\documentclass[phd,lfcs,twoside,logo]{infthesis}

\usepackage{makeidx}
\makeindex

\usepackage[ddmmyyyy]{datetime}
\usepackage[numbers]{natbib}
\usepackage{color}
\usepackage[usenames,dvipsnames]{xcolor}
\usepackage{hyperref}
\usepackage{amssymb}
\usepackage{amsmath}
\usepackage{amsthm}
\usepackage{proof}
\usepackage{stmaryrd}
\usepackage{verbatim}
\usepackage[all]{xy}
\usepackage{multirow}
\usepackage{bigints}
\usepackage{enumerate}
\usepackage{nccmath}
\usepackage{csquotes}

\newtheorem{theorem}{Theorem}[section]
\newtheorem{proposition}[theorem]{Proposition}
\newtheorem{corollary}[theorem]{Corollary}

\newtheorem{example}[theorem]{Example}

\theoremstyle{definition}
\newtheorem{definition}[theorem]{Definition}

\newcommand{\dcomment}[1]{\colorbox{lightgray}{{\fontsize{0.32cm}{0.32cm}$#1$}}}
\newcommand{\dhide}[1]{\colorbox{white}{{\fontsize{0.33cm}{0.32cm}$#1$}}}
\newcommand{\dscomment}[1]{\colorbox{lightgray}{{\fontsize{0.2cm}{0.32cm}$#1$}}}
\newcommand{\dshide}[1]{\colorbox{white}{{\fontsize{0.22cm}{0.32cm}$#1$}}}

\newcommand{\vertbar}{\,\vert\,}

\newcommand{\SCCompC}{SCCompC}

\newcommand{\mkrule}[2]{\infer{#1}{#2}}
\newcommand{\mkrulelabel}[3]{\infer[#3]{#1}{#2}}

\newcommand{\defeq}{\stackrel{\mbox{\tiny def}}=}

\newcommand{\return}[1]{\mathtt{return~}#1}
\newcommand{\doto}[4]{#1 \mathtt{~to~} #2 \mathtt{~in}_{#3} \mathtt{~} #4}

\newcommand{\inl}[2]{\mathtt{inl}_{#1} \mathtt{~} #2}
\newcommand{\inr}[2]{\mathtt{inr}_{#1} \mathtt{~} #2}
\newcommand{\seminl}[2]{\mathsf{inl}_{#1} \mathtt{~} #2}
\newcommand{\seminr}[2]{\mathsf{inr}_{#1} \mathtt{~} #2}

\newcommand{\absurd}[2]{\mathtt{case~} #2 \mathtt{~of}_{#1} \mathtt{~} ()}

\newcommand{\case}[4]{\mathtt{case~} #1 \mathtt{~of}_{#2} \mathtt{~} (#3,#4)}
\newcommand{\casesf}[4]{\mathsf{case~} #1 \mathsf{~of}_{#2} \mathsf{~} (#3,#4)}

\newcommand{\algop}{\mathtt{op}}
\newcommand{\sigalgop}{\mathsf{op}}

\newcommand{\Nat}{\mathsf{Nat}}

\newcommand{\Exception}{\mathsf{Exc}}
\newcommand{\State}{\mathsf{St}}
\newcommand{\Location}{\mathsf{Loc}}
\newcommand{\Value}{\mathsf{Val}}
\newcommand{\Character}{\mathsf{Chr}}
\newcommand{\Updates}{\mathsf{Upd}}

\newcommand{\zero}{\mathtt{zero}}
\newcommand{\suc}[1]{\mathtt{succ~} #1}
\newcommand{\succc}{\mathtt{succ}}
\newcommand{\natrec}[4]{\mathtt{nat\text{-}elim}_{#1}(#2, #3,#4)}

\newcommand{\refl}[2]{\mathtt{refl}_{} \mathtt{~} #2}
\newcommand{\pathind}[6]{\mathtt{eq\text{-}elim}_{{#1}}(#2,#3,#4,#5,#6)}

\newcommand{\funsection}{\mathsf{s}}

\newcommand{\thunk}[1]{\mathtt{thunk~} #1}

\newcommand{\force}[2]{\mathtt{force}_{#1} \mathtt{~} #2}

\newcommand{\dtensorlet}[3]{#2 \mathtt{~to~} #1 \mathtt{~in~} #3}

\newcommand{\lj}[2]{#1\vdash #2}
\newcommand{\zj}[3]{#1\vdash #2:#3}
\newcommand{\vj}[3]{#1\vdash #2:#3}
\newcommand{\cj}[3]{#1\vdash #2: #3}
\newcommand{\hj}[4]{#1 \vertbar #2 \vdash #3:#4}

\newcommand{\ljeq}[3]{#1\vdash #2 = #3}

\newcommand{\veq}[4]{#1 \vdash #2 = #3:#4}
\newcommand{\ceq}[4]{#1 \vdash #2 = #3:#4}
\newcommand{\heq}[5]{#1 \vertbar #2 \vdash #3 = #4:#5}

\newcommand{\pair}[2]{\langle #1,#2 \rangle}

\newcommand{\pmatch}[4]{\mathtt{pm~} #1 \mathtt{~as~} #2 \mathtt{~in}_{#3} \mathtt{~} #4}
\newcommand{\pmatchsf}[4]{\mathsf{pm~} #1 \mathsf{~as~} #2 \mathsf{~in}_{#3} \mathsf{~} #4}

\newcommand{\runas}[4]{#1~\mathtt{as}~#2~\mathtt{in}_{#3}~#4}

\newcommand{\id}{\mathsf{id}} 
\newcommand{\comp}{\circ}

\newcommand{\Id}{\mathsf{Id}}

\newcommand{\sem}[1]{\llbracket #1 \rrbracket}

\newcommand{\efftrans}[2]{\llparenthesis \, #1 \, \rrparenthesis_{#2}}

\newcommand{\fst}[1]{\mathtt{fst}\,#1}
\newcommand{\snd}[1]{\mathtt{snd}\,#1}
\newcommand{\semfst}[1]{\mathsf{fst}\,#1}
\newcommand{\semsnd}[1]{\mathsf{snd}\,#1}

\newcommand{\Set}{\mathsf{Set}}

\newcommand{\Mod}{\mathsf{Mod}}
\newcommand{\CPO}{\mathsf{CPO}}
\newcommand{\Law}{\mathsf{Law_c}}

\newcommand{\Fam}{\mathsf{Fam}}
\newcommand{\CFam}{\mathsf{CFam}}

\newcommand{\ul}[1]{\underline{#1}}

\newcommand{\ia}[1]{\{#1\}}

\newcommand{\sproj}[4]{\mathsf{proj}_{#1;#2 : #3;#4}}
\newcommand{\ssubst}[5]{\mathsf{subst}_{#1;#2 : #3;#4;#5}}

\newcommand{\pl}[1]{{\sc{#1}}}

\addtocontents{toc}{\protect\setcounter{tocdepth}{2}}

\title{
Fibred Computational Effects
}
\author{Danel Ahman}

\abstract{%
We study the interplay between \emph{dependent types} and \emph{computational effects}, 
two important areas of modern programming language research. On the one hand, dependent types underlie proof assistants such as Coq and functional programming languages such as Agda, Idris, and F*, providing programmers a means for encoding detailed specifications of program behaviour using types. On the other hand, computational effects, such as exceptions, nondeterminism, state, I/O, probability, etc., are integral to all widely-used programming languages, ranging from imperative languages, such as C, to functional languages, such as ML and Haskell. Separately, dependent types and computational effects both come with rigorous mathematical foundations, dependent types in the effect-free setting and computational effects in the simply typed setting. Their \emph{combination}, however, has received much less attention and no similarly exhaustive theory has been developed. 
In this thesis we address this shortcoming by providing a comprehensive treatment of the combination of  these two fields, and demonstrating that they admit a mathematically elegant and natural combination.

Specifically, we develop a core effectful dependently typed language, eMLTT, based on Martin-L\"{o}f's intensional type theory and a clear separation between (effect-free) values and (possibly effectful) computations familiar from simply typed languages such as Levy's Call-By-Push-Value and Egger et al.'s Enriched Effect Calculus. 
 A novel feature of our language is the \emph{computational $\Sigma$-type}, which we use to give a uniform treatment of type-dependency in sequential composition.
In addition, we define and study a class of category-theoretic models, called \emph{fibred adjunction models}, that are suitable for defining a sound and complete interpretation of eMLTT. Specifically, fibred adjunction models naturally combine standard category-theoretic models of dependent types (split closed comprehension categories) with those of computational effects (adjunctions). We discuss and study various examples of these models, including a domain-theoretic model so as to extend eMLTT with general recursion.

We also investigate a dependently typed generalisation of the algebraic treatment of computational effects by showing how to extend eMLTT with \emph{fibred algebraic effects} and their \emph{handlers}.
In particular, we specify fibred algebraic effects using a dependently typed generalisation of Plotkin and Pretnar's effect theories, enabling us to capture precise notions of computation such as state with location-dependent store types and dependently typed update monads.
For handlers, we observe that their conventional term-level definition leads to unsound program equivalences becoming derivable in languages that include a notion of homomorphism, such as eMLTT. To solve this problem, we propose a novel type-based treatment of handlers via a new computation type,  the \emph{user-defined algebra type}, which pairs a value type (the carrier) with a family of value terms (the operations). This type internalises Plotkin and Pretnar's insight that handlers denote algebras for a given equational theory of computational effects. We demonstrate the generality of our type-based treatment of handlers by showing that their conventional term-level presentation can be routinely derived, and this treatment provides a useful mechanism for reasoning about effectful computations. Finally, we show that these extensions of eMLTT  can be soundly interpreted in a fibred adjunction model based on the families of sets fibration and models of Lawvere theories.
}

\begin{document}

\begin{preliminary}

\maketitle

\begin{laysummary}
\emph{Dependent types} provide a lightweight and modular means to integrate programming and formal program verification. In particular, the types of programs written in dependently typed programming languages (Agda, Idris, F*, etc.) can be used to express specifications of program correctness. These specifications can vary from being as simple as requiring the divisor in the division function to be non-zero, to as complex as specifying the correctness of compilers of industrial-strength languages. Successful compilation of a program then guarantees that it satisfies its type-based specification.

While dependent types allow many runtime errors to be eliminated by rejecting erroneous programs at compile-time, dependently typed languages are yet to gain popularity in the wider programming community. One reason for this is their limited support for \emph{computational effects}, an integral part of all widely used programming languages, ranging from imperative languages, such as C, to functional languages, such as ML and Haskell.
For example, in addition to simply turning their inputs to outputs, programs written in these programming languages can raise exceptions, access computer's memory, communicate over a network, render images on a screen, etc.

Therefore, if 
dependently typed programming languages are to truly live up to their promise of seamlessly integrating  programming and formal program verification, we must first understand how to properly account for computational effects in such languages.
While there already exists work on this topic, ingredients needed for a comprehensive theory are generally missing. For example, foundations are often not settled; available effects may be limited; or effects may not be treated systematically.

In this thesis we address these shortcomings by providing a comprehensive treatment of the combination of dependent types and general computational effects. Specifically, we i) define a core effectful dependently typed programming language; ii) study its category-theoretic denotational semantics;
and iii) demonstrate how to extend the algebraic treatment of computational effects (including the handlers of algebraic effects) to the dependently typed setting, enabling us to uniformly specify a wide range of computational effects in terms of operations and equations. 
In particular, in this thesis we demonstrate that  dependent types and computational effects admit a mathematically natural combination, in which well-known concepts and results from the simply typed setting can be reused and adapted, but which also reveals new and interesting programming language features and corresponding mathematical structures. 

\end{laysummary}

\begin{acknowledgements}
I would like to thank the many people who have been important  
to my PhD studies. This long journey would not have been 
possible without their support and guidance.

First and foremost, I would like to express my sincerest gratitude to 
my supervisor Gordon Plotkin for all the guidance, support, and
encouragement he provided during my time in Edinburgh. 
Although we often found ourselves working in different geographic locations 
and timezones, he always found time to comment on my work and 
come up with useful suggestions, and provide a great deal of invaluable feedback.

I would also like to thank my second supervisor Alex Simpson for discussions 
and suggestions concerning my research in the earlier stages of my PhD studies. 
I am also grateful to Ian Stark for taking over the second supervisor's duties when 
Alex moved to Ljubljana. I would also like to thank Phil Wadler for agreeing to be 
on my yearly review meeting panels, and for all the feedback and constructive criticism he 
provided. 

I am also grateful to Paul Levy and James Cheney for agreeing to 
be my examiners, for spending many hours of their time carefully reading through a thesis of this length, 
and for all their feedback that helped to greatly improve 
the final version of this thesis.

I am very grateful to LFCS and its members for providing an excellent research 
environment. In particular, I would like to thank my fellow PhD students of IF 5.32, past and present, for many 
interesting discussions about work and life in general: 
Alyssa Alcorn, Simon Fowler, Weili Fu, Jiansen He, Ben Kavanagh, Craig McLaughlin, Fabian Nagel, 
Shayan Najd, Jack Williams, and Jakub Zalewski. I am also grateful for the support and friendship of many other LFCS students: Daniel Hillerstr\"{o}m, Theodoros Kapourniotis, Karoliina Lehtinen, Kristjan Liiva, Einar Pius, Panagiotis Stratis, and Marcin Szymczak.
I would also like to thank the occupants of IF 5.28, past and present, for having time 
to discuss various aspects of research, both mine and theirs: Bob Atkey, Brian Campbell, Sam Lindley, James McKinna, and Garrett Morris.

I would also like to thank the members of the MSP group at the University of Strathclyde for many interesting visits, seminars, and reading groups; these provided a useful and much needed distraction  from my studies. In particular, I would like to thank Neil Ghani with whom I and Gordon co-authored the FoSSaCS'16 paper on dependent types and computational effects which became the basis of the work presented in this thesis. I would also like to thank Fredrik Forsberg and Conor McBride for many interesting discussions about dependent types, both conventional and cubical. More generally, I would like to thank all the various Scottish programming languages and semantics research groups for making Scotland such a wonderful research environment, in particular, by organising events such as SPLS, ScotCats, and CLAP Scotland.

I am also grateful to the past and present members of the Logic and Semantics Group at the 
Institute of Cybernetics (now at the Department of 
Software Science) at the Tallinn University of Technology for hosting my visits, and for organising great events such as the Estonian Winter School in Computer Science and the Estonian Computer Science Theory Days. Specifically, I would like to thank James Chapman and Tarmo Uustalu for their encouragement and support regarding my PhD research, and also for our continued  collaborations on directed containers and related topics.

I would also like to thank Mihai Budiu and Nikhil Swamy for inviting me to do internships 
at Microsoft Research, and Gordon for putting me in contact with them in the 
first place. During these two internships, I learnt a lot about conducting research and working in a large   corporate environment. Both internships also greatly broadened my knowledge about the more practical aspects of programming language research.

I am also grateful to Ohad Kammar, Justus Matthiesen, and Kayvan Memarian for many 
interesting discussions about programming language research, hiking, cycling, and life in general, 
and for accommodating me during my visits to Cambridge.

Special thanks are reserved to the many people I shared the Bruntsfield Gardens flat during my four and half year stay in Edinburgh, and who helped to make it a true home away from home: Barbara Balazs, Krzysztof Geras, Michaela Keil, Stephen McGroarty, Zs\'{o}fia Nem\'{e}nyi, Toomas Remmelg, and Michael Wilson. 

I am also forever indebted to Ege Ello and Veiko Vostrjakov who have offered immense emotional support over the past years, provided me with a place to stay when visiting Estonia, and more generally have taken me in as a member of their family.

Finally, I would like to acknowledge the financial support of the University of Edinburgh (through the Principal's Career Development PhD Scholarship) and the Archimedes Foundation (in collaboration with the Estonian Ministry of Education and Research, through the scholarship program Kristjan Jaak). I am also grateful for the travel funding provided by the LFCS, the Archimedes Foundation, the ACM SIGPLAN PAC, the ERDF funded Coinduction project, and the EUTypes Cost Action. I am also grateful for C\u{a}t\u{a}lin Hri\c{t}cu and the ERC SECOMP project for funding me while this thesis was under examination  and during the preparation of its final version.
\end{acknowledgements}

\standarddeclaration


\tableofcontents


\end{preliminary}




\chapter{Introduction}
\label{chap:introduction}

In this thesis we study the interplay between dependent types and computational effects, 
two important areas of modern programming language research. 

On the one hand, \emph{dependent types} underlie modern proof assistants such as Coq~\cite{Coq:Manual}, and  programming languages such as Agda~\cite{Norell:Thesis}, Idris~\cite{Brady:Idris}, and F*~\cite{Swamy:FStar}. In particular, dependent types provide a lightweight and modular means to formally specify and verify properties of programs using their types. These specifications can vary from being as simple as requiring the divisor in the division function to be non-zero, to as complex as specifying the correctness of compilers of industrial-strength languages~\cite{CompCert}.

\index{ I@$I/O$ (input/output)}
On the other hand, \emph{computational effects}, such as exceptions, nondeterminism, state, interactive I/O, etc., are an integral part of all widely used programming languages, ranging from imperative languages, such as C~\cite{Kernighan:CLanguage}, to functional languages, such as ML~\cite{ML:Standard} and Haskell~\cite{Marlow:HaskellReport}. While some computational effects can be \emph{represented} in languages that do not support them off the shelf, e.g., stateful programs can be naturally encoded as functions $\State \to A \times \State$, having \emph{primitive support} for computational effects such as state means that compilers can perform effect-dependent optimisations (e.g., see~\cite{Kammar:Thesis}) which can lead to programs being executed more efficiently.

Consequently, if dependently typed languages are to truly live up to their promise of providing a lightweight means for integrating formal verification and practical (functional) programming, we must first understand how to correctly account for computational effects in this setting.
At the moment, the level of support for them varies greatly in existing languages. 
For example, Agda does not offer any principled support except for a very basic foreign function interface to Haskell; Idris represents computational effects using a domain specific language that is elaborated to the underlying pure language; and F* includes primitive support for state and exceptions via extraction to OCaml, and allows some other effects to be represented using monads defined in a simply typed definition language.
As a testament to the benefits of including computational effects primitively in the design of dependently typed languages, F*, with its support to program with and reason about primitively supported effects such as state, has a central role in Microsoft's Everest project~\cite{Everest:Project} that aims to deliver a high-performance, standards-compliant, verified implementation of the full HTTPS ecosystem.

While the Everest project demonstrates the potentially groundbreaking impact that effectful dependently typed programming languages could have in the years to come, the intersection of these two fields still lacks a general and exhaustive treatment. This is in stark contrast to our rigorous understanding of computational effects in the simply typed setting.
We address this shortcoming by providing a comprehensive language-based, category-theoretic, and algebraic treatment of the combination of dependent types and computational effects. 
Our goal is to establish the following claim: 

\vspace{0.2cm}

\begin{displayquote}
\textit{Dependent types and computational effects admit a natural combination.}
\end{displayquote}

It is worth noting that compared to the kinds of computational effects supported by languages such as Idris and F*, we take a step back and investigate more foundational questions in the design and semantics of effectful dependently typed languages, leaving questions about a general treatment of more expressive type-and-effect systems (such as the ones used in Idris and F*) for future work---see Sections~\ref{sect:fibredparametrisedeffects} and~\ref{sect:fibDijkstramonads}.

\section{Two guiding questions}
\label{sect:twoguidingquestions}

That the above claim is non-trivial was already recognised by Moggi~\cite{Moggi:NotionsofComputationandMonads}. In this thesis we identify two key questions that one needs to answer in order to provide a general treatment of the combination of dependent types and computational effects: 
\begin{itemize}
\item Should one allow effectful computations to appear in types?
\item How should one treat type-dependency in sequential composition?
\end{itemize}
We discuss both of these questions and some possible answers to them in detail below, separately highlighting the answers that form the basis of the effectful dependently typed language we develop in this thesis, and its denotational semantics. For both questions, our choice of answers is based on being able to maximally reuse and naturally combine respective existing work on dependent types and computational effects.

We assume that the reader has basic knowledge of both dependent types and computational effects. For a good overview of dependent types and their denotational semantics, we suggest~\cite{Hofmann:SyntaxAndSemantics,Jacobs:Book,Streicher:Semantics}. 
For computational effects, there are a variety of sources one can consult, ranging from the seminal monad-based work of Moggi~\cite{Moggi:ComputationalLambdaCalculus,Moggi:NotionsofComputationandMonads}, to the more  recent adjunction-based work of Levy~\cite{Levy:CBPV}, to the algebraic treatment pioneered by Plotkin and Power~\cite{Plotkin:SemanticsForAlgOperations,Plotkin:NotionsOfComputation}. For functional programmers, a good overview of the algebraic treatment and a source of further references is Pretnar's tutorial~\cite{Pretnar:Tutorial}. 
We give a short overview of these three approaches 
in Section~\ref{sect:modelsofeffects}.

\subsection*{Should one allow effectful computations to appear in types?}

To make the first question more concrete, let us consider the prototypical example of a dependent type, namely, the type of vectors 
of values from some type $A$, written $\mathsf{Vec}~A~M$, where $M$ is a term of type $\mathsf{Nat}$. We can then rephrase the question as follows:
\begin{itemize}
\item Should one allow $M$ to raise an effect in $\mathsf{Vec}~A~M$, e.g., perform I/O?
\end{itemize}

In practice, the answer to this question depends on the kinds of computational effects one considers. In particular, it depends on whether we can expect to evaluate a closed $M$ to a natural number at compile-time, which is crucial to typecheck the constructors of $\mathsf{Vec}~A~M$ and to compare two such types for equality.

For computational effects that do not involve interaction with the runtime environment, $M$ does not need to be restricted. For example, such effects include local names~\cite{Pitts:NominalMLTT,Cheney:DepNomTypeTheory} and  recursion~\cite{Casinghino:CombiningProofs}. While $M$ might diverge in the latter case, making typechecking undecidable, it is important that its evaluation does not get stuck because of, for example, the need to perform I/O. On the other hand, if one wants to accommodate a wide range of different computational effects, including both local names and I/O, $M$ should be restricted to a value so as to guarantee that it evaluates to a natural number during typechecking.
Furthermore, while there exists semantics for these two specific computational effects (see~\cite{Pitts:NominalMLTT} and~\cite{Palmgren:DomainInterpretation}, respectively), we do not know of a denotational semantics that would account for type-dependency on an unrestricted $M$. 

It is worthwhile to note that the problem with general $M$ only arises when one takes a coarse-grained language, such as Moggi's computational $\lambda$-calculus~\cite{Moggi:ComputationalLambdaCalculus}, as a basis for building an effectful dependently typed language. In particular, in such languages neither the types nor the typing judgements contain information about whether and which computational effects a term might perform. As a result, a closed term of type $\mathsf{Nat}$ can ``surprisingly" perform I/O or raise an exception, instead of evaluating to a natural number. While this style of programming is convenient in many situations, it contradicts the spirit of dependently typed programming, namely, that types ought to be as precise descriptions of program properties  as possible. For example, when pattern-matching on a vector of type $\mathsf{Vec}~A~5$, 
the typechecker can readily use the knowledge that this vector must be exactly of length $5$. 
Thus, when combining dependent types and computational effects, one naturally expects similar precision to also apply to the types of effectful computations, i.e., we had better not be able to assign just the type $\mathsf{Nat}$ to a program that can potentially perform I/O or raise an exception. 

\vspace{0.25cm}

\noindent
\textbf{Our solution:} Guided by the above discussion, we allow types to depend only on values in order to support a wide range of computational effects. While this decision might seem limiting at first, we recover the ability to depend on effectful computations via thunks and handlers (see Section~\ref{section:usinghandlersforreasoning} for examples of this). 
To ensure that types depend only on values, we make a clear distinction between value types $A$ and computation types $\ul{C}$, and between value terms $V$ and computation terms $M$, as is done in simply typed effectful languages such as Levy's Call-By-Push-Value (CBPV)~\cite{Levy:CBPV}, 
\index{ CBPV@CBPV (Call-By-Push-Value)}
and Egger, M{\o}gelberg, and Simpson's Enriched Effect Calculus (EEC)~\cite{Egger:EnrichedEffectCalculus}. 
\index{ EEC@EEC (Enriched Effect Calculus)}

Specifically, the well-formed types of the effectful dependently typed language we develop in this thesis are defined using judgements $\lj \Gamma A$ and $\lj \Gamma \ul{C}$, where the variables in contexts $\Gamma$ are required to range exclusively over value types. 
As a result,
our language lends itself to a very natural denotational semantics that combines standard category-theoretic models of dependent types 
and the corresponding generalisation of standard adjunction-based models of computational effects.

It is worth noting that we could have chosen other effectful simply typed languages as the basis of our effectful dependently typed language, such as Moggi's monadic metalanguage~\cite{Moggi:NotionsofComputationandMonads} or Levy's fine-grain call-by-value language~\cite[Appendix~A.3.2]{Levy:CBPV}. Each of these languages distinguishes between effect-free values and possibly effectful computations. The former does so by assigning effectful computations monadic types $TA$, while the latter makes this distinction already in the grammar of terms. However, as we next discuss, by basing our work in this thesis on CBPV and EEC, we are able to give a more uniform treatment of type-dependency in sequential composition.

\subsection*{How should one treat type-dependency in sequential composition?}

In order to make the above question more concrete, we recall the typing rule for the \linebreak 

\pagebreak
\noindent
sequential composition of effectful computations from CBPV:
\vspace{0.15cm}
\[
\mkrule
{\cj \Gamma {\doto M {x \!:\! A} {} N} {\ul{C}}}
{
\cj \Gamma M FA
\quad
\cj {\Gamma, x \!:\! A} N \ul{C}
}
\vspace{-0.1cm}
\]
It is important to observe that in the dependently typed setting, this typing rule is no longer correct because the second premise allows the value variable $x$ to appear freely in $\ul{C}$, meaning that $x$ can also appear free in the conclusion where it ought to be bound.
Based on this observation, we rephrase our second guiding question as follows:
\begin{itemize}
\item How should one fix this typing rule for sequential composition so that the value variable $x$ would not appear free in its conclusion?
\end{itemize}

As already suggested by Levy~\cite[Section~12.4.1]{Levy:CBPV}, the most straightforward solution to this problem would be to not allow the variable $x$ to appear free in the computation type $\ul{C}$, i.e., require that $\ul{C}$ is well-formed in $\Gamma$ and change the typing rule to
\vspace{-0.05cm}
\[
\mkrulelabel
{\cj \Gamma {\doto M {x \!:\! A} {} N} {\ul{C}}}
{
\cj \Gamma M FA
\quad
\lj \Gamma \ul{C}
\quad
\cj {\Gamma, x \!:\! A} N \ul{C}
}
{(*)}
\vspace{-0.05cm}
\]
On the one hand, this solution would have two important advantages: i) it solves the above-mentioned problem with minimal changes; and ii) sequential composition typed using this rule can be given a  denotational semantics using split fibred adjunctions, naturally generalising the semantics of CBPV's simply typed sequential composition. On the other hand, this solution can be somewhat restrictive in some situations. For example, when $M$ involves opening a file and the return values of $M$ model whether the file was opened successfully or not, the computation type $\ul{C}$ and the computation terms  allowed to inhabit it could  crucially depend on the return values of $M$. For instance, if the given file was not opened successfully in $M$, we might want to use dependency on the variable $x$ in $\ul{C}$ to prohibit reading from and writing to the given file in $N$.

In recent unpublished manu\-scripts~\cite{Vakar:EffectfulDepTypes,Vakar:FrameworkForDepEffs}, V{\'{a}}k{\'{a}}r has proposed an alternative solution in which $\ul{C}$ is allowed to depend on thunks of computations. In particular, V{\'{a}}k{\'{a}}r studies a dependently typed CBPV in which sequential composition is typed as
\vspace{-0.05cm}
\[
\mkrule
{\cj {\Gamma_1, \Gamma_2[\thunk M/y]} {\doto {M} {x \!:\! A} {} {N}} {\ul{C}[\thunk M/y]}}
{
\begin{array}{c}
\cj {\Gamma_1} M {FA}
\quad
\lj {\Gamma_1, y \!:\! U\!FA, \Gamma_2} {\ul{C}}
\\[-0.5mm]
\cj {\Gamma_1, x \!:\! A, \Gamma_2[\thunk\! (\return x)/y]} {N} {\ul{C}[\thunk\! (\return x)/y]}
\end{array}
}
\vspace{-0.1cm}
\]
However, while this rule solves the above-mentioned problem with the typing rule of sequential composition, its thunks-based type-dependency introduces new problems in the presence of algebraic effects. We discuss these problems further in Section~\ref{sect:relatedwork}.

Finally, if we would have chosen Moggi's monadic metalanguage as the basis of our  language, instead of CBPV and EEC, we would have had the option to simply use the standard (value) $\Sigma$-type to fix the typing rule of sequential composition. In particular, given two terms $\cj \Gamma M TA$ and $\cj {\Gamma, x \!:\! A} N TB$, we could have considered ``closing-off" the type of the sequential composition of $M$ and $N$ as $T(\Sigma\, x \!:\! A.\, B)$.
However, it is not immediate whether this proposed use of the $\Sigma$-type is the right solution to this problem, 
e.g., why not use $\Sigma\, x \!:\! A.\, TB$, or why use the $\Sigma$-type in the first place? 

\vspace{0.25cm}

\noindent
\textbf{Our solution:}
Guided by the above discussion, we choose to use the restricted rule $(*)$ to type sequential composition for the reasons listed earlier, namely, because it solves the above-mentioned problem with minimal changes and the resulting language lends itself to a denotational semantics that naturally generalises that of simply typed CBPV.
We overcome the restrictive nature of this rule (see earlier discussion) by ``closing-off" $\ul{C}$ using a $\Sigma$-type.
However, as $\ul{C}$ is a computation type, we cannot use the standard (value) $\Sigma$-type from Martin-L\"{o}f's type theory (MLTT)~\cite{MartinLof:IntuitionisticTT}. 
\index{ M@MLTT (Martin-L\"{o}f's type theory)}
Instead, we introduce a computational variant of it, written $\Sigma\, x \!:\! A .\, \ul{C}$, to complement the typing rule $(*)$. 
In particular, by combing the typing rule $(*)$ with the computational $\Sigma$-type, we can derive
\vspace{-0.1cm}
\[
\mkrule
{\cj \Gamma {\doto M {x \!:\! A} {} {\langle x , N \rangle}} {\Sigma\, x \!:\! A .\, \ul{C}}}
{
\quad
\cj \Gamma M FA
\quad
\lj \Gamma {\Sigma\, x \!:\! A .\, \ul{C}}
\quad
\mkrule
{\cj {\Gamma, x \!:\! A} {\langle x , N \rangle} {\Sigma\, x \!:\! A .\, \ul{C}}}
{
\mkrule
{\cj {\Gamma, x \!:\! A} x A}
{}
\quad
\cj {\Gamma, x \!:\! A} N \ul{C}
}
}
\]
where the type of $N$ is allowed to depend on the values returned by $M$, but we ``close it off" using the introduction form for $\Sigma\, x \!:\! A .\, \ul{C}$, given by computational pairing, before concluding the derivation by applying the typing rule $(*)$ for sequential composition. 

Our use of the computational $\Sigma$-type is inspired by the algebraic treatment of computational effects in which computational effects are specified using equational theories, whose algebras then model computation types.
To explain this further, let us consider the effect of accessing read-only memory that stores a single bit. Following~\cite{Staton:Instances}, this effect can be represented using a binary operation $?$ and the equations:
\vspace{-0.45cm}
\[
M ~?~ M = M
\qquad
(M_1 ~?~ M_2) ~?~ (N_1 ~?~ N_2) = M_1 ~?~ N_2
\]
The idea is that $M ~?~ N$ is a computation that first reads the bit in the store and then continues by following either $M$ or $N$, depending on whether the bit was $0$ or $1$.

Next, let us consider the following program: 
\[
\doto {((\return 4) ~ ? ~ (\return 2))} {x \!:\! \Nat} {} {N}
\]
When we examine how this program would evaluate, assuming that $\cj {x \!:\! \Nat} {N} {\ul{C}}$ and $\lj {x \!:\! \Nat} \ul{C}$, we see that after reading the bit in the store, we continue either by evaluating $N[4/x]$, that has type $\ul{C}[4/x]$, \emph{or} by evaluating $N[2/x]$, that has type $\ul{C}[2/x]$, leading us to naturally conclude that the whole program denotes an element of the coproduct of algebras denoted by $\ul{C}[4/x]$ and $\ul{C}[2/x]$. The elements of this coproduct are equivalence classes of binary computation trees whose leaves are given by elements of the algebras denoted by $\ul{C}[4/x]$ and $\ul{C}[2/x]$. For example, the given program denotes the tree
\[
\xymatrix@C=0.75em@R=2.5em@M=0.25em{
& ? \ar@{-}[dl] \ar@{-}[dr] &
\\
N[4/x] & & N[2/x]
}
\]

As the same pattern also reoccurs with computational effects other than reading a bit and dependency on arbitrary value types, we introduce the computational $\Sigma$-type as a uniform means to account for general type-dependency in sequential composition.
In particular, if $\lj {x \!:\! A} \ul{C}$ denotes an $A$-indexed family of algebras, then the computational $\Sigma$-type $\Sigma\, x \!:\! A .\, \ul{C}$ denotes an $A$-indexed coproduct of algebras $\ul{C}[a_i/x]$.
However, as already demonstrated in the above derivation of ${\doto M {x \!:\! A} {} {\langle x , N \rangle}}$, we separate concerns by not making the computational $\Sigma$-type part of the typing rule for sequential composition, but instead equip it with its own introduction and elimination forms, given by pairing and pattern-matching, analogously to the (value) $\Sigma$-type from MLTT.

This general treatment also justifies our earlier proposal of using the (value) $\Sigma$-type to ``close-off"  free variables to type sequential composition in a dependently typed version of Moggi's monadic metalanguage. In particular, based on the above discussion, an embedding of Moggi's monadic metalanguage in ours by taking $TA \defeq U\!FA$, and a computation type isomorphism $\Gamma \vdash F(\Sigma\, x \!:\! A .\, B) \cong \Sigma\, x \!:\! A .\, FB$ provable in our language, we get that the canonical treatment of type-dependency in sequential composition for Moggi's metalanguage amounts to using the following derivable typing rule:
\vspace{0.25cm}
\[
\mkrule
{\cj \Gamma {\doto M {x \!:\! A} {} {(\doto {N} {y \!:\! B} {} {\return \langle x , y \rangle}})} {T(\Sigma\, x \!:\! A.\, B)}}
{
\cj \Gamma M TA
\quad
\cj {\Gamma, x \!:\! A} N TB
}
\vspace{-0.1cm}
\]

The above discussion also shows us why using $\Sigma\, x \!:\! A.\, TB$ to type sequential composition in a dependently typed version of Moggi's monadic metalanguage would not have the desired effect. In particular, if $\Sigma\, x \!:\! A.\, TB$ were used to type the sequential composition of ${\cj \Gamma M TA}$ and ${\cj {\Gamma, x \!:\! A} N TB}$, then $M$ would need to return the same exact value of type $A$ in each of its branches. From an algebraic perspective, if $B$ denotes an $A$-indexed family of sets, then ${\Sigma\, x \!:\! A.\, TB}$ would denote an $A$-indexed coproduct of sets of computation trees, where each tree in an $a$'th component of this coproduct would have all its leaves given by elements of the set denoted by $B[a/x]$. 
As a result, the example program ${\doto {((\return 4) ~ ? ~ (\return 2))} {x \!:\! \Nat} {} {N}}$, where now \linebreak ${\cj {x \!:\! \Nat} N TB}$, 
could not be modelled as an element of the set denoted by $\Sigma\, x \!:\! \Nat.\, TB$. 

While useful for providing a general treatment of type-dependency in sequential composition, we note that we have yet to find interesting examples involving computation types $\Sigma\, x \!:\! A .\, \ul{C}$ where $\ul{C}$ would not be of the form $FB$. In particular, a natural implementation of the dependently typed parsing example we alluded to in~\cite{Ahman:FibredEffects} also turns out to only require types of the form $FB$ (and $\Sigma\, x \!:\! A .\, FB$), e.g., as sketched in Appendix~\ref{chap:appendixC1}  using a shallow embedding of Moggi's monadic metalanguage in Agda~\cite{Norell:Thesis}.
However, we would like to draw the reader's attention to that in the context of more expressive typing disciplines than considered in this thesis, the combination of the computational $\Sigma$-type and computation types of the form $F_W B$ (where $F$ is now indexed by a value term $W$) can give rise to interesting consequences, e.g., as discussed in Section~\ref{sect:fibredparametrisedeffects}.

\section{Contributions}
\label{sect:contributions}

The main contributions of this thesis are:

\begin{itemize}
\item An \emph{effectful dependently typed language}, called eMLTT, that naturally combines intensional MLTT 
with general computational effects, based on a clear separation between values and computations. The most notable feature of eMLTT is the computational $\Sigma$-type that provides a uniform treatment of type-dependency in the sequential composition of effectful computations.
\item A \emph{class of category-theoretic models}, called fibred adjunction models, with respect to which eMLTT is both sound and complete. These models naturally combine standard category-theoretic models of dependent types (split closed comprehension categories) and computational effects (adjunctions). Further, they
\begin{itemize}
\item provide evidence that one can keep using monads and adjunctions to model computational effects in the dependently typed setting;
\item demonstrate that computational $\Pi$- and $\Sigma$-types can be modelled analogously to their value counterparts, as adjoints to weakening functors; and 
\item provide a category-theoretically natural axiomatisation of structures needed for modelling type-dependency in the elimination forms of value types. 
\end{itemize}
\item A \emph{collection of natural examples} of fibred adjunction models, based on
\begin{itemize}
\item simple fibrations and models of EEC (i.e., enriched adjunctions);  
\item the families of sets fibration and lifting of adjunctions; 
\item the Eilenberg-Moore resolutions of split fibred monads, where we 
give sufficient conditions for the Eilenberg-Moore fibration to support computational $\Pi$- and $\Sigma$-types, generalising known results about the existence of products and coproducts in the Eilenberg-Moore category of a monad; and
\item the fibration of continuous families of $\omega$-complete partial orders and lifting of $\CPO$-enriched adjunctions, so as to accommodate general recursion.
\end{itemize}
\item An extension of eMLTT with \emph{algebraic effects} and their \emph{handlers}, including
\begin{itemize}
\item a notion of fibred effect theory that allows computational effects to be specified using dependently typed operation symbols (and equations), enabling one to capture precise notions of computation, such as state with location-dependent store types and dependently typed update monads;
\item an observation that naively following the literature and defining handlers at the term level leads to unsound program equivalences becoming derivable in languages involving a notion of homomorphism, such as eMLTT;
\item a novel computation type, called the user-defined algebra type, that pairs a value type (the carrier) to a family of value terms (the operations), allowing us to safely extend eMLTT with handlers of fibred algebraic effects;
\item a demonstration that the conventional term-level presentation of handlers can be routinely derived from our type-based treatment; and
\item a proof that this extended language can be soundly interpreted in a fibred adjunction model based on the families of sets fibration and models of a countable Lawvere theory we derive from the given fibred effect theory.
\end{itemize}
\item A demonstration that our type-based treatment of handlers provides a useful mechanism for \emph{reasoning about effectful computations}, e.g., allowing us to 
\begin{itemize}
\item lift predicates given on return values to predicates on computations; 
\item define Dijkstra's weakest precondition predicate transformers; and 
\item specify detailed patterns of allowed (I/O-)effects in computations. 
\end{itemize}
\end{itemize}

\section{Organisation}

\noindent
\textbf{Chapter~\ref{chap:introduction}} is this introduction which also includes an overview of related work.

\vspace{0.05cm}

\noindent
\textbf{Chapter~\ref{chap:preliminaries}} recalls some preliminaries of category-theoretic models of computational effects and dependent types that are needed to understand the rest of the thesis.

\vspace{0.05cm}

\noindent
\textbf{Chapter~\ref{chap:syntax}} introduces and studies our effectful dependently typed language eMLTT.

\vspace{0.05cm}

\noindent
\textbf{Chapter~\ref{chap:fibadjmodels}} defines and studies fibred adjunction models, including various examples.

\vspace{0.05cm}

\noindent
\textbf{Chapter~\ref{chap:interpretation}} defines and studies the interpretation of eMLTT in fibred adjunction models. 

\vspace{0.05cm}

\noindent
\textbf{Chapter~\ref{chap:fibalgeffects}} extends eMLTT with fibred algebraic effects.

\vspace{0.05cm}

\noindent
\textbf{Chapter~\ref{chap:handlers}} extends eMLTT with handlers of fibred algebraic effects.

\vspace{0.05cm}

\noindent
\textbf{Chapter~\ref{chap:conclusions}} concludes the thesis and discusses some possible future work directions.

\vspace{0.05cm}

\noindent
\textbf{Appendix~\ref{chap:appendixC1}} presents an example of dependently typed monadic parsing.

\vspace{0.05cm}

\noindent
\textbf{Appendices~\ref{chap:appendixC4},~\ref{chap:appendixC5}}, and\textbf{~\ref{chap:appendixC6}} contain detailed proofs of results in Chapters~\ref{chap:fibadjmodels},~\ref{chap:interpretation}, and~\ref{chap:fibalgeffects}.

\section{Related work}
\label{sect:relatedwork}

In this section we give an overview of existing work on combining dependent types and computational effects. In particular, these works either 
\begin{itemize}
\item develop new dependently typed languages in which computational effects are included primitively in the design of the language; or 
\item use domain specific languages to represent computational effects in existing dependently typed languages.
\end{itemize}

Compared to our treatment of the combination of dependent types and computational effects, all these languages lack ingredients needed for a general theory, e.g., the foundations are often not settled; the available effects may be limited; or they may lack a systematic treatment of (equational) effect specification. However, it is also important to note that by being more specialised than eMLTT, and focussed on specific computational effects and effect-typing disciplines, some of these languages support more sophisticated types for effectful computations than eMLTT (see below). 

\subsubsection*{Dependently typed CBPV}

A dependently typed CBPV was already briefly discussed as a potential future work direction in Levy's original work~\cite[Section~12.4.1]{Levy:CBPV}. In particular, similarly to us, Levy recognises that one cannot use CBPV's typing rule for sequential composition in the dependently typed setting. 
To solve this problem, Levy suggests the same solution that we have adopted in eMLTT, i.e., to type sequential composition as follows:
\vspace{0.2cm}
\[
\mkrule
{\cj \Gamma {\doto M {x \!:\! A} {} N} {\ul{C}}}
{
\cj \Gamma M FA
\quad
\lj \Gamma \ul{C}
\quad
\cj {\Gamma, x \!:\! A} N \ul{C}
}
\vspace{-0.1cm}
\]
However, Levy does not investigate this dependently typed version of CBPV further. In particular, he does not consider the computational $\Sigma$-type or any other means to overcome the restrictive nature of this typing rule. 
On the other hand, he highlights an important drawback of allowing type-dependency only on values and typing sequential composition using this rule. Namely, there is no obvious translation from a dependently typed $\lambda$-calculus into this version of CBPV, in contrast to the simply typed setting where there exists two canonical translations (call-by-value and call-by-name).

The closest work to ours 
appears in recent unpublished manu\-scripts by V{\'{a}}k{\'{a}}r~\cite{Vakar:EffectfulDepTypes,Vakar:FrameworkForDepEffs}, which appeared independently of the author's paper~\cite{Ahman:FibredEffects} that large parts of this thesis are based on. In particular, V{\'{a}}k{\'{a}}r develops a dependently typed version of CBPV, gives it a denotational semantics based on indexed categories and indexed adjunctions between them, and an abstract machine based operational semantics. Compared to eMLTT, V{\'{a}}k{\'{a}}r's language lacks the computational $\Sigma$-type, he only considers algebraic operations whose possible continuations are listed explicitly, and he does not provide any treatment of handlers in the dependently typed setting. 
On the other hand, V{\'{a}}k{\'{a}}r fixes the typing rule for sequential composition differently from us and Levy, by
\[
\mkrule
{\cj {\Gamma_1, \Gamma_2[\thunk M/y]} {\doto {M} {x \!:\! A} {} {N}} {\ul{C}[\thunk M/y]}}
{
\begin{array}{c}
\cj {\Gamma_1} M {FA}
\quad
\lj {\Gamma_1, y \!:\! U\!FA, \Gamma_2} {\ul{C}}
\\
\cj {\Gamma_1, x \!:\! A, \Gamma_2[\thunk\! (\return x)/y]} {N} {\ul{C}[\thunk\! (\return x)/y]}
\end{array}
}
\]

However, while this typing rule enables V{\'{a}}k{\'{a}}r to define call-by-value and call-by-name translations from a dependently typed $\lambda$-calculus into his language, he observes that not all computationally natural monads support this typing rule for sequential composition, when considering  liftings of their Eilenberg-Moore resolutions to families of sets. For example, while the exceptions monad supports this typing rule, the standard monads for reading, writing, state, and continuations fail to do so. Significant problems also arise in the presence of algebraic effects due to the thunks-based type-dependency in his proposed typing rule for sequential composition. For example, in order to prove that the subject reduction property holds of his language in the presence of algebraic effects that involve equations, such as nondeterminism and state, V{\'{a}}k{\'{a}}r needed to further 
extend his language with the following
typing rules (for all algebraic operations)\footnote{In other words, without these additional typing rules for algebraic operations, one would be unable to assign a single canonical computation type to the left- and right-hand sides of definitional algebraicity equations of the form $\doto {\algop(M_1, \ldots, M_n)} {x \!:\! A} {} {N} = \algop(\doto {M_1} {x \!:\! A} {} {N}, \ldots, \doto {M_n} {x \!:\! A} {} {N})$.}:
\vspace{-0.1cm}
\[
\mkrule
{\cj \Gamma {M} {\ul{C}[\thunk\! (\algop(M_1, \ldots, M_n))/x]}}
{\cj \Gamma {M} {\ul{C}[\thunk M_i/x]}}
\]
It is however not immediate if and how these typing rules could be adapted to general algebraic operations that involve variable bindings, such as the ones we use to extend eMLTT in Chapters~\ref{chap:fibalgeffects} and~\ref{chap:handlers}. 
In summary, we conjecture that in order to make V{\'{a}}k{\'{a}}r's proposed approach work in general, the computation type $\ul{C}$ should be allowed to depend directly on computations of type $FA$ rather than their thunks. We discuss some possibilities for extending eMLTT with this kind of type-dependency in Section~\ref{sect:typedependencyonfeffects}.

\subsubsection*{Linear dependent types}

While not directly addressing the combination of dependent types and computational effects, the recent works of Krishnaswami et al.~\cite{Krishnaswami:LinearDependentTypes} and V{\'{a}}k{\'{a}}r~\cite{Vakar:LinearLF} on integrating dependent and linear types have many features in common with our work. In particular, they develop languages that contain both intuitionistic and linear fragments,  based on adjunction models of linear logic, with both kinds of types allowed to depend only on intuitionistic variables. Compared to eMLTT, where the contexts of (linear) computation variables contain exactly one variable, the contexts of linear variables in their work can contain arbitrary number of variables, which can be used in any order.
Further, while the type of homomorphisms between computation types (the \emph{homomorphic function type} $\ul{C} \multimap \ul{D}$) has to be treated as a value type in eMLTT, so as to be able to capture a wide range of computational effects, the adjunctions used to model linear logic enable them to treat the corresponding type of linear functions as a linear type. 
Krishnaswami et al.~also investigate extending their language with computational effects via a state monad on the linear types, allowing them to support state with strong (type-changing) updates and to  encode the effectful primitives of Hoare Type Theory.

\subsubsection*{Hoare Type Theory}

Hoare Type Theory (HTT)~\cite{Nanevski:HTT} was developed by Nanevski et al.~to allow programmers to specify and verify stateful computations in a dependently typed setting using the Hoare type $\{P\} x \!:\! A \{Q\}$, where $P$ and $Q$ are logical pre- and postcondition formulae over the heap, and $A$ is the type of values returned by a computation of this type. 
The resulting language is a dependently typed version of Moggi's monadic metalanguage, with the ordinary monad replaced by the Hoare type, and with the typing rules of relevant terms adapted accordingly.
HTT resolves the problem with type-dependency in sequential composition by restricting the free variables in the return type, and by using existential quantification to ``close-off" the free variable in the logical formulae. In particular, using idealised syntax, HTT's typing rule for sequential composition is
\vspace{0.15cm}
\[
\mkrule
{\cj \Gamma {\doto M {x \!:\! A} {} N} {\{P\} y \!:\! B \{\exists\, x \!:\! A .\, R\}}}
{
\cj \Gamma M \{P\} x \!:\! A \{Q\}
\quad
\lj \Gamma B
\quad
\cj {\Gamma, x \!:\! A} N \{Q\} y \!:\! B \{R\}
}
\vspace{-0.1cm}
\]
HTT has been given both an abstract machine based operational semantics~\cite{Nanevski:HTT} and a realisability based denotational semantics~\cite{Petersen:HTT}. The latter is organised using a split fibration of uniform families of chain-complete partial equivalence relations, a split comprehension category of uniform families of assemblies, and a split fibred reflection between them, based on Jacobs's fibrational models of higher-order dependent predicate logic and full higher-order dependent type theory, see~\cite[Sections~11.2 and~11.6]{Jacobs:Book}.

\subsubsection*{F*}

Swamy et al.'s F*~\cite{Swamy:FStar} is a closely related language to HTT.
As well as state, F* supports other computational effects such as exceptions and divergence, and their combinations, organised in a lattice of effects.
Compared to the pre- and postconditions based reasoning in HTT, F* instead uses weakest precondition predicate transformers, structured as Dijkstra monads $T\, A\,\, w\!p$, to reason about the behaviour of effectful computations. Here, $A$ is the type of values returned by a computation of this type and $w\!p$ is the weakest precondition transformer. 
For example, for the state effect one has
\[
w\!p : (A \to \State \to \mathsf{Type}) \to \State \to \mathsf{Type}
\]
Namely, $w\!p$ transforms a type-theoretic predicate on return values and final states to a predicate on  initial states. Specifically, in a total correctness setting, being able to assign the type $T\, A\,\, w\!p$ to a stateful computation $M$ guarantees that if $M$ is executed in a state $V_S$ that satisfies $w\!p\,\, V_{\!Q}\,\, V_{\!S}$ (for some postcondition $V_{\!Q}$), then the execution of $M$ produces a value $V$ and a state $V'_{\!S}$ that satisfy $V_{\!Q}\,\, V\,\, V'_{S}$.
In a recent joint work by the author and Hri{\c t}cu et al.~\cite{Ahman:DM4Free}, F* has been extended with a means for representing computational effects and their combinations using monads defined in a simply typed definition language, with the corresponding predicate transformers and Dijkstra monads derived automatically using a selective CPS-transformation; this includes global state, exceptions, and continuations, but currently excludes I/O and probability.

F* resolves the problem with type-dependency in sequential composition by restricting the free variables in the return type, and by using the monad structure of the weakest precondition predicate transformers. In particular, using idealised syntax, F* includes the following typing rule for sequential composition: 
\vspace{0.15cm}
\[
\mkrule
{\cj \Gamma {\doto M {x \!:\! A} {} N} {T~A~(\doto {w\!p_1} {x \!:\! A} {} {w\!p_2})}}
{
\cj \Gamma M T~A~w\!p_1
\quad
\lj \Gamma B
\quad
\cj {\Gamma, x \!:\! A} N T~B~w\!p_2
}
\vspace{-0.1cm}
\]
While F* has been equipped with an operational semantics, a (category-theoretic) denotational semantics and a general algebraic account of it remain open problems. 
Finally, it is worth noting that while the treatment of computational effects in F* is based on weakest precondition predicate transformers, its end users are usually presented with a HTT-style programming interface that uses pre- and postconditions.

\subsubsection*{Dependent types and local names}

Pitts et al.~\cite{Pitts:NominalMLTT} and Cheney~\cite{Cheney:DepNomTypeTheory} have successfully combined dependent types with another important notion of computation, namely, local names.
A significant difference between these works and eMLTT, and the languages described above, is that the languages developed Pitts et al.~and Cheney do not include a distinct layer of effectful computations, i.e., return and sequential composition. 
As a result, types can depend directly on terms that contain name abstractions and concretion.
This is possible because local names, when considered as a computational effect, do not require interaction with the runtime environment, and so a closed term of type $A$ will always evaluate to a value of type $A$ in these languages during typechecking.
Pitts et al.~show how to define a sound interpretation of their language in a category with families (CwF, see~\cite{Hofmann:SyntaxAndSemantics}) of nominal sets.
Meanwhile, Cheney develops a sound and complete normalisation algorithm for the equational theory of his language. 

A somewhat different approach to combining dependent types with local names has been taken by Sch\"{o}pp and Stark~\cite{Schopp:DTTWithNames} (also~\cite{Schopp:Thesis}) who extend an extensional MLTT with ideas from the logic of bunched implications~\cite{Pym:BunchedImplications}. In particular, they derive the operations for local names, such as name abstraction, from a central notion of freshness that they formalise using monoidal versions of $\Sigma$- and $\Pi$-types. Similarly to the work of both Pitts et al.~and Cheney, the resulting language does not include a distinct layer of effectful computations. Sch\"{o}pp and Stark  give their language a denotational semantics based on split closed comprehension categories that have affine monoidal bases, and that additionally support monoidal versions of split dependent sums and products.

\subsubsection*{Dependent types and general recursion}

Casinghino et al.~\cite{Casinghino:CombiningProofs} have studied combining dependent types with another important computational effect, namely, general recursion and the possibility of divergence. 
While their language also does not include combinators for returning values and sequential composition, similarly to the work on local names discussed earlier, their language includes separate typing judgements to distinguish between code that is guaranteed to terminate (used for logical reasoning) and code that can potentially diverge (used for programming). 
Casinghino et al.~equip their language with an operational semantics but they do not investigate a corresponding denotational semantics. 
 
A domain-theoretic denotational semantics has been developed by Palmgren and Stoltenberg-Hansen~\cite{Palmgren:DomainInterpretation} for intensional MLTT that supports recursion via an iteration type $\Omega$. In contrast to Casinghino et al.'s work, this variant of MLTT  does not distinguish between terminating and diverging code. As a result, it is inconsistent as a logic because all its types are inhabited, including the empty type $0$.
In comparison, when we extend eMLTT with general recursion in Section~\ref{sect:continuousfamilies}, 
we take care to ensure that recursion can only be used in computation terms, thus ensuring that the pure fragment of eMLTT can be used to reason about programs. This is similar to the distinction between terminating and possibly diverging code in Casinghino et al.'s work.

It is worth noting that Agda~\cite{Norell:Thesis}, Idris~\cite{Brady:Idris}, and F* also support general recursion. 
While the first two use syntactic termination checkers to ensure the totality of general recursive definitions and therefore the consistency of logical reasoning, F* supports a semantic termination check based on a well-founded partial order on its terms.

\subsubsection*{Representing computational effects using interaction structures}

Regarding the use of domain specific languages to represent computational effects in existing dependently typed languages, a general treatment has been developed by Hancock and Setzer~\cite{Hancock:InteractivePrograms}, who embed Moggi's monadic metalanguage in type theory. They achieve generality by specifying computational effects using interaction structures; these are an abstract representation of single-sorted signatures, also known in the literature under the name of containers~\cite{Abbott:Containers}. The corresponding monad is then generated freely on the polynomial endofunctor induced by the given interaction structure. 
It is worth noting that while this work is general enough to capture all single-sorted signatures, it does not support the equational specification of computational effects.

\subsubsection*{Representing effects using monads on indexed sets and parameterised monads}

We conclude our overview of related work by discussing the work of McBride~\cite{McBride:Kleisli} and Brady~\cite{Brady:Effects}, who also use existing languages to monadically represent computational effects. Compared to Hancock and Setzer's work, McBride and Brady also support sophisticated effect-typing disciplines to specify and verify properties of effectful programs, e.g., that one can read from a file only after it has been opened. McBride's representation of computational effects and the corresponding effect-typing is based on monads on indexed sets, using a custom version of Haskell as the underlying language. 
Brady, on the other hand, uses a natural dependently typed generalisation of Atkey's parameterised monads (see Section~\ref{sect:fibredparametrisedeffects}) to represent computational effects in Idris, and to use pre- and postconditions to track the ``worlds" of computation, e.g., whether a file is open or closed. For example, the type of the function representing sequential composition in Brady's work can be described using the following rule:
\vspace{0.15cm}
\[
\mkrule
{\cj \Gamma {\doto M {x \!:\! A} {} N} {T_{W1,W3}~B}}
{
\cj \Gamma M T_{W1,W2}~A
\quad
\lj \Gamma B
\quad
\vj \Gamma {W_3} {\mathsf{World}}
\quad
\cj {\Gamma, x \!:\! A} N T_{W2,W3}~B
}
\vspace{-0.1cm}
\]

More recently, Brady has extended his (dependently typed parameterised) monads with additional type-dependency~\cite{Brady:ResourceDependentEffects}, by allowing the postcondition worlds to depend on return values, thus alleviating the limitation that $W_2$ is not allowed to depend on the return values of $M$ (via $x$) in the above typing of sequential composition. 
However, as part of our exploratory investigations into extending eMLTT with dependently typed effect typing, we observed that there does not seem to be a category-theoretically natural axiomatisation of the corresponding adjunctions. Instead, our preliminary work has shown that this more dependently typed version of Brady's monad turns out to be simply the composite of the less dependently typed parameterised adjoints and our computational $\Sigma$-type. We discuss this observation in more detail in Section~\ref{sect:fibredparametrisedeffects}.


\chapter{Semantic preliminaries}
\label{chap:preliminaries}

To make our work accessible to a wider audience, we begin by recalling some preliminaries of category-theoretic models of computational effects (monads, adjunctions, and Lawvere theories) 
and dependent types (split fibrations and split comprehension categories). 
We assume familiarity with basic category theoretical concepts such as categories, functors, and natural transformations---we refer the reader to Mac Lane's book~\cite{MacLane:CatWM} for an in-depth overview. 
Later in the thesis, we also assume familiarity with basic enriched category theory---see Kelly's book~\cite{Kelly:EnrichedCats} for an in-depth overview. 

We also note that throughout this chapter (and more generally, 
throughout this entire thesis), we assume the Axiom of Choice in results involving sets and functions.

\section{Models of computational effects}
\label{sect:modelsofeffects}

We begin by recalling the definitions and key properties of category-theoretic structures used for modelling computational effects: monads, adjunctions, and Lawvere theories, including their relationships. For more details, see~\cite{MacLane:CatWM,Barr:Toposes,Manes:AlgTheories,Borceux:HandbookVol2,Power:CountableTheories}. 

\subsection{Monads}

The uniform category-theoretic study of computational effects dates back to the seminal work of Moggi~\cite{Moggi:ComputationalLambdaCalculus,Moggi:NotionsofComputationandMonads}, who recognised that all computational effects commonly used in programming languages can be modelled using (strong) monads. 

\begin{definition}
\label{def:monad}
\index{monad}
\index{ T@$\mathbf{T}$ (monad)}
\index{ T@$(T,\eta,\mu)$ (monad)}
\index{ e@$\eta$ (unit of a monad)}
\index{ m@$\mu$ (multiplication of a monad)}
Given a category $\mathcal{V}$, a \emph{monad} $\mathbf{T} = (T,\eta,\mu)$ 
on $\mathcal{V}$ is given by a functor $T : \mathcal{V} \longrightarrow \mathcal{V}$ and two natural transofrmations, the \emph{unit} $\eta : \id_{\mathcal{V}} \longrightarrow T$ and the \emph{multiplication} $\mu : T \comp T \longrightarrow T$, 
subject to the following commuting diagrams:
\[
\xymatrix@C=3em@R=3em@M=0.5em{
T \ar[r]^-{\eta \,\comp\, T} \ar[dr]_-{\id_T\!\!\!} & T \comp T \ar[d]_-{\mu} & T \ar[l]_-{T \,\comp\, \eta} \ar[dl]^-{\id_T}
\\
& T
}
\qquad\qquad
\xymatrix@C=3em@R=3em@M=0.5em{
T \comp T \comp T \ar[r]^-{T \,\comp\, \mu} \ar[d]_-{\mu \,\comp\, T} & T \comp T \ar[d]^-{\mu}
\\
T \comp T \ar[r]_-{\mu} & T
}
\]
\end{definition}

Through the work of Wadler~\cite{Wadler:Monads}, who popularised the use of monads as a convenient means to structure functional programs, and the subsequent adoption of monads as a uniform mechanism to include computational effects in Haskell, functional programmers are probably more familiar with the Kleisli triple presentation of monads.

\begin{definition}
\index{Kleisli!-- triple}
\index{ f@$f^\dagger$ (Kleisli extension)}
\index{Kleisli!-- extension}
\index{ ob@$\mathsf{ob}(\hspace{-0.05cm}\mathcal{V})$ (class of objects of a category $\mathcal{V}$)}
Given a category $\mathcal{V}$, a \emph{Kleisli triple} $(T,\eta,(-)^\dagger)$ on $\mathcal{V}$ is given by a mapping $T : \mathsf{ob}(\!\mathcal{V}) \longrightarrow \mathsf{ob}(\!\mathcal{V})$, a family of morphisms $\eta_A : A \longrightarrow T(A)$ (for all $A$ in $\mathcal{V}$) and morphisms $f^\dagger : T(A) \longrightarrow T(B)$ (for all $f : A \longrightarrow T(B)$ in $\mathcal{V}$) 
such that
\[
\eta_A^\dagger = \id_{T(A)}
\qquad
f^\dagger \comp \eta_A = f
\qquad
g^\dagger \comp f^\dagger = (g^\dagger \comp f)^\dagger
\]
for all $f : A \longrightarrow T(B)$ and $g : B \longrightarrow T(C)$.
\end{definition}

As is well-known, these two definitions are in fact equivalent.

\begin{proposition}[{\cite[Theorem~3.18]{Manes:AlgTheories}}]
Monads and Kleisli triples on a category $\mathcal{V}$ are in a 1-to-1 correspondence, with $f^\dagger$ and $\mu$ defined respectively as follows:
\[
f^\dagger \defeq \mu_B \comp T(f)
\qquad
\mu_A \defeq \id_{T(A)}^\dagger
\]
\end{proposition}

Moggi's insight was that for a suitable monad $(T,\eta,\mu)$, the object $T(A)$ can be used to model effectful computations that return values modelled by $A$;
the unit $\eta$ can be used to model the effect-free computation that returns a value and does not perform any effects; and the multiplication $\mu$ (or, equivalently, the Kleisli extension $(-)^\dagger$) can be used to model the sequential composition of effectful computations, e.g., the $\mathtt{let}$-expression in the ML-family of languages.
Based on this monadic approach to  denotational semantics, Moggi also developed two simply typed languages to provide a formal basis for proving equivalences between effectful programs, namely, the computational $\lambda$-calculus~\cite{Moggi:ComputationalLambdaCalculus} and the monadic metalanguage~\cite{Moggi:NotionsofComputationandMonads}. These languages were later refined by Levy to a single fine-grain call-by-value language~\cite[Appendix~A.3.2]{Levy:CBPV}. 

\index{ Set@$\Set$ (category of sets and functions)}
Below we list some computational effects that Moggi considered and recall the underlying functors of the corresponding monads (for simplicity, on the category $\Set$):  
\begin{itemize}
\item for \emph{exceptions}, the monad is given on objects by $T_{\text{EXC}}(A) \defeq A + E$;
\index{monad!exceptions --}
\item for \emph{nondeterminism}, the monad is given on objects by $T_{\text{ND}}(A) \defeq \mathcal{P}_{\text{fin}}^+(A)$;
\index{monad!nondeterminism --}
\item for \emph{global state}, the monad is given on objects by $T_{\text{GS}}(A) \defeq S \Rightarrow (A \times S)$;
\index{monad!global state}
\index{ A@$A \times B$ (Cartesian product of $A$ and $B$)}
\index{ A@$A \Rightarrow B$ (exponential object)}
\item for \emph{I/O}, the monad is given on objects by $T_{\text{I/O}}(A) \defeq \mu\, X .\, A + (I \Rightarrow X) + (O \times X)$; and 
\index{monad!I/O --}
\index{ m@$\mu X .\, F(X)$ (least fixed point of the endofunctor $F$)}
\item for \emph{continuations}, the monad is given on objects by $T_{\text{CONT}}(A) \defeq (A \Rightarrow R) \Rightarrow R$, 
\index{monad!continuations --}
\end{itemize}
where $E$ is a set of exception names, $S$ of store values, $I$ of input values, $O$ of output values,  and $R$ of results. We omit the definitions of $\eta$ and $\mu$ for each of these monads; they can be  readily found in~\cite{Moggi:NotionsofComputationandMonads}. These monads also generalise straightforwardly to categories $\mathcal{V}$ other than $\Set$, as long as $\mathcal{V}$ has appropriate structure, e.g., coproducts for the exceptions monad, and Cartesian products and exponentials for global state.

It is worth noting that the global state monad only models state where the store is changed by overwriting. In contrast, the author has also studied models of a more fine-grained notion of state where the store is changed by applying (potentially small) updates to it. 
For this notion of state, the store is modelled using a set $S$, the updates using a monoid $(P,\mathsf{o},\oplus)$, and the interaction of the two using an action $\downarrow$ of the monoid on $S$.
The corresponding \emph{update monad} is given on objects by $T_{\text{UPD}}(A) \defeq S \Rightarrow (P \times A)$. 
\index{monad!update --}
For more details about update monads, see the author's joint paper with Uustalu~\cite{Ahman:UpdateMonads}.

This paper also describes a natural \emph{dependently typed generalisation} of update monads, where one uses a dependently typed generalisation of a monoid, a \emph{directed container} $(S,P,\downarrow,\mathsf{o},\oplus)$~\cite{Ahman:Dcontainers}. In short, in $(S,P,\downarrow,\mathsf{o},\oplus)$, $P$ is not a set but instead an $S$-indexed family of sets, with $\downarrow$, $\mathsf{o}$, and $\oplus$ typed accordingly, thus providing fine-grained control over which updates are applicable to specific stores---see Example~\ref{ex:fibsigofdeptypedupdatemonad} for more details.
The corresponding dependently typed update monad is then given on objects by\footnote{Given a set $X$ and an $X$-indexed family of sets $Y$, then $\bigsqcap_{x \in X} Y_x$ is the $X$-indexed product of $Y_x$'s.} $T_{\text{DUPD}}(A) \defeq \bigsqcap_{s \in S} (P_s \times A)$.
\index{monad!update --!dependently typed --}
We discuss the equational presentations of (dependently typed) update monads in Examples~\ref{ex:fibsigofupdatemonad},~\ref{ex:fibsigofdeptypedupdatemonad},~\ref{ex:fibtheoryofupdatemonads},  and~\ref{ex:fibtheoryofdeptypedupdatemonads}. 

Moggi also observed that monads by themselves are not sufficient to model computations in non-empty contexts of variables. Correspondingly, one requires the given monad $(T,\eta,\mu)$ to also be strong, so as to ensure that every morphism of the form $A \times B \longrightarrow T(C)$ canonically induces a morphism of the form $A \times T(B) \longrightarrow T(C)$.

\begin{definition}
\label{def:strengthofmonad}
\index{monad!strong --}
\index{ sigma@$\sigma$ (strength of a monad)}
A monad $\mathbf{T} = (T,\eta,\mu)$ on a category $\mathcal{V}$ with Cartesian products is said to be \emph{strong} if it is equipped with a natural transformation 
\[
\sigma : (-) \times T(=) \longrightarrow T((-) \times (=))
\]
making the following four diagrams commute:
\[
\xymatrix@C=3em@R=4em@M=0.5em{
1 \times T(A) \ar[r]^-{\sigma_{I,A}} \ar[dr]_-{\lambda_{T(A)}} & T(1 \times A) \ar[d]^-{T(\lambda_A)}
\\
& T(A)
}
\]

\vspace{0.1cm}

\[
\xymatrix@C=3em@R=4em@M=0.5em{
(A \times B) \times T(C) \ar[d]_-{\alpha_{A,B,T(C)}} \ar[rr]^-{\sigma_{A \,\times\, B,C}} && T((A \times B) \times C) \ar[d]^-{T(\alpha_{A,B,C})}
\\
A \times (B \times T(C)) \ar[r]_-{\id_A \times\, \sigma_{B,C}} & A \times T(B \times C) \ar[r]_-{\sigma_{A,B \,\times\, C}} & T(A \times (B \times C))
}
\]

\vspace{0.1cm}

\[
\xymatrix@C=3em@R=4em@M=0.5em{
A \times B \ar[r]^-{\id_A \times\, \eta_B} \ar[dr]_-{\eta_{A \,\times\, B}} & A \times T(B) \ar[d]^-{\sigma_{A,B}}
\\
& T(A \times B)
}
\]

\vspace{0.1cm}

\[
\xymatrix@C=3em@R=4em@M=0.5em{
A \times T(T(B)) \ar[d]_-{\id_A \times\, \mu_B} \ar[r]^-{\sigma_{A,T(B)}} & T(A \times T(B)) \ar[r]^-{T(\sigma_{A,B})} & T(T(A \times B)) \ar[d]^-{\mu_{A \,\times\, B}}
\\
A \times T(B) \ar[rr]_-{\sigma_{A,B}} && T(A \times B)
}
\]
where $\lambda_A : I \times A \overset{\cong}{\longrightarrow} A$ and $\alpha_{A,B,C} : (A \times B) \times C \overset{\cong}{\longrightarrow} A \times (B \times C)$ are components of the canonical natural isomorphisms induced by the Cartesian monoidal structure of $\mathcal{V}$.
The notion of strength easily generalises to arbitrary monoidal categories, see~\cite{Kock:StrongMonads}.
\end{definition}

We conclude by recalling a well-known result that every monad on $\Set$ has a unique strength $\sigma$, given by $\sigma_{A,B} \defeq \langle a , d \rangle \mapsto T(b \mapsto \langle a , b \rangle)(d)$. One way to show this result is to first observe that a strong monad on a monoidal closed category $\mathcal{V}$ can be equivalently characterised as a $\mathcal{V}$-enriched monad on $\mathcal{V}$ (see~\cite{Kock:StrongMonads}), and as it happens, every monad on $\Set$ is trivially $\Set$-enriched. The uniqueness of $\sigma$ then follows from $\Set$ having enough points, in that for any two functions $f,g \!:\! A \longrightarrow B$, we have that $(\forall h \!:\! 1 \longrightarrow A .\, f \comp h = g \comp h)$ implies $f = g$ (see~\cite[Proposition~3.4]{Moggi:NotionsofComputationandMonads} for more details).

\subsection{Adjunctions}
\label{sect:adjunctionsbackground}

A decade after Moggi's seminal work, Levy~\cite{Levy:CBPV} gave a more fine-grained analysis of effects based on adjunctions, using them to account for the clear separation between values and computations in his Call-By-Push-Value (CBPV) language. Adjunctions were also important in Egger et al.'s~\cite{Egger:EnrichedEffectCalculus} subsequent work on the linear aspects of effects, and for giving a denotational semantics to their Enriched Effect Calculus (EEC).

\begin{definition}
\label{def:adjunction}
\index{adjunction}
\index{adjoint!left --}
\index{adjoint!right --}
\index{ F@$F \dashv\, U$ (adjunction)}
\index{ e@$\eta$ (unit of an adjunction)}
\index{ e@$\varepsilon$ (counit of an adjunction)}
An \emph{adjunction} $F \dashv\, U : \mathcal{C} \longrightarrow \mathcal{V}$ between categories $\mathcal{V}$ and $\mathcal{C}$ is given by two functors, the \emph{left adjoint} $F : \mathcal{V} \longrightarrow \mathcal{C}$ and the \emph{right adjoint} $U : \mathcal{C} \longrightarrow \mathcal{V}$, and two natural transformations, the \emph{unit} $\eta : \id_{\mathcal{V}} \longrightarrow U \comp F$ and the \emph{counit} $\varepsilon : F \comp U \longrightarrow \id_{\mathcal{C}}$, 
subject to the following two commuting diagrams:
\[
\xymatrix@C=3em@R=3em@M=0.5em{
U \ar[r]^-{\eta \,\comp\, U} \ar[dr]_-{\id_U\!\!} & U \comp F \comp U \ar[d]^-{U \,\comp\, \varepsilon}
&
F \ar[r]^-{F \,\comp\, \eta} \ar[dr]_-{\id_F\!\!} & F \comp U \comp F \ar[d]^-{\varepsilon \,\comp\, F}
\\
& U
&
& F
}
\]
\end{definition}

As a convention, we write $A,B,\ldots$ for the objects of $\mathcal{V}$ and $\ul{C},\ul{D},\ldots$ for the objects of $\mathcal{C}$. While this notation coincides with our notation for value and computation types, we make sure that it is clear from the context whether $A$ and $\ul{C}$ mean types or objects.

Similarly to monads, there are other, equivalent ways in which one can define adjunctions. We recall the other commonly used definition based on hom-sets.

\begin{definition}
\label{def:adjunctionhomsets}
\index{adjunction!hom-set presentation of an --}
\index{ @$\cong$ (isomorphism)}
\index{ V@$\mathcal{V}(A,B)$ (hom-set between objects $A$ and $B$ in $\mathcal{V}$)}
A \emph{hom-set presentation} of an adjunction ${F \dashv\, U : \mathcal{C} \longrightarrow \mathcal{V}}$  consists of two functors, the left adjoint ${F : \mathcal{V} \longrightarrow \mathcal{C}}$ and the right adjoint ${U : \mathcal{C} \longrightarrow \mathcal{V}}$, and an isomorphism of hom-sets $\mathcal{C}(FA,\ul{C}) \cong \mathcal{V}(A,U\ul{C})$ that is natural in both $A$ and $\ul{C}$.
\end{definition}

A useful property of adjoints is that they are unique up-to-isomorphism. 

\begin{proposition}[{\cite[Section~IV.1]{MacLane:CatWM}}]
\label{prop:adjointsareunique}
\index{ @$\cong$ (natural isomorphism)}
Given $F \dashv\, U : \mathcal{C} \longrightarrow \mathcal{V}$ and $F' \dashv\, U : \mathcal{C} \longrightarrow \mathcal{V}$, then there exists a natural isomorphism $F \cong F'$. Analogously, given $F \dashv\, U : \mathcal{C} \longrightarrow \mathcal{V}$ and $F \dashv\, U' : \mathcal{C} \longrightarrow \mathcal{V}$, then there exists a natural isomorphism $U \cong U'$. 
\end{proposition}

\index{ V@$\mathcal{V}^\text{op}$ (opposite category of $\mathcal{V}$)}
Analogously to using monads for giving denotational semantics to effectful languages, adjunctions by themselves are not sufficient to model CBPV and EEC's terms in non-empty contexts. To this end, one requires the adjunction $F \dashv\, U : \mathcal{C} \longrightarrow \mathcal{V}$ to be \linebreak $\Set^{\mathcal{V}^\text{op}}$-enriched in the models of CBPV and $\mathcal{V}$-enriched in the models of EEC.

Next, we recall the close relationship between adjunctions and monads.

\begin{proposition}[{\cite[Section~VI.1]{MacLane:CatWM}}]
\label{prop:monadfromadjunction}
Given an adjunction $F \dashv\, U : \mathcal{C} \longrightarrow \mathcal{V}$ between categories $\mathcal{V}$ and $\mathcal{C}$, we get a monad $(U \comp F, \eta, U \comp \varepsilon \comp F)$ on the category $\mathcal{V}$.
\end{proposition}

\begin{definition}
\label{def:resolutionofamonad}
\index{resolution!-- of a monad}
Given a monad $(T,\eta,\mu)$ on a category $\mathcal{V}$, a \emph{resolution} of $(T,\eta,\mu)$ is given by a category $\mathcal{C}$ and an adjunction $F \dashv\, U : \mathcal{C} \longrightarrow \mathcal{V}$ such that $(T,\eta,\mu)$ coincides with the monad canonically derived from this adjunction, as given in Proposition~\ref{prop:monadfromadjunction}.
\end{definition}

While there does not exist a unique resolution of a monad, it is well-known that there exist two canonical resolutions: the \emph{Kleisli} and \emph{Eilenberg-Moore} resolutions. In fact, these two resolutions turn out to be the initial and terminal object in the category of resolutions of a monad, respectively---see~\cite[Chapter~VI]{MacLane:CatWM} for more details. These resolutions are also a common source of models of languages such as CBPV and EEC.

\begin{definition}
\label{def:kleisliresolution}
\index{resolution!-- of a monad!Kleisli --}
\index{ V@$\mathcal{V}_{\mathbf{T}}$ (Kleisli category of a monad $\mathbf{T}$ on $\mathcal{V}$)}
\index{category!Kleisli --}
\index{ F@$F_{\mathbf{T}} \dashv U_{\mathbf{T}}$ (Kleisli resolution)}
Given a monad $\mathbf{T} = (T,\eta,\mu)$ on a category $\mathcal{V}$, its \emph{Kleisli resolution} is given by a category $\mathcal{V}_{\mathbf{T}}$ 
and an adjunction $F_{\mathbf{T}} \dashv U_{\mathbf{T}} : \mathcal{V}_{\mathbf{T}} \longrightarrow \mathcal{V}$, 
where the objects of $\mathcal{V}_{\mathbf{T}}$ are the objects of $\mathcal{V}$; and the morphisms $A \longrightarrow B$ in $\mathcal{V}_{\mathbf{T}}$ are the morphisms $A \longrightarrow T(A)$ in $\mathcal{V}$. The left and right adjoints are defined as follows:
\[
\begin{array}{c}
F_{\mathbf{T}}(A) \defeq A 
\qquad
F_{\mathbf{T}}(f) \defeq \eta_B \comp f
\qquad
U_{\mathbf{T}}(A) \defeq T(A)
\qquad
U_{\mathbf{T}}(h) \defeq \mu_B \comp T(h)
\end{array}
\]
where $f : A \longrightarrow B$ in $\mathcal{V}$ and $h : A \longrightarrow B$ in $\mathcal{V}_{\mathbf{T}}$.
\end{definition}

\begin{definition}
\label{def:EMresolution}
\index{resolution!-- of a monad!Eilenberg-Moore --}
\index{ EM@EM (Eilenberg-Moore)}
\index{ V@$\mathcal{V}^{\mathbf{T}}$ (Eilenberg-Moore category of a monad $\mathbf{T}$ on $\mathcal{V}$)}
\index{category!Eilenberg-Moore --}
\index{ F@$F^{\mathbf{T}} \dashv U^{\mathbf{T}}$ (Eilenberg-Moore resolution)}
Given a monad $\mathbf{T} = (T,\eta,\mu)$ on a category $\mathcal{V}$, its \emph{Eilenberg-Moore (EM-) resolution} is given by a category $\mathcal{V}^{\mathbf{T}}$ 
and an adjunction $F^{\mathbf{T}} \dashv U^{\mathbf{T}} : \mathcal{V}^{\mathbf{T}} \longrightarrow \mathcal{V}$. 
The objects of $\mathcal{V}^{\mathbf{T}}$ are given by pairs $(A,\alpha)$ of an object $A$ in $\mathcal{V}$ and a morphism \linebreak $\alpha : T(A) \longrightarrow A$ in $\mathcal{V}$ such that the following two diagrams commute:
\[
\xymatrix@C=3em@R=3em@M=0.5em{
A \ar[dr]_-{\id_A} \ar[r]^-{\eta_A} & T(A) \ar[d]^-{\alpha}
&
T(T(A)) \ar[d]_-{\mu_A} \ar[r]^-{T(\alpha)} & T(A) \ar[d]^-{\alpha}
\\
&
A
&
T(A) \ar[r]_-{\alpha} & A
}
\]
A morphism $h : (A,\alpha) \longrightarrow (B,\beta)$ in $\mathcal{V}^{\mathbf{T}}$ is given by a morphism $h : A \longrightarrow B$ in $\mathcal{V}$ such that the following diagram commutes:
\[
\xymatrix@C=3em@R=3em@M=0.5em{
T(A) \ar[r]^-{T(h)} \ar[d]_-{\alpha} & T(B) \ar[d]^-{\beta}
\\
A \ar[r]_-{h} & B
}
\]
The left and right adjoints are defined as follows:
\[
F^{\mathbf{T}} (A) \defeq (T(A),\mu_A)
\qquad
F^{\mathbf{T}} (f) \defeq T(f)
\qquad
U^{\mathbf{T}} (A,\alpha) \defeq A
\qquad
U^{\mathbf{T}} (h) \defeq h
\]
where $f : A \longrightarrow B$ in $\mathcal{V}$ and $h : (A,\alpha) \longrightarrow (B,\beta)$ in $\mathcal{V}^{\mathbf{T}}$.
\end{definition}

The category $\mathcal{V}^{\mathbf{T}}$ is called the \emph{Eilenberg-Moore (EM-) category} of the monad $\mathbf{T}$. Its objects are commonly known as the \emph{Eilenberg-Moore (EM-) algebras}
\index{Eilenberg-Moore!-- algebra}
 of $\mathbf{T}$ and its morphisms as the EM-algebra \emph{homomorphisms}.
For a given EM-algebra $(A,\alpha)$, 
\index{ A@$(A,\alpha)$ (Eilenberg-Moore algebra)}
the object $A$ is typically called the \emph{carrier}, 
\index{Eilenberg-Moore!-- algebra!carrier of an --}
and the morphism $\alpha$ the \emph{structure map}.
\index{Eilenberg-Moore!-- algebra!structure map of an --}

It is worth noting that some computationally important monads can be naturally decomposed into resolutions other than their Kleisli and EM-resolutions. 
Below we assume that the monads in question are given on some Cartesian closed category $\mathcal{V}$.

\begin{proposition}
\label{prop:statemonadresolution}
\index{resolution!-- of the state monad}
The global state monad, given by $T_{\text{GS}}(A) \defeq S \Rightarrow (A \times S)$, can be decomposed into the resolution given by $(-) \times S \dashv\, S \Rightarrow (-) : \mathcal{V} \longrightarrow \mathcal{V}$.
\end{proposition}

\begin{proposition}
\label{prop:continuationsmonadresolution}
\index{resolution!-- of the continuations monad}
The continuations monad, given by $T_{\text{CONT}}(A) \defeq (A \Rightarrow R) \Rightarrow R$, can be decomposed into the resolution given by $(-) \Rightarrow R \dashv\, (-) \Rightarrow R : \mathcal{V}^{\text{op}} \longrightarrow \mathcal{V}$.
\end{proposition}

We conclude our discussion about monads and adjunctions by recalling some known results about the existence of products and coproducts in the EM-category of a monad. 
We later use these results and their natural fibrational generalisations as a basis for constructing examples of models of eMLTT---see Section~\ref{sect:examplesoffibadjmodels} for details.

In the interest of generality, we state these existence results in terms of limits and colimits, from which the results for Cartesian products and coproducts follow as simple corollaries. To this end, we first recall the definitions of limits and colimits.

\begin{definition}
\index{diagram of given shape}
\index{ D@$\mathcal{D}$ (shape of a diagram)}
\index{ J@$J$ (diagram)}
Given any category $\mathcal{V}$ and a small category $\mathcal{D}$, we say that a functor $J : \mathcal{D} \longrightarrow \mathcal{V}$ is a \emph{diagram of shape $\mathcal{D}$}.
\end{definition}

\begin{definition}
\label{def:cone}
\index{cone}
Given a diagram $J : \mathcal{D} \longrightarrow \mathcal{V}$ and an object $A$ in $\mathcal{V}$, we say that a natural transformation $\alpha : \Delta(A) \longrightarrow J$ is a \emph{cone over $J$}. We call $A$ the \emph{vertex} of $\alpha$.
\index{cone!vertex of a --}
\end{definition}

\index{functor!diagonal --}
\index{ D@$\Delta$ (diagonal functor)}
In the above definition, $\Delta : \mathcal{V} \longrightarrow \mathcal{V}^{\mathcal{D}}$ 
is the standard \emph{diagonal functor} that maps an object $A$ in $\mathcal{V}$ to the constant functor that maps every $D$ in $\mathcal{D}$ to the given $A$ in $\mathcal{V}$.

\begin{definition}
\index{morphism!-- of cones}
Given a diagram $J : \mathcal{D} \longrightarrow \mathcal{V}$, and two cones $\alpha : \Delta(A) \longrightarrow J$ and $\beta : \Delta(B) \longrightarrow J$, we say that a morphism $h : A \longrightarrow B$ in $\mathcal{V}$ is \emph{a morphism of cones} from $\alpha$ to $\beta$ if for all objects $D$ in $\mathcal{D}$, we have 
$
\beta_D \comp h = \alpha_D
$.
\end{definition}

\begin{definition}
\index{limit}
\index{ proj@$\mathsf{pr}^{J}$ (limit of $J$)}
\index{ lim@$\mathsf{lim}(J)$ (vertex of the limit of $J$)}
Given a diagram $J : \mathcal{D} \longrightarrow \mathcal{V}$, a \emph{limit of $J$} is the terminal cone over $J$, which we write as  $\mathsf{pr}^{J} : \Delta(\mathsf{lim}(J)) \longrightarrow J$. 
For any other cone $\alpha : \Delta(A) \longrightarrow J$, we write $\langle \alpha \rangle$ for the unique mediating morphism of cones from $\alpha$ to $\mathsf{pr}^{J}$.
\index{ a@$\langle \alpha \rangle$ (unique mediating morphism of cones from $\alpha$ to $\mathsf{pr}^{J}$)}
\end{definition}

\begin{definition}
If the category $\mathcal{V}$ has limits for all diagrams  $J : \mathcal{D} \longrightarrow \mathcal{V}$, we say that $\mathcal{V}$ has \emph{limits of shape} $\mathcal{D}$. Further, if the category $\mathcal{V}$ has limits of all shapes $\mathcal{D}$, we say that $\mathcal{V}$ has all \emph{small limits} and that $\mathcal{V}$ is \emph{complete}.
\index{category!complete --}
\index{limit!small --}
\end{definition}

\begin{definition}
\index{pullback}
A \emph{pullback} of morphisms $f : A \longrightarrow C$ and $g : B \longrightarrow C$ in $\mathcal{V}$ is the limit of a diagram $J : \mathcal{D} \longrightarrow \mathcal{V}$, where $\mathcal{D}$ is given by morphisms ${i : D_1 \longrightarrow D_3}$ and ${j : D_2 \longrightarrow D_3}$; and $J$ is given by  $J(i) \defeq f$ and $J(j) \defeq g$.
\end{definition}

\index{pullback!-- square}
\index{ @$\lrcorner$ (notation for pullback squares)}
As standard, we denote the existence of the pullback of $f : A \longrightarrow C$ and $g : B \longrightarrow C$ in $\mathcal{V}$ using a diagram of the following form, commonly called a \emph{pullback square}:
\[
\xymatrix@C=3.5em@R=3.5em@M=0.5em{
\mathsf{lim}(J) \ar[r]^{\mathsf{pr}^{J}_{D_1}} \ar[d]_{\mathsf{pr}^{J}_{D_2}}^<{\,\big\lrcorner} & A \ar[d]^{f}
\\
B \ar[r]_{g} & C
}
\]
As also standard, we often leave the diagram $J$ and the corresponding terminal cone implicit, and instead write pullback squares as in Definition~\ref{def:kernelpair} below.

\begin{definition}
\label{def:kernelpair}
\index{kernel pair}
A \emph{kernel pair} of a morphism $f : A \longrightarrow B$ is a pair of morphisms $g,h : C \longrightarrow A$ that form the pullback of $f$ and $f$, as illustrated in the following diagram:
\[
\xymatrix@C=3.5em@R=3.5em@M=0.5em{
C \ar[r]^{h} \ar[d]_{g}^<{\,\big\lrcorner} & A \ar[d]^{f}
\\
A \ar[r]_{f} & B
}
\]
\end{definition}

\begin{definition}
\index{cocone}
Given a diagram $J : \mathcal{D} \longrightarrow \mathcal{V}$ and an object $A$ in $\mathcal{V}$, we say that a natural transformation $\alpha : J \longrightarrow \Delta(A)$ is a \emph{cocone over $J$}. We call $A$ the \emph{vertex} of $\alpha$.
\index{cocone!vertex of a --}
\end{definition}

\begin{definition}
\index{morphism!-- of cocones}
Given a diagram $J : \mathcal{D} \longrightarrow \mathcal{V}$, and two cocones $\alpha : J \longrightarrow \Delta(A)$ and $\beta : J \longrightarrow \Delta(B)$, we say that a morphism $h : A \longrightarrow B$ in $\mathcal{V}$ is \emph{a morphism of cocones} from $\alpha$ to $\beta$ if for all objects $D$ in $\mathcal{D}$, we have 
$
h \comp \alpha_D = \beta_D
$.
\end{definition}

\begin{definition}
\label{def:colimits}
\index{colimit}
\index{ in@$\mathsf{in}^{J}$ (colimit of $J$)}
\index{ colim@$\mathsf{colim}(J)$ (vertex of the colimit of $J$)}
Given a diagram $J : \mathcal{D} \longrightarrow \mathcal{V}$, a \emph{colimit of $J$} is the initial cocone over $J$, which we write as $\mathsf{in}^{J} : J \longrightarrow \Delta(\mathsf{colim}(J))$. 
For any other cocone $\alpha : J \longrightarrow \Delta(A)$, we write $[\alpha]$ for the unique mediating morphism of cocones from $\mathsf{in}^{J}$ to $\alpha$.
\index{ a@$[\alpha]$ (unique mediating morphism of cocones from $\mathsf{in}^{J}$ to $\alpha$)}
\end{definition}

\begin{definition}
If the category $\mathcal{V}$ has colimits for all diagrams  $J : \mathcal{D} \longrightarrow \mathcal{V}$, we say that $\mathcal{V}$ has \emph{colimits of shape} $\mathcal{D}$. Further, if the category $\mathcal{V}$ has colimits of all shapes $\mathcal{D}$, we say that $\mathcal{V}$ has all \emph{small colimits} and that $\mathcal{V}$ is \emph{cocomplete}.
\index{category!cocomplete --}
\index{colimit!small --}
\end{definition}

\begin{definition}
\index{pushout}
A \emph{pushout} of morphisms $f : A \longrightarrow B$ and $g : A \longrightarrow C$ in $\mathcal{V}$ is the colimit of a diagram $J : \mathcal{D} \longrightarrow \mathcal{V}$, where $\mathcal{D}$ is given by morphisms ${i : D_1 \longrightarrow D_2}$ and ${j : D_1 \longrightarrow D_3}$; and $J$ is given by  $J(i) \defeq f$ and $J(j) \defeq g$.
\end{definition}

\begin{definition}
\index{coequalizer}
A \emph{coequalizer} of a parallel pair of morphisms $f,g : A \longrightarrow B$ in $\mathcal{V}$ is the colimit of a diagram $J : \mathcal{D} \longrightarrow \mathcal{V}$, where $\mathcal{D}$ consists of a parallel pair of morphisms $i,j : D_1 \longrightarrow D_2$; and $J$ is given by $J(i) \defeq f$ and $J(j) \defeq g$.
\end{definition}

\begin{definition}
\index{section}
A \emph{section} of a morphism $f : A \longrightarrow B$ is a morphism $g : B \longrightarrow A$ that is a right inverse to $f$, i.e., $f \comp g = \id_B$.
\end{definition}

\begin{definition}
\index{coequalizer!reflexive --}
A \emph{reflexive coequalizer} is a coequalizer of a parallel pair of morphisms $f,g : A \longrightarrow B$ that have a common section. \end{definition}

\index{ 2@$\mathbf{2}$ (discrete two-object category)}
We now list the results about the existence of limits and colimits in the EM-category of a monad.
The cases of Cartesian products and coproducts follow as simple corollaries to these results if we take $\mathcal{D} \defeq \mathbf{2}$, where $\mathbf{2}$ is the discrete two-object category. 

\begin{proposition}[{\cite[Proposition~4.3.1]{Borceux:HandbookVol2}}]
\label{prop:limitsinEMcategory}
Given a monad $\mathbf{T}$ on a category $\mathcal{V}$ and a diagram $J : \mathcal{D} \longrightarrow \mathcal{V}^{\mathbf{T}}$, then there exists a limit of $J$ if there exists a limit of the composite diagram $U^{\mathbf{T}} \comp J$. In particular, if $\mathcal{V}$ has all limits of shape $\mathcal{D}$, then $\mathcal{V}^{\mathbf{T}}$ also has all limits of shape $\mathcal{D}$.
\end{proposition}

\begin{proposition}[{\cite[Proposition~4.3.2]{Borceux:HandbookVol2}}]
\label{prop:colimitsinEMcategory1}
Given a monad $\mathbf{T} = (T,\eta,\mu)$ on a category $\mathcal{V}$ and a diagram $J : \mathcal{D} \longrightarrow \mathcal{V}^{\mathbf{T}}$ such that there exists a colimit of the composite functor $U^{\mathbf{T}} \comp J$ which is preserved by $T$, then there exists a colimit of $J$ and it is preserved by $U^{\mathbf{T}}$. In particular, if $\mathcal{V}$ has all colimits of shape $\mathcal{D}$ and they are preserved by $T$, then $\mathcal{V}^{\mathbf{T}}$ also has all colimits of shape $\mathcal{D}$ and they are preserved by $U^{\mathbf{T}}$.
\end{proposition}

\begin{proposition}[{\cite[Corollary~2]{Linton:Coequalizers}}]
\label{prop:colimitsinEMcategory2}
Given a monad $\mathbf{T}$ on a cocomplete category $\mathcal{V}$, then $\mathcal{V}^{\mathbf{T}}$ is cocomplete if it has reflexive coequalizers. 
\end{proposition}

\begin{proposition}[{\cite[Theorem~4.3.5 (i)]{Borceux:HandbookVol2}}]
\label{prop:EMcategoryiscompletecocompleteandregular}
Given a monad $\mathbf{T}$ on a complete, cocomplete, and regular category $\mathcal{V}$ in which every regular epimorphism has a section, then $\mathcal{V}^{\mathbf{T}}$ is complete, cocomplete, and regular.
\end{proposition}

In particular, recall that a category is \emph{regular} 
\index{category!regular --}
when i) every morphism in it has a kernel pair, ii) every kernel pair has a coequalizer, and iii) the pullback of a regular epimorphism 
\index{epimorphism!regular --}
(a coequalizer of some parallel pair of morphisms) along any morphism exists and is again a regular epimorphism---see~\cite[Chapter~2]{Borceux:HandbookVol2} for more details.

Finally, it is worth highlighting a useful result about the EM-categories of monads on $\Set$ that follows as a straightforward corollary to Propositions~\ref{prop:limitsinEMcategory} and~\ref{prop:EMcategoryiscompletecocompleteandregular}. We use this result in Section~\ref{sect:fibadjmodelsfromfamiliesofsets} to construct examples of models of eMLTT.

\begin{proposition}
\label{prop:EMcategoryonSetiscompleteandcocomplete}
For any monad $\mathbf{T}$ on $\Set$, $\Set^{\mathbf{T}}$ is both complete and cocomplete.
\end{proposition}

\begin{proof}
Completeness of $\Set^{\mathbf{T}}$ follows directly from Proposition~\ref{prop:limitsinEMcategory} because it is well-known that $\Set$ is complete. Cocompleteness of $\Set^{\mathbf{T}}$ then follows from Proposition~\ref{prop:EMcategoryiscompletecocompleteandregular} because it is well-known that $\Set$ is also cocomplete, and that by the Axiom of Choice, every epimorphism (i.e., surjective function) in $\Set$ has a section. Further, it is also well-know that $\Set$ is a regular category (e.g., see~\cite[Example~2.4.2]{Borceux:HandbookVol2}).
\end{proof}

\subsection{Algebraic treatment of computational effects}
\label{sect:algebraictreatmentofeffects}

While Moggi's seminal work shows that strong monads provide a uniform means to model basic combinators on effectful computations (returning values and sequential composition), it leaves two important questions unanswered: 
\begin{itemize}
\item Given any programming language with its computational effects, which monad should we use to model this language?
\item Given monads for two or more computational effects, how should we combine them into a monad for the combination of these effects?
\end{itemize}
In particular, recall that the monads Moggi considered were defined on a case-by-case basis for specific computational effects. Further, while his subsequent work with Cenciarelli~\cite{Cenciarelli:Modularity} on monad transformers (pointed endofunctors on the category of strong monads) provides some means of modularity for combining monads, these transformers are defined on a similar case-by-case basis for specific computational effects.

\index{algebraic effect}
An elegant answer to both questions is provided by the algebraic treatment of computational effects, as originally proposed and developed by Plotkin and Power~\cite{Plotkin:SemanticsForAlgOperations,Plotkin:NotionsOfComputation,Plotkin:AlgOperations}. In this  approach, one represents computational effects algebraically using a set of operation symbols (representing the sources of effects) and a set of equations (describing their computational properties). Consequently, computational effects that fit this approach are commonly called \emph{algebraic}. 
A good source of examples of algebraic effects is~\cite{Plotkin:HandlingEffects}. 
We also discuss examples of common algebraic effects in Section~\ref{sect:fibeffecttheories}.

Plotkin and Power's key insight was that computational effects themselves, when represented using operations and equations, canonically determine the monads and adjunctions that one can use to model effectful languages. 
For example, the global state monad is determined by operations for reading from and writing to the store, together with a set of  equations describing the natural computational behaviour of reading and writing, as given in~\cite{Plotkin:NotionsOfComputation}. 
In the same way one can recover most of Moggi's monads, with the notable exception of the continuation monad that is not algebraic~\cite{Hyland:Continuations}.

Focussing on operations and equations also enables computational effects and the corresponding monads to be composed modularly. For example, Hyland et al.~\cite{Hyland:CombiningEffects} explain various commonly used monad transformers in terms of two canonical constructions on equational theories, the \emph{sum} and \emph{tensor product} of equational theories. For both constructions, the set of operation symbols of the resulting theory is given by the union of the sets of operation symbols of the given theories, and the set of equations of the resulting theory includes the union of the sets of equations of the given theories. For the tensor product, the set of equations of the resulting theory additionally includes equations ensuring that operations from different given theories commute with each other. The algebraic treatment of computational effects has also resulted in a general account of effect handlers~\cite{Plotkin:HandlingEffects}, and has been successfully applied to effect-dependent optimisations~\cite{Kammar:AlgebraicFoundations}, operational semantics~\cite{Plotkin:TensorsOfModels,Saleh:OpSemantics}, and a logic of effects~\cite{Plotkin:Logic}.

In more detail, Plotkin and Power modelled algebraic computational effects using equationally presented  Lawvere theories, making use of the wealth of existing theory and the particularly good properties of Lawvere theories and their models. Informally, a Lawvere theory is an abstract, category-theoretic  description of the clone of equational theories.
Plotkin and Power's original work has since been extended to account for more general notions of algebraic effects. For example: i) countable Lawvere theories~\cite{Power:CountableTheories} enable one to model natural number valued global state, ii) enriched Lawvere theories~\cite{Power:EnrichedLawvereTheories,Hyland:CombiningEffects,Hyland:DiscreteLawTh} enable one to model partiality and recursion, and iii) indexed Lawvere theories~\cite{Power:IndexedLawvereTheories} and parameterised algebraic theories~\cite{Staton:Instances} enable one to model local computational effects, such as the allocation of fresh names and references. 

In this thesis, we use \emph{countable Lawvere theories} for modelling algebraic effects. 
We recall their definition and some key properties below---see~\cite{Power:CountableTheories} for more details.
In particular, on the one hand, countable Lawvere theories are general enough to be used to model the corresponding extensions of eMLTT we discuss in Chapters~\ref{chap:fibalgeffects} and~\ref{chap:handlers}. On the other hand, they are sufficiently  concrete to be accessible to a wide audience.

\begin{definition}
\label{def:countablelawveretheory}
\index{Lawvere theory!countable --}
\index{ I@$I$ (countable Lawvere theory)}
\index{ L@$\mathcal{L}$ (countable Lawvere theory)}
\index{category!skeleton of the -- of countable sets}
\index{ ab@$\aleph_{\hspace{-0.05cm}1}$ (skeleton of the category of countable sets)}
A \emph{countable Lawvere theory} is given by a small category $\mathcal{L}$ with countable products and a strict countable-product preserving identity-on-objects functor  $I : \aleph_{\!\!1}^{\text{op}} \longrightarrow \mathcal{L}$, 
where $\aleph_{\!\!1}$ is the skeleton of the category of countable sets and all functions between them (countable coproducts in $\aleph_{\!\!1}$ are given by cardinal sum).
\end{definition}

It is worthwhile to note that the objects of $\mathcal{L}$ are exactly those of $\aleph_{\!\!1}^{\text{op}}$, or equivalently, those of $\aleph_{\!\!1}$. In more concrete terms, every object of $\mathcal{L}$ is either a natural number or the distinguished symbol $\omega$ denoting the cardinality of countable sets. 

In the rest of this thesis we let $n,m,\ldots$ to range over the objects of $\aleph_{\!\!1}$; we ensure that it is clear from the context whether $n$ denotes a natural number or the symbol $\omega$.

\begin{definition}
\index{morphism!-- of countable Lawvere theories}
A \emph{morphism} of countable Lawvere theories, from $I_1 : \aleph_{\!\!1}^{\text{op}} \longrightarrow \mathcal{L}_1$ to \linebreak $I_2 : \aleph_{\!\!1}^{\text{op}} \longrightarrow \mathcal{L}_2$, is given by a strict countable-product preserving functor from $\mathcal{L}_1$ to $\mathcal{L}_2$ that commutes with the functors $I_1$ and $I_2$. This gives us the category $\Law$.
\index{ Law@$\Law$ (category of countable Lawvere theories)}
\end{definition}

\begin{definition}
\index{Lawvere theory!countable --!model of a --}
A \emph{model} of a countable Lawvere theory $I : \aleph_{\!\!1}^{\text{op}} \longrightarrow \mathcal{L}$ in a category $\mathcal{V}$ with countable products is a countable-product preserving functor $\mathcal{M} : \mathcal{L} \longrightarrow \mathcal{V}$.
\index{ M@$\mathcal{M}$ (model of a countable Lawvere theory)}
\end{definition}

\begin{definition}
\index{morphism!-- of models of a countable Lawvere theory}
\index{ Mod@$\Mod(\hspace{-0.05cm}\mathcal{L},\mathcal{V})$ (category of models  in $\mathcal{V}$ of a countable Lawvere theory $\mathcal{L}$)}
A \emph{morphism} of models $\mathcal{M}_1 : \mathcal{L} \longrightarrow \mathcal{V}$ and $\mathcal{M}_2 : \mathcal{L} \longrightarrow \mathcal{V}$ of a countable Lawvere theory $I : \aleph_{\!\!1}^{\text{op}} \longrightarrow \mathcal{L}$ in a category $\mathcal{V}$ with countable products is given by a natural transformation from $\mathcal{M}_1$ to $\mathcal{M}_2$. This gives us a category $\Mod(\!\mathcal{L},\mathcal{V})$.
\end{definition}

The category $\Mod(\!\mathcal{L},\mathcal{V})$ comes equipped with a canonical forgetful functor to $\mathcal{V}$, whose left adjoint, if it exists, enables us to derive the corresponding monad on $\mathcal{V}$.

\begin{definition}
The canonical \emph{forgetful functor} $U_{\!\mathcal{L}} : \Mod(\!\mathcal{L},\mathcal{V}) \longrightarrow \mathcal{V}$ is given by 
\[
U_{\!\mathcal{L}}(\mathcal{M}) \defeq \mathcal{M}(1)
\qquad
U_{\!\mathcal{L}}(\alpha) \defeq \alpha_1
\]
\end{definition}

\begin{proposition}
If the forgetful functor $U_{\!\mathcal{L}}$ has a left adjoint $F_{\mathcal{L}}$, then it exhibits $\Mod(\!\mathcal{L},\mathcal{V})$ as equivalent to the EM-category for the monad given by  $T_{\mathcal{L}} \defeq U_{\!\mathcal{L}} \comp F_{\mathcal{L}}$.
\index{ T@$(T_{\mathcal{L}},\eta_{\mathcal{L}},\mu_{\mathcal{L}})$ (monad derived from a countable Lawvere theory $\mathcal{L}$)}
\index{ F@$F_{\hspace{-0.05cm}\mathcal{L}} \dashv\, U_{\hspace{-0.05cm}\mathcal{L}}$ (canonical adjunction for models of a countable Lawvere theory $\mathcal{L}$)}
\end{proposition}

\begin{proposition}
\label{prop:leftadjointexistswhenlocallycountablypresentable}
The left adjoint $F_{\mathcal{L}}$ exists when $\mathcal{V}$ is locally countably presentable.
\end{proposition}

\index{category!locally countably presentable --}
\index{object!countably presentable --}
\index{colimit!countably directed --}
We recall that $\mathcal{V}$ is \emph{locally countably presentable} if it is cocomplete and there is a set $\mathcal{A}$ of countably presentable objects such that every object in $\mathcal{V}$ is a  countably directed colimit of objects in $\mathcal{A}$. We also recall that $A$ is a \emph{countably presentable object} if its hom-functor $\mathcal{V}(A,-) : \mathcal{V} \longrightarrow \Set$ preserves countably directed colimits. Finally, we recall that a \emph{countably directed colimit} is the colimit of a diagram $J : \mathcal{D} \longrightarrow \mathcal{V}$ whose shape $\mathcal{D}$ is given by a partial order in which every countable subset has an upper bound.
See~\cite[Chapter~1]{Adamek:LocallyPresentableCats} for more details about locally presentable categories. 

In particular, we recall from op.~cit. that $\Set$ is locally countably presentable. 

\begin{corollary}
Given any countable Lawvere theory $I : \aleph_{\!\!1}^{\text{op}} \longrightarrow \mathcal{L}$, there exists an adjunction $F_{\mathcal{L}} \dashv\, U_{\!\mathcal{L}} : \Mod(\!\mathcal{L},\Set) \longrightarrow \Set$.
\end{corollary}

We later use these adjunctions as a basis for giving a denotational semantics to the extensions of eMLTT with fibred algebraic effects and their handlers, see Chapters~\ref{chap:fibalgeffects} and~\ref{chap:handlers} for details. In order to show that these models of eMLTT support the computational $\Sigma$- and $\Pi$-types, we recall another important property of $\Mod(\!\mathcal{L},\Set)$ from~\cite{Power:CountableTheories}.

\begin{proposition}
\label{prop:modelsoflawveretheoriesinsetcocomplete}
For any countable Lawvere theory $I : \aleph_{\!\!1}^{\text{op}} \longrightarrow \mathcal{L}$, the category $\Mod(\!\mathcal{L},\Set)$ is both complete and cocomplete.
\end{proposition}

Next, we show how to specify countable Lawvere theories using operation symbols and equations, the key idea underlying the algebraic treatment of computational effects, as discussed earlier. We follow Plotkin and Pretnar~\cite{Plotkin:HandlingEffects} in using countable equational theories~\cite{Gratzer:UniversalAlgebra} for freely generating countable Lawvere theories. Equivalently, and more abstractly, one could also specify countable Lawvere theories using single-sorted countable-product sketches~\cite{Barr:CatThForCS}, e.g., as discussed by Power in~\cite{Power:CountableTheories}.
We choose to use countable equational theories instead of the corresponding sketches with the aim of making our work more accessible to a functional programming audience.

\begin{definition}
\index{signature!countable --}
\index{ S@$\mathbb{S}$ (countable signature)}
A \emph{countable signature} is given by set $\mathbb{S}$ of operation symbols $\mathsf{op}$, 
\index{ op@$\mathsf{op}$ (operation symbol in a countable signature)}
and an assignment $\mathsf{op} : n$ of an arity to each $\mathsf{op}$ in $\mathbb{S}$, where $n$ is an object of $\aleph_{\!\!1}$.
\end{definition}

\begin{definition}
\index{ t@$t,u,\ldots$ (terms derived from a countable signature)}
\index{term!-- derived from a countable signature}
\index{ x@$x,y,\ldots$ (variables of terms derived from a countable signature)}
\index{ @$\leq$ (less-than-equal order on natural numbers)}
Given a countably infinite set of variables ranged over by $x,y,\ldots$, 
the set of \emph{terms} $t,u,\ldots$ 
derivable from a countable signature $\mathbb{S}$ is given by the grammar
\[
\begin{array}{r c l @{\qquad\qquad}l}
t & ::= & x \,\,\,\vertbar\,\,\, \mathsf{op}(t_i)_{1 \leq i \leq n}
\end{array}
\] 
where $n$ is the arity of the operation symbol $\mathsf{op}$, given either by a natural number or the distinguished symbol $\omega$\footnote{In this thesis, we use the convention that if the arity $n$ of $\sigalgop$ is the distinguished symbol $\omega$, then the notation $1 \leq i \leq n$ stands for $i \in \mathbb{N}$, where $\mathbb{N}$ is the set of natural numbers.}. As a convention, if $n = 1$, we write $\mathsf{op}(t)$ for $\mathsf{op}(t_i)_{1 \leq i \leq 1}$.
\end{definition}

We can then define substitution by straightforward structural recursion.

\begin{definition}
\label{def:eqtheorysimultaneoussubstitution}
\index{ t@$t[u/x]$ (substitution of $u$ for $x$ in $t$)}
Given terms $t$, $u_1$, \ldots, $u_n$, and variables $x_1$, \ldots, $x_n$, then the \emph{simultaneous substitution} of $u_1$, \ldots, $u_n$ for $x_1$, \ldots, $x_n$ in $t$, written $t[u_1/x_1, \ldots, u_n/x_n]$, or $t[\overrightarrow{u_i}/\overrightarrow{x_i}]$ for short, is defined by recursion on the structure of $t$ as follows:
\[
\begin{array}{l c l}
x_i[\overrightarrow{u_i}/\overrightarrow{x_i}] & \defeq & u_i
\\
y[\overrightarrow{u_i}/\overrightarrow{x_i}] & \defeq & y \qquad\qquad\qquad\qquad\qquad\qquad (\text{if~} y \not\in \{x_1, \ldots, x_n\})
\\
\mathsf{op}(t_j)_{1 \leq j \leq m}[\overrightarrow{u_i}/\overrightarrow{x_i}] & \defeq & \mathsf{op}(t_j[\overrightarrow{u_i}/\overrightarrow{x_i}])_{1 \leq j \leq m}
\end{array}
\]
\end{definition}

\begin{definition}
A \emph{context} $\Delta$ is a countable list of distinct variables.
\index{ D@$\Delta$ (context of variables for well-formed terms derived from a countable signature)}
\index{context!-- of variables for well-formed terms derived from a countable signature}
\end{definition}

\begin{definition}
We say that a term $t$ derived from $\mathbb{S}$ is \emph{well-formed} in context $\Delta$ when there exists a derivation for the judgement $\lj \Delta t$ using the following rules:
\index{term!-- derived from a countable signature!well-formed --}
\[
\mkrule
{\lj \Delta x}
{x \in \Delta}
\qquad
\mkrulelabel
{\lj \Delta {\mathsf{op}(t_i)_{1 \leq i \leq n}}}
{\lj \Delta t_i \qquad (1 \leq i \leq n)}
{(\mathsf{op} : n \in \mathbb{S})}
\]
\end{definition}

It is then straightforward to show that these derivations are closed under the standard rules of weakening, exchange of variables, and substitution.

\begin{proposition}
\mbox{}
\begin{itemize}
\item Given a term $t$ and a variable $x$ such that  $\lj {\Delta} t$ and $x \not\in \Delta$, then we have $\lj {\Delta, x} t$.
\item Given a term $t$ such that $\lj {\Delta, x, y, \Delta'} t$, then we have $\lj {\Delta, y, x, \Delta'} t$.
\item Given a term $t$ and terms $u_1, \ldots, u_n$ such that $\lj {\overrightarrow{x_i}} t$ and $\lj {\Delta} {u_i}$ (for all $x_i$), then we have $\lj {\Delta} {t[\overrightarrow{u_i}/\overrightarrow{x_i}]}$.
\end{itemize}
\end{proposition}

\begin{proof}
All three cases are proved by induction on the given derivation.
\end{proof}

\begin{definition}
\label{def:countableequationaltheory}
\index{theory!countable equational --}
\index{ T@$\mathbb{T}$ (countable equational theory)}
A \emph{countable equational theory} $\mathbb{T}$ is given by a countable signature $\mathbb{S}$ and a set $\mathbb{E}$ 
\index{ E@$\mathbb{E}$ (set of equations of a countable equational theory)}
of equations $\ljeq \Delta t u$ between well-formed terms $\lj \Delta t$ and $\lj \Delta u$, closed under the rules of reflexivity, symmetry, transitivity, replacement, and substitution:
\[
\begin{array}{c}
\mkrule
{\ljeq \Delta t t}
{\lj \Delta t}
\qquad
\mkrule
{\ljeq \Delta t u}
{\ljeq \Delta u t}
\qquad
\mkrule
{\ljeq \Delta {t_1} {t_3}}
{\ljeq \Delta {t_1} {t_2} \quad \ljeq \Delta {t_2} {t_3}}
\\[4mm]
\mkrule
{\ljeq \Delta {t[\overrightarrow{u_i}/\overrightarrow{x_i}]} {t[\overrightarrow{u'_i}/\overrightarrow{x_i}]}}
{\lj {\overrightarrow{x_i}} t \quad \ljeq \Delta {u_i} {u'_i} \quad (\text{for all } x_i)}
\qquad
\mkrule
{\ljeq \Delta {t[\overrightarrow{u_i}/\overrightarrow{x_i}]} {t'[\overrightarrow{u_i}/\overrightarrow{x_i}]}}
{\ljeq {\overrightarrow{x_i}} t t' \quad \lj \Delta {u_i} \quad (\text{for all } x_i)}
\end{array}
\]
\end{definition}

Next, we show how to construct a countable Lawvere theory from a given countable equational theory, 
based on the intuition that a countable Lawvere theory $I : \aleph_{\!\!1}^{\text{op}} \longrightarrow \mathcal{L}$ is an abstract, category-theoretic description of a clone of  countable equational theories. 
In particular, one should think of morphisms $n \longrightarrow 1$ in $\mathcal{L}$ as terms in $n$ variables. 

\begin{definition}
\label{def:lawveretheoryfromequationaltheory}
Given a countable equational theory $\mathbb{T} = (\mathbb{S},\mathbb{E})$, we define a category $\mathcal{L}_{\mathbb{T}}$, 
\index{ L@$\mathcal{L}_{\mathbb{T}}$ (countable Lawvere theory derived from a countable equational theory $\mathbb{T}$)}
whose 
\begin{itemize}
\item objects $n$ are those of $\aleph_{\!\!1}$ (i.e., $n$ is either a natural number or the distinguished symbol $\omega$ denoting the cardinality of countable sets);  
\item morphisms $n \longrightarrow m$ are given by $m$-tuples $(\lj {\overrightarrow{x_i}} {t_{\!j}})_{1 \leq j \leq m}$ of equivalence classes of terms in $n$ variables (for convenience, we refer to these equivalence classes via their representatives, i.e., we write $\lj {\overrightarrow{x_i}} {t_{\!j}}$ for the equivalence class $[\lj {\overrightarrow{x_i}} {t_{\!j}}]$);
\item identity morphisms are given by tuples of variables, i.e., given an object $n$ in $\mathcal{L}_{\mathbb{T}}$, the identity morphism $\id_n : n \longrightarrow n$ is given by the tuple $(\lj {\overrightarrow{x_i}} {x_{\!j}})_{1 \leq j \leq n}$; and
\item composition of morphisms is given by substitution, i.e., given two morphisms $n_1 \longrightarrow n_2$ and $n_2 \longrightarrow n_3$, given by the tuples $(\lj {\overrightarrow{x_i}} {t_{\!j}})_{1 \leq j \leq n_2}$ and $(\lj {\overrightarrow{y_{\!j}}} {u_k})_{1 \leq k \leq n_3}$, then their composite is given by the tuple $(\lj {\overrightarrow{x_i}} {u_k[\overrightarrow{t_{\!j}}/\overrightarrow{y_{\!j}}]})_{1 \leq k \leq n_3}$.
\end{itemize}
\end{definition}

\begin{proposition}
\label{prop:clonehascountableproducts}
The category $\mathcal{L}_{\mathbb{T}}$ has countable products.
\end{proposition}

\begin{proof}
Given a countable set $\mathbb{I}$ and objects $n_i$ in $\mathcal{L}_{\mathbb{T}}$, for all $i$ in $\mathbb{I}$, their countable product $\bigsqcap_{i \in \mathbb{I}} n_i$ is given by their cardinal sum $+_{\!i \in \mathbb{I}}\, n_i$. 
\index{ @$+_{i \in \mathbb{I}}$ (cardinal sum)}
By the Axiom of (Countable) Choice, the cardinal sum of countably many objects $\omega$ is again $\omega$, and therefore an object of $\mathcal{L}_{\mathbb{T}}$. 
\index{ projb@$\mathsf{proj}_{j}$ ($j$'th projection for countable products)}
The $j$'th projection $\mathsf{proj}_{j} : \bigsqcap_{i \in \mathbb{I}} n_i \longrightarrow n_{\!j}$ is given by the following $n_{\!j}$-tuple of variables:
\[
(\lj {\overrightarrow{x_l}} {x_{+_{\!(1 \,\leq\, l \,\leq\, j \,-\, 1)}\,\, n_l \,+\, k}})_{1 \,\leq\, k \,\leq\, n_{\!j}}
\]
Given morphisms $m \longrightarrow n_i$, represented by tuples of terms 
$(\lj {\overrightarrow{y_k}} {t_{i,j}})_{1 \,\leq\, j \,\leq\, n_i}$, 
for all $i$ in $\mathbb{I}$, the unique mediating morphism $m \longrightarrow \bigsqcap_{i \in \mathbb{I}} n_i$ is given by the $+_{\!i \in \mathbb{I}}\, n_i$-tuple of terms
\[
(\lj {\overrightarrow{y_l}} {t_{f(k)}})_{1 \,\leq\, k \,\leq\, +_{\!i \in \mathbb{I}}\, n_i}
\]
where the auxiliary function $f$ is given by
\[
f \defeq j \mapsto (i , j - +_{\!(1 \,\leq\, k \,\leq\, i \,-\, 1)}\,\, n_k) 
\qquad
\quad
(\text{when } +_{\!(1 \,\leq\, k \,\leq\, i \,-\, 1)}\, n_k < j \leq +_{\!(1 \,\leq\, k \,\leq\, i)}\,\, n_k)
\]
The proof that these definitions indeed equip $\mathcal{L}_{\mathbb{T}}$ with countable products is straightforward. It involves unfolding the definition of composition of morphisms in $\mathcal{L}_{\mathbb{T}}$ and then using standard properties of substitution. We omit the details of this proof.
\end{proof}

\begin{proposition}
\label{prop:lawveretheoryfromequationaltheory}
\index{ I@$I_{\mathbb{T}}$ (countable Lawvere theory derived from a countable equational theory $\mathbb{T}$)}
The functor $I_{\mathbb{T}} : \aleph_{\!\!1}^{\text{op}} \longrightarrow \mathcal{L}_{\mathbb{T}}$, given by
\[
I_{\mathbb{T}}(n) \defeq n
\qquad
I_{\mathbb{T}}(f) \defeq (\lj {\overrightarrow{x_i}} {x_{f(j)}})_{1 \,\leq\, j \,\leq\, m} : n \longrightarrow m \qquad (\text{where } f : n \longrightarrow m \in \aleph_{\!\!1}^{\text{op}})
\]
is a countable Lawvere theory.
\end{proposition}

\begin{proof}
First, we prove that $I_{\mathbb{T}}$ is indeed a functor, by proving the following equations:
\begin{fleqn}[0.75cm]
\begin{align*}
& I_{\mathbb{T}}(\id_n) \qquad\qquad\qquad\qquad\qquad
&
& I_{\mathbb{T}}(g \comp f)
\\
=\,\, & (\lj {\overrightarrow{x_i}} {x_{\id_n(j)}})_{1 \,\leq\, j \,\leq\, n} 
&
=\,\, & (\lj {\overrightarrow{x_i}} {x_{f(g(k))}})_{1 \,\leq\, k \,\leq\, n_3} 
\\
=\,\, & (\lj {\overrightarrow{x_i}} {x_{\!j}})_{1 \,\leq\, j \,\leq\, n}
&
=\,\, & (\lj {\overrightarrow{x_i}} {y_{g(k)}[\overrightarrow{x_{f(j)}}/\overrightarrow{y_{\!j}}]})_{1 \,\leq\, k \,\leq\, n_3} 
\\
=\,\, & \id_{I_{\mathbb{T}}(n)}
&
=\,\, & (\lj {\overrightarrow{y_{\!j}}} {y_{g(k)}})_{1 \,\leq\, k \,\leq\, n_3} \comp (\lj {\overrightarrow{x_i}} {x_{f(j)}})_{1 \,\leq\, j \,\leq\, n_2}
\\
&&
=\,\, & I_{\mathbb{T}}(g) \comp I_{\mathbb{T}}(f)
\end{align*}
\end{fleqn}
where $f : n_1 \longrightarrow n_2$ and $g : n_2 \longrightarrow n_3$ are morphisms in $\aleph_{\!\!1}^{\text{op}}$.

We also need to show that $I_{\mathbb{T}}$ strictly preserves countable products in $\aleph_{\!\!1}^{\text{op}}$. Specifically, we need to prove that the following three equations hold:
\index{ Product@$\bigsqcap_{i \in \mathbb{I}}$ (countable product)}
\[
I_{\mathbb{T}}(\bigsqcap_{i \in \mathbb{I}} n_i) = \bigsqcap_{i \in \mathbb{I}} (I_{\mathbb{T}}(n_i))
\qquad
I_{\mathbb{T}}(\mathsf{proj}_j) = \mathsf{proj}_j
\qquad
I_{\mathbb{T}}(\langle f_i \rangle_{i \in \mathbb{I}}) = \langle I_{\mathbb{T}}(f_i) \rangle_{i \in \mathbb{I}}
\]

To this end, we recall from~\cite{Power:CountableTheories} that the countable products in $\aleph_{\!\!1}^{\text{op}}$ are given by countable coproducts in $\aleph_{\!\!1}$, which are in turn given by the cardinal sum of objects in $\aleph_{\!\!1}$.
In particular, given a countable set $\mathbb{I}$ and objects $n_i$ in $\aleph_{\!\!1}^{\text{op}}$ (for all $i$ in $\mathbb{I}$), their countable product $\bigsqcap_{i \in \mathbb{I}} n_i$ is given by $+_{\!i \in \mathbb{I}}\,\, n_i$. Consequently, we can show that 
\[
\begin{array}{c}
I_{\mathbb{T}}(\bigsqcap_{i \in \mathbb{I}} n_i) = \bigsqcap_{i \in \mathbb{I}} n_i = +_{\!i \in \mathbb{I}}\,\, n_i = \bigsqcap_{i \in \mathbb{I}} n_i = \bigsqcap_{i \in \mathbb{I}} (I_{\mathbb{T}}(n_i))
\end{array}
\]

Next, we recall that the $j$'th projection morphism $\mathsf{proj}_j : \bigsqcap_{i \in \mathbb{I}} n_i \longrightarrow n_{\!j}$ in $\aleph_{\!\!1}^{\text{op}}$ is given by the corresponding $j$'th injection function in $\aleph_{\!\!1}$, namely, by a function $n_{\!j} \longrightarrow +_{\!i \in \mathbb{I}}\,\, n_i$ given by $k \mapsto +_{\!(1 \,\leq\, l \,\leq\, j \,-\, 1)}\,\, n_l \,+\, k$. Consequently, we can show that
\[
\begin{array}{c}
I_{\mathbb{T}}(\mathsf{proj}_j) = (\lj {\overrightarrow{x_l}} {x_{+_{\!(1 \,\leq\, l \,\leq\, j \,-\, 1)}\,\, n_l \,+\, k}})_{1 \,\leq\, k \,\leq\, n_{\!j}} = \mathsf{proj}_j : \bigsqcap_{i \in \mathbb{I}} (I_{\mathbb{T}}(n_i))\longrightarrow I_{\mathbb{T}}(n_{\!j})
\end{array}
\]

Finally, we recall that given a countable set $\mathbb{I}$ and morphisms $f_i : m \longrightarrow n_i$ in $\aleph_{\!\!1}^{\text{op}}$ (for all $i$ in $\mathbb{I}$), the unique mediating morphism $\langle f_i \rangle_{i \in \mathbb{I}} : m \longrightarrow \bigsqcap_{i \in \mathbb{I}} n_i$ 
\index{ f@$\langle f_i \rangle_{i \in \mathbb{I}}$ (unique mediating morphism for countable products)}
in $\aleph_{\!\!1}^{\text{op}}$ is given by the corresponding unique mediating morphism for countable coproducts in $\aleph_{\!\!1}$, namely, by a function $[f_i]_{i \in \mathbb{I}} : +_{\!i \in \mathbb{I}}\,\, n_i \longrightarrow m$ defined as
\index{ f@$[f_i]_{i \in \mathbb{I}}$ (unique mediating morphism for countable coproducts)} 
\[
[f_i]_{i \in \mathbb{I}} \defeq 
j \mapsto f_{\!i}(j - +_{\!(1 \,\leq\, k \,\leq\, i \,-\, 1)}\,\, n_k)
\qquad
\quad\!\!\!
(\text{when } +_{\!(1 \,\leq\, k \,\leq\, i-1)}\, n_k < j \leq +_{\!(1 \,\leq\, k \,\leq\, i)}\,\, n_k)
\]
Now, if we write $t_{i,f_i(k)}$ for ${y_{f_i(k)}}$, and unfold the definitions of $I_{\mathbb{T}}(f_i)$ and $\langle I_{\mathbb{T}}(f_i) \rangle_{i \in \mathbb{I}}$, we see that showing 
$
I_{\mathbb{T}}(\langle f_i \rangle_{i \in \mathbb{I}}) = \langle I_{\mathbb{T}}(f_i) \rangle_{i \in \mathbb{I}}$
amounts to proving the following equation:
\[
(\lj {\overrightarrow{y_l}} {y_{[ f_i ]_{i \in \mathbb{I}}(k)}})_{1 \,\leq\, k \,\leq\, +_{\!i \in \mathbb{I}}\,\, n_i} = (\lj {\overrightarrow{y_l}} {t_{f(k)}})_{1 \,\leq\, k \,\leq\, +_{\!i \in \mathbb{I}}\, n_i}
\]
where the auxiliary function $f$ is defined as in the proof of Proposition~\ref{prop:clonehascountableproducts}. We prove this equation by showing that for every $k$, the $k$'th terms in the two tuples are equal. 
As $1 \leq k \leq +_{\!i \in \mathbb{I}}\, n_i$, there must be a $i$ such that $+_{\!(1 \,\leq\, l \,\leq\, i-1)}\, n_l < k \leq +_{\!(1 \,\leq\, l \,\leq\, i)}\,\, n_l$. Based on this observation, we can show that the $k$'th terms in these tuples are equal:
\[
{y_{[ f_i ]_{i \in \mathbb{I}}(k)}} 
= 
{y_{\!f_{\!i}(k \,-\, +_{\!(1 \,\leq\, l \,\leq\, i \,-\,1)}\,\, n_l)}} 
= 
t_{i, f_{\!i}(k \,-\, +_{\!(1 \,\leq\, l \,\leq\, i \,-\, 1)}\,\, n_l)} 
= 
t_{f(k)}
\]
\end{proof}

We conclude this section by formally presenting the equational theory of global state 
\index{theory!countable equational --!-- of global state}
which we used as an informal example of algebraic effects towards the beginning of this section. In particular, given a countable set $S$ of store values, the countable equational theory of global state is given by an $\vertbar S \vertbar\!$-ary\footnote{As standard in the literature, we write $\vertbar X \vertbar$ for the \emph{cardinality} of a given set $X$.} operation symbol $\mathsf{get} : \!\vertbar S \vertbar$ and an $\vertbar S \vertbar\!$-indexed family of unary operation symbols $\mathsf{put}_s : 1$, and the following equations:
\[
\begin{array}{c}
\ljeq {x} {\mathsf{get}(\mathsf{put}_s(x))_{1 \,\leq\, s \,\leq\, \vertbar S \vertbar}} {x}
\\[2mm]
\ljeq {\overrightarrow{x_s}} {\mathsf{put}_{s'}(\mathsf{get}(x_s)_{1 \,\leq\, s \,\leq\, \vertbar S \vertbar})} {\mathsf{put}_{s'}(x_{s'})}
\\[2mm]
\ljeq {x} {\mathsf{put}_{s}(\mathsf{put}_{s'}(x))} {\mathsf{put}_{s'}(x)}
\end{array}
\]
closed under the rules of reflexivity, symmetry, transitivity, replacement, and substitution. 
The monad one obtains from the corresponding countable Lawvere theory is the standard one for global state, given on objects by $T_{\text{GS}}(A) \defeq \vert S \vert \Rightarrow (A \times \vert S \vert)$, see~\cite{Plotkin:NotionsOfComputation}. 

\section{Fibred category theory}
\label{sect:fibrationsbasics}

In this section we recall some basic definitions and results from fibred category theory which we use through Chapters~\ref{chap:fibadjmodels}--\ref{chap:handlers}
for giving a denotational semantics to eMLTT and its extensions. A much more detailed overview of fibred category theory, including its use in modelling various type theories and logics, can be found in~\cite{Jacobs:Book}. While the results we present in this section are well-known (see~op.~cit.), we spell out some of the proofs to introduce the reader to the style of proofs used in fibred category theory.

We have chosen to work with fibred category theory because it provides a natural framework for developing denotational semantics of dependently typed languages. In particular, i) functors model type-dependency; ii) split fibrations model substitution; and iii) the notion of comprehension models context extension. However, it is worth noting that the ideas we develop in this thesis also apply to other models of dependent types, such as categories with families, categories with attributes, and contextual categories\footnote{In the field of homotopy type theory, the latter are also known under the name of C-systems~\cite{Voevodsky:CSystems}.}. We suggest~\cite{Hofmann:SyntaxAndSemantics,Pitts:CategoricalLogic} for an overview of these models of dependent types.

We begin our overview of fibred category theory with some common terminology.

\begin{definition}
Given a functor ${p : \mathcal{V} \longrightarrow \mathcal{B}}$, we say that an object $A$ in $\mathcal{V}$ is \emph{over} an object $X$ in $\mathcal{B}$ when $p(A) = X$. Analogously, we say that a morphism $f : A \longrightarrow B$ in $\mathcal{V}$ is \emph{over} a morphism $g : X \longrightarrow Y$ in $\mathcal{B}$ when $p(A) = X$, $p(B) = Y$, and $p(f) = g$.
\end{definition}

\begin{definition}
\index{morphism!vertical --}
Given a functor ${p : \mathcal{V} \longrightarrow \mathcal{B}}$, we say that a morphism $f : A \longrightarrow B$ in $\mathcal{V}$ is \emph{vertical} when $p(A) = p(B) = X$ and $p(f) = \id_{X}$. 
\end{definition}

\begin{definition}
\index{category!fibre --}
Given a functor ${p : \mathcal{V} \longrightarrow \mathcal{B}}$ and an object $X$ in $\mathcal{B}$, we write $\mathcal{V}_X$ for the \emph{fibre (category)} over $X$, 
\index{ V@$\mathcal{V}_X$ (fibre (category) over $X$)}
i.e., for the subcategory of $\mathcal{V}$ consisting of objects over $X$ and vertical morphisms over $\id_X$.
\end{definition}

Next, we define two important concepts in fibred category theory: Cartesian morphisms and fibrations. We also recall some basic but useful facts about these concepts.

\begin{definition}
\index{morphism!Cartesian --}
Given a functor ${p : \mathcal{V} \longrightarrow \mathcal{B}}$, a morphism ${f : A \longrightarrow B}$ in $\mathcal{V}$ is said to be \emph{Cartesian} over a morphism ${g : X \longrightarrow Y}$ in $\mathcal{B}$ if ${p(f) = g}$, and if for all ${i : C \longrightarrow B}$ in $\mathcal{V}$ and ${j : p(C) \longrightarrow X}$ in $\mathcal{B}$ such that ${p(i) = g \comp j}$, there exists a unique mediating morphism $h : C \longrightarrow A$ over $j$ such that $f \,\comp\, h = i$, as illustrated in the following  diagram:
\[
\xymatrix@C=3em@R=3em@M=0.5em{
C \ar@/^2pc/[rr]^{i} \ar@{-->}[r]_{h} & A \ar[r]_{f} & B && \text{in} & \mathcal{V} \ar[d]^{p}
\\
p(C) \ar[r]^{j} \ar@/_2pc/[rr]_{p(i)}  & X \ar[r]^{g = p(f)} & Y && \text{in} & \mathcal{B}
}
\]
\end{definition}

Throughout the rest of this thesis, we often do not mention the morphism $g$ explicitly because it is equal to $p(f)$. In that case, we simply say that $f$ is a Cartesian morphism. In addition, we often omit the lower part of such diagrams and only work with the top part when the morphism $j$ is clear from the surrounding context.

An important property of Cartesian morphisms worth noting is that they are unique up-to a unique isomorphism, as made precise in the next proposition.

\begin{proposition}[{\cite[Exercise~1.1.1 (i)]{Jacobs:Book}}]
\label{prop:cartesianmorphismsareunique}
Given a functor ${p : \mathcal{V} \longrightarrow \mathcal{B}}$, and two Cartesian morphisms $f : A \longrightarrow C$ and $g : B \longrightarrow C$ such that $p(f) = p(g)$, then there is a unique vertical isomorphism $\psi_{f,g} : A \overset{\cong}{\longrightarrow} B$ such that $f = g \comp \psi_{f,g}$.
\index{ psi@$\psi_{f,g}$ (vertical isomorphism witnessing the uniqueness of Cartesian morphisms)}
\end{proposition}

\begin{proof}
To improve readability, we let $X \defeq p(A)$. As a consequence, also $p(B) = X$.

Then, we define $\psi_{f,g} : A \longrightarrow B$ as the unique mediating morphism over $\id_X$ in 
\[
\xymatrix@C=5em@R=3em@M=0.5em{
A \ar@{-->}[r]_{\psi_{f,g}} \ar@/^2pc/[rr]^{f} & B \ar[r]_-{g} & C
}
\]
and $\psi^{-1}_{f,g} : B \longrightarrow A$ as the unique mediating morphism over $\id_X$ in the diagram
\[
\xymatrix@C=5em@R=3em@M=0.5em{
B \ar@{-->}[r]_{\psi^{-1}_{f,g}} \ar@/^2pc/[rr]^{g} & A \ar[r]_-{f} & C
}
\]

Clearly, both $\psi^{-1}_{f,g} \comp \psi_{f,g} : A \longrightarrow A$ and $\psi_{f,g} \comp \psi^{-1}_{f,g} : B \longrightarrow B$ are vertical over \linebreak $\id_X :  X \longrightarrow X$, and they are determined uniquely.
Therefore, it  remains to show that
\[
\psi^{-1}_{f,g} \comp \psi_{f,g} = \id_A
\qquad
\psi_{f,g} \comp \psi^{-1}_{f,g} = \id_B
\]
which we do by using the universal properties of the Cartesian morphisms $f$ and $g$, respectively. In particular, we observe that the following two diagrams commute:
\[
\xymatrix@C=7em@R=2em@M=0.5em{
&&& \ar@{}[d]_{\dcomment{\text{def. of } \psi_{f,g}} \qquad\qquad\qquad\qquad\qquad}
\\
\ar@{}[d]^>>>{\qquad\qquad\qquad\,\,\, \dcomment{\text{composition}}} & B \ar[dr]^-{\psi^{-1}_{f,g}} \ar@/^1.5pc/[rrd]^{g} & & \ar@{}[d]_{\dcomment{\text{def. of } \psi^{-1}_{f,g}} \qquad\qquad\qquad}
\\
A \ar[rr]_{\psi^{-1}_{f,g} \,\comp\, \psi_{f,g}} \ar[ur]^-{\psi_{f,g}} \ar@/^7pc/[rrr]^{f} && A \ar[r]_-{f} & C
}
\]
\[
\xymatrix@C=7em@R=2em@M=0.5em{
&&& \ar@{}[d]_{\dcomment{\text{def. of } \psi^{-1}_{f,g}} \qquad\qquad\qquad\qquad\qquad}
\\
\ar@{}[d]^>>>{\qquad\qquad\qquad\,\,\, \dcomment{\text{composition}}} & A \ar[dr]^-{\psi_{f,g}} \ar@/^1.5pc/[rrd]^{f} & & \ar@{}[d]_{\dcomment{\text{def. of } \psi_{f,g}} \qquad\qquad\qquad}
\\
B \ar[rr]_{\psi_{f,g} \,\comp\, \psi^{-1}_{f,g}} \ar[ur]^-{\psi^{-1}_{f,g}} \ar@/^7pc/[rrr]^{g} && B \ar[r]_-{g} & C
}
\]
from which it follows that the composite morphisms $\psi^{-1}_{f,g} \comp \psi_{f,g}$ and $\psi_{f,g} \comp \psi^{-1}_{f,g}$ are equal to the unique mediating morphisms over $\id_X$ induced by $f$ and $g$, respectively. Namely, the commutativity of these two diagrams shows that these composite morphisms satisfy the same universal properties that uniquely determine these mediating morphisms. However, as the identity morphisms $\id_{A}$ and $\id_{B}$ also satisfy the same universal properties, these composite morphisms are in fact equal to $\id_{A}$ and $\id_{B}$.
\end{proof}

\begin{proposition}[{\cite[Exercise~1.1.4 (ii)]{Jacobs:Book}}]
\label{prop:cartesianmorphismscompose}
The composition of two Cartesian morphisms is itself a Cartesian morphism.
\end{proposition}

\begin{proof}
According to the definition of Cartesian morphisms, given a functor ${p \!:\! \mathcal{V} \!\longrightarrow\! \mathcal{B}}$, two Cartesian morphisms $f : A \longrightarrow B$ and $g : B \longrightarrow C$, a morphism ${i : D \longrightarrow C}$ in $\mathcal{V}$, and a morphism ${j : p(D) \longrightarrow p(A)}$ in $\mathcal{B}$ such that ${p(i) = p(g) \comp p(f) \comp j}$, we need to construct a unique mediating morphism $h : D \longrightarrow A$ over $j$ such that $g \comp f \comp h = i$, as in 
\[
\xymatrix@C=3em@R=3em@M=0.5em{
D \ar@/^2pc/[rrr]^{i} \ar@{-->}[r]_{h} & A \ar[r]_{f} & B \ar[r]_-{g} & C
}
\]

First, we use the universal property of the Cartesian morphism $g : B \longrightarrow C$
to construct a unique mediating morphism $h' : D \longrightarrow B$ over $p(f) \comp j$, 
as illustrated below:
\[
\xymatrix@C=3em@R=3em@M=0.5em{
D \ar@/^2pc/[rrr]^{i} \ar@{-->}[rr]_{h'} & & B \ar[r]_-{g} & C
}
\]

Next, we use the universal property of the Cartesian morphism $f : A \longrightarrow B$
to construct a unique morphism $h : D \longrightarrow A$ over $j$, 
as illustrated below:
\[
\xymatrix@C=3em@R=3em@M=0.5em{
D \ar@/^2pc/[rrr]^{h'} \ar@{-->}[rr]_{h} & & A \ar[r]_-{f} & B
}
\]
After combining $g \comp h' = i$ and $f \comp h = h'$, we see that $h$ also satisfies $g \comp f \comp h = i$.

Finally, we need to show that $h$ is the unique morphism over $j$ satisfying $g \comp f \comp h = i$. This  follows straightforwardly from the definitions of $h'$ and $h$. Namely, given any other morphism $h'' : D \longrightarrow A$ over $j$ such that $g \comp f \comp h'' = i$, we first get $f \comp h'' = h'$ by using the uniqueness of $h'$, and then $h'' = h$ by using the uniqueness of $h$.
\end{proof}

\begin{definition}
\index{fibration}
\index{category!total --}
\index{category!base --}
\index{ p@$p,q,\ldots$ (fibrations)}
A functor ${p : \mathcal{V} \longrightarrow \mathcal{B}}$ is called a \emph{fibration} if for every object $B$ in $\mathcal{V}$ and every morphism $g : X \longrightarrow p(B)$ in $\mathcal{B}$, there exists a morphism $f : A \longrightarrow B$ that is Cartesian over $g$. We refer to $\mathcal{V}$ as the \emph{total} category and to $\mathcal{B}$ as the \emph{base} category.
\end{definition}

\begin{definition}
\index{fibration!cloven --}
\index{ f@$\overline{f}(A)$ (chosen Cartesian morphism in a cloven fibration)}
A fibration ${p : \mathcal{V} \longrightarrow \mathcal{B}}$ is said to be \emph{cloven} if it comes with a choice of Cartesian morphisms.
As standard, we write ${\overline{f}(A) : f^*(A) \longrightarrow A}$ for the \emph{chosen} Cartesian morphism (and $f^*(A)$ for its domain) over a morphism ${f : X \longrightarrow p(A)}$ in $\mathcal{B}$. 
\end{definition}

See Examples~\ref{ex:codomainfibration}--\ref{ex:simplefibration} below for common cloven fibrations. 

\begin{definition}
\label{def:uniquemediatingmorphismforCartesianmorphism}
\index{ f@$f^\dagger$ (unique mediating morphism induced by the chosen Cartesian morphism $\overline{p(f)}(B)$)}
Given a cloven fibration ${p : \mathcal{V} \longrightarrow \mathcal{B}}$ and a morphism $f : A \longrightarrow B$ in $\mathcal{V}$, then we write $f^\dagger : A \longrightarrow (p(f))^*(B)$ for the unique mediating morphism induced by the universal property of the Cartesian morphism $\overline{p(f)}(B) : (p(f))^*(B) \longrightarrow B$, as in
\[
\xymatrix@C=3em@R=3em@M=0.5em{
A \ar@/^2pc/[rr]^{f} \ar@{-->}[r]_-{f^\dagger} & (p(f))^*(B) \ar[r]_-{\overline{p(f)}(B)} & B
}
\]
\end{definition}

\begin{proposition}[{\cite[Section~1.4]{Jacobs:Book}}]
\label{prop:clovenfibration}
\index{functor!reindexing --}
\index{ f@$f^*$ (reindexing functor)}
Given a cloven fibration ${p : \mathcal{V} \longrightarrow \mathcal{B}}$, then any morphism ${f : X \longrightarrow Y}$ in $\mathcal{B}$ induces a \emph{reindexing functor} ${f^* : \mathcal{V}_Y \longrightarrow \mathcal{V}_X}$, which maps an object $A$ to the domain $f^*(A)$ of the chosen Cartesian morphism $\overline{f}(A)$ over $f$; and a morphism $g : A \longrightarrow B$ to the unique mediating morphism induced by $\overline{f}(B)$, as in 
\[
\xymatrix@C=3.5em@R=3.5em@M=0.5em{
f^*(A) \ar@{-->}[d]_{f^*(g)} \ar[r]^-{\overline{f}(A)} & A \ar[d]^{g}
\\
f^*(B) \ar[r]_-{\overline{f}(B)} & B
}
\]
\end{proposition}

\begin{proposition}[{\cite[Section~1.4]{Jacobs:Book}}]
\label{prop:reindexingfunctornaturalisos}
Given a cloven fibration ${p : \mathcal{V} \longrightarrow \mathcal{B}}$, the reindexing functors ${f^* : \mathcal{V}_Y \longrightarrow \mathcal{V}_X}$ satisfy the following two natural isomorphisms:
\[
{(\id_X)^* \cong \id_{\mathcal{V}_X}}
\qquad
{(h \comp g)^* \cong g^* \comp h^*}
\]
where $g : X \longrightarrow Y$ and $h : Y \longrightarrow Z$.
\end{proposition}

\begin{definition}
\index{fibration!split --}
A cloven fibration is said to be \emph{split} if the isomorphisms given in Proposition~\ref{prop:reindexingfunctornaturalisos} are identities, i.e., when ${(\id_X)^* = \id_{\mathcal{V}_X}}$ and ${(h \comp g)^* = g^* \comp h^*}$.
\end{definition}

\begin{definition}
For any category-theoretic structure $\circledast$, such as products, coproducts, etc., a split fibration $p : \mathcal{V} \longrightarrow \mathcal{B}$ is said to have \emph{split fibred $\circledast$} if every fibre $\mathcal{V}_X$ has $\circledast$ and this structure is preserved on-the-nose by reindexing functors.
\end{definition}

\begin{example}
\label{ex:codomainfibration}
\index{fibration!codomain --}
\index{functor!codomain --}
\index{ cod@$\mathsf{cod}_\mathcal{B}$ (codomain fibration)}
\index{ B@$\mathcal{B}^\to$ (total category of a codomain fibration)}
A prototypical example of a cloven fibration is given by the \emph{codomain} functor $\mathsf{cod}_\mathcal{B} : \mathcal{B}^\to \longrightarrow \mathcal{B}$, 
for any category $\mathcal{B}$ with pullbacks. 
In this case, given an object $f : X \to Y$ in $\mathcal{B}^\to$ and a morphism  $g : Z \longrightarrow Y$ in $\mathcal{B}$, the chosen Cartesian morphism over $g$ is given by the following pullback square:
\[
\xymatrix@C=3.5em@R=3.5em@M=0.5em{
g^*(X) \ar[r] \ar[d]_{g^*(f)}^<{\,\big\lrcorner} & X \ar[d]^{f}
\\
Z \ar[r]_{g} & Y
}
\]
\index{category!arrow --}
\index{ B@$\mathcal{B}^\to$ (arrow category)}
\!Here, $\mathcal{B}^\to$ is the \emph{arrow category} of $\mathcal{B}$. Its objects are given by morphisms $f : X \longrightarrow Y$ of $\mathcal{B}$; and its morphisms from $f : X_1 \longrightarrow Y_1$ to $g : X_2 \longrightarrow Y_2$ are given by pairs $(h_1,h_2)$ of morphisms $h_1 : X_1 \longrightarrow X_2$ and $h_2 : Y_1 \longrightarrow Y_2$ in $\mathcal{B}$ such that $g \comp h_1 = h_2 \comp f$.
\end{example}

While $\mathsf{cod}_\mathcal{B}$ is cloven, it is well-known that it fails to be split because pullback squares are closed under composition only up-to-isomorphism, and not up-to-equality. 

\begin{example}
\label{ex:familiesfibration}
\index{fibration!families --}
\index{functor!families --}
Another common example of a cloven fibration is given by the \emph{$\mathcal{V}$-valued families} functor $\mathsf{fam}_{\mathcal{V}}: \Fam(\!\mathcal{V}\,) \longrightarrow \mathcal{\Set}$, for any category $\mathcal{V}$. 
\index{ fam@$\mathsf{fam}_{\mathcal{V}}$ ($\mathcal{V}$-valued families fibration)}
Here, the objects of $\Fam(\!\mathcal{V}\,)$ 
\index{ Fam@$\Fam(\hspace{-0.05cm}\mathcal{V})$ (total category of a $\mathcal{V}$-valued families fibration)}
are pairs $(X,A)$ of a set $X$ and a functor $A : X \longrightarrow \mathcal{V}$, i.e., an $X$-indexed family of objects of $\mathcal{V}$. Similarly, a morphism from $(X,A)$ to $(Y,B)$ is given by a pair $(f,g)$ of a function $f : X \longrightarrow Y$ and a natural transformation $g : A \longrightarrow B \comp f$, i.e., an $X$-indexed family of morphisms $\{g_x : A(x) \longrightarrow B(f(x))\}_{x \in X}$ in $\mathcal{V}$.
For an object $(Y,A)$ in $\Fam(\!\mathcal{V}\,)$ and a function $f : X \longrightarrow Y$, the chosen Cartesian morphism over $f$ is
\[
\overline{f}(Y,A) \defeq (f , \{\id_{A(f(x))}\}_{x \in X}) : (X , A \comp f) \longrightarrow (Y,A)
\]
A typical example of families fibrations is the \emph{families of sets fibration} with $\mathcal{V} \defeq \Set$.
\end{example}

Compared to the codomain fibrations, the families fibrations are split because composition of morphisms  is strictly associative, see~\cite[Section~1.4]{Jacobs:Book} for more details.

\begin{example}
\label{ex:simplefibration}
\index{fibration!simple --}
The third and final class of examples of cloven fibrations we consider in this section is given by the \emph{simple fibration} construction on any category $\mathcal{V}$ that has Cartesian products, see~\cite[Definition~1.3.1]{Jacobs:Book}. In particular, we can construct a category $\mathsf{s}(\!\mathcal{V}\,)$ 
\index{ s@$\mathsf{s}(\hspace{-0.05cm}\mathcal{V})$ (total category of a simple fibration)}
whose objects are given by pairs $(X,A)$ of objects of $\mathcal{V}$, and whose morphisms $(X,A) \longrightarrow (Y,B)$ are given by pairs $(f,g)$ of  morphisms $f : X \!\longrightarrow\! Y$ and $g : X \times A \!\longrightarrow\! B$ in $\mathcal{V}$. The \emph{simple fibration $\mathsf{s}_{\mathcal{V}} \!:\! \mathsf{s}(\!\mathcal{V}\,) \!\longrightarrow\! \mathcal{V}$} is then given by the functor
\[
\mathsf{s}_{\mathcal{V}}(X,A) \defeq X
\qquad
\mathsf{s}_{\mathcal{V}}(f,g) \defeq f
\]
Given a morphism $f : X \longrightarrow Y$ in $\mathcal{V}$ and an object $(Y,A)$ in $\mathsf{s}(\mathcal{V})$, the Cartesian morphism over $f$ can be shown to be given by $\overline{f}(Y,A) \defeq (f,\mathsf{snd}) : (X,A) \longrightarrow (Y,A)$.
\index{ s@$\mathsf{s}_{\mathcal{V}}$ (simple fibration built from a Cartesian category $\mathcal{V}$)}
\end{example}

Analogously to the families fibrations, the simple fibrations are also split. 
In fact, one can view the simple fibrations as a non-indexed version of the families fibrations. 

As the main use of fibred category theory in this thesis is to give a denotational semantics to eMLTT and its extensions, we only focus on split fibrations and constructions on them that preserve Cartesian morphisms on-the-nose. Informally, the on-the-nose preservation of Cartesian morphisms corresponds to the up-to-equality preservation of type- and term-formers by substitution in dependently typed languages. Therefore, we only consider split versions of fibred functors, fibred natural transformations, fibred adjunctions, etc. The non-split variants of these constructions can be easily recovered by relaxing the preservation conditions for reindexing so that they hold up-to-isomorphism rather than equality.
In addition, it is well-known that one can transform every (possibly non-split) fibration into an equivalent split fibration---see~\cite[Lemma~5.2.4, Corollary~5.2.5]{Jacobs:Book} for details of this construction.

Next, we equip split fibrations over some base category $\mathcal{B}$ with the structure of a 2-category, given by split fibred functors and split fibred natural transformations.

\begin{definition}
\index{functor!split fibred --}
Given two split fibrations ${p : \mathcal{V} \longrightarrow \mathcal{B}}$ and ${q : \mathcal{C} \longrightarrow \mathcal{B}}$, a \emph{split fibred functor} ${F : p \longrightarrow q}$ is given by a functor ${F : \mathcal{V} \longrightarrow \mathcal{C}}$ such that the diagram
\[
\xymatrix@C=3em@R=3em@M=0.5em{
\mathcal{V} \ar[dr]_{p} \ar[rr]^{F} & & \mathcal{C} \ar[dl]^{q}
\\
& \mathcal{B} 
}
\]
commutes and $F$ preserves the chosen Cartesian morphisms on-the-nose.
\end{definition}

\begin{proposition}
Given split fibrations ${p : \mathcal{V} \longrightarrow \mathcal{B}}$ and ${q : \mathcal{C} \longrightarrow \mathcal{B}}$, a split fibred functor ${F : p \longrightarrow q}$, an object $B$ in $\mathcal{V}$, and a morphism $f : X \longrightarrow p(A)$ in $\mathcal{B}$, then 
\[
f^*(F(A)) = F(f^*(A))
\]
\end{proposition}

\begin{proof}
This equality follows directly from the on-the-nose preservation of the chosen Cartesian morphisms by $F$, i.e., from $\overline{f}(F(A))$ and $F(\overline{f}(A))$ being equal.
\end{proof}

In the diagrammatic proofs we present in the rest of this thesis, we represent such equalities on objects using morphisms which we write as $f^*(F(A)) \overset{=}{\longrightarrow} F(f^*(A))$. 
\begin{definition}
\index{natural transformation!split fibred --}
Given two split fibrations ${p : \mathcal{V} \longrightarrow \mathcal{B}}$ and ${q : \mathcal{C} \longrightarrow \mathcal{B}}$, and two split fibred functors ${F : p \longrightarrow q}$ and ${G : p \longrightarrow q}$, a \emph{split fibred natural transformation} ${\alpha : F \longrightarrow G}$ is given by a natural transformation ${\alpha : F \longrightarrow G}$, whose every component $\alpha_A : F(A) \longrightarrow G(A)$ is vertical over $\id_{p(A)}$.
\end{definition}

\begin{proposition}
\label{prop:fibrednaturaltransformationspreserved}
Given two split fibrations ${p : \mathcal{V} \longrightarrow \mathcal{B}}$ and ${q : \mathcal{C} \longrightarrow \mathcal{B}}$, two split fibred functors ${F : p \longrightarrow q}$ and ${G : p \longrightarrow q}$, and a split fibred natural transformation $\alpha : F \longrightarrow G$, then the components of $\alpha$ are preserved by reindexing, i.e., we have
\[
f^*(\alpha_A) = \alpha_{f^*(A)}
\]
in $\mathcal{C}_{X}$, for any object $A$ of $\mathcal{V}$ and any morphism $f : X \longrightarrow p(A)$ in $\mathcal{B}$.
\end{proposition}

\begin{proof}
First, we observe that the following two diagrams commute in $\mathcal{C}$:
\[
\xymatrix@C=5.25em@R=3.5em@M=0.5em{
\ar@{}[d]^{\qquad\qquad\qquad\quad \dcomment{\text{def. of } f^*(\alpha_A)}} & F(A) \ar@/^2pc/[drr]^{\alpha_A} & & \ar@{}[d]_>>>{\dcomment{G \text{ is split fibred}} \qquad\qquad\qquad\,\,\,\,\,\,}
\\
f^*(F(A)) \ar[r]_-{f^*(\alpha_A)} \ar[ur]^{\overline{f}(F(A))} & f^*(G(A)) \ar[r]_-{=} \ar@/^2.7pc/[rr]^{\overline{f}(G(A))} & G(f^*(A)) \ar[r]_-{G(\overline{f}(A))} & G(A)
}
\]
\[
\xymatrix@C=5.25em@R=3.5em@M=0.5em{
\ar@{}[d]^>>>{\qquad\qquad\,\,\, \dcomment{F \text{ is split fibred}}} & F(A) \ar@/^2pc/[drr]^{\alpha_A} & & \ar@{}[d]_{\dcomment{\text{nat. of } \alpha} \qquad\qquad\qquad\qquad\quad}
\\
f^*(F(A)) \ar[r]_-{=} \ar[ur]^{\overline{f}(F(A))} & F(f^*(A)) \ar[u]_-{F(\overline{f}(A))} \ar[r]_-{\alpha_{f^*(A)}} & G(f^*(A)) \ar[r]_-{G(\overline{f}(A))} & G(A)
}
\vspace{0.25cm}
\]
As we also know that $q(f^*(\alpha_A)) = q(\alpha_{f^*(A)}) = \id_{X}$, the universal property of the Cartesian morphism $G(\overline{f}(A))$ tells us that the vertical morphisms $f^*(\alpha_A)$ and $\alpha_{f^*(A)}$ are both equal to the unique mediating morphism over $\id_{X}$ induced by  $\alpha_A \comp \overline{f}(F(A))$. 
\end{proof}

\begin{proposition}[{\cite[Section~1.7]{Jacobs:Book}}]
\index{ Fib@$\mathsf{Fib}_{\mathsf{split}}(\mathcal{B})$ (2-category of split fibrations with a base category $\mathcal{B}$, split fibred functors, and split fibred natural transformations)}
Split fibrations with a base category $\mathcal{B}$, split fibred functors, and split fibred natural transformations form the 2-category $\mathsf{Fib}_{\mathsf{split}}(\mathcal{B})$.
\end{proposition}

The denotational semantics of eMLTT and its extensions is based on split fibred adjunctions.
These are defined in $\mathsf{Fib}_{\mathsf{split}}(\mathcal{B})$ analogously to how ordinary adjunctions are defined in the 2-category $\mathsf{Cat}$ of categories, functors, and natural transformations. 
\index{ Cat@$\mathsf{Cat}$ (2-category of categories, functors, and natural transformations)}

\begin{definition}
\index{adjunction!split fibred --}
\index{ F@$F \dashv\, U$ (split fibred adjunction)}
\index{ e@$\eta$ (unit of a split fibred adjunction)}
\index{ e@$\varepsilon$ (counit of a split fibred adjunction)}
Given two split fibrations ${p : \mathcal{V} \longrightarrow \mathcal{B}}$ and ${q : \mathcal{C} \longrightarrow \mathcal{B}}$, a \emph{split fibred adjunction} ${F \dashv\, U : q \longrightarrow p}$ 
is given by two split fibred functors ${F : p \longrightarrow q}$ and ${U : q \longrightarrow p}$, and two split fibred natural transformations  ${\eta : \id_\mathcal{V} \longrightarrow U \comp F}$ and ${\varepsilon : F \comp U \longrightarrow \id_\mathcal{C}}$, subject to the standard two unit-counit laws (see Definition~\ref{def:adjunction}).
\end{definition}

\begin{proposition}
Given two split fibrations ${p : \mathcal{V} \longrightarrow \mathcal{B}}$ and ${q : \mathcal{C} \longrightarrow \mathcal{B}}$, and a \emph{split fibred adjunction} ${F \dashv\, U : q \longrightarrow p}$, then, for every object $X$ in $\mathcal{B}$, the restriction of $F$ and $U$ to the fibres over $X$ determines an adjunction ${F_X \dashv\, U_X : \mathcal{C}_X \longrightarrow \mathcal{V}_X}$.
\end{proposition}

\begin{proof}
The adjunction ${F_X \dashv\, U_X : \mathcal{C}_X \longrightarrow \mathcal{V}_X}$ follows directly from $F$ and $U$ being split fibred functors, and the components of $\eta$ and $\varepsilon$ being vertical morphisms.
\end{proof}

As we know by definition that $F_X(A) = F(A)$ and $F_X(f) = F(f)$, and similarly for $U_X$, we often omit the subscripts in $F_X$ and $U_X$ when $X$ is clear from the context.

Next, we define split fibred monads. Similarly to split fibred adjunctions, these are defined in  $\mathsf{Fib}_{\mathsf{split}}(\mathcal{B})$ analogously to how ordinary monads are defined in $\mathsf{Cat}$.

\begin{definition}
\label{def:fibredmonad}
\index{monad!split fibred --}
\index{ T@$\mathbf{T}$ (split fibred monad)}
\index{ T@$(T,\eta,\mu)$ (split fibred monad)}
\index{ e@$\eta$ (unit of a split fibred monad)}
\index{ m@$\mu$ (multiplication of a split fibred monad)}
A \emph{split fibred monad} $\mathbf{T} = (T,\eta,\mu)$ on a split fibration $p : \mathcal{V} \longrightarrow \mathcal{B}$ is given by a split fibred functor $T : p \longrightarrow p$, and split fibred natural transformations $\eta : \id_p \longrightarrow T$ and $\mu : T \comp T \longrightarrow T$, subject to standard monad laws (see Definition~\ref{def:monad}).
\end{definition}

Analogously, we can also define a split fibred variant of resolutions of monads.

\begin{definition}
\label{def:fibredresolution}
\index{resolution!-- of a monad!split fibred --}
Given a split fibred monad $(T,\eta,\mu)$ on a split fibration $p : \mathcal{V} \longrightarrow \mathcal{B}$,  its \emph{split fibred resolution} is given by a split fibration $q : \mathcal{C} \longrightarrow \mathcal{B}$ and a split fibred adjunction $F \dashv\, U : q \longrightarrow p$ such that $(T,\eta,\mu)$ coincides with the split fibred monad canonically derived from this split fibred adjunction (this monad is derived analogously to the ordinary, non-fibred case discussed in Proposition~\ref{prop:monadfromadjunction}).
\end{definition}

Analogously to monads in $\mathsf{Cat}$, there are again two canonical split fibred resolutions of a split fibred monad, the Kleisli and Eilenberg-Moore resolutions. As before, these are the initial and terminal objects in the category of split fibred resolutions. As we only use the split fibred Eilenberg-Moore resolution in this thesis, we omit the definition of the Kleisli resolution---it can be found in~\cite[Exercise~1.7.9 (i)]{Jacobs:Book}.

\begin{proposition}[{\cite[Exercise~1.7.9 (ii)]{Jacobs:Book}}]
\label{prop:definingthesplitEMfibration}
\index{Eilenberg-Moore!split fibred -- resolution}
Given a split fibred monad $\mathbf{T} = (T,\eta,\mu)$ on a split fibration $p : \mathcal{V} \longrightarrow \mathcal{B}$, its \emph{split fibred Eilenberg-Moore resolution} is given by a split fibration $p^{\mathbf{T}} : \mathcal{V}^{\mathbf{T}} \longrightarrow \mathcal{B}$ and a split fibred adjunction $F^{\mathbf{T}} \dashv\, U^{\mathbf{T}} : p^{\mathbf{T}} \longrightarrow p$, where the category $\mathcal{V}^{\mathbf{T}}$ and the adjunction $F^{\mathbf{T}} \dashv\,  U^{\mathbf{T}} : \mathcal{V}^{\mathbf{T}} \longrightarrow \mathcal{V}$ are defined as if we were constructing the EM-resolution of the monad $(T,\eta,\mu)$ on $\mathcal{V}$ (see Definition~\ref{def:EMresolution}).
\end{proposition}

\index{fibration!Eilenberg-Moore --}\index{Eilenberg-Moore!-- fibration}
\index{ p@$p^{\mathbf{T}}$ (split Eilenberg-Moore fibration of a split fibred monad $\mathbf{T}$ on $p$)}
In Proposition~\ref{prop:definingthesplitEMfibration}, the functor $p^{\mathbf{T}} : \mathcal{V}^{\mathbf{T}} \longrightarrow \mathcal{B}$ is given by $p^{\mathbf{T}}(A,\alpha) \defeq p(A)$ and $p^{\mathbf{T}}(h) \defeq p(h)$. We call $p^{\mathbf{T}}$ the \emph{split Eilenberg-Moore (EM-) fibration}
of $\mathbf{T}$. The chosen Cartesian morphism in $p^{\mathbf{T}}$ over a morphism $f : X \longrightarrow p^{\mathbf{T}}(B,\beta)$ in $\mathcal{B}$ is given by
\[
\overline{f}(B,\beta) \defeq \overline{f}(B) : (f^*(B), f^*(\beta)) \longrightarrow (B,\beta)
\]

We conclude our overview of fibred category theory by discussing structures that are commonly used to model the core features of dependently typed languages.

The general idea behind modelling a dependent type $\lj \Gamma A$ in a split fibration \linebreak $p : \mathcal{V} \longrightarrow \mathcal{B}$ is to interpret the context $\Gamma$ as an object $\sem{\Gamma}$ in the base category $\mathcal{B}$ and the dependent type $A$ as an object $\sem{A}$ in the total category $\mathcal{V}$, such that $p(\sem{A}) = \sem{\Gamma}$. 

Regardless of the particular grammar of types, a crucial step in the definition of the interpretation of contexts $\Gamma$ (lists of distinct variables $x$ annotated with types $A$) involves defining the interpretation of extended contexts $\Gamma, x \!:\! A$. In fibrational models of dependently typed languages, $\Gamma, x \!:\! A$ is most naturally interpreted using the notion of comprehension, which we define below, in terms of a terminal object functor for $p$.

\begin{definition}
\label{def:terminalobjectfunctor}
\index{functor!split terminal object --}
\index{ 1@$1$ (split terminal object functor)}
A \emph{split terminal object functor} for a split fibration $p : \mathcal{V} \longrightarrow \mathcal{B}$ is given by a functor $1 : \mathcal{B} \longrightarrow \mathcal{V}$ that is a split fibred right adjoint to $p$ in $\mathsf{Fib}_{\mathsf{split}}(\mathcal{B})$, i.e., 
\[
\xymatrix@C=3em@R=3em@M=0.5em{
\mathcal{V} \ar[dr]_{p} \ar@/^1pc/[rr]^{p} & \bot & \mathcal{B} \ar[dl]^{\id_{\mathcal{B}}} \ar@/^1pc/[ll]^{1}
\\
& \mathcal{B}
}
\]
\end{definition}

Below we note that the existence of a such terminal object functor equips every fibre $\mathcal{V}_X$ with a terminal object $1_X$, and these are preserved on-the-nose by reindexing.
\index{ 1@$1_X$ (terminal object in $\mathcal{V}_X$)}

\begin{proposition}
\label{prop:fibredterminalobjects}
If $p : \mathcal{V} \longrightarrow \mathcal{B}$ is a split fibration, then $p$ comes equipped with a split terminal object functor $1 : \mathcal{B} \longrightarrow \mathcal{V}$ if and only if every fibre of $p$ has a terminal object and these terminal objects are preserved on-the-nose by reindexing.
\end{proposition}

\begin{proof}
For a detailed proof, we refer the reader to~\cite[Lemma~1.8.8]{Jacobs:Book}, where a non-split version of this proposition is proved. The proof of this split version is proved analogously, but using the additional information that for any morphism $f : X \longrightarrow Y$ in $\mathcal{B}$, we have $f^*(1_Y) = 1_X$. Here, we simply sketch the definitions one uses to prove both directions of this proposition.
First, in the \emph{if}-direction, we define the terminal object functor $1 : \mathcal{B} \longrightarrow \mathcal{V}$ by mapping an object $X$ in $\mathcal{B}$ to the terminal object $1_X$ in $\mathcal{V}_X$; and by mapping a morphism $f : X \longrightarrow Y$ in $\mathcal{B}$ to the composite morphism $1_X \overset{=}{\longrightarrow} f^*(1_Y) \overset{\overline{f}(1_Y)}{\longrightarrow} 1_Y$. In the opposite direction, we define the terminal object $1_X$ in $\mathcal{V}_X$ to be $1(X)$. The on-the-nose preservation of terminal objects by reindexing follows from the on-the-nose preservation of Cartesian morphisms by $1 : \mathcal{B} \longrightarrow \mathcal{V}$.
\end{proof}

\index{ 1@$\mathbf{1}$ (trivial one object category)}
As noted by Jacobs~\cite[Section~1.8]{Jacobs:Book}, this characterisation is a fibred analogue of a category $\mathcal{V}$ having a terminal object if and only if the unique functor $!_{\mathcal{V}} : \mathcal{V} \longrightarrow \mathbf{1}$ has a right adjoint. In $\mathsf{Fib}_{\mathsf{split}}(\mathcal{B})$, the terminal object is given by $\id_{\mathcal{B}} : \mathcal{B} \longrightarrow \mathcal{B}$.

Based on this correspondence, we use the convention of writing $1_X$ for $1(X)$.

\begin{definition}
\index{category!split comprehension -- with unit}
\index{functor!comprehension --}
\index{ @$\ia -$ (comprehension functor)}
A split fibration $p : \mathcal{V} \longrightarrow \mathcal{B}$ is called a \emph{split comprehension \linebreak category with unit} if i) $p$ comes equipped with a split terminal object functor \linebreak $1 : \mathcal{B} \longrightarrow \mathcal{V}$; and ii) this terminal object functor has a (not necessarily fibred) right adjoint $\ia - : \mathcal{V} \longrightarrow \mathcal{B}$ in $\mathsf{Cat}$, called the \emph{comprehension} functor, as illustrated below:
\[
\xymatrix@C=0.001em@R=2em@M=0.5em{
&\mathcal{V} \ar@/_3pc/[dd]_{p} \ar@/^3pc/[dd]^{\ia{-}}
\\
\dashv & & \!\!\!\!\dashv
\\
&\mathcal{B} \ar[uu]^{1}
}
\]
\end{definition}

\begin{proposition}[{\cite[Section~10.4]{Jacobs:Book}}]
\label{prop:comprehensioncategorywithunit}
Given a split comprehension category with unit $p : \mathcal{V} \longrightarrow \mathcal{B}$, then there exists a functor $\mathcal{P} : \mathcal{V} \longrightarrow \mathcal{B}^\to$ such that $p = \mathsf{cod}_{\mathcal{B}} \comp \mathcal{P}$ and $\mathcal{P}$ sends the chosen Cartesian morphisms in $\mathcal{V}$ to pullback squares in $\mathcal{B}^\to$. A functor with these properties is called a \emph{comprehension category}.
\index{category!comprehension --}
\index{ P@$\mathcal{P}$ (comprehension category)}
\end{proposition}

We recall from~\cite[Section~10.4]{Jacobs:Book} that the functor $\mathcal{P} : \mathcal{V} \longrightarrow \mathcal{B}^\to$ is given on objects by mapping an object $A$ in $\mathcal{V}$ to the morphism $\ia A \overset{=}{\longrightarrow} p(1_{\ia A}) \overset{p(\varepsilon_A^{1 \dashv \ia -})}{\longrightarrow} p(A)$, and by mapping a morphism $f : A \longrightarrow B$ in $\mathcal{V}$ to the following commuting diagram:
\[
\xymatrix@C=5em@R=4em@M=0.5em{
\ia{A} \ar[r]^{\ia f} \ar[d]_{=}^{\qquad\,\,\,\, \dcomment{p \comp 1 = \id_{\mathcal{B}}}} & \ia{B} \ar[d]^{=} 
\\
p(1_{\ia A}) \ar[d]_{p(\varepsilon_A^{1 \dashv \ia -})}^{\qquad \dcomment{\text{nat. of } \varepsilon^{1 \dashv \ia -}}} \ar[r]^{p(1(\ia f))} & p(1_{\ia A}) \ar[d]^{p(\varepsilon_B^{1 \dashv \ia -})}
\\
p(A) \ar[r]_{p(f)} & p(B)
}
\]

\begin{definition}
\index{morphism!projection --}
\index{functor!weakening --}
Given a split comprehension category with unit $p : \mathcal{V} \longrightarrow \mathcal{B}$ and an object $A$ in $\mathcal{V}$, the morphism $\mathcal{P}(A) : \ia A \longrightarrow p(A)$ is called a \emph{projection morphism} and commonly written as $\pi_A$. 
\index{ product@$\pi_A$ (projection morphism)}
The reindexing functor $\pi_A^*$ is called a \emph{weakening functor}. 
\index{ product@$\pi_A^*$ (weakening functor)}
\end{definition}

\begin{definition}
\index{category!split comprehension -- with unit!full --}
A  split comprehension category with unit $p : \mathcal{V} \longrightarrow \mathcal{B}$ is said to be \emph{full} if the corresponding comprehension category $\mathcal{P} : \mathcal{V} \longrightarrow \mathcal{B}^\to$ is full and faithful.
\end{definition}

Returning to the interpretation of dependently typed languages (such as MLTT), we now briefly describe how to interpret well typed terms in a full split comprehension category with unit $p : \mathcal{V} \longrightarrow \mathcal{B}$. In the literature, a well typed term $\vj \Gamma V A$ is usually interpreted either i) as a \emph{global element} $1_{\sem{\Gamma}} \longrightarrow \sem{A}$ 
\index{global element}
of $\sem{A}$ in $\mathcal{V}_{\sem{\Gamma}}$, or ii) as a \emph{section} 
of the projection morphism $\pi_{\sem{A}} : \ia{\sem{A}} \longrightarrow \sem{\Gamma}$ in $\mathcal{B}$.
However, as is well known, these two ways of interpreting terms are interchangeable, see Proposition~\ref{prop:globalelementssections} below. Therefore, one  often switches between i) and ii) when working with the denotations of terms. In particular, ii) corresponds to the fact that the fully-faithfulness of $\mathcal{P}$ allows us to consider $\pi_{\sem{A}} : \ia{\sem{A}} \longrightarrow \sem{\Gamma}$ in $\mathcal{B}$ as an equivalent denotation of a type $\lj \Gamma A$. 

\begin{proposition}
\label{prop:globalelementssections}
Given a split comprehension category with unit $p : \mathcal{V} \longrightarrow \mathcal{B}$ and an object $A$ in $\mathcal{V}$, then there exists an isomorphism
\[
\mathcal{V}_{p(A)}(1_{p(A)} , A) \cong \{f : p(A) \longrightarrow \ia A \vertbar \pi_A \comp f = \id_{p(A)}\}
\]
\end{proposition}

\begin{proof}
This proposition is a special case of~\cite[Lemma~10.4.9 (i)]{Jacobs:Book}, whose proof is omitted in op. cit. Here we give the proof of the above isomorphism explicitly.

First, given a global element $f : 1_{p(A)} \longrightarrow A$ of $A$ in $\mathcal{V}_{p(A)}$, 
\index{global element}
we define the corresponding section 
$\funsection(f) : p(A) \longrightarrow \ia A$ in $\mathcal{B}$ as the following composite morphism: 
\index{ s@$\funsection(f)$ (section corresponding to a vertical global element $f$)}
\[
\xymatrix@C=5em@R=3em@M=0.5em{
p(A) \ar[r]^-{\eta^{1 \dashv \ia -}_{p(A)}} & \ia {1_{p(A)}} \ar[r]^-{\ia {f}} & \ia A
}
\]
The required equation $\pi_A \comp \funsection(f) = \id_{p(A)}$ then follows from the commutativity of the following diagram:
\[
\xymatrix@C=8em@R=4em@M=0.5em{
p(A) \ar@/^3pc/[rr]^-{\mathsf{s}(f)}_*+<0.75em>{\dcomment{\text{def. of } \mathsf{s}(f)}} \ar[d]_{=}^<<<<<<<<{\qquad\qquad\,\, \dcomment{p \comp 1 = \id_{\mathcal{B}}}} \ar[r]^-{\eta^{1 \dashv \ia -}_{p(A)}} & \ia {1_{p(A)}} \ar[d]_{=} \ar[r]^-{\ia {f}} \ar@{}[dd]^{\qquad\qquad\quad\,\,\, \dcomment{\mathcal{P}(f)}} & \ia A \ar[d]^{=} \ar@/^3pc/[dd]^-{\pi_{A}}_>>>>>>>>{\dcomment{\text{def. of } \pi_A}\,\,\,\,\,}
\\
p(1_{p(A)}) \ar[dr]_{\id_{p(1_{p(A)})}} \ar[r]^-{p(1(\eta^{1 \dashv \ia -}_{p(A)}))} \ar@{}[d]^<<<<<<{\qquad\qquad\qquad\qquad\!\!\! \dcomment{1 \dashv \ia -}} & p(1_{\ia {1_{p(A)}}}) \ar[d]^{p(\varepsilon^{1 \dashv \ia -}_{1_{p(A)}})} & p(1_{\ia A}) \ar@/_2.5pc/[d]_-{p(\varepsilon^{1 \dashv \ia -}_A)}
\\
& p(1_{p(A)}) \ar[r]_{=} & p(A)
}
\]

Next, given a morphism $f : p(A) \longrightarrow \ia A$ in $\mathcal{V}$ such that $\pi_A \comp f = \id_{p(A)}$, we define the corresponding global element $\mathsf{s}^{-1}(f) : 1_{p(A)} \longrightarrow A$ in $\mathcal{V}$ as the composite 
\index{ s@$\mathsf{s}^{-1}(f)$ (vertical global element corresponding to a section $f$)}
\[
\xymatrix@C=5em@R=2em@M=0.5em{
1_{p(A)} \ar[r]^-{1(f)} & 1_{\ia A} \ar[r]^-{\varepsilon^{1 \dashv \ia -}_A} & A
}
\]
and show that $\mathsf{s}^{-1}(f)$ is in $\mathcal{V}_{p(A)}$ by proving that the following diagram commutes:
\[
\xymatrix@C=5em@R=4em@M=0.5em{
p(1_{p(A)}) \ar@/^3.25pc/[rr]^-{p(\mathsf{s}^{-1}(f))}_*+<1em>{\dcomment{\text{def. of } \mathsf{s}^{-1}(f)}} \ar[d]_{=}^{\qquad\,\,\,\,\,\, \dcomment{p \comp 1 = \id_{\mathcal{B}}}} \ar[r]^-{p(1(f))} & p(1_{\ia A}) \ar[d]_{=}^{\qquad\,\,\,\,\, \dcomment{\text{def. of } \pi_A}} \ar[r]^-{p(\varepsilon^{1 \dashv \ia -}_A)} & p(A) \ar[d]^{\id_{p(A)}}
\\
p(A) \ar@/_3.25pc/[rr]_-{\id_{p(A)}}^*+<1.25em>{\dcomment{f \text{ is a section of } \pi_A}} \ar[r]_{f} & \ia {A} \ar[r]_-{\pi_A} & p(A)
}
\]

Next, we show that the equation $\mathsf{s}^{-1}(\mathsf{s}(f)) = f$ holds for all $f : 1_{p(A)} \longrightarrow A$ in $\mathcal{V}_{p(A)}$, by proving that the following diagram commutes:
\[
\xymatrix@C=5em@R=4em@M=0.5em{
1_{p(A)} \ar[dr]_-{1(\eta^{1 \dashv \ia -}_{p(A)})} \ar@/^4pc/[rrr]^-{\mathsf{s}^{-1}(\mathsf{s}(f))}_*+<1.75em>{\dcomment{\text{def. of } \mathsf{s}^{-1}(\mathsf{s}(f))}} \ar@/_2.5pc/[ddr]_{\id_{1_{p(A)}}} \ar[rr]_-{1(\mathsf{s}(f))} \ar@{}[d]^{\qquad\qquad\qquad\,\,\, \dcomment{\text{def. of } \mathsf{s}(f)}} & & 1_{\ia A} \ar[r]_-{\varepsilon^{1 \dashv \ia -}_A} & A 
\\
& 1_{\ia {1_{p(A)}}} \ar[ur]_-{1 (\ia f)} \ar[d]^-{\varepsilon^{1 \dashv \ia -}_{1_{p(A)}}}_<<<{\dcomment{1 \dashv \ia -} \quad\!\!\!} & 
\\
& 1_{p(A)} \ar@/_2.5pc/[uurr]_-{f}^>>>>>>>>>>>>>>>>>>>>>>>>>{\dcomment{\text{nat. of } \varepsilon^{1 \dashv \ia -}}\qquad\quad} & &
}
\]

Finally, we show that the equation $\mathsf{s}(\mathsf{s}^{-1}(f)) = f$ holds for all $f : p(A) \longrightarrow \ia A$ in $\mathcal{V}$ with $\pi_A \comp f = \id_{p(A)}$, by proving that the following diagram commutes:
\[
\xymatrix@C=5em@R=3em@M=0.5em{
p(A) \ar@/_2.5pc/[ddrr]_-{f}^>>>>>>>>>>>>>>>>>>>>>>>>>{\qquad\quad\dcomment{\text{nat. of } \eta^{1 \dashv \ia -}}} \ar@/^4pc/[rrr]^-{\mathsf{s}(\mathsf{s}^{-1}(f))}_*+<1.75em>{\dcomment{\text{def. of } \mathsf{s}(\mathsf{s}^{-1}(f))}}  \ar[r]_-{\eta^{1 \dashv \ia -}_{p(A)}} & \ia {1_{p(A)}} \ar[dr]_{\ia {1(f)}} \ar@{}[d]^{\qquad\qquad\qquad \dcomment{\text{def. of } \mathsf{s}^{-1}(f)}} \ar[rr]_-{\ia {\mathsf{s}^{-1}(f)}} & & \ia A 
\\
& & \ia {1_{\ia A}} \ar[ur]_-{\ia {\varepsilon^{1 \dashv \ia -}_A}} & 
\\
& & \ia {A} \ar@/_3.5pc/[uur]_{\id_{\ia A}} \ar[u]^-{\eta^{1 \dashv \ia -}_{\ia A}}_>>>{\quad\!\!\!\!\! \dcomment{1 \dashv \ia -}}
}
\]
\end{proof}

To make better use of this interchangeability of the global elements $f :  1_{p(A)} \longrightarrow A$ in the fibres and the sections $\mathsf{s}(f) : p(A) \longrightarrow \ia A$ in the base category, we now describe a construction for $\mathsf{s}(f)$ that corresponds to applying a reindexing functor to $f$.

\begin{proposition}
\label{prop:reindexinginthebasecategory}
Given a split comprehension category with unit $p : \mathcal{V} \longrightarrow \mathcal{B}$, a global element $f : 1_{p(A)} \longrightarrow A$ of $A$ in $\mathcal{V}_{p(A)}$ and a morphism $g : X \longrightarrow p(A)$ in $\mathcal{B}$, then
\[
\mathsf{s}^{-1}(h) = g^*(f)
\]
where $h : X \longrightarrow \ia {g^*(A)}$ is the unique mediating morphism in the following pullback situation:
\[
\xymatrix@C=5em@R=4em@M=0.5em{
& p(A) \ar@/^1.5pc/[dr]^-{\mathsf{s}(f)} &
\\
X \ar@/_1.5pc/[dr]_{\id_X} \ar@/^1.5pc/[ur]^-{g} \ar@{-->}[r]^-{h} & \ia {g^*(A)} \ar[d]_{\pi_{g^*(A)}}^<{\,\big\lrcorner} \ar[r]^-{\ia {\overline{g}(A)}} & \ia A \ar[d]^{\pi_A}_{\dcomment{\mathcal{P}(\overline{g}(A))}\qquad\,\,\,\,\,\,\,}
\\
& X \ar[r]_-{g} & p(A)
}
\]
The composite morphisms that make up the outer perimeter from $X$ to $p(A)$ are equal because of Proposition~\ref{prop:globalelementssections}, namely, because we know that $\pi_A \comp \mathsf{s}(f) = \id_{p(A)}$.
\end{proposition}

\begin{proof}
In order to prove the required equation
\[
\mathsf{s}^{-1}(h) = g^*(f)
\]
we instead prove an auxiliary equation 
\[
h = \mathsf{s}(g^*(f))
\]
from which the required equation follows because $\mathsf{s}$ and $\mathsf{s}^{-1}$ form an isomorphism. 

We show that this auxiliary equation holds by observing that the morphism $\mathsf{s}(g^*(f))$ satisfies the same universal property as the unique mediating morphism $h$ given above in the proposition. In particular, we first show that the following diagram commutes:
\vspace{0.6cm}
\[
\xymatrix@C=6em@R=4em@M=0.5em{
p(A) \ar@/^8pc/[ddrrr]^-{\mathsf{s}(f)}_<<<<<<<<<{\quad\qquad\qquad\qquad\qquad\qquad\qquad\dcomment{\text{def. of } \mathsf{s}(f)}} \ar[r]_-{\eta^{1 \,\dashv\, \ia -}_{p(A)}} & \ia {1_{p(A)}} \ar@/^3.5pc/[ddrr]_-{\ia f}
\\
&& \ia {g^*(1_{p(A)})} \ar[ul]_-{\ia {\overline{g}(1_{p(A)})}} &
\\
X \ar@/_3.25pc/[rr]_-{\mathsf{s}(g^*(f))}^*+<1em>{\dcomment{\text{def. of } \mathsf{s}(g^*(f))}} \ar[uu]^-{g}_-{\qquad\dcomment{\text{nat. of } \eta^{1 \,\dashv\, \ia -}}} \ar[r]^-{\eta^{1 \,\dashv\, \ia -}_{X}} & \ia {1_X} \ar[uu]_-{\ia {1(g)}}_-{\qquad\quad\!\!\!\!\dcomment{1 \text{ is split fibred}}} \ar[ur]_-{=}_>>>>>>>>>{\qquad\qquad\dcomment{\text{def. of } g^*(f)}}\ar[r]^-{\ia {g^*(f)}} & \ia {g^*(A)} \ar[r]_-{\ia {\overline{g}(A)}} & \ia A
}
\vspace{0.5cm}
\]
Next, we note that $\pi_{g^*(A)} \comp \mathsf{s}(g^*(f)) = \id_X$ by Proposition~\ref{prop:globalelementssections}.
As a result, we get that $\mathsf{s}(g^*(f))$ is equal to the unique mediating morphism $h : X \longrightarrow \ia {g^*(A)}$.
\end{proof}

In addition to allowing us to translate between the global elements $f :  1_{p(A)} \longrightarrow A$ in the fibres and the sections $\mathsf{s}(f) : p(A) \longrightarrow \ia A$ in the base category, the unit $\eta^{1 \,\dashv\, \ia -}$ of the adjunction $1 \dashv \ia -$ has a further useful property, namely, it is in fact a natural isomorphism, as noted in~\cite[Section~10.4]{Jacobs:Book} and proved in detail below.

\begin{proposition}[{\cite[Exercise~10.4.7 (i)]{Jacobs:Book}}] 
\label{prop:compcatunitiso}
Given a split comprehension category with unit $p : \mathcal{V} \longrightarrow \mathcal{B}$ and 
an object $X$ in $\mathcal{B}$, then the component $\eta^{1 \,\dashv\, \ia -}_X : X \longrightarrow \ia {1_X}$ of the unit $\eta^{1 \,\dashv\, \ia -}$ is an isomorphism, with its inverse given by $\pi_{1_X} : \ia {1_X} \longrightarrow X$.
\end{proposition}

\begin{proof}
A neat way to prove this proposition is to note that $\eta^{1 \,\dashv\, \ia -}$ must be a natural isomorphism to start with because the left adjoint $1$ in $1 \dashv \ia -$ is fully-faithful. In more detail, as highlighted in~\cite[Section~10.4]{Jacobs:Book}, the fact that $p \dashv 1$ is a fibred adjunction, means that $p \comp 1 = \id_{\mathcal{B}}$, resulting in the counit $\varepsilon^{\,p \,\,\dashv\, 1}$ being  identity and therefore also a natural isomorphism. However, it is well-known that the counit of an adjunction is a natural isomorphism if and only if the right adjoint is fully-faithful, e.g., see~\cite[Section~IV.4]{MacLane:CatWM}. In the context of this proposition, this means that $1$ must be fully-faithful. The dual of this fact states that the unit of an adjunction is a natural isomorphism if and only if the left adjoint is fully-faithful. Therefore, as we know that $1$ is fully-faithful, and it is the left adjoint in $1 \dashv \ia -$, the unit $\eta^{1 \,\dashv\, \ia -}$ must be a natural isomorphism.

Now, as we know that for each object $X$ in $\mathcal{B}$, $\eta^{1 \,\dashv\, \ia -}_X$ must have an inverse $(\eta^{1 \,\dashv\, \ia -}_X)^{-1}$, we are left with showing that $(\eta^{1 \,\dashv\, \ia -}_X)^{-1} = \pi_{1_X}$. To this end, we first show that $\pi_{1_X}$ is the left inverse of $\eta^{1 \,\dashv\, \ia -}_X$, i.e., that $\pi_{1_X} \comp \eta^{1 \,\dashv\, \ia -}_X = \id_X$. This equation follows straightforwardly  from the commutativity of the following diagram:
\[
\xymatrix@C=5em@R=5em@M=0.5em{
X \ar[d]_{=}^{\qquad\,\,\dcomment{p \comp 1 = \id_{\mathcal{B}}}} \ar[r]^-{\eta^{1 \,\dashv\, \ia -}_X} & \ia {1_X} \ar[d]_{=}^{\qquad\,\,\,\dcomment{\text{def. of } \pi_{1_X}}} \ar[r]^-{\pi_{1_X}} & X \ar[d]_{=}
\\
p(1_X) \ar[r]^-{p(1(\eta^{1 \,\dashv\, \ia -}_X))} \ar@/_3.25pc/[rr]_{\id_{p(1_X)}}^*+<1em>{\dcomment{1 \,\dashv\, \ia -}} & p(1_{\ia {1_X}}) \ar[r]^-{p(\varepsilon^{1 \,\dashv\, \ia -}_{1_X})} & p(1_X)
}
\]
Finally, we show that the equation $(\eta^{1 \,\dashv\, \ia -}_X)^{-1} = \pi_{1_X}$ holds by observing that
\[
(\eta^{1 \,\dashv\, \ia -}_X)^{-1} = \id_X \comp (\eta^{1 \,\dashv\, \ia -}_X)^{-1} = \pi_{1_X} \comp \eta^{1 \,\dashv\, \ia -}_X \comp (\eta^{1 \,\dashv\, \ia -}_X)^{-1} = \pi_{1_X} \comp \id_{\ia {1_X}} = \pi_{1_X}
\]
\end{proof}


\chapter[eMLTT: Martin-L\"{o}f's type theory with fibred computational effects]{eMLTT: Martin-L\"{o}f's type theory \\with fibred computational effects}
\label{chap:syntax}

\index{ e@eMLTT (our effectful dependently typed language)}
In this chapter we introduce and study eMLTT---our take on intensional MLTT with general computational effects. 
Specifically, eMLTT combines dependently typed programming in MLTT with features familiar from simply typed languages with computational effects such as CBPV and EEC. 
Similarly to CBPV and EEC, eMLTT makes a clear distinction between values (i.e., effect-free programs) and computations (i.e., potentially effectful programs), at both the level of types and the level of terms, with both kinds of types only allowed to depend on values---see the discussion in Section~\ref{sect:twoguidingquestions}.

In Section~\ref{sect:syntax}, we present the syntax of eMLTT; in Section~\ref{sect:judgements}, we 
equip eMLTT with a type system and define its equational theory; and in Section~\ref{sect:metatheory}, we establish some basic meta-theoretic properties of eMLTT, including the closure of well-formed expressions under weakening and substitution. We conclude this chapter by discussing syntax that is not part of the definition of eMLTT but that is nevertheless derivable. In Section~\ref{sect:derivableeliminationforms}, we show how to eliminate various value types into computations;  and in Section~\ref{sect:derivableequations}, we derive some standard equations that are familiar from other computational languages, such as the unit and associativity laws for sequential composition.

\section{Syntax}
\label{sect:syntax}

\index{ x@$x,y,\ldots$ (value variables)}
\index{ z@$z, \ldots$ (computation variables)}
We begin by assuming two disjoint and countably infinite sets of \emph{value variables} and 
\emph{computation variables}, respectively. 
We use $x,y,\ldots$ to range over value variables and 
\index{variable!value --}
$z, \ldots$ to range over computation variables. 
\index{variable!computation --}
The former are treated intuitionistically, as in MLTT, and enjoy structural properties of weakening and contraction. The latter are treated linearly, as in EEC, so as to ensure that effectful computations are not duplicated or discarded arbitrarily---an important property for the correct formulation of the elimination rule for the computational $\Sigma$-type, as discussed later in this section.

\index{ A@$A,B,\ldots$ (value types)}
\index{ C@$\ul{C},\ul{D},\ldots$ (computation types)}
\index{ V@$V,W,\ldots$ (value terms)}
\index{ M@$M,N,\ldots$ (computation terms)}
\index{ K@$K,L,\ldots$ (homomorphism terms)}
For types, we use $A,B,\ldots$ to range over \emph{value types}; 
and $\ul{C},\ul{D},\ldots$ to range over \emph{computation types}. 
For terms, we use $V,W,\ldots$ to range over \emph{value terms}; 
$M,N,\ldots$ to range over \emph{computation terms}; 
and $K,L,\ldots$ to range over \emph{homomorphism terms}. 
As is common for dependently typed languages, eMLTT's types and terms are given by a mutually inductive definition---see Definitions~\ref{def:types} and~\ref{def:terms}, respectively.
We discuss these different kinds of types and terms after their respective definitions.

\begin{definition}
\label{def:types}
\index{type!value --}
eMLTT's \emph{value} and \emph{computation types} are given by
\[
\begin{array}{r c l @{\qquad\qquad}l}
\index{ Nat@$\Nat$ (type of natural numbers)}
A & ::= & \Nat & \text{type of natural numbers}
\\
\index{ 1@$1$ (unit type)}
& \vertbar & 1 & \text{unit type}
\\
\index{ Sigma@$\Sigma \, x \hspace{-0.05cm}:\hspace{-0.05cm} A .\, B$ (value $\Sigma$-type)}
& \vertbar & \Sigma \, x \!:\! A .\, B & \text{value $\Sigma$-type}
\\
\index{ Product@$\Pi \, x \hspace{-0.05cm}:\hspace{-0.05cm} A .\, B$ (value $\Pi$-type)}
& \vertbar & \Pi \, x \!:\! A .\, B & \text{value $\Pi$-type}
\\
\index{ 0@$0$ (empty type)}
& \vertbar & 0 & \text{empty type}
\\
\index{ A@$A + B$ (coproduct type)}
& \vertbar & A + B & \text{coproduct type}
\\
\index{ V@$V =_A W$ (propositional equality)}
& \vertbar & V =_A W & \text{propositional equality}
\\
\index{ UC@$U \ul{C}$ (type of thunked computations)}
& \vertbar & U \ul{C} & \text{type of thunked computations}
\\
\index{ C@$\ul{C} \multimap \ul{D}$ (homomorphic function type)}
& \vertbar & \ul{C} \multimap \ul{D} & \text{homomorphic function type}
\index{type!computation --}
\\[5mm]
\index{ FA@$FA$ (type of computations that return values of type $A$)}
\ul{C} & ::= & FA & \text{type of computations that return values of type $A$}
\\
\index{ Sigma@$\Sigma \, x \hspace{-0.05cm}:\hspace{-0.05cm} A .\, \ul{C}$ (computational $\Sigma$-type)}
& \vertbar & \Sigma \, x \!:\! A .\, \ul{C} & \text{computational $\Sigma$-type}
\\
\index{ Product@$\Pi \, x \hspace{-0.05cm}:\hspace{-0.05cm} A .\, \ul{C}$ (computational $\Pi$-type)}
& \vertbar & \Pi \, x \!:\! A .\, \ul{C} & \text{computational $\Pi$-type}
\end{array}
\vspace{0.2cm}
\]
where 
\begin{itemize}
\item in $\Sigma \, x \!:\! A .\, B$ and $\Pi \, x \!:\! A .\, B$, the value variable $x$ is bound in $B$; and
\item in $\Sigma \, x \!:\! A .\, \ul{C}$ and $\Pi \, x \!:\! A .\, \ul{C}$, the value variable $x$ is bound in $\ul{C}$.
\end{itemize}
\end{definition}

We write $FVV(A)$ and $FVV(\ul{C})$ for the sets of \emph{free value variables} of a value type $A$ and a computation type $\ul{C}$, respectively.

eMLTT's \emph{value types} coincide with the types of MLTT, except for the type $U\ul{C}$ of thunks of computations of type $\ul{C}$ and the homomorphic function type $\ul{C} \multimap \ul{D}$; these are respectively based on analogous types in CBPV and EEC. The inhabitants of the former are suspended computations of type $\ul{C}$. The inhabitants of the latter are effectful functions that accept computations of type $\ul{C}$ as their arguments and additionally guarantee that the argument computation ``happens first" in function application---see the homomorphic lambda abstraction in Definition~\ref{def:terms}. 
To keep the presentation focussed on the computational aspects of eMLTT, we omit general inductive types and use the type of natural numbers as a representative example. If needed, general inductive types can be added using standard techniques, e.g., using $W$\!-types~\cite{MartinLof:Bibliopolis}. 
As is common in presentations of dependently typed languages, we use simply typed notation for $\Sigma\, x \!:\! A .\, B$ and $\Pi\, x \!:\! A .\, B$ when $x \not\in FVV(B)$, writing $A \times B$ and $A \to B$, respectively. 
\index{ A@$A \to B$ (non-dependent value function type)}
\index{ A@$A \times B$ (non-dependent value product type)}

eMLTT's \emph{computation types} include the type $FA$ of computations that return values of type $A$. Computation types also include computational variants of the $\Sigma$- and $\Pi$-types, written $\Sigma\, x \!:\! A .\, \ul{C}$ and $\Pi\, x \!:\! A .\, \ul{C}$. The former is a natural dependently typed generalisation of EEC's computational tensor type. The latter is a natural dependently typed generalisation of EEC and CBPV's computational function type. Based on this relationship, we use simply typed notation for $\Sigma\, x \!:\! A .\, \ul{C}$ and $\Pi\, x \!:\! A .\, \ul{C}$ when $x \not\in FVV(\ul{C})$, writing  $A \otimes\, \ul{C}$ and $A \to \ul{C}$, respectively. 
\index{ A@$A \to \ul{C}$ (non-dependent computational function type)}%
\index{ A@$A \otimes\, \ul{C}$ (non-dependent computational tensor type)}%
Compared to EEC and CBPV, we omit binary (and nullary) coproducts and products of 
computation types. In the models we study in this thesis, these are special cases 
of $\Sigma\, x \!:\! A .\, \ul{C}$ and $\Pi\, x \!:\! A .\, \ul{C}$. In order to define them 
in terms of $\Sigma\, x \!:\! A .\, \ul{C}$ and $\Pi\, x \!:\! A .\, \ul{C}$ in the language itself, 
one needs to extend eMLTT, e.g., either with large elimination forms or universes (see the next paragraph).

We note that as our focus is on the general principles of combining  
dependent types and computational effects, we treat type-dependency 
abstractly in most parts of this thesis, leaving the exact means through 
which one defines dependent types implicit (with the exception of  
propositional equality). Analogously to accommodating general 
inductive types, one can easily extend eMLTT using standard techniques. For example,  
one can extend its value and computation types 
with large elimination forms, such as 
\[
\begin{array}{r c l}
A & ::= & \ldots \,\,\,\vertbar\,\,\, \pmatch V {(x_1 \!:\! A_1, x_2 \!:\! A_2)} {} A
\\[2mm]
\ul{C} & ::= & \ldots \,\,\,\vertbar\,\,\, \pmatch V {(x_1 \!:\! A_1, x_2 \!:\! A_2)} {} \ul{C}
\end{array}
\]
and analogously for eliminating other value types. 
Alternatively, one could also extend eMLTT with universes 
of value and computation types, e.g., as discussed and used in Section~\ref{section:usinghandlersforreasoning}.
Finally, we note that in Section~\ref{sect:typedependencyonfeffects} we 
discuss a possible way to also accommodate type-dependency 
on computations directly, rather than only via thunks.

We also highlight that eMLTT is a minor extension of the language the author studied in the paper~\cite{Ahman:FibredEffects} large parts of this thesis are based on. eMLTT additionally includes the empty type, the coproduct type, and the homomorphic function type. The first two extensions better align it with dependently typed  languages such as Agda and Idris, and enable us to specify signatures of fibred algebraic effects in Chapter~\ref{chap:fibalgeffects}; the latter extension enables us to eliminate various value types into computations.
We also note that in Section~\ref{sect:continuousfamilies} we extend the core language presented in this chapter with general recursion, and in Chapters~\ref{chap:fibalgeffects} and~\ref{chap:handlers} with  algebraic effects and their handlers.

\begin{definition}
\label{def:terms}
eMLTT's \emph{value}, \emph{computation}, and \emph{homomorphism terms} are given by
\index{term!value --}
\[
\begin{array}{r c l @{\qquad\qquad}l}
V & ::= & x & \text{value variable}
\\
& \vertbar & \zero & \text{zero}
\\
& \vertbar & \suc V & \text{successor}
\\
& \vertbar & \natrec {x.\,A} {V_z} {y_1.\, y_2.\, V_s} {V} & \text{primitive recursion}
\\
& \vertbar & \star & \text{unit}
\\
& \vertbar & \langle V , W \rangle_{(x : A).\, B}  & \text{pairing}
\\
& \vertbar & \pmatch V {(x_1 \!:\! A_1, x_2 \!:\! A_2)} {y.\, B} W & \text{pattern-matching}
\\
& \vertbar & \lambda \, x \!:\! A .\, V & \text{lambda abstraction}
\\
& \vertbar & V(W)_{(x : A).\,B}  & \text{function application}
\\
& \vertbar & \absurd {x.\,A} V & \text{empty case analysis}
\\
& \vertbar & \inl {A + B} V & \text{left injection}
\\
& \vertbar & \inr {A + B} V & \text{right injection}
\\
& \vertbar & \mathtt{case~} V \mathtt{~of}_{x.\,B} \mathtt{~} ({\inl {\!} {\!\!(y_1 \!:\! A_1)} \mapsto W_1}, & \text{binary case analysis}
\\[-1mm]
&& \hspace{2.55cm} {\inr {\!} {\!\!(y_2 \!:\! A_2)} \mapsto W_2})
\\
& \vertbar & \refl A V & \text{reflexivity of}
\\[-2mm]
& & & \text{propositional equality}
\\
& \vertbar & \pathind A {x_1.\, x_2.\, x_3.\, B} {y.\, W} {V_1} {V_2} {V_p} & \text{elimination of}
\\[-2mm]
& & & \text{propositional equality}
\\
& \vertbar & \thunk M & \text{thunked computation}
\\
& \vertbar & \lambda\, z \!:\! \ul{C} .\, K & \text{homomorphic}
\\[-2mm]
& & & \text{lambda abstraction}
\\[5mm]
\index{term!computation --}
M & ::= & \return V & \text{returning a value}
\\
& \vertbar & \doto M {x \!:\! A} {\ul{C}} N & \text{sequential composition}
\\
& \vertbar & \langle V , M \rangle_{(x : A).\, \ul{C}} & \text{computational pairing}
\\
& \vertbar & \doto M {(x \!:\! A, z \!:\! \ul{C})} {\ul{D}} K & \text{computational}
\\[-2mm]
& & & \text{pattern-matching}
\\
& \vertbar & \lambda \, x \!:\! A .\, M & \text{computational}
\\[-2mm]
& & & \text{lambda abstraction}
\end{array}
\]

\[
\begin{array}{r c l @{\qquad\qquad}l}
& \vertbar & M(V)_{(x : A).\,\ul{C}}  & \text{computational}
\\[-2mm]
& & & \text{function application}
\\
& \vertbar & \force {\ul{C}} V & \text{forcing a}
\\[-2mm]
& & & \text{thunked computation}
\\
& \vertbar & V(M)_{\ul{C}, \ul{D}} & \text{homomorphic}
\\[-2mm]
& & & \text{function application}
\\[5mm]
\index{term!homomorphism --}
K & ::= & z & \text{computation variable}
\\
& \vertbar & \doto K {x \!:\! A} {\ul{C}} M & \text{sequential composition}
\\
& \vertbar & \langle V , K \rangle_{(x : A).\, \ul{C}} & \text{computational pairing}
\\
& \vertbar & \doto K {(x \!:\! A, z \!:\! \ul{C})} {\ul{D}} L & \text{computational}
\\[-2mm]
&&& \text{pattern-matching}
\\
& \vertbar & \lambda \, x \!:\! A .\, K & \text{computational}
\\[-2mm]
&&& \text{lambda abstraction}
\\
& \vertbar & K(V)_{(x : A).\, \ul{C}}  & \text{computational}
\\[-2mm]
&&& \text{function application}
\\
& \vertbar & V(K)_{\ul{C}, \ul{D}} & \text{homomorphic}
\\[-2mm]
&&& \text{function application}
\end{array}
\]
where
\begin{itemize}
\item in $\natrec {x.\,A} {V_z} {y_1.\, y_2.\, V_s} {V}$, the value variable $x$ is bound in $A$, and the value variables $y_1$ and $y_2$ are bound in $V_s$;
\item in $\langle V , W \rangle_{(x : A).\, B}$, the value variable $x$ is bound in $B$;
\item in $\pmatch V {(x_1 \!:\! A_1, x_2 \!:\! A_2)} {y.\, B} W$, the value variable $x_1$ is bound in $A_2$ and $W$, the value variable $x_2$ is bound in $W$, and the value variable $y$ is bound in $B$;
\item in $\lambda \, x \!:\! A .\, V$, the value variable $x$ is bound in $V$;
\item in $V(W)_{(x : A).\,B}$, the value variable $x$ is bound in $B$;
\item in $\absurd {x.\,A} V$, the value variable $x$ is bound in $A$
\item in $\case V {x.\, B} {\inl {\!} {\!\!(y_1 \!:\! A_1)} \mapsto W_1} {\inr {\!} {\!\!(y_2 \!:\! A_2)} \mapsto W_2}$, the value variable $x$ is bound in $B$, the value variable $y_1$ is bound in $W_1$, and the value variable $y_2$ is bound in $W_2$;
\item in $\pathind A {x_1.\, x_2.\, x_3.\, B} {y.\, W} {V_1} {V_2} {V_p}$, the value variables $x_1$, $x_2$,  and $x_3$ are bound in $B$, and the value variable $y$ is bound in $W$;
\item in $\lambda\, z \!:\! \ul{C} .\, K$, the computation variable $z$ is bound in $K$;
\item in $\doto M {x \!:\! A} {\ul{C}} N$, the value variable $x$ is bound in $N$;
\item in $\langle V , M \rangle_{(x : A).\, \ul{C}}$, the value variable $x$ is bound in $\ul{C}$;
\item in $\doto M {(x \!:\! A, z \!:\! \ul{C})} {\ul{D}} K$, the value variable $x$ is bound in $\ul{C}$ and $K$, and the computation variable $z$ is bound in $K$;
\item in $\lambda \, x \!:\! A .\, M$, the value variable $x$ is bound in $M$;
\item in $M(V)_{(x : A).\,\ul{C}}$, the value variable $x$ is bound in $\ul{C}$;
\item in $\doto K {x \!:\! A} {\ul{C}} M$, the value variable $x$ is bound in $M$;
\item in $\langle V , K \rangle_{(x : A).\, \ul{C}}$, the value variable $x$ is bound in $\ul{C}$;
\item in $\doto K {(x \!:\! A, z \!:\! \ul{C})} {\ul{D}} L$, the value variable $x$ is bound in $\ul{C}$ and $L$, and the computation variable $z$ is bound in $L$;
\item in $\lambda \, x \!:\! A .\, K$, the value variable $x$ is bound in $K$; and
\item in $K(V)_{(x : A).\, \ul{C}}$, the value variable $x$ is bound in $\ul{C}$.
\end{itemize}
\end{definition}

\index{ fst@$\fst$ (first projection, derived from pattern-matching)}
\index{ snd@$\snd$ (second projection, derived from pattern-matching)}
For better readability, we sometimes use left and right \emph{projections} instead of pattern-matching in our examples. These projections are derived from pattern-matching as
\[
\begin{array}{c}
\fst V \defeq \pmatch {V} {(x_1 \!:\! A , x_2 \!:\! B)} {} {x_1}
\qquad
\snd V \defeq \pmatch {V} {(x_1 \!:\! A , x_2 \!:\! B)} {} {x_2}
\end{array}
\]

We write $FVV(V)$, $FVV(M)$, and $FVV(K)$ 
for the sets of \emph{free value variables} of a value term $V$, a computation term $M$, and a homomorphism term $K$, respectively. In particular, the free value variables of terms also include the free variables of type annotations. Regarding\, \emph{free computation variables}, we have the following properties:

\begin{proposition}
\mbox{}
\begin{enumerate}
\item[(a)] Value types $A$, computation types $\ul{C}$, value terms $V$, and computation terms $M$ do not contain free computation variables.
\item[(b)] Every homomorphism term $K$ contains exactly one \emph{free computation variable} which we write  as $FCV(K)$.
\index{ FCV@$FCV(K)$ (free computation variable of $K$)}
\end{enumerate}
\end{proposition}

\begin{proof}
We prove $(a)$ and $(b)$ by simultaneous induction: the former by induction on the structure of $A$, $\ul{C}$, $V$, and $M$, and the latter by induction on the structure of $K$.
\end{proof}

eMLTT's \emph{value terms} coincide with the terms of MLTT, except for thunked computations 
$\thunk M$ and the homomorphic lambda abstraction $ \lambda\, z \!:\! \ul{C} .\, K$. 
In contrast to CBPV, eMLTT's value terms include elimination forms for value types  
in order to accommodate effect-free programs that the types of eMLTT could depend on.

eMLTT's \emph{computation terms} include standard combinators for programming with computational effects, namely, returning values and sequential composition of computations. They also include introduction and elimination forms for the computational $\Sigma$- and $\Pi$-types, forcing of thunked computations, and homomorphic function applications. 
We deliberately use similar notation for sequential composition and computational pattern-matching---compare $\doto M {x \!:\! A} {} N$ and $\doto M {(x \!:\! A, z \!:\! \ul{C})} {} K$---so as to emphasise that  the effects of $M$ are performed before the effects of $N$ and $K$, respectively. Further, notice that in order to account for the fact that $M$ produces a pair of a value and a computation, the second term in computational pattern-matching is necessarily a homomorphism term. 

eMLTT's \emph{homomorphism terms} are analogous to EEC's linear terms. Similarly to computation terms, they also include sequential composition, and introduction and elimination forms for the computational $\Sigma$- and $\Pi$-types. However, unlike computation terms, they do not include returning values and forcing of thunked computations but instead include computation variables $z$ that are required to be used linearly. Further, the definition of homomorphism terms also requires computation variables to be used so that effects of a computation bound to $z$ always ``happen first" in a term containing it. For example, when eliminating a computational pair $\pair V M$, the effects of $M$ are guaranteed to be performed before the effects of $K$ in ${\dtensorlet {(x \!:\! A , z \!:\! \ul{C})} {\pair V M} K}$.

This linear use of computation variables, together with leaving out returning values and forcing thunked computations, ensures that homomorphism terms denote algebra homomorphisms in the examples based on Eilenberg-Moore algebras of monads (see Section~\ref{sect:fibredmonadsandEMfibs}), or on algebraic effects (see Section~\ref{sect:fibalgeffectsmodel}): hence their name. 
A similar form of linearity is also present in CBPV with stacks, as defined in~\cite[\S 2.3.4]{Levy:CBPV}. Indeed,  homomorphism terms can be viewed as a programmer-friendly syntax for dependently typed CBPV stack terms (or equivalently, for one-hole evaluation contexts). 

Later, in Section~\ref{sect:alternativepresentations}, we also briefly discuss an alternative presentation of eMLTT in which one omits computation variables and homomorphism terms, and instead uses value variables and computation terms, in combination with equational proof obligations ensuring that the value variables are used analogously to computation variables.

It is worth observing that compared to the corresponding terms in CBPV and EEC, eMLTT's computation terms (resp.~homomorphism terms) do not include elimination forms for value types as they can be easily derived from the corresponding elimination forms included in eMLTT's value terms using thunking and forcing (resp.~homomorphic lambda abstraction and function application). We show this in Section~\ref{sect:derivableeliminationforms}.

The reader should also note that eMLTT's terms are decorated with more type annotations than one usually expects to see in a high-level (dependently typed) programming language. We later use these annotations to define the interpretation of eMLTT in fibred adjunction models by recursion on (raw) types and terms rather than the (non-unique) typing derivations. This is a standard technique in the literature  to avoid having to define the interpretation  simultaneously with proving its coherence, see~\cite{Streicher:Semantics,Hofmann:SyntaxAndSemantics}.
Nevertheless, we conjecture that this annotated syntax is equivalent to the corresponding unannotated syntax, based on analogous results for MLTT, e.g., see~\cite{Streicher:Semantics,Castellan:Report}. We leave a formal proof of this conjecture for future work. However, for better readability, we often omit these annotations in our examples and informal discussion.

We conclude this section by establishing some useful properties of substitution in eMLTT. 
In order to keep our definitions and propositions concise and readable, we refer to types and terms collectively as \emph{expressions} and use $E, \ldots$ to range over them. In detail, expressions are given by the following grammar:
\index{expression}
\index{ E@$E, \ldots$ (expressions, collective name for types and terms of eMLTT)}
\[
E \,\,\,\,::=\,\,\,\, A \,\,\,\,\vertbar\,\,\,\, \ul{C} \,\,\,\,\vertbar\,\,\,\, V \,\,\,\,\vertbar\,\,\,\, M \,\,\,\,\vertbar\,\,\,\, K
\]

\index{variable!-- convention}
When working with expressions involving bound variables, we follow the standard conventions:
i) we identify expressions that differ only in the names of bound variables, i.e., we identify expressions up-to $\alpha$-equivalence; and ii) we assume that in any mathematical context (definitions, theorems, proofs, etc.), the bound variables of expressions are chosen to be different from the free variables appearing in that context.

\begin{definition}
\label{def:substvaluevariables}
\index{substitution!-- of a value term}
The \emph{substitution of a value term $V$ for a value variable $x$ in an expression $E$}, written $E[V/x]$, 
\index{ E@$E[V/x]$ (substitution of $V$ for $x$ in $E$)}
is defined by recursion on the structure of $E$ as follows:
\[
\begin{array}{l c l}
\Nat[V/x] & \defeq & \Nat
\\
& \ldots &
\\
x[V/x] & \defeq & V
\\
y[V/x] & \defeq & y \qquad\qquad\qquad\qquad\qquad\qquad\qquad\qquad\,\,\,\, (\text{if~} x \neq y)
\\
& \ldots &
\\
(\return W)[V/x] & \defeq & \return (W[V/x])
\\
(\doto M {y \!:\! A} {\ul{C}} N)[V/x] & \defeq & \doto {M[V/x]} {y \!:\! A[V/x]} {\ul{C}[V/x]} {N[V/x]}
\\
& \ldots &
\end{array}
\]

\[
\hspace{-3.9cm}
\begin{array}{l c l}
(K(W)_{(y : A).\, \ul{C}})[V/x] & \defeq & (K[V/x])(W[V/x])_{(y : A[V/x]).\, \ul{C}[V/x]}
\\
(W(K)_{\ul{C}, \ul{D}})[V/x] & \defeq & (W[V/x])(K[V/x])_{\ul{C}[V/x], \ul{D}[V/x]}
\end{array}
\]
where the bound value variables are assumed to be different from the value variable $x$ we are substituting $W$ for, according to the variable conventions we have adopted.
\end{definition}

Later, in Section~\ref{sect:completeness}, we also demonstrate that this definition of unary 
substitutions naturally generalises to a definition of simultaneous substitutions. 
We then use simultaneous substitutions in op.~cit.~when constructing the classifying categorical model of eMLTT, and in Chapters~\ref{chap:fibalgeffects} and~\ref{chap:handlers} when 
extending eMLTT with algebraic effects and their handlers. Meanwhile, we found it more 
convenient to work with unary substitutions when presenting the well-formed syntax and meta-theory of eMLTT (Sections~\ref{sect:judgements} and~\ref{sect:metatheory}), and its denotational semantics and the soundness proof (Sections~\ref{sect:interpretation} and~\ref{sect:soundness}).

\begin{proposition}
Given a value variable $x$, a value term $V$, and an expression $E$, then $E[V/x]$ is the same kind of expression as $E$, e.g., if $E$ is a computation term, then  $E[V/x]$ is also a computation term.
\end{proposition}

\begin{proof}
By induction on the structure of $E$.
\end{proof}

\begin{proposition}
\label{prop:freevariablesofsubsstitution}
Given a value variable $x$, a value term $V$, and an expression $E$, then $FVV(E[V/x]) \subseteq (FVV(E) - \{x\}) \,\cup\, FVV(V)$.
\end{proposition}

\begin{proof}
By induction on the structure of $E$.
\end{proof}

\begin{proposition}
\label{prop:valuesubstlemma1}
Given a value variable $x$, a value term $V$, and an expression $E$ such that $x \not\in FVV(E)$, then $E[V/x] = E$.
\index{ FVV@$FVV(E)$ (set of free value variables of $E$)}
\end{proposition}

\begin{proof}
By induction on the structure of $E$.
\end{proof}

\begin{proposition}
\label{prop:valuesubstlemma2}
Given a value variable $x$ and an expression $E$, then $E[x/x] = E$.
\end{proposition}

\begin{proof}
By induction on the structure of $E$.
\end{proof}

\begin{proposition}
\label{prop:valuesubstlemma3}
Given value variables $x$ and $y$, value terms $W_1$ and $W_2$, and an expression $E$ such that $x \neq y$ and $x \not\in FVV(W)$, then $E[V/x][W/y] = E[W/y][V[W/y]/x]$.
\end{proposition}

\begin{proof}
By induction on the structure of $E$.
\end{proof}

\begin{definition}
\index{substitution!-- of a computation term}
The \emph{substitution of a computation term $M$ for a computation variable $FCV(K) = z$ in a homomorphism term $K$}, written $K[M/z]$, is defined by recursion on the structure of $K$ as follows:
\index{ K@$K[M/z]$ (substitution of $M$ for $z$ in $K$)}
\[
\begin{array}{l c l}
z[M/z] & \defeq & M
\\
(\doto K {x \!:\! A} {\ul{C}} N)[M/z] & \defeq & \doto {K[M/z]} {x \!:\! A} {\ul{C}} N
\\
(\langle V , K \rangle_{(x : A).\, \ul{C}})[M/z] & \defeq & \langle V , K[M/z] \rangle_{(x : A).\, \ul{C}}
\\
(\doto K {(x \!:\! A, z' \!:\! \ul{C})} {\ul{D}} L)[M/z] & \defeq & \doto {K[M/z]} {(x \!:\! A, z' \!:\! \ul{C})} {\ul{D}} L
\\
(\lambda \, x \!:\! A .\, K)[M/z] & \defeq & \lambda \, x \!:\! A .\, K[M/z]
\\
(K(V)_{(x : A).\, \ul{C}})[M/z] & \defeq & (K[M/z])(V)_{(x : A).\, \ul{C}}
\\
(V(K)_{\ul{C}, \ul{D}})[M/z] & \defeq & V(K[M/z])_{\ul{C}, \ul{D}}
\end{array}
\]
\end{definition}
\vspace{0.2cm}

\begin{proposition}
Given a homomorphism term $K$ with $FCV(K) = z$ and a computation term $M$, then $K[M/z]$ is a computation term.
\end{proposition}

\begin{proof}
By induction on the structure of $K$.
\end{proof}

\begin{proposition}
\label{prop:compsubstvaluesubst}
Given a homomorphism term $K$ with $FCV(K) = z$, a computation term $M$, a value variable $x$, and a value term $V$, then $K[M/z][V/x] = K[V/x][M[V/x]/z]$.
\end{proposition}

\begin{proof}
By induction on the structure of $K$.
\end{proof}

\begin{definition}
\index{substitution!-- of a homomorphism term}
The \emph{substitution of a homomorphism term $K$ \,for a computation variable $FCV(L) = z$ in a homomorphism term $L$}, written $L[K/z]$, is defined by recursion on the structure of $L$ as follows:
\index{ L@$L[K/z]$ (substitution of $K$ for $z$ in $L$)}
\[
\begin{array}{l c l}
z[K/z] & \defeq & K
\\
(\doto L {x \!:\! A} {\ul{C}} N)[K/z] & \defeq & \doto {L[K/z]} {x \!:\! A} {\ul{C}} N
\\
(\langle V , L \rangle_{(x : A).\, \ul{C}})[K/z] & \defeq & \langle V , L[K/z] \rangle_{(x : A).\, \ul{C}}
\\
(\doto {L_1} {(x \!:\! A, z' \!:\! \ul{C})} {\ul{D}} {L_2})[K/z] & \defeq & \doto {L_1[K/z]} {(x \!:\! A, z' \!:\! \ul{C})} {\ul{D}} {L_2}
\\
(\lambda \, x \!:\! A .\, L)[K/z] & \defeq & \lambda \, x \!:\! A .\, L[K/z]
\\
(L(V)_{(x : A).\, \ul{C}})[K/z] & \defeq & (L[K/z])(V)_{(x : A).\, \ul{C}}
\\
(V(L)_{\ul{C}, \ul{D}})[K/z] & \defeq & V(L[K/z])_{\ul{C}, \ul{D}}
\end{array}
\]
\end{definition}
\vspace{0.2cm}

\begin{proposition}
Given homomorphism terms $K$ and $L$ with $FCV(K) = z$, then $K[L/z]$ is also a homomorphism term.
\end{proposition}

\begin{proof}
By induction on the structure of $K$.
\end{proof}

\begin{proposition}
\label{prop:compvariableofhomsubst}
Given homomorphism terms $K$ and $L$ with $FCV(L) = z$, then $FCV(L[K/z]) = FCV(K)$.
\end{proposition}

\begin{proof}
By induction on the structure of $L$.
\end{proof}

\begin{proposition}
\label{prop:homsubstlemma}
Given a homomorphism term $K$ with $FCV(K) \!=\! z$, then $K[z/z] \!=\! K$.
\end{proposition}

\begin{proof}
By induction on the structure of $K$.
\end{proof}

\begin{proposition}
\label{prop:hompsubstvaluesubst}
Given homomorphism terms $K$ and $L$ with $FCV(L) = z$, a value variable $x$, and a value term $V$, then $L[K/z][V/x] = L[V/x][K[V/x]/z]$.
\end{proposition}

\begin{proof}
By induction on the structure of $K$.
\end{proof}

\begin{proposition}
\label{prop:hompsubstcompsubst}
Given homomorphism terms $K$ and $L$ with $FCV(L) = z_1$ and \linebreak $FCV(K) = z_2$, and a computation term $M$, then $L[K/z_1][M/z_2] = L[K[M/z_2]/z_1]$.
\end{proposition}

\begin{proof}
By induction on the structure of $K_1$.
\end{proof}

\begin{proposition}
\label{prop:hompsubsthomsubst}
Given homomorphism terms $K_1$, $K_2$, and $K_3$ with $FCV(K_1) = z_1$ and $FCV(K_2) = z_2$, then $K_1[K_2/z_1][K_3/z_2] = K_1[K_2[K_3/z_2]/z_1]$.
\end{proposition}

\begin{proof}
By induction on the structure of $K_1$.
\end{proof}

\section{Well-formed syntax and equational theory}
\label{sect:judgements}

In this section we present the well-formed syntax of eMLTT and its equational theory.

First, we define the notions of value and computation contexts, and prove some basic properties about the former.

\begin{definition}
\label{def:contexts}
\index{context!value --}
\index{ G@$\Gamma$ (value context)}
A \emph{value context} $\Gamma$ is a finite list $x_1 \!:\! A_1, \ldots, x_n \!:\! A_n$ of pairs of value variables $x_i$ and value types $A_i$ such that all the value variables $x_i$ are distinct. We write $\diamond$ for the \emph{empty value context} and $V\!ars(\Gamma)$ for the \emph{set of value variables} in $\Gamma$.
\index{ @$\diamond$ (empty value context)}
\index{ Vars@$Vars(\Gamma)$ (set of variables in $\Gamma$)}
\end{definition}

\begin{definition}
\index{context!computation --}
A \emph{computation context} $z \!:\! \ul{C}$ is a pair of a computation variable $z$ and a computation type $\ul{C}$.
\index{ z@$z : \ul{C}$ (computation context)}
\end{definition}

\begin{definition}
Given two value contexts $\Gamma_1$ and $\Gamma_2$, we say that $\Gamma_1$ and $\Gamma_2$ are \emph{disjoint} when $V\!ars(\Gamma_1) \,\cap\, V\!ars(\Gamma_2) = \emptyset$.
\end{definition}

\begin{definition}
Given two disjoint value contexts $\Gamma_1$ and $\Gamma_2$, their \emph{concatenation} is written $\Gamma_1,\Gamma_2$, and defined by recursion on the length of $\Gamma_2$:
\index{ G@$\Gamma_1,\Gamma_2$ (concatenation of value contexts)}
\[
\begin{array}{l c l}
\Gamma_1,\diamond & \defeq & \Gamma_1
\\
\Gamma_1, (\Gamma_2,x \!:\! A) & \defeq & (\Gamma_1,\Gamma_2), x \!:\! A
\end{array}
\]
where the second case is well-defined because of the disjointness assumption.
\end{definition}

\begin{proposition}
\label{prop:unionofvariablesinconctexts}
Given that $\Gamma_1,\Gamma_2$ exists, then $V\!ars(\Gamma_1,\Gamma_2) = V\!ars(\Gamma_1) \,\cup\, V\!ars(\Gamma_2)$.
\end{proposition}

\begin{proof}
By induction on the length of $\Gamma_2$.
\end{proof}

Next, we note that the notion of substitution of value terms for value variables extends straightforwardly from value types to value contexts:

\begin{definition}
\label{def:contextsubstitution}
The \emph{substitution of a value term $V$ for a value variable $x$ in a value context $\Gamma = y_1 \!:\! A_1, \ldots, y_n \!:\! A_n$}, written $\Gamma[V/x]$, is defined by 
\index{ G@$\Gamma[V/x]$ (substitution of $V$ for $x$ in $\Gamma$)}
\[
\Gamma[V/x] \defeq y_1 \!:\! A_1[V/x], \ldots, y_n \!:\! A_n[V/x]
\]
\end{definition}

\begin{proposition}
\label{prop:contextsubstlemma}
Propositions~\ref{prop:valuesubstlemma1},~\ref{prop:valuesubstlemma2}, and~\ref{prop:valuesubstlemma3} extend to Definition~\ref{def:contextsubstitution}.
\end{proposition}

\begin{proof}
By applying the cases of value types of Propositions~\ref{prop:valuesubstlemma1},~\ref{prop:valuesubstlemma2}, and~\ref{prop:valuesubstlemma3} to each of the value types $A_i$ in the given value context $\Gamma = y_1 \!:\! A_1, \ldots, y_n \!:\! A_n$.
\end{proof}

Finally, we define the well-formed syntax of eMLTT and its equational theory.
To this end, it is worth recalling that eMLTT is based on MLTT with intensional propositional equality. As a consequence, we do not include an $\eta$-equation for the elimination form of propositional equality. 
We also do not include an $\eta$-equation for primitive recursion. In both cases, we do so to avoid a known source of undecidability for the equational theory of eMLTT, see~\cite{Hofmann:Thesis} and~\cite{Okada:Rewriting}, respectively. While we do not investigate normalisation of eMLTT's equational theory in this thesis, it would be an important property for future implementations of eMLTT (see Section~\ref{sect:normalisationandimplementation}). 

In addition, as discussed in Section~\ref{sect:twoguidingquestions}, the rules for sequential composition (both for computation and homomorphism terms) disallow the type of the second computation to depend on the variable bound by sequential composition. As also discussed in Section~\ref{sect:twoguidingquestions}, we can uniformly recover this type-dependency using the computational $\Sigma$-type. A similar restriction on type-dependency also appears in the rules for computational pattern-matching; this type-dependency can be recovered analogously to the case of sequential composition. These restrictions on type-dependency enable us to give eMLTT a denotational semantics using a natural fibrational generalisation of the adjunction-based semantics of CBPV and EEC---see Chapters~\ref{chap:fibadjmodels} and~\ref{chap:interpretation} for details.

\begin{definition}
\label{def:judgements}
\index{well-formed syntax}
The \emph{well-formed syntax} of eMLTT and its \emph{equational theory} are defined using the following judgement forms:
\[
\begin{array}{l @{\qquad\qquad}l}
\index{context!value --!well-formed --}
\vdash \Gamma & \text{well-formed value context}
\\
\vdash \Gamma_1 = \Gamma_2 & \text{definitionally equal value contexts}
\\
\index{type!value --!well-formed --}
\lj \Gamma A & \text{well-formed value type}
\\
\ljeq \Gamma A B & \text{definitionally equal value types}
\\
\index{type!computation --!well-formed --}
\lj \Gamma \ul{C} & \text{well-formed computation type}
\\
\ljeq \Gamma {\ul{C}} {\ul{D}} & \text{definitionally equal computation types}
\\
\index{term!value --!well-typed --}
\vj \Gamma V A & \text{well-typed value term}
\\
\veq \Gamma V W A & \text{definitionally equal value terms}
\\
\index{term!computation --!well-typed --}
\cj \Gamma M \ul{C} & \text{well-typed computation term}
\\
\ceq \Gamma M N \ul{C} & \text{definitionally equal computation terms}
\\
\index{term!homomorphism --!well-typed --}
\hj \Gamma {z \!:\! \ul{C}} K \ul{D} & \text{well-typed homomorphism term}
\\
\heq \Gamma {z \!:\! \ul{C}} K L \ul{D} & \text{definitionally equal homomorphism terms}
\end{array}
\]
\end{definition}
\vspace{0.2cm}

\noindent 
These judgements are defined mutually inductively, using the rules given below.
We have organised these rules so that closely-related rules for different judgements are grouped together. For example, we group the formation rule for the type of natural numbers together with the corresponding typing rules and definitional equations.

\paragraph*{Well-formed value contexts} \mbox{}

\noindent
Formation rules for value contexts:
\vspace{0.15cm}
\[
\mkrule
{\vdash \diamond}
{}
\qquad
\mkrule
{\vdash \Gamma, x \!:\! A}
{\vdash \Gamma \quad \lj \Gamma A \quad x \not\in V\!ars(\Gamma)}
\]

\noindent
Rules for definitionally equal value contexts:
\vspace{0.15cm}
\[
\mkrule
{\ljeq {} \diamond \diamond}
{}
\qquad
\mkrule
{\ljeq {} {\Gamma_1, x \!:\! A} {\Gamma_2, x \!:\! B}}
{\ljeq {} {\Gamma_1} {\Gamma_2} \quad \ljeq {\Gamma_1} A B \quad x \not\in V\!ars(\Gamma_1) \quad x \not\in V\!ars(\Gamma_2)}
\]

\paragraph*{Context and type conversions} \mbox{}

\noindent
Context conversion rules for types:
\vspace{0.15cm}
\[
\begin{array}{c}
\mkrule
{\lj {\Gamma_2} A}
{\lj {\Gamma_1} A \quad \ljeq {} {\Gamma_1} {\Gamma_2}}
\qquad
\mkrule
{\lj {\Gamma_2} {\ul{C}}}
{\lj {\Gamma_1} {\ul{C}} \quad \ljeq {} {\Gamma_1} {\Gamma_2}}
\\[6mm]
\mkrule
{\ljeq {\Gamma_2} A B}
{\ljeq {\Gamma_1} A B \quad \ljeq {} {\Gamma_1} {\Gamma_2}}
\qquad
\mkrule
{\ljeq {\Gamma_2} {\ul{C}} {\ul{D}}}
{\ljeq {\Gamma_1} {\ul{C}} {\ul{D}} \quad \ljeq {} {\Gamma_1} {\Gamma_2}}
\end{array}
\]

\noindent
Context and type conversion rules for terms:
\vspace{0.15cm}
\[
\begin{array}{c}
\mkrule
{\vj {\Gamma_2} V B}
{\vj {\Gamma_1} V A \quad \ljeq {} {\Gamma_1} {\Gamma_2} \quad \ljeq {\Gamma_1} A B}
\qquad
\mkrule
{\cj {\Gamma_2} M {\ul{D}}}
{\cj {\Gamma_1} M {\ul{C}} \quad \ljeq {} {\Gamma_1} {\Gamma_2} \quad \ljeq {\Gamma_1} {\ul{C}} {\ul{D}}}
\\[6mm]
\mkrule
{\hj {\Gamma_2} {z \!:\! \ul{C}_2} K {\ul{D}_2}}
{\hj {\Gamma_1} {z \!:\! \ul{C}_1} K {\ul{D}_1} \quad \ljeq {} {\Gamma_1} {\Gamma_2} \quad \ljeq {\Gamma_1} {\ul{C}_1} {\ul{C}_2} \quad \ljeq {\Gamma_1} {\ul{D}_1} {\ul{D}_2}}
\\[6mm]
\mkrule
{\veq {\Gamma_2} V W B}
{\veq {\Gamma_1} V W A \quad \ljeq {} {\Gamma_1} {\Gamma_2} \quad \ljeq {\Gamma_1} A B}
\\[6mm]
\mkrule
{\ceq {\Gamma_2} M N {\ul{D}}}
{\ceq {\Gamma_1} M N {\ul{C}} \quad \ljeq {} {\Gamma_1} {\Gamma_2} \quad \ljeq {\Gamma_1} {\ul{C}} {\ul{D}}}
\\[6mm]
\mkrule
{\heq {\Gamma_2} {z \!:\! \ul{C}_2} K L {\ul{D}_2}}
{\heq {\Gamma_1} {z \!:\! \ul{C}_1} K L {\ul{D}_1} \quad \ljeq {} {\Gamma_1} {\Gamma_2} \quad \ljeq {\Gamma_1} {\ul{C}_1} {\ul{C}_2} \quad \ljeq {\Gamma_1} {\ul{D}_1} {\ul{D}_2}}
\end{array}
\]

\paragraph*{Reflexivity, symmetry and transitivity} \mbox{}

\noindent
Rules for reflexivity:
\vspace{0.15cm}
\[
\begin{array}{c}
\mkrule
{\ljeq \Gamma A A}
{\lj \Gamma A}
\qquad
\mkrule
{\ljeq \Gamma {\ul{C}} {\ul{C}}}
{\lj \Gamma {\ul{C}}}
\\[6mm]
\mkrule
{\veq \Gamma V V A}
{\vj \Gamma V A}
\qquad
\mkrule
{\ceq \Gamma M M {\ul{C}}}
{\cj \Gamma M {\ul{C}}}
\qquad
\mkrule
{\heq \Gamma {z \!:\! \ul{C}} K K {\ul{D}}}
{\hj \Gamma {z \!:\! \ul{C}} K {\ul{D}}}
\end{array}
\]

\noindent
Rules for symmetry:
\vspace{0.15cm}
\[
\begin{array}{c}
\mkrule
{\ljeq \Gamma A B}
{\ljeq \Gamma B A}
\qquad
\mkrule
{\ljeq \Gamma {\ul{C}} {\ul{D}}}
{\ljeq \Gamma {\ul{D}} {\ul{C}}}
\\[6mm]
\mkrule
{\veq \Gamma V W A}
{\veq \Gamma W V A}
\qquad
\mkrule
{\ceq \Gamma M N {\ul{C}}}
{\ceq \Gamma N M {\ul{C}}}
\qquad
\mkrule
{\heq \Gamma {z \!:\! \ul{C}} K L {\ul{D}}}
{\heq \Gamma {z \!:\! \ul{C}} L K {\ul{D}}}
\end{array}
\]

\noindent
Rules for transitivity:
\vspace{0.15cm}
\[
\begin{array}{c}
\mkrule
{\ljeq \Gamma {A_1} {A_3}}
{\ljeq \Gamma {A_1} {A_2} \quad \ljeq \Gamma {A_2} {A_3}}
\qquad
\mkrule
{\ljeq \Gamma {\ul{C}_1} {\ul{C}_3}}
{\ljeq \Gamma {\ul{C}_1} {\ul{C}_2} \quad \ljeq \Gamma {\ul{C}_2} {\ul{C}_3}}
\\[6mm]
\mkrule
{\veq \Gamma {V_1} {V_3} A}
{\veq \Gamma {V_1} {V_2} A \quad \veq \Gamma {V_2} {V_3} A}
\qquad
\mkrule
{\ceq \Gamma {M_1} {M_3} {\ul{C}}}
{\ceq \Gamma {M_1} {M_2} {\ul{C}} \quad \ceq \Gamma {M_2} {M_3} {\ul{C}}}
\\[6mm]
\mkrule
{\heq \Gamma {z \!:\! \ul{C}} {K_1} {K_3} {\ul{D}}}
{\heq \Gamma {z \!:\! \ul{C}} {K_1} {K_2} {\ul{D}} \quad \heq \Gamma {z \!:\! \ul{C}} {K_2} {K_3} {\ul{D}}}
\end{array}
\]

\paragraph*{Replacement} \mbox{}

\noindent
Replacement rules for value and computation types:
\vspace{0.25cm}
\[
\begin{array}{c}
\mkrule
{\ljeq {\Gamma_1, \Gamma_2[V_1/x]} {B[V_1/x]} {B[V_2/x]}}
{\lj {\Gamma_1, x \!:\! A, \Gamma_2} B \quad \veq {\Gamma_1} {V_1} {V_2} A}
\qquad
\mkrule
{\ljeq {\Gamma_1, \Gamma_2[V_1/x]} {\ul{C}[V_1/x]} {\ul{C}[V_2/x]}}
{\lj {\Gamma_1, x \!:\! A, \Gamma_2} \ul{C} \quad \veq {\Gamma_1} {V_1} {V_2} A}
\end{array}
\]

\vspace{0.25cm}
\noindent
Replacement rules for value, computation, and homomorphism terms:
\vspace{0.25cm}
\[
\begin{array}{c}
\mkrule
{\veq {\Gamma_1, \Gamma_2[V_1/x]} {W[V_1/x]} {W[V_2/x]} {B[V_1/x]}}
{\vj {\Gamma_1, x \!:\! A, \Gamma_2} W B \quad \veq {\Gamma_1} {V_1} {V_2} A}
\\[6mm]
\mkrule
{\ceq {\Gamma_1, \Gamma_2[V_1/x]} {M[V_1/x]} {M[V_2/x]} {\ul{C}[V_1/x]}}
{\cj {\Gamma_1, x \!:\! A, \Gamma_2} M {\ul{C}} \quad \veq {\Gamma_1} {V_1} {V_2} A}
\\[6mm]
\mkrule
{\heq {\Gamma_1, \Gamma_2[V_1/x]} {z \!:\! \ul{C}[V_1/x]} {K[V_1/x]} {K[V_2/x]} {\ul{D}[V_1/x]}}
{\hj {\Gamma_1, x \!:\! A, \Gamma_2} {z \!:\! \ul{C}} K {\ul{D}} \quad \veq {\Gamma_1} {V_1} {V_2} A}
\\[6mm]
\mkrule
{\ceq {\Gamma} {K[M/z]} {K[N/z]} {\ul{D}}}
{\hj {\Gamma} {z \!:\! \ul{C}} K {\ul{D}} \quad \ceq {\Gamma} {M} {N} {\ul{C}}}
\\[6mm]
\mkrule
{\heq {\Gamma} {z_2 \!:\! \ul{C}} {K[L_1/z_1]} {K[L_2/z_1]} {\ul{D}_2}}
{\hj {\Gamma} {z_1 \!:\! \ul{D}_1} K {\ul{D}_2} \quad \heq {\Gamma} {z_2 \!:\! \ul{C}} {L_1} {L_2} {\ul{D}_1}}
\end{array}
\]

\paragraph*{Variables} \mbox{}

\noindent
Typing rules for value and computation variables:
\[
\mkrule
{\vj {\Gamma_1, x \!:\! A, \Gamma_2} x A}
{\lj {} {\Gamma_1, x \!:\! A, \Gamma_2}}
\qquad
\mkrule
{\hj {\Gamma} {z \!:\! \ul{C}} {z} {\ul{C}}}
{
\lj \Gamma {\ul{C}}
}
\]

\paragraph*{Natural numbers} \mbox{}

\noindent
Formation rule for the type of natural numbers:
\[
\mkrule
{\lj \Gamma {\Nat}}
{\vdash \Gamma}
\]

\noindent
Typing rules for zero, successor, and primitive recursion:
\[
\begin{array}{c}
\mkrule
{\vj \Gamma {\zero} {\Nat}}
{\vdash \Gamma}
\qquad
\mkrule
{\vj \Gamma {\suc V} {\Nat}}
{\vj \Gamma V \Nat}
\\[5mm]
\mkrule
{\vj \Gamma {\natrec {x.\,A} {V_z} {y_1.\,y_2.\,V_s} {V}} {A[V/x]}}
{
\begin{array}{c}
\lj {\Gamma, x \!:\! \Nat} A \quad \vj \Gamma V \Nat 
\\[-1mm]
\vj \Gamma {V_z} {A[\zero/x]} \quad \vj {\Gamma, y_1 \!:\! \Nat, y_2 \!:\! A[y_1/x]} {V_s} {A[\suc y_1/x]}
\end{array}
}
\end{array}
\]

\noindent
Congruence rules for successor and primitive recursion:
\[
\begin{array}{c}
\mkrule
{\veq \Gamma {\suc V} {\suc W} {\Nat}}
{\veq \Gamma V W {\Nat}}
\\[5mm]
\mkrule
{\veq \Gamma {\natrec {x.\,A} {V_z} {y_1.\,y_2.\,V_s} {V}} {\natrec {x.\,B} {W_z} {y_1.\,y_2.\,W_s} {W}} {A[V/x]}}
{
\begin{array}{c}
\ljeq {\Gamma, x \!:\! \Nat} {A} {B}
\quad
\veq \Gamma V W {\Nat}
\\[-1mm]
\veq \Gamma {V_z} {W_z} {A[\zero/x]}
\quad
\veq {\Gamma, y_1 \!:\! \Nat, y_2 \!:\! A[y_1/x]} {V_s} {W_s} {A[\suc y_1/x]}
\end{array}
}
\end{array}
\]

\noindent
$\beta$-equations for primitive recursion:
\[
\begin{array}{c}
\mkrule
{\veq \Gamma {\natrec {x.\,A} {V_z} {y_1.\,y_2.\,V_s} {\zero}} {V_z} {A[\zero/x]}}
{\lj {\Gamma, x \!:\! \Nat} A \quad \vj \Gamma {V_z} {A[\zero/x]} \quad \vj {\Gamma, y_1 \!:\! \Nat, y_2 \!:\! A[y_1/x]} {V_s} {A[\suc y_1/x]}}
\\[5mm]
\mkrule
{
\begin{array}{c@{~} c@{~} l}
\Gamma & \vdash & {\natrec {x.\,A} {V_z} {y_1.\,y_2.\,V_s} {\suc V}} 
\\[-1mm]
& = & {V_s[V/y_1][\natrec {x.A} {V_z} {y_1.\,y_2.\,V_s} {V}/y_2]} : {A[\suc V/x]}
\end{array}
}
{
\begin{array}{c}
\lj {\Gamma, x \!:\! \Nat} A \quad \vj \Gamma V \Nat 
\\[-1mm]
\vj \Gamma {V_z} {A[\zero/x]} \quad \vj {\Gamma, y_1 \!:\! \Nat, y_2 \!:\! A[y_1/x]} {V_s} {A[\suc y_1/x]}
\end{array}
}
\end{array}
\]

\paragraph*{Unit type} \mbox{}

\noindent
Formation rule for the unit type:
\[
\mkrule
{\lj \Gamma 1}
{\vdash \Gamma}
\]

\noindent
Typing rule for the unit:
\[
\mkrule
{\vj \Gamma \star 1}
{\vdash \Gamma}
\]

\noindent
$\eta$-equation for the unit:
\[
\mkrule
{\veq \Gamma V \star 1}
{\vj \Gamma V 1}
\] 

\paragraph*{Value $\Sigma$-type} \mbox{}

\noindent
Formation rule for the value $\Sigma$-type:
\[
\mkrule
{\lj \Gamma {\Sigma \, x \!:\! A .\, B}}
{\lj {\Gamma, x \!:\! A} B}
\]

\noindent
Typing rules for pairing and pattern-matching:
\[
\begin{array}{c}
\mkrule
{\vj \Gamma {\langle V , W \rangle_{(x : A).\,B}} {\Sigma \, x \!:\! A .\, B}}
{\vj \Gamma V A \quad \lj {\Gamma, x \!:\! A} B \quad \vj \Gamma W {B[V/x]}}
\\[5mm]
\mkrule
{\vj \Gamma {\pmatch V {(x_1 \!:\! A_1, x_2 \!:\! A_2)} {y.\,B} W} {B[V/y]}}
{
\begin{array}{c}
\lj {\Gamma, y \!:\! \Sigma \, x_1 \!:\! A_1 .\, A_2} B 
\\[-1mm]
\vj \Gamma V {\Sigma \, x_1 \!:\! A_1 .\, A_2} \quad \vj {\Gamma, x_1 \!:\! A_1, x_2 \!:\! A_2} {W} {B[\langle x_1 , x_2 \rangle_{(x_1 : A_1).\,A_2} /y]}
\end{array}
}
\end{array}
\]

\noindent
Congruence rules for the value $\Sigma$-type, pairing, and pattern-matching:
\[
\begin{array}{c}
\mkrule
{\ljeq \Gamma {\Sigma \, x \!:\! A_1 .\, B_1} {\Sigma \, x \!:\! A_2 .\, B_2}}
{\ljeq \Gamma {A_1} {A_2} \quad \ljeq {\Gamma, x \!:\! A_1} {B_1} {B_2}}
\\[6mm]
\mkrule
{\veq \Gamma {\langle V_1 , W_1 \rangle_{(x : A_1).\, B_1}} {\langle V_2 , W_2 \rangle_{(x : A_2).\, B_2}} {\Sigma \, x \!:\! A_1 . B_1}}
{\ljeq \Gamma {A_1} {A_2} \quad \ljeq {\Gamma, x \!:\! A_1} {B_1} {B_2} \quad \veq \Gamma {V_1} {V_2} {A_1} \quad \veq \Gamma {W_1} {W_2} {B_1[V_1/x]}}
\\[5mm]
\mkrule
{
\begin{array}{c@{~} c@{~} l}
\Gamma & \vdash & {\pmatch {V_1} {(x_1 \!:\! A_{11}, x_2 \!:\! A_{21})} {y.\, B_1} {W_1}} 
\\[-1mm]
& = & {\pmatch {V_2} {(x_1 \!:\! A_{12}, x_2 \!:\! A_{22})} {y.\, B_2} {W_2}} : {B_1[V_1/y]}
\end{array}
}
{
\begin{array}{c}
\ljeq \Gamma {A_{11}} {A_{12}} \quad \ljeq {\Gamma, x_1 \!:\! A_{11}} {A_{21}} {A_{22}} \quad \ljeq {\Gamma, y \!:\! \Sigma \, x_1 \!:\! A_{11} .\, A_{21}} {B_1} {B_2} 
\\[-1mm]
\veq \Gamma {V_1} {V_2} {\Sigma \, x_1 \!:\! A_{11} .\, A_{21}} \quad \veq {\Gamma, x_1 \!:\! A_{11}, x_2 \!:\! A_{21}} {W_1} {W_2} {B_1[\langle x_1 , x_2 \rangle_{(x_1 : A_{11}).\,A_{21}} /y]}
\end{array}
}
\end{array}
\]

\noindent
$\beta$- and $\eta$-equations  for pattern-matching:
\[
\begin{array}{c}
\mkrule
{
\begin{array}{c@{~} c@{~} l}
\Gamma & \vdash & {\pmatch {\langle V_1 , V_2 \rangle_{(x_1 : A_1).\,A_2}} {(x_1 \!:\! A_1, x_2 \!:\! A_2)} {y.\,B} W} 
\\
& = & {W[V_1/x_1][V_2/x_2]} : {B[\langle V_1 , V_2 \rangle_{(x_1 : A_1).\,A_2}/y]}
\end{array}
}
{
\begin{array}{c}
\lj {\Gamma, y \!:\! \Sigma \, x_1 \!:\! A_1 .\, A_2} B
\quad
\vj \Gamma {V_1} {A_1} \quad \vj \Gamma {V_2} {A_2[V_1/x_1]}
\\[-1mm]
\vj {\Gamma, x_1 \!:\! A_1, x_2 \!:\! A_2} {W} {B[\langle x_1 , x_2 \rangle_{(x_1 : A_1).\,A_2} /y]}
\end{array}
}
\\[9mm]
\mkrule
{\veq \Gamma {\pmatch V {(x_1 \!:\! A_1, x_2 \!:\! A_2)} {y_1.\,B} {W[{\langle x_1 , x_2 \rangle_{(x_1 : A_1).\,A_2}}/y_2]}} {W[V/y_2]} {B[V/y_1]}}
{
\begin{array}{c}
\lj \Gamma {A_1} \quad \lj {\Gamma, x_1 \!:\! A_1} {A_2}
\quad
\vj \Gamma V {\Sigma \, x_1 \!:\! A_1 .\, A_2}  
\\[-1mm]
\lj {\Gamma, y_1 \!:\! \Sigma \, x_1 \!:\! A_1 .\, A_2} B \quad \vj {\Gamma, y_2 \!:\! \Sigma \, x_1 \!:\! A_1 .\, A_2} {W} {B[y_2/y_1]}
\end{array}
}
\end{array}
\]

\paragraph*{Value $\Pi$-type} \mbox{}

\noindent
Formation rule for the value $\Pi$-type:
\[
\mkrule
{\lj \Gamma {\Pi \, x \!:\! A .\, B}}
{\lj {\Gamma, x \!:\! A} B}
\]

\noindent
Typing rules for lambda abstraction and function application:
\[
\mkrule
{\vj \Gamma {\lambda \, x \!:\! A .\, V} {\Pi \, x \!:\! A .\, B}}
{\vj {\Gamma, x \!:\! A} V B}
\qquad
\mkrule
{\vj \Gamma {V(W)_{(x : A).\, B}} {B[W/x]}}
{\lj {\Gamma, x \!:\! A} {B} \quad \vj \Gamma V {\Pi \, x \!:\! A .\, B} \quad \vj \Gamma W A}
\]

\noindent
Congruence rules for the value $\Pi$-type, lambda abstraction, and function application:
\[
\begin{array}{c}
\mkrule
{\ljeq \Gamma {\Pi \, x \!:\! A_1 .\, B_1} {\Pi \, x \!:\! A_2 .\, B_2}}
{\ljeq \Gamma {A_1} {A_2} \quad \ljeq {\Gamma, x \!:\! A_1} {B_1} {B_2}}
\\[6mm]
\mkrule
{\veq \Gamma {\lambda \, x \!:\! A_1 .\, V_1} {\lambda \, x \!:\! A_2 .\, V_2} {\Pi \, x \!:\! A_1 .\, B}}
{\ljeq \Gamma {A_1} {A_2} \quad \veq {\Gamma, x \!:\! A_1} {V_1} {V_2} B}
\\[6mm]
\mkrule
{\veq \Gamma {V_1(W_1)_{(x : A_1).\, B_1}} {V_2(W_2)_{(x : A_2).\, B_2}} {B_1[W_1/x]}}
{\ljeq \Gamma {A_1} {A_2} \quad \ljeq {\Gamma, x \!:\! A_1} {B_1} {B_2} \quad \veq \Gamma {V_1} {V_2} {\Pi \, x \!:\! A_1 .\, B_1} \quad \veq {\Gamma} {W_1} {W_2} {A_1}}
\end{array}
\]

\noindent
$\beta$- and $\eta$-equations for lambda abstraction and function application:
\[
\begin{array}{c}
\mkrule
{\veq \Gamma {(\lambda \, x \!:\! A .\, V)(W)_{(x : A) .\, B}} {V[W/x]} {B[W/x]}}
{\vj {\Gamma, x \!:\! A} V B \quad \vj \Gamma W A}
\\[6mm]
\mkrule
{\veq \Gamma {V} {\lambda \, x \!:\! A .\, V(x)_{(x : A) .\, B}} {\Pi \, x \!:\! A .\, B}}
{\lj {\Gamma, x \!:\! A} {B} \quad \vj \Gamma V {\Pi \, x \!:\! A .\, B}}
\end{array}
\]

\paragraph*{Empty type} \mbox{}

\noindent
Formation rule for the empty type:
\[
\mkrule
{\lj \Gamma 0}
{\vdash \Gamma}
\]

\noindent
Typing rule for empty case analysis:
\[
\mkrule
{\vj \Gamma {\absurd {x.\, A} V} {A[V/x]}}
{\lj {\Gamma, x \!:\! 0} A \quad \vj \Gamma V 0}
\]

\noindent
Congruence rule for empty case analysis:
\[
\mkrule
{\veq \Gamma {\absurd {x.\, A_1} {V_1}} {\absurd {x.\, A_2} {V_2}} {A_1[V_1/x]}}
{\ljeq {\Gamma, x \!:\! 0} {A_1} {A_2} \quad \veq \Gamma {V_1} {V_2} 0}
\]

\noindent
$\eta$-equation for empty case analysis:
\[
\mkrule
{\veq \Gamma {\absurd {x.\, A} V} {W[V/x]} {A[V/x]}}
{\lj {\Gamma, x \!:\! 0} A \quad \vj \Gamma V 0 \quad \vj {\Gamma, x \!:\! 0} W A}
\]

\paragraph*{Coproduct type} \mbox{}

\noindent
Formation rule for the coproduct type:
\[
\mkrule
{\lj \Gamma {A + B}}
{\lj \Gamma A \quad \lj \Gamma B}
\]

\noindent
Typing rules for the left and right injections, and binary case analysis:
\[
\begin{array}{c}
\mkrule
{\vj \Gamma {\inl {A + B} V} {A + B}}
{\vj \Gamma V A}
\qquad
\mkrule
{\vj \Gamma {\inr {A + B} V} {A + B}}
{\vj \Gamma V B}
\\[5mm]
\mkrule
{\vj \Gamma {\case {V} {x.\, B} {\inl {\!} {\!\!(y_1 \!:\! A_1)} \mapsto W_1} {\inr {\!} {\!\!(y_2 \!:\! A_2)} \mapsto W_2}} {B[V/x]}}
{
\begin{array}{c}
\lj {\Gamma, x \!:\! A_1 + A_2} B \quad \vj \Gamma V {A_1 + A_2} 
\\[-1mm]
\vj {\Gamma, y_1 \!:\! A_1} {W_1} {B[\inl {A_1 + A_2} y_1/x]} \quad \vj {\Gamma, y_2 \!:\! A_2} {W_2} {B[\inr {A_1 + A_2} y_2/x]}
\end{array}
}
\end{array}
\]

\noindent
Congruence rules for the coproduct type, left and right injections, and binary case analysis:
\[
\begin{array}{c}
\mkrule
{\ljeq \Gamma {A_1 + B_1} {A_2 + B_2}}
{\ljeq \Gamma {A_1} {A_2} \quad \ljeq \Gamma {B_1} {B_2}}
\\[6mm]
\mkrule
{\veq \Gamma {\inl {A_1 + B_1} {V_1}} {\inl {A_2 + B_2} {V_2}} {A_1 + B_1}}
{\ljeq \Gamma {A_1} {A_2} \quad \ljeq \Gamma {B_1} {B_2} \quad \veq \Gamma {V_1} {V_2} {A_1}}
\\[6.5mm]
\mkrule
{\veq \Gamma {\inr {A_1 + B_1} {V_1}} {\inr {A_2 + B_2} {V_2}} {A_1 + B_1}}
{\ljeq \Gamma {A_1} {A_2} \quad \ljeq \Gamma {B_1} {B_2} \quad \veq \Gamma {V_1} {V_2} {B_1}}
\\[5mm]
\mkrule
{
\begin{array}{c@{~} c@{~} l}
\Gamma & \vdash & {\case {V_1} {x.\, B_1} {\inl {\!} {\!\!(y_1 \!:\! A_{11})} \mapsto W_{11}} {\inr {\!} {\!\!(y_2 \!:\! A_{21})} \mapsto W_{21}}}
\\ 
& = & {\case {V_2} {x.\, B_2} {\inl {\!} {\!\!(y_1 \!:\! A_{12})} \mapsto W_{12}} {\inr {\!} {\!\!(y_2 \!:\! A_{22})} \mapsto W_{22}}} : {B_1[V_1/x]}
\end{array}
}
{
\begin{array}{c}
\ljeq \Gamma {A_{11}} {A_{12}} \quad \ljeq \Gamma {A_{21}} {A_{22}} \quad \ljeq {\Gamma, x \!:\! A_{11} + A_{21}} {B_1} {B_2} 
\\[-1mm]
\veq \Gamma {V_1} {V_2} {A_{11} + A_{21}}  \quad \veq {\Gamma, y_1 \!:\! A_{11}} {W_{11}} {W_{12}} {B_1[\inl {A_{11} + A_{21}} y_1/x]} 
\\[-1mm]
\veq {\Gamma, y_2 \!:\! A_{12}} {W_{21}} {W_{22}} {B_1[\inr {A_{12} + A_{22}} y_2/x]} 
\end{array}
}
\end{array}
\]

\noindent
$\beta$- and $\eta$-equations for binary case analysis:
\[
\begin{array}{c}
\mkrule
{
\begin{array}{c@{~} c@{~} l}
\Gamma & \vdash & {\case {(\inl {A_1 + A_2} V)} {x.\, B} {\inl {\!} {\!\!(y_1 \!:\! A_1)} \mapsto W_1} {\inr {\!} {\!\!(y_2 \!:\! A_2)} \mapsto W_2}} 
\\
& = & {W_1[V/y_1]} : {B[\inl {A_1 + A_2} V/x]}
\end{array}
}
{
\begin{array}{c}
\lj {\Gamma, x \!:\! A_1 + A_2} B \quad \vj \Gamma V {A_1} 
\\
\vj {\Gamma, y_1 \!:\! A_1} {W_1} {B[\inl {A_1 + A_2} y_1/x]} \quad \vj {\Gamma, y_2 \!:\! A_2} {W_2} {B[\inr {A_1 + A_2} y_2/x]}
\end{array}
}
\\[9mm]
\mkrule
{
\begin{array}{c@{~} c@{~} l}
\Gamma & \vdash & {\case {(\inr {A_1 + A_2} V)} {x.\, B} {\inl {\!} {\!\!(y_1 \!:\! A_1)} \mapsto W_1} {\inr {\!} {\!\!(y_2 \!:\! A_2)} \mapsto W_2}} 
\\
& = & {W_2[V/y_2]} : {B[\inr {A_1 + A_2} V/x]}
\end{array}
}
{
\begin{array}{c}
\lj {\Gamma, x \!:\! A_1 + A_2} B \quad \vj \Gamma V {A_2} 
\\[-1mm]
\vj {\Gamma, y_1 \!:\! A_1} {W_1} {B[\inl {A_1 + A_2} y_1/x]} \quad \vj {\Gamma, y_2 \!:\! A_2} {W_2} {B[\inr {A_1 + A_2} y_2/x]}
\end{array}
}
\\[9mm]
\mkrule
{
\begin{array}{c}
\hspace{-3.7cm} \Gamma \vdash {\mathtt{case~} V \mathtt{~of}_{x_1.\,B} \mathtt{~} ({\inl {\!} {\!\!(y_1 \!:\! A_1)} \mapsto W[\inl {A_1 + A_2} y_1/x_2]} , }
\\[-1mm]
\hspace{3.35cm} {\inr {\!} {\!\!(y_2 \!:\! A_2)} \mapsto W[\inr {A_1 + A_2} y_2/x_2]}) = {W[V/x_2]} : {B[V/x_1]}
\end{array}
}
{
\begin{array}{c}
\lj {\Gamma, x_1 \!:\! A_1 + A_2} B
\quad
\vj \Gamma V {A_1 + A_2} \quad \vj {\Gamma, x_2 \!:\! A_1 + A_2} {W} {B[x_2/x_1]}
\end{array} 
}
\end{array}
\]

\paragraph*{Propositional equality} \mbox{}

\noindent
Formation rule for propositional equality:
\[
\mkrule
{\lj \Gamma {V =_A W}}
{\vj \Gamma V A \quad \vj \Gamma W A}
\]

\noindent
Typing rules for the introduction and elimination forms of propositional equality:
\[
\begin{array}{c}
\mkrule
{\vj \Gamma {\refl A V} {V =_A V}}
{\vj \Gamma V A}
\\[3mm]
\mkrule
{\vj \Gamma {\pathind A {x_1.\, x_2.\, x_3.\, B} {y.\, W} {V_1} {V_2} {V_p}} {B[V_1/x_1][V_2/x_2][V_p/x_3]}}
{
\begin{array}{c}
\lj \Gamma A \quad \lj {\Gamma, x_1 \!:\! A, x_2 \!:\! A, x_3 \!:\! x_1 =_A x_2} {B} \quad \vj \Gamma {V_1} A \quad \vj \Gamma {V_2} A 
\\[-1mm]
\vj \Gamma {V_p} {V_1 =_A V_2} \quad \vj {\Gamma, y \!:\! A} {W} {B[y/x_1][y/x_2][\refl A y/x_3]}
\end{array}
}
\end{array}
\]

\noindent
Congruence rules for propositional equality, and its introduction and elimination forms:
\vspace{0.1cm}
\[
\begin{array}{c}
\mkrule
{\ljeq \Gamma {(V_1 =_{A_1} {W_1})} {(V_2 =_{A_2} {W_2})}}
{\ljeq \Gamma {A_1} {A_2} \quad \veq \Gamma {V_1} {V_2} {A_1} \quad \veq \Gamma {W_1} {W_2} {A_1}}
\\[6mm]
\mkrule
{\veq \Gamma {\refl {A_1} {V_1}} {\refl {A_2} {V_2}} {V_1 =_{A} {V_1}}}
{\veq \Gamma {V_1} {V_2} {A}}
\\[4mm]
\mkrule
{
\begin{array}{c@{~} c@{~} l}
\Gamma & \vdash & {\pathind {A_1} {x_1.\, x_2.\, x_3.\, B_1} {y.\, W_1} {V_{11}} {V_{21}} {V_{p1}}} 
\\
& = & {\pathind {A_2} {x_1.\, x_2.\, x_3.\, B_2} {y.\, W_2} {V_{12}} {V_{22}} {V_{p2}}} : {B_1[V_{11}/x_1][V_{21}/x_2][V_{p1}/x_3]}
\end{array}
}
{
\begin{array}{c}
\ljeq \Gamma {A_1} {A_2} \quad \ljeq {\Gamma, x_1 \!:\! A_1, x_2 \!:\! A_1, x_3 \!:\! x_1 =_{A_1} x_2} {B_1} {B_2} 
\\[-1mm]
\veq \Gamma {V_{11}} {V_{12}} {A_1} \quad \veq \Gamma {V_{21}} {V_{22}} {A_1} \quad \veq \Gamma {V_{p1}} {V_{p2}} {V_{11} =_{A_1} V_{21}} 
\\[-1mm]
\veq {\Gamma, y \!:\! A_1} {W_1} {W_2} {B_1[y/x_1][y/x_2][\refl {A_1} y/x_3]}
\end{array}
}
\end{array}
\]

\noindent
$\beta$-equation for the elimination form of propositional equality:
\vspace{-0.1cm}
\[
\mkrule
{\veq \Gamma {\pathind A {x_1.\, x_2.\, x_3.\, B} {y.\, W} {V} {V} {\refl A V}} {W[V/y]} {B[V/x_1][V/x_2][\refl A V/x_3]}}
{
\begin{array}{c}
\lj \Gamma A \quad \lj {\Gamma, x_1 \!:\! A, x_2 \!:\! A, x_3 \!:\! x_1 =_A x_2} {B}
\\[-1mm]
\vj \Gamma {V} A \quad \vj {\Gamma, y \!:\! A} {W} {B[y/x_1][y/x_2][\refl A y/x_3]}
\end{array}
}
\]

\paragraph*{Thunked computations} \mbox{}

\noindent
Formation rule for the type of thunked computations:
\[
\mkrule
{\lj \Gamma {U\ul{C}}}
{\lj \Gamma {\ul{C}}}
\]

\noindent
Typing rules for thunking and forcing:
\[
\mkrule
{\vj \Gamma {\thunk M} {U\ul{C}}}
{\cj \Gamma M {\ul{C}}}
\qquad
\mkrule
{\cj \Gamma {\force {\ul{C}} V} {\ul{C}}}
{\vj \Gamma V {U\ul{C}}}
\]

\noindent
Congruence rules for the type of thunked computations, thunking, and forcing:
\[
\begin{array}{c}
\mkrule
{\ljeq \Gamma {U\ul{C}_1} {U\ul{C}_2}}
{\ljeq \Gamma {\ul{C}_1} {\ul{C}_2}}
\\[6mm]
\mkrule
{\veq \Gamma {\thunk M_1} {\thunk M_2} {U\ul{C}}}
{\ceq \Gamma {M_1} {M_2} \ul{C}}
\\[6mm]
\mkrule
{\ceq \Gamma {\force {\ul{C}_1} {V_1}} {\force {\ul{C}_2} V_2} {\ul{C}_1}}
{\ljeq \Gamma {\ul{C}_1} {\ul{C}_2} \quad \veq \Gamma {V_1} {V_2} {U\ul{C}_1}}
\end{array}
\]

\noindent
Equations relating thunking and forcing:
\[
\begin{array}{c}
\mkrule
{\veq \Gamma {\thunk (\force {\ul{C}} V)} {V} {U\ul{C}}}
{\vj \Gamma V {U\ul{C}}}
\qquad
\mkrule
{\ceq \Gamma {\force {\ul{C}} {(\thunk M)}} {M} {\ul{C}}}
{\cj \Gamma M {\ul{C}}}
\end{array}
\]

\paragraph*{Homomorphic function type} \mbox{}

\noindent
Formation rule for the homomorphic function type:
\[
\mkrule
{\lj \Gamma {\ul{C} \multimap \ul{D}}}
{\lj \Gamma {\ul{C}} \quad \lj \Gamma {\ul{D}}}
\]

\noindent
Typing rules for the homomorphic lambda abstraction and function application:
\[
\begin{array}{c}
\mkrule
{\vj \Gamma {\lambda \, z \!:\! \ul{C} .\, K} {\ul{C} \multimap \ul{D}}}
{\hj \Gamma {z \!:\! \ul{C}} K {\ul{D}}}
\\[6mm]
\mkrule
{\cj \Gamma {V(M)_{\ul{C}, \ul{D}}} {\ul{D}}}
{\vj \Gamma V {\ul{C} \multimap \ul{D}} \quad \cj \Gamma M \ul{C}}
\qquad
\mkrule
{\hj \Gamma {z \!:\! \ul{C}} {V(K)_{\ul{D}_1, \ul{D}_2}} {\ul{D}_2}}
{\vj \Gamma V {\ul{D}_1 \multimap \ul{D}_2} \quad \hj \Gamma {z \!:\! \ul{C}} K {\ul{D}_1}}
\vspace{0.1cm}
\end{array}
\]

\noindent
Congruence rules for the homomorphic function type, lambda abstraction, and function application:
\[
\begin{array}{c}
\mkrule
{\ljeq \Gamma {\ul{C}_1 \multimap \ul{D}_1} {\ul{C}_2 \multimap \ul{D}_2}}
{\ljeq \Gamma {\ul{C}_1} {\ul{C}_2} \quad \ljeq \Gamma {\ul{D}_1} {\ul{D}_2}}
\\[6mm]
\mkrule
{\veq \Gamma {\lambda \, z \!:\! \ul{C}_1 .\, K_1} {\lambda \, z \!:\! \ul{C}_2 .\, K_2} {\ul{C}_1 \multimap \ul{D}}}
{\ljeq {\Gamma} {\ul{C}_1} {\ul{C}_2} \quad \heq \Gamma {z \!:\! \ul{C}_1} {K_1} {K_2} {\ul{D}}}
\\[6mm]
\mkrule
{\ceq \Gamma {V_1(M_1)_{\ul{C}_1, \ul{D}_1}} {V_2(M_2)_{\ul{C}_2, \ul{D}_2}} {\ul{D}_1}}
{\ljeq {\Gamma} {\ul{C}_1} {\ul{C}_2} \quad \ljeq {\Gamma} {\ul{D}_1} {\ul{D}_2} \quad \veq \Gamma {V_1} {V_2} {\ul{C}_1 \multimap \ul{D}_1} \quad \ceq \Gamma {M_1} {M_2} {\ul{C}_1}}
\\[6mm]
\mkrule
{\heq \Gamma {z \!:\! \ul{C}} {V_1(K_1)_{\ul{D}_{11}, \ul{D}_{21}}} {V_2(K_2)_{\ul{D}_{12}, \ul{D}_{22}}} {\ul{D}_{21}}}
{\ljeq {\Gamma} {\ul{D}_{11}} {\ul{D}_{12}} \quad \ljeq {\Gamma} {\ul{D}_{21}} {\ul{D}_{22}} \quad \veq \Gamma {V_1} {V_2} {\ul{D}_{11} \multimap \ul{D}_{21}} \quad \heq \Gamma {z \!:\! \ul{C}} {K_1} {K_2} {\ul{D}_{11}}}
\end{array}
\]

\noindent
$\beta$- and $\eta$-equations for homomorphic lambda abstraction and function application:
\[
\begin{array}{c}
\mkrule
{\ceq \Gamma {(\lambda \, z \!:\! \ul{C} .\, K)(M)_{\ul{C}, \ul{D}}} {K[M/z]} {\ul{D}}}
{\cj \Gamma M \ul{C} \quad \hj \Gamma {z \!:\! \ul{C}} K {\ul{D}}}
\\[6mm]
\mkrule
{\heq \Gamma {z_1 \!:\! \ul{C}} {(\lambda \, z_2 \!:\! \ul{D}_1 .\, L)(K)_{\ul{D}_1, \ul{D}_2}} {L[K/z_2]} {\ul{D}_2}}
{\hj \Gamma {z_1 \!:\! \ul{C}} K {\ul{D}_1} \quad \hj \Gamma {z_2 \!:\! \ul{D}_1} L {\ul{D}_2}}
\\[6mm]
\mkrule
{\veq \Gamma V {\lambda\, z \!:\! \ul{C} .\, V(z)_{\ul{C}, \ul{D}}} {\ul{C} \multimap \ul{D}}}
{\vj \Gamma V {\ul{C} \multimap \ul{D}}}
\end{array}
\]

\paragraph*{Type of computations that return values of a give value type} \mbox{}

\noindent
Formation rule for the type of computations that return values of a give value type:
\[
\mkrule
{\lj \Gamma {FA}}
{\lj \Gamma A}
\]

\noindent
Typing rules for returning a value and sequential composition:
\[
\begin{array}{c}
\mkrule
{\cj \Gamma {\return V} {FA}}
{\vj \Gamma V A}
\\[6mm]
\mkrule
{\cj \Gamma {\doto M {x \!:\! A} {\ul{C}} N} {\ul{C}}}
{\cj \Gamma M {FA} \quad \lj \Gamma {\ul{C}} \quad \cj {\Gamma, x \!:\! A} {N} {\ul{C}}}
\\[6mm]
\mkrule
{\hj \Gamma {z \!:\! \ul{C}} {\doto K {x \!:\! A} {\ul{D}} M} {\ul{D}}}
{\hj \Gamma {z \!:\! \ul{C}} K {FA} \quad \lj \Gamma {\ul{D}} \quad \cj {\Gamma, x \!:\! A} {M} {\ul{D}}}
\end{array}
\]

\noindent
Congruence rules for the type of computations that return values of a given value type, returning a value,  and sequential composition:
\[
\begin{array}{c}
\mkrule
{\ljeq \Gamma {FA_1} {FA_2}}
{\ljeq \Gamma {A_1} {A_2}}
\\[6mm]
\mkrule
{\ceq \Gamma {\return {V_1}} {\return {V_2}} {FA}}
{\veq \Gamma {V_1} {V_2} A}
\\[6mm]
\mkrule
{\ceq \Gamma {\doto {M_1} {x \!:\! A_1} {\ul{C}_1} {N_1}} {\doto {M_2} {x \!:\! A_2} {\ul{C}_2} {N_2}} {\ul{C}_1}}
{\ljeq \Gamma {A_1} {A_2} \quad \ceq \Gamma {M_1} {M_2} {FA_1} \quad \ljeq \Gamma {\ul{C}_1} {\ul{C}_2} \quad \ceq {\Gamma, x \!:\! A_1} {N_1} {N_2} {\ul{C}_1}}
\\[6mm]
\mkrule
{\heq \Gamma {z \!:\! \ul{C}} {\doto {K_1} {x \!:\! A_1} {\ul{D}_1} {M_1}} {\doto {K_2} {x \!:\! A_2} {\ul{D}_2} {M_2}} {\ul{D}_1}}
{\ljeq \Gamma {A_1} {A_2} \quad \heq \Gamma {z \!:\! \ul{C}} {K_1} {K_2} {FA_1} \quad \ljeq \Gamma {\ul{D}_1} {\ul{D}_2} \quad \ceq {\Gamma, x \!:\! A_1} {M_1} {M_2} {\ul{D}_1}}
\end{array}
\]

\pagebreak

\noindent
$\beta$- and $\eta$-equations for sequential composition:
\[
\begin{array}{c}
\mkrule
{\ceq \Gamma {\doto {\return V} {x \!:\! A} {\ul{C}} M} {M[V/x]} {\ul{C}}}
{\vj \Gamma V {A} \quad \lj \Gamma {\ul{C}} \quad \cj {\Gamma, x \!:\! A} {M} {\ul{C}}}
\\[6mm]
\mkrule
{\ceq \Gamma {\doto M {x \!:\! A} {\ul{C}} {K[\return x/z]}} {K[M/z]} {\ul{C}}}
{\cj \Gamma M {FA} \quad \lj \Gamma {\ul{C}} \quad \hj {\Gamma} {z \!:\! FA} {K} {\ul{C}}}
\\[6mm]
\mkrule
{\heq \Gamma {z_1 \!:\! \ul{C}} {\doto K {x \!:\! A} {\ul{D}} {L[\return x/z_2]}} {L[K/z_2]} {\ul{D}}}
{\hj \Gamma {z_1 \!:\! \ul{C}} K {FA} \quad \lj \Gamma {\ul{D}} \quad \hj {\Gamma} {z_2 \!:\! FA} {L} {\ul{D}}}
\end{array}
\]

\paragraph*{Computational $\Sigma$-type} \mbox{}

\noindent
Formation rule for the computational $\Sigma$-type:
\[
\begin{array}{c}
\mkrule
{\lj \Gamma {\Sigma \, x \!:\! A .\, \ul{C}}}
{\lj {\Gamma, x \!:\! A} {\ul{C}}}
\end{array}
\]

\noindent
Typing rules for computational pairing and pattern-matching: 
\[
\begin{array}{c}
\mkrule
{\cj \Gamma {\langle V , M \rangle_{(x : A).\, \ul{C}}} {\Sigma \, x \!:\! A .\, \ul{C}}}
{\vj \Gamma V A \quad \lj {\Gamma, x \!:\! A} {\ul{C}} \quad \cj \Gamma M {\ul{C}[V/x]}}
\\[6mm]
\mkrule
{\cj \Gamma {\doto M {(x \!:\! A, z \!:\! \ul{C})} {\ul{D}} K} {\ul{D}}}
{\cj \Gamma M {\Sigma \, x \!:\! A .\, \ul{C}} \quad \lj \Gamma {\ul{D}} \quad \hj {\Gamma, x \!:\! A} {z \!:\! \ul{C}} {K} {\ul{D}}}
\\[6mm]
\mkrule
{\hj \Gamma {z \!:\! \ul{C}} {\langle V , K \rangle_{(x : A).\, \ul{D}}} {\Sigma \, x \!:\! A .\, \ul{D}}}
{\vj \Gamma V A \quad \lj {\Gamma, x \!:\! A} {\ul{D}} \quad \hj \Gamma {z \!:\! \ul{C}} K {\ul{D}[V/x]}}
\\[6mm]
\mkrule
{\hj \Gamma {z_1 \!:\! \ul{C}} {\doto K {(x \!:\! A, z_2 \!:\! \ul{D}_1)} {\ul{D}_2} L} {\ul{D}_2}}
{\hj \Gamma {z_1 \!:\! \ul{C}} K {\Sigma \, x \!:\! A .\, \ul{D}_1} \quad \lj \Gamma {\ul{D}_2} \quad \hj {\Gamma, x \!:\! A} {z_2 \!:\! \ul{D}_1} {L} {\ul{D}_2}}
\end{array}
\]

\noindent
Congruence rules for the computational $\Sigma$-type, computational pairing, and pattern-matching:
\[
\begin{array}{c}
\mkrule
{\ljeq \Gamma {\Sigma \, x \!:\! A_1 .\, \ul{C}_1} {\Sigma \, x \!:\! A_2 .\, \ul{C}_2}}
{\ljeq \Gamma {A_1} {A_2} \quad \ljeq {\Gamma, x \!:\! A_1} {\ul{C}_1} {\ul{C}_2}}
\\[6mm]
\mkrule
{\ceq \Gamma {\langle V_1 , M_1 \rangle_{(x : A_1).\, \ul{C}_1}} {\langle V_2 , M_2 \rangle_{(x : A_2).\, \ul{C}_2}} {\Sigma \, x \!:\! A_1 . \ul{C}_1}}
{\ljeq \Gamma {A_1} {A_2} \quad \ljeq {\Gamma, x \!:\! A_1} {\ul{C}_1} {\ul{C}_2} \quad \veq \Gamma {V_1} {V_2} {A_1} \quad \ceq \Gamma {M_1} {M_2} {\ul{C}_1[V_1/x]}}
\\[6mm]
\mkrule
{\ceq \Gamma {\doto {M_1} {(x \!:\! A_1, z \!:\! \ul{C}_1)} {\ul{D}_1} {K_1}} {\doto {M_2} {(x \!:\! A_2, z \!:\! \ul{C}_2)} {\ul{D}_2} {K_2}} {\ul{D}_1}}
{
\begin{array}{c}
\ljeq \Gamma {A_1} {A_2} \quad \ljeq {\Gamma, x \!:\! A_1} {\ul{C}_1} {\ul{C}_2} \quad \ljeq \Gamma {\ul{D}_1} {\ul{D}_2}
\\[-1mm]
\ceq \Gamma {M_1} {M_2} {\Sigma \, x \!:\! A_1 .\, \ul{C}_1} \quad \heq {\Gamma, x \!:\! A_1} {z \!:\! \ul{C}_1} {K_1} {K_2} {\ul{D}_1}
\end{array}
}
\end{array}
\]

\[
\begin{array}{c}
\mkrule
{\heq \Gamma {z \!:\! \ul{C}} {\langle V_1 , K_1 \rangle_{(x : A_1).\, \ul{D}_1}} {\langle V_2 , K_2 \rangle_{(x : A_2).\, \ul{D}_2}} {\Sigma \, x \!:\! A_1 . \ul{D}_1}}
{\ljeq \Gamma {A_1} {A_2} \quad \ljeq {\Gamma, x \!:\! A_1} {\ul{D}_1} {\ul{D}_2} \quad \veq \Gamma {V_1} {V_2} {A_1} \quad \heq \Gamma {z \!:\! \ul{C}} {K_1} {K_2} {\ul{D}_1[V_1/x]}}
\\[5mm]
\mkrule
{\heq \Gamma {z_1 \!:\! \ul{C}} {\doto {K_1} {(x \!:\! A_1, z_2 \!:\! \ul{D}_{11})} {\ul{D}_{21}} {L_1}} {\doto {K_2} {(x \!:\! A_2, z_2 \!:\! \ul{D}_{12})} {\ul{D}_{22}} {L_2}} {\ul{D}_{21}}}
{
\begin{array}{c}
\ljeq \Gamma {A_1} {A_2} \quad \ljeq {\Gamma, x \!:\! A_1} {\ul{D}_{11}} {\ul{D}_{12}} \quad \ljeq \Gamma {\ul{D}_{21}} {\ul{D}_{22}}
\\[-1mm]
\heq \Gamma {z_1 \!:\! \ul{C}} {K_1} {K_2} {\Sigma \, x \!:\! A_1 .\, \ul{D}_{11}} \quad \heq {\Gamma, x \!:\! A_1} {z_2 \!:\! \ul{D}_{11}} {L_1} {L_2} {\ul{D}_{21}}
\end{array}
}
\end{array}
\]

\noindent
$\beta$- and $\eta$-equations for computational pattern-matching:
\[
\begin{array}{c}
\mkrule
{\ceq \Gamma {\doto {\langle V , M \rangle_{(x : A).\, \ul{C}}} {(x \!:\! A, z \!:\! \ul{C})} {\ul{D}} K} {K[V/x][M/z]} {\ul{D}}}
{\vj \Gamma V A \quad \cj \Gamma M {\ul{C}[V/x]} \quad \lj \Gamma {\ul{D}} \quad \hj {\Gamma, x \!:\! A} {z \!:\! \ul{C}} {K} {\ul{D}}}
\\[6mm]
\mkrule
{\ceq \Gamma {\doto M {(x \!:\! A, z_1 \!:\! \ul{C})} {\ul{D}} {K[{\langle x , z_1 \rangle_{(x : A).\, \ul{C}}}/z_2]}} {K[M/z_2]} {\ul{D}}}
{\lj {\Gamma, x \!:\! A} {\ul{C}} \quad \cj \Gamma M {\Sigma \, x \!:\! A .\, \ul{C}} \quad \lj \Gamma {\ul{D}} \quad \hj {\Gamma} {z_2 \!:\! \Sigma \, x \!:\! A .\, \ul{C}} {K} {\ul{D}}}
\\[6mm]
\mkrule
{\heq \Gamma {z_1 \!:\! \ul{C}} {\doto {\langle V , K \rangle_{(x : A).\, \ul{D}_1}} {(x \!:\! A, z_2 \!:\! \ul{D}_1)} {\ul{D}_2} L} {L[V/x][K/z_2]} {\ul{D}_2}}
{\vj \Gamma V A \quad \hj \Gamma {z_1 \!:\! \ul{C}} K {\ul{D}_1[V/x]} \quad \lj \Gamma {\ul{D}_2} \quad \hj {\Gamma, x \!:\! A} {z_2 \!:\! \ul{D}_1} {L} {\ul{D}_2}}
\\[6mm]
\mkrule
{\heq \Gamma {z_1 \!:\! \ul{C}} {\doto K {(x \!:\! A, z_2 \!:\! \ul{D}_1)} {\ul{D}_2} {L[{\langle x , z_2 \rangle_{(x : A).\, \ul{D}_1}}/z_3]}} {L[K/z_3]} {\ul{D}_2}}
{\lj {\Gamma, x \!:\! A} {\ul{D}_1} \quad \hj \Gamma {z_1 \!:\! \ul{C}} K {\Sigma \, x \!:\! A .\, \ul{D}_1} \quad \lj \Gamma {\ul{D}_2} \quad \hj {\Gamma} {z_3 \!:\! \Sigma \, x \!:\! A .\, \ul{D}_1} {K} {\ul{D}_2}}
\end{array}
\]

\paragraph*{Computational $\Pi$-type} \mbox{}

\noindent
Formation rule for the computational $\Pi$-type:
\[
\mkrule
{\lj \Gamma {\Pi \, x \!:\! A .\, \ul{C}}}
{\lj {\Gamma, x \!:\! A} \ul{C}}
\]

\noindent
Typing rules for computational lambda abstraction and function application:
\[
\begin{array}{c}
\mkrule
{\cj \Gamma {\lambda \, x \!:\! A .\, M} {\Pi \, x \!:\! A .\, \ul{C}}}
{\cj {\Gamma, x \!:\! A} M \ul{C}}
\qquad
\mkrule
{\cj \Gamma {M(V)_{(x : A).\, \ul{C}}} {\ul{C}[V/x]}}
{\lj {\Gamma, x \!:\! A} {\ul{C}} \quad \cj \Gamma M {\Pi \, x \!:\! A .\, \ul{C}} \quad \vj \Gamma V A}
\\[6mm]
\mkrule
{\hj \Gamma {z \!:\! \ul{C}} {\lambda \, x \!:\! A .\, K} {\Pi \, x \!:\! A .\, \ul{D}}}
{\lj \Gamma {\ul{C}} \quad \hj {\Gamma, x \!:\! A} {z \!:\! \ul{C}} K \ul{D}}
\qquad
\mkrule
{\hj \Gamma {z \!:\! \ul{C}} {K(V)_{(x : A).\, \ul{D}}} {\ul{D}[V/x]}}
{\lj {\Gamma, x \!:\! A} {\ul{D}} \quad \hj \Gamma {z \!:\! \ul{C}} K {\Pi \, x \!:\! A .\, \ul{D}} \quad \vj \Gamma V A}
\end{array}
\]

\noindent
Congruence rules for the computational $\Pi$-type, lambda abstraction, and function application:
\[
\begin{array}{c}
\mkrule
{\ljeq \Gamma {\Pi \, x \!:\! A_1 .\, \ul{C}_1} {\Pi \, x \!:\! A_2 .\, \ul{C}_2}}
{\ljeq \Gamma {A_1} {A_2} \quad \ljeq {\Gamma, x \!:\! A_1} {\ul{C}_1} {\ul{C}_2}}
\\[6mm]
\mkrule
{\ceq \Gamma {\lambda \, x \!:\! A_1 .\, M_1} {\lambda \, x \!:\! A_2 .\, M_2} {\Pi \, x \!:\! A_1 .\, \ul{C}}}
{\ljeq \Gamma {A_1} {A_2} \quad \ceq {\Gamma, x \!:\! A_1} {M_1} {M_2} {\ul{C}}}
\end{array}
\]

\[
\begin{array}{c}
\mkrule
{\ceq \Gamma {M_1(V_1)_{(x : A_1).\, \ul{C}_1}} {M_2(V_2)_{(x : A_2).\, \ul{C}_2}} {\ul{C}_1[V_1/x]}}
{\ljeq \Gamma {A_1} {A_2} \quad \ljeq {\Gamma, x \!:\! A_1} {\ul{C}_1} {\ul{C}_2} \quad \ceq \Gamma {M_1} {M_2} {\Pi \, x \!:\! A_1 .\, \ul{C}_1} \quad \veq {\Gamma} {V_1} {V_2} {A_1}}
\\[6mm]
\mkrule
{\heq \Gamma {z \!:\! \ul{C}} {\lambda \, x \!:\! A_1 .\, K_1} {\lambda \, x \!:\! A_2 .\, K_2} {\Pi \, x \!:\! A_1 .\, \ul{D}}}
{\ljeq \Gamma {A_1} {A_2} \quad \lj \Gamma {\ul{C}} \quad \heq {\Gamma, x \!:\! A_1} {z \!:\! \ul{C}} {K_1} {K_2} {\ul{D}}}
\\[6mm]
\mkrule
{\heq \Gamma {z \!:\! \ul{C}} {K_1(V_1)_{(x : A_1).\, \ul{D}_1}} {K_2(V_2)_{(x : A_2).\, \ul{D}_2}} {\ul{D}_1[V_1/x]}}
{\ljeq \Gamma {A_1} {A_2} \quad \ljeq {\Gamma, x \!:\! A_1} {\ul{D}_1} {\ul{D}_2} \quad \ceq \Gamma {K_1} {K_2} {\Pi \, x \!:\! A_1 .\, \ul{D}_1} \quad \veq {\Gamma} {V_1} {V_2} {A_1}}
\end{array}
\]

\vspace{0.25cm}

\noindent
$\beta$- and $\eta$-equations for computational lambda abstraction and function application:
\vspace{0.05cm}
\[
\begin{array}{c}
\mkrule
{\ceq \Gamma {(\lambda \, x \!:\! A .\, M)(V)_{(x : A) .\, \ul{C}}} {M[V/x]} {\ul{C}[V/x]}}
{\cj {\Gamma, x \!:\! A} M \ul{C} \quad \vj \Gamma V A}
\\[6mm]
\mkrule
{\ceq \Gamma {M} {\lambda \, x \!:\! A .\, M(x)_{(x : A) .\, \ul{C}}} {\Pi \, x \!:\! A .\, \ul{C}}}
{\lj {\Gamma, x \!:\! A} {\ul{C}} \quad \cj \Gamma M {\Pi \, x \!:\! A .\, \ul{C}}}
\\[6mm]
\mkrule
{\heq \Gamma {z \!:\! \ul{C}} {(\lambda \, x \!:\! A .\, K)(V)_{(x : A) .\, \ul{D}}} {K[V/x]} {\ul{D}[V/x]}}
{\lj \Gamma {\ul{C}} \quad \hj {\Gamma, x \!:\! A} {z \!:\! \ul{C}} K \ul{D} \quad \vj \Gamma V A}
\\[6mm]
\mkrule
{\heq \Gamma {z \!:\! \ul{C}} {K} {\lambda \, x \!:\! A .\, K(x)_{(x : A) .\, \ul{D}}} {\Pi \, x \!:\! A .\, \ul{D}}}
{\lj {\Gamma, x \!:\! A} {\ul{D}} \quad \hj \Gamma {z \!:\! \ul{C}} K {\Pi \, x \!:\! A .\, \ul{D}}}
\end{array}
\]

\vspace{0.1cm}

\subsection*{Convention for proving definitional equations}

In this thesis we use the following convention: when we say that we prove a definitional equation $\veq \Gamma {V_1} {V_n} A$, we formally mean constructing a corresponding derivation using the rules given above. In order to improve the readability of these proofs, we omit the full derivations and instead present the proofs as sequences of equations
\[
\Gamma \vdash V_1 = V_2 = \ldots = V_{n-1} = V_n : A
\]
where each individual equation corresponds to a derivation of  $\veq \Gamma {V_i} {V_j} A$, and the sequence as a whole corresponds to the derivation of $\veq \Gamma {V_1} {V_n} A$ by repeated use of the transitivity rule on these individual derivations. These individual derivations often consist of an application of a  $\beta$- or $\eta$-rule under some number of congruence rules.

\section{Meta-theory}
\label{sect:metatheory}

In this section we prove various meta-theoretical properties of the well-formed syntax of eMLTT we introduced in the previous section; these include closure under weakening and substitution, and that well-typed terms are assigned only well-formed types. 

We begin with some properties of well-formed value contexts. First, we show that all value types in a well-formed value context are themselves well-formed.

\begin{proposition}
\label{prop:wellformedtypesinwellformedcontexts}
Given value contexts $\Gamma_1$ and $\Gamma_2$, a value variable $x$, and a value type $A$ such that $\vdash \Gamma_1, x \!:\! A, \Gamma_2$, then we also have $\lj {\Gamma_1} A$.
\end{proposition}

\begin{proof}
By induction on the length of $\Gamma_2$ and by observing that in each of the cases the derivation of $\vdash \Gamma_1, x \!:\! A, \Gamma_2$ has to end with the context extension rule. 
\end{proof}

Next, we observe that our axiomatisation of definitionally equal value contexts in terms of a reflexivity rule for the empty context and a congruence rule for context extension gives rise to an equivalence relation. 

\begin{proposition}
The following rules are admissible for value contexts:
\[
\mkrule
{\ljeq {} \Gamma \Gamma}
{\vdash \Gamma}
\qquad
\mkrule
{\ljeq {} {\Gamma_1} {\Gamma_2}}
{\ljeq {} {\Gamma_2} {\Gamma_1}}
\qquad
\mkrule
{\ljeq {} {\Gamma_1} {\Gamma_3}}
{\ljeq {} {\Gamma_1} {\Gamma_2} \quad \ljeq {} {\Gamma_2} {\Gamma_3}}
\]
\end{proposition}

\begin{proof}
We prove reflexivity by induction on the derivation of $\vdash \Gamma$, symmetry by induction on the derivation of $\ljeq {} {\Gamma_2} {\Gamma_1}$, and transitivity by induction on the sum of the heights of the derivations of $\ljeq {} {\Gamma_1} {\Gamma_2}$ and $\ljeq {} {\Gamma_2} {\Gamma_3}$. We use the reflexivity, 
symmetry, transitivity, and context conversion rules for value types where needed.
\end{proof}

Next, we show that definitionally equal value contexts have the same structure.

\begin{proposition}
\label{prop:setsofvariablesofequalcontexts}
Given $\Gamma_1$ and $\Gamma_2$ such that ${\ljeq {} {\Gamma_1} {\Gamma_2}}$, then $V\!ars(\Gamma_1) = V\!ars(\Gamma_2)$; and moreover, these value variables  
are given in the same exact order in both contexts.
\end{proposition}

\begin{proof}
By induction on the derivation of $\ljeq {} {\Gamma_1} {\Gamma_2}$.
\end{proof}

\begin{proposition}
\label{prop:equalcontextconcatenation1}
Given value contexts $\Gamma_1$ and $\Gamma_2$ and $\Gamma_3$ such that $\ljeq {} {\Gamma_1, \Gamma_2} {\Gamma_3}$, then there exist value contexts $\Gamma_4$ and $\Gamma_5$ such that $\Gamma_3 = \Gamma_4, \Gamma_5$ and $\ljeq {} {\Gamma_1} {\Gamma_4}$.
\end{proposition}

\begin{proof}
By induction on the length of $\Gamma_2$. 
\end{proof}

\begin{corollary}
\label{cor:equalcontextconcatenation2}
Given value contexts $\Gamma_1$ and $\Gamma_2$ and $\Gamma_3$, a value variable $x$, and a value type $A$ such that $\ljeq {} {\Gamma_1, x \!:\! A, \Gamma_2} {\Gamma_3}$, then there exist value contexts $\Gamma_4$ and $\Gamma_5$, and a value type $B$ such that $\Gamma_3 = \Gamma_4, x \!:\! B, \Gamma_5$ and $\ljeq {} {\Gamma_1} {\Gamma_4}$ and $\ljeq {\Gamma_1} {A} {B}$.
\end{corollary}

\begin{proof}
We use Proposition~\ref{prop:equalcontextconcatenation1} with value contexts $\Gamma_1, x \!:\! A$ and $\Gamma_2$ and $\Gamma_3$, and observe that the last rule used in the derivation of $\vdash {\Gamma_1, x \!:\! A} = \Gamma_4$ has to be the congruence rule for definitionally equal value contexts, giving us the required value type $B$.
\end{proof}

\begin{proposition}
\label{prop:contextextensionwithcontext}
Given value contexts $\Gamma_1$ and $\Gamma_2$ and $\Gamma$ such that $\ljeq {} {\Gamma_1} {\Gamma_2}$ and $\lj {} {\Gamma_1,\Gamma}$, then $\ljeq {} {\Gamma_1,\Gamma} {\Gamma_2,\Gamma}$.
\end{proposition}

\begin{proof}
By induction on the length of $\Gamma$. 
\end{proof}

Next, we show that the free value variables of a well-formed expression are contained in the value context in which the given expression is well-formed.

\begin{proposition} \mbox{}
\label{prop:freevariablesofwellformedexpressions}
\begin{enumerate}[(a)]
\item Given $\lj \Gamma A$, then $FVV(A) \subseteq V\!ars(\Gamma)$.
\item Given $\ljeq \Gamma A B$, then $FVV(A) \subseteq V\!ars(\Gamma)$ and $FVV(B) \subseteq V\!ars(\Gamma)$.
\item Given $\lj \Gamma \ul{C}$, then $FVV(\ul{C}) \subseteq V\!ars(\Gamma)$.
\item Given $\ljeq \Gamma {\ul{C}} {\ul{D}}$, then $FVV(\ul{C}) \subseteq V\!ars(\Gamma)$ and $FVV(\ul{D}) \subseteq V\!ars(\Gamma)$.
\item Given $\vj \Gamma V A$, then $FVV(V) \subseteq V\!ars(\Gamma)$ and $FVV(A) \subseteq V\!ars(\Gamma)$.
\item Given $\veq \Gamma V W A$, then $FVV(V) \subseteq V\!ars(\Gamma)$, $FVV(W) \subseteq V\!ars(\Gamma)$, and \\$FVV(A) \subseteq V\!ars(\Gamma)$.
\item Given $\cj \Gamma M \ul{C}$, then $FVV(M) \subseteq V\!ars(\Gamma)$ and $FVV(\ul{C}) \subseteq V\!ars(\Gamma)$.
\item Given $\ceq \Gamma M N \ul{C}$, then $FVV(M) \subseteq V\!ars(\Gamma)$, $FVV(N) \subseteq V\!ars(\Gamma)$, and \\ $FVV(\ul{C}) \subseteq V\!ars(\Gamma)$.
\item Given $\hj \Gamma {z \!:\! \ul{C}} K \ul{D}$, then $FVV(\ul{C}) \subseteq V\!ars(\Gamma)$, $FVV(K) \subseteq V\!ars(\Gamma)$, and \\ $FVV(\ul{D}) \subseteq V\!ars(\Gamma)$.
\item Given $\heq \Gamma {z \!:\! \ul{C}} K L \ul{D}$, then $FVV(\ul{C}) \subseteq V\!ars(\Gamma)$, $FVV(K) \subseteq V\!ars(\Gamma)$, \\ $FVV(L) \subseteq V\!ars(\Gamma)$, and $FVV(\ul{D}) \subseteq V\!ars(\Gamma)$.
\end{enumerate}
\end{proposition}

\begin{proof}
We prove $(a)$--$(j)$ simultaneously, by induction on the derivations of the given judgements, using Proposition~\ref{prop:freevariablesofsubsstitution} for rules involving the substitution of value terms for value variables.
\end{proof}

Analogously, we can also show that the free computation variable of a well-typed homomorphism term matches the computation variable mentioned in its typing judgement, and analogously for definitional equations between homomorphism terms:

\pagebreak

\begin{proposition} \mbox{}
\label{prop:freecomputationvariables}
\begin{enumerate}[(a)]
\item Given $\hj \Gamma {z \!:\! \ul{C}} K {\ul{D}}$, then $FCV(K) = z$.
\item Given $\heq \Gamma {z \!:\! \ul{C}} K L {\ul{D}}$, then $FCV(K) = z$ and $FCV(L) = z$.
\end{enumerate}
\end{proposition}

\begin{proof}
We first prove $(a)$ and then $(b)$, by induction on the given derivations of \linebreak $\hj \Gamma {z \!:\! \ul{C}} K {\ul{D}}$ and  $\heq \Gamma {z \!:\! \ul{C}} K L {\ul{D}}$, respectively. In the cases that involve substituting homomorphism terms for computation variables, we use Proposition~\ref{prop:compvariableofhomsubst}.
\end{proof}

We are now ready to prove that the judgements of eMLTT are closed under weakening of value contexts and under substitution of value terms for value variables. 

\begin{theorem}[Weakening] 
\label{thm:weakening}
\index{weakening theorem!syntactic --}
Assuming that $x \not\in V\!ars(\Gamma_1,\Gamma_2)$ and $\lj {\Gamma_1} A$, we have:
\label{thm:weakening}
\begin{enumerate}[(a)]
\item Given $\vdash {\Gamma_1,\Gamma_2}$, then $\vdash {\Gamma_1,x \!:\! A,\Gamma_2}$.
\item Given $\ljeq{} {\Gamma_1,\Gamma_2} {\Gamma_3,\Gamma_4}$, then $\ljeq {} {\Gamma_1,x \!:\! A,\Gamma_2} {\Gamma_3,x \!:\! A,\Gamma_4}$.
\item Given $\lj {\Gamma_1,\Gamma_2} B$, then $\lj {\Gamma_1,x \!:\! A,\Gamma_2} B$.
\item Given $\ljeq {\Gamma_1,\Gamma_2} {B_1} {B_2}$, then $\ljeq {\Gamma_1,x \!:\! A,\Gamma_2} {B_1} {B_2}$.
\item Given $\lj {\Gamma_1,\Gamma_2} \ul{C}$, then $\lj {\Gamma_1,x \!:\! A,\Gamma_2} \ul{C}$.
\item Given $\ljeq {\Gamma_1,\Gamma_2} {\ul{C}} {\ul{D}}$, then $\ljeq {\Gamma_1,x \!:\! A,\Gamma_2} {\ul{C}} {\ul{D}}$.
\item Given $\vj {\Gamma_1,\Gamma_2} V B$, then $\vj {\Gamma_1,x \!:\! A,\Gamma_2} V B$.
\item Given $\veq {\Gamma_1,\Gamma_2} V W B$, then $\veq {\Gamma_1,x \!:\! A,\Gamma_2} V W B$.
\item Given $\cj {\Gamma_1,\Gamma_2} M \ul{C}$, then $\cj {\Gamma_1,x \!:\! A,\Gamma_2} M \ul{C}$.
\item Given $\ceq {\Gamma_1,\Gamma_2} M N \ul{C}$, then $\ceq {\Gamma_1,x \!:\! A,\Gamma_2} M N \ul{C}$.
\item Given $\hj {\Gamma_1,\Gamma_2} {z \!:\! \ul{C}} K \ul{D}$, then $\hj {\Gamma_1,x \!:\! A,\Gamma_2} {z \!:\! \ul{C}} K \ul{D}$.
\item Given $\heq {\Gamma_1,\Gamma_2} {z \!:\! \ul{C}} K L \ul{D}$, then $\heq {\Gamma_1,x \!:\! A,\Gamma_2} {z \!:\! \ul{C}} K L \ul{D}$.
\end{enumerate}
\end{theorem}

\begin{proof}
We prove this theorem simultaneously with Theorem~\ref{thm:substitution}.
We prove all cases simultaneously: $(a)$--$(b)$ by induction on the length of $\Gamma_2$ 
and $(c)$--$(l)$ by induction on the derivations of the given judgements. 
We sketch the proofs of two cases that need extra work, namely, those involving context and type conversions, and substitution. 

\vspace{0.2cm}

\noindent \textbf{Context and type conversion rule for computation terms:}
In this case, the given derivation ends with
\[
\mkrule
{\cj {\Gamma_1,\Gamma_2} M {\ul{D}}}
{\cj {\Gamma_3} M {\ul{C}} \quad \vdash \Gamma_3 = \Gamma_1,\Gamma_2 \quad \ljeq {\Gamma_3} {\ul{C}} {\ul{D}}}
\]
and we are required to construct a derivation of ${\cj {\Gamma_1, x \!:\! A,\Gamma_2} M {\ul{D}}}$.

First, by using Proposition~\ref{prop:equalcontextconcatenation1}, together with the symmetry of definitionally equal value contexts, we get that $\Gamma_3 = \Gamma_4,\Gamma_5$ and $\ljeq {} {\Gamma_1} {\Gamma_4}$, for some $\Gamma_4$ and $\Gamma_5$. 

Next, by combining this definitional equation with the context and type conversion rule for value types, we get a derivation of $\lj {\Gamma_4} A$.

Next, we observe that our assumptions give us that $x \not\in V\!ars(\Gamma_1,\Gamma_2)$, and Proposition~\ref{prop:setsofvariablesofequalcontexts} gives us that $V\!ars(\Gamma_1,\Gamma_2) = V\!ars(\Gamma_3) = V\!ars(\Gamma_4,\Gamma_5)$. 

Therefore, we can use the induction hypothesis on the derivation of $\cj {\Gamma_4,\Gamma_5} M {\ul{C}}$ to get a derivation of $\cj {\Gamma_4,x \!:\! A,\Gamma_5} M {\ul{C}}$. Analogously, we can use $(b)$ on the derivation of $\ljeq {} {\Gamma_4,\Gamma_5} {\Gamma_1,\Gamma_2}$ to get a derivation of $\ljeq {} {\Gamma_4,x \!:\! A,\Gamma_5} {\Gamma_1,x \!:\! A,\Gamma_2}$, and $(f)$ on the derivation of $\ljeq {\Gamma_4,\Gamma_5} {\ul{C}} {\ul{D}}$ to get a derivation of $\ljeq {\Gamma_4,x \!:\! A,\Gamma_5} {\ul{C}} {\ul{D}}$.

As a result, we can now construct the required derivation, as shown below:
\[
\mkrule
{\cj {\Gamma_1, x \!:\! A,\Gamma_2} M {\ul{D}}}
{\cj {\Gamma_4,x \!:\! A,\Gamma_5} M {\ul{C}} \quad \vdash \Gamma_4,x \!:\! A,\Gamma_5 = \Gamma_1, x \!:\! A,\Gamma_2 \quad \ljeq {\Gamma_4,x \!:\! A,\Gamma_5} {\ul{C}} {\ul{D}}}
\]

Other cases involving context and type conversion rules are proved analogously.

\vspace{0.2cm}

\noindent \textbf{Replacement rule for computation terms:}
In this case, the given derivation ends with
\[
\mkrule
{\ceq {\Gamma_3, \Gamma_4[V_1/y]} {M[V_1/y]} {M[V_2/y]} {\ul{C}[V_1/y]}}
{\vj {\Gamma_3, y \!:\! B, \Gamma_4} M {\ul{C}} \quad \veq {\Gamma_3} {V_1} {V_2} {B}}
\]
with $\Gamma_1,\Gamma_2 = \Gamma_3, \Gamma_4[V_1/y]$. 

We now have two possibilities to consider, according to how $\Gamma_1$ and $\Gamma_3$ overlap. 
In both cases, we note that our adopted variable convention does not apply because $y$ is not a bound value variable. As a result, it is not guaranteed that $x$ and $y$ are different. To overcome this, we choose a fresh $y'$ satisfying $y' \not\in V\!ars(\Gamma_3, y \!:\! B, \Gamma_4)$ and $y' \neq x$.

\vspace{0.2cm}
\noindent
\emph{Case for $V\!ars(\Gamma_3) \subseteq V\!ars(\Gamma_1)$}: In this case, the value context $\Gamma_4[V_1/y]$ is of the form $\Gamma_{41}[V_1/y], \Gamma_{42}[V_1/y]$, with $\Gamma_1 = \Gamma_3,\Gamma_{41}[V_1/y]$ and $\Gamma_2 = \Gamma_{42}[V_1/y]$, and we are required to construct a derivation of
\[
\ceq {\Gamma_3, \Gamma_{41}[V_1/y], x \!:\! A, \Gamma_{42}[V_1/y]} {M[V_1/y]} {M[V_2/y]} {\ul{C}[V_1/y]}
\]

Based on that ${y' \not\in V\!ars(\Gamma_3, y \!:\! B, \Gamma_4)}$, we can use $(i)$ on the given derivation of \linebreak$\cj {\Gamma_3, y \!:\! B, \Gamma_{41}, \Gamma_{42}} M {\ul{C}}$ to get a derivation of $\cj {\Gamma_3, y' \!:\! B, y \!:\! B, \Gamma_{41}, \Gamma_{42}} M {\ul{C}}$, \linebreak on which we can in turn use $(i)$ of Theorem~\ref{thm:substitution} to get a derivation of \linebreak $\cj {\Gamma_{3}, y' \!:\! B, \Gamma_{41}[y'/y], \Gamma_{42}[y'/y]} {M[y'/y]} {\ul{C}[y'/y]}$.

Next, we observe that $x \not\in V\!ars(\Gamma_3, y' \!:\! B, \Gamma_{41}[y'/y], \Gamma_{42}[y'/y])$.
Therefore, we can use $(i)$ on the derivation of $\cj {\Gamma_3, y' \!:\! B, \Gamma_{41}[y'/y], \Gamma_{42}[y'/y]} {M[y'/y]} {\ul{C}[y'/y]}$ to get a derivation of $\cj {\Gamma_3, y' \!:\! B, \Gamma_{41}[y'/y], x \!:\! A, \Gamma_{42}[y'/y]} {M[y'/y]} {\ul{C}[y'/y]}$. 

Now, by applying the replacement rule for computation terms on this derivation, we get a derivation ending with
\[
\mkrulelabel
{
\begin{array}{c@{~} c@{~} l}
{\Gamma_3, \Gamma_{41}[y'/y][V_1/y'], x \!:\! A[V_1/y'], \Gamma_{42}[y'/y][V_1/y']} & \vdash & {M[y'/y][V_1/y']}
\\
& = & {M[y'/y][V_2/y']} : {\ul{C}[y'/y][V_1/y']}
\end{array}
}
{\cj {\Gamma_3, y' \!:\! B, \Gamma_{41}[y'/y],x \!:\! A,\Gamma_{42}[y'/y]} {M[y'/y]} {\ul{C}[y'/y]} \quad \veq {\Gamma_3} {V_1} {V_2} {B}}
{(*)}
\]

Next, according to Proposition~\ref{prop:freevariablesofwellformedexpressions}, we know that $FVV(V_1) \subseteq V\!ars(\Gamma_3)$ and $FVV(V_2) \subseteq V\!ars(\Gamma_3)$. Therefore, we have that $y \not\in FVV(V_1)$ and $y \not\in FVV(V_2)$. 
\linebreak
Furthermore, as a consequence of the way we have chosen $y'$, we also know that $y' \not\in FVV(\Gamma_{41})$, $y' \not\in FVV(\Gamma_{42})$, $y' \not\in FVV(M)$, $y' \not\in FVV(A)$, and $y' \not\in FVV(\ul{C})$. 

By combining these observations with Definition~\ref{def:substvaluevariables} and Propositions~\ref{prop:valuesubstlemma1},~\ref{prop:valuesubstlemma3}, and~\ref{prop:contextsubstlemma}, we can  show that the following equations hold:
\[
\begin{array}{c}
\Gamma_{41}[y'/y][V_1/y'] = \Gamma_{41}[V_1/y'][y'[V_1/y']/y] = \Gamma_{41}[y'[V_1/y']/y] = \Gamma_{41}[V_1/y]
\\[1mm]
A[V_1/y'] = A
\\[1mm]
\Gamma_{42}[y'/y][V_1/y'] = \Gamma_{42}[V_1/y'][y'[V_1/y']/y] = \Gamma_{42}[y'[V_1/y']/y] = \Gamma_{42}[V_1/y]
\\[1mm]
M[y'/y][V_1/y'] = M[V_1/y'][y'[V_1/y']/y] = M[y'[V_1/y']/y] = M[V_1/y]
\\[1mm]
M[y'/y][V_2/y'] = M[V_2/y'][y'[V_2/y']/y] = M[y'[V_2/y']/y] = M[V_2/y]
\\[1mm]
\ul{C}[y'/y][V_1/y'] = \ul{C}[V_1/y'][y'[V_1/y']/y] = \ul{C}[y'[V_1/y']/y] = \ul{C}[V_1/y]
\end{array}
\]

As a result, the derivation we constructed above that ends with $(*)$ gives us the required derivation of
$
\ceq {\Gamma_3, \Gamma_{41}[V_1/y], x \!:\! A, \Gamma_{42}[V_1/y]} {M[V_1/y]} {M[V_2/y]} {\ul{C}[V_1/y]}
$.

\vspace{0.2cm}
\noindent
\emph{Case for $V\!ars(\Gamma_3) \not\subseteq V\!ars(\Gamma_1)$}: In this case, the value context $\Gamma_3$ is of the form $\Gamma_{31}, \Gamma_{32}$, with $\Gamma_1 = \Gamma_{31}$ and $\Gamma_2 = \Gamma_{32},\Gamma_4[V_1/y]$, and we are required to construct a derivation of 
\[
\ceq {\Gamma_{31}, x \!:\! A,\Gamma_{32}, \Gamma_4[V_1/y]} {M[V_1/y]} {M[V_2/y]} {\ul{C}[V_1/y]}
\]

Based on that $y' \not\in V\!ars(\Gamma_3, y \!:\! B, \Gamma_4)$, we can use $(i)$ on the derivation of \linebreak$\cj {\Gamma_{31},\Gamma_{32}, y \!:\! B, \Gamma_4} M {\ul{C}}$ to get a derivation of $\cj {\Gamma_{31},\Gamma_{32}, y' \!:\! B, y \!:\! B, \Gamma_4} M {\ul{C}}$, \linebreak on which we can in turn use $(i)$ of Theorem~\ref{thm:substitution} to get a derivation of \linebreak $\cj {\Gamma_{31},\Gamma_{32}, y' \!:\! B, \Gamma_4[y'/y]} {M[y'/y]} {\ul{C}[y'/y]}$.

Next, we observe that $x \not\in V\!ars(\Gamma_{31},\Gamma_{32}, y' \!:\! B, \Gamma_4[y'/y])$.
Therefore, we can use $(i)$ on the derivation of $\cj {\Gamma_{31},\Gamma_{32}, y' \!:\! B, \Gamma_4[y'/y]} {M[y'/y]} {\ul{C}[y'/y]}$ to get a derivation of $\cj {\Gamma_{31}, x \!:\! A,\Gamma_{32}, y' \!:\! B, \Gamma_4[y'/y]} {M[y'/y]} {\ul{C}[y'/y]}$. 
Furthermore, as we also know that $x \not\in V\!ars(\Gamma_{31},\Gamma_{32})$, we can use $(h)$ on the derivation of $\veq {\Gamma_{31},\Gamma_{32}} {V_1} {V_2} {B}$ to get a derivation of $\veq {\Gamma_{31},x \!:\! A,\Gamma_{32}} {V_1} {V_2} {B}$. 

Now, by applying the replacement rule for computation terms on these derivations we get a derivation ending with 
\[
\mkrulelabel
{\ceq {\Gamma_{31}, x \!:\! A,\Gamma_{32}, \Gamma_4[y'/y][V_1/y']} {M[y'/y][V_1/y']} {M[y'/y][V_2/y']} {\ul{C}[y'/y][V_1/y']}}
{\cj {\Gamma_{31}, x \!:\! A,\Gamma_{32}, y' \!:\! B, \Gamma_4[y'/y]} {M[y'/y]} {\ul{C}[y'/y]} \quad \veq {\Gamma_{31}, x \!:\! A,\Gamma_{32}} {V_1} {V_2} {B}}
{(**)}
\]

Next, according to Proposition~\ref{prop:freevariablesofwellformedexpressions}, we know that $FVV(V_1) \subseteq V\!ars(\Gamma_{31},\Gamma_{32})$ and $FVV(V_2) \subseteq V\!ars(\Gamma_{31},\Gamma_{32})$. Therefore, we have that $y \not\in FVV(V_1)$ and $y \not\in FVV(V_2)$. Furthermore, as a consequence of the way we have picked $y'$, we also know that  \linebreak $y' \not\in FVV(\Gamma_4)$, $y' \not\in FVV(M)$, and $y' \not\in FVV(\ul{C})$. 

By combining these observations with Definition~\ref{def:substvaluevariables} and Propositions~\ref{prop:valuesubstlemma1},~\ref{prop:valuesubstlemma3}, and~\ref{prop:contextsubstlemma},  we can  show that the following equations hold: 
\[
\begin{array}{c}
\Gamma_4[y'/y][V_1/y'] = \Gamma_4[V_1/y'][y'[V_1/y']/y] = \Gamma_4[y'[V_1/y']/y] = \Gamma_4[V_1/y]
\\[1mm]
M[y'/y][V_1/y'] = M[V_1/y'][y'[V_1/y']/y] = M[y'[V_1/y']/y] = M[V_1/y]
\\[1mm]
M[y'/y][V_2/y'] = M[V_2/y'][y'[V_2/y']/y] = M[y'[V_2/y']/y] = M[V_2/y]
\\[1mm]
\ul{C}[y'/y][V_1/y'] = \ul{C}[V_1/y'][y'[V_1/y']/y] = \ul{C}[y'[V_1/y']/y] = \ul{C}[V_1/y]
\end{array}
\]

As a result, the derivation we constructed above that ends with $(**)$ gives us the required derivation for
$
\ceq {\Gamma_{31}, x \!:\! A,\Gamma_{32}, \Gamma_4[V_1/y]} {M[V_1/y]} {M[V_2/y]} {\ul{C}[V_1/y]}
$.

Other cases that involve substitution in rule conclusions are proved analogously.
\end{proof}

\begin{theorem}[Value term substitution] 
\label{thm:substitution}
\index{substitution theorem!syntactic --!-- for value terms}
Assuming $\vj {\Gamma_1} V A$, we have:
\begin{enumerate}[(a)]
\item Given $\vdash {\Gamma_1, x \!:\! A,\Gamma_2}$, then $\vdash {\Gamma_1,\Gamma_2[V/x]}$.
\item Given $\ljeq{} {\Gamma_1, x \!:\! A,\Gamma_2} {\Gamma_3, x \!:\! A',\Gamma_4}$, then $\ljeq {} {\Gamma_1,\Gamma_2[V/x]} {\Gamma_3,\Gamma_4[V/x]}$.
\item Given $\lj {\Gamma_1, x \!:\! A,\Gamma_2} B$, then $\lj {\Gamma_1,\Gamma_2[V/x]} B[V/x]$.
\item Given $\ljeq {\Gamma_1,x \!:\! A,\Gamma_2} {B_1} {B_2}$, then $\ljeq {\Gamma_1,\Gamma_2[V/x]} {B_1[V/x]} {B_2[V/x]}$.
\item Given $\lj {\Gamma_1,x \!:\! A,\Gamma_2} \ul{C}$, then $\lj {\Gamma_1,\Gamma_2[V/x]} \ul{C}[V/x]$.
\item Given $\ljeq {\Gamma_1,x \!:\! A,\Gamma_2} {\ul{C}} {\ul{D}}$, then $\ljeq {\Gamma_1,\Gamma_2[V/x]} {\ul{C}[V/x]} {\ul{D}[V/x]}$.
\item Given $\vj {\Gamma_1,x \!:\! A,\Gamma_2} W B$, then $\vj {\Gamma_1,\Gamma_2[V/x]} {W[V/x]} {B[V/x]}$.
\item Given $\veq {\Gamma_1,x \!:\! A,\Gamma_2} {W_1} {W_2} B$, then $\veq {\Gamma_1,\Gamma_2[V/x]} {W_1[V/x]} {W_2[V/x]} {B[V/x]}$.
\item Given $\cj {\Gamma_1,x \!:\! A,\Gamma_2} M \ul{C}$, then $\cj {\Gamma_1,\Gamma_2[V/x]} {M[V/x]} {\ul{C}[V/x]}$.
\item Given $\ceq {\Gamma_1,x \!:\! A,\Gamma_2} M N \ul{C}$, then $\ceq {\Gamma_1,\Gamma_2[V/x]} {M[V/x]} {N[V/x]} {\ul{C}[V/x]}$.
\item Given $\hj {\Gamma_1,x \!:\! A,\Gamma_2} {z \!:\! \ul{C}} K \ul{D}$, then $\hj {\Gamma_1,\Gamma_2[V/x]} {z \!:\! \ul{C}[V/x]} {K[V/x]} {\ul{D}[V/x]}$.
\item Given $\heq {\Gamma_1,x \!:\! A,\Gamma_2} {z \!:\! \ul{C}} K L \ul{D}$, then 

\hfill$\heq {\Gamma_1,\Gamma_2[V/x]} {z \!:\! \ul{C}[V/x]} {K[V/x]} {L[V/x]} {\ul{D}[V/x]}$.
\end{enumerate}
\end{theorem}

\begin{proof}
We prove this theorem simultaneously with Theorem~\ref{thm:weakening}.
We prove all cases simultaneously: $(a)$--$(b)$ by induction on the length of $\Gamma_2$ and $(c)$--$(l)$ by induction on the derivations of the given judgements. We sketch the proofs of three cases that need extra work, namely, those involving context and type conversions, and substitution.

\vspace{0.2cm}

\noindent \textbf{Context and type conversion rule for computation terms:}
In this case, the given derivation ends with
\[
\mkrule
{\cj {\Gamma_1, x \!:\! A,\Gamma_2} M {\ul{D}}}
{\cj {\Gamma_3} M {\ul{C}} \quad \vdash \Gamma_3 = \Gamma_1, x \!:\! A,\Gamma_2 \quad \ljeq {\Gamma_3} {\ul{C}} {\ul{D}}}
\]
and we are required to construct a derivation of ${\cj {\Gamma_1, \Gamma_2[V/x]} {M[V/x]} {\ul{D}[V/x]}}$.

First, using Corollary~\ref{cor:equalcontextconcatenation2}, together with the symmetry of definitionally equal value contexts, we get that $\Gamma_3 = \Gamma_4, x \!:\! B,\Gamma_5$ and $\ljeq{} {\Gamma_1} {\Gamma_4}$ and $\ljeq {\Gamma_1} {A} {B}$, for some value contexts $\Gamma_4$ and $\Gamma_5$, and a value type $B$. 

Next, by combining these definitional equations with the context and type conversion rule for value terms, we get a derivation of $\vj {\Gamma_4} V {B}$.  

As a result, we can use the induction hypothesis on the given derivation of \linebreak $\cj {\Gamma_4, x \!:\! B,\Gamma_5} M {\ul{C}}$ to get a derivation of $\cj {\Gamma_4,\Gamma_5[V/x]} {M[V/x]} {\ul{C}[V/x]}$. Analogously, we can use $(b)$ on the derivation of $\ljeq {} {\Gamma_4, x \!:\! B,\Gamma_5} {\Gamma_1, x \!:\! A,\Gamma_2}$ to get a derivation of  $\ljeq {} {\Gamma_4,\Gamma_5[V/x]} {\Gamma_1,\Gamma_2[V/x]}$, and $(d)$ on the derivation of $\ljeq {\Gamma_4, x \!:\! B,\Gamma_5} {\ul{C}} {\ul{D}}$ to get a derivation of $\ljeq {\Gamma_4,\Gamma_5[V/x]} {\ul{C}[V/x]} {\ul{D}[V/x]}$.

As a result, we can now construct the required derivation, as shown below:
\[
\mkrule
{\cj {\Gamma_1, \Gamma_2[V/x]} {M[V/x]} {\ul{D}[V/x]}}
{
\begin{array}{c}
(1)
\quad
\vdash \Gamma_4,\Gamma_5[V/x] = \Gamma_1,\Gamma_2[V/x] \quad \ljeq {\Gamma_4,\Gamma_5[V/x]} {\ul{C}[V/x]} {\ul{D}[V/x]}
\end{array}
}
\]
where $(1)$ denotes a derivation of
\[
\cj {\Gamma_4,\Gamma_5[V/x]} {M[V/x]} {\ul{C}[V/x]} 
\]

Other cases involving context and type conversion rules are proved analogously.

\vspace{0.2cm}

\noindent \textbf{Typing rule for computational pairing:}
In this case, the given derivation ends with
\[
\mkrule
{\cj {\Gamma_1, x \!:\! A,\Gamma_2} {\langle W , M \rangle_{(y : B).\, \ul{C}}} {\Sigma \, y \!:\! B .\, \ul{C}}}
{\vj {\Gamma_1, x \!:\! A,\Gamma_2} W B \quad \lj {\Gamma_1, x \!:\! A,\Gamma_2, y \!:\! B} {\ul{C}} \quad \cj {\Gamma_1, x \!:\! A,\Gamma_2} M {\ul{C}[W/y]}}
\]
and we are required to construct a derivation of 
\[
\cj {\Gamma_1, \Gamma_2[V/x]} {\langle W[V/x] , M[V/x] \rangle_{(y : B[V/x]).\, \ul{C}[V/x]}} {\Sigma \, y \!:\! B[V/x] .\, \ul{C}[V/x]}
\]

First, we use $(g)$ on the derivation of $\vj {\Gamma_1, x \!:\! A,\Gamma_2} W A$ to get a derivation \linebreak of $\vj {\Gamma_1,\Gamma_2[V/x]} {W[V/x]} {A[V/x]}$. 
Analogously, we use $(e)$ on the given derivation of $\lj {\Gamma_1, x \!:\! A,\Gamma_2, y \!:\! B} {\ul{C}}$ to get a derivation of $\lj {\Gamma_1,\Gamma_2[V/x], y \!:\! B[V/x]} {\ul{C}[V/x]}$. 

Next, we use the induction hypothesis on the derivation of $\cj {\Gamma_1, x \!:\! A,\Gamma_2} M {\ul{C}[W/x]}$ to get a derivation of $\cj {\Gamma_1, \Gamma_2[V/x]} {M[V/x]} {\ul{C}[W/y][V/x]}$, which is the same as a derivation of $\cj {\Gamma_1, \Gamma_2[V/x]} {M[V/x]} {\ul{C}[V/x][W[V/x]/y]}$ due to Proposition~\ref{prop:valuesubstlemma3}. 
In particular, based on our adopted variable convention, we have $y \not\in V\!ars(\Gamma_1, x \!:\! A, \Gamma_2)$, from which it follows that $y \not\in V\!ars(\Gamma_1)$.
Further, according to Proposition~\ref{prop:freevariablesofwellformedexpressions}, we also know that $FVV(V) \subseteq V\!ars(\Gamma_1)$. Therefore, $y \not\in FVV(V)$.
Combining these observations with Proposition~\ref{prop:valuesubstlemma3}, we get that $\ul{C}[W/y][V/x] = \ul{C}[V/x][W[V/x]/y]$.

Finally, we can construct the required derivation by using the typing rule for computational pairing with the derivations we have constructed above, as shown below:
\[
\mkrule
{\cj {\Gamma_1, \Gamma_2[V/x]} {\langle W[V/x] , M[V/x] \rangle_{(y : B[V/x]).\, \ul{C}[V/x]}} {\Sigma \, y \!:\! B[V/x] .\, \ul{C}[V/x]}}
{
\vj {\Gamma_1,\Gamma_2[V/x]} {W[V/x]} {B[V/x]} \quad (1) \quad \cj {\Gamma_1,\Gamma_2[V/x]} {M[V/x]} {\ul{C}[V/x][W[V/x]/y]}
}
\]
where $(1)$ denotes a derivation of 
\[
\lj {\Gamma_1,\Gamma_2[V/x], y \!:\! B[V/x]} {\ul{C}[V/x]} 
\]

Other cases that involve substitution in rule premises are proved analogously.

\vspace{0.2cm}

\noindent \textbf{Replacement rule for computation terms:}
In this case, the given derivation ends with
\[
\mkrule
{\ceq {\Gamma_3, \Gamma_4[V_1/y]} {M[V_1/y]} {M[V_2/y]} {\ul{C}[V_1/y]}}
{\cj {\Gamma_3, y \!:\! B, \Gamma_4} M {\ul{C}} \quad \veq {\Gamma_3} {V_1} {V_2} {B}}
\]
where $\Gamma_1, x \!:\! A,\Gamma_2 = \Gamma_3, \Gamma_4[V_1/y]$. 

Similarly to the analogous case in the proof of Theorem~\ref{thm:weakening}, we now have two possibilities to consider, depending on whether $x \in V\!ars(\Gamma_3)$ or $x \in V\!ars(\Gamma_4[V_1/y])$. 

\vspace{0.2cm}
\noindent
\emph{Case for $x \in V\!ars(\Gamma_3)$}: In this case, the value context $\Gamma_3$ is of the form $\Gamma_{31}, x \!:\! A, \Gamma_{32}$, with $\Gamma_1 = \Gamma_{31}$ and $\Gamma_2 = \Gamma_{31}, \Gamma_4[V_1/y]$, and we are required to construct a derivation of 
\[
\ceq {\Gamma_3, \Gamma_4[V_1/y][V/x]} {M[V_1/y][V/x]} {M[V_2/y][V/x]} {\ul{C}[V_1/y][V/x]}
\]

First, we use $(i)$ on the derivation of $\cj {\Gamma_{31}, x \!:\! A, \Gamma_{32}, y \!:\! B, \Gamma_4} M {\ul{C}}$ to get a \linebreak derivation of $\cj {\Gamma_{31}, \Gamma_{32}[V/x], y \!:\! B[V/x], \Gamma_4[V/x]} {M[V/x]} {\ul{C}[V/x]}$. 

Next, we use $(h)$ on the derivation of $\veq {\Gamma_{31}, x \!:\! A, \Gamma_{32}} {V_1} {V_2} {B}$ to get a derivation of $\veq {\Gamma_{31}, \Gamma_{32}[V/x]} {V_1[V/x]} {V_2[V/x]} {B[V/x]}$. 

Using these derivations, we can use the replacement rule for computation terms to get a derivation ending with
\[
\mkrulelabel
{
\begin{array}{c@{~} c@{~} l}
{\Gamma_3, \Gamma_4[V/x][V_1[V/x]/y]} & \vdash & {M[V/x][V_1[V/x]/y]} 
\\[-1mm]
& = & {M[V/x][V_2[V/x]/y]} : {\ul{C}[V/x][V_1[V/x]/y]}
\end{array}
}
{\cj {\Gamma_3, y \!:\! B[V/x], \Gamma_4[V/x]} {M[V/x]} {\ul{C}} \quad \veq {\Gamma_3[V/x]} {V_1[V/x]} {V_2[V/x]} {B[V/x]}}
{(*)}
\]

Next, according to Proposition~\ref{prop:freevariablesofwellformedexpressions}, we know that $FVV(V) \subseteq V\!ars(\Gamma_{31})$. Therefore, as $y \not\in V\!ars(\Gamma_{31})$, we also know that $y \not\in FVV(V)$. 

By combining these observations with Propositions~\ref{prop:valuesubstlemma3} and~\ref{prop:contextsubstlemma}, we can show that the following equations hold: 
\[
\begin{array}{c}
\Gamma_4[V/x][V_1[V/x]/y] = \Gamma_4[V_1/y][V/x]
\\[1mm]
M[V/x][V_1[V/x]/y] = M[V_1/y][V/x]
\\[1mm]
M[V/x][V_2[V/x]/y] = M[V_2/y][V/x]
\\[1mm]
\ul{C}[V/x][V_1[V/x]/y] = \ul{C}[V_1/y][V/x]
\end{array}
\]

As a result, the derivation we constructed above that ends with $(*)$ gives us the required derivation of 
$
\ceq {\Gamma_3, \Gamma_4[V_1/y][V/x]} {M[V_1/y][V/x]} {M[V_2/y][V/x]} {\ul{C}[V_1/y][V/x]}
$.

\vspace{0.2cm}
\noindent
\emph{Case for $x \in V\!ars(\Gamma_4[V_1/y])$}: In this case, the value context $\Gamma_4[V_1/y]$ is of the form $\Gamma_{41}[V_1/y], x \!:\! A[V_1/y], \Gamma_{42}[V_1/y]$, with $\Gamma_1 = \Gamma_3, \Gamma_{41}[V_1/y]$ and $\Gamma_2 = \Gamma_{42}[V_1/y]$, and we are required to construct a derivation of 
\[
\ceq {\Gamma_3, \Gamma_{41}[V_1/y],\Gamma_{42}[V_1/y][V/x]} {M[V_1/y][V/x]} {M[V_2/y][V/x]} {\ul{C}[V_1/y][V/x]}
\]

First, we use $(i)$ on the derivation of $\cj {\Gamma_3, y \!:\! B, \Gamma_{41}, x \!:\! A, \Gamma_{42}} M {\ul{C}}$ to get a derivation of $\cj {\Gamma_3, y \!:\! B, \Gamma_{41}, \Gamma_{42}[V/x]} {M[V/x]} {\ul{C}[V/x]}$. 

Using this derivation, together with the given derivation of $\veq {\Gamma_3} {V_1} {V_2} {B}$, we can  use the replacement rule for computation terms to get a derivation ending with
\[
\mkrulelabel
{\ceq {\Gamma_3, \Gamma_{41}[V_1/y],\Gamma_{42}[V/x][V_1/y]} {M[V/x][V_1/y]} {M[V/x][V_2/y]} {\ul{C}[V/x][V_1/y]}}
{\cj {\Gamma_3, y \!:\! B, \Gamma_{41},\Gamma_{42}[V/x]} {M[V/x]} {\ul{C}[V/x]} \quad \veq {\Gamma_3} {V_1} {V_2} {B}}
{(**)}
\]

Next, according to Proposition~\ref{prop:freevariablesofwellformedexpressions}, we know that $FVV(V) \subseteq V\!ars(\Gamma_3, \Gamma_{41}[V_1/y])$. Therefore, as $y \not\in V\!ars(\Gamma_3, \Gamma_{41}[V_1/y])$, we also know that $y \not\in FVV(V)$. 

In addition, according to Proposition~\ref{prop:freevariablesofwellformedexpressions}, we know that $FVV(V_1) \subseteq V\!ars(\Gamma_3)$ and $FVV(V_2) \subseteq V\!ars(\Gamma_3)$. Therefore, as $x \not\in V\!ars(\Gamma_3)$, we also have that $x \not\in FVV(V_1)$ and $x \not\in FVV(V_2)$. 

By combining these observations with Propositions~\ref{prop:valuesubstlemma1},~\ref{prop:valuesubstlemma3}, and~\ref{prop:contextsubstlemma}, we can show that the following equations hold:
\[
\begin{array}{c}
\Gamma_{42}[V/x][V_1/y] = \Gamma_{42}[V/x][V_1[V/x]/y] = \Gamma_{42}[V_1/y][V/x]
\\[1mm]
M[V/x][V_1/y] = M[V/x][V_1[V/x]/y] = M[V_1/y][V/x]
\\[1mm]
M[V/x][V_2/y] = M[V/x][V_2[V/x]/y] = M[V_2/y][V/x]
\\[1mm]
\ul{C}[V/x][V_1/y] = \ul{C}[V/x][V_1[V/x]/y] = \ul{C}[V_1/y][V/x]
\end{array}
\]

As a result, the above derivation ending with $(**)$ gives us the required derivation of 
$
\ceq {\Gamma_3, \Gamma_{41}[V_1/y],\Gamma_{42}[V_1/y][V/x]} {M[V_1/y][V/x]} {M[V_2/y][V/x]} {\ul{C}[V_1/y][V/x]}
$.

Other cases involving substitution in rule conclusions are proved similarly. In the cases that involve substituting computation terms and homomorphism terms for computation variables, the proofs use Propositions~\ref{prop:compsubstvaluesubst} and~\ref{prop:hompsubstvaluesubst}, respectively.
\end{proof}

In addition to the substitution theorem for value variables, we also prove two substitution theorems for computation variables, one for substituting computation terms and one for substituting homomorphism terms.

\begin{theorem}[Computation term substitution] 
\label{thm:compsubstitution}
\index{substitution theorem!syntactic --!-- for computation terms}
Assuming $\cj {\Gamma} M {\ul{C}}$, we have:
\begin{enumerate}[(a)]
\item Given $\hj {\Gamma} {z \!:\! \ul{C}} K \ul{D}$, then $\cj {\Gamma} {K[M/z]} {\ul{D}}$.
\item Given $\heq {\Gamma} {z \!:\! \ul{C}} K L \ul{D}$, then $\ceq {\Gamma} {K[M/z]} {L[M/z]} {\ul{D}}$.
\end{enumerate}
\end{theorem}

\begin{proof}
We first prove $(a)$ and then $(b)$, by induction on the given derivations of \linebreak $\hj {\Gamma} {z \!:\! \ul{C}} K \ul{D}$ and $\heq {\Gamma} {z \!:\! \ul{C}} K L \ul{D}$, respectively. 
We sketch the proof of a case that needs extra work compared to other cases, namely, the case that involves substitution.

\vspace{0.2cm}

\noindent \textbf{Replacement rule for homomorphism terms:}
In this case, the given derivation ends with
\[
\mkrule
{\heq {\Gamma} {z_2 \!:\! \ul{C}} {K[L_1/z_1]} {K[L_2/z_1]} {\ul{D}_2}}
{\hj {\Gamma} {z_1 \!:\! \ul{D}_1} K {\ul{D}_2} \quad \heq {\Gamma} {z_2 \!:\! \ul{C}} {L_1} {L_2} {\ul{D}_1}}
\]
and we are required to build a derivation of $\ceq {\Gamma} {K[L_1/z_1][M/z_2]} {K[L_2/z_1][M/z_2]} {\ul{D}_2}$.

First, we use the induction hypothesis on the derivation of  $\heq {\Gamma} {z_2 \!:\! \ul{C}} {L_1} {L_2} {\ul{D}_1}$ to get a derivation of $\ceq {\Gamma} {L_1[M/z_2]} {L_2[M/z_2]} {\ul{D}_1}$. 

Next, using this derivation, together with the derivation of $\hj {\Gamma} {z_1 \!:\! \ul{D}_1} K {\ul{D}_2}$, we can use the replacement rule for homomorphism terms to get a derivation ending with
\[
\mkrulelabel
{\ceq {\Gamma} {K[L_1[M/z_2]/z_1]} {K[L_2[M/z_2]/z_1]} {\ul{D}_2}}
{\hj {\Gamma} {z_1 \!:\! \ul{D}_1} K {\ul{D}_2} \quad \ceq {\Gamma} {L_1[M/z_2]} {L_2[M/z_2]} {\ul{D}_1}}
{(*)}
\]

Next, we can use Proposition~\ref{prop:hompsubstcompsubst} to show that the following equations hold: 
\[
\begin{array}{c}
K[L_1[M/z_2]/z_1] = K[L_1/z_1][M/z_2]
\qquad
K[L_2[M/z_2]/z_1] = K[L_2/z_1][M/z_2]
\end{array}
\]

As a result,  the derivation we constructed above that ends with $(*)$ gives us the required derivation of 
$\ceq {\Gamma} {K[L_1/z_1][M/z_2]} {K[L_2/z_1][M/z_2]} {\ul{D}_2}$.

Other cases involving substitution are proved similarly. In the cases that involve substituting value terms for value variables, the proofs use Propositions~\ref{prop:valuesubstlemma1} and~\ref{prop:compsubstvaluesubst}.
\end{proof}

\begin{theorem}[Homomorphism term substitution] 
\label{thm:homsubstitution}
\index{substitution theorem!syntactic --!-- for homomorphism terms}
Assuming $\hj {\Gamma} {z_1 \!:\! \ul{C}} K {\ul{D}_1}$, we have:
\begin{enumerate}[(a)]
\item Given $\hj {\Gamma} {z_2 \!:\! \ul{D}_1} L {\ul{D}_2}$, then $\hj {\Gamma} {z_1 \!:\! \ul{C}} {L[K/z]} {\ul{D}_2}$.
\item Given $\heq {\Gamma} {z_2 \!:\! \ul{D}_1} {L_1} {L_2} {\ul{D}_2}$, then $\heq {\Gamma} {z_1 \!:\! \ul{C}} {L_1[K/z]} {L_2[K/z]} {\ul{D}_2}$.
\end{enumerate}
\end{theorem}

\begin{proof}
We first prove $(a)$ and then $(b)$, by induction on the derivations of \linebreak $\hj {\Gamma} {z_2 \!:\! \ul{D}_1} L {\ul{D}_2}$ and $\heq {\Gamma} {z_2 \!:\! \ul{D}_1} {L_1} {L_2} {\ul{D}_2}$, respectively. 
We sketch the proof of a case that needs extra work compared to other cases, namely, the case that involves substitution.

\vspace{0.2cm}

\noindent \textbf{Replacement rule for homomorphism terms:}
In this case, the given derivation ends with
\[
\mkrule
{\heq {\Gamma} {z_2 \!:\! \ul{D}_1} {L[L_1/z_1]} {L[L_2/z_1]} {\ul{D}_2}}
{\hj {\Gamma} {z_3 \!:\! \ul{D}_3} L {\ul{D}_2} \quad \heq {\Gamma} {z_2 \!:\! \ul{D}_1} {L_1} {L_2} {\ul{D}_3}}
\]
and we are required to construct a derivation of 
\[
{\heq {\Gamma} {z_1 \!:\! \ul{C}} {L[L_1/z_1][K/z_2]} {L[L_2/z_1][K/z_2]} {\ul{D}_2}}
\]

First, we use the induction hypothesis on the derivation of  ${\heq {\Gamma} {z_2 \!:\! \ul{D}_1} {L_1} {L_2} {\ul{D}_3}}$ to get a derivation of $\heq {\Gamma} {z_1 \!:\! \ul{C}} {L_1[K/z_2]} {L_2[K/z_2]} {\ul{D}_1}$. 

Next, using this derivation, together with the derivation of $\hj {\Gamma} {z_3 \!:\! \ul{D}_3} L {\ul{D}_2}$, we can use the replacement rule for homomorphism terms to get a derivation ending with
\[
\mkrulelabel
{\heq {\Gamma} {z_1 \!:\! \ul{C}} {L[L_1[K/z_2]/z_1]} {L[L_2[K/z_2]/z_1]} {\ul{D}_2}}
{\hj {\Gamma} {z_3 \!:\! \ul{D}_3} L {\ul{D}_2} \quad \heq {\Gamma} {z_1 \!:\! \ul{C}} {L_1[K/z_2]} {L_2[K/z_2]} {\ul{D}_3}}
{(*)}
\]

Next, we can use Proposition~\ref{prop:hompsubsthomsubst} to show that the following equations hold: 
\[
\begin{array}{c}
L[L_1[K/z_2]/z_1] = L[L_1/z_1][K/z_2]
\qquad
L[L_2[K/z_2]/z_1] = L[L_2/z_1][K/z_2]
\end{array}
\]

As a result, the derivation we constructed above that ends with $(*)$ gives us the required derivation of 
$\heq {\Gamma} {z_1 \!:\! \ul{C}} {L[L_1/z_1][K/z_2]} {L[L_2/z_1][K/z_2]} {\ul{D}_2}$.

Other cases involving substitution are proved similarly. In the cases that involve substituting value terms for value variables, the proofs use Propositions~\ref{prop:valuesubstlemma1} and~\ref{prop:hompsubstvaluesubst}.
\end{proof}

In the next proposition we show that the value types that are assigned to the same variable in definitionally equal contexts are themselves definitionally equal.

\begin{proposition}
\label{prop:findingvariableinequalcontexts}
Given value contexts $\Gamma_1$ and $\Gamma_2$, a value variable $x$, and a value type $A$ such that $\vdash \Gamma_1 = \Gamma_2$ and $x \!:\! A \in \Gamma_1$, then there exists a value type $B$ such that $x \!:\! B \in \Gamma_2$ and $\ljeq {\Gamma_1} A B$.
\end{proposition}

\begin{proof}
We prove this proposition by induction on the derivation of $\vdash \Gamma_1 = \Gamma_2$, using Theorem~\ref{thm:weakening} when the last variable of $\Gamma_1$ and $\Gamma_2$ is different from $x$ in order to weaken the definitional equation given by the induction hypothesis.
\end{proof}

Next, we prove inversion lemmas for well-formed expressions, showing that the corresponding formation and typing rules can be inverted.

\begin{proposition} 
\label{prop:valuetypeinversion}
The following inversion principles are valid for value types:
\begin{enumerate}[(a)]
\item Given $\lj \Gamma {\Sigma\, x \!:\! A .\, B}$, then $\lj \Gamma A$ and $\lj {\Gamma, x \!:\! A} B$.
\item Given $\lj \Gamma {\Pi\, x \!:\! A .\, B}$, then $\lj \Gamma A$ and $\lj {\Gamma, x \!:\! A} B$.
\item Given $\lj \Gamma {A + B}$, then $\lj \Gamma A$ and $\lj {\Gamma} B$.
\item Given $\lj \Gamma {V =_A W}$, then $\lj \Gamma A$ and $\vj {\Gamma} V A$ and $\vj {\Gamma} W A$.
\item Given $\lj \Gamma {U\ul{C}}$, then $\lj \Gamma \ul{C}$.
\item Given $\lj \Gamma {\ul{C} \multimap \ul{D}}$, then $\lj \Gamma \ul{C}$ and $\lj \Gamma \ul{D}$.
\end{enumerate}
\end{proposition}

\begin{proof}
We prove $(a)$--$(f)$ independently of each other, by induction on the given derivations. 

First, by recalling the rules that define the judgement $\lj \Gamma A$ from Definition~\ref{def:judgements}, we can see that the given derivation can only end with the corresponding type formation rule or with the context conversion rule for value types.

If the given derivation ends with a type formation rule, the required derivations follow immediately from the premises of that rule.

If the given derivation ends with the context conversion rule, we apply the context conversion rule on the derivations given to us by the induction hypotheses. For example, in the case of the context conversion rule for $(a)$, the given derivation ends with
\[
\mkrule
{\lj {\Gamma_2} {\Sigma\, x \!:\! A .\, B}}
{\vdash \Gamma_1 = \Gamma_2 \quad \lj {\Gamma_1} {\Sigma\, x \!:\! A .\, B}}
\]
By using the induction hypothesis on the derivation of $\lj {\Gamma_1} {\Sigma\, x \!:\! A .\, B}$, we get derivations of $\lj {\Gamma_1} {A}$ and $\lj {\Gamma_1, x \!:\! A} {B}$. We can then use the context conversion rule with $\vdash \Gamma_1 = \Gamma_2$ on these derivations to get the required derivations of $\lj {\Gamma_2} {A}$ and $\lj {\Gamma_2, x \!:\! A} {B}$.
\end{proof}

\begin{proposition} 
\label{prop:computationtypeinversion}
The following inversion principles are valid for computation types:
\begin{enumerate}[(a)]
\item Given $\lj \Gamma {FA}$, then $\lj \Gamma A$.
\item Given $\lj \Gamma {\Sigma\, x \!:\! A .\, \ul{C}}$, then $\lj \Gamma A$ and $\lj {\Gamma, x \!:\! A} \ul{C}$.
\item Given $\lj \Gamma {\Pi\, x \!:\! A .\, \ul{C}}$, then $\lj \Gamma A$ and $\lj {\Gamma, x \!:\! A} \ul{C}$.
\end{enumerate}
\end{proposition}

\begin{proof}
We prove $(a)$--$(c)$ independently of each other, by induction on the given derivations. As in the proof of Proposition~\ref{prop:valuetypeinversion}, we again observe that the given derivations can only end with a corresponding type formation rule, or with a context conversion rule. Both cases are treated analogously to the proof of Proposition~\ref{prop:valuetypeinversion}.
\end{proof}

\begin{proposition}
\label{prop:valueterminversion}
The following inversion principles are valid for value terms:
\begin{enumerate}[(a)]
\item Given $\vj {\Gamma} {x} A$, then $x \!:\! B \in \Gamma$ and $\ljeq \Gamma A B$, for some value type $B$.
\item Given $\vj \Gamma {\zero} A$, then $\ljeq \Gamma A {\Nat}$.
\item Given $\vj \Gamma {\suc V} A$, then $\ljeq \Gamma A {\Nat}$ and $\vj \Gamma V \Nat$.
\item Given $\vj \Gamma {\natrec {x.A} {V_z} {y_1.y_2.V_s} V} {B}$, then $\ljeq \Gamma B {A[V/x]}$ and 

\vspace{-0.2cm}
\hfill $\vj \Gamma {V_z} {\Nat}$ and $\vj {\Gamma, y_1 \!:\! \Nat, y_2 \!:\! A[y_1/x]} {V_s} {A[\suc y_1/x]}$ and $\vj \Gamma V \Nat$.
\item[] \ldots
\item[(q)] Given $\vj \Gamma {\lambda\, x \!:\! \ul{C} .\, K} {A}$, then $\ljeq \Gamma A {\ul{C} \multimap \ul{D}}$ and $\hj {\Gamma} {z \!:\! \ul{C}} K {\ul{D}}$, 

\vspace{-0.2cm}
\hfill for some computation type $\ul{D}$ with $\lj {\Gamma} {\ul{D}}$.

\end{enumerate}
\end{proposition}

\begin{proof}
We prove $(a)$--$(q)$ independently of each other, by induction on the given derivations. 

First, by recalling the rules that define the judgement $\vj \Gamma V A$ from Definition~\ref{def:judgements}, we can see that the given derivation can only end with the corresponding typing rule, or with a context and type conversion rule.

If the given derivation ends with a typing rule, the required derivations of the subterms of the given term follow immediately from the premises of that rule. Further, we prove the required definitional equations using the reflexivity of $\ljeq \Gamma A B$. 

If the given derivation ends with a context and type conversion rule, we apply the context and type conversion rule on the derivations given by the induction hypotheses.  For example, in the case of the context and type conversion rule for $(q)$, the given derivation ends with 
\[
\mkrule
{\vj {\Gamma_2} {\lambda\, z \!:\! \ul{C} .\, K} {B}}
{\vj {\Gamma_1} {\lambda\, z \!:\! \ul{C} .\, K} {A} \quad \vdash \Gamma_1 = \Gamma_2 \quad \ljeq {\Gamma_1} {A} {B}}
\]
Now, by using the induction hypothesis on the derivation of $\vj {\Gamma_1} {\lambda\, z \!:\! \ul{C} .\, K} {A}$, we get a type $\ul{D}$, and derivations of $\lj {\Gamma_1} {\ul{D}}$, $\ljeq {\Gamma_1} {A} {\ul{C} \multimap \ul{D}}$, and $\hj {\Gamma_1} {z \!:\! \ul{C}} K {\ul{D}}$. We can \linebreak then use the context and type conversion rules on these derivations, in combination with the assumed derivations of $\vdash \Gamma_1 = \Gamma_2$ and $\ljeq {\Gamma_1} {A} {B}$, to get the required derivations of $\lj {\Gamma_2} {\ul{D}}$ and $\ljeq {\Gamma_2} {B} {\ul{C} \multimap \ul{D}}$ and $\hj {\Gamma_2} {z \!:\! \ul{C}} K {\ul{D}}$.

In the induction step case for $(a)$, we further use Proposition~\ref{prop:findingvariableinequalcontexts} to show that the types assigned to $x$ in definitionally equal contexts are definitionally equal.
\end{proof}

\begin{proposition}
\label{prop:compterminversion}
The following inversion principles are valid for computation terms:
\begin{enumerate}[(a)]
\item Given $\cj \Gamma {\return V} \ul{C}$, then $\ljeq \Gamma {\ul{C}} {FA}$ and $\vj \Gamma V A$, for some value type $A$.
\item Given $\cj \Gamma {\doto M {x \!:\! A} {\ul{C}} {N}} \ul{D}$, then $\ljeq \Gamma {\ul{D}} {\ul{C}}$ and $\cj \Gamma M {FA}$ and $\cj {\Gamma, x \!:\! A} N {\ul{C}}$.
\item Given $\cj \Gamma {\langle V , M \rangle_{(x : A).\ul{C}}} {\ul{D}}$, then $\ljeq \Gamma {\ul{D}} {\Sigma\, x \!:\! A .\, \ul{C}}$ and $\vj \Gamma V A$ and $\cj \Gamma M {\ul{C}[V/x]}$.
\item Given $\cj \Gamma {\doto {M} {(x \!:\! A , z \!:\! \ul{C}_1)} {\ul{C}_2} {K}} {\ul{D}}$, then $\ljeq \Gamma {\ul{D}} {\ul{C}_2}$ and 

\vspace{-0.2cm}
\hfill $\cj \Gamma M {\Sigma\, x \!:\! A .\, \ul{C}_1}$ and $\hj {\Gamma, x \!:\! A} {z \!:\! \ul{C}_1} K {\ul{C}_2}$.
\item Given $\cj \Gamma {\lambda\, x \!:\! A .\, M} {\ul{D}}$, then $\ljeq \Gamma {\ul{D}} {\Pi\, x \!:\! A .\, \ul{C}}$ and $\vj {\Gamma, x \!:\! A} M {\ul{C}}$, 

\vspace{-0.2cm}
\hfill for some computation type $\ul{C}$ with $\lj {\Gamma, x \!:\! A} {\ul{C}}$.
\item Given $\cj \Gamma {M(V)_{(x : A) .\, \ul{C}}} {\ul{D}}$, then $\ljeq \Gamma {\ul{D}} {\ul{C}[V/x]}$ and $\cj \Gamma M {\Pi\, x \!:\! A .\, \ul{C}}$ and $\vj \Gamma V A$.
\item Given $\cj \Gamma {\force {\ul{C}} V} {\ul{D}}$, then $\ljeq \Gamma {\ul{D}} {\ul{C}}$ and $\vj \Gamma V {U\ul{C}}$.
\item Given $\cj \Gamma {V(M)_{\ul{C}_1, \ul{C}_2}} {\ul{D}}$, then $\ljeq \Gamma {\ul{D}} {\ul{C}_2}$ and $\vj \Gamma V {\ul{C}_1 \multimap \ul{C}_2}$ and $\cj \Gamma M {\ul{C}_1}$.
\end{enumerate}
\end{proposition}

\begin{proof}
We prove $(a)$--$(h)$ independently of each other, by induction on the given derivations. Similarly to the proof of Proposition~\ref{prop:valueterminversion}, we again observe that the given derivation can only end with a corresponding typing rule, or with a context and type conversion rule. Both cases are treated analogously to the proof of Proposition~\ref{prop:valueterminversion}.
\end{proof}

\begin{proposition}
\label{prop:homterminversion}
The following inversion principles are valid for homomorphism terms:
\begin{enumerate}[(a)]
\item Given $\hj \Gamma {z_1 \!:\! \ul{C}} {z_2} \ul{D}$, then $\ljeq \Gamma {\ul{D}} {\ul{C}}$ and $z_1 = z_2$.
\item Given $\hj \Gamma {z \!:\! \ul{C}} {\doto K {x \!:\! A} {\ul{D}_1} {N}} \ul{D}_2$, then $\ljeq \Gamma {\ul{D}_2} {\ul{D}_1}$ and 

\vspace{-0.2cm}
\hfill $\hj \Gamma {z \!:\! \ul{C}} K {FA}$ and $\cj {\Gamma, x \!:\! A} N {\ul{D}_1}$.
\item Given $\hj \Gamma {z \!:\! \ul{C}} {\langle V , K \rangle_{(x : A).\ul{D}_1}} {\ul{D}_2}$, then $\ljeq \Gamma {\ul{D}_2} {\Sigma\, x \!:\! A .\, \ul{D}_1}$ and 

\vspace{-0.2cm}
\hfill $\vj \Gamma V A$ and $\hj \Gamma {z \!:\! \ul{C}} K {\ul{D}_1[V/x]}$.
\item Given $\hj \Gamma {z_1 \!:\! \ul{C}} {\doto {K} {(x \!:\! A , z_2 \!:\! \ul{D}_1)} {\ul{D}_2} {L}} {\ul{D}_3}$, then $\ljeq \Gamma {\ul{D}_3} {\ul{D}_2}$ and 

\vspace{-0.2cm}
\hfill $\hj \Gamma {z_1 \!:\! \ul{C}} K {\Sigma\, x \!:\! A .\, \ul{D}_1}$ and $\hj {\Gamma, x \!:\! A} {z_2 \!:\! \ul{D}_1} L {\ul{D}_2}$.
\item Given $\hj \Gamma {z \!:\! \ul{C}} {\lambda\, x \!:\! A .\, K} {\ul{D}_2}$, then $\ljeq \Gamma {\ul{D}_2} {\Pi\, x \!:\! A .\, \ul{D}_1}$ and $\hj {\Gamma, x \!:\! A} {z \!:\! \ul{C}} K {\ul{D}_1}$, 

\vspace{-0.2cm}
\hfill for some computation type $\ul{D}_1$ with $\lj {\Gamma, x \!:\! A} {\ul{D}_1}$.
\item Given $\hj \Gamma {z \!:\! \ul{C}} {K(V)_{(x : A) .\, \ul{D}_1}} {\ul{D}_2}$, then $\ljeq \Gamma {\ul{D}_2} {\ul{D}_1[V/x]}$ and 

\vspace{-0.2cm}
\hfill $\hj \Gamma {z \!:\! \ul{C}} K {\Pi\, x \!:\! A .\, \ul{D}_1}$ and $\vj \Gamma V A$.
\item Given $\hj \Gamma {z \!:\! \ul{C}} {V(K)_{\ul{D}_1, \ul{D}_2}} {\ul{D}_3}$, then $\ljeq \Gamma {\ul{D}_3} {\ul{D}_2}$ and 

\vspace{-0.2cm}
\hfill $\vj \Gamma V {\ul{D}_1 \multimap \ul{D}_2}$ and $\hj \Gamma {z \!:\! \ul{C}} K {\ul{D}_1}$.
\end{enumerate}
\end{proposition}

\begin{proof}
We prove $(a)$--$(g)$ independently of each other, by induction on the given derivations. Similarly to the proof of Proposition~\ref{prop:valueterminversion}, we again observe that the given derivation can  only end with a corresponding typing rule, or with a context and type conversion rule. Both cases are treated analogously to the proof of Proposition~\ref{prop:valueterminversion}.
\end{proof}

As a direct consequence of these three inversion lemmas, we can prove that the type assignment in eMLTT is unique up to definitional equations between types.

\pagebreak

\begin{proposition} \mbox{}
\label{prop:uniquenessoftypeassignement}
\begin{enumerate}[(a)]
\item Given $\vj \Gamma V A_1$ and $\vj \Gamma V A_2$, then $\ljeq \Gamma {A_1} {A_2}$.
\item Given $\cj \Gamma M \ul{C}_1$ and $\cj \Gamma M \ul{C}_2$, then $\ljeq \Gamma {\ul{C}_1} {\ul{C}_2}$.
\item Given $\hj \Gamma {z \!:\! \ul{C}} K \ul{D}_1$ and $\hj \Gamma {z \!:\! \ul{C}} K \ul{D}_2$, then $\ljeq \Gamma {\ul{D}_1} {\ul{D}_2}$.
\end{enumerate}
\end{proposition}

\begin{proof}
We prove $(a)$--$(c)$ simultaneously: $(a)$ by induction on the structure of $V$, $(b)$ by induction on the structure of $M$, and $(c)$ by induction on the structure of $K$. In each of these cases, we use the corresponding case of Proposition~\ref{prop:valueterminversion},~\ref{prop:compterminversion}, or~\ref{prop:homterminversion}.

For example, in the case when $V$ is a value variable $x$, the case $(a)$ of Proposition~\ref{prop:valueterminversion} gives us that $x \!:\! B \in \Gamma$, for some value type $B$, together with derivations of $\ljeq \Gamma {A_1} B$ and $\ljeq \Gamma {A_2} B$. By combining these derivations using the symmetry and transitivity rules for definitionally equal value types, we get the required derivation of $\ljeq \Gamma {A_1} {A_2}$.

As another example, in the case when $M$ is of the form $\doto {N_1} {x \!:\! A} {\ul{C}} {N_2}$, the case $(b)$ of Proposition~\ref{prop:compterminversion} gives us derivations of $\ljeq \Gamma {\ul{D}_1} {\ul{C}}$ and $\ljeq \Gamma {\ul{D}_2} {\ul{C}}$. By combining these derivations using the symmetry and transitivity rules for definitionally equal computation types, we get the required derivation of $\ljeq \Gamma {\ul{D}_1} {\ul{D}_2}$.

As a final example, in the case when $K$ is of the form $\lambda\, x \!:\! B .\, L$, the case $(e)$ of Proposition~\ref{prop:homterminversion} gives us derivations of $\ljeq \Gamma {\ul{D}_1} {\Pi\, x \!:\! B .\, \ul{D}_3}$ and $\ljeq \Gamma {\ul{D}_2} {\Pi\, x \!:\! B .\, \ul{D}_4}$, \linebreak together with derivations of $\hj {\Gamma, x \!:\! B} {z \!:\! \ul{C}} L {\ul{D}_3}$ and $\hj {\Gamma, x \!:\! B} {z \!:\! \ul{C}} L {\ul{D}_4}$, for some computation types $\ul{D}_3$ and $\ul{D}_4$. Next, by using the induction hypothesis on the latter two derivations, we get a derivation of $\ljeq {\Gamma, x \!:\! B} {\ul{D}_3} {\ul{D}_4}$. Further, by using the \linebreak congruence rule for the computational $\Pi$-type on this derivation, we get a derivation of \linebreak $\ljeq \Gamma {\Pi\, x \!:\! B .\, \ul{D}_3} {\Pi\, x \!:\! B .\, \ul{D}_4}$. Finally, by combining this derivation with the derivations of $\ljeq \Gamma {\ul{D}_1} {\Pi\, x \!:\! B .\, \ul{D}_3}$ and $\ljeq \Gamma {\ul{D}_2} {\Pi\, x \!:\! B .\, \ul{D}_4}$ using the symmetry and transitivity rules for definitionally equal computation types, we get the required derivation of $\ljeq \Gamma {\ul{D}_1} {\ul{D}_2}$.
\end{proof}

We conclude this section by showing that the judgements of well-formed expressions and definitional equations only involve well-formed contexts, well-formed types, and well-typed terms.

\begin{proposition} \mbox{}
\label{prop:wellformedcomponentsofjudgements}
\begin{enumerate}[(a)]
\item Given $\lj {\Gamma} A$, then $\vdash \Gamma$.
\item Given $\ljeq {\Gamma} {A} {B}$, then $\lj \Gamma A$ and $\lj \Gamma B$.
\item Given $\lj {\Gamma} \ul{C}$, then $\vdash \Gamma$.
\item Given $\ljeq {\Gamma} {\ul{C}} {\ul{D}}$, then $\lj \Gamma \ul{C}$ and $\lj \Gamma \ul{D}$.
\item Given $\vj {\Gamma} V A$, then $\lj \Gamma A$.
\item Given $\veq {\Gamma} V W A$, then $\vj \Gamma V A$ and $\vj \Gamma W A$.
\item Given $\cj {\Gamma} M \ul{C}$, then $\lj \Gamma \ul{C}$.
\item Given $\ceq {\Gamma} M N \ul{C}$, then $\cj \Gamma M \ul{C}$ and $\cj \Gamma N \ul{C}$.
\item Given $\hj {\Gamma} {z \!:\! \ul{C}} K \ul{D}$, then $\lj \Gamma \ul{C}$ and $\lj \Gamma \ul{D}$.
\item Given $\heq {\Gamma} {z \!:\! \ul{C}} K L \ul{D}$, then $\hj {\Gamma} {z \!:\! \ul{C}} K \ul{D}$ and $\hj {\Gamma} {z \!:\! \ul{C}} L \ul{D}$.
\end{enumerate}
\end{proposition}

\begin{proof}
We prove $(a)$--$(j)$ simultaneously, by induction on the given derivations. Below we sketch the proofs of four cases that demonstrate the use of the weakening and substitution theorems, the inversion lemmas from Proposition~\ref{prop:valuetypeinversion}, and the well-formedness assumptions for type annotations.

\vspace{0.2cm}

\noindent \textbf{Typing rule for value variables:}
In this case, the given derivation ends with
\[
\mkrule
{\vj {\Gamma_1, x \!:\! A, \Gamma_2} {x} {A}}
{\vdash \Gamma_1, x \!:\! A, \Gamma_2}
\]
and we are required to construct a derivation of $\lj {\Gamma_1, x \!:\! A, \Gamma_2} A$.

First, we use Proposition~\ref{prop:wellformedtypesinwellformedcontexts} to get a derivation of $\lj {\Gamma_1} A$. Next, we use $(c)$ of Theorem~\ref{thm:weakening} on this derivation of $\lj {\Gamma_1} A$ to get a derivation of  $\lj {\Gamma_1, x \!:\! A} A$. 

Finally, we turn this derivation of $\lj {\Gamma_1, x \!:\! A} A$ into the derivation of $\lj {\Gamma_1, x \!:\! A, \Gamma_2} A$ by induction on the length of $\Gamma_2$, using $(c)$ of Theorem~\ref{thm:weakening}, and by observing that in each of the cases the last rule used in the derivation of $\vdash \Gamma_1, x \!:\! A, \Gamma_2$ has to be the context extension rule.

\vspace{0.2cm}

\noindent \textbf{Typing rule for forcing a thunked computation:}
In this case, the given derivation ends with
\[
\mkrule
{\cj \Gamma {\force {\ul{C}} V} {\ul{C}}}
{\vj \Gamma V U\ul{C}}
\]
and we are required to construct a derivation of $\lj \Gamma \ul{C}$.

We observe that by using $(e)$ on the derivation of $\vj \Gamma V U\ul{C}$, we get a derivation $\lj {\Gamma} U\ul{C}$. As a result, by using Proposition~\ref{prop:valuetypeinversion} on this derivation of $\lj {\Gamma} U\ul{C}$, we get the required derivation of $\lj {\Gamma} \ul{C}$.

\vspace{0.2cm}

\noindent \textbf{Typing rule for computational pairing:}
In this case, the given derivation ends with
\[
\mkrule
{\cj {\Gamma} {\langle W , M \rangle_{(x : A).\, \ul{C}}} {\Sigma \, x \!:\! A .\, \ul{C}}}
{\vj {\Gamma} W B \quad \lj {\Gamma, x \!:\! A} {\ul{C}} \quad \cj {\Gamma} M {\ul{C}[W/y]}}
\]
and we are required to construct derivations of $\vdash \Gamma$ and $\lj \Gamma {\Sigma \, x \!:\! A .\, \ul{C}}$.

We observe that by using the derivation of $\lj {\Gamma, x \!:\! A} {\ul{C}}$, we can construct the required derivation of $\lj \Gamma {\Sigma \, x \!:\! A .\, \ul{C}}$ simply by using the corresponding type formation rule.

\vspace{0.2cm}

\noindent \textbf{Replacement rule for computation terms:}
In this case, the given derivation ends with
\[
\mkrule
{\ceq {\Gamma_1, \Gamma_2[V_1/x]} {M[V_1/x]} {M[V_2/x]} {\ul{C}[V_1/x]}}
{\cj {\Gamma_1, x \!:\! A, \Gamma_2} M {\ul{C}} \quad \veq {\Gamma_1} {V_1} {V_2} {A}}
\]
and we are required to construct derivations of $\cj {\Gamma_1, \Gamma_2[V_1/x]} {M[V_1/x]} {\ul{C}[V_1/x]}$ and \linebreak$\cj {\Gamma_1, \Gamma_2[V_1/x]} {M[V_2/x]} {\ul{C}[V_1/x]}$.

First, we use $(g)$ on the given derivation of $\cj {\Gamma_1, x \!:\! A, \Gamma_2} M {\ul{C}}$ to get derivations of \linebreak $\vdash \Gamma_1, x \!:\! A, \Gamma_2$ and $\lj {\Gamma_1, x \!:\! A, \Gamma_2} {\ul{C}}$. 

Next, we use $(f)$ on the given derivation of $\veq {\Gamma_1} {V_1} {V_2} {A}$ to get derivations of $\vj {\Gamma_1} {V_1} {A}$ and $\vj {\Gamma_1} {V_2} {A}$. 

Further, we use $(i)$ of Theorem~\ref{thm:substitution} on the derivations of $\cj {\Gamma_1, y \!:\! A, \Gamma_2} M {\ul{C}}$, $\vj {\Gamma_1} {V_1} {A}$, and $\vj {\Gamma_1} {V_2} {A}$ to get derivations of $\cj {\Gamma_1, \Gamma_2[V_1/x]} {M[V_1/x]} {\ul{C}[V_1/x]}$ and $\cj {\Gamma_1, \Gamma_2[V_2/x]} {M[V_2/x]} {\ul{C}[V_2/x]}$. 

Finally, we construct the required derivation of $\cj {\Gamma_1, \Gamma_2[V_1/x]} {M[V_2/x]} {\ul{C}[V_1/x]}$ by combining the context and type conversion rule for computation terms with replacement rules for value contexts and computation types, as shown below
\[
\mkrule
{\cj {\Gamma_1, \Gamma_2[V_1/x]} {M[V_2/x]} {\ul{C}[V_1/x]}}
{
(1)
\quad
\mkrule
{\ljeq {} {\Gamma_1, \Gamma_2[V_2/x]} {\Gamma_1, \Gamma_2[V_1/x]}}
{(2)}
\quad
\mkrule
{\ljeq {\Gamma_1, \Gamma_2[V_2/x]} {\ul{C}[V_2/x]} {\ul{C}[V_1/x]}}
{(3)}
}
\]
where $(1)$ is given by
\[
\cj {\Gamma_1, \Gamma_2[V_2/x]} {M[V_2/x]} {\ul{C}[V_2/x]}
\]
and $(2)$ by
\[
\mkrule
{\ljeq {} {\Gamma_1, \Gamma_2[V_2/x]} {\Gamma_1, \Gamma_2[V_1/x]}}
{
\mkrule
{\ljeq {} {\Gamma_1, x \!:\! A, \Gamma_2} {\Gamma_1, x \!:\! A, \Gamma_2}}
{\vdash {\Gamma_1, x \!:\! A, \Gamma_2}}
\quad 
\mkrule
{\veq {\Gamma_1} {V_2} {V_1} {A}}
{\veq {\Gamma_1} {V_1} {V_2} {A}}
}
\]
and $(3)$ by
\[
\mkrule
{\ljeq {\Gamma_1, \Gamma_2[V_2/x]} {\ul{C}[V_2/x]} {\ul{C}[V_1/x]}}
{
\mkrule
{\ljeq {\Gamma_1, x \!:\! A, \Gamma_2} {\ul{C}} {\ul{C}}}
{\lj {\Gamma_1, x \!:\! A, \Gamma_2} {\ul{C}}}
\quad
\mkrule
{\veq {\Gamma_1} {V_2} {V_1} {A}}
{\veq {\Gamma_1} {V_1} {V_2} {A}}
}
\]

\end{proof}

\begin{proposition}
Given $\vdash {\Gamma_1} =  \Gamma_2$, then $\vdash \Gamma_1$ and $\vdash \Gamma_2$.
\end{proposition}

\begin{proof}
We prove this proposition by induction on the derivation of $\vdash {\Gamma_1} =  \Gamma_2$. 

In the case of the congruence rule for context extension, when the equation is of the form $\vdash {\Gamma_1, x \!:\! A} = \Gamma_2, x \!:\! B$, we use $(b)$ of Proposition~\ref{prop:wellformedcomponentsofjudgements} to get derivations of $\lj {\Gamma_1} A$ and $\lj {\Gamma_1} B$ from the derivation of $\ljeq {\Gamma_1} A B$. We then use the context conversion rule on $\lj {\Gamma_1} B$ to get a derivation of $\lj {\Gamma_2} B$. Finally, we use the context extension rule on $\lj {\Gamma_1} A$ and $\lj {\Gamma_2} B$ to get the required derivations of $\lj {} {\Gamma_1, x \!:\! A}$ and $\lj {} {\Gamma_2, x \!:\! B}$.
\end{proof}

\section{Derivable elimination forms}
\label{sect:derivableeliminationforms}

In this section we show how to eliminate value types into computation and homomorphism terms. In addition, we show that the corresponding $\beta$- and $\eta$-equations hold.

Recall that the type of natural numbers, the value $\Sigma$-type, the empty type, the coproduct type,  and propositional equality are only eliminated into value terms in eMLTT. While being able to eliminate these types into value terms is necessary to accommodate effect-free programs on which eMLTT's types could depend on, it is also desirable to be able to eliminate these types into computation and homomorphism terms, e.g., to use primitive recursion in effectful programs.
Below we show that such elimination forms are derivable using thunking and forcing for computation terms, and homomorphic lambda abstraction and function application for homomorphism terms.

First, we show how to derive the elimination forms for computation terms.

\begin{definition} The \emph{computation term variant of}
\label{def:derivablecomputationtermvariantsofeliminators}
\begin{itemize}
\item \emph{primitive recursion} is defined as
\[
\begin{array}{c}
\hspace{-7.5cm} \natrec {x.\,\ul{C}} {M_z} {y_1.\, y_2.\, M_s} {V} \defeq 
\\[1mm]
\hspace{3cm} \force {\ul{C}[V/x]} (\natrec {x.\,U\ul{C}} {\thunk M_z} {y_1.\, y_2.\, \thunk M_s} {V})
\end{array}
\]
\item \emph{pattern-matching} is defined as
\[
\begin{array}{c}
\hspace{-7cm} \pmatch V {(x_1 \!:\! A_1, x_2 \!:\! A_2)} {y.\, \ul{C}} M  \defeq
\\[1mm]
\hspace{3cm} \force {\ul{C}[V/y]} (\pmatch V {(x_1 \!:\! A_1, x_2 \!:\! A_2)} {y.\, U\ul{C}} {\thunk M})
\end{array}
\]
\item \emph{empty case analysis} is defined as
\[
\absurd {x.\,\ul{C}} V \defeq \force {\ul{C}[V/x]} (\absurd {x.\,U\ul{C}} V)
\]
\item \emph{binary case analysis} is defined as
\[
\begin{array}{c}
\hspace{-3.6cm} \case {V} {y.\, \ul{C}} {\inl {\!} {\!\!(x_1 \!:\! A_1)} \mapsto M} {\inr {\!} {\!\!(x_2 \!:\! A_2)} \mapsto N} \defeq
\\[1mm]
\hspace{2cm} \force {\ul{C}[V/y]} (\case {V} {y.\, U\ul{C}} {\inl {\!} {\!\!(x_1 \!:\! A_1)} \mapsto \thunk M} {
\\[-1mm]
\hspace{7.15cm} \inr {\!} {\!\!(x_2 \!:\! A_2)} \mapsto \thunk N})
\end{array}
\]
\item \emph{elimination of propositional equality} is defined as
\[
\begin{array}{c}
\hspace{-6cm} \pathind A {x_1.\, x_2.\, x_3.\, \ul{C}} {y.\, M} {V_1} {V_2} {V_p} \defeq 
\\[1mm]
\hspace{0.9cm} \force {\ul{C}[V_1/x_1][V_2/x_2][V_p/x_3]} (\pathind A {x_1.\, x_2.\, x_3.\, U\ul{C}} {y.\, \thunk M} {V_1} {V_2} {V_p})
\end{array}
\]
\end{itemize}
\end{definition}

Next, we show that these derived elimination forms are well-typed. 

\begin{proposition}
The following typing rules are derivable for the computation term variant of:
\begin{itemize}
\item primitive recursion
\[
\mkrule
{\cj \Gamma {\natrec {x.\,\ul{C}} {M_z} {y_1.\, y_2.\, M_s} {V}} {\ul{C}[V/x]}}
{
\begin{array}{c}
\lj {\Gamma, x \!:\! \Nat} {\ul{C}} \quad \vj \Gamma V \Nat 
\\[-1mm]
\cj \Gamma {M_z} {\ul{C}[\zero/x]} \quad \cj {\Gamma, y_1 \!:\! \Nat, y_2 \!:\! U\ul{C}[y_1/x]} {M_z} {\ul{C}[\suc y_1/x]}
\end{array}
}
\]
\item pattern-matching
\[
\mkrule
{\cj \Gamma {\pmatch V {(x_1 \!:\! A_1, x_2 \!:\! A_2)} {y.\,\ul{C}} M} {\ul{C}[V/y]}}
{
\begin{array}{c}
\lj {\Gamma, y \!:\! \Sigma \, x_1 \!:\! A_1 .\, A_2} {\ul{C}}
\\[-1mm]
\vj \Gamma V {\Sigma \, x_1 \!:\! A_1 .\, A_2} \quad \cj {\Gamma, x_1 \!:\! A_1, x_2 \!:\! A_2} {M} {\ul{C}[\langle x_1 , x_2 \rangle /y]}
\end{array}
}
\]
\item empty case analysis
\[
\mkrule
{\cj \Gamma {\absurd {x.\,\ul{C}} V} {\ul{C}[V/x]}}
{
\vj {\Gamma} V 0 \quad \lj {\Gamma, x \!:\! 0} \ul{C}
}
\]
\item binary case analysis
\[
\mkrule
{\cj \Gamma {\case {V} {y.\, \ul{C}} {\inl {\!} {\!\!(x_1 \!:\! A_1)} \mapsto M} {\inr {\!} {\!\!(x_2 \!:\! A_2)} \mapsto N}} {\ul{C}[V/y]}}
{
\begin{array}{c}
\lj {\Gamma, y \!:\! A_1 + A_2} \ul{C} \quad \vj \Gamma V {A_1 + A_2} 
\\[-1mm]
\cj {\Gamma, x_1 \!:\! A_1} {M} {\ul{C}[\inl {A_1 + A_2} x_1/y]} \quad \cj {\Gamma, x_2 \!:\! A_2} {N} {\ul{C}[\inr {A_1 + A_2} x_2/y]}
\end{array}
}
\]
\item elimination of propositional equality 
\[
\mkrule
{\cj \Gamma {\pathind A {x_1.\, x_2.\, x_3.\, \ul{C}} {y.\, M} {V_1} {V_2} {V_p}} {\ul{C}[V_1/x_1][V_2/x_2][V_p/x_3]}}
{
\begin{array}{c}
\lj \Gamma A \quad \lj {\Gamma, x_1 \!:\! A, x_2 \!:\! A, x_3 \!:\! x_1 =_A x_2} {\ul{C}} \quad \vj \Gamma {V_1} A \quad \vj \Gamma {V_2} A 
\\[-1mm]
\vj \Gamma {V_p} {V_1 =_A V_2} \quad \cj {\Gamma, y \!:\! A} {M} {\ul{C}[y/x_1][y/x_2][\refl A y/x_3]}
\end{array}
}
\]
\end{itemize}
\end{proposition}

\begin{proof}
We prove this proposition by constructing the corresponding typing derivations. 

For example, the typing derivation for primitive recursion is given by
\[
\mkrule
{\cj \Gamma {\natrec {x.\,\ul{C}} {M_z} {y_1.\, y_2.\, M_s} {V}} {\ul{C}[V/x]}}
{
\mkrule
{\cj \Gamma {\force {\ul{C}} (\natrec {x.\,U\ul{C}} {\thunk M_z} {y_1.\, y_2.\, \thunk M_s} {V})} {\ul{C}[V/x]}}
{
\mkrule
{\vj \Gamma {\natrec {x.\,U\ul{C}} {\thunk M_z} {y_1.\, y_2.\, \thunk M_s} {V}} {U\ul{C}[V/x]}}
{
\mkrule
{\lj {\Gamma, x \!:\! \Nat} {U\ul{C}}}
{\lj {\Gamma, x \!:\! \Nat} {\ul{C}}}
\quad
\vj \Gamma V \Nat
\quad
(1)
\quad
(2)
}
}
}
\]
where $(1)$ is given by
\[
\mkrule
{\vj \Gamma {\thunk M_z} {U\ul{C}[\zero/x]}}
{\cj \Gamma {M_z} {\ul{C}[\zero/x]}}
\]
and $(2)$ by
\[
\mkrule
{\vj {\Gamma, y_1 \!:\! \Nat, y_2 \!:\! U\ul{C}[y_1/x]} {\thunk M_z} {U\ul{C}[\suc y_1/x]}}
{\cj {\Gamma, y_1 \!:\! \Nat, y_2 \!:\! U\ul{C}[y_1/x]} {M_z} {\ul{C}[\suc y_1/x]}}
\]
Derivations of the other typing rules listed above are constructed similarly.
\end{proof}

We finish our discussion about these derived elimination forms by proving that they satisfy computation term variants of the $\beta$- and $\eta$-equations given in Definition~\ref{def:judgements}.

\begin{proposition}
\label{prop:derivablecomputationtermvariantsofeliminatorsequations}
The following $\beta$- and $\eta$-equations are derivable for the computation term variant of:
\begin{itemize}
\item primitive recursion
\[
\mkrule
{\ceq \Gamma {\natrec {x.\,\ul{C}} {M_z} {y_1.\,y_2.\,M_s} {\zero}} {M_z} {\ul{C}[\zero/x]}}
{
\begin{array}{c}
\lj {\Gamma, x \!:\! \Nat} {\ul{C}} 
\\[-1mm]
\cj \Gamma {M_z} {\ul{C}[\zero/x]} \quad \cj {\Gamma, y_1 \!:\! \Nat, y_2 \!:\! U\ul{C}[y_1/x]} {M_s} {\ul{C}[\suc y_1/x]}
\end{array}
}
\]
\[
\mkrule
{
\begin{array}{c@{~} c@{~} l}
\Gamma & \vdash & {\natrec {x.\,\ul{C}} {M_z} {y_1.\,y_2.\,M_s} {\suc V}} 
\\
& = & {M_s[V/y_1][\thunk (\natrec {x.\ul{C}} {M_z} {y_1.\,y_2.\,M_s} {V})/y_2]} : {\ul{C}[\suc V/x]}
\end{array}
}
{
\begin{array}{c}
\lj {\Gamma, x \!:\! \Nat} \ul{C} \quad \vj \Gamma V \Nat 
\\[-1mm]
\cj \Gamma {M_z} {\ul{C}[\zero/x]} \quad \cj {\Gamma, y_1 \!:\! \Nat, y_2 \!:\! U\ul{C}[y_1/x]} {M_s} {\ul{C}[\suc y_1/x]}
\end{array}
}
\]
\item pattern-matching
\[
\mkrule
{
\begin{array}{c@{~} c@{~} l}
\Gamma & \vdash & {\pmatch {\langle V_1 , V_2 \rangle} {(x_1 \!:\! A_1, x_2 \!:\! A_2)} {y.\,\ul{C}} M} 
\\
& = & {M[V_1/x_1][V_2/x_2]} : {\ul{C}[\langle V_1 , V_2 \rangle/y]}
\end{array}
}
{
\begin{array}{c}
\lj {\Gamma, y \!:\! \Sigma \, x_1 \!:\! A_1 .\, A_2} \ul{C}
\\[-1mm]
\vj \Gamma {V_1} {A_1} \quad \vj \Gamma {V_2} {A_2[V_1/x_1]}  \quad \cj {\Gamma, x_1 \!:\! A_1, x_2 \!:\! A_2} {M} {\ul{C}[\langle x_1 , x_2 \rangle /y]}
\end{array}
}
\]

\[
\mkrule
{\veq \Gamma {\pmatch V {(x_1 \!:\! A_1, x_2 \!:\! A_2)} {y_1.\,\ul{C}} {M[{\langle x_1 , x_2 \rangle}/y_2]}} {M[V/y_2]} {\ul{C}[V/y_1]}}
{
\begin{array}{c}
\lj \Gamma {A_1} \quad \lj {\Gamma, x_1 \!:\! A_1} {A_2} \quad \vj \Gamma V {\Sigma \, x_1 \!:\! A_1 .\, A_2}
\\[-1mm]
\lj {\Gamma, y_1 \!:\! \Sigma \, x_1 \!:\! A_1 .\, A_2} \ul{C} \quad \cj {\Gamma, y_2 \!:\! \Sigma \, x_1 \!:\! A_1 .\, A_2} {M} {\ul{C}[y_2/y_1]}
\end{array}
}
\]
\item empty case analysis
\[
\mkrule
{\ceq \Gamma {\absurd {x.\, \ul{C}} V} {M[V/x]} {\ul{C}[V/x]}}
{\lj {\Gamma, x \!:\! 0} \ul{C} \quad \vj \Gamma V 0 \quad \cj {\Gamma, x \!:\! 0} M \ul{C}}
\]
\item binary case analysis
\[
\mkrule
{
\begin{array}{c@{~} c@{~} l}
\Gamma & \vdash & {\case {(\inl {A_1 + A_2} V)} {y.\, \ul{C}} {\inl {\!} {\!\!(x_1 \!:\! A_1)} \mapsto M} {\inr {\!} {\!\!(x_2 \!:\! A_2)} \mapsto N}} 
\\
& = & {M[V/x_1]} : {\ul{C}[\inl {A_1 + A_2} V/y]}
\end{array}
}
{
\begin{array}{c}
\lj {\Gamma, y \!:\! A_1 + A_2} \ul{C} \quad \vj \Gamma V {A_1} 
\\[-1mm]
\vj {\Gamma, x_1 \!:\! A_1} {M} {\ul{C}[\inl {A_1 + A_2} x_1/y]} \quad \cj {\Gamma, x_2 \!:\! A_2} {N} {\ul{C}[\inr {A_1 + A_2} x_2/y]}
\end{array}
}
\]

\[
\mkrule
{
\begin{array}{c@{~} c@{~} l}
\Gamma & \vdash & {\case {(\inr {A_1 + A_2} V)} {y.\, \ul{C}} {\inl {\!} {\!\!(x_1 \!:\! A_1)} \mapsto M} {\inr {\!} {\!\!(x_2 \!:\! A_2)} \mapsto N}} 
\\
& = & {N[V/x_2]} : {\ul{C}[\inr {A_1 + A_2} V/y]}
\end{array}
}
{
\begin{array}{c}
\lj {\Gamma, y \!:\! A_1 + A_2} \ul{C} \quad \vj \Gamma V {A_2} 
\\[-1mm]
\vj {\Gamma, x_1 \!:\! A_1} {M} {\ul{C}[\inl {A_1 + A_2} x_1/y]} \quad \cj {\Gamma, x_2 \!:\! A_2} {N} {\ul{C}[\inr {A_1 + A_2} x_2/y]}
\end{array}
}
\]

\[
\mkrule
{
\begin{array}{c@{~} c@{~} l}
\Gamma & \vdash & {\mathtt{case~} V \mathtt{~of}_{y_1.\,\ul{C}} \mathtt{~} ({\inl {\!} {\!\!(x_1 \!:\! A_1)} \mapsto M[\inl {A_1 + A_2} x_1/y_2]} , }
\\[-1mm]
&& \hspace{2.7cm} {\inr {\!} {\!\!(x_2 \!:\! A_2)} \mapsto M[\inr {A_1 + A_2} x_2/y_2]}) 
\\
& = & {M[V/y_2]} : {\ul{C}[V/y_1]}
\end{array}
}
{
\begin{array}{c}
\lj {\Gamma, y_1 \!:\! A_1 + A_2} \ul{C}
\quad
\vj \Gamma V {A_1 + A_2} \quad \cj {\Gamma, y_2 \!:\! A_1 + A_2} {M} {\ul{C}[y_2/y_1]}
\end{array} 
}
\]
\item elimination of propositional equality
\[
\mkrule
{
\begin{array}{c@{~} c@{~} l}
\Gamma & \vdash & {\pathind A {x_1.\, x_2.\, x_3.\, \ul{C}} {y.\, M} {V} {V} {\refl A V}} 
\\
& = & {M[V/y]} : {\ul{C}[V/x_1][V/x_2][\refl A V/x_3]}
\end{array}
}
{
\begin{array}{c}
\lj \Gamma A \quad \lj {\Gamma, x_1 \!:\! A, x_2 \!:\! A, x_3 \!:\! x_1 =_A x_2} {\ul{C}}
\\[-1mm]
\vj \Gamma {V} A \quad \cj {\Gamma, y \!:\! A} {M} {\ul{C}[y/x_1][y/x_2][\refl A y/x_3]}
\end{array}
}
\]
\end{itemize}
\end{proposition}

\begin{proof}
We prove this proposition by using the corresponding $\beta$- and $\eta$-equations for the value term variants of these elimination forms, together with the equations for thunking and forcing. 

For example, the first $\beta$-equation for primitive recursion is proved as follows:
\begin{fleqn}[0.3cm]
\begin{align*}
\Gamma \,\vdash\,\, & \natrec {x.\,\ul{C}} {M_z} {y_1.\, y_2.\, M_s} {\zero}
\\
=\,\, & \force {\ul{C}[\zero/x]} \big(\natrec {x.\,U\ul{C}} {\thunk M_z} {y_1.\, y_2.\, \thunk M_s} {\zero}\big)
\\
=\,\, & \force {\ul{C}[\zero/x]} (\thunk M_z)
\\
=\,\, & M_z \,:\, \ul{C}[\zero/x]
\end{align*}
\end{fleqn}
and the second $\beta$-equation as follows:
\begin{fleqn}[0.1cm]
\begin{align*}
\Gamma \,\vdash\,\, & \natrec {x.\,\ul{C}} {M_z} {y_1.\, y_2.\, M_s} {\suc V}
\\
=\,\, & \force {\ul{C}[\suc V/x]} \big(\natrec {x.\,U\ul{C}} {\thunk M_z} {y_1.\, y_2.\, \thunk M_s} {\suc V}\big)
\\
=\,\, & \force {\ul{C}[\suc V/x]} \big((\thunk M_s)[V/y_1]
\\[-2mm]
& \hspace{5.08cm} [\natrec {x.\,U\ul{C}} {\thunk M_z} {y_1.\, y_2.\, \thunk M_s} {V}/y_2]\big)
\\
=\,\, & \force {\ul{C}[\suc V/x]} \big(\thunk (M_s[V/y_1]
\\[-2mm]
& \hspace{4.93cm} [\natrec {x.\,U\ul{C}} {\thunk M_z} {y_1.\, y_2.\, \thunk M_s} {V}/y_2])\big)
\\
=\,\, & M_s[V/y_1][\natrec {x.\,U\ul{C}} {\thunk M_z} {y_1.\, y_2.\, \thunk M_s} {V}/y_2]
\\
=\,\, & M_s[V/y_1]
\\[-2mm]
& \hspace{0.5cm} [\thunk \big(\force {\ul{C}[V/x]} \big(\natrec {x.\,U\ul{C}} {\thunk M_z} {y_1.\, y_2.\, \thunk M_s} {V}\big)\big)/y_2]
\\
=\,\, & M_s[V/y_1][\thunk (\natrec {x.\,\ul{C}} {M_z} {y_1.\, y_2.\, M_s} {V})/y_2] \,:\, \ul{C}[\suc V/x]
\end{align*}
\end{fleqn}

\vspace{0.3cm}

As another example, the $\eta$-equation for pattern-matching is proved as follows:
\begin{fleqn}[0.3cm]
\begin{align*}
\Gamma \,\vdash\,\, & \pmatch V {(x_1 \!:\! A_1, x_2 \!:\! A_2)} {y.\,\ul{C}} {M[{\langle x_1 , x_2 \rangle}/y]}
\\
=\,\, & \force {\ul{C}[V/x]} \big(\pmatch V {(x_1 \!:\! A_1, x_2 \!:\! A_2)} {y.\, \ul{C}} {(\thunk M[{\langle x_1 , x_2 \rangle}/y])}\big)
\\
=\,\, & \force {\ul{C}[V/x]} \big(\pmatch V {(x_1 \!:\! A_1, x_2 \!:\! A_2)} {y.\, \ul{C}} {(\thunk M)[{\langle x_1 , x_2 \rangle}/y]}\big)
\\
=\,\, & \force {\ul{C}[V/x]} ((\thunk M)[V/y])
\\
=\,\, & \force {\ul{C}[V/x]} (\thunk M[V/y])
\\
=\,\, & M[V/y] \,:\, \ul{C}[V/y]
\end{align*}
\end{fleqn}
The other $\beta$- and $\eta$-equations listed above are proved similarly.
\end{proof}

Next, we show how to derive the elimination forms for homomorphism terms.

\pagebreak

\begin{definition} The \emph{homomorphism term variant of}
\begin{itemize}
\item \emph{primitive recursion} is defined as
\[
\begin{array}{c}
\hspace{-8.5cm} \natrec {\ul{C}} {K_z} {y .\, K_s} {V} \defeq 
\\[1mm]
\hspace{3cm} \big(\natrec {x_1.\,\ul{C} \,\multimap\, \ul{C}} {\lambda\, z \!:\! \ul{C} .\, K_z} {y.\, x_2.\, \lambda\, z \!:\! \ul{C} .\, K_s[x_2\, z/z]} {V}\big)\,\, z
\end{array}
\]
given that $FCV(K_z) = FCV(K_s) = z$, and where $x_1$ and $x_2$ are chosen fresh;
\item \emph{pattern-matching} is defined as
\[
\begin{array}{c}
\hspace{-6.5cm} \pmatch V {(y_1 \!:\! A, y_2 \!:\! B)} {y_3.\, \ul{C} , y_4 .\, \ul{D}} K \defeq 
\\[1mm]
\hspace{1.4cm} \big(\pmatch V {(y_1 \!:\! A, y_2 \!:\! B)} {x.\, \ul{C}[x/y_3] \,\multimap\, \ul{D}[x/y_4]} {(\lambda\, z \!:\! \ul{C}[\langle y_1 , y_2 \rangle/y_3] .\, K)\big) \,\, z}
\end{array}
\]
given that $FCV(K) = z$, and where $x$ is chosen fresh;
\item \emph{empty case analysis} is defined as
\[
\absurd {y_1.\,\ul{C},y_2.\,\ul{D}} {V} \defeq \big(\absurd {x.\,\ul{C}[x/y_1] \,\multimap\, \ul{D}[x/y_2]} V\big)\,\, z
\]
where $x$ is chosen fresh;
\item \emph{binary case analysis} is defined as
\[
\begin{array}{c}
\hspace{-3.2cm} \case {V} {y_1.\, \ul{C},y_2.\, \ul{D}} {\inl {\!} {\!\!(y_3 \!:\! A)} \mapsto K} {\inr {\!} {\!\!(y_4 \!:\! B)} \mapsto L} \defeq
\\[1mm]
\hspace{-0.09cm} \big(\case {V} {x .\, \ul{C}[x/y_1] \,\multimap\, \ul{D}[x/y_2]} {\inl {\!} {\!\!(y_3 \!:\! A)} \mapsto \lambda\, z \!:\! \ul{C}[\inl {A + B} y_3/y_1] .\, K} {
\\[-1mm]
\hspace{5.25cm} \inr {\!} {\!\!(y_4 \!:\! B)} \mapsto \lambda\, z \!:\! \ul{C}[\inr {A + B} y_4/y_1] .\, L} \big) \,\, z
\end{array}
\]
given that $FCV(K) = FCV(L) = z$, and where $x$ is chosen fresh; and 
\item \emph{elimination of propositional equality} is defined as
\[
\begin{array}{c}
\hspace{-3.6cm} \pathind A {y_1.\, y_2.\, y_3.\, \ul{C} , y_4.\, y_5.\, y_6.\, \ul{D}} {y_7.\, K} {V_1} {V_2} {V_p} \defeq 
\\[1mm]
\hspace{1cm} \big(\pathind A {x_1.\, x_2.\, x_3.\, \ul{C}[x_1/y_1][x_2/y_2][x_3/y_3] \multimap \ul{D}[x_1/y_4][x_2/y_5][x_3/y_6]} {
\\[-1mm]
\hspace{2.2cm} y_7 .\, \lambda\, z \!:\! \ul{C}[y_7/y_1][y_7/y_2][\refl A y_7/y_3] .\, K} {V_1} {V_2} {V_p} \big)\,\, z_2
\end{array}
\]
given that $FCV(K) = z$, and where $x_1$, $x_2$, and $x_3$ are chosen fresh.
\end{itemize}
\end{definition}

\pagebreak

\begin{proposition}
The following typing rules are derivable for the homomorphism term variant of:
\begin{itemize}
\item primitive recursion
\[
\mkrule
{\hj \Gamma {z \!:\! \ul{C}} {\natrec {\ul{C}} {K_z} {y .\, K_s} {V}} {\ul{C}}}
{
\lj {\Gamma} {\ul{C}} \quad \vj \Gamma V \Nat 
\quad
\hj \Gamma {z \!:\! \ul{C}} {K_z} {\ul{C}} \quad \hj {\Gamma, y \!:\! \Nat} {z \!:\! \ul{C}} {K_s} {\ul{C}}
}
\]
\item pattern-matching
\[
\mkrule
{\hj \Gamma {z \!:\! \ul{C}[V/y_3]} {\pmatch V {(y_1 \!:\! A, y_2 \!:\! B)} {y_3.\, \ul{C} , y_4 .\, \ul{D}} K} {\ul{D}[V/y_4]}}
{
\begin{array}{c}
\lj {\Gamma, y_3 \!:\! \Sigma \, y_1 \!:\! A .\, B} {\ul{C}} \quad \lj {\Gamma, y_4 \!:\! \Sigma \, y_1 \!:\! A .\, B} {\ul{D}}
\\[-1mm]
\vj \Gamma V {\Sigma \, y_1 \!:\! A .\, B} \quad \hj {\Gamma, y_1 \!:\! A, y_2 \!:\! B} {z \!:\! \ul{C}[\langle y_1 , y_2 \rangle /y_3]} {K} {\ul{D}[\langle y_1 , y_2 \rangle /y_4]}
\end{array}
}
\]
\item empty case analysis
\[
\mkrule
{\hj \Gamma {z \!:\! \ul{C}[V/y_1]} {\absurd {y_1.\,\ul{C},y_2.\,\ul{D}} V} {\ul{D}[V/y_2]}}
{
\vj {\Gamma} V 0 \quad \lj {\Gamma, y_1 \!:\! 0} \ul{C} \quad \lj {\Gamma, y_2 \!:\! 0} \ul{D}
}
\]
\item binary case analysis
\[
\mkrule
{
\begin{array}{c}
\hspace{-1.5cm}\Gamma \vertbar {z \!:\! \ul{C}[V/y_1]} \vdash \mathtt{case~} V \mathtt{~of}_{y_1.\, \ul{C},y_2.\, \ul{D}} \mathtt{~} (\inl {\!} {\!\!(y_3 \!:\! A)} \mapsto K , 
\\[-1mm]
\hspace{6.24cm} \inr {\!} {\!\!(y_4 \!:\! B)} \mapsto L) : \ul{D}[V/y_2]
\end{array}
}
{
\begin{array}{c}
\lj {\Gamma, y_1 \!:\! A + B} \ul{C} \quad \lj {\Gamma, y_2 \!:\! A + B} \ul{D}\quad \vj \Gamma V {A + B} 
\\[-1mm]
\hj {\Gamma, y_3 \!:\! A} {z \!:\! \ul{C}[\inl {A + B} y_3/y_1]} {K} {\ul{D}[\inl {A + B} y_3/y_2]} 
\\[-1mm]
\hj {\Gamma, y_4 \!:\! B} {z \!:\! \ul{C}[\inr {A + B} y_4/y_1]} {L} {\ul{D}[\inr {A + B} y_4/y_2]}
\end{array}
}
\]
\item elimination of propositional equality 
\[
\mkrule
{
\begin{array}{c}
\hspace{-8cm} \Gamma \vertbar {z \!:\! \ul{C}[V_1/y_1][V_2/y_2][V_p/y_3]} \vdash 
\\[-1mm]
\hspace{0.5cm} {\pathind A {y_1.\, y_2.\, y_3.\, \ul{C} , y_4.\, y_5.\, y_6.\, \ul{D}} {y_7.\, K} {V_1} {V_2} {V_p}} 
\\[-1mm]
\hspace{8.8cm} : {\ul{D}[V_1/y_4][V_2/y_5][V_p/y_6]}
\end{array}
}
{
\begin{array}{c}
\lj \Gamma A \quad \lj {\Gamma, y_1 \!:\! A, y_2 \!:\! A, y_3 \!:\! y_1 =_A y_2} {\ul{C}} \quad \lj {\Gamma, y_4 \!:\! A, y_5 \!:\! A, y_6 \!:\! y_4 =_A y_5} {\ul{D}} 
\\[-1mm]
\vj \Gamma {V_1} A \quad \vj \Gamma {V_2} A \quad \vj \Gamma {V_p} {V_1 =_A V_2}
\\[-1mm]
\hj {\Gamma, y_7 \!:\! A} {z \!:\! \ul{C}[y_7/y_1][y_7/y_2][\refl A y_7/y_3]} {K} {\ul{D}[y_7/y_4][y_7/y_5][\refl A y_7/y_6]}
\end{array}
}
\]
\end{itemize}
\end{proposition}

\begin{proof}
We prove this proposition by constructing the corresponding typing derivations. 

For example, the typing derivation for the homomorphism term variant of pattern-matching is constructed as follows:
\[
\mkrule
{\hj \Gamma {z \!:\! \ul{C}[V/y_3]} {\pmatch V {(y_1 \!:\! A, y_2 \!:\! B)} {y_3.\, \ul{C} , y_4 .\, \ul{D}} K} {\ul{D}[V/y_4]}}
{
\mkrule
{
\begin{array}{c}
\hspace{-11.5cm} \Gamma \vertbar {z \!:\! \ul{C}[V/y_3]} \vdash 
\\[-1mm]
\hspace{0.5cm} {\big(\pmatch V {(y_1 \!:\! A, y_2 \!:\! B)} {x.\, \ul{C}[x/y_3] \,\multimap\, \ul{D}[x/y_4]} {(\lambda\, z \!:\! \ul{C}[\langle y_1 , y_2 \rangle/y_3] .\, K)}\big) \,\, z} : {\ul{D}[V/y_4]}
\end{array}
}
{
(1)
\qquad\qquad\qquad
\mkrule
{\hj \Gamma {z \!:\! \ul{C}[V/y_3]} {z} {\ul{C}[V/y_3]}}
{
\mkrule
{\lj \Gamma {\ul{C}[V/y_3]}}
{\vj \Gamma V {\Sigma\, y_1 \!:\! A .\, B} \quad \lj {\Gamma, y_3 \!:\! \Sigma\, y_1 \!:\! A .\, B} {\ul{C}}}
}
}
}
\]
where $(1)$ is given by
\[
\mkrule
{
\begin{array}{c}
\hspace{-2cm} \Gamma \vdash {\pmatch V {(y_1 \!:\! A, y_2 \!:\! B)} {x.\, \ul{C}[x/y_3] \,\multimap\, \ul{D}[x/y_4]} {(\lambda\, z \!:\! \ul{C}[\langle y_1 , y_2 \rangle/y_3] .\, K)}} 
\\[-1mm]
\hspace{9.7cm} : {\ul{C}[V/y_3] \,\multimap\, \ul{D}[V/y_4]}
\end{array}
}
{
\vj \Gamma V {\Sigma \, y_1 \!:\! A .\, B} \quad (2) \quad (3)
}
\]
and $(2)$ by
\[
\mkrule
{\lj {\Gamma, x \!:\! \Sigma\, y_1 \!:\! A .\, B} {\ul{C}[x/y_3] \,\multimap\, \ul{D}[x/y_4]}}
{
\mkrule
{\lj {\Gamma, x \!:\! \Sigma\, y_1 \!:\! A .\, B} {\ul{C}[x/y_3]}}
{
\mkrule
{\lj {\Gamma, x \!:\! \Sigma\, y_1 \!:\! A .\, B, y_3 \!:\! \Sigma\, y_1 \!:\! A .\, B} {\ul{C}} \quad (4)}
{\lj {\Gamma, y_3 \!:\! \Sigma\, y_1 \!:\! A .\, B} {\ul{C}} \quad 
\mkrule
{x \not\in V\!ars(\Gamma) \cup \{y_3\}}
{x \text{~is chosen fresh}}}
}
\quad
\mkrule
{\lj {\Gamma, x \!:\! \Sigma\, y_1 \!:\! A .\, B} {\ul{D}[x/y_4]}}
{
\mkrule
{\lj {\Gamma, x \!:\! \Sigma\, y_1 \!:\! A .\, B, y_4 \!:\! \Sigma\, y_1 \!:\! A .\, B} {\ul{D}} \quad (4)}
{\lj {\Gamma, y_4 \!:\! \Sigma\, y_1 \!:\! A .\, B} {\ul{D}} \quad 
\mkrule
{x \not\in V\!ars(\Gamma) \cup \{y_4\}}
{x \text{~is chosen fresh}}
}
}
}
\]
and $(3)$ by
\[
\mkrule
{\vj {\Gamma, y_1 \!:\! A , y_2 \!:\! B} {\lambda\, z \!:\! \ul{C}[\langle y_1 , y_2 \rangle/y_3] .\, K} {\ul{C}[\langle y_1 , y_2 \rangle/y_3] \multimap \ul{D}[\langle y_1 , y_2 \rangle/y_4]}}
{\hj {\Gamma, y_1 \!:\! A, y_2 \!:\! B} {z \!:\! \ul{C}[\langle y_1 , y_2 \rangle /y_3]} {K} {\ul{D}[\langle y_1 , y_2 \rangle /y_4]}}
\]
and $(4)$ by
\[
\mkrule
{\vj {\Gamma, x \!:\! \Sigma\, y_1 \!:\! A .\, B} x {\Sigma\, y_1 \!:\! A .\, B}}
{
\mkrule
{\vdash {\Gamma, x \!:\! \Sigma\, y_1 \!:\! A .\, B}}
{
\hspace{0.45cm}
\vdash \Gamma
\quad
\lj \Gamma {\Sigma\, y_1 \!:\! A .\, B
\quad
\mkrule
{x \not\in V\!ars(\Gamma)}}
{x \text{~is chosen fresh}}
}
}
\]
Derivations of the other typing rules listed above are constructed similarly.
\end{proof}

\begin{proposition}
The following $\beta$- and $\eta$-equations are derivable for the homomorphism term variant of:
\begin{itemize}
\item primitive recursion
\[
\mkrule
{\heq \Gamma {z \!:\! \ul{C}} {\natrec {\ul{C}} {K_z} {y .\, K_s} {\zero}} {K_z} {\ul{C}}}
{
\lj {\Gamma} {\ul{C}} \quad \hj \Gamma {z \!:\! \ul{C}} {K_z} {\ul{C}} \quad \hj {\Gamma, y \!:\! \Nat} {z \!:\! \ul{C}} {K_s} {\ul{C}}
}
\]
\[
\mkrule
{
\begin{array}{c}
\Gamma \vertbar {z \!:\! \ul{C}} \vdash {\natrec {\ul{C}} {K_z} {y .\, K_s} {\suc V}} = 
{K_s[V/y][\natrec {\ul{C}} {K_z} {y .\, K_s} {V}/z]} : {\ul{C}}
\end{array}
}
{
\lj {\Gamma} {\ul{C}} \quad \vj \Gamma V \Nat
\quad
\hj \Gamma {z \!:\! \ul{C}} {K_z} {\ul{C}} \quad \hj {\Gamma, y \!:\! \Nat} {z \!:\! \ul{C}} {K_s} {\ul{C}}
}
\]
\item pattern-matching
\[
\mkrule
{
\begin{array}{c@{~} c@{~} l}
\Gamma \vertbar {z \!:\! \ul{C}[\langle V_1 , V_2 \rangle/y_3]} & \vdash & {\pmatch {\langle V_1 , V_2 \rangle} {(y_1 \!:\! A, y_2 \!:\! B)} {y_3.\, \ul{C} , y_4 .\, \ul{D}} K} 
\\[-1mm]
& = & K[V_1/y_1][V_2/y_2] : {\ul{D}[\langle V_1 , V_2 \rangle/y_4]}
\end{array}
}
{
\begin{array}{c}
\lj {\Gamma, y_3 \!:\! \Sigma \, y_1 \!:\! A .\, B} {\ul{C}} \quad \lj {\Gamma, y_4 \!:\! \Sigma \, y_1 \!:\! A .\, B} {\ul{D}} \quad \vj \Gamma {V_1} {A}
\\[-1mm]
\vj \Gamma {V_2} {B[V_1/y_1]} \quad \hj {\Gamma, y_1 \!:\! A, y_2 \!:\! B} {z \!:\! \ul{C}[\langle y_1 , y_2 \rangle /y_3]} {K} {\ul{D}[\langle y_1 , y_2 \rangle /y_4]}
\end{array}
}
\]

\[
\mkrule
{
\begin{array}{c@{~} c@{~} l}
\Gamma \vertbar {z \!:\! \ul{C}[V/y_3]} & \vdash & {\pmatch V {(y_1 \!:\! A, y_2 \!:\! B)} {y_3.\, \ul{C} , y_4 .\, \ul{D}} K[\pair {y_1} {y_2}/y_5]} 
\\[-1mm]
& = & K[V/y_5] : {\ul{D}[V/y_4]}
\end{array}
}
{
\begin{array}{c}
\lj {\Gamma, y_3 \!:\! \Sigma \, y_1 \!:\! A .\, B} {\ul{C}} \quad \lj {\Gamma, y_4 \!:\! \Sigma \, y_1 \!:\! A .\, B} {\ul{D}} 
\\[-1mm]
\vj \Gamma V {\Sigma \, y_1 \!:\! A .\, B} \quad \hj {\Gamma, y_5 \!:\! \Sigma\, y_1 \!:\! A .\, B} {z \!:\! \ul{C}[y_5 /y_3]} {K} {\ul{D}[y_5 /y_4]}
\end{array}
}
\]
\item empty case analysis
\[
\mkrule
{\heq \Gamma {z \!:\! \ul{C}[V/y_1]} {\absurd {y_1.\,\ul{C},y_2.\,\ul{D}} V} {K[V/y_3]} {\ul{D}[V/y_2]}}
{
\lj {\Gamma, y_1 \!:\! 0} \ul{C} \quad \lj {\Gamma, y_2 \!:\! 0} \ul{D}
\quad
\vj {\Gamma} V 0 \quad \hj {\Gamma, y_3 \!:\! 0} {z \!:\! \ul{C}[y_3/y_1]} K {\ul{D}[y_3/y_2]}
}
\]
\item binary case analysis
\[
\mkrule
{
\begin{array}{c@{~} c@{~} l}
\Gamma \vertbar {z \!:\! \ul{C}[\inl {A + B} V/y_1]} & \vdash & \case {(\inl {A + B} V)} {y_1.\, \ul{C},y_2.\, \ul{D}} {\inl {\!} {\!\!(y_3 \!:\! A)} \mapsto  K} {\\[-1mm] && \hspace{5.15cm} \inr {\!} {\!\!(y_4 \!:\! B)} \mapsto L}
\\
& = & K[V/y_3] : {\ul{D}[\inl {A + B} V/y_2]}
\end{array}
}
{
\begin{array}{c}
\lj {\Gamma, y_1 \!:\! A + B} \ul{C} \quad \lj {\Gamma, y_2 \!:\! A + B} \ul{D} \quad \vj \Gamma V {A} 
\\[-1mm]
\hj {\Gamma, y_3 \!:\! A} {z \!:\! \ul{C}[\inl {A + B} y_3/y_1]} {K} {\ul{D}[\inl {A + B} y_3/y_2]} 
\\[-1mm]
\hj {\Gamma, y_4 \!:\! B} {z \!:\! \ul{C}[\inr {A + B} y_4/y_1]} {L} {\ul{D}[\inr {A + B} y_4/y_2]}
\end{array}
}
\]

\[
\mkrule
{
\begin{array}{c@{~} c@{~} l}
\Gamma \vertbar {z \!:\! \ul{C}[\inr {A + B} V/y_1]} & \vdash & \case {(\inr {A + B} V)} {y_1.\, \ul{C},y_2.\, \ul{D}} {\inl {\!} {\!\!(y_3 \!:\! A)} \mapsto K} {\\[-1mm] && \hspace{5.15cm} \inr {\!} {\!\!(y_4 \!:\! B)} \mapsto L} 
\\
& = & L[V/y_4] : {\ul{D}[\inr {A + B} V/y_2]}
\end{array}
}
{
\begin{array}{c}
\lj {\Gamma, y_1 \!:\! A + B} \ul{C} \quad \lj {\Gamma, y_2 \!:\! A + B} \ul{D} \quad \vj \Gamma V {B} 
\\[-1mm]
\hj {\Gamma, y_3 \!:\! A} {z \!:\! \ul{C}[\inl {A + B} y_3/y_1]} {K} {\ul{D}[\inl {A + B} y_3/y_2]} 
\\[-1mm]
\hj {\Gamma, y_4 \!:\! B} {z \!:\! \ul{C}[\inr {A + B} y_4/y_1]} {L} {\ul{D}[\inr {A + B} y_4/y_2]}
\end{array}
}
\]

\[
\mkrule
{
\begin{array}{c@{~} c@{~} l}
\Gamma \vertbar {z \!:\! \ul{C}[V/y_1]} & \vdash & \mathtt{case~} V \mathtt{~of}_{y_1.\, \ul{C},y_2.\, \ul{D}} \mathtt{~} (\inl {\!} {\!\!(y_3 \!:\! A)} \mapsto K[\inl {A + B} y_3/y_5],
\\[-1mm]
&& \hspace{3.425cm} \inr {\!} {\!\!(y_4 \!:\! B)} \mapsto K[\inr {A + B} y_4/y_5]) 
\\
& = & K[V/y_4] : {\ul{D}[V/y_2]}
\end{array}
}
{
\begin{array}{c}
\lj {\Gamma, y_1 \!:\! A + B} \ul{C} \quad \lj {\Gamma, y_2 \!:\! A + B} \ul{D} 
\\[-1mm]
\vj \Gamma V {A + B} \quad \hj {\Gamma, y_5 \!:\! A + B} {z \!:\! \ul{C}[y_5/y_1]} {K} {\ul{D}[y_5/y_2]} 
\end{array}
}
\]
\pagebreak
\item elimination of propositional equality
\[
\mkrule
{
\begin{array}{c}
\hspace{-6.5cm} \Gamma \vertbar {z \!:\! \ul{C}[V/y_1][V/y_2][\refl A V/y_3]} \vdash 
\\[-1mm]
\hspace{3cm} {\pathind A {y_1.\, y_2.\, y_3.\, \ul{C} , y_4.\, y_5.\, y_6.\, \ul{D}} {y_7.\, K} {V} {V} {\refl A V}} 
\\
\hspace{-0.3cm} = K[V/y_7] : {\ul{D}[V/y_4][V/y_5][\refl A V/y_6]}
\end{array}
}
{
\begin{array}{c}
\lj \Gamma A
\quad
\vj \Gamma {V} A
\\[-1mm]
\lj {\Gamma, y_1 \!:\! A, y_2 \!:\! A, y_3 \!:\! y_1 =_A y_2} {\ul{C}} 
\quad 
\lj {\Gamma, y_4 \!:\! A, y_5 \!:\! A, y_6 \!:\! y_4 =_A y_5} {\ul{D}} 
\\[-1mm]
\hj {\Gamma, y_7 \!:\! A} {z \!:\! \ul{C}[y_7/y_1][y_7/y_2][\refl A y_7/y_3]} {K} {\ul{D}[y_7/y_4][y_7/y_5][\refl A y_7/y_6]}
\end{array}
}
\]
\end{itemize}
\end{proposition}

\begin{proof}
We prove this proposition by using the corresponding $\beta$- and $\eta$-equations for the value term variants of these elimination rules, together with the $\beta$-equation for homomorphic lambda abstraction and homomorphic function application. 

For example, the first $\beta$-equation for primitive recursion is proved as follows:
\begin{fleqn}[0.3cm]
\begin{align*}
\Gamma \vertbar {z \!:\! \ul{C}} \,\vdash\,\, & \natrec {\ul{C}} {K_z} {y .\, K_s} {\zero}
\\
=\,\, & \big(\natrec {x_1.\,\ul{C} \,\multimap\, \ul{C}} {\lambda\, z \!:\! \ul{C} .\, K_z} {y.\, x_2.\, \lambda\, z \!:\! \ul{C} .\, K_s[x_2\, z/z]} {\zero}\big)\,\, z
\\
=\,\, & (\lambda\, z \!:\! \ul{C} .\, K_z)\,\, z
\\
=\,\, & K_z : \ul{C}
\end{align*}
\end{fleqn}
and the second $\beta$-equation as follows:
\begin{fleqn}[0.3cm]
\begin{align*}
\Gamma \vertbar {z \!:\! \ul{C}} \,\vdash\,\, & \natrec {\ul{C}} {K_z} {y .\, K_s} {\suc V , z_3}
\\
=\,\, & \big(\natrec {x_1.\,\ul{C} \,\multimap\, \ul{C}} {\lambda\, z \!:\! \ul{C} .\, K_z} {y.\, x_2.\, \lambda\, z \!:\! \ul{C} .\, K_s[x_2\, z/z]} {\suc V}\big)\,\, z
\\
=\,\, & \big((\lambda\, z \!:\! \ul{C} .\, K_s[x_2\, z/z])[V/y]
\\[-2mm]
& \hspace{0.3cm} [\natrec {x_1.\,\ul{C} \,\multimap\, \ul{C}} {\lambda\, z \!:\! \ul{C} .\, K_z} {y.\, x_2.\, \lambda\, z \!:\! \ul{C} .\, K_s[x_2\, z/z]} {V}/x_2]\big)\,\, z
\\
=\,\, & K_s[x_2\, z/z][V/y][\natrec {x_1.\,\ul{C} \,\multimap\, \ul{C}} {\lambda\, z \!:\! \ul{C} .\, K_z} {y.\, x_2.\, \lambda\, z \!:\! \ul{C} .\, K_s[x_2\, z/z]} {V}/x_2]
\\
=\,\, & K_s[V/y][x_2\, z/z][\natrec {x_1.\,\ul{C} \,\multimap\, \ul{C}} {\lambda\, z \!:\! \ul{C} .\, K_z} {y.\, x_2.\, \lambda\, z \!:\! \ul{C} .\, K_s[x_2\, z/z]} {V}/x_2]
\\
=\,\, & K_s[V/y][\natrec {\ul{C}} {K_z} {y .\, K_s} {V}/z] : \ul{C}
\end{align*}
\end{fleqn}

As another example, the $\eta$-equation for pattern-matching is proved as follows:
\begin{fleqn}[0.2cm]
\begin{align*}
\Gamma \vertbar {z \!:\! \ul{C}[V/y_3]} \,\vdash\,\, & {\pmatch V {(y_1 \!:\! A, y_2 \!:\! B)} {y_3.\, \ul{C} , y_4 .\, \ul{D}} (K[\pair {y_1} {y_2}/y_5])}
\\
=\,\, & \big(\pmatch V {(y_1 \!:\! A, y_2 \!:\! B)} {x.\, \ul{C}[x/y_3] \,\multimap\, \ul{D}[x/y_4]} {
\\ 
& \hspace{5cm} (\lambda\, z \!:\! \ul{C}[\langle y_1 , y_2 \rangle/y_3] .\, K[\pair {y_1} {y_2}/y_5])\big) \,\, z}
\\
=\,\, & (\lambda\, z \!:\! \ul{C}[V/y_3] .\, K[V/y_5])\,\, z
\\
=\,\, & K[V/y_5] : \ul{D}[V/y_4]
\end{align*}
\end{fleqn}
The other $\beta$- and $\eta$-equations listed above are proved similarly.
\end{proof}

\section{Derivable equations}
\label{sect:derivableequations}

We conclude this chapter by presenting some useful derivable equations.

We begin with equations that are familiar from other (monadic) effectful languages, such as Moggi's computational $\lambda$-calculus~\cite{Moggi:ComputationalLambdaCalculus}. Specifically, we show how to derive the right unit and associativity equations for for sequential composition. Note that the left unit equation is already included in the definition of eMLTT's equational theory.

\begin{proposition}
\label{prop:monadicequations}
The following right unit and associativity equations are derivable for sequential composition:
\[
\mkrule
{\ceq \Gamma {\doto {M} {x \!:\! A} {} {\return x}} {M} {FA}}
{\cj \Gamma M {FA} \quad x \not\in V\!ars(\Gamma)}
\]

\[
\mkrule
{\heq \Gamma {z \!:\! \ul{C}} {\doto {K} {x \!:\! A} {} {\return x}} {K} {FA}}
{\hj \Gamma {z \!:\! \ul{C}} K {FA} \quad x \not\in V\!ars(\Gamma)}
\]

\[
\mkrule
{\ceq \Gamma {\doto {M_1} {x \!:\! A} {} {(\doto {M_2} {y \!:\! B} {} {M_3})}} {\doto {(\doto {M_1} {x \!:\! A} {} {M_2})} {y \!:\! B} {} {M_3}} {\ul{C}}}
{\cj \Gamma {M_1} {FA} \quad \cj {\Gamma, x \!:\! A} {M_2} {FB} \quad \cj {\Gamma, y \!:\! B} {M_3} {\ul{C}} \quad \lj \Gamma {\ul{C}} \quad x \neq y}
\]

\[
\mkrule
{\heq \Gamma {z \!:\! \ul{C}} {\doto {K} {x \!:\! A} {} {(\doto {M} {y \!:\! B} {} {N})}} {\doto {(\doto {K} {x \!:\! A} {} {M})} {y \!:\! B} {} {N}} {\ul{D}}}
{\hj \Gamma {z \!:\! \ul{C}} {K} {FA} \quad \cj {\Gamma, x \!:\! A} {M} {FB} \quad \cj {\Gamma, y \!:\! B} {N} {\ul{D}} \quad \lj \Gamma {\ul{D}} \quad x \neq y}
\]
\end{proposition}

\begin{proof}
All four equations are proved using the $\beta$- and $\eta$-equations for sequential composition.
In particular, the two right unit equations are proved as follows:
\begin{fleqn}[0.3cm]
\begin{align*}
\Gamma \,\vdash\,\, & \doto {M} {x \!:\! A} {} {\return x}
&
\Gamma \vertbar {z \!:\! \ul{C}} \,\vdash\,\, & \doto {K} {x \!:\! A} {} {\return x}
\\
=\,\, & \doto {M} {x \!:\! A} {} {z[\return x]/z}
&
=\,\, & \doto {K} {x \!:\! A} {} {z[\return x]/z}
\\
=\,\, & z[M/z]
&
=\,\, & z[K/z]
\\
=\,\, & M : FA
&
=\,\, & K : FA
\end{align*}
\end{fleqn}
the associativity equation for computation terms is proved as follows:
\begin{fleqn}[0.3cm]
\begin{align*}
\Gamma \,\vdash\,\, & \doto {M_1} {x \!:\! A} {} {(\doto {M_2} {y \!:\! B} {} {M_3})}
\\
=\,\, & \doto {M_1} {x \!:\! A} {} {\big(\doto {(\doto {\return x} {x' \!:\! A} {} {M_2[x'/x]})} {y \!:\! B} {} {M_3}\big)}
\\
=\,\, & \doto {M_1} {x \!:\! A} {} {\big(\doto {(\doto {z} {x' \!:\! A} {} {M_2[x'/x]})} {y \!:\! B} {} {M_3}\big)[\return x/z]}
\\
=\,\, & \doto {(\doto {M_1} {x' \!:\! A} {} {M_2[x'/x]})} {y \!:\! B} {} {M_3}
\\
=\,\, & \doto {(\doto {M_1} {x \!:\! A} {} {M_2})} {y \!:\! B} {} {M_3} : \ul{C}
\end{align*}
\end{fleqn}
and the associativity equation for homomorphism terms is proved as follows:
\begin{fleqn}[0.3cm]
\begin{align*}
\Gamma \vertbar {z \!:\! \ul{C}} \,\vdash\,\, & \doto {K} {x \!:\! A} {} {(\doto {M} {y \!:\! B} {} {N})}
\\
=\,\, & \doto {K} {x \!:\! A} {} {\big(\doto {(\doto {\return x} {x' \!:\! A} {} {M[x'/x]})} {y \!:\! B} {} {N}\big)}
\\
=\,\, & \doto {K} {x \!:\! A} {} {\big(\doto {(\doto {z} {x' \!:\! A} {} {M[x'/x]})} {y \!:\! B} {} {N}\big)[\return x/z]}
\\
=\,\, & \doto {(\doto {K} {x' \!:\! A} {} {M[x'/x]})} {y \!:\! B} {} {N}
\\
=\,\, & \doto {(\doto {K} {x \!:\! A} {} {M})} {y \!:\! B} {} {N} : \ul{D}
\end{align*}
\end{fleqn}
\end{proof}

Next, we observe that we can also derive analogous right unit and associativity equations for computational pattern-matching.

\begin{proposition}
\label{prop:comppatternmatchingequations}
The following right unit and associativity equations are derivable for computational pattern-matching:
\[
\mkrule
{\ceq \Gamma {\doto {M} {(x \!:\! A , z \!:\! \ul{C})} {} {\pair x z}} {M} {\Sigma\, x \!:\! A .\, \ul{C}}}
{\cj \Gamma M {\Sigma\, x \!:\! A .\, \ul{C}}}
\vspace{0.25cm}
\]

\[
\mkrule
{\heq \Gamma {z_1 \!:\! \ul{C}} {\doto {K} {(x \!:\! A , z_2 \!:\! \ul{D})} {} {\pair x {z_2}}} {K} {\Sigma\, x \!:\! A .\, \ul{D}}}
{\hj \Gamma {z_1 \!:\! \ul{C}} K {\Sigma\, x \!:\! A .\, \ul{D}}}
\vspace{0.25cm}
\]

\[
\mkrule
{
\begin{array}{c@{~} c@{~} l}
\Gamma & \vdash & {\doto {M} {(x \!:\! A , z_1 \!:\! \ul{C}_1)} {} {(\doto {K} {(y \!:\! B , z_2 \!:\! \ul{C}_2)} {} {L})}} 
\\[-1mm]
& = & {\doto {(\doto {M} {(x \!:\! A , z_1 \!:\! \ul{C}_1)} {} {K})} {(y \!:\! B , z_2 \!:\! \ul{C}_2)} {} {L}} : {\ul{D}}
\end{array}
}
{
\cj \Gamma {M} {\Sigma\, x \!:\! A .\, \ul{C}_1} \quad
\hj {\Gamma, x \!:\! A} {z_1 \!:\! \ul{C}_1} {K} {\Sigma\, y \!:\! B .\, \ul{C}_2} 
\quad
\hj {\Gamma, y \!:\! B} {z_2 \!:\! \ul{C}_2} {L} {\ul{D}}
\quad
\lj \Gamma {\ul{D}}
}
\]

\[
\mkrule
{
\begin{array}{c@{~} c@{~} l}
\Gamma \vertbar {z_1 \!:\! \ul{C}} & \vdash & {\doto {K_1} {(x \!:\! A , z_2 \!:\! \ul{D}_1)} {} {(\doto {K_2} {(y \!:\! B , z_3 \!:\! \ul{D}_2)} {} {K_3})}} 
\\[-1mm]
& = & {\doto {(\doto {K_1} {(x \!:\! A , z_2 \!:\! \ul{D}_1)} {} {K_2})} {(y \!:\! B , z_3 \!:\! \ul{D}_2)} {} {K_3}} : {\ul{D}_3}
\end{array}
}
{
\begin{array}{c}
\hj \Gamma {z_1 \!:\! \ul{C}} {K_1} {\Sigma\, x \!:\! A .\, \ul{D}_1} \quad
\hj {\Gamma, x \!:\! A} {z_2 \!:\! \ul{D}_1} {K_2} {\Sigma\, y \!:\! B .\, \ul{D}_2} 
\\[-1mm]
\hj {\Gamma, y \!:\! B} {z_3 \!:\! \ul{D}_2} {K_3} {\ul{D}_3}
\quad
\lj \Gamma {\ul{D}_3}
\end{array}
}
\]
\end{proposition}

\begin{proof}
These four equations are proved similarly to the equations given in Proposition~\ref{prop:monadicequations}, using the $\beta$- and $\eta$-equations for computational pattern-matching.
\end{proof}

Next, we can show that sequential composition commutes with other computational term formers from the left.

\begin{proposition}
\label{prop:seqcompdistributivity}
Sequential composition commutes with computational pairing, \linebreak computational pattern-matching, lambda abstraction, computational function appli-\linebreak 

\pagebreak
\noindent
cation, and homomorphic function application from the left:
\[
\mkrule
{\ceq \Gamma {\doto {M} {x \!:\! A} {} {\pair V N}} {\pair V {\doto M {x \!:\! A} {} N}} {\Sigma \, y \!:\! B .\, \ul{C}}}
{\cj \Gamma M {FA} \quad \vj \Gamma V B \quad \cj {\Gamma, x \!:\! A} N {\ul{C}[V/y]} \quad \lj {\Gamma, y \!:\! B} \ul{C}}
\]

\vspace{0.1cm}

\[
\mkrule
{\heq \Gamma {z \!:\! \ul{C}} {\doto {K} {x \!:\! A} {} {\pair V M}} {\pair V {\doto K {x \!:\! A} {} M}} {\Sigma \, y \!:\! B .\, \ul{D}}}
{\hj \Gamma {z \!:\! \ul{C}} K {FA} \quad \vj \Gamma V B \quad \cj {\Gamma, x \!:\! A} M {\ul{D}[V/y]} \quad \lj {\Gamma, y \!:\! B} \ul{D}}
\]

\vspace{0.1cm}

\[
\mkrule
{
\begin{array}{c@{~} c@{~} l}
\Gamma & \vdash & {\doto M {x \!:\! A} {} {(\doto N {(y \!:\! B , z \!:\! \ul{C})} {} {K})}} 
\\[-1mm]
& = & {\doto {(\doto M {x \!:\! A} {} N)} {(y \!:\! B , z \!:\! \ul{C})} {} {K}} : {\ul{D}}
\end{array}
}
{\cj \Gamma M {FA} \quad \cj {\Gamma, x \!:\! A} N {\Sigma\, y \!:\! B .\, \ul{C}} \quad \lj {\Gamma, y \!:\! B} {\ul{C}} \quad \hj {\Gamma, y \!:\! B} {z \!:\! \ul{C}} K {\ul{D}} \quad \lj \Gamma \ul{D}}
\]

\vspace{0.1cm}

\[
\mkrule
{
\begin{array}{c@{~} c@{~} l}
\Gamma \vertbar {z_1 \!:\! \ul{C}} & \vdash & {\doto K {x \!:\! A} {} {(\doto M {(y \!:\! B , z_2 \!:\! \ul{D}_1)} {} {L})}} 
\\[-2mm]
& = & {\doto {(\doto K {x \!:\! A} {} M)} {(y \!:\! B , z_2 \!:\! \ul{D}_1)} {} {L}} : {\ul{D}_2}
\end{array}
}
{
\begin{array}{c}
\hj \Gamma {z_1 \!:\! \ul{C}} K {FA} \quad \cj {\Gamma, x \!:\! A} M {\Sigma\, y \!:\! B .\, \ul{D}_1}
\\[-1mm]
\lj {\Gamma, y \!:\! B} {\ul{D}_1} \quad \hj {\Gamma, y \!:\! B} {z_2 \!:\! \ul{C}} L {\ul{D}_2} \quad \lj \Gamma \ul{D}_2
\end{array}
}
\]

\vspace{0.1cm}

\[
\mkrule
{\ceq \Gamma {\doto M {x \!:\! A} {} {(\lambda\, y \!:\! B .\, N)}} {\lambda \, y \!:\! B .\, (\doto M {x \!:\! A} {} N)} {\Pi\, y \!:\! B .\, \ul{C}}}
{\cj \Gamma M {FA} \quad \cj {\Gamma, x \!:\! A, y \!:\! B} N {\ul{C}} \quad \lj \Gamma B \quad \lj {\Gamma, y \!:\! B} \ul{C}}
\]

\vspace{0.1cm}

\[
\mkrule
{\heq \Gamma {z \!:\! \ul{C}} {\doto K {x \!:\! A} {} {(\lambda\, y \!:\! B .\, M)}} {\lambda \, y \!:\! B .\, (\doto K {x \!:\! A} {} M)} {\Pi\, y \!:\! B .\, \ul{D}}}
{\hj \Gamma {z \!:\! \ul{C}} K {FA} \quad \cj {\Gamma, x \!:\! A, y \!:\! B} M {\ul{D}} \quad \lj \Gamma B \quad \lj {\Gamma, y \!:\! B} \ul{D}}
\]

\vspace{0.1cm}

\[
\mkrule
{\ceq \Gamma {\doto M {x \!:\! A} {} {N\, V}} {(\doto M {x \!:\! A} {} N)\, V} {\ul{C}[V/x]}}
{\cj \Gamma M {FA} \quad \cj {\Gamma, x \!:\! A} N {\Pi\, y \!:\! B .\, \ul{C}} \quad \vj \Gamma V B \quad \lj {\Gamma, y \!:\! B} {\ul{C}}}
\]

\vspace{0.1cm}

\[
\mkrule
{\heq \Gamma {z \!:\! \ul{C}} {\doto K {x \!:\! A} {} {M\, V}} {(\doto K {x \!:\! A} {} M)\, V} {\ul{D}[V/x]}}
{\hj \Gamma {z \!:\! \ul{C}} K {FA} \quad \cj {\Gamma, x \!:\! A} M {\Pi\, y \!:\! B .\, \ul{D}} \quad \vj \Gamma V B \quad \lj {\Gamma, y \!:\! B} {\ul{D}}}
\]

\vspace{0.1cm}

\[
\mkrule
{\ceq \Gamma {\doto M {x \!:\! A} {} {V\, N}} {V\, (\doto M {x \!:\! A} {} {N})} {\ul{D}}}
{\cj \Gamma M {FA} \quad \vj \Gamma {V} {\ul{C} \multimap \ul{D}} \quad \cj {\Gamma, x \!:\! A} N {\ul{C}}}
\]

\vspace{0.1cm}

\[
\mkrule
{\heq \Gamma {z \!:\! \ul{C}} {\doto K {x \!:\! A} {} {V\, M}} {V\, (\doto K {x \!:\! A} {} {M})} {\ul{D}_2}}
{\hj \Gamma {z \!:\! \ul{C}} K {FA} \quad \vj \Gamma {V} {\ul{D}_1 \multimap \ul{D}_2} \quad \cj {\Gamma, x \!:\! A} M {\ul{D}_1}}
\]
\end{proposition}

\begin{proof}
All these equations are proved using the $\beta$- and $\eta$-equations for sequential composition, following a similar pattern to the proofs of Propositions~\ref{prop:monadicequations} and~\ref{prop:comppatternmatchingequations}. 

\pagebreak

For example, commutativity with homomorphic function application is proved as follows:
\begin{fleqn}[0.3cm]
\begin{align*}
\Gamma \vertbar {z \!:\! \ul{C}} \,\vdash\,\, & \doto K {x \!:\! A} {} {V\, M}
\\
=\,\, & \doto K {x \!:\! A} {} {V\, (\doto {\return x} {y \!:\! A} {} {M[y/x]})}
\\
=\,\, & \doto K {x \!:\! A} {} {V\, (\doto {z} {y \!:\! A} {} {M[y/x]})[\return x/z]}
\\
=\,\, & \doto K {x \!:\! A} {} {\big(V\, (\doto {z} {y \!:\! A} {} {M[y/x]})\big)[\return x/z]}
\\
=\,\, & V\, (\doto K {y \!:\! A} {} {M[y/x]})
\\
=\,\, & V\, (\doto K {x \!:\! A} {} {M}) : \ul{D}_2
\end{align*}
\end{fleqn}
\end{proof}

Analogously to sequential composition, computational pattern-matching also commutes with other computational term-formers from the left, as shown next.

\begin{proposition}
\label{prop:comppatternmatchingdistributivity}
Computational pattern-matching commutes with sequential composition, computational pairing, computational lambda abstraction, computational function application, and homomorphic function application from the left:
\vspace{0.2cm}
\[
\mkrule
{\ceq \Gamma {\doto M {(x \!:\! A , z \!:\! \ul{C})} {} {(\doto K {y \!:\! B} {} {N})}} {\doto {(\doto M {(x \!:\! A , z \!:\! \ul{C})} {} {K})} {y \!:\! B} {} {N}} {\ul{D}}}
{\cj \Gamma M {\Sigma\, x \!:\! A .\, \ul{C}} \quad \hj {\Gamma, x \!:\! A} {z \!:\! \ul{C}} {K} {FB} \quad \cj {\Gamma, y \!:\! B} N {\ul{D}} \quad \lj \Gamma \ul{D}}
\]

\vspace{0.1cm}

\[
\mkrule
{
\begin{array}{c@{~} c@{~} l}
\Gamma \vertbar {z_1 \!:\! \ul{C}} & \vdash & {\doto K {(x \!:\! A , z_2 \!:\! \ul{D}_1)} {} {(\doto L {y \!:\! B} {} {M})}} 
\\[-1mm]
& = & {\doto {(\doto K {(x \!:\! A , z_2 \!:\! \ul{D}_1)} {} {L})} {y \!:\! B} {} {M}} : {\ul{D}_2}
\end{array}
}
{\hj \Gamma {z_1 \!:\! \ul{C}} K {\Sigma\, x \!:\! A .\, \ul{D}_1} \quad \hj {\Gamma, x \!:\! A} {z_2 \!:\! \ul{D}_1} {L} {FB} \quad \cj {\Gamma, y \!:\! B} M {\ul{D}_2} \quad \lj \Gamma \ul{D}_2} 
\]

\vspace{0.1cm}

\[
\mkrule
{\ceq \Gamma {\doto M {(x \!:\! A , z \!:\! \ul{C})} {} {\pair V K}} {\pair V {\doto M {(x \!:\! A , z \!:\! \ul{C})} {} {K}}} {\Sigma\, y \!:\! B .\, \ul{D}}}
{\cj \Gamma M {\Sigma\, x \!:\! A .\, \ul{C}} \quad \vj \Gamma V B \quad \hj {\Gamma, x \!:\! A} {z \!:\! \ul{C}} {K} {\ul{D}[V/y]} \quad \lj {\Gamma, y \!:\! B} \ul{D}}
\]

\vspace{0.1cm}

\[
\mkrule
{\heq \Gamma {z_1 \!:\! \ul{C}} {\doto K {(x \!:\! A , z_2 \!:\! \ul{D}_1)} {} {\pair V L}} {\pair V {\doto K {(x \!:\! A , z_2 \!:\! \ul{D}_1)} {} {L}}} {\Sigma\, y \!:\! B .\, \ul{D}_2}}
{\hj \Gamma {z_1 \!:\! \ul{C}} K {\Sigma\, x \!:\! A .\, \ul{D}_1} \quad \vj \Gamma V B \quad \hj {\Gamma, x \!:\! A} {z \!:\! \ul{D}_1} {L} {\ul{D}_2[V/y]} \quad \lj {\Gamma, y \!:\! B} \ul{D}_2}
\]

\vspace{0.1cm}

\[
\mkrule
{\ceq \Gamma {\doto M {(x \!:\! A , z \!:\! \ul{C})} {} {(\lambda\, y \!:\! B .\, K)}} {\lambda\, y \!:\! B .\, (\doto M {(x \!:\! A , z \!:\! \ul{C})} {} {K})} {\Pi\, y \!:\! B .\, \ul{D}}}
{\cj \Gamma M {\Sigma\, x \!:\! A .\, \ul{C}} \quad \hj {\Gamma, x \!:\! A, y \!:\! B} {z \!:\! \ul{C}} {K} {\ul{D}} \quad \lj {\Gamma, y \!:\! B} \ul{D}}
\]

\vspace{0.1cm}

\[
\mkrule
{
\begin{array}{c@{~} c@{~} l}
\Gamma \vertbar {z_1 \!:\! \ul{C}} & \vdash & {\doto K {(x \!:\! A , z_2 \!:\! \ul{D}_1)} {} {(\lambda\, y \!:\! B .\, L)}} 
\\[-1mm]
& = & {\lambda\, y \!:\! B .\, (\doto K {(x \!:\! A , z_2 \!:\! \ul{D}_1)} {} {L})} : {\Pi\, y \!:\! B .\, \ul{D}_2}
\end{array}
}
{\hj \Gamma {z_1 \!:\! \ul{C}} K {\Sigma\, x \!:\! A .\, \ul{D}_1} \quad \hj {\Gamma, x \!:\! A, y \!:\! B} {z_2 \!:\! \ul{D}_1} {L} {\ul{D}_2} \quad \lj {\Gamma, y \!:\! B} \ul{D}_2}
\]

\vspace{0.1cm}

\[
\mkrule
{\ceq \Gamma {\doto M {(x \!:\! A , z \!:\! \ul{C})} {} {K\,\, V}} {(\doto M {(x \!:\! A , z \!:\! \ul{C})} {} {K})\,\, V} {\ul{D}[V/y]}}
{\cj \Gamma M {\Sigma\, x \!:\! A .\, \ul{C}} \quad \hj {\Gamma, x \!:\! A} {z \!:\! \ul{C}} K {\Pi\, y \!:\! B .\, \ul{D}} \quad \vj \Gamma V B \quad \lj {\Gamma, y \!:\! B} \ul{D}}
\]

\vspace{0.1cm}

\[
\mkrule
{\heq \Gamma {z_1 \!:\! \ul{C}} {\doto K {(x \!:\! A , z_2 \!:\! \ul{D}_1)} {} {L\,\, V}} {(\doto K {(x \!:\! A , z_2 \!:\! \ul{D}_1)} {} {L})\,\, V} {\ul{D}_2[V/y]}}
{\hj \Gamma {z_1 \!:\! \ul{C}} K {\Sigma\, x \!:\! A .\, \ul{D}_1} \quad \hj {\Gamma, x \!:\! A} {z_2 \!:\! \ul{D}_1} L {\Pi\, y \!:\! B .\, \ul{D}_2} \quad \vj \Gamma V B \quad \lj {\Gamma, y \!:\! B} \ul{D}_2}
\]

\vspace{0.1cm}

\[
\mkrule
{\ceq \Gamma {\doto M {(x \!:\! A , z \!:\! \ul{C})} {} {V\, K}} {V\, (\doto M {(x \!:\! A , z \!:\! \ul{C})} {} {K})} {\ul{D}_2}}
{\cj \Gamma M {\Sigma\, x \!:\! A .\, \ul{C}} \quad \vj \Gamma V {\ul{D}_1 \multimap \ul{D}_2} \quad \hj {\Gamma, x \!:\! A} {z \!:\! \ul{C}} K {\ul{D}_1}}
\]

\vspace{0.1cm}

\[
\mkrule
{\heq \Gamma {z_1 \!:\! \ul{C}} {\doto K {(x \!:\! A , z_2 \!:\! \ul{D}_1)} {} {V\, L}} {V\, (\doto K {(x \!:\! A , z_2 \!:\! \ul{D}_1)} {} {L})} {\ul{D}_3}}
{\hj \Gamma {z_1 \!:\! \ul{C}} K {\Sigma\, x \!:\! A .\, \ul{D}_1} \quad \vj \Gamma V {\ul{D}_2 \multimap \ul{D}_3} \quad \hj {\Gamma, x \!:\! A} {z_2 \!:\! \ul{D}_1} L {\ul{D}_2}}
\]
\end{proposition}

\begin{proof}
All these equations are proved by using the $\beta$- and $\eta$-equations for computational pattern-matching, following a similar pattern to the proof of Proposition~\ref{prop:seqcompdistributivity}. 

For example, commutativity with sequential composition is proved as follows:
\begin{fleqn}[0.2cm]
\begin{align*}
\Gamma \,\vdash\,\, & \doto M {(x \!:\! A,z \!:\! \ul{C})} {} {(\doto K {y \!:\! B} {} {N})}
\\
=\,\, & \doto M {(x \!:\! A,z \!:\! \ul{C})} {} {\big(\doto {\big(\doto {\langle x , z \rangle} {(x' \!:\! A, z' \!:\! \ul{C}[x'/x])} {} {K[x'/x][z'/z]}\big)} {y \!:\! B} {} {N}\big)}
\\
=\,\, & \doto M {(x \!:\! A,z \!:\! \ul{C})} {} {\big(\\[-1mm] & \hspace{2.45cm} \doto {\big(\doto {z''} {(x' \!:\! A, z' \!:\! \ul{C}[x'/x])} {} {K[x'/x][z'/z]}\big)[\langle x , z \rangle/z'']} {y \!:\! B} {} {N}\big)}
\\
=\,\, & \doto M {(x \!:\! A,z \!:\! \ul{C})} {} {\big(\\[-1mm] & \hspace{2.45cm} \doto {\big(\doto {z''} {(x' \!:\! A, z' \!:\! \ul{C}[x'/x])} {} {K[x'/x][z'/z]}\big)} {y \!:\! B} {} {N}\big)[\langle x , z \rangle/z'']}
\\
=\,\, & \doto {\big(\doto {M} {(x' \!:\! A, z' \!:\! \ul{C}[x'/x])} {} {K[x'/x][z'/z]}\big)} {y \!:\! B} {} {N}
\\
=\,\, & \doto {(\doto {M} {(x \!:\! A, z \!:\! \ul{C})} {} {K})} {y \!:\! B} {} {N} : \ul{D}
\end{align*}
\end{fleqn}

As another example, commutativity with homomorphic function application is proved as follows:
\begin{fleqn}[0.3cm]
\begin{align*}
\Gamma \,\vdash\,\, & \doto M {(x \!:\! A , z \!:\! \ul{C})} {} {V\, K}
\\
=\,\, & \doto M {(x \!:\! A , z \!:\! \ul{C})} {} {V\, \big(\doto {\langle x , z \rangle} {(y \!:\! A, z' \!:\! \ul{C}[z'/z])} {} {K[y/x][z'/z]}\big)}
\\
=\,\, & \doto M {(x \!:\! A , z \!:\! \ul{C})} {} {\big(V\, \big(\doto {z''} {(y \!:\! A, z' \!:\! \ul{C}[z'/z])} {} {K[y/x][z'/z]}\big)\big)[\langle x , z \rangle/z'']}
\\
=\,\, & V\, \big(\doto {M} {(y \!:\! A, z' \!:\! \ul{C}[z'/z])} {} {K[y/x][z'/z]}\big)
\\
=\,\, & V\, (\doto {M} {(x \!:\! A, z \!:\! \ul{C})} {} {K}) : \ul{D}_2
\end{align*}
\end{fleqn}
\end{proof}


\chapter{Fibred adjunction models}
\label{chap:fibadjmodels}

In this chapter we discuss the category-theoretic structures we use in Chapter~\ref{chap:interpretation} to give eMLTT a denotational semantics. Specifically, we work in the setting of fibred category theory because it provides a natural framework for developing the semantics of dependently typed languages, where i) functors model type-dependency, ii) split fibrations model substitution, and iii) split closed comprehension categories model $\Sigma$- and $\Pi$-types. See Section~\ref{sect:fibrationsbasics} for a brief overview of fibred category theory. It is important to note that the ideas we develop in this chapter can also be expressed in terms of other equivalent category-theoretic models of dependent types, such as contextual categories~\cite{Streicher:Semantics}, categories with families~\cite{Hofmann:SyntaxAndSemantics}, or categories with attributes~\cite{Pitts:CategoricalLogic}.

Specifically, in Section~\ref{sect:fibadjmodelsstructure}, we study the category-theoretic structures needed to model value and computation $\Sigma$- and $\Pi$-types, the empty type, the coproduct type, the type of natural numbers, intensional propositional equality, and the homomorphic function type. It is worth highlighting that in the case of the empty type, the coproduct type, and type of natural numbers, we identify category-theoretically more natural axiomatisations than commonly used in the semantics of dependently typed languages.

In Section~\ref{sect:fibadjmodels}, we combine these category-theoretic structures into a class of models suitable for giving a denotational semantics to eMLTT, called \emph{fibred adjunction models}. These models are a natural fibrational generalisation of adjunction-based models of simply typed computational languages such as CBPV and EEC.
Finally, in Section~\ref{sect:examplesoffibadjmodels}, we discuss some examples of these models, arising from i) identity adjunctions, ii) simple fibrations and models of EEC, iii) families fibrations and lifting of adjunctions, iv) the Eilenberg-Moore fibrations of split fibred monads, and v) the fibration of continuous families of $\omega$-complete partial orders and lifting of CPO-enriched Eilenberg-Moore adjunctions (so as to extend eMLTT with general recursion).

\section{Category theory for modelling eMLTT}
\label{sect:fibadjmodelsstructure}

In this section we discuss the category-theoretic structures that we use in Chapter~\ref{chap:interpretation} to give eMLTT a sound and complete denotational semantics.
Similarly to the overview of fibred category theory we gave in Section~\ref{sect:fibrationsbasics}, we only discuss split versions of these structures. The non-split versions can be recovered by relaxing the preservation conditions for reindexing so that they hold up-to-isomorphism rather than equality.

\subsection{$\Pi$- and $\Sigma$-types}
\label{sect:modelsofdependentsunctionandsumtypes}

We begin by discussing the structures we use to model the value $\Pi$- and $\Sigma$-types. 
As standard in categorical semantics of dependent types, we use well-behaved right and left adjoints to weakening functors to model these types, e.g., see~\cite[Section~10.5]{Jacobs:Book}.

\begin{definition}
\label{def:splitdependentproducts}
\index{split dependent!-- products}
\index{ Product@$\Pi_A$ (split dependent product)}
A split comprehension category with unit $p : \mathcal{V} \longrightarrow \mathcal{B}$ is said to have  \emph{split dependent products} if every weakening functor $\pi_A^* : \mathcal{V}_{p(A)} \longrightarrow \mathcal{V}_{\ia A}$ has a right adjoint $\Pi_A : \mathcal{V}_{\ia A} \longrightarrow \mathcal{V}_{p(A)}$ such that the split Beck-Chevalley condition holds: for any Cartesian morphism $f : A \longrightarrow B$ in $\mathcal{V}$, the canonical natural transformation
\[
\hspace{-0.15cm}
\xymatrix@C=1.15em@R=1.25em@M=0.5em{
(p(f))^* \comp \Pi_B \ar[rrrr]^-{\eta^{\pi^*_A \,\dashv\, \Pi_A} \,\comp\, (p(f))^* \,\comp\, \Pi_B} &&&& \Pi_A \comp \pi^*_A \comp (p(f))^* \comp \Pi_B \ar[r]^-{=} & \Pi_A \comp (p(f) \comp \pi_A)^* \comp \Pi_B \ar[d]^{=}
\\
\Pi_A \comp \ia f^* &&&& \Pi_A \comp \ia f ^* \comp \pi^*_B \comp \Pi_B \ar[llll]^-{\Pi_A \,\comp\, \ia f^* \,\comp\, \varepsilon^{\pi^*_B \,\dashv\, \Pi_B}} & \Pi_A \comp (\pi_B \comp \ia f)^* \comp \Pi_B \ar[l]^-{=}
}
\]
is required to be an identity. In particular, we must have $(p(f))^* \comp \Pi_B = \Pi_A \comp \ia f^*$.
\end{definition}

\begin{definition}
\label{def:weaksplitdependentsums}
\index{split dependent!weak -- sums}
\index{ Sigma@$\Sigma_A$ (weak split dependent sum)}
A split comprehension category with unit $p : \mathcal{V} \longrightarrow \mathcal{B}$ is said to have  \emph{weak split dependent sums} if every weakening functor $\pi_A^* : \mathcal{V}_{p(A)} \longrightarrow \mathcal{V}_{\ia A}$ has a left adjoint $\Sigma_A : \mathcal{V}_{\ia A} \longrightarrow \mathcal{V}_{p(A)}$ such that the split Beck-Chevalley condition holds: for any Cartesian morphism $f : A \longrightarrow B$ in $\mathcal{V}$, the canonical natural transformation
\[
\hspace{-0.15cm}
\xymatrix@C=1.25em@R=1.25em@M=0.5em{
\Sigma_A \comp \ia f^* \ar[rrrr]^-{\Sigma_A \,\comp\, \ia f^* \,\comp\, \eta^{\Sigma_B \,\dashv\, \pi^*_B}} &&&& \Sigma_A \comp \ia f^* \comp \pi^*_B \comp \Sigma_B \ar[r]^-{=} & \Sigma_A \comp (\pi_B \comp \ia f)^* \comp \Sigma_B \ar[d]^{=}
\\
(p(f))^* \comp \Sigma_B &&&& \Sigma_A \comp \pi^*_A \comp (p(f))^* \comp \Sigma_B \ar[llll]^-{\varepsilon^{\Sigma_A \,\dashv\, \pi^*_A} \,\comp\, (p(f))^* \,\comp\, \Sigma_B} & \Sigma_A \comp (p(f) \comp \pi_A)^* \comp \Sigma_B \ar[l]^-{=}
}
\]
is required to be an identity. In particular, we must have $\Sigma_A \comp \ia f^* = (p(f))^* \comp \Sigma_B$.
\end{definition}

Observe that these Beck-Chevalley conditions seem to guarantee that only the $\Pi_A$- and $\Sigma_A$-functors are preserved by reindexing. However, as we show below, they in fact ensure that the units and counits of the corresponding adjunctions are also preserved. 

\begin{proposition}
\label{prop:BCfordepproducts}
Given a split comprehension category with unit $p : \mathcal{V} \longrightarrow \mathcal{B}$ with split dependent products, and a Cartesian morphism $f : A \longrightarrow B$ in $\mathcal{V}$, then we have 
\[
(p(f))^* \comp \eta^{\pi^*_B \,\dashv\, \Pi_B} = \eta^{\pi^*_A \,\dashv\, \Pi_A} \comp (p(f))^*
\qquad
\ia f ^* \comp \varepsilon^{\pi^*_B \,\dashv\, \Pi_B} = \varepsilon^{\pi^*_A \,\dashv\, \Pi_A} \comp \ia f ^*
\]
\end{proposition}

\begin{proof}
These two equations follow from the commutativity of the next two diagrams.
\vspace{0.25cm}
\[
\xymatrix@C=0.6em@R=5em@M=0.5em{
(p(f))^* \ar[r]^-{\eta^{\pi^*_A \,\dashv\, \Pi_A} \,\comp\, (p(f))^*} \ar@/_6pc/[dd]_-{(p(f))^* \comp \,\eta^{\pi^*_B \,\dashv\, \Pi_B}}^<<<<<<<<<<<<<<<{\quad\dcomment{\text{nat. of } \eta^{\pi^*_A \,\dashv\, \Pi_A}}} & \Pi_A \comp \pi^*_A \comp (p(f))^* \ar@/^6.5pc/[ddd]^>>>>>>>>>>>>>>>>>>>>>>>>>>{=} \ar[dl]_{\Pi_A \,\comp\, \pi^*_A \,\comp\, (p(f))^* \,\comp\, \eta^{\pi^*_B \,\dashv\, \Pi_B}\quad}^{\qquad\quad\!\!\!\dcomment{\pi^*_A \comp (p(f))^* = \ia f ^* \comp \pi^*_B}}
\\
\Pi_A \comp \pi^*_A \comp (p(f))^* \comp \Pi_B \comp \pi^*_B \ar[r]^-{=} & \Pi_A \comp \ia f ^* \comp \pi^*_B \comp \Pi_B \comp \pi^*_B \ar@/_2pc/[d]_>>>>>>>{\Pi_A \,\comp\, \ia f ^* \,\comp\, \varepsilon^{\pi^*_B \,\dashv\, \Pi_B} \,\comp\, \pi^*_B}^>>>>>>{\,\,\,\,\,\,\dcomment{\pi^*_B \,\dashv\, \Pi_B}}
\\
(p(f))^* \comp \Pi_B \comp \pi^*_B \ar[r]_-{=} \ar[u]^-{\dhide{\eta^{\pi^*_A \,\dashv\, \Pi_A} \comp (p(f))^* \comp \Pi_B \comp \pi^*_B}\!\!}_>>>>>>{\quad\quad\!\!\dcomment{\text{split Beck-Chevalley}}} & \Pi_A \comp \ia f ^* \comp \pi^*_B  & 
\\
& \Pi_A \comp \ia f ^* \comp \pi^*_B \ar@/_3.5pc/[uu]_>>>>>>>{\!\!\!\!\dhide{\Pi_A \comp \ia f ^* \comp \pi^*_B \comp \eta^{\pi^*_B \,\dashv\, \Pi_B}}} \ar[u]^{\id_{\Pi_A \comp\, \ia f ^* \,\comp\, \pi^*_B}}
}
\]

\vspace{0.15cm}

\[
\xymatrix@C=0.6em@R=5em@M=0.5em{
\ia f ^* \comp \pi^*_B \comp \Pi_B \ar@{<-}@/_7pc/[ddd]_>>>>>>>>>>>>>>>>>>>>>>>>>>>{=} \ar[r]^-{\ia f ^* \,\comp\, \varepsilon^{\pi^*_B \,\dashv\, \Pi_B}} & \ia f ^* \ar@{<-}@/^5.5pc/[dd]^-{\varepsilon^{\pi^*_A \,\dashv\, \Pi_A} \,\comp\, \ia f ^*}_<<<<<<<<<<<<<<<{\dcomment{\text{nat. of } \varepsilon^{\pi^*_A \,\dashv\, \Pi_A}}\quad}
\\
\pi^*_A \comp \Pi_A \comp \pi^*_A \comp (p(f))^* \comp \pi_B \ar[r]^-{=} \ar@/_4pc/[dd]_<<<<<<<<<{\dhide{\varepsilon^{\pi^*_A \,\dashv\, \Pi_A} \comp \pi^*_A \comp (p(f))^* \comp \Pi_B}\!\!\!\!}^<<<<<<<<<<<<{\,\,\,\,\,\,\,\dcomment{\pi^*_A \dashv \Pi_A}} & \pi^*_A \comp \Pi_A \comp \ia f ^* \comp \pi^*_B \comp \Pi_B \ar[ul]_{\quad\,\,\,\varepsilon^{\pi^*_A \,\dashv\, \Pi_A} \,\comp\, \ia f ^* \,\comp\, \pi^*_B \,\comp\, \Pi_B}^{\dcomment{\pi^*_A \comp (p(f))^* = \ia f ^* \comp \pi^*_B}\qquad\quad} \ar[d]^-{\!\!\dhide{\pi^*_A \comp \Pi_A \comp \ia f ^* \comp \varepsilon^{\pi^*_B \,\dashv\, \Pi_B}}}_>>>>>>{\dcomment{\text{split Beck-Chevalley}}\qquad\!\!}
\\
\pi^*_A \comp (p(f))^* \comp \Pi_B \ar[d]^{\id_{\pi^*_A \comp\, (p(f))^* \comp\, \Pi_B}} \ar[r]_-{=} \ar@/_2pc/[u]_>>>>>>>{\pi^*_A \,\comp\, \eta^{\pi^*_A \,\dashv\, \Pi_A} \,\comp\, (p(f))^* \,\comp\, \Pi_B} & \pi^*_A \comp \Pi_A \comp \ia f ^*
\\
\pi^*_A \comp (p(f))^* \comp \Pi_B
}
\]

\pagebreak

In both diagrams, we write $\pi^*_A \comp (p(f))^* = \ia f ^* \comp \pi^*_B$  for the composite equation
\[
\pi^*_A \comp (p(f))^* = (p(f) \comp \pi_A)^* = (\pi_B \comp \ia f)^* = \ia f ^* \comp \pi^*_B
\]
where the middle equation holds because $(\ia f , p(f)) : \pi_A \longrightarrow \pi_B$ is a morphism in $\mathcal{B}^\to$ given by $\mathcal{P}(f)$---see Proposition~\ref{prop:comprehensioncategorywithunit} for the definition of $\mathcal{P} : \mathcal{V} \longrightarrow \mathcal{B}^\to$.
\end{proof}

\begin{proposition}
\label{prop:BCfordepsums}
Given a split comprehension category with unit $p : \mathcal{V} \longrightarrow \mathcal{B}$ with weak split dependent sums, and a Cartesian morphism $f : A \longrightarrow B$ in $\mathcal{V}$, then we have 
\[
\ia f ^* \comp \eta^{\Sigma_B \,\dashv\, \pi^*_B} = \eta^{\Sigma_A \,\dashv\, \pi^*_A} \comp \ia f ^*
\qquad
(p(f))^* \comp \varepsilon^{\Sigma_B \,\dashv\, \pi^*_B} = \varepsilon^{\Sigma_A \,\dashv\, \pi^*_A} \comp (p(f))^*
\]
\end{proposition}

\begin{proof}
By straightforward diagram chasing, analogously to Proposition~\ref{prop:BCfordepproducts}.
\end{proof}

Analogously to~\cite[Section~10.5]{Jacobs:Book}, we also require these split dependent sums to be strong, as made precise in Definition~\ref{def:strongsums}, so as to be able to model the type-dependency appearing in the typing rule of the elimination form for the value $\Sigma$-type. 

\begin{definition}
\label{def:strongsums}
\index{split dependent!strong -- sums}
\index{ Sigma@$\Sigma_A$ (strong split dependent sum)}
\index{ k@$\kappa_{A,B}$ (isomorphism witnessing the strength of split dependent sums)}
A split comprehension category with unit $p : \mathcal{V} \longrightarrow \mathcal{B}$ is said to have \emph{strong split dependent sums} if it has weak split dependent sums, as defined above, and if for every two objects $A$ in $\mathcal{V}$ and $B$ in $\mathcal{V}_{\ia{A}}$, the canonical composite morphism 
\[
\xymatrix@C=5em@R=1em@M=0.5em{
\kappa_{A,B} \quad \defeq \quad \ia B \ar[r]^-{\ia {\eta_B^{\Sigma_A \,\dashv\, \pi_A^*}}} & \ia {\pi^*_A(\Sigma_A (B))} \ar[r]^-{\ia {\overline{\pi_A}(\Sigma_A (B))}} & \ia {\Sigma_A (B)}
}
\]
is an isomorphism. 
\end{definition}

Following~\cite[Section~10.5]{Jacobs:Book}, we next combine split dependent products and strong split dependent sums into a structure that forms the basis of our semantics of eMLTT's value fragment. In particular, this structure allows us to model the core features of eMLTT such as the empty value context, the extension of value contexts, and the value $\Sigma$- and $\Pi$-types, including the corresponding introduction and elimination forms. 

\begin{definition}
\label{def:sccompc}
\index{category!split closed comprehension --}
\index{ SCCompC@\SCCompC\, (split closed comprehension category)}
\index{ 1@$1$ (terminal object)}
\index{object!terminal --}
A \emph{split closed comprehension category} (\SCCompC) is a full split comprehension category with unit that has both split dependent products and strong split dependent sums, and a terminal object $1$ in its base category.
\end{definition}

Similarly to how we defined the product type $A \times B$ and the function type $A \to B$ as non-dependent versions of $\Sigma\, x \!:\! A .\, B$ and $\Pi\, x \!:\! A .\, B$ in Chapter~\ref{chap:syntax}, the split dependent products and strong split dependent sums make each fibre of a given \SCCompC\, into a Cartesian closed category (CCC), as illustrated in the next proposition.

\pagebreak

\begin{proposition}
\label{prop:sccompc}
\index{ CCC@CCC (Cartesian closed category)}
\index{category!Cartesian closed --}
\index{ A@$A \times_X B$ (Cartesian product of $A$ and $B$ in $\mathcal{V}_X$)}
\index{ A@$A \Rightarrow_X B$ (exponential object in $\mathcal{V}_X$)}
Given a \SCCompC ~$p : \mathcal{V} \longrightarrow \mathcal{B}$, then every fibre $\mathcal{V}_X$ is a CCC with
\[
A \times_X B \defeq \Sigma_A (\pi_A^*(B))
\qquad
A \Rightarrow_X B \defeq \Pi_A (\pi_A^*(B))
\]
and this structure is preserved on-the-nose by reindexing, i.e., $p$ is a split fibred CCC.
\end{proposition}

We omit the details of the proof of this proposition and refer the reader to the proof of the non-split version of this proposition given in~\cite[Proposition~10.5.4]{Jacobs:Book}.

However, as the fibre-wise Cartesian products $A \times_X B$ play an important role in the interpretation of sequential composition in~Section~\ref{sect:interpretation}, we spell out the definitions of the corresponding vertical projection and pairing morphisms.

\index{projection!first --}
\index{ fst@$\semfst$ (first projection for Cartesian products)}
We obtain the \emph{first projection} $\semfst : A \times_X B \longrightarrow A$ by using the fully-faithfulness of $\mathcal{P} : \mathcal{V} \longrightarrow \mathcal{B}^{\to}$ on the following morphism between $\pi_{\Sigma_A(\pi_A^*(B))}$ and $\pi_A$ in $\mathcal{B}^\to$:
\[
\xymatrix@C=8.5em@R=6em@M=0.5em{
\ia {\Sigma_A (\pi_A^*(B))} \ar[ddd]_{\pi_{\Sigma_A(\pi_A^*(B))}}^>>>>>>>>>>>{\qquad\qquad\qquad\qquad\dcomment{\text{id. law}}} \ar[r]^-{\kappa^{-1}_{A,\pi_A^*(B)}} \ar@/_3pc/[ddr]_{\id_{\ia {\Sigma_A (\pi_A^*(B))}}\!\!}^<<<<<<<<{\qquad\dcomment{\kappa_{A,\pi_A^*(B)} \text{ is an iso.}}} & \ia {\pi_A^*(B)} \ar[r]^-{\pi_{\pi_A^*(B)}} \ar@/_7.5pc/[dd]_{\kappa_{A,\pi_A^*(B)}}^<<<<<<<<<<<<<<<<<<{\quad\!\!\!\!\!\!\!\dcomment{\text{def. of } \kappa_{A,\pi_A^*(B)}}} \ar[d]^{\ia {\eta^{\Sigma_A \,\dashv\, \pi^*_A}_{\pi_A^*(B)}}} & \ia A \ar[ddd]^{\pi_A}_>>>>>>>>>>>>>>>>>>>>>>>>>>>>>>>>{\dcomment{\mathcal{P}(\overline{\pi_A}(\Sigma_A(\pi_A^*(B))))}\qquad}
\\
& \ia {\pi^*_A(\Sigma_A(\pi_A^*(B)))} \ar[d]^{\ia {\overline{\pi_A}(\Sigma_A(\pi_A^*(B)))}} \ar[ur]_{\quad\,\,\,\pi_{\pi^*_A(\Sigma_A(\pi_A^*(B)))}}^>>>>>>>>>>>>>>>>>>>>>>{\dcomment{\mathcal{P}(\eta^{\Sigma_A \,\dashv\, \pi^*_A}_{\pi_A^*(B)})}\quad\,\,\,\,\,\,\,\,\,\,\,}
\\
& \ia {\Sigma_A(\pi_A^*(B))} \ar[dr]_{\pi_{\Sigma_A(\pi_A^*(B))}}
\\
X \ar[rr]_{\id_X} & & X
}
\]

\index{projection!second --}
\index{ snd@$\semsnd$ (second projection for Cartesian products)}
We obtain the \emph{second projection} $\semsnd : A \times_X B \!\longrightarrow\! B$ by defining 
\[
\semsnd \defeq \varepsilon^{\Sigma_A \,\dashv\, \pi^*_A}_B
\]

\index{ f@$\pair f g$ (unique mediating (pairing) morphism for Cartesian products)}
\index{morphism!unique mediating (pairing) --}
Finally, given two vertical morphisms $f : C \longrightarrow A$ and $g : C \longrightarrow B$ in $\mathcal{V}_X$, we obtain the unique mediating (\emph{pairing}) morphism $\pair f g : C \longrightarrow A \times_X B$  using the fully-faithfulness of $\mathcal{P}$ on the following morphism between $\pi_C$ and $\pi_{\Sigma_A(\pi^*_A(B))}$ in $\mathcal{B}^\to$:

\[
\hspace{-0.1cm}
\xymatrix@C=4.4em@R=4em@M=0.5em{
\ia{C} \ar[dddd]_{\pi_C}^>>>>>>>>>>>>>>>>>>>>>>>>>>>>>>>{\qquad\dcomment{\mathcal{P}(f)}} \ar[dddr]_{\ia f}^<<<<<<<<<<{\quad\,\,\,\,\,\,\,\dcomment{\text{def. of } h}} \ar[r]^-{h} & \ia {\pi^*_A(B)} \ar[dd]_{\id_{\ia {\pi^*_A(B)}}}^>>>>>>>>>>>>>>>>>>>>{\qquad\dcomment{\Sigma_A \dashv \pi^*_A}} \ar[dr]_{\ia {\eta^{\Sigma_A \,\dashv\, \pi^*_A}_{\pi^*_A(B)}}}^<<<<<<<<<<<<<{\qquad\dcomment{\text{def. of } \kappa_{A,\pi^*_A(B)}}} \ar[rr]^-{\kappa_{A,\pi^*_A(B)}}&  & \ia {\Sigma_A(\pi^*_A(B))} \ar[dddd]^{\pi_{\Sigma_A(\pi^*_A(B))}}_{\dcomment{\mathcal{P}(\varepsilon^{\Sigma_A \,\dashv\, \pi^*_A}_B)}\qquad\,\,\,\,} \ar[ddl]^{\!\!\!\!\!\!\ia {\varepsilon^{\Sigma_A \,\dashv\, \pi^*_A}_B}}_>{\dcomment{\text{def. of } \pi^*_A(\varepsilon^{\Sigma_A \,\dashv\, \pi^*_A}_B)}}
\\
& & \ia {\pi^*_A(\Sigma_A(\pi^*_A(B)))} \ar[dl]_{\ia {\pi^*_A(\varepsilon^{\Sigma_A \,\dashv\, \pi^*_A}_B)}\quad\!\!\!} \ar[ur]^{\ia{\overline{\pi_A}(\Sigma_A(\pi^*_A(B)))}\quad}
\\
& \ia {\pi^*_A(B)} \ar[d]^{\pi_{\pi^*_{A}(B)}} \ar[r]_{\ia {\overline{\pi_A}(B)}} & \ia {B} \ar[ddr]^{\pi_B}_<<<<<<<<<<<<<<<{\dcomment{\mathcal{P}(\overline{\pi_A}(B))}\qquad\qquad}
\\
& \ia A \ar[dl]^{\pi_A} \ar[drr]^{\pi_A}_{\dcomment{\text{id. law}}\qquad\qquad\qquad\qquad\qquad\qquad} & &
\\
X \ar[rrr]_{\id_X} & & & X 
}
\]
where $h$ is the unique mediating morphism in the following pullback situation:
\[
\xymatrix@C=4em@R=4em@M=0.5em{
\ia C \ar@/_2pc/[dr]_{\ia f} \ar@/^3pc/[rr]^{\ia g} \ar@{-->}[r]^-{h} & \ia {\pi^*_A(B)} \ar[d]_{\pi_{\pi^*_A(B)}}^<{\,\big\lrcorner} \ar[r]^-{\ia {\overline{\pi_A}(B)}} & \ia B \ar[d]^{\pi_B}_{\dcomment{\mathcal{P}(\overline{\pi_A}(B))}\quad\,\,\,\,\,\,\,\,\,}
\\
& \ia A \ar[r]_-{\pi_A} & X
}
\]
and where $\pi_A \comp \ia f = \pi_C = \pi_B \comp \ia g$ follow from $\mathcal{P}(f)$ and $\mathcal{P}(g)$, respectively.

Next, we prove two useful results (Proposition~\ref{prop:semsubstintoweakenedterm} and Corollary~\ref{cor:semsubstintoweakenedterm2}) that we later use to relate different ways of modelling non-dependent substitution. 
At a high level, Proposition~\ref{prop:semsubstintoweakenedterm} says that given value terms $\vj \Gamma V A$ and $\vj {\Gamma, x \!:\! A} W B$, where the value type $B$ does not depend on $x$, i.e., $\lj \Gamma B$, we can model the substitution $\vj \Gamma {W[V/x]} B$ either by applying a reindexing functor, or by composing morphisms.

\begin{proposition}
\label{prop:semsubstintoweakenedterm}
Given a full split comprehension category with unit $p : \mathcal{V} \longrightarrow \mathcal{B}$ with strong split dependent sums, an object $X$ in $\mathcal{B}$, objects $A$ and $B$ in $\mathcal{V}_X$, a morphism $f : 1_X \longrightarrow A$ in $\mathcal{V}_{X}$, and a morphism $g : 1_{\ia A} \longrightarrow \pi^*_A(B)$ in $\mathcal{V}_{\ia A}$, then
\[
\xymatrix@C=4em@R=2em@M=0.5em{
1_X \ar[r]^-{=} \ar[d]_-{f} & (\mathsf{s}(f))^*(1_{\ia A}) \ar[r]^-{(\mathsf{s}(f))^*(g)} & (\mathsf{s}(f))^*(\pi^*_A(B)) \ar[r]^-{=} & B
\\
A \ar[r]_-{\langle \id_{A} , !_A \rangle} & \Sigma_A(\pi^*_A(1_X)) \ar[r]_-{=} & \Sigma_A(1_{\ia A}) \ar[r]_-{\Sigma_A(g)} & \Sigma_A(\pi^*_A(B)) \ar[u]_-{\varepsilon^{\Sigma_A \,\dashv\, \pi^*_A}_{B}}
}
\]
commutes in $\mathcal{V}_X$.
\end{proposition}

\begin{proof}
This diagram commutes because we have
\[
\xymatrix@C=4em@R=4em@M=0.5em{
1_X \ar[r]^-{=} \ar[dd]_-{f}^-{\!\!\!\!\!\!\!\!\quad\qquad\dcomment{(a)}} \ar[dr]_-{1(\mathsf{s}(f))} & (\mathsf{s}(f))^*(1_{\ia A}) \ar[r]^-{(\mathsf{s}(f))^*(g)} \ar[d]^-{\overline{\mathsf{s}(f)}(1_{\ia A})}_<<<{\dcomment{1 \text{ is s. fib.}}\quad\!\!\!}^>>{\qquad\dcomment{\text{def. of } (\mathsf{s}(f))^*(g)}} & (\mathsf{s}(f))^*(\pi^*_A(B)) \ar[r]^-{=} \ar[d]_-{\overline{\mathsf{s}(f)}(\pi^*_A(B))}^<<<<{\!\!\!\quad\dcomment{p \text{ is a s. fib.}}} & B
\\
& 1_{\ia{A}} \ar[r]_-{g} \ar[dl]^<<<<<<<<{\varepsilon^{1 \,\dashv\, \ia -}_{A}}^<<<<<<<<{\qquad\qquad\qquad\qquad\dcomment{\text{using the fully-faithfulness of } \mathcal{P} \text{ on } (b)}} & \pi^*_A(B) \ar[ur]_-{\overline{\pi_A}(B)}
\\
A \ar[r]_-{\langle \id_{A} , !_A \rangle} & \Sigma_A(\pi^*_A(1_X)) \ar[r]_-{=} & \Sigma_A(1_{\ia A}) \ar[r]_-{\Sigma_A(g)} & \Sigma_A(\pi^*_A(B)) \ar[uu]_-{\varepsilon^{\Sigma_A \,\dashv\, \pi^*_A}_{B}}
}
\]
where $(a)$ commutes because we have
\[
\xymatrix@C=6em@R=3em@M=0.5em{
1_X \ar@/^3pc/[rr]^-{1(\mathsf{s}(f))}_*+<1em>{\dcomment{\text{def. of } \mathsf{s}(f)}} \ar[r]_-{1(\eta^{1 \,\dashv\, \ia -}_X)} \ar@/_2.5pc/[ddr]_{f} \ar@/_1pc/[dr]_-{\id_{1_X}} & 1_{\ia {1_X}} \ar[r]_-{1(\ia {f})} \ar[d]^-{\varepsilon^{1 \,\dashv\, \ia -}_{1_X}}_-{\dcomment{1 \,\dashv\, \ia -}\,\,\,} & 1_{\ia A} \ar@/^2.5pc/[ddl]^-{\varepsilon^{1 \,\dashv\, \ia -}_A}_-{\dcomment{\text{nat. of } \varepsilon^{1 \,\dashv\, \ia -}}\,\,\,}
\\
& 1_X \ar[d]^-{f}_<<{\dcomment{\text{id. law}}\qquad}
\\
& A
}
\]
Further, we note that $(b)$ refers to a diagram in $\mathcal{B}^{\to}$ between $\pi_{1_{\ia A}} : \ia {1_{\ia A}} \longrightarrow \ia {A}$ and $\pi_B : \ia {B} \longrightarrow X$ that commutes because i) we have the following sequence of equations:
\[
\begin{array}{c}
\hspace{-4.25cm}
p(\overline{\pi_A}(B)) \comp p(g) = p(\overline{\pi_A}(B)) \comp \id_{\ia A} = \pi_A = p(\varepsilon^{1 \,\dashv\, \ia -}_A) =
\\[1mm]
\hspace{2.5cm}
\id_{X} \comp p(\varepsilon^{1 \,\dashv\, \ia -}_A)  = p(\varepsilon^{\Sigma_A \,\dashv\, \pi^*_A}_B) \comp p(\Sigma_A(g)) \comp p(\langle \id_A , !_A \rangle) \comp p (\varepsilon^{1 \,\dashv\, \ia -}_A)
\end{array}
\]
for the morphisms between the codomains of $\pi_{1_{\ia A}}$ and $\pi_B$; and ii) we can 
show for the morphisms between the domains of $\pi_{1_{\ia A}}$ and $\pi_B$ that the following diagram commutes:
\[
\hspace{-0.25cm}
\xymatrix@C=0.25em@R=5em@M=0.5em{
\ia {1_{\ia A}} \ar[dd]_-{\ia {\varepsilon^{1 \,\dashv\, \ia -}_A}} \ar[rrrr]^-{\ia g} \ar[drr]^-{\!\!\!\eta^{1 \,\dashv\, \ia -}_{\ia {1_{\ia A}}}}_-{\dcomment{\text{iso.}}\!\!} \ar@/^5.25pc/[ddrrr]^-{\!\!\!\!\!\!\!\id_{\ia {1_{\ia A}}}}_-{\dcomment{1 \,\dashv\, \ia -}\,\,\,} &&&& \ia {\pi^*_A(B)} \ar[r]^-{\ia {\overline{\pi_A}(B)}} & \ia B
\\
&& \ia {1_{\ia {1_{\ia A}}}} \ar@/_1.5pc/[dr]_<<{\ia {1(\ia {\varepsilon^{1 \,\dashv\, \ia -}_A})}\!} \ar@/^2pc/[dr]_>>>>>>>>>>>{\ia {\varepsilon^{1 \,\dashv\, \ia -}_{1_{\ia A}}}\!\!\!}_>>>>>>>>>>>>>{\dcomment{\mathcal{P}(1(\ia {\varepsilon^{1\,\dashv\, \ia -}_A}))}\qquad\qquad\qquad\qquad\,\,\,\,}_<<<<<<<{\dcomment{1 \text{ is  s. f.}}} \ar@/^2pc/[ull]^-{\pi_{1_{\ia {1_{\ia A}}}}\!\!\!\!\!} & 
\\
\ia A \ar[dd]_-{\langle \id_A , !_A \rangle}^-{\quad\!\!\dcomment{\text{def. of } \langle \id_A , !_A \rangle}} \ar@/_0.75pc/[rrr]_-{\eta^{1 \,\dashv\, \ia -}_{\ia A}} \ar[drr]_-{h} &&& \ia {1_{\ia A}} \ar@/_1.5pc/[lll]_-{\pi_{1_{\ia A}}}^-{\dcomment{\eta^{1 \,\dashv\, \ia -}_{\ia A} \text{is an iso.}}} \ar[dl]^-{=}_>>>>>>>>>{\dcomment{(c)}\qquad\!\!\!\!} \ar[r]_-{\!\!\ia g}
&
\ia {\pi_A^*(B)}
\ar[dddd]_-{\ia {\eta^{\Sigma_A \,\dashv\, \pi^*_A}_{\pi^*_A(B)}}}
\ar@/^3pc/[uu]^-{\id_{\ia {\pi^*_A(B)}}}^>>>>{\dcomment{\id_{\ia {\pi^*_A(B)}} \,\comp\, \ia g = \ia g \,\comp \, \id_{\ia {1_{\ia A}}}}\quad\!\!\!}
\\
&& \ia {\pi^*_A(1_X)} \ar[dll]_-{\kappa_{A,\pi^*_A(1_X)}} \ar[dd]^-{\ia {\eta^{\Sigma_A \,\dashv\, \pi^*_A}_{\pi^*_A(1_X)}}}_-{\dcomment{\text{def. of } \kappa_{A,\pi^*_A(1_X)}} \quad\!\!\!\!}^>>>>>>>>>>{\qquad\qquad\dcomment{\text{nat. of } \eta^{\Sigma_A \,\dashv\, \pi^*_A}}}
\\
\ia {\Sigma_A(\pi^*_A(1_X))} \ar[ddd]_-{=} && 
\\
&& \ia {\pi^*_A(\Sigma_A(\pi^*_A(1_X)))} \ar[ull]^<<<<<<<<<{\ia {\overline{\pi_A}(\Sigma_A(\pi^*_A(1_X)))}\quad\!\!\!\!\!} \ar[d]_-{=}_-{\dcomment{1 \text{ is split fibred}}\qquad\,\,\,}
\\
&& \ia {\pi^*_A(\Sigma_A(1_{\ia A}))} \ar[dll]^-{\,\,\,\ia {\overline{\pi_A}(\Sigma_A(1_{\ia A}))}}^-{\qquad\qquad\qquad\qquad\qquad\dcomment{\text{def. of } \pi^*_A(\Sigma_A(g))}} \ar[rr]_-{\ia {\pi^*_A(\Sigma_A(g))}} && \ia {\pi^*_A(\Sigma_A(\pi^*_A(B)))} \ar[dr]_-{\ia {\overline{\pi_A}(\Sigma_A(\pi^*_A(B)))}} \ar@/_3pc/[uuuuuu]_<<<<<<<<<<<<<<<<<{\ia {\pi^*_A(\varepsilon^{\Sigma \,\dashv\, \pi^*_A}_B)}}^>>>>>>>>>>>>>>>>>>>>>>>>{\dcomment{\Sigma_A \,\dashv\, \pi^*_A}\quad\!\!\!\!\!}
\\
\ia {\Sigma_A(1_{\ia A})} \ar[rrrrr]_-{\ia {\Sigma_A(g)}} &&&&& \ia {\Sigma_A(\pi^*_A(B))} \ar[uuuuuuu]_-{\ia {\varepsilon^{\Sigma \,\dashv\, \pi^*_A}_B}}^>>>>>>{\dcomment{\text{def. of } \pi^*_A(\varepsilon^{\Sigma \,\dashv\, \pi^*_A})}\,\,\,}
}
\]
In the previous diagram, $h$ is defined as the unique mediating morphism into the pullback square given by $\mathcal{P}(\overline{\pi_A}(1_X))$, for $\ia {!_A} : \ia {A} \longrightarrow \ia {1_{X}}$ and $\id_{\ia A} : \ia {A} \longrightarrow \ia {A}$. 

Finally, we show that $(c)$ commutes by observing that $\eta^{1 \,\dashv\, \ia -}_{\ia A}$ satisfies the same universal property as $h$, as shown in the following diagram:
\[
\xymatrix@C=9em@R=6em@M=0.5em{
& \ia {1_X} \ar[d]^-{\pi_{1_X}}_>{\dcomment{\mathcal{P}(!_A)}\qquad}^>{\,\,\,\,\,\,\,\dcomment{\eta^{1 \,\dashv\, \ia -}_X \text{ is an iso.}}} \ar@/^3.5pc/[dddr]^-{\id_{\ia {1_X}}}
\\
& X \ar@/^2pc/[ddr]_<<<<<<<<{\eta^{1 \,\dashv\, \ia -}_{X}}_<<<<<<<<<<<<<<<<<<{\dcomment{\text{nat. of } \eta^{1 \,\dashv\, \ia -}}\qquad\qquad}
\\
& \ia {1_{\ia A}} \ar[d]_-{=}^{\,\,\,\,\quad\dcomment{1 \text{ is split fibred}}}_>>>>{\dcomment{1 \text{ is s. fib.}}\,\,\,\,} \ar@/_5pc/[dd]_-{\pi_{1_{\ia A}}}_>>>>>>>>>>>>>>>>>>{\dcomment{\eta^{1 \,\dashv\, \ia -}_{\ia A} \text{is an iso.}}\,\,} \ar@/^2pc/[dr]^<<<<<<{\ia {1(\pi_A)}}
\\
\ia {A} \ar[ur]^-{\eta^{1 \,\dashv\, \ia -}_{\ia A}\!\!\!\!\!\!\!\!} \ar@/_2.75pc/[dr]_{\id_{\ia A}} \ar@/^1.5pc/[uur]_-{\pi_A} \ar@/^3.5pc/[uuur]^-{\ia {!_A}} & \ia {\pi^*_A(1_X)} \ar[d]_-{\pi_{\pi^*_A(1_X)}}^<<{\,\,\,\big\lrcorner} \ar[r]^-{\ia {\overline{\pi_A}(1_X)}} & \ia {1_X} \ar[d]^-{\pi_A}_-{\dcomment{\mathcal{P}(\overline{\pi_A}(1_X))}\qquad\qquad\quad\!\!\!\!\!\!\!\!\!\!}
\\
& \ia A \ar[r]_-{\pi_A} & X
}
\]
\end{proof}

Corollary~\ref{cor:semsubstintoweakenedterm2} follows from  Proposition~\ref{prop:semsubstintoweakenedterm} by setting $B \defeq U\ul{C}$. Intuitively, it says that given a value term $\vj \Gamma V A$ and a computation term $\cj {\Gamma, x \!:\! A} M \ul{C}$, where the computation type $\ul{C}$ does not depend on $x$, we can model the substitution $\cj \Gamma {M[V/x]} {\ul{C}}$ either by applying a reindexing functor, or by composing vertical morphisms.

\begin{corollary}
\label{cor:semsubstintoweakenedterm2}
Given a full split comprehension category with unit $p : \mathcal{V} \longrightarrow \mathcal{B}$ \linebreak with strong split dependent sums, a split fibration $q : \mathcal{C} \longrightarrow \mathcal{B}$, a split fibred functor $U : q \longrightarrow p$, an object $X$ in $\mathcal{B}$, an object $A$ in $\mathcal{V}_X$, an object $\ul{C}$ in $\mathcal{C}_X$, a morphism $f : 1_X \longrightarrow A$ in $\mathcal{V}_{X}$, and a morphism $g : 1_{\ia A} \longrightarrow U(\pi^*_A(\ul{C}))$ in $\mathcal{V}_{\ia A}$, then 
\[
\hspace{-0.25cm}
\xymatrix@C=2.25em@R=3em@M=0.5em{
1_X \ar[r]^-{=} \ar[d]_-{f} & (\mathsf{s}(f))^*(1_{\ia A}) \ar[r]^-{(\mathsf{s}(f))^*(g)} & (\mathsf{s}(f))^*(U(\pi^*_A(\ul{C}))) \ar[r]^-{=} & (\mathsf{s}(f))^*(\pi^*_A(U(\ul{C}))) \ar[d]^-{=}
\\
A \ar[d]_-{\langle \id_{A} , !_A \rangle} & & & U(\ul{C})
\\
\Sigma_A(\pi^*_A(1_X)) \ar[r]_-{=} & \Sigma_A(1_{\ia A}) \ar[r]_-{\Sigma_A(g)} & \Sigma_A(U(\pi^*_A(\ul{C}))) \ar[r]_-{=} & \Sigma_A(\pi^*_A(U(\ul{C}))) \ar[u]_-{\varepsilon^{\Sigma_A \,\dashv\, \pi^*_A}_{B}}
}
\]
commutes in $\mathcal{V}_X$.
\end{corollary}

Next, we define the structures we use to model the computational $\Pi$- and $\Sigma$-types. Similarly to their value counterparts, we also model these types using well-behaved right and left adjoints to weakening functors, but in a different fibration. These definitions are based on $\mathcal{P}$-products and -coproducts discussed in~\cite[Definition~9.3.5]{Jacobs:Book}.

\begin{definition}
\label{def:splitdependentcompproducts}
\index{split dependent!-- $p$-products}
\index{ Product@$\Pi_A$ (split dependent $p$-product)}
Given a split comprehension category with unit $p : \mathcal{V} \longrightarrow \mathcal{B}$, a split fibration $q : \mathcal{C} \longrightarrow \mathcal{B}$ is said to have  \emph{split dependent $p$-products} if every weakening functor $\pi_A^* : \mathcal{C}_{p(A)} \longrightarrow \mathcal{C}_{\ia A}$ has a right adjoint $\Pi_A : \mathcal{C}_{\ia A} \longrightarrow \mathcal{C}_{p(A)}$ such that the split Beck-Chevalley condition holds: for any Cartesian morphism $f : A \longrightarrow B$ in $\mathcal{V}$, the canonical natural transformation
\[
\xymatrix@C=1em@R=1.5em@M=0.5em{
(p(f))^* \comp \Pi_B \ar[rrrr]^-{\eta^{\pi^*_A \,\dashv\, \Pi_A} \,\comp\, (p(f))^* \,\comp\, \Pi_B} &&&& \Pi_A \comp \pi^*_A \comp (p(f))^* \comp \Pi_B \ar[r]^-{=} & \Pi_A \comp (p(f) \comp \pi_A)^* \comp \Pi_B \ar[d]^{=}
\\
\Pi_A \comp \ia f^* &&&& \Pi_A \comp \ia f ^* \comp \pi^*_B \comp \Pi_B \ar[llll]^-{\Pi_A \,\comp\, \ia f^* \,\comp\, \varepsilon^{\pi^*_B \,\dashv\, \Pi_B}} & \Pi_A \comp (\pi_B \comp \ia f)^* \comp \Pi_B \ar[l]^-{=}
}
\]
is required to be an identity. In particular, we must have $(p(f))^* \comp \Pi_B = \Pi_A \comp \ia f^*$.
\end{definition}

Observe that while the projection morphism $\pi_A : \ia A \longrightarrow p(A)$ in $\mathcal{B}$ is still induced by the split comprehension category with unit $p$, as in Definition~\ref{def:splitdependentproducts}, the weakening functor $\pi_A^* : \mathcal{C}_{p(A)} \longrightarrow \mathcal{C}_{\ia A}$ is now induced by reindexing along $\pi_A$ in $q$. 

\begin{definition}
\label{def:splitdependentcompsums}
\index{split dependent!-- $p$-sums}
\index{ Sigma@$\Sigma_A$ (split dependent $p$-sum)}
Given a split comprehension category with unit $p : \mathcal{V} \longrightarrow \mathcal{B}$,  a split fibration $q : \mathcal{C} \longrightarrow \mathcal{B}$ is said to have \emph{split dependent $p$-sums} if every weakening functor $\pi_A^* : \mathcal{C}_{p(A)} \longrightarrow \mathcal{C}_{\ia A}$ has a left adjoint $\Sigma_A : \mathcal{C}_{\ia A} \longrightarrow \mathcal{C}_{p(A)}$ such that the split Beck-Chevalley condition holds: for any Cartesian morphism $f : A \longrightarrow B$ in $\mathcal{V}$, the canonical natural transformation
\[
\hspace{-0.1cm}
\xymatrix@C=1.25em@R=1.5em@M=0.5em{
\Sigma_A \comp \ia f^* \ar[rrrr]^-{\Sigma_A \,\comp\, \ia f^* \,\comp\, \eta^{\Sigma_B \,\dashv\, \pi^*_B}} &&&& \Sigma_A \comp \ia f^* \comp \pi^*_B \comp \Sigma_B \ar[r]^-{=} & \Sigma_A \comp (\pi_B \comp \ia f)^* \comp \Sigma_B \ar[d]^{=}
\\
(p(f))^* \comp \Sigma_B &&&& \Sigma_A \comp \pi^*_A \comp (p(f))^* \comp \Sigma_B \ar[llll]^-{\varepsilon^{\Sigma_A \,\dashv\, \pi^*_A} \,\comp\, (p(f))^* \,\comp\, \Sigma_B} & \Sigma_A \comp (p(f) \comp \pi_A)^* \comp \Sigma_B \ar[l]^-{=}
}
\]
is required to be an identity. In particular, we must have $\Sigma_A \comp \ia f^* = (p(f))^* \comp \Sigma_B$.
\end{definition}

Observe that compared to the split dependent sums of $p$ (see Definition~\ref{def:strongsums}), we do not attempt to define a notion of strength for the split dependent $p$-sums of $q$. We do so because the typing rule of the elimination form for the computational $\Sigma$-type does not involve type-dependency, compared to the elimination form for the value $\Sigma$-type.

Analogously to Propositions~\ref{prop:BCfordepproducts} and~\ref{prop:BCfordepsums}, these split Beck-Chevalley conditions again ensure that the units and counits of these adjunctions are preserved by reindexing.

\begin{proposition}
\label{prop:BCfordepcompproducts}
Given a split comprehension category with unit $p : \mathcal{V} \longrightarrow \mathcal{B}$, a split fibration $q : \mathcal{C} \longrightarrow \mathcal{B}$ with split dependent $p$-products, and a Cartesian morphism $f : A \longrightarrow B$ in $\mathcal{V}$, then we have 
\[
(p(f))^* \comp \eta^{\pi^*_B \,\dashv\, \Pi_B} = \eta^{\pi^*_A \,\dashv\, \Pi_A} \comp (p(f))^*
\qquad
\ia f ^* \comp \varepsilon^{\pi^*_B \,\dashv\, \Pi_B} = \varepsilon^{\pi^*_A \,\dashv\, \Pi_A} \comp \ia f ^*
\]
\end{proposition}

\begin{proposition}
\label{prop:BCfordepcompsums}
Given a split comprehension category with unit $p : \mathcal{V} \longrightarrow \mathcal{B}$, a split fibration $q : \mathcal{C} \longrightarrow \mathcal{B}$ with split dependent $p$-sums, and a Cartesian morphism $f : A \longrightarrow B$ in $\mathcal{V}$, then we have 
\[
\ia f ^* \comp \eta^{\Sigma_B \,\dashv\, \pi^*_B} = \eta^{\Sigma_A \,\dashv\, \pi^*_A} \comp \ia f ^*
\qquad
(p(f))^* \comp \varepsilon^{\Sigma_B \,\dashv\, \pi^*_B} = \varepsilon^{\Sigma_A \,\dashv\, \pi^*_A} \comp (p(f))^*
\]
\end{proposition}

\begin{proof}
Propositions~\ref{prop:BCfordepcompproducts} and~\ref{prop:BCfordepcompsums} are proved by straightforward diagram chasing, analogously to the proof of Proposition~\ref{prop:BCfordepproducts}.
\end{proof}

We conclude this section by showing that if the split comprehension category with unit $p : \mathcal{V} \longrightarrow \mathcal{B}$ and the split fibration $q : \mathcal{C} \longrightarrow \mathcal{B}$ are connected by a split fibred adjunction $F \dashv\, U : q \longrightarrow p$, then $U$ preserves split dependent products and $F$ preserves split dependent sums.

\begin{proposition}
\label{prop:UandFpreserveSigmaPi}
Given a split comprehension category with unit $p : \mathcal{V} \longrightarrow \mathcal{B}$ \linebreak with split dependent products (resp.~weak split dependent sums), a split fibration \linebreak $q : \mathcal{C} \longrightarrow \mathcal{B}$ with split dependent $p$-products (resp.~split dependent $p$-sums), and a split fibred adjunction $F \dashv\, U : q \longrightarrow p$, then we have the natural isomorphism
\[
U \comp \Pi_A \cong \Pi_A \comp U : \mathcal{C}_{\ia A} \longrightarrow \mathcal{V}_{p(A)}
\qquad
(\text{resp.~} F \comp \Sigma_A \cong \Sigma_A \comp F : \mathcal{V}_{\ia A} \longrightarrow \mathcal{C}_{p(A)})
\]
for all objects $A$ in $\mathcal{V}$.
\end{proposition}

\begin{proof}

Both natural isomorphisms follow straightforwardly from the fact that adjoints are unique up-to a unique natural isomorphism. 

Specifically, in order to prove that the left-hand natural isomorphism involving $U$ and $\Pi_A$ exists, we first observe that we have the following two composite adjunctions:
\[
\xymatrix@C=3em@R=1.5em@M=0.5em{
\mathcal{V}_{p(A)} \ar@/^2pc/[rr]^{F} & \dhide{\bot} & \mathcal{C}_{p(A)} \ar@/^2pc/[rr]^{\pi^*_A} \ar@/^2pc/[ll]^{U} & \dhide{\bot} & \mathcal{C}_{\ia A} \ar@/^2pc/[ll]^{\Pi_A}
}
\]
\[
\xymatrix@C=3em@R=1.5em@M=0.5em{
\mathcal{V}_{p(A)} \ar@/^2pc/[rr]^{\pi^*_A} & \dhide{\bot} & \mathcal{V}_{\ia A} \ar@/^2pc/[rr]^{F} \ar@/^2pc/[ll]^{\Pi_A} & \dhide{\bot} & \mathcal{C}_{\ia A} \ar@/^2pc/[ll]^{U}
}
\vspace{0.2cm}
\]
We also recall that $F$ is a split fibred functor, meaning that $\pi^*_A \comp F = F \comp \pi^*_A$. 

By combining these two observations, we get that both $U \comp \Pi_A$ and $\Pi_A \comp U$ are right adjoints to $\pi^*_A \comp F$ (or, equivalently, to $F \comp \pi^*_A$). Therefore, as right adjoints are unique up-to a unique natural isomorphism, we get that $U \comp \Pi_A \cong \Pi_A \comp U$. In detail, this natural isomorphism is given by the following two vertical natural transformations:
\[
\hspace{-0.1cm}
\xymatrix@C=2em@R=2em@M=0.5em{
U \comp \Pi_A \ar[rr]^-{\eta^{\pi^*_A \,\dashv\, \Pi_A} \,\comp\, U \,\comp\, \Pi_A} && \Pi_A \comp \pi^*_A \comp U \comp \Pi_A \ar[r]^{=} & \Pi_A \comp U \comp \pi^*_A \comp \Pi_A \ar[rr]^-{\Pi_A \,\comp\, U \,\comp\, \varepsilon^{\pi^*_A \,\dashv\, \Pi_A}} && \Pi_A \comp U
}
\]
and
\[
\hspace{-0.15cm}
\xymatrix@C=7em@R=2em@M=0.5em{
\Pi_A \comp U \ar[r]^-{\eta^{F \,\dashv\, U} \,\comp\, \Pi_A \,\comp\, U} & U \comp F \comp \Pi_A \comp U \ar[r]^-{U \,\comp\, \eta^{\pi^*_A \,\dashv\, \Pi_A} \,\comp\, F \,\comp\, \Pi_A \,\comp\, U} & U \comp \Pi_A \comp \pi^*_A \comp F \comp \Pi_A \comp U \ar[d]^{=}
\\
U \comp \Pi_A & U \comp \Pi_A \comp F \comp U \ar[l]^-{U \,\comp\, \Pi_A \,\comp\, \varepsilon^{F \,\dashv\, U}} & U \comp \Pi_A \comp F \comp \pi^*_A \comp \Pi_A \comp U \ar[l]^-{U \,\comp\, \Pi_A \,\comp\, F \,\comp\, \varepsilon^{\pi^*_A \,\dashv\, \Pi_A} \,\comp\, U}
}
\]
We denote this natural isomorphism by $\zeta_{\Pi, A} : U \comp \Pi_A \overset{\cong}{\,\longrightarrow\,} \Pi_A \comp U$.
\index{ zeta@$\zeta_{\Pi, A}$ (natural isomorphism witnessing that $U$ preserves split dependent products)}

The other natural isomorphism is constructed similarly, by combining the adjunctions $\Sigma_A \dashv \pi^*_A$ with the fact that $U$ is a split fibred functor. In detail, it is given by
\[
\hspace{-0.2cm}
\xymatrix@C=7.25em@R=2em@M=0.5em{
F \comp \Sigma_A \ar[r]^-{F \,\comp\, \Sigma_A \,\comp\, \eta^{F \,\dashv\, U}} & F \comp \Sigma_A \comp U \comp F \ar[r]^-{F \,\comp\, \Sigma_A \,\comp\, U \,\comp\, \eta^{\Sigma_A \,\dashv\, \pi^*_A} \,\comp\, F} & F \comp \Sigma_A \comp U \comp \pi^*_A \comp \Sigma_A \comp F \ar[d]^-{=}
\\
\Sigma_A \comp F & F \comp U \comp \Sigma_A \comp F \ar[l]^-{\varepsilon^{F \,\dashv\, U} \,\comp\, \Sigma_A \,\comp\, F} & F \comp \Sigma_A\comp \pi^*_A \comp U \comp \Sigma_A \comp F \ar[l]^-{F \,\comp\, \varepsilon^{\Sigma_A \,\dashv\, \pi^*_A} \,\comp\, U \,\comp\, \Sigma_A \,\comp\, F}
}
\]
and
\[
\hspace{-0.1cm}
\xymatrix@C=2.25em@R=2em@M=0.5em{
\Sigma_A \comp F \ar[rr]^-{\Sigma_A \,\comp\, F \,\comp\, \eta^{\Sigma_A \,\dashv\, \pi^*_A}} && \Sigma_A \comp F \comp \pi^*_A \comp \Sigma_A \ar[r]^{=} & \Sigma_A \comp \pi^*_A \comp F \comp \Sigma_A \ar[rr]^-{\varepsilon^{\Sigma_A \,\dashv\, \pi^*_A} \,\comp\, F \,\comp\, \Sigma_A} && F \comp \Sigma_A
}
\]
We denote this natural isomorphism by $\zeta_{\Sigma,A} : F \comp \Sigma_A \overset{\cong}{\,\longrightarrow\,} \Sigma_A \comp F$.
\index{ zeta@$\zeta_{\Sigma,A}$ (natural isomorphism witnessing that $F$ preserves split dependent sums)}
\end{proof}

\noindent We now show that the corresponding units and counits are also preserved by $U$ and $F$.

\begin{proposition}
\label{prop:PiUnitCounitPreservedByU}
Given a split comprehension category with unit $p : \mathcal{V} \longrightarrow \mathcal{B}$ with split dependent products, a split fibration $q : \mathcal{C} \longrightarrow \mathcal{B}$ with split dependent $p$-products, and a split fibred adjunction $F \dashv\, U : q \longrightarrow p$, then the next two diagrams commute.
\[
\hspace{-0.1cm}
\xymatrix@C=3.75em@R=3em@M=0.5em{
U \ar[r]^-{\eta^{\pi^*_A \,\dashv\, \Pi_A} \,\comp\, U} \ar[d]_-{U \,\comp\, \eta^{\pi^*_A \,\dashv\, \Pi_A}} & \Pi_A \comp \pi^*_A \comp U
\\
U \comp \Pi_A \comp \pi^*_A \ar[r]_-{\zeta_{\Pi,A} \,\comp\, \pi^*_A} & \Pi_A \comp U \comp \pi^*_A \ar[u]_-{=}
}
\quad
\xymatrix@C=3.75em@R=3em@M=0.5em{
\pi^*_A \comp \Pi_A \comp U \ar[r]^-{\varepsilon^{\pi^*_A \,\dashv\, \Pi_A} \,\comp\, U} \ar[d]_-{\pi^*_A \,\comp\, \zeta^{-1}_{\Pi,A}} & U
\\
\pi^*_A \comp U \comp \Pi_A \ar[r]_-{=} & U \comp \pi^*_A \comp \Pi_A \ar[u]_-{U \,\comp\, \varepsilon^{\pi^*_A \,\dashv\, \Pi_A}}
}
\]
\end{proposition}

\begin{proof}
We show the commutativity of these two diagrams by straightforward diagram chasing.
For example, the left-hand square commutes because we have
\[
\hspace{-0.1cm}
\xymatrix@C=6em@R=4em@M=0.5em{
U 
\ar[r]^-{\eta^{\pi^*_A \,\dashv\, \Pi_A} \,\comp\, U} \ar@/_6pc/[dd]_-{U \,\comp\, \eta^{\pi^*_A \,\dashv\, \Pi_A}}
\ar@{}[d]^-{\!\!\!\!\!\!\!\!\!\!\!\!\!\!\!\!\!\!\!\!\!\!\!\!\!\!\!\!\!\!\dcomment{\text{nat. of } \eta^{\pi^*_A \,\dashv\, \Pi_A}}}
& 
\Pi_A \comp \pi^*_A \comp U
\ar[dl]_<<<<<<<<<<<<{\Pi_A \,\comp\, \pi^*_A \,\comp\, U \,\comp\, \eta^{\pi^*_A \,\dashv\, \Pi_A}\quad}
\ar[d]_>>>{\Pi_A \,\comp\, U \,\comp\, \pi^*_A \,\comp\, \eta^{\pi^*_A \,\dashv\, \Pi_A}}_<<<<<<<<{\dcomment{U \text{ is s. fib.}}\quad}^>>>>{\quad\dcomment{\pi^*_A \dashv \Pi_A}}
\\
\Pi_A \comp \pi^*_A \comp U \comp \Pi_A \comp \pi^*_A
\ar[r]_-{=} 
& 
\Pi_A \comp U \comp \pi^*_A \comp \Pi_A \comp \pi^*_A
\ar[d]_-{\Pi_A \,\comp\, U \,\comp\, \varepsilon^{\pi^*_A \,\dashv\, \Pi_A} \,\comp\, \pi^*_A}
\\
U \comp \Pi_A \comp \pi^*_A  
\ar[r]_-{\zeta_{\Pi,A} \,\comp\, \pi^*_A}
\ar[u]_-{\eta^{\pi^*_A \,\dashv\, \Pi_A} \,\comp\, U \,\comp\, \Pi_A \,\comp\, \pi^*_A}_<<{\,\,\,\,\,\,\quad\qquad\qquad\dcomment{\text{def. of } \zeta_{\Pi,A}}}
& 
\Pi_A \comp U \comp \pi^*_A
\ar@/_6pc/[uu]_-{=}
}
\]
The commutativity of the second square is proved analogously.
\end{proof}

\begin{proposition}
\label{prop:SigmaUnitCounitPreservedByF}
Given a split comprehension category with unit $p : \mathcal{V} \longrightarrow \mathcal{B}$ with weak split dependent sums, a split fibration $q : \mathcal{C} \longrightarrow \mathcal{B}$ with split dependent $p$-sums, and a split fibred adjunction $F \dashv\, U : q \longrightarrow p$, then the next two diagrams  commute.
\[
\xymatrix@C=3.75em@R=3em@M=0.5em{
F 
\ar[r]^-{\eta^{\Sigma_A \,\dashv\, \pi^*_A} \,\comp\, F}
\ar[d]_-{F \,\comp\, \eta^{\Sigma_A \,\dashv\, \pi^*_A}}
& 
\pi^*_A \comp \Sigma_A \comp F
\\
F \comp \pi^*_A \comp \Sigma_A
\ar[r]_-{=} 
& 
\pi^*_A \comp F \comp \Sigma_A
\ar[u]_-{\pi^*_A \,\comp\, \zeta_{\Sigma,A}}
}
\quad
\xymatrix@C=3.75em@R=3em@M=0.5em{
\Sigma_A \comp \pi^*_A \comp F
\ar[r]^-{\varepsilon^{\Sigma_A \,\dashv\, \pi^*_A} \,\comp\, F} \ar[d]_-{=}
&
F
\\
\Sigma_A \comp F \comp \pi^*_A
\ar[r]_-{\zeta^{-1}_{\Sigma,A} \,\comp\, \pi^*_A}
&
F  \comp \Sigma_A \comp \pi^*_A
\ar[u]_-{F \,\comp\, \varepsilon^{\Sigma_A \,\dashv\, \pi^*_A}}
}
\]
\end{proposition}

\begin{proof}
This proposition is proved analogously to Proposition~\ref{prop:PiUnitCounitPreservedByU}, also by straightforward diagram chasing, and by unfolding the definitions of $\zeta_{\Sigma,A}$ and $\zeta_{\Sigma,A}^{-1}$.
\end{proof}

\subsection{Empty type and coproduct type}
\label{sect:fibadjmodelscolimits}

As their syntax suggests, the empty type $0$ and the coproduct type $A + B$ are respectively most naturally modelled in terms of split fibred initial objects and split fibred binary coproducts in some split fibration $p : \mathcal{V} \longrightarrow \mathcal{B}$. However, it is important to observe that assuming such split fibred structure by itself does not suffice to model the dependently typed elimination forms for these types, i.e., the empty and binary case analysis.

For the coproduct type, an appropriate fibrational structure has been characterised by Jacobs in~\cite[Exercise~10.5.6]{Jacobs:Book}. Specifically, Jacobs requires a certain mediating functor, induced by the injections of the split fibred coproducts, to be \emph{fully-faithful}.

In this section we observe that the structure Jacobs suggested for modelling the dependently typed  elimination form for the coproduct type is in fact an instance of a more general phenomenon. Namely, we show that Jacobs's ideas apply to arbitrary split fibred colimits, including split fibred initial objects, enabling us to also model the empty type and the empty case analysis. In addition, we demonstrate that in fact one does not need to separately assume the existence of split fibred colimits of a given shape before imposing the fully-faithfulness condition---every split fibred cocone of a given shape for which the fully-faithfulness condition on the induced mediating functor holds turns out to be a split fibred colimit of that shape. We refer the reader to Section~\ref{sect:adjunctionsbackground} for the definitions of shapes, diagrams, cones, cocones, limits, and colimits.

We begin by defining a notion of strong colimits, based on the fully-faithfulness condition that Jacobs proposed for fibred coproducts in~\cite[Exercise~10.5.6]{Jacobs:Book}.

\begin{definition}
\label{def:strongcolimits}
\index{colimit!strong --}
\index{ in@$\mathsf{\ul{in}}^{J}$ (strong colimit of $J$)}
\index{ colim@$\mathsf{\ul{colim}}(J)$ (vertex of the strong colimit of $J$)}
\index{ J@$\widehat{J}$ ($\mathsf{Cat}$-valued diagram derived from $J$)}
Let us assume a small category $\mathcal{D}$ and a full split comprehension category with unit $p : \mathcal{V} \longrightarrow \mathcal{B}$.
Then, given an object $X$ in $\mathcal{B}$, we say that the fibre $\mathcal{V}_X$ has \emph{strong colimits of shape $\mathcal{D}$} if for every diagram $J : \mathcal{D} \longrightarrow \mathcal{V}_X$,  there exists a cocone $\mathsf{\ul{in}}^{J} : J \longrightarrow \Delta(\mathsf{\ul{colim}}(J))$ over $J$ such that the unique mediating functor $\langle \ia {\mathsf{\ul{in}}^J_D}^*_{D \in \mathcal{D}} \rangle : \mathcal{V}_{\ia {\mathsf{\ul{colim}}(J)}} \longrightarrow \mathsf{lim}(\widehat{J})$, induced by the universal property of the limit $\mathsf{pr}^{\widehat{J}} : \Delta(\mathsf{lim}(\widehat{J})) \longrightarrow \widehat{J}$, is fully-faithful. Here, the diagram $\widehat{J} : \mathcal{D}^{\text{op}} \longrightarrow \mathsf{Cat}$ is given by 
\[
\widehat{J}(D) \defeq \mathcal{V}_{\ia {J(D)}}
\qquad
\widehat{J}(g) \defeq \ia {J(g)}^*
\]
\end{definition}

More specifically, the functor $\langle \ia {\mathsf{\ul{in}}^J_D}^*_{D \in \mathcal{D}} \rangle$ arises as the unique mediating morphism in $\mathsf{Cat}$ for $\mathsf{lim}(\widehat{J})$ because the reindexing functors $\ia {\mathsf{\ul{in}}^J_D}^*$ form a cone over $\widehat{J}$. In particular, for all morphisms $g : D_i \longrightarrow D_j$ in $\mathcal{D}$, the outer triangle commutes in 
\[
\xymatrix@C=1.5em@R=2em@M=0.5em{
& & \mathcal{V}_{\ia {\mathsf{\ul{colim}}(J)}} \ar@/_2pc/[dddll]_-{\ia {\mathsf{\ul{in}}^{J}_{D_j}} ^*} \ar@/^2pc/[dddrr]^-{\ia {\mathsf{\ul{in}}^{J}_{D_i}} ^*} \ar@{-->}[dd]^{\langle \ia {\mathsf{\ul{in}}^J_D}^*_{D \in \mathcal{D}} \rangle}
\\
\\
& & \mathsf{lim}(\widehat{J}) \ar[dl]_{\mathsf{pr}^{\widehat{J}}_{D_j}} \ar[dr]^{\mathsf{pr}^{\widehat{J}}_{D_i}}
\\
\mathcal{V}_{\ia {J(D_j)}} \ar[r]_-{=} & \widehat{J}(D_j) \ar[rr]_{\ia {J(g)} ^*} & & \widehat{J}(D_i) \ar[r]_-{=} &  \mathcal{V}_{\ia {J(D_i)}}
}
\]
because we have the following sequence of equations: 
\[
\ia {\mathsf{\ul{in}}^{J}_{D_i}} ^* = \ia {\mathsf{\ul{in}}^{J}_{D_j} \comp\, J(g)} ^* = \ia {J(g)} ^* \comp \ia {\mathsf{\ul{in}}^{J}_{D_j}} ^*
\]
where the left-hand equation holds because $\mathsf{\ul{colim}}(J)$ is the vertex of the cocone $\mathsf{\ul{in}}^{J}$.

\begin{definition} 
\label{def:strongsplitfibredcolims}
\index{colimit!split fibred strong --}
\index{ in@$\mathsf{\ul{in}}^{J}$ (split fibred strong colimit of $J$)}
\index{ colim@$\mathsf{\ul{colim}}(J)$ (vertex of the split fibred strong colimit of $J$)}
A full split comprehension category with unit $p : \mathcal{V} \longrightarrow \mathcal{B}$ has \emph{split fibred strong colimits of shape $\mathcal{D}$} if every fibre of $p$ has strong colimits of shape $\mathcal{D}$ and this structure is preserved on-the-nose by reindexing, i.e., given any morphism $f : X \longrightarrow Y$ in $\mathcal{B}$ and any diagram $J : \mathcal{D} \longrightarrow \mathcal{V}_Y$, then we must have 
\[
f^*(\mathsf{\ul{colim}}(J)) = \mathsf{\ul{colim}}(f^* \comp\, J)
\qquad
f^*(\mathsf{\ul{in}}^{J}_D) =  \mathsf{\ul{in}}^{f^* \comp\, J}_D: f^*(J(D)) \longrightarrow \mathsf{\ul{colim}}(f^* \comp\, J)
\]
\end{definition}

It is instructive to see what the above characterisation means for modelling the empty type and the coproduct type in a full split comprehension category with unit $p$.

\index{object!split fibred strong initial --}
\index{ 0@$\mathbf{0}$ (empty category)}
\index{ 0@$0_X$ (initial object in $\mathcal{V}_X$)}
\index{ @$?_A$ (unique morphism in $\mathcal{V}_{\ia {0_X}}$ from $1_{\ia {0_X}}$ to $A$)}
To model the \emph{empty type}, we require $p$ to have split fibred strong colimits of shape $\mathbf{0}$, i.e., we require $p$ to have split fibred strong initial objects. 
\index{initial object!split fibred strong --}
Writing $0_X$ for the split fibred strong initial object in the fibre $\mathcal{V}_X$, its strength ensures that in the fibre $\mathcal{V}_{\ia {0_X}}$ there is exactly one morphism between any two objects. In particular, when later defining the interpretation of eMLTT, we make use of the fact that there is a unique vertical morphism from $1_{\ia {0_X}}$ to any other object $A$ in $\mathcal{V}_{\ia {0_X}}$, written ${?_A : 1_{\ia {0_X}} \longrightarrow A}$.

\index{ A@$A +_X B$ (binary coproduct of $A$ and $B$ in $\mathcal{V}_X$)}
\index{injection!left --}
\index{injection!right --}
\index{ inl@$\mathsf{inl}$ (left injection for binary coproducts)}
\index{ inr@$\mathsf{inr}$ (right injection for binary coproducts)}
To model the \emph{coproduct type}, we require $p$ to have split fibred strong colimits of shape $\mathbf{2}$, i.e., we require $p$ to have split fibred strong  coproducts. 
\index{coproduct!split fibred strong --}
Writing ${A_1 +_X A_2}$ for the split fibred strong binary coproduct of $A_1$ and $A_2$ in $\mathcal{V}_X$, its strength ensures that vertical morphisms of the form $B_1 \longrightarrow B_2$ in $\mathcal{V}_{\ia {A_1 +_X A_2}}$ are in one-to-one \linebreak correspondence with pairs of vertical morphisms $\ia {\mathsf{inl}} ^*(B_1) \longrightarrow \ia {\mathsf{inl}} ^*(B_2)$ and \linebreak $\ia {\mathsf{inr}} ^*(B_1) \longrightarrow \ia {\mathsf{inr}} ^*(B_2)$ in $\mathcal{V}_{\ia {A_1}}$ and $\mathcal{V}_{\ia {A_2}}$, respectively, where we write \linebreak $\mathsf{inl}$ for $\mathsf{\ul{in}}^J_0 : A_1 \longrightarrow A_1 +_X A_2$ and $\mathsf{inr}$ for $\mathsf{\ul{in}}^J_1 : A_2 \longrightarrow A_1 +_X A_2$, with $J(0) = A_1$ and \linebreak $J(1) = A_2$.
This one-to-one correspondence gives us a dependent case analysis principle for $A_1 +_X A_2$, arising as a special case of a corresponding dependent elimination principle for arbitrary split fibred strong colimits whose existence we show next.

\begin{proposition}
\label{prop:indexedelimcolimits}
\index{ f@$[f_D]_{D \in \mathcal{D}}$ (dependently typed elimination principle for split fibred strong colimits)}
Let us assume a full split comprehension category with unit \linebreak $p : \mathcal{V} \longrightarrow \mathcal{B}$ that has split fibred strong colimits of shape $\mathcal{D}$, a diagram of the form \linebreak $J : \mathcal{D} \longrightarrow \mathcal{V}_X$, and an object $A$ in $\mathcal{V}_{\ia {\mathsf{\ul{colim}}(J)}}$. Then, given a family of vertical morphisms $f_{D} : 1_{\ia {J(D)}} \longrightarrow \ia {\mathsf{\ul{in}}^{J}_D} ^*(A)$, for all objects $D$ in $\mathcal{D}$, such that for all morphisms \linebreak$g : D_i \longrightarrow D_{\!j}$ in $\mathcal{D}$ we have $\ia {J(g)} ^*(f_{D_{\!j}}) = f_{D_i}$, there exists a unique vertical morphism $[f_D]_{D \in \mathcal{D}} : 1_{\ia {\mathsf{\ul{colim}}(J)}} \longrightarrow A$ in $\mathcal{V}_{\ia {\mathsf{\ul{colim}}(J)}}$ satisfying the following ``$\beta$-equations":
\[
\ia {\mathsf{\ul{in}}^{J}_{D_i}}^*([f_D]_{D \in \mathcal{D}}) = f_{D_i}  : 1_{\ia {J(D_i)}} \longrightarrow \ia {\mathsf{\ul{in}}^{J}_{D_i}} ^*(A)
\]
for all objects $D_i$ in $\mathcal{D}$.
\end{proposition}

\begin{proof}
We postpone the lengthy details of this proof to Appendix~\ref{sect:proofofprop:indexedelimcolimits}, where  
much of the space is taken up by straightforward but laborious diagram chasing.
At a high level, the proof is based on using the universal property of the limit $\mathsf{pr}^{\widehat{J}} : \Delta(\mathsf{lim}(\widehat{J})) \longrightarrow \widehat{J}$ and the fully-faithfulness of the induced functor $\langle \ia {\mathsf{\ul{in}}^J_D}^*_{D \in \mathcal{D}} \rangle : \mathcal{V}_{\ia {\mathsf{\ul{colim}}(J)}} \longrightarrow \mathsf{lim}(\widehat{J})$. 
\end{proof}

\index{ f@$[f,g]$ (unique copairing of vertical morphisms)}
In particular, when we define the interpretation of eMLTT's coproduct type in Chapter~\ref{chap:interpretation}, we write $[f,g] : 1_{\ia {A_1 +_X A_2}} \longrightarrow B$ for the corresponding unique copairing of any two vertical morphisms  $f : 1_{\ia {A_1}} \longrightarrow \ia {\mathsf{inl}} ^*(B)$ and $g : 1_{\ia {A_2}} \longrightarrow \ia {\mathsf{inr}} ^*(B)$.

Finally, notice that we have suggestively written the cocone in Definitions~\ref{def:strongcolimits} and~\ref{def:strongsplitfibredcolims} as $\mathsf{\ul{in}}^{J} : J \longrightarrow \Delta(\mathsf{\ul{colim}}(J))$.
As the notation suggests, and as promised earlier, it turns out that the fully-faithfulness condition ensures 
that $\mathsf{\ul{in}}^{J}$ forms a colimit of $J$ in $\mathcal{V}_X$ in the standard sense.
In particular, the next proposition generalises an analogous result for strong fibred coproducts in $\mathsf{cod}_\mathcal{B}$, as given in~\cite[Exercise~10.5.6 (ii)]{Jacobs:Book}.

\begin{proposition}
\label{prop:fibredcolimits}
Let us assume a full split comprehension category with unit \linebreak $p : \mathcal{V} \longrightarrow \mathcal{B}$ that has split fibred strong colimits of shape $\mathcal{D}$. Then, given a diagram of the form $J : \mathcal{D} \longrightarrow \mathcal{V}_X$, the cocone $\mathsf{\ul{in}}^{J} : J \longrightarrow \Delta(\mathsf{\ul{colim}}(J))$, induced by the existence of split fibred strong colimits of shape $\mathcal{D}$, is a colimit of $J$ in $\mathcal{V}_X$ in the standard sense, i.e., the cocone $\mathsf{\ul{in}}^{J} : J \longrightarrow \Delta(\mathsf{\ul{colim}}(J))$ is initial amongst the cocones over $J$ in $\mathcal{V}_X$.
\end{proposition}

\begin{proof}
We prove this proposition by appropriately instantiating Proposition~\ref{prop:indexedelimcolimits}. 
In particular, given another cocone $\alpha : J \longrightarrow \Delta(A)$ in $\mathcal{V}_X$, we choose the object in $\mathcal{V}_{\ia {\mathsf{\ul{colim}}(J)}}$ to be $\pi^*_{\mathsf{\ul{colim}}(J)}(A)$ 
and derive each $f_D$ from the corresponding component $\alpha_D$ of the given cocone $\alpha$.
We postpone the details of this proof to Appendix~\ref{sect:proofofprop:fibredcolimits}.
\end{proof}

Further, observe that according to Definition~\ref{def:strongsplitfibredcolims}, if the given full split comprehension category with unit has split fibred strong colimits, the colimiting cocones in the fibres are preserved on-the-nose by reindexing. To add to this, we show below that the unique mediating morphisms are also preserved on-the-nose by reindexing.

\begin{proposition}
\index{ f@$f^*(\alpha)$ (component-wise reindexing of the cocone $\alpha$)}
Given a full split comprehension category with unit $p : \mathcal{V} \longrightarrow \mathcal{B}$ 
that has split fibred strong colimits of shape $\mathcal{D}$, a diagram $J : \mathcal{D} \longrightarrow \mathcal{V}_Y$, a cocone $\alpha : J \longrightarrow \Delta(A)$ in $\mathcal{V}_Y$, and a morphism $f : X \longrightarrow Y$ in $\mathcal{B}$, then we have 
\[
f^*([\alpha]) = [f^*(\alpha)]
\]
where $f^*(\alpha)$ is a cocone with $(f^*(\alpha))_D \defeq f^*(\alpha_D)$.
Analogously, the unique morphisms arising from Proposition~\ref{prop:indexedelimcolimits} are also preserved on-the-nose by reindexing, i.e., 
\[
\ia {\overline{f}(\mathsf{\ul{colim}}(J))}^*([f_D]_{D \in \mathcal{D}}) = [\ia {\overline{f}(J(D))}^*(f_D)]_{D \in \mathcal{D}}
\]
\end{proposition}

\begin{proof}
As $[\alpha]$ is a morphism of cocones from $\mathsf{\ul{in}}^J$ to $\alpha$, we know that
\[
[\alpha] \comp \mathsf{\ul{in}}^J_D = \alpha_D
\]
for all $D$ in $\mathcal{D}$. Next, using the functoriality of the reindexing functor $f^*$, we get that
\[
f^*([\alpha]) \comp f^*(\mathsf{\ul{in}}^J_D) = f^*(\alpha_D)
\]
Now, as we have assumed split fibred strong colimits of shape $\mathcal{D}$, we have that
\[
f^*(\mathsf{\ul{in}}^J_D) = \mathsf{\ul{in}}^{f^* \comp\, J}_D
\]
from which we get that
\[
f^*([\alpha]) \comp \mathsf{\ul{in}}^{f^* \comp\, J}_D = f^*(\alpha_D)
\]
meaning that $f^*([\alpha])$ is a morphism of cocones from $\mathsf{\ul{in}}^{f^* \comp\, J}$ to $f^*(\alpha)$. But as we know that $\mathsf{\ul{in}}^{f^* \comp\, J}$ is the colimit of $f^* \comp J$, there is exactly one such morphism of cocones, namely, $[f^*(\alpha)]$. Therefore, we have successfully shown that $f^*([\alpha]) = [f^*(\alpha)]$.

The proof that the unique morphisms $[f_D]_{D \in \mathcal{D}}$ arising from Proposition~\ref{prop:indexedelimcolimits} are also preserved on-the-nose by reindexing proceeds similarly: we show that the morphism $\ia {\overline{f}(\mathsf{\ul{colim}}(J))}^*([f_D]_{D \in \mathcal{D}})$ satisfies the same universal property as the unique morphism $[\ia {\overline{f}(J(D))}^*(f_D)]_{D \in \mathcal{D}}$, i.e., we show for all $D$ in $\mathcal{D}$ that we have 
\[
\ia {\mathsf{\ul{in}}^{f^* \comp\, J}_D}^*(\ia {\overline{f}(\mathsf{\ul{colim}}(J))}^*([f_D]_{D \in \mathcal{D}}))
=
\ia {\overline{f}(J(D))}^*(f_D)
\]
which follows by straightforward diagram chasing, 
based on $p$ being a split fibration, and using the equations given in Definition~\ref{def:strongsplitfibredcolims} and the definition of $f^*(\mathsf{\ul{in}}^J_D)$.
\end{proof}

\subsection{Natural numbers}

We recall that in the paper~\cite{Ahman:FibredEffects} on which this thesis is based on, the semantics of the type of natural numbers was given somewhat synthetically, by reading the semantic axiomatisation directly off the corresponding typing rules. 
Similar syntax-based axiomatisations appear for natural numbers also elsewhere in the literature, e.g., in~\cite{Atkey:DepTypes}.

It is worth noting that while such syntax-based axiomatisation provides the structure one needs to interpret the type of natural numbers and the corresponding dependently typed elimination form, it is not immediate how it relates to the existing work on fibrational models of the induction principle for natural numbers in predicate logic, which corresponds to the dependently typed elimination form via the Curry-Howard correspondence.
Specifically, in fibrational models of predicate logic, the induction principle for an inductive type is commonly modelled by giving an algebra for the lifting of the endofunctor whose least fixed point defines the inductive type in question, e.g., as studied by Hermida and Jacobs~\cite{Hermida:fibinduction}, and Ghani et al.~\cite{Ghani:FibredInduction}. 

In this section we propose a category-theoretically more natural characterisation of the structure we used in~\cite{Ahman:FibredEffects} for modelling the type of natural numbers, inspired by the above-mentioned fibrational treatment of the induction principle for natural numbers.

\begin{definition}
\label{def:strongsplitfibredweaknaturals}
\index{weak split fibred strong natural numbers}
\index{ zero@$\mathsf{zero}$ (zero morphism of weak split fibred strong natural numbers)}
\index{ succ@$\mathsf{succ}$ (successor morphism of weak split fibred strong natural numbers)}
\index{ N@$\mathbb{N}$ (weak split fibred strong natural numbers)}
\index{ rec@$\mathsf{rec}(f_z,f_s)$ (elimination principle for weak split fibred strong natural numbers)}
Given a full split comprehension category with unit $p : \mathcal{V} \longrightarrow \mathcal{B}$ such that $\mathcal{B}$ has a terminal object,
we say that $p$ has \emph{weak split fibred strong natural numbers} if there exists a distinguished object $\mathbb{N}$ in $\mathcal{V}_1$, together with a pair of vertical morphisms
\[
\xymatrix@C=7em@R=1em@M=0.5em{
1_1 \ar[r]^-{\mathsf{zero}} & \mathbb{N} & \mathbb{N} \ar[l]_-{\mathsf{succ}}
}
\]
such that for any object $X$ in $\mathcal{B}$ and any pair of morphisms
\[
\hspace{-0.5cm}
\xymatrix@C=7em@R=1em@M=0.5em{
1_{\ia {1_X}} \ar[r]^-{f_z} & A & A \ar[l]_-{f_s}
}
\]
in $\mathcal{V}$, with 
\[
p(A) = \ia {!_X^*(\mathbb{N})}
\qquad
p(f_z) = \ia {!_X^*(\mathsf{zero})}
\qquad
p(f_s) = \ia {!_X^*(\mathsf{succ})}
\]
there exists a (not necessarily unique) section $\mathsf{rec}(f_z,f_s)$ of 
$\pi_A : \ia {A} \longrightarrow \ia {!_X^*(\mathbb{N})}$, making the following two squares commute:
\[
\xymatrix@C=5em@R=5em@M=0.5em{
\ia {1_X} \ar[r]^-{\ia {!_X^*(\mathsf{zero})}} \ar[d]_-{\eta^{1 \,\dashv\, \ia -}_{\ia {1_X}}} & \ia {!_X^*(\mathbb{N})} \ar[d]_-{\mathsf{rec}(f_z,f_s)} & \ia {!_X^*(\mathbb{N})} \ar[l]_-{\ia {!_X^*(\mathsf{succ})}} \ar[d]^-{\mathsf{rec}(f_z,f_s)}
\\
\ia {1_{\ia {1_X}}} \ar[r]_-{\ia {f_z}} & \ia {A} & \ia {A} \ar[l]^-{\ia {f_s}}
}
\]
\end{definition}

\index{ NNO@NNO (natural numbers object)}
As a direct consequence of the above definition, we can show that every fibre of $p$ has a weak natural numbers object (NNO) that also supports a dependently typed elimination principle in the sense of the axiomatisation used in~\cite{Ahman:FibredEffects}, as shown next.

\begin{proposition}
\label{prop:fibredNNO}
Let us assume a full split comprehension category with unit \linebreak $p : \mathcal{V} \longrightarrow \mathcal{B}$ such that $\mathcal{B}$ has a terminal object and $p$ has weak split fibred strong 
natural numbers. Then, each fibre of $p$ has a weak NNO and this structure is preserved on-the-nose by reindexing.
\end{proposition}

\begin{proof}
Due to its length, we postpone the proof of Proposition~\ref{prop:fibredNNO} to Appendix~\ref{sect:proofofprop:fibredNNO}. 
Here, we only note that given an object $X$ in $\mathcal{B}$, a weak NNO in $\mathcal{V}_X$ can be given by
\[
\xymatrix@C=7em@R=6em@M=0.5em{
1_X \ar[r]^-{!_X^*(\mathsf{zero})} & !_X^*(\mathbb{N}) & !_X^*(\mathbb{N}) \ar[l]_-{!_X^*(\mathsf{succ})}
}
\]
\end{proof}

At this point, we would also like to report on a small oversight in~\cite{Ahman:FibredEffects}. Namely, 
the semantic ``$\beta$-equation" that corresponds to the application of the elimination form for natural numbers to the successor should have of course been given by
\[
\ia {!_X^*(\mathsf{succ})}^* (\mathsf{i}_A(f_z,f_s)) 
=
(\mathsf{s}(\mathsf{i}_A(f_z,f_s)))^*(f_s) 
\]
Taking this oversight into account, we show that the axiomatisations are equivalent.

\begin{proposition}
\label{prop:equivalenceofnaturalnumbersinthesisandpaper}
\index{ i@$\mathsf{i}_A(f_z,f_s)$ (elimination principle for weak split fibred strong natural numbers)}
Let us assume a full split comprehension category with unit \linebreak $p : \mathcal{V} \longrightarrow \mathcal{B}$ such that $\mathcal{B}$ has a terminal object. Then, $p$ having weak split fibred strong natural numbers is equivalent to $p$ supporting weak natural numbers as in~\cite{Ahman:FibredEffects}, i.e., for every object $X$ in $\mathcal{B}$, every object $A$ in $\mathcal{V}_{\ia {!_X^*(\mathbb{N})}}$, every morphism 
\[
f_z : 1_X \longrightarrow (\funsection(!_X^*(\mathsf{zero})))^*(A)
\]
in $\mathcal{V}_X$, and every morphism 
\[
f_s : 1_{\ia A} \longrightarrow \pi_A^*(\ia {!_X^*(\mathsf{succ})}^* (A))
\]
in $\mathcal{V}_{\ia A}$, there exists a morphism 
\[
\mathsf{i}_A(f_z,f_s) : 1_{\ia {!^*_X(\mathbb{N})}} \longrightarrow A
\]
in $\mathcal{V}_{\ia {!^*_X(\mathbb{N})}}$ such that 
\[
\begin{array}{c}
(\funsection(!_X^*(\mathsf{zero})))^*(\mathsf{i}_A(f_z,f_s)) = f_z
\\[3mm]
\ia {!_X^*(\mathsf{succ})}^* (\mathsf{i}_A(f_z,f_s)) 
=
(\mathsf{s}(\mathsf{i}_A(f_z,f_s)))^*(f_s) 
\end{array}
\vspace{0.2cm}
\]
\end{proposition}

\begin{proof}
We postpone the straightforward but somewhat lengthy details of this proof to Appendix~\ref{sect:proofofprop:equivalenceofnaturalnumbersinthesisandpaper}, where much of the space is taken up by laborious diagram chasing.
\end{proof}

\subsection{Propositional equality}

We recall that for dependently typed languages that support extensional propositional equality, i.e., languages that include an $\eta$-equation for propositional equality, the required category-theoretical structure is most naturally characterised by requiring all contraction functors (defined later in this section) to have well-behaved left adjoints, e.g., as discussed in~\cite[Section~10.5]{Jacobs:Book}.
While we could use this adjunction-based characterisation of models of extensional propositional equality to prove the soundness of the interpretation of eMLTT, we would not be able to prove the completeness of the interpretation in Section~\ref{sect:completeness} because eMLTT's propositional equality is intensional.

Therefore, in order to be able to later prove the completeness of the interpretation of eMLTT, we characterise the structure needed to model its intensional propositional equality similarly axiomatically as in the paper~\cite{Ahman:FibredEffects} on which this thesis is based on.

\begin{definition}
\label{def:diagonalmorphisminbase}
\index{morphism!diagonal --}
\index{ d@$\delta_A$ (diagonal morphism)}
Given a split comprehension category with unit $p : \mathcal{V} \longrightarrow \mathcal{B}$ and an object $A$ in $\mathcal{V}$, the unique mediating morphism $\delta_A : \ia {A} \longrightarrow \ia {\pi_A^*(A)}$ induced by the pullback situation below is called a \emph{diagonal morphism}.
\[
\xymatrix@C=4em@R=4em@M=0.5em{
\ia A \ar@/_2pc/[dr]_{\id_{\ia A}} \ar@/^3pc/[rr]^{\id_{\ia A}} \ar@{-->}[r]^-{\delta_A} & \ia {\pi^*_A(A)} \ar[d]_{\pi_{\pi^*_A(A)}}^<{\,\big\lrcorner} \ar[r]^-{\ia {\overline{\pi_A}(A)}} & \ia A \ar[d]^{\pi_A}_{\dcomment{\mathcal{P}(\overline{\pi_A}(A))}\quad\,\,\,\,\,\,\,\,\,}
\\
& \ia A \ar[r]_-{\pi_A} & p(A)
}
\]
\end{definition}

\begin{definition}
\index{functor!contraction --}
\index{ d@$\delta^*_A$ (contraction functor)}
Given a split comprehension category with unit $p : \mathcal{V} \longrightarrow \mathcal{B}$ and an object $A$ in $\mathcal{V}$, the functor $\delta^*_A : \mathcal{V}_{\ia {\pi^*_A(A)}} \longrightarrow \mathcal{V}_{\ia{A}}$ is called a \emph{contraction functor}.
\end{definition}

\begin{definition}
\label{def:strongpropequality}
\index{split intensional propositional equality}
\index{ Id@$\Id_A$ (split intensional propositional equality)}
\index{ r@$\mathsf{r}_A$ (reflexivity of split intensional propositional equality)}
\index{ i@$\mathsf{i}_{A,B}(f)$ (elimination principle for split intensional propositional equality)}
Given a split comprehension category with unit $p : \mathcal{V} \longrightarrow \mathcal{B}$, we say that $p$ supports \emph{split intensional propositional equality} if for every $A$ in $\mathcal{V}$, there exists an object $\Id_A$ in $\mathcal{V}_{\ia {\pi^*_A(A)}}$ and a morphism $\mathsf{r}_A : 1_{\ia A} \longrightarrow \delta^*_A(\Id_A)$ in $\mathcal{V}_{\ia {A}}$, such that for every object $B$ in $\mathcal{V}_{\ia {\Id_A}}$ and morphism $f : 1_{\ia A} \longrightarrow (\mathsf{s}(\mathsf{r}_A))^*(\ia {\overline{\delta_A}(\Id_A)} ^* (B))$ in $\mathcal{V}_{\ia {A}}$, there exists a morphism $\mathsf{i}_{A,B}(f) : 1_{\ia {\Id_A}} \longrightarrow B$ in $\mathcal{V}_{\ia {\Id_A}}$, satisfying 
\[
(\mathsf{s}(\mathsf{r}_A))^*(\ia {\overline{\delta_A}(\Id_A)} ^* (\mathsf{i}_{A,B}(f))) = f
\]
such that for any Cartesian morphism $g : A \longrightarrow A'$ in $\mathcal{V}$, the following equations hold:
\[
\begin{array}{c}
{\ia {g'}}^*(\Id_{A'}) = \Id_A
\\[3mm]
\ia g^*(\mathsf{r}_{A'}) = \mathsf{r}_{A}
\\[3mm]
\ia {\overline{\ia {g'}}(\Id_{A'})}^*(\mathsf{i}_{A',B}(f)) = \mathsf{i}_{A,\ia {\overline{\ia {g'}}(\Id_{A'})}^*(B)}(\ia g^*(f))
\end{array}
\]
Here, the morphism $g' : \pi^*_A(A) \longrightarrow \pi^*_{A'}(A')$ in $\mathcal{V}$ is induced by the universal property of the Cartesian morphism $\overline{\pi_{A'}}(A') : \pi^*_{A'}(A') \longrightarrow A'$, as illustrated in the next diagram.
\[
\xymatrix@C=3em@R=3em@M=0.5em{
& A \ar[dr]^-{g}
\\
\pi^*_A(A) \ar@{-->}[r]_-{g'} \ar[ur]^-{\overline{\pi_A}(A)} & \pi^*_{A'}(A') \ar[r]_-{\overline{\pi_{A'}}(A')} & A'
\\
\ia{A} \ar[r]^-{\ia g} \ar[dr]_-{\pi_A} \ar@{}[d]^<<<<<<{\qquad\qquad\quad\,\,\,\,\,\dcomment{\mathcal{P}(g)}} & \ia {A'} \ar[r]^-{\pi_{A'}} & p(A')
\\
& p(A) \ar[ur]_-{p(g)}
}
\]
\end{definition}

It is worth noting that the second equation $\ia g^*(\mathsf{r}_{A'}) = \mathsf{r}_{A}$ is well-formed because the morphisms $\ia {g'} \comp \delta_A : \ia {A} \longrightarrow \ia {\pi^*_{A'}(A')}$ and $\delta_{A'} \comp \ia g : \ia {A} \longrightarrow \ia {\pi^*_{A'}(A')}$ are equal. In particular, these morphisms satisfy the same universal property as the unnamed unique mediating morphism in the following pullback situation:
\[
\xymatrix@C=4em@R=4em@M=0.5em{
\ia A \ar@/_2pc/[dr]_{\ia g} \ar@/^3pc/[rr]^{\ia g} \ar@{-->}[r]^-{} & \ia {\pi^*_{A'}(A')} \ar[d]_{\pi_{\pi^*_{A'}(A')}}^<{\,\big\lrcorner} \ar[r]^-{\ia {\overline{\pi_{A'}}(A')}} & \ia {A'} \ar[d]^{\pi_{A'}}_{\dcomment{\mathcal{P}(\overline{\pi_{A'}}(A'))}\quad\,\,\,\,\,\,\,\,\,}
\\
& \ia {A'} \ar[r]_-{\pi_{A'}} & p(A')
}
\]

Finally, the third equation $\ia {\overline{\ia {g'}}(\Id_{A'})}^*(\mathsf{i}_{A',B}(f)) = \mathsf{i}_{A,\ia {\overline{\ia {g'}}(\Id_{A'})}^*(B)}(\ia g^*(f))$ is well-formed because the following diagram commutes:
\[
\xymatrix@C=5em@R=5em@M=0.5em{
\ia {A}
\ar[rr]^-{\ia {g}}
\ar@/_2pc/[d]_>>>>>{\eta^{1 \,\dashv\, \ia -}_{\ia A}\!\!}
\ar@/_5pc/[dd]_<<<<<<{\mathsf{s}(\mathsf{r}_A)\!\!}
&
&
\ia {A'}
\ar@/^2pc/[d]^>>>>>{\!\!\eta^{1 \,\dashv\, \ia -}_{\ia {A'}}}_-{\dcomment{\mathcal{P}(\overline{\ia g}(1_{\ia {A'}}))}\qquad\qquad\qquad\qquad\qquad\quad\,\,\,}_-{\dcomment{\text{iso.}}\quad}
\ar@/^5pc/[dd]^<<<<<<{\!\!\mathsf{s}(\mathsf{r}_{A'})}
\\
\ia {1_{\ia A}}
\ar[r]^-{=}
\ar@/_2pc/[u]_-{\pi_{1_{\ia A}}}^-{\dcomment{\text{iso.}}\quad}
\ar[d]_-{\ia {\mathsf{r}_A}}_<<<<{\dcomment{\text{def.}}\quad}^-{\,\,\,\qquad\dcomment{\text{property of } \mathsf{r_A}}}
&
\ia {\ia {g}^*(1_{\ia {A'}})}
\ar[r]^-{\ia {\overline{\ia g}(1_{\ia {A'}})}}
\ar[d]_-{\ia {\ia g^*(\mathsf{r}_{A'})}}^-{\qquad\quad\dcomment{\text{def. of } \ia g^*(\mathsf{r}_{A'})}}
&
\ia {1_{\ia {A'}}}
\ar@/^2pc/[u]^-{\pi_{1_{\ia {A'}}}}
\ar[d]^-{\ia {\mathsf{r}_{A'}}}^<<<<{\quad\dcomment{\text{def.}}}
\\
\ia {\delta_A^*(\Id_A)}
\ar[r]_-{=}
\ar[d]_{\ia {\overline{\delta_A}(\Id_A)}}^-{\qquad\qquad\dcomment{\text{property of } \Id_A}}^-{\,\,\,\quad\qquad\qquad\qquad\qquad\qquad\qquad\dcomment{p \text{ is a split fibration}}}
&
\ia {\ia g^*(\delta_{A'}^*(\Id_{A'}))}
\ar[r]_-{\ia {\overline{\ia g}(\delta_{A'}^*(\Id_{A'}))}}
&
\ia {\delta_{A'}^*(\Id_{A'})}
\ar[d]^-{\ia {\overline{\delta_{A'}}(\Id_{A'})}}
\\
\ia {\Id_A}
\ar[r]_-{=}
&
\ia {\ia {g'}^*(\Id_{A'})}
\ar[r]_-{\ia {\overline{\ia {g'}}(\Id_{A'})}}
&
\ia {\Id_{A'}}
}
\]

\subsection{Homomorphic function type}
\label{sect:shallowenrichment}

Analogously to EEC in the simply typed setting, the syntax of eMLTT suggests that the homomorphic function type $\ul{C} \multimap \ul{D}$ ought to be modelled in terms of enrichment. In particular, we seem to need a fibre-wise enrichment of the fibrations we use for modelling computation types in the fibrations we use for modelling value types, such that the enriched structure is preserved by reindexing in some appropriate sense. 

While informally such fibre-wise enrichment might seem straightforward, then formally the situation turns out to be much more involved. In particular, even if all the fibres are enriched, the total category of the fibration we use for modelling computation types also includes non-vertical morphisms, which would need to be compatible with the vertical morphisms, now modelled using enrichment and hom-objects. 
Correspondingly, the fibration would need to be given by a functor that is enriched when restricted to fibres.
But for this to be even possible, the base category would also need to be an enriched category, with its enrichment compatible with that of the fibres. 

As a result, the situation where some parts of the fibration are enriched and others are not seems overly complicated and somewhat ad-hoc, particularly, when compared to the arguably very natural and elegant models of eMLTT without the homomorphic function type, as studied in~\cite{Ahman:FibredEffects}. Ideally, one would like the models of eMLTT with the homomorphic function type to be only a small variation of the models given in op. cit. 

As asking for fibre-wise enrichment does not seem to give a satisfactory semantic structure for modelling the homomorphic function type, one could wonder why not use the existing work on enriched fibrations, such as~\cite{Shulman:EnrichedIndexedCategories} and~\cite[Section~8.1]{Vasilakopoulou:Thesis}? Unfortunately, while this existing work gives two systematic approaches to combining enrichment and fibrations, neither fits well into the setting we are working in. 

On the one hand, compared to the models of eMLTT without the homomorphic function type from~\cite{Ahman:FibredEffects}, trying to adapt~\cite{Shulman:EnrichedIndexedCategories} to our setting would lead us to having to require the base categories of the fibrations we use to additionally have finite products. While this would not be a significant problem in itself, the notion of enriched fibration one gets by applying the Grothendieck construction (see~\cite[Definition~1.10.1]{Jacobs:Book}) to the enriched indexed categories developed in~\cite{Shulman:EnrichedIndexedCategories} would be significantly more involved compared to the ordinary (unenriched) fibrations that are used to model eMLTT in~\cite{Ahman:FibredEffects}. 

On the other hand, trying to adapt~\cite[Section~8.1]{Vasilakopoulou:Thesis} to our setting would involve imposing even more substantial conditions on the base categories of the fibrations we use. In particular, we would need to require the  base categories of the fibrations we work with to be self-enriched. Again, imposing such condition would be a significant change from the kinds of fibrations we used to model eMLTT in~\cite{Ahman:FibredEffects}.

Having discussed some approaches that do not work, we now explain one that does work and that we use for modelling the homomorphic function type in Chapter~\ref{chap:interpretation}. While we still follow the intuition that the fibrations we use for modelling computation types should be fibre-wise enriched in the fibrations we use for modelling value types, the enrichment-like structure we use is sufficiently relaxed to make the compatibility issues between vertical and non-vertical morphisms disappear. In particular, we continue to use (unenriched) fibrations to model computation types, but additionally require the existence of fibre-wise ``hom-objects", given by functors of the form $\mathcal{C}^{\text{op}}_X \times \mathcal{C}_X \longrightarrow \mathcal{V}_X$, satisfying certain compatibility conditions, as made precise below.

\index{ @$\int$ (Grothendieck construction)}
Before we define the relaxed notion of enrichment suitable for modelling eMLTT's homomorphic function type, we first note that 
given any split fibration $q : \mathcal{C} \longrightarrow \mathcal{B}$, we can construct a new split fibration
\vspace{-0.25cm}
\[
\begin{array}{c}
r : \bigintsss (X \mapsto \mathcal{C}^{\text{op}}_X \times \mathcal{C}_X) \longrightarrow \mathcal{B}
\end{array}
\]
by applying the
\index{Grothendieck construction}
\index{category!split $\mathcal{B}$-indexed --}
\!\emph{Grothendieck construction} to the split $\mathcal{B}$-indexed category\footnote{A \emph{split $\mathcal{B}$-indexed category} is given by a functor $\mathcal{B}^{\text{op}} \longrightarrow \mathsf{Cat}$, see~\cite[Definition~1.4.4 (ii)]{Jacobs:Book}.} given by 
\[
X \mapsto \mathcal{C}^{\text{op}}_X \times \mathcal{C}_X
\qquad
f \mapsto (f^*)^{\text{op}} \times f^*
\]

Concretely, the objects of the total category $\bigintsss (X \mapsto \mathcal{C}^{\text{op}}_X \times \mathcal{C}_X)$ are triples $(X , \ul{C}, \ul{D})$, where $X$ is an object of $\mathcal{B}$, and $\ul{C}$ and $\ul{D}$ are objects of $\mathcal{C}_X$. A morphism from $(X , \ul{C}_1, \ul{D}_1)$ to $(Y , \ul{C}_2, \ul{D}_2)$ is given by a triple $(f,h,k)$, where $f : X \longrightarrow Y$ is a morphism in $\mathcal{B}$, and $h : f^*(\ul{C}_2) \!\longrightarrow\! \ul{C}_1$ and $k : \ul{D}_1 \longrightarrow f^*(\ul{D}_2)$ are morphisms in $\mathcal{C}_{X}$.  $r$ is then given by 
\[
r(X , \ul{C}, \ul{D}) \defeq X \qquad r(f,h,k) \defeq f
\] 
Finally, we note that $r$ is split and the chosen Cartesian morphisms are of the form 
\[
(f , \id_{f^*(\ul{C})} , \id_{f^*(\ul{D})}) : (X , f^*(\ul{C}),f^*(\ul{D})) \longrightarrow (Y,\ul{C},\ul{D})
\]

It is informative to observe that while the above definition of $r$ is convenient for us to work with, it can also be characterised in more abstract terms. Namely, the split $\mathcal{B}$-indexed category given by $X \mapsto \mathcal{C}^{\text{op}}_X \times \mathcal{C}_X$ is the Cartesian product (in the $2$-category of split $\mathcal{B}$-indexed categories) of the split $\mathcal{B}$-indexed categories given by $X \mapsto \mathcal{C}^{\text{op}}_X$ and $X \mapsto \mathcal{C}_X$, of which the former is the opposite  of the latter, e.g., as discussed in \cite[Definition~1.10.10]{Jacobs:Book}. As a result, based on the fact that the Grothendieck construction forms an equivalence of categories between split $\mathcal{B}$-indexed categories and split fibrations with base category $\mathcal{B}$ (see \cite[Proposition~1.10.9]{Jacobs:Book}), and that it takes the split $\mathcal{B}$-indexed category given by $X \mapsto \mathcal{C}^{\text{op}}_X$ to the opposite $q^{\text{op}}$ of the split fibration $q$ (see \cite[Exercise~1.10.9]{Jacobs:Book}), the split fibration $r$ can equivalently be characterised as the Cartesian product $q^{\text{op}} \times q$ of the split fibrations $q^{\text{op}}$ and $q$, in the $2$-category $\mathsf{Fib}_{\mathsf{split}}(\mathcal{B})$.

\index{ q@$q^{\text{op}}$ (opposite of the split fibration $q$)}

We now define the relaxed notion of enrichment suitable for modelling eMLTT.

\begin{definition}
\label{def:shallowfibredenrichment}
\index{split fibred pre-enrichment}
\index{ @$\multimap$ (split fibred functor witnessing split fibred pre-enrichment)}
\index{ xi@$\xi_{X,\ul{C},\ul{D}}$ (isomorphism witnessing split fibred pre-enrichment)}
\index{ C@$\ul{C} \multimap_X \ul{D}$ (shorthand for $\multimap (X,\ul{C},\ul{D})$)}
Given two split fibrations $p : \mathcal{V} \longrightarrow \mathcal{B}$ and $q : \mathcal{C} \longrightarrow \mathcal{B}$ such that $p$ has split fibred terminal objects, we say that $q$ admits \emph{split fibred pre-enrichment} in $p$ if there exists a split fibred functor $\multimap\,\, : r \longrightarrow p$, as depicted in
\[
\xymatrix@C=1.7em@R=3em@M=0.5em{
\bigintsss (X \mapsto \mathcal{C}^{\text{op}}_X \times \mathcal{C}_X) \ar[rrr]^-{\multimap} \ar[dr]_-{r} &&& \mathcal{V} \ar[dll]^-{p}
\\
& \mathcal{B} &
}
\]
together with a family of isomorphisms (where we write $\ul{C} \multimap_X \ul{D}$ for $\multimap (X,\ul{C},\ul{D})$)
\[
\xi_{X,\ul{C},\ul{D}} : \mathcal{V}_X(1_X , \ul{C} \multimap_X \ul{D}) \overset{\cong}{\longrightarrow} \mathcal{C}_X(\ul{C},\ul{D})
\]
that are natural in both $\ul{C}$ and $\ul{D}$, and preserved on-the-nose by reindexing, as respectively illustrated by the commutativity of the two squares in the following diagram:
\[
\xymatrix@C=9em@R=2.25em@M=0.5em{
\mathcal{V}_X(1_X , \ul{C}_1 \multimap_X \ul{D}_1) \ar[r]^-{\xi_{X,\ul{C}_1,\ul{D}_1}} \ar[d]_-{\mathcal{V}_X(1_X , h \,\multimap_{\id_X}\, k)}
&
\mathcal{C}_X(\ul{C}_1, \ul{D}_1) \ar[d]^-{\mathcal{C}_X(h, k)}
\\
\mathcal{V}_X(1_X , f^*(\ul{C}_2) \multimap_X f^*(\ul{D}_2)) \ar[r]^-{\xi_{X,f^*(\ul{C}_2),f^*(\ul{D}_2)}}
&
\mathcal{C}_X(f^*(\ul{C}_2), f^*(\ul{D}_2))
\\
\mathcal{V}_X(f^*(1_Y) , f^*(\ul{C}_2 \multimap_Y \ul{D}_2)) \ar[u]^-{=}
\\
\mathcal{V}_Y(1_Y , \ul{C}_2 \multimap_Y \ul{D}_2) \ar[u]^-{f^*} \ar[r]_-{\xi_{Y,\ul{C}_2,\ul{D}_2}}
&
\mathcal{C}_Y(\ul{C}_2, \ul{D}_2) \ar[uu]_-{f^*}
}
\]
for every morphism
$
(f,h,k) : (X,\ul{C}_1,\ul{D}_1) \longrightarrow (Y,\ul{C}_2,\ul{D}_2)
$
in $\bigintsss  (X \mapsto \mathcal{C}^{\text{op}}_X \times \mathcal{C}_X)$.
\end{definition}

\index{ C@$\ul{C} \multimap \ul{D}$ (shorthand for $\ul{C} \multimap_X \ul{D}$)}
To improve the readability of our proofs, we sometimes omit the subscript on the functor $\multimap$ when it is clear from the context, i.e., we write $\ul{C} \multimap \ul{D}$ for $\ul{C} \multimap_X \ul{D}$. 

\section{Fibred adjunction models}
\label{sect:fibadjmodels}

In this short section we combine the category-theoretic structures we discussed in Section~\ref{sect:fibadjmodelsstructure}
into a class of categorical models suitable for interpreting eMLTT, called \emph{fibred adjunction models}. We use the same name for this class of models as we did in~\cite{Ahman:FibredEffects} for a more restricted class of models because the core of the models remains the same.

\begin{definition}
\label{def:fibadjmodels}
\index{fibred adjunction model}
A \emph{fibred adjunction model} is given by 
\begin{itemize}
\item a split closed comprehension category $p : \mathcal{V} \longrightarrow \mathcal{B}$, 
\item a split fibration $q : \mathcal{C} \longrightarrow \mathcal{B}$, and
\item a split fibred adjunction $F \dashv\, U : q \longrightarrow p$
\end{itemize}
such that
\begin{itemize}
\item $q$ has split dependent $p$-products (as in Definition~\ref{def:splitdependentcompproducts}),
\item $q$ has split dependent $p$-sums (as in Definition~\ref{def:splitdependentcompsums}), 
\item $p$ has split fibred strong colimits of shape $\mathbf{0}$ and $\mathbf{2}$ (as in Definition~\ref{def:strongsplitfibredcolims}), 
\item $p$ has weak split fibred strong natural numbers (as in Definition~\ref{def:strongsplitfibredweaknaturals}), 
\item $p$ has split intensional propositional equality (as in Definition~\ref{def:strongpropequality}), and
\item $q$ admits split fibred pre-enrichment in $p$ (as in Definition~\ref{def:shallowfibredenrichment}),
\end{itemize}
as depicted in
\vspace{-2cm}
\[
\xymatrix@C=4em@R=5em@M=0.5em{
\ar@{}[dd]^-{\!\!\quad\qquad\qquad\perp}
\\
\mathcal{V} \ar@/_1.75pc/[d]_-{p} \ar@{}[d]_-{\dashv\,\,\,\,\,} \ar@{}[d]^-{\,\,\,\,\,\,\,\dashv} \ar@/^1.75pc/[d]^-{\ia {-}} \ar@/^1.25pc/[rr]^-{F} &  &  \mathcal{C} \ar@/^1.25pc/[ll]^-{U} \ar@/^1pc/[dll]^-{q}
\\
\mathcal{B} \ar[u]_-{\!1}
}
\vspace{0.25cm}
\]
\end{definition}

\noindent
In the rest of this thesis, we assume that whenever we work with fibred adjunction models, their structure is given using the notation used in Definition~\ref{def:fibadjmodels} above, e.g., we use $p$ for the split closed comprehension category, $F \dashv\, U$ for the adjunction, etc.

\section{Examples of fibred adjunction models}
\label{sect:examplesoffibadjmodels}

We now discuss some examples of fibred adjunction models.

\subsection{Identity adjunctions}

Given an \SCCompC\, $p : \mathcal{V} \longrightarrow \mathcal{B}$ with split fibred strong colimits of shape $\mathbf{0}$ and $\mathbf{2}$, weak split fibred strong natural numbers, and split intensional propositional equality, we can always pick the \emph{identity adjunction} $\id_{\mathcal{V}} \dashv \id_{\mathcal{V}} : \mathcal{V} \longrightarrow \mathcal{V}$ 
\index{adjunction!identity --}
to get an ``effect-free" fibred adjunction model, by letting $q \defeq p$ and  observing that $\id_{\mathcal{V}}$ is trivially split fibred. Further, observe that the split dependent $p$-products and split dependent $p$-sums are given in $q$ by the corresponding structure in $p$. Finally, we can define the split fibred pre-enrichment of $q$ in $p$ using the fact that $p$ is a split fibred CCC, i.e., we let $A \multimap_X B \defeq A \Rightarrow_X B$. We summarise this discussion in the next theorem.

\begin{theorem}
\label{thm:effectfreefibadjmodel}
\index{fibred adjunction model!-- built from identity adjunction}
Given an \SCCompC\, $p : \mathcal{V} \longrightarrow \mathcal{B}$ with split fibred strong colimits of shape $\mathbf{0}$ and $\mathbf{2}$, weak split fibred strong natural numbers, and split intensional propositional equality, the identity adjunction $\id_{\mathcal{V}} \dashv\, \id_{\mathcal{V}} : \mathcal{V} \longrightarrow \mathcal{V}$ gives rise to an ``effect-free" fibred adjunction model.
\end{theorem}

\subsection[Simple fibrations and models of EEC\raisebox{0.75pt}{+}]{Simple fibrations and models of EEC\raisebox{1.75pt}{+}}
\label{sect:fibadjmodelsfromeecmodels}

\index{ EEC@EEC\raisebox{0.75pt}{+} (extension of EEC with finite products)}
Our second example of fibred adjunction models is based on the models of EEC\raisebox{0.75pt}{+}, where EEC\raisebox{0.75pt}{+} stands for an extension of EEC with finite coproducts, see~\cite[Definition~6.6]{Egger:EnrichedEffectCalculus}. The resulting fibred adjunction models are a restricted form of models defined in Definition~\ref{def:fibadjmodels} in that they do not support  propositional equality.

\begin{definition}
\index{model of EEC\raisebox{0.75pt}{+}}
A \emph{model of EEC\raisebox{0.75pt}{+} with weak natural numbers} is given by a $\mathcal{V}$-enriched adjunction $F \dashv\, U : \mathcal{C} \longrightarrow \mathcal{V}$, where $\mathcal{V}$ is a CCC that also has finite coproducts and a weak NNO, and where $\mathcal{C}$ 
is $\mathcal{V}$-enriched, having all $\mathcal{V}$-tensors and $\mathcal{V}$-cotensors.
\end{definition}

In this example we use $X,Y,A,B,\ldots$ and $f,g,\ldots$ to range over the objects and morphisms of $\mathcal{V}$, and $\ul{C}, \ul{D}, \ldots$ and $h, k, \ldots$ to range over the objects and morphisms of $\mathcal{C}$, respectively. We denote the Cartesian closed structure of $\mathcal{V}$ by $A \times B$ and $A \Rightarrow B$, 
the $A$-fold $\mathcal{V}$-tensors of $\mathcal{C}$ by $A \otimes\, \ul{C}$, and the $A$-fold $\mathcal{V}$-cotensors of $\mathcal{C}$ by $A \Rightarrow \ul{C}$. 
\index{ A@$A \otimes\, \ul{C}$ ($A$-fold $\mathcal{V}$-tensor)}
\index{ A@$A \Rightarrow \ul{C}$ ($A$-fold $\mathcal{V}$-cotensor)}

\index{a@$A$-fold $\mathcal{V}$-tensor}
\index{a@$A$-fold $\mathcal{V}$-cotensor}
We recall from~\cite{Kelly:EnrichedCats} that the universal properties of the $A$-fold \emph{$\mathcal{V}$-tensors} and \emph{$\mathcal{V}$-cotensors} of $\mathcal{C}$ are characterised as the following two $\mathcal{V}$-isomorphisms:
\[
\mathcal{C}(A \otimes\, \ul{C},\ul{D}) \cong A \Rightarrow \mathcal{C}(\ul{C},\ul{D})
\qquad
\mathcal{C}(\ul{C},A \Rightarrow \ul{D}) \cong A \Rightarrow \mathcal{C}(\ul{C},\ul{D})
\]

In order to improve the readability of this example, and to simplify the associated proofs, we present this fibred adjunction model using the internal language of the models of EEC\raisebox{0.75pt}{+}, namely, a variant\footnote{Compared to the syntax used to present EEC\raisebox{0.75pt}{+} in~\cite{Egger:EnrichedEffectCalculus}, we use eMLTT's syntax for its elimination forms. Furthermore, we write $F$ for the EEC type former $!$ and make the type former $U$ explicit.} of the syntax of EEC\raisebox{0.75pt}{+}.
This syntactic presentation is justified by the soundness and completeness results proved in~\cite[Theorem~7.1]{Egger:EnrichedEffectCalculus}. 
Specifically, we represent morphisms $f : X \longrightarrow Y$ of $\mathcal{V}$ as EEC\raisebox{0.75pt}{+}'s non-linear terms $\zj {x \!:\! X} {f(x)} Y$, and morphisms $h : \ul{C} \longrightarrow \ul{D}$ of $\mathcal{C}$ as EEC\raisebox{0.75pt}{+}'s linear terms $\zj {z \!:\! \ul{C}} {h(z)} {\ul{D}}$.

We proceed by defining the \SCCompC\, part of this example of fibred adjunction models, based on the simple fibration construction discussed in Example~\ref{ex:simplefibration}. In particular, we let $p \defeq \mathsf{s}_{\mathcal{V}}$, which gives us an \SCCompC\, because $\mathsf{s}_{\mathcal{V}}$ can be easily seen to be split, and because we have the following result regarding closed comprehension categories (this closed comprehension category structure is also easily seen to be split).

\begin{proposition}[{\cite[Theorem~10.5.5 (i)]{Jacobs:Book}}]
The simple fibration $\mathsf{s}_{\mathcal{V}} : \mathsf{s}(\!\mathcal{V}) \longrightarrow \mathcal{V}$ is a closed comprehension category if and only if $\mathcal{V}$ is a CCC.
\end{proposition}

In particular, the corresponding terminal object functor $1 : \mathcal{V} \longrightarrow \mathsf{s}(\!\mathcal{V})$ and comprehension functor $\ia - : \mathsf{s}(\!\mathcal{V}) \longrightarrow \mathcal{V}$ are given by
\[
1(X) \defeq (X,1)
\qquad
\ia {(X,A)} \defeq X \times\, A
\]

The split dependent products and strong split dependent sums are given by 
\[
\Pi_{(X,A)}(X \times A,B) \defeq (X, A \Rightarrow B)
\qquad
\Sigma_{(X,A)}(X \times A, B) \defeq (X, A \times B)
\]
with the strength of the latter witnessed by isomorphisms $(X \times A) \times B \,\cong\, X \times (A \times B)$. 

Next, we note that $\mathsf{s}_{\mathcal{V}}$ also has other structure we require from $p$ in Definition~\ref{def:fibadjmodels}, except for split intensional propositional equality, as mentioned earlier.

\begin{proposition}
$\mathsf{s}_{\mathcal{V}}$ has split fibred strong colimits of shape $\mathbf{0}$ and $\mathbf{2}$, and weak split fibred strong natural numbers.
\end{proposition}

\begin{proof}
The split fibred strong colimits of shape $\mathbf{0}$ and $\mathbf{2}$ are given in terms of the initial object and binary coproducts in $\mathcal{V}$, i.e., 
\[
0_X \defeq (X , 0) 
\qquad
(X,A) +_X (X,B) \defeq (X, A + B)
\]
and the weak split fibred strong natural numbers in terms of the weak NNO in $\mathcal{V}$, i.e., 
\[
\begin{array}{c}
\mathbb{N} \defeq (1, \mathbf{N})
\\[1mm]
\mathsf{zero} \defeq \big(\id_1 , (\zj {x \!:\! 1 \times 1} {\mathsf{z}\, (\star)} {\mathbf{N}})\big)
\qquad
\mathsf{succ} \defeq \big(\id_1, (\zj {x \!:\! 1 \times \mathbf{N}} {\mathsf{s}\, (\snd x)} {\mathbf{N}})\big)
\end{array}
\]
where $\mathsf{z} : 1 \longrightarrow \mathbf{N}$ and $\mathsf{s} : \mathbf{N} \longrightarrow \mathbf{N}$ are the zero and successor morphisms associated with the weak NNO $\mathbf{N}$ assumed to exist in $\mathcal{V}$.
\index{ z@$\mathsf{z}$ (zero morphism associated with a weak NNO)}
\index{ s@$\mathsf{s}$ (successor morphism associated with a weak NNO)}

The proofs that these definitions give rise to the required structure consist of straightforward reasoning in the equational theory of EEC\raisebox{0.75pt}{+}. We thus omit these proofs.
\end{proof}

We proceed by observing that it is possible to extend the simple fibration construction to an enriched (effectful) setting. In particular, we can construct a category $\mathsf{s}(\!\mathcal{V},\mathcal{C})$ whose objects are given by pairs $(X,\ul{C})$ of an object $X$ of $\mathcal{V}$ and an object $\ul{C}$ of $\mathcal{C}$, and whose morphisms $(X,\ul{C}) \longrightarrow (Y,\ul{D})$ are given by pairs $(f,h)$ of a morphism $f : X \longrightarrow Y$ in $\mathcal{V}$ and a morphism $h : X \otimes\, \ul{C} \longrightarrow \ul{D}$ in $\mathcal{C}$. 
Analogously to the simple fibration construction, we can define a split fibration whose total category is $\mathsf{s}(\!\mathcal{V},\mathcal{C})$.

\begin{proposition}
\index{fibration!simple $\mathcal{V}$-enriched --}
\index{ s@$\mathsf{s}_{\mathcal{V},\mathcal{C}}$ (simple $\mathcal{V}$-enriched fibration built from a $\mathcal{V}$-enriched category $\mathcal{C}$)}
\index{ s@$\mathsf{s}(\hspace{-0.05cm}\mathcal{V},\mathcal{C})$ (total category of a simple $\mathcal{V}$-enriched fibration)}
The functor $\mathsf{s}_{\mathcal{V}, \mathcal{C}} : \mathsf{s}(\!\mathcal{V},\mathcal{C}) \longrightarrow \mathcal{V}$, given by
\[
\mathsf{s}_{\mathcal{V},\mathcal{C}}(X,\ul{C}) \defeq X
\qquad
\mathsf{s}_{\mathcal{V},\mathcal{C}}(f,h) \defeq f
\]
is a split fibration, called the \emph{simple $\mathcal{V}$-enriched fibration}.
\end{proposition}

\begin{proof}
It is straightforward to show that $\mathsf{s}_{\mathcal{V},\mathcal{C}}$ preserves identities and composition---these properties follow from routine reasoning in the equational theory of EEC\raisebox{0.75pt}{+}.

Given a morphism $f : X \longrightarrow Y$ and an object $(Y,\ul{D})$ in $\mathsf{s}(\!\mathcal{V},\mathcal{C})$, the chosen Cartesian morphism over $f$ can be shown to be given by
\[
\overline{f}(Y,\ul{D}) \defeq \big(f, (\zj {z \!:\! X \otimes\, \ul{D}} {\doto z {(x,z')} {} {z'}} {\ul{D}})\big) : (X,\ul{D}) \longrightarrow (Y,\ul{D})
\]
As was the case for the simple fibration $\mathsf{s}_{\mathcal{V}}$, it is also easy to verify that $\mathsf{s}_{\mathcal{V},\mathcal{C}}$ is split.
\end{proof}

Next, we show that the $\mathcal{V}$-enriched adjunction $F \dashv\, U : \mathcal{C} \longrightarrow \mathcal{V}$ can be lifted to a split fibred adjunction between the two simple fibration constructions.

\begin{proposition}
\index{ F@$\widehat{F}$ (lifting of the functor $F$)}
\index{ U@$\widehat{U}$ (lifting of the functor $U$)}
\index{adjunction!lifting of --}
The $\mathcal{V}$-enriched adjunction $F \dashv\, U : \mathcal{C} \longrightarrow \mathcal{V}$ lifts to a split fibred adjunction $\widehat{F} \dashv\, \widehat{U} : \mathsf{s}_{\mathcal{V},\mathcal{C}} \longrightarrow \mathsf{s}_{\mathcal{V}}$.
\end{proposition}

\begin{proof}
The functors $\widehat{F}$ and $\widehat{U}$ are given on objects by
\[
\widehat{F}(X,A) \defeq (X,F(A))
\qquad
\widehat{U}(X,\ul{C}) \defeq (X,U(\ul{C}))
\]
and on morphisms by
\[
\begin{array}{c}
\widehat{F}(f,g) \defeq \big(f,\big(\zj {z \!:\! X \otimes\, F(A)} {\doto z {(x,z')} {} {F_{A,B}(\lambda\, y \!:\! A .\, g
\, \langle x,y \rangle)(z')}} {F(B)}\big)\big)
\\[2mm]
\widehat{U}(f,h) \defeq \big(f, \big(\zj {x \!:\! X \times U(\ul{C})} {U_{\ul{C},\ul{D}}(\lambda\, z \!:\! \ul{C} .\, h\, \langle \fst x, z \rangle)(\snd x)} {U(\ul{D})}\big)\big)
\end{array}
\]
where the two morphisms
\[
\zj {x \!:\! A \Rightarrow B} {F_{A,B}(x)} {F(A) \multimap F(B)}
\qquad
\zj {x \!:\! \ul{C} \multimap \ul{D}} {U_{\ul{C},\ul{D}}(x)} {U(\ul{C}) \Rightarrow U(\ul{D})}
\]
are given by the $\mathcal{V}$-enrichment of $F$ and $U$, respectively.

It is straightforward to show that $\widehat{F}$ and $\widehat{U}$ preserve identities and composition---these properties follow from routine reasoning in the equational theory of EEC\raisebox{0.75pt}{+}, using the preservation of identities and composition by $F$ and $U$, respectively.
We omit these proofs but show how to prove that both $\widehat{F}$ and $\widehat{U}$ preserve Cartesian morphisms on-the-nose, so as to illustrate the kinds of equational reasoning the proofs in this example are based on. 
Specifically, given a morphism $f : X \longrightarrow Y$ in $\mathcal{V}$, we have 
\begin{fleqn}[0.3cm]
\begin{align*}
& \widehat{F}(\overline{f}(Y,B))
\\
=\,\, & \widehat{F}\big(f, \big(\zj {x \!:\! X \times B} {\snd\, x} {B}\big)\big)
\\
=\,\, &
\big(f, \big(\zj {z \!:\! X \otimes\, F(B)} {\doto z {(x,z')} {} {F_{B,B}(\lambda\, y \!:\! B .\, \snd \langle x,y \rangle\big)(z')}} {F(B)}\big)\big)
\\
=\,\, &
\big(f, \big(\zj {z \!:\! X \otimes\, F(B)} {\doto z {(x,z')} {} {F_{B,B}(\lambda\, y \!:\! B .\, y)(z')}} {F(B)}\big)\big)
\\
=\,\, &
\big(f, \big(\zj {z \!:\! X \otimes\, F(B)} {\doto z {(x,z')} {} {(\lambda\, z'' \!:\! F(B) .\, z'')(z')}} {F(B)}\big)\big)
\\
=\,\, &
\big(f, \big(\zj {z \!:\! X \otimes\, F(B)} {\doto z {(x,z')} {} {z'}} {F(B)}\big)\big)
\\
=\,\, & \overline{f}(Y,F(B))
\\
=\,\, & \overline{f}(\widehat{F}(Y,B))
\end{align*}
\end{fleqn}
and 
\begin{fleqn}[0.3cm]
\begin{align*}
& \widehat{U}(\overline{f}(Y,\ul{D}))
\\
=\,\, & \widehat{U}\big(f, \big(\zj {z \!:\! X \otimes\, \ul{D}} {\doto z {(x,z')} {} {z'}} {\ul{D}}\big)\big)
\\
=\,\, &
\big(f, \big(\zj {x \!:\! X \times U(\ul{D})} {U_{\ul{D},\ul{D}}(\lambda\, z \!:\! \ul{D} .\, \doto {\langle \fst x, z \rangle} {(x,z')} {} {z'})(\snd x)} {U(\ul{D})}\big)\big)
\\
=\,\, &
\big(f, \big(\zj {x \!:\! X \times U(\ul{D})} {U_{\ul{D},\ul{D}}(\lambda\, z \!:\! \ul{D} .\, z)(\snd x)} {U(\ul{D})}\big)\big)
\\
=\,\, &
\big(f, \big(\zj {x \!:\! X \times U(\ul{D})} {(\lambda\, y \!:\! U(\ul{D}) .\, y)(\snd x)} {U(\ul{D})}\big)\big)
\\
=\,\, &
\big(f, \big(\zj {x \!:\! X \times U(\ul{D})} {\snd\, x} {U(\ul{D})}\big)\big)
\\
=\,\, & \overline{f}(Y,U(\ul{D}))
\\
=\,\, & \overline{f}(\widehat{U}(Y,\ul{D}))
\end{align*}
\end{fleqn}

The unit and counit of the adjunction $\widehat{F} \dashv\, \widehat{U}$ are given by components
\[
\begin{array}{c}
\eta_{(X,A)} \defeq \big(\id_X, \big(\zj {x \!:\! X \times A} {\eta^{F \,\dashv\,\, U}_A(\snd\, x)} {U(F(A))}\big)\big) 
\\[2mm]
\varepsilon_{(X,\ul{C})} \defeq \big(\id_X, \big(\zj {z \!:\! X \otimes\, F(U(\ul{C}))} {\doto z {(x,z')} {} {\varepsilon_{\ul{C}}^{F \,\dashv\,\, U}(z')}} {\ul{C}}\big)\big) 
\end{array}
\]
where the two morphisms
\[
\zj {x \!:\! A} {\eta^{F \,\dashv\,\, U}_A(x)} {U(F(A))}
\qquad
\zj {z \!:\! F(U(\ul{C}))} {\varepsilon_{\ul{C}}^{F \,\dashv\,\, U}(z)} {\ul{C}}
\]
are given by the components of the unit and counit of the assumed adjunction $F \dashv U$.

The naturality of $\eta$ and $\varepsilon$, and the two unit-counit laws are proved by straightforward equational reasoning in the equational theory of EEC\raisebox{0.75pt}{+}, using the naturality of $\eta^{F \,\dashv\, U}$ and $\varepsilon^{F \,\dashv\, U}$, and the commutativity of the corresponding unit-counit triangles.
\end{proof}

We proceed by showing that $\mathsf{s}_{\mathcal{V},\mathcal{C}}$ has split dependent $\mathsf{s}_{\mathcal{V}}$-products and $\mathsf{s}_{\mathcal{V}}$-sums.

\begin{proposition}
\label{prop:eecsplitdependentproducts}
$\mathsf{s}_{\mathcal{V},\mathcal{C}}$ has split dependent $\mathsf{s}_{\mathcal{V}}$-products.
\end{proposition}

\begin{proof}
The functor 
\[
\Pi_{(X,A)} : \mathsf{s}(\!\mathcal{V},\mathcal{C})_{X \times A} \longrightarrow \mathsf{s}(\!\mathcal{V},\mathcal{C})_X
\]
is given on objects by 
\[
\begin{array}{c}
\Pi_{(X,A)}(X \times A, \ul{C}) \defeq (X,A \Rightarrow \ul{C})
\end{array}
\]
and on morphisms by
\[
\begin{array}{c}
\hspace{-10.5cm} \Pi_{(X,A)}(\id_{X \times A},h) \defeq 
\\[-0.5mm]
\hspace{2cm} \big(\id_X, \big(\zj {z \!:\! X \otimes (A \Rightarrow \ul{C})} {\doto z {(x,z')} {} {\lambda\, y \!:\! A .\, h\, \langle \langle x , y \rangle , z' \rangle}} {A \Rightarrow \ul{D}}\big)\big)
\end{array}
\]
where $h : (X \times A) \otimes\, \ul{C} \longrightarrow \ul{D}$.

The unit and counit of the adjunction $\pi^*_{(X,A)} \dashv \Pi_{(X,A)}$ are given by components
\[
\begin{array}{c}
\eta_{(X, \ul{C})} \defeq \big(\id_X, \big(\zj {z \!:\! X \otimes\, \ul{C}} {\doto {z} {(x,z')} {} {\lambda\, y \!:\! A .\, z'}} {A \Rightarrow \ul{C}}\big)\big) 
\\[2mm]
\varepsilon_{(X \times A,\ul{C})} \defeq \big(\id_{X \times A}, \big(\zj {z \!:\! (X \times A) \otimes (A \Rightarrow \ul{C})} {\doto {z} {(x,z')} {} {z'\, (\snd x)}} {\ul{C}}\big)\big) 
\end{array}
\vspace{0.15cm}
\]

The well-definedness of $\Pi_{(X,A)}$, the naturality of $\eta$ and $\varepsilon$, and the corresponding unit-counit laws are proved by straightforward equational reasoning in the equational theory of EEC\raisebox{0.75pt}{+}. We omit the details of these proofs but show how to prove that the split Beck-Chevalley condition holds. 
Specifically, given a Cartesian morphism 
\[
\overline{f}(Y,B) \defeq \big(f, (\vj {x \!:\! X \times B} {\snd x} {B})\big) : (X,B) \longrightarrow (Y,B)
\]
in $\mathsf{s}_{\mathcal{V}}$, we show that the canonical natural transformation given in Definition~\ref{def:splitdependentcompproducts} is an identity. 

In particular, for the fibred adjunction model we are constructing in this example, the components of the canonical natural transformation given in Definition~\ref{def:splitdependentcompproducts} can be shown to be given by the composition of morphisms of the following form:
\[
\begin{array}{c}
\hspace{-3cm} \big(\id_X, \big(\zj {z \!:\! X \otimes (B \Rightarrow \ul{C})} {\doto z {(x,z')} {} {\lambda\, y \!:\! B .\, z'}} {B \Rightarrow (B \Rightarrow \ul{C})} \big)\big) 
\\[-1mm]
\hspace{7.75cm} : (X,B \Rightarrow \ul{C}) \longrightarrow (X,B \Rightarrow (B \Rightarrow \ul{C}))
\end{array}
\]
and
\[
\begin{array}{c}
\hspace{-1cm} \big(\id_X, \big(\zj {z'' \!:\! X \otimes (B \Rightarrow (B \Rightarrow \ul{C}))} {\doto {z''} {(x',z''')} {} {\lambda\, y' \!:\! B .\, (z'''\, y')\,\, y'}} {B \Rightarrow \ul{C}}\big)\big)
\\[-1mm]
\hspace{7.75cm} : (X,B \Rightarrow (B \Rightarrow \ul{C})) \longrightarrow (X,B \Rightarrow \ul{C})
\end{array}
\]
which we can then show to be equal to the identity morphism $\id_{(X,B \Rightarrow \ul{C})}$ by 
\begin{fleqn}[0.3cm]
\begin{align*}
& \big(\id_X, \big(\zj {z'' \!:\! X \otimes (B \Rightarrow (B \Rightarrow \ul{C}))} {\doto {z''} {(x',z''')} {} {\lambda\, y' \!:\! B .\, (z'''\, y')\,\, y'}} {B \Rightarrow \ul{C}}\big)\big) \,\, \comp \,\,
\\[-1mm]
& \hspace{2.9cm} \big(\id_X, \big(\zj {z \!:\! X \otimes (B \Rightarrow \ul{C})} {\doto z {(x,z')} {} {\lambda\, y \!:\! B .\, z'}} {B \Rightarrow (B \Rightarrow \ul{C})} \big)\big)
\\
=\,\, & 
\big(\id_X, \big(\zj {z \!:\! X \otimes (B \Rightarrow \ul{C})} {\doto z {(x'',z'''')} {} {\\[-1.5mm] & \hspace{0.15cm} \big(\doto {\langle x'' , (\doto {\langle x'', z'''' \rangle} {(x,z')} {} {\lambda\, y \!:\! B .\, z' )\rangle}} {(x',z''')} {} {\lambda\, y' \!:\! B .\, (z'''\, y')\,\, y'}\big)}} {B \Rightarrow \ul{C}}\big)\big)
\\
=\,\, &
\big(\id_X, \big(\zj {z \!:\! X \otimes (B \Rightarrow \ul{C})} {\doto z {(x'',z'''')} {} {\\[-1.5mm] & \hspace{3.9cm} \big(\doto {\langle x'' , \lambda\, y \!:\! B .\, z'''' \rangle} {(x',z''')} {} {\lambda\, y' \!:\! B .\, (z'''\, y')\,\, y'}\big)}} {B \Rightarrow \ul{C}}\big)\big)
\\
=\,\, &
\big(\id_X, \big(\zj {z \!:\! X \otimes (B \Rightarrow \ul{C})} {\doto z {(x'',z'''')} {} {\lambda\, y' \!:\! B .\, ((\lambda\, y \!:\! B .\, z'''')\,\, y')\,\, y'}} {B \Rightarrow \ul{C}}\big)\big)
\\
=\,\, &
\big(\id_X, \big(\zj {z \!:\! X \otimes (B \Rightarrow \ul{C})} {\doto z {(x'',z'''')} {} {\lambda\, y' \!:\! B .\, z''''\, y'}} {B \Rightarrow \ul{C}}\big)\big)
\\
=\,\, &
\big(\id_X, \big(\zj {z \!:\! X \otimes (B \Rightarrow \ul{C})} {\doto z {(x'',z'''')} {} {z''''}} {B \Rightarrow \ul{C}}\big)\big)
\\
=\,\, &
\id_{(X,B \Rightarrow \ul{C})}
\end{align*}
\end{fleqn}
from which it then follows that the corresponding canonical natural transformation is an identity, as required.
\end{proof}

\begin{proposition}
$\mathsf{s}_{\mathcal{V},\mathcal{C}}$ has split dependent $\mathsf{s}_{\mathcal{V}}$-sums.
\end{proposition}

\begin{proof}
The functor 
\[
\Sigma_{(X,A)} : \mathsf{s}(\!\mathcal{V},\mathcal{C})_{X \times A} \longrightarrow \mathsf{s}(\!\mathcal{V},\mathcal{C})_X
\]
is given on objects by
\[
\begin{array}{c}
\Sigma_{(X,A)}(X \times A, \ul{C}) \defeq (X, A \otimes\, \ul{C})
\end{array}
\]
and on morphisms by
\[
\begin{array}{c}
\hspace{-11cm} \Sigma_{(X,A)}(\id_{X \times A},h) \defeq 
\\
\hspace{0.5cm} \big(\id_X, \big(\zj {z \!:\! X \otimes (A \otimes\, \ul{C})} {\doto {z} {(x,z')} {} {(\doto {z'} {(y,z'')} {} {\big\langle y , h\, \langle \langle x , y \rangle , z'' \rangle\big\rangle})}} {A \otimes\, \ul{D}}\big)\big)
\end{array}
\]
where $h : (X \times A) \otimes\, \ul{C} \longrightarrow \ul{D}$.

The unit and counit of the adjunction $\Sigma_{(X,A)} \dashv \pi^*_{(X,A)}$ are given by components
\[
\begin{array}{c}
\eta_{(X \times A , \ul{C})} \defeq \big(\id_{X \times A}, \big(\zj {z \!:\! (X \times A) \otimes\, \ul{C}} {\doto {z} {(x,z')} {} {\langle \snd x , z' \rangle}} {A \otimes\, \ul{C}}\big) \big)
\\[2mm]
\varepsilon_{(X,\ul{C})} \defeq \big(\id_X, \big(\zj {z \!:\! X \otimes (A \otimes\, \ul{C})} {\doto z {(x,z')} {} {(\doto {z'} {(y,z'')} {} {z''})}} {\ul{C}}\big)\big)
\end{array}
\vspace{0.15cm}
\]

The well-definedness of $\Sigma_{(X,A)}$, the naturality of $\eta$ and $\varepsilon$, and the corresponding unit-counit laws are proved by straightforward equational reasoning in the equational theory of EEC\raisebox{0.75pt}{+}. The proof that the split Beck-Chevalley condition holds is analogous to the corresponding proof we gave for the split dependent $\mathsf{s}_{\mathcal{V}}$-products earlier.
\end{proof}

Finally, we show that $\mathsf{s}_{\mathcal{V},\mathcal{C}}$ admits split fibred pre-enrichment in $\mathsf{s}_{\mathcal{V}}$.

\begin{proposition}
$\mathsf{s}_{\mathcal{V},\mathcal{C}}$ admits split fibred pre-enrichment in $\mathsf{s}_{\mathcal{V}}$.
\end{proposition}

\begin{proof}
The functor
\[
\multimap\,\, : \int (X \mapsto \mathcal{C}^{\text{op}}_X \times \mathcal{C}_X) \longrightarrow \mathcal{V} 
\]
is given on objects by
\[
\begin{array}{c}
\multimap (X, (X,\ul{C}), (X,\ul{D})) \defeq (X, \ul{C} \multimap \ul{D})
\end{array}
\]
and on morphisms by 
\[
\begin{array}{c}
\hspace{-9.5cm} \multimap (f, (\id_X, h), (\id_X, k)) \defeq 
\\
\hspace{0.75cm} \big(f, \big(\zj {x \!:\! X \times (\ul{C}_1 \multimap \ul{D}_1)} {\lambda\, z \!:\! \ul{C}_2 .\, k\, \langle \fst x , (\snd x)(h\, \langle \fst x , z \rangle) \rangle} {\ul{C}_2 \multimap \ul{D}_2}\big) \big)
\end{array}
\]
where $h : X \otimes\, \ul{C}_2 \longrightarrow \ul{C}_1$ and $k : X \otimes\, \ul{D}_1 \longrightarrow \ul{D}_2$. 

The isomorphisms $\xi_{X,(X,\ul{C}),(X,\ul{D})}$ between hom-sets are witnessed by functions
\[
\begin{array}{c}
\xi_{X,(X,\ul{C}),(X,\ul{D})}(\id_X, f) \defeq \big(\id_X, \big(\zj {z \!:\! X \otimes\, \ul{C}} {\doto {z} {(x,z')} {} {(f\, \langle x , \star \rangle)\, z'}} {\ul{D}}\big)\big)
\\[2mm]
\xi_{X,(X,\ul{C}),(X,\ul{D})}^{-1}(\id_X, h) \defeq \big(\id_X, \big(\zj {x \!:\! X \times 1} {\lambda\, z \!:\! \ul{C} .\, h\, \langle \fst x , z \rangle} {\ul{C} \multimap \ul{D}}\big) \big)
\end{array}
\]
where $f : X \times 1 \longrightarrow \ul{C} \multimap \ul{D}$ and $h : X \otimes\, \ul{C} \longrightarrow \ul{D}$.

The well-definedness of the functor $\multimap$, the naturality of $\xi$ and $\xi^{-1}$ in $(X,\ul{C})$ and $(X,\ul{D})$, and their preservation under reindexing are proved by straightforward equational reasoning in the equational theory of EEC\raisebox{0.75pt}{+}. We thus omit these proofs.
\end{proof}

We conclude this example by summarising the above results in the next theorem.

\begin{theorem}
\label{thm:eecfibadjmodels}
\index{fibred adjunction model!-- built from model of EEC\raisebox{0.75pt}{+}}
Given a model $F \dashv\, U : \mathcal{C} \longrightarrow \mathcal{V}$ of EEC\raisebox{0.75pt}{+} with weak natural numbers, we get a fibred adjunction model (without split intensional propositional equality) by letting $p \defeq \mathsf{s}_{\mathcal{V}}$ and $q \defeq \mathsf{s}_{\mathcal{V},\mathcal{C}}$, and by using the lifted adjunction $\widehat{F} \dashv\, \widehat{U} : \mathsf{s}_{\mathcal{V},\mathcal{C}} \longrightarrow \mathsf{s}_{\mathcal{V}}$.
\end{theorem}

\subsection{Families of sets fibration and liftings of adjunctions}
\label{sect:fibadjmodelsfromfamiliesofsets}

Our third example of fibred adjunction models is based on the \emph{families of sets fibration} 
$\mathsf{fam}_{\Set} : \Fam(\Set) \longrightarrow \Set$, a prototypical model of dependent types. This split fibration is a $\Set$-valued instance of the families fibrations we discussed  in Example~\ref{ex:familiesfibration}. 

\index{ fam@$\mathsf{fam}_{\Set}$ (families of sets fibration)}
\index{fibration!families of sets --}
First, we define the \SCCompC\, part of the fibred adjunction model by letting \linebreak $p \defeq \mathsf{fam}_{\Set}$. This gives us an \SCCompC\, because of the following well-known result. 

\begin{proposition}[{\cite[Section~10.5]{Jacobs:Book}}]
\label{prop:familiesofsetsissccompc}
$\mathsf{fam}_{\Set}$ is an SCCompC.
\end{proposition}

In particular, the corresponding terminal object functor $1 : \Set \longrightarrow \Fam(\Set)$ and  comprehension functor $\ia - : \Fam(\Set) \longrightarrow \Set$ are given by
\[
1(X) \defeq (X , x \mapsto 1) \qquad
\ia {(X,A)} \defeq \bigsqcup_{x \in X } A(x)
\]
where $1$ is the terminal object in $\Set$, i.e., a one-element set.

\index{coproduct!set-indexed --}
\index{product!set-indexed --}
\index{ Coproduct@$\bigsqcup_{x \in  X}$ (set-indexed coproduct)}
\index{ Product@$\bigsqcap_{x \in  X}$ (set-indexed product)}
\index{ x@$\langle x , a \rangle$ ($x$'th injection into a set-indexed coproduct)}
The split dependent products and strong split dependent sums are given by
\[
\Pi_{(X,A)}(\bigsqcup_{x \in X}\, A(x), B) \defeq (X , x \mapsto \bigsqcap_{a \in A(x)} B \,\langle x , a \rangle)
\]
\[
\Sigma_{(X,A)}(\bigsqcup_{x \in X}\, A(x), B) \defeq (X , x \mapsto \bigsqcup_{a \in A(x)} B \,\langle x , a \rangle)
\]
where $\bigsqcap_{x \in  X}\, A(x)$ and $\bigsqcup_{x \in  X}\, A(x)$ denote $X$-indexed products and coproducts, respectively; and where $\langle x , a \rangle$ denotes the $x$'th injection into $\bigsqcup_{x \in  X}\, A(x)$. 

Next, we note that $\mathsf{fam}_{\Set}$ also has all other structure we require in Definition~\ref{def:fibadjmodels}.

\begin{proposition}
$\mathsf{fam}_{\Set}$ has split fibred strong colimits of shape $\mathbf{0}$ and $\mathbf{2}$, weak split fibred strong natural numbers, and split intensional propositional equality.
\end{proposition}

\begin{proof}
All the structure mentioned in this proposition is given pointwise in terms of the corresponding set-theoretic structure. 

First, the split fibred strong colimits of shape $\mathbf{0}$ and $\mathbf{2}$ can be shown to be given by
\[
0_X \defeq (X , x \mapsto 0) 
\qquad
(X,A) +_X (X,B) \defeq (X, x \mapsto A(x) + B(x))
\]
where $0$ is the initial object in $\Set$, i.e., the empty set; and $A(x) + B(x)$ is the coproduct of the sets $A(x)$ and $B(x)$, i.e., the disjoint union of $A(x)$ and $B(x)$.

\index{ N@$\mathbf{N}$ (set of natural numbers)}
\index{ z@$\mathsf{z}$ (zero function associated with the set of natural numbers)}
\index{ s@$\mathsf{s}$ (successor function associated with the set of natural numbers)}
Second, the weak split fibred strong natural numbers can be shown to be given by
\[
\mathbb{N} \defeq (1, \star \mapsto \mathbf{N})
\qquad
\mathsf{zero} \defeq (\id_1 , \{\mathsf{z}\}_{\star \in 1})
\qquad
\mathsf{succ} \defeq (\id_1, \{\mathsf{s}\}_{\star \in 1})
\]
where $\mathsf{z} : 1 \longrightarrow \mathbf{N}$ and $\mathsf{s} : \mathbf{N} \longrightarrow \mathbf{N}$ are the zero and successor functions associated with the set $\mathbf{N}$ of natural numbers.

Finally, split intensional propositional equality can be shown to be given by
\[
\Id_{(X,A)} \defeq \big(\bigsqcup_{\langle x , a \rangle \in\, \bigsqcup_{x \in X}\, A(x)}\, A(x) , \langle \langle x , a \rangle , a' \rangle \mapsto \ia {\star \vertbar a = a'}\big)
\]

Showing that these definitions indeed determine the required structure in $\mathsf{fam}_{\Set}$ amounts to straightforward set-theoretic reasoning, using the universal properties of the set-theoretic structure used in these definitions.
\end{proof}

Next, we recall a well-known result about lifting adjunctions to families fibrations.

\begin{proposition}[{\cite[Example~1.8.7 (i)]{Jacobs:Book}}]
\label{prop:liftingadjunctionstofamilies}
\index{adjunction!lifting of --}
\index{ F@$\widehat{F}$ (lifting of the functor $F$)}
\index{ U@$\widehat{U}$ (lifting of the functor $U$)}
Every adjunction $F \dashv\, U : \mathcal{C} \longrightarrow \mathcal{V}$ lifts pointwise to a split fibred adjunction $\widehat{F} \dashv\, \widehat{U} : \mathsf{fam}_{\mathcal{C}} \!\longrightarrow\! \mathsf{fam}_{\mathcal{V}}$ as follows:
\[
\widehat{F}(X,A) \defeq (X, x \mapsto F(A(x)))
\qquad
\widehat{U}(X,\ul{C}) \defeq (X, x \mapsto U(\ul{C}(x)))
\]
\end{proposition}

Given an adjunction $F \dashv\, U : \mathcal{C} \longrightarrow \Set$, we next give sufficient conditions   for $\mathsf{fam}_{\mathcal{C}}$ to have split dependent $\mathsf{fam}_{\Set}$-products and split dependent $\mathsf{fam}_{\Set}$-sums.

\begin{proposition}
Given an adjunction $F \dashv\, U : \mathcal{C} \longrightarrow \Set$, then if $\mathcal{C}$ has set-indexed products, the split fibration $\mathsf{fam}_{\mathcal{C}}$ has split dependent $\mathsf{fam}_{\Set}$-products.
\end{proposition}

\begin{proof}
The split dependent $\mathsf{fam}_{\Set}$-products are defined analogously to how the split dependent products are defined in $\mathsf{fam}_{\Set}$, i.e., they are given on objects by
\[
\Pi_{(X,A)}(\bigsqcup_{x \in X}\, A(x), \ul{C}) \defeq (X , x \mapsto \bigsqcap_{a \in A(x)}\, \ul{C}\, \langle x , a \rangle)
\]
Defining $\Pi_{(X,A)}$ on morphisms and showing the existence of the corresponding adjunction $\pi_{(X,A)}^* \dashv \Pi_{(X,A)}$ amounts to straightforward set-theoretic reasoning.
\end{proof}

\begin{proposition}
Given an adjunction $F \dashv\, U : \mathcal{C} \longrightarrow \Set$, then if $\mathcal{C}$ has set-indexed coproducts, the split fibration $\mathsf{fam}_{\mathcal{C}}$ has split dependent $\mathsf{fam}_{\Set}$-sums.
\end{proposition}

\begin{proof}
The split dependent $\mathsf{fam}_{\Set}$-sums are defined analogously to how strong split dependent sums are defined in $\mathsf{fam}_{\Set}$, i.e., they are given on objects by
\[
\Sigma_{(X,A)}(\bigsqcup_{x \in X}\, A(x), \ul{C}) \defeq (X , x \mapsto \bigsqcup_{a \in A(x)}\, \ul{C}\, \langle x , a \rangle)
\]
Defining $\Sigma_{(X,A)}$ on morphisms and showing the existence of the corresponding adjunction $\Sigma_{(X,A)} \dashv \pi_{(X,A)}^*$ amounts to straightforward set-theoretic reasoning.
\end{proof}

Finally, we show that $\mathsf{fam}_{\mathcal{C}}$ admits split fibred pre-enrichment in $\mathsf{fam}_{\Set}$.

\begin{proposition}
\label{prop:familiesofsetsshallwoenrichment}
Given an adjunction $F \dashv\, U : \mathcal{C} \longrightarrow \Set$, the split fibration $\mathsf{fam}_{\mathcal{C}}$ admits split fibred pre-enrichment in $\mathsf{fam}_{\Set}$.
\end{proposition}

\begin{proof}
We define the functor 
\[
\multimap \,\,: \int (X \mapsto \Fam_X(\mathcal{C})^{\text{op}} \times \Fam_X(\mathcal{C})) \longrightarrow \Fam(\Set)
\]
pointwise by using the $\Set$-enrichment of $\mathcal{C}$, i.e., we define it as
\[
\begin{array}{c}
\multimap (X,(X,\ul{C}),(X,\ul{D})) \defeq (X,x \mapsto \mathcal{C}(\ul{C}(x),\ul{D}(x)))
\\[1mm]
\multimap (f,(\id_X,h),(\id_X,k)) \defeq (f, \{l_x \mapsto k_x \comp l_x \comp h_x\}_{x \in X})
\end{array}
\]
where $h_x : \ul{C}_2(f(x)) \longrightarrow \ul{C}_1(x)$, $k_x : \ul{D}_1(x) \longrightarrow \ul{D}_2(f(x))$,  and $l_x : \ul{C}_1(x) \longrightarrow \ul{D}_1(x)$. We omit the straightforward proofs showing that $\multimap$ preserves identities and composition but show how to prove that it preserves Cartesian morphisms on-the-nose:
\begin{fleqn}[0.3cm]
\begin{align*}
& \multimap (f,(\id_X,\{\id_{\ul{C}(x)}\}_{x \in X}),(\id_X,\{\id_{\ul{D}(x)}\}_{x \in X})) 
\\
=\,\, &
(f , \{l_x \mapsto \id_{\ul{D}(x)} \comp l_x \comp \id_{\ul{C}(x)}\}_{x \in X})
\\
=\,\, &
(f , \{l_x \mapsto l_x\}_{x \in X})
\\
=\,\, &
(f , \{\id_{\mathcal{C}(\ul{C}(x),\ul{D}(x))}\}_{x \in X})
\end{align*}
\end{fleqn}

The isomorphisms $\xi_{X,(X,\ul{C}),(X,\ul{D})}$ between hom-sets are witnessed by functions
\[
\begin{array}{c}
\xi_{X,(X,\ul{C}),(X,\ul{D})} (\id_X, f) \defeq (\id_X, \{f_x(\star)\}_{x \in X})
\\[1mm]
\xi_{X,(X,\ul{C}),(X,\ul{D})}^{-1} (\id_X, h) \defeq (\id_X, \{\star \mapsto h_x\}_{x \in X})
\end{array}
\]
where $f_x : 1 \longrightarrow \mathcal{C}(\ul{C}(x),\ul{D}(x))$ and $h_x : \ul{C}(x) \longrightarrow \ul{D}(x)$. 

The naturality of $\xi$ and $\xi^{-1}$ in $(X,\ul{C})$ and $(X,\ul{D})$, and their preservation under reindexing are proved using straightforward set-theoretic reasoning. 
\end{proof}

We summarise these results in the next theorem.

\begin{theorem}
\label{thm:liftedfibradjmodels}
\index{fibred adjunction model!-- built from families fibration}
Given an adjunction $F \dashv\, U : \mathcal{C} \longrightarrow \Set$ such that $\mathcal{C}$ has set-indexed products and set-indexed coproducts, then the families fibrations $\mathsf{fam}_{\Set}$ and $\mathsf{fam}_{\mathcal{C}}$, together with the lifting $\widehat{F} \dashv\, \widehat{U}$ of $F \dashv\, U$, give rise to a fibred adjunction model.
\end{theorem}

We conclude this section by highlighting some concrete examples of fibred adjunction models built from the families of sets fibration $\mathsf{fam}_{\Set}$. These examples follow as corollaries to Theorem~\ref{thm:liftedfibradjmodels} by instantiating the adjunction $F \dashv\, U$ appropriately. 

The first two instances of Theorem~\ref{thm:liftedfibradjmodels} are based on the decompositions of two of Moggi's monads---the global state monad and the continuations monad---into resolutions other than their Eilenberg-Moore resolutions.

\begin{corollary}
Given a set $S$, the adjunction $(-) \times S \dashv S \Rightarrow (-) : \Set \longrightarrow \Set$ gives rise to a fibred adjunction model. 
\end{corollary}

\begin{corollary}
\label{cor:continuationsmonad}
Given a set $R$, the adjunction $(-) \Rightarrow R \dashv (-) \Rightarrow R : \Set^{\text{op}} \longrightarrow \Set$ gives rise to a fibred adjunction model. 
\end{corollary}

For Corollary~\ref{cor:continuationsmonad} to be an instance of Theorem~\ref{thm:liftedfibradjmodels}, it suffices to recall that  $\Set^{\text{op}}$ trivially has set-indexed products and coproducts---these are given by the set-indexed coproducts and set-indexed products in $\Set$, respectively.

The next instance of Theorem~\ref{thm:liftedfibradjmodels} arises from the algebraic treatment of computational effects, namely, from countable Lawvere theories (see Section~\ref{sect:algebraictreatmentofeffects}).

\begin{corollary}
\label{cor:modelsoflawveretheories}
Given a countable Lawvere theory $I : \aleph_{\!1}^{\text{op}} \longrightarrow \mathcal{L}$, the free model adjunction $F_{\mathcal{L}} \dashv\, U_{\mathcal{L}} : \mathsf{Mod}(\mathcal{L},\Set) \longrightarrow \Set$ gives rise to a fibred adjunction model. 
\end{corollary}

For Corollary~\ref{cor:modelsoflawveretheories} to be an instance of Theorem~\ref{thm:liftedfibradjmodels}, it suffices to recall that $\mathsf{Mod}(\mathcal{L},\Set)$ is both complete and cocomplete (see Proposition~\ref{prop:modelsoflawveretheoriesinsetcocomplete}), meaning that $\mathsf{Mod}(\mathcal{L},\Set)$ has all set-indexed products and set-indexed coproducts, as required.

The final instance of Theorem~\ref{thm:liftedfibradjmodels} we present is based on one of the two standard ways of decomposing monads into adjunctions---the Eilenberg-Moore resolution.

\begin{corollary}
\label{cor:emadjunctionsofmonadsonset}
Given a monad $\mathbf{T} = (T,\eta,\mu)$ on $\Set$, its Eilenberg-Moore resolution $F^{\mathbf{T}} \dashv\, U^{\mathbf{T}} : \Set^{\mathbf{T}} \longrightarrow \Set$ gives rise to a fibred adjunction model. 
\end{corollary}

For Corollary~\ref{cor:emadjunctionsofmonadsonset} to be an instance of Theorem~\ref{thm:liftedfibradjmodels}, it suffices to recall that $\Set^{\mathbf{T}}$ is both complete and cocomplete for any monad $\mathbf{T}$ on $\Set$ (see Proposition~\ref{prop:EMcategoryiscompletecocompleteandregular}), meaning that $\Set^{\mathbf{T}}$ has all set-indexed products and set-indexed coproducts. 

\subsection{Eilenberg-Moore fibrations of fibred monads}
\label{sect:fibredmonadsandEMfibs}

We continue our overview of examples of fibred adjunction models by investigating the conditions under which the Eilenberg-Moore fibration $p^{\mathbf{T}}$ of a split fibred monad ${\mathbf{T}} =(T,\eta,\mu)$ supports split dependent $p$-products and split dependent $p$-sums.
To this end, we generalise some well-known results about the existence of limits and colimits in the EM-category of a monad (see Section~\ref{sect:modelsofeffects} for an overview) from products and coproducts to split dependent $p$-products and split dependent $p$-sums, respectively. 

We begin by recalling a useful fact about the EM-algebras of a split fibred monad. This result later enables us to define split dependent $p$-products and $p$-sums in $p^{\mathbf{T}}$ using the functoriality of the corresponding structure in $p$.

\begin{proposition}[{\cite[Exercise~1.7.9 (ii)]{Jacobs:Book}}]
\label{prop:verticalEMalgebras}
The structure map of every EM-algebra $(A,\alpha)$ of a split fibred monad $\mathbf{T} = (T,\eta,\mu)$ on a split fibration $p : \mathcal{V} \longrightarrow \mathcal{B}$ is vertical.
\end{proposition}

\begin{proof}
The proof of this proposition is straightforward. All one needs to do is to consider the diagram relating $\eta$ and $\alpha$, and apply the functor $p$ to it, i.e., we have
\[
p(\alpha) = p(\alpha) \comp \id_{p(A)} = p(\alpha) \comp p(\eta_A) = p(\alpha \comp \eta_A) = p(\id_A) = \id_{p(A)}
\]
where $p(A) = p(T(A))$ holds because $T$ is assumed to be a fibred functor.
\end{proof}

Another observation we make about split fibred monads is that every split fibred monad on a split comprehension category with unit and strong split dependent sums comes equipped with a dependent notion of strength. 
We use this dependent strength to impose one of the conditions under which $p^{\mathbf{T}}$ has split dependent $p$-sums.

\begin{proposition}
\label{prop:strengthofsplitfibredmonads}
\index{monad!split fibred --!dependent strength of a --}
\index{ sigma@$\sigma_A$ (dependent strength of a split fibred monad)}
Given a split comprehension category with unit $p : \mathcal{V} \longrightarrow \mathcal{B}$ with strong split dependent sums and a split fibred monad $\mathbf{T} = (T,\eta,\mu)$ on it, then there exists a family of natural transformations 
\[
\sigma_A : \Sigma_A \comp T \longrightarrow T \comp \Sigma_A \qquad\qquad\qquad (A \in \mathcal{V})
\]
collectively called the \emph{dependent strength} of $\mathbf{T}$, satisfying the following diagrams:

\vspace{0.5cm}

\[
\xymatrix@C=4em@R=5em@M=0.5em{
\Sigma_{1_{p(A)}}(\pi_{1_{p(A)}}^*(T(A))) \ar[r]^-{=} \ar[dr]_-{\varepsilon^{\Sigma_{1_{p(A)}} \dashv\,\, \pi_{1_{p(A)}}^*}_{T(A)}} & \Sigma_{1_{p(A)}}(T(\pi_{1_{p(A)}}^*(A))) \ar[r]^-{\sigma_{1_{p(A)},\pi_{1_{p(A)}}^*(A)}} & T(\Sigma_{1_{p(A)}}(\pi_{1_{p(A)}}^*(A))) \ar[dl]^-{T(\varepsilon^{\Sigma_{1_{p(A)}} \dashv\,\, \pi_{1_{p(A)}}^*}_A)}_-{(1)\qquad\qquad\qquad\quad\,\,\,\,}
\\
& T(A)
}
\]

\vspace{0.3cm}

\[
\xymatrix@C=0.8em@R=5em@M=0.5em{
\Sigma_{\Sigma_{A}(B)}(T(C)) \ar[rr]^-{\sigma_{\Sigma_{A}(B),T(C)}} \ar[d]_-{\alpha_{A,B,T(C)}} & & T(\Sigma_{\Sigma_{A}(B)}(C)) \ar[dd]^-{T(\alpha_{A,B,C})}_-{(2)\qquad\qquad\qquad\qquad\quad\,\,\,\,\,}
\\
\Sigma_{A}(\Sigma_{B}(\kappa_{A,B}^*(T(C)))) \ar[d]_-{=}
\\
\Sigma_{A}(\Sigma_{B}(T(\kappa_{A,B}^*(C)))) \ar[dr]_-{\Sigma_{A}(\sigma_{B,\kappa_{A,B}^*(C)})\quad\,\,\,\,} &  & T(\Sigma_{A}(\Sigma_{B}(\kappa_{A,B}^*(C))))
\\
& \Sigma_{A}(T(\Sigma_{B}(\kappa_{A,B}^*(C)))) \ar[ur]_-{\,\,\,\,\quad\sigma_{A,\Sigma_{B}(\kappa_{A,B}^*(C))}}
}
\]

\vspace{0.3cm}

\[
\xymatrix@C=10em@R=6em@M=0.5em{
\Sigma_A(B) \ar[r]^-{\Sigma_A(\eta_B)} \ar[dr]_-{\eta_{\Sigma_A(B)}} & \Sigma_A(T(B)) \ar[d]^-{\sigma_{A,B}}_<<<<<<<<{(3)\qquad\quad}
\\
& T(\Sigma_A(B))
}
\]

\vspace{0.3cm}

\[
\xymatrix@C=5em@R=6em@M=0.5em{
\Sigma_A(T(T(B))) \ar[r]^-{\sigma_{A,T(B)}} \ar[d]_-{\Sigma_A(\mu_B)} & T(\Sigma_A(T(B))) \ar[r]^-{T(\sigma_{A,B})} & T(T(\Sigma_A(B))) \ar[d]^-{\mu_{\Sigma_A(B)}}_-{(4)\qquad\qquad\qquad\qquad\qquad\quad\!\!\!\!}
\\
\Sigma_A(T(B)) \ar[rr]_-{\sigma_{A,B}} && T(\Sigma_A(B))
}
\]

\noindent where 
\[
\alpha_{A,B} : \Sigma_{\Sigma_A(B)} \longrightarrow \Sigma_A \comp \Sigma_B \comp \kappa^*_{A,B} \qquad\qquad\qquad (A \in \mathcal{V}, B \in \mathcal{V}_{\ia A})
\]
\index{ ab@$\alpha_{A,B}$ (natural associativity isomorphism)}
is a family of natural associativity isomorphisms, given by the following composites:
\[
\hspace{-0.075cm}
\xymatrix@C=7.25em@R=4em@M=0.5em{
\Sigma_{\Sigma_A(B)} \ar[r]^-{=} & \Sigma_{\Sigma_A(B)} \comp (\kappa^{-1}_{A,B})^* \comp \kappa^*_{A,B} \ar[d]^-{\Sigma_{\Sigma_A(B)} \,\comp\, (\kappa^{-1}_{A,B})^* \,\comp\, \eta^{\Sigma_A \,\comp\, \Sigma_B \,\dashv\, \pi^*_B \,\comp\, \pi^*_A} \,\comp\, \kappa^*_{A,B}}
\\
& \Sigma_{\Sigma_A(B)} \comp (\kappa^{-1}_{A,B})^* \comp \pi^*_B \comp \pi^*_A \comp \Sigma_A \comp \Sigma_B \comp \kappa^*_{A,B} \ar[d]^-{=}
\\
\Sigma_A \comp \Sigma_B \comp \kappa^*_{A,B} & \Sigma_{\Sigma_A(B)} \comp \pi^*_{\Sigma_A(B)} \comp \Sigma_A \comp \Sigma_B \comp \kappa^*_{A,B} \ar[l]^-{\varepsilon^{\Sigma_{\Sigma_A(B)} \,\dashv\, \pi^*_{\Sigma_A(B)}} \,\comp\, \Sigma_A \,\comp\, \Sigma_B \,\comp\, \kappa^*_{A,B}}
}
\]
\end{proposition}

We note that the second equality morphism used in the definition of $\alpha_{A,B}$ follows from the commutativity of the following diagram:
\[
\xymatrix@C=6em@R=7em@M=0.5em{
\ia {B} \ar[r]_-{\ia {\eta^{\Sigma_A \,\dashv\, \pi^*_A}_B}} \ar[dr]_-{\pi_B} \ar@/^2.25pc/[rr]^-{\kappa_{A,B}}_{\dcomment{\text{def. of } \kappa_{A,B}}} & \ia {\pi^*_A(\Sigma_A(B))} \ar[r]_-{\ia {\overline{\pi_A}(\Sigma_A(B))}} \ar[d]^-{\pi_{\pi^*_A(\Sigma_A(B))}}_<<<<<<<<<<{\dcomment{\mathcal{P}(\eta^{\Sigma_A \,\dashv\, \pi^*_A}_B)}\,\,\,\,\,\,} & \ia {\Sigma_A(B)} \ar[d]^-{\pi_{\Sigma_A(B)}}_-{\dcomment{\mathcal{P}(\overline{\pi_A}(\Sigma_A(B)))}\quad\,\,\,\,\,} \ar@/_5.15pc/[ll]_-{\kappa^{-1}_{A,B}}^*+<0.5em>{\dcomment{\kappa_{A,B} \text{ is an iso.}}}
\\
& \ia A \ar[r]_-{\pi_A} & p(A)
}
\]

\begin{proof}
Due to its length, we postpone the proof of Proposition~\ref{prop:strengthofsplitfibredmonads} to Appendix~\ref{sect:proofofprop:strengthofsplitfibredmonads} and only note here that each of the natural transformations $\sigma_A$ is given by the composite 
\[
\xymatrix@C=2.15em@R=5em@M=0.5em{
\Sigma_A \comp T \ar[rr]^-{\Sigma_A \,\comp\, T \,\comp\, \eta^{\Sigma_A \,\dashv\, \pi^*_A}} && \Sigma_A \comp T \comp \pi^*_A \comp \Sigma_A \ar[r]^-{=} & \Sigma_A \comp \pi^*_A \comp T \comp \Sigma_A \ar[rr]^-{\varepsilon^{\Sigma_A \,\dashv\, \pi^*_A} \,\comp\, T \,\comp\, \Sigma_A} && T \comp \Sigma_A
}
\]
\end{proof}

We now proceed by investigating the conditions under which split dependent $p$-products and split dependent $p$-sums exist in $p^{\mathbf{T}}$.

On the one hand, it can be easily seen that the EM-fibration of a split fibred monad on a split comprehension category with unit $p : \mathcal{V} \longrightarrow \mathcal{B}$ with split dependent products always has split dependent $p$-products. 

\begin{theorem}
\label{thm:dependentproductsinEMfibration}
\index{Eilenberg-Moore!-- fibration!split dependent $p$-products in --}
Given a split comprehension category with unit $p : \mathcal{V} \longrightarrow \mathcal{B}$ with split dependent products and a split fibred monad  $\mathbf{T} = (T,\eta,\mu)$ on it, then the corresponding EM-fibration $p^{\mathbf{T}} : \mathcal{V}^{\mathbf{T}} \longrightarrow \mathcal{B}$ has split dependent $p$-products.
\end{theorem}

\begin{proof}
\index{ Product@$\Pi^{\mathbf{T}}_A$ (split dependent $p$-product in $p^{\mathbf{T}}$)}
\index{ b@$\beta_{\Pi^{\mathbf{T}}_A}$ (structure map of $\Pi^{\mathbf{T}}_A(B,\beta)$)}
Due to its length, we postpone the proof of Theorem~\ref{thm:dependentproductsinEMfibration} to Appendix~\ref{sect:proofofthm:dependentproductsinEMfibration} and only note here that the split dependent $p$-products are given in the EM-fibration $p^{\mathbf{T}}$ by functors $\Pi^{\mathbf{T}}_A : \mathcal{V}^{\mathbf{T}}_{\ia A} \longrightarrow \mathcal{V}^{\mathbf{T}}_{p(A)}$ that are defined on an object $(B,\beta)$ of $\mathcal{V}^{\mathbf{T}}_{\ia A}$ by 
\[
\Pi^{\mathbf{T}}_A(B,\beta) \defeq (\Pi_A(B), \beta_{\Pi^{\mathbf{T}}_A})
\]
where the structure map $\beta_{\Pi^{\mathbf{T}}_A} : T(\Pi_A(B)) \longrightarrow \Pi_A(B)$ is given by the morphism
\[
\xymatrix@C=5em@R=0.25em@M=0.5em{
T(\Pi_A(B)) \ar[r]^-{\eta^{\pi^*_A \,\dashv\, \Pi_A}_{T(\Pi_A(B))}} & \Pi_A(\pi^*_A(T(\Pi_A(B)))) \ar[dr]^-{=}
\\
& & \Pi_A(T(\pi^*_A(\Pi_A(B)))) \ar[dl]^-{\Pi_A(T(\varepsilon^{\pi^*_A \,\dashv\, \Pi_A}_B))}
\\
\Pi_A(B) & \Pi_A(T(B)) \ar[l]^-{\Pi_A(\beta)}
}
\]
using the split dependent products in $p$, i.e., the adjunction $\pi^*_A \dashv \Pi_A : \mathcal{V}_{\ia A} \longrightarrow \mathcal{V}_{p(A)}$. We use Proposition~\ref{prop:verticalEMalgebras} in the definition of $\beta_{\Pi^{\mathbf{T}}_A}$ to ensure that $\beta$ is vertical.
\end{proof}

As split dependent products can be viewed as a natural generalisation of set-indexed products to products indexed by an arbitrary object in the total category $\mathcal{V}$, then it is not surprising that Theorem~\ref{thm:dependentproductsinEMfibration} and its proof are similar to the set-indexed products instance of Proposition~\ref{prop:limitsinEMcategory}.

For constructing a fibred adjunction model based on the EM-fibration of a split fibred monad, 
we have the following corollary to Theorem~\ref{thm:dependentproductsinEMfibration}.

\begin{corollary}
Given an SCCompC $p : \mathcal{V} \longrightarrow \mathcal{B}$ and a split fibred monad on it, then the corresponding EM-fibration has split dependent $p$-products.
\end{corollary}

On the other hand, the situation with the split dependent $p$-sums in the EM-fibration $p^{\mathbf{T}}$ of a split fibred monad $\mathbf{T}$ is analogous to the existence of coproducts in the EM-category of a monad. In particular, generally they can not be defined directly in terms of the split dependent sums in $p$. However, analogously to the existence of coproducts in the EM-category of a monad, under certain conditions the split dependent $p$-sums in $p^{\mathbf{T}}$ do exist and can be defined in terms of the split dependent sums in $p$. 

In this thesis,  
we investigate two such conditions for $p^{\mathbf{T}}$: i) when $\mathbf{T}$ preserves the split dependent sums in $p$ via its dependent strength (see Theorem~\ref{thm:dependentsumsinEMfibrationwhenmonadpreservesthem} below); and ii) when $p^{\mathbf{T}}$ has split fibred reflexive coequalizers (see Theorem~\ref{thm:dependentsumsinEMfibration} below). 

\begin{theorem}
\label{thm:dependentsumsinEMfibrationwhenmonadpreservesthem}
\index{Eilenberg-Moore!-- fibration!split dependent $p$-sums in --}
Given a split comprehension category with unit $p : \mathcal{V} \longrightarrow \mathcal{B}$ with strong split dependent sums and a split fibred monad $\mathbf{T} = (T,\eta,\mu)$ on it, then the corresponding EM-fibration $p^{\mathbf{T}} : \mathcal{V}^{\mathbf{T}} \!\longrightarrow\! \mathcal{B}$ has split dependent $p$-sums if the dependent strength of $\mathbf{T}$ is given by a family of natural isomorphisms, i.e., if for every $A$ in $\mathcal{V}$, $\sigma_A : \Sigma_A \comp T \longrightarrow T \comp \Sigma_A$ is a natural isomorphism.
Furthermore, these split dependent $p$-sums are preserved on-the-nose by $U^{\mathbf{T}}$, i.e., we have $U^{\mathbf{T}}(\Sigma^{\mathbf{T}}_A(B,\beta)) = \Sigma_A(U^{\mathbf{T}}(B,\beta))$.
\end{theorem}

\begin{proof}
Due to its length, we postpone the proof of Theorem~\ref{thm:dependentsumsinEMfibrationwhenmonadpreservesthem} to Appendix~\ref{sect:proofofthm:dependentsumsinEMfibrationwhenmonadpreservesthem} and only note here that the split dependent $p$-sums are given in the EM-fibration $p^{\mathbf{T}}$ by functors $\Sigma^{\mathbf{T}}_A : \mathcal{V}^{\mathbf{T}}_{\ia A} \longrightarrow \mathcal{V}^{\mathbf{T}}_{p(A)}$ that are defined on an object $(B,\beta)$ of $\mathcal{V}^{\mathbf{T}}_{\ia A}$ by 
\[
\Sigma^{\mathbf{T}}_A(B,\beta) \defeq (\Sigma_A(B), \beta_{\Sigma^{\mathbf{T}}_A})
\]
\index{ Sigma@$\Sigma^{\mathbf{T}}_A$ (split dependent $p$-sum in $p^{\mathbf{T}}$)}
\index{ b@$\beta_{\Sigma^{\mathbf{T}}_A}$ (structure map of $\Sigma^{\mathbf{T}}_A(B,\beta)$)}
where the structure map $\beta_{\Sigma^{\mathbf{T}}_A} : T(\Sigma_A(B)) \longrightarrow \Sigma_A(B)$ is given by the morphism
\[
\xymatrix@C=5em@R=5em@M=0.5em{
T(\Sigma_A(B)) \ar[r]^-{\sigma^{-1}_{A,B}} & \Sigma_A(T(B)) \ar[r]^-{\Sigma_A(\beta)} & \Sigma_A(B)
}
\]
using the split dependent sums in $p$, i.e., the adjunction $\Sigma_A \dashv \pi^*_A : \mathcal{V}_{p(A)} \longrightarrow \mathcal{V}_{\ia A}$.
\end{proof}

We note that Theorem~\ref{thm:dependentsumsinEMfibrationwhenmonadpreservesthem} is similar to Proposition~\ref{prop:colimitsinEMcategory1} where the existence of colimits in the EM-category of a monad followed from the preservation of colimits by that monad. 
However, in contrast to Proposition~\ref{prop:colimitsinEMcategory1}, our preservation condition is formulated using a specific isomorphism, based on the dependent strength of $\mathbf{T}$. We note that this particular choice of the preservation isomorphism is crucial for our proof of this theorem to go through---see Appendix~\ref{sect:proofofthm:dependentsumsinEMfibrationwhenmonadpreservesthem} for details.

For constructing a fibred adjunction model based on the EM-fibration of a split fibred monad, we have the following corollary to Theorem~\ref{thm:dependentsumsinEMfibrationwhenmonadpreservesthem}.

\begin{corollary}
Given an SCCompC $p : \mathcal{V} \longrightarrow \mathcal{B}$ and a split fibred monad on it, then the corresponding EM-fibration has split dependent $p$-sums if the dependent strength of this split fibred monad is given by a family of natural isomorphisms.
\end{corollary}

\begin{theorem}
\label{thm:dependentsumsinEMfibration}
\index{Eilenberg-Moore!-- fibration!split dependent $p$-sums in --}
Given a split comprehension category with unit $p : \mathcal{V} \longrightarrow \mathcal{B}$ with strong split dependent sums and a split fibred monad $\mathbf{T} = (T,\eta,\mu)$ on it, then the corresponding EM-fibration $p^{\mathbf{T}} : \mathcal{V}^{\mathbf{T}} \longrightarrow \mathcal{B}$ has split dependent $p$-sums if $p^{\mathbf{T}}$ has split fibred reflexive coequalizers.
\end{theorem}

\begin{proof}
\index{ e@$e_{A,(B,\beta)}$ (reflexive coequalizer used to define $\Sigma^{\mathbf{T}}_A(B,\beta)$)}
Due to its length, we postpone the proof of Theorem~\ref{thm:dependentsumsinEMfibration} to Appendix~\ref{sect:proofofthm:dependentsumsinEMfibration} and only note here that the split dependent $p$-sums are given in the EM-fibration $p^{\mathbf{T}}$ by functors $\Sigma^{\mathbf{T}}_A : \mathcal{V}^{\mathbf{T}}_{\ia A} \longrightarrow \mathcal{V}^{\mathbf{T}}_{p(A)}$ that are defined on an object $(B,\beta)$ of $\mathcal{V}^{\mathbf{T}}_{\ia A}$ as the reflexive coequalizer
\[
\xymatrix@C=5em@R=5em@M=0.5em{
(T(\Sigma_A(B)), \mu_{\Sigma_A(B)}) \ar[r]^-{e_{A,(B,\beta)}} & \Sigma^{\mathbf{T}}_A(B,\beta)
}
\]
of the following pair of morphisms in $\mathcal{V}^{\mathbf{T}}_{p(A)}$:
\[
\xymatrix@C=5em@R=5em@M=0.5em{
(T(\Sigma_A(T(B))), \mu_{\Sigma_A(T(B))}) \ar[r]^-{T(\Sigma_A(\beta))} & (T(\Sigma_A(B)), \mu_{\Sigma_A(B)})
}
\]
and
\[
\xymatrix@C=6em@R=3em@M=0.5em{
(T(\Sigma_A(T(B))), \mu_{\Sigma_A(T(B))}) \ar[r]^-{T(\Sigma_A(T(\eta^{\Sigma_A \,\dashv\, \pi^*_A}_B)))} & (T(\Sigma_A(T(\pi^*_A(\Sigma_A(B))))), \mu_{\Sigma_A(T(\pi^*_A(\Sigma_A(B))))}) \ar[d]^-{=}
\\
& (T(\Sigma_A(\pi^*_A(T(\Sigma_A(B))))), \mu_{\Sigma_A(\pi^*_A(T(\Sigma_A(B))))}) \ar[d]^-{T(\varepsilon^{\Sigma_A \,\dashv\, \pi^*_A}_{T(\Sigma_A(B))})}
\\
(T(\Sigma_A(B)), \mu_{\Sigma_A(B)}) & \ar[l]^-{\mu_{\Sigma_A(B)}} (T(T(\Sigma_A(B))), \mu_{T(\Sigma_A(B))})
}
\]
using the split dependent sums in $p$, i.e., the adjunction $\Sigma_A \dashv \pi^*_A : \mathcal{V}_{p(A)} \longrightarrow \mathcal{V}_{\ia A}$. 
\end{proof}

We note that Theorem~\ref{thm:dependentsumsinEMfibration} is similar to Proposition~\ref{prop:colimitsinEMcategory2} where the existence of coproducts in the EM-category of a monad followed from the existence of reflexive coequalizers. 
In particular, as split dependent sums can be viewed as a natural generalisation of set-indexed coproducts to coproducts indexed by an arbitrary object in the total category $\mathcal{V}$, it is not surprising that Theorem~\ref{thm:dependentsumsinEMfibration} and its proof are similar to Proposition~\ref{prop:colimitsinEMcategory2}.

For constructing a fibred adjunction model based on the EM-fibration of a split fibred monad, we have the following corollary to Theorem~\ref{thm:dependentsumsinEMfibration}.

\begin{corollary}
Given an SCCompC $p : \mathcal{V} \longrightarrow \mathcal{B}$ and a split fibred monad on it, then the corresponding EM-fibration has split dependent $p$-sums if the EM-fibration of this split fibred monad has split fibred reflexive coequalizers.
\end{corollary}

\subsection{Continuous families fibration and general recursion}
\label{sect:continuousfamilies}

We conclude our overview of examples of fibred adjunction models by presenting a domain-theoretic generalisation of the families of sets fibration based models from Section~\ref{sect:fibadjmodelsfromfamiliesofsets}. Furthermore, we show how to use this domain-theoretic  model to give a denotational semantics to an extension of eMLTT with general recursion.

To improve the readability of this section, we have split it into three parts: i) in Section~\ref{sect:domaintheorypreliminaries}, we recall some preliminaries of domain theory; ii) in Section~\ref{sect:domaintheoreticfibredadjunctionmodel} we construct a domain-theoretic fibred adjunction model based on continuous families; and iii) in Section~\ref{sect:extensionofeMLTTwithrecursion}, we present an extension of eMLTT with general recursion, and show how to interpret it in our domain-theoretic fibred adjunction model.

\subsubsection{Domain-theoretic preliminaries}
\label{sect:domaintheorypreliminaries}

In this section we give a brief overview of basic domain-theoretic concepts and results that we later use to construct a domain-theoretic fibred adjunction model. We refer the reader to~\cite{Gierz:ContinuousLattices,Plotkin:PisaNotes,Abramsky:DomainTheory} for a more detailed overview of the relevant domain theory. 

\begin{definition}
\index{partial order!$\omega$-complete --}
\index{ cpo@cpo ($\omega$-complete partial order)}
\index{ x@$\langle x_n \rangle$ (increasing $\omega$-chain $x_1 \leq_X x_2 \leq_X \ldots$)}
\index{ V@$\bigvee_{\hspace{-0.05cm}n} \,x_n$ (least upper bound of an increasing $\omega$-chain $\langle x_n \rangle$)}
\index{ @$\leq_X$ (partial order of a cpo $X$)}
\index{ X@$\vert X \vert$ (underlying set of a cpo $X$)}
\index{ X@$(\vert X \vert , \leq_X)$ (cpo)}
An \emph{$\omega$-complete partial order} (cpo) is a partial order $X = (\vert X \vert , \leq_X)$ which has least upper bounds $\bigvee_{\!\!n} \,x_n$ of all increasing $\omega$-chains $\langle x_n \rangle \defeq x_1 \leq_X x_2 \leq_X \ldots$
\end{definition}

\begin{definition}
\index{function!continuous --}
A function $f : X \longrightarrow Y$ between cpos is \emph{continuous} if it is monotone and preserves the least upper bounds of increasing $\omega$-chains, i.e., when we have
\[
\begin{array}{c}
x_1 \leq_X x_2 \implies f(x_1) \leq_Y f(x_2)
\qquad
f(\bigvee_{\!\!n} \,x_n) = \bigvee_{\!\!n}\, f(x_n)
\end{array}
\]
\end{definition}

\begin{proposition}
\index{category!-- of cpos and continuous functions}
\index{ CPO@$\CPO$ (category of cpos and continuous functions)}
Cpos and continuous functions form the category $\CPO$.
\end{proposition}

\begin{definition}
\index{partial order!$\omega$-complete --!discrete --}
A cpo $X$ is \emph{discrete} if its partial order is given by equality.
\end{definition}

In particular, observe that every set $A$ trivially determines a discrete cpo $(A,=)$. 

\begin{proposition}[{\cite[Section~2]{Plotkin:PisaNotes}}]
$\CPO$ is Cartesian closed.
\end{proposition}

In particular, the terminal object $1$ is given by the unique cpo on a singleton set; the Cartesian product $X \times Y$ is given by the set $\vert X \vert \times \vert Y \vert$ and the component-wise  order
\[
\langle x_1 , y_1 \rangle \leq_{X \times Y} \langle x_2 , y_2 \rangle \text{~~iff~~} x_1 \leq_X x_2 ~~\wedge~~ y_1 \leq_Y y_2
\]
and the exponential $X \Rightarrow Y$ is given by the set of continuous functions from $X$ to $Y$ with the pointwise order
\[
f \leq_{X \Rightarrow Y} g \text{~~iff~~} \forall x \in \vert X \vert .\, f(x) \leq_Y g(x)
\]

\begin{proposition}[{\cite[Section~3]{Plotkin:PisaNotes}}]
\label{prop:CPOhasfinitecoproducts}
$\CPO$ has finite coproducts. 
\end{proposition}

In particular, the initial object $0$ is given by the unique cpo on the empty set; and the binary coproduct $X + Y$ is given by the set $\vert X \vert + \vert Y \vert$ with the disjoint order
\[
\mathsf{inl}~x_1 \leq_{X + Y} \mathsf{inl}~x_2 \text{~~iff~~} x_1 \leq_X x_2
\qquad
\qquad
\mathsf{inr}~y_1 \leq_{X + Y} \mathsf{inr}~y_2 \text{~~iff~~} y_1 \leq_Y y_2
\]

An important variation of the category $\CPO$ we use is the category $\CPO^{\text{EP}}$ of embedding-projection pairs. For us, the embedding-projection pairs are essential to accommodate the contravariance arising in the definition of split dependent products.

\begin{definition}
\index{embedding-projection pair}
\index{ f@$(f^e,f^p)$ (embedding-projection pair)}
An \emph{embedding-projection pair} $(f^e,f^p) : X \longrightarrow Y$ between cpos $X$ and $Y$ is given by continuous functions $f^e : X \longrightarrow Y$ and $f^p : Y \longrightarrow X$, satisfying
\[
f^p \comp f^e = \id_X
\qquad
f^e \comp f^p \leq_Y \id_Y
\]
\end{definition}

\begin{proposition}
\index{category!-- of embedding-projection pairs between cpos}
\index{ CPO@$\CPO^{\text{EP}}$ (category of embedding-projection pairs between cpos)}
Cpos and embedding-projection pairs form the category $\CPO^{\text{EP}}$.
\end{proposition}

An important property of the category $\CPO^{\text{EP}}$ that we use pervasively in our proofs is the following instance of the well-known limit-colimit coincidence property for embed\-ding-projection pairs---see~\cite[Theorem~2]{Smyth:RecDomEqs} for more details.

\begin{proposition}
\label{prop:limitcolimitcoincidenceforcfam}
\index{limit-colimit coincidence}
Given a cpo $X$, a functor $A : X \longrightarrow \CPO^{\text{EP}}$, and an increasing $\omega$-chain $\langle x_n \rangle$, then the cocone $\alpha : J \longrightarrow \Delta(A(\bigvee_{\!\!n}\, x_n))$, given by components 
\[
\begin{array}{c}
\alpha_{x_n} \defeq (A(x_n \leq_X \bigvee_{\!\!n}\, x_n)^e , A(x_n \leq_X \bigvee_{\!\!n}\, x_n)^p) : A(x_n) \longrightarrow A(\bigvee_{\!\!n}\, x_n)
\end{array}
\]
is a colimit of the diagram $J : \langle x_n \rangle \longrightarrow \CPO^{\text{EP}}$, given by $J(x_n) \defeq A(x_n)$, if and only if 
\[
\begin{array}{c}
\bigvee_{\!\!n}\, (A(x_n \leq_X \bigvee_{\!\!n}\, x_n)^e \comp A(x_n \leq_X \bigvee_{\!\!n}\, x_n)^p) = \id_{A(\bigvee_{\!\!n}\, x_n)} 
\end{array}
\]
\end{proposition}

\subsubsection{A domain-theoretic fibred adjunction model}
\label{sect:domaintheoreticfibredadjunctionmodel}

The domain-theoretic fibred adjunction model we construct below is based on lifting the EM-resolution of a suitable $\CPO$-enriched monad $\mathbf{T} = (T,\eta,\mu)$ on $\CPO$ to a split fibred adjunction. It is well-known that the corresponding EM-category $\CPO^{\mathbf{T}}$ and the adjunction $F^{\mathbf{T}} \dashv\, U^{\mathbf{T}} : \CPO^{\mathbf{T}} \longrightarrow \CPO$ are both  $\CPO$-enriched. In particular, the hom-cpo $\CPO^{\mathbf{T}}((A,\alpha),(B,\beta))$ is given by the cpo of continuous functions from $A$ to $B$ that additionally satisfy the condition of being an EM-algebra homomorphism. 

\index{ O@$\Omega_{(A,\alpha)}$ (least zero-ary operation)}
In order to be able to model general recursion, we assume that the monad $\mathbf{T}$ supports a least zero-ary operation, in the sense of~\cite[Section~6]{Plotkin:SemanticsForAlgOperations}. In more detail, this means that for every EM-algebra $(A,\alpha)$, there must exist an element $\Omega_{(A,\alpha)}$ in $\vert A \vert$ such that $\Omega_{(A,\alpha)} \leq_A a$, for all elements $a$ in $\vert A \vert$; and the continuous EM-algebra homomorphisms must strictly preserve these least elements.

Further, in order to be able to define split dependent sums, we assume that $\CPO^{\mathbf{T}}$ has reflexive coequalizers---see Proposition~\ref{prop:cfamEMalgebrassplitdependentsums} for how they are used.

We note that a monad $\mathbf{T}_{\!\mathcal{L}}$ satisfying these requirements is induced by any  discrete $\CPO$-enriched countable Lawvere theory $\mathcal{L}$ (see~\cite{Hyland:DiscreteLawTh} for details) that includes a least zero-ary operation. In particular, the monad is given by the functor $T_{\!\mathcal{L}} \defeq U_{\!\mathcal{L}} \comp F_{\mathcal{L}}$, and the category $\CPO^{\mathbf{T}_{\!\mathcal{L}}}$ is complete and cocomplete, and thus has reflexive coequalizers.

\index{partial order!directed-complete --}
\index{ dcpo@dcpo (directed-complete partial order)}
We begin by defining a split fibration that is suitable for modelling eMLTT's value types, the \emph{fibration $\mathsf{cfam}_{\CPO} : \CFam(\CPO) \longrightarrow \CPO$ of continuous families}. This fibration is based on the analogous fibration of continuous families of directed-complete partial orders (dcpos) studied in~\cite[Section~10.6]{Jacobs:Book}. We discuss the reasons why we use continuous families of cpos instead of dcpos in Proposition~\ref{prop:dcposarenotpresentable}.

\begin{definition}
\label{def:catofcontfamilies}
\index{category!-- of continuous families of cpos}
\index{functor!continuous --}
\index{ CFam@$\CFam(\CPO)$ (total category of the fibration of continuous families of cpos)}
The objects of the category $\CFam(\CPO)$ are given by pairs $(X,A)$ of a cpo $X$ and a continuous functor $A : X \longrightarrow \CPO^{\text{EP}}$, i.e., a functor that preserves colimits of $\omega$-chains when we treat the cpo $X$ as a category. A morphism from $(X,A)$ to $(Y,B)$ is given by a pair $(f,\{g_x\}_{x \in \vert X \vert})$ of a continuous function $f : X \longrightarrow Y$ and a family of continuous functions $\{g_x : A(x) \longrightarrow B(f(x))\}_{x \in \vert X \vert}$, satisfying
\vspace{-0.15cm}
\[
\hspace{-0.175cm}
\begin{array}{c}
x_1 \leq_X x_2 \implies B(f(x_1) \leq_Y f(x_2))^e \comp g_{x_1} \leq _{\CPO(A(x_1),B(f(x_2)))} 
g_{x_2} \comp A(x_1 \leq_X x_2)^e
\\[2mm]
\langle x_n \rangle \text{~is incr.~} \omega \text{-chain} \implies g_{\bigvee_{\!n} x_n} = \bigvee_{\!n} \big(B(f(x_n) \leq_Y f(\bigvee_{\!n} x_n))^e \comp g_{x_n} \comp A(x_n \leq_X \bigvee_{\!n} x_n)^p\big)
\end{array}
\vspace{-0.1cm}
\]
These conditions express that $g$ is a continuously indexed natural transformation.
\end{definition}

For better readability, we often define the continuous functors $A : X \longrightarrow \CPO^{\text{EP}}$ using the $\mapsto$ notation, leaving the action of $A$ on $\leq_X$ implicit when it is clear from the context. To keep this example concise, we also omit details of some other definitions when they are analogous to those for the families of sets fibration $\mathsf{fam}_{\Set}$ at the level of the underlying sets, and when the additional order-theoretic proof obligations are straightforward (e.g., we only give the object parts of $\Pi_{(X,A)}$ and $\Sigma_{(X,A)}$). Further, we also omit the laborious but straightforward proofs of the propositions given below.

\begin{proposition}
\label{prop:continuousfamiliesfibration}
\index{fibration!-- of continuous families of cpos}
\index{ cfam@$\mathsf{cfam}_{\CPO}$ (fibration of continuous families of cpos)}
The functor $\mathsf{cfam}_{\CPO} : \CFam(\CPO) \longrightarrow \CPO$ of \emph{continuous families of cpos}, given by
\[
\mathsf{cfam}_{\CPO}(X,A) \defeq X
\qquad
\mathsf{cfam}_{\CPO}(f,\{g_x\}_{x \in \vert X \vert}) \defeq f
\]
is a split fibration. In fact, it is a full split comprehension category with unit.
\end{proposition}

We refer the reader to the detailed proof of the dcpo-version of this proposition given in~\cite[Lemma~10.6.2]{Jacobs:Book}.

In particular, given a continuous function $f : X \longrightarrow Y$, the chosen Cartesian morphism over $f$ in $\mathsf{cfam}_{\CPO}$ is given as in $\mathsf{fam}_{\Set}$, i.e., by
\[
\overline{f}(Y,A) \defeq (f , \{\id_{A(f(x))}\}_{x \in X}) : (X , A \comp f) \longrightarrow (Y,A)
\]

Further, the corresponding terminal object functor $1 : \CPO \longrightarrow \CFam(\CPO)$ and  comprehension functor $\ia - : \CFam(\CPO) \longrightarrow \CPO$ are given by
\[
1(X) \defeq (X , x \mapsto (1,=)) \qquad
\ia {(X,A)} \defeq \bigsqcup_{X}\, A
\]
where the latter is defined using a \emph{cpo-indexed coproduct}, given by
\index{coproduct!cpo-indexed --}
\index{ Coproduct@$\bigsqcup_{X}$ (cpo-indexed coproduct)}
\[
\bigsqcup_{X}\, A \defeq (\bigsqcup_{x \in \vert X \vert}\!\! \vert A(x) \vert , \leq_{\bigsqcup_{X} A})
\]
and where the partial order $\leq_{\bigsqcup_{X} A}$ is given by
\[
\langle x_1,a_1 \rangle \leq_{\bigsqcup_{X} A} \langle x_2 , a_2 \rangle \text{~~iff~~} x_1 \leq_X x_2 ~~\wedge~~ A(x_1 \leq_X x_2)^e(a_1) \leq_{A(x_2)} a_2
\]

\index{ x@$\langle x , a \rangle$ ($x$'th injection into a cpo-indexed coproduct)}
These cpo-indexed coproducts $\bigsqcup_{X}\, A$ come equipped with natural continuous injection and copairing morphisms, given by the injection and copairing morphisms of the underlying set-indexed coproducts of sets. Analogously to the set-indexed coproducts of sets, we write $\langle x , a \rangle$ for the $x$'th injection into the $X$-indexed coproduct $\bigsqcup_{X}\, A$. 

Next, we show that $\mathsf{cfam}_{\CPO}$ has the structure one needs to model a version of eMLTT in which propositional equality is restricted to be over terms whose types denote families of discrete cpos---see the discussion later in this section.

As mentioned earlier, our proofs in this section rely on the limit-colimit coincidence for embedding-projection pairs (see~Proposition~\ref{prop:limitcolimitcoincidenceforcfam}). In particular, we use Proposition~\ref{prop:limitcolimitcoincidenceforcfam} pervasively to show that the second components of the objects of $\CFam(\CPO)$ we define below are continuous functors. An analogous property for dcpos had an important role in the corresponding proofs given in~\cite[Section~10.6]{Jacobs:Book}.

\begin{proposition}
\label{prop:cfamhassepproductsandsums}
$\mathsf{cfam}_{\CPO}$ has split dependent products and strong split dependent sums.
\end{proposition}

In particular, the corresponding functors are given on objects by 
\[
\Pi_{(X,A)}(\bigsqcup_{X}\, A , B) \defeq (X , x \mapsto \bigsqcap_{A(x)} B\, \langle x , - \rangle)
\]
\[
\Sigma_{(X,A)}(\bigsqcup_{X}\, A , B) \defeq (X , x \mapsto \bigsqcup_{A(x)} B\, \langle x , - \rangle)
\]
where the former is defined using a \emph{cpo-indexed product}, given by
\index{product!cpo-indexed --}
\index{ Product@$\bigsqcap_{X}$ (cpo-indexed product)}
\[
\bigsqcap_{A(x)} B\, \langle x , - \rangle \defeq (\{f : A(x) \longrightarrow \bigsqcup_{A(x)} B\, \langle x , - \rangle \vertbar \mathsf{fst} \comp f = \id_{A(x)}\} , \leq_{\bigsqcap_{A(x)} B\, \langle x , - \rangle})
\]
and where the partial order $\leq_{\bigsqcap_{A(x)} B\, \langle x , - \rangle}$ is given pointwise by
\[
f_1 \leq_{\bigsqcap_{A(x)} B\, \langle x , - \rangle} f_2 \text{~~iff~~} \forall a \in \vert A(x) \vert .\, f_1(a) \leq_{\bigsqcup_{A(x)} B\, \langle x , - \rangle} f_2(a)
\]

It is worth noting that the action of $x \mapsto \bigsqcap_{A(x)} B\, \langle x , - \rangle$ on the partial order $\leq_X$ crucially relies on $A$ being $\CPO^{\text{EP}}$-valued rather than $\CPO$-valued. In particular, the projections enable us to account for the contravariance arising from $\bigsqcap_{A(x)} B\, \langle x , - \rangle$. 

Dcpo-versions of these definitions and the corresponding proofs can be found in~\cite[Section~10.6]{Jacobs:Book}. 

Next, by combining Propositions~\ref{prop:continuousfamiliesfibration} and~\ref{prop:cfamhassepproductsandsums}, we get the following corollary.

\begin{corollary}
$\mathsf{cfam}_{\CPO}$ is a \SCCompC.
\end{corollary}

Next, we note that $\mathsf{cfam}_{\CPO}$ also has all other structure we require from $p$ in Definition~\ref{def:fibadjmodels}, except for split intensional propositional equality, which we here restrict to be over continuous families of discrete cpos, as explained below.

\begin{proposition}
$\mathsf{cfam}_{\CPO}$ has split fibred strong colimits of shape $\mathbf{0}$ and $\mathbf{2}$, and weak split fibred strong natural numbers.
\end{proposition}

\index{ N@$\mathbf{N}_=$ (discrete cpo on the set of natural numbers)}
\index{ z@$\mathsf{z}$ (continuous zero function associated with the discrete cpo of natural numbers)}
\index{ s@$\mathsf{s}$ (continuous successor function associated with the discrete cpo of natural numbers)}
In particular, these split fibred strong colimits can be shown to be given by
\[
0_X \defeq (X , x \mapsto 0)
\qquad
(X,A) +_X (X,B) \defeq (X , x \mapsto A(x) + B(x))
\]
and the weak split fibred strong natural numbers can be shown to be given by
\[
\mathbb{N} \defeq (1, \star \mapsto \mathbf{N}_=)
\qquad
\mathsf{zero} \defeq (\id_1 , \{\mathsf{z}\}_{\star \in 1})
\qquad
\mathsf{succ} \defeq (\id_1, \{\mathsf{s}\}_{\star \in 1})
\]
where $\mathsf{z} : 1 \longrightarrow \mathbf{N}_=$ and $\mathsf{s} : \mathbf{N}_= \longrightarrow \mathbf{N}_=$ are the continuous zero and successor functions associated with the discrete cpo $\mathbf{N}_=$ on the set $\mathbf{N}$ of natural numbers.

As mentioned earlier, for split intensional propositional equality, we only consider continuous families $A : X \longrightarrow \CPO^{\text{EP}}$ where each $A(x)$ is a discrete cpo, so as to guarantee that the second component of $\Id_{(X,A)}$ we define below is a continuous functor. 

\begin{proposition}
$\mathsf{cfam}_{\CPO}$ has split intensional propositional equality over continuous families of discrete cpos.
\end{proposition}

In particular, the discreteness assumption allows us to define the split intensional propositional equality analogously to $\mathsf{fam}_{\Set}$, i.e., by
\[
\Id_{(X,A)} \defeq \big(\{\pi^*_{(X,A)}(X,A)\}, \langle \langle x , a \rangle , a' \rangle \mapsto (\{\star \vertbar a = a'\}, =)\big)
\]

Next, we define a fibration suitable for modelling computation types, the \emph{fibration $\mathsf{cfam}_{\CPO^{\mathbf{T}}} : \CFam(\CPO^{\mathbf{T}}) \longrightarrow \CPO$ of continuous families of continuous EM-algebras}, for the monad $\mathbf{T} = (T,\eta,\mu)$ we assumed earlier in this section.
This is a natural domain-theoretic generalisation of the families of EM-algebras fibration used in Section~\ref{sect:fibadjmodelsfromfamiliesofsets}.

\begin{definition}
\index{category!-- of continuous families of continuous EM-algebras}
\index{ CFam@$\CFam(\CPO^{\mathbf{T}})$ (total category of the fibration of continuous families of continuous EM-algebras)}
The objects of the category $\CFam(\CPO^{\mathbf{T}})$ are given by pairs $(X,\ul{C})$ of a cpo $X$ and a continuous functor $\ul{C} : X \longrightarrow (\CPO^{\mathbf{T}})^{\text{EP}}$, i.e., a functor that preserves colimits of $\omega$-chains when we treat $X$ as a category. A morphism from $(X,\ul{C})$ to $(Y,\ul{D})$ is given by a pair $(f,\{h_x\}_{x \in \vert X \vert})$ of a continuous function $f : X \longrightarrow Y$ and a family of continuous EM-algebra homomorphisms $\{h_x : \ul{C}(x) \longrightarrow \ul{D}(f(x))\}_{x \in \vert X \vert}$, satisfying
\vspace{-0.15cm}
\[
\hspace{-0.2cm}
\begin{array}{c}
x_1 \leq_X x_2 \implies \ul{D}(f(x_1) \leq_Y f(x_2))^e \comp h_{x_1} \leq _{\CPO^{\mathbf{T}}(\ul{C}(x_1),\ul{D}(f(x_2)))}
h_{x_2} \comp \ul{C}(x_1 \leq_X x_2)^e
\\[2mm]
\langle x_n \rangle \text{~is incr.~} \omega \text{-chain} \implies h_{\bigvee_{\!n} x_n} = \bigvee_{\!n} \big(\ul{D}(f(x_n) \leq_Y f(\bigvee_{\!n} x_n))^e \comp h_{x_n} \comp \ul{C}(x_n \leq_X \bigvee_{\!n} x_n)^p\big)
\end{array}
\vspace{-0.1cm}
\]
These conditions express that $h$ is a continuously indexed natural transformation.
\end{definition}

\begin{proposition}
\index{fibration!-- of continuous families of continuous EM-algebras}
\index{ cfam@$\mathsf{cfam}_{\CPO^{\mathbf{T}}}$ (fibration of continuous families of continuous EM-algebras)}
The functor $\mathsf{cfam}_{\CPO^{\mathbf{T}}} : \CFam(\CPO^{\mathbf{T}}) \longrightarrow \CPO$ of \emph{continuous families of continuous EM-algebras}, given by
\[
\mathsf{cfam}_{\CPO^{\mathbf{T}}}(X,\ul{C}) \defeq X
\qquad
\mathsf{cfam}_{\CPO^{\mathbf{T}}}(f,\{h_x\}_{x \in \vert X \vert}) \defeq f
\]
is a split fibration.
\end{proposition}

\index{ CPO@$(\CPO^{\mathbf{T}})^{\text{EP}}$ (category of embedding-projection pairs between continuous EM-algebras)}
\index{category!-- of embedding-projection pairs between continuous EM-algebras}
We note that the category $(\CPO^{\mathbf{T}})^{\text{EP}}$ is given by \emph{embedding-projection pairs} of continuous homomorphisms between continuous EM-algebras for the monad $\mathbf{T}$. 

It is worth noting that an analogue of Proposition~\ref{prop:limitcolimitcoincidenceforcfam} also holds for $(\CPO^{\mathbf{T}})^{\text{EP}}$.

\begin{proposition}
\label{prop:limitcolimitcoincidenceforEMalgebras}
\index{limit-colimit coincidence}
Given a cpo $X$, a functor $\ul{C} : X \longrightarrow (\CPO^{\mathbf{T}})^{\text{EP}}$, and an increasing $\omega$-chain $\langle x_n \rangle$, then the cocone $\alpha : J \longrightarrow \Delta(\ul{C}(\bigvee_{\!\!n}\, x_n))$, given by components 
\[
\begin{array}{c}
\alpha_{x_n} \defeq (\ul{C}(x_n \leq_X \bigvee_{\!\!n}\, x_n)^e , \ul{C}(x_n \leq_X \bigvee_{\!\!n}\, x_n)^p) : \ul{C}(x_n) \longrightarrow \ul{C}(\bigvee_{\!\!n}\, x_n)
\end{array}
\]
is a colimit of $J : \langle x_n \rangle \longrightarrow (\CPO^{\mathbf{T}})^{\text{EP}}$, given by $J(x_n) \defeq \ul{C}(x_n)$, if and only if 
\[
\begin{array}{c}
\bigvee_{\!\!n}\, (\ul{C}(x_n \leq_X \bigvee_{\!\!n}\, x_n)^e \comp \ul{C}(x_n \leq_X \bigvee_{\!\!n}\, x_n)^p) = \id_{\ul{C}(\bigvee_{\!\!n}\, x_n)} 
\end{array}
\]
\end{proposition}

Analogously to Proposition~\ref{prop:limitcolimitcoincidenceforcfam}, Proposition~\ref{prop:limitcolimitcoincidenceforEMalgebras} also follows from the well-known limit-colimit coincidence property for embed\-ding-projection pairs---see~\cite[Theorem~2]{Smyth:RecDomEqs} for details.
We use it pervasively to show that the second components of the objects of $\CFam(\CPO^{\mathbf{T}})$ we define below are continuous functors. 

Next, analogously to lifting adjunctions to the families fibrations (see Proposition~\ref{prop:liftingadjunctionstofamilies}), we can lift the $\CPO$-enriched EM-adjunction ${F^{\mathbf{T}} \dashv\, U^{\mathbf{T}} : \CPO^{\mathbf{T}} \longrightarrow \CPO}$ to a corresponding split fibred adjunction $\widehat{F^{\mathbf{T}}} \dashv\, \widehat{U^{\mathbf{T}}} : \mathsf{cfam}_{\CPO^{\mathbf{T}}} \longrightarrow \mathsf{cfam}_{\CPO}$.

\begin{proposition}
\index{ F@$\widehat{F^{\mathbf{T}}}$ (lifting of the functor $F^{\mathbf{T}}$)}
\index{ U@$\widehat{U^{\mathbf{T}}}$ (lifting of the functor $U^{\mathbf{T}}$)}
\index{adjunction!lifting of --}
The two split fibred functors ${\widehat{F^{\mathbf{T}}} : \mathsf{cfam}_{\CPO} \longrightarrow \mathsf{cfam}_{\CPO^{\mathbf{T}}}}$ and \linebreak $\widehat{U^{\mathbf{T}}} : \mathsf{cfam}_{\CPO^{\mathbf{T}}} \longrightarrow \mathsf{cfam}_{\CPO}$, given by a pointwise construction, i.e., by
\[
\widehat{F^{\mathbf{T}}}(X,A) \defeq (X , x \mapsto F^{\mathbf{T}}(A(x)))
\qquad
\widehat{U^{\mathbf{T}}}(X,\ul{C}) \defeq (X , x \mapsto U^{\mathbf{T}}(\ul{C}(x)))
\]
constitute a split fibred adjunction $\widehat{F^{\mathbf{T}}} \dashv\, \widehat{U^{\mathbf{T}}} : \mathsf{cfam}_{\CPO^{\mathbf{T}}} \longrightarrow \mathsf{cfam}_{\CPO}$.
\end{proposition}

Next, we show that $\mathsf{cfam}_{\CPO^{\mathbf{T}}}$ has split dependent $\mathsf{cfam}_{\CPO}$-products and -sums. 

\begin{proposition}
$\mathsf{cfam}_{\CPO^{\mathbf{T}}}$ has split dependent $\mathsf{cfam}_{\CPO}$-products.
\end{proposition}

The corresponding functor $\Pi_{(X,A)} : \CFam_{\ia {(X,A)}}(\CPO^{\mathbf{T}}) \longrightarrow \CFam_X(\CPO^{\mathbf{T}})$ is defined on objects as 
\[
\Pi_{(X,A)}(\bigsqcup_{X}\, A , \ul{C}) \defeq (X , x \mapsto \bigsqcap_{A(x)} \ul{C}\, \langle x , - \rangle)
\]
using a \emph{cpo-indexed product of continuous EM-algebras}, given by 
\index{product!cpo-indexed --}
\index{ Product@$\bigsqcap_{X}$ (cpo-indexed product)}
\[
\bigsqcap_{A(x)} \ul{C}\, \langle x , - \rangle \defeq (\,\bigsqcap_{A(x)} (U^{\mathbf{T}} \comp\, \ul{C}\, \langle x , - \rangle) , \alpha)
\]
where the continuous structure map 
\[
\alpha : T(\bigsqcap_{A(x)} (U^{\mathbf{T}} \comp\, \ul{C}\, \langle x , - \rangle)) \longrightarrow \bigsqcap_{A(x)} (U^{\mathbf{T}} \comp\, \ul{C}\, \langle x , - \rangle)
\]
is given by
\[
\alpha \defeq \langle \alpha_{\ul{C} \langle x , a \rangle} \comp T(\mathsf{proj}_a) \rangle_{a \in \vert A(x) \vert}
\]
using the continuous pairing and projection morphisms associated with the cpo-indexed products in $\CPO$, and where $\alpha_{\ul{C}(x)}$ denotes the structure map of $\ul{C}(x)$. 

Observe that the use of cpo-indexed products in $\CPO$ to define cpo-indexed products in $\CPO^{\mathbf{T}}$ is analogous to how set-indexed products in $\mathcal{V}$ are used to define set-indexed products in $\mathcal{V}^{\mathbf{T}}$---see Proposition~\ref{prop:limitsinEMcategory} for more details.

\begin{proposition}
\label{prop:cfamEMalgebrassplitdependentsums}
$\mathsf{cfam}_{\CPO^{\mathbf{T}}}$ has split dependent $\mathsf{cfam}_{\CPO}$-sums.
\end{proposition}

The corresponding functor $\Sigma_{(X,A)} : \CFam_{\ia {(X,A)}}(\CPO^{\mathbf{T}}) \longrightarrow \CFam_X(\CPO^{\mathbf{T}})$ is defined on objects as 
\[
\Sigma_{(X,A)}(\bigsqcup_{X}\, A , \ul{C}) \defeq (X , x \mapsto \bigsqcup_{A(x)} \ul{C}\, \langle x , - \rangle)
\]
using a \emph{cpo-indexed coproduct of continuous EM-algebras}, given by 
the reflexive coequalizer $e : F^{\mathbf{T}} (\bigsqcup_{A(x)} (U^{\mathbf{T}} \comp\, \ul{C}\, \langle x , - \rangle)) \longrightarrow \bigsqcup_{A(x)} \ul{C}\, \langle x , - \rangle$ of the pair of morphisms
\index{coproduct!cpo-indexed --}
\index{ Coproduct@$\bigsqcup_{X}$ (cpo-indexed coproduct)}
\index{ e@$e$ (reflexive coequalizer used to define a cpo-indexed coproduct of continuous EM-algebras)}
\vspace{-0.25cm}
\[
\hspace{-0.15cm}
\xymatrix@C=3em@R=2em@M=0.5em{
F^{\mathbf{T}} (\bigsqcup_{A(x)} (T \comp U^{\mathbf{T}} \comp\, \ul{C}\, \langle x , - \rangle)) \ar@<-0.8ex>[rrr]_-{\mu_{\bigsqcup_{A(x)} (U^{\mathbf{T}} \comp\,\, \ul{C}\, \langle x , - \rangle)} \comp F^{\mathbf{T}}([T(\mathsf{inj}_a)]_{a \in \vert A(x) \vert})} \ar@<0.8ex>[rrr]^-{F^{\mathbf{T}} ([\mathsf{inj}_a \,\,\comp\,\, \alpha_{\ul{C}\, \langle x , a \rangle}]_{a \in \vert A(x) \vert})} &&& F^{\mathbf{T}} (\bigsqcup_{A(x)} (U^{\mathbf{T}} \comp\, \ul{C}\, \langle x , - \rangle))
}
\]
whose common section is given by 
\[
\xymatrix@C=4.5em@R=2em@M=0.5em{
F^{\mathbf{T}} (\bigsqcup_{A(x)} (U^{\mathbf{T}} \comp\, \ul{C}\, \langle x , - \rangle)) \ar[rr]^-{F^T ([\mathsf{inj}_a \,\,\comp \,\, \eta_{U^{\mathbf{T}}(\ul{C}\, \langle x , a \rangle)}]_{a \in \vertbar A(x) \vertbar})} 
&&
F^{\mathbf{T}} (\bigsqcup_{A(x)} (T \comp U^{\mathbf{T}} \comp\, \ul{C}\, \langle x , - \rangle))
}
\]
On morphisms, we define $\Sigma_{(X,A)}$ using the universal property of reflexive coequalizers. 

Observe that the use of reflexive coequalizers to define cpo-indexed coproducts in $\CPO^{\mathbf{T}}$ is analogous to the use of reflexive coequalizers to define set-indexed coproducts in $\mathcal{V}^{\mathbf{T}}$---see Proposition~\ref{prop:colimitsinEMcategory2} for more details.
Similar use of (split fibred) reflexive coequalizers also appears in our definition of split dependent sums in the EM-fibrations of split fibred monads---see the proof of Theorem~\ref{thm:dependentsumsinEMfibration} for details, e.g., for the definition of $\Sigma_{(X,A)}$ on morphisms using the universal property of reflexive coequalizers.

The final ingredient for constructing a fibred adjunction model based on $\mathsf{cfam}_{\CPO}$ and $\mathsf{cfam}_{\CPO^{\mathbf{T}}}$ is the split fibred pre-enrichment of $\mathsf{cfam}_{\CPO^{\mathbf{T}}}$ in $\mathsf{cfam}_{\CPO}$.

\begin{proposition}
$\mathsf{cfam}_{\CPO^{\mathbf{T}}}$ admits split fibred pre-enrichment in $\mathsf{cfam}_{\CPO}$.
\end{proposition}

The corresponding functor 
\[
\multimap \,\,: \int (X \mapsto \CFam_X({\CPO^{\mathbf{T}}})^{\text{op}} \times \CFam_X({\CPO^{\mathbf{T}}})) \longrightarrow \CFam(\CPO)
\]
is given on objects by 
\[
\begin{array}{c}
\multimap (X,(X,\ul{C}),(X,\ul{D})) \defeq (X,x \mapsto \CPO^{\mathbf{T}}(\ul{C}(x),\ul{D}(x)))
\end{array}
\]
using the $\CPO$-enrichment of $\CPO^{\mathbf{T}}$ discussed earlier in this section.

We summarise these results in the next theorem.

\begin{theorem}
\label{thm:continuousfamiliesfibadjmodel}
Given a $\CPO$-enriched monad $\mathbf{T} = (T,\eta,\mu)$ on $\CPO$ that supports a least zero-ary operation, in the sense of~\cite[Section~6]{Plotkin:SemanticsForAlgOperations}, such that $\CPO^{\mathbf{T}}$ has reflexive coequalizers, then the fibrations $\mathsf{cfam}_{\CPO}$ and $\mathsf{cfam}_{\CPO^{\mathbf{T}}}$ give rise to a split fibred adjunction model where propositional equality is restricted to families of discrete cpos.
\end{theorem}

It is worth noting that the existence of the least zero-ary operation is only required in order to use this domain-theoretic fibred adjunction model to give a denotational semantics to an extension of eMLTT with general recursion. This requirement can be dropped when giving a denotational semantics to eMLTT as defined in Chapter~\ref{chap:syntax}.

A good source of such fibred adjunction models is the algebraic treatment of computational effects, as made precise in the following corollary to Theorem~\ref{thm:continuousfamiliesfibadjmodel}.

\begin{corollary}
\label{cor:fibredadjunctionmodelsfromdiscretelawveretheories}
Given a discrete $\CPO$-enriched countable Lawvere theory $\mathcal{L}$ \linebreak (see~\cite{Hyland:DiscreteLawTh}) that includes a least zero-ary operation, the corresponding $\CPO$-enriched monad on $T \defeq U_{\!\mathcal{L}} \comp F_{\mathcal{L}}$ gives us a fibred adjunction model where propositional equality is restricted to families of discrete cpos. Here, $F_{\mathcal{L}} \dashv\, U_{\!\mathcal{L}} : \Mod(\!\mathcal{L},\CPO) \longrightarrow \CPO$.
\end{corollary}

In particular, in future work we plan to extend fibred algebraic effects and their handlers with inequations based on $\mathsf{cfam}_{\CPO}$ and discrete $\CPO$-enriched countable Lawvere theories, analogously to how $\mathsf{fam}_{\Set}$ and countable Lawvere theories are used to model equationally presented fibred algebraic effects and their handlers in Chapters~\ref{chap:fibalgeffects} and~\ref{chap:handlers}. We recall from~\cite{Hyland:DiscreteLawTh} that a key prerequisite for this to work, i.e., for the corresponding left adjoint $F_{\mathcal{L}}$ to exist, is that $\CPO$ is locally countably presentable (see~\cite[Example~1.18 (2)]{Adamek:LocallyPresentableCats}). 
Unfortunately, this is not the case for the category of dcpos. 

\begin{proposition}[{\cite[Example~1.18 (5)]{Adamek:LocallyPresentableCats}}]
\label{prop:dcposarenotpresentable}
The category of dcpos and continuous functions between them is not locally (countably) presentable. 
\end{proposition}

The failure of the category of dcpos to be locally countably presentable is the main reason why we use cpos instead of dcpos in this section, compared to the domain-theoretic fibrational model of dependent types given in~\cite[Section~10.6]{Jacobs:Book}.

We conclude this section by explaining why we use $\mathsf{cfam}_{\CPO}$ instead of the other natural candidate, the codomain fibration $\mathsf{cod}_{\CPO} : \CPO^{\to} \longrightarrow \CPO$. 

On the one hand, $\mathsf{cod}_{\CPO}$ is not split, but this can be overcome because every fibration is equivalent to a split one, see~\cite[Corollary~5.2.5]{Jacobs:Book}. On the other hand, for $\mathsf{cod}_{\CPO}$ to be a even a non-split CCompC, $\CPO$ must be locally Cartesian closed, see~\cite[Theorem~10.5.5 (ii)]{Jacobs:Book}. However, as the next result shows, this is not the case.

\begin{theorem}
$\CPO$ is not locally Cartesian closed.
\end{theorem}

\begin{proof}
Recall that for $\CPO$ to be locally Cartesian closed, every base change functor $f^* : \CPO/Y \longrightarrow \CPO/X$ must have a right adjoint, meaning that $f^*$ itself has to be a left adjoint and thus it has to preserve colimits. In particular, $f^*$ has to preserve epimorphisms because they can be characterised as certain colimits. Specifically, it is well-known that $g : A \longrightarrow B$ is an epimorphism when $\id_B \comp g$ and $\id_B \comp g$ form a pushout.

Below we show that this is not the case in $\CPO$ by giving an example of a particular base change functor and an epimorphism it does not preserve. The proof is essentially based on the fact that not all epimorphisms in $\CPO$ are given by surjective functions.

\index{category!slice --}
\index{ CPO@$\CPO/X$ (slice category of $\CPO$ over an object $X$)}
Here, $\CPO/X$ is the \emph{slice category} of $\CPO$ over $X$. Its objects are given by continuous functions with codomain $X$.  A morphism in $\CPO/X$ from $f : Y \longrightarrow X$ to $g : Z \longrightarrow X$ is given by a continuous function $h : Y \longrightarrow Z$ such that $g \comp h = f$.

\index{ N@$\mathbf{N}_{\omega}$ (cpo of natural numbers with the $\leq$ order and a top element)}
We write $\mathbf{N}_=$ for the discrete cpo on the set of natural numbers and $\mathbf{N}_{\omega}$ for the cpo on the set of natural numbers extended with a top element $\omega$, where $\leq_{\mathbf{N}_{\omega}}$ is given by
\[
n \leq_{\mathbf{N}_{\omega}} m \text{~~iff~~} n,m \text{~are natural numbers} ~~\wedge~~ n \leq m
\qquad\quad
n \leq_{\mathbf{N}_{\omega}} \omega \text{~~for all~} n
\]

Next, recall that given a continuous function $f : X \longrightarrow Y$, the base change functor $f^* : \CPO/Y \longrightarrow \CPO/X$ is given by sending a continuous function $g : Z \longrightarrow Y$ to the continuous function $f^*(g) : f^*(Z) \longrightarrow X$ in the following pullback square:
\[
\xymatrix@C=3em@R=3em@M=0.5em{
f^*(Z) \ar[r] \ar[d]_{f^*(g)}^<{\,\big\lrcorner} & Z \ar[d]^{g}
\\
X \ar[r]_{f} & Y
}
\]
On morphisms of $\CPO/Y$, $f^*$ is defined using the universal property of pullbacks.

The particular epimorphism of interest to us in $\CPO$ is $e : \mathbf{N}_= \longrightarrow \mathbf{N}_{\omega}$, given by mapping $n$ to $n$. 
It is easy to see that $e$ is an epimorphism: given two continuous functions $h_1 : \mathbf{N}_{\omega} \longrightarrow Y$ and $h_2 : \mathbf{N}_{\omega} \longrightarrow Y$, for some $Y$, such that $h_1 \comp e = h_2 \comp e$, then it suffices to show that $h_1(n) = h_2(n)$, for all natural numbers $n$, and that $h_1(\omega) = h_2(\omega)$, for the top element $\omega$. The proofs for these equations are straightforward:
\[
\begin{array}{c}
h_1(n) = h_1(e(n)) = h_2(e(n)) = h_2(n)
\\[3mm]
h_1(\omega) = h_1(\bigvee_{\!\!n}\, n) = 
\bigvee_{\!\!n}\, h_1(e(n)) = \bigvee_{\!\!n}\, h_2(e(n)) = 
h_2(\bigvee_{\!\!n}\, n) = h_2(\omega) 
\end{array}
\]
using the fact that $\omega$ is the least upper bound of the $\omega$-chain $\langle n \rangle \defeq 0 \leq 1 \leq \ldots$ 

Importantly for us, $e$ also gives us an epimorphism in the slice category $\CPO/{\mathbf{N}_{\omega}}$: 
\[
\xymatrix@C=3em@R=2em@M=0.5em{
\mathbf{N}_= \ar[rr]^-{e} \ar[dr]_-{e} & & \mathbf{N}_{\omega} \ar[dl]^-{\id_{\mathbf{N}_{\omega}}}
\\
& \mathbf{N}_{\omega}
}
\]

Now, assuming a non-empty cpo $X$, let us consider the base change functor \linebreak $f_{\omega}^* : \CPO/{\mathbf{N}_{\omega}} \longrightarrow \CPO/X$ for a constant function $f_\omega : X \longrightarrow \mathbf{N}_{\omega}$ that is given by mapping every $x$ to $\omega$. If $\CPO$ were locally Cartesian closed, this base change functor must preserve colimits, in particular, the epimorphism in $\CPO/{\mathbf{N}_{\omega}}$  given by $e$. 

When we apply $f_\omega^*$ to this epimorphism, we get the following morphism in $\CPO/X$:
\[
\xymatrix@C=3em@R=2em@M=0.5em{
f_\omega^*(\mathbf{N}_=) \ar[rr]^-{g} \ar[dr]_-{f_\omega^*(e)} & & f_\omega^*(\mathbf{N}_\omega) \ar[dl]^-{f_\omega^*(\id_{\mathbf{N}_\omega})}
\\
& X
}
\]
where $g : f_\omega^*(\mathbf{N}_=) \longrightarrow f_\omega^*(\mathbf{N}_\omega)$ is the result of the action of $f_\omega^*$ on the morphism $e$ in $\mathbf{N}_\omega$.
By spelling out the definition of (the chosen) pullbacks in $\CPO$, we see that 
\[
\begin{array}{c}
f_\omega^*(\mathbf{N}_=) = (\{\langle x , n \rangle \vertbar f_\omega(x) = e(n) \} , \leq) = (\{\langle x , n \rangle \vertbar \omega = n \} , \leq) = (\emptyset, =)
\\[3mm]
f_\omega^*(\mathbf{N}_\omega) = (\{\langle x , n \rangle \vertbar f_\omega(x) = n \}, \leq') = (\{\langle x , \omega \rangle \vertbar x \in \vert X \vert \}, \leq')
\end{array}
\]
from which it follows that $g : f_\omega^*(\mathbf{N}_=) \longrightarrow f_\omega^*(\mathbf{N}_\omega)$ is not an epimorphism in $\CPO/X$. 

For example, take $X$ to be the discrete cpo on the set $\{a,b\}$ and $Y$ to be the discrete cpo on the set $\{a,b,c\}$. Then, if we consider functions $h_1 : f_\omega^*(\mathbf{N}_\omega) \longrightarrow Y$ and \linebreak $h_2 : f_\omega^*(\mathbf{N}_\omega) \longrightarrow Y$, given by $h_1(a) \defeq h_2(a) \defeq a$, $h_1(b) \defeq b$, and $h_2(b) \defeq c$, we have
\[
\xymatrix@C=3em@R=3em@M=0.5em{
f_\omega^*(\mathbf{N}_=) \ar[rr]^-{g} \ar@/_1pc/[drr]_-{f_\omega^*(e)} 
& & 
f_\omega^*(\mathbf{N}_\omega) \ar[d]^-{f_\omega^*(\id_{\mathbf{N}_\omega})}
\ar@<.75ex>[rr]^-{h_1}
\ar@<-.75ex>[rr]_-{h_2}
& &
Y
\ar@/^1pc/[dll]^-{\quad a \,\mapsto\, a \,,\, b \,\mapsto\, b \,,\, c \,\mapsto\, b}
\\
& & X
}
\]
where $h_1 \comp g = h_2 \comp g$ holds vacuously. However, we do not have that $h_1 = h_2$.
\end{proof}

\subsubsection{Extension of eMLTT with general recursion}
\label{sect:extensionofeMLTTwithrecursion}

We now show how the domain-theoretic fibred adjunction model we constructed in the previous section can be used to model an extension of eMLTT with general recursion.

We note that the following discussion ought to be preceded by the definition of the interpretation of eMLTT in fibred adjunction models given in Section~\ref{sect:interpretation}. Therefore, we advise the reader to first read Section~\ref{sect:interpretation} and then the rest of this section.

\index{general recursion}
\index{type!value --!discrete --}
\index{fixed point operation}
\index{ A@$A_{\mathsf{disc}},B_{\mathsf{disc}},\ldots$ (discrete value types)}
The version of eMLTT we consider in this section includes two changes compared to the definition given in Chapter~\ref{chap:syntax}. First, we restrict the types appearing in propositional equality to those that denote continuous families of \emph{discrete} cpos. More precisely, we only allow $V =_{A_{\mathsf{disc}}} W$ where $A_{\mathsf{disc}}$ is given by the following grammar:
\[
\begin{array}{r c l}
A_{\mathsf{disc}},B_{\mathsf{disc}} & ::= & \Nat \vertbar 1 \vertbar \Sigma\, x \!:\! A_{\mathsf{disc}} .\, B_{\mathsf{disc}} \vertbar \Pi\, x \!:\! A .\, B_{\mathsf{disc}} \vertbar 0 \vertbar A_{\mathsf{disc}} + B_{\mathsf{disc}} \vertbar V =_{A_{\mathsf{disc}}} W
\end{array}
\]
Second, we extend the grammar of eMLTT's computation terms with a \emph{fixed point operation}  $\mu\, x \!:\! U\ul{C} .\, M$, with the corresponding typing rule given by
\[
\mkrule
{\cj \Gamma {\mu\, x \!:\! U\ul{C} .\, M} {\ul{C}}}
{
\lj \Gamma \ul{C}
\quad
\cj {\Gamma, x \!:\! U\ul{C}} M \ul{C}
}
\]
Further, we extend eMLTT's equational theory with a congruence equation
\[
\mkrule
{\ceq \Gamma {\mu\, x \!:\! U\ul{C} .\, M} {\mu\, x \!:\! U\ul{D} .\, N} \ul{D}}
{\ljeq \Gamma {\ul{C}} {\ul{D}}
\quad
\ceq {\Gamma, x \!:\! U\ul{C}} M N {\ul{C}}}
\]
and an equation that describes the unfolding of fixed points:
\[
\mkrule
{\ceq \Gamma {\mu\, x \!:\! U\ul{C} .\, M} {M[\thunk (\mu\, x \!:\! U\ul{C} .\, M)/x]} {\ul{C}}}
{
\lj \Gamma \ul{C}
\quad
\cj {\Gamma, x \!:\! U\ul{C}} M \ul{C}
}
\]

We note that the meta-theory we established for eMLTT in Section~\ref{sect:metatheory} straightforwardly extends to this version of eMLTT. In particular, the fixed point operation and the corresponding definitional equations are treated analogously to other computation terms and definitional equations that involve variable bindings and type annotations. 

\index{interpretation function}
\index{ @$\sem{-}$ (interpretation function)}
We can interpret this version of eMLTT in the fibred adjunction model defined in Theorem~\ref{thm:continuousfamiliesfibadjmodel} by extending the interpretation of eMLTT given in Section~\ref{sect:interpretation} with
\[
\mkrule
{
\begin{array}{l c l}
\sem{\Gamma; \mu\, x \!:\! U\ul{C} .\, M}_1 & \defeq & \id_{\sem{\Gamma}}
\\
(\sem{\Gamma; \mu\, x \!:\! U\ul{C} .\, M}_2)_\gamma(\star) & \defeq & \mu\, (c  \mapsto (\sem{\Gamma, x \!:\! U\ul{C};M}_2)_{\langle \gamma , c \rangle}(\star))
\end{array}
}
{
\begin{array}{c}
\sem{\Gamma; \ul{C}}_1 = \sem{\Gamma} \in \CPO
\qquad
\sem{\Gamma; \ul{C}}_2(\gamma) \in (\CPO^{\mathbf{T}})^{\text{EP}}
\\
\sem{\Gamma, x \!:\! U\ul{C};M}_1 = \id_{\sem{\Gamma, x : U\ul{C}}}
\qquad
(\sem{\Gamma, x \!:\! U\ul{C};M}_2)_{\langle \gamma , c \rangle} : 1 \longrightarrow U^{\mathbf{T}}(\sem{\Gamma;\ul{C}}_2(\gamma))
\end{array}
}
\]
where $c$ is an element of the set $\vert U^{\mathbf{T}}(\sem{\Gamma; \ul{C}}_2(\gamma)) \vert$.

For better readability, we use subscripts to denote the first and second components of the objects and morphisms in $\CFam(\CPO)$ and $\CFam(\CPO^{\mathbf{T}})$. This notation is analogous to the conventions we adopt in Sections~\ref{sect:fibalgeffectsmodel} and~\ref{sect:interpretingemlttwithhandlers} for $\Fam(\Set)$.

It is then straightforward to show that the soundness results presented in Section~\ref{sect:soundness} remain true for this extension of eMLTT, as discussed in detail below. 

First, the least fixed points $\mu\, (c  \mapsto (\sem{\Gamma, x \!:\! U\ul{C};M}_2)_{\langle \gamma , c \rangle}(\star))$ are guaranteed to exist because our assumptions about $\mathbf{T}$ ensure that every $U^{\mathbf{T}}(\sem{\Gamma;\ul{C}}_2(\gamma))$ is a pointed cpo. 

Next, showing that $\sem{\Gamma; \mu\, x \!:\! U\ul{C} .\, M}$ is indeed a morphism from $(\sem{\Gamma}, \gamma \mapsto 1)$ to $(\sem{\Gamma}, \gamma \mapsto U^{\mathbf{T}}(\sem{\Gamma; \ul{C}}_2(\gamma)))$ in $\CFam_{\sem{\Gamma}}(\CPO)$ amounts to showing that $\sem{\Gamma; \mu\, x \!:\! U\ul{C} .\, M}_2$ is a continuously indexed natural transformation in the sense of Definition~\ref{def:catofcontfamilies}. 

For the naturality of $\sem{\Gamma; \mu\, x \!:\! U\ul{C} .\, M}_2$, we have to prove that $\gamma_1 \leq_{\sem{\Gamma}} \gamma_2$ implies
\[
\begin{array}{c}
(U^{\mathbf{T}} \comp \sem{\Gamma;\ul{C}}_2)(\gamma_1 \leq_{\sem{\Gamma}} \gamma_2)^e\big(\mu\, (c_1 \mapsto (\sem{\Gamma, x \!:\! U\ul{C};M}_2)_{\langle \gamma_1 , c_1 \rangle}(\star) )\big)
\\[1.5mm]
\leq_{U^{\mathbf{T}}(\sem{\Gamma;\ul{C}}_2(\gamma_2))}
\\[1.5mm]
\mu\, (c_2 \mapsto (\sem{\Gamma, x \!:\! U\ul{C};M}_2)_{\langle \gamma_2 , c_2 \rangle}(\star))
\end{array}
\]

We prove this inequation by first recalling a standard result in domain theory that the least fixed point operation is itself continuous, e.g., see~\cite[Section~2]{Plotkin:PisaNotes}. Then, using the fact that $\sem{\Gamma, x \!:\! U\ul{C};M}$ is assumed to be a morphism in $\CFam_{\sem{\Gamma, x : U\ul{C}}}(\CPO)$, i.e., \linebreak $\sem{\Gamma, x \!:\! U\ul{C};M}_2$ is natural in the sense of Definition~\ref{def:catofcontfamilies}, we get the inequation
\[
\hspace{-0.05cm}
\begin{array}{c}
\mu\, \big(c_2 \mapsto (U^{\mathbf{T}} \comp \sem{\Gamma;\ul{C}}_2)(\gamma_1 \leq_{\sem{\Gamma}} \gamma_2)^e\big((\sem{\Gamma, x \!:\! U\ul{C};M}_2)_{\langle \gamma_1 , (U^{\mathbf{T}} \comp \sem{\Gamma;\ul{C}}_2)(\gamma_1 \leq_{\sem{\Gamma}} \gamma_2)^p(c_1) \rangle}(\star)\big)\big)
\\[1.5mm]
\leq_{U^{\mathbf{T}}(\sem{\Gamma;\ul{C}}_2(\gamma_2))}
\\[1.5mm]
\mu\, (c_2 \mapsto (\sem{\Gamma, x \!:\! U\ul{C};M}_2)_{\langle \gamma_2 , c_2 \rangle}(\star))
\end{array}
\]
meaning that we are left with proving that the following inequation holds:
\[
\hspace{-0.05cm}
\begin{array}{c}
(U^{\mathbf{T}} \comp \sem{\Gamma;\ul{C}}_2)(\gamma_1 \leq_{\sem{\Gamma}} \gamma_2)^e\big(\mu\, (c_1 \mapsto (\sem{\Gamma, x \!:\! U\ul{C};M}_2)_{\langle \gamma_1 , c_1 \rangle}(\star) )\big)
\\[1.5mm]
\leq_{U^{\mathbf{T}}(\sem{\Gamma;\ul{C}}_2(\gamma_2))}
\\[1.5mm]
\mu\, \big(c_2 \mapsto (U^{\mathbf{T}} \comp \sem{\Gamma;\ul{C}}_2)(\gamma_1 \leq_{\sem{\Gamma}} \gamma_2)^e\big((\sem{\Gamma, x \!:\! U\ul{C};M}_2)_{\langle \gamma_1 , (U^{\mathbf{T}} \comp \sem{\Gamma;\ul{C}}_2)(\gamma_1 \leq_{\sem{\Gamma}} \gamma_2)^p(c_1) \rangle}(\star)\big)\big)
\end{array}
\]

We prove this last inequation below, using a natural deduction style presentation. In order to improve the readability of this proof, we use the following auxiliary notation:
\[
\begin{array}{c}
E \defeq (U^{\mathbf{T}} \comp \sem{\Gamma;\ul{C}}_2)(\gamma_1 \leq_{\sem{\Gamma}} \gamma_2)^e
\\[1.5mm]
P \defeq (U^{\mathbf{T}} \comp \sem{\Gamma;\ul{C}}_2)(\gamma_1 \leq_{\sem{\Gamma}} \gamma_2)^p
\\[1.5mm]
f\, \langle \gamma , c \rangle \defeq (\sem{\Gamma, x \!:\! U\ul{C};M}_2)_{\langle \gamma , c \rangle}(\star)
\end{array}
\]
and omit the subscripts on $\leq$. The above inequation is then proved as follows:
\[
\mkrulelabel
{
E\big(\mu\, (c_1 \mapsto f\, \langle \gamma_1 , c_1 \rangle)\big) 
\leq 
\mu\, (c_2 \mapsto E(f\, \langle \gamma_1, P(c_2) \rangle))
}
{
\mkrulelabel
{
E\big(\mu\, (c_1 \mapsto f\, \langle \gamma_1 , c_1 \rangle)\big) 
\leq 
E\big(f\, \big\langle \gamma_1 , P(\mu\, (c_2 \mapsto E(f\, \langle \gamma_1, P(c_2) \rangle))) \big\rangle\big)
}
{
\mkrulelabel
{
\mu\, (c_1 \mapsto f\, \langle \gamma_1 , c_1 \rangle)
\leq 
f\, \big\langle \gamma_1 , P(\mu\, (c_2 \mapsto E(f\, \langle \gamma_1, P(c_2) \rangle))) \big\rangle
}
{
\mkrulelabel
{
\begin{array}{c}
f\, \big\langle \gamma_1 , f\, \big\langle \gamma_1 , P\big(\mu\, (c_2 \mapsto E(f\, \langle \gamma_1, P(c_2) \rangle))\big) \big\rangle \big\rangle
\\[-1.5mm]
\leq
\\[-1.5mm] 
f\, \big\langle \gamma_1 , P(\mu\, (c_2 \mapsto E(f\, \langle \gamma_1, P(c_2) \rangle))) \big\rangle
\end{array}
}
{
\mkrulelabel
{
\begin{array}{c}
f\, \big\langle \gamma_1 , f\, \big\langle \gamma_1 , P\big(\mu\, (c_2 \mapsto E(f\, \langle \gamma_1, P(c_2) \rangle))\big) \big\rangle \big\rangle
\\[-1.5mm]
\leq
\\[-1.5mm] 
f\, \big\langle \gamma_1 , P\big(E\big(f\, \big\langle \gamma_1, P\big(\mu\, (c_2 \mapsto E(f\, \langle \gamma_1, P(c_2) \rangle))\big) \big\rangle\big)\big) \big\rangle
\end{array}
}
{
\begin{array}{c}
f\, \big\langle \gamma_1 , f\, \big\langle \gamma_1 , P\big(\mu\, (c_2 \mapsto E(f\, \langle \gamma_1, P(c_2) \rangle))\big) \big\rangle \big\rangle
\\[-1.5mm]
=
\\[-1.5mm] 
f\, \big\langle \gamma_1 , f\, \big\langle \gamma_1, P\big(\mu\, (c_2 \mapsto E(f\, \langle \gamma_1, P(c_2) \rangle))\big) \big\rangle \big\rangle
\end{array}
}
{(e)}
}
{(d)}
}
{(c)}
}
{(b)}
}
{(a)}
\]
where $(a)$, $(c)$, and $(d)$ follow from properties of least fixed points; $(b)$ holds because $E$ is  monotone; and $(e)$ holds because $E$ and $P$ form an embedding-projection pair.

For showing that $\sem{\Gamma; \mu\, x \!:\! U\ul{C} .\, M}_2$ is continuously indexed, we have to prove  
\[
\begin{array}{c}
\mu\, (c \mapsto (\sem{\Gamma, x \!:\! U\ul{C};M}_2)_{\langle \bigvee_{\!n}\! \gamma_n , c \rangle}(\star))
\\
=
\\
\bigvee_{\!n}\, \big((U^{\mathbf{T}} \comp \sem{\Gamma;\ul{C}}_2)(\gamma_n \leq_{\sem{\Gamma}} \bigvee_{\!n} \gamma_n)^e\big(\mu\, (c_n \mapsto (\sem{\Gamma, x \!:\! U\ul{C};M}_2)_{\langle \gamma_n , c_n \rangle}(\star))\big)\big)
\end{array}
\]

We prove this equation by again using the fact that the least fixed point operation is itself continuous and the fact that $\sem{\Gamma, x \!:\! U\ul{C};M}$ is assumed to be a morphism in $\CFam_{\sem{\Gamma, x : U\ul{C}}}(\CPO)$. These observations give us the following equations:
\[
\begin{array}{c}
\mu\, (c \mapsto (\sem{\Gamma, x \!:\! U\ul{C};M}_2)_{\langle \bigvee_{\!n}\! \gamma_n , c \rangle}(\star))
\\[3mm]
=
\\[-3mm]
\hspace{-13.25cm}
\mu\, \big(c \mapsto 
\\
\hspace{0.25cm}
\bigvee_{\!n}\, (U^{\mathbf{T}} \comp \sem{\Gamma;\ul{C}}_2)(\gamma_n \leq_{\sem{\Gamma}} \bigvee_{\!n} \gamma_n)^e\big((\sem{\Gamma, x \!:\! U\ul{C};M}_2)_{\langle \gamma_n , (U^{\mathbf{T}} \comp \sem{\Gamma;\ul{C}}_2)(\gamma_n \leq_{\sem{\Gamma}} \bigvee_{\!n} \gamma_n)^p(c) \rangle}(\star)\big)\big)
\\[3mm]
=
\\[-3mm]
\hspace{-12.65cm}
\bigvee_{\!n} \big(\mu\, \big(c \mapsto 
\\
\hspace{0.6cm}
(U^{\mathbf{T}} \comp \sem{\Gamma;\ul{C}}_2)(\gamma_n \leq_{\sem{\Gamma}} \bigvee_{\!n} \gamma_n)^e\big((\sem{\Gamma, x \!:\! U\ul{C};M}_2)_{\langle \gamma_n , (U^{\mathbf{T}} \comp \sem{\Gamma;\ul{C}}_2)(\gamma_n \leq_{\sem{\Gamma}} \bigvee_{\!n} \gamma_n)^p(c) \rangle}(\star)\big)\big)\big)
\end{array}
\]
meaning that we are left with showing that the following equation holds for all $n$:
\[
\begin{array}{c}
\hspace{-13.25cm}
\mu\, \big(c \mapsto 
\\
\hspace{0.75cm}
(U^{\mathbf{T}} \comp \sem{\Gamma;\ul{C}}_2)(\gamma_n \leq_{\sem{\Gamma}} \bigvee_{\!n} \gamma_n)^e\big((\sem{\Gamma, x \!:\! U\ul{C};M}_2)_{\langle \gamma_n , (U^{\mathbf{T}} \comp \sem{\Gamma;\ul{C}}_2)(\gamma_n \leq_{\sem{\Gamma}} \bigvee_{\!n} \gamma_n)^p(c) \rangle}(\star)\big)\big)
\\
=
\\
(U^{\mathbf{T}} \comp \sem{\Gamma;\ul{C}}_2)(\gamma_n \leq_{\sem{\Gamma}} \bigvee_{\!n} \gamma_n)^e\big(\mu\, (c_n \mapsto (\sem{\Gamma, x \!:\! U\ul{C};M}_2)_{\langle \gamma_n , c_n \rangle}(\star))\big)
\end{array}
\]

We prove this last equation by showing that we have inequations in both directions. Similarly to the naturality proof, we use auxiliary notation in this proof, given by
\[
\begin{array}{c}
E \defeq (U^{\mathbf{T}} \comp \sem{\Gamma;\ul{C}}_2)(\gamma_n \leq_{\sem{\Gamma}} \bigvee_{\!n} \gamma_n)^e
\\[1.5mm]
P \defeq (U^{\mathbf{T}} \comp \sem{\Gamma;\ul{C}}_2)(\gamma_n \leq_{\sem{\Gamma}} \bigvee_{\!n} \gamma_n)^p
\\[1.5mm]
f\, \langle \gamma , c \rangle \defeq (\sem{\Gamma, x \!:\! U\ul{C};M}_2)_{\langle \gamma , c \rangle}(\star)
\end{array}
\]

In the left-to-right direction, we have
\[
\mkrulelabel
{
\mu\, (c \mapsto E(f\, \langle \gamma_n, P(c) \rangle))
\leq 
E\big(\mu\, (c_n \mapsto f\, \langle \gamma_n , c_n \rangle)\big) 
}
{
\mkrulelabel
{
E\big(f\, \big\langle \gamma_n, P\big(E\big(\mu\, (c_n \mapsto f\, \langle \gamma_n , c_n \rangle)\big)\big) \big\rangle\big)
\leq 
E\big(\mu\, (c_n \mapsto f\, \langle \gamma_n , c_n \rangle)\big) 
}
{
\mkrulelabel
{
E\big(f\, \big\langle \gamma_n, \mu\, (c_n \mapsto f\, \langle \gamma_n , c_n \rangle) \big\rangle\big)
\leq 
E\big(\mu\, (c_n \mapsto f\, \langle \gamma_n , c_n \rangle)\big) 
}
{
\mkrulelabel
{
f\, \big\langle \gamma_n, \mu\, (c_n \mapsto f\, \langle \gamma_n , c_n \rangle) \big\rangle
\leq 
\mu\, (c_n \mapsto f\, \langle \gamma_n , c_n \rangle)
}
{
f\, \big\langle \gamma_n, \mu\, (c_n \mapsto f\, \langle \gamma_n , c_n \rangle) \big\rangle
=
f\, \big\langle \gamma_n, \mu\, (c_n \mapsto f\, \langle \gamma_n , c_n \rangle) \big\rangle
}
{(d)}
}
{(c)}
}
{(b)}
}
{(a)}
\]
where $(a)$ and $(d)$ follow from the properties of least fixed points; $(b)$ holds because $E$ and $P$ form an embedding-projection pair; and $(c)$ holds because $E$ is monotone. 

In the right-to-left direction, we have to prove that the following inequation holds:
\[
E\big(\mu\, (c_n \mapsto f\, \langle \gamma_n , c_n \rangle)\big) 
\leq 
\mu\, (c \mapsto E(f\, \langle \gamma_n, P(c) \rangle))
\]
As the proof of this inequation has the same structure as the proof of the corresponding inequation in the earlier naturality proof for $\sem{\Gamma; \mu\, x \!:\! U\ul{C} .\, M}_2$, we omit its proof here.

Finally, it is easy to see that the interpretation validates the fixed point unfolding equation, namely, because we have interpreted $\mu\, x \!:\! U\ul{C} .\, M$ using a least fixed point.


\chapter{Denotational semantics of eMLTT}
\label{chap:interpretation}

In this chapter we show how to interpret eMLTT in the fibred adjunction models we defined in Chapter~\ref{chap:fibadjmodels}; and we prove that this interpretation is sound and complete.

\section{Interpreting eMLTT in fibred adjunction models}
\label{sect:interpretation}

Following the standard approach in the literature on dependently typed languages, e.g., as advocated by Streicher~\cite{Streicher:Semantics} and Hoffmann~\cite{Hofmann:Thesis}, we define the interpretation function as an \emph{a priori} partial mapping $\sem{-}$ from the raw (i.e., not necessarily well-formed) expressions of eMLTT into a given fibred adjunction model. It is only afterwards that we prove in the soundness theorem that $\sem{-}$ is defined on well-formed expressions and, furthermore, that it validates the equational theory of eMLTT. In order to be able to define the interpretation function as a partial mapping, we have decorated the syntax of eMLTT with a range of additional type annotations, as discussed in Section~\ref{sect:syntax}.

Analogously to the work of Streicher and Hoffmann, the main reason for defining $\sem{-}$ as a partial mapping is to avoid the coherence issues that arise when trying to define $\sem{-}$ directly on the derivations of well-formed expressions. In particular, as the typing derivations of eMLTT are not unique, due to the context and type conversion rules (see Section~\ref{sect:judgements}), defining the interpretation on the derivations of well-formed types and terms would require us to also simultaneously prove the coherence of the interpretation.
Furthermore, as the context and type conversion rules contain definitional equations, 
defining the interpretation on derivations would also require us to simultaneously prove that the interpretation validates the equational theory of eMLTT.

Throughout this section, we assume given a fibred adjunction model using the notation of Definition~\ref{def:fibadjmodels}, i.e., given by $p : \mathcal{V} \longrightarrow \mathcal{B}$, $q : \mathcal{C} \longrightarrow \mathcal{B}$, and $F \dashv\, U : q \longrightarrow p$.

We begin by defining a notion of size for eMLTT's expressions and value contexts.

\begin{definition}
\index{size!-- of expression}
\index{ size@$\mathsf{size}(E)$ (size of an expression $E$)}
The \emph{size} of an expression $E$, written $\mathsf{size}(E)$, is defined by recursion on the structure of $E$ as follows:
\[
\begin{array}{l c l}
\mathsf{size}(\Nat) & \defeq & 1
\\[-1.5mm]
& \ldots &
\\[1.5mm]
\mathsf{size}(x) & \defeq & 1
\\[-1.5mm]
& \ldots &
\\[1.5mm]
\mathsf{size}(\return V) & \defeq & \mathsf{size}(V) + 1
\\
\mathsf{size}(\doto M {y \!:\! A} {\ul{C}} N) & \defeq & \mathsf{size}(M) + \mathsf{size}(A) + \mathsf{size}(\ul{C}) + \mathsf{size}(N) + 1
\\[-1.5mm]
& \ldots &
\\[1.5mm]
\mathsf{size}(K(V)_{(y : A).\, \ul{C}}) & \defeq & \mathsf{size}(K) + \mathsf{size}(V) + \mathsf{size}(A) + \mathsf{size}(\ul{C}) + 1
\\
\mathsf{size}(V(K)_{\ul{C}, \ul{D}}) & \defeq & \mathsf{size}(V) + \mathsf{size}(K) + \mathsf{size}(\ul{C}) + \mathsf{size}(\ul{D}) + 1
\end{array}
\]
\end{definition}

\begin{definition}
\index{size!-- of value context}
\index{ size@$\mathsf{size}(\Gamma)$ (size of a value context $\Gamma$)}
Given a value context $\Gamma$, its \emph{size}, written $\mathsf{size}(\Gamma)$, is defined as
\[
\mathsf{size}(\diamond) \,\,\,\defeq\,\,\, 0
\qquad
\mathsf{size}(\Gamma, x \!:\! A) \,\,\,\defeq\,\,\, \mathsf{size}(\Gamma) + \mathsf{size}(A)
\]
\end{definition}

Using this notion of size, we now define the partial interpretation function $\sem{-}$.

\begin{definition}
\index{interpretation function}
\index{ @$\sem{-}$ (interpretation function)}
The \emph{a priori} partial \emph{interpretation function} $\sem{-}$ is defined by induction on the sum of the sizes of its arguments (see below) such that, if defined, it maps
\begin{itemize}
\item a value context $\Gamma$ to an object $\sem{\Gamma}$ in $\mathcal{B}$, 
\item a pair of a value context $\Gamma$ and a value type $A$ to an object $\sem{\Gamma;A}$ in $\mathcal{V}_{\sem{\Gamma}}$, 
\item a pair of a value context $\Gamma$ and a computation type $\ul{C}$ to an object $\sem{\Gamma;\ul{C}}$ in $\mathcal{C}_{\sem{\Gamma}}$, 
\item a pair of a value context $\Gamma$ and a value term $V$ to an object $A$ in $\mathcal{V}_{\sem{\Gamma}}$ and a morphism $\sem{\Gamma;V} : 1_{\sem{\Gamma}} \longrightarrow A$ in $\mathcal{V}_{\sem{\Gamma}}$, 
\item a pair of a value context $\Gamma$ and a computation term $M$ to an object $\ul{C}$ in $\mathcal{C}_{\sem{\Gamma}}$ and a morphism $\sem{\Gamma;M} : 1_{\sem{\Gamma}} \longrightarrow U(\ul{C})$ in $\mathcal{V}_{\sem{\Gamma}}$, and
\item a quadruple of a value context $\Gamma$, a computation variable $z$, a computation type \linebreak $\ul{C}$, and a homomorphism term $K$ to an object $\ul{D}$ in $\mathcal{C}_{\sem{\Gamma}}$ and a morphism \linebreak $\sem{\Gamma;z \!:\! \ul{C};K} : \sem{\Gamma;\ul{C}} \longrightarrow \ul{D}$ in $\mathcal{C}_{\sem{\Gamma}}$.
\end{itemize}

Compared to how we defined $\sem{-}$ in~\cite{Ahman:FibredEffects}, we give its definition here in natural deduction style instead of using the Kleene equality $\simeq$.
This makes it easier for the reader to follow the details of the definition, such as the domains and codomains of the interpretation of the subterms of a given term. 
Specifically, we define $\sem{-}$ using rules whose premises describe the conditions we require to hold for the corresponding conclusions to be defined. For example, in the premise of the case for function application $V(W)_{(x : A) .\, B}$, we require the application of $\sem{-}$ on the given function term $V$ to be defined and its codomain to be an application of the $\Pi_{\sem{\Gamma;A}}$-functor. 
Observe that it is precisely these kinds of conditions that make the definition of $\sem{-}$ partial. 

In terms of notation, we write $\sem{\Gamma;A} \in \mathcal{V}_{\sem{\Gamma}}$ to mean that $\sem{\Gamma;A}$ is defined and given by an object in $\mathcal{V}_{\sem{\Gamma}}$, and similarly for computation types. Analogously, we write $\sem{\Gamma;V} : 1_{\sem{\Gamma}} \longrightarrow A$ to mean that $\sem{\Gamma;V}$ is defined and given by a morphism $1_{\sem{\Gamma}} \longrightarrow A$ in $\mathcal{V}_{\sem{\Gamma}}$, for some object $A$, and  similarly for computation and homomorphism terms.

To improve the readability of the definition of $\sem{-}$, we leave some premises implicit if they can be inferred from others. For example, when we write $\sem{\Gamma;V} : 1_{\sem{\Gamma}} \longrightarrow \sem{\Gamma;A}$ in the premise of a rule, we implicitly assume that $\sem{\Gamma} \in \mathcal{B}$ and $\sem{\Gamma;A} \in \mathcal{V}_{\sem{\Gamma}}$.

We now give the rules that define $\sem{-}$.

\paragraph*{Value contexts}
\mbox{}\\
\[
{\sem{\diamond} \defeq 1}
\qquad
\mkrule
{\sem{\Gamma, x \!:\! A} \defeq \ia {\sem{\Gamma;A}}}
{
\begin{array}{c}
\sem{\Gamma;A} \in \mathcal{V}_{\sem{\Gamma}} \quad x \not\in V\!ars(\Gamma)
\end{array}
}
\]
\end{definition}

\paragraph*{Value types}
\mbox{}\\
\[
\mkrule
{\sem{\Gamma;\Nat} \defeq !^*_{\sem{\Gamma}}(\mathbb{N})}
{
\begin{array}{c}
\sem{\Gamma} \in \mathcal{B}
\end{array}
}
\qquad
\mkrule
{\sem{\Gamma;1} \defeq 1_{\sem{\Gamma}}}
{
\begin{array}{c}
\sem{\Gamma} \in \mathcal{B}
\end{array}
}
\]

\vspace{0.05cm}

\[
\mkrule
{\sem{\Gamma;\Sigma\, x \!:\! A .\, B} \defeq \Sigma_{\sem{\Gamma;A}} (\sem{\Gamma, x \!:\! A;B})}
{
\begin{array}{c}
\sem{\Gamma;A} \in \mathcal{V}_{\sem{\Gamma}} \quad \sem{\Gamma, x \!:\! A;B} \in \mathcal{V}_{\ia {\sem{\Gamma;A}}}
\end{array}
}
\]

\vspace{0.05cm}

\[
\mkrule
{\sem{\Gamma;\Pi\, x \!:\! A .\, B} \defeq \Pi_{\sem{\Gamma;A}} (\sem{\Gamma, x \!:\! A;B})}
{
\begin{array}{c}
\sem{\Gamma;A} \in \mathcal{V}_{\sem{\Gamma}} \quad \sem{\Gamma, x \!:\! A;B} \in \mathcal{V}_{\ia {\sem{\Gamma;A}}}
\end{array}
}
\]

\vspace{0.05cm}

\[
\mkrule
{\sem{\Gamma;0} \defeq 0_{\sem{\Gamma}}}
{
\begin{array}{c}
\sem{\Gamma} \in \mathcal{B}
\end{array}
}
\qquad
\mkrule
{\sem{\Gamma;A + B} \defeq \sem{\Gamma;A} +_{\sem{\Gamma}} \sem{\Gamma;B}}
{
\begin{array}{c}
\sem{\Gamma;A} \in \mathcal{V}_{\sem{\Gamma}} \quad \sem{\Gamma;B} \in \mathcal{V}_{\sem{\Gamma}}
\end{array}
}
\]

\vspace{-0.25cm}

\[
\mkrule
{\sem{\Gamma; V =_A W} \defeq h^*(\Id_{\sem{\Gamma;A}})}
{
\begin{array}{c}
\sem{\Gamma;V} : 1_{\sem{\Gamma}} \longrightarrow \sem{\Gamma;A} \quad \sem{\Gamma;W} : 1_{\sem{\Gamma}} \longrightarrow \sem{\Gamma;A}
\end{array}
}
\]
where $h$ is the unique mediating morphism in the following pullback situation:
\[
\xymatrix@C=5em@R=6em@M=0.5em{
\sem{\Gamma} \ar@/_2pc/[dr]_{\mathsf{s}(\sem{\Gamma;V})} \ar@/^3.5pc/[rr]^{\mathsf{s}(\sem{\Gamma;W})} \ar@{-->}[r]_-{h} & \ia {\pi^*_{\sem{\Gamma;A}}(\sem{\Gamma;A})} \ar[d]_{\pi_{\pi^*_{\sem{\Gamma;A}}({\sem{\Gamma;A}})}}^<{\,\,\,\big\lrcorner} \ar[r]_-{\ia {\overline{\pi_{\sem{\Gamma;A}}}({\sem{\Gamma;A}})}} & \ia {\sem{\Gamma;A}} \ar[d]^{\pi_{\sem{\Gamma;A}}}_{\dcomment{\mathcal{P}(\overline{\pi_{\sem{\Gamma;A}}}({\sem{\Gamma;A}}))}\quad\,\,\,\,\,\,\,\,\,}
\\
& \ia {\sem{\Gamma;A}} \ar[r]_-{\pi_{\sem{\Gamma;A}}} & \sem{\Gamma}
}
\]

\vspace{0.05cm}

\[
\mkrule
{\sem{\Gamma;U\ul{C}} \defeq U(\sem{\Gamma;\ul{C}})}
{
\begin{array}{c}
\sem{\Gamma;\ul{C}} \in \mathcal{C}_{\sem{\Gamma}}
\end{array}
}
\]

\vspace{-0.25cm}

\[
\mkrule
{\sem{\Gamma;\ul{C} \multimap \ul{D}} \defeq \sem{\Gamma;\ul{C}} \multimap_{\sem{\Gamma}} \sem{\Gamma;\ul{D}}}
{
\begin{array}{c}
\sem{\Gamma;\ul{C}} \in \mathcal{C}_{\sem{\Gamma}} \quad \sem{\Gamma;\ul{D}} \in \mathcal{C}_{\sem{\Gamma}}
\end{array}
}
\]

\paragraph*{Computation types}
\mbox{}\\
\[
\mkrule
{\sem{\Gamma;FA} \defeq F(\sem{\Gamma;A})}
{
\begin{array}{c}
\sem{\Gamma;A} \in \mathcal{V}_{\sem{\Gamma}}
\end{array}
}
\]

\vspace{-0.25cm}

\[
\mkrule
{\sem{\Gamma;\Sigma\, x \!:\! A .\, \ul{C}} \defeq \Sigma_{\sem{\Gamma;A}} (\sem{\Gamma, x \!:\! A;\ul{C}})}
{
\begin{array}{c}
\sem{\Gamma;A} \in \mathcal{V}_{\sem{\Gamma}} \quad \sem{\Gamma, x \!:\! A;\ul{C}} \in \mathcal{C}_{\ia {\sem{\Gamma;A}}}
\end{array}
}
\]

\vspace{-0.25cm}

\[
\mkrule
{\sem{\Gamma;\Pi\, x \!:\! A .\, \ul{C}} \defeq \Pi_{\sem{\Gamma;A}} (\sem{\Gamma, x \!:\! A;\ul{C}})}
{
\begin{array}{c}
\sem{\Gamma;A} \in \mathcal{V}_{\sem{\Gamma}} \quad \sem{\Gamma, x \!:\! A;\ul{C}} \in \mathcal{C}_{\ia {\sem{\Gamma;A}}}
\end{array}
}
\]

\paragraph*{Value variables (case 1)}
\mbox{}\\
\[
\mkrule
{
\xymatrix@C=3em@R=2em@M=0.5em{
\txt<25pc>{$\sem{\Gamma, x \!:\! A;x} $\\$ \defeq $\\$ 1_{\ia {\sem{\Gamma;A}}}$}
\ar[dd]_-{\eta^{\Sigma_{\sem{\Gamma;A}} \,\dashv\, \pi^*_{\sem{\Gamma;A}}}_{1_{\ia {\sem{\Gamma;A}}}}}
\\
\\
\pi^*_{\sem{\Gamma;A}}(\Sigma_{\sem{\Gamma;A}}(1_{\ia {\sem{\Gamma;A}}})) 
\ar[d]_-{=}
\\
\pi^*_{\sem{\Gamma;A}}(\Sigma_{\sem{\Gamma;A}}(\pi^*_{\sem{\Gamma;A}}(1_{\sem{\Gamma}}))) 
\ar[dd]_-{\pi^*_{\sem{\Gamma;A}}(\mathsf{fst})}
\\
\\
\pi^*_{\sem{\Gamma;A}}(\sem{\Gamma;A})
}
}
{
\begin{array}{c}
\sem{\Gamma;A} \in \mathcal{V}_{\sem{\Gamma}} \quad x \not\in V\!ars(\Gamma)
\end{array}
}
\]

\paragraph*{Value variables (case 2)}
\mbox{}\\
\[
\mkrule
{
\xymatrix@C=3em@R=2em@M=0.5em{
\txt<25pc>{$\sem{\Gamma_1, x \!:\! A_1, \Gamma_2, y \!:\! A_2; x} $\\$ \defeq $\\$
1_{\sem{\Gamma_1, x : A_1, \Gamma_2, y : A_2}}$}
\ar[d]_-{=}
\\
1_{\ia {\sem{\Gamma_1, x : A_1, \Gamma_2; A_2}}}
\ar[d]_-{=}
\\
\pi^*_{\sem{\Gamma_1, x : A_1, \Gamma_2;A_2}}(1_{\sem{\Gamma_1, x : A_1, \Gamma_2}}) 
\ar[dd]_-{\pi^*_{\sem{\Gamma_1, x : A_1, \Gamma_2;A_2}}(\sem{\Gamma_1, x : A_1, \Gamma_2; x})}
\\
\\
\pi^*_{\sem{\Gamma_1, x : A_1, \Gamma_2;A_2}}(B)
}
}
{
\begin{array}{c}
y \not\in V\!ars(\Gamma_1, x \!:\! A_1, \Gamma_2)
\\[1mm]
\sem{\Gamma_1, x \!:\! A_1, \Gamma_2; A_2} \in \mathcal{V}_{\sem{\Gamma_1, x : A_1, \Gamma_2}} \quad
\sem{\Gamma_1, x \!:\! A_1, \Gamma_2; x} : 1_{\sem{\Gamma_1, x : A_1, \Gamma_2}} \longrightarrow B
\end{array}
}
\]

\pagebreak

\paragraph*{Zero}
\mbox{}\\
\[
\mkrule
{\sem{\Gamma;\zero} \defeq 1_{\sem{\Gamma}} \overset{!^*_{\sem{\Gamma}}(\mathsf{zero})}{\,-\!\!\!\!-\!\!\!\!-\!\!\!\!-\!\!\!\!-\!\!\!\!-\!\!\!\!\longrightarrow\,} !^*_{\sem{\Gamma}}(\mathbb{N})}
{
\begin{array}{c}
\sem{\Gamma} \in \mathcal{B}
\end{array}
}
\]

\paragraph*{Successor}
\mbox{}\\
\[
\mkrule
{\sem{\Gamma;\succc\, V} \defeq 1_{\sem{\Gamma}} \overset{\sem{\Gamma;V}}{\,-\!\!\!\!-\!\!\!\!-\!\!\!\!-\!\!\!\!-\!\!\!\!-\!\!\!\!\longrightarrow\,} !^*_{\sem{\Gamma}}(\mathbb{N}) \overset{!^*_{\sem{\Gamma}}(\mathsf{succ})}{\,-\!\!\!\!-\!\!\!\!-\!\!\!\!-\!\!\!\!-\!\!\!\!-\!\!\!\!\longrightarrow\,} !^*_{\sem{\Gamma}}(\mathbb{N})}
{
\begin{array}{c}
\sem{\Gamma;V} : 1_{\sem{\Gamma}} \longrightarrow !^*_{\sem{\Gamma}}(\mathbb{N})
\end{array}
}
\]

\paragraph*{Primitive recursion}
\mbox{}\\
\[
\mkrule
{
\xymatrix@C=3em@R=2em@M=0.5em{
\txt<25pc>{$\sem{\Gamma;\natrec {x.\,A} {V_z} {y_1.\, y_2.\, V_s} {V}} $\\$ \defeq $\\$ 1_{\sem{\Gamma}}$} 
\ar[d]_-{=} 
\\
(\mathsf{s}(\sem{\Gamma;V}))^*(1_{\ia {!^*_{\sem{\Gamma}}(\mathbb{N})}}) 
\ar[dd]_-{(\mathsf{s}(\sem{\Gamma;V}))^*(\mathsf{i}_{\sem{\Gamma, x : \Nat;A}}(\sem{\Gamma, V_z}, \sem{\Gamma, y_1 : \Nat, y_2 : A[y_1/x];V_s}))}
\\
\\
(\mathsf{s}(\sem{\Gamma;V}))^*(\sem{\Gamma, x \!:\! \Nat;A})
}
}
{
\begin{array}{c}
\sem{\Gamma;V} : 1_{\sem{\Gamma}} \longrightarrow !^*_{\sem{\Gamma}}(\mathbb{N}) \quad \sem{\Gamma;V_z} : 1_{\sem{\Gamma}} \longrightarrow (\funsection(!_X^*(\mathsf{zero})))^*(\sem{\Gamma, x \!:\! \Nat; A}) 
\\[1mm]
\sem{\Gamma, y_1 \!:\! \Nat, y_2 \!:\! A[y_1/x];V_s} : 1_{\ia {\sem{\Gamma, x : \Nat; A}}} \longrightarrow \pi_{\sem{\Gamma, x : \Nat; A}}^*(\ia {!_X^*(\mathsf{succ})}^* ({\sem{\Gamma, x \!:\! \Nat; A}}))
\end{array}
}
\]

\paragraph*{Unit}
\mbox{}\\
\[
\mkrule
{\sem{\Gamma;\star} \defeq 1_{\sem{\Gamma}} \overset{\id_{1_{\sem{\Gamma}}}}{\,-\!\!\!\!-\!\!\!\!-\!\!\!\!-\!\!\!\!\longrightarrow\,} 1_{\sem{\Gamma}}}
{
\begin{array}{c}
\sem{\Gamma} \in \mathcal{B}
\end{array}
}
\]

\paragraph*{Pairing}
\mbox{}\\
\[
\mkrule
{
\xymatrix@C=3em@R=1.75em@M=0.5em{
\txt<25pc>{$\sem{\Gamma; \langle V , W \rangle_{(x : A) .\, B} } $\\$ \defeq $\\$ 1_{\sem{\Gamma}}$}
\ar[dd]_-{\sem{\Gamma;W}}
\\
\\
(\mathsf{s}(\sem{\Gamma;V}))^*(\sem{\Gamma, x \!:\! A; B}) 
\ar[dd]_-{(\mathsf{s}(\sem{\Gamma;V}))^*(\eta^{\Sigma_{\sem{\Gamma;A}} \,\dashv\, \pi^*_{\sem{\Gamma;A}}}_{\sem{\Gamma, x : A;B}})}
\\
\\
(\mathsf{s}(\sem{\Gamma;V}))^*(\pi^*_{\sem{\Gamma;A}}(\Sigma_{\sem{\Gamma;A}}(\sem{\Gamma, x \!:\! A; B}))) 
\ar[d]_-{=}
\\
\Sigma_{\sem{\Gamma;A}}(\sem{\Gamma, x \!:\! A; B})
}
} 
{
\begin{array}{c}
\sem{\Gamma;V} : 1_{\sem{\Gamma}} \longrightarrow \sem{\Gamma;A} \quad \sem{\Gamma;W} : 1_{\sem{\Gamma}} \longrightarrow (\mathsf{s}(\sem{\Gamma;V}))^*(\sem{\Gamma, x \!:\! A; B})
\end{array}
}
\]

\paragraph*{Pattern-matching}
\mbox{}\\
\[
\mkrule
{
\xymatrix@C=3em@R=1.75em@M=0.5em{
\txt<25pc>{$\sem{\Gamma;\pmatch V {(x_1 \!:\! A_1, x_2 \!:\! A_2)} {y.\, B} W} $\\$ \defeq $\\$ 
1_{\sem{\Gamma}}$}
\ar[d]_-{=} 
\\
(\mathsf{s}(\sem{\Gamma;V}))^*((\kappa^{-1})^*(1_{\ia {\sem{\Gamma, x_1 : A_1;A_2}}})) 
\ar[dd]_-{(\mathsf{s}(\sem{\Gamma;V}))^*((\kappa^{-1})^*(\sem{\Gamma, x_1 \!:\! A_1, x_2 \!:\! A_2;W}))}
\\
\\
(\mathsf{s}(\sem{\Gamma;V}))^*((\kappa^{-1})^*(\kappa^*(\sem{\Gamma, y \!:\! (\Sigma\, x_1 \!:\! A_1 .\, A_2);B}))) 
\ar[d]_-{=}
\\
(\mathsf{s}(\sem{\Gamma;V}))^*(\sem{\Gamma, y \!:\! (\Sigma\, x_1 \!:\! A_1 .\, A_2);B})
}
}
{
\begin{array}{c}
\sem{\Gamma;V} : 1_{\sem{\Gamma}} \longrightarrow \Sigma_{\sem{\Gamma;A_1}} (\sem{\Gamma, x_1 \!:\! A_2;A_2})
\\[1mm]
\sem{\Gamma, x_1 \!:\! A_1, x_2 \!:\! A_2;W} : 1_{\ia {\sem{\Gamma, x_1 : A_1;A_2}}} \longrightarrow \kappa_{\sem{\Gamma;A_1},\sem{\Gamma, x_1 \!:\! A_1;A_2}}^*(\sem{\Gamma, y \!:\! (\Sigma\, x_1 \!:\! A_1 .\, A_2);B})
\end{array}
}
\]
where we omit the subscripts in $\kappa_{\sem{\Gamma;A_1},\sem{\Gamma, x_1 \!:\! A_1;A_2}}$ and $\kappa_{\sem{\Gamma;A_1},\sem{\Gamma, x_1 \!:\! A_1;A_2}}^{-1}$ in the conclusion for better readability.

\paragraph*{Lambda abstraction}
\mbox{}\\
\[
\mkrule
{
\xymatrix@C=3em@R=2em@M=0.5em{
\txt<25pc>{$\sem{\Gamma; \lambda\, x \!:\! A .\, V} $\\$ \defeq $\\$ 1_{\sem{\Gamma}}$}
\ar[dd]_-{\eta^{\pi^*_{\sem{\Gamma;A}} \,\dashv\, \Pi_{\sem{\Gamma;A}}}_{1_{\sem{\Gamma}}}}
\\
\\
\Pi_{\sem{\Gamma;A}}(\pi^*_{\sem{\Gamma;A}}(1_{\sem{\Gamma}})) 
\ar[d]_-{=} 
\\
\Pi_{\sem{\Gamma;A}}(1_{\ia {\sem{\Gamma;A}}}) 
\ar[dd]_-{\Pi_{\sem{\Gamma;A}}(\sem{\Gamma, x : A;V})}
\\
\\
\Pi_{\sem{\Gamma;A}}(B)
}
}
{
\begin{array}{c}
\sem{\Gamma;A} \in \mathcal{V}_{\sem{\Gamma}} \quad \sem{\Gamma, x \!:\! A; V} : 1_{\ia {\sem{\Gamma;A}}} \longrightarrow B
\end{array}
}
\]

\paragraph*{Function application}
\mbox{}\\
\[
\mkrule
{
\xymatrix@C=3em@R=2em@M=0.5em{
\txt<25pc>{$\sem{\Gamma; V(W)_{(x : A) .\, B}} $\\$ \defeq $\\$ 1_{\sem{\Gamma}}$}
\ar[d]_-{=}
\\
(\mathsf{s}(\sem{\Gamma;W}))^*(\pi^*_{\sem{\Gamma;A}}(1_{\sem{\Gamma}})) 
\ar[dd]_-{(\mathsf{s}(\sem{\Gamma;W}))^*(\pi^*_{\sem{\Gamma;A}}(\sem{\Gamma;V}))}
\\
\\
(\mathsf{s}(\sem{\Gamma;W}))^*(\pi^*_{\sem{\Gamma;A}}(\Pi_{\sem{\Gamma;A}}(\sem{\Gamma, x \!:\! A;B}))) 
\ar[dd]_-{(\mathsf{s}(\sem{\Gamma;W}))^*(\varepsilon^{\pi^*_{\sem{\Gamma;A}} \,\dashv\, \Pi_{\sem{\Gamma;A}}}_{\sem{\Gamma, x : A;B}})}
\\
\\
(\mathsf{s}(\sem{\Gamma;W}))^*(\sem{\Gamma, x \!:\! A;B})
}
}
{
\begin{array}{c}
\sem{\Gamma;V} : 1_{\sem{\Gamma}} \longrightarrow \Pi_{\sem{\Gamma;A}}(\sem{\Gamma, x \!:\! A;B}) \quad \sem{\Gamma;W} :  1_{\sem{\Gamma}} \longrightarrow {\sem{\Gamma;A}}
\end{array}
}
\]

\paragraph*{Empty case analysis}
\mbox{}\\
\[
\mkrule
{
\xymatrix@C=3em@R=2em@M=0.5em{
\txt<25pc>{$\sem{\Gamma;\absurd {x.\,A} V} $\\$ \defeq $\\$ 1_{\sem{\Gamma}}$}
\ar[d]_-{=}
\\
(\mathsf{s}(\sem{\Gamma;V}))^*(1_{\ia {0_{\sem{\Gamma}}}}) 
\ar[dd]_-{(\mathsf{s}(\sem{\Gamma;V}))^*(?_{\sem{\Gamma, x : 0;A}})}
\\
\\
(\mathsf{s}(\sem{\Gamma;V}))^*(\sem{\Gamma, x \!:\! 0;A})
}
}
{
\begin{array}{c}
\sem{\Gamma, x \!:\! 0;A} \in \mathcal{V}_{\ia {0_{\sem{\Gamma}}}} \quad \sem{\Gamma;V} : 1_{\sem{\Gamma}} \longrightarrow 0_{\sem{\Gamma}}
\end{array}
}
\]

\paragraph*{Binary case analysis}
\mbox{}\\
\[
\mkrule
{
\xymatrix@C=3em@R=2em@M=0.5em{
\txt<25pc>{$\sem{\Gamma; \mathtt{case~} V \mathtt{~of}_{x.\,B} \mathtt{~} ({\inl {\!} {\!\!(y_1 \!:\! A_1)} \mapsto W_1} , {\inr {\!} {\!\!(y_2 \!:\! A_2)} \mapsto W_2})} $\\$ \defeq $\\$ 1_{\sem{\Gamma}}$}
\ar[d]_-{=}
\\
(\mathsf{s}(\sem{\Gamma;V}))^*(1_{\ia {\sem{\Gamma;A_1} +_{\sem{\Gamma}} \sem{\Gamma;A_2}}}) 
\ar[dd]_-{(\mathsf{s}(\sem{\Gamma;V}))^*([\sem{\Gamma, y_1 : A_1;W_1},\sem{\Gamma, y_2 : A_2;W_2}])}
\\
\\
(\mathsf{s}(\sem{\Gamma;V}))^*(\sem{\Gamma, x \!:\! A_1 + A_2;B})
}
}
{
\begin{array}{c}
\sem{\Gamma;V} : 1_{\sem{\Gamma}} \longrightarrow \sem{\Gamma;A_1} +_{\sem{\Gamma}} \sem{\Gamma;A_2} 
\\[1mm]
\sem{\Gamma, y_1 \!:\! A_1;W_1} : 1_{\ia {\sem{\Gamma;A_1}}} \longrightarrow \ia {\mathsf{inl}}^*(\sem{\Gamma, x \!:\! A_1 + A_2;B})
\\[1mm]
\sem{\Gamma, y_2 \!:\! A_2;W_2} : 1_{\ia {\sem{\Gamma;A_2}}} \longrightarrow \ia {\mathsf{inr}}^*(\sem{\Gamma, x \!:\! A_1 + A_2;B})
\end{array}
}
\]

\paragraph*{Left injection}
\mbox{}\\
\[
\mkrule
{\sem{\Gamma;\inl {A + B} V} \defeq 1_{\sem{\Gamma}} \overset{\sem{\Gamma;V}}{\,-\!\!\!\!-\!\!\!\!-\!\!\!\!-\!\!\!\!\longrightarrow\,} \sem{\Gamma;A} \overset{\mathsf{inl}}{\,-\!\!\!\!-\!\!\!\!\longrightarrow\,} \sem{\Gamma;A} +_{\sem{\Gamma}} \sem{\Gamma;B}}
{
\begin{array}{c}
\sem{\Gamma;V} : 1_{\sem{\Gamma}} \longrightarrow \sem{\Gamma;A} \quad \sem{\Gamma;B} \in \mathcal{V}_{\sem{\Gamma}}
\end{array}
}
\]

\paragraph*{Right injection}
\mbox{}\\
\[
\mkrule
{\sem{\Gamma;\inr {A + B} V} \defeq 1_{\sem{\Gamma}} \overset{\sem{\Gamma;V}}{\,-\!\!\!\!-\!\!\!\!-\!\!\!\!-\!\!\!\!\longrightarrow\,} \sem{\Gamma;B} \overset{\mathsf{inr}}{\,-\!\!\!\!-\!\!\!\!\longrightarrow\,} \sem{\Gamma;A} +_{\sem{\Gamma}} \sem{\Gamma;B}}
{
\begin{array}{c}
\sem{\Gamma;A} \in \mathcal{V}_{\sem{\Gamma}} \quad \sem{\Gamma;V} : 1_{\sem{\Gamma}} \longrightarrow \sem{\Gamma;B}
\end{array}
}
\]

\paragraph*{Reflexivity of propositional equality}
\mbox{}\\
\[
\mkrule
{
\sem{\Gamma;\refl A V} \defeq 1_{\sem{\Gamma}} \overset{=}{\,\longrightarrow\,} (\mathsf{s}(\sem{\Gamma;V}))^*(1_{\ia {A}}) \overset{(\mathsf{s}(\sem{\Gamma;V}))^*(\mathsf{r}_{A})}{\,-\!\!\!\!-\!\!\!\!-\!\!\!\!-\!\!\!\!-\!\!\!\!-\!\!\!\!-\!\!\!\!-\!\!\!\!-\!\!\!\!-\!\!\!\!\longrightarrow\,} (\mathsf{s}(\sem{\Gamma;V}))^*(\delta^*_{A}(\Id_{A}))
}
{
\begin{array}{c}
\sem{\Gamma;V} : 1_{\sem{\Gamma}} \longrightarrow A
\end{array}
}
\]

\paragraph*{Elimination of propositional equality}
\mbox{}\\
\[
\mkrule
{
\xymatrix@C=3em@R=2em@M=0.5em{
\txt<25pc>{$\sem{\Gamma;\pathind A {x_1.\, x_2.\, x_3.\, B} {y.\, W} {V_1} {V_2} {V_p}} $\\$ \defeq $\\$ 1_{\sem{\Gamma}}$}
\ar[d]_-{=}
\\
(\mathsf{s}(\sem{\Gamma;V_p}))^*(\ia {\overline{h}(\Id_{\sem{\Gamma;A}})}^*(1_{\ia {\Id_{\sem{\Gamma;A}}}})) 
\ar[dd]_-{(\mathsf{s}(\sem{\Gamma;V_p}))^*(\ia {\overline{h}(\Id_{\sem{\Gamma;A}})}^*(\mathsf{i}(\sem{\Gamma, y : A;W})))}
\\
\\
(\mathsf{s}(\sem{\Gamma;V_p}))^*(\ia {\overline{h}(\Id_{\sem{\Gamma;A}})}^*(\sem{\Gamma, x_1 \!:\! A, x_2 \!:\! A, x_3 \!:\! (x_1 =_A x_2);B}))
}
}
{
\begin{array}{c}
\sem{\Gamma;V_1} : 1_{\sem{\Gamma}} \longrightarrow \sem{\Gamma;A} 
\quad
\sem{\Gamma;V_2} : 1_{\sem{\Gamma}} \longrightarrow \sem{\Gamma;A} 
\quad
\sem{\Gamma;V_p} : 1_{\sem{\Gamma}} \longrightarrow h^*(\Id_{\sem{\Gamma;A}})
\\[1mm]
\hspace{-9.25cm}
\sem{\Gamma, y \!:\! A;W} : 1_{\ia {\Id_{\sem{\Gamma;A}}}} \longrightarrow 
\\
\hspace{3cm}
(\mathsf{s}(\mathsf{r}_{\sem{\Gamma;A}}))^*(\ia {\overline{\delta_{\sem{\Gamma;A}}}(\Id_{\sem{\Gamma;A}})}^*(\sem{\Gamma, x_1 \!:\! A, x_2 \!:\! A, x_3 \!:\! (x_1 =_A x_2);B}))
\end{array}
}
\vspace{0.1cm}
\]
where we omit the subscripts in $\mathsf{i}_{\sem{\Gamma;A},\sem{\Gamma, x_1 \!:\! A, x_2 \!:\! A, x_3 \!:\! (x_1 =_A x_2);B}}$ in the conclusion; and \linebreak

\pagebreak
\noindent
where $h$ is the unique mediating morphism in the following pullback situation:
\vspace{0.1cm}
\[
\xymatrix@C=5em@R=6em@M=0.5em{
\sem{\Gamma} \ar@/_2pc/[dr]_{\mathsf{s}(\sem{\Gamma;V_1})} \ar@/^3.5pc/[rr]^{\mathsf{s}(\sem{\Gamma;V_2})} \ar@{-->}[r]_-{h} & \ia {\pi^*_{\sem{\Gamma;A}}(\sem{\Gamma;A})} \ar[d]_{\pi_{\pi^*_{\sem{\Gamma;A}}({\sem{\Gamma;A}})}}^<{\,\,\,\big\lrcorner} \ar[r]_-{\ia {\overline{\pi_{\sem{\Gamma;A}}}({\sem{\Gamma;A}})}} & \ia {\sem{\Gamma;A}} \ar[d]^{\pi_{\sem{\Gamma;A}}}_{\dcomment{\mathcal{P}(\overline{\pi_{\sem{\Gamma;A}}}({\sem{\Gamma;A}}))}\quad\,\,\,\,\,\,\,\,\,\,\,\,\,}
\\
& \ia {\sem{\Gamma;A}} \ar[r]_-{\pi_{\sem{\Gamma;A}}} & \sem{\Gamma}
}
\]

\paragraph*{Thunking a computation}
\mbox{}\\
\[
\mkrule
{\sem{\Gamma;\thunk M} \defeq 1_{\sem{\Gamma}} \overset{\sem{\Gamma;M}}{\,-\!\!\!\!-\!\!\!\!-\!\!\!\!-\!\!\!\!\longrightarrow\,} U(\ul{C})}
{
\begin{array}{c}
\sem{\Gamma;M} : 1_{\sem{\Gamma}} \longrightarrow U(\ul{C})
\end{array}
}
\]

\paragraph*{Homomorphic lambda abstraction}
\mbox{}\\
\[
\mkrule
{\sem{\Gamma;\lambda\, z \!:\! \ul{C} .\, K} \defeq 1_{\sem{\Gamma}} \overset{\xi^{-1}_{\sem{\Gamma},\sem{\Gamma;\ul{C}},\ul{D}}(\sem{\Gamma;z : \ul{C};K})}{\,-\!\!\!\!-\!\!\!\!-\!\!\!\!-\!\!\!\!-\!\!\!\!-\!\!\!\!-\!\!\!\!-\!\!\!\!-\!\!\!\!-\!\!\!\!-\!\!\!\!-\!\!\!\!-\!\!\!\!-\!\!\!\!-\!\!\!\!\longrightarrow\,} \sem{\Gamma;\ul{C}} \multimap_{\sem{\Gamma}} \ul{D} }
{
\begin{array}{c}
\sem{\ul{C}} \in \mathcal{V}_{\sem{\Gamma}} \quad \sem{\Gamma;z \!:\! \ul{C};K} : \sem{\Gamma;\ul{C}} \longrightarrow \ul{D}
\end{array}
}
\]

\paragraph*{Returning a value}
\mbox{}\\
\[
\mkrule
{\sem{\Gamma;\return V} \defeq 1_{\sem{\Gamma}} \overset{\sem{\Gamma;V}}{\,-\!\!\!\!-\!\!\!\!-\!\!\!\!\longrightarrow\,} A \overset{\eta^{F \,\dashv\, U}_A}{\,-\!\!\!\!-\!\!\!\!-\!\!\!\!-\!\!\!\!\longrightarrow\,} U(F(A))}
{
\begin{array}{c}
\sem{\Gamma;V} : 1_{\sem{\Gamma}} \longrightarrow A
\end{array}
}
\]

\paragraph*{Sequential composition}
\mbox{}\\
\[
\mkrule
{
\xymatrix@C=3em@R=2em@M=0.5em{
\txt<25pc>{$\sem{\Gamma;\doto M {x \!:\! A} {\ul{C}} N } $\\$ \defeq $\\$ 1_{\sem{\Gamma}}$}
\ar[dd]_-{\sem{\Gamma;M}}
\\
\\
U(F(\sem{\Gamma;A})) 
\ar[dd]_-{U(F(\langle \id_{\sem{\Gamma;A}}, ! \rangle))}
\\
\\
U(F(\Sigma_{\sem{\Gamma;A}}(\pi^*_{\sem{\Gamma;A}}(1_{\sem{\Gamma}})))) 
\ar[d]_-{=} 
\\
U(F(\Sigma_{\sem{\Gamma;A}}(1_{\ia {\sem{\Gamma;A}}}))) 
\ar[dd]_-{U(F(\Sigma_{\sem{\Gamma;A}}(\sem{\Gamma, x : A;N})))}
\\
\\
U(F(\Sigma_{\sem{\Gamma;A}}(U(\pi^*_{\sem{\Gamma;A}}(\sem{\Gamma;\ul{C}}))))) 
\ar[d]_-{=}
\\
U(F(\Sigma_{\sem{\Gamma;A}}(\pi^*_{\sem{\Gamma;A}}(U(\sem{\Gamma;\ul{C}}))))) 
\ar[dd]_-{U(F(\varepsilon^{\Sigma_{\sem{\Gamma;A}} \,\dashv\, \pi^*_{\sem{\Gamma;A}}}_{U(\sem{\Gamma;\ul{C}})}))}
\\
\\
U(F(U(\sem{\Gamma;\ul{C}}))) 
\ar[dd]_-{U(\varepsilon^{F \,\dashv\, U}_{\sem{\Gamma;\ul{C}}})}
\\
\\
U(\sem{\Gamma;\ul{C}})
}
}
{
\begin{array}{c}
\sem{\Gamma;M} : 1_{\sem{\Gamma}} \longrightarrow \sem{\Gamma;A} \quad \sem{\Gamma, x \!:\! A;N} : 1_{\ia {\sem{\Gamma;A}}} \longrightarrow U(\pi^*_{\sem{\Gamma;A}}(\sem{\Gamma;\ul{C}}))
\end{array}
}
\]

\pagebreak

\pagebreak

\paragraph*{Computational pairing}
\mbox{}\\
\[
\mkrule
{
\xymatrix@C=3em@R=1.6em@M=0.5em{
\txt<25pc>{$\sem{\Gamma;\langle V , M \rangle_{(x : A).\, \ul{C}}} $\\$ \defeq $\\$ 1_{\sem{\Gamma}}$}
\ar[dd]_-{\sem{\Gamma;M}}
\\
\\
U((\mathsf{s}(\sem{\Gamma;V}))^*(\sem{\Gamma, x \!:\! A;\ul{C}})) 
\ar[dd]_-{U((\mathsf{s}(\sem{\Gamma;V}))^*(\eta^{\Sigma_{\sem{\Gamma;A}} \,\dashv\, \pi^*_{\sem{\Gamma;A}}}_{\sem{\Gamma, x \!:\! A;\ul{C}}}))}
\\
\\
U((\mathsf{s}(\sem{\Gamma;V}))^*(\pi^*_{\sem{\Gamma;A}}(\Sigma_{\sem{\Gamma;A}}(\sem{\Gamma, x \!:\! A;\ul{C}})))) 
\ar[d]_-{=}
\\
U(\Sigma_{\sem{\Gamma;A}}(\sem{\Gamma, x \!:\! A;\ul{C}}))
}
}
{
\begin{array}{c}
\sem{\Gamma;V} : 1_{\sem{\Gamma}} \longrightarrow \sem{\Gamma;A} \quad \sem{\Gamma;M} : 1_{\sem{\Gamma}} \longrightarrow U((\mathsf{s}(\sem{\Gamma;V}))^*(\sem{\Gamma, x \!:\! A;\ul{C}}))
\end{array}
}
\]

\paragraph*{Computational pattern-matching}
\mbox{}\\
\[
\mkrule
{
\xymatrix@C=3em@R=1.6em@M=0.5em{
\txt<25pc>{$\sem{\Gamma;\doto M {(x \!:\! A, z \!:\! \ul{C})} {\ul{D}} K} $\\$ \defeq $\\$ 1_{\sem{\Gamma}}$} 
\ar[dd]_-{\sem{\Gamma;M}}
\\
\\
U(\Sigma_{\sem{\Gamma;A}}(\sem{\Gamma, x \!:\! A;\ul{C}})) 
\ar[dd]_-{U(\Sigma_{\sem{\Gamma;A}}(\sem{\Gamma, x : A;z : \ul{C};K}))}
\\
\\
U(\Sigma_{\sem{\Gamma;A}}(\pi^*_{\sem{\Gamma;A}}(\sem{\Gamma;\ul{D}}))) 
\ar[dd]_-{U(\varepsilon^{\Sigma_{\sem{\Gamma;A}} \,\dashv\, \pi^*_{\sem{\Gamma;A}}}_{\sem{\Gamma;\ul{D}}})}
\\
\\
U(\sem{\Gamma;\ul{D}})
}
}
{
\begin{array}{c}
\sem{\Gamma;M} : 1_{\sem{\Gamma}} \longrightarrow U(\Sigma_{\sem{\Gamma;A}}(\sem{\Gamma, x \!:\! A;\ul{C}})) 
\\
\sem{\Gamma, x \!:\! A;z \!:\! \ul{C};K} : \sem{\Gamma, x \!:\! A; \ul{C}} \longrightarrow \pi^*_{\sem{\Gamma;A}}(\sem{\Gamma;\ul{D}})
\end{array}
}
\]

\paragraph*{Computational lambda abstraction}
\mbox{}\\[1cm]
\[
\mkrule
{
\xymatrix@C=3em@R=3em@M=0.5em{
\txt<25pc>{$\sem{\Gamma;\lambda \, x \!:\! A .\, M} $\\$ \defeq $\\$ 1_{\sem{\Gamma}}$}
\ar[dd]_-{\eta^{\pi^*_{\sem{\Gamma;A}} \,\dashv\, \Pi_{\sem{\Gamma;A}}}_{1_{\sem{\Gamma}}}}
\\
\\
\Pi_{\sem{\Gamma;A}}(\pi^*_{\sem{\Gamma;A}}(1_{\sem{\Gamma}})) 
\ar[d]_-{=} 
\\
\Pi_{\sem{\Gamma;A}}(1_{\ia {\sem{\Gamma;A}}}) 
\ar[dd]_-{\Pi_{\sem{\Gamma;A}}(\sem{\Gamma, x : A;M})}
\\
\\
\Pi_{\sem{\Gamma;A}}(U(\ul{C})) 
\ar[dd]_-{(\zeta^{-1}_{\Pi,\sem{\Gamma;A}})_{\ul{C}}}
\\
\\
U(\Pi_{\sem{\Gamma;A}}(\ul{C}))
}
}
{
\begin{array}{c}
\sem{\Gamma;A} \in \mathcal{V}_{\sem{\Gamma}} \quad \sem{\Gamma, x \!:\! A;M} : 1_{\ia {\sem{\Gamma;A}}} \longrightarrow U(\ul{C})
\end{array}
}
\vspace{1cm}
\]
where $\zeta_{\Pi,\sem{\Gamma;A}} : U \comp \Pi_{\sem{\Gamma;A}} \overset{\cong}{\,\longrightarrow\,} \Pi_{\sem{\Gamma;A}} \comp U$ is the natural isomorphism defined in Proposition~\ref{prop:UandFpreserveSigmaPi}.

\newpage

\paragraph*{Computational function application}
\mbox{}\\
\[
\mkrule
{
\xymatrix@C=3em@R=2em@M=0.5em{
\txt<25pc>{$\sem{\Gamma;M(V)_{(x : A).\,\ul{C}}} $\\$ \defeq $\\$ 1_{\sem{\Gamma}}  $}
\ar[d]_-{=}
\\
(\mathsf{s}(\sem{\Gamma;V}))^*(\pi^*_{\sem{\Gamma;A}}(1_{\sem{\Gamma}})) 
\ar[dd]_-{(\mathsf{s}(\sem{\Gamma;V}))^*(\pi^*_{\sem{\Gamma;A}}(\sem{\Gamma;M}))} 
\\
\\
(\mathsf{s}(\sem{\Gamma;V}))^*(\pi^*_{\sem{\Gamma;A}}(U(\Pi_{\sem{\Gamma;A}}(\sem{\Gamma, x \!:\! A; \ul{C}})))) 
\ar[d]_-{=}
\\
U((\mathsf{s}(\sem{\Gamma;V}))^*(\pi^*_{\sem{\Gamma;A}}(\Pi_{\sem{\Gamma;A}}(\sem{\Gamma, x \!:\! A; \ul{C}})))) 
\ar[dd]_-{U((\mathsf{s}(\sem{\Gamma;V}))^*(\varepsilon^{\pi^*_{\sem{\Gamma;A}} \,\dashv\, \Pi_{\sem{\Gamma;A}}}_{\sem{\Gamma, x \!:\! A; \ul{C}}}))}
\\
\\
U((\mathsf{s}(\sem{\Gamma;V}))^*(\sem{\Gamma, x \!:\! A; \ul{C}}))
}
}
{
\begin{array}{c}
\sem{\Gamma;M} : 1_{\sem{\Gamma}} \longrightarrow U(\Pi_{\sem{\Gamma;A}}(\sem{\Gamma, x \!:\! A; \ul{C}})) \quad \sem{\Gamma;V} : 1_{\sem{\Gamma}} \longrightarrow \sem{\Gamma;A}
\end{array}
}
\]

\paragraph*{Forcing a thunked computation}
\mbox{}\\
\[
\mkrule
{\sem{\Gamma;\force {\ul{C}} V} \defeq 1_{\sem{\Gamma}} \overset{\sem{\Gamma;V}}{\,-\!\!\!\!-\!\!\!\!-\!\!\!\!-\!\!\!\!\longrightarrow\,} U(\sem{\Gamma;\ul{C}})}
{
\begin{array}{c}
\sem{\Gamma;V} : 1_{\sem{\Gamma}} \longrightarrow U(\sem{\Gamma;\ul{C}})
\end{array}
}
\]

\paragraph*{Homomorphic function application}
\mbox{}\\
\[
\mkrule
{\sem{\Gamma;V(M)_{\ul{C}, \ul{D}}} \defeq 1_{\sem{\Gamma}} \overset{\sem{\Gamma;M}}{\,-\!\!\!\!-\!\!\!\!-\!\!\!\!-\!\!\!\!\longrightarrow\,} U(\sem{\Gamma;\ul{C}}) \overset{U(\xi_{\sem{\Gamma},\sem{\Gamma;\ul{C}},\sem{\Gamma;\ul{D}}}(\sem{\Gamma;V}))}{\,-\!\!\!\!-\!\!\!\!-\!\!\!\!-\!\!\!\!-\!\!\!\!-\!\!\!\!-\!\!\!\!-\!\!\!\!-\!\!\!\!-\!\!\!\!-\!\!\!\!-\!\!\!\!-\!\!\!\!-\!\!\!\!-\!\!\!\!-\!\!\!\!\longrightarrow\,} U(\sem{\Gamma;\ul{D}})}
{
\begin{array}{c}
\sem{\Gamma;V} : 1_{\sem{\Gamma}} \longrightarrow \sem{\Gamma;\ul{C}} \multimap_{\sem{\Gamma}} \sem{\Gamma;\ul{D}} \quad \sem{\Gamma;M} : 1_{\sem{\Gamma}} \longrightarrow U(\sem{\Gamma;\ul{C}})
\end{array}
}
\]

\paragraph*{Computation variables}
\mbox{}\\
\[
\mkrule
{\sem{\Gamma;z \!:\! \ul{C};z} \defeq \sem{\Gamma;\ul{C}} \overset{\id_{\sem{\Gamma;\ul{C}}}}{\,-\!\!\!\!-\!\!\!\!-\!\!\!\!-\!\!\!\!-\!\!\!\!\longrightarrow\,} \sem{\Gamma;\ul{C}}}
{
\begin{array}{c}
\sem{\Gamma;\ul{C}} \in \mathcal{C}_{\sem{\Gamma}}
\end{array}
}
\]

\pagebreak

\paragraph*{Sequential composition}
\mbox{}\\[0.3cm]
\[
\mkrule
{
\xymatrix@C=3em@R=2em@M=0.5em{
\txt<25pc>{$\sem{\Gamma;z \!:\! \ul{C};\doto K {x \!:\! A} {\ul{D}} M} $\\$ \defeq $\\$ \sem{\Gamma;\ul{C}}$}
\ar[dd]_-{\sem{\Gamma;z : \ul{C};K}}
\\
\\
F(\sem{\Gamma;A}) 
\ar[dd]_-{F(\langle \id_{\sem{\Gamma;A}} , !\rangle)}
\\
\\
F(\Sigma_{\sem{\Gamma;A}}(\pi^*_{\sem{\Gamma;A}}(1_{\sem{\Gamma}}))) 
\ar[d]_-{=}
\\
F(\Sigma_{\sem{\Gamma;A}}(1_{\ia {\sem{\Gamma;A}}})) 
\ar[dd]_-{F(\Sigma_{\sem{\Gamma;A}}(\sem{\Gamma, x : A ; M}))}
\\
\\
F(\Sigma_{\sem{\Gamma;A}}(U(\pi^*_{\sem{\Gamma;A}}(\sem{\Gamma;\ul{D}}))))
\ar[d]_-{=}
\\
F(\Sigma_{\sem{\Gamma;A}}(\pi^*_{\sem{\Gamma;A}}(U(\sem{\Gamma;\ul{D}})))) 
\ar[dd]_-{F(\varepsilon^{\Sigma_{\sem{\Gamma;A}} \,\dashv\, \pi^*_{\sem{\Gamma;A}}}_{U(\sem{\Gamma;\ul{D}})})}
\\
\\
F(U(\sem{\Gamma;\ul{D}})) 
\ar[dd]_-{\varepsilon^{F \,\dashv\, U}_{\sem{\Gamma;\ul{D}}}}
\\
\\
\sem{\Gamma;\ul{D}}
}
}
{
\begin{array}{c}
\sem{\Gamma;\ul{C}} \in \mathcal{C}_{\sem{\Gamma}} 
\quad
\sem{\Gamma;z \!:\! \ul{C};K} : \sem{\Gamma;\ul{C}} \longrightarrow F(\sem{\Gamma;A}) 
\\
\sem{\Gamma, x \!:\! A ; M} : 1_{\ia {\sem{\Gamma;A}}} \longrightarrow U(\pi^*_{\sem{\Gamma;A}}(\sem{\Gamma;\ul{D}}))
\end{array}
}
\]

\newpage

\paragraph*{Computational pairing}
\mbox{}\\
\[
\mkrule
{
\xymatrix@C=3em@R=1.5em@M=0.5em{
\txt<25pc>{$\sem{\Gamma;z \!:\! \ul{C};\langle V , K \rangle_{(x : A) . \ul{D}}} $\\$ \defeq $\\$ \sem{\Gamma;\ul{C}}$} 
\ar[dd]_-{\sem{\Gamma;z : \ul{C};K}}
\\
\\
(\mathsf{s}(\sem{\Gamma;V}))^*(\sem{\Gamma, x \!:\! A;\ul{D}}) 
\ar[dd]_-{(\mathsf{s}(\sem{\Gamma;V}))^*(\eta^{\Sigma_{\sem{\Gamma;A}} \,\dashv\, \pi^*_{\sem{\Gamma;A}}}_{\sem{\Gamma, x : A;\ul{D}}})}
\\
\\
(\mathsf{s}(\sem{\Gamma;V}))^*(\pi^*_{\sem{\Gamma;A}}(\Sigma_{\sem{\Gamma;A}}(\sem{\Gamma, x \!:\! A;\ul{D}}))) 
\ar[d]_-{=} 
\\
\Sigma_{\sem{\Gamma;A}}(\sem{\Gamma, x \!:\! A;\ul{D}})
}
}
{
\begin{array}{c}
\sem{\Gamma;\ul{C}} \in \mathcal{C}_{\sem{\Gamma}}
\quad
\sem{\Gamma;V} : 1_{\sem{\Gamma}} \longrightarrow \sem{\Gamma;A} 
\\
\sem{\Gamma;z \!:\! \ul{C};K} : \sem{\Gamma;\ul{C}} \longrightarrow (\mathsf{s}(\sem{\Gamma;V}))^*(\sem{\Gamma, x \!:\! A;\ul{D}})
\end{array}
}
\]

\paragraph*{Computational pattern-matching}
\mbox{}\\
\[
\mkrule
{
\xymatrix@C=3em@R=1.5em@M=0.5em{
\txt<25pc>{$\sem{\Gamma;z_1 \!:\! \ul{C};\doto K {(x \!:\! A, z_2 \!:\! \ul{D}_1)} {\ul{D}_2} L} $\\$ \defeq $\\$ \sem{\Gamma;\ul{C}}$}
\ar[dd]_-{\sem{\Gamma;z_1 : \ul{C};K}}
\\
\\
\Sigma_{\sem{\Gamma;A}}(\sem{\Gamma, x \!:\! A; \ul{D}_1}) 
\ar[dd]_-{\Sigma_{\sem{\Gamma;A}}(\sem{\Gamma, x : A; z_2 : \ul{D}_1;L})}
\\
\\
\Sigma_{\sem{\Gamma;A}}(\pi^*_{\sem{\Gamma;A}}(\sem{\Gamma;\ul{D}_2})) 
\ar[dd]_-{\varepsilon^{\Sigma_{\sem{\Gamma;A}} \,\dashv\, \pi^*_{\sem{\Gamma;A}}}_{\sem{\Gamma;\ul{D}_2}}}
\\
\\
\sem{\Gamma;\ul{D}_2}
}
} 
{
\begin{array}{c}
\sem{\Gamma;\ul{C}} \in \mathcal{C}_{\sem{\Gamma}} \quad 
\sem{\Gamma;z_1 \!:\! \ul{C};K} : \sem{\Gamma;\ul{C}} \longrightarrow \Sigma_{\sem{\Gamma;A}}(\sem{\Gamma, x \!:\! A; \ul{D}_1}) 
\\
\sem{\Gamma, x \!:\! A; z_2 \!:\! \ul{D}_1;L} : \sem{\Gamma, x \!:\! A; \ul{D}_1} \longrightarrow \pi^*_{\sem{\Gamma;A}}(\sem{\Gamma;\ul{D}_2})
\end{array}
}
\]

\paragraph*{Computational lambda abstraction}
\mbox{}\\
\[
\mkrule
{
\xymatrix@C=3em@R=2em@M=0.5em{
\txt<25pc>{$\sem{\Gamma;z \!:\! \ul{C};\lambda \, x \!:\! A .\, K} $\\$ \defeq $\\$ \sem{\Gamma;\ul{C}}$}
\ar[dd]_-{\eta^{\pi^*_{\sem{\Gamma;A}} \,\dashv\, \Pi_{\sem{\Gamma;A}}}_{\sem{\Gamma;\ul{C}}}}
\\
\\
\Pi_{\sem{\Gamma;A}}(\pi^*_{\sem{\Gamma;A}}(\sem{\Gamma;\ul{C}})) 
\ar[dd]_-{\Pi_{\sem{\Gamma;A}}(\sem{\Gamma, x : A; z : \ul{C}; K})}
\\
\\
\Pi_{\sem{\Gamma;A}}(\ul{D})
}
}
{
\begin{array}{c}
\sem{\Gamma;\ul{C}} \in \mathcal{C}_{\sem{\Gamma}} \quad \sem{\Gamma, x \!:\! A; z \!:\! \ul{C}; K} : \pi^*_{\sem{\Gamma;A}}(\sem{\Gamma;\ul{C}}) \longrightarrow \ul{D}
\end{array}
}
\]

\paragraph*{Computational function application}
\mbox{}\\
\[
\mkrule
{
\xymatrix@C=3em@R=2em@M=0.5em{
\txt<25pc>{$\sem{\Gamma;z \!:\! \ul{C};K(V)_{(x : A).\, \ul{D}}} $\\$ \defeq $\\$ \sem{\Gamma;\ul{C}}$}
\ar[d]_-{=}
\\
(\mathsf{s}(\sem{\Gamma;V}))^*(\pi^*_{\sem{\Gamma;A}}(\sem{\Gamma;\ul{C}})) 
\ar[dd]_-{(\mathsf{s}(\sem{\Gamma;V}))^*(\pi^*_{\sem{\Gamma;A}}(\sem{\Gamma;z : \ul{C};K}))}
\\
\\
(\mathsf{s}(\sem{\Gamma;V}))^*(\pi^*_{\sem{\Gamma;A}}(\Pi_{\sem{\Gamma;A}} (\sem{\Gamma, x \!:\! A; \ul{D}}))) 
\ar[dd]_-{(\mathsf{s}(\sem{\Gamma;V}))^*(\varepsilon^{\pi^*_{\sem{\Gamma;A}} \,\dashv\, \Pi_{\sem{\Gamma;A}}}_{\sem{\Gamma, x : A; \ul{D}}})}
\\
\\
(\mathsf{s}(\sem{\Gamma;V}))^* (\sem{\Gamma, x \!:\! A; \ul{D}})
}
}
{
\begin{array}{c}
\sem{\Gamma;\ul{C}} \in \mathcal{C}_{\sem{\Gamma}} 
\quad
\sem{\Gamma;V} : 1_{\sem{\Gamma}} \longrightarrow \sem{\Gamma;A}
\\
\sem{\Gamma;z \!:\! \ul{C};K} : \sem{\Gamma;\ul{C}} \longrightarrow \Pi_{\sem{\Gamma;A}} (\sem{\Gamma, x \!:\! A; \ul{D}})
\end{array}
}
\]

\paragraph*{Homomorphic function application}
\mbox{}\\
\[
\mkrule
{\sem{\Gamma;z \!:\! \ul{C};V(K)_{\ul{D}_1, \ul{D}_2}} \defeq \sem{\Gamma;\ul{C}} \overset{\sem{\Gamma; z : \ul{C};K}}{\,-\!\!\!\!-\!\!\!\!-\!\!\!\!-\!\!\!\!-\!\!\!\!-\!\!\!\!\longrightarrow\,} \sem{\Gamma;\ul{D}_1} \overset{\xi_{\sem{\Gamma}, \sem{\Gamma;\ul{D}_1}, \sem{\Gamma;\ul{D}_2}}(\sem{\Gamma;V})}{\,-\!\!\!\!-\!\!\!\!-\!\!\!\!-\!\!\!\!-\!\!\!\!-\!\!\!\!-\!\!\!\!-\!\!\!\!-\!\!\!\!-\!\!\!\!-\!\!\!\!-\!\!\!\!-\!\!\!\!-\!\!\!\!-\!\!\!\!-\!\!\!\!\longrightarrow\,} \sem{\Gamma;\ul{D}_2}}
{
\begin{array}{c}
\sem{\Gamma;V} : 1_{\sem{\Gamma}} \longrightarrow \sem{\Gamma;\ul{D}_1} \multimap_{\sem{\Gamma}} \sem{\Gamma;\ul{D}_2} \quad \sem{\Gamma;z \!:\! \ul{C};K} : \sem{\Gamma;\ul{C}} \longrightarrow \sem{\Gamma;\ul{D}_1}
\end{array}
}
\]

\section{Soundness}
\label{sect:soundness}

In this section we show that the interpretation of eMLTT we defined in Section~\ref{sect:interpretation} is sound.
In particular, we prove that $\sem{-}$ is defined on well-formed expressions, and that it validates the equational theory of eMLTT. We state and prove this result in Theorem~\ref{thm:soundness}. However, before we do so, we first define semantic notions of weakening and substitution, and relate them to their syntactic counterparts, analogously to the soundness proofs for the denotational semantics of MLTT given in~\cite{Streicher:Semantics,Hofmann:Thesis}.

First, we observe that if we assume that $\sem{\Gamma_1,\Gamma_2} \in \mathcal{B}$, then $\Gamma_1,\Gamma_2$ must be a valid value context to begin with, and thus $\Gamma_1$ and $\Gamma_2$ must be disjoint according to the definition of value contexts, namely, because the variables in $\Gamma_1,\Gamma_2$ are distinct. As a consequence, we do not need to include explicit disjointness requirements on value contexts in the propositions and theorems we prove in the rest of this section.

Next, we define the semantic notions of weakening and substitution.
 
\begin{definition}
\label{def:semweakening}
\index{morphism!semantic projection --}
\index{ projb@$\sproj {\Gamma_1} x A {\Gamma_2}$ (semantic projection morphism)}
Given value contexts $\Gamma_1$ and $\Gamma_2$, a value type $A$, and a value variable $x$ such that $\sem{\Gamma_1,\Gamma_2} \in \mathcal{B}$ and $\sem{\Gamma_1, x \!:\! A, \Gamma_2} \in \mathcal{B}$, we define the \emph{semantic projection morphisms} as the \emph{a priori} partially defined family of morphisms
\[
\sproj {\Gamma_1} x A {\Gamma_2} : \sem{\Gamma_1, x \!:\! A, \Gamma_2} \longrightarrow \sem{\Gamma_1,\Gamma_2}
\]
that are defined by induction on the size of $\Gamma_2$, as follows:
\[
\begin{array}{c}
\sproj {\Gamma_1} x A {\diamond} \defeq \ia {\sem{\Gamma_1;A}} \overset{\pi_{\sem{\Gamma_1;A}}}{\,-\!\!\!\!-\!\!\!\!-\!\!\!\!\longrightarrow\,} \sem{\Gamma_1}
\\[5mm]
\sproj {\Gamma_1} x A {\Gamma_2, y : B} \defeq \ia {\sproj {\Gamma_1} x A {\Gamma_2} ^*(\sem{\Gamma_1, \Gamma_2;B})} \overset{\ia {\overline{\sproj {\Gamma_1} x A {\Gamma_2}}(\sem{\Gamma_1, \Gamma_2;B})}}{\,-\!\!\!\!-\!\!\!\!-\!\!\!\!-\!\!\!\!-\!\!\!\!-\!\!\!\!-\!\!\!\!-\!\!\!\!-\!\!\!\!-\!\!\!\!-\!\!\!\!-\!\!\!\!-\!\!\!\!-\!\!\!\!-\!\!\!\!\longrightarrow\,} \ia {\sem{\Gamma_1, \Gamma_2;B}}
\end{array}
\]
where the base case is always defined because the assumption $\sem{\Gamma_1, x \!:\! A} \in \mathcal{B}$ allows us to deduce $\sem{\Gamma_1;A} \in \mathcal{V}_{\sem{\Gamma_1}}$ from it; and
 the step case is only defined when we have 
 \[
 \sem{\Gamma_1, x \!:\! A, \Gamma_2;B} = \sproj {\Gamma_1} x A {\Gamma_2} ^*(\sem{\Gamma_1, \Gamma_2;B})
 \]
\end{definition}

\begin{definition}
\label{def:semsubstitution}
\index{morphism!semantic substitution --}
\index{ subst@$\ssubst {\Gamma_1} x A {\Gamma_2} V$ (semantic substitution morphism)}
Given value contexts $\Gamma_1$ and $\Gamma_2$, a value type $A$, a value variable $x$, and a value term $V$ such that $\sem{\Gamma_1, x \!:\! A, \Gamma_2} \in \mathcal{B}$ and $\sem{\Gamma_1,\Gamma_2[V/x]} \in \mathcal{B}$, and such that $\sem{\Gamma_1;V} : 1_{\sem{\Gamma_1}} \longrightarrow \sem{\Gamma_1;A}$, we define the \emph{semantic substitution morphisms} as the \emph{a priori} partially defined family of morphisms
\[
\ssubst {\Gamma_1} x A {\Gamma_2} V : \sem{\Gamma_1,\Gamma_2[V/x]} \longrightarrow \sem{\Gamma_1,x \!:\! A, \Gamma_2}
\]
that are defined by induction on the size of $\Gamma_2$, as follows:
\[
\begin{array}{c}
\ssubst {\Gamma_1} x A {\diamond} V \defeq \sem{\Gamma_1} \overset{\sem{\Gamma;V}}{\,-\!\!\!\!-\!\!\!\!-\!\!\!\!\longrightarrow\,} \ia {\sem{\Gamma_1;A}}
\\[5mm]
\hspace{-11cm}
\ssubst {\Gamma_1} x A {\Gamma_2, y : B} V \defeq 
\\
\ia {\ssubst {\Gamma_1} x A {\Gamma_2} V^*(\sem{\Gamma_1, x \!:\! A, \Gamma_2;B})} \overset{\ia {\overline{\ssubst {\Gamma_1} x A {\Gamma_2} V}(\sem{\Gamma_1, x : A, \Gamma_2;B})}}{\,-\!\!\!\!-\!\!\!\!-\!\!\!\!-\!\!\!\!-\!\!\!\!-\!\!\!\!-\!\!\!\!-\!\!\!\!-\!\!\!\!-\!\!\!\!-\!\!\!\!-\!\!\!\!-\!\!\!\!-\!\!\!\!-\!\!\!\!-\!\!\!\!-\!\!\!\!-\!\!\!\!-\!\!\!\!\longrightarrow\,} 
\ia {\sem{\Gamma_1, x \!:\! A, \Gamma_2;B}}
\end{array}
\]
where the base case is always defined due to the assumption $\sem{\Gamma_1;V} : 1_{\sem{\Gamma_1}} \longrightarrow \sem{\Gamma_1;A}$; and 
where the step case is only defined when we have
\[
\sem{\Gamma_1, \Gamma_2[V/x]; B[V/x]} = \ssubst {\Gamma_1} x A {\Gamma_2} V^*(\sem{\Gamma_1, x \!:\! A, \Gamma_2;B})
\]
\end{definition}

Intuitively, the family of morphisms $\sproj {\Gamma_1} x A {\Gamma_2}$ corresponds to projecting out the value context $\Gamma_1,\Gamma_2$ from $\Gamma_1, x \!:\! A, \Gamma_2$;
and the family of morphisms $\ssubst {\Gamma_1} x A {\Gamma_2} V$ corresponds to substituting the value term $V$ for the value variable $x$ in $\Gamma_1, x \!:\! A, \Gamma_2$.
We make this intuition formal in the semantic weakening and substitution lemmas below.
Simultaneously with these lemmas, we also prove that the \emph{a priori} partially defined families of morphisms $\sproj {\Gamma_1} x A {\Gamma_2}$ and $\ssubst {\Gamma_1} x A {\Gamma_2} V$ are in fact defined for all $\Gamma_2$.

\begin{proposition}
\label{prop:semweakening1}
Given value contexts $\Gamma_1$ and $\Gamma_2$, a value type $A$, and a value variable $x$ such that $\sem{\Gamma_1,\Gamma_2} \in \mathcal{B}$ and $\sem{\Gamma_1, x \!:\! A, \Gamma_2} \in \mathcal{B}$, then the \emph{a priori} partially defined semantic projection morphism $\sproj {\Gamma_1} x A {\Gamma_2} : \sem{\Gamma_1, x \!:\! A, \Gamma_2} \longrightarrow \sem{\Gamma_1,\Gamma_2}$
is defined.
\end{proposition}

\begin{proof}
We prove this proposition simultaneously with Proposition~\ref{prop:semweakening2}. 
The proof is straightforward---it proceeds induction on the size of $\Gamma_2$. As mentioned in the definition of $\sproj {\Gamma_1} x A {\Gamma_2}$, the base case is always defined because in this case we assume that $\sem{\Gamma_1, x \!:\! A} \in \mathcal{B}$, from which it follows that $\sem{\Gamma_1;A} \in \mathcal{V}_{\sem{\Gamma_1}}$ by inspecting the definition of $\sem{-}$ for $\Gamma_1, x \!:\! A$. For showing that the step case is defined, we use $(a)$ of Proposition~\ref{prop:semweakening2}, which gives us $\sem{\Gamma_1, x \!:\! A, \Gamma_2;B} = \sproj {\Gamma_1} x A {\Gamma_2} ^*(\sem{\Gamma_1, \Gamma_2;B}) \in \mathcal{V}_{\sem{\Gamma_1, x \!:\! A, \Gamma_2}}$.
\end{proof}

\begin{proposition}[Semantic weakening]
\label{prop:semweakening2}
\index{weakening theorem!semantic --}
Given value contexts $\Gamma_1$ and $\Gamma_2$, a value type $A$, and a value variable $x$ such that $\sem{\Gamma_1,\Gamma_2} \in \mathcal{B}$ and $\sem{\Gamma_1, x \!:\! A, \Gamma_2} \in \mathcal{B}$, then we have: 
\begin{enumerate}[(a)]
\item Given a value type $B$ such that $\sem{\Gamma_1,\Gamma_2;B} \in \mathcal{V}_{\sem{\Gamma_1,\Gamma_2}}$, then 
\[
\sem{\Gamma_1, x \!:\! A,\Gamma_2;B} = \sproj {\Gamma_1} x A {\Gamma_2}^*(\sem{\Gamma_1,\Gamma_2;B}) \in \mathcal{V}_{\sem{\Gamma_1, x : A,\Gamma_2}}
\]
\item Given a computation type $\ul{C}$ such that $\sem{\Gamma_1,\Gamma_2;\ul{C}} \in \mathcal{C}_{\sem{\Gamma_1,\Gamma_2}}$, then
\[
\sem{\Gamma_1, x \!:\! A,\Gamma_2;\ul{C}} = \sproj {\Gamma_1} x A {\Gamma_2}^*(\sem{\Gamma_1,\Gamma_2;\ul{C}}) \in \mathcal{C}_{\sem{\Gamma_1, x : A,\Gamma_2}}
\]
\item Given a value term $V$ such that $\sem{\Gamma_1,\Gamma_2;V} : 1_{\sem{\Gamma_1,\Gamma_2}} \longrightarrow B$, then
\[
\sem{\Gamma_1, x \!:\! A,\Gamma_2;V} = \sproj {\Gamma_1} x A {\Gamma_2}^*(\sem{\Gamma_1,\Gamma_2;V}) : 1_{\sem{\Gamma_1, x : A,\Gamma_2}} \longrightarrow \sproj {\Gamma_1} x A {\Gamma_2}^*(B)
\]
\item Given a computation term $M$ such that $\sem{\Gamma_1,\Gamma_2;M} : 1_{\sem{\Gamma_1,\Gamma_2}} \longrightarrow U(\ul{C})$, then 
\[
\sem{\Gamma_1, x \!:\! A,\Gamma_2;M} = \sproj {\Gamma_1} x A {\Gamma_2}^*(\sem{\Gamma_1,\Gamma_2;M}) : 1_{\sem{\Gamma_1, x : A,\Gamma_2}} \longrightarrow U(\sproj {\Gamma_1} x A {\Gamma_2}^*(\ul{C}))
\]
\item Given a computation variable $z$, a computation type $\ul{C}$, and a homomorphism term $K$ such that $\sem{\Gamma_1, \Gamma_2; z \!:\! \ul{C}; K} : \sem{\Gamma_1,\Gamma_2;\ul{C}} \longrightarrow \ul{D}$ in $\mathcal{C}_{\sem{\Gamma_1,\Gamma_2}}$, then

\[
\begin{array}{c}
\hspace{-4.5cm}
\sem{\Gamma_1, x \!:\! A, \Gamma_2; z \!:\! \ul{C}; K} = \sproj {\Gamma_1} x A {\Gamma_2}^*(\sem{\Gamma_1, \Gamma_2; z \!:\! \ul{C}; K}) 
\\
\hspace{5.5cm}
: \sproj {\Gamma_1} x A {\Gamma_2}^*(\sem{\Gamma_1, \Gamma_2; \ul{C}}) \longrightarrow \sproj {\Gamma_1} x A {\Gamma_2}^*(\ul{D})
\end{array}
\]
\end{enumerate}
where we use the notation
\[
\sem{\Gamma_1, x \!:\! A,\Gamma_2;B} = \sproj {\Gamma_1} x A {\Gamma_2}^*(\sem{\Gamma_1,\Gamma_2;B}) \in \mathcal{V}_{\sem{\Gamma_1, x : A,\Gamma_2}}
\]
to mean that $\sem{\Gamma_1, x \!:\! A,\Gamma_2;B}$ is defined and that it is equal to $\sproj {\Gamma_1} x A {\Gamma_2}^*(\sem{\Gamma_1,\Gamma_2;B})$ as an object of $\mathcal{V}_{\sem{\Gamma_1, x : A,\Gamma_2}}$. We also use analogous notation for terms and morphisms.
\end{proposition}

\begin{proof}
We prove this proposition simultaneously with Proposition~\ref{prop:semweakening1}. We prove $(a)$--$(e)$ simultaneously, by induction on the sum of the sizes of the arguments to $\sem{-}$. We postpone the straightforward but laborious details of this proof to Appendix~\ref{sect:proofofprop:semweakening2}. 

In the setting of contextual categories, a detailed proof of this proposition can be found for MLTT in~\cite[Chapter~III]{Streicher:Semantics}.
\end{proof}

\begin{proposition}
\label{prop:semsubstitution1}
Given value contexts $\Gamma_1$ and $\Gamma_2$, a value type $A$, a value variable $x$, and a value term $V$ such that $\sem{\Gamma_1, x \!:\! A, \Gamma_2} \in \mathcal{B}$ and $\sem{\Gamma_1,\Gamma_2[V/x]} \in \mathcal{B}$, and \linebreak such that $\sem{\Gamma_1;V} : 1_{\sem{\Gamma_1}} \longrightarrow \sem{\Gamma_1;A}$, then the \emph{a priori} partially defined semantic substitution morphism 
$\ssubst {\Gamma_1} x A {\Gamma_2} V : \sem{\Gamma_1,\Gamma_2[V/x]} \longrightarrow \sem{\Gamma_1,x \!:\! A, \Gamma_2}$
is defined.
\end{proposition}

\begin{proof}
We prove this proposition simultaneously with Proposition~\ref{prop:semsubstitution2}. The proof is straightforward and very similar to the proof of Proposition~\ref{prop:semweakening1}---it proceeds by  induction on the size of $\Gamma_2$. As remarked in the definition of $\ssubst {\Gamma_1} x A {\Gamma_2} V$, the base case is always defined because we assume that $\sem{\Gamma_1,\Gamma_2;V} : 1_{\Gamma_1} \longrightarrow \sem{\Gamma_1;A}$. For showing that the step case is always defined, we use $(a)$ of Proposition~\ref{prop:semsubstitution2}, which gives us that $\sem{\Gamma_1,\Gamma_2[V/x];B[V/x]} = \ssubst {\Gamma_1} x A {\Gamma_2} V^*(\sem{\Gamma_1, x \!:\! A,\Gamma_2;B}) \in \mathcal{V}_{\sem{\Gamma_1,\Gamma_2[V/x]}}$.
\end{proof}

\begin{proposition}[Semantic value term substitution]
\label{prop:semsubstitution2}
\index{substitution theorem!semantic --!-- for value terms}
Given value contexts $\Gamma_1$ and $\Gamma_2$, a value type $A$, a value variable $x$, and a value term $V$ such that ${\sem{\Gamma_1, x \!:\! A, \Gamma_2} \in \mathcal{B}}$ and ${\sem{\Gamma_1,\Gamma_2[V/x]} \in \mathcal{B}}$, and such that $\sem{\Gamma_1;V} : 1_{\sem{\Gamma_1}} \longrightarrow \sem{\Gamma_1;A}$, then we have:
\begin{enumerate}[(a)]
\item Given a value type $B$ such that $\sem{\Gamma_1, x \!:\! A,\Gamma_2;B} \in \mathcal{V}_{\sem{\Gamma_1, x : A,\Gamma_2}}$, then 
\[
\sem{\Gamma_1,\Gamma_2[V/x];B[V/x]} = \ssubst {\Gamma_1} x A {\Gamma_2} V^*(\sem{\Gamma_1, x \!:\! A,\Gamma_2;B}) \in \mathcal{V}_{\sem{\Gamma_1,\Gamma_2[V/x]}}
\]
\item Given a computation type $\ul{C}$ such that $\sem{\Gamma_1, x \!:\! A,\Gamma_2;\ul{C}} \in \mathcal{C}_{\sem{\Gamma_1, x : A,\Gamma_2}}$, then 
\[
\sem{\Gamma_1,\Gamma_2[V/x];\ul{C}[V/x]} = \ssubst {\Gamma_1} x A {\Gamma_2} V^*(\sem{\Gamma_1, x \!:\! A,\Gamma_2;\ul{C}}) \in \mathcal{C}_{\sem{\Gamma_1,\Gamma_2[V/x]}}
\]
\item Given a value term $W$ such that $\sem{\Gamma_1, x \!:\! A,\Gamma_2;W} : 1_{\sem{\Gamma_1, x : A,\Gamma_2}} \longrightarrow B$, then 
\[
\begin{array}{c}
\hspace{-3.65cm}
\sem{\Gamma_1,\Gamma_2[V/x];W[V/x]} = \ssubst {\Gamma_1} x A {\Gamma_2} V^*(\sem{\Gamma_1, x \!:\! A,\Gamma_2;W}) 
\\
\hspace{7.5cm}
: 1_{\sem{\Gamma_1,\Gamma_2[V/x]}} \longrightarrow \ssubst {\Gamma_1} x A {\Gamma_2} V^*(B)
\end{array}
\]
\item Given a computation term $M$ such that $\sem{\Gamma_1, x \!:\! A,\Gamma_2;M} : 1_{\sem{\Gamma_1, x : A,\Gamma_2}} \longrightarrow U(\ul{C})$, then 
\[
\begin{array}{c}
\hspace{-3.5cm}
\sem{\Gamma_1,\Gamma_2[V/x];M[V/x]} = \ssubst {\Gamma_1} x A {\Gamma_2} V^*(\sem{\Gamma_1, x \!:\! A,\Gamma_2;M}) 
\\
\hspace{6.8cm}
: 1_{\sem{\Gamma_1,\Gamma_2[V/x]}} \longrightarrow U(\ssubst {\Gamma_1} x A {\Gamma_2} V^*(\ul{C}))
\end{array}
\]
\item Given a computation variable $z$, a computation type $\ul{C}$, and a homomorphism term $K$ such that $\sem{\Gamma_1, x \!:\! A,\Gamma_2; z \!:\! \ul{C};K} : \sem{\Gamma_1, x : A,\Gamma_2;\ul{C}} \longrightarrow \ul{D}$ in $\mathcal{C}_{\sem{\Gamma_1, x : A,\Gamma_2}}$, then 
\[
\begin{array}{c}
\hspace{-1.2cm}
\sem{\Gamma_1,\Gamma_2[V/x];z \!:\! \ul{C}[V/x];K[V/x]} = \ssubst {\Gamma_1} x A {\Gamma_2} V^*(\sem{\Gamma_1, x \!:\! A,\Gamma_2;z \!:\! \ul{C};K}) 
\\
\hspace{4cm}
: \ssubst {\Gamma_1} x A {\Gamma_2} V^*(\sem{\Gamma_1, x \!:\! A,\Gamma_2;\ul{C}}) \longrightarrow \ssubst {\Gamma_1} x A {\Gamma_2} V^*(\ul{D})
\end{array}
\]
\end{enumerate}
\end{proposition}

\begin{proof}
We prove this proposition simultaneously with Proposition~\ref{prop:semsubstitution1}.
We prove $(a)$--$(e)$ simultaneously, by induction on the sum of the sizes of the arguments to $\sem{-}$. We omit the lengthy proof of this proposition because it is analogous to the proof of Proposition~\ref{prop:semweakening2}, due to the similar use of the comprehension functor $\ia -$ and Cartesian morphisms in the definitions of $\ssubst {\Gamma_1} x A {\Gamma_2} V$ and $\sproj {\Gamma_1} x A {\Gamma_2}$. 

\pagebreak

Analogously to Proposition~\ref{prop:semweakening2}, in the setting of contextual categories, a detailed proof of this proposition can be found for MLTT in~\cite[Chapter~III]{Streicher:Semantics}.
\end{proof}

Next, we show that the semantic projection and substitution morphisms commute with each other.

\begin{proposition}
\label{prop:semweakeningandsubstitutioncommuting}
Given value contexts $\Gamma_1$ and $\Gamma_2$, value variables $x$ and $y$, value types $A$ and $B$, and a value term $V$ such that $\sem{\Gamma_1,\Gamma_2[V/y]} \in \mathcal{B}$, $\sem{\Gamma_1, y \!:\! B, \Gamma_2} \in \mathcal{B}$, $\sem{\Gamma_1, x \!:\! A,\Gamma_2[V/y]} \in \mathcal{B}$, $\sem{\Gamma_1, x \!:\! A, y \!:\! B,\Gamma_2} \in \mathcal{B}$, and $\sem{\Gamma_1; V} : 1_{\sem{\Gamma_1}} \longrightarrow \sem{\Gamma_1;B}$, then 
\[
\ssubst {\Gamma_1} {y} {B} {\Gamma_2} {V} \comp \sproj {\Gamma_1} {x} {A} {\Gamma_1[V/y]}
=
\sproj {\Gamma_1} {x} {A} {y : B, \Gamma_2} \comp \ssubst {\Gamma_1, x : A} y B {\Gamma_2} V 
\]
\end{proposition}

\begin{proof}
We prove this proposition by induction on the length of $\Gamma_2$. Both the base case and the step case of induction are proved by straightforward diagram chasing. We postpone the details of this proof to Appendix~\ref{sect:proofofprop:semweakeningandsubstitutioncommuting}.
\end{proof}

Next, we recall from Section~\ref{sect:fibadjmodelsstructure} that the motivation for requiring the split dependent sums to be strong is to be able to model the type-dependency in the elimination form for $\Sigma\, x \!:\! A .\, B$. We make this informal motivation precise in the next proposition.

\begin{proposition}
\label{prop:reindexingalongkappaandpairing}
Given a value context $\Gamma$, value variables $x_1$, $x_2$, and $y$, and \linebreak value types $A_1$, $A_2$, and $B$ such that $x_2 \not\in V\!ars(\Gamma) \cup \{y\}$, $\sem{\Gamma} \in \mathcal{B}$, $\sem{\Gamma;A} \in \mathcal{V}_{\sem{\Gamma}}$, \linebreak $\sem{\Gamma, x_1 \!:\! A_1;A_2} \in \mathcal{V}_{\sem{\Gamma, x_1 \!:\! A_1}}$, and $\sem{\Gamma, y \!:\! (\Sigma\, x_1 \!:\! A_1 .\, A_2) ; B} \in \mathcal{V}_{\sem{\Gamma, y : (\Sigma\, x_1 : A_1 .\, A_2)}}$, then we have
\[
\sem{\Gamma, x_1 \!:\! A_1, x_2 \!:\! A_2, B[\langle x_1, x_2 \rangle/y]} = \kappa_{\sem{\Gamma; A_1},\sem{\Gamma, x_1 : A_1; A_2}}^*(\sem{\Gamma, y \!:\! (\Sigma\, x_1 \!:\! A_1 .\, A_2); B}) 
\]
\end{proposition}

\begin{proof}
We begin by noting that both sides of this equation can be rewritten. 

On the one hand, the left-hand side of this equation can be rewritten as
\[
\begin{array}{c}
\hspace{-8cm}
(\mathsf{s}(\sem{\Gamma, x_1 \!:\! A_1, x_2 \!:\! A_2; \langle x_1 , x_2 \rangle}))^*(
\\
\hspace{-4cm}
\ia{\overline{\pi_{\sem{\Gamma, x_1 : A_1 ; A_2}}}(\sem{\Gamma, x_1 \!:\! A_1; \Sigma\, x_1 \!:\! A_1 .\, A_2})}^*(
\\
\hspace{4cm}
\ia{\overline{\pi_{\sem{\Gamma;A_1}}}(\sem{\Gamma; \Sigma\, x_1 \!:\! A_1 .\, A_2})}^*(\sem{\Gamma, y \!:\! (\Sigma\, x_1 \!:\! A_1 .\, A_2); B})))
\end{array}
\]
based on Propositions~\ref{prop:semweakening2} and~\ref{prop:semsubstitution2}, and the definition of morphisms $\sproj {\Gamma_1} {x} {A} {\Gamma_2}$.

On the other hand, the right-hand side of this equation can be rewritten as
\[
\ia {\eta^{\Sigma_{\sem{\Gamma; A_1}} \,\dashv\, \pi^*_{\sem{\Gamma; A_1}}}_{\sem{\Gamma, x_1 : A_1; A_2}}}^*(\ia {\overline{\pi_{\sem{\Gamma;A_1}}}(\sem{\Gamma; \Sigma\, x_1 \!:\! A_1 .\, A_2})}^*(\sem{\Gamma, y \!:\! (\Sigma\, x_1 \!:\! A_1 .\, A_2); B}))
\]
based on the definitions of $\kappa_{\sem{\Gamma; A_1},\sem{\Gamma, x_1 : A_1; A_2}}$ and $\sem{\Gamma;\Sigma\, x_1 \!:\! A_1 .\, A_2}$.

Now, as a result of $p : \mathcal{V} \longrightarrow \mathcal{B}$ being a split fibration, it suffices to show 
\[
\begin{array}{c}
\ia{\overline{\pi_{\sem{\Gamma, x_1 : A_1 ; A_2}}}(\sem{\Gamma, x_1 \!:\! A_1; \Sigma\, x_1 \!:\! A_1 .\, A_2})} 
\comp \mathsf{s}(\sem{\Gamma, x_1 \!:\! A_1, x_2 \!:\! A_2; \langle x_1 , x_2 \rangle})
\\
=
\\
\ia {\eta^{\Sigma_{\sem{\Gamma; A_1}} \,\dashv\, \pi^*_{\sem{\Gamma; A_1}}}_{\sem{\Gamma, x_1 : A_1; A_2}}}
\end{array}
\]
for the required equation to be true. We show that these two morphisms are equal by straightforward diagram chasing. We postpone these details to Appendix~\ref{sect:proofofprop:reindexingalongkappaandpairing}.
\end{proof}

In addition to relating the substitution of value terms for value variables to its semantic counterpart, we also need to do the same for the substitution of computation and homomorphism terms for computation variables. To this end, we show in the next two propositions that these two kinds of substitution correspond to composition.

\begin{proposition}[Semantic computation term substitution]
\label{prop:semsubstitution3}
\index{substitution theorem!semantic --!-- for computation terms}
Given a value context $\Gamma$, a computation variable $z$, a computation type $\ul{C}$, a computation term $M$, and a homomorphism term $K$ such that $\sem{\Gamma;M} : 1_{\sem{\Gamma}} \longrightarrow U(\sem{\Gamma;\ul{C}})$ and $\sem{\Gamma;z \!:\! \ul{C};K} : \sem{\Gamma;\ul{C}} \longrightarrow \ul{D}$, then we have 
\[
\sem{\Gamma;K[M/z]} \quad = \quad 1_{\sem{\Gamma}} \overset{\sem{\Gamma;M}}{\,-\!\!\!\!-\!\!\!\!-\!\!\!\!-\!\!\!\!-\!\!\!\!\longrightarrow\,} U(\sem{\Gamma;\ul{C}}) \overset{U(\sem{\Gamma;z : \ul{C};K})}{\,-\!\!\!\!-\!\!\!\!-\!\!\!\!-\!\!\!\!-\!\!\!\!-\!\!\!\!-\!\!\!\!-\!\!\!\!-\!\!\!\!-\!\!\!\!\longrightarrow\,} U(\ul{D})
\]
\end{proposition}

\begin{proof}
We prove this proposition by induction on the sum of the sizes of $\Gamma$, $\ul{C}$ and $K$. 

First, by inspecting the definitions of substitution and $\sem{-}$ for homomorphism terms, we see that in most cases, the part of $K[M/z]$ that contains $M$ (i.e., the part of $K$ that contains $z$) is interpreted as first in a sequence of morphisms. As a result, the proofs for these cases consist of using the induction hypothesis on the part of $K[M/z]$ that contains $M$, the functoriality of $U$, and the definition of $\sem{-}$ under $U$. 

The only exception to this general pattern is the case for computational lambda abstraction, where the part of $K[M/z]$ that contains $M$ is not interpreted first in a sequence of morphisms and, moreover, this part of $K[M/z]$ is interpreted under the $\Pi_{\sem{\Gamma;A}}$-functor. Therefore, we present a detailed proof of this case below. 

We also present detailed proofs of the cases for computation variables and sequential composition as representative examples of the other more straightforward cases.

\vspace{0.1cm}
\noindent
\textbf{Computation variables:}
In this case, we need to show that 
\[
\sem{\Gamma;z[M/z]} \quad = \quad 1_{\sem{\Gamma}} \overset{\sem{\Gamma;M}}{\,-\!\!\!\!-\!\!\!\!-\!\!\!\!-\!\!\!\!-\!\!\!\!\longrightarrow\,} U(\sem{\Gamma;\ul{C}}) \overset{U(\sem{\Gamma;z : \ul{C};z})}{\,-\!\!\!\!-\!\!\!\!-\!\!\!\!-\!\!\!\!-\!\!\!\!-\!\!\!\!-\!\!\!\!-\!\!\!\!-\!\!\!\!-\!\!\!\!\longrightarrow\,} U(\sem{\Gamma;\ul{C}})
\]

\pagebreak
\noindent
First, by inspecting the definition of substitution for $z$, we get that
\[
z[M/z] = M
\]
Secondly, by inspecting the definition of $\sem{-}$ for $z$, we get that
\[
\sem{\Gamma;z : \ul{C};z} = \id_{\sem{\Gamma;\ul{C}}} : \sem{\Gamma;\ul{C}} \longrightarrow \sem{\Gamma;\ul{C}}
\]
Therefore, we are left with having to show that
\[
\sem{\Gamma;M} \quad = \quad 1_{\sem{\Gamma}} \overset{\sem{\Gamma;M}}{\,-\!\!\!\!-\!\!\!\!-\!\!\!\!-\!\!\!\!-\!\!\!\!\longrightarrow\,} U(\sem{\Gamma;\ul{C}}) \overset{U(\id_{\sem{\Gamma;\ul{C}}})}{\,-\!\!\!\!-\!\!\!\!-\!\!\!\!-\!\!\!\!-\!\!\!\!-\!\!\!\!-\!\!\!\!-\!\!\!\!-\!\!\!\!-\!\!\!\!\longrightarrow\,} U(\sem{\Gamma;\ul{C}})
\]
which follows from the functoriality of $U$ and the properties of composition.

\vspace{0.2cm}
\noindent
\textbf{Sequential composition:}
In this case, we need to show that
\[
\begin{array}{c}
\hspace{-8.5cm}
\sem{\Gamma;(\doto K {x \!:\! A} {\ul{D}} N)[M/z]} = 
\\[1mm]
\hspace{3cm}
1_{\sem{\Gamma}} \overset{\sem{\Gamma;M}}{\,-\!\!\!\!-\!\!\!\!-\!\!\!\!-\!\!\!\!-\!\!\!\!\longrightarrow\,} U(\sem{\Gamma;\ul{C}}) \overset{U(\sem{\Gamma;z : \ul{C};\doto K {x : A} {\ul{D}} N})}{\,-\!\!\!\!-\!\!\!\!-\!\!\!\!-\!\!\!\!-\!\!\!\!-\!\!\!\!-\!\!\!\!-\!\!\!\!-\!\!\!\!-\!\!\!\!-\!\!\!\!-\!\!\!\!-\!\!\!\!-\!\!\!\!-\!\!\!\!-\!\!\!\!-\!\!\!\!-\!\!\!\!\longrightarrow\,} U(\sem{\Gamma;\ul{D}})
\end{array}
\]
First, by inspecting the definition of $\sem{-}$ for $\doto K {x \!:\! A} {\ul{D}} N$, we get that 
\[
\begin{array}{c}
\sem{\Gamma;z \!:\! \ul{C};K} : \sem{\Gamma;\ul{C}} \longrightarrow F(\sem{\Gamma;A})
\\[2mm]
\sem{\Gamma, x \!:\! A; N} : 1_{\sem{\Gamma, x : A}} \longrightarrow U(\pi^*_{\sem{\Gamma;A}}(\sem{\Gamma;\ul{D}}))
\end{array}
\]
Next, by inspecting the definition of substitution for $\doto K {x \!:\! A} {\ul{D}} N$, we get that
\[
(\doto K {x \!:\! A} {\ul{D}} N)[M/z] = \doto {K[M/z]} {x \!:\! A} {\ul{D}} N
\]
Finally, we show that the required equation holds by proving that the following diagram commutes:
\vspace{-0.1cm}
\[
\xymatrix@C=2.5em@R=3em@M=0.3em{
\ar@{}[d]^-{\,\,\qquad\qquad\qquad\quad\dcomment{\text{use of the i.h. on } K}}
\\
1_{\sem{\Gamma}} \ar@/_9pc/[ddddddr]_>>>>>>>>>>>>>>>>>>{\sem{\Gamma;\doto {K[M/z]} {x : A} {\ul{D}} N}\quad\,\,} \ar[r]_-{\sem{\Gamma;K[M/z]}} \ar@/^4pc/[rr]^-{\sem{\Gamma;M}}
& 
U(F(\sem{\Gamma;A})) \ar[d]^-{U(F(\langle \id_{\sem{\Gamma;A}} , ! \rangle))}
&
U(\sem{\Gamma;\ul{C}}) \ar[l]^-{U(\sem{\Gamma; z : \ul{C}; K})} \ar@/^9pc/[ddddddl]^>>>>>>>>>>>>>>>>>>>>{\quad U(\sem{\doto {K} {x : A} {\ul{D}} N})}
\\
& 
U(F(\Sigma_{\sem{\Gamma;A}}(\pi^*_{\sem{\Gamma;A}}(1_{\sem{\Gamma}})))) \ar[d]^-{=}^-{\,\,\,\,\qquad\dcomment{\text{def. of } \sem{\doto {K} {x \!:\! A} {\ul{D}} N}}}
\\
& 
U(F(\Sigma_{\sem{\Gamma;A}}(1_{\ia {\sem{\Gamma;A}}}))) \ar[d]^-{U(F(\Sigma_{\sem{\Gamma;A}}(\sem{\Gamma, x : A;N})))}_-{\dcomment{\text{def. of } \sem{\Gamma;\doto {K[M/z]} {x \!:\! A} {\ul{D}} N}}\,\,\,\,\quad}
\\
& 
U(F(\Sigma_{\sem{\Gamma;A}}(U(\pi^*_{\sem{\Gamma;A}}(\sem{\Gamma;\ul{D}}))))) \ar[d]^-{=}^-{\quad\qquad\dcomment{\text{functoriality of } U}}
\\
& 
U(F(\Sigma_{\sem{\Gamma;A}}(\pi^*_{\sem{\Gamma;A}}(U(\sem{\Gamma;\ul{D}}))))) \ar[d]^-{U(F(\varepsilon^{\Sigma_{\sem{\Gamma;A}} \,\dashv\, \pi^*_{\sem{\Gamma;A}}}_{U(\sem{\Gamma;\ul{D}})}))}
\\
& 
U(F(U(\sem{\Gamma;\ul{D}}))) \ar[d]^-{U(\varepsilon^{F \,\dashv\, U}_{\sem{\Gamma;\ul{D}}})}
\\
& 
U(\sem{\Gamma;\ul{D}})
}
\]

\vspace{0.2cm}
\noindent
\textbf{Computational lambda abstraction:}
In this case, we need to show that 
\[
\begin{array}{c}
\hspace{-9.5cm}
\sem{\Gamma;(\lambda \, x \!:\! A .\, K)[M/z]} = 
\\[1mm]
\hspace{2.5cm}
1_{\sem{\Gamma}} \overset{\sem{\Gamma;M}}{\,-\!\!\!\!-\!\!\!\!-\!\!\!\!-\!\!\!\!-\!\!\!\!\longrightarrow\,} U(\sem{\Gamma;\ul{C}}) \overset{U(\sem{\Gamma;z : \ul{C};\lambda \, x : A .\, K})}{\,-\!\!\!\!-\!\!\!\!-\!\!\!\!-\!\!\!\!-\!\!\!\!-\!\!\!\!-\!\!\!\!-\!\!\!\!-\!\!\!\!-\!\!\!\!-\!\!\!\!-\!\!\!\!-\!\!\!\!-\!\!\!\!-\!\!\!\!-\!\!\!\!-\!\!\!\!-\!\!\!\!\longrightarrow\,} U(\Pi_{\sem{\Gamma;A}}(\ul{D}))
\end{array}
\]
First, by inspecting the definition of $\sem{-}$ for $\lambda \, x \!:\! A .\, K$, we get that
\[
\sem{\Gamma, x \!:\! A;z \!:\! \ul{C};K} : \pi^*_{\sem{\Gamma;A}}(\sem{\Gamma;\ul{C}}) \longrightarrow \ul{D}
\]
Next, by inspecting the definition of substitution for $\lambda \, x \!:\! A .\, K$, we get that
\[
(\lambda \, x \!:\! A .\, K)[M/z] = \lambda \, x \!:\! A .\, (K[M/z])
\]
Finally, we show that the required equation holds by proving that the following diagram commutes:
\[
\hspace{-0.25cm}
\xymatrix@C=7em@R=2.75em@M=0.3em{
1_{\sem{\Gamma}} 
\ar[r]^-{\sem{\Gamma;M}} \ar[ddd]^>>>>>>>>>>>>{\eta^{\pi^*_{\sem{\Gamma;A}} \,\dashv\, \Pi_{\sem{\Gamma;A}}}_{1_{\sem{\Gamma}}}}
\ar@/_4.5pc/[ddddddd]_<<<<<<{\sem{\Gamma;\lambda\, x : A .\, K[M/z]}}
\ar@{}[ddd]^>>>{\!\qquad\qquad\dcomment{\text{nat. of } \eta^{\pi^*_{\sem{\Gamma;A}} \,\dashv\, \Pi_{\sem{\Gamma;A}}}}}
& 
U(\sem{\Gamma;\ul{C}})
\ar@/_12pc/[ddd]_<<<<<<<<<<<<<<<<<{\eta^{\pi^*_{\sem{\Gamma;A}} \,\dashv\, \Pi_{\sem{\Gamma;A}}}_{U(\sem{\Gamma;\ul{C}})}\!\!\!\!\!\!\!\!\!\!\!\!}
\ar[d]_>>>>{U(\eta^{\pi^*_{\sem{\Gamma;A}} \,\dashv\, \Pi_{\sem{\Gamma;A}}}_{\sem{\Gamma;\ul{C}}})}
\ar@/^12pc/[ddddddd]^<<<<<<<<{U(\sem{\Gamma; z : \ul{C}; \lambda \, x : A .\, K})}_-{\dcomment{\text{def.}}\quad\!\!\!}
\\
\ar@{}[dd]_-{\dcomment{\text{def.}}\quad}
& 
\txt<5pc>{$U(\Pi_{\sem{\Gamma;A}}($ $\pi^*_{\sem{\Gamma;A}}(\sem{\Gamma;\ul{C}})))$} \ar[d]_-{(\zeta_{\Pi;\sem{\Gamma;A}})_{\pi^*_{\sem{\Gamma;A}}(\sem{\Gamma;\ul{C}})}}
\ar@/^8.5pc/[ddddd]_-{=}
\ar@{}[dd]_-{\dcomment{\text{Proposition~\ref{prop:PiUnitCounitPreservedByU}}}\qquad\quad}
\\
& 
\txt<5pc>{$\Pi_{\sem{\Gamma;A}}(U($ $\pi^*_{\sem{\Gamma;A}}(\sem{\Gamma;\ul{C}})))$} \ar[d]_-{=}^-{\,\,\,\quad\dcomment{\zeta \text{ is nat. iso.}}}
\\
\Pi_{\sem{\Gamma;A}}(\pi^*_{\sem{\Gamma;A}}(1_{\sem{\Gamma}}))
\ar[d]^-{=}^-{\qquad\qquad\dcomment{\text{Proposition~\ref{prop:semweakening2} } (d)}}
\ar[r]_-{\Pi_{\sem{\Gamma;A}}(\pi^*_{\sem{\Gamma;A}}(\sem{\Gamma;M}))}
&
\txt<5pc>{$\Pi_{\sem{\Gamma;A}}(\pi^*_{\sem{\Gamma;A}}($ $U(\sem{\Gamma;\ul{C}})))$}
\ar[d]_-{=}^-{\,\,\,\dcomment{\text{Proposition~\ref{prop:semweakening2} } (a)}}
\\
\Pi_{\sem{\Gamma;A}}(1_{\ia {\sem{\Gamma;A}}})
\ar[d]^>>>>{\Pi_{\sem{\Gamma;A}}(\sem{\Gamma, x : A;K[M/z]})}^<<<{\quad\qquad\dcomment{\text{use of the i.h. on } K}}
\ar[r]^-{\Pi_{\sem{\Gamma;A}}(\sem{\Gamma, x : A;M})}
&
\Pi_{\sem{\Gamma;A}}(U(\sem{\Gamma, x \!:\! A;\ul{C}}))
\ar@/^2pc/[dl]^>>>>>>>>>>{\,\,\,\quad\qquad\Pi_{\sem{\Gamma;A}}(U(\sem{\Gamma, x : A; z : \ul{C};K}))}
\ar[dd]^<<<<<{(\zeta^{-1}_{\Pi,\sem{\Gamma;A}})_{\sem{\Gamma, x : A;\ul{C}}}}
\\
\Pi_{\sem{\Gamma;A}}(U(\ul{D}))
\ar[dd]^-{(\zeta^{-1}_{\Pi,\sem{\Gamma;A}})_{\ul{D}}}^<<<<<<<<<<{\quad\qquad\qquad\dcomment{\text{nat. of } \zeta^{-1}_{\Pi,\sem{\Gamma;A}}}}
\\
& 
U(\Pi_{\sem{\Gamma;A}}(\sem{\Gamma, x \!:\! A;\ul{C}}))
\ar[d]_-{U(\Pi_{\sem{\Gamma;A}}(\sem{\Gamma, x : A;z : \ul{C};K}))}
\\
U(\Pi_{\sem{\Gamma;A}}(\ul{D}))
&
U(\Pi_{\sem{\Gamma;A}}(\ul{D})) \ar[l]^-{\id_{U(\Pi_{\sem{\Gamma;A}}(\ul{D}))}}
}
\vspace{-1cm}
\]
\end{proof}

\begin{proposition}[Semantic homomorphism term substitution]
\label{prop:semsubstitution4}
\index{substitution theorem!semantic --!-- for homomorphism terms}
Given a value \linebreak context $\Gamma$, computation variables $z_1$ and $z_2$, computation types $\ul{C}_1$ and $\ul{C}_2$, and \linebreak homomorphism terms $K$ and $L$ such that $\sem{\Gamma;z_1 \!:\! \ul{C}_1 ;K} : \sem{\Gamma;\ul{C}_1} \longrightarrow \sem{\Gamma;\ul{C}_2}$ and \linebreak $\sem{\Gamma;z_2 \!:\! \ul{C}_2;L} : \sem{\Gamma;\ul{C}_2} \longrightarrow \ul{D}$, then we have 
\[
\sem{\Gamma;z_1 \!:\! \ul{C}_1;L[K/z_2]} \quad = \quad \sem{\Gamma;\ul{C}_1} \overset{\sem{\Gamma;z_1 : \ul{C}_1;K}}{\,-\!\!\!\!-\!\!\!\!-\!\!\!\!-\!\!\!\!-\!\!\!\!-\!\!\!\!-\!\!\!\!-\!\!\!\!-\!\!\!\!-\!\!\!\!\longrightarrow\,} \sem{\Gamma;\ul{C}_2} \overset{\sem{\Gamma;z_2 : \ul{C}_2;L})}{\,-\!\!\!\!-\!\!\!\!-\!\!\!\!-\!\!\!\!-\!\!\!\!-\!\!\!\!-\!\!\!\!-\!\!\!\!-\!\!\!\!-\!\!\!\!\longrightarrow\,} \ul{D}
\]
\end{proposition}

\begin{proof}
We omit the proof of this proposition because it proceeds analogously to the proof of Proposition~\ref{prop:semsubstitution3} discussed above---by using the induction hypothesis on the part of $L[K/z]$ that contains $K$ and the definition of $\sem{-}$ for homomorphism terms.
\end{proof}

We also note that Propositions~\ref{prop:semweakening2} and~\ref{prop:semsubstitution2} also 
extend to several value variables and value terms, as respectively shown in Propositions~\ref{prop:semweakening5} and~\ref{prop:semsubstitution5} below. 

\begin{proposition}
\label{prop:semweakening5}
\index{ projb@$\mathsf{proj}$ (composition of semantic projection morphisms)}
Given value contexts $\Gamma_1$ and $\Gamma_2$ and $\Gamma_3$ (for simplicity, we assume that $\Gamma_2 = x_1 \!:\! A_1, \ldots, x_n \!:\! A_n$) such that $\sem{\Gamma_1,\Gamma_3} \in \mathcal{B}$ and $\sem{\Gamma_1, x_1 \!:\! A_1, \ldots, x_i \!:\! A_i, \Gamma_3} \in \mathcal{B}$, for all $1 \leq i \leq n$, then, using the following abbreviation:
\[
\mathsf{proj} \defeq \sproj {\Gamma_1} {x_1} {A_1} {\Gamma_3} \,\comp\, \ldots \,\comp\, \sproj {\Gamma_1, x_1 : A_1, \ldots, x_{n-1} : A_{n-1}} {x_n} {A_n} {\Gamma_3}
\]
we have:
\begin{enumerate}[(a)]
\item Given a value type $A$ such that $\sem{\Gamma_1,\Gamma_3;A} \in \mathcal{V}_{\sem{\Gamma_1,\Gamma_3}}$, then 
\[
\sem{\Gamma_1,\Gamma_2,\Gamma_3;A} = \mathsf{proj}^*(\sem{\Gamma_1,\Gamma_3;A}) \in \mathcal{V}_{\sem{\Gamma_1,\Gamma_2,\Gamma_3}}
\]
\item Given a computation type $\ul{C}$ such that $\sem{\Gamma_1,\Gamma_3;\ul{C}} \in \mathcal{C}_{\sem{\Gamma_1,\Gamma_3}}$, then
\[
\sem{\Gamma_1,\Gamma_2,\Gamma_3;\ul{C}} = \mathsf{proj}^*(\sem{\Gamma_1,\Gamma_3;\ul{C}}) \in \mathcal{C}_{\sem{\Gamma_1,\Gamma_2,\Gamma_3}}
\]
\item Given a value term $V$ such that $\sem{\Gamma_1,\Gamma_3;V} : 1_{\sem{\Gamma_1,\Gamma_3}} \longrightarrow B$, then
\[
\sem{\Gamma_1,\Gamma_2,\Gamma_3;V} = \mathsf{proj}^*(\sem{\Gamma_1,\Gamma_3;V}) : 1_{\sem{\Gamma_1,\Gamma_2,\Gamma_3}} \longrightarrow\,\, \mathsf{proj}^*(B)
\]
\item Given a computation term $M$ such that $\sem{\Gamma_1,\Gamma_3;M} : 1_{\sem{\Gamma_1,\Gamma_3}} \longrightarrow U(\ul{C})$, then 
\[
\sem{\Gamma_1,\Gamma_2,\Gamma_3;M} = \mathsf{proj}^*(\sem{\Gamma_1,\Gamma_3;M}) : 1_{\sem{\Gamma_1,\Gamma_2,\Gamma_3}} \longrightarrow U(\mathsf{proj}^*(\ul{C}))
\]
\item Given a computation variable $z$, a computation type $\ul{C}$, and a homomorphism term $K$ such that $\sem{\Gamma_1,\Gamma_3; z \!:\! \ul{C}; K} : \sem{\Gamma_1,\Gamma_3;\ul{C}} \longrightarrow \ul{D}$ in $\mathcal{C}_{\sem{\Gamma_1,\Gamma_3}}$, then
\[
\sem{\Gamma_1,\Gamma_2,\Gamma_3; z \!:\! \ul{C}; K} = \mathsf{proj}^*(\sem{\Gamma_1,\Gamma_3; z \!:\! \ul{C}; K}) : \mathsf{proj}^*(\sem{\Gamma_1,\Gamma_3; \ul{C}}) \longrightarrow\,\, \mathsf{proj}^*(\ul{D})
\]
\end{enumerate}
\end{proposition}

\begin{proof}
We prove $(a)$--$(e)$ independently, by induction on the length of $\Gamma_2$. As all the cases are similar, we only consider $(b)$ in detail as a representative example.

\vspace{0.1cm}

\noindent\textit{Base case} (with $\Gamma_2 = \diamond$): 
This case is trivial because we need to show 
\[
\sem{\Gamma_1,\Gamma_3; \ul{C}} = \sem{\Gamma_1,\Gamma_3; \ul{C}} \in \mathcal{C}_{\sem{\Gamma_1,\Gamma_3}}
\]
which follows directly from our assumptions.

\vspace{0.1cm}

\noindent\textit{Step case} (with $\Gamma_2 = x_1 \!:\! A_1, \Gamma$):
First, we note that according to our assumptions, we know that $\sem{\Gamma_1,\Gamma_3} \in \mathcal{B}$ and $\sem{\Gamma_1, x_1 \!:\! A_1, \Gamma_3} \in \mathcal{B}$, which means that we can use $(b)$ of Propositon~\ref{prop:semweakening2} to get 
\[
\sem{\Gamma_1, x_1 \!:\! A_1,\Gamma_3;\ul{C}} = \sproj {\Gamma_1} {x_1} {A_1} {\Gamma_3}^*(\sem{\Gamma_1,\Gamma_3;\ul{C}}) \in \mathcal{C}_{\sem{\Gamma_1, x_1 : A_1,\Gamma_3}}
\]

Next, we use the induction hypothesis on $\sem{\Gamma_1, x_1 \!:\! A_1,\Gamma_3;\ul{C}}$, with the three contexts chosen to be $\Gamma_1, x_1 \!:\! A_1$ and $\Gamma$ and $\Gamma_3$, to get
\[
\sem{\Gamma_1, x_1 \!:\! A_1,\Gamma,\Gamma_3;\ul{C}} = \mathsf{proj}'^*(\sem{\Gamma_1, x_1 \!:\! A_1,\Gamma_3;\ul{C}}) \in \mathcal{C}_{\sem{\Gamma_1, x_1 : A_1,\Gamma,\Gamma_3}}
\]
where
\[
\mathsf{proj'} \defeq \sproj {\Gamma_1, x_1 : A_1} {x_2} {A_2} {\Gamma_3} \,\comp\, \ldots \,\comp\, \sproj {\Gamma_1, x_1 : A_1, \ldots, x_{n-1} : A_{n-1}} {x_n} {A_n} {\Gamma_3}
\]

Next, by combining the previous two equations with $\Gamma_2 = x_1 \!:\! A_1, \Gamma$, we get
\[
\sem{\Gamma_1,\Gamma_2,\Gamma_3; \ul{C}} = \mathsf{proj}'^*(\sproj {\Gamma_1} {x_1} {A_1} {\Gamma_3}^*(\sem{\Gamma_1,\Gamma_3;\ul{C}})) \in \mathcal{C}_{\sem{\Gamma_1,\Gamma_2,\Gamma_3}}
\]

Finally, by observing that $\sproj {\Gamma_1} {x_1} {A_1} {\Gamma_3} \comp \mathsf{proj}' = \mathsf{proj}$, in combination with the fact that $p$ is a split fibration, we get the required equation
\[
\sem{\Gamma_1,\Gamma_2,\Gamma_3; \ul{C}} = \mathsf{proj}^*(\sem{\Gamma_1,\Gamma_3;\ul{C}}) \in \mathcal{C}_{\sem{\Gamma_1,\Gamma_2,\Gamma_3}}
\]
\end{proof}

\begin{proposition}
\label{prop:semsubstitution5}
\index{ subst@$\mathsf{subst}$ (composition of semantic substitution morphisms)}
Given value contexts $\Gamma_1$ and $\Gamma_2$ and $\Gamma_3$ (where, for simplicity, we assume that $\Gamma_2 = x_1 \!:\! A_1, \ldots, x_n \!:\! A_n$) and value terms $V_i$ (for all $1 \leq i \leq n$) such that 
 $\sem{\Gamma_1,\Gamma_2,\Gamma_3} \in \mathcal{B}$ and 
\[
\begin{array}{c}
\sem{\Gamma_1, \Gamma_{2i}[V_1/x_1] \ldots [V_{i-1}/x_{i-1}], \Gamma_3[V_1/x_1] \ldots [V_{i-1}/x_{i-1}]} \in \mathcal{B}
\\[2mm]
\sem{\Gamma_1; V_i} : 1_{\sem{\Gamma_1}} \longrightarrow \sem{\Gamma_1; A_i[V_1/x_1] \ldots [V_{i-1}/x_{i-1}]}
\end{array}
\]
where $\Gamma_{2i} = x_i \!:\! A_i, \ldots, x_n \!:\! A_n$, then, using the following abbreviation:
\[
\begin{array}{c}
\hspace{-6.5cm}
\mathsf{subst} \,\,\defeq\,\, \ssubst {\Gamma_1} {x_1} {A_1} {x_2 : A_2, \ldots, x_n : A_n,\Gamma_3} {V_1} 
\\
\hspace{4cm}
\comp\, \ldots  \,\comp\, \ssubst {\Gamma_1} {x_n} {A_n[V_1/x_1] \ldots [V_{n-1}/x_{n-1}]} {\Gamma_3[V_1/x_1] \ldots [V_{n-1}/x_{n-1}]} {V_n}
\end{array}
\]
we have:
\begin{enumerate}[(a)]
\item Given a value type $B$ such that $\sem{\Gamma_1, \Gamma_2,\Gamma_3;B} \in \mathcal{V}_{\sem{\Gamma_1, \Gamma_2,\Gamma_3}}$, then 
\[
\begin{array}{c}
\hspace{-5cm}
\sem{\Gamma_1,\Gamma_3[V_1/x_1] \ldots [V_n/x_n];B[V_1/x_1] \ldots [V_n/x_n]} 
\\
\hspace{4.75cm}
= \mathsf{subst}^*(\sem{\Gamma_1, \Gamma_2,\Gamma_3;B}) \in \mathcal{V}_{\sem{\Gamma_1,\Gamma_3[V_1/x_1] \ldots [V_n/x_n]}}
\end{array}
\]
\item Given a computation type $\ul{C}$ such that $\sem{\Gamma_1, \Gamma_2,\Gamma_3;\ul{C}} \in \mathcal{C}_{\sem{\Gamma_1, \Gamma_2,\Gamma_3}}$, then 
\[
\begin{array}{c}
\hspace{-5cm}
\sem{\Gamma_1,\Gamma_3[V_1/x_1] \ldots [V_n/x_n];\ul{C}[V_1/x_1] \ldots [V_n/x_n]} 
\\
\hspace{4.75cm}
= \mathsf{subst}^*(\sem{\Gamma_1, \Gamma_2,\Gamma_3;\ul{C}}) \in \mathcal{C}_{\sem{\Gamma_1,\Gamma_3[V_1/x_1] \ldots [V_n/x_n]}}
\end{array}
\]
\item Given a value term $W$ such that $\sem{\Gamma_1, \Gamma_2,\Gamma_3;W} : 1_{\sem{\Gamma_1, \Gamma_2,\Gamma_3}} \longrightarrow B$, then 
\[
\begin{array}{c}
\hspace{-4.85cm}
\sem{\Gamma_1,\Gamma_3[V_1/x_1] \ldots [V_n/x_n];W[V_1/x_1] \ldots [V_n/x_n]} 
\\
\hspace{2.25cm}
= \mathsf{subst}^*(\sem{\Gamma_1, \Gamma_2,\Gamma_3;W}) 
: 1_{\sem{\Gamma_1,\Gamma_3[V_1/x_1] \ldots [V_n/x_n]}} \longrightarrow \mathsf{subst}^*(B)
\end{array}
\]
\item Given a computation term $M$ such that $\sem{\Gamma_1, \Gamma_2,\Gamma_3;M} : 1_{\sem{\Gamma_1, \Gamma_2,\Gamma_3}} \longrightarrow U(\ul{C})$, then 
\[
\begin{array}{c}
\hspace{-4.85cm}
\sem{\Gamma_1,\Gamma_3[V_1/x_1] \ldots [V_n/x_n];M[V_1/x_1] \ldots [V_n/x_n]} 
\\
\hspace{1.6cm}
= \mathsf{subst}^*(\sem{\Gamma_1, \Gamma_2,\Gamma_3;M}) 
: 1_{\sem{\Gamma_1,\Gamma_3[V_1/x_1] \ldots [V_n/x_n]}} \longrightarrow U(\mathsf{subst}^*(\ul{C}))
\end{array}
\]
\item Given a computation variable $z$, a computation type $\ul{C}$, and a homomorphism term $K$ such that $\sem{\Gamma_1, \Gamma_2,\Gamma_3; z \!:\! \ul{C};K} : \sem{\Gamma_1, \Gamma_2,\Gamma_3;\ul{C}} \longrightarrow \ul{D}$ in $\mathcal{C}_{\sem{\Gamma_1, \Gamma_2,\Gamma_3}}$, then 
\[
\begin{array}{c}
\hspace{-1.2cm}
\sem{\Gamma_1,\Gamma_3[V_1/x_1] \ldots [V_n/x_n];z \!:\! \ul{C}[V_1/x_1] \ldots [V_n/x_n];K[V_1/x_1] \ldots [V_n/x_n]} 
\\
\hspace{1cm}
= \mathsf{subst}^*(\sem{\Gamma_1,\Gamma_2,\Gamma_3;z \!:\! \ul{C};K}) 
: \mathsf{subst}^*(\sem{\Gamma_1, \Gamma_2,\Gamma_3;\ul{C}}) \longrightarrow \mathsf{subst}^*(\ul{D})
\end{array}
\]
\end{enumerate}
\end{proposition}

\begin{proof}
We prove $(a)$--$(e)$ independently, by induction on the length of $\Gamma_2$. As all the cases are similar, we only consider $(b)$ in detail as a representative example below.

\vspace{0.1cm}

\noindent\textit{Base case} (with $\Gamma_2 = \diamond$): 
This case is trivial because we need to show 
\[
\sem{\Gamma_1,\Gamma_3; \ul{C}} = \sem{\Gamma_1,\Gamma_3; \ul{C}} \in \mathcal{C}_{\sem{\Gamma_1,\Gamma_3}}
\]
which follows directly from our assumptions.

\vspace{0.1cm}

\noindent\textit{Step case} (with $\Gamma_2 = x_1 \!:\! A_1, \Gamma$):
To begin with, we note that according to our assumptions, we know that ${\sem{\Gamma_1, x_1 \!:\! A_1, \Gamma, \Gamma_3} \in \mathcal{B}}$, $\sem{\Gamma_1, \Gamma[V_1/x_1],\Gamma_3[V_1/x_1]} \in \mathcal{B}$, and $\sem{\Gamma_1; V_1} : 1_{\sem{\Gamma_1}} \longrightarrow \sem{\Gamma_1;A_1}$, which means that we can use Proposition~\ref{prop:semsubstitution2} to get 
\[
\begin{array}{c}
\hspace{-7.5cm}
\sem{\Gamma_1, \Gamma[V_1/x_1], \Gamma_3[V_1/x_1]; \ul{C}[V_1/x_1]} 
\\
\hspace{2.5cm}
= \ssubst {\Gamma_1} {x_1} {A_1} {\Gamma,\Gamma_3} {V_1}^*(\sem{\Gamma_1, x_1 \!:\! A_1, \Gamma, \Gamma_3; \ul{C}}) \in \mathcal{C}_{\sem{\Gamma_1, \Gamma[V_1/x_1], \Gamma_3[V_1/x_1]}}
\end{array}
\]

Next, we use the induction hypothesis on $\sem{\Gamma_1, \Gamma[V_1/x_1], \Gamma_3[V_1/x_1]; \ul{C}[V_1/x_1]}$, with the three contexts chosen to be $\Gamma_1$ and $\Gamma[V_1/x_1]$ and $\Gamma_3[V_1/x_1]$, to get
\[
\begin{array}{c}
\hspace{-5cm}
\sem{\Gamma_1,\Gamma_3[V_1/x_1] \ldots [V_n/x_n];\ul{C}[V_1/x_1] \ldots [V_n/x_n]} 
\\
\hspace{1.75cm}
= \mathsf{subst}'^*(\sem{\Gamma_1, \Gamma[V_1/x_1],\Gamma_3[V_1/x_1];\ul{C}[V_1/x_1]}) \in \mathcal{C}_{\sem{\Gamma_1,\Gamma_3[V_1/x_1] \ldots [V_n/x_n]}}
\end{array}
\]
where
\[
\begin{array}{c}
\hspace{-2.5cm}
\mathsf{subst'} \,\,\defeq\,\, \ssubst {\Gamma_1} {x_2} {A_2[V_1/x_1]} {x_3 : A_3[V_1/x_1], \ldots, x_n : A_n[V_1/x_1],\Gamma_3[V_1/x_1]} {V_2} 
\\
\hspace{4cm}
\comp \ldots  \comp \ssubst {\Gamma_1} {x_n} {A_n[V_1/x_1] \ldots [V_{n-1}/x_{n-1}]} {\Gamma_3[V_1/x_1] \ldots [V_{n-1}/x_{n-1}]} {V_n}
\end{array}
\]

Next, combining these two equations (where $\Gamma_2 = x_1 \!:\! A_1, \Gamma$), we get 
\[
\begin{array}{c}
\hspace{-5cm}
\sem{\Gamma_1,\Gamma_3[V_1/x_1] \ldots [V_n/x_n];\ul{C}[V_1/x_1] \ldots [V_n/x_n]} 
\\
\hspace{1.75cm}
= \mathsf{subst}'^*(\ssubst {\Gamma_1} {x_1} {A_1} {\Gamma,\Gamma_3} {V_1}^*(\sem{\Gamma_1, \Gamma_2, \Gamma_3; \ul{C}})) \in \mathcal{C}_{\sem{\Gamma_1,\Gamma_3[V_1/x_1] \ldots [V_n/x_n]}}
\end{array}
\]

Finally, by observing that $\ssubst {\Gamma_1} {x_1} {A_1} {\Gamma,\Gamma_3} {V_1} \comp \mathsf{subst}' = \mathsf{subst}$, in combination with the fact that $p$ is a split fibration, we get the required equation
\[
\begin{array}{c}
\hspace{-5cm}
\sem{\Gamma_1,\Gamma_3[V_1/x_1] \ldots [V_n/x_n];\ul{C}[V_1/x_1] \ldots [V_n/x_n]} 
\\
\hspace{4.75cm}
= \mathsf{subst}^*(\sem{\Gamma_1, \Gamma_2,\Gamma_3;\ul{C}}) \in \mathcal{C}_{\sem{\Gamma_1,\Gamma_3[V_1/x_1] \ldots [V_n/x_n]}}
\end{array}
\]
\end{proof}

Further, for the soundness results proved in Sections~\ref{sect:fibalgeffectsmodel} and~\ref{sect:interpretingemlttwithhandlers}, it is also useful to note that some special cases of Proposition~\ref{prop:semweakening5} admit more concise characterisations. In particular, we consider two such cases, where the type or term to be weakened is given in i) the empty value context or ii) a context containing only one variable.

\begin{proposition}
\label{prop:semweakening3}
Given a value context $\Gamma$ such that $\sem{\Gamma} \in \mathcal{B}$, then we have: 
\begin{enumerate}[(a)]
\item Given a value type $A$ such that $\sem{\diamond;A} \in \mathcal{V}_1$, then 
\[
\sem{\Gamma;A} =\,\, !_{\sem{\Gamma}}^*(\sem{\diamond;A}) \in \mathcal{V}_{\sem{\Gamma}}
\]
\item Given a computation type $\ul{C}$ such that $\sem{\diamond;\ul{C}} \in \mathcal{C}_1$, then
\[
\sem{\Gamma;\ul{C}} =\,\, !_{\sem{\Gamma}}^*(\sem{\diamond;\ul{C}}) \in \mathcal{C}_{\sem{\Gamma}}
\]
\item Given a value term $V$ such that $\sem{\diamond;V} : 1_1 \longrightarrow B$, then
\[
\sem{\Gamma;V} =\,\, !_{\sem{\Gamma}}^*(\sem{\diamond;V}) : 1_{\sem{\Gamma}} \longrightarrow\,\, !_{\sem{\Gamma}}^*(B)
\]
\item Given a computation term $M$ such that $\sem{\diamond;M} : 1_1 \longrightarrow U(\ul{C})$, then 
\[
\sem{\Gamma;M} =\,\, !_{\sem{\Gamma}}^*(\sem{\diamond;M}) : 1_{\sem{\Gamma}} \longrightarrow U(!_{\sem{\Gamma}}^*(\ul{C}))
\]
\item Given a computation variable $z$, a computation type $\ul{C}$, and a homomorphism term $K$ such that $\sem{\diamond; z \!:\! \ul{C}; K} : \sem{\diamond;\ul{C}} \longrightarrow \ul{D}$ in $\mathcal{C}_1$, then
\[
\sem{\Gamma; z \!:\! \ul{C}; K} =\,\, !_{\sem{\Gamma}}^*(\sem{\diamond; z \!:\! \ul{C}; K}) :\,\, !_{\sem{\Gamma}}^*(\sem{\diamond; \ul{C}}) \longrightarrow\,\, !_{\sem{\Gamma}}^*(\ul{D})
\]
\end{enumerate}
\end{proposition}

\begin{proof}
We prove $(a)$--$(e)$ independently, with all cases following the same pattern. 
As all the cases are similar, we only consider $(b)$ in detail as a representative example below.
For simplicity, we assume that $\Gamma = x_1 \!:\! A_1, \ldots, x_n \!:\! A_n$. 

First, we note that from the assumption $\sem{\Gamma} \in \mathcal{B}$, it follows that $\sem{\Gamma'} \in \mathcal{B}$ for every prefix $\Gamma'$ of $\Gamma$. 

As a result, we can use Proposition~\ref{prop:semweakening5}, with the three contexts chosen to be $\diamond$ and $\Gamma$ and $\diamond$, to show that $\sem{\Gamma;\ul{C}}$ is equal to 
\[
(\sproj {\diamond} {x_1} {A_1} {\diamond} \,\comp\,  \ldots \,\comp\,  \sproj {x_1 : A_1, \ldots, x_{n-1} : A_{n-1}} {x_n} {A_n} {\diamond})^*(\sem{\diamond; \ul{C}})
\]

Next, according to the definition of semantic projection morphisms, the domain of the morphism $\sproj {x_1 : A_1, \ldots, x_{n-1} : A_{n-1}} {x_n} {A_n} {\diamond}$ is $\sem{\Gamma}$. Similarly, the codomain of the morphism $\sproj {\diamond} {x_1} {A_1} {\diamond}$ is $\sem{\diamond}$, which is equal to the terminal object $1$ by definition. 

As a result, we have
\[
\sproj {\diamond} {x_1} {A_1} {\diamond} \,\comp\, \ldots \,\comp\, \sproj {x_1 : A_1, \ldots, x_{n-1} : A_{n-1}} {x_n} {A_n} {\diamond} : \sem{\Gamma} \longrightarrow 1
\]

Finally, using the universal property of the terminal object $1$, this composite must be equal to the unique such morphism, namely, to $!_{\sem{\Gamma}} : \sem{\Gamma}\longrightarrow 1$, meaning that  
\[
\sem{\Gamma;\ul{C}} =\,\, !^*_{\sem{\Gamma}}(\sem{\diamond; \ul{C}}) \in \mathcal{C}_{\sem{\Gamma}}
\vspace{-0.5cm}
\]
\end{proof}

\begin{proposition}
\label{prop:semweakening4}
Given a value context $\Gamma$, a value variable $x$, and a value type $A$ such that $x \not\in V\!ars(\Gamma)$ and $\sem{\Gamma} \in \mathcal{B}$, then we have: 
\begin{enumerate}[(a)]
\item Given a value type $B$ such that $\sem{x \!:\! A;B} \in \mathcal{V}_{\ia {\sem{\diamond; A}}}$, then 
\[
\sem{\Gamma, x \!:\! A;B} = \ia {\overline{!_{\sem{\Gamma}}}(\sem{\diamond; A})}^*(\sem{x \!:\! A; B}) \in \mathcal{V}_{\ia {!_{\sem{\Gamma}}^*(\sem{\diamond;A})}}
\]
\item Given a computation type $\ul{C}$ such that $\sem{x \!:\! A;\ul{C}} \in \mathcal{C}_{\ia {\sem{\diamond; A}}}$, then
\[
\sem{\Gamma, x \!:\! A;\ul{C}} = \ia {\overline{!_{\sem{\Gamma}}}(\sem{\diamond; A})}^*(\sem{x \!:\! A; \ul{C}}) \in \mathcal{C}_{\ia {!_{\sem{\Gamma}}^*(\sem{\diamond;A})}}
\]
\item Given a value term $V$ such that $\sem{x \!:\! A;V} : 1_{\ia {\sem{\diamond; A}}} \longrightarrow B$, then
\[
\sem{\Gamma, x \!:\! A;V} = \ia {\overline{!_{\sem{\Gamma}}}(\sem{\diamond; A})}^*(\sem{x \!:\! A;V}) : 1_{\ia {!_{\sem{\Gamma}}^*(\sem{\diamond;A})}} \longrightarrow \ia {\overline{!_{\sem{\Gamma}}}(\sem{\diamond; A})}^*(B)
\]
\item Given a computation term $M$ such that $\sem{x \!:\! A;M} : 1_{\ia {\sem{\diamond; A}}} \longrightarrow U(\ul{C})$, then 
\[
\sem{\Gamma, x \!:\! A;M} = \ia {\overline{!_{\sem{\Gamma}}}(\sem{\diamond; A})}^*(\sem{x \!:\! A;M}) : 1_{\ia {!_{\sem{\Gamma}}^*(\sem{\diamond;A})}} \longrightarrow U(\ia {\overline{!_{\sem{\Gamma}}}(\sem{\diamond; A})}^*(\ul{C}))
\]
\item Given a computation variable $z$, a computation type $\ul{C}$, and a homomorphism term $K$ such that $\sem{x \!:\! A; z \!:\! \ul{C}; K} : \sem{x\!:\! A;\ul{C}} \longrightarrow \ul{D}$ in $\mathcal{C}_{\ia {\sem{\diamond; A}}}$, then
\[
\begin{array}{c}
\hspace{-5.6cm}
\sem{\Gamma, x \!:\! A; z \!:\! \ul{C}; K} = \ia {\overline{!_{\sem{\Gamma}}}(\sem{\diamond; A})}^*(\sem{x \!:\! A; z \!:\! \ul{C}; K}) 
\\
\hspace{5.1cm}
: \ia {\overline{!_{\sem{\Gamma}}}(\sem{\diamond; A})}^*(\sem{x \!:\! A; \ul{C}}) \longrightarrow \ia {\overline{!_{\sem{\Gamma}}}(\sem{\diamond; A})}^*(\ul{D})
\end{array}
\]
\end{enumerate}
\end{proposition}

\begin{proof}
We prove $(a)$--$(e)$ independently, with all cases following the same pattern. 
As all the cases are similar, we only consider $(b)$ in detail as a representative example.
For simplicity, we assume that $\Gamma = x_1 \!:\! A_1, \ldots, x_n \!:\! A_n$. 

First, we note that from the assumption $\sem{\Gamma} \in \mathcal{B}$, it follows that $\sem{\Gamma'} \in \mathcal{B}$ for every prefix $\Gamma'$ of $\Gamma$. 

As a result, we can use Proposition~\ref{prop:semweakening5}, with the three contexts chosen to be $\diamond$ and $\Gamma$ and $x \!:\! A$, to show that $\sem{\Gamma, x \!:\! A;\ul{C}}$ is equal to 
\[
(\sproj {\diamond} {x_1} {A_1} {x : A} \comp \ldots \comp \sproj {x_1 : A_1, \ldots, x_{n-1} : A_{n-1}} {x_n} {A_n} {x : A})^*(\sem{x \!:\! A; \ul{C}})
\]

Next, according to the definition of semantic projection morphisms and the functoriality of $\ia -$, the above reindexing functor is equal to reindexing along 
\[
\ia {\overline{\sproj \diamond {x_{1}} {A_{1}} {\diamond}}(\sem{\diamond ; A})
\comp \ldots \comp 
\overline{\sproj {x_1 : A_1, \ldots, x_{n-1} : A_{n-1}} {x_n} {A_n} {\diamond}}(\sem{x_1 \!:\! A_1, \ldots, x_{n-1} \!:\! A_{n-1} ; A})}
\]
which is equal to reindexing along the morphism that results from applying $\ia -$ to 
\[
\overline{\sproj \diamond {x_{1}} {A_{1}} {\diamond}}(\sem{\diamond ; A})
\comp \ldots \comp 
\overline{\sproj {x_1 : A_1, \ldots, x_{n-1} : A_{n-1}} {x_n} {A_n} {\diamond}}(\sem{x_1 \!:\! A_1, \ldots, x_{n-1} \!:\! A_{n-1} ; A})
\]

Next, by observing that this last morphism is the composition of Cartesian morphisms, we can make the following three observations: i) the domain of this composite morphism is $\sproj {x_1 : A_1, \ldots, x_{n-1} : A_{n-1}} {x_n} {A_n} {\diamond}^*(\sem{x_1 : A_1, \ldots, x_{n-1} : A_{n-1} ; A})$, which, according to Proposition~\ref{prop:semweakening2}, is the same as $\sem{\Gamma; A}$, and which, according to Proposition~\ref{prop:semweakening3} is the same as $!_{\sem{\Gamma}}^*(\sem{\diamond; A})$; ii) the codomain of this composite morphism is $\sem{\diamond; A}$; and iii) this composite morphism is itself Cartesian, according to Proposition~\ref{prop:cartesianmorphismscompose}. 

As a result, according to the choice of Cartesian morphisms in $p$, this composite Cartesian morphism must be equal to $\overline{!_{\sem{\Gamma}}}(\sem{\diamond; A}) : \,\, !_{\sem{\Gamma}}^*(\sem{\diamond; A}) \longrightarrow \sem{\diamond; A}$. 

Finally, after applying $\ia -$ to this last morphism, we get that reindexing along 
\[
\ia {\overline{\sproj \diamond {x_{1}} {A_{1}} {\diamond}}(\sem{\diamond ; A})
\comp \ldots \comp 
\overline{\sproj {x_1 : A_1, \ldots, x_{n-1} : A_{n-1}} {x_n} {A_n} {\diamond}}(\sem{x_1 \!:\! A_1, \ldots, x_{n-1} \!:\! A_{n-1} ; A})}
\]
is equal to reindexing along $\ia {\overline{!_{\sem{\Gamma}}}(\sem{\diamond; A})}$, meaning that 
\[
\sem{\Gamma, x \!:\! A; \ul{C}} = \ia {\overline{!_{\sem{\Gamma}}}(\sem{\diamond; A})}^*(\sem{x \!:\! A; \ul{C}}) \in \mathcal{C}_{\ia {!_{\sem{\Gamma}}^*(\sem{\diamond;A})}}
\vspace{-0.25cm}
\]
\end{proof}

We now state and prove the soundness theorem\footnote{Using the terminology of~\cite{Streicher:Semantics}, Theorem~\ref{thm:soundness} could also be called the correctness theorem.} for the interpretation of eMLTT in fibred adjunction models. In particular, we show that for well-formed types and well-typed terms, the \emph{a priori} partially defined interpretation function $\sem -$ is in fact always defined, and that it maps definitionally equal contexts,  types, and terms to equal objects and morphisms.
As noted in the beginning of this section, the proof of this theorem crucially relies on the semantic weakening and substitution lemmas we proved above.

\begin{theorem}[Soundness]
\label{thm:soundness}
\index{soundness theorem}
\mbox{}
\begin{enumerate}[(a)]
\item Given $\vdash \Gamma$, then $\sem{\Gamma} \in \mathcal{B}$.
\item Given $\lj \Gamma A$, then $\sem{\Gamma;A} \in \mathcal{V}_{\sem{\Gamma}}$.
\item Given $\lj \Gamma \ul{C}$, then $\sem{\Gamma;\ul{C}} \in \mathcal{C}_{\sem{\Gamma}}$.
\item Given $\vj \Gamma V A$, then $\sem{\Gamma;V} : 1_{\sem{\Gamma}} \longrightarrow \sem{\Gamma;A}$.
\item Given $\cj \Gamma M \ul{C}$, then $\sem{\Gamma;M} : 1_{\sem{\Gamma}} \longrightarrow U(\sem{\Gamma;\ul{C}})$.
\item Given $\hj \Gamma {z \!:\! \ul{C}} K \ul{D}$, then $\sem{\Gamma;z \!:\! \ul{C}; K} : \sem{\Gamma;\ul{C}} \longrightarrow \sem{\Gamma;\ul{D}}$.
\item Given $\vdash {\Gamma_1} = {\Gamma_2}$, then $\sem{\Gamma_1} = \sem{\Gamma_2} \in \mathcal{B}$.
\item Given $\ljeq \Gamma A B$, then $\sem{\Gamma;A} = \sem{\Gamma;B} \in \mathcal{V}_{\sem{\Gamma}}$.
\item Given $\ljeq \Gamma {\ul{C}} {\ul{D}}$, then $\sem{\Gamma;\ul{C}} = \sem{\Gamma;\ul{D}} \in \mathcal{C}_{\sem{\Gamma}}$.
\item Given $\veq \Gamma V W A$, then $\sem{\Gamma;V} = \sem{\Gamma;W} : 1_{\sem{\Gamma}} \longrightarrow \sem{\Gamma;A}$.
\item Given $\ceq \Gamma M N \ul{C}$, then $\sem{\Gamma;M} = \sem{\Gamma;N} : 1_{\sem{\Gamma}} \longrightarrow U(\sem{\Gamma;\ul{C}})$.
\item Given $\heq \Gamma {z \!:\! \ul{C}} K L \ul{D}$, then $\sem{\Gamma;z \!:\! \ul{C};K} = \sem{\Gamma; z \!:\! \ul{C}; L} : \sem{\Gamma;\ul{C}} \longrightarrow \sem{\Gamma;\ul{D}}$.
\end{enumerate}
\end{theorem}

\begin{proof}
We prove $(a)$--$(l)$ simultaneously, by induction on the given derivations, using Propositions~\ref{prop:semweakening2},~\ref{prop:semsubstitution2},~\ref{prop:semsubstitution3},  and~\ref{prop:semsubstitution4} to relate syntactic weakening and substitution to reindexing along semantic projection and projection morphisms. 

Similarly to the proofs of  Propositions~\ref{prop:semweakening2} and~\ref{prop:semsubstitution2}, in the setting of contextual categories, detailed proofs of the cases that involve the MLTT fragment of eMLTT can be found in~\cite[Chapter~III]{Streicher:Semantics}.
We thus omit the proofs of these cases.

We illustrate the eMLTT-specific cases of $(b)$ and $(c)$ by giving a detailed proof for the formation rule for the computational $\Sigma$-type. 

We omit most of the cases of $(e)$ and $(f)$ (i.e., the cases concerning the computational $\Sigma$- and $\Pi$-types) because their proofs are analogous to the detailed proofs given for the corresponding terms in the MLTT fragment of eMLTT in~\cite[Chapter~III]{Streicher:Semantics}. For $(e)$ and $(f)$, we only present the proof for the typing rule for sequential composition. 

Regarding $(k)$ and $(l)$, we again omit most of the cases and only present detailed proofs for the $\beta$- and $\eta$-equations for homomorphic lambda abstraction and function application, and for sequential composition. It is worth noting that the proofs for the cases of $(k)$ and $(l)$ that involve the computational $\Sigma$- and $\Pi$-types follow directly from the properties of the corresponding adjunctions $\Sigma_A \dashv \pi^*_{A}$ and  $\pi^*_{A} \dashv \Pi_A$, respectively.

\vspace{0.2cm}
\noindent
\textbf{Computational $\Sigma$-type:}
In this case, the given derivation ends with
\[
\mkrule
{\lj \Gamma {\Sigma\, x \!:\! A .\, \ul{C}}}
{\lj \Gamma A 
\qquad
\lj {\Gamma, x \!:\! A} \ul{C}
}
\]
and we need to show that
\[
\sem{\Gamma;\Sigma\, x \!:\! A .\, \ul{C}} \in \mathcal{C}_{\sem{\Gamma}}
\]
First, by using $(b)$ and the induction hypothesis on the two assumed derivations, we get that
\[
\sem{\Gamma;A} \in \mathcal{V}_{\sem{\Gamma}}
\qquad
\sem{\Gamma, x \!:\! A;\ul{C}} \in \mathcal{C}_{\sem{\Gamma, x : A}}
\]
Next, by inspecting the definition of $\sem{-}$ for $\Gamma, x \!:\! A$, we get that
\[
\sem{\Gamma, x : A} = \ia {\sem{\Gamma;A}}
\]
which means that we can use the existence of split dependent $p$-sums to get that
\[
\Sigma_{\sem{\Gamma;A}} (\sem{\Gamma, x \!:\! A;\ul{C}}) \in \mathcal{C}_{\sem{\Gamma}}
\]
Finally, the required object in $\mathcal{C}_{\sem{\Gamma}}$ exists because by the definition of $\sem{-}$ we have that
\[
\sem{\Gamma;\Sigma\, x \!:\! A .\, \ul{C}} = \Sigma_{\sem{\Gamma;A}} (\sem{\Gamma, x \!:\! A;\ul{C}})
\]

\vspace{0.2cm}
\noindent
\textbf{Typing rule for sequential composition for computation terms:}
In this case, the given derivation ends with
\[
\mkrule
{\cj \Gamma {\doto M {x \!:\! A} {\ul{C}} N} {\ul{C}}}
{\cj \Gamma M {FA} \quad \lj \Gamma {\ul{C}} \quad \cj {\Gamma, x \!:\! A} N {\ul{C}}}
\]
and we need to show that
\[
\sem{\Gamma;\doto M {x \!:\! A} {\ul{C}} N} : 1_{\sem{\Gamma}} \longrightarrow U(\sem{\Gamma;\ul{C}})
\]
First, by using the induction hypothesis on the two assumed derivations, we get that
\[
\sem{\Gamma;M} : 1_{\sem{\Gamma}} \longrightarrow U(\sem{\Gamma;FA})
\qquad
\sem{\Gamma, x \!:\! A;N} : 1_{\sem{\Gamma, x : A}} \longrightarrow U(\sem{\Gamma, x \!:\! A; \ul{C}})
\]
Next, by inspecting the definition of $\sem{-}$ for $\Gamma, x \!:\! A$ and $FA$, we get that
\[
\sem{\Gamma;M} : 1_{\sem{\Gamma}} \longrightarrow U(F(\sem{\Gamma;A}))
\qquad
\sem{\Gamma, x \!:\! A;N} : 1_{\ia {\sem{\Gamma;A}}} \longrightarrow U(\sem{\Gamma, x \!:\! A; \ul{C}})
\]
Further, by using $(b)$ of Proposition~\ref{prop:semweakening2} and the definition of $\sproj \Gamma x A \diamond$, we get that
\[
\sem{\Gamma, x \!:\! A;N} : 1_{\ia {\sem{\Gamma;A}}} \longrightarrow U(\pi^*_{\sem{\Gamma;A}}(\sem{\Gamma; \ul{C}}))
\]
Finally, by inspecting the definition of $\sem{-}$ for $\doto M {x \!:\! A} {\ul{C}} N$, we see that $\sem{\Gamma;M}$ and $\sem{\Gamma, x \!:\! A;N}$ satisfy the corresponding pre-conditions. Therefore, we get that
\[
\sem{\Gamma;\doto M {x \!:\! A} {\ul{C}} N} : 1_{\sem{\Gamma}} \longrightarrow U(\sem{\Gamma;\ul{C}})
\]

\vspace{0.2cm}
\noindent
\textbf{$\beta$-equation for homomorphic function application for computation terms:}
In this case, the given derivation ends with
\[
\mkrule
{\ceq \Gamma {(\lambda \, z \!:\! \ul{C} .\, K)(M)_{\ul{C}, \ul{D}}} {K[M/z]} {\ul{D}}}
{\cj \Gamma M \ul{C} \quad \hj \Gamma {z \!:\! \ul{C}} K {\ul{D}}}
\]
and we need to show that
\[
\sem{\Gamma;(\lambda \, z \!:\! \ul{C} .\, K)(M)_{\ul{C}, \ul{D}}} = \sem{\Gamma;K[M/z]} : 1_{\sem{\Gamma}} \longrightarrow U(\sem{\Gamma;\ul{D}})
\]
By using $(e)$ and $(f)$ on the two assumed derivations, we get that
\[
\sem{\Gamma;M} : 1_{\sem{\Gamma}} \longrightarrow U(\sem{\Gamma;\ul{C}})
\qquad
\sem{\Gamma;z \!:\! \ul{C};K} : \sem{\Gamma;\ul{C}} \longrightarrow \sem{\Gamma;\ul{D}}
\]
The required equation then follows from the commutativity of the following diagram:
\[
\xymatrix@C=16em@R=8em@M=0.3em{
1_{\sem{\Gamma}} \ar@/^4pc/[dr]^-{\,\,\,\,\,\,\,\,\sem{\Gamma;K[M/z]}} \ar[d]_-{\sem{\Gamma;M}}^<<<<<<<<<<<<<{\qquad\qquad\dcomment{\text{Proposition~\ref{prop:semsubstitution3}}}} \ar@/_4pc/[dd]_-{\sem{\Gamma;(\lambda \, z : \ul{C} .\, K)(M)_{\ul{C}, \ul{D}}}} \ar@{}[dd]^-{\qquad\qquad\qquad\dcomment{\xi_{\sem{\Gamma},\sem{\Gamma;\ul{C}},\sem{\Gamma;\ul{D}}} \text{ is an iso.}}}
\\
U(\sem{\Gamma;\ul{C}}) \ar@/^2pc/[r]^-{U(\sem{\Gamma;z : \ul{C}; K})} \ar@/_2pc/[r]_-{U(\xi_{\sem{\Gamma},\sem{\Gamma;\ul{C}},\sem{\Gamma;\ul{D}}}(\xi^{-1}_{\sem{\Gamma},\sem{\Gamma;\ul{C}},\sem{\Gamma;\ul{D}}}(\sem{\Gamma; z : \ul{C}; K})))}\ar[d]^>>>>>>>>>{U(\xi_{\sem{\Gamma},\sem{\Gamma;\ul{C}},\sem{\Gamma;\ul{D}}}(\sem{\Gamma;\lambda \, z : \ul{C} .\, K}))}_<<<<<<<<<{\dcomment{\text{def.}}\quad\!\!\!\!}^-{\qquad\qquad\dcomment{\text{def. of } \sem{\Gamma;\lambda \, z \!:\! \ul{C} .\, K}}} & U(\sem{\Gamma;\ul{D}})
\\
U(\sem{\Gamma;\ul{D}}) \ar@/_4pc/[ur]_-{\id_{\sem{\Gamma;\ul{C}}}}
}
\]
The case for the $\beta$-equation for homomorphic lambda abstraction for homomorphism terms is proved analogously.

\vspace{0.2cm}
\noindent
\textbf{$\eta$-equation for homomorphic function application:}
In this case, the given derivation ends with
\[
\mkrule
{\veq \Gamma V {\lambda\, z \!:\! \ul{C} .\, V(z)_{\ul{C}, \ul{D}}} {\ul{C} \multimap \ul{D}}}
{\vj \Gamma V {\ul{C} \multimap \ul{D}}}
\]
and we need to show that
\[
\sem{\Gamma;V} = \sem{\Gamma;\lambda\, z \!:\! \ul{C} .\, V(z)_{\ul{C}, \ul{D}}} : 1_{\sem{\Gamma}} \longrightarrow \sem{\Gamma;\ul{C}} \multimap_{\sem{\Gamma}} \sem{\Gamma;\ul{D}}
\]
First, by using $(d)$ on the assumed derivation, we get that
\[
\sem{\Gamma;V} : 1_{\sem{\Gamma}} \longrightarrow \sem{\Gamma;\ul{C} \multimap \ul{D}}
\]
Further, by inspecting the definition of $\sem{-}$ for $\ul{C} \multimap \ul{D}$, we get that
\[
\sem{\Gamma;V} : 1_{\sem{\Gamma}} \longrightarrow \sem{\Gamma;\ul{C}} \multimap_{\sem{\Gamma}} \sem{\Gamma;\ul{D}}
\]
Next, by inspecting the definition of $\sem{-}$ for $z$, we know that
\[
\sem{\Gamma;z \!:\! \ul{C};z} = \id_{\sem{\Gamma;\ul{C}}} : \sem{\Gamma;\ul{C}} \longrightarrow \sem{\Gamma;\ul{C}}
\]
Finally, the required equation follows from the commutativity of the following diagram:
\vspace{-0.15cm}
\[
\hspace{-0.25cm}
\xymatrix@C=15em@R=12em@M=0.3em{
1_{\sem{\Gamma}} \ar@/^1.25pc/[r]^{\sem{\Gamma;V}} \ar@/_2pc/[r]_-{\xi^{-1}_{\sem{\Gamma},\sem{\Gamma;\ul{C}},\sem{\Gamma;\ul{D}}}(\xi_{\sem{\Gamma},\sem{\Gamma;\ul{C}},\sem{\Gamma;\ul{D}}}(\sem{\Gamma;V}))} \ar@/_1.5pc/[d]_<<<<<<{\sem{\Gamma;\lambda\, z : \ul{C} .\, V(z)_{\ul{C}, \ul{D}}}}^-{\,\,\,\,\,\dcomment{\text{def.}}}^>>>>>>>>>>>>>>>>>>>>{\quad\qquad\qquad\qquad\dcomment{\xi_{\sem{\Gamma},\sem{\Gamma;\ul{C}},\sem{\Gamma;\ul{D}}} \text{ is an iso.}}} \ar@/^1.5pc/[d]^<<<<<<<<<<<<<<<<<<<<<<<<<<<<<<<<<<<<<<<<{\xi^{-1}_{\sem{\Gamma},\sem{\Gamma;\ul{C}},\sem{\Gamma;\ul{D}}}(\sem{\Gamma;z : \ul{C};V(z)_{\ul{C}, \ul{D}}})}^<<<<<<<<<<<<<<<<<<<<<<<<<<<<<<<<<{\dcomment{\text{def. of } \sem{\Gamma;z \!:\! \ul{C};V(z)_{\ul{C}, \ul{D}}}}} \ar@/_2.5pc/[dr]^<<<<<<<<<<<<<<<<<<<<<<<<<<<<<<<<{\!\!\quad\qquad\xi^{-1}_{\sem{\Gamma},\sem{\Gamma;\ul{C}},\sem{\Gamma;\ul{D}}}(\xi_{\sem{\Gamma},\sem{\Gamma;\ul{C}},\sem{\Gamma;\ul{D}}}(\sem{\Gamma;V}) \,\comp\, \sem{\Gamma;z : \ul{C};z})}^<<<<<<<<<<<<<<<<<<<<<<{\qquad\qquad\dcomment{\text{def. of } \sem{\Gamma;z \!:\! \ul{C}; z}}}^<<<<<<<<<<<<<<<<<<<<<<{\qquad\qquad\qquad\qquad\qquad\qquad\dcomment{\text{comp. with id.}}} & \sem{\Gamma;\ul{C}} \multimap \sem{\Gamma;\ul{D}}
\\
\sem{\Gamma;\ul{C}} \multimap \sem{\Gamma;\ul{D}} \ar[r]_-{\id_{\sem{\Gamma;\ul{C}} \multimap \sem{\Gamma;\ul{D}}}} & \sem{\Gamma;\ul{C}} \multimap \sem{\Gamma;\ul{D}} \ar@/_3pc/[u]_<<<<<<{\id_{\sem{\Gamma;\ul{C}} \multimap \sem{\Gamma;\ul{D}}}}
}
\]

\pagebreak

\noindent
\textbf{$\beta$-equation for sequential composition for computation terms:}
In this case, the given derivation ends with 
\[
\mkrule
{\ceq \Gamma {\doto {\return V} {x \!:\! A} {\ul{C}} M} {M[V/x]} {\ul{C}}}
{\vj \Gamma V {A} \quad \lj \Gamma {\ul{C}} \quad \cj {\Gamma, x \!:\! A} {M} {\ul{C}}}
\]
and we need to show that
\[
\sem{\Gamma;\doto {\return V} {x \!:\! A} {\ul{C}} M} = \sem{\Gamma;M[V/x]} : 1_{\sem{\Gamma}} \longrightarrow U(\sem{\Gamma;\ul{C}})
\]
First, by using $(d)$ and $(e)$ on the assumed derivations, we get that
\[
\sem{\Gamma;V} : 1_{\sem{\Gamma}} \longrightarrow \sem{\Gamma;A}
\qquad
\sem{\Gamma, x \!:\! A;M} : 1_{\sem{\Gamma, x : A}} \longrightarrow U(\sem{\Gamma, x \!:\! A;\ul{C}})
\]
Next, by inspecting the definition of $\sem{-}$ for $\Gamma, x \!:\! A$, we get that
\[
\sem{\Gamma, x \!:\! A;M} : 1_{\ia {\sem{\Gamma;A}}} \longrightarrow U(\sem{\Gamma, x \!:\! A;\ul{C}})
\] 
Further, by using $(b)$ of Proposition~\ref{prop:semweakening2} and the definition of $\sproj \Gamma x A \diamond$, we get that
\[
\sem{\Gamma, x \!:\! A;M} : 1_{\ia {\sem{\Gamma;A}}} \longrightarrow U(\pi^*_{\sem{\Gamma;A}}(\sem{\Gamma;\ul{C}}))
\] 
Finally, the required equation follows from the commutativity of the following diagram:

\pagebreak

\mbox{}

\vspace{0.5cm}

\[
\hspace{-0.6cm}
\xymatrix@C=1.5em@R=2.5em@M=0.3em{
1_{\sem{\Gamma}} 
\ar[rr]^-{\sem{\Gamma;M[V/x]}}
\ar[dr]^-{=}^>>>>>>>>>>{\quad\qquad\qquad\qquad\dcomment{\text{Proposition~\ref{prop:semsubstitution2}}}} \ar[dd]^>>>>>{\!\!\sem{\Gamma;V}}_>>>>>{\dcomment{\text{def.}}\,\,}
\ar@/_3pc/[ddddr]^<<<<<<<<<<<<<<<{\eta^{F \,\dashv\, U}_{1_{\sem{\Gamma}}}}^>>>>{\,\,\,\qquad\dcomment{\text{nat. of } \eta^{F \,\dashv\, U}}}
\ar@/_8pc/[dddddddddd]_<<<<{\sem{\Gamma;\doto {\return V} {x : A} {\ul{C}} M}}
\ar@/_2.5pc/[dddd]_>>>>>>{\sem{\Gamma; \return V}}
&&
U(\sem{\Gamma;\ul{C}})
\ar[dddddddddd]_-{\id_{U(\sem{\Gamma;\ul{C}})}}
\\
& 
(\mathsf{s}(\sem{\Gamma;V}))^*(1_{\ia {\sem{\Gamma;A}}})
\ar[d]_-{(\mathsf{s}(\sem{\Gamma;V}))^*(\sem{\Gamma, x : A;M})}
\\
\sem{\Gamma;A}
\ar[dd]^<<<<<<{\eta^{F \,\dashv\, U}_{\sem{\Gamma;A}}}^>>>>>>>>{\!\!\!\!\qquad\dcomment{\text{nat. of } \eta^{F \,\dashv\, U}}}
& 
(\mathsf{s}(\sem{\Gamma;V}))^*(U(\pi^*_{\sem{\Gamma;A}}(\sem{\Gamma;\ul{C}}))) 
\ar@/_2pc/[uur]^-{=}
\ar@/^5pc/[ddddd]_<<<<<<{\eta^{F \,\dashv\, U}_{(\mathsf{s}(\sem{\Gamma;V}))^*(U(\pi^*_{\sem{\Gamma;A}}(\sem{\Gamma;\ul{C}})))}\!\!\!\!\!}
\\
&
\\
U(F(\sem{\Gamma;A}))
\ar[d]^-{U(F(\langle \id_{\sem{\Gamma;A}} , ! \rangle))}
&
U(F(1_{\sem{\Gamma}}))
\ar[l]_-{U(F(\sem{\Gamma;V}))}
\ar[d]_-{=}
\\
U(F(\Sigma_{\sem{\Gamma;A}}(\pi^*_{\sem{\Gamma;A}}(1_{\sem{\Gamma}}))))
\ar[d]^-{=}_-{\dcomment{\text{def.}}\qquad\quad}
&
\txt<7pc>{$U(F((\mathsf{s}(\sem{\Gamma;V}))^*($\\$1_{\ia {\sem{\Gamma;A}}})))$}
\ar[dd]_<<<<<{U(F((\mathsf{s}(\sem{\Gamma;V}))^*(\sem{\Gamma, x : A;M})))}
\\
U(F(\Sigma_{\sem{\Gamma;A}}(1_{\ia {\sem{\Gamma;A}}})))
\ar[d]^-{U(F(\Sigma_{\sem{\Gamma;A}}(\sem{\Gamma, x : A; M})))}
\\
\txt<7pc>{$U(F(\Sigma_{\sem{\Gamma;A}}(U($ $\pi^*_{\sem{\Gamma;A}}(\sem{\Gamma;\ul{C}})))))$}
\ar[d]^-{=}^-{\,\,\,\,\,\,\,\qquad\quad\dcomment{\text{Corollary~\ref{cor:semsubstintoweakenedterm2}}}}
&
\txt<8pc>{$U(F((\mathsf{s}(\sem{\Gamma;V}))^*($\\$U(\pi^*_{\sem{\Gamma;A}}(\sem{\Gamma;\ul{C}})))))$}
\ar[d]_-{=}
\\
\txt<7pc>{$U(F(\Sigma_{\sem{\Gamma;A}}(\pi^*_{\sem{\Gamma;A}}($ $U(\sem{\Gamma;\ul{C}})))))$}
\ar[d]^-{U(F(\varepsilon^{\Sigma_{\sem{\Gamma;A}} \,\dashv\, \pi^*_{\sem{\Gamma;A}}}_{U(\sem{\Gamma;\ul{C}})}))}
&
\txt<8pc>{$U(F((\mathsf{s}(\sem{\Gamma;V}))^*($\\$\pi^*_{\sem{\Gamma;A}}(U(\sem{\Gamma;\ul{C}})))))$}
\ar@/^2pc/[dl]^-{=}^-{\qquad\qquad\qquad\qquad\qquad\qquad\dcomment{F \,\dashv\, U}}
\\
\txt<5pc>{$U(F(U($\\$\sem{\Gamma;\ul{C}})))$}
\ar[d]^-{U(\varepsilon^{F \,\dashv\, U}_{\sem{\Gamma;\ul{C}}})}
\\
U(\sem{\Gamma;\ul{C}})
&&
U(\sem{\Gamma;\ul{C}})
\ar[ll]^-{\id_{U(\sem{\Gamma;\ul{C}})}}
}
\]

\pagebreak
 
\noindent
\textbf{$\eta$-equation for sequential composition for computation terms:}
In this case, the given derivation ends with 
\[
\mkrule
{\ceq \Gamma {\doto M {x \!:\! A} {\ul{C}} {K[\return x/z]}} {K[M/z]} {\ul{C}}}
{\cj \Gamma M {FA} \quad \lj \Gamma {\ul{C}} \quad \hj {\Gamma} {z \!:\! FA} {K} {\ul{C}}}
\]
and we need to show that
\[
\sem{\Gamma;\doto M {x \!:\! A} {\ul{C}} {K[\return x/z]}} = \sem{\Gamma;K[M/z]} : 1_{\sem{\Gamma}} \longrightarrow U(\sem{\Gamma;\ul{C}})
\]
First, by using $(e)$ and $(f)$ on the assumed derivations, we get that
\[
\sem{\Gamma;M} : 1_{\sem{\Gamma}} \longrightarrow U(\sem{\Gamma;FA})
\qquad
\sem{\Gamma;z \!:\! FA;K} : \sem{\Gamma;FA} \longrightarrow \sem{\Gamma;\ul{C}}
\]
Further, by using the definition of $\sem{-}$ for $FA$, we get that
\[
\sem{\Gamma;M} : 1_{\sem{\Gamma}} \longrightarrow U(F(\sem{\Gamma;A}))
\qquad
\sem{\Gamma;z \!:\! FA;K} : F(\sem{\Gamma;A}) \longrightarrow \sem{\Gamma;\ul{C}}
\]
Next, by using $(e)$ of Proposition~\ref{prop:semweakening2} and the definition of $\sproj \Gamma x A \diamond$, we get that
\[
\sem{\Gamma, x \!:\! A; z \!:\! FA; K} = \pi^*_{\sem{\Gamma;A}}(\sem{\Gamma, z \!:\! FA;K}) : \pi^*_{\sem{\Gamma;A}}(F(\sem{\Gamma;A})) \longrightarrow \pi^*_{\sem{\Gamma;A}}(\sem{\Gamma;\ul{C}})
\]
Finally, the required equation follows from the commutativity of the following diagram:

\mbox{}

\vspace{0.2cm}

\[
\hspace{0.5cm}
\xymatrix@C=2em@R=3.5em@M=0.3em{
1_{\sem{\Gamma}} 
\ar[rr]^-{\sem{\Gamma;K[M/z]}}
\ar[dr]_-{\sem{\Gamma;M}}
\ar@/_8pc/[ddddddddd]
\ar@{}[d]^<<<<{\qquad\qquad\qquad\qquad\dcomment{\text{Proposition~\ref{prop:semsubstitution3}}}}
&
\ar@{}[dd]_>>>>>>>>>>>>>{\sem{\Gamma; \doto M {x : A} {\ul{C}} {K[\return x/z]}}\qquad\quad}_>>>>>>>>>>>>>>>>>>>>{\dcomment{\text{def.}}\qquad\qquad\qquad\quad}
& 
U(\sem{\Gamma;\ul{C}})
\ar@/^5.25pc/[ddddddddd]
\\
& U(F(\sem{\Gamma;A})) 
\ar[ur]_-{\quad U(\sem{\Gamma; z : FA; K})}
\ar@/^1pc/[dl]^<<<<<{\,U(F(\langle \id_{\sem{\Gamma;A}}, ! \rangle))}
\\
\txt<7pc>{$U(F(\Sigma_{\sem{\Gamma;A}}($\\$\pi^*_{\sem{\Gamma;A}}(1_{\sem{\Gamma}}))))$}
\ar@/_3pc/[ddd]_-{=}
&
\txt<7pc>{$U(F(\Sigma_{\sem{\Gamma;A}}(\pi^*_{\sem{\Gamma;A}}($\\$\Sigma_{\sem{\Gamma;A}}(1_{\ia {\sem{\Gamma;A}}})))))$}
\ar@/^2pc/[d]^-{U(F(\Sigma_{\sem{\Gamma;A}}(\pi^*_{\sem{\Gamma;A}}(\mathsf{fst}))))}_-{\dcomment{\text{def.}}\quad}
&
\\
&
\txt<7pc>{$U(F(\Sigma_{\sem{\Gamma;A}}($\\$\pi^*_{\sem{\Gamma;A}}(\sem{\Gamma;A}))))$}
\ar[dr]^-{U(F(\Sigma_{\sem{\Gamma;A}}(\eta^{F \,\dashv\, U}_{\pi^*_{\sem{\Gamma;A}}(\sem{\Gamma;A})})))}
\\
& 
&
\txt<9pc>{$U(F(\Sigma_{\sem{\Gamma;A}}(U($\\$F(\pi^*_{\sem{\Gamma;A}}(\sem{\Gamma;A}))))))$}
\ar[dd]^-{=}_<<<<<<<<<{\dcomment{\text{def. of } \sem{\Gamma, x \!:\! A;\return x}}\qquad\qquad\quad}
\\
U(F(\Sigma_{\sem{\Gamma;A}}(1_{\ia {\sem{\Gamma;A}}})))
\ar[d]^-{U(F(\Sigma_{\sem{\Gamma;A}}(\sem{\Gamma, x : A; K[\return x / z]})))}^>{\qquad\qquad\qquad\dcomment{\text{Proposition~\ref{prop:semsubstitution3}}}}
\ar@/^2pc/[drr]^<<<<<<<<<<<<<<<{\qquad\quad U(F(\Sigma_{\sem{\Gamma;A}}(\sem{\Gamma, x : A; \return x})))}
\ar[uur]_-{U(F(\Sigma_{\sem{\Gamma;A}}(\sem{\Gamma, x : A;x})))}
\ar[uuur]^>>>>>>>>>>>{U(F(\Sigma_{\sem{\Gamma;A}}(\eta^{\Sigma_{\sem{\Gamma;A}} \,\dashv\, \pi^*_{\sem{\Gamma;A}}}_{1_{\ia {\sem{\Gamma;A}}}})))\!\!\!}
\\
\txt<7pc>{$U(F(\Sigma_{\sem{\Gamma;A}}(U($\\$\pi^*_{\sem{\Gamma;A}}(\sem{\Gamma;\ul{C}})))))$}
\ar[d]_-{=}
\ar@{}[dd]^-{\quad\qquad\qquad\qquad\qquad\qquad\dcomment{(*)}}
&&
\txt<9pc>{$U(F(\Sigma_{\sem{\Gamma;A}}(U($\\$\pi^*_{\sem{\Gamma;A}}(F(\sem{\Gamma;\ul{C}}))))))$}
\ar[ll]^-{U(F(\Sigma_{\sem{\Gamma;A}}(U(\pi^*_{\sem{\Gamma;A}}(\sem{\Gamma; z : FA; K})))))}
\\
\txt<8.5pc>{$U(F(\Sigma_{\sem{\Gamma;A}}($\\$\pi^*_{\sem{\Gamma;A}}(U(\sem{\Gamma;\ul{C}})))))$}
\ar[d]^-{U(F(\varepsilon^{\Sigma_{\sem{\Gamma;A}} \,\dashv\, \pi^*_{\sem{\Gamma;A}}}_{U(\sem{\Gamma;\ul{C}})}))}
&&
\\
\txt<7pc>{$U(F(U($\\$\sem{\Gamma;\ul{C}})))$}
\ar[d]^-{U(\varepsilon^{F \,\dashv\, U}_{\sem{\Gamma;\ul{C}}})}
&&
\\
U(\sem{\Gamma;\ul{C}})
&&
U(\sem{\Gamma;\ul{C}})
\ar@{}[uuu]_>>>>>>>>>>>>>>>>>>>>>>>{\id_{U(\sem{\Gamma;\ul{C}})}}
\ar[ll]_-{\id_{U(\sem{\Gamma;\ul{C}})}}
}
\]

\noindent
where we show that the subdiagram marked with $(*)$ commutes as follows:
\[
\hspace{-0.5cm}
\xymatrix@C=1em@R=4em@M=0.3em{
&
U(F(\sem{\Gamma;A}))
\ar[dl]_-{U(F(\langle \id_{\sem{\Gamma;A}} , ! \rangle))}
\ar[ddr]^-{\id_{U(F(\sem{\Gamma;A}))}}
\ar[ddd]^-{U(F(\eta^{F \,\dashv\, U}_{\sem{\Gamma;A}}))}^>>>>>>>>>>>>>>>>>>>>>>>>>>{\qquad\dcomment{F \,\dashv\, U}}
\\
\txt<7pc>{$U(F(\Sigma_{\sem{\Gamma;A}}($\\$\pi^*_{\sem{\Gamma;A}}(1_{\sem{\Gamma}}))))$}
\ar[d]_-{=}^-{\,\,\,\,\dcomment{\text{functoriality of } U \,\comp\, F \text{ on } (**)}}
\\
U(F(\Sigma_{\sem{\Gamma;A}}(1_{\ia {\sem{\Gamma;A}}})))
\ar[d]_-{U(F(\Sigma_{\sem{\Gamma;A}}(\eta^{\Sigma_{\sem{\Gamma;A}} \,\dashv\, \pi^*_{\sem{\Gamma;A}}}_{1_{\ia {\sem{\Gamma;A}}}})))}
&
&
U(F(\sem{\Gamma;A}))
\ar[d]^-{U(\sem{\Gamma; z : FA;K})}_>>>>>{\dcomment{\text{nat. of } \varepsilon^{F \,\dashv\, U}}\quad}
\\
\txt<7pc>{$U(F(\Sigma_{\sem{\Gamma;A}}(\pi^*_{\sem{\Gamma;A}}($\\$\Sigma_{\sem{\Gamma;A}}(1_{\ia {\sem{\Gamma;A}}})))))$}
\ar[d]_-{U(F(\Sigma_{\sem{\Gamma;A}}(\pi^*_{\sem{\Gamma;A}}(\mathsf{fst}))))}
&
U(F(UF(\sem{\Gamma;A})))
\ar@/^2.5pc/[dr]
\ar[ur]^-{U(\varepsilon^{F \,\dashv\, U}_{F(\sem{\Gamma;A})})}
&
U(\sem{\Gamma;\ul{C}})
\\
\txt<7pc>{$U(F(\Sigma_{\sem{\Gamma;A}}($\\$\pi^*_{\sem{\Gamma;A}}(\sem{\Gamma;A}))))$}
\ar[d]_-{U(F(\Sigma_{\Gamma;A}(\eta^{F \,\dashv\, U}_{\pi^*_{\sem{\Gamma;A}}(\sem{\Gamma;A})})))}^-{\qquad\!\dcomment{U , F \text{ are s. fib.}}}
&
\txt<9pc>{$U(F(\Sigma_{\sem{\Gamma;A}}(\pi^*_{\sem{\Gamma;A}}($\\$U(F(\sem{\Gamma;\ul{A}}))))))$}
\ar@/^2.5pc/[dr]_>>>>>>>{U(F(\Sigma_{\sem{\Gamma;A}}(\pi^*_{\sem{\Gamma;A}}(U(\sem{\Gamma;z : FA;K})))))\!\!\!}
\ar[u]^-{U(F(\varepsilon^{\Sigma_{\sem{\Gamma;A}} \,\dashv\, \pi^*_{\sem{\Gamma;A}}}_{U(F(\sem{\Gamma;A}))}))}_<<<<{\qquad\,\,\, U(F(U(\sem{\Gamma; z : FA; K})))}_>>>>>>>>>>>{\,\,\,\,\dcomment{\text{nat. of } \varepsilon^{\Sigma_{\sem{\Gamma;A}} \,\dashv\, \pi^*_{\sem{\Gamma;A}}}}}
&
U(F(U(\sem{\Gamma;\ul{C}})))
\ar[u]_-{U(\varepsilon^{F \,\dashv\, U}_{\sem{\Gamma;\ul{C}}})}
\\
\txt<9pc>{$U(F(\Sigma_{\sem{\Gamma;A}}(U($\\$F(\pi^*_{\sem{\Gamma;A}}(\sem{\Gamma;A}))))))$}
\ar[d]_-{=}
\ar@/^2.5pc/[ur]^-{=}
&
&
\txt<8.5pc>{$U(F(\Sigma_{\sem{\Gamma;A}}($\\$\pi^*_{\sem{\Gamma;A}}(U(\sem{\Gamma;\ul{C}})))))$}
\ar[u]_-{U(F(\varepsilon^{\Sigma_{\sem{\Gamma;A}} \,\dashv\, \pi^*_{\sem{\Gamma;A}}}_{U(\sem{\Gamma;\ul{C}})}))}
\\
\txt<9pc>{$U(F(\Sigma_{\sem{\Gamma;A}}(U($\\$\pi^*_{\sem{\Gamma;A}}(F(\sem{\Gamma;\ul{A}}))))))$}
\ar[rr]_-{U(F(\Sigma_{\Gamma;A}(U(\pi^*_{\Gamma;A}(\sem{\Gamma;z : FA;K})))))}
\ar[uur]_-{=}
&&
\txt<7pc>{$U(F(\Sigma_{\sem{\Gamma;A}}(U($\\$\pi^*_{\sem{\Gamma;A}}(\sem{\Gamma;\ul{C}})))))$}
\ar[u]_-{=}^-{\dcomment{U \text{ is split fibred}}\qquad\qquad\qquad}
}
\]
\pagebreak

\noindent
and where the diagram we refer to as $(**)$ commutes because we have
\[
\xymatrix@C=1em@R=5em@M=0.3em{
\\
\Sigma_{\sem{\Gamma;A}}(\pi^*_{\sem{\Gamma;A}}(1_{\sem{\Gamma}}))
\ar[d]_-{=}^-{\qquad\qquad\qquad\dcomment{\mathsf{fst} \,\comp\, \langle \id_{\sem{\Gamma;A}} , ! \rangle = \id_{\sem{\Gamma;A}}}}
&
&
\sem{\Gamma;A}
\ar[ll]_-{\langle \id_{\sem{\Gamma;A}} , ! \rangle}
\ar[ddd]^-{\eta^{F \,\dashv\, U}_{\sem{\Gamma;A}}}_>>>>>>{\dcomment{\text{nat. of } \varepsilon^{\Sigma_{\sem{\Gamma;A}} \,\dashv\, \pi^*_{\sem{\Gamma;A}}}}\qquad\quad}
\ar@/^2pc/@{<-}[ddl]^-{\id_{\sem{\Gamma;A}}}
\\
\Sigma_{\sem{\Gamma;A}}(1_{\ia {\sem{\Gamma;A}}})
\ar[d]_-{\Sigma_{\sem{\Gamma;A}}(\eta^{\Sigma_{\sem{\Gamma;A}} \,\dashv\, \pi^*_{\sem{\Gamma;A}}}_{1_{\ia {\sem{\Gamma;A}}}})}^<<<<<{\!\!\!\!\!\!\quad\dcomment{\Sigma_{\sem{\Gamma;A}} \,\dashv\, \pi^*_{\sem{\Gamma;A}}}}^>>>>{\qquad\qquad\dcomment{\text{nat. of } \varepsilon^{\Sigma_{\sem{\Gamma;A}} \,\dashv\, \pi^*_{\sem{\Gamma;A}}}}}
\ar[r]^-{\id_{\Sigma_{\sem{\Gamma;A}}(1_{\ia {\sem{\Gamma;A}}})}}
&
\Sigma_{\sem{\Gamma;A}}(1_{\ia {\sem{\Gamma;A}}})
\ar@/^2.5pc/[d]^-{\mathsf{fst}}
\\
\Sigma_{\sem{\Gamma;A}}(\pi^*_{\sem{\Gamma;A}}(\Sigma_{\sem{\Gamma;A}}(1_{\ia {\sem{\Gamma;A}}})))
\ar[d]_-{\Sigma_{\sem{\Gamma;A}}(\pi^*_{\sem{\Gamma;A}}(\mathsf{fst}))}
\ar[ur]_>>>>>>>{\varepsilon^{\Sigma_{\sem{\Gamma;A}} \,\dashv\, \pi^*_{\sem{\Gamma;A}}}_{\Sigma_{\sem{\Gamma;A}}(1_{\ia {\sem{\Gamma;A}}})}}
&
\sem{\Gamma;A}
\\
\Sigma_{\sem{\Gamma;A}}(\pi^*_{\sem{\Gamma;A}}(\sem{\Gamma;A}))
\ar[d]_-{\Sigma_{\sem{\Gamma;A}}(\eta^{F \,\dashv\, U}_{\pi^*_{\sem{\Gamma;A}}(\sem{\Gamma;A})})}^>>>>{\,\,\,\,\,\,\,\dcomment{\eta^{F \,\dashv\, U} \text{ is a split fib. nat. transformation}}}
\ar[ur]_-{\varepsilon^{\Sigma_{\sem{\Gamma;A}} \,\dashv\, \pi^*_{\sem{\Gamma;A}}}_{\sem{\Gamma;A}}}
\ar[drr]^-{\quad\Sigma_{\sem{\Gamma;A}}(\pi^*_{\sem{\Gamma;A}}(\eta^{F \,\dashv\, U}_{\sem{\Gamma;A}}))}
&
&
U(F(\sem{\Gamma;A}))
\\
\Sigma_{\sem{\Gamma;A}}(U(F(\pi^*_{\sem{\Gamma;A}}(\sem{\Gamma;A}))))
\ar[rr]_-{=}
&&
\Sigma_{\sem{\Gamma;A}}(\pi^*_{\sem{\Gamma;A}}(U(F(\sem{\Gamma;A}))))
\ar[u]_-{\varepsilon^{\Sigma_{\sem{\Gamma;A}} \,\dashv\, \pi^*_{\sem{\Gamma;A}}}_{U(F(\sem{\Gamma;A}))}}
}
\]
\end{proof}

\section{Completeness}
\label{sect:completeness}

We now demonstrate that eMLTT is complete for fibred adjunction models.
We do so by proving that the well-formed syntax of eMLTT itself forms a fibred adjunction model, the \emph{classifying fibred adjunction model}. We construct this model by building on and extending the classifying \SCCompC\, construction for MLTT, as sketched in~\cite[Sections~10.3--10.5]{Jacobs:Book}. 
More specifically, we show in this section how to use  
well-formed\linebreak 

\pagebreak

\noindent contexts, types, and terms to construct the categorical structure depicted in
\vspace{-2.1cm}
\[
\xymatrix@C=4em@R=5em@M=0.5em{
\ar@{}[dd]^-{\!\!\quad\qquad\qquad\perp}
\\
\mathcal{V} \ar@/_1.75pc/[d]_-{p} \ar@{}[d]_-{\dashv\,\,\,\,\,} \ar@{}[d]^-{\,\,\,\,\,\,\,\dashv} \ar@/^1.75pc/[d]^-{\ia {-}} \ar@/^1.25pc/[rr]^-{F} &  &  \mathcal{C} \ar@/^1.25pc/[ll]^-{U} \ar@/^1pc/[dll]^-{q}
\\
\mathcal{B} \ar[u]_-{\!1}
}
\]
together with the structure we use to model eMLTT's value and computation types.

In order to make our discussion about the computational fragment of this classifying model easier to follow, we begin by recalling the core details of the classifying \SCCompC\, construction from op. cit., comprising the adjunctions $p \dashv 1$ and $1 \dashv \ia -$. 

To this end, we first show how to extend the unary substitutions we defined in Definition~\ref{def:substvaluevariables} (and the corresponding results thereafter)
to simultaneous substitutions. 
Of course, unary substitutions are just a special case of these simultaneous substitutions.

\begin{definition}
\label{def:simultaneoussubstvaluevariables}
\index{substitution!simultaneous --}
The $n$-ary \emph{simultaneous substitution} of value terms $V_1, \ldots, V_n$ 
for distinct value variables $x_1, \ldots, x_n$ in an expression $E$, written $E[V_1/x_1, \ldots, V_n/x_n]$ 
(or $E[\overrightarrow{V_i}/\overrightarrow{x_i}]$ for short), 
\index{ E@$E[V_1/x_1, \ldots, V_n/x_n]$ (simultaneous substitution)}
\index{ E@$E[\overrightarrow{V_i}/\overrightarrow{x_i}]$ (simultaneous substitution)}
is defined by recursion on the structure of $E$ as follows:
\[
\begin{array}{l c l}
\Nat[\overrightarrow{V_i}/\overrightarrow{x_i}] & \defeq & \Nat
\\
& \ldots &
\\
x_i[\overrightarrow{V_i}/\overrightarrow{x_i}] & \defeq & V_i
\\
y[\overrightarrow{V_i}/\overrightarrow{x_i}] & \defeq & y \qquad\qquad\qquad\qquad\qquad\qquad (\text{if~} y \not\in \{x_1, \ldots, x_n\})
\\
& \ldots &
\\
(\return W)[\overrightarrow{V_i}/\overrightarrow{x_i}] & \defeq & \return (W[\overrightarrow{V_i}/\overrightarrow{x_i}])
\\
(\doto M {y \!:\! A} {\ul{C}} N)[\overrightarrow{V_i}/\overrightarrow{x_i}] & \defeq & \doto {M[\overrightarrow{V_i}/\overrightarrow{x_i}]} {y \!:\! A[\overrightarrow{V_i}/\overrightarrow{x_i}]} {\ul{C}[\overrightarrow{V_i}/\overrightarrow{x_i}]} {N[\overrightarrow{V_i}/\overrightarrow{x_i}]}
\\
& \ldots &
\\
(K(W)_{(y : A).\, \ul{C}})[\overrightarrow{V_i}/\overrightarrow{x_i}] & \defeq & (K[\overrightarrow{V_i}/\overrightarrow{x_i}])(W[\overrightarrow{V_i}/\overrightarrow{x_i}])_{(y : A[\overrightarrow{V_i}/\overrightarrow{x_i}]).\, \ul{C}[\overrightarrow{V_i}/\overrightarrow{x_i}]}
\\
(W(K)_{\ul{C}, \ul{D}})[\overrightarrow{V_i}/\overrightarrow{x_i}] & \defeq & (W[\overrightarrow{V_i}/\overrightarrow{x_i}])(K[\overrightarrow{V_i}/\overrightarrow{x_i}])_{\ul{C}[\overrightarrow{V_i}/\overrightarrow{x_i}], \ul{D}[\overrightarrow{V_i}/\overrightarrow{x_i}]}
\end{array}
\]
where, according to our adopted variable conventions, the bound value variables are assumed to be different from the variables $x_1, \ldots, x_n$ we are substituting $V_1, \ldots, V_n$ for.
\end{definition}

Below, we list some useful properties of simultaneous substitutions that we use 
for constructing the classifying fibred adjunction model for eMLTT; many of them are natural generalisations of the properties we proved for unary substitutions in Chapter~\ref{chap:syntax}.

\pagebreak 

\begin{proposition}
\label{prop:freevariablesofsubsstitutionsimultaneous}
Given an expression $E$, then 
\index{ FVV@$FVV(E)$ (set of free value variables of $E$)}
\[
FVV(E[\overrightarrow{V_i}/\overrightarrow{x_i}]) \subseteq (FVV(E) - \{x_1, \ldots, x_n\}) \,\cup\, FVV(V_1) \,\cup\, \ldots \,\cup\, FVV(V_n)
\]
\end{proposition}

\begin{proof}
By induction on the structure of $E$.
\end{proof}

\begin{proposition}
\label{prop:valuesubstlemma1simultaneous}
Given an expression $E$ such that $x_i \not\in FVV(E)$, then 
\[
E[V_1/x_1, \ldots, V_i/x_i, \ldots, V_n/x_n] = E[V_1/x_1, \ldots, V_{i-1}/x_{i-1}, V_{i+1}/x_{i+1}, \ldots, V_n/x_n]
\]
\end{proposition}

\begin{proof}
By induction on the structure of $E$.
\end{proof}

\begin{proposition}
\label{prop:valuesubstlemma2simultaneous}
Given an expression $E$, then 
\[
E[V_1/x_1, \ldots, x_i/x_i, \ldots, V_n/x_n] = E[V_1/x_1, \ldots, V_{i-1}/x_{i-1}, V_{i+1}/x_{i+1}, \ldots, V_n/x_n]
\]
\end{proposition}

\begin{proof}
By induction on the structure of $E$.
\end{proof}

\begin{proposition}
\label{prop:valuesubstlemma3simultaneous}
Given an expression $E$ such that $\{x_1, \ldots, x_n\} \cap \{y_1, \ldots, y_m\} = \emptyset$ and 
$\{x_1, \ldots, x_n\} \cap (FVV(W_1) \cup \ldots \cup FVV(W_m)) = \emptyset$, then 
\[
E[\overrightarrow{V_i}/\overrightarrow{x_i}][\overrightarrow{W_j}/\overrightarrow{y_j}] = 
E[\overrightarrow{W_j}/\overrightarrow{y_j}][\overrightarrow{V_i[\overrightarrow{W_j}/\overrightarrow{y_j}]}/\overrightarrow{x_i}]
\]
\end{proposition}

\begin{proof}
By induction on the structure of $E$.
\end{proof}

\begin{proposition}
\label{prop:simultaneoussubstlemma2}
Given an expression $E$ such that $\{x_1, \ldots, x_n\} \cap \{y_1, \ldots, y_m\} = \emptyset$ 
and $\{y_1, \ldots, y_m\} \cap (FVV(V_1) \cup \ldots \cup FVV(V_n)) = \emptyset$, then 
\[
E[\overrightarrow{V_i}/\overrightarrow{x_i}][\overrightarrow{W_j}/\overrightarrow{y_j}] = E[\overrightarrow{V_i}/\overrightarrow{x_i},\overrightarrow{W_j}/\overrightarrow{y_j}]
\]
\end{proposition}

\begin{proof}
By induction on the structure of $E$. 
\end{proof}

\begin{proposition}
\label{prop:simultaneoussubstlemma1}
Given an expression $E$ such that $FVV(E) \subseteq \{x_1, \ldots, x_n\}$, then 
\[
E[\overrightarrow{V_i}/\overrightarrow{x_i}][\overrightarrow{W_j}/\overrightarrow{y_j}] = E[\overrightarrow{V_i[\overrightarrow{W_j}/\overrightarrow{y_j}]}/\overrightarrow{x_i}]
\]
\end{proposition}

\begin{proof}
By induction on the structure of $E$. 
\end{proof}

\begin{proposition}
\label{prop:simultaneoussubstlemma3}
Given an expression $E$ such that $\{x'_1, \ldots, x'_n\} \cap FVV(E) = \emptyset$, then 
\[
E[\overrightarrow{V_i}/\overrightarrow{x_i}] = E[\overrightarrow{x'_i}/\overrightarrow{x_i}][\overrightarrow{V_i}/\overrightarrow{x'_i}]
\]
\end{proposition}

\begin{proof}
By induction on the structure of $E$.
\end{proof}

\begin{proposition}
Given a homomorphism term $K$ with $FCV(K) = z$, then 
\[
K[M/z][\overrightarrow{V_i}/\overrightarrow{x_i}] = K[\overrightarrow{V_i}/\overrightarrow{x_i}][M[\overrightarrow{V_i}/\overrightarrow{x_i}]/z]
\]
\end{proposition}

\begin{proof}
By induction on the structure of $K$.
\end{proof}

\begin{proposition}
Given homomorphism terms $K$ and $L$ with $FCV(L) = z$, then 
\[
L[K/z][\overrightarrow{V_i}/\overrightarrow{x_i}] = L[\overrightarrow{V_i}/\overrightarrow{x_i}][K[\overrightarrow{V_i}/\overrightarrow{x_i}]/z]
\]
\end{proposition}

\begin{proof}
By induction on the structure of $K$.
\end{proof}

Finally, we show that in addition to unary substitutions (Theorem~\ref{thm:substitution}), eMLTT's judgements are 
are also closed under simultaneous substitutions of 
Definition~\ref{def:simultaneoussubstvaluevariables}.

\begin{theorem}[Simultaneous value term substitution] 
\label{thm:simultaneoussubstitution}
\index{substitution theorem!syntactic --!-- for value terms}
Given $\Gamma_2 = x_1 \!:\! A_1, \ldots, x_n \!:\! A_n$ and value terms 
$\vj {\Gamma_1} {V_1} {A_1}$, $\ldots$, 
$\vj {\Gamma_1} {V_n} {A_n[V_1/x_1, \ldots, V_{n-1}/x_{n-1}]}$, then we have:
\begin{enumerate}[(a)]
\item Given $\lj {\Gamma_2} B$, then $\lj {\Gamma_1} B[\overrightarrow{V_i}/\overrightarrow{x_i}]$.
\item Given $\ljeq {\Gamma_2} {B_1} {B_2}$, then $\ljeq {\Gamma_1} {B_1[\overrightarrow{V_i}/\overrightarrow{x_i}]} {B_2[\overrightarrow{V_i}/\overrightarrow{x_i}]}$.
\item Given $\lj {\Gamma_2} \ul{C}$, then $\lj {\Gamma_1} \ul{C}[\overrightarrow{V_i}/\overrightarrow{x_i}]$.
\item Given $\ljeq {\Gamma_2} {\ul{C}} {\ul{D}}$, then $\ljeq {\Gamma_1} {\ul{C}[\overrightarrow{V_i}/\overrightarrow{x_i}]} {\ul{D}[\overrightarrow{V_i}/\overrightarrow{x_i}]}$.
\item Given $\vj {\Gamma_2} W B$, then $\vj {\Gamma_1} {W[\overrightarrow{V_i}/\overrightarrow{x_i}]} {B[\overrightarrow{V_i}/\overrightarrow{x_i}]}$.
\item Given $\veq {\Gamma_2} {W_1} {W_2} B$, then $\veq {\Gamma_1} {W_1[\overrightarrow{V_i}/\overrightarrow{x_i}]} {W_2[\overrightarrow{V_i}/\overrightarrow{x_i}]} {B[\overrightarrow{V_i}/\overrightarrow{x_i}]}$.
\item Given $\cj {\Gamma_2} M \ul{C}$, then $\cj {\Gamma_1} {M[\overrightarrow{V_i}/\overrightarrow{x_i}]} {\ul{C}[\overrightarrow{V_i}/\overrightarrow{x_i}]}$.
\item Given $\ceq {\Gamma_2} M N \ul{C}$, then $\ceq {\Gamma_1} {M[\overrightarrow{V_i}/\overrightarrow{x_i}]} {N[\overrightarrow{V_i}/\overrightarrow{x_i}]} {\ul{C}[\overrightarrow{V_i}/\overrightarrow{x_i}]}$.
\item Given $\hj {\Gamma_2} {z \!:\! \ul{C}} K \ul{D}$, then $\hj {\Gamma_1} {z \!:\! \ul{C}[\overrightarrow{V_i}/\overrightarrow{x_i}]} {K[\overrightarrow{V_i}/\overrightarrow{x_i}]} {\ul{D}[\overrightarrow{V_i}/\overrightarrow{x_i}]}$.
\item Given $\heq {\Gamma_2} {z \!:\! \ul{C}} K L \ul{D}$, then $\heq {\Gamma_1} {z \!:\! \ul{C}[\overrightarrow{V_i}/\overrightarrow{x_i}]} {K[\overrightarrow{V_i}/\overrightarrow{x_i}]} {L[\overrightarrow{V_i}/\overrightarrow{x_i}]} {\ul{D}[\overrightarrow{V_i}/\overrightarrow{x_i}]}$.
\end{enumerate}
\end{theorem}

\begin{proof}
We prove $(a)$--$(j)$ using the combination of Theorems~\ref{thm:weakening} and~\ref{thm:substitution}, and other results we established earlier. For example, for $(c)$ the proof proceeds as follows.

To begin with, we use Proposition~\ref{prop:wellformedcomponentsofjudgements} with 
$\lj {\Gamma_2} \ul{C}$ to get that $\lj {} {\Gamma_2}$, whose derivation also gives us that 
$\lj {x_1 \!:\! A_1, \ldots, x_{i-1} \!:\! A_{i-1}} A_i$, for all $1 \leq i \leq n$. 
In addition, we use Proposition~\ref{prop:freevariablesofwellformedexpressions}
to get the inclusion $FVV(\ul{C}) \subseteq V\!ars(\Gamma_2) = \{x_1, \ldots, x_n\}$.

Next, we choose distinct value variables $x'_1, \ldots, x'_n$ such that they are disjoint from 
the variables of $\Gamma_1$ and $\Gamma_2$. We write $\widehat{\Gamma_2}$ for the 
``fresh" version of $\Gamma_2$, given by
\[
x'_1 \!:\! A_1, x'_2 \!:\! A_2[x'_1/x_1], \ldots, x'_n \!:\! A_n[x'_1/x_1]\ldots[x'_{n-1}/x_{n-1}]
\]
with Theorems~\ref{thm:weakening} and~\ref{thm:substitution}
allowing us to show that for all $1 \leq i \leq n$, we have 
\[
\lj {x'_1 \!:\! A_1, \ldots, x'_{i-1} \!:\! A_{i-1}[x'_1/x_1] \ldots [x'_{i-2}/x_{i-2}]} {A_i[x'_1/x_1] \ldots [x'_{i-1}/x_{i-1}]}
\]
and as $\{x_1, \ldots, x_n\} \cap \{x'_1, \ldots, x'_n\} = \emptyset$, we can use 
Proposition~\ref{prop:simultaneoussubstlemma2} to show that
\[
\lj {x'_1 \!:\! A_1, \ldots, x'_{i-1} \!:\! A_{i-1}[x'_1/x_1, \ldots ,x'_{i-2}/x_{i-2}]} {A_i[x'_1/x_1, \ldots ,x'_{i-1}/x_{i-1}]}
\]

Furthermore, by using Proposition~\ref{prop:freevariablesofwellformedexpressions}, we get that
$FVV(A_i) \subseteq \{x_1, \ldots, x_{i-1}\}$, for all $1 \leq i \leq n$, and thus 
we have for all $1 \leq i \leq n$ that $\{x'_1, \ldots, x'_n\} \cap FVV(A_i) = \emptyset$. 
As a consequence, we can use Proposition~\ref{prop:simultaneoussubstlemma3} with $A_i$ (for all $1 \leq i \leq n$) to get that
\[
A_i[V_1/x_1, \ldots, V_{i-1}/x_{i-1}] = A_i[x'_1/x_1, \ldots, x'_{i-1}/x_{i-1}][V_1/x'_1, \ldots, V_{i-1}/x'_{i-1}]
\]
and then Proposition~\ref{prop:simultaneoussubstlemma2} (as $FVV(V_i) \subseteq V\!ars(\Gamma_1)$ by Proposition~\ref{prop:freevariablesofwellformedexpressions}) to get that 
\[
\begin{array}{c}
A_i[x'_1/x_1, \ldots, x'_{i-1}/x_{i-1}][V_1/x'_1, \ldots, V_{i-1}/x'_{i-1}]
\\
=
\\ 
A_i[x'_1/x_1, \ldots, x'_{i-1}/x_{i-1}][V_1/x'_1]\ldots[V_{i-1}/x'_{i-1}]
\end{array}
\]
from which it follows that the assumed derivations of $V_i$ are also derivations of
\[
\vj {\Gamma_1} {V_i} {A_i[x'_1/x_1, \ldots, x'_{i-1}/x_{i-1}][V_1/x'_1]\ldots[V_{i-1}/x'_{i-1}]}
\]

Next, we repeatedly use Theorem~\ref{thm:weakening} to get a derivation of 
$\lj {\widehat{\Gamma_2},\Gamma_2} {\ul{C}}$ from that of $\lj {\Gamma_2} {\ul{C}}$, and then  
Theorem~\ref{thm:substitution} to get a derivation of 
$\lj {\widehat{\Gamma_2}} {\ul{C}[x'_1/x_1]\ldots[x'_n/x_n]}$. \linebreak 
However, as $\{x_1, \ldots, x_n\} \cap \{x'_1, \ldots, x'_n\} = \emptyset$, we can use 
Proposition~\ref{prop:simultaneoussubstlemma2} to get that 
\[
{\ul{C}[x'_1/x_1]\ldots[x'_n/x_n]} = {\ul{C}[x'_1/x_1, \ldots, x'_n/x_n]}
\]

Finally, we repeatedly use Theorem~\ref{thm:weakening} to get  
$\lj {\Gamma_1, \widehat{\Gamma_2}} {\ul{C}[x'_1/x_1, \ldots, x'_n/x_n]}$ from the derivation of 
$\lj {\widehat{\Gamma_2}} {\ul{C}[x'_1/x_1, \ldots, x'_n/x_n]}$, and then 
Theorem~\ref{thm:substitution} to get that 
\[
\lj {\Gamma_1} {\ul{C}[x'_1/x_1, \ldots, x'_n/x_n]}[V_1/x'_1]\ldots[V_n/x'_n]
\]
However, as we know that $FVV(\ul{C}) \subseteq \{x_1, \ldots, x_n\}$ and $FVV(V_i) \subseteq V\!ars(\Gamma_1)$, for all $1 \leq i \leq n$, 
then we can use 
Propositions~\ref{prop:simultaneoussubstlemma1} and~\ref{prop:valuesubstlemma1} to respectively get that 
\[
\ul{C}[x'_1/x_1, \ldots, x'_n/x_n][V_1/x'_1]\ldots[V_n/x'_n]
=
\ul{C}[\overrightarrow{x'_i[V_1/x'_1]\ldots[V_n/x'_n]}/\overrightarrow{x_i}]
=
\ul{C}[\overrightarrow{V_i}/\overrightarrow{x_i}]
\]
giving us the required derivation of
\[
\lj {\Gamma_1} {\ul{C}[V_1/x_1, \ldots, V_n/x_n]}
\]
\end{proof}

\subsection*{Base category $\mathcal{B}$ of value contexts}

The objects of $\mathcal{B}$ are given by equivalence classes $[\, \vdash \Gamma\, ]$ of well-formed value contexts $\vdash \Gamma$, where the equivalence relation is given by 
\[
\mkrule
{\vdash \Gamma_1 ~\equiv~ \vdash \Gamma_2}
{\ljeq {} {\Gamma_1} {\Gamma_2}}
\]

In order to improve the readability of the material presented in this section, we follow the standard convention of referring to the various equivalence classes we use by their representatives, i.e., we write $\vdash \Gamma$ for $[\, \vdash \Gamma\, ]$. To see that this simplification is valid, we observe that by definition the well-formed types, terms, and definitional equations are closed under context and type conversions.
Furthermore, according to Theorems~\ref{thm:weakening},~\ref{thm:substitution},~\ref{thm:compsubstitution}, and~\ref{thm:simultaneoussubstitution}, well-formed contexts, types, terms, and definitional equations are also closed under weakening and substitution. 

To further improve the readability of this section, we also omit the turnstile symbol when referring to  well-formed contexts, and simply write $\Gamma$ instead of $\vdash \Gamma$ (and $[\,\vdash \Gamma\,]$).

Given well-formed value contexts $\Gamma_1$ and $\Gamma_2$ such that $\Gamma_2 = x_1 \!:\! A_1, \ldots, x_n \!:\! A_n$, a morphism $\Gamma_1 \longrightarrow \Gamma_2$ in $\mathcal{B}$ is given by an equivalence class of tuples $( V_1, \ldots, V_n )$ of well-typed value terms, where $\vj {\Gamma_1} {V_i} {A_i[V_1/x_1, \ldots, V_{i-1}/x_{i-1}]}$, for all $1 \leq i \leq n$; and where the equivalence relation on such tuples of value terms is given component-wise:
\[
\mkrule
{{(V_1, \ldots, V_n)} \equiv {(W_1, \ldots, W_n)} : \Gamma_1 \longrightarrow \Gamma_2}
{\veq {\Gamma_1} {V_i} {W_i} {A_i[V_1/x_1, \ldots, V_{i-1}/x_{i-1}]} \qquad (1 \leq i \leq n)}
\]
Throughout this section, we often abbreviate tuples $(V_1, \ldots, V_n)$ of value terms as $\overrightarrow{V_i}$.
\index{ V@$\overrightarrow{V_i}$ (shorthand for $(V_1, \ldots, V_n)$)}

Next, the composition of any two morphisms of the form
\[
(V_1, \ldots, V_n) : \Gamma_1 \longrightarrow \Gamma_2
\qquad
(W_1, \ldots, W_m) : \Gamma_2 \longrightarrow \Gamma_3
\]
is given by simultaneous substitution of value terms for value variables, namely, by
\[
(W_1, \ldots, W_m) \comp (V_1, \ldots, V_n) \defeq (W_1[\overrightarrow{V_i}/\overrightarrow{x_i}], \ldots, W_m[\overrightarrow{V_i}/\overrightarrow{x_i}]) 
\]
assuming that $\Gamma_2 = x_1 \!:\! A_1, \ldots, x_n \!:\! A_n$.

Further, given any well-formed value context $\Gamma$ such that $\Gamma = x_1 \!:\! A_1, \ldots, x_n \!:\! A_n$, the identity morphism $\id_{\Gamma}$ is given by the variables of $\Gamma$, namely, by  
\[
\id_\Gamma \defeq (x_1, \ldots, x_n) : \Gamma \longrightarrow \Gamma
\]

The associativity and identity laws for composition follow from the properties we established about simultaneous substitutions in 
the beginning of this section.

Finally, this category also has a terminal object, given by the empty context $\diamond$, with the corresponding unique morphisms $\Gamma \longrightarrow \diamond$ given by the empty tuple of value terms.

\subsection*{Category $\mathcal{V}$ of value types}

The objects of $\mathcal{V}$ are given by equivalence classes of well-formed value types $\lj \Gamma A$. 
Given two value types $\lj {\Gamma_1} A$ and $\lj {\Gamma_2} B$ such that $\Gamma_2 =  x_1 \!:\! A_1, \ldots, x_n \!:\! A_n$, a morphism $\lj {\Gamma_1} A \longrightarrow \lj {\Gamma_2} B$ is given by an equivalence class of tuples $(V_1, \ldots, V_n, x.\,V)$, where $\vj {\Gamma_1} {V_i} {A_i[V_1/x_1, \ldots, V_{i-1}/x_{i-1}]}$, for all $1 \leq i \leq n$, as above, and where \linebreak $\vj {\Gamma_1, x \!:\! A} V {B[\overrightarrow{V_i}/\overrightarrow{x_i}]}$. In $(V_1, \ldots, V_n, x.\,V)$, the value variable $x$ is bound in the value term $V$. The equivalence relation is again given component-wise, namely, by
\[
\mkrule
{{(V_1, \ldots, V_n, x.\,V)} \equiv {(W_1, \ldots, W_n, x.\,W)} : \lj {\Gamma_1} A \longrightarrow \lj {\Gamma_2} B}
{
\begin{array}{c}
\veq {\Gamma_1} {V_i} {W_i} {A_i[V_1/x_1, \ldots, V_{i-1}/x_{i-1}]} \quad (1 \leq i \leq n)
\\
\veq {\Gamma_1, x \!:\! A} {V} {W} {B[V_1/x_1, \ldots, V_n/x_n]}
\end{array}
}
\]

Analogously to $\mathcal{B}$, the composition of morphisms is given by simultaneous substitution of value terms for value variables. 
Further, given any value type $\lj \Gamma A$ such that $\Gamma = x_1 \!:\! A_1, \ldots, x_n \!:\! A_n$, the identity morphism $\id_{\lj {\Gamma\,} {\,A}}$ is given by a tuple of variables: 
\[
\id_{\lj {\Gamma\,} {\,A}} \defeq (x_1, \ldots, x_n, x.\,x) : \lj \Gamma A \longrightarrow \lj \Gamma A
\]

Finally, the associativity and identity laws for composition follow from the properties we established 
about simultaneous substitutions in the beginning of this section.

\subsection*{Split fibration $p : \mathcal{V} \longrightarrow \mathcal{B}$}

We define the functor $p$ by mapping a well-formed value type to its context, given by
\[
p(\lj \Gamma A) \defeq \Gamma
\qquad
p(V_1, \ldots, V_n, x.\,V) \defeq (V_1, \ldots, V_n)
\]
We omit the proofs showing that $p$ preserves identity morphisms and composition of morphisms. We also omit analogous proofs for all other functors defined in this section. 
 
Given a morphism $\overrightarrow{V_i} : \Gamma_1 \longrightarrow \Gamma_2$ and a value type  $\lj {\Gamma_2} A$, the morphism
\[
(\overrightarrow{V_i}, x.\,x) : \lj {\Gamma_1} {A[\overrightarrow{V_i}/\overrightarrow{x_i}]} \longrightarrow \lj {\Gamma_2} A
\]
is Cartesian over $\overrightarrow{V_i} : \Gamma_1 \longrightarrow p(\lj {\Gamma_2} A)$, with the unique mediating morphism in
\[
\xymatrix@C=4.25em@R=3em@M=0.5em{
\lj {\Gamma_3} B \ar@/^2pc/[rr]^-{(\overrightarrow{W_i}, x.W)} \ar@{-->}[r] & \lj {\Gamma_1} {A[\overrightarrow{V_i}/\overrightarrow{x_i}]} \ar[r]_-{(\overrightarrow{V_i}, x.\,x)} & \lj {\Gamma_2} A & \text{in} & \mathcal{V} \ar[d]^-{p}
\\
\Gamma_3 \ar[r]^-{\overrightarrow{V'_j}} \ar@/_2pc/[rr]_{p(\overrightarrow{W_i}, x.W)}  & \Gamma_1 \ar[r]^{\overrightarrow{V_i}} & \Gamma_2 & \text{in} & \mathcal{B}
}
\]
given by
\[
(V'_1, \ldots, V'_m, x.\,W) : \lj {\Gamma_3} B \longrightarrow \lj {\Gamma_1} {A[\overrightarrow{V_i}/\overrightarrow{x_i}]}
\]

This morphism is well-formed because the commutativity of the lower diagram in $\mathcal{B}$ means that we must have
\[
\veq {\Gamma_3} {W_i} {V_i[\overrightarrow{V'_j}/\overrightarrow{y_j}]} {A_i[W_1/x_1, \ldots, W_{i-1}/x_{i-1}]}
\]
for all $1 \leq i \leq n$. 

In addition, the two value types that are assigned to $W$ in the two morphisms containing it are definitionally equal because of the properties we established about simultaneous substitutions 
in the beginning of this section. Concretely, we have
\[
\ljeq {\Gamma_3} {A[\overrightarrow{V_i}/\overrightarrow{x_i}][\overrightarrow{V'_j}/\overrightarrow{y_j}]} {A[V_1[\overrightarrow{V'_j}/\overrightarrow{y_j}]/x_1, \ldots, V_n[\overrightarrow{V'_j}/\overrightarrow{y_j}]/x_n]}
\]

The uniqueness of $(\overrightarrow{V'_j}, x.\,W) : \lj {\Gamma_3} B \longrightarrow \lj {\Gamma_1} {A[\overrightarrow{V_i}/\overrightarrow{x_i}]}$ is also easy to prove. Specifically, given any other morphism $(\overrightarrow{W'_j}, x.\, W') : \lj {\Gamma_3} B \longrightarrow \lj {\Gamma_1} {A[\overrightarrow{V_i}/\overrightarrow{x_i}]}$ that makes the previous two diagrams commute, this commutativity means that we have
\[
\veq {\Gamma_3, x \!:\! B} {W'} {W} {A[\overrightarrow{V_i}/\overrightarrow{x_i}]}
\]
and that for all $1 \leq j \leq m$, we also have
\[
\veq {\Gamma_3} {W'_j} {V'_j} {A'_i[V'_1/y_1, \ldots, V'_{j-1}/y_{j-1}]}
\]
As a result, we get the required equation
\[
(W'_1, \ldots, W'_m, x.\, W') = (V'_1, \ldots, V'_m, x.\,W)
\]

The definition of the chosen Cartesian morphisms also tells us how the induced reindexing functors $(V_1, \ldots, V_n)^* : \mathcal{V}_{\Gamma_2} \longrightarrow \mathcal{V}_{\Gamma_1}$ are defined. They are given on objects by
\vspace{-0.25cm}
\[
(\overrightarrow{V_i})^*(\lj {\Gamma_2} A) \defeq \lj {\Gamma_1} A[\overrightarrow{V_i}/\overrightarrow{x_i}]
\]
and on morphisms $(\overrightarrow{x_i}, x.\,V) : \lj {\Gamma_2} {A} \longrightarrow \lj {\Gamma_2} {B}$ by
\[
(\overrightarrow{V_i})^*(\overrightarrow{x_i}, x.\,V) \defeq (\overrightarrow{y_j}, x.\,V[\overrightarrow{V_i}/\overrightarrow{x_i}])
: \lj {\Gamma_1} A[\overrightarrow{V_i}/\overrightarrow{x_i}] \longrightarrow \lj {\Gamma_1} B[\overrightarrow{V_i}/\overrightarrow{x_i}]
\]

Finally, we show that $p$ is a split fibration. On the one hand, we observe that for any well-formed value context $\Gamma$ such that $\Gamma = x_1 \!:\! A_1, \ldots, x_n \!:\! A_n$ we have
\[
\id_{\Gamma}^*(\lj \Gamma A) = (\overrightarrow{x_i})^*(\lj \Gamma A) = \lj \Gamma A[\overrightarrow{x_i}/\overrightarrow{x_i}] = \lj \Gamma A
\]
On the other hand, given any two morphisms $\overrightarrow{V_i} : \Gamma_1 \longrightarrow \Gamma_2$ and $\overrightarrow{W_j} : \Gamma_2 \longrightarrow \Gamma_3$ in $\mathcal{B}$ such  that $\Gamma_2 = x_1 \!:\! B_1, \ldots, x_n \!:\! B_n$ and $\Gamma_3 = y_1 \!:\! B'_1, \ldots, y_m \!:\! B'_m$, we have
\begin{fleqn}[0.3cm]
\begin{align*}
& (\overrightarrow{W_j} \comp \overrightarrow{V_i})^*(\lj {\Gamma_3} A) 
\\
=\,\, &
(W_1[\overrightarrow{V_i}/\overrightarrow{x_i}], \ldots, W_m[\overrightarrow{V_i}/\overrightarrow{x_i}])^*(\lj {\Gamma_3} A)
\\
=\,\, & \lj {\Gamma_1} {A[W_1[\overrightarrow{V_i}/\overrightarrow{x_i}]/y_1, \ldots, W_m[\overrightarrow{V_i}/\overrightarrow{x_i}]/y_m]}
\\
=\,\, & \lj {\Gamma_1} {A[\overrightarrow{W_j}/\overrightarrow{y_j}][\overrightarrow{V_i}/\overrightarrow{x_i}]}
\\
=\,\, & (\overrightarrow{V_i})^*(\lj {\Gamma_2} {A[\overrightarrow{W_j}/\overrightarrow{y_j}]})
\\
=\,\, & (\overrightarrow{V_i})^*((\overrightarrow{W_j})^*(\lj {\Gamma_3} A))
\end{align*}
\end{fleqn}

\subsection*{Split fibred terminal object functor $1 : \mathcal{B} \longrightarrow \mathcal{V}$}

We define the terminal object functor $1$ in terms of the unit type, by 
\[
1 (\Gamma) \defeq \lj \Gamma 1
\qquad
1 (V_1, \ldots, V_n) \defeq (V_1, \ldots, V_n, x.\,x)
\]

We proceed by showing that $1$ is split fibred. On the one hand, we trivially have that $p \comp 1 = \id_{\mathcal{B}}$. On the other hand, we also see that $1$ preserves Cartesian morphisms on-the-nose---in $\id_{\mathcal{B}} : \mathcal{B} \longrightarrow \mathcal{B}$, every morphism in the total category is Cartesian.

The unit and counit of the adjunction $p \dashv 1$ are given by components
\[
\eta_{\lj {\Gamma\,} {\,A}} \defeq (\overrightarrow{x_i}, x.\,\star) : \lj \Gamma A \longrightarrow \lj \Gamma 1
\qquad
\varepsilon_{\Gamma} \defeq \overrightarrow{x_i} : \Gamma \longrightarrow \Gamma 
\]
assuming that $\Gamma = x_1 \!:\! A_1, \ldots, x_n \!:\! A_n$.

We omit the proofs of the naturality of $\eta$ and $\varepsilon$. 
We also omit the naturality proofs for all other natural transformations we define in the rest of this section. Finally, we note that the two unit-counit laws hold because both $\eta$ and $\varepsilon$ are defined as identities on contexts; and every well-formed value term of type $1$ is definitionally equal to $\star$.

\subsection*{Comprehension functor $\ia - : \mathcal{V} \longrightarrow \mathcal{B}$}

We define $\ia -$ in terms of context extension.
To facilitate this, we fix a choice of a fresh value variable $\mathsf{fresh}(X)$ for every finite set 
$X$ of value variables, and then define
\index{ f@$\mathsf{fresh}(X)$ (a choice of a fresh value variable for a finite set $X$ of value variables)}
\[
\ia {\lj \Gamma A} \defeq \Gamma, x \!:\! A
\qquad
\ia {(\overrightarrow{V_i}, y.\, V)} \defeq (\overrightarrow{V_i}, V[y_1/y]) : \Gamma_1, y_1 \!:\! A \longrightarrow \Gamma_2, y_2 \!:\! B
\]
where $(\overrightarrow{V_i}, y.\, V) : \lj {\Gamma_1} A \longrightarrow \lj {\Gamma_2} B$ and 
$x \defeq \mathsf{fresh}(V\!ars(\Gamma))$, with $y_1$ and $y_2$ chosen similarly.
In the rest of this chapter, we often leave the uses of $\mathsf{fresh}(X)$ implicit.

The unit and counit of the adjunction $1 \dashv \ia -$ are given by components
\[
\eta_{\Gamma} \defeq (\overrightarrow{x_i}, \star) : \Gamma \longrightarrow \Gamma, x \!:\! 1
\qquad
\varepsilon_{\lj {\Gamma\,} {\,A}} \defeq (\overrightarrow{x_i}, y.\, x) : \lj {\Gamma, x \!:\! A} 1 \longrightarrow \lj \Gamma A
\]
assuming that $\Gamma = x_1 \!:\! A_1, \ldots, x_n \!:\! A_n$. 

We conclude by showing that the two unit-counit laws hold for these $\eta$ and $\varepsilon$.

On the one hand, we observe that the first unit-counit triangle
\[
\xymatrix@C=5em@R=5em@M=0.5em{
\ia - \ar[r]^-{\eta \,\comp\, \ia -} \ar[dr]_{\id_{\ia -}} & \ia - \comp 1 \comp \ia - \ar[d]^-{\ia - \,\comp\, \varepsilon}
\\
& \ia -
}
\]
can be rewritten for each well formed value type $\lj \Gamma A$ (with $\Gamma = x_1 \!:\! A_1, \ldots, x_n \!:\! A_n$) as follows:
\[
\xymatrix@C=5em@R=5em@M=0.5em{
\Gamma, x \!:\! A \ar[r]^-{(\overrightarrow{x_i}, x, \star)} \ar[dr]_{(\overrightarrow{x_i}, x)} & \Gamma, x \!:\! A, y \!:\! 1 \ar[d]^-{(\overrightarrow{x_i}, x)}
\\
& \Gamma, x \!:\! A
}
\]
It is now easy to verify that this triangle commutes, simply by using the simultaneous substitution based definition of the composition of morphisms in $\mathcal{B}$. 

On the other hand, we observe that the second unit-counit triangle
\[
\xymatrix@C=5em@R=5em@M=0.5em{
1 \ar[r]^-{1 \,\comp\, \eta} \ar[dr]_{\id_1} & 1 \comp \ia - \comp 1 \ar[d]^-{\varepsilon \,\comp\, 1}
\\
& 1
}
\]
can be rewritten for each well-formed value context $\Gamma = x_1 \!:\! A_1, \ldots, x_n \!:\! A_n$ as follows:
\[
\xymatrix@C=5em@R=5em@M=0.5em{
\lj \Gamma 1 \ar[r]^-{(\overrightarrow{x_i}, \star, y. y)} \ar[dr]_{(\overrightarrow{x_i}, x. x)} & \lj {\Gamma, x \!:\! 1} 1 \ar[d]^-{(\overrightarrow{x_i}, y'\!\!. x)}
\\
& \lj \Gamma 1
}
\]
Similarly to the other unit-counit triangle, it is now easy to verify that this triangle commutes, simply by using the definition of the composition of morphisms in $\mathcal{B}$, and the fact that every well-typed term of type $1$ is definitionally equal to $\star$.

\subsection*{The induced split full comprehension category $\mathcal{P} : \mathcal{V} \longrightarrow \mathcal{B}^{\to}$}

We begin by recalling that we showed how the comprehension category $\mathcal{P} : \mathcal{V} \longrightarrow \mathcal{B}^{\to}$ is derived from the adjunction $1 \dashv\, \ia -$ in Proposition~\ref{prop:comprehensioncategorywithunit}. In this classifying fibred adjunction model, the functor $\mathcal{P} : \mathcal{V} \longrightarrow \mathcal{B}^{\to}$ can be shown to be given on objects by
\[
\mathcal{P}(\lj \Gamma A) \defeq (x_1, \ldots, x_n) : \Gamma, x \!:\! A \longrightarrow \Gamma
\]
assuming that $\Gamma = x_1 \!:\! A_1, \ldots, x_n \!:\! A_n$, with $x \defeq \mathsf{fresh}(V\!ars(\Gamma))$; and on morphisms by
\[
\xymatrix@C=6em@R=4em@M=0.5em{
\Gamma_1, y_1 \!:\! A 
\ar[d]_-{\txt<14pc>{$\mathcal{P}(\overrightarrow{V_j}, x.\, V) \qquad \defeq $}}_-{\overrightarrow{x_i}}
\ar[r]^-{(\overrightarrow{V_j}, V[y_1/x])}
&
\Gamma_2, y_2 \!:\! B
\ar[d]^-{\overrightarrow{x'_j}}
\\
\Gamma_1
\ar[r]_-{\overrightarrow{V_j}}
&
\Gamma_2
}
\]
where $(\overrightarrow{V_j}, x.\, V) : \lj {\Gamma_1} A \longrightarrow \lj {\Gamma_2} B$. Further, for better readability, we  
assume in the previous diagram that $\Gamma_1 = x_1 \!:\! A_1, \ldots, x_n \!:\! A_n$ and $\Gamma_2 = x'_1 \!:\! B_1, \ldots, x'_m \!:\! B_m$. 

In order to show that this comprehension category is full, we need to prove that the functor $\mathcal{P}$ is fully-faithful. We do so by first constructing a mapping 
\[
\mathcal{P}^{-1}_{\lj {\Gamma_1} A, \lj {\Gamma_2} B} : \mathcal{B}^{\to}(\mathcal{P}(\lj {\Gamma_1} A) , \mathcal{P}(\lj {\Gamma_2} B)) \longrightarrow \mathcal{V}(\lj {\Gamma_1} A , \lj {\Gamma_2} B)
\]
for any two well-formed value types $\lj {\Gamma_1} A$ and $\lj {\Gamma_2} B$. In particular, given a morphism 
\[
\xymatrix@C=6em@R=4em@M=0.5em{
\Gamma_1, y_1 \!:\! A 
\ar[d]_-{\overrightarrow{x_i}}
\ar[r]^-{(\overrightarrow{W_j}, W)}
&
\Gamma_2, y_2 \!:\! B
\ar[d]^-{\overrightarrow{x'_j}}
\\
\Gamma_1
\ar[r]_-{\overrightarrow{V_j}}
&
\Gamma_2
}
\]
in $\mathcal{B}^{\to}(\mathcal{P}(\lj {\Gamma_1} A) , \mathcal{P}(\lj {\Gamma_2} B))$, we first observe that the commutativity of the above square entails $\veq {\Gamma_1} {W_j} {V_j} {B_j[\overrightarrow{W_k}/\overrightarrow{x'_k}]}$, for all $1 \leq j \leq m$. Therefore, we can  define 
\[
\mathcal{P}^{-1}_{\lj {\Gamma_1} A, \lj {\Gamma_2} B}((\overrightarrow{W_j}, V),\overrightarrow{V_j}) \defeq (\overrightarrow{V_j}, y_1.\, W)
\]

Now, verifying that $\mathcal{P}$ is fully-faithful is straightforward: on the one hand, we have 
\begin{fleqn}[0.3cm]
\begin{align*}
& \mathcal{P}(\mathcal{P}^{-1}_{\lj {\Gamma_1} A, \lj {\Gamma_2} B}((\overrightarrow{W_j}, V),\overrightarrow{V_j})) 
\\
=\,\, &
\mathcal{P}(\overrightarrow{V_j}, y_1.\, W)
\\
=\,\, &
((\overrightarrow{V_j},W[y_1/y_1]) , \overrightarrow{V_j})
\\
=\,\, &
((\overrightarrow{V_j},W) , \overrightarrow{V_j})
\\
=\,\, &
((\overrightarrow{W_j},W) , \overrightarrow{V_j})
\end{align*}
\end{fleqn}
and on the other hand, we have 
\begin{fleqn}[0.3cm]
\begin{align*}
& \mathcal{P}^{-1}_{\lj {\Gamma_1} A, \lj {\Gamma_2} B}(\mathcal{P}(\overrightarrow{V_j}, x .\, V))
\\
=\,\, &
\mathcal{P}^{-1}_{\lj {\Gamma_1} A, \lj {\Gamma_2} B}((\overrightarrow{V_j}, V[y_1/x]),(\overrightarrow{V_j}))
\\
=\,\, &
(\overrightarrow{V_j}, y_1.\,V[y_1/x])
\\
=\,\, &
(\overrightarrow{V_j}, x .\,V)
\end{align*}
\end{fleqn}

\subsection*{Category $\mathcal{C}$ of computation types}

The category $\mathcal{C}$ of computation types is defined similarly to the category $\mathcal{V}$ of value types. First, the objects of $\mathcal{C}$ are given by well-formed computation types $\lj \Gamma \ul{C}$. Secondly, given two well-formed computation types $\lj {\Gamma_1} \ul{C}$ and $\lj {\Gamma_2} \ul{D}$ such that \linebreak $\Gamma_2 = x_1 \!:\! A_1, \ldots, x_n \!:\! A_n$, a morphism $\lj {\Gamma_1} \ul{C} \longrightarrow \lj {\Gamma_2} \ul{D}$ is given by an equivalence \linebreak class of tuples $(V_1, \ldots, V_n, z.\, K)$ of well-typed value and homomorphism terms, \linebreak where $\vj {\Gamma_1} {V_i} {A_i[V_1/x_1, \ldots, V_{i-1}/x_{i-1}]}$, for all $1 \leq i \leq n$, as before; and where \linebreak $\hj {\Gamma_1} {z \!:\! \ul{C}} {K} {\ul{D}[\overrightarrow{V_i}/\overrightarrow{x_i}]}$. In $(V_1, \ldots, V_n, z.\,K)$, the computation variable $z$ is bound in the homomorphism term $K$.
The equivalence relation on such tuples of well-typed value and homomorphism terms is again given component-wise, namely, by
\[
\mkrule
{{(V_1, \ldots, V_n, z.\,K)} \equiv {(W_1, \ldots, W_n, z.\,L)} : \lj {\Gamma_1} \ul{C} \longrightarrow \lj {\Gamma_2} \ul{D}}
{
\begin{array}{c}
\veq {\Gamma_1} {V_i} {W_i} {A_i[V_1/x_1, \ldots, V_{i-1}/x_{i-1}]} \qquad (1 \leq i \leq n)
\\
\heq {\Gamma_1} {z \!:\! \ul{C}} {K} {L} {\ul{D}[V_1/x_1, \ldots, V_n/x_n]}
\end{array}
}
\]

Analogously to $\mathcal{V}$, the composition of morphisms is again defined using simultaneous substitutions, but this time by also using the substitution of homomorphism terms for computation variables. In detail, the composition of any two morphisms
\[
(V_1, \ldots, V_n, z_1.\, K) : \lj {\Gamma_1} {\ul{C}_1} \longrightarrow \lj {\Gamma_2} {\ul{C}_2}
\qquad
(W_1, \ldots, W_m, z_2.\, L) : \lj {\Gamma_2} {\ul{C}_2} \longrightarrow \lj {\Gamma_3} {\ul{C}_3}
\]
is given by
\[
\hspace{-0.1cm}
(W_1, \ldots, W_m, z_2.\, L) \comp (V_1, \ldots, V_n, z_1.\, K) \defeq (W_1[\overrightarrow{V_i}/\overrightarrow{x_i}], \ldots, W_m[\overrightarrow{V_i}/\overrightarrow{x_i}], z_1.\, L[\overrightarrow{V_i}/\overrightarrow{x_i}][K/z_2])
\]
where we assume that $\Gamma_2 =  x_1 \!:\! A_1, \ldots, x_n \!:\! A_n$. 

For any well-formed computation type $\lj \Gamma \ul{C}$, the identity morphism $\id_{\lj {\Gamma\,} {\,\ul{C}}}$ is again given by a tuple of variables, namely, by
\[
\id_{\lj {\Gamma\,} {\,\ul{C}}} \defeq (x_1, \ldots, x_n, z.\, z) : \lj \Gamma \ul{C} \longrightarrow \lj \Gamma \ul{C}
\]
assuming that $\Gamma = x_1 \!:\! A_1, \ldots, x_n \!:\! A_n$.

Finally, analogously to $\mathcal{V}$, the associativity and identity laws for composition follow  from the properties we established about simultaneous substitutions of value terms for value variables in the beginning of this section, 
and the results we established in Section~\ref{sect:syntax} about substituting homomorphism terms for computation variables.

\subsection*{Split fibration $q : \mathcal{C} \longrightarrow \mathcal{B}$}

We define the functor $q$ by mapping a well-formed computation type to its context:
\[
q(\lj \Gamma A) \defeq \Gamma
\qquad
q(V_1, \ldots, V_n, z.\, K) \defeq (V_1, \ldots, V_n)
\]
where $(V_1, \ldots, V_n, z.\, K) : \lj {\Gamma_1} {\ul{C}} \longrightarrow \lj {\Gamma_2} {\ul{D}}$.

Given a morphism $\overrightarrow{V_i} : \Gamma_1 \longrightarrow \Gamma_2$ in $\mathcal{B}$ and a well-formed computation type $\lj {\Gamma_2} {\ul{C}}$, we note that the morphism given by
\[
(\overrightarrow{V_i}, z.\,z) : \lj {\Gamma_1} {\ul{C}[V_1/x_1, \ldots, V_n/x_n]} \longrightarrow \lj {\Gamma_2} {\ul{C}}
\]
is Cartesian over $\overrightarrow{V_i} : \Gamma_1 \longrightarrow q(\lj {\Gamma_2} \ul{C})$, with the unique mediating morphism in
\[
\xymatrix@C=4.25em@R=3em@M=0.5em{
\lj {\Gamma_3} \ul{D} \ar@/^2pc/[rr]^-{(\overrightarrow{W_i}, z.K)} \ar@{-->}[r] & \lj {\Gamma_1} {\ul{C}[\overrightarrow{V_i}/\overrightarrow{x_i}]} \ar[r]_-{(\overrightarrow{V_i}, z.\,z)} & \lj {\Gamma_2} \ul{C} & \text{in} & \mathcal{V} \ar[d]^-{p}
\\
\Gamma_3 \ar[r]^-{\overrightarrow{V'_j}} \ar@/_2pc/[rr]_{q(\overrightarrow{W_i}, z.K)}  & \Gamma_1 \ar[r]^{\overrightarrow{V_i}} & \Gamma_2 & \text{in} & \mathcal{B}
}
\]
given by
\[
(\overrightarrow{V'_j}, z.\,K) : \lj {\Gamma_3} \ul{D} \longrightarrow \lj {\Gamma_1} {\ul{C}[\overrightarrow{V_i}/\overrightarrow{x_i}]}
\]

Analogously to the chosen Cartesian morphisms in $p$, this mediating morphism is well-defined because of the commutativity of the lower diagram in $\mathcal{B}$; and because of the properties we established about simultaneous substitutions of value terms for value variables in the beginning of this section, and the results we established in Section~\ref{sect:syntax} about substituting homomorphism terms for computation variables.
The uniqueness of this mediating morphism is then proved analogously to the uniqueness of the corresponding mediating morphisms in $p$, as discussed in detail earlier in this section.

The induced reindexing functors $(V_1, \ldots, V_n)^* : \mathcal{C}_{\Gamma_2} \longrightarrow \mathcal{C}_{\Gamma_1}$ are given analogously to the corresponding functors for $p$: on objects, they are given by
\[
(\overrightarrow{V_i})^*(\lj {\Gamma_2} \ul{C}) \defeq \lj {\Gamma_1} {\ul{C}[\overrightarrow{V_i}/\overrightarrow{x_i}]}
\]
and on morphisms $(x_1, \ldots, x_n, z.\, K) : \lj {\Gamma_2} {\ul{C}} \longrightarrow \lj {\Gamma_2}  {\ul{D}}$, they are given by
\[
(\overrightarrow{V_i})^*(x_1, \ldots, x_n, z .\, K) \defeq (y_1, \ldots, y_m, z .\, K[\overrightarrow{V_i}/\overrightarrow{x_i}])
\]

Finally, we note that analogously to the fibration $p$ defined earlier, $q$ is also split.

\subsection*{$p$ has split dependent products and strong split dependent sums}

We omit a detailed discussion about these properties of $p$ because we discuss analogous properties for $q$ in detail later in this section.
We only note that the split dependent products are defined in terms of the value $\Pi$-type $\Pi\, x\!:\! A .\, B$; and the strong split dependent sums in terms of the value $\Sigma$-type $\Sigma\, x \!:\! A .\, B$. The strength of the latter is witnessed by isomorphisms $\kappa_{(\Gamma \,\vdash\, A),\, ({\Gamma, x : A} \,\vdash\, B)} : \Gamma, x \!:\! A, y \!:\! B \overset{\cong}{\longrightarrow} \Gamma, y' \!:\! \Sigma\, x \!:\! A .\, B$, given by 
\[
\kappa_{(\Gamma \,\vdash\, A),\, ({\Gamma, x : A} \,\vdash\, B)} \defeq (\overrightarrow{x_i}, \langle x , y \rangle)
\qquad
\kappa_{(\Gamma \,\vdash\, A),\, ({\Gamma, x : A} \,\vdash\, B)}^{-1} \defeq (\overrightarrow{x_i}, \fst y' , \snd y')
\]
assuming that $\Gamma = x_1 \!:\! A_1, \ldots, x_n \!:\! A_n$.

\subsection*{$p$ has strong fibred colimits of shape $\mathbf{0}$}

Given any diagram of the form $J : \mathbf{0} \longrightarrow \mathcal{V}_\Gamma$, we define the vertex $\mathsf{\ul{colim}}(J)$ as
\[
\mathsf{\ul{colim}}(J) \defeq \lj \Gamma 0
\]
and note that the cocone $\mathsf{\ul{in}}^J : J \longrightarrow \Delta(\mathsf{\ul{colim}}(J))$ is given vacuously as a natural transformation between functors with empty domains. It is easy to verify that both $\mathsf{\ul{colim}}(J)$ and $\mathsf{\ul{in}}^J$ are preserved by reindexing, i.e., substituting  value terms for value variables.

Next, we observe that for any diagram of the form $J : \mathbf{0} \longrightarrow \mathcal{V}_\Gamma$, we have 
\[
\mathsf{lim}(\widehat{J}) = \mathbf{1}
\]
where the diagram $\widehat{J} : \mathbf{0}^{\text{op}} \longrightarrow \mathsf{Cat}$ is derived from $J$ trivially. 

As a consequence of the above observation about $\mathsf{lim}(\widehat{J})$, the unique mediating functor $\langle \{\mathsf{\ul{in}}^J_D\}^*_{D \,\in\, \mathbf{0}} \rangle : \mathcal{V}_{\ia {\mathsf{\ul{colim}}(J)}} \longrightarrow \mathsf{lim}(\widehat{J})$ turns out to be the trivially constant functor with codomain $\mathbf{1}$. Therefore, showing that $\langle \{\mathsf{\ul{in}}^J_D\}^*_{D \,\in\, \mathbf{0}} \rangle$ is fully-faithful simplifies to showing that there is exactly one morphism between every pair of objects in $\mathcal{V}_{\Gamma,\, x : 0}$, which is straightforward. In particular, assuming  $\Gamma = x_1 \!:\! A_1, \ldots, x_n \!:\! A_n$, any morphism
\[
(\overrightarrow{x_i}, x, y.\,V) : \lj {\Gamma, x \!:\! 0} A \longrightarrow \lj {\Gamma, x \!:\! 0} B
\]
in $\mathcal{V}_{\Gamma,\, x : 0}$ can be shown to be equal to
\[
(\overrightarrow{x_i}, x, y.\,\absurd {} x)
\]
using the $\eta$-equation for empty case analysis. 

\subsection*{$p$ has strong fibred colimits of shape $\mathbf{2}$}

Given any diagram of the form $J : \mathbf{2} \longrightarrow \mathcal{V}_\Gamma$, we define the vertex $\mathsf{\ul{colim}}(J)$ as
\[
\mathsf{\ul{colim}}(J) \defeq \lj \Gamma {J(0) + J(1)}
\]
and the cocone $\mathsf{\ul{in}}^J : J \longrightarrow \Delta(\mathsf{\ul{colim}}(J))$ by 
\[
\begin{array}{c}
\mathsf{\ul{in}}^J_0 \defeq (\overrightarrow{x_i}, x.\, \inl {} x) : \lj \Gamma {J(0)} \longrightarrow \lj \Gamma {J(0) + J(1)}
\\[1mm]
\mathsf{\ul{in}}^J_1 \defeq (\overrightarrow{x_i}, x.\, \inr {} x) : \lj \Gamma {J(1)} \longrightarrow \lj \Gamma {J(0) + J(1)}
\end{array}
\]
assuming that $\Gamma = x_1 \!:\! A_1, \ldots, x_n \!:\! A_n$.
Similarly to the strong fibred colimits of shape $\mathbf{0}$, it is again easy to verify that both $\mathsf{\ul{colim}}(J)$ and $\mathsf{\ul{in}}^J$ are preserved by reindexing.

Next, we observe that for any diagram of the form $J : \mathbf{2} \longrightarrow \mathcal{V}_\Gamma$, we have 
\[
\mathsf{lim}(\widehat{J}) = \mathcal{V}_{\Gamma,\, y_0 : J(0)} \times \mathcal{V}_{\Gamma,\, y_1 : J(1)}
\]
where the diagram $\widehat{J} : \mathbf{2}^{\text{op}} \longrightarrow \mathsf{Cat}$ is derived from $J$ as in Definition~\ref{def:strongcolimits}. 

As a consequence of this observation about $\mathsf{lim}(\widehat{J})$, the unique mediating functor \linebreak $\langle \{\mathsf{\ul{in}}^J_D\}^*_{D \,\in\, \mathbf{2}} \rangle : \mathcal{V}_{\ia {\mathsf{\ul{colim}}(J)}} \longrightarrow \mathsf{lim}(\widehat{J})$ sends any morphism
\[
(\overrightarrow{x_i}, x, y .\, V) : \lj {\Gamma, x \!:\! J(0) + J(1)}{A} \longrightarrow \lj {\Gamma, x \!:\! J(0) + J(1)}{B} 
\]
in $\mathcal{V}_{\Gamma,\, x : J(0) \,+\, J(1)}$ to a pair of morphisms
\[
\begin{array}{c}
(\overrightarrow{x_i}, y_0, y .\, V[\inl {} {y_0}/x]) : \lj {\Gamma, y_0 \!:\! J(0)}{A[\inl {} {y_0}/x]} \longrightarrow \lj {\Gamma, y_0 \!:\! J(0)}{B[\inl {} {y_0}/x]}
\\[1mm]
(\overrightarrow{x_i}, y_1, y .\, V[\inr {} {y_1}/x]) : \lj {\Gamma, y_1 \!:\! J(1)}{A[\inr {} {y_1}/x]} \longrightarrow \lj {\Gamma, y_1 \!:\! J(1)}{B[\inr {} {y_1}/x]}
\end{array}
\]
in $\mathcal{V}_{\Gamma,\, y_0 : J(0)}$ and $\mathcal{V}_{\Gamma,\, y_1 : J(1)}$, respectively.
We note that for better readability, we use different names ($y_0$ and $y_1$) for the variable $x \defeq \mathsf{fresh}(V\!ars(\Gamma))$ in the last two morphisms.

In order to show that $\langle \{\mathsf{\ul{in}}^J_D\}^*_{D \,\in\, \mathbf{2}} \rangle$ is fully-faithful, we first define the mapping of morphisms in the reverse direction: we send any pair of morphisms
\[
\begin{array}{c}
(\overrightarrow{x_i}, y_0, y .\, W_0) : \lj {\Gamma, y_0 \!:\! J(0)}{A[\inl {} {y_0}/x]} \longrightarrow \lj {\Gamma, y_0 \!:\! J(0)}{B[\inl {} {y_0}/x]}
\\[1mm]
(\overrightarrow{x_i}, y_1, y .\, W_1) : \lj {\Gamma, y_1 \!:\! J(1)}{A[\inr {} {y_1}/x]} \longrightarrow \lj {\Gamma, y_1 \!:\! J(1)}{B[\inr {} {y_1}/x]}
\end{array}
\]
in $\mathcal{V}_{\Gamma,\, y_0 : J(0)}$ and $\mathcal{V}_{\Gamma,\, y_1 : J(1)}$, respectively, to a morphism
\[
\begin{array}{c}
\hspace{-1.25cm} \big(\overrightarrow{x_i}, x, y.\big(\mathtt{case~} x \mathtt{~of}_{} \mathtt{~} ({\inl {\!} {\!\!(y_0 \!:\! J(0))} \mapsto \lambda y' \!:\! A[\inl {} {y_0}/x] .\, W_0[y'/y]} , 
\\[-1mm]
\hspace{3.025cm} {\inr {\!} {\!\!(y_1 \!:\! J(1))} \mapsto \lambda y' \!:\! A[\inr {} {y_1}/x] .\, W_1[y'/y]})\big)\,\, y\big)
\end{array}
\]
in $\mathcal{V}_{\Gamma,\, x : J(0) \,+\, J(1)}$, with $y'$ chosen fresh. 

Now, to show that the round-trip (on morphisms) from $\mathcal{V}_{\Gamma,\, x : J(0) \,+\, J(1)}$ to the product $\mathcal{V}_{\Gamma,\, y_0 : J(0)} \times \mathcal{V}_{\Gamma,\, y_1 : J(1)}$ and back is an identity, it suffices to observe that any morphism 
\[
(\overrightarrow{x_i}, x, y .\, V) : \lj {\Gamma, x \!:\! J(0) + J(1)}{A} \longrightarrow \lj {\Gamma, x \!:\! J(0) + J(1)}{B} 
\]
in $\mathcal{V}_{\Gamma,\, x : J(0) \,+\, J(1)}$ can be shown to be equal to
\[
\begin{array}{c}
\hspace{-0.85cm} \big(\overrightarrow{x_i}, x, y.\big(\mathtt{case~} x \mathtt{~of}_{} \mathtt{~} ({\inl {\!} {\!\!(y_0 \!:\! J(0))} \mapsto \lambda y' \!:\! A[\inl {} {y_0}/x] .\, V[y'/y][\inl {} {y_0}/x]} , 
\\[-1mm]
\hspace{3.425cm} {\inr {\!} {\!\!(y_1 \!:\! J(1))} \mapsto \lambda y' \!:\! A[\inr {} {y_1}/x] .\, V[y'/y][\inr {} {y_1}/x]})\big)\,\, y\big)
\end{array}
\]
using 
the $\eta$-equation for binary case analysis. 

Showing that the other round-trip (on morphisms) from $\mathcal{V}_{\Gamma,\, y_0 : J(0)} \times \mathcal{V}_{\Gamma,\, y_1 : J(1)}$ to $\mathcal{V}_{\Gamma,\, x : J(0) \,+\, J(1)}$ and back is an identity is similarly straightforward, by using the respective $\beta$-equations for binary case analysis and function application.

\subsection*{$p$ has weak split fibred strong natural numbers}

We define the weak split fibred strong natural numbers in $p$ by 
\[
\begin{array}{c}
\mathbb{N} \defeq \lj \diamond \Nat
\\[1mm]
\mathsf{zero} \defeq (x.\, \zero) : \lj \diamond 1 \longrightarrow \lj \diamond \Nat
\\[1mm]
\mathsf{succ} \defeq (x.\, \suc x) : \lj \diamond \Nat \longrightarrow \lj \diamond \Nat
\end{array}
\]
Next, if we assume that $\Gamma =  x_1 \!:\! A_1, \ldots, x_n \!:\! A_n$, then given any pair of morphisms 
\[
\begin{array}{c}
f_z \defeq (\overrightarrow{x_i}, \zero, y.\, V_z) : \lj {\Gamma, x \!:\! 1} 1 \longrightarrow \lj {\Gamma, x \!:\! \Nat} A
\\[1mm]
f_s \defeq (\overrightarrow{x_i}, \suc x, y.\, V_s) : \lj {\Gamma, x \!:\! \Nat} A \longrightarrow \lj {\Gamma, x \!:\! \Nat} A
\end{array}
\]
in $\mathcal{V}$, the mediating morphism $\mathsf{rec}(f_z,f_s) : \Gamma, x \!:\! \Nat \longrightarrow \Gamma, x \!:\! \Nat, y \!:\! A$ is defined as
\[
\mathsf{rec}(f_z,f_s) \defeq 
\big(\overrightarrow{x_i}, x, \natrec {} {V_z[\star/x][\star/y]} {y_1.\, y_2.\, V_s[y_1/x][y_2/y]} {x}\big)  
\]
with $y_1$ and $y_2$ chosen fresh.

As $\mathsf{rec}(f_z,f_s)$ is clearly a section of $(\overrightarrow{x_i}, x) : \Gamma, x \!:\! \Nat, y \!:\! A \longrightarrow \Gamma, x \!:\! \Nat$, we are left with showing that the diagram of ``$\beta$-equations" given in Definition~\ref{def:strongsplitfibredweaknaturals} commutes. To this end, we show that the two squares in the above-mentioned diagram commute, by first observing that these squares can be rewritten in this classifying model as
\[
\xymatrix@C=6em@R=4em@M=0.5em{
\Gamma, x \!:\! 1 \ar[r]^-{(\overrightarrow{x_i}, \zero)} \ar[d]_-{(\overrightarrow{x_i}, x\!, \star)} & \Gamma, x \!:\! \Nat \ar[d]^-{\mathsf{rec}(f_z,f_s)}
\\
\Gamma, x \!:\! 1, y \!:\! 1 \ar[r]_-{(\overrightarrow{x_i}, \zero, V_z)} & \Gamma, x \!:\! \Nat, y \!:\! A
}
\]
and
\[
\xymatrix@C=6em@R=4em@M=0.5em{
\Gamma, x \!:\! \Nat \ar[d]_-{\mathsf{rec}(f_z,f_s)} & \Gamma, x \!:\! \Nat \ar[l]_-{(\overrightarrow{x_i}, \suc x)} \ar[d]^-{\mathsf{rec}(f_z,f_s)}
\\
\Gamma, x \!:\! \Nat, y \!:\! A & \Gamma, x \!:\! \Nat, y \!:\! A \ar[l]^-{(\overrightarrow{x_i}, \suc x, V_s)}
}
\vspace{0.25cm}
\]
It is now straightforward to show that these squares commute, by
\begin{fleqn}[0.3cm]
\begin{align*}
& \big(\overrightarrow{x_i}, x, \natrec {} {V_z[\star/x][\star/y]} {y_1.\, y_2.\, V_s[y_1/x][y_2/y]} {x}\big) \comp (\overrightarrow{x_i}, \zero)
\\
=\,\, & \big(\overrightarrow{x_i}, \zero, \natrec {} {V_z[\star/x][\star/y]} {y_1.\, y_2.\, V_s[y_1/x][y_2/y]} {\zero}\big)
\\
=\,\, &
(\overrightarrow{x_i}, \zero, V_z[\star/x][\star/y])
\\
=\,\, &
(\overrightarrow{x_i}, \zero, V_z[\star/y])
\\
=\,\, &
(\overrightarrow{x_i} , \zero, V_z) \comp (\overrightarrow{x_i} , x\!, \star)
\end{align*}
\end{fleqn}
and
\begin{fleqn}[0.3cm]
\begin{align*}
& \big(\overrightarrow{x_i}, x, \natrec {} {V_z[\star/x][\star/y]} {y_1.\, y_2.\, V_s[y_1/x][y_2/y]} {x}\big) \comp (\overrightarrow{x_i}, \suc x)
\\
=\,\, & \big(\overrightarrow{x_i}, \suc x, \natrec {} {V_z[\star/x][\star/y]} {y_1.\, y_2.\, V_s[y_1/x][y_2/y]} {\suc x}\big)
\\
=\,\, &
(\overrightarrow{x_i}, \suc x, V_s[y_1/x][y_2/y][x/y_1]
\\[-2mm]
&
\hspace{2.4cm} \big[\natrec {} {V_z[\star/x][\star/y]} {y_1.\, y_2.\, V_s[y_1/x][y_2/y]} {x}/y_2\big])
\\
=\,\, &
(\overrightarrow{x_i}, \suc x, V_s\big[\natrec {} {V_z[\star/x][\star/y]} {y_1.\, y_2.\, V_s[y_1/x][y_2/y]} {x}/y\big])
\\
=\,\, &
(\overrightarrow{x_i}, \suc x, V_s) \comp \big(\overrightarrow{x_i}, x, \natrec {} {V_z[\star/x][\star/y]} {y_1.\, y_2.\, V_s[y_1/x][y_2/y]} {x}\big)
\end{align*}
\end{fleqn}

\subsection*{$p$ has split intensional propositional equality}

We define the object $\Id_{\lj {\Gamma\,} {\,A}}$ in $\mathcal{V}_{\Gamma, x : A, y : A}$ as
\[
\Id_{\lj {\Gamma\,} {\,A}} \defeq \lj {\Gamma, x \!:\! A, y \!:\! A} {x =_A y}
\]
and the morphism $\mathsf{r}_A : \lj {\Gamma, x \!:\! A} 1 \longrightarrow \lj {\Gamma, x \!:\! A} {x =_A x}$ as
\[
\mathsf{r}_A \defeq (\overrightarrow{x_i}, x, y.\, \refl {} x)
\]

Next, given a well-formed value type $\lj {\Gamma, x_1 \!:\! A, x_2 \!:\! A, x_3 \!:\! x_1 =_A x_2} {B}$ and a morphism 
\[
f \defeq (\overrightarrow{x_i} , x , y.\, V) : \lj {\Gamma, x \!:\! A} 1 \longrightarrow \lj {\Gamma, x \!:\! A} {B[x/x_1][x/x_2][\refl {} x/x_3]}
\]
we define the morphism 
\[
\begin{array}{c}
\hspace{-9cm}
\mathsf{i}_{\lj {\Gamma\,} {\,A} \,,\, \lj {\Gamma, x_1 : A, x_2 : A, x_3 : x_1 =_A x_2\,} {\,B}}(f) 
\\
\hspace{2cm}
: \lj {\Gamma, x_1 \!:\! A, x_2 \!:\! A, x_3 \!:\! x_1 =_A x_2} 1 \longrightarrow \lj {\Gamma, x_1 \!:\! A, x_2 \!:\! A, x_3 \!:\! x_1 =_A x_2} B
\end{array}
\]
as
\[
(\overrightarrow{x_i} , x_1, x_2, x_3 , y .\, \pathind A {x_1.\, x_2.\, x_3.\, B} {x.\, V} {x_1} {x_2} {x_3})
\]
We note that for better readability, we write $x_1$ and $x_2$ for the freshly chosen value variables $x \defeq \mathsf{fresh}(V\!ars(\Gamma))$ and $y \defeq \mathsf{fresh}(V\!ars(\Gamma) \cup \{x\})$ in the above definition.

Next, we note that in this classifying model, the equation relating the morphisms $\mathsf{r}_A$ and $\mathsf{i}_{\lj {\Gamma\,} {\,A} \,,\, \lj {\Gamma, x_1 \!:\! A, x_2 \!:\! A, x_3 \!:\! x_1 =_A x_2\,} {\,B}}(f)$, as given in Definition~\ref{def:strongpropequality}, amounts to showing
\[
\begin{array}{c}
(\overrightarrow{x_i} , x , y .\, \pathind A {x_1.\, x_2.\, x_3.\, B} {x.\, V} {x} {x} {\refl {} x}) = (\overrightarrow{x_i} , x , y.\, V)
\end{array}
\]
which follows straightforwardly from the $\beta$-equation for propositional equality.

Finally, we note that all the structure we defined above is also preserved on-the-nose by reindexing, as required in Definition~\ref{def:strongpropequality}, because the type- and term-formers used in these definitions are all preserved on-the-nose by substitution.

\subsection*{Split fibred adjunction $F \dashv\, U$}

We define the functor $F : \mathcal{V} \longrightarrow \mathcal{C}$ on objects using the type of free computations:
\[
F(\lj \Gamma A) \defeq \lj \Gamma FA
\]
and on morphisms using sequential composition:
\[
F(\overrightarrow{V_i}, x.\, V) \defeq (\overrightarrow{V_i}, z.\, \doto z {x \!:\! A} {} {\return V})
\]
where $(\overrightarrow{V_i}, x.\, V) : \lj {\Gamma_1} {A} \longrightarrow \lj {\Gamma_2} B$.

We proceed by showing that $F$ is split fibred. On the one hand, we observe that $F$ does not alter the context $\Gamma$ of the given well-formed value type $\lj \Gamma A$ (or a morphism between them). On the other hand, $F$ preserves Cartesian morphisms on-the-nose:
\[
F(\overrightarrow{V_i}, x.\, x) = (\overrightarrow{V_i}, z.\, \doto z {x \!:\! A[\overrightarrow{V_i}/\overrightarrow{x_i}]} {} {\return x}) = (\overrightarrow{V_i}, z.\, z)
\]
where $(\overrightarrow{V_i}, x.\, x) : \lj {\Gamma_1} {A[V_1/x_1, \ldots, V_n/x_n]} \longrightarrow \lj {\Gamma_2} A$.

We define the functor $U$ on objects using the type of thunked computations:
\[
U(\lj \Gamma \ul{C}) \defeq \lj \Gamma U\ul{C}
\]
and on morphisms using thunking and forcing:
\[
U(\overrightarrow{V_i}, z.\, K) \defeq (\overrightarrow{V_i}, x.\, \thunk (K[\force {\ul{C}} x/z]))
\]
with $x$ chosen fresh; and 
where $(\overrightarrow{V_i}, z.\, K) : \lj {\Gamma_1} {\ul{C}} \longrightarrow \lj {\Gamma_2} {\ul{D}}$.

Showing that $U$ is split fibred is also straightforward. On the one hand, $U$ does not alter the context $\Gamma$ of the given well-formed computation type $\lj \Gamma {\ul{C}}$ (or a morphism between them). On the other hand, $U$ preserves Cartesian morphisms on-the-nose:
\[
U(\overrightarrow{V_i}, z.\, z) = (\overrightarrow{V_i}, x.\, \thunk (z[\force {\ul{C}} x / z])) = (\overrightarrow{V_i}, x.\,  \thunk (\force {\ul{C}} x)) = (\overrightarrow{V_i}, x.\, x)
\]
where $(\overrightarrow{V_i}, z.\, z) : \lj {\Gamma_1} {\ul{C}[V_1/x_1, \ldots, V_n/x_n]} \longrightarrow \lj {\Gamma_2} \ul{C}$.

Next, the unit and counit of the adjunction $F \dashv\, U$ are given by components
\[
\begin{array}{c}
\eta_{\lj {\Gamma\,} {\,A}} \defeq (\overrightarrow{x_i}, x.\, \thunk (\return x)) : \lj \Gamma A \longrightarrow \lj \Gamma {UFA}
\\[2mm]
\varepsilon_{\lj {\Gamma\,} {\,\ul{C}}} \defeq (\overrightarrow{x_i}, z.\, \doto z {y \!:\! U\ul{C}} {} \force {\ul{C}} y) : \lj \Gamma {FU\ul{C}} \longrightarrow \lj \Gamma \ul{C}
\end{array}
\]
assuming that $\Gamma = x_1 \!:\! A_1, \ldots, x_n \!:\! A_n$, and where $x$,$y$, and $z$ are chosen fresh. 

Finally, we show that the two unit-counit laws hold. 

On the one hand, assuming that $\Gamma = x_1 \!:\! A_1, \ldots, x_n \!:\! A_n$, we observe that the triangle
\[
\xymatrix@C=5em@R=5em@M=0.5em{
U \ar[r]^-{\eta \,\comp\, U} \ar[dr]_{\id_{U}} & U \comp F \comp U \ar[d]^-{U \,\comp\, \varepsilon}
\\
& U
}
\]
can be rewritten for each well-formed computation type $\lj \Gamma \ul{C}$ as follows:
\[
\xymatrix@C=5em@R=5em@M=0.5em{
\lj \Gamma U\ul{C} \ar[r]^-{f} \ar[dr]_{(\overrightarrow{x_i}, x. x)} & \lj \Gamma UFU\ul{C} \ar[d]^-{g}
\\
& \lj \Gamma U\ul{C}
}
\]
where the morphisms $f$ and $g$ are given by
\[
\begin{array}{c}
f \defeq (\overrightarrow{x_i}, x.\,  \thunk (\return x))
\\[2mm]
g \defeq \big(\overrightarrow{x_i}, x' .\,  \thunk \big(\doto {(\force {FU\ul{C}} x')} {y \!:\! U\ul{C}} {} \force {\ul{C}} y\big)\big)
\end{array}
\]
It is now straightforward to show that this triangle commutes, by
\begin{fleqn}[0.3cm]
\begin{align*}
&
\big(\overrightarrow{x_i}, x' .\,  \thunk \big(\doto {(\force {FU\ul{C}} x')} {y \!:\! U\ul{C}} {} \force {\ul{C}} y\big)\big) \comp (\overrightarrow{x_i}, x.\,  \thunk (\return x))
\\
=\,\, &
\big(\overrightarrow{x_i}, x.\,  \thunk \big(\doto {(\force {FU\ul{C}} (\thunk (\return x)))} {y \!:\! U\ul{C}} {} \force {\ul{C}} y\big)\big)
\\
=\,\, & 
\big(\overrightarrow{x_i}, x.\,  \thunk \big(\doto {(\return x)} {y \!:\! U\ul{C}} {} \force {\ul{C}} y\big)\big)
\\
=\,\, & 
(\overrightarrow{x_i}, x.\,  \thunk (\force {\ul{C}} x))
\\
=\,\, & 
(\overrightarrow{x_i}, x.\,  x)
\end{align*}
\end{fleqn}

On the other hand, assuming that $\Gamma = x_1 \!:\! A_1, \ldots, x_n \!:\! A_n$, we observe that the  triangle
\[
\xymatrix@C=5em@R=5em@M=0.5em{
F \ar[r]^-{F \,\comp\, \eta} \ar[dr]_{\id_F} & F \comp U \comp F \ar[d]^-{\varepsilon \,\comp\, F}
\\
& F
}
\]
can be rewritten for each well-formed value type $\lj \Gamma A$ as follows
\[
\hspace{-0.1cm}
\xymatrix@C=5em@R=5em@M=0.5em{
\lj \Gamma FA \ar[r]^-{h} \ar[dr]_{(\overrightarrow{x_i}, z. z)} & \lj \Gamma FUFA \ar[d]^-{k}
\\
& \lj \Gamma FA
}
\]
where the morphisms $h$ and $k$ are given by
\[
\begin{array}{c}
h \defeq \big(\overrightarrow{x_i}, z.\,  \doto z {x \!:\! A} {} {\return (\thunk (\return x))}\big)
\\[2mm]
k \defeq (\overrightarrow{x_i}, z'\! .\,  \doto {z'} {y \!:\! UFA} {} {\force {FA} y})
\end{array}
\]
It is now straightforward to show that this triangle commutes, by
\begin{fleqn}[0.3cm]
\begin{align*}
&
(\overrightarrow{x_i}, z'\!.\,  \doto {z'} {y : UFA} {} {\force {FA} y}) \,\,\comp 
\\[-2mm]
&
\hspace{4.7cm} \big(\overrightarrow{x_i}, z.\,  \doto z {x : A} {} {\return (\thunk (\return x))}\big)
\\
=\,\, &
\big(\overrightarrow{x_i}, z.\,  \doto {\big(\doto z {x : A} {} {\return (\thunk (\return x))}\big)} {y : UFA} {} {\force {FA} y}\big)
\\
=\,\,&
\big(\overrightarrow{x_i}, z .\,  \doto z {x : A} {} {\big(\doto {\return (\thunk (\return x))} {y : UFA} {} {\force {FA} y}\big)}\big)
\\
=\,\,&
\big(\overrightarrow{x_i}, z .\,  \doto z {x : A} {} {\force {FA} (\thunk (\return x))}\big)
\\
=\,\,&
(\overrightarrow{x_i}, z .\,  \doto z {x : A} {} {\return x})
\\
=\,\,&
(\overrightarrow{x_i}, z .\,  z)
\end{align*}
\end{fleqn}

\subsection*{$q$ has split dependent $p$-products}

We define the functor 
$
\Pi_{\lj {\Gamma\,} {\,A}} : \mathcal{C}_{\Gamma, x : A} \longrightarrow \mathcal{C}_{\Gamma}
$
on objects by 
\[
\Pi_{\lj {\Gamma\,} {\,A}}(\lj {\Gamma, x \!:\! A} \ul{C}) \defeq \lj \Gamma {\Pi\, x \!:\! A .\, \ul{C}}
\]
and on morphisms by
\[
\Pi_{\lj {\Gamma\,} {\,A}}(\overrightarrow{x_i}, x, z.\,  K) \defeq (\overrightarrow{x_i}, z'\!.\,  \lambda\, x \!:\! A .\, K[z'\, x/z])
\]
with $z'$ chosen fresh.

Next, we note that in this classifying model, the projection morphisms are given by
\[
\pi_{\lj {\Gamma\,} {\,A}} \defeq \overrightarrow{x_i}: \Gamma, x \!:\! A \longrightarrow \Gamma
\]
assuming that $\Gamma = x_1 \!:\! A_1, \ldots, x_n \!:\! A_n$, with $x \defeq \mathsf{fresh}(V\!ars(\Gamma))$. 

As a result, the weakening functors $\pi^*_{\lj {\Gamma\,} {\,A}}$
can be shown to be given by syntactic weakening, both on objects and morphisms:
\[
\pi^*_{\lj {\Gamma\,} {\,A}}(\lj \Gamma \ul{C}) \defeq \lj {\Gamma, x \!:\! A} \ul{C}
\qquad
\pi^*_{\lj {\Gamma\,} {\,A}}(\overrightarrow{x_i}, z.\, K) \defeq (\overrightarrow{x_i}, x, z.\, K)
\]

Next, the unit and counit of the adjunction $\pi^*_{\lj {\Gamma\,} {\,A}} \dashv\, \Pi_{\lj {\Gamma\,} {\,A}}$ are given by components
\[
\begin{array}{c}
\eta_{\lj {\Gamma\,} {\,\ul{C}}} \defeq (\overrightarrow{x_i}, z.\,  \lambda\, x \!:\! A .\, z) : \lj \Gamma \ul{C} \longrightarrow \lj \Gamma {\Pi\, x \!:\! A .\, \ul{C}}
\\[2mm]
\varepsilon_{\lj {\Gamma, x : A\,} {\ul{C}}} \defeq (\overrightarrow{x_i}, x, z .\,  z\,\, x) : \lj {\Gamma, x \!:\! A} {\Pi\, y \!:\! A .\, \ul{C}[y/x]} \longrightarrow \lj {\Gamma, x \!:\! A} {\ul{C}}
\end{array}
\]
with $z$ chosen fresh in both definitions. 

Next, we show that the split Beck-Chevalley condition holds. First, we observe that the components of the corresponding natural transformation are given by morphisms
\[
\begin{array}{c}
\hspace{-7cm}
\big(\overrightarrow{y_{\!j}}, z.\, \lambda x \!:\! A[\overrightarrow{V_i}/\overrightarrow{x_i}] .\, ((\lambda y \!:\! A[\overrightarrow{V_i}/\overrightarrow{x_i}] .\, z)\, x)\, x\big) 
\\[1mm]
\hspace{2.5cm}
: \lj {\Gamma_1} {\Pi\, x \!:\! A[\overrightarrow{V_i}/\overrightarrow{x_i}] .\, \ul{C}[\overrightarrow{V_i}/\overrightarrow{x_i}]} \longrightarrow \lj {\Gamma_1} {\Pi\, x \!:\! A[\overrightarrow{V_i}/\overrightarrow{x_i}] .\, \ul{C}[\overrightarrow{V_i}/\overrightarrow{x_i}]}
\end{array}
\]
for any Cartesian morphism $(\overrightarrow{V_i},x.\, x) : \lj {\Gamma_1} {A[\overrightarrow{V_i}/\overrightarrow{x_i}]} \longrightarrow \lj {\Gamma_2} {A}$ and $\lj {\Gamma_2, x \!:\! A} {\ul{C}}$.
Then, it is easy to verify that the above components are equal to identity, i.e., $(\overrightarrow{y_{\!j}}, z.\, z)$; namely, by using the $\beta$- and $\eta$-equations for computational function application.

Finally, we show that the two unit-counit laws hold. 

On the one hand, assuming that $\Gamma = x_1 \!:\! A_1, \ldots, x_n \!:\! A_n$, we observe that the triangle
\[
\xymatrix@C=5em@R=5em@M=0.5em{
\Pi_{\lj {\Gamma\,} {\,A}} \ar[r]^-{\eta \,\comp\, \Pi_{\lj {\Gamma\,} {\,A}}} \ar[dr]_{\id_{\Pi_{\lj {\Gamma\,} {\,A}}}} & \Pi_{\lj {\Gamma\,} {\,A}} \comp \pi^*_{\lj {\Gamma\,} {\,A}} \comp \Pi_{\lj {\Gamma\,} {\,A}} \ar[d]^-{\Pi_{\lj {\Gamma\,} {\,A}} \,\comp\, \varepsilon}
\\
& \Pi_{\lj {\Gamma\,} {\,A}}
}
\]
can be rewritten for each computation type $\lj {\Gamma, x \!:\! A} \ul{C}$ as follows
\[
\xymatrix@C=5em@R=5em@M=0.5em{
\lj {\Gamma} \Pi\, x \!:\! A .\, \ul{C} \ar[r]^-{(\overrightarrow{x_i}, z.  \lambda x : A . z)} \ar[dr]_{(\overrightarrow{x_i}, z. z)} & \lj {\Gamma} {\Pi\, x \!:\! A .\, \Pi\, y \!:\! A .\, \ul{C}[y/x]} \ar[d]^-{(\overrightarrow{x_i}, z'\! .  \lambda x : A . (z'\, x)\,x)}
\\
& \lj {\Gamma} \Pi\, x \!:\! A .\, \ul{C}
}
\]
which commutes because we have
\begin{fleqn}[0.3cm]
\begin{align*}
& (\overrightarrow{x_i}, z'\! .\,  \lambda\, x \!:\! A .\, (z'\, x)\, x) \comp (\overrightarrow{x_i}, z.\,  \lambda\, x \!:\! A .\, z) 
\\
=\,\, &
(\overrightarrow{x_i}, z .\,  \lambda\, x \!:\! A .\, ((\lambda\, y \!:\! A .\, z)\, x)\, x)
\\
=\,\, &
(\overrightarrow{x_i}, z .\,  \lambda\, x \!:\! A .\, (z[x/y])\, x)
\\
=\,\, &
(\overrightarrow{x_i}, z .\,  \lambda\, x \!:\! A .\, z\,\, x)
\\
=\,\, &
(\overrightarrow{x_i}, z .\,  z)
\end{align*}
\end{fleqn}

On the other hand, assuming that $\Gamma = x_1 \!:\! A_1, \ldots, x_n \!:\! A_n$, we observe that the triangle
\[
\xymatrix@C=5em@R=5em@M=0.5em{
\pi^*_{\lj {\Gamma\,} {\,A}} \ar[r]^-{\pi^*_{\lj {\Gamma\,} {\,A}} \,\comp\, \eta} \ar[dr]_{\id_{\pi^*_{\lj {\Gamma\,} {\,A}}}} & \pi^*_{\lj {\Gamma\,} {\,A}} \comp \Pi_{\lj {\Gamma\,} {\,A}} \comp \pi^*_{\lj {\Gamma\,} {\,A}} \ar[d]^-{\varepsilon \,\comp\, \pi^*_{\lj {\Gamma\,} {\,A}}}
\\
& \pi^*_{\lj {\Gamma\,} {\,A}}
}
\]
can be rewritten for each computation type $\lj \Gamma \ul{C}$ as follows:
\[
\xymatrix@C=5em@R=5em@M=0.5em{
\lj {\Gamma, x \!:\! A} \ul{C} \ar[r]^-{(\overrightarrow{x_i}, x, z. \lambda y : A . z)} \ar[dr]_{(\overrightarrow{x_i}, x, z.z)} & \lj {\Gamma, x \!:\! A} {\Pi\, y \!:\! A .\, \ul{C}} \ar[d]^-{(\overrightarrow{x_i}, x, z'\!. z'\, x)}
\\
& \lj {\Gamma, x \!:\! A} \ul{C}
}
\]
which commutes because we have
\[
(\overrightarrow{x_i}, x, z'\! .\, z'\, x) \comp (\overrightarrow{x_i}, x, z.\,  \lambda\, y \!:\! A .\, z)
=
(\overrightarrow{x_i}, x, z.\,  (\lambda\, y \!:\! A .\, z)\, x)
=
(\overrightarrow{x_i}, x, z.\, z[x/y])
=
(\overrightarrow{x_i}, x, z.\, z)
\]

\subsection*{$q$ has split dependent $p$-sums}

We define the functor 
$
\Sigma_{\lj {\Gamma\,} {\,A}} : \mathcal{C}_{\Gamma, x : A} \longrightarrow \mathcal{C}_{\Gamma}
$
on objects by 
\[
\Sigma_{\lj {\Gamma\,} {\,A}}(\lj {\Gamma, x \!:\! A} \ul{C}) \defeq \lj \Gamma {\Sigma\, x \!:\! A .\, \ul{C}}
\]
and on morphisms by
\[
\Sigma_{\lj {\Gamma\,} {\,A}}(\overrightarrow{x_i}, x, z.\, K) \defeq (\overrightarrow{x_i}, z'\! .\, \doto {z'} {(x \!:\! A, z'' \!:\! \ul{C})} {} {\langle x , K[z''/z] \rangle})
\]
with $z'$ and $z''$ chosen fresh. 

Next, the unit and counit of the adjunction $\Sigma_{\lj {\Gamma\,} {\,A}} \dashv\, \pi^*_{\lj {\Gamma\,} {\,A}}$ are given by components
\[
\begin{array}{c}
\eta_{\lj {\Gamma, x : A\,} {\,\ul{C}}} \defeq (\overrightarrow{x_i}, x, z.  \langle x , z \rangle) : \lj {\Gamma, x \!:\! A} \ul{C} \longrightarrow \lj {\Gamma, x \!:\! A} {\Sigma\, y \!:\! A .\, \ul{C}[y/x]}
\\[2mm]
\varepsilon_{\lj {\Gamma\,} {\ul{C}}} \defeq (\overrightarrow{x_i}, z .\,  \doto {z} {(x \!:\! A, z' \!:\! \ul{C})} {} {z'}) : \lj {\Gamma} {\Sigma\, x \!:\! A .\, \ul{C}} \longrightarrow \lj {\Gamma} {\ul{C}}
\end{array}
\]
with $z$ and $z'$ chosen fresh.

Next, we show that the split Beck-Chevalley condition holds. First, we observe that the components of the corresponding natural transformation are given by morphisms
\[
\begin{array}{c}
\hspace{-5.5cm}
(\overrightarrow{y_{\!j}}, z.\, \doto {z} {(x ,z')} {} {(\doto {\langle x , \langle x , z' \rangle \rangle} {(y,z'')} {} {z''})}) 
\\[1mm]
\hspace{2.5cm}
: \lj {\Gamma_1} {\Sigma\, x \!:\! A[\overrightarrow{V_i}/\overrightarrow{x_i}] .\, \ul{C}[\overrightarrow{V_i}/\overrightarrow{x_i}]} \longrightarrow \lj {\Gamma_1} {\Sigma\, x \!:\! A[\overrightarrow{V_i}/\overrightarrow{x_i}] .\, \ul{C}[\overrightarrow{V_i}/\overrightarrow{x_i}]}
\end{array}
\]
for any Cartesian morphism $(\overrightarrow{V_i},x.\, x) : \lj {\Gamma_1} {A[\overrightarrow{V_i}/\overrightarrow{x_i}]} \longrightarrow \lj {\Gamma_2} {A}$ and $\lj {\Gamma_2, x \!:\! A} {\ul{C}}$.
Then, it is easy to verify that the above components are equal to identity, i.e., $(\overrightarrow{y_{\!j}}, z.\, z)$; namely, by using the $\beta$- and $\eta$-equations for computational pattern-matching.

Finally, we show that the two unit-counit laws hold. 

On the one hand, assuming that $\Gamma = x_1 \!:\! A_1, \ldots, x_n \!:\! A_n$, we observe that the triangle
\[
\xymatrix@C=5em@R=5em@M=0.5em{
\pi^*_{\lj {\Gamma\,} {\,A}} \ar[r]^-{\eta \,\comp\, \pi^*_{\lj {\Gamma\,} {\,A}}} \ar[dr]_{\id_{\pi^*_{\lj {\Gamma\,} {\,A}}}} & \pi^*_{\lj {\Gamma\,} {\,A}} \comp \Sigma_{\lj {\Gamma\,} {\,A}} \comp \pi^*_{\lj {\Gamma\,} {\,A}} \ar[d]^-{\pi^*_{\lj {\Gamma\,} {\,A}} \,\comp\, \varepsilon}
\\
& \pi^*_{\lj {\Gamma\,} {\,A}}
}
\]
can be rewritten for each computation type $\lj \Gamma \ul{C}$ as follows:
\[
\xymatrix@C=5em@R=5em@M=0.5em{
\lj {\Gamma, x \!:\! A} \ul{C} \ar[r]^-{(\overrightarrow{x_i}, x, z. \langle x , z \rangle)} \ar[dr]_{(\overrightarrow{x_i}, x, z.z)} & \lj {\Gamma, x \!:\! A} {\Sigma\, y \!:\! A .\, \ul{C}} \ar[d]^-{h}
\\
& \lj {\Gamma, x \!:\! A} \ul{C}
}
\]
where the morphism $h$ is given by
\[
h \defeq \big(\overrightarrow{x_i}, x, z'\! .\, \doto {z'} {(y : A, z'' : \ul{C})} {} {z''}\big)
\]
It is now straightforward to show that this triangle commutes, by
\begin{fleqn}[0.3cm]
\begin{align*}
& 
\big(\overrightarrow{x_i}, x, z'\! .\, \doto {z'} {(y : A, z'' : \ul{C})} {} {z''}\big) \comp (\overrightarrow{x_i}, x, z. \langle x , z \rangle)
\\
=\,\, &
\big(\overrightarrow{x_i}, x, z. \doto {\langle x , z \rangle} {(y : A, z'' : \ul{C})} {} {z''}\big)
\\
=\,\, &
(\overrightarrow{x_i}, x, z.\, z''[z/z''])
\\
=\,\, &
(\overrightarrow{x_i}, x, z.\, z)
\end{align*}
\end{fleqn}

On the other hand, assuming that $\Gamma = x_1 \!:\! A_1, \ldots, x_n \!:\! A_n$, we observe that the triangle
\[
\xymatrix@C=5em@R=5em@M=0.5em{
\Sigma_{\lj {\Gamma\,} {\,A}} \ar[r]^-{\Sigma_{\lj {\Gamma\,} {\,A}} \,\comp\, \eta} \ar[dr]_{\id_{\Sigma_{\lj {\Gamma\,} {\,A}}}} & \Sigma_{\lj {\Gamma\,} {\,A}} \comp \pi^*_{\lj {\Gamma\,} {\,A}} \comp \Sigma_{\lj {\Gamma\,} {\,A}} \ar[d]^-{\varepsilon \,\comp\, \Sigma_{\lj {\Gamma\,} {\,A}}}
\\
& \Sigma_{\lj {\Gamma\,} {\,A}}
}
\]
can be rewritten for every computation type $\lj {\Gamma, x \!:\! A} \ul{C}$ as follows:
\[
\xymatrix@C=5em@R=5em@M=0.5em{
\lj \Gamma {\Sigma\, x \!:\! A .\, \ul{C}} \ar[r]^-{k} \ar[dr]_{(\overrightarrow{x_i}, z.z)} & \lj \Gamma {\Sigma\, y \!:\! A .\, \Sigma\, x \!:\! A.\, \ul{C}} \ar[d]^-{l}
\\
& \lj \Gamma {\Sigma\, x \!:\! A .\, \ul{C}}
}
\]
where the morphisms $k$ and $l$ are given by
\[
\begin{array}{c}
k \defeq \big(\overrightarrow{x_i}, z.\, \doto {z} {(x \!:\! A, z' \!:\! \ul{C})} {} {\langle x , \langle x , z' \rangle \rangle}\big)
\\[2mm]
l \defeq \big(\overrightarrow{x_i}, z''.\, \doto {z''} {(y \!:\! A, z''' \!:\! \Sigma\, x \!:\! A .\, \ul{C})} {} {z'''}\big)
\end{array}
\]
It is now straightforward to show that this triangle commutes, by
\begin{fleqn}[0.3cm]
\begin{align*}
&
\big(\overrightarrow{x_i}, z''.\, \doto {z''} {(y \!:\! A, z''' \!:\! \Sigma\, x \!:\! A .\, \ul{C})} {} {z'''}\big) \comp \big(\overrightarrow{x_i}, z.\, \doto {z} {(x \!:\! A, z' \!:\! \ul{C})} {} {\langle x , \langle x , z' \rangle \rangle}\big)
\\
=\,\, &
\big(\overrightarrow{x_i}, z.\, \doto {\big(\doto {z} {(x \!:\! A, z' \!:\! \ul{C})} {} {\langle x , \langle x , z' \rangle \rangle}\big)} {(y \!:\! A, z''' \!:\! \Sigma\, x \!:\! A .\, \ul{C})} {} {z'''}\big)
\\
=\,\, &
\big(\overrightarrow{x_i}, z.\, \doto {z} {(x \!:\! A, z' \!:\! \ul{C})} {} {\big(\doto {\langle x , \langle x , z' \rangle \rangle} {(y \!:\! A, z''' \!:\! \Sigma\, x \!:\! A .\, \ul{C})} {} {z'''}\big)}\big)
\\
=\,\, &
\big(\overrightarrow{x_i}, z.\, \doto {z} {(x \!:\! A, z' \!:\! \ul{C})} {} {z'''[\langle x , z' \rangle/z''']}\big)
\\
=\,\, &
\big(\overrightarrow{x_i}, z.\, \doto {z} {(x \!:\! A, z' \!:\! \ul{C})} {} {\langle x , z' \rangle}\big)
\\
=\,\, &
(\overrightarrow{x_i}, z.\, z)
\end{align*}
\end{fleqn}

\subsection*{$q$ admits split fibred pre-enrichment in $p$}

We define the functor $\multimap \,\,: \bigintsss (\Gamma \mapsto \mathcal{C}^{\text{op}}_\Gamma \times \mathcal{C}_\Gamma) \longrightarrow \mathcal{V}$ on objects by
\[
\multimap (\Gamma, \lj \Gamma \ul{C}, \lj \Gamma \ul{D}) \defeq \lj \Gamma {\ul{C} \multimap \ul{D}}
\]
and on morphisms by
\[
\multimap (\overrightarrow{V_i}, z.\, K, z'\! .\, L) \defeq (\overrightarrow{V_i}, x.\, \lambda\, z'' \!:\! \ul{C}_2[\overrightarrow{V_i}/\overrightarrow{x_i}] .\, L[x\, K[z''/z]/z'])
\]
where $\overrightarrow{V_i} : \Gamma_1 \longrightarrow \Gamma_2$, and $\hj {\Gamma_1} {z \!:\! \ul{C}_2[\overrightarrow{V_i}/\overrightarrow{x_i}]} K \ul{C}_1$, and $\hj {\Gamma_1} {z' \!:\! \ul{D}_1} L \ul{D}_2[\overrightarrow{V_i}/\overrightarrow{x_i}]$; where $x$ and $z''$ are chosen fresh; and where we assume that $\Gamma_2 = x_1 \!:\! A_1, \ldots, x_n \!:\! A_n$. 

It is easy to verify that the functor $\multimap$ is split fibred. On the one hand, $\multimap$ does not alter the context part of the given objects and morphisms. On the other hand, $\multimap$ preserves Cartesian morphisms on-the-nose because we have 
\begin{fleqn}[0.3cm]
\begin{align*}
& \multimap (\overrightarrow{V_i}, z.\, z, z'\! .\, z') 
\\
=\,\, &
(\overrightarrow{V_i}, x.\, \lambda\, z'' \!:\! \ul{C}_2[\overrightarrow{V_i}/\overrightarrow{x_i}] .\, z'[x\,\, z[z''\!/z]/z']) 
\\
=\,\, &
(\overrightarrow{V_i}, x.\, \lambda\, z'' \!:\! \ul{C}_2[\overrightarrow{V_i}/\overrightarrow{x_i}] .\, x\,\, z[z''\!/z]) 
\\
=\,\, &
(\overrightarrow{V_i}, x.\, \lambda\, z'' \!:\! \ul{C}_2[\overrightarrow{V_i}/\overrightarrow{x_i}] .\, x\,\, z'')
\\
=\,\, &
(\overrightarrow{V_i}, x.\, x)
\end{align*}
\end{fleqn}

Finally, the isomorphisms $\xi_{\Gamma,\lj {\Gamma\,} {\,\ul{C}}, \lj {\Gamma\,} {\,\ul{D}}}$ between hom-sets are witnessed by the following functions:
\[
\begin{array}{c}
\xi_{\Gamma,\lj {\Gamma\,} {\,\ul{C}}, \lj {\Gamma\,} {\,\ul{D}}}(\overrightarrow{x_i}, x.\, V) \defeq (\overrightarrow{x_i}, z.\, (V[\star/x])\, z)
\\[1mm]
\xi^{-1}_{\Gamma,\lj {\Gamma\,} {\,\ul{C}}, \lj {\Gamma\,} {\,\ul{D}}}(\overrightarrow{x_i}, z.\, K) \defeq (\overrightarrow{x_i}, y.\, \lambda\, z' \!:\! \ul{C} .\, K[z'/z])
\end{array}
\]
where $z$ is chosen fresh in the former; and $y$ and $z'$ are chosen fresh in the latter. 

\subsection*{The completeness theorem}

To begin with, we first summarise the above definitions and results in the next theorem.

\begin{theorem}
\label{thm:classifyingmodel}
\index{fibred adjunction model!classifying --}
The above definitions, based on the well-formed syntax of eMLTT, constitute a fibred adjunction model, called the \emph{classifying fibred adjunction model}.
\end{theorem}

Next, we show that  
the interpretation function $\sem{-}$ maps eMLTT's types and terms to 
their respective equivalence classes 
in this classifying fibred adjunction model.

\pagebreak

\begin{proposition}
\label{prop:classifyingmodelinterpretation2} 
Assuming that $\Gamma = x_1 \!:\! A_1, \ldots, x_n \!:\! A_n$, we have:
\mbox{}
\begin{enumerate}[(a)]
\item If    
$\sem{\Gamma} = \Gamma' \in \mathcal{B}$, then $\Gamma' = x'_1 \!:\! A'_1, \ldots, x'_n \!:\! A'_n$ and for all $1 \leq i \leq n$, we have 
\[
\ljeq {x'_1 \!:\! A'_1, \ldots, x'_{i-1} \!:\! A'_{i-1}} {A'_i} {A_i[x'_1/x_1, \ldots, x'_{i-1}/x_{i-1}]}
\]
\item If  
$\sem{\Gamma;A} = \lj {\sem{\Gamma}} A' \in \mathcal{V}_{\sem{\Gamma}}$, then $\ljeq {\sem{\Gamma}} {A'} {A[x'_1/x_1, \ldots, x'_n/x_n]}$.
\item If   
$\sem{\Gamma;\ul{C}} = \lj {\sem{\Gamma}} \ul{C}' \in \mathcal{C}_{\sem{\Gamma}}$, then $\ljeq {\sem{\Gamma}} {\ul{C}'} {\ul{C}[x'_1/x_1, \ldots, x'_n/x_n]}$.
\item If $\sem{\Gamma;V} = (x'_1, \ldots, x'_n, y.\,V') : \lj {\sem{\Gamma}} 1 \longrightarrow \lj {\sem{\Gamma}} {A'}$, then  
\[
\veq {\sem{\Gamma}} {V'} {V[x'_1/x_1,\ldots,x'_n/x_n]} {A'}
\]
\item If $\sem{\Gamma;M} = (x'_1, \ldots, x'_n, y.\,V_M) : \lj {\sem{\Gamma}} 1 \longrightarrow \lj {\sem{\Gamma}} {U\ul{C}'}$, then  
\[
\veq {\sem{\Gamma}} {\force {\ul{C}'} V_M} {M[x'_1/x_1,\ldots,x'_n/x_n]} {\ul{C}'}
\]
\item If $\sem{\Gamma;z \!:\! \ul{C};K} = (x'_1, \ldots, x'_n, z.\, K') : \lj {\Gamma} {\ul{C}} \longrightarrow \lj {\Gamma} {\ul{D}'}$, then 
\[
\heq {\sem{\Gamma}} {z \!:\! \ul{C}[x'_1/x_1,\ldots,x'_n/x_n]} {K'} {K[x'_1/x_1,\ldots,x'_n/x_n]} {\ul{D}'}
\]
\end{enumerate}

\noindent
\end{proposition}

\begin{proof}
We prove $(a)$--$(f)$ simultaneously, by induction on sum of the sizes of 
the arguments to $\sem{-}$. 
As two representative examples, we present the cases corresponding to the computational 
$\Sigma$-type and the sequential composition of computation terms.

\vspace{0.1cm}
\noindent
\textbf{The computational $\Sigma$-type:} 
In this case, we have that $\ul{C} = {\Sigma\, x \!:\! A .\, \ul{D}}$.

First, by inspecting the definition of $\sem{-}$ for the computational $\Sigma$-type, we get that 
\[
\sem{\Gamma;{\Sigma\, x \!:\! A .\, \ul{D}}} = \lj {\sem{\Gamma}} {{\Sigma\, x'' \!:\! A' .\, \ul{D}'}} \in \mathcal{C}_{\sem{\Gamma}}
\]
with $\sem{\Gamma;A} = \lj {\sem{\Gamma}} {A'} \in \mathcal{V}_{\sem{\Gamma}}$ and $\sem{\Gamma,x\!:\! A;\ul{D}} = \lj {\sem{\Gamma},x''\!:\! A'} \ul{D}' \in \mathcal{C}_{\sem{\Gamma},x'' :  A'}$.

As a result, we can use $(b)$ and the induction hypothesis to get that 
\[
\ljeq {\sem{\Gamma}} {A'} {A[x'_1/x_1, \ldots, x'_n/x_n]} 
\qquad
\ljeq {\sem{\Gamma},x'' :  A'} {\ul{D}'} {\ul{D}[x'_1/x_1, \ldots, x'_n/x_n,x''/x]}
\]

Finally, by using the congruence equation for the computational $\Sigma$-type, we get that
\[
\ljeq {\sem{\Gamma}} {\Sigma\, x'' \!:\! A' .\, \ul{D}'} {\Sigma\, x'' \!:\! A[x'_1/x_1, \ldots, x'_n/x_n] .\, \ul{D}[x'_1/x_1, \ldots, x'_n/x_n,x''/x]}
\]
which, according to our chosen variable conventions, is in fact the proof of 
\[
\ljeq {\sem{\Gamma}} {\Sigma\, x'' \!:\! A' .\, \ul{D}'} {\Sigma\, x \!:\! A[x'_1/x_1, \ldots, x'_n/x_n] .\, \ul{D}[x'_1/x_1, \ldots, x'_n/x_n]}
\]
and which, according to the definition of simultaneous substitutions, is the proof of  
\[
\ljeq {\sem{\Gamma}} {\Sigma\, x'' \!:\! A' .\, \ul{D}'} {(\Sigma\, x \!:\! A .\, \ul{D})[x'_1/x_1, \ldots, x'_n/x_n]}
\]

\vspace{0.1cm}
\noindent
\textbf{Sequential composition:}
In this case, we have that $M = {\doto {N_1} {x \!:\! A} {\ul{C}} {N_2}}$. 

First, by unfolding the definition of $\sem{-}$ for sequential composition, we get that 
\[
\begin{array}{c}
\sem{\Gamma;\doto {N_1} {x \!:\! A} {\ul{C}} {N_2}}
\\
=
\\
\big(\overrightarrow{x'_i}, y_{11}.\, \thunk \big(\doto {(\force {} {y_{11}})} {y_{12} \!:\! U\ul{C}'} {} {(\force {} {y_{12}})}\big)\big)
\\
\comp
\\
\hspace{-6.1cm}
\big(\overrightarrow{x'_i}, y_7.\, \thunk \big(\doto {(\force {} y_7)} {y_8 \!:\! A' \times U\ul{C}'} {} {\\[-1mm] \hspace{5.3cm} \return (\pmatch {y_8} {(y_9 \!:\! A',y_{10} \!:\! U\ul{C}')} {} {y_{10}})}\big)\big)
\\
\comp
\\
\hspace{-6.5cm}
\big(\overrightarrow{x'_i}, y_4.\, \thunk \big(\doto {(\force {} y_4)} {y_5 \!:\! A' \times 1} {} {\\[-1mm] \hspace{5cm} \return (\pmatch {y_5} {(x'' \!:\! A',y_6 \!:\! 1)} {} {\langle x'' , V_{N_2} \rangle})}\big)\big)
\\
\comp 
\\
\big(\overrightarrow{x'_i}, y_2.\, \thunk \big(\doto {(\force {} y_2)} {y_3 \!:\! A'} {} {\return \langle y_3, \star \rangle}\big)\big)
\\
\comp
\\
(\overrightarrow{x'_i}, y_1.\, V_{N_1})
\\[0.5mm]
=
\\[-1mm]
\big(\overrightarrow{x'_i}, y_1.\, V_{(\doto {N_1\!} {\!x : A\!} {\ul{C}} {\!N_2})}\big)
\end{array}
\]
as a morphism $\lj {\sem{\Gamma}} 1 \longrightarrow \lj {\sem{\Gamma}} {U\ul{C}'}$ in $\mathcal{V}_{\sem{\Gamma}}$, with  
\[ 
\begin{array}{c}
\sem{\Gamma;A} = \lj {\sem{\Gamma}} {A'} \in \mathcal{V}_{\sem{\Gamma}}
\qquad
\sem{\Gamma;\ul{C}} = \lj {\sem{\Gamma}} \ul{C}' \in \mathcal{C}_{\sem{\Gamma}}
\\[2mm]
\sem{\Gamma;N_1} = (x'_1, \ldots, x'_n, y_1.\,V_{N_1}) : \lj {\sem{\Gamma}} 1 \longrightarrow \lj {\sem{\Gamma}} {A'} 
\\[2mm]
\sem{\Gamma, x\!:\! A;N_2} = (x'_1, \ldots, x'_n, x'', y_6.\,V_{N_2}) : \lj {\sem{\Gamma}, x'' : A'} 1 \longrightarrow \lj {\sem{\Gamma},x'' \!:\! A'} {U\ul{C}'}
\end{array}
\]

As a result, we can use $(b)$, $(c)$, and the induction hypothesis to get that 
\[
\begin{array}{c}
\ljeq {\sem{\Gamma}} {A'} {A[\overrightarrow{x'_i}/\overrightarrow{x_i}]}
\qquad
\ljeq {\sem{\Gamma}} {\ul{C}'} {\ul{C}[\overrightarrow{x'_i}/\overrightarrow{x_i}]}
\\[2mm]
\veq {\sem{\Gamma}} {\force {\ul{FA}'} V_{N_1}} {N_1[\overrightarrow{x'_i}/\overrightarrow{x_i}]} {FA'}
\\[2mm]
\veq {\sem{\Gamma},x''\!:\! A'} {\force {\ul{C}'} V_{N_2}} {N_2[\overrightarrow{x'_i}/\overrightarrow{x_i},x''/x]} {\ul{C}'}
\end{array}
\]

Finally, we note that the required equation
\[
\veq {\sem{\Gamma}} {\force {\ul{C}'} V_{(\doto {N_1\!} {\!x : A\!} {\ul{C}} {\!N_2})}} {(\doto {N_1} {x \!:\! A} {\ul{C}} {N_2})[\overrightarrow{x'_i}/\overrightarrow{x_i}]} {\ul{C}'}
\]
follows by straightforward equational reasoning, based on the definitional equations we just derived, in combination with the unfolding of $\sem{\Gamma;\doto {N_1} {x \!:\! A} {\ul{C}} {N_2}}$.
\end{proof}

Finally, we prove the completeness of fibred adjunction models for eMLTT.

\begin{theorem}[Completeness] 
\label{thm:completeness}
\index{completeness theorem}
If we assume given contexts $\Gamma_1 = y_1 \!:\! B_1, \ldots, y_n \!:\! B_n$ and 
$\Gamma_2 = y'_1 \!:\! B'_1, \ldots, y'_m \!:\! B'_m$, then we have:
\begin{enumerate}[(a)]
\item If $\sem{\Gamma_1;A} = \sem{\Gamma_2;B}$ in all fibred adjunction models, then $n = m$ and  
\[
\begin{array}{c}
\ljeq \Gamma {A[x_1/y_1,\ldots,x_n/y_n]} {B[x_1/y'_1,\ldots,x_n/y'_n]}
\end{array}
\]
for some $\Gamma = x_1\!:\! A_1, \ldots, x_n \!:\! A_n$ such that for all $1 \leq i \leq n$, we have 
\[
\ljeq {x_1 \!:\! A_1, \ldots, x_{i-1} \!:\! A_{i-1}} {A_i} {B_i[x_1/y_1, \ldots, x_{i-1}/y_{i-1}]} = {B'_i[x_1/y'_1, \ldots, x_{i-1}/y'_{i-1}]}
\]
\item If $\sem{\Gamma_1;\ul{C}} = \sem{\Gamma_2;\ul{D}}$ in all fibred adjunction models, then $n = m$ and  
\[
\begin{array}{c}
\ljeq \Gamma {\ul{C}[x_1/y_1,\ldots,x_n/y_n]} {\ul{D}[x_1/y'_1,\ldots,x_n/y'_n]}
\end{array}
\]
for some $\Gamma = x_1\!:\! A_1, \ldots, x_n \!:\! A_n$ such that for all $1 \leq i \leq n$, we have 
\[
\ljeq {x_1 \!:\! A_1, \ldots, x_{i-1} \!:\! A_{i-1}} {A_i} {B_i[x_1/y_1, \ldots, x_{i-1}/y_{i-1}]} = {B'_i[x_1/y'_1, \ldots, x_{i-1}/y'_{i-1}]}
\]
\item If $\sem{\Gamma_1;V} = \sem{\Gamma_2;W}$ in all fibred adjunction models, then $n = m$ and  
\[
\begin{array}{c}
\veq \Gamma {V[x_1/y_1,\ldots,x_n/y_n]} {W[x_1/y'_1,\ldots,x_n/y'_n]} {A}
\end{array}
\]
for some $A$ and $\Gamma = x_1\!:\! A_1, \ldots, x_n \!:\! A_n$ such that for all $1 \leq i \leq n$, we have 
\[
\ljeq {x_1 \!:\! A_1, \ldots, x_{i-1} \!:\! A_{i-1}} {A_i} {B_i[x_1/y_1, \ldots, x_{i-1}/y_{i-1}]} = {B'_i[x_1/y'_1, \ldots, x_{i-1}/y'_{i-1}]}
\]
\item If $\sem{\Gamma_1;M} = \sem{\Gamma_2;N}$ in all fibred adjunction models, then $n = m$ and  
\[
\begin{array}{c}
\ceq \Gamma {M[x_1/y_1,\ldots,x_n/y_n]} {N[x_1/y'_1,\ldots,x_n/y'_n]} {\ul{C}}
\end{array}
\]
for some $\ul{C}$ and $\Gamma = x_1\!:\! A_1, \ldots, x_n \!:\! A_n$ such that for all $1 \leq i \leq n$, we have 
\[
\ljeq {x_1 \!:\! A_1, \ldots, x_{i-1} \!:\! A_{i-1}} {A_i} {B_i[x_1/y_1, \ldots, x_{i-1}/y_{i-1}]} = {B'_i[x_1/y'_1, \ldots, x_{i-1}/y'_{i-1}]}
\]
\item If $\sem{\Gamma_1;z_1\!:\!\ul{C}_1;K} = \sem{\Gamma_2;z_2\!:\!\ul{C}_2;L}$ in all fibred adjunction models, then $n = m$ and  
\[
\begin{array}{c}
\heq \Gamma {z_1 \!:\! \ul{C}_1[x_1/y_1,\ldots,x_n/y_n]} {K[x_1/y_1,\ldots,x_n/y_n]} {L[x_1/y'_1,\ldots,x_n/y'_n][z_1/z_2]} {\ul{D}}
\end{array}
\]
for some $\ul{D}$ and $\Gamma = x_1\!:\! A_1, \ldots, x_n \!:\! A_n$ such that for all $1 \leq i \leq n$, we have 
\[
\ljeq {x_1 \!:\! A_1, \ldots, x_{i-1} \!:\! A_{i-1}} {A_i} {B_i[x_1/y_1, \ldots, x_{i-1}/y_{i-1}]} = {B'_i[x_1/y'_1, \ldots, x_{i-1}/y'_{i-1}]}
\]
\end{enumerate}
\noindent
\end{theorem}

\begin{proof}
We prove $(a)$--$(e)$ simultaneously, following the same general pattern of using the 
interpretation in the classifying fibred adjunction model, in combination with Proposition~\ref{prop:classifyingmodelinterpretation2}.
As a representative example, we present the proof of $(d)$ below.

First, we observe that if $\sem{\Gamma_1;M} = \sem{\Gamma_2;N}$ in all fibred adjunction models, 
then \linebreak $\sem{\Gamma_1;M} = \sem{\Gamma_2;N}$ in the classifying fibred adjunction model. 
As a result, we have 
\[
\begin{array}{c}
\sem{\Gamma_1;M} = (x_1,\ldots, x_n, x.\, V_M) : \lj {\sem{\Gamma_1}} 1 \longrightarrow \lj {\sem{\Gamma_1}} U\ul{D}
\\[2mm]
\sem{\Gamma_2;N} = (x_1,\ldots, x_n, x.\, V_N) : \lj {\sem{\Gamma_2}} 1 \longrightarrow \lj {\sem{\Gamma_2}} U\ul{D}'
\end{array}
\]
such that (as contexts, types, and terms are identified in the classifying fibred adjunction model when they are definitionally equal---see the definitions of $\mathcal{B}$, $\mathcal{V}$, and $\mathcal{C}$)
\[
\ljeq {} {\sem{\Gamma_1}} {\sem{\Gamma_2}}
\qquad
\ljeq {\sem{\Gamma_1}} {U\ul{D}} {U\ul{D}'}
\qquad
\veq {\sem{\Gamma_1}, x \!:\! 1} {V_M} {V_N} {U\ul{D}}
\]
from which it follows that $n = m$. As a result, we can consider $\sem{\Gamma_1;M}$ and $\sem{\Gamma_1;N}$ as 
\[
\begin{array}{c}
\sem{\Gamma_1;M} = (x_1,\ldots, x_n, x.\, V_M) : \lj {\sem{\Gamma_1}} 1 \longrightarrow \lj {\sem{\Gamma_1}} U\ul{D}
\\[2mm]
\sem{\Gamma_2;N} = (x_1,\ldots, x_n, x.\, V_N) : \lj {\sem{\Gamma_1}} 1 \longrightarrow \lj {\sem{\Gamma_1}} U\ul{D}
\end{array}
\]
and choose $\ul{C} \defeq \ul{D}$.

Next, we choose $\Gamma \defeq \sem{\Gamma_1}$ and then use $(a)$ of Proposition~\ref{prop:classifyingmodelinterpretation2} to get that 
\[
\ljeq {x_1 \!:\! A_1, \ldots, x_{i-1} \!:\! A_{i-1}} {A_i} {B'_i[x_1/y'_1, \ldots, x_{i-1}/y'_{i-1}]}
\]
for all $1 \leq i \leq n$, which, when combined with Proposition~\ref{prop:valuesubstlemma2simultaneous}, gives us that 
\[
\ljeq {x_1 \!:\! A_1, \ldots, x_{i-1} \!:\! A_{i-1}} {A_i} {B_i[x_1/y_1, \ldots, x_{i-1}/y_{i-1}]} = {B'_i[x_1/y'_1, \ldots, x_{i-1}/y'_{i-1}]}
\]
for all $1 \leq i \leq n$.

Next, by using $(e)$ of Proposition~\ref{prop:classifyingmodelinterpretation2}, we get that 
\[
\begin{array}{c}
\veq {\sem{\Gamma_1}} {\force {\ul{D}} V_M} {M[x_1/y_1,\ldots,x_n/y_n]} {\ul{D}}
\\[2mm]
\veq {\sem{\Gamma_1}} {\force {\ul{D}} V_N} {N[x_1/y'_1,\ldots,x_n/y'_n]} {\ul{D}}
\end{array}
\]

Next, we use Propositions~\ref{prop:freevariablesofwellformedexpressions} and~\ref{prop:wellformedcomponentsofjudgements} to get that $x \not\in FVV(V_M)$, $x \not\in FVV(V_N)$, and $x \not\in FVV(\ul{D})$. As a result, we get a proof of the following definitional equation:
\[
\veq {\sem{\Gamma_1}} {V_M} {V_N} {\ul{D}}
\]
by substituting $\star$ for $x$ in $\veq {\sem{\Gamma_1}, x \!:\! 1} {V_M} {V_N} {\ul{D}}$, and by using Proposition~\ref{prop:valuesubstlemma1}.

Finally, the required equation now follows by using the rules of symmetry and transitivity, and the congruence rule for forcing thunked computations, giving us 
\[
\ceq {\sem{\Gamma_1}} {M[x_1/y_1,\ldots,x_n/y_n]} {N[x_1/y'_1,\ldots,x_n/y'_n]} {\ul{D}}
\]
\end{proof}


\chapter[eMLTT$_{\!\mathcal{T}_{\text{eff}}}$: an extension of eMLTT with fibred algebraic effects]{eMLTT$_{\mathcal{T}_{\text{eff}}}$: an extension of eMLTT \\with fibred algebraic effects}
\label{chap:fibalgeffects}

\index{ e@eMLTT$_{\mathcal{T}_{\text{eff}}}$ (extension of eMLTT with fibred algebraic effects)}
While eMLTT makes it clear how to account for type-dependency in composite effectful dependently typed programs (using the combination of sequential composition and computational $\Sigma$-types), it  provides programmers with no way to use specific computational effects in their code, such as exceptions, nondeterminism, state, I/O, etc. In this chapter, we address this limitation by extending eMLTT with corresponding language primitives and definitional equations. This extension of eMLTT is based on the algebraic treatment of computational effects---see Section~\ref{sect:algebraictreatmentofeffects} for an overview.  Thus it allows us to uniformly capture a wide range of computational effects in eMLTT.

In Section~\ref{sect:fibeffecttheories}, we define a notion of fibred effect theory so as to specify computational effects using operations and equations. 
Unlike the existing work on algebraic effects, our operation symbols are dependently typed, enabling us to capture precise notions of computation, such as state with location-dependent store types and dependently typed update monads. 
In Section~\ref{sect:fibalgeffectsineMLTT}, we show how to extend eMLTT with computational effects specified by a given fibred effect theory ${\mathcal{T}_{\text{eff}}}$. In particular, we extend its computation terms with algebraic operations, and its equational theory with the corresponding  definitional equations. We call the resulting language eMLTT$_{\mathcal{T}_{\text{eff}}}$. 

In Section~\ref{sect:emlttalgeffectsmetatheory}, we show how to extend the meta-theory of eMLTT to eMLTT$_{\mathcal{T}_{\text{eff}}}$. \linebreak
In Section~\ref{sect:derivableequationsforeMLTTwithfibalgeffects}, we present some useful definitional equations derivable in eMLTT$_{\mathcal{T}_{\text{eff}}}$. In Section~\ref{sect:fibalgeffectsmodel}, we equip eMLTT$_{\mathcal{T}_{\text{eff}}}$ with a denotational semantics,  showing how to define a sound interpretation of it in a fibred adjunction model based on the families of sets fibration and models of a countable Lawvere theory we derive from $\mathcal{T}_{\text{eff}}$. Finally, in Section~\ref{ref:genericeffects}, we briefly discuss an equivalent extension of eMLTT with generic effects.

\section{Fibred algebraic effects}
\label{sect:fibeffecttheories}

In this section we develop a means to uniformly specify a wide range of computational effects, ranging from well-known examples such as exceptions, nondeterminism, state, IO, etc. to a less well-known example of (dependently typed) update monads. 

Following the work of Plotkin and Pretnar in the simply typed setting~\cite{Plotkin:HandlingEffects}, we develop a notion of fibred effect theory so as to specify computational effects in terms of operation symbols and equations. 
The former denote the sources of computational effects, with the latter describing their computational properties. 
Following Plotkin and Pretnar, we begin by defining a notion of \emph{fibred effect signature} (given by a finite set of dependently typed operation symbols) and then extend it to a notion of \emph{fibred effect theory} (given by extending a fibred effect signature with a finite set of equations).
To emphasise the dependently typed nature of our operation symbols, we refer to the computational effects specified by these theories as \emph{fibred algebraic effects}. 
\index{algebraic effect!fibred --}

\subsection{Fibred effect signatures}

As mentioned earlier, our fibred effect signatures consist of operation symbols that are dependently typed. We specify these dependent types internally in a certain fragment of eMLTT, consisting of  \emph{pure value types} and \emph{pure value terms}, as defined below. 

\begin{definition}
\index{type!value --!pure --}
An eMLTT value type is said to be \emph{pure} if it is constructed only from $\Nat$, $1$, $\Sigma \, x \!:\! A .\, B $, $\Pi \, x \!:\! A .\, B$, $0$, $A + B$, and $V =_A W$, with $A$ pure in propositional equality.
\end{definition}

In other words, a value type is pure exactly when it contains neither $U$ nor $\multimap$.
It is also worth noting that pure value types are very similar to the discrete value types we used in Section~\ref{sect:extensionofeMLTTwithrecursion}---the only difference being in the argument type of the value $\Pi$-type. For pure types, the argument type has to be pure, whereas for discrete types the argument type could be arbitrary, as discreteness is determined by the result type.

\begin{definition}
\index{term!value --!pure --}
An eMLTT value term is said to be \emph{pure} if it does not contain $\thunk{}$ terms and homomorphic lambda abstractions, and all its type annotations are pure.
\end{definition}

Based on this fragment of eMLTT, we now define a notion of fibred effect signature.

\begin{definition}
\index{signature!fibred effect --}
\index{ I@$I$ (input type of an operation symbol)}
\index{ O@$O$ (output type of an operation symbol)}
\index{ op@$\sigalgop$ (operation symbol in a fibred effect signature)}
\index{type!input --}
\index{type!output --}
\index{ S@$\mathcal{S}_{\text{eff}}$ (fibred effect signature)}
A \emph{fibred effect signature} $\mathcal{S}_{\text{eff}}$ is a finite set of typed operation symbols
\[
\sigalgop : (x \!:\! I) \longrightarrow O
\]
where $\lj {\diamond} I$ and $\lj {x \!:\! I} O$ are well-formed pure value types, called the \emph{input} and \emph{output} type of $\sigalgop$, respectively.
\end{definition}

Analogously to types and terms that involve variable bindings, the variable $x$ is bound in $O$ in the type of $\sigalgop : (x \!:\! I) \longrightarrow O$; and we do not distinguish between $\alpha$-equivalent types of operation symbols. We also assume that in any mathematical context, the bound variable $x$ in the type of $\sigalgop$ is always chosen to be different from the free variables of that context.
As a further simplification, if the variable $x$ is not free in $O$, we omit the variable binding and simply write the type of $\sigalgop$ simply as $I \longrightarrow O$.

Intuitively, in models where dependent value types denote families of sets (see the model of eMLTT$_{\mathcal{T}_{\text{eff}}}$ given in Section~\ref{sect:fibalgeffectsmodel}), one thinks of  $\sigalgop : (x \!:\! I) \longrightarrow O$ as describing an $I$-indexed family of algebraic operations $\sigalgop_i$, each of whose arity is the cardinality of the set denoted by $O[i/x]$. 
From a computational perspective, the input type of an operation should be understood as specifying the values used to parameterise the corresponding effect, e.g., the memory locations to be accessed; and the output type of an operation as specifying the values that are produced by performing the corresponding effect, e.g., for the $\mathsf{get}$ operation, the value stored in the memory. 
Based on these intuitions, $I$ could also be called a \emph{parameter} type and $O$ an \emph{arity} type, e.g., as in \cite{Plotkin:HandlingEffects}.

We now give some examples of fibred effect signatures for important computational effects, 
starting with ones based on simply typed effect signatures from~\cite{Plotkin:HandlingEffects}.

\begin{example}[Exceptions]
\label{ex:fibsigofexceptions}
\index{signature!fibred effect --!-- of exceptions}
\index{ Exc@$\Exception$ (type of exception names)}
Assuming given a well-formed pure value type $\lj \diamond \Exception$ of exception names, the signature $\mathcal{S}_{\text{EXC}}$ of exceptions is given by one operation symbol
\[
\mathsf{raise} : \Exception \longrightarrow 0
\]

The idea is that $\mathsf{raise}$ denotes the effect of raising an exception corresponding to a given value of type $\Exception$. The output type of $\mathsf{raise}$ is the empty type $0$ because after raising an exception in a program, there is no further continuation to be evaluated.
\end{example}

\begin{example}[Binary nondeterminism]
\label{ex:fibsigofnondeterminism}
\index{signature!fibred effect --!-- of binary nondeterminism}
The signature $\mathcal{S}_{\text{ND}}$ of binary nondeterminism is given by one operation symbol
\[
\mathsf{choose} : 1 \longrightarrow 1 + 1
\]

The idea is that $\mathsf{choose}$ denotes the effect of nondeterministically making a binary
choice, with the outcome witnessed by returning either the value $\inl {} \star$ or $\inr {} \star$. \end{example}

\begin{example}[Global state]
\label{ex:fibsigofstate}
\index{signature!fibred effect --!-- of global state}
\index{ St@$\State$ (type of store values)}
Assuming given a well-formed pure value type $\lj \diamond \State$ of store values, the signature $\mathcal{S}_{GS}$ of global state is given by two operation symbols
\[
\mathsf{get} : 1 \longrightarrow \State 
\qquad
\mathsf{put} : \State \longrightarrow 1 
\]

The idea is that $\mathsf{get}$ denotes the effect of reading and returning the current value of the store; and $\mathsf{put}$ denotes the effect of setting the store to a given value of type $\State$. 
\end{example}

Observe that in the previous example, $\mathsf{get}$ and $\mathsf{put}$ operate on the whole state. Below, we consider a common variation of the signature of global state that incorporates multiple memory locations. However, in contrast to the simply typed effect signature for global state with locations, where all locations have to store values of the same type, e.g., see~\cite{Plotkin:HandlingEffects},
the presence of dependent types allows us to make the type of store values dependent on locations, giving a more realistic presentation of global state. 

\begin{example}[Global state with locations]
\label{ex:fibsigofstatewithlocations}
\index{signature!fibred effect --!-- of global state with locations}
\index{ Loc@$\Location$ (type of memory locations)}
\index{ Val@$\Value$ (type of values stored at memory locations)}
Assuming well-formed pure value types 
\[
\lj \diamond \Location
\qquad
\lj {x \!:\! \Location} \Value
\]
of memory locations and values stored at them, respectively, the signature $\mathcal{S}_{GSL}$ of global state with locations is  given by two operation symbols
\[
\mathsf{get} : (x \!:\! \Location) \longrightarrow \Value
\qquad
\mathsf{put} : \Sigma\, x \!:\! \Location .\, \Value \longrightarrow 1
\]

Observe that compared to the operation symbols given in Example~\ref{ex:fibsigofstate}, this $\mathsf{get}$ and $\mathsf{put}$ take the memory location to be accessed as an additional value argument.
\end{example}

In the simply typed setting, where $\Value$ would not be allowed to depend on $\Location$, this signature would need to be given either i) by restricting all locations to store values of the same type (as already suggested earlier), or ii) by families of operation symbols
\[
\mathsf{get}_V : 1 \longrightarrow \Value_V
\qquad
\mathsf{put}_V : \Value_V \longrightarrow 1
\]
where $\mathsf{get}$, $\mathsf{put}$, and $\Value$ are all indexed by closed normal forms $V$ of type $\Location$. 

However, if we were to extend a simply typed programming language with primitives corresponding to the second approach, we must bear in mind that in state-manipulating programs it is often desirable to use $\mathsf{get}$ and $\mathsf{put}$ with non-normal and open arguments of type $\Location$. While this could be achieved to some extent using case analysis on the given value argument of type $\Location$, the lack of dependent types means that the corresponding derived operation would have the following imprecise type:
\[
\mathsf{get} : \Location \longrightarrow \Value_{V_1} ~+~ \ldots ~+~ \Value_{V_n}
\]
where we use $V_1, \ldots, V_n$ to range over the closed normal forms of type $\Location$.

As a final example of well-known and important computational effects, we present the fibred effect signature of interactive character input/output. This signature does not use any type-dependency and is therefore exactly the same as the one given in~\cite{Plotkin:HandlingEffects}.

\begin{example}[Input/output]
\label{ex:fibsigofIO}
\index{signature!fibred effect --!-- of input/output}
\index{ Chr@$\Character$ (type of characters)}
Assuming a well-formed pure value type $\lj \diamond \Character$ of characters, the signature $\mathcal{S}_{\text{I/O}}$ of input/output is given by two operation symbols
\[
\mathsf{read} : 1 \longrightarrow \Character
\qquad
\mathsf{write} : \Character \longrightarrow 1
\]

The idea is that $\mathsf{read}$ denotes the effect of reading a character from the terminal; and $\mathsf{write}$ denotes the effect of writing the given character to the terminal. Note that the same signature can also be used to describe input/output over, say, a network.
\end{example}

Observe that the signature of input/output given in Example~\ref{ex:fibsigofIO} is essentially the same as the signature of global state given in Example~\ref{ex:fibsigofstate}, modulo the names of the operation symbols and their types; these two computational effects differ in the equations one  imposes on them---see Examples~\ref{ex:fibtheoryofglobalstate} and~\ref{ex:fibtheoryofIO} for details. 

Further, observe that analogously to the signature $\mathcal{S}_{GSL}$ of global state with locations, one can also extend $\mathcal{S}_{\text{I/O}}$ with type-dependency by considering multiple terminals (or network channels) and making the values read from and written to terminals (resp. network channels) dependent on terminal names (resp. channel names).

In addition to these well-known effect theories from the algebraic effects literature, we also want to draw the reader's attention to a less well-known example of global state, in which the store is changed not by overwriting but by applying (potentially small) updates to it. This notion of global state is modelled by   update monads that were introduced and thoroughly studied by the author in a joint paper with Uustalu~\cite{Ahman:UpdateMonads}. 

\begin{example}[Update monads]
\label{ex:fibsigofupdatemonad}
\index{signature!fibred effect --!-- of an update monad}
\index{ Upd@$(\Updates,\mathsf{o},\oplus)$ (monoid of updates)}
\index{ @$\downarrow$ (action of the monoid of updates on the set of store values)}
\index{monoid}
Assuming given two well-formed pure value types
\[
\lj {\diamond} \State
\qquad
\lj {\diamond} \Updates
\]
of store values and store updates, together with well-typed closed pure value terms
\[
\downarrow\, : \State \to \Updates \to \State
\qquad
\mathsf{o} : \Updates
\qquad
\oplus : \Updates \to \Updates \to \Updates
\]
satisfying the following five closed equations (in the equational theory of eMLTT):
\[
V \downarrow \mathsf{o} = V 
\qquad
V \downarrow (W_1 \oplus W_2) = (V \downarrow W_1) \downarrow W_2
\]
\[
W \oplus \mathsf{o} = W
\qquad
\mathsf{o} \oplus W = W
\qquad
(W_1 \oplus W_2) \oplus W_3 = W_1 \oplus (W_2 \oplus W_3)
\]
the signature $\mathcal{S}_{\text{UPD}}$ of a (simply typed) update monad is given by two operation symbols
\[
\mathsf{lookup} : 1 \longrightarrow \State
\qquad
\mathsf{update} : \Updates \longrightarrow 1
\]

For better readability, we omit the empty contexts and $\,\vdash$ in the typing of $\downarrow$, $\mathsf{o}$, and $\oplus$, and in the equations. Further, we omit the types of the equations and assume that all value terms are well-typed according to the typing of $\downarrow$, $\mathsf{o}$, $\oplus$. To improve the readability further, we also use infix notation when applying $\downarrow$ and $\oplus$ to their arguments.

The idea is that $(\Updates, \mathsf{o}, \oplus)$ forms a monoid of updates which can be applied to the store values via its action $\downarrow$ on $\State$; $\mathsf{lookup}$ denotes the effect of reading the current value of the store; and $\mathsf{update}$ denotes the effect of applying the update given by a value argument of type $\Updates$ to the current store. Regarding the monoid, the intuition is that $\mathsf{o}$ denotes the ``do nothing" update and $\oplus$ is used to combine successive updates.
\end{example}

While (simply typed) update monads are useful for modelling state changes by (small) incremental updates, their simply typed nature means that one must be able to meaningfully describe the action of all possible updates on all possible store values---see the type of $\downarrow$ given in Example~\ref{ex:fibsigofupdatemonad}. To address this limitation, we introduced a dependently typed generalisation of update monads in the above-mentioned joint paper with Uustalu. These monads are parameterised not by a monoid and its action on the store values, but instead by a dependently typed generalisation of monoids and their actions, in which the type of updates is allowed to depend on the type of store values, enabling us to precisely specify which updates are applicable to particular store values. 

This dependently typed generalisation of monoids and their actions is known in the literature under the name of \emph{directed containers}---see the author's joint paper with Chapman and Uustalu~\cite{Ahman:Dcontainers} for more details and their original use for modelling tree-like datastructures with a well-behaved notion of sub-datastructure.

\begin{example}[Dependently typed update monads]
\label{ex:fibsigofdeptypedupdatemonad}
\index{signature!fibred effect --!-- of a dependently typed update monad}
\index{directed container}
\index{ St@$(\State,\Updates,\downarrow,\mathsf{o},\oplus)$ (directed container of store values and updates)}
Assuming given two well-formed pure value types
\[
\lj {\diamond} \State
\qquad
\lj {x \!:\! \State} \Updates
\]
of store values and store updates, together with well-typed closed pure value terms
\[
\begin{array}{c}
\downarrow\, : \Pi\, x \!:\! \State .\, \Updates \to \State
\qquad
\mathsf{o} : \Pi\, x \!:\! \State .\, \Updates
\\[1mm]
\oplus : \Pi\,x \!:\! \State .\, \Pi y \!:\! \Updates .\, \Updates[x \downarrow y/x] \to \Updates
\end{array}
\]
satisfying the following five closed equations (in the equational theory of eMLTT):
\[
\begin{array}{c}
V \downarrow (\mathsf{o}\,\, V) = V 
\qquad
V \downarrow (W_1 \oplus_{V} W_2) = (V \downarrow W_1) \downarrow W_2
\\[2mm]
W \oplus_{V} (\mathsf{o}\,\, (V \downarrow W)) = W
\qquad
(\mathsf{o}\,\, V) \oplus_V W = W
\\[2mm]
(W_1 \oplus_V W_2) \oplus_V W_3 = W_1 \oplus_V (W_2 \oplus_{V \downarrow W_1} W_3)
\end{array}
\]
the signature $\mathcal{S}_{\text{DUPD}}$ of a dependently typed update monad is given by two operation symbols
\[
\mathsf{lookup} : 1 \longrightarrow \State
\qquad
\mathsf{update} : \Pi\, x \!:\! \State .\, \Updates \longrightarrow 1
\]

In addition to the presentational conventions used in Example~\ref{ex:fibsigofupdatemonad}, we further improve readability by writing the first argument to $\oplus$ as a subscript in the equations.
\end{example}

In~\cite[Examples~10 and~11]{Ahman:UpdateMonads}, it is demonstrated that dependently typed update monads can be used for natural state-based modelling of \emph{non-overflowing  buffers} and \emph{non-under\-flowing stacks}, by ensuring that the size of the data written to a buffer does not exceed the remaining free space, and by not allowing an empty stack to be popped.

It is worth noting that differently from~\cite{Ahman:UpdateMonads}, where algebras of dependently typed update monads are studied using a single operation symbol, typed as 
\[
\mathsf{act} : \Pi\, x \!:\! \State .\, \Updates \longrightarrow \State
\]
we present dependently typed update monads here using two operation symbols, analogously to how we have presented the global state and simply typed update monads in the previous examples.  We omit the details of the equivalence of these presentations and instead refer the reader to~\cite[Section~2.3]{Ahman:UpdateMonads} where the relationship between the corresponding one and two operation presentations is discussed for the simply typed case---the equivalence for the dependently typed case is proved analogously. 

We note that in~\cite{Ahman:UpdateMonads} the two operation presentation was considered only for simply typed update monads because it followed naturally from the analysis of simply typed update monads as compatible compositions of reader and writer monads. 
For the dependently typed generalisation of update monads, it is currently not known whether it is possible to build them naturally as a composition of two or more ordinary monads. 
In particular, we only know how to build dependently typed update monads from reader and writer monad like relative monads~\cite{Altenkirch:RelMon2}, as discussed in~\cite[Section~3]{Ahman:UpdateMonads}.

\subsection{Fibred effect theories}

Next, again following~\cite{Plotkin:HandlingEffects}, we describe the computational behaviour of the  effects specified by a fibred effect signature using equations between \emph{effect terms}. 

\index{variable!effect --}
\index{ w@$w, \ldots$ (effect variables)}
Specifically, assuming a countably infinite set of \emph{effect variables} that is disjoint from the sets of value and computation variables, and ranged over by $w, \ldots$, the effect terms $T$ are built from these effect variables, algebraic operations corresponding to the operation symbols in the given fibred effect signature, and elimination forms for pure value types, as defined in Definition~\ref{def:fibeffectterms}. 
Intuitively, these effect terms denote the computation trees one can build from the operation symbols in the given signature.

\begin{definition}
\label{def:fibeffectterms}
\index{term!effect --}
\index{ T@$T,\ldots$ (effect terms)}
The \emph{effect terms} derivable from a fibred effect signature $\mathcal{S}_{\text{eff}}$ are given by the following grammar:
\[
\begin{array}{r c l @{\qquad\qquad\qquad\quad}l}
T & ::= & w\, (V) &
\\
& \vertbar & \sigalgop_V(y .\, T) & (\sigalgop : (x \!:\! I) \longrightarrow O \in \mathcal{S}_{\text{eff}})
\\
& \vertbar & \pmatchsf V {(x_1 \!:\! A_1, x_2 \!:\! A_2)} {} T
\\
& \vertbar & \multicolumn{2}{l}{\mathsf{case~} V \mathsf{~of}_{} \mathsf{~} ({\seminl {\!} {\!\!(x_1 \!:\! A_1)} \mapsto T_1}, {\seminr {\!} {\!\!(x_2 \!:\! A_2)} \mapsto T_2})}
\end{array}
\]
where all value types and terms are assumed to be pure; and where
\begin{itemize}
\item in $\sigalgop_V(y .\, T)$, the value variable $y$ is bound in $T$;
\item in $\pmatchsf V {(x_1 \!:\! A_1, x_2 \!:\! A_2)} {} T$, the value variable $x_1$ is bound in $A_2$ and $T$, and the value variable $x_2$ is bound in $T$; and
\item in $\casesf V {} {\seminl {\!} {\!\!(x_1 \!:\! A_1)} \mapsto T_1} {\seminr {\!} {\!\!(x_2 \!:\! A_2)} \mapsto T_2}$, the value variable $x_1$ is bound in $T_1$ and the value variable $x_2$ is bound in $T_2$.
\end{itemize}
\end{definition}

\index{ FEV@$F\hspace{-0.05cm}EV(T)$ (set of free effect variables of $T$)}
We write $F\!EV(T)$ for the set of \emph{free effect variables} of the effect term $T$.

It is worth noting that effect variables can appear only in applied form. This is so because in this CPS-style calculus of effect terms, effect variables denote continuations that have to be applied to values before they can be used in computation. See the rules given in Definition~\ref{def:wellformedeffectterms} for details about how these value terms have to be typed.

Furthermore, observe that the definition of effect terms does not include elimination forms for neither propositional equality nor the type of natural numbers. On the one hand, as the former is only useful in  terms that get assigned dependent types, we do not need it for effect terms which will not be assigned types at all. On the other hand, while the latter would allow us to build compound computation trees using primitive recursion, we are unaware of interesting computational effects whose specification in terms of operations and equations would require the use of primitive recursion, even in the simply typed setting. However, observe that one is still free to use either of these elimination forms in the pure value terms  that appear in effect terms. 

We proceed by defining well-formed effect terms using the judgement $\lj {\Gamma \vertbar \Delta} T$, where $\Gamma$ is a pure value context and $\Delta$ is an effect context.

\begin{definition}
\index{context!value --!pure --}
An eMLTT value context $\Gamma$ is said to be \emph{pure} if $A_i$ is a pure value type for every $x_i \!:\! A_i$ in $\Gamma$.
\end{definition}

\begin{definition}
\index{context!effect --}
\index{ D@$\Delta$ (effect context)}
An \emph{effect context} $\Delta$ is a finite list $w_1 \!:\! A_1, \ldots, w_n \!:\! A_n$ of pairs of effect variables $w_{i}$ and pure value types $A_{i}$, such that all the effect variables $w_{i}$ are distinct. 
\end{definition}

\begin{definition}
\label{def:wellformedeffectcontext}
\index{context!effect --!well-formed --}
An effect context $\Delta$ is said to be \emph{well-formed} in a pure value context $\Gamma$, written $\lj \Gamma \Delta$, if $\vdash \Gamma$ and if we have $\lj \Gamma {A_i}$, for every $w_{i} \!:\! A_{i}$ in $\Delta$.
\end{definition}

\begin{definition}
\label{def:wellformedeffectterms}
\index{term!effect --!well formed --}
\emph{Well-formed} effect terms are given by the following rules:

\vspace{0.15cm}

\[
\mkrule
{\lj {\Gamma \vertbar \Delta_1, w \!:\! A, \Delta_2} {w\,(V)}}
{\lj \Gamma {\Delta_1, w \!:\! A, \Delta_2} \quad \vj \Gamma V A}
\]

\vspace{0.05cm}

\[
\mkrulelabel
{\lj {\Gamma \vertbar \Delta} {\sigalgop_V(y .\, T)}}
{
\lj \Gamma \Delta 
\quad 
\vj \Gamma V I 
\quad 
\lj {\Gamma, y \!:\! O[V/x] \vertbar \Delta} {T}
}
{(\sigalgop : (x \!:\! I) \longrightarrow O \in \mathcal{S}_{\text{eff}})}
\]

\vspace{0.05cm}

\[
\mkrule
{\lj {\Gamma \vertbar \Delta} {\pmatchsf V {(x_1 \!:\! A_1, x_2 \!:\! A_2)} {} T}}
{
\lj \Gamma \Delta 
\quad
\vj \Gamma V \Sigma\, x_1 \!:\! A_1 .\, A_2
\quad
\lj {\Gamma, x_1 \!:\! A_1, x_2 \!:\! A_2 \vertbar \Delta} T
}
\]

\vspace{0.05cm}

\[
\mkrule
{\lj {\Gamma \vertbar \Delta} {\mathsf{case~} V \mathsf{~of}_{} \mathsf{~} ({\seminl {\!} {\!\!(x_1 \!:\! A_1)} \mapsto T_1}, {\seminr {\!} {\!\!(x_2 \!:\! A_2)} \mapsto T_2})}}
{
\lj \Gamma \Delta 
\quad
\vj \Gamma V A_1 + A_2
\quad
\lj {\Gamma, x_1 \!:\! A_1 \vertbar \Delta} T_1
\quad
\lj {\Gamma, x_2 \!:\! A_2 \vertbar \Delta} T_2
}
\]
\end{definition}

Next, we prove two meta-theoretical results about effect terms that are analogous to Propositions~\ref{prop:freevariablesofwellformedexpressions} and~\ref{prop:wellformedcomponentsofjudgements} which we established for eMLTT in Chapter~\ref{chap:syntax}.

\begin{proposition}
\label{prop:freevariablesofeffectterms}
Given a well-formed effect term $\lj {\Gamma \vertbar \Delta} T$, then 
\[
FVV(T) \subseteq V\!ars(\Gamma)
\qquad
F\!EV(T) \subseteq V\!ars(\Delta)
\]
\end{proposition}

\begin{proof}
We prove this proposition by induction on the derivation of $\lj {\Gamma \vertbar \Delta} T$. We use Proposition~\ref{prop:freevariablesofwellformedexpressions} to derive inclusions $FVV(A) \subseteq V\!ars(\Gamma)$ and $FVV(V) \subseteq V\!ars(\Gamma)$ for well-formed pure value types $\lj \Gamma A$ and well-typed pure value terms $\vj \Gamma V A$.
\end{proof}

\begin{proposition}
\label{prop:wellformedfibredeffecttermhaswellformedcontext}
Given a well-formed effect term $\lj {\Gamma \vertbar \Delta} T$, then $\lj \Gamma \Delta$.
\end{proposition}

\begin{proof}
By induction on the derivation of $\lj {\Gamma \vertbar \Delta} T$.
\end{proof}

Further, when we combine Proposition~\ref{prop:wellformedfibredeffecttermhaswellformedcontext} with Definition~\ref{def:wellformedeffectcontext}, we get the following corollary.

\begin{corollary}
Given a well-formed effect term $\lj {\Gamma \vertbar \Delta} T$, then $\vdash \Gamma$.
\end{corollary}

We are now ready to define the notion of fibred effect theory so as to specify both the side-effect causing dependently typed effects and their computational behaviour.

\begin{definition}
\index{theory!fibred effect --}
\index{ T@$\mathcal{T}_{\text{eff}}$ (fibred effect theory)}
\index{ E@$\mathcal{E}_{\text{eff}}$ (set of equations of a fibred effect theory)}
\index{ S@$(\mathcal{S}_{\text{eff}},\mathcal{E}_{\text{eff}})$ (fibred effect theory)}
A \emph{fibred effect theory} $\mathcal{T}_{\text{eff}}$ is given by a fibred effect signature $\mathcal{S}_{\text{eff}}$ and a finite set $\mathcal{E}_{\text{eff}}$ of equations $\ljeq {\Gamma \vertbar \Delta} {T_1} {T_2}$, where $\lj {\Gamma \vertbar \Delta} {T_1}$ and $\lj {\Gamma \vertbar \Delta} {T_2}$ are two well-formed  effect terms derived from $\mathcal{S}_{\text{eff}}$. 
\end{definition}

We conclude this section by revisiting the examples of computational effects we discussed earlier and equip the corresponding signatures with equations, where appropriate. We follow~\cite{Plotkin:HandlingEffects} for the fibred effect signatures given in Examples~\ref{ex:fibsigofexceptions}--\ref{ex:fibsigofIO}, and the joint paper with Uustalu~\cite{Ahman:UpdateMonads} for the equational presentation of (dependently typed) update monads. 

To improve the readability of our examples, we omit the value argument $V$ in the effect term $\sigalgop_V(y .\, T)$ when the input type of $\sigalgop$ is $1$. For the same reason, we also omit the variable binding in the effect term $\sigalgop_V(y .\, T)$ when the output type of $\sigalgop$ is $1$.

\begin{example}[Exceptions]
\label{ex:fibtheoryofexceptions}
\index{theory!fibred effect --!-- of exceptions}
The fibred effect theory $\mathcal{T}_{\text{EXC}}$ of exceptions is given by the signature $\mathcal{S}_{\text{EXC}}$ and no equations.
\end{example}

\begin{example}[Binary nondeterminism]
\label{ex:fibtheoryofnondeterminism}
\index{theory!fibred effect --!-- of binary nondeterminism}
The fibred effect theory $\mathcal{T}_{\text{ND}}$ of binary nondeterminism is given by the signature $\mathcal{S}_{\text{ND}}$ and the following three equations:
\[
\ljeq {\diamond \vertbar w \!:\! 1} {\mathsf{choose}(x .\, w\, (\star))} {w\, (\star)}
\vspace{0.15cm}
\]

\[
\begin{array}{c@{~} c@{~} l}
{\diamond \vertbar w_1 \!:\! 1, w_2 \!:\! 1} & \vdash & {\mathsf{choose}(x .\, \mathsf{case~} x \mathsf{~of}_{} \mathsf{~} ({\seminl {\!} {\!\!(x_1 \!:\! 1)} \mapsto w_1\, (\star)}, {\seminr {\!} {\!\!(x_2 \!:\! 1)} \mapsto w_2\, (\star)}))}
\\[-0.5mm]
& = & {\mathsf{choose}(x .\, \mathsf{case~} x \mathsf{~of}_{} \mathsf{~} ({\seminl {\!} {\!\!(x_1 \!:\! 1)} \mapsto w_2\, (\star)}, {\seminr {\!} {\!\!(x_2 \!:\! 1)} \mapsto w_1\, (\star)}))}
\end{array}
\vspace{0.15cm}
\]

\[
\begin{array}{c}
\hspace{-10cm}
{\diamond \vertbar w_1 \!:\! 1, w_2 \!:\! 1, w_3 \!:\! 1} \vdash 
\\[-1mm]
\hspace{0.2cm}
\mathsf{choose}\big(x .\, \mathsf{case~} x \mathsf{~of}_{} \mathsf{~} \big(\seminl {\!} {\!\!(x_1 \!:\! 1)} \mapsto 
\mathsf{choose}(x'\!  .\, \mathsf{case~} x' \mathsf{~of}_{} \mathsf{~} ({\seminl {\!} {\!\!(x_3 \!:\! 1)} \mapsto w_1\, (\star)}, 
\\[-0.5mm]
\hspace{9.7cm}
{\seminr {\!} {\!\!(x_4 \!:\! 1)} \mapsto w_2\, (\star)})), 
\\[-1.5mm]
\hspace{-1.75cm}
{\seminr {\!} {\!\!(x_2 \!:\! 1)} \mapsto w_3\, (\star)}\big)\big) 
\\[1mm]
\hspace{-5.95cm}
= \mathsf{choose}\big(x  .\, \mathsf{case~} x \mathsf{~of}_{} \mathsf{~} \big(\seminl {\!} {\!\!(x_1 \!:\! 1)} \mapsto w_1\, (\star), 
\\[-0.5mm]
\hspace{3.7cm}
\seminr {\!} {\!\!(x_2 \!:\! 1)} \mapsto \mathsf{choose}(x'\! .\, \mathsf{case~} x' \mathsf{~of}_{} \mathsf{~} ({\seminl {\!} {\!\!(x_3 \!:\! 1)} \mapsto w_2\, (\star)}, 
\\[-0.5mm]
\hspace{10.05cm}
{\seminr {\!} {\!\!(x_4 \!:\! 1)} \mapsto w_3\, (\star)}))\big)\big)
\end{array}
\]

The idea is that nondeterministic choices are not observable if the continuation does not depend on the choice (1st equation); the choices are fair (2nd equation); and different nondeterministic choices are independent of each other (3rd equation).
\end{example}

\pagebreak

\begin{example}[Global state]
\label{ex:fibtheoryofglobalstate}
\index{theory!fibred effect --!-- of global state}
The fibred effect theory $\mathcal{T}_{\text{GS}}$ of global state is given by the signature $\mathcal{S}_{\text{GS}}$ and the following three equations:
\[
{\diamond \vertbar w \!:\! 1} \vdash \mathsf{get}(x .\, \mathsf{put}_x(w\, (\star))) = w\, (\star)
\]

\[
{x \!:\! \State \vertbar w \!:\! \State} \vdash \mathsf{put}_x(\mathsf{get}(y .\, w\, (y))) = \mathsf{put}_x(w\, (x))
\vspace{0.15cm}
\]

\[
{x \!:\! \State, y \!:\! \State \vertbar w \!:\! 1} \vdash \mathsf{put}_x(\mathsf{put}_y(w\, (\star))) = \mathsf{put}_y(w\, (\star))
\]

These equations describe the expected behaviour $\mathsf{get}$ and $\mathsf{put}$: trivial store changes are not observable (1st equation); $\mathtt{get}$ returns the most recent value the store has been set to (2nd equation); and $\mathtt{put}$ overwrites the contents of the store (3rd equation).
\end{example}

\begin{example}[Global state with locations]
\label{ex:fibtheoryofglobalstatewithlocations}
\index{theory!fibred effect --!-- of global state with locations}
The fibred effect theory $\mathcal{T}_{\text{GSL}}$ of global state with locations is given by the signature $\mathcal{S}_{\text{GSL}}$ and the following five equations:
\[
{x \!:\! \Location \vertbar w \!:\! 1} \vdash \mathsf{get}_x(y .\, \mathsf{put}_{\langle x , y \rangle}(w\, (\star))) = w\, (\star)
\vspace{0.15cm}
\]

\[
{x \!:\! \Location, y \!:\! \Value \vertbar w \!:\! \Value} \vdash \mathsf{put}_{\langle x , y \rangle}(\mathsf{get}_x(y' .\, w\, (y'))) = \mathsf{put}_{\langle x , y \rangle}(w\, (y))
\vspace{0.15cm}
\]

\[
{x \!:\! \Location, y_1 \!:\! \Value, y_2 \!:\! \Value \vertbar w \!:\! 1} \vdash \mathsf{put}_{\langle x , y_1 \rangle}(\mathsf{put}_{\langle x , y_2 \rangle}(w\, (\star))) = \mathsf{put}_{\langle x , y_2 \rangle}(w\, (\star))
\vspace{0.15cm}
\]

\[
\begin{array}{c}
\hspace{-6.75cm}
{x_1 \!:\! \Location, x_2 \!:\! \Location \vertbar w \!:\! \Value[x_1/x] \times \Value[x_2/x]} \vdash 
\\[-0.5mm]
\hspace{0.75cm}
\mathsf{get}_{x_1}(y_1.\, \mathsf{get}_{x_2}(y_2 .\, w\, (\langle y_1 , y_2 \rangle))) = 
\mathsf{get}_{x_2}(y_2 .\, \mathsf{get}_{x_1}(y_1 .\, w\, (\langle y_1 , y_2 \rangle)))
\qquad (x_1 \neq x_2)
\end{array}
\vspace{0.15cm}
\]

\[
\begin{array}{c}
\hspace{-5.45cm}
{x_1 \!:\! \Location, x_2 \!:\! \Location, y_1 \!:\! \Value[x_1/x], y_2 \!:\! \Value[x_2/x] \vertbar w \!:\! 1} \vdash
\\[-0.5mm]
\hspace{2.25cm}
\mathsf{put}_{\langle x_1, y_1 \rangle}(\mathsf{put}_{\langle x_2, y_2 \rangle}(w\, (\star))) = \mathsf{put}_{\langle x_2, y_2 \rangle}(\mathsf{put}_{\langle x_1, y_1 \rangle}(w\, (\star)))
\qquad (x_1 \neq x_2)
\end{array}
\vspace{0.15cm}
\]

Observe that the first three equations are $\Location$-indexed variants of the equations from Example~\ref{ex:fibtheoryofglobalstate}. The last two equations describe that $\mathsf{get}$ and $\mathsf{put}$ effects for different locations commute with each other. To this end, the last two equations both come with a side-condition requiring the locations denoted by $x_1$ and $x_2$ to be different. 

\end{example}

Similarly to~\cite{Plotkin:HandlingEffects}, this notation for side-conditions is an informal short-hand for a formal presentation based on using case analysis. Specifically, we assume a decidable (for simplicity, boolean-valued) equality on locations, given by a closed well-typed pure value term  $ {\mathtt{eq}} : {\Location \times \Location \to 1 + 1}$, and then write the right-hand sides of these equations using case analysis. For example, the last equation is formally written as
\[
\begin{array}{c}
\hspace{-5cm}
{x_1 \!:\! \Location, x_2 \!:\! \Location, y_1 \!:\! \Value[x_1/x], y_2 \!:\! \Value[x_2/x] \vertbar w \!:\! 1} \vdash
\\[0.5mm]
\hspace{-5.35cm}
\mathsf{put}_{\langle x_1, y_1 \rangle}(\mathsf{put}_{\langle x_2, y_2 \rangle}(w\, (\star))) 
\\[1mm]
\hspace{0.1cm}
= \mathsf{case~} (\mathtt{eq}~ \langle x_1, x_2 \rangle) \mathsf{~of}_{} \mathsf{~} ({\seminl {\!} {\!\!(x'_1 \!:\! 1)} \mapsto \mathsf{put}_{\langle x_1, y_1 \rangle}(\mathsf{put}_{\langle x_2, y_2 \rangle}(w\, (\star)))}, 
\\[-0.5mm]
\hspace{4.2cm}
{\seminr {\!} {\!\!(x'_2 \!:\! 1)} \mapsto \mathsf{put}_{\langle x_2, y_2 \rangle}(\mathsf{put}_{\langle x_1, y_1 \rangle}(w\, (\star)))})
\end{array}
\]

\begin{example}[Input/output]
\label{ex:fibtheoryofIO}
\index{theory!fibred effect --!-- of input/output}
The fibred effect theory $\mathcal{T}_{\text{I/O}}$ of input/output is given by the signature $\mathcal{S}_{\text{I/O}}$ and no equations.
\end{example}

\begin{example}[Update monads]
\label{ex:fibtheoryofupdatemonads}
\index{theory!fibred effect --!-- of an update monad}
The fibred effect theory $\mathcal{T}_{\text{UPD}}$ of an update monad is given by the signature $\mathcal{S}_{\text{UPD}}$ and the following three equations:
\[
{\diamond \vertbar w \!:\! 1} \vdash \mathsf{lookup}(x .\, \mathsf{update}_{\mathsf{o}}(w\, (\star))) = w\, (\star)
\vspace{0.15cm}
\]

\[
\begin{array}{c@{~} c@{~} l}
{x \!:\! \Updates \vertbar w \!:\! \State \times \State} & \vdash & \mathsf{lookup}(y .\, \mathsf{update}_x(\mathsf{lookup}(y' .\, w\, (\langle y , y' \rangle))) 
\\[-0.5mm]
& = & \mathsf{lookup}(y .\, \mathsf{update}_x(w\, (\langle y , y \downarrow x \rangle))))
\end{array}
\vspace{0.15cm}
\]

\[
{x \!:\! \Updates, y \!:\! \Updates \vertbar w \!:\! 1} \vdash \mathsf{update}_x(\mathsf{update}_y(w\, (\star))) = \mathsf{update}_{x \, \oplus \, y}(w\, (\star))
\]

These equations are similar to those given for global state in Example~\ref{ex:fibtheoryofglobalstate}, but instead of describing only overwriting-based store manipulations, they describe store manipulations using the action $\downarrow$ of the monoid $(\Updates, \mathsf{o}, \oplus)$ on store values. Further, observe how $\oplus$ is used to combine consecutive updates in the third equation.
\end{example}

In~\cite{Ahman:UpdateMonads}, we also consider other,  equivalent sets of equations for the algebras of simply typed update monads, based on the different ways they can constructed from other monads, e.g., as a compatible composition of reader and writer monads.

\begin{example}[Dependently typed update monads]
\label{ex:fibtheoryofdeptypedupdatemonads}
\index{theory!fibred effect --!-- of a dependently typed update monad}
The fibred effect theory $\mathcal{T}_{\text{DUPD}}$ of a dependently typed update monad is given by the signature $\mathcal{S}_{\text{DUPD}}$ and the following three equations:
\[
{\diamond \vertbar w \!:\! 1} \vdash \mathsf{lookup}(x .\, \mathsf{update}_{\lambda\, y : \State .\, \mathsf{o}\, y}(w\, (\star))) = w\, (\star)
\vspace{0.15cm}
\]

\[
\begin{array}{c@{~} c@{~} l}
{x \!:\! (\Pi\, x' \!:\! \State .\, \Updates[x'/x]) \vertbar w \!:\! \State \times \State} & \vdash & \mathsf{lookup}(y .\, \mathsf{update}_{x}(\mathsf{lookup}(y' .\, w\, (\langle y , y' \rangle)))  
\\[-0.5mm]
& = & \mathsf{lookup}(y .\, \mathsf{update}_x(w\, (\langle y , y \downarrow (x\,\, y) \rangle))))
\end{array}
\vspace{0.15cm}
\]

\[
\begin{array}{c}
\hspace{-4.8cm}
{x \!:\! (\Pi\, x' \!:\! \State .\, \Updates[x'/x]), y \!:\! (\Pi\, y' \!:\! \State .\, \Updates[y'/x]) \vertbar w \!:\! 1} \vdash 
\\[-0.5mm]
\hspace{2.5cm}
\mathsf{update}_x(\mathsf{update}_y(w\, (\star))) = \mathsf{update}_{\lambda\, x''.\, (x\,\, x'') \, \oplus_{x''} \, (y\,\, (x'' \,\downarrow\, (x\,\, x'')))}(w\, (\star))
\end{array}
\]

These three equations are analogous to the equations given for simply typed update monads in Example~\ref{ex:fibtheoryofupdatemonads}, except for  
the 1st and 3rd equation now having to account for the input type of $\mathsf{update}$ being $\Pi\, x \!:\! \State.\, \Updates$ instead of simply $\Updates$. 
\end{example}

\section{Extending eMLTT with fibred algebraic effects} 
\label{sect:fibalgeffectsineMLTT}

In this section we show how to extend eMLTT with fibred algebraic effects given by a fibred effect theory $\mathcal{T}_{\text{eff}} = (\mathcal{S}_{\text{eff}},\mathcal{E}_{\text{eff}})$. We call the resulting language eMLTT$_{\mathcal{T}_{\text{eff}}}$. 

\begin{definition}
\label{def:extensionofeMLTTsyntaxwithfibalgeffects}
\index{extension of eMLTT!-- with fibred algebraic effects}
\index{algebraic operation}
The syntax of eMLTT$_{\mathcal{T}_{\text{eff}}}$ is given by extending eMLTT's computation terms with \emph{algebraic operations}:
\[
\begin{array}{r c l @{\qquad\qquad}l}
M & ::= & \ldots \,\,\,\vertbar\,\,\, \algop^{\ul{C}}_V(y .\, M)
\end{array}
\]
for all operation symbols $\sigalgop : (x \!:\! I) \longrightarrow O$ in $\mathcal{S}_{\text{eff}}$ and all  computation types $\ul{C}$.
\end{definition}

In $\algop^{\ul{C}}_V(y .\, M)$, the value variable $y$ is bound in $M$.
Similarly to effect terms, we omit the variable binding in $\algop^{\ul{C}}_V(y .\, M)$ for better readability when the output type of $\sigalgop$ is $1$. Analogously, 
we also omit the value argument $V$ when the input type of $\sigalgop$ is $1$. 

The different kinds of substitution we defined for eMLTT extend straightforwardly to eMLTT$_{\mathcal{T}_{\text{eff}}}$: we extend the (simultaneous) substitution of value terms with
\[
(\algop^{\ul{C}}_V(y .\, M))[\overrightarrow{W}/\overrightarrow{x}] \defeq \algop^{\ul{C}[\overrightarrow{W}/\overrightarrow{x}]}_{V[\overrightarrow{W}/\overrightarrow{x}]}(y .\, M[\overrightarrow{W}/\overrightarrow{x}])
\]
and keep the substitution of computation and homomorphism terms for computation variables unchanged. 
The properties of substitution we established for eMLTT in Sections~\ref{sect:syntax} and~\ref{sect:completeness} also extend straightforwardly to eMLTT$_{\mathcal{T}_{\text{eff}}}$---the  proof principles remain unchanged, and the cases for the algebraic operations are treated analogously to other computation terms that involve variable bindings and type annotations.

Unless stated otherwise, the types and terms we use in the rest of this chapter are those of eMLTT$_{\mathcal{T}_{\text{eff}}}$.
This also includes the definitions of pure value types and pure value terms appearing in effect terms because every pure eMLTT value type (resp. term) can be trivially considered as a pure eMLTT$_{\mathcal{T}_{\text{eff}}}$ value type (resp. term).

Next, we extend the typing rules and equational theory of eMLTT with fibred algebraic effects. However, 
before doing so, we first need to define a translation of effect terms into eMLTT$_{\mathcal{T}_{\text{eff}}}$. While it might be more natural to translate effect terms into computation terms, we have decided to translate them into value terms instead. We do so to avoid having to define another similar translation in Chapter~\ref{chap:handlers}. We note that this choice does not restrict the definitional equations between computation terms that one can derive from the equations given in $\mathcal{E}_{\text{eff}}$, as illustrated later in this section.

In order to simplify the presentation of eMLTT$_{\text{eff}}$, we assume that
\[
\Gamma = x_1 \!:\! A_1, \ldots, x_n \!:\! A_n
\qquad
\Delta = w_1 \!:\! A'_1, \ldots, w_m \!:\! A'_m
\] 
throughout this section.
In order to further improve the readability, we use vector notation for sets of value terms, i.e., we write $\overrightarrow{V_i}$ for a set of value terms $\{V_1, \ldots, V_n\}$. 
\index{ V@$\overrightarrow{V_i}$ (shorthand for $\{V_1, \ldots, V_n\}$)}

We also note that we only translate well-formed effect terms $\lj {\Gamma \vertbar \Delta} {T}$ because it makes it easier to account for the substitution of value terms for effect variables in the definition of the translation. In particular, the later results refer to value terms substituted for all effect variables in $\Delta$, not just for the free variables appearing in $T$.

\begin{definition}
\label{def:transofefftermstovalueterms}
\index{translation of effect terms into value terms}
\index{ T@$\efftrans T {A; \overrightarrow{V_i}; \overrightarrow{V'_{j}}; \overrightarrow{W_{\sigalgop}}}$ (translation of effect terms into value terms)}
Given a well-formed effect term $\lj {\Gamma \vertbar \Delta} T$ derived from $\mathcal{S}_{\text{eff}}$, a value type $A$, value terms $V_{i}$ (for all $x_i \!:\! A_i$ in $\Gamma$), value terms $V'_{\!j}$ (for all $w_{\!j} \!:\! A'_{\!j}$ in $\Delta$), and value terms $W_{\sigalgop}$ (for all $\sigalgop : (x \!:\! I) \longrightarrow O$ in $\mathcal{S}_{\text{eff}}$), the \emph{translation} of the effect term $T$ into a value term $\efftrans T {A; \overrightarrow{V_i}; \overrightarrow{V'_{\!j}}; \overrightarrow{W_{\sigalgop}}}$ is defined by recursion on the structure of $T$ as follows:
\[
\begin{array}{l c l}
\efftrans {w_{\!j}\, (V)} {} & \defeq & V'_{\!j}\,\, (V[\overrightarrow{V_i}/\overrightarrow{x_i}])
\\[2mm]
\efftrans {\sigalgop_V(y .\, T)} {} & \defeq & W_{\sigalgop}\,\, \langle V[\overrightarrow{V_i}/\overrightarrow{x_i}] , \lambda\, y \!:\! O[V[\overrightarrow{V_i}/\overrightarrow{x_i}]/x] .\, \efftrans {T} {}\rangle
\\[2mm]
\multicolumn{3}{l}{\efftrans {\pmatchsf V {(y_1 \!:\! B_1, y_2 \!:\! B_2)} {} T} {}}
\\
\multicolumn{3}{l}{\hspace{3cm} \defeq \,\,\,\,\,\, \pmatch {V[\overrightarrow{V_i}/\overrightarrow{x_i}]} {(y_1 \!:\! B_1[\overrightarrow{V_i}/\overrightarrow{x_i}], y_2 \!:\! B_2[\overrightarrow{V_i}/\overrightarrow{x_i}])} {y .\, A} {\efftrans T {}} }
\\[2mm]
\multicolumn{3}{l}{\efftrans {\mathsf{case~} V \mathsf{~of}_{} \mathsf{~} ({\seminl {\!} {\!\!(y_1 \!:\! B_1)} \mapsto T_1}, {\seminr {\!} {\!\!(y_2 \!:\! B_2)} \mapsto T_2})} {}}
\\
\multicolumn{3}{l}{\hspace{4cm} \defeq \,\,\,\,\,\, \mathtt{case~} V[\overrightarrow{V_i}/\overrightarrow{x_i}] \mathtt{~of}_{y.\,A} \mathtt{~} ({\inl {\!} {\!\!(y_1 \!:\! B_1[\overrightarrow{V_i}/\overrightarrow{x_i}])} \mapsto \efftrans {T_1} {}},} \\[-0.5mm]
\multicolumn{3}{l}{\hfill {\inr {\!} {\!\!(y_2 \!:\! B_2[\overrightarrow{V_i}/\overrightarrow{x_i}])} \mapsto \efftrans {T_2} {}})}
\end{array}
\]
where in the last two cases the value variable $y$ is chosen fresh. While in the above we omit the subscripts on the translation for better readability, it is important to note that in the cases where the given effect term $T$ involves variable bindings, the set of value terms $\overrightarrow{V_i}$ is extended with the corresponding value variables in the right-hand side. For example, the right-hand side of the algebraic operations case is formally written as
\[
W_{\sigalgop}\,\, \langle V[\overrightarrow{V_i}/\overrightarrow{x_i}] , \lambda\, y \!:\! O[V[\overrightarrow{V_i}/\overrightarrow{x_i}]/x] .\, \efftrans {T} {A; \overrightarrow{V_i},\, y; \overrightarrow{V'_{\!j}}; \overrightarrow{W_{\sigalgop}}}\rangle
\]
\end{definition}

Later, in Proposition~\ref{prop:welltypednessoftranslatingeffectterms}, we show that under appropriate well-formedness assumptions about the value type $A$ and the value terms $V_i$, $V'_{\!j}$,  and $W_{\sigalgop}$, the translation of the effect term $T$ results in a well-typed value term $\efftrans T {A; \overrightarrow{V_i}; \overrightarrow{V'_{\!j}}; \overrightarrow{W_{\sigalgop}}}$ of type $A$.

Using this translation, we can now extend the typing rules and definitional equations of eMLTT with fibred algebraic effects.

\begin{definition}
\label{def:extensionofeMLTTwithfibalgeffects}
\index{well-formed syntax}
The \emph{well-formed syntax} of eMLTT$_{\mathcal{T}_{\text{eff}}}$
is given by extending the typing rules for eMLTT's well-typed  computation terms with
\vspace{0.2cm}
\[
\mkrulelabel
{\cj \Gamma {\algop^{\ul{C}}_V(y .\, M)} {\ul{C}}}
{
\vj \Gamma V I
\quad
\lj \Gamma \ul{C}
\quad
\cj {\Gamma, y \!:\! O[V/x]} {M} {\ul{C}}
}
{(\sigalgop : (x \!:\! I) \longrightarrow O \in \mathcal{S}_{\text{eff}})}
\]
and the equational theory of eMLTT with rules for
\begin{itemize}
\item congruence equations:
\[
\hspace{-0.5cm}
\mkrulelabel
{\ceq \Gamma {\algop^{\ul{C}}_V(y .\, M)} {{\algop^{\ul{D}}_W(y .\, N)}} {\ul{C}}}
{
\veq \Gamma V W I
\quad
\ljeq \Gamma {\ul{C}} {\ul{D}}
\quad
\ceq {\Gamma, y \!:\! O[V/x]} {M} {N} {\ul{C}}
}
{(\sigalgop : (x \!:\! I) \longrightarrow O \in \mathcal{S}_{\text{eff}})}
\]
\item general algebraicity\footnote{We are using the terminology of~\cite[Section~5.3]{Pretnar:Thesis}.} equation:
\index{algebraicity equation!general --}
\[
\mkrulelabel
{\ceq \Gamma {K[\algop^{\ul{C}}_V(y . M)/z]} {\algop^{\ul{D}}_V(y . K[M/z])} {\ul{D}}}
{
\vj \Gamma V {I} 
\quad 
\cj {\Gamma, y \!:\! O[V/x]} M {\ul{C}} 
\quad 
\hj \Gamma {z \!:\! \ul{C}} K {\ul{D}}}
{(\sigalgop : (x \!:\! I) \longrightarrow O \in \mathcal{S}_{\text{eff}})}
\]
\item equations of the given fibred effect theory:
\[
\mkrulelabel
{
\veq {\Gamma'} {\efftrans {T_1} {U\ul{C}; \overrightarrow{V_i}; \overrightarrow{V'_{\!j}}; \overrightarrow{W_{\sigalgop}}}} {\efftrans {T_2} {U\ul{C}; \overrightarrow{V_i}; \overrightarrow{V'_{\!j}}; \overrightarrow{W_{\sigalgop}}}} {U\ul{C}}
}
{
\begin{array}{c@{\qquad\quad} l}
\lj {\Gamma'} \ul{C}
\\[-0.5mm]
\vj {\Gamma'} {V_i} {A_i[V_1/x_1, \ldots, V_{i-1}/x_{i-1}]} & (1 \leq i \leq n)
\\[0.5mm]
\vj {\Gamma'} {V'_{\!j}} {A'_{\!j}[\overrightarrow{V_i}/\overrightarrow{x_i}] \to U\ul{C}} & (1 \leq j \leq m)
\end{array}
}
{(\ljeq {\Gamma \vertbar \Delta} {T_1\!} {\!T_2} \in \!\mathcal{E}_{\text{eff}})}
\]
with the well-typed value terms $\vj {\Gamma'} {W_{\sigalgop}} {(\Sigma\, x \!:\! I .\, O \to U\ul{C}) \to U\ul{C}}$ given by
\[
\begin{array}{c}
\hspace{-8cm}
W_{\sigalgop} \defeq \lambda x' \!:\! (\Sigma\, x \!:\! I.\, O \to U\ul{C}).\, 
\\
\hspace{1.75cm}
\pmatch {x'} {(x \!:\! I, y \!:\! O \to U\ul{C})} {x''\!.\, U\ul{C}} {\thunk (\algop^{\ul{C}}_{x}(y'.\, \force {\ul{C}} {(y\,\, y')}))} 
\end{array}
\]
for each operation symbol $\sigalgop : (x \!:\! I) \longrightarrow O$ in $\mathcal{S}_{\text{eff}}$, with $x''$  chosen fresh.
\end{itemize}
\end{definition}

Finally, is worth noting that we include the equations of the given fibred effect theory in eMLTT$_{\mathcal{T}_{\text{eff}}}$ as definitional equations between value terms, rather than as equations between computation terms. This is analogous to how we have defined the translation of effect terms, namely,  into value terms rather than computation terms.
Nevertheless, the expected equations between computation terms are still derivable, using thunking and forcing. For example, we can derive the following definitional equation:
\[
\ceq {\Gamma} {\mathtt{get}^{\ul{C}}(x .\, \mathtt{put}^{\ul{C}}_x(M))} {M} {\ul{C}}
\]
from the equation 
\[
{\diamond \vertbar w \!:\! 1} \vdash \mathsf{get}(x .\, \mathsf{put}_x(w\, (\star))) = w\, (\star)
\]
given in the global state theory $\mathcal{T}_{\text{GS}}$ as follows:

\begin{fleqn}[0.3cm]
\begin{align*}
\Gamma \,\vdash\,\, & \mathtt{get}^{\ul{C}}(x .\, \mathtt{put}^{\ul{C}}_x(M))
\\
=\,\, & \force {\ul{C}} (\thunk (\mathtt{get}^{\ul{C}}(x .\, \force {\ul{C}}(\thunk ( \mathtt{put}^{\ul{C}}_x(\force {\ul{C}} (\thunk M)))))))
\\
=\,\, & \force {\ul{C}} \big(\big(\lambda x' \!:\! 1 \times (\State \to U\ul{C}) .\, 
\\[-1mm]
& \hspace{0.5cm} \pmatch {x'} {(x'_1 \!:\! 1, x'_2 \!:\! \State \to U\ul{C})} {} {\thunk (\mathtt{get}^{\ul{C}}(x .\, \force {\ul{C}} (x'_2\,\, x)))\big)}\,  
\\[-1mm]
& \hspace{5.05cm} \langle \star , \lambda\, x \!:\! \State .\, \thunk (\mathtt{put}^{\ul{C}}_x(\force {\ul{C}} (\thunk M))) \rangle \big)
\\
=\,\, & \force {\ul{C}} \big(\big(\lambda x' \!:\! 1 \times (\State \to U\ul{C}) .\, 
\\[-1mm]
& \hspace{1cm} \pmatch {x'} {(x'_1 \!:\! 1, x'_2 \!:\! \State \to U\ul{C})} {} {\thunk (\mathtt{get}^{\ul{C}}(x .\, \force {\ul{C}} (x'_2\,\, x)))\big)}\,  
\\[-1mm]
& \hspace{2cm} \big\langle \star , \lambda\, x \!:\! \State .\, (\lambda x'' \!:\! \State \times (1 \to U\ul{C}) .\, 
\\[-1mm]
& \hspace{3cm} \pmatch {x''} {(x''_1 \!:\! \State, x''_2 \!:\! 1 \to U\ul{C})} {} {\\& \hspace{3.9cm} \thunk (\mathtt{put}^{\ul{C}}_{x''_1}(\force {\ul{C}} (x''_2\,\, \star)))})\, \langle x , \lambda\, x''' \!:\! 1 .\, \thunk M \rangle \big\rangle \big)
\\
=\,\, & \force {\ul{C}} \big(\big(\lambda x' \!:\! 1 \times (\State \to U\ul{C}) .\, 
\\[-1mm]
& \hspace{0.35cm} \pmatch {x'} {(x'_1 \!:\! 1, x'_2 \!:\! \State \to U\ul{C})} {} {\thunk (\mathtt{get}^{\ul{C}}(x .\, \force {\ul{C}} (x'_2\,\, x)))\big)}\,  
\\[-1mm]
& \hspace{0.8cm} \big\langle \star , \lambda\, x \!:\! \State .\, (\lambda x'' \!:\! \State \times (1 \to U\ul{C}) .\, 
\\[-1mm]
& \hspace{1.55cm} \pmatch {x''} {(x''_1 \!:\! \State, x''_2 \!:\! 1 \to U\ul{C})} {} {\\[-1mm]& \hspace{2.25cm} \thunk (\mathtt{put}^{\ul{C}}_{x''_1}(\force {\ul{C}} (x''_2\,\, \star)))}) \, \langle x , \lambda\, x''' \!:\! 1 .\, (\lambda\, y \!:\! 1 .\, \thunk M)\, \star \rangle \big\rangle \big)
\\
=\,\, & \force {\ul{C}} \big(\efftrans {\mathtt{get}(x .\, \mathtt{put}_x(w\, (\star)))} {U\ul{C}; \emptyset; \lambda\, y : 1. \thunk M; \overrightarrow{W_{\sigalgop}}} \big)
\\
=\,\, & \force {\ul{C}} \big(\efftrans {w\, (\star)} {U\ul{C}; \emptyset; \lambda\, y : 1. \thunk M; \overrightarrow{W_{\sigalgop}}} \big)
\\
=\,\, & \force {\ul{C}} ((\lambda\, y : 1.\, \thunk M)\, \star)
\\
=\,\, & \force {\ul{C}} (\thunk M)
\\
=\,\, & M : \ul{C}
\end{align*}
\end{fleqn}
where the well-typed value terms $W_{\mathsf{get}}$ and $W_{\mathsf{put}}$ are respectively given by
\[
\begin{array}{c}
\hspace{-8.5cm}
W_{\mathsf{get}} \defeq \lambda x' \!:\! 1 \times (\State \to U\ul{C}).\, 
\\[-1mm]
\hspace{2cm}
\pmatch {x'} {(x'_1 \!:\! 1, x'_2 \!:\! \State \to U\ul{C})} {} {\thunk (\mathsf{get}^{\ul{C}}(x.\, \force {\ul{C}} {(x'_2\,\, x)}))}
\\[2mm]
\hspace{-8.4cm}
W_{\mathsf{put}} \defeq \lambda x'' \!:\! \State \times (1 \to U\ul{C}).\, 
\\[-1mm]
\hspace{2.1cm}
\pmatch {x''} {(x''_1 \!:\! \State, x''_2 \!:\! 1 \to U\ul{C})} {} {\thunk (\mathsf{put}^{\ul{C}}_{x''_1}(\force {\ul{C}} {(x''_2\,\, \star)}))}
\end{array}
\]

\section{Meta-theory} 
\label{sect:emlttalgeffectsmetatheory}

In this section we show how to extend the meta-theory we established for eMLTT in Section~\ref{sect:metatheory} (and in the beginning of Section~\ref{sect:completeness}) to eMLTT$_{\mathcal{T}_{\text{eff}}}$. While some of these results extend straightforwardly from eMLTT to eMLTT$_{\mathcal{T}_{\text{eff}}}$ (either the proof remains the same or it can be easily adapted), others require little more work. In particular, as the definitional equations now involve the translation of effect terms, some of the results below need to be now proved in conjunction with corresponding results about the translation. We omit the propositions and theorems whose proofs extend straightforwardly to eMLTT$_{\mathcal{T}_{\text{eff}}}$ and only comment on those whose proofs are more involved.

\subsection*{Extending Proposition~\ref{prop:freevariablesofwellformedexpressions} to eMLTT$_{\!\mathcal{T}_{\text{eff}}}$}

We begin by recalling that in Proposition~\ref{prop:freevariablesofwellformedexpressions} we showed that the free value variables of well-formed eMLTT expressions and definitional equations are included in their respective value contexts. For example, given $\ceq \Gamma M N \ul{C}$, we showed that
\[
FVV(M) \subseteq V\!ars(\Gamma)
\qquad
FVV(N) \subseteq V\!ars(\Gamma)
\qquad
FVV(\ul{C}) \subseteq V\!ars(\Gamma)
\]

When extending Proposition~\ref{prop:freevariablesofwellformedexpressions} to eMLTT$_{\mathcal{T}_{\text{eff}}}$, we keep the basic proof principle the same: we prove $(a)$--$(j)$ for the different kinds of types, terms, and definitional equations simultaneously, by induction on the given derivations. 

The new cases for algebraic operations, and the corresponding congruence and general algebraicity equations are proved analogously to other computation terms and definitional equations that involve variable bindings and type annotations. However, in order to account for the case that corresponds to the third group of definitional equations given in Definition~\ref{def:extensionofeMLTTwithfibalgeffects}, we need to prove the eMLTT$_{\mathcal{T}_{\text{eff}}}$ version of Proposition~\ref{prop:freevariablesofwellformedexpressions} simultaneously with Proposition~\ref{prop:effecttermtranslationinclusionofvariables} below.

\begin{proposition}
\label{prop:effecttermtranslationinclusionofvariables}
Given a well-formed effect term $\lj {\Gamma \vertbar \Delta} T$ derived from $\mathcal{S}_{\text{eff}}$, a value type $A$, value terms $V_{i}$ (for all $x_i \!:\! A_i$ in $\Gamma$), value terms $V'_{\!j}$ (for all $w_{\!j} \!:\! A'_{\!j}$ in $\Delta$), and value terms $W_{\sigalgop}$ (for all $\sigalgop : (x \!:\! I) \longrightarrow O$ in $\mathcal{S}_{\text{eff}}$), then we have
\[
FVV(\efftrans T {A; \overrightarrow{V_i}; \overrightarrow{V'_{\!j}}; \overrightarrow{W_{\sigalgop}}}) 
\vspace{-0.15cm}
\]
\[
\subseteq
\vspace{-0.1cm}
\]
\[
FVV(A) \,\cup\!\! \bigcup_{V_i \in \overrightarrow{V_i}} \! FVV(V_i) \,\cup\!\!\! \bigcup_{V'_{\!j} \in \overrightarrow{V'_{\!j}}} \! FVV(V'_{\!j}) \,\cup\!\!\!\!\!\!\! \bigcup_{W_{\sigalgop} \in \overrightarrow{W_{\sigalgop}}} \!\!\!\!\!\! FVV(W_{\sigalgop}) 
\]
\end{proposition}

\begin{proof}
We prove this proposition by induction on the derivation of $\lj {\Gamma \vertbar \Delta} T$, using the simultaneously proved eMLTT$_{\mathcal{T}_{\text{eff}}}$ version of Proposition~\ref{prop:freevariablesofwellformedexpressions} to show inclusions for the sets of free value variables of the pure value types and pure value terms appearing in $T$. The proof also uses the eMLTT$_{\mathcal{T}_{\text{eff}}}$ version of Proposition~\ref{prop:freevariablesofsubsstitutionsimultaneous} that shows how the sets of free value variables are computed for expressions involving substitution. 

As a representative example, we consider the case of algebraic operations, for which we need to show that the following inclusion holds:
\[
FVV(W_{\sigalgop}\,\, \langle V[\overrightarrow{V_i}/\overrightarrow{x_i}] , \lambda\, y \!:\! O[V[\overrightarrow{V_i}/\overrightarrow{x_i}]/x] .\, \efftrans {T} {A; \overrightarrow{V_i},\, y; \overrightarrow{V'_{\!j}}; \overrightarrow{W_{\sigalgop}}}\rangle)
\vspace{-0.3cm}
\]
\[
\subseteq
\vspace{-0.1cm}
\]
\[
FVV(A) \,\cup\!\! \bigcup_{V_i \in \overrightarrow{V_i}} \! FVV(V_i) \,\cup\!\!\! \bigcup_{V'_{\!j} \in \overrightarrow{V'_{\!j}}} \! FVV(V'_{\!j}) \,\cup\!\!\!\!\!\!\! \bigcup_{W_{\sigalgop} \in \overrightarrow{W_{\sigalgop}}} \!\!\!\!\!\! FVV(W_{\sigalgop}) 
\]

First, according to how the set of free value variables of a value term is computed, we know that the following sequence of equations holds:
\[
\begin{array}{c}
FVV(W_{\sigalgop}\,\, \langle V[\overrightarrow{V_i}/\overrightarrow{x_i}] , \lambda\, y \!:\! O[V[\overrightarrow{V_i}/\overrightarrow{x_i}]/x] .\, \efftrans {T} {A; \overrightarrow{V_i} ,\, y; \overrightarrow{V'_{\!j}}; \overrightarrow{W_{\sigalgop}}}\rangle)
\\
=
\\[1mm]
FVV(W_{\sigalgop}) \,\cup\, FVV(V[\overrightarrow{V_i}/\overrightarrow{x_i}]) \,\cup\, FVV\big(\lambda\, y \!:\! O[V[\overrightarrow{V_i}/\overrightarrow{x_i}]/x] .\, \efftrans {T} {A; \overrightarrow{V_i},\, y; \overrightarrow{V'_{\!j}}; \overrightarrow{W_{\sigalgop}}}\big)
\\
=
\\[1mm]
\hspace{-7.225cm}
FVV(W_{\sigalgop}) \,\cup\, FVV(V[\overrightarrow{V_i}/\overrightarrow{x_i}]) \,\,\,\cup\, 
\\
\hspace{3.35cm}
FVV(O[V[\overrightarrow{V_i}/\overrightarrow{x_i}]/x]) \,\cup\, \big(FVV(\efftrans {T} {A; \overrightarrow{V_i} ,\, y; \overrightarrow{V'_{\!j}}; \overrightarrow{W_{\sigalgop}}}) - \ia y \big)
\end{array}
\]

Next, by using Proposition~\ref{prop:freevariablesofsubsstitutionsimultaneous} with  $V[\overrightarrow{V_i}/\overrightarrow{x_i}]$, we get that
\[
FVV(V[\overrightarrow{V_i}/\overrightarrow{x_i}]) \subseteq (FVV(V) - \overrightarrow{x_i}) \,\,\cup \bigcup_{V_i \in \overrightarrow{V_i}} \! FVV(V_i)
\]

However, as we know that $\vj \Gamma V I$, we can use $(e)$ of the simultaneously proved eMLTT$_{\mathcal{T}_{\text{eff}}}$ version of Proposition~\ref{prop:freevariablesofwellformedexpressions} on this derivation to get $FVV(V) \subseteq V\!ars(\Gamma)$. 

Further, according to the definition of the translation of effect terms into value terms, we also know that $\overrightarrow{x_i} = V\!ars(\Gamma)$. As a result, we can conclude that
\[
FVV(V[\overrightarrow{V_i}/\overrightarrow{x_i}]) \subseteq (FVV(V) - V\!ars(\Gamma)) \,\cup\!\! \bigcup_{V_i \in \overrightarrow{V_i}} \!  FVV(V_i) = \bigcup_{V_i \in \overrightarrow{V_i}} \!  FVV(V_i)
\] 
Using these same arguments with $O[V[\overrightarrow{V_i}/\overrightarrow{x_i}]/x]$, we also get that
\[
FVV(O[V[\overrightarrow{V_i}/\overrightarrow{x_i}]/x]) \subseteq \bigcup_{V_i \in \overrightarrow{V_i}} \!  FVV(V_i)
\]

\pagebreak

Next, by using the induction hypothesis on $\lj {\Gamma, y \!:\! O[V/x] \vertbar \Delta} T$, we get
\[
FVV(\efftrans T {A; \overrightarrow{V_i},\, y; \overrightarrow{V'_{\!j}}; \overrightarrow{W_{\sigalgop}}}) 
\]
\[
\subseteq 
\]
\[
FVV(A) \,\cup\!\!\!\! \bigcup_{V_i \in \overrightarrow{V_i},\, y} \!\!\! FVV(V_i) \,\cup\!\!\!  \bigcup_{V'_{\!j} \in \overrightarrow{V'_{\!j}}} \!\! FVV(V'_{\!j}) \,\cup\!\!\!\!\!\!\! \bigcup_{W_{\sigalgop} \in \overrightarrow{W_{\sigalgop}}} \!\!\!\!\!\! FVV(W_{\sigalgop}) 
\]

Finally, by observing that $y$ is fresh according to our chosen variable convention, we can combine the  inclusions we proved above to get the required inclusion
\[
FVV(W_{\sigalgop}\,\, \langle V[\overrightarrow{V_i}/\overrightarrow{x_i}] , \lambda\, y \!:\! O[V[\overrightarrow{V_i}/\overrightarrow{x_i}]/x] .\, \efftrans {T} {A; \overrightarrow{V_i},\, y; \overrightarrow{V'_{\!j}}; \overrightarrow{W_{\sigalgop}}}\rangle)
\vspace{-0.25cm}
\]
\[
\subseteq
\]
\[
FVV(A) \,\cup\!\! \bigcup_{V_i \in \overrightarrow{V_i}} \! FVV(V_i) \,\cup\!\!\! \bigcup_{V'_{\!j} \in \overrightarrow{V'_{\!j}}} \! FVV(V'_{\!j}) \,\cup\!\!\!\!\!\!\! \bigcup_{W_{\sigalgop} \in \overrightarrow{W_{\sigalgop}}} \!\!\!\!\!\! FVV(W_{\sigalgop}) 
\vspace{-0.25cm}
\]
\end{proof}

We now return to the eMLTT$_{\mathcal{T}_{\text{eff}}}$ version of Proposition~\ref{prop:freevariablesofwellformedexpressions} and the case of its proof that corresponds to the third group of definitional equations given in Definition~\ref{def:extensionofeMLTTwithfibalgeffects}. 

Specifically, in this case the given derivation ends with
\[
\mkrulelabel
{
\veq {\Gamma'} {\efftrans {T_1} {U\ul{C}; \overrightarrow{V_i}; \overrightarrow{V'_{\!j}}; \overrightarrow{W_{\sigalgop}}}} {\efftrans {T_2} {U\ul{C}; \overrightarrow{V_i}; \overrightarrow{V'_{\!j}}; \overrightarrow{W_{\sigalgop}}}} {U\ul{C}}
}
{
\begin{array}{c@{\qquad\qquad} l}
\lj {\Gamma'} \ul{C}
\\
\vj {\Gamma'} {V_i} {A_i[V_1/x_1, \ldots, V_{i-1}/x_{i-1}]} & (1 \leq i \leq n)
\\
\vj {\Gamma'} {V'_{\!j}} {A'_{\!j}[\overrightarrow{V_i}/\overrightarrow{x_i}] \to U\ul{C}} & (1 \leq j \leq m)
\end{array}
}
{(\ljeq {\Gamma \vertbar \Delta} {T_1\!} {\!T_2} \in \!\mathcal{E}_{\text{eff}})}
\]
with the well-typed value terms $\vj {\Gamma'} {W_{\sigalgop}} {(\Sigma\, x \!:\! I .\, O \to U\ul{C}) \to U\ul{C}}$ defined as
\[
\begin{array}{c}
\hspace{-8.75cm}
W_{\sigalgop} \defeq \lambda x' \!:\! (\Sigma\, x \!:\! I.\, O \to U\ul{C}).\, 
\\
\hspace{2cm}
\pmatch {x'} {(x \!:\! I, y \!:\! O \to U\ul{C})} {x''\!.\, U\ul{C}} {\thunk (\algop^{\ul{C}}_{x}(y'.\, \force {\ul{C}} {(y\,\, y')}))} 
\end{array}
\]
for each the operation symbols $\sigalgop : (x \!:\! I) \longrightarrow O$ in $\mathcal{S}_{\text{eff}}$, 
and we need to show that
\[
\begin{array}{c}
FVV(U\ul{C}) \subseteq V\!ars(\Gamma')
\\[2.5mm]
FVV(\efftrans {T_1} {U\ul{C}; \overrightarrow{V_i}; \overrightarrow{V'_{\!j}}; \overrightarrow{W_{\sigalgop}}}) \subseteq V\!ars(\Gamma')
\\[3mm]
FVV(\efftrans {T_2} {U\ul{C}; \overrightarrow{V_i}; \overrightarrow{V'_{\!j}}; \overrightarrow{W_{\sigalgop}}}) \subseteq V\!ars(\Gamma')
\end{array}
\]

First, we use $(c)$ on $\lj {\Gamma'} {\ul{C}}$ to get the inclusion $FVV(\ul{C}) \subseteq V\!ars(\Gamma')$, from which then $FVV(U\ul{C}) \subseteq V\!ars(\Gamma')$ follows trivially because we have $FVV(U\ul{C}) = FVV(\ul{C})$. 

Next, by using $(e)$ on the other derivations in the premise of the given rule, we get $FVV(V_i) \subseteq V\!ars(\Gamma')$, $FVV(V'_{\!j}) \subseteq V\!ars(\Gamma')$, and $FVV(W_{\sigalgop}) \subseteq V\!ars(\Gamma')$ for all value terms $V_i$, $V'_{\!j}$, and $W_{\sigalgop}$ mentioned in the subscripts of the translations of $T_1$ and $T_2$. 

Finally, combining the above inclusions with the inclusion given by the simultaneously proved Proposition~\ref{prop:effecttermtranslationinclusionofvariables}, the required three inclusions follow straightforwardly.

\subsection*{Extending Theorem~\ref{thm:substitution} (Value term substitution) to eMLTT$_{\!\mathcal{T}_{\text{eff}}}$}
\index{substitution theorem!syntactic --!-- for value terms}

We begin by recalling that in Theorem~\ref{thm:substitution} we showed that the substitution rule is admissible in eMLTT for substituting value terms for value variables. When extending Theorem~\ref{thm:substitution} to eMLTT$_{\mathcal{T}_{\text{eff}}}$, we keep the basic proof principle the same: we prove $(a)$--$(l)$ for different kinds of types, terms, and definitional equations simultaneously, with $(a)$--$(b)$ proved by induction on the length of the given value context $\Gamma_2$, and $(c)$--$(l)$  by induction on the given derivations; and this theorem as a whole is proved simultaneously with the eMLTT$_{\mathcal{T}_{\text{eff}}}$ version of the weakening theorem (Theorem~\ref{thm:weakening}). 

The new cases for algebraic operations, and the corresponding congruence and general algebraicity equations are analogous to other computation terms and definitional equations that involve variable bindings and type annotations. However, in order to account for the case that corresponds to the third group of equations given in Definition~\ref{def:extensionofeMLTTwithfibalgeffects}, we additionally need to prove Proposition~\ref{prop:effecttermstranslationsubstitution} below. It is worth noting that this  proposition does not need to be proved simultaneously with the eMLTT$_{\mathcal{T}_{\text{eff}}}$ version of Theorem~\ref{thm:substitution}, the latter simply uses it in its proof. 
It is also worth highlighting that as a consequence of extending Theorem~\ref{thm:substitution} to eMLTT$_{\!\mathcal{T}_{\text{eff}}}$, 
the analogous theorem about simultaneous substitutions (Theorem~\ref{thm:simultaneoussubstitution}) also holds for eMLTT$_{\!\mathcal{T}_{\text{eff}}}$.

\begin{proposition}
\label{prop:effecttermstranslationsubstitution}
Given an effect term $\lj {\Gamma \vertbar \Delta} T$ derived from $\mathcal{S}_{\text{eff}}$, a value type $A$, value terms $V_{i}$ (for all $x_i \!:\! A_i$ in $\Gamma$), value terms $V'_{\!j}$ (for all $w_{\!j} \!:\! A'_{\!j}$ in $\Delta$), value terms $W_{\sigalgop}$ (for all $\sigalgop : (x \!:\! I) \longrightarrow O$ in $\mathcal{S}_{\text{eff}}$), a value variable $y$, and a value term $W$, then 
\[
\efftrans T {A; \overrightarrow{V_i}; \overrightarrow{V'_{\!j}}; \overrightarrow{W_{\sigalgop}}}[W/y] = \efftrans T {A[W/y]; \overrightarrow{V_i[W/y]}; \overrightarrow{V'_{\!j}[W/y]}; \overrightarrow{W_{\sigalgop}[W/y]}}
\]
\end{proposition}

\begin{proof}
We prove this proposition by induction on the derivation of $\lj {\Gamma \vertbar \Delta} T$. 

As a representative example, we consider the case of algebraic operations, for which we need to show that the following equation holds:
\[
\begin{array}{c}
(W_{\sigalgop}\,\, \langle V[\overrightarrow{V_i}/\overrightarrow{x_i}] , \lambda\, y' \!:\! O[V[\overrightarrow{V_i}/\overrightarrow{x_i}]/x] .\, \efftrans {T} {A; \overrightarrow{V_i},\, y'; \overrightarrow{V'_{\!j}}; \overrightarrow{W_{\sigalgop}}}\rangle)[W/y] 
\\[-1mm]
=
\\[-1mm]
\hspace{-8.5cm}
(W_{\sigalgop}[W/y])\, \langle V[\overrightarrow{V_i[W/y]}/\overrightarrow{x_i}] , 
\\
\hspace{2.75cm}
\lambda\, y' \!:\! O[V[\overrightarrow{V_i[W/y]}/\overrightarrow{x_i}]/x] .\, \efftrans {T} {A[W/y]; \overrightarrow{V_i[W/y]}, y'; \overrightarrow{V'_{\!j}[W/y]}; \overrightarrow{W_{\sigalgop}[W/y]}}\rangle
\end{array}
\]

First, by using the definition of substitution, we get that
\[
\begin{array}{c}
(W_{\sigalgop}\,\, \langle V[\overrightarrow{V_i}/\overrightarrow{x_i}] , \lambda\, y' \!:\! O[V[\overrightarrow{V_i}/\overrightarrow{x_i}]/x] .\, \efftrans {T} {A; \overrightarrow{V_i} ,\, y'; \overrightarrow{V'_{\!j}}; \overrightarrow{W_{\sigalgop}}}\rangle)[W/y]
\\[-1mm]
=
\\[1mm]
(W_{\sigalgop}[W/y])\, \langle V[\overrightarrow{V_i}/\overrightarrow{x_i}][W/y] , \lambda\, y' \!:\! O[V[\overrightarrow{V_i}/\overrightarrow{x_i}]/x][W/y] .\, \efftrans {T} {A; \overrightarrow{V_i},\, y'; \overrightarrow{V'_{\!j}}; \overrightarrow{W_{\sigalgop}}}[W/y]\rangle
\end{array}
\]

Next, by using the eMLTT$_{\mathcal{T}_{\text{eff}}}$ version of Proposition~\ref{prop:freevariablesofwellformedexpressions} on the assumed derivation of $\vj \Gamma V I$, we get  $FVV(V) \subseteq V\!ars(\Gamma) = \{x_1, \ldots, x_n\}$. 
Based on this inclusion, we can use the eMLTT$_{\mathcal{T}_{\text{eff}}}$ version of Proposition~\ref{prop:simultaneoussubstlemma1} to prove the following equation:
\[
V[\overrightarrow{V_i}/\overrightarrow{x_i}][W/y] = V[\overrightarrow{V_i[W/y]}/\overrightarrow{x_i}]
\]
Further, using the eMLTT$_{\mathcal{T}_{\text{eff}}}$ versions of Propositions~\ref{prop:freevariablesofwellformedexpressions} and~\ref{prop:simultaneoussubstlemma1}, we also get that
\[
O[V[\overrightarrow{V_i}/\overrightarrow{x_i}]/x][W/y] = O[V[\overrightarrow{V_i}/\overrightarrow{x_i}][W/y]/x] = O[V[\overrightarrow{V_i[W/y]}/\overrightarrow{x_i}]/x]
\]

Next, by using the induction hypothesis on $\lj {\Gamma, y' \!:\! O[V/x] \vertbar \Delta} T$, we get that 
\[
\efftrans {T} {A; \overrightarrow{V_i}\!, y'; \overrightarrow{V'_{\!j}}; \overrightarrow{W_{\sigalgop}}}[W/y] = \efftrans {T} {A[W/y]; \overrightarrow{V_i[W/y]}, y'[W/y]; \overrightarrow{V'_{\!j}[W/y]}; \overrightarrow{W_{\sigalgop}[W/y]}}
\]
However, as $y \neq y'$ due to our adopted variable conventions, the above is equivalent to  
\[
\efftrans {T} {A; \overrightarrow{V_i}\!, y'; \overrightarrow{V'_{\!j}}; \overrightarrow{W_{\sigalgop}}}[W/y] = \efftrans {T} {A[W/y]; \overrightarrow{V_i[W/y]}, y'; \overrightarrow{V'_{\!j}[W/y]}; \overrightarrow{W_{\sigalgop}[W/y]}}
\]

Finally, when we combine the above equations, we get the required equation
\[
\begin{array}{c}
(W_{\sigalgop}\,\, \langle V[\overrightarrow{V_i}/\overrightarrow{x_i}] , \lambda\, y' \!:\! O[V[\overrightarrow{V_i}/\overrightarrow{x_i}]/x] .\, \efftrans {T} {A; \overrightarrow{V_i}\!, y'; \overrightarrow{V'_{\!j}}; \overrightarrow{W_{\sigalgop}}}\rangle)[W/y] 
\\[-1mm]
=
\\[-1mm]
\hspace{-8.5cm}
(W_{\sigalgop}[W/y])\, \langle V[\overrightarrow{V_i[W/y]}/\overrightarrow{x_i}] , 
\\
\hspace{2.75cm}
\lambda\, y' \!:\! O[V[\overrightarrow{V_i[W/y]}/\overrightarrow{x_i}]/x] .\, \efftrans {T} {A[W/y]; \overrightarrow{V_i[W/y]}, y'; \overrightarrow{V'_{\!j}[W/y]}; \overrightarrow{W_{\sigalgop}[W/y]}}\rangle
\end{array}
\]
\end{proof}

We now return to the eMLTT$_{\mathcal{T}_{\text{eff}}}$ version of Theorem~\ref{thm:substitution} and the case of its proof that corresponds to the third group of definitional equations given in Definition~\ref{def:extensionofeMLTTwithfibalgeffects}. 

Specifically, in this case we are given
$
\vj {\Gamma'_1} W A
$
and
\[
\mkrulelabel
{
\veq {\Gamma'_1, y \!:\! A, \Gamma'_2} {\efftrans {T_1} {U\ul{C}; \overrightarrow{V_i}; \overrightarrow{V'_{\!j}}; \overrightarrow{W_{\sigalgop}}}} {\efftrans {T_2} {U\ul{C}; \overrightarrow{V_i}; \overrightarrow{V'_{\!j}}; \overrightarrow{W_{\sigalgop}}}} {U\ul{C}}
}
{
\begin{array}{c@{\qquad\quad} l}
\lj {\Gamma'_1, y \!:\! A, \Gamma'_2} \ul{C}
\\
\vj {\Gamma'_1, y \!:\! A, \Gamma'_2} {V_i} {A_i[V_1/x_1, \ldots, V_{i-1}/x_{i-1}]} & (1 \leq i \leq n)
\\
\vj {\Gamma'_1, y \!:\! A, \Gamma'_2} {V'_{\!j}} {A'_{\!j}[\overrightarrow{V_i}/\overrightarrow{x_i}] \to U\ul{C}} & (1 \leq j \leq m)
\end{array}
}
{(\ljeq {\Gamma \vertbar \Delta} {T_1\!} {\!T_2} \in \!\mathcal{E}_{\text{eff}})}
\]
with the value terms $\vj {\Gamma'_1, y \!:\! A, \Gamma'_2} {W_{\sigalgop}} {(\Sigma\, x \!:\! I .\, O \to U\ul{C}) \to U\ul{C}}$ defined as
\[
\begin{array}{c}
\hspace{-8.75cm}
W_{\sigalgop} \defeq \lambda x' \!:\! (\Sigma\, x \!:\! I.\, O \to U\ul{C}).\, 
\\
\hspace{2cm}
\pmatch {x'} {(x \!:\! I, y' \!:\! O \to U\ul{C})} {x''\!.\, U\ul{C}} {\thunk (\algop^{\ul{C}}_{x}(y''.\, \force {\ul{C}} {(y'\,\, y'')}))} 
\end{array}
\]
for each $\sigalgop : (x \!:\! I) \longrightarrow O$ in $\mathcal{S}_{\text{eff}}$, 
and we need to prove the following equation:
\[
\veq {\Gamma'_1, \Gamma'_2[W/y]} {\efftrans {T_1} {U\ul{C}; \overrightarrow{V_i}; \overrightarrow{V'_{\!j}}; \overrightarrow{W_{\sigalgop}}}[W/y]} {\efftrans {T_2} {U\ul{C}; \overrightarrow{V_i}; \overrightarrow{V'_{\!j}}; \overrightarrow{W_{\sigalgop}}}[W/y]} {U\ul{C}[W/y]}
\]

First, we use $(e)$ on the assumed derivation of $\lj {\Gamma'_1, y \!:\! A, \Gamma'_2} {\ul{C}}$ to get a derivation \linebreak of $\lj {\Gamma'_1, \Gamma'_2[W/y]} {\ul{C}[W/y]}$. 

Next, by using $(g)$ on the other assumptions of the given rule, in combination with 
the properties of (simultaneous) substitution we established in Sections~\ref{sect:syntax} and~\ref{sect:completeness}, and the eMLTT$_{\mathcal{T}_{\text{eff}}}$ version of Proposition~\ref{prop:freevariablesofwellformedexpressions} proved earlier, we get derivations of 
\[
\begin{array}{c}
\vj {\Gamma'_1, \Gamma'_2[W/y]} {V_i[W/y]} {A_i[V_1[W/y]/x_1, \ldots, V_{i-1}[W/y]/x_{i-1}]}
\\[2mm]
\vj {\Gamma'_1, \Gamma'_2[W/y]} {V'_{\!j}[W/y]} {A'_{\!j}[\overrightarrow{V_i[W/y]}/\overrightarrow{x_i}] \to U\ul{C}[W/y]}\end{array}
\]
for all $1 \leq i \leq n$ and $1 \leq j \leq m$, respectively.

Next, we use the rule for the third group of definitional equations given in Definition~\ref{def:extensionofeMLTTwithfibalgeffects}, together with the derivations we have constructed above, to prove
\[
\begin{array}{c}
\hspace{-5cm}
{\Gamma'_1, \Gamma'_2[W/y]} \vdash {\efftrans {T_1} {U\ul{C}[W/y]; \overrightarrow{V_i[W/y]}; \overrightarrow{V'_{\!j}[W/y]}; \overrightarrow{W_{\sigalgop}[W/y]}}} = 
\\
\hspace{6cm}
{\efftrans {T_2} {U\ul{C}[W/y]; \overrightarrow{V_i[W/y]}; \overrightarrow{V'_{\!j}[W/y]}; \overrightarrow{W_{\sigalgop}[W/y]}}} : {U\ul{C}[W/y]}
\end{array}
\]

Finally, we use Proposition~\ref{prop:effecttermstranslationsubstitution} to turn the this proof into the required proof of
\[
\veq {\Gamma'_1, \Gamma'_2[W/y]} {\efftrans {T_1} {U\ul{C}; \overrightarrow{V_i}; \overrightarrow{V'_{\!j}}; \overrightarrow{W_{\sigalgop}}}[W/y]} {\efftrans {T_2} {U\ul{C}; \overrightarrow{V_i}; \overrightarrow{V'_{\!j}}; \overrightarrow{W_{\sigalgop}}}[W/y]} {U\ul{C}[W/y]}
\]

\subsection*{Extending Proposition~\ref{prop:wellformedcomponentsofjudgements} to eMLTT$_{\!\mathcal{T}_{\text{eff}}}$}

We begin by recalling that in Proposition~\ref{prop:wellformedcomponentsofjudgements} we showed that the judgements of well-formed expressions and definitional equations only involve well-formed contexts and types, and well-typed terms. For example, given $\ceq \Gamma M N {\ul{C}}$, we showed that
\[
\cj \Gamma M \ul{C}
\qquad
\cj \Gamma N \ul{C}
\]

When extending Proposition~\ref{prop:wellformedcomponentsofjudgements} to eMLTT$_{\mathcal{T}_{\text{eff}}}$, we keep the basic proof principle the same: we prove $(a)$--$(j)$ for different kinds of types, terms, and definitional equations simultaneously, by induction on the given derivations, using the eMLTT$_{\mathcal{T}_{\text{eff}}}$ versions of the weakening and substitution theorems, where required.

The new cases for algebraic operations,  and the corresponding congruence and general algebraicity equations are analogous to other computation terms and definitional equations that involve variable bindings and type annotations. However, in order to account for the case that corresponds to the third group of equations given in Definition~\ref{def:extensionofeMLTTwithfibalgeffects}, we additionally need to prove Proposition~\ref{prop:welltypednessoftranslatingeffectterms} below. It is worth noting that this  proposition does not need to be proved simultaneously with the eMLTT$_{\mathcal{T}_{\text{eff}}}$ version of Proposition~\ref{prop:wellformedcomponentsofjudgements}, the latter simply uses it in its proof.

\begin{proposition}
\label{prop:welltypednessoftranslatingeffectterms}
\index{translation of effect terms into value terms! well-typedness of --}
Given a well-formed effect term $\lj {\Gamma \vertbar \Delta} T$ derived from $\mathcal{S}_{\text{eff}}$, a value type $A$, value terms $V_{i}$ (for all $x_i \!:\! A_i$ in $\Gamma$), value terms $V'_{\!j}$ (for all $w_{\!j} \!:\! A'_{\!j}$ in $\Delta$), value terms $W_{\sigalgop}$ (for all $\sigalgop : (x \!:\! I) \longrightarrow O$ in $\mathcal{S}_{\text{eff}}$), and a value context $\Gamma'$ such that
\begin{itemize}
\item $\vdash \Gamma'$, 
\item $\lj {\Gamma'} A$, 
\item $\vj {\Gamma'} {V_i} {A_i[V_1/x_1, \ldots, V_{i-1}/x_{i-1}]}$,  
\item $\vj {\Gamma'} {V'_{\!j}} {A'_{\!j}[V_1/x_1, \ldots, V_n/x_n] \to A}$, and 
\item $\vj {\Gamma'} {W_{\sigalgop}} {(\Sigma\, x \!:\! I .\, O \to A) \to A}$,  
\end{itemize}
then 
the result of the translation of $T$ is well-typed as $\vj {\Gamma'} {\efftrans T {A; \overrightarrow{V_i}; \overrightarrow{V'_{\!j}}; \overrightarrow{W_{\sigalgop}}}} {A}$.
\end{proposition}

\begin{proof}
We prove this proposition by induction on the derivation of $\lj {\Gamma \vertbar \Delta} T$, using the eMLTT$_{\mathcal{T}_{\text{eff}}}$ versions of the weakening and substitution theorems, where necessary.

As a representative example, we consider the case of algebraic operations, for which we need to construct a derivation of
\[
\vj {\Gamma'} {W_{\sigalgop}\,\, \langle V[\overrightarrow{V_i}/\overrightarrow{x_i}] , \lambda\, y \!:\! O[V[\overrightarrow{V_i}/\overrightarrow{x_i}]/x] .\, \efftrans {T} {A; \overrightarrow{V_i}\!, y; \overrightarrow{V'_{\!j}}; \overrightarrow{W_{\sigalgop}}}\rangle} {A}
\]

First, by using the eMLTT$_{\mathcal{T}_{\text{eff}}}$ version of Theorem~\ref{thm:simultaneoussubstitution} (simultaneous value term substitution) with the derivation of $\vj \Gamma V I$, we get a derivation of
\[
\vj {\Gamma'} {V[\overrightarrow{V_i}/\overrightarrow{x_i}]} {I[\overrightarrow{V_i}/\overrightarrow{x_i}]}
\]
However, as $\lj {\diamond} I$, we can use the eMLTT$_{\mathcal{T}_{\text{eff}}}$ version of Proposition~\ref{prop:freevariablesofwellformedexpressions} to get that $FVV(I) = \emptyset$, and 
thus we can use the eMLTT$_{\mathcal{T}_{\text{eff}}}$ version of Propopsition~\ref{prop:valuesubstlemma1simultaneous} to get 
\[
\vj {\Gamma'} {V[\overrightarrow{V_i}/\overrightarrow{x_i}]} {I}
\]
As a consequence, we can use the eMLTT$_{\mathcal{T}_{\text{eff}}}$ version of Theorem~\ref{thm:simultaneoussubstitution} (simultaneous value term substitution) with the derivation of $\lj {x \!:\! I} O$, to also get a derivation of
\[
\lj {\Gamma'} {O[V[\overrightarrow{V_i}/\overrightarrow{x_i}]/x]}
\]

Next, as $y$ is fresh by our adopted variable conventions, we have a derivation of $\lj {} {\Gamma', y \!:\! O[V[\overrightarrow{V_i}/\overrightarrow{x_i}]/x]}$ 
and thus we can use the induction hypothesis to get 
\[
\vj {\Gamma', y \!:\! O[V[\overrightarrow{V_i}/\overrightarrow{x_i}]/x]} {\efftrans {T} {A; \overrightarrow{V_i}\!, y; \overrightarrow{V'_{\!j}}; \overrightarrow{W_{\sigalgop}}}} {A}
\]

Next, by using the typing rule for lambda abstraction, we get a derivation of
\[
\vj {\Gamma'} {\lambda\, y \!:\! O[V[\overrightarrow{V_i}/\overrightarrow{x_i}]/x] .\, \efftrans {T} {A; \overrightarrow{V_i}\!, y; \overrightarrow{V'_{\!j}}; \overrightarrow{W_{\sigalgop}}}} {O[V[\overrightarrow{V_i}/\overrightarrow{x_i}]/x] \to A}
\]

Finally, by using the typing rules for function application and pairing, together with the derivations constructed above, we get the required derivation of
\[
\vj {\Gamma'} {W_{\sigalgop}\,\, \langle V[\overrightarrow{V_i}/\overrightarrow{x_i}] , \lambda\, y \!:\! O[V[\overrightarrow{V_i}/\overrightarrow{x_i}]/x] .\, \efftrans {T} {A; \overrightarrow{V_i}\!, y; \overrightarrow{V'_{\!j}}; \overrightarrow{W_{\sigalgop}}}\rangle} {A}
\]
\end{proof}

We now return to the eMLTT$_{\mathcal{T}_{\text{eff}}}$ version of Proposition~\ref{prop:wellformedcomponentsofjudgements} and the case of its proof that corresponds to the third group of definitional equations given in Definition~\ref{def:extensionofeMLTTwithfibalgeffects}.

Specifically, in this case the given derivation ends with
\[
\mkrulelabel
{
\veq {\Gamma'} {\efftrans {T_1} {U\ul{C}; \overrightarrow{V_i}; \overrightarrow{V'_{\!j}}; \overrightarrow{W_{\sigalgop}}}} {\efftrans {T_2} {U\ul{C}; \overrightarrow{V_i}; \overrightarrow{V'_{\!j}}; \overrightarrow{W_{\sigalgop}}}} {U\ul{C}}
}
{
\begin{array}{c@{\qquad\quad} l}
\lj {\Gamma'} \ul{C}
\\
\vj {\Gamma'} {V_i} {A_i[V_1/x_1, \ldots, V_{i-1}/x_{i-1}]} & (1 \leq i \leq n)
\\
\vj {\Gamma'} {V'_{\!j}} {A'_{\!j}[\overrightarrow{V_i}/\overrightarrow{x_i}] \to U\ul{C}} & (1 \leq j \leq m)
\end{array}
}
{(\ljeq {\Gamma \vertbar \Delta} {T_1\!} {\!T_2} \in \!\mathcal{E}_{\text{eff}})}
\]

\pagebreak
\noindent
and we need to construct derivations for
\[
\vj {\Gamma'} {\efftrans {T_1} {U\ul{C}; \overrightarrow{V_i}; \overrightarrow{V'_{\!j}}; \overrightarrow{W_{\sigalgop}}}} {U\ul{C}}
\qquad
\vj {\Gamma'} {\efftrans {T_2} {U\ul{C}; \overrightarrow{V_i}; \overrightarrow{V'_{\!j}}; \overrightarrow{W_{\sigalgop}}}} {U\ul{C}}
\]
both of which follow immediately from Proposition~\ref{prop:welltypednessoftranslatingeffectterms} proved above.

\section{Derivable equations}
\label{sect:derivableequationsforeMLTTwithfibalgeffects}

In this short section we present some useful equations that are derivable in eMLTT$_{\mathcal{T}_{\text{eff}}}$, in addition to the equations that we showed to be derivable in eMLTT in Section~\ref{sect:derivableequations}. Namely, by using the general algebraicity equation from Definition~\ref{def:extensionofeMLTTwithfibalgeffects}, we can derive more specialised algebraicity equations that  describe the commutativity of algebraic operations with specific computation terms. Many of these equations appear in languages with algebraic effects based on CBPV without stacks, e.g., see~\cite{Kammar:AlgebraicFoundations,Pretnar:Thesis}.

\begin{proposition}
\label{prop:specialisedalgebraicity}
\index{algebraicity equation!specialised --}
The following definitional equations between computation terms are derivable in eMLTT$_{\mathcal{T}_{\text{eff}}}$, for every operation symbol $\sigalgop : (x \!:\! I) \longrightarrow O$ in $\mathcal{S}_{\text{eff}}$:

\[
\mkrule
{\ceq \Gamma {\doto {\algop^{FA}_V(y .\, M)} {y' \!:\! A} {\ul{C}} {N}} {\algop^{\ul{C}}_V(y.\, \doto M {y' \!:\! A} {\ul{C}} {N})} {\ul{C}}}
{
\vj \Gamma V I
\quad
\cj {\Gamma, y \!:\! O[V/x]} M {FA}
\quad
\lj \Gamma \ul{C}
\quad
\cj {\Gamma, y' \!:\! A} N \ul{C}
}
\vspace{0.35cm}
\]

\[
\mkrule
{\ceq \Gamma {\langle W , \algop^{\ul{C}[W/y']}_V(y .\, M) \rangle_{(y' : A).\, \ul{C}}} {\algop^{\Sigma\, y' : A .\, \ul{C}}_V(y.\, \langle W , M \rangle_{(y' : A).\, \ul{C}})} {\Sigma\, y' \!:\! A .\, \ul{C}}}
{
\vj \Gamma V I
\quad
\vj \Gamma W A
\quad
\lj {\Gamma, y' \!:\! A} \ul{C}
\quad
\cj {\Gamma, y \!:\! O[V/x]} M {\ul{C}[W/y']}
}
\vspace{0.35cm}
\]

\[
\mkrule
{\ceq \Gamma {\doto {\algop^{\Sigma\, y' : A .\, \ul{C}}_V(y .\, M)} {(y' \!:\! A, z \!:\! \ul{C})} {\ul{D}} K} {\algop^{\ul{D}}_V(y .\, \doto M {(y' \!:\! A, z \!:\! \ul{C})} {\ul{D}} K)} {\ul{D}}}
{
\vj \Gamma V I
\quad
\cj {\Gamma, y \!:\! O[V/x]} M {\Sigma\, y' \!:\! A .\, \ul{C}}
\quad
\lj \Gamma \ul{D}
\quad
\hj {\Gamma, y' \!:\! A} {z \!:\! \ul{C}} K \ul{D} 
}
\vspace{0.35cm}
\]

\[
\mkrule
{\ceq \Gamma {\lambda \, y' \!:\! A .\, \algop^{\ul{C}}_V(y.\, M)} {\algop^{\Pi\, y' : A .\, \ul{C}}_V(y .\, \lambda\, y' \!:\! A .\, M)} {\Pi\, y' \!:\! A .\, \ul{C}}}
{
\vj \Gamma V I
\quad
\lj {\Gamma, y' \!:\! A} {\ul{C}}
\quad
\cj {\Gamma, y \!:\! O[V/x], y' \!:\! A} {M} {\ul{C}} 
}
\vspace{0.35cm}
\]

\[
\mkrule
{\ceq \Gamma {(\algop^{\Pi\, y' : A .\, \ul{C}}_V(y.\, M))(W)_{(y' : A).\,\ul{C}}} {\algop^{\ul{C}[W/y']}_V(y.\, M(W)_{(y' : A).\,\ul{C}})} {\ul{C}[W/y']}}
{
\vj \Gamma V I
\quad
\vj \Gamma W A
\quad
\lj {\Gamma, y' \!:\! A} \ul{C}
\quad
\cj {\Gamma, y \!:\! O[V/x]} M {\Pi\, y' \!:\! A .\, \ul{C}}
}
\vspace{0.35cm}
\]

\[
\mkrule
{\ceq \Gamma {W(\algop^{\ul{C}}_V(y.\, M))_{\ul{C}, \ul{D}}} {\algop^{\ul{D}}_V(y.\, W(M)_{\ul{C}, \ul{D}})} {\ul{D}}}
{
\vj \Gamma V I
\quad
\vj \Gamma W {\ul{C} \multimap \ul{D}}
\quad
\cj {\Gamma, y \!:\! O[V/x]} M \ul{C}
}
\]
\end{proposition}

\pagebreak

\begin{proof}
All six equations are proved similarly, by using the general algebraicity equation from Definition~\ref{def:extensionofeMLTTwithfibalgeffects} in combination with the definition of substituting  computation terms for computation variables, e.g., the first equation is proved as follows:
\begin{fleqn}[0.3cm]
\begin{align*}
\Gamma \,\vdash\,\, & \doto {\algop^{FA}_V(y .\, M)} {y' \!:\! A} {\ul{C}} {N}
\\
=\,\, & (\doto {z} {y' \!:\! A} {\ul{C}} {N})[\algop^{FA}_V(y .\, M)/z]
\\
=\,\, & \algop^{\ul{C}}_V(y. (\doto {z} {y' \!:\! A} {\ul{C}} {N})[M/z])
\\
=\,\, & \algop^{\ul{C}}_V(y. \doto {z[M/z]} {y' \!:\! A} {\ul{C}} {N})
\\
=\,\, & \algop^{\ul{C}}_V(y. \doto {M} {y' \!:\! A} {\ul{C}} {N}) : \ul{C}
\vspace{-1cm}
\end{align*}
\end{fleqn}
\end{proof}

\section{Interpreting eMLTT$_{\!\mathcal{T}_{\text{eff}}}$ in a fibred adjunction model}
\label{sect:fibalgeffectsmodel}

In this section we equip eMLTT$_{\mathcal{T}_{\text{eff}}}$ with a denotational semantics by showing how to interpret it in a fibred adjunction model based on the prototypical model of dependent types, the families of sets fibration. More precisely, the fibred adjunction model we use is based on the lifting of the adjunction $F_{\!\mathcal{L}_{\mathcal{T}_{\text{eff}}}} \!\dashv\, U_{\!\mathcal{L}_{\mathcal{T}_{\text{eff}}}} : \Mod(\!\mathcal{L}_{\mathcal{T}_{\text{eff}}},\Set) \longrightarrow \Set$ to families fibrations (see Theorem~\ref{thm:liftedfibradjmodels} and Corollary~\ref{cor:modelsoflawveretheories}), where $\mathcal{L}_{\mathcal{T}_{\text{eff}}}$ is a countable Lawvere theory that we derive from the given fibred effect theory $\mathcal{T}_{\text{eff}} = (\mathcal{S}_{\text{eff}},\mathcal{E}_{\text{eff}})$. 

It is worth noting that compared to the level of generality with which we investigated the denotational semantics of eMLTT in Chapter~\ref{chap:interpretation}, 
we only study the interpretation of eMLTT$_{\mathcal{T}_{\text{eff}}}$ in one specific fibred adjunction model. We leave an investigation of a more general class of models for future work. In particular, we expect that our future study of fibred notions of universal algebra and Lawvere theories (see Section~\ref{sect:fiblawveretheories}) will also provide us with a good general notion of a model for eMLTT$_{\mathcal{T}_{\text{eff}}}$. Further, defining the interpretation of eMLTT$_{\mathcal{T}_{\text{eff}}}$ in a model based on families of sets should make it more accessible to the less categorically inclined audience of this thesis.

In order to reuse the established theory concerning countable Lawvere theories (see Section~\ref{sect:algebraictreatmentofeffects}), we restrict our attention to fibred effect theories which we call \emph{countable}, by imposing conditions on how certain contexts and types must be interpreted in the families of sets fibration $\mathsf{fam}_{\Set} : \Fam(\Set) \longrightarrow \Set$, using the interpretation function $\sem{-}$ that we defined for eMLTT in Section~\ref{sect:interpretation}. Note that as the input and output types of operation symbols are given by well-formed pure eMLTT value types, the soundness results we established in Section~\ref{sect:soundness} ensure that $\sem{-}$ is defined on them.

\index{ F@$\sem{\Gamma; A}_{1}$ (first component of the object $\sem{\Gamma; A}$)}
\index{ F@$\sem{\Gamma; A}_{2}$ (second component of the object $\sem{\Gamma; A}$)}
For better readability, we use the convention that if $\sem{\Gamma; A} = (X, B)$ in $\Fam(\Set)$, then we write $\sem{\Gamma; A}_{1}$ for the set $X$ and $\sem{\Gamma; A}_{2}$ for the functor $B : X \longrightarrow \Set$, and similarly for the interpretation $\sem{\Gamma;\ul{C}}$ of computation types in $\Fam(\Mod(\!\mathcal{L}_{\mathcal{T}_{\text{eff}}},\Set))$. Analogously, if $\sem{\Gamma;V} = (f,g)$ in $\Fam(\Set)$, then we write $\sem{\Gamma;V}_1$ for $f$ and $\sem{\Gamma;V}_2$ for $g$, and similarly for the interpretation of computation and homomorphism terms.
\index{ F@$\sem{\Gamma; V}_{1}$ (first component of the morphism $\sem{\Gamma; V}$)}
\index{ F@$\sem{\Gamma; V}_{2}$ (second component of the morphism $\sem{\Gamma; V}$)}

\begin{definition}
\label{def:countableeffecthteory}
\index{theory!fibred effect --!countable --}
\index{signature!fibred effect --!countable --}
A fibred effect theory $\mathcal{T}_{\text{eff}} = (\mathcal{S}_{\text{eff}}, \mathcal{E}_{\text{eff}})$ is \emph{countable} if $\sem{x \!:\! I; O}_{2}$ is a family of countable sets, for all $\sigalgop : (x \!:\! I) \longrightarrow O$ in $\mathcal{S}_{\text{eff}}$, and if $\sem{\Gamma; A'_{\!j}}_{2}$ is a family \linebreak of countable sets, for all equations $\ljeq {\Gamma \vertbar \Delta} {T_1} {T_2}$ in $\mathcal{E}_{\text{eff}}$ and variables $w_{\!j} \!:\! A'_{\!j}$ in $\Delta$.
\end{definition}

In the rest of this section, we assume that the given fibred effect theory $\mathcal{T}_{\text{eff}}$ is countable.

Next, we show how to derive the countable Lawvere theory $\mathcal{L}_{\mathcal{T}_{\text{eff}}}$ from the given fibred effect theory  $\mathcal{T}_{\text{eff}}$. In particular, we first show that $\mathcal{T}_{\text{eff}}$ gives rise to a countable equational theory $\mathbb{T}_{\!\mathcal{T}_{\text{eff}}} = (\mathbb{S}_{\!\mathcal{T}_{\text{eff}}},\mathbb{E}_{\!\mathcal{T}_{\text{eff}}})$, from which we can then derive the countable Lawvere theory $\mathcal{L}_{\mathcal{T}_{\text{eff}}}$ following Definition~\ref{def:lawveretheoryfromequationaltheory} and Proposition~\ref{prop:lawveretheoryfromequationaltheory}. 

The construction of $\mathbb{T}_{\mathcal{T}_{\text{eff}}}$ is based on the intuitive reading of operation symbols discussed in Section~\ref{sect:fibeffecttheories}, i.e., every operation symbol $\sigalgop : (x \!:\! I) \longrightarrow O$ can be viewed as an $\sem{\diamond;I}_2(\star)$-indexed family of operation symbols in the ordinary universal-algebraic sense. 

We recall that an analogous expansion of effect theories to countable equational theories was also used by Plotkin and Pretnar in their work on handlers in the simply typed setting, see~\cite[Section~4]{Plotkin:HandlingEffects}. In this thesis, we want to emphasise that countable fibred effect theories, despite their additional type-dependency, can be also naturally expanded into  countable equational theories.

\begin{definition}
\index{ S@$\mathbb{S}_{\mathcal{T}_{\text{eff}}}$ (countable signature derived from a countable fibred effect theory $\mathcal{T}_{\text{eff}}$)}
The \emph{countable signature} $\mathbb{S}_{\mathcal{T}_{\text{eff}}}$ is given by operation symbols $\mathsf{op}_i$, each with arity $\vert \sem{x \!:\! I; O}_{2}\, \langle \star , i \rangle \vert$, for all operation symbols $\sigalgop : (x \!:\! I) \longrightarrow O$ in $\mathcal{S}_{\text{eff}}$ and all $i$ in $\sem{\diamond; I}_{2}(\star)$, where, as standard, we use $\vert X \vert$ to denote the cardinality of the set $X$.
\end{definition}

As we have assumed $\mathcal{T}_{\text{eff}}$ to be countable, the previous definition is well-formed because the arities $\vert \sem{x \!:\! I; O}_{2}\, \langle \star , i \rangle \vert$ of operations $\sigalgop_i$ are guaranteed to be countable.

\begin{proposition}
\index{ T@$T^\gamma$ (term of a countable equational theory, derived from an effect term $T$)}
\index{ D@$\Delta^\gamma$ (context of a countable equational theory, derived from an effect context $\Delta$)}
Every well-formed effect term $\lj {\Gamma \vertbar \Delta} T$ derived from ${\mathcal{S}_{\text{eff}}}$ determines a set of well-formed terms $\lj {\Delta^\gamma} {T^\gamma}$ derivable from $\mathbb{S}_{\mathcal{T}_{\text{eff}}}$, for all $\gamma$ in $\sem{\Gamma}$, where $\Delta^\gamma$ is a context of variables $x^{a}_{w_{\!j}}$ for all $w_{\!j} \!:\! A'_{\!j}$ in $\Delta$ and $a$ in $\sem{\Gamma; A'_{\!j}}_{2}(\gamma)$.
\end{proposition}

\begin{proof}
First, we note that as we have assumed $\mathcal{T}_{\text{eff}}$ to be countable, every $\sem{\Gamma; A'_{\!j}}_{2}$ is a family of countable sets. As a result, every context $\Delta^\gamma$ is a countable list of variables.

Next, the terms $T^\gamma$ are computed by recursion on the structure of $T$, as follows: 
\[
\begin{array}{l c l}
(w_{\!j}(V))^\gamma & \defeq & x_{w_{\!j}}^{(\sem{\Gamma;V}_{2})_\gamma(\star)}
\\[4mm]
(\sigalgop_V(y.\, T))^\gamma & \defeq & \mathsf{op}_{i}(T^{\langle \gamma , o \rangle})_{1 \,\leq\, o \,\leq\, \vert \sem{x : I; O}_{2}\, \langle \star , i \rangle \vert} \hfill (\text{where } i \defeq (\sem{\Gamma;V}_{2})_\gamma(\star))
\\[4mm]
\multicolumn{3}{l}{(\pmatchsf V {(y_1 \!:\! B_1, y_2 \!:\! B_2)} {} T)^{\gamma} \,\,\,\,\,\, \defeq \,\,\,\,\,\, T^{\langle \langle \gamma , b_1 \rangle , b_2 \rangle} \qquad (\text{when } (\sem{\Gamma;V}_2)_\gamma(\star) = \langle b_1 , b_2 \rangle)}
\\[4mm]
\multicolumn{3}{l}{(\mathsf{case~} V \mathsf{~of}_{} \mathsf{~} ({\seminl {\!} {\!\!(y_1 \!:\! B_1)} \mapsto T_1}, {\seminr {\!} {\!\!(y_2 \!:\! B_2)} \mapsto T_2}))^{\gamma} \,\,\,\,\,\, \defeq \,\,\,\,\,\, T_1^{\langle \gamma , b \rangle}}
\\[-1mm]
&& \hfill (\text{when } (\sem{\Gamma;V}_2)_\gamma(\star) = \mathsf{inl}\, b)
\\[4mm]
\multicolumn{3}{l}{(\mathsf{case~} V \mathsf{~of}_{} \mathsf{~} ({\seminl {\!} {\!\!(y_1 \!:\! B_1)} \mapsto T_1}, {\seminr {\!} {\!\!(y_2 \!:\! B_2)} \mapsto T_2}))^{\gamma} \,\,\,\,\,\, \defeq \,\,\,\,\,\, T_2^{\langle \gamma , b \rangle}}
\\[-1mm]
&& \hfill (\text{when } (\sem{\Gamma;V}_2)_\gamma(\star) = \mathsf{inr}\, b)
\end{array}
\]

Observe that for better readability, we use $o$ to denote both a natural number between $1$ and $\vert \sem{x : I; O}_{2}\, \langle \star , i \rangle \vert$, and the corresponding element of $\sem{x : I; O}_{2}\, \langle \star , i \rangle$.

Finally, we can construct the required derivations of well-formed terms $\lj {\Delta^\gamma} {\!T^\gamma}$ \linebreak by straightforward induction on the given derivation of the effect term $\lj {\Gamma \vertbar \Delta} \!T$.
\end{proof}

\begin{definition}
\label{def:countableeqthfromeffth}
\index{ T@$\mathbb{T}_{\hspace{-0.05cm}\mathcal{T}_{\text{eff}}}$ (countable equational theory derived from a countable fibred effect theory $\mathcal{T}_{\text{eff}}$)}
\index{ E@$\mathbb{E}_{\hspace{-0.05cm}\mathcal{T}_{\text{eff}}}$ (set of equations of a countable equational theory derived from a countable fibred effect theory $\mathcal{T}_{\text{eff}}$)}
The \emph{countable equational theory} $\mathbb{T}_{\!\mathcal{T}_{\text{eff}}}$ is given by the countable signature $\mathbb{S}_{\mathcal{T}_{\text{eff}}}$ and a set $\mathbb{E}_{\!\mathcal{T}_{\text{eff}}}$, which is the least set containing equations $\ljeq {\Delta^\gamma} {T^\gamma_1} {T^\gamma_2}$, for all  $\ljeq {\Gamma \vertbar \Delta} {T_1} {T_2}$ in $\mathcal{E}_{\text{eff}}$ and all $\gamma$ in $\sem{\Gamma}$, that is closed under the rules of reflexivity, symmetry, transitivity, replacement, and substitution (see Definition~\ref{def:countableequationaltheory}).
\end{definition}

\index{ L@$\mathcal{L}_{\mathcal{T}_{\text{eff}}}$ (countable Lawvere theory derived from a countable fibred effect theory $\mathcal{T}_{\text{eff}}$)}
\index{ I@$I_{\mathcal{T}_{\text{eff}}}$ (countable Lawvere theory derived from a countable fibred effect theory $\mathcal{T}_{\text{eff}}$)}
Now, we know from Definition~\ref{def:lawveretheoryfromequationaltheory} and Proposition~\ref{prop:lawveretheoryfromequationaltheory} that there exists a category $\mathcal{L}_{\mathcal{T}_{\text{eff}}}$ and a corresponding countable Lawvere theory $I_{\mathcal{T}_{\text{eff}}} : \aleph^{\text{op}}_{\!1} \longrightarrow \mathcal{L}_{\mathcal{T}_{\text{eff}}}$, both built from $\mathbb{T}_{\!\mathcal{T}_{\text{eff}}}$. There also exists an adjunction $F_{\!\mathcal{L}_{\mathcal{T}_{\text{eff}}}} \!\dashv\, U_{\!\mathcal{L}_{\mathcal{T}_{\text{eff}}}} : \Mod(\!\mathcal{L}_{\mathcal{T}_{\text{eff}}},\Set) \longrightarrow \Set$. 

Using 
Corollary~\ref{cor:modelsoflawveretheories}, we can lift this adjunction to a split fibred one between the families fibrations $\mathsf{fam}_{\Set}$ and $\mathsf{fam}_{\Mod(\!\mathcal{L}_{\mathcal{T}_{\text{eff}}},\Set)}$, giving us a fibred adjunction model
\vspace{-2.5cm}
\[
\xymatrix@C=0.5em@R=7em@M=0.5em{
\ar@{}[dd]^-{\!\!\quad\qquad\qquad\qquad\qquad\perp}
\\
\Fam(\Set) \ar@/_2.5pc/[d]_-{\mathsf{fam}_{\Set}} \ar@{}[d]_-{\dashv\,\,\,\,\,\,\,\,\,\,} \ar@{}[d]^-{\,\,\,\,\,\,\,\,\,\,\dashv} \ar@/^2.5pc/[d]^-{\ia {-}} \ar@/^1.25pc/[rrrrrrrrrr]^-{\widehat{F_{\!\mathcal{L}_{\mathcal{T}_{\text{eff}}}}}} &  &&&&&&&&& \,\,\,\,\,\,\,\,\,\,\,\,\,\,\,\,\,\,\,\, \ar@/^1.25pc/[llllllllll]^-{\widehat{U_{\!\mathcal{L}_{\mathcal{T}_{\text{eff}}}}}}  & \hspace{-1.5cm} \Fam(\Mod(\!\mathcal{L}_{\mathcal{T}_{\text{eff}}},\Set)) \ar@/^2pc/[dlllllllllll]^-{\!\!\!\!\!\!\quad\qquad\mathsf{fam}_{\Mod(\!\mathcal{L}_{\mathcal{T}_{\text{eff}}},\Set)}}
\\
\mathcal{\Set} \ar[u]_-{1}
}
\]
suitable for modelling eMLTT. In the rest of this section, we show that this fibred adjunction model is also suitable for defining a sound interpretation for eMLTT$_{\mathcal{T}_{\text{eff}}}$. 

\pagebreak

\begin{definition}
\index{interpretation function}
\index{ @$\sem{-}$ (interpretation function)}
We extend the \emph{interpretation} of eMLTT to eMLTT$_{\mathcal{T}_{\text{eff}}}$ by defining it on algebraic operations $\algop^{\ul{C}}_V(y .\, M)$, for each $\sigalgop : (x \!:\! I) \longrightarrow O$ in $\mathcal{S}_{\text{eff}}$, as follows:
\[
\mkrule
{
\xymatrix@C=7em@R=7.5em@M=0.5em{
\txt<10pc>{$\sem{\Gamma; \algop^{\ul{C}}_V(y .\, M)}_1 $\\$ \defeq $\\$ \sem{\Gamma}$}
\ar[ddd]_-{\id_{\sem{\Gamma}}}
&
\txt<10pc>{$(\sem{\Gamma; \algop^{\ul{C}}_V(y .\, M)}_2)_{\gamma} $\\$ \defeq $\\$ 1$}
\ar[d]_-{\langle \id_1 \rangle_{o \in \sem{\Gamma; O[V/x]}_2(\gamma)}}
\\
&
\bigsqcap_{o \in \sem{\Gamma; O[V/x]}_2(\gamma)} 1
\ar[d]_-{\bigsqcap_{o \in \sem{\Gamma; O[V/x]}_2(\gamma)} ((\sem{\Gamma, y : O[V/x]; M}_2)_{\langle \gamma , o \rangle})}
\\
&
\bigsqcap_{o \in \sem{\Gamma; O[V/x]}_2(\gamma)} (U_{\!\mathcal{L}_{\mathcal{T}_{\text{eff}}}}(\sem{\Gamma;\ul{C}}_2(\gamma)))
\ar[d]_-{\sigalgop^{\sem{\Gamma;\ul{C}}_2(\gamma)}_{(\sem{\Gamma; V}_2)_\gamma(\star)}}
\\
\sem{\Gamma}
&
U_{\!\mathcal{L}_{\mathcal{T}_{\text{eff}}}}(\sem{\Gamma;\ul{C}}_2(\gamma))
}
}
{
\begin{array}{c}
\sem{\Gamma;V}_1 = \id_{\sem{\Gamma}} : \sem{\Gamma} \longrightarrow \sem{\Gamma}
\\[2mm]
(\sem{\Gamma;V}_2)_{\gamma} : 1 \longrightarrow \sem{\diamond; I}_2(\star)
\\[2mm]
\sem{\Gamma;O[V/x]}_1 = \sem{\Gamma} \in \Set
\\[2mm]
\sem{\Gamma;O[V/x]}_2(\gamma) = \sem{\Gamma, x \!:\! I; O}_2\, \langle \gamma , (\sem{\Gamma;V}_2)_\gamma(\star) \rangle  \in \Set
\\[2mm]
\sem{\Gamma, y \!:\! O[V/x]; M}_1 =  \id_{\bigsqcup_{\gamma \in \sem{\Gamma}} (\sem{\Gamma;O[V/x]}_2(\gamma))} : \sem{\Gamma, y \!:\! O[V/x]} \longrightarrow \sem{\Gamma, y \!:\! O[V/x]}
\\[2mm]
(\sem{\Gamma, y \!:\! O[V/x]; M}_2)_{\langle \gamma , o \rangle} : 1 \longrightarrow U_{\!\mathcal{L}_{\mathcal{T}_{\text{eff}}}}(\sem{\Gamma; \ul{C}}_2(\gamma))
\end{array}
}
\]
\index{ op@$\sigalgop^{\sem{\Gamma;\ul{C}}_2(\gamma)}_{(\sem{\Gamma; V}_2)_\gamma(\star)}$ (algebraic operation)}
where the function (\emph{algebraic operation}) $\sigalgop^{\sem{\Gamma;\ul{C}}_2(\gamma)}_{(\sem{\Gamma; V}_2)_\gamma(\star)}$ is defined using the countable-product preservation property of $\sem{\Gamma;\ul{C}}_2(\gamma)$ as the following composite function:
\[
\hspace{-3cm}
\xymatrix@C=2em@R=3.75em@M=0.5em{
\bigsqcap_{o \in \sem{\Gamma; O[V/x]}_2(\gamma)} (U_{\!\mathcal{L}_{\mathcal{T}_{\text{eff}}}}(\sem{\Gamma;\ul{C}}_2(\gamma)))
\ar[d]_-{=}
\\
\bigsqcap_{o \in \sem{\Gamma; O[V/x]}_2(\gamma)} ((\sem{\Gamma;\ul{C}}_2(\gamma))(1))
\ar[d]_-{\cong}
\\
(\sem{\Gamma;\ul{C}}_2(\gamma))(+_{o \in \sem{\Gamma; O[V/x]}_2(\gamma)}\, 1)
\ar[d]_-{=}
\\
(\sem{\Gamma;\ul{C}}_2(\gamma))(\vert \sem{\Gamma; O[V/x]}_2(\gamma) \vert)
\ar[d]_-{(\sem{\Gamma;\ul{C}}_2(\gamma))(\lj {\overrightarrow{x_o}\,} {\,\sigalgop_{(\sem{\Gamma;V}_2)_\gamma(\star)}(x_o)_{1 \,\leq\, o \,\leq\, \vert \sem{\Gamma; O[V/x]}_2(\gamma) \vert}})}
\\
(\sem{\Gamma;\ul{C}}_2(\gamma))(1)
\ar[d]_-{=}
\\
U_{\!\mathcal{L}_{\mathcal{T}_{\text{eff}}}}(\sem{\Gamma;\ul{C}}_2(\gamma))
}
\]
\end{definition}

As the interpretation of eMLTT$_{\mathcal{T}_{\text{eff}}}$ is defined \emph{a priori} partially, analogously to the interpretation of eMLTT, we again have to separately show that $\sem{-}$ is defined on all well-formed contexts, types, and terms; and that it validates the equational theory of eMLTT$_{\mathcal{T}_{\text{eff}}}$. 
While most of the results proved for eMLTT in Section~\ref{sect:soundness} (the propositions relating  weakening and substitution to reindexing along semantic projection and substitution morphisms) extend to the interpretation of eMLTT$_{\mathcal{T}_{\text{eff}}}$ without any substantial additional work, the soundness theorem (Theorem~\ref{thm:soundness}) needs more attention. 

In particular, in order to prove the soundness theorem for eMLTT$_{\mathcal{T}_{\text{eff}}}$, we first need to relate two ways of interpreting well-formed effect terms. Specifically,  we relate the interpretations of i) the translation of an effect term $\lj {\Gamma \vertbar \Delta} T$ and ii) the corresponding terms $\lj {\Delta^\gamma} {T^\gamma}$ derivable from the countable signature $\mathbb{S}_{\mathcal{T}_{\text{eff}}}$, as discussed next. 

We note that in order to conveniently reuse the next proposition in Chapter~\ref{chap:handlers} to prove the soundness of the interpretation of the user-defined algebra type, we state it in terms of the given fibred effect signature $\mathcal{S}_{\text{eff}}$ rather than the fibred effect theory $\mathcal{T}_{\text{eff}}$.

\begin{proposition}
\label{prop:relatingsemanticsoffibeffectterms}
Given a well-formed effect term $\lj {\Gamma \vertbar \Delta} T$ derived from $\mathcal{S}_{\text{eff}}$, a computation type $\ul{C}$, value terms $V_{i}$ (for all $x_i \!:\! A_i$ in $\Gamma$), value terms $V'_{\!j}$ (for all $w_{\!j} \!:\! A'_{\!j}$ in $\Delta$), value terms $W_{\sigalgop}$ (for all $\sigalgop : (x \!:\! I) \longrightarrow O$ in $\mathcal{S}_{\text{eff}}$), and a value context $\Gamma'$ such that
\begin{itemize}
\item $\sem{\Gamma'} \in \Set$, 
\item $\sem{\Gamma';V_i}_1 = \id_{\sem{\Gamma'}}  : \sem{\Gamma'} \longrightarrow \sem{\Gamma'}$, and 
\item $(\sem{\Gamma';V_i}_2)_{\gamma\,'} : 1 \longrightarrow $
\\[-7.5mm]

\hspace{1cm} $\sem{x_1 \!:\! A_1, \ldots, x_{i-1} \!:\! A_{i - 1};A_i}_2\, \langle \langle \langle \star , (\sem{\Gamma'; V_1}_2)_{\gamma\,'}(\star) \rangle , \ldots \rangle , (\sem{\Gamma'; V_{i - 1}}_2)_{\gamma\,'}(\star) \rangle$, 
\end{itemize}
together with a value type $A$ and a family of models $\mathcal{M}_{\gamma\,'} : \mathcal{L}_{\mathcal{T}^{d}_{\text{eff}}} \longrightarrow \Set$ (for all $\gamma\,'$ in $\sem{\Gamma'}$) of the Lawvere theory $I_{\mathcal{T}^{d}_{\text{eff}}} : \aleph_{\!\!1}^{\text{op}} \longrightarrow \mathcal{L}_{\mathcal{T}^{d}_{\text{eff}}}$ (where ${\mathcal{T}^{d}_{\text{eff}}} \defeq (\mathcal{S}_{\text{eff}}, \emptyset)$) such that
\index{ T@${\mathcal{T}^{d}_{\text{eff}}}$ (fibred effect theory $(\mathcal{S}_{\text{eff}}, \emptyset)$)}
\begin{itemize}
\item $\sem{\Gamma';A}_1 = \sem{\Gamma'}$, 
\item $\sem{\Gamma';A}_2(\gamma\,') = \mathcal{M}_{\gamma\,'}(1)$, 
\item $\sem{\Gamma';V'_{\!j}}_1 = \id_{\sem{\Gamma'}}  : \sem{\Gamma'} \longrightarrow \sem{\Gamma'}$, 
\item $(\sem{\Gamma';V'_{\!j}}_2)_{\gamma\,'} : 1 \longrightarrow \bigsqcap_{a \in \sem{\Gamma;A'_{\!j}}_2(\gamma)}(\sem{\Gamma';A}_2(\gamma\,'))$, 
\item $\sem{\Gamma';W_{\sigalgop}}_1 = \id_{\sem{\Gamma'}} : \sem{\Gamma'} \longrightarrow \sem{\Gamma'}$, and
\item $(\sem{\Gamma';W_{\sigalgop}}_2)_{\gamma\,'} = \bigsqcap_{\langle i , f \rangle} \sigalgop^{\mathcal{M}_{\gamma\,'}}_{i} \,\comp\,\, \bigsqcap_{\langle i , f \rangle} (\star \mapsto f) \,\comp\,\, \langle \id_1 \rangle_{\langle i , f \rangle}$
\\[-7.5mm]

\hspace{3.5cm} $: 1 \longrightarrow \bigsqcap_{\langle i , f \rangle \in \bigsqcup_{i \in \sem{\diamond;I}_2(\star)} \bigsqcap_{o \in {\sem{x : I ; O}_2\, \langle \star , i \rangle}} (\sem{\Gamma';A}_2(\gamma\,'))} (\sem{\Gamma';A}_2(\gamma\,'))$, 
\end{itemize}
then 
\[
\sem{\Gamma'; \efftrans T {\!\!A; \overrightarrow{V_i}; \overrightarrow{V'_{\!j}}; \overrightarrow{W_{\sigalgop}}}}_1 = \id_{\sem{\Gamma'}} : \sem{\Gamma'} \longrightarrow \sem{\Gamma'}
\]
and, for all $\gamma\,'$ in $\sem{\Gamma'}$, the function 
\[
(\sem{\Gamma'; \efftrans T {\!\!A; \overrightarrow{V_i}; \overrightarrow{V'_{\!j}}; \overrightarrow{W_{\sigalgop}}}}_2)_{\gamma\,'} : 1 \longrightarrow \sem{\Gamma';A}_2(\gamma\,')
\]
is defined and equal to the following composite function:
\[
\xymatrix@C=4em@R=5em@M=0.5em{
1 
\ar[rr]^-{\langle (\sem{\Gamma';V'_{\!j}}_2)_{\gamma\,'} \rangle_{w_{\!j} : A'_{\!j} \in \Delta}}
&&
\bigsqcap_{w_{\!j} : A'_{\!j} \in \Delta} \bigsqcap_{a \in {\sem{\Gamma;A'_{\!j}}_2(\gamma)}} (\sem{\Gamma';A}_2(\gamma\,'))
\ar[d]^-{\cong}
\\
\sem{\Gamma';A}_2(\gamma\,')
&
\mathcal{M}_{\gamma\,'}(1)
\ar[l]^-{=}
&
\mathcal{M}_{\gamma\,'}(\vert \Delta^\gamma \vert)
\ar[l]^-{\mathcal{M}_{\gamma\,'}(\lj {\Delta^\gamma\,\,} {\,T^\gamma})}
}
\]
where $\vert \Delta^\gamma \vert$ denotes the length of the context $\Delta^\gamma$; and where we use the abbreviation
\[
\gamma \,\defeq \langle \langle \langle \star , (\sem{\Gamma'; V_1}_2)_{\gamma\,'}(\star) \rangle , \ldots \rangle , (\sem{\Gamma'; V_n}_2)_{\gamma\,'}(\star) \rangle
\]
\end{proposition}

\begin{proof}
We prove this proposition by induction on the given derivation of $\lj {\Gamma \vertbar \Delta} T$, using the eMLTT$_{\mathcal{T}_{\text{eff}}}$ versions of Propositions~\ref{prop:semweakening2},~\ref{prop:semsubstitution2},~\ref{prop:semweakening5}, and~\ref{prop:semsubstitution5} to relate syntactic weakening and substitution to their semantic counterparts. 
We postpone the straightforward but lengthy details of this proof to Appendix~\ref{sect:proofofprop:relatingsemanticsoffibeffectterms}.
\end{proof}

We are now ready to prove the soundness of the interpretation of eMLTT$_{\mathcal{T}_{\text{eff}}}$ in the fibred adjunction model given by the split fibred adjunction $\widehat{F_{\!\mathcal{L}_{\mathcal{T}_{\text{eff}}}}} \dashv\, \widehat{U_{\!\mathcal{L}_{\mathcal{T}_{\text{eff}}}}}$.

\subsection*{Extending Theorem~\ref{thm:soundness} (Soundness) to eMLTT$_{\!\mathcal{T}_{\text{eff}}}$}
\index{soundness theorem}

We begin by recalling that in Theorem~\ref{thm:soundness} we showed that the \emph{a priori} partially defined interpretation function $\sem{-}$  is defined on well-formed types and contexts, and well-typed terms, and that it maps definitionally equal contexts, types, and terms to equal objects and morphisms. For example, given $\ceq \Gamma M N \ul{C}$, we showed that
\[
\sem{\Gamma;M} 
=
\sem{\Gamma;N} 
: 1_{\sem{\Gamma}} \longrightarrow \widehat{U_{\mathcal{L}_{\mathcal{T}_{\text{eff}}}}}(\sem{\Gamma;\ul{C}})
\]

When extending Theorem~\ref{thm:soundness} to eMLTT$_{\mathcal{T}_{\text{eff}}}$, we keep the basic proof principle the same: $(a)$--$(l)$ are proved simultaneously, by induction on the given derivations, using the eMLTT$_{\mathcal{T}_{\text{eff}}}$ versions of Propositions~\ref{prop:semweakening2},~\ref{prop:semsubstitution2},~\ref{prop:semsubstitution3}, and~\ref{prop:semsubstitution4} to relate weakening and substitution to their semantic counterparts. As mentioned earlier, these propositions extend straightforwardly from eMLTT to eMLTT$_{\mathcal{T}_{\text{eff}}}$.

The case for algebraic operations is analogous to other computation terms that involve variable bindings and type annotations. Namely, the premises of the typing rule for algebraic operations and the induction hypotheses are enough to satisfy the conditions required for $\sem{\Gamma;\algop^{\ul{C}}_V(y.\, M)} : 1_{\sem{\Gamma}} \longrightarrow \widehat{U_{\mathcal{L}_{\mathcal{T}_{\text{eff}}}}}(\sem{\Gamma;\ul{C}})$ to be defined. 

The case for the congruence rule for algebraic operations is also straightforward. Similarly to the typing rule for algebraic operations, the premises of this rule and the induction hypotheses are enough to ensure that we have $\sem{\Gamma;\algop^{\ul{C}}_V(y.\, M)} = \sem{\Gamma;\algop^{\ul{D}}_W(y.\, N)}$.

Finally, we discuss the cases corresponding to the general algebraicity equation and the equations involving the translation of effect terms into value terms in detail.

\vspace{0.2cm}

\noindent\textbf{General algebraicity equation:} In this case, the given derivation ends with
\vspace{0.2cm}
\[
\mkrulelabel
{\ceq \Gamma {K[\algop^{\ul{C}}_V(y . M)/z]} {\algop^{\ul{D}}_V(y . K[M/z])} {\ul{D}}}
{
\vj \Gamma V {I} 
\quad 
\cj {\Gamma, y \!:\! O[V/x]} M {\ul{C}} 
\quad 
\hj \Gamma {z \!:\! \ul{C}} K {\ul{D}}}
{(\sigalgop : (x \!:\! I) \longrightarrow O \in \mathcal{S}_{\text{eff}})}
\]
and we need to show that
\[
\sem{\Gamma;{K[\algop^{\ul{C}}_V(y . M)/z]}} = \sem{\Gamma;{\algop^{\ul{D}}_V(y . K[M/z])}} : 1_{\sem{\Gamma}} \longrightarrow \widehat{U_{\mathcal{L}_{\mathcal{T}_{\text{eff}}}}}(\sem{\Gamma;\ul{D}})
\]
which, for the fibred adjunction model we are working with, is equivalent to showing 
\[
\sem{\Gamma;{K[\algop^{\ul{C}}_V(y . M)/z]}}_1 = \sem{\Gamma;{\algop^{\ul{D}}_V(y . K[M/z])}}_1 = \id_{\sem{\Gamma}} : \sem{\Gamma} \longrightarrow \sem{\Gamma}
\]
and, for all $\gamma$ in $\sem{\Gamma}$, that
\[
(\sem{\Gamma;{K[\algop^{\ul{C}}_V(y . M)/z]}}_2)_\gamma = (\sem{\Gamma;{\algop^{\ul{D}}_V(y . K[M/z])}}_2)_\gamma : 1 \longrightarrow U_{\!\mathcal{L}_{\mathcal{T}_{\text{eff}}}}(\sem{\Gamma;\ul{D}}_2(\gamma))
\]

First, we use $(d)$ on the given derivation of $\vj \Gamma V {I}$, $(e)$ on the given derivation of \linebreak $\cj {\Gamma, y \!:\! O[V/x]} M {\ul{C}}$, and $(f)$ on the given derivation of $\hj \Gamma {z \!:\! \ul{C}} K {\ul{D}}$, in combination with the propositions that relate weakening and substitution to reindexing along semantic projection and substitution morphisms, to get 
\[
\begin{array}{c}
\sem{\Gamma;V}_1 = \id_{\sem{\Gamma}} : \sem{\Gamma} \longrightarrow \sem{\Gamma}
\\[3mm]
(\sem{\Gamma;V}_2)_\gamma : 1 \longrightarrow \sem{\diamond; I}_2(\star)
\\[3mm]
\sem{\Gamma, y \!:\! O[V/x]; M}_1 = \id_{\bigsqcup_{\gamma \in \sem{\Gamma}} (\sem{x : I; O}_2\, \langle \star , (\sem{\Gamma;V}_2)_\gamma(\star) \rangle)} : \sem{\Gamma, y \!:\! O[V/x]} \longrightarrow \sem{\Gamma, y \!:\! O[V/x]}
\\[3mm]
(\sem{\Gamma, y \!:\! O[V/x]; M}_2)_{\langle \gamma, o \rangle} : 1 \longrightarrow U_{\!\mathcal{L}_{\mathcal{T}_{\text{eff}}}}(\sem{\Gamma;\ul{C}}_2(\gamma))
\\[3mm]
\sem{\Gamma; z \!:\! \ul{C}; K}_1 = \id_{\sem{\Gamma}} : \sem{\Gamma} \longrightarrow \sem{\Gamma}
\\[3mm]
(\sem{\Gamma; z \!:\! \ul{C}; K}_2)_\gamma : \sem{\Gamma;\ul{C}}_2(\gamma) \longrightarrow \sem{\Gamma;\ul{D}}_2(\gamma)
\end{array}
\]

Next, we observe that the first required equation
\[
\sem{\Gamma;{K[\algop^{\ul{C}}_V(y . M)/z]}}_1 = \sem{\Gamma;{\algop^{\ul{D}}_V(y . K[M/z])}}_1 = \id_{\sem{\Gamma}} : \sem{\Gamma} \longrightarrow \sem{\Gamma}
\]
follows straightforwardly by unfolding the definition of $\sem -$ for both sides of the equation, and by noting that the first components of all involved morphisms are identities.

Finally, for all $\gamma$ in $\sem{\Gamma}$, the second required equation
\[
(\sem{\Gamma;{K[\algop^{\ul{C}}_V(y . M)/z]}}_2)_\gamma = (\sem{\Gamma;{\algop^{\ul{D}}_V(y . K[M/z])}}_2)_\gamma : 1 \longrightarrow U_{\!\mathcal{L}_{\mathcal{T}_{\text{eff}}}}(\sem{\Gamma;\ul{D}}_2(\gamma))
\]
follows from the commutativity of the diagram below, where the two top-to-bottom composite morphisms, along the perimeter of the diagram, can be shown to be equal to the two sides of the above equation, using the definition of $\sem -$ and Proposition~\ref{prop:semsubstitution3}.

\[
\hspace{-0.5cm}
\scriptsize
\xymatrix@C=5em@R=4em@M=0.5em{
&
1
\ar[d]^-{\langle \id_1 \rangle_{o \in \sem{\Gamma; O [V/x]}_2(\gamma)}}
\ar@/_3pc/[dddl]_-{(\sem{\Gamma;\algop^{\ul{C}}_V(y.\, M)}_2)_\gamma}^<<<<<<<<<<<<<<<<<{\qquad\dscomment{\text{def. of } \sem{-}}}
&
\\
&
\bigsqcap_o 1
\ar[d]_-{\bigsqcap_o ((\sem{\Gamma, y; M}_2)_{\langle \gamma , o \rangle})}^-{\qquad\dscomment{\text{weakening}}}
\ar@/^2.5pc/[ddr]^-{\quad\bigsqcap_o ((\sem{\Gamma, y ; K[M/z]}_2)_{\langle \gamma , o \rangle})}_>>>>>>>>>>>{\dscomment{\text{Proposition~\ref{prop:semsubstitution3}}}\quad\,\,\,\,\,}
\\
&
\bigsqcap_o (U_{\!\mathcal{L}_{\mathcal{T}_{\text{eff}}}}(\sem{\Gamma;\ul{C}}_2(\gamma)))
\ar@/_1.5pc/[dd]_-{=}
\ar[dl]^-{\sigalgop^{\sem{\Gamma;\ul{C}}_2(\gamma)}_{(\sem{\Gamma;V}_2)_\gamma(\star)}}
\ar@/_1pc/[dr]_-{\bigsqcap_o (U_{\!\mathcal{L}_{\mathcal{T}_{\text{eff}}}}((\sem{\Gamma; z : \ul{C}; K}_2)_\gamma))\qquad\quad}
\\
U_{\!\mathcal{L}_{\mathcal{T}_{\text{eff}}}}(\sem{\Gamma;\ul{C}}_2(\gamma))
\ar[dddddd]_-{U_{\!\mathcal{L}_{\mathcal{T}_{\text{eff}}}}((\sem{\Gamma; z : \ul{C}; K}_2)_\gamma)}
&
&
\bigsqcap_o (U_{\!\mathcal{L}_{\mathcal{T}_{\text{eff}}}}(\sem{\Gamma;\ul{D}}_2(\gamma)))
\ar[d]^-{=}_<<<<{\dscomment{\text{def. of } U_{\!\mathcal{L}_{\mathcal{T}_{\text{eff}}}}}\qquad\qquad\qquad\qquad}
\ar@/^5.5pc/[dddddd]
\\
&
\bigsqcap_o ((\sem{\Gamma;\ul{C}}_2(\gamma))(1))
\ar[d]_-{\cong}_-{\dscomment{\text{def. of } \sigalgop^{\sem{\Gamma;\ul{C}}_2(\gamma)}_{(\sem{\Gamma;V}_2)_\gamma(\star)}}\qquad\quad}^-{\qquad\qquad\dscomment{\text{preservation of countable products}}}
\ar[r]^-{\bigsqcap_o (((\sem{\Gamma; z : \ul{C}; K}_2)_\gamma)_1)}
&
\bigsqcap_o ((\sem{\Gamma;\ul{D}}_2(\gamma))(1))
\ar@/^1pc/[dd]^-{\cong}^>>>>>{\qquad\dscomment{\text{def.}}}
\\
&
(\sem{\Gamma;\ul{C}}_2(\gamma))(\vert \sem{\Gamma;O[V/x]}_2(\gamma) \vert)
\ar[ddd]^<<<<<<<<<<<<<<<<<<<<<{(\sem{\Gamma;\ul{C}}_2(\gamma))(\lj {\overrightarrow{x_o}\,\,} {\,\sigalgop_{(\sem{\Gamma;V})_\gamma(\star)}(x_o)_o})}
\ar[dr]^-{\,\,\,\,\,\,\qquad((\sem{\Gamma; z : \ul{C}; K}_2)_\gamma)_{\vert \sem{\Gamma;O[V/x]}_2(\gamma) \vert}}
&
\\
&
&
(\sem{\Gamma;\ul{D}}_2(\gamma))(\vert \sem{\Gamma;O[V/x]}_2(\gamma) \vert)
\ar[dd]_>>>>>>{(\sem{\Gamma;\ul{D}}_2(\gamma))(\lj {\overrightarrow{x_o}\,\,} {\,\sigalgop_{(\sem{\Gamma;V})_\gamma(\star)}(x_o)_o})}^-{\!\!\!\qquad\sigalgop^{\sem{\Gamma;\ul{D}}_2(\gamma)}_{(\sem{\Gamma;V}_2)_\gamma(\star)}}_-{\dscomment{\text{nat. of } (\sem{\Gamma; z : \ul{C}; K}_2)_\gamma}\qquad\qquad\,\,\,\,\,\,\,\,\,}
\\
&
&
\\
&
(\sem{\Gamma;\ul{C}}_2(\gamma))(1)
\ar[r]_-{((\sem{\Gamma; z : \ul{C}; K}_2)_\gamma)_1}
\ar@/^2pc/[uuuuul]^-{=}
&
(\sem{\Gamma;\ul{D}}_2(\gamma))(1)
\ar[d]^-{=}_-{\dscomment{\text{def. of } U_{\!\mathcal{L}_{\mathcal{T}_{\text{eff}}}}}\qquad\qquad\qquad\qquad\qquad\qquad\qquad\qquad}
\\
U_{\!\mathcal{L}_{\mathcal{T}_{\text{eff}}}}(\sem{\Gamma;\ul{D}}_2(\gamma))
\ar[rr]_-{\id_{U_{\!\mathcal{L}_{\mathcal{T}_{\text{eff}}}}(\sem{\Gamma;\ul{D}}_2(\gamma))}}
&
&
U_{\!\mathcal{L}_{\mathcal{T}_{\text{eff}}}}(\sem{\Gamma;\ul{D}}_2(\gamma))
}
\]

\vspace{0.75cm}

\noindent\textbf{Equations involving the translation of effect terms:} In this case, the given derivation ends with
\[
\mkrulelabel
{
\veq {\Gamma'} {\efftrans {T_1} {U\ul{C}; \overrightarrow{V_i}; \overrightarrow{V'_{\!j}}; \overrightarrow{W_{\sigalgop}}}} {\efftrans {T_2} {U\ul{C}; \overrightarrow{V_i}; \overrightarrow{V'_{\!j}}; \overrightarrow{W_{\sigalgop}}}} {U\ul{C}}
}
{
\begin{array}{c@{\qquad\quad} l}
\lj {\Gamma'} \ul{C}
\\
\vj {\Gamma'} {V_i} {A_i[V_1/x_1, \ldots, V_{i-1}/x_{i-1}]} & (1 \leq i \leq n)
\\
\vj {\Gamma'} {V'_{\!j}} {A'_{\!j}[\overrightarrow{V_i}/\overrightarrow{x_i}] \to U\ul{C}} & (1 \leq j \leq m)
\end{array}
}
{(\ljeq {\Gamma \vertbar \Delta} {T_1} {T_2} \in \!\mathcal{E}_{\text{eff}})}
\]
where 
\[
\begin{array}{c}
\hspace{-8cm}
W_{\sigalgop} \defeq \lambda x' \!:\! (\Sigma\, x \!:\! I.\, O \to U\ul{C}).\, 
\\
\hspace{1.75cm}
\pmatch {x'} {(x \!:\! I, y \!:\! O \to U\ul{C})} {x''\!.\, U\ul{C}} {\thunk (\algop^{\ul{C}}_{x}(y'.\, \force {\ul{C}} {(y\,\, y')}))} 
\end{array}
\]
and we need to show 
\[
\sem{\Gamma';\efftrans {T_1} {U\ul{C}; \overrightarrow{V_i}; \overrightarrow{V'_{\!j}}; \overrightarrow{W_{\sigalgop}}}} = \sem{\Gamma'; \efftrans {T_2} {U\ul{C}; \overrightarrow{V_i}; \overrightarrow{V'_{\!j}}; \overrightarrow{W_{\sigalgop}}}} : 1_{\sem{\Gamma}} \longrightarrow \widehat{U_{\mathcal{L}_{\mathcal{T}_{\text{eff}}}}}(\sem{\Gamma';\ul{C}})
\]
which, for the fibred adjunction model we are working with, is equivalent to showing 
\[
\sem{\Gamma';\efftrans {T_1} {U\ul{C}; \overrightarrow{V_i}; \overrightarrow{V'_{\!j}}; \overrightarrow{W_{\sigalgop}}}}_1 = \sem{\Gamma'; \efftrans {T_2} {U\ul{C}; \overrightarrow{V_i}; \overrightarrow{V'_{\!j}}; \overrightarrow{W_{\sigalgop}}}}_1 = \id_{\sem{\Gamma'}} : \sem{\Gamma'} \longrightarrow \sem{\Gamma'}
\]
and, for all $\gamma\,'$ in $\sem{\Gamma'}$, that 
\[
(\sem{\Gamma';\efftrans {T_1} {U\ul{C}; \overrightarrow{V_i}; \overrightarrow{V'_{\!j}}; \overrightarrow{W_{\sigalgop}}}}_2)_{\gamma\,'} = (\sem{\Gamma'; \efftrans {T_2} {U\ul{C}; \overrightarrow{V_i}; \overrightarrow{V'_{\!j}}; \overrightarrow{W_{\sigalgop}}}}_2)_{\gamma\,'} : 1 \longrightarrow U_{\!\mathcal{L}_{\mathcal{T}_{\text{eff}}}}(\sem{\Gamma;\ul{C}}_2(\gamma))
\]

First, we use $(c)$ on the given derivation of $\lj {\Gamma'} \ul{C}$ and $(d)$ on the given derivations of  $\vj {\Gamma'} {V_i} {A_i[V_1/x_1, \ldots, V_{i-1}/x_{i-1}]}$ and $\vj {\Gamma'} {V'_{\!j}} {A'_{\!j}[\overrightarrow{V_i}/\overrightarrow{x_i}] \to U\ul{C}}$, for all $1 \leq i \leq n$ and $1 \leq j \leq m$, in combination with the propositions that relate weakening and substitution to reindexing along semantic projection and substitution morphisms, to get 
\[
\begin{array}{c}
\sem{\Gamma';\ul{C}}_1 = \sem{\Gamma'}
\\[3mm]
\sem{\Gamma';\ul{C}}_2 : \sem{\Gamma'} \longrightarrow \Mod(\!\mathcal{L}_{\mathcal{T}_{\text{eff}}},\Set)
\\[3mm]
\sem{\Gamma';V_i}_1 = \id_{\sem{\Gamma'}}  : \sem{\Gamma'} \longrightarrow \sem{\Gamma'}
\\[3mm]
\hspace{-10.5cm}
(\sem{\Gamma';V_i}_2)_{\gamma\,'} : 1 \longrightarrow 
\\[-1mm]
\hspace{2cm}
\sem{x_1 \!:\! A_1, \ldots, x_{i-1} \!:\! A_{i - 1};A_i}_2\, \langle \langle \langle \star , (\sem{\Gamma'; V_1}_2)_{\gamma\,'}(\star) \rangle , \ldots \rangle , (\sem{\Gamma'; V_{i - 1}}_2)_{\gamma\,'}(\star) \rangle
\\[3mm]
\sem{\Gamma';V'_{\!j}}_1 = \id_{\sem{\Gamma'}}  : \sem{\Gamma'} \longrightarrow \sem{\Gamma'}
\\[3mm]
(\sem{\Gamma';V'_{\!j}}_2)_{\gamma\,'} : 1 \longrightarrow \bigsqcap_{a \in \sem{\Gamma;A'_{\!j}}_2(\gamma)}(U_{\!\mathcal{L}_{\mathcal{T}_{\text{eff}}}}(\sem{\Gamma';\ul{C}}_2(\gamma\,')))
\end{array}
\]
where
\[
\gamma \defeq \langle \langle \langle \star , (\sem{\Gamma'; V_1}_2)_{\gamma\,'}(\star) \rangle , \ldots \rangle , (\sem{\Gamma'; V_n}_2)_{\gamma\,'}(\star) \rangle
\]
In particular, we prove equations involving simultaneous substitutions, e.g.,   
\[
\begin{array}{c}
\sem{\Gamma';A_i[V_1/x_1, \ldots, V_{i-1}/x_{i-1}]}_2(\gamma\,')
\\
=
\\
\sem{x_1 \!:\! A_1, \ldots, x_{i-1} \!:\! A_{i - 1};A_i}_2\, \langle \langle \langle \star , (\sem{\Gamma'; V_1}_2)_{\gamma\,'}(\star) \rangle , \ldots \rangle , (\sem{\Gamma'; V_{i - 1}}_2)_{\gamma\,'}(\star) \rangle
\end{array}
\]
by first noting that analogously to the proof of Theorem~\ref{thm:simultaneoussubstitution}, we can first show that 
\[
A_i[V_1/x_1, \ldots, V_{i-1}/x_{i-1}] = A_i[x'_1/x_1]\ldots[x'_{i-1}/x_{i-1}][V_1/x'_1]\ldots[V_{i-1}/x'_{i-1}]
\]

\pagebreak
\noindent
for freshly chosen value variables $x'_1, \ldots, x'_{i-1}$, and then use the eMLTT$_{\mathcal{T}_{\text{eff}}}$ versions of the propositions
that relate weakening and (unary) substitution to reindexing along semantic projection and substitution morphisms (in particular, see Proposition~\ref{prop:semsubstitution5}).

Next, by letting ${\mathcal{T}^{d}_{\text{eff}}} \defeq (\mathcal{S}_{\text{eff}}, \emptyset)$, we get that  $\mathbb{E}_{\!\mathcal{T}^{d}_{\text{eff}}} \subseteq \mathbb{E}_{\!\mathcal{T}_{\text{eff}}}$, based on the definitions of $\mathbb{E}_{\!\mathcal{T}^{d}_{\text{eff}}}$ and $\mathbb{E}_{\!\mathcal{T}_{\text{eff}}}$. Using this inclusion, we get a morphism of countable Lawvere theories from $I_{\mathcal{T}^d_{\text{eff}}} : \aleph_{\!\!1}^{\text{op}} \longrightarrow \mathcal{L}_{\mathcal{T}^d_{\text{eff}}}$ to $I_{\mathcal{T}_{\text{eff}}} : \aleph_{\!\!1}^{\text{op}} \longrightarrow \mathcal{L}_{\mathcal{T}_{\text{eff}}}$, by defining the corresponding functor $\mathcal{L}_{\mathcal{T}^d_{\text{eff}}} \longrightarrow \mathcal{L}_{\mathcal{T}_{\text{eff}}}$ as identity on objects and by sending every tuple of terms (i.e., a morphism in $\mathcal{L}_{\mathcal{T}^d_{\text{eff}}}$) to its equivalence class (i.e., a morphism in $\mathcal{L}_{\mathcal{T}_{\text{eff}}}$). 

It is well-known that any morphism of countable Lawvere theories induces a functor between the corresponding categories of models, defined by composition of countable product preserving functors,  and going in the opposite direction. Concretely, for the purposes of this thesis, there exists a functor $\Mod(\!\mathcal{L}_{\mathcal{T}_{\text{eff}}},\Set) \longrightarrow \Mod(\!\mathcal{L}_{\mathcal{T}^d_{\text{eff}}},\Set)$, meaning that every model of $I_{\mathcal{T}_{\text{eff}}} : \aleph_{\!\!1}^{\text{op}} \!\longrightarrow\! \mathcal{L}_{\mathcal{T}_{\text{eff}}}$ is also a model of $I_{\mathcal{T}^d_{\text{eff}}} : \aleph_{\!\!1}^{\text{op}} \!\longrightarrow\! \mathcal{L}_{\mathcal{T}^d_{\text{eff}}}$. 

In particular, the above observation means that $\sem{\Gamma';\ul{C}}_2(\gamma\,') : \mathcal{L}_{\mathcal{T}_{\text{eff}}} \longrightarrow \Set$ is also a model of $I_{\mathcal{T}^d_{\text{eff}}} : \aleph_{\!\!1}^{\text{op}} \longrightarrow \mathcal{L}_{\mathcal{T}^d_{\text{eff}}}$, for all $\gamma\,'$ in $\sem{\Gamma'}$.

Another observation we make is that by unfolding the definition of $\sem{-}$ for lambda abstractions, pattern-matching, thunking, algebraic operations, and forcing, we get 
\[
\begin{array}{c}
\hspace{-4cm}
(\sem{\Gamma';W_{\sigalgop}}_2)_{\gamma\,'} = \bigsqcap_{\langle i , f \rangle} \sigalgop^{\mathcal{M}_{\gamma\,'}}_{i} \,\comp\,\, \bigsqcap_{\langle i , f \rangle} (\star \mapsto f) \,\comp\,\, \langle \id_1 \rangle_{\langle i , f \rangle}
\\[1mm]
\hspace{1.5cm}: 1 \longrightarrow \bigsqcap_{\langle i , f \rangle \in \bigsqcup_{i \in \sem{\diamond;I}_2(\star)} \bigsqcap_{o \in {\sem{x : I ; O}_2\, \langle \star , i \rangle}} (U_{\!\mathcal{L}_{\mathcal{T}_{\text{eff}}}}(\sem{\Gamma';\ul{C}}_2(\gamma\,')))} (U_{\!\mathcal{L}_{\mathcal{T}_{\text{eff}}}}(\sem{\Gamma';\ul{C}}_2(\gamma\,')))
\end{array}
\]

As a consequence of these observations, we can now use Proposition~\ref{prop:relatingsemanticsoffibeffectterms} to prove the required equations. In more detail, the first required equation
\[
\sem{\Gamma';\efftrans {T_1} {U\ul{C}; \overrightarrow{V_i}; \overrightarrow{V'_{\!j}}; \overrightarrow{W_{\sigalgop}}}}_1 = \sem{\Gamma'; \efftrans {T_2} {U\ul{C}; \overrightarrow{V_i}; \overrightarrow{V'_{\!j}}; \overrightarrow{W_{\sigalgop}}}}_1 = \id_{\sem{\Gamma'}} : \sem{\Gamma'} \longrightarrow \sem{\Gamma'}
\]
follows immediately from Proposition~\ref{prop:relatingsemanticsoffibeffectterms}. To prove the second required equation
\[
(\sem{\Gamma';\efftrans {T_1} {U\ul{C}; \overrightarrow{V_i}; \overrightarrow{V'_{\!j}}; \overrightarrow{W_{\sigalgop}}}}_2)_{\gamma\,'} = (\sem{\Gamma'; \efftrans {T_2} {U\ul{C}; \overrightarrow{V_i}; \overrightarrow{V'_{\!j}}; \overrightarrow{W_{\sigalgop}}}}_2)_{\gamma\,'} : 1 \longrightarrow U_{\!\mathcal{L}_{\mathcal{T}_{\text{eff}}}}(\sem{\Gamma;\ul{C}}_2(\gamma))
\]
for all $\gamma\,'$ in $\sem{\Gamma'}$, we combine Proposition~\ref{prop:relatingsemanticsoffibeffectterms} with the following commuting diagram:
\[
\xymatrix@C=2em@R=3em@M=0.5em{
1
\ar[d]_-{\langle (\sem{\Gamma';V'_{\!j}}_2)_{\gamma\,'} \rangle_{w_{\!j} : A'_{\!j} \in \Delta}}
\\
\bigsqcap_{w_{\!j} : A'_{\!j} \in \Delta} \bigsqcap_{a \in {\sem{\Gamma;A'_{\!j}}_2(\gamma)}} (U_{\!\mathcal{L}_{\mathcal{T}_{\text{eff}}}}(\sem{\Gamma';\ul{C}}_2(\gamma\,')))
\ar[d]_-{\cong}
\\
(\sem{\Gamma';\ul{C}}_2(\gamma\,'))(\vert \Delta^\gamma \vert)
\ar@/_4pc/[dd]_-{(\sem{\Gamma';\ul{C}}_2(\gamma\,'))(\lj {\Delta^\gamma\,\,} {\,T_1^\gamma})}^-{\,\,\,\,\,\,\,\quad\dcomment{\ljeq {\Delta^\gamma} {T_1^\gamma} {T_2^\gamma}}}
\ar@/^4pc/[dd]^-{(\sem{\Gamma';\ul{C}}_2(\gamma\,'))(\lj {\Delta^\gamma\,\,} {\,T_2^\gamma})}
\\
\\
(\sem{\Gamma';\ul{C}}_2(\gamma\,'))(1)
\ar[d]_-{=}
\\
U_{\!\mathcal{L}_{\mathcal{T}_{\text{eff}}}}(\sem{\Gamma';\ul{C}}_2(\gamma\,'))
}
\]
where the equation $\ljeq {\Delta^\gamma} {T_1^\gamma} {T_2^\gamma}$ follows from the assumed equation $\ljeq {\Gamma \vertbar \Delta} {T_1} {T_2}$, based on the way the set of equations $\mathbb{E}_{\!\mathcal{T}_{\text{eff}}}$ is derived from $\mathcal{S}_{\text{eff}}$ in Definition~\ref{def:countableeqthfromeffth}.

\section{Generic effects}
\label{ref:genericeffects}

\index{generic effect}
\index{ g@$\mathtt{gen}_{\sigalgop}$ (generic effect)}
It is worth noting that instead of extending eMLTT with algebraic operations $\algop^{\ul{C}}_V(y.\, M)$, for all operation symbols $\sigalgop : (x \!:\! I) \longrightarrow O$ in the given fibred effect signature $\mathcal{S}_{\text{eff}}$, we could have alternatively extended eMLTT with \emph{generic effects}, analogously to the simply typed setting~\cite{Plotkin:AlgOperations}. More precisely, we could have extended eMLTT's computation terms with function constants $\mathtt{gen}_{\sigalgop} : \Pi\, x \!:\! I .\, FO$, for all $\sigalgop : (x \!:\! I) \longrightarrow O$ in $\mathcal{S}_{\text{eff}}$.

While algebraic operations are more convenient to reason about (equationally), generic effects are closer to the language primitives that programmers are familiar, e.g., from ML-style languages. 
However, analogously to the simply typed setting, these two ways of extending eMLTT are in fact equivalent, as we show below. 

On the one hand, we can define generic effects using algebraic operations as
\[
\mathtt{gen}_{\sigalgop} \defeq  \lambda\, x \!:\! I .\, \algop^{FO}_x(y.\, \return y)
\]
Clearly, the right-hand side has type $\Pi\, x \!:\! I .\, FO$ in the empty context, by simply using the typing rules for lambda abstraction, algebraic operations, and returning a value.

On the other hand, we can define algebraic operations using generic effects as
\[
\algop^{\ul{C}}_V(y.\, M) \defeq \doto {(\mathtt{gen}_{\sigalgop}\, V)} {y \!:\! O[V/x]} {} {M}
\]
Clearly, if $\vj \Gamma V I$, $\lj \Gamma {\ul{C}}$, and $\cj {\Gamma, y \!:\! O[V/x]} M {\ul{C}}$, the right-hand side is well-typed in $\Gamma$ at computation type $\ul{C}$, by simply using the typing rule for sequential composition.

\begin{proposition}
These two definitions of generic effects and algebraic operations in terms of each other constitute an isomorphism. 
\end{proposition}

\begin{proof}
The proof is exactly the same as one would have in the simply typed setting, modulo the possibility of $O$ depending on values of type $I$.

In one direction, we have
\begin{fleqn}[0.3cm]
\begin{align*}
\Gamma \,\vdash\,\, & \doto {\big((\lambda\, x \!:\! I .\, \algop^{FO}_x(y.\, \return y))\, V\big)} {y \!:\! O[V/x]} {} {M}
\\
=\,\, & \doto {\big((\lambda\, x \!:\! I .\, \algop^{FO}_x(y.\, \return y))\, V\big)} {y' \!:\! O[V/x]} {} {M[y'/y]}
\\
=\,\, & \doto {\algop^{FO[V/x]}_V(y.\, \return y)} {y' \!:\! O[V/x]} {} {M[y'/y]}
\\
=\,\, & \algop^{\ul{C}}_V\big(y.\, \doto {(\return y)} {y' \!:\! O[V/x]} {} {M[y'/y]}\big)
\\
=\,\, & \algop^{\ul{C}}_V(y.\, M) : \ul{C}
\end{align*}
\end{fleqn}

In the other direction, we have 
\begin{fleqn}[0.3cm]
\begin{align*}
\Gamma \,\vdash\,\, & \lambda\, x \!:\! I .\, \doto {(\mathtt{gen}_{\sigalgop}\, x)} {y \!:\! O} {} {\return y}
\\
=\,\, & \lambda\, x \!:\! I .\, \doto {(\mathtt{gen}_{\sigalgop}\, x)} {y \!:\! O} {} {z[\return y/z]}
\\
=\,\, & \lambda\, x \!:\! I .\, \mathtt{gen}_{\sigalgop}\, x
\\
=\,\, & \mathtt{gen}_{\sigalgop} : \Pi\, x \!:\! I .\, FO
\end{align*}
\end{fleqn}
\end{proof}


\chapter[eMLTT$_{\!\mathcal{T}_{\text{eff}}}^{\mathcal{H}}$: an extension of eMLTT$_{\!\mathcal{T}_{\text{eff}}}$ with handlers]{eMLTT$_{\!\mathcal{T}_{\text{eff}}}^{\mathcal{H}}$: an extension of eMLTT$_{\!\mathcal{T}_{\text{eff}}}$ with handlers}
\label{chap:handlers}

\index{ e@eMLTT$_{\mathcal{T}_{\text{eff}}}^{\mathcal{H}}$ (extension of eMLTT$_{\mathcal{T}_{\text{eff}}}$ with handlers of fibred algebraic effects)}
In this chapter we show how to extend eMLTT$_{\mathcal{T}_{\text{eff}}}$ with handlers of fibred algebraic effects. Our work builds on the pioneering work of Plotkin and Pretnar who generalised exception handlers to all algebraic effects in the simply typed setting~\cite{Plotkin:HandlingEffects}.
They also showed how handlers can be used to neatly implement relabelling and restriction in Milner's CCS, timeouts, rollbacks, stream redirection, etc., 
paving the way for handlers to become a practical modular programming language abstraction.

In Section~\ref{sect:handlersoverview}, we recall the conventional definition of handlers and their use in programming languages.
Next, in Section~\ref{sect:problemwithhandlers}, we make an important observation that will be key for the rest of this chapter. Namely, we observe that using the conventional term-level definition of handlers to extend eMLTT$_{\mathcal{T}_{\text{eff}}}$ leads to unsound program equivalences becoming derivable. We solve this problem in Section~\ref{sect:extendingemlttwithuserdefinedalgebras} by giving handlers a novel type-based treatment via a new computation type, the \emph{user-defined algebra type}, which pairs a value type (the carrier) with a family of value terms (the operations). This type internalises Plotkin and Pretnar's insight that handlers denote algebras for a given equational theory of effects. We call this extended language eMLTT$_{\mathcal{T}_{\text{eff}}}^{\mathcal{H}}$. 

We demonstrate the generality of our type-based treatment of handlers by showing in Section~\ref{sect:derivingconventionalhandlers} that their conventional term-level presentation can be routinely derived, and demonstrating in Section~\ref{section:usinghandlersforreasoning} that the type-based treatment provides a useful mechanism for reasoning about effectful computations.
Next, in Section~\ref{section:handlersmetatheory}, we study the meta-theory of eMLTT$_{\mathcal{T}_{\text{eff}}}^{\mathcal{H}}$, and in Section~\ref{sect:derivableisomorphismswithhandlers}, we present some useful derivable equations. Finally, in Section~\ref{sect:interpretingemlttwithhandlers}, we equip eMLTT$_{\mathcal{T}_{\text{eff}}}^{\mathcal{H}}$ with a denotational semantics. In particular, we show how to define a sound interpretation of it in the same fibred adjunction model we used in Secton~\ref{sect:fibalgeffectsmodel} for giving a denotational semantics to eMLTT$_{\mathcal{T}_{\text{eff}}}$.

\section{Handlers of algebraic effects}
\label{sect:handlersoverview}

Handlers of algebraic effects were introduced by Plotkin and Pretnar~\cite{Plotkin:HandlingEffects,Plotkin:Handlers} as a natural generalisation of exception handlers to all algebraic effects. 
Building on the algebraic treatment of computational effects, Plotkin and Pretnar's key insight was to understand exception handlers as defining new algebras for the equational theory of exceptions. 
Taking this insight as a starting point, they then generalised handlers to arbitrary algebraic effects given by (countable) equational theories, where
\begin{itemize}
\item a \emph{handler} defines a new, user-defined algebra for the given equational theory by providing redefinitions of all its algebraic operations; and
\item the \emph{handling} construct denotes the application of the unique mediating homomorphism between the free algebra and the one denoted by the given handler. 
\end{itemize}

\index{handler}
\index{handling construct}
Plotkin and Pretnar formalise these ideas by extending Levy's CBPV with algebraic effects and their handlers, as explained below. They also give their language a denotational semantics using the  adjunction determined by the category of models of the given equational theory of computational effects. In particular, given an effect theory\footnote{In the sense of~\cite{Plotkin:HandlingEffects}, i.e., a non-dependent version of the fibred effect theories defined in Section~\ref{sect:fibeffecttheories}.}  $\mathcal{T}_{\text{eff}} = (\mathcal{S}_{\text{eff}},\mathcal{E}_{\text{eff}})$, they extend CBPV's computation terms with the following term former that combines a handler $\{\mathtt{op}_x(x') \mapsto N_{\sigalgop}\}_{\sigalgop \,\in\, \mathcal{S}_{\text{eff}}}$ with the handling construct:
\[
\mkrule
{\Gamma \vdash {M \mathtt{~handled~with~} \{\mathtt{op}_x(x') \mapsto N_{\sigalgop}\}_{\sigalgop \,\in\, \mathcal{S}_{\text{eff}}} \mathtt{~to~} y \!:\! A \mathtt{~in~} N_{\mathsf{ret}}} : \ul{C}}
{
\cj \Gamma M {FA}
\quad
\{\cj {\Gamma, x \!:\! I, x' \!:\! O \to U\ul{C}} {N_{\sigalgop}} {\ul{C}}\}_{\sigalgop \,:\, I \longrightarrow O \,\in\, \mathcal{S}_{\text{eff}}}
\quad
\cj {\Gamma, y \!:\! A} {N_{\mathsf{ret}}} {\ul{C}}
}
\]
and the equational theory of computation terms with two $\beta$-equations, given by
\[
\begin{array}{c@{\,\,} c@{\,\,} l}
\Gamma & \vdash & {(\algop^{FA}_V(y'\!.\, M)) \mathtt{~handled~with~} \{\mathtt{op}_x(x') \mapsto N_{\sigalgop}\}_{\sigalgop \,\in\, \mathcal{S}_{\text{eff}}} \mathtt{~to~} y \!:\! A \mathtt{~in~} N_{\mathsf{ret}}}
\\[0.5mm]
& = & {N_{\sigalgop}[V/x][\lambda\, y' \!:\! O[V/x] .\, \thunk H/x']} : \ul{C}
\end{array}
\]
and
\[
\begin{array}{c@{\,\,} c@{\,\,} l}
\Gamma & \vdash & {(\return V) \mathtt{~handled~with~} \{\mathtt{op}_x(x') \mapsto N_{\sigalgop}\}_{\sigalgop \,\in\, \mathcal{S}_{\text{eff}}} \mathtt{~to~} y \!:\! A \mathtt{~in~} N_{\mathsf{ret}}} 
\\[0.5mm]
& = & {N_{\mathsf{ret}}[V/y]} : \ul{C}
\end{array}
\]
where, for better readability, we abbreviate the handling construct as
\[
H \,\defeq\, M \mathtt{~handled~with~} \{\mathtt{op}_x(x') \mapsto N_{\sigalgop}\}_{\sigalgop \,\in\, \mathcal{S}_{\text{eff}}} \mathtt{~to~} y \!:\! A \mathtt{~in~} N_{\mathsf{ret}}
\]

It is worth noting that while these $\beta$-equations capture the intuition that the handling construct denotes the application of the mediating homomorphism between the free algebra denoted by $FA$ and the algebra defined by the family of terms $N_{\sigalgop}$, they do not capture the idea that the handling construct denotes the unique such homomorphism. 
To capture the uniqueness of the handling construct, one would also need to extend Plotkin and Pretnar's version of CBPV with a corresponding $\eta$-equation, e.g., as considered in~\cite[Section~6]{Ahman:NBE} for Levy's fine-grain call-by-value language.

From a programming language perspective, the first $\beta$-equation describes that handling consists of traversing the given program and replacing each algebraic operation with the corresponding user-defined term $N_{\sigalgop}$.
The second $\beta$-equation describes that when handling reaches return values (the end of the given program we are handling), a substitution instance of the specified continuation $N_{\mathsf{ret}}$ is evaluated next.

As mentioned earlier, handlers can be used to neatly implement timeouts, rollbacks, stream redirection, etc., see~\cite[Section~3]{Plotkin:HandlingEffects} for details of these and other examples. 

\index{ Terminal@$\mathsf{Terminal}$ (type of terminal names)}
For example, let us consider an extension of the theory of input/output from Example~\ref{ex:fibtheoryofIO} with multiple terminals, where the operation symbols are typed as follows:
\[
\mathsf{read} : \mathsf{Terminal} \longrightarrow \Character
\qquad
\mathsf{write} : \mathsf{Terminal} \times \Character \longrightarrow 1
\]
Now, assuming two distinguished terminal names $V_t$ and $V_{t'}$, we can neatly redirect the output on $V_t$ to $V_{t'}$ by handling a given program using the following handler:
\[
\begin{array}{l c l}
\mathtt{read}_x(x') & \mapsto & \mathtt{read}_x(y' .\, \force {} (x'\, y'))
\\[2mm]
\mathtt{write}_x(x') & \mapsto & \mathtt{if}~ (\mathtt{eq}~(\fst x)~V_t)~\mathtt{then}~(\mathtt{write}_{\langle V_{t'}  , \snd x \rangle}(\force {} (x'\, \star)))
\\
&& \hspace{3.165cm} \mathtt{else}~(\mathtt{write}_{x}(\force {} (x'\, \star)))
\end{array}
\]

More recently, handlers have also gained popularity as a practical and modular programming language abstraction, allowing programmers to write their programs generically in terms of algebraic operations,  and then use handlers to modularly provide different fit-for-purpose implementations of these generic programs. 
A prototypical example of this approach involves implementing the global state operations using the natural representation of stateful programs as state-passing functions $\State \to A \times \State$.

To facilitate this style of programming, Kammar et al.~\cite{Kammar:Handlers} have extended Haskell, OCaml, SML, and Racket with algebraic effects and their handlers, implemented using free monads and (delimited) continuations. Further, Bauer and Pretnar~\cite{Bauer:AlgebraicEffects,Bauer:EffectSystem} have built an entire ML-like language, called Eff, around this style of programming. This style of programming has also been successfully combined with row-based type-and-effect systems, as demonstrated by Hillerstr\"{o}m and Lindley~\cite{Hillerstrom:Liberating}, and Leijen~\cite{Leijen:Handlers}. Handlers are also central to the ongoing effort to extend OCaml with shared memory multicore parallelism (see~\cite{MulticoreOCaml} for details of the Multicore OCaml project), providing a convenient  means for programmers to implement their own fit-for-purpose schedulers.

\index{handler!multi--}
Recently, Lindley et al.~\cite{Lindley:DoBeDoBeDo} have also investigated a generalisation of handlers, called \emph{multihandlers}, that allow multiple computations to be handled simultaneously. A binary instance of this generalisation was discussed by Plotkin in an earlier invited talk~\cite{Plotkin:BinaryHandlers}. In particular, Plotkin showed how to define binary handlers in terms of (standard) unary handlers, and how to use them to implement interleaving concurrency.

\section{Problems with the term-level definition of handlers}
\label{sect:problemwithhandlers}

In this section we make an important observation that will be key to our work regarding an extension of eMLTT$_{\mathcal{T}_{\text{eff}}}$ with handlers of fibred algebraic effects. In particular, we observe that naively following the existing work on handlers to extend eMLTT$_{\mathcal{T}_{\text{eff}}}$ (or any other language with a notion of homomorphism, such as CPBV with stack terms or EEC) would lead to unsound program equivalences becoming derivable. 

More concretely, let us assume we were to extend eMLTT$_{\mathcal{T}_{\text{eff}}}$ with handlers \`a la Plotkin and Pretnar~\cite{Plotkin:HandlingEffects} by extending eMLTT$_{\mathcal{T}_{\text{eff}}}$'s computation terms and the corresponding equational theory as discussed in Section~\ref{sect:handlersoverview}. While this extension suffices for CBPV without stack terms, as considered by Plotkin and Pretnar~\cite{Plotkin:HandlingEffects}, and Kammar et al.~\cite{Kammar:Handlers}, languages that also include a notion of homomorphism (e.g., CBPV with stack terms, EEC, and eMLTT$_{\mathcal{T}_{\text{eff}}}$) ought to be extended further. 
Specifically, in addition to only extending computation terms, one should also extend the corresponding notion of homomorphism with the handling construct. In particular, for eMLTT$_{\mathcal{T}_{\text{eff}}}$ this would mean extending homomorphism terms with the following term former:
\[
{K \mathtt{~handled~with~} \{\mathtt{op}_x(x') \mapsto N_{\sigalgop}\}_{\sigalgop \,\in\, \mathcal{S}_{\text{eff}}} \mathtt{~to~} y \!:\! A \mathtt{~in~} N_{\mathsf{ret}}}
\]

We highlight two reasons for needing such terms when extending eMLTT$_{\mathcal{T}_{\text{eff}}}$
with handlers \`a la Plotkin and Pretnar. First, as the handling construct naturally denotes the application of the mediating homomorphism between a free algebra and the algebra defined by the family of terms $N_{\sigalgop}$, it is natural to also make it into a homomorphism in the language, thus making the language more complete with respect to its models. 
Second, making the handling construct into a homomorphism term is also important to ensure that effectful programs could be combined modularly, e.g., to be able to write
\[
{\doto {M} {(y \!:\! A_1, z \!:\! FA_2)} {} {\big(z} \mathtt{~handled~with~} \{\mathtt{op}_x(x') \mapsto N_{\sigalgop}\}_{\sigalgop \,\in\, \mathcal{S}_{\text{eff}}} \mathtt{~to~} y' \!:\! B \mathtt{~in~} N_{\mathsf{ret}}}\big)
\]
where the term being handled is given by the computation variable $z$.

Unfortunately, if we were to follow this approach for extending eMLTT$_{\mathcal{T}_{\text{eff}}}$ with handlers of fibred algebraic effects, it becomes possible to derive unsound program equivalences in the resulting language, such as the following equation for input/output:
\[
\ceq {\Gamma} {\mathtt{write}^{F1}_{\mathsf{a}}(\return *)} {\mathtt{write}^{F1}_{\mathsf{b}}(\return *)} {F1}
\]

This problem arises from the type of the handling construct not containing any information about the specific handler being used. In particular, recall that a key property of homomorphism terms is that their interaction with algebraic operations is determined exclusively by their types---see the general algebraicity equation given in Definition~\ref{def:extensionofeMLTTwithfibalgeffects}. Unfortunately, this property is not true for the handling construct. Specifically, 
the handling construct gives rise to a critical pair in the equational theory: 
\[
{\algop^{FA}_V(y.\, M) \mathtt{~handled~with~} \{\mathtt{op}_x(x') \mapsto N_{\sigalgop}\}_{\sigalgop \,\in\, \mathcal{S}_{\text{eff}}} \mathtt{~to~} y' \!:\! A \mathtt{~in~} N_{\mathsf{ret}}}
\]
matches both the $\beta$-equation for handlers given in the previous section and the general algebraicity equation given in Definition~\ref{def:extensionofeMLTTwithfibalgeffects}. It is easy to show that this critical pair is not convergent. 
For example, let us consider the following handler for input/output:
\[
\begin{array}{l c l}
\mathtt{read}_x(x') & \mapsto & \mathtt{read}(y' .\, \force {} (x'\, y'))
\\[1mm]
\mathtt{write}_x(x') & \mapsto & \mathtt{write}_{\mathtt{b}}(\force {} (x'\,\star))
\end{array}
\]

On the one hand, the $\beta$-equation for the handling construct allows us to derive
\begin{fleqn}[0.3cm]
\begin{align*}
\Gamma \,\vdash\,\, & {(\mathtt{write}^{F1}_{\mathtt{a}}(\return *)) \mathtt{~handled~with~} \{\mathtt{op}_x(x') \mapsto N_{\sigalgop}\}_{\sigalgop \,\in\, \mathcal{S}_{\text{I/O}}} \mathtt{~to~} y \!:\! 1 \mathtt{~in~} N_{\mathsf{ret}}}
\\
=\,\, & \mathtt{write}^{F1}_{\mathtt{b}}\big({(\return *) \mathtt{~handled~with~} \{\mathtt{op}_x(x') \mapsto N_{\sigalgop}\}_{\sigalgop \,\in\, \mathcal{S}_{\text{I/O}}} \mathtt{~to~} y \!:\! 1 \mathtt{~in~} N_{\mathsf{ret}}}\big)
\\
=\,\, & \mathtt{write}^{F1}_{\mathtt{b}}(\return *) : F1
\end{align*}
\end{fleqn}
assuming that the handler in question 
is defined as above, and $N_{\mathsf{ret}} \defeq \return *$. 

On the other hand, the general algebraicity equation allows us to derive
\begin{fleqn}
\begin{align*}
\hspace{-0.05cm}
\Gamma \,\vdash\,\, & {(\mathtt{write}^{F1}_{\mathtt{a}}(\return *)) \mathtt{~handled~with~} \{\mathtt{op}_x(x') \mapsto N_{\sigalgop}\}_{\sigalgop \,\in\, \mathcal{S}_{\text{I/O}}} \mathtt{~to~} y \!:\! 1 \mathtt{~in~} N_{\mathsf{ret}}}
\\
=\,\, & {\big(z \mathtt{~handled~with~} \{\mathtt{op}_x(x') \mapsto N_{\sigalgop}\}_{\sigalgop \,\in\, \mathcal{S}_{\text{I/O}}} \mathtt{~to~} y \!:\! 1 \mathtt{~in~} N_{\mathsf{ret}}\big)[\mathtt{write}^{F1}_{\mathtt{a}}(\return *)/z]}
\\
=\,\, & \mathtt{write}^{F1}_{\mathtt{a}}\big(\big(z \mathtt{~handled~with~} \{\mathtt{op}_x(x') \mapsto N_{\sigalgop}\}_{\sigalgop \,\in\, \mathcal{S}_{\text{I/O}}} \mathtt{~to~} y \!:\! 1 \mathtt{~in~} N_{\mathsf{ret}}\big)[\return */z]\big)
\\
=\,\, & \mathtt{write}^{F1}_{\mathtt{a}}\big((\return *) \mathtt{~handled~with~} \{\mathtt{op}_x(x') \mapsto N_{\sigalgop}\}_{\sigalgop \,\in\, \mathcal{S}_{\text{I/O}}} \mathtt{~to~} y \!:\! 1 \mathtt{~in~} N_{\mathsf{ret}}\big)
\\
=\,\, & \mathtt{write}^{F1}_{\mathtt{a}}(\return *) : F1
\end{align*}
\end{fleqn}
allowing us to conclude that the following unsound definitional equation is derivable:
\[
\ceq {\Gamma} {\mathtt{write}^{F1}_{\mathsf{a}}(\return *)} {\mathtt{write}^{F1}_{\mathsf{b}}(\return *)} {F1}
\]

From a semantic perspective, the above discussion also exposes a conflict between the term-level definition of handlers, and Plotkin and Pretnar's semantic insight that they ought to denote algebras for a given equational theory of computational effects.

The reason why Plotkin and Pretnar were able to define a sound interpretation for their language 
is because they were using CBPV without stack terms, i.e., without a notion of homomorphism. As CBPV's computation terms are interpreted as elements of the carriers of the algebras denoted by their types, the  interpretation of their language and its soundness were not affected by the type of the handling construct not mentioning the corresponding handler. In particular, the carrier of the algebra denoted by the type of the handling construct is the same as the carrier of the algebra denoted by the corresponding handler. The lack of a notion of homomorphism also meant that their equational theory did not have critical pairs arising from the handling construct.

\section{Extending eMLTT$_{\!\mathcal{T}_{\text{eff}}}$ with a type-based treatment of handlers}
\label{sect:extendingemlttwithuserdefinedalgebras}

As demonstrated in the previous section, we cannot naively follow Plotkin and Pretnar's approach to extend eMLTT$_{\mathcal{T}_{\text{eff}}}$ with handlers of fibred algebraic effects by defining them at the term level for both computation and homomorphism terms. Instead, we either have to i) change the existing equational theory of eMLTT$_{\mathcal{T}_{\text{eff}}}$'s homomorphism terms (e.g., as investigated by Levy for exception handlers in CBPV with stacks; however, in which case the homomorphism terms would not denote homomorphisms any more---see~\cite{Levy:MonadsForExceptions}); or 
ii) find an alternative solution that would allow handlers to be soundly accommodated in eMLTT$_{\mathcal{T}_{\text{eff}}}$ without changing its existing definition. 

In this thesis, we follow ii) and  
accommodate handlers  in eMLTT$_{\mathcal{T}_{\text{eff}}}$ via a novel type-level extension that internalises Plotkin and Pretnar's semantic insight that handlers of algebraic effects denote algebras for the corresponding equational theories. Specifically, we extend eMLTT$_{\mathcal{T}_{\text{eff}}}$ with a novel computation type that pairs a value type (the carrier) with a family of appropriately typed value terms (the operations), denoting a user-defined algebra for the given fibred effect theory $\mathcal{T}_{\text{eff}}  = (\mathcal{S}_{\text{eff}},\mathcal{E}_{\text{eff}})$. 

\begin{definition}
\label{def:extensionofeMLTTsyntaxwithhandlers}
\index{extension of eMLTT!-- with handlers of fibred algebraic effects}
\index{type!user-defined algebra --}
\index{ A@$\langle A , \{V_{\sigalgop}\}_{\sigalgop \in \mathcal{S}_{\text{eff}}} \rangle$ (user-defined algebra type)}
\index{composition operation}
The syntax of eMLTT$_{\mathcal{T}_{\text{eff}}}^{\mathcal{H}}$ is given by extending eMLTT$_{\mathcal{T}_{\text{eff}}}$'s computation types with the \emph{user-defined algebra type}:
\[
\begin{array}{r c l @{\qquad\qquad}l}
\ul{C} & ::= & \ldots \,\,\,\vertbar\,\,\, \langle A , \{V_{\sigalgop}\}_{\sigalgop \in \mathcal{S}_{\text{eff}}} \rangle
\end{array}
\]
and computation and homomorphism terms with \emph{composition operations}:
\[
\begin{array}{r c l @{\qquad\qquad}l}
M & ::= & \ldots \,\,\,\vertbar\,\,\, \runas M {x \!:\! U\ul{C}} {\ul{D}} {N}
\\[2mm]
K & ::= & \ldots \,\,\,\vertbar\,\,\, \runas K {x \!:\! U\ul{C}} {\ul{D}} {M}
\end{array}
\]
\end{definition}

In both $\runas M {x \!:\! U\ul{C}} {\ul{D}} {N}$ and $\runas K {x \!:\! U\ul{C}} {\ul{D}} {N}$, the value variable $x$ is bound in $N$. Similarly to other terms, we often omit the type annotation $\ul{D}$ for better readability---these annotations exist in order to be able define the interpretation of eMLTT$_{\mathcal{T}_{\text{eff}}}^{\mathcal{H}}$ as a partial mapping from raw expressions to a suitable fibred adjunction model.

As a special case, these composition operations act as \emph{elimination forms} for the user-defined algebra type, i.e., when $\ul{C} = \langle A , \{V_{\sigalgop}\}_{\sigalgop \in \mathcal{S}_{\text{eff}}} \rangle$. In principle, we could have restricted them to only the user-defined algebra type, but then we would not have been able to derive a useful type isomorphism (see Proposition~\ref{prop:typeisomorphismforuserdefinedalgebras}) that allows us to coerce computations between $\ul{C}$ and the corresponding user-defined algebra type, namely, 
\[
\begin{array}{c}
\hspace{-8.75cm}
{\ul{C}} \cong 
\langle U\ul{C} , \{\lambda\, y \!:\! (\Sigma\, x \!:\! I .\, O \to U\ul{C}) .\, 
\\[-1mm]
\hspace{2.6cm}
\pmatch y {(x \!:\! I, x' \!:\! O \to U\ul{C})} {} {\thunk (\algop^{\ul{C}}_{x}(y'\!.\, \force {\ul{C}} (x'\,y')))}  \}_{\sigalgop \in \mathcal{S}_{\text{eff}}} \rangle
\end{array}
\]

We further note that computation terms of type $\langle A , \{V_{\sigalgop}\}_{\sigalgop \in \mathcal{S}_{\text{eff}}} \rangle$ are \emph{introduced} by forcing values of type $A$, i.e., thunked computations of type $U\langle A , \{V_{\sigalgop}\}_{\sigalgop \in \mathcal{S}_{\text{eff}}} \rangle$. 

Conceptually, these composition operations are a special kind of explicit substitution of thunked computations for value variables, e.g., the definitional equations accompanying these terms allow us to prove the following definitional equation: 
\[
\ceq \Gamma {\runas M {x \!:\! U\ul{C}} {\ul{D}} {N}} {N[\thunk M / x]} {\ul{D}}
\]
As such, the value variable $x$ refers to the whole of (the thunk of) the computation term $M$, compared to, e.g., sequential composition $\doto M {x \!:\! A} {} N$, where $x$ refers only to the return value computed by $M$. Therefore, we use $\mathtt{as}$ (running $M$ \emph{as} if it was $x$) instead of $\mathtt{to}$ (running $M$ \emph{to} produce a value for $x$) for the composition operations.

As already hinted above, there is more to these composition operations than just substituting  thunked computations for value variables. In particular, the typing rules for $\runas M {x \!:\! U\ul{C}} {\ul{D}} {N}$ and $\runas K {x \!:\! U\ul{C}} {\ul{D}} {N}$ (see Definition~\ref{def:extensionofeMLTTwithhandlers} below) require that the value variable $x$ is used as if it was a computation variable, in that $x$ must not be duplicated or discarded arbitrarily. 
However, rather than extending eMLTT$_{\mathcal{T}_{\text{eff}}}$ with some form of linear typing for such $x$, we impose these requirements via equational proof obligations by requiring that  $N$ commutes with algebraic operations (when substituted for $x$ using thunks). This ensures that $N$ behaves as if it was a homomorphism term, meaning that the effects in $M$ and $K$ are guaranteed to happen before those in $N$.

The different kinds of substitution defined for eMLTT$_{\mathcal{T}_{\text{eff}}}$ extend straightforwardly to eMLTT$_{\mathcal{T}_{\text{eff}}}^{\mathcal{H}}$: we extend the (simultaneous) substitution of value terms with
\[
\begin{array}{l c l}
\langle A , \{V_{\sigalgop}\}_{\sigalgop \in \mathcal{S}_{\text{eff}}} \rangle[\overrightarrow{W}/\overrightarrow{x}] & \defeq & \langle A[\overrightarrow{W}/\overrightarrow{x}] , \{V_{\sigalgop}[\overrightarrow{W}/\overrightarrow{x}]\}_{\sigalgop \in \mathcal{S}_{\text{eff}}} \rangle
\\[2mm]
(\runas M {y \!:\! U\ul{C}} {\ul{D}} {N})[\overrightarrow{W}/\overrightarrow{x}] & \defeq & 
\runas {M[\overrightarrow{W}/\overrightarrow{x}]} {y \!:\! U\ul{C}[\overrightarrow{W}/\overrightarrow{x}]} {\ul{D}[\overrightarrow{W}/\overrightarrow{x}]} {N[\overrightarrow{W}/\overrightarrow{x}]}
\\[2mm]
(\runas K {y \!:\! U\ul{C}} {\ul{D}} {M})[\overrightarrow{W}/\overrightarrow{x}] & \defeq & 
\runas {K[\overrightarrow{W}/\overrightarrow{x}]} {y \!:\! U\ul{C}[\overrightarrow{W}/\overrightarrow{x}]} {\ul{D}[\overrightarrow{W}/\overrightarrow{x}]} {M[\overrightarrow{W}/\overrightarrow{x}]}
\end{array}
\]
the substitution of computation terms for computation variables with
\[
\begin{array}{l c l}
(\runas K {x \!:\! U\ul{C}} {\ul{D}} {N})[M/z] & \defeq & 
\runas {K[M/z]} {x \!:\! U\ul{C}} {\ul{D}} {N}
\end{array}
\]
and the substitution of homomorphism terms for computation variables with
\[
\begin{array}{l c l}
(\runas L {x \!:\! U\ul{C}} {\ul{D}} {N})[K/z] & \defeq & 
\runas {L[K/z]} {x \!:\! U\ul{C}} {\ul{D}} {N}
\end{array}
\]

The properties of substitution we established for eMLTT in Sections~\ref{sect:syntax} and~\ref{sect:completeness} also extend straightforwardly from eMLTT and eMLTT$_{\mathcal{T}_{\text{eff}}}$ to eMLTT$_{\mathcal{T}_{\text{eff}}}^{\mathcal{H}}$. Specifically, the proof principles remain unchanged: the user-defined algebra type and the value terms appearing in it are treated analogously to propositional equality and the terms appearing in it; and the composition operations are treated analogously to other computation and homomorphism terms that involve variable bindings and type annotations.

Unless stated otherwise, the types and terms we use in the rest of this chapter are those of eMLTT$_{\mathcal{T}_{\text{eff}}}^{\mathcal{H}}$.
This also includes the definitions of pure value types and pure value terms appearing in effect terms because every pure eMLTT value type (resp. term) can be trivially considered as a pure eMLTT$_{\mathcal{T}_{\text{eff}}}^{\mathcal{H}}$ value type (resp. term).

Next, we extend the typing rules and definitional equations of eMLTT$_{\mathcal{T}_{\text{eff}}}$ with the user-defined algebra type and the composition operations. Similarly to eMLTT$_{\mathcal{T}_{\text{eff}}}$, the new rules involve the translation of well-formed effect terms $\lj {\Gamma \vertbar \Delta} T$ into value terms $\efftrans T {A; \overrightarrow{V_i}; \overrightarrow{V'_{\!j}}; \overrightarrow{W_{\sigalgop}}}$. The definition of this translation remains unchanged because it only depends on the structure of $T$ and does not inspect the subscripts $A$, $\overrightarrow{V_i}$, $\overrightarrow{V'_{\!j}}$, and $\overrightarrow{W_{\sigalgop}}$.

Similarly to Chapter~\ref{chap:fibalgeffects}, we assume 
\[
\Gamma = x_1 \!:\! A_1, \ldots, x_n \!:\! A_n
\qquad
\Delta = w_1 \!:\! A'_1, \ldots, w_m \!:\! A'_m
\] 
throughout this chapter, so as to simplify the presentation of typing rules, definitional equations, and the meta-theory of eMLTT$_{\mathcal{T}_{\text{eff}}}^{\mathcal{H}}$. As in Chapter~\ref{chap:fibalgeffects}, we use vector notation for sets of value terms, e.g., we use $\overrightarrow{V_i}$ to denote a set of value terms $\{V_1, \ldots, V_n\}$.

\begin{definition}
\label{def:extensionofeMLTTwithhandlers}
\index{well-formed syntax}
The \emph{well-formed syntax} of eMLTT$_{\mathcal{T}_{\text{eff}}}^{\mathcal{H}}$ is given by extending the typing rules and definitional equations of eMLTT$_{\mathcal{T}_{\text{eff}}}$ with 
\begin{itemize}
\item a formation rule for the user-defined algebra type:
\[
\mkrule
{
\lj {\Gamma'} {\langle A , \{V_{\sigalgop}\}_{\sigalgop \in \mathcal{S}_{\text{eff}}} \rangle}
}
{
\begin{array}{c}
\lj {\Gamma'} A
\quad
\vj {\Gamma'} {V_{\sigalgop}} {(\Sigma\, x \!:\! I . O \to A) \to A}
\\[2mm]
\hspace{-3.9cm} \veq {\Gamma'} {\overrightarrow{\lambda\, x'_i \!:\! \widehat{A_i} .}\, \overrightarrow{\lambda\, x_{w_{\!j}} \!:\! \widehat{A'_j} \to A .}\, \efftrans {T_1} {A; \overrightarrow{x'_i}; \overrightarrow{x_{w_{\!j}}}; \overrightarrow{V_{\sigalgop}}} \\ \hspace{0.45cm} } {\overrightarrow{\lambda\, x'_i \!:\! \widehat{A_i} .}\, \overrightarrow{\lambda\, x_{w_{\!j}} \!:\! \widehat{A'_j} \to A .}\, \efftrans {T_2} {A; \overrightarrow{x'_i}; \overrightarrow{x_{w_{\!j}}}; \overrightarrow{V_{\sigalgop}}}\,} {\,\overrightarrow{\Pi x'_i \!:\! \widehat{A_i} .}\, \overrightarrow{\widehat{A'_j} \to A} \to A} 
\\[3mm]
(\text{for all } \sigalgop : (x \!:\! I) \longrightarrow O \in \mathcal{S}_{\text{eff}}
\text{ and }
\ljeq {\Gamma \vertbar \Delta} {T_1} {T_2} \in \mathcal{E}_{\text{eff}})
\end{array}
}
\]
where $\widehat{A_i} \defeq A_i[x'_1/x_1, \ldots, x'_{i-1}/x_{i-1}]$ and $\widehat{A'_j} \defeq A'_j[x'_1/x_1, \ldots, x'_n/x_n]$; and where we write $\overrightarrow{\lambda x'_i \!:\! \widehat{A_i} .}$, $\overrightarrow{\lambda x_{w_{\!j}} \!:\! \widehat{A'_j} \to A .}$, $\overrightarrow{\Pi x'_i \!:\! \widehat{A_i}}$, and $\overrightarrow{\widehat{A'_j} \to A}$ for sequences of lambda abstractions and sequences of  (dependent) function types, respectively.
\index{ A@$\widehat{A_i}$ ($A_i$, with its variables $x_i$ replaced with fresh ones $x'_i$)}
\index{ lambda@$\overrightarrow{\lambda\, x_{w_j} \hspace{-0.05cm}:\hspace{-0.05cm} \widehat{A'_j} \to A .}$ (sequence of lambda abstractions)}
\index{ A@$\overrightarrow{\widehat{A'_j} \to A}$ (sequence of function types)}
\item typing rules for the two composition operations:
\[
\mkrule
{
\cj {\Gamma} {\runas M {y \!:\! U\ul{C}} {\ul{D}} {N}} {\ul{D}}
}
{
\begin{array}{c@{\qquad} c}
\cj \Gamma M \ul{C} 
\quad
\lj \Gamma \ul{D}
\quad
\cj {\Gamma, y \!:\! U\ul{C}} N \ul{D}
\\[2mm]
\hspace{-0.95cm}
\ceq \Gamma {\lambda\, x \!:\! I .\, \lambda\, x' \!:\! O \to U\ul{C} .\, N[\thunk (\algop^{\ul{C}}_x(y'\!.\, \force {\ul{C}} (x'\, y')))/y] \\ \hspace{0.25cm}} { \lambda\, x \!:\! I .\, \lambda\, x' \!:\! O \to U\ul{C} .\, \algop^{\ul{D}}_x(y'\!.\, N[x'\, y'/y])} {\Pi\, x \!:\! I .\, (O \to U\ul{C}) \to \ul{D}}
\\[-1mm]
& \hspace{-3.2cm} (\sigalgop : (x \!:\! I) \longrightarrow O \in \mathcal{S}_{\text{eff}}) 
\end{array}
}
\]

\vspace{-0.15cm}

\[
\mkrule
{
\hj {\Gamma} {z \!:\! \ul{C}} {\runas K {y : U\ul{D}_1} {\ul{D}_2} {M}} {\ul{D}_2}
}
{
\begin{array}{c@{\qquad} c}
\hj \Gamma{z \!:\! \ul{C}} K \ul{D}_1 
\quad
\lj \Gamma \ul{D}_2
\quad
\cj {\Gamma, y \!:\! U\ul{D}_1} M \ul{D}_2
\\[2mm]
\hspace{-1.1cm}
\ceq \Gamma {\lambda\, x \!:\! I .\, \lambda\, x' \!:\! O \to U\ul{D}_1 .\, M[\thunk (\algop^{\ul{D}_1}_x(y'\!.\, \force {\ul{D}_1} (x'\, y')))/y] \\ \hspace{0.25cm}} { \lambda\, x \!:\! I .\, \lambda\, x' \!:\! O \to U\ul{D}_1 .\, \algop^{\ul{D}_2}_x(y'\!.\, M[x'\, y'/y])} {\Pi\, x \!:\! I .\, (O \to U\ul{D}_1) \to \ul{D}_2}
\\[-1mm]
& \hspace{-4cm} (\sigalgop : (x \!:\! I) \longrightarrow O \in \mathcal{S}_{\text{eff}})
\end{array}
}
\]
\item congruence rules for the user-defined algebra type and the composition operations:
\[
\mkrule
{
\ljeq {\Gamma'} {\langle A , \{V_{\sigalgop}\}_{\sigalgop \in \mathcal{S}_{\text{eff}}} \rangle} {\langle B , \{W_{\sigalgop}\}_{\sigalgop \in \mathcal{S}_{\text{eff}}} \rangle}
}
{
\begin{array}{c}
\ljeq {\Gamma'} A B
\quad
\veq {\Gamma'} {V_{\sigalgop}} {W_{\sigalgop}} {(\Sigma\, x \!:\! I . O \to A) \to A}
\\[2mm]
\hspace{-3.9cm} \veq {\Gamma'} {\overrightarrow{\lambda\, x'_i \!:\! \widehat{A_i} .}\, \overrightarrow{\lambda\, x_{w_{\!j}} \!:\! \widehat{A'_j} \to A .}\, \efftrans {T_1} {A; \overrightarrow{x'_i}; \overrightarrow{x_{w_{\!j}}}; \overrightarrow{V_{\sigalgop}}} \\ \hspace{0.45cm} } {\overrightarrow{\lambda\, x'_i \!:\! \widehat{A_i} .}\, \overrightarrow{\lambda\, x_{w_{\!j}} \!:\! \widehat{A'_j} \to A .}\, \efftrans {T_2} {A; \overrightarrow{x'_i}; \overrightarrow{x_{w_{\!j}}}; \overrightarrow{V_{\sigalgop}}}\,} {\,\overrightarrow{\Pi x'_i \!:\! \widehat{A_i} .}\, \overrightarrow{\widehat{A'_j} \to A} \to A}
\\[3mm]
(\text{for all } \sigalgop : (x \!:\! I) \longrightarrow O \in \mathcal{S}_{\text{eff}}
\text{ and }
\ljeq {\Gamma \vertbar \Delta} {T_1} {T_2} \in \mathcal{E}_{\text{eff}})
\end{array}
}
\]

\vspace{-0.15cm}

\[
\mkrule
{
\ceq {\Gamma} {\runas {M_1} {y \!:\! U\ul{C}_1} {\ul{D}_1} {N_1}} {\runas {M_2} {y \!:\! U\ul{C}_2} {\ul{D}_2} {N_2}} {\ul{D}_1}
}
{
\begin{array}{c@{\qquad} c}
\ljeq \Gamma {\ul{C}_1} {\ul{C}_2}
\quad
\ceq \Gamma {M_1} {M_2} \ul{C}_1 
\\[1mm]
\ljeq \Gamma {\ul{D}_1} {\ul{D}_2}
\quad
\ceq {\Gamma, y \!:\! U\ul{C}_1} {N_1} {N_2} \ul{D}_1
\\[2mm]
\hspace{-1.15cm}
\ceq \Gamma {\lambda\, x \!:\! I .\, \lambda\, x' \!:\! O \to U\ul{C}_1 .\, N_1[\thunk (\algop^{\ul{C}_1}_x(y'\!.\, \force {\ul{C}_1} (x'\, y')))/y] \\ \hspace{0.25cm}} { \lambda\, x \!:\! I .\, \lambda\, x' \!:\! O \to U\ul{C}_1 .\, \algop^{\ul{D}_1}_x(y'\!.\, N_1[x'\, y'/y])} {\Pi\, x \!:\! I .\, (O \to U\ul{C}_1) \to \ul{D}_1}
\\[-1mm]
& \hspace{-4cm} (\sigalgop : (x \!:\! I) \longrightarrow O \in \mathcal{S}_{\text{eff}})
\end{array}
}
\]

\vspace{0.05cm}

\[
\hspace{-0.25cm}
\mkrule
{
\heq {\Gamma} {z \!:\! \ul{C}} {\runas {K} {y \!:\! U\ul{D}_{11}} {\ul{D}_{21}} {M}} {\runas {L} {y \!:\! U\ul{D}_{12}} {\ul{D}_{22}} {N}} {\ul{D}_{21}}
}
{
\begin{array}{c@{\qquad} c}
\ljeq \Gamma {\ul{D}_{11}} {\ul{D}_{12}}
\quad
\heq \Gamma {z \!:\! \ul{C}} {K} {L} \ul{D}_{11} 
\\[1mm]
\ljeq \Gamma {\ul{D}_{21}} {\ul{D}_{22}}
\quad
\ceq {\Gamma, y \!:\! U\ul{D}_{11}} {M} {N} \ul{D}_{21}
\\[2mm]
\hspace{-1.15cm}
\ceq \Gamma {\lambda\, x \!:\! I .\, \lambda\, x' \!:\! O \to U\ul{D}_{11} .\, M[\thunk (\algop^{\ul{D}_{11}}_x(y'\!.\, \force {\ul{D}_{11}} (x'\, y')))/y] \\ \hspace{0.4cm}} { \lambda\, x \!:\! I .\, \lambda\, x' \!:\! O \to U\ul{D}_{11} .\, \algop^{\ul{D}_{21}}_x(y'\!.\, M[x'\, y'/y])} {\Pi\, x \!:\! I .\, (O \to U\ul{D}_{11}) \to \ul{D}_{21}}
\\[-1mm]
& \hspace{-4.5cm} (\sigalgop : (x \!:\! I) \longrightarrow O \in \mathcal{S}_{\text{eff}})
\end{array}
}
\]
\item a $\beta$-equation for the user-defined algebra type:
\vspace{0.1cm}
\[
\mkrule
{
\ljeq {\Gamma} {U\langle A , \{V_{\sigalgop}\}_{\sigalgop \in \mathcal{S}_{\text{eff}}} \rangle} {A}
}
{
\lj \Gamma \langle A , \{V_{\sigalgop}\}_{\sigalgop \in \mathcal{S}_{\text{eff}}} \rangle
}
\]

\item $\beta$- and $\eta$-equations for the composition operation for computation terms:
\[
\mkrule
{
\ceq {\Gamma} {\runas {(\force {\ul{C}} V)} {y \!:\! U\ul{C}} {\ul{D}} {M}} {M[V/y]} {\ul{D}}
}
{
\begin{array}{c@{\qquad} c}
\vj \Gamma V U\ul{C}
\quad
\lj \Gamma \ul{D}
\quad
\cj {\Gamma, y \!:\! U\ul{C}} M \ul{D}
\\[2mm]
\hspace{-0.95cm}
\ceq \Gamma {\lambda\, x \!:\! I .\, \lambda\, x' \!:\! O \to U\ul{C} .\, M[\thunk (\algop^{\ul{C}}_x(y'\!.\, \force {\ul{C}} (x'\, y')))/y] \\ \hspace{0.25cm}} { \lambda\, x \!:\! I .\, \lambda\, x' \!:\! O \to U\ul{C} .\, \algop^{\ul{D}}_x(y'\!.\, M[x'\, y'/y])} {\Pi\, x \!:\! I .\, (O \to U\ul{C}) \to \ul{D}}
\\[-1mm]
& \hspace{-3.3cm} (\sigalgop : (x \!:\! I) \longrightarrow O \in \mathcal{S}_{\text{eff}}) 
\end{array}
}
\]

\vspace{0.05cm}

\[
\mkrule
{
\ceq {\Gamma} {\runas {M} {y \!:\! U\ul{C}} {\ul{D}} {K[\force {\ul{C}} y/z]}} {K[M/z]} {\ul{D}}
}
{
\cj \Gamma M \ul{C} 
\quad
\hj {\Gamma} {z \!:\! \ul{C}} K \ul{D}
}
\]
\item an $\eta$-equation for the composition operation for homomorphism terms:
\vspace{0.1cm}
\[
\mkrule
{
\heq {\Gamma} {z_1 \!:\! \ul{C}} {\runas {K} {y \!:\! U\ul{D}_1} {\ul{D}_2} {L[\force {\ul{D}_1} y/z_2]}} {L[K/z_2]} {\ul{D}_2}
}
{
\hj \Gamma {z_1 \!:\! \ul{C}} K \ul{D}_1 
\quad
\hj {\Gamma} {z_2 \!:\! \ul{D}_1} L \ul{D}_2
}
\]
\item an $\eta$-equation for algebraic operations at the user-defined algebra type:
\[
\mkrule
{
\begin{array}{r@{\,\,} l}
\ceq \Gamma {& \algop^{\langle A , \{V_{\sigalgop}\}_{\sigalgop \in \mathcal{S}_{\text{eff}}} \rangle}_V(y.\, M) \\} { & \force {\langle A , \{V_{\sigalgop}\}_{\sigalgop \in \mathcal{S}_{\text{eff}}} \rangle} (V_{\sigalgop}\, \langle V , \lambda\, y \!:\! O[V/x] .\, \thunk M \rangle)} {\langle A , \{V_{\sigalgop}\}_{\sigalgop \in \mathcal{S}_{\text{eff}}} \rangle}
\end{array}
}
{
\begin{array}{c}
\vj \Gamma V I 
\quad
\lj \Gamma \langle A , \{V_{\sigalgop}\}_{\sigalgop \in \mathcal{S}_{\text{eff}}} \rangle
\quad
\cj {\Gamma, y \!:\! O[V/x]} M {\langle A , \{V_{\sigalgop}\}_{\sigalgop \in \mathcal{S}_{\text{eff}}} \rangle}
\end{array}
}
\]
\end{itemize}
\end{definition}

Observe that the $\beta$-equation for the user-defined algebra type captures the intuition that the value type $A$ denotes the carrier of the algebra denoted by $\langle A , \{V_{\sigalgop}\}_{\sigalgop \in \mathcal{S}_{\text{eff}}} \rangle$.
Analogously, the $\eta$-equation for algebraic operations captures the intuition that the value terms $V_{\sigalgop}$ denote the operations of the algebra denoted by $\langle A , \{V_{\sigalgop}\}_{\sigalgop \in \mathcal{S}_{\text{eff}}} \rangle$.

It is also worthwhile to note that the equational theory of eMLTT$_{\mathcal{T}_{\text{eff}}}^{\mathcal{H}}$ does not include an $\eta$-equation for the user-defined algebra type, namely, 
\[
\hspace{-0.25cm}
\mkrule
{
\begin{array}{c}
\hspace{-7cm}
\ljeq {\Gamma} {\ul{C}} {\langle U\ul{C} , \{\lambda\, y \!:\! (\Sigma\, x \!:\! I .\, O \to U\ul{C}) .\, 
\\[-1mm]
\hspace{3cm}
\pmatch y {x \!:\! I, x' \!:\! O \to U\ul{C}} {} {\algop^{\ul{C}}_{x}(y'\!.\, \force {\ul{C}} (x'\, y'))}  \}_{\sigalgop \in \mathcal{S}_{\text{eff}}} \rangle}
\end{array}
}
{
\lj \Gamma \ul{C}
}
\]

We omit this equation because it does not hold in the natural fibred adjunction model we use for giving a denotational semantics to eMLTT$_{\mathcal{T}_{\text{eff}}}^{\mathcal{H}}$ in Section~\ref{sect:interpretingemlttwithhandlers}, based on models of countable Lawvere theories. 
However, it is also important to note that this equation would hold in a variant of that fibred adjunction model, based on models of countable equational theories. This illustrates that while the two categories 
of models might be equivalent as categories, they differ in the strict equations that they support.

Instead, as promised earlier, we can derive a type isomorphism that allows us to coerce computations between $\ul{C}$ and the corresponding user-defined algebra type.

\begin{proposition}
\label{prop:typeisomorphismforuserdefinedalgebras}
Given $\lj \Gamma \ul{C}$, we can derive a computation type isomorphism
\[
\Gamma \vdash \ul{C} \cong \langle U\ul{C} , \{V_{\sigalgop}\}_{\sigalgop \in \mathcal{S}_{\text{eff}}} \rangle
\]
where, for all $\sigalgop : (x \!:\! I) \longrightarrow O$ in $\mathcal{S}_{\text{eff}}$, the value terms $V_{\sigalgop}$ are given by
\[
\lambda\, y \!:\! (\Sigma\, x \!:\! I .\, O \to U\ul{C}) .\, 
\pmatch y {(x \!:\! I, x' \!:\! O \to U\ul{C})} {} {\thunk (\algop^{\ul{C}}_{x}(y'\!.\, \force {\ul{C}} (x'\,y')))}
\]
\end{proposition}

\begin{proof}
This type isomorphism is witnessed by the well-typed homomorphism terms
\[
\begin{array}{c}
\hj \Gamma {z \!:\! \ul{C}} {\runas z {y \!:\! U\ul{C}} {} {\force {\langle U\ul{C} , \{V_{\sigalgop}\}_{\sigalgop \in \mathcal{S}_{\text{eff}}} \rangle} {y}}} {\langle U\ul{C} , \{V_{\sigalgop}\}_{\sigalgop \in \mathcal{S}_{\text{eff}}} \rangle}
\\[3mm]
\hj \Gamma {z \!:\! \langle U\ul{C} , \{V_{\sigalgop}\}_{\sigalgop \in \mathcal{S}_{\text{eff}}} \rangle} {\runas z {y \!:\! U\langle U\ul{C} , \{V_{\sigalgop}\}_{\sigalgop \in \mathcal{S}_{\text{eff}}} \rangle} {} {\force {\ul{C}} {y}}} {\ul{C}}
\end{array}
\]

The proofs that both composites of these terms are definitionally equal to $z$ (i.e., to identity) are straightforward, using the $\beta$- and $\eta$-equations for composition operations.
\end{proof}

We conclude this section by making a simple yet useful observation about our equational proof obligations that allows 
many of them to be proved at little extra cost.

\begin{proposition}
Given a homomorphism term $\hj \Gamma {z \!:\! \ul{C}} K {\ul{D}}$, then we have 
\[
\begin{array}{c}
\hspace{-0.9cm}
\ceq \Gamma {\lambda\, x \!:\! I .\, \lambda\, x' \!:\! O \to U\ul{C} .\, K[\force {\ul{C}} y/z][\thunk (\algop^{\ul{C}}_x(y'\!.\, \force {\ul{C}} (x'\, y')))/y] \\ \hspace{0.25cm}} { \lambda\, x \!:\! I .\, \lambda\, x' \!:\! O \to U\ul{C} .\, \algop^{\ul{D}}_x(y'\!.\, K[\force {\ul{C}} y/z][x'\, y'/y])} {\Pi\, x \!:\! I .\, (O \to U\ul{C}) \to \ul{D}}
\end{array}
\]
for all operation symbols $\sigalgop : (x \!:\! I) \longrightarrow O$ in $\mathcal{S}_{\text{eff}}$.
\end{proposition}

\begin{proof}
By straightforward equational reasoning, using the definitions of different kinds of substitution, and the general algebraicity equation given in Definition~\ref{def:extensionofeMLTTwithfibalgeffects}.
\end{proof}

\section{Deriving the conventional presentation of handlers}
\label{sect:derivingconventionalhandlers}

In this section we show how to derive the conventional term-level presentation of handlers (as discussed in Section~\ref{sect:handlersoverview}) in eMLTT$_{\mathcal{T}_{\text{eff}}}^{\mathcal{H}}$, so as to provide programmers with a familiar syntax for programming with handlers within computation terms.

\index{handling construct}
In detail, we define the handling construct 
\[
{M \mathtt{~handled~with~} \{\mathtt{op}_x(x') \mapsto N_{\sigalgop}\}_{\sigalgop \,\in\, \mathcal{S}_{\text{eff}}} \mathtt{~to~} y \!:\! A \mathtt{~in~} N_{\mathsf{ret}}}
\]
using sequential composition as the following composite computation term:
\[
\force {\ul{C}} (\thunk (\doto M {y \!:\! A} {} {\force {\langle U\ul{C} , \{V_{\sigalgop}\}_{\sigalgop \,\in\, \mathcal{S}_{\text{eff}}} \rangle} {(\thunk N_{\mathsf{ret}})}}))
\]
where, for all $\sigalgop : (x \!:\! I) \longrightarrow O$ in $\mathcal{S}_{\text{eff}}$, the value terms $V_{\sigalgop}$ are given by
\[
V_{\sigalgop} \defeq \lambda y' \!:\! (\Sigma\, x \!:\! I .\, O \to U\ul{C}) .\, \pmatch {y'} {(x \!:\! I, x' \!:\! O \to U\ul{C})} {} {\thunk N_{\sigalgop}}
\]

Next, we show that the corresponding typing rule is derivable. Compared to the typing rule considered by Plotkin and Pretnar, ours includes explicit equational proof obligations so as to ensure that the computation type $\langle U\ul{C} , \{V_{\sigalgop}\}_{\sigalgop \,\in\, \mathcal{S}_{\text{eff}}} \rangle$ is well-formed.

\begin{proposition}
\label{prop:handlertyping}
The following typing rule is derivable:
\[
\mkrule
{
\cj {\Gamma'} {M \mathtt{~handled~with~} \{\mathtt{op}_x(x') \mapsto N_{\sigalgop}\}_{\sigalgop \,\in\, \mathcal{S}_{\text{eff}}} \mathtt{~to~} y \!:\! A \mathtt{~in~} N_{\mathsf{ret}}} {\ul{C}}
}
{
\begin{array}{c}
\cj {\Gamma'} M FA
\quad
\lj {\Gamma'} \ul{C}
\quad
\cj {\Gamma', y \!:\! A} {N_{\mathsf{ret}}} {\ul{C}}
\quad
\cj {\Gamma', x \!:\! I, x' \!:\! O \to U\ul{C}} {N_{\sigalgop}} {\ul{C}}
\\[2mm]
\hspace{-3.9cm} \veq {\Gamma'} {\overrightarrow{\lambda\, x'_i \!:\! \widehat{A_i} .}\, \overrightarrow{\lambda\, x_{w_{\!j}} \!:\! \widehat{A'_j} \to A .}\, \efftrans {T_1} {A; \overrightarrow{x'_i}; \overrightarrow{x_{w_{\!j}}}; \overrightarrow{V_{\sigalgop}}} \\ \hspace{0.45cm} } {\overrightarrow{\lambda\, x'_i \!:\! \widehat{A_i} .}\, \overrightarrow{\lambda\, x_{w_{\!j}} \!:\! \widehat{A'_j} \to A .}\, \efftrans {T_2} {A; \overrightarrow{x'_i}; \overrightarrow{x_{w_{\!j}}}; \overrightarrow{V_{\sigalgop}}}\,} {\,\overrightarrow{\Pi x'_i \!:\! \widehat{A_i} .}\, \overrightarrow{\widehat{A'_j} \to A} \to A}
\\[3mm]
(\text{for all } \sigalgop : (x \!:\! I) \longrightarrow O \in \mathcal{S}_{\text{eff}}
\text{ and }
\ljeq {\Gamma \vertbar \Delta} {T_1} {T_2} \in \mathcal{E}_{\text{eff}})
\end{array}
}
\]
where, for all $\sigalgop : (x \!:\! I) \longrightarrow O$ in $\mathcal{S}_{\text{eff}}$, the value terms $V_{\sigalgop}$ are given by
\[
V_{\sigalgop} \defeq \lambda y' \!:\! (\Sigma\, x \!:\! I .\, O \to U\ul{C}) .\, \pmatch {y'} {(x \!:\! I, x' \!:\! O \to U\ul{C})} {} {\thunk N_{\sigalgop}}
\]
and where the value types $\widehat{A_i}$ and $\widehat{A'_j}$ are defined as in Definition~\ref{def:extensionofeMLTTwithhandlers}.
\end{proposition} 

\begin{proof}
The derivation of this typing rule is constructed straightforwardly. The derivation consists of using the respective typing rules for forcing of thunked computations, thunking of computations, and sequential 
composition of computation terms. 
\end{proof}

Further, we can also show that the corresponding $\beta$-equations are derivable.

\begin{proposition}
\label{prop:handlerequations}
The following two definitional $\beta$-equations are derivable:
\[
\mkrule
{
\begin{array}{r@{\,\,} l}
\ceq {\Gamma'} {& (\algop^{FA}_V(y'\!.\, M)) \mathtt{~handled~with~} \{\mathtt{op}_x(x') \mapsto N_{\sigalgop}\}_{\sigalgop \,\in\, \mathcal{S}_{\text{eff}}} \mathtt{~to~} y \!:\! A \mathtt{~in~} N_{\mathsf{ret}} \\} { & N_{\sigalgop}[V/x][\lambda\, y' \!:\! O[V/x] .\, \thunk H/x']} {\ul{C}}
\end{array}
}
{
\begin{array}{c}
\vj {\Gamma'} V I
\quad
\cj {\Gamma', y' \!:\! O[V/x]} M FA
\quad
\lj {\Gamma'} \ul{C}
\\
\cj {\Gamma', y \!:\! A} {N_{\mathsf{ret}}} {\ul{C}}
\quad
\cj {\Gamma', x \!:\! I, x' \!:\! O \to U\ul{C}} {N_{\sigalgop}} {\ul{C}}
\\[2mm]
\hspace{-3.9cm} \veq {\Gamma'} {\overrightarrow{\lambda\, x'_i \!:\! \widehat{A_i} .}\, \overrightarrow{\lambda\, x_{w_{\!j}} \!:\! \widehat{A'_j} \to A .}\, \efftrans {T_1} {A; \overrightarrow{x'_i}; \overrightarrow{x_{w_{\!j}}}; \overrightarrow{V_{\sigalgop}}} \\ \hspace{0.45cm} } {\overrightarrow{\lambda\, x'_i \!:\! \widehat{A_i} .}\, \overrightarrow{\lambda\, x_{w_{\!j}} \!:\! \widehat{A'_j} \to A .}\, \efftrans {T_2} {A; \overrightarrow{x'_i}; \overrightarrow{x_{w_{\!j}}}; \overrightarrow{V_{\sigalgop}}}\,} {\,\overrightarrow{\Pi x'_i \!:\! \widehat{A_i} .}\, \overrightarrow{\widehat{A'_j} \to A} \to A}
\\[3mm]
(\text{for all } \sigalgop : (x \!:\! I) \longrightarrow O \in \mathcal{S}_{\text{eff}}
\text{ and }
\ljeq {\Gamma \vertbar \Delta} {T_1} {T_2} \in \mathcal{E}_{\text{eff}})
\end{array}
}
\]

\vspace{0.01cm}

\[
\mkrule
{
\begin{array}{r@{\,\,} l}
\ceq {\Gamma'} {& (\return V) \mathtt{~handled~with~} \{\mathtt{op}_x(x') \mapsto N_{\sigalgop}\}_{\sigalgop \,\in\, \mathcal{S}_{\text{eff}}} \mathtt{~to~} y \!:\! A \mathtt{~in~} N_{\mathsf{ret}} \\} { & N_{\mathsf{ret}}[V/y]} {\ul{C}}
\end{array}
}
{
\begin{array}{c}
\vj {\Gamma'} V A
\quad
\lj {\Gamma'} \ul{C}
\quad
\cj {\Gamma', y \!:\! A} {N_{\mathsf{ret}}} {\ul{C}}
\quad
\cj {\Gamma', x \!:\! I, x' \!:\! O \to U\ul{C}} {N_{\sigalgop}} {\ul{C}}
\\[2mm]
\hspace{-3.9cm} \veq {\Gamma'} {\overrightarrow{\lambda\, x'_i \!:\! \widehat{A_i} .}\, \overrightarrow{\lambda\, x_{w_{\!j}} \!:\! \widehat{A'_j} \to A .}\, \efftrans {T_1} {A; \overrightarrow{x'_i}; \overrightarrow{x_{w_{\!j}}}; \overrightarrow{V_{\sigalgop}}} \\ \hspace{0.45cm} } {\overrightarrow{\lambda\, x'_i \!:\! \widehat{A_i} .}\, \overrightarrow{\lambda\, x_{w_{\!j}} \!:\! \widehat{A'_j} \to A .}\, \efftrans {T_2} {A; \overrightarrow{x'_i}; \overrightarrow{x_{w_{\!j}}}; \overrightarrow{V_{\sigalgop}}}\,} {\,\overrightarrow{\Pi x'_i \!:\! \widehat{A_i} .}\, \overrightarrow{\widehat{A'_j} \to A} \to A}
\\[3mm]
(\text{for all } \sigalgop : (x \!:\! I) \longrightarrow O \in \mathcal{S}_{\text{eff}}
\text{ and }
\ljeq {\Gamma \vertbar \Delta} {T_1} {T_2} \in \mathcal{E}_{\text{eff}})
\end{array}
}
\]
where we abbreviate the handling construct in the first equation as
\[
H \,\defeq\, M \mathtt{~handled~with~} \{\mathtt{op}_x(x') \mapsto N_{\sigalgop}\}_{\sigalgop \,\in\, \mathcal{S}_{\text{eff}}} \mathtt{~to~} y \!:\! A \mathtt{~in~} N_{\mathsf{ret}}
\]
\end{proposition}

\begin{proof}
These two definitional $\beta$-equations are proved as follows:
\begin{fleqn}[0.3cm]
\begin{align*}
\Gamma' \,\vdash\,\, & (\algop^{FA}_V(y'\!.\, M)) \mathtt{~handled~with~} \{\mathtt{op}_x(x') \mapsto N_{\sigalgop}\}_{\sigalgop \,\in\, \mathcal{S}_{\text{eff}}} \mathtt{~to~} y \!:\! A \mathtt{~in~} N_{\mathsf{ret}}
\\[1mm]
=\,\, & \force {\ul{C}} \big(\thunk \big(
\\[-1.5mm]
& \hspace{2.25cm} \doto {(\algop^{FA}_V(y'\!.\, M))} {y \!:\! A} {} {\force {\langle U\ul{C} , \{V_{\sigalgop}\}_{\sigalgop \,\in\, \mathcal{S}_{\text{eff}}} \rangle} {\thunk (N_{\mathsf{ret}})}}\big)\big)
\\[2mm]
=\,\, & \force {\ul{C}} \big(\thunk \big(\algop^{\langle U\ul{C} , \{V_{\sigalgop}\}_{\sigalgop \,\in\, \mathcal{S}_{\text{eff}}} \rangle}_V(y'\!.\, 
\\[-2mm]
& \hspace{3.95cm} \doto M {y \!:\! A} {} {\force {\langle U\ul{C} , \{V_{\sigalgop}\}_{\sigalgop \,\in\, \mathcal{S}_{\text{eff}}} \rangle} {(\thunk N_{\mathsf{ret}})}})\big)\big)
\\
=\,\, & \force {\ul{C}} \big(\thunk \big(\\[-1.5mm] & \hspace{2cm} \force {\langle U\ul{C} , \{V_{\sigalgop}\}_{\sigalgop \,\in\, \mathcal{S}_{\text{eff}}} \rangle} {\big(V_{\sigalgop}\, \big\langle V , \lambda\, y' \!:\! O[V/x] .\, \thunk (\\[-1mm] & \hspace{3.6cm} \doto M {y \!:\! A} {} {\force {\langle U\ul{C} , \{V_{\sigalgop}\}_{\sigalgop \,\in\, \mathcal{S}_{\text{eff}}} \rangle} {(\thunk N_{\mathsf{ret}})}}) \big\rangle\big)}\big)\big)
\\[1mm]
=\,\, & \force {\ul{C}} \big(V_{\sigalgop}\, \big\langle V , \lambda\, y' \!:\! O[V/x] .\, \thunk (\\[-1.5mm] & \hspace{3.95cm} \doto M {y \!:\! A} {} {\force {\langle U\ul{C} , \{V_{\sigalgop}\}_{\sigalgop \,\in\, \mathcal{S}_{\text{eff}}} \rangle} {(\thunk N_{\mathsf{ret}})}}) \big\rangle\big)
\\[1mm]
=\,\, & \force {\ul{C}} \big(\thunk \big(N_{\sigalgop}[V/x]\big[\lambda\, y' \!:\! O[V/x] .\, \\[-2mm] & \hspace{2cm} \thunk (\doto M {y \!:\! A} {} {\force {\langle U\ul{C} , \{V_{\sigalgop}\}_{\sigalgop \,\in\, \mathcal{S}_{\text{eff}}} \rangle} {(\thunk N_{\mathsf{ret}})}})/x'\big]\big)\big)
\\[1mm]
=\,\, & N_{\sigalgop}[V/x]\big[\lambda\, y' \!:\! O[V/x] .\, \\[-2mm] & \hspace{2.35cm} \thunk (\doto M {y \!:\! A} {} {\force {\langle U\ul{C} , \{V_{\sigalgop}\}_{\sigalgop \,\in\, \mathcal{S}_{\text{eff}}} \rangle} {(\thunk N_{\mathsf{ret}})}})/x'\big]
\\[1mm]
=\,\, & N_{\sigalgop}[V/x]\big[\lambda\, y' \!:\! O[V/x] .\, \thunk ( \force {\ul{C}} (\thunk (
\\[-2mm] & \hspace{3.35cm} \doto M {y \!:\! A} {} {\force {\langle U\ul{C} , \{V_{\sigalgop}\}_{\sigalgop \,\in\, \mathcal{S}_{\text{eff}}} \rangle} {(\thunk N_{\mathsf{ret}})}})))/x'\big]
\\[1mm]
=\,\, & N_{\sigalgop}[V/x]\big[\lambda\, y' \!:\! O[V/x] .\, \thunk (
\\[-2mm] & \hspace{2.1cm}
M \mathtt{~handled~with~} \{\mathtt{op}_x(x') \mapsto N_{\sigalgop}\}_{\sigalgop \,\in\, \mathcal{S}_{\text{eff}}} \mathtt{~to~} y \!:\! A \mathtt{~in~} N_{\mathsf{ret}})/x'\big] : \ul{C}
\end{align*}
\end{fleqn}
and

\begin{fleqn}[0.3cm]
\begin{align*}
\Gamma' \,\vdash\,\, & (\return V) \mathtt{~handled~with~} \{\mathtt{op}_x(x') \mapsto N_{\sigalgop}\}_{\sigalgop \,\in\, \mathcal{S}_{\text{eff}}} \mathtt{~to~} y \!:\! A \mathtt{~in~} N_{\mathsf{ret}}
\\
=\,\, & \force {\ul{C}} \big(\thunk \big(
\\[-2mm]
& \hspace{2cm} \doto {(\return V)} {y \!:\! A} {} {\force {\langle U\ul{C} , \{V_{\sigalgop}\}_{\sigalgop \,\in\, \mathcal{S}_{\text{eff}}} \rangle} {(\thunk N_{\mathsf{ret}})}}\big)\big)
\\
=\,\, & \force {\ul{C}} \big(\thunk \big(\force {\langle U\ul{C} , \{V_{\sigalgop}\}_{\sigalgop \,\in\, \mathcal{S}_{\text{eff}}} \rangle} {(\thunk N_{\mathsf{ret}}[V/y])}\big)\big)
\\
=\,\, & \force {\ul{C}} {(\thunk N_{\mathsf{ret}}[V/y])}
\\
=\,\, & N_{\mathsf{ret}}[V/y] : \ul{C}
\\[-1.25cm]
\end{align*}
\end{fleqn}
\end{proof}

By being able to derive the conventional term-level presentation of handlers, we can also straightforwardly accommodate all the typical example uses of handlers proposed by Plotkin and Pretnar~\cite{Plotkin:HandlingEffects}, e.g., implementing timeouts, rollbacks, stream redirection, etc. 
We refer the reader to op.~cit. for a detailed overview of these examples.

It is worth noting that the problems discussed in Section~\ref{sect:problemwithhandlers} do not arise in this extension of eMLTT$_{\mathcal{T}_{\text{eff}}}$ because we can not define the handling construct
\[
{K \mathtt{~handled~with~} \{\mathtt{op}_x(x') \mapsto N_{\sigalgop}\}_{\sigalgop \,\in\, \mathcal{S}_{\text{eff}}} \mathtt{~to~} y \!:\! A \mathtt{~in}_{\ul{C}} \mathtt{~} N_{\mathsf{ret}}}
\]
satisfying analogous equations to those given in 
Definition~\ref{prop:handlerequations}. Intuitively, we can not derive such terms because the nature of  homomorphism terms does not allow us to temporarily forget about the (algebra) structure of $\ul{C}$ and instead work with $U\ul{C}$, e.g., as used in the definition of 
the computation term variant of the handling construct above.

Finally, we recall that Plotkin and Pretnar do not enforce the correctness of their handlers during typechecking because it is in general an undecidable problem, see~\cite[\S6]{Plotkin:HandlingEffects} for details. In other words, they do not require the user-defined computation terms $N_{\sigalgop}$ to satisfy the equations given in $\mathcal{E}_{\text{eff}}$. In comparison, we include the corresponding proof obligations in eMLTT$_{\mathcal{T}_{\text{eff}}}^{\mathcal{H}}$'s typing rules and definitional equations because in this thesis we only study a declarative presentation of eMLTT and its extensions.

We plan to address the issue of algorithmic typechecking in future extensions of this work. 
For example, we could develop a normaliser for eMLTT$_{\mathcal{T}_{\text{eff}}}^{\mathcal{H}}$ that is optimised for important fibred effect theories 
(e.g., for global state, as studied in~\cite[\S5.2]{Ahman:NBE}), 
and require  
programmers to manually prove equations that cannot be established automatically. To facilitate the latter, we could change eMLTT$_{\mathcal{T}_{\text{eff}}}^{\mathcal{H}}$ to use propositional equalities in proof obligations instead of definitional equations---see Section~\ref{sect:normalisationandimplementation}.

\section{Using handlers to reason about algebraic effects}
\label{section:usinghandlersforreasoning}

In this section we demonstrate that our type-based treatment of handlers provides a useful mechanism for reasoning about effectful computations, giving us an alternative to defining predicates on effectful computations using propositional equality on thunks.

To facilitate such reasoning, we first extend eMLTT$_{\mathcal{T}_{\text{eff}}}^{\mathcal{H}}$ with \emph{universes}. To keep to the declarative presentation we are using for eMLTT and its extensions, we extend eMLTT$_{\mathcal{T}_{\text{eff}}}^{\mathcal{H}}$ with universes \emph{\`a la Tarski}\footnote{This terminology was originally proposed by Martin-L\"{o}f in~\cite{MartinLof:Bibliopolis}, due to the similarity between the explicit decoding function and Tarski's definition of truth~\cite{Tarski:ConceptOfTruth}.} by making the decoding function explicit. 

In detail, we extend eMLTT$_{\mathcal{T}_{\text{eff}}}^{\mathcal{H}}$ with i) \emph{universes} of codes of types and ii) \emph{decoding functions} that provide a meaning to these codes by ``interpreting" them as corresponding types.
As eMLTT$_{\mathcal{T}_{\text{eff}}}^{\mathcal{H}}$ includes both value and computation types, it is natural to include two kinds of universes, albeit we only use the former in our examples. 

Specifically, we extend eMLTT$_{\mathcal{T}_{\text{eff}}}^{\mathcal{H}}$'s types with
\[
\begin{array}{r c l @{\qquad\qquad}l}
A & ::= & \ldots & 
\\
& \vertbar & \mathsf{VU} & \text{universe of codes of value types}
\\
& \vertbar & \mathsf{CU} & \text{universe of codes of computation types}
\\
& \vertbar & \mathsf{El}~ V & \text{decoding function for the codes of value types}
\\[5mm]
\ul{C} & ::= & \ldots & 
\\
& \vertbar & \mathsf{El}~ V & \text{decoding function for the codes of computation types}
\end{array}
\]
\index{universe}
\index{ El@$\mathsf{El}$ (decoding function for codes of types)}
\index{ VU@$\mathsf{VU}$ (universe of codes of value types)}
\index{ CU@$\mathsf{CU}$ (universe of codes of computation types)}

Observe that we consider only one universe level for both value and computation types---this can be straightforwardly extended to a hierarchy of universes using standard techiques, e.g., as discussed in~\cite{Palmgren:Universes}.

The concrete codes of eMLTT$_{\mathcal{T}_{\text{eff}}}^{\mathcal{H}}$'s value and computation types are  given by terms of type $\mathsf{VU}$ and $\mathsf{CU}$, respectively. Specifically, we extend eMLTT$_{\mathcal{T}_{\text{eff}}}^{\mathcal{H}}$'s value terms with
\[
\begin{array}{r c l @{\qquad\qquad}l}
V & ::= & \ldots & 
\\
& \vertbar & \mathtt{nat\text{-}code} \hspace{2.5cm} & 
\multirow{8}{*}{
\hspace{-0.7cm}
\vbox{
\hsize = 3cm
\[
\left. \begin{array}{l}
\vspace{-0.7cm}
\,\\
\,\\
\,\\
\,\\
\,\\
\,\\
\,\\
\,\\
\,
\end{array} \right\} \begin{array}{l} \text{codes of value types} \end{array}
\]
}
}
\\
& \vertbar & \mathtt{unit\text{-}code}
\\
& \vertbar & \mathtt{v\text{-}sigma\text{-}code}(V,x.\,W)
\\
& \vertbar & \mathtt{v\text{-}pi\text{-}code}(V,x.\,W)
\\
& \vertbar & \mathtt{empty\text{-}code}
\\
& \vertbar & \mathtt{sum\text{-}code}(V,W)
\\
& \vertbar & \mathtt{eq\text{-}code}(V,W_1,W_2)
\\
& \vertbar & \mathtt{u\text{-}code~} V
\\
& \vertbar & \mathtt{hom\text{-}code}(V,W)
\\[5mm]
& \vertbar & \mathtt{f\text{-}code~} V & 
\multirow{3}{*}{
\hspace{-1.07cm}
\vbox{
\hsize = 5cm
\[
\left. \begin{array}{l}
\vspace{-0.7cm}
\,\\
\,\\
\,\\
\,
\end{array} \right\} \begin{array}{l} \text{codes of} \\[-2mm] \text{computation types} \end{array}
\]
}
}
\\
& \vertbar & \mathtt{c\text{-}sigma\text{-}code}(V,x.\,W)
\\
& \vertbar & \mathtt{c\text{-}pi\text{-}code}(V,x.\,W)
\\
& \vertbar & \mathtt{u\text{-}alg\text{-}code}(V,\{W_{\sigalgop}\}_{\sigalgop \in \mathcal{S}_{\text{eff}}})
\end{array}
\vspace{0.25cm}
\]

We also extend the well-formed syntax of eMLTT$_{\mathcal{T}_{\text{eff}}}^{\mathcal{H}}$ with rules of the form:
\vspace{0.25cm}

\[
\mkrule
{\lj \Gamma {\mathsf{VU}}}
{\lj {} \Gamma}
\qquad
\mkrule
{\lj \Gamma {\mathsf{El}~V}}
{\vj \Gamma V {\mathsf{VU}}}
\qquad
\mkrule
{\lj \Gamma {\mathsf{El}~V}}
{\vj \Gamma V {\mathsf{CU}}}
\]

\vspace{0.01cm}

\[
\mkrule
{\vj \Gamma {\mathtt{hom\text{-}code}(V,W)} {\mathsf{VU}}}
{\vj \Gamma {V} {\mathsf{CU}} \quad \vj \Gamma {W} {\mathsf{CU}}}
\]

\vspace{0.01cm}

\[
\mkrule
{\vj \Gamma {\mathtt{u\text{-}code~} V} {\mathsf{VU}}}
{\vj \Gamma {V} {\mathsf{CU}}}
\qquad
\mkrule
{\vj \Gamma {\mathtt{f\text{-}code~} V} {\mathsf{CU}}}
{\vj \Gamma V {\mathsf{VU}}}
\]

\vspace{0.01cm}

\[
\mkrule
{\vj \Gamma {\mathtt{c\text{-}sigma\text{-}code}(V,x.\,W)} {\mathsf{CU}}}
{\vj \Gamma V {\mathsf{VU}} \quad \vj {\Gamma, x \!:\! \mathsf{El}~ V} {W} {\mathsf{CU}}}
\qquad
\mkrule
{\vj \Gamma {\mathtt{c\text{-}pi\text{-}code}(V,x.\,W)} {\mathsf{CU}}}
{\vj \Gamma V {\mathsf{VU}} \quad \vj {\Gamma, x \!:\! \mathsf{El}~ V} {W} {\mathsf{CU}}}
\]

The behaviour of the two decoding functions, both written $\mathsf{El}~ V$, is described using definitional equations between value and computation types, e.g., as given by
\vspace{0.25cm}
\[
\mkrule
{\ljeq \Gamma {\mathsf{El}~(\mathtt{hom\text{-}code}(V,W))} {(\mathsf{El}~V) \multimap (\mathsf{El}~W)}}
{\vj \Gamma {V} {\mathsf{CU}} \quad \vj \Gamma {W} {\mathsf{CU}}}
\]

\vspace{0.01cm}

\[
\mkrule
{\ljeq \Gamma {\mathsf{El}~(\mathtt{u\text{-}code~} V)} {U(\mathsf{El}~V)}}
{\vj \Gamma {V} {\mathsf{CU}}}
\qquad
\mkrule
{\ljeq \Gamma {\mathsf{El}~(\mathtt{f\text{-}code~} V)} {F(\mathsf{El}~V)}}
{\vj \Gamma V {\mathsf{VU}}}
\]

\vspace{0.01cm}

\[
\mkrule
{\ljeq \Gamma {\mathsf{El}~(\mathtt{c\text{-}sigma\text{-}code}(V,x.\,W))} {\Sigma \, x \!:\! (\mathsf{El}~V).\, \mathsf{El}~W}}
{\vj \Gamma V {\mathsf{VU}} \quad \vj {\Gamma, x \!:\! \mathsf{El}~V} {W} {\mathsf{CU}}}
\]

\vspace{0.01cm}

\[
\mkrule
{\ljeq \Gamma {\mathsf{El}~(\mathtt{c\text{-}pi\text{-}code}(V,x.\,W))} {\Pi \, x \!:\! (\mathsf{El}~V) .\, \mathsf{El}~W}}
{\vj \Gamma V {\mathsf{VU}} \quad \vj {\Gamma, x \!:\! \mathsf{El}~V} {W} {\mathsf{CU}}}
\]

Using these universes (in particular, the value universe $\mathsf{VU}$), we can now define predicates on (thunks of) effectful computations of type $F\!A$ as value terms of the form $\vj \Gamma V {U\!FA \to \mathsf{VU}}$, with the aim of using them to refine (thunks of) effectful computations using the value $\Sigma$-type, as $\Sigma\, x \!:\! U\!FA .\, \mathsf{El}~(V\,x)$. In more detail, we define these predicates by i) equipping the universe $\mathsf{VU}$ (or a value type we define using it) with an appropriate \emph{algebra} for the given fibred effect theory, and by ii) using a combination of thunking-forcing and sequential composition to \emph{handle} the given computation of type $F\!A$ with the above-mentioned algebra on $\mathsf{VU}$ (or on a value type we define using it).

Below we consider two kinds of examples of defining predicates on computations using our type-based treatment of handlers: i) lifting predicates from return values to predicates on computations; and ii) specifying patterns of allowed (I/O-)effects. 

\subsection{Lifting predicates from return values to computations}
\label{section:liftingpredicatesexamples}

Lifting predicates from return values to computations
is easiest when the fibred effect theory in question does not contain equations, because then we do not have to prove equational proof obligations for the user-defined algebra type. Therefore, let us first consider the \emph{theory $\mathcal{T}_{\text{I/O}}$ of input/output} from Examples~\ref{ex:fibsigofIO} and~\ref{ex:fibtheoryofIO}---other equation-free fibred algebraic effects can be reasoned about similarly, e.g., exceptions.

In particular, we lift a given predicate $\vj \Gamma {V_{\!P}} {A \to \mathsf{VU}}$ on return values to a predicate $\vj \Gamma {V_{\!\widehat{P}}} {U\!FA \to \mathsf{VU}}$ on (thunks of) computations by 
\[
\begin{array}{c}
V_{\!\widehat{P}} \defeq \lambda\, y \!:\! U\!F\!A .\, 
\thunk\! \big(\doto {(\force {F\!A} y)} {y' \!:\! A} {} {\force {\langle \mathsf{VU} , \{V_{\sigalgop}\}_{\sigalgop \in \mathcal{S}_{\text{I/O}}} \rangle} (V_{\!P}\, y')}\big)
\end{array}
\]
where the value terms $V_{\mathsf{read}}$ and $V_{\mathsf{write}}$ are given by 
\[
\begin{array}{l@{~} c @{~} l}
V_{\mathsf{read}} & \defeq & \lambda\, y \!:\! (\Sigma\, x \!:\! 1 . \Character \to \mathsf{VU}) .\, \mathtt{v\text{-}sigma\text{-}code}(\mathtt{chr\text{-}code},y'\!.\, (\snd y)\, y')
\\[2mm]
V_{\mathsf{write}} & \defeq & \lambda\, y \!:\! (\Sigma\, x \!:\! \Character .\, 1 \to \mathsf{VU}) .\, (\snd y)\, \star
\end{array}
\]
and where $\mathtt{chr\text{-}code}$ is the code of the assumed value type $\Character$ of characters, i.e., 
\[
\ljeq \Gamma {\mathsf{El}~\mathtt{chr\text{-}code}} {\Character}
\]

On closer inspection, we can see that the predicate $V_{\!\widehat{P}}$ agrees with the possibility modality from Evaluation Logic~\cite{PittsAM:evall}, in that a computation satisfies $V_{\!\widehat{P}}$ if there \emph{exists} a return value that satisfies $V_{\!P}$. For example, to prove that $V_{\!\widehat{P}}$ holds of a computation \linebreak term $\mathtt{read}^{FA}(y.\, \mathtt{write}^{FA}_{V}(\return W))$, we need to construct an inhabitant for the right-hand side of the following derivable definitional equation between value types:
\[
\Gamma \vdash \mathsf{El}~\big(V_{\!\widehat{P}}\,\, (\thunk\! (\mathtt{read}^{FA}(y.\, \mathtt{write}^{FA}_{V}(\return W))))\big) = \Sigma\, y \!:\! \Character .\, \mathsf{El}~ (V_{\!P}\,\, W)
\]
If we replace $\mathtt{v\text{-}sigma\text{-}code}$ with $\mathtt{v\text{-}pi\text{-}code}$ in the definition of $V_{\mathsf{read}}$, we get a predicate that holds if \emph{all} the return values of the given computation satisfy $V_{\!P}$. 

As a second example, we consider a fibred effect theory that does include equations, namely, the \emph{theory $\mathcal{T}_{\text{GS}}$ of global state} from Examples~\ref{ex:fibsigofstate} and~\ref{ex:fibtheoryofglobalstate}. 

In particular, given a predicate $\vj \Gamma {V_{\!Q}} {A \to \State \to \mathsf{VU}}$ on return values  and \emph{final} store values, we define a predicate $\vj {\Gamma} {V_{\!\widehat{Q}}} {U\!FA \to \State \to \mathsf{VU}}$ on (thunks of) computations and \emph{initial} store values by
\[
\begin{array}{c}
\hspace{-3.75cm}
V_{\!\widehat{Q}} \defeq \lambda\, y \!:\! U\!F\!A .\, \lambda\, x_{S} \!:\! \State .\, 
\fst \big(\big(\thunk\! \big(\doto {(\force {F\!A} y)} {y' \!:\! A} {} {\\ \hspace{4.25cm}  \force {\langle \State \,\to\, (\mathsf{VU} \times \State) , \{V_{\sigalgop}\}_{\sigalgop \in \mathcal{S}_{\text{GS}}} \rangle} (\lambda\, x'_{S} \!:\! \State .\, \langle V_{\!Q}\,\, y'\, x'_{S} , x'_{S} \rangle)}\big)\big)\,\,  x_{S}\big)
\end{array}
\]
where the value terms $V_{\mathsf{get}}$ and $V_{\mathsf{put}}$ are  defined using the natural representation of stateful programs as state-passing functions ${\State \,\to\, (\mathsf{VU} \times \State)}$, e.g., $V_{\mathsf{put}}$ is defined as 
\[
\begin{array}{l@{~} c@{~} l}
V_{\mathsf{put}} & \defeq & \lambda\, y \!:\! (\Sigma\, x \!:\! \State.\, 1 \to (\State \,\to\, (\mathsf{VU} \times \State))) .\, \lambda\, x_S \!:\! \State .\,
\\[-1mm]
\multicolumn{3}{c}{\hspace{4.8cm}\pmatch y {(x \!:\! \State, x' \!:\! 1 \to (\State \,\to\, (\mathsf{VU} \times \State)))} {} {x'\, \star\,\, x}}
\end{array}
\]
In other words, $V_{\mathsf{get}}$ and $V_{\mathsf{put}}$ are defined as if they were operations of the free algebra on $\mathsf{VU}$ for the equational theory corresponding to the fibred effect theory $\mathcal{T}_{\text{GS}}$.

On closer inspection, we can see that $V_{\!\widehat{Q}}$ corresponds to Dijkstra's weakest precondition semantics of stateful programs~\cite{Dijkstra:GCommands}, as made precise in the next proposition.

\begin{proposition}
The following definitional equations are derivable in eMLTT$_{\mathcal{T}_{\text{GS}}}^{\mathcal{H}}$:
\[
\mkrule
{\Gamma \vdash V_{\!\widehat{Q}}\,\, (\thunk\! (\return V))\,\, V_S = V_{\!Q}\,\, V\,\, V_S : \mathsf{VU}}
{\vj \Gamma {V_{\!Q}} {A \to \State \to \mathsf{VU}} \quad \vj \Gamma V A \quad \vj \Gamma {V_S} \State}
\vspace{0.25cm}
\]

\[
\mkrule
{\Gamma \vdash V_{\!\widehat{Q}}\,\, (\thunk\! (\mathtt{get}^{F\!A}(y.\, M)))\,\, V_S = V_{\!\widehat{Q}}\,\, (\thunk M[V_S/y])\,\, V_S : \mathsf{VU}}
{\vj \Gamma {V_{\!Q}} {A \to \State \to \mathsf{VU}} \quad \cj {\Gamma, y \!:\! \State} {M} {FA} \quad \vj \Gamma {V_S} \State}
\vspace{0.25cm}
\]
\[
\mkrule
{\Gamma \vdash V_{\!\widehat{Q}}\,\, (\thunk\! (\mathtt{put}^{F\!A}_{V'_S}(M)))\,\, V_S = V_{\!\widehat{Q}}\,\, (\thunk M)\,\, V'_S : \mathsf{VU}}
{\vj \Gamma {V_{\!Q}} {A \to \State \to \mathsf{VU}} \quad \cj {\Gamma} {M} {FA} \quad \vj \Gamma {V_S} \State \quad \vj \Gamma {V'_S} {\State}}
\]
\end{proposition}

\begin{proof}
All three equations are proved by straightforward equational reasoning, e.g., 
\begin{fleqn}[0.3cm]
\begin{align*}
\Gamma \,\vdash\,\, & V_{\!\widehat{Q}}\,\, (\thunk\! (\mathtt{put}^{F\!A}_{V'_S}(M)))\,\, V_S
\\[1mm]
=\,\, & \fst \big(\big(\thunk\! \big(\doto {(\force {F\!A} (\thunk\! (\mathtt{put}^{F\!A}_{V'_S}(M))))} {y' \!:\! A} {} {\\[-2mm] & \hspace{3.5cm}  \force {\langle \mathsf{S} \,\to\, \mathsf{VU} \times \mathsf{S} , \{V_{\sigalgop}\}_{\sigalgop \in \mathcal{S}_{\text{GS}}} \rangle} (\lambda\, x'_{S} \!:\! \State .\, \langle V_{\!Q}\,\, y'\, x'_{S} , x'_{S} \rangle)}\big)\big)\,\,  V_S\big)
\\[1mm]
=\,\, & \fst \big(\big(\thunk\! \big(\doto {\mathtt{put}^{F\!A}_{V'_S}(M)} {y' \!:\! A} {} {\\[-2mm] & \hspace{3.5cm} \force {\langle \mathsf{S} \,\to\, \mathsf{VU} \times \mathsf{S} , \{V_{\sigalgop}\}_{\sigalgop \in \mathcal{S}_{\text{GS}}} \rangle} (\lambda\, x'_{S} \!:\! \State .\, \langle V_{\!Q}\,\, y'\, x'_{S} , x'_{S} \rangle)}\big)\big)\,\, V_S\big)
\\[2.5mm]
=\,\, & \fst \big(\big(\thunk\! \big(\doto {\mathtt{put}^{\!\langle \mathsf{S} \,\to\, \mathsf{VU} \times \mathsf{S} , \{V_{\sigalgop}\}_{\sigalgop \in \mathcal{S}_{\text{GS}}} \rangle}_{V'_S}(M} {y' \!:\! A} {} {\\[-2mm] & \hspace{3.35cm} \force {\langle \mathsf{S} \,\to\, \mathsf{VU} \times \mathsf{S} , \{V_{\sigalgop}\}_{\sigalgop \in \mathcal{S}_{\text{GS}}} \rangle} (\lambda\, x'_{S} \!:\! \State .\, \langle V_{\!Q}\,\, y'\, x'_{S} , x'_{S} \rangle))}\big)\big)\,\, V_S\big)
\\[1mm]
=\,\, & \fst \big(\big(\thunk\! \big(\force {\langle \mathsf{S} \,\to\, \mathsf{VU} \times \mathsf{S} , \{V_{\sigalgop}\}_{\sigalgop \in \mathcal{S}_{\text{GS}}} \rangle}\big(\\[-1mm] & \hspace{1.5cm} \lambda\, x_S \!:\! \State .\, \pmatch {\big\langle V'_S , \lambda\, y \!:\! 1 .\, \thunk (\doto M {y' \!:\! A} {} {\\[-2mm] & \hspace{3cm} \force {\langle \mathsf{S} \,\to\, \mathsf{VU} \times \mathsf{S} , \{V_{\sigalgop}\}_{\sigalgop \in \mathcal{S}_{\text{GS}}} \rangle} (\lambda\, x'_{S} \!:\! \State .\, \langle V_{\!Q}\,\, y'\, x'_{S} , x'_{S} \rangle)})  \big\rangle} {\\[-1mm] & \hspace{4.85cm} (x \!:\! \State, x' \!:\! 1 \to (\State \,\to\, \mathsf{VU} \times \State))} {} {x'\, \star\,\, x}\big)\big)\big)\,\, V_S\big)
\\[1mm]
=\,\, & \fst \big(\big( \lambda\, x_S \!:\! \State .\, \pmatch {\big\langle V'_S , \lambda\, y \!:\! 1 .\, \thunk\! (\doto M {y' \!:\! A} {} {\\[-2mm] & \hspace{3cm} \force {\langle \mathsf{S} \,\to\, \mathsf{VU} \times \mathsf{S} , \{V_{\sigalgop}\}_{\sigalgop \in \mathcal{S}_{\text{GS}}} \rangle} (\lambda\, x'_{S} \!:\! \State .\, \langle V_{\!Q}\,\, y'\, x'_{S} , x'_{S} \rangle)})  \big\rangle} {\\[-1mm] & \hspace{5.2cm} (x \!:\! \State, x' \!:\! 1 \to (\State \,\to\, \mathsf{VU} \times \State))} {} {x'\, \star\,\, x}\big)\,\, V_S\big)
\\[1mm]
=\,\, & \fst \big(\pmatch {\big\langle V'_S , \lambda\, y \!:\! 1 .\, \thunk\! (\doto M {y' \!:\! A} {} {\\[-2mm] & \hspace{2.5cm} \force {\langle \mathsf{S} \,\to\, \mathsf{VU} \times \mathsf{S} , \{V_{\sigalgop}\}_{\sigalgop \in \mathcal{S}_{\text{GS}}} \rangle} (\lambda\, x'_{S} \!:\! \State .\, \langle V_{\!Q}\,\, y'\, x'_{S} , x'_{S} \rangle)})  \big\rangle} {\\[-1mm] & \hspace{5.85cm} (x \!:\! \State, x' \!:\! 1 \to (\State \,\to\, \mathsf{VU} \times \State))} {} {x'\, \star\,\, x}\big)
\\
=\,\, & \fst \big(\big(\thunk\! \big(\doto {M} {y' \!:\! A} {} {\\[-2mm] & \hspace{3.5cm} \force {\langle \mathsf{S} \,\to\, \mathsf{VU} \times \mathsf{S} , \{V_{\sigalgop}\}_{\sigalgop \in \mathcal{S}_{\text{GS}}} \rangle} (\lambda\, x'_{S} \!:\! \State .\, \langle V_{\!Q}\,\, y'\, x'_{S} , x'_{S} \rangle)}\big)\big)\,\, V'_S\big)
\\[1mm]
=\,\, & \fst \big(\big(\thunk\! \big(\doto {(\force {F\!A} (\thunk M))} {y' \!:\! A} {} {\\[-2mm] & \hspace{3.5cm}  \force {\langle \mathsf{S} \,\to\, \mathsf{VU} \times \mathsf{S} , \{V_{\sigalgop}\}_{\sigalgop \in \mathcal{S}_{\text{GS}}} \rangle} (\lambda\, x'_{S} \!:\! \State .\, \langle V_{\!Q}\,\, y'\, x'_{S} , x'_{S} \rangle)}\big)\big)\,\,  V'_S\big)
\\[1mm]
=\,\, & V_{\!\widehat{Q}}\,\, (\thunk M)\,\, V'_S : \mathsf{VU}
\end{align*}
\end{fleqn}
The proofs of the other two equations follow a similar pattern.
\end{proof}

We leave comparing our handler-based definition of Dijkstra's weakest precondition semantics to the CPS-translation based definition used in \pl{F*}~\cite{Ahman:DM4Free} for future work.

\subsection{Specifying patterns of allowed effects}

Analogously to lifting predicates from return values to computations, specifying patterns 
of allowed effects is easiest when the given fibred effect theory does not contain equations.
Therefore, let us again consider the fibred effect theory  $\mathcal{T}_{\text{I/O}}$ of input/output. 

As a first example, we define a very coarse grained predicate $V_{\mathsf{no\text{-}w}}$ on the allowed I/O-effects, namely, one that \emph{disallows all writes}. This predicate is defined as follows: 
\[
V_{\mathsf{no\text{-}w}} \defeq \lambda\, y \!:\! U\!F\!A .\, \thunk\! \big(\doto {(\force {F\!A} y)} {y' \!:\! A} {} {\force {\langle \mathsf{VU} , \{V_{\sigalgop}\}_{\sigalgop \in \mathcal{S}_{\text{I/O}}} \rangle} \mathtt{unit\text{-}code}}\big)
\]
where the value terms $V_{\mathsf{read}}$ and $V_{\mathsf{write}}$ are given by 
\[
\begin{array}{l@{~} c @{~} l}
V_{\mathsf{read}} & \defeq & \lambda\, y \!:\! (\Sigma\, x \!:\! 1 .\, \Character \to \mathsf{VU}) .\, \mathtt{v\text{-}pi\text{-}code}(\mathtt{chr\text{-}code},y'\!.\, (\snd y)\,\, y')
\\[2mm]
V_{\mathsf{write}} & \defeq & \lambda\, y \!:\! (\Sigma\, x \!:\! \Character .\, 1 \to \mathsf{VU}) .\, \mathtt{empty\text{-}code}
\end{array}
\]

For example, the computation term $\mathtt{read}^{F\!A}(x.\,\mathtt{write}_V^{F\!A}(M))$ does not satisfy $V_{\mathsf{no\text{-}w}}$ (if $\Gamma$ is consistent) because of the following derivable value type isomorphism:
\[
\Gamma \vdash \mathsf{El}(V_{\mathsf{no\text{-}w}} (\thunk \! (\mathtt{read}^{F\!A}(y.\,\mathtt{write}_V^{F\!A}(M))))) = \Pi\, y \!:\! 1 \!+\! 1 .\, 0 \cong 0
\]
 
Next, we show how to 
specify more complex patterns of allowed I/O-effects in the style of session types~\cite{Honda:LangPrimitives}. To this end, for our second example, let us assume an inductive type $\lj \diamond \mathsf{Protocol}$, defined using three constructors with the following types:
\[
\begin{array}{c}
\mathtt{e} : \mathsf{Protocol}
\qquad
\mathtt{r} : (\Character \to \mathsf{Protocol}) \to \mathsf{Protocol}
\\[2mm]
\mathtt{w} : (\Character \to \mathsf{VU}) \times \mathsf{Protocol} \to \mathsf{Protocol}
\end{array}
\]

Intuitively, $\mathtt{e}$ stands for the end of communication; $\mathtt{r}$ specifies that the next allowed I/O-effect has to be a read; and $\mathtt{w}$ specifies that the next I/O-effect has to be a write. 

Note that both $\mathtt{r}$ and $\mathtt{w}$ take a $\mathsf{Protocol}$-valued argument. This argument specifies the pattern of I/O-effects that are allowed after performing a read or write effect, respectively. Further, observe that for $\mathtt{r}$, the $\mathsf{Protocol}$-valued argument can depend on the character being read from the input. It is also worth noting that $\mathtt{w}$ takes a second argument (with type $\Character \to \mathsf{VU}$). This argument denotes a predicate on the values of type $\Character$ that are allowed to be written to the output by the corresponding write effect.

Then, given some particular protocol $\Gamma \vdash V_{\mathsf{pr}} : \mathsf{Protocol}$, we define a predicate
\[
\begin{array}{c}
\hspace{-5.5cm}
V_{\widehat{\mathsf{pr}}} \defeq \lambda\, y \!:\! U\!F\!A .\, \big(\thunk\! \big(\doto {(\force {F\!A} y)} {y' \!:\! A} {} {\\[-1.5mm] \hspace{7cm} \force {\langle \mathsf{Protocol} \to \mathsf{VU} , \{V_{\sigalgop}\}_{\sigalgop \in \mathcal{S}_{\text{I/O}}} \rangle} V_{\mathsf{ret}}}\big)\big)\, V_{\mathsf{pr}}
\end{array}
\]
where the value terms $V_{\mathsf{ret}}$, $V_{\mathsf{read}}$, and $V_{\mathsf{write}}$ are defined as follows (for better readability, we opt to give their definitions by pattern-matching on their respective arguments):
\[
\begin{array}{l@{~} l@{~} l@{~~} c@{~~} l}
V_{\mathsf{ret}}& & \mathtt{e} & \defeq & \mathtt{unit\text{-}code}
\\[2mm]
V_{\mathsf{read}}& \langle V , V_{\mathsf{rk}} \rangle& (\mathtt{r}~ V'_{\mathsf{pr}}) & \defeq & \mathtt{v\text{-}pi\text{-}code}(\mathtt{chr\text{-}code},y.\, (V_{\mathsf{rk}}\,\, y)\,\, (V'_{\mathsf{pr}}\,\, y))
\\[2mm]
V_{\mathsf{write}}& \langle V , V_{\mathsf{wk}} \rangle& (\mathtt{w}~ \langle V_{\!P} , V'_{\mathsf{pr}} \rangle) & \defeq & \mathtt{v\text{-}sigma\text{-}code}(V_{\!P}\,\, V, y.\, V_{\mathsf{wk}}\,\, \star\,\, V'_{\mathsf{pr}})
\end{array}
\]
with all other cases defined to be equal to $\mathtt{empty\text{-}code}$; and where 
\[
\begin{array}{c}
\Gamma \vdash V_{\mathsf{rk}} : \Character \to \mathsf{Protocol} \to \mathsf{VU}
\qquad
\Gamma \vdash V_{\mathsf{wk}} : 1 \to \mathsf{Protocol} \to \mathsf{VU}
\end{array}
\]
are the respective continuations of the algebraic operations denoted by $V_{\mathsf{read}}$ and $V_{\mathsf{write}}$.

The high-level idea is that $V_{\widehat{\mathsf{pr}}}$ computes to $\mathtt{empty\text{-}code}$ if the given computation does not conform to the pattern of I/O-effects specified by the given protocol $V_{\mathsf{pr}}$. On the other hand, if the given computation happens to conform to the given protocol, $V_{\widehat{\mathsf{pr}}}$ will compute to a representation of a sequence of value $\Pi$- and $\Sigma$-types (ending with $1$) for which one can easily construct an inhabitant, and thus prove that $V_{\widehat{\mathsf{pr}}}$ holds.

We conclude by noting that one can easily combine $V_{\mathsf{no\text{-}w}}$ and $V_{\widehat{\mathsf{pr}}}$ with predicates from Section~\ref{section:liftingpredicatesexamples}, by replacing $\mathtt{unit\text{-}code}$ with a predicate $V_{\!P}$ on return values.

\section{Meta-theory} 
\label{section:handlersmetatheory}

In this section we show how to extend the meta-theory of eMLTT and eMLTT$_{\!\mathcal{T}_{\text{eff}}}$ to eMLTT$_{\!\mathcal{T}_{\text{eff}}}^{\mathcal{H}}$, analogously to how we extended the meta-theory of eMLTT to eMLTT$_{\!\mathcal{T}_{\text{eff}}}$ in Section~\ref{sect:emlttalgeffectsmetatheory}. Similarly to eMLTT$_{\!\mathcal{T}_{\text{eff}}}$, many of the results from Section~\ref{sect:metatheory} (and from the beginning of Section~\ref{sect:completeness}) extend straightforwardly to eMLTT$_{\!\mathcal{T}_{\text{eff}}}^{\mathcal{H}}$, with either the proof remaining the same or it can be easily adapted for eMLTT$_{\!\mathcal{T}_{\text{eff}}}^{\mathcal{H}}$. Analogously to  Section~\ref{sect:emlttalgeffectsmetatheory}, we omit the proofs of the propositions and theorems that extend to eMLTT$_{\!\mathcal{T}_{\text{eff}}}^{\mathcal{H}}$ straightforwardly and only comment on the more involved proofs.

\subsection*{Extending Theorem~\ref{thm:substitution} (Value term substitution) to eMLTT$_{\!\mathcal{T}_{\text{eff}}}^{\mathcal{H}}$}

\index{substitution theorem!syntactic --!-- for value terms}

We begin by recalling that in Theorem~\ref{thm:substitution} we showed that the substitution rule is admissible in eMLTT for substituting value terms for value variables. When extending Theorem~\ref{thm:substitution} to eMLTT$_{\mathcal{T}_{\text{eff}}}^{\mathcal{H}}$, we keep the basic proof principle the same: we prove $(a)$--$(l)$ for different kinds of types, terms, and definitional equations simultaneously, with $(a)$--$(b)$ proved by induction on the length of the given value context $\Gamma_2$, and $(c)$--$(l)$  by induction on the given derivations; and this theorem as a whole is proved simultaneously with the eMLTT$_{\mathcal{T}_{\text{eff}}}^{\mathcal{H}}$ version of the weakening theorem (Theorem~\ref{thm:weakening}). 

The cases for the terms and definitional equations introduced by eMLTT$_{\mathcal{T}_{\text{eff}}}$ are proved analogously to Section~\ref{sect:fibalgeffectsineMLTT}; this also includes additionally proving an eMLTT$_{\mathcal{T}_{\text{eff}}}^{\mathcal{H}}$ version of Proposition~\ref{prop:effecttermstranslationsubstitution}, so as to account for substituting value terms for value variables in the translation of effect terms. The new cases for the types, terms, and definitional equations introduced by eMLTT$_{\mathcal{T}_{\text{eff}}}^{\mathcal{H}}$ are proved analogously to other types, terms, and definitional equations that involve variable bindings and type annotations.

For example, in the case corresponding to the formation rule for the user-defined algebra type, the given derivation ends with 
\[
\hspace{-0.15cm}
\mkrule
{
\lj {\Gamma_1,y \!:\! B, \Gamma_2} {\langle A , \{V_{\sigalgop}\}_{\sigalgop \in \mathcal{S}_{\text{eff}}} \rangle}
}
{
\begin{array}{c}
\lj {\Gamma_1,y \!:\! B, \Gamma_2} A
\quad
\vj {\Gamma_1,y \!:\! B, \Gamma_2} {V_{\sigalgop}} {(\Sigma\, x \!:\! I . O \to A) \to A}
\\[2mm]
\hspace{-4cm} \veq {\Gamma_1,y \!:\! B, \Gamma_2} {\overrightarrow{\lambda\, x'_i \!:\! \widehat{A_i} .}\, \overrightarrow{\lambda\, x_{w_{\!j}} \!:\! \widehat{A'_j} \to A .}\, \efftrans {T_1} {A; \overrightarrow{x'_i}; \overrightarrow{x_{w_{\!j}}}; \overrightarrow{V_{\sigalgop}}} \\ \hspace{1.9cm} } {\overrightarrow{\lambda\, x'_i \!:\! \widehat{A_i} .}\, \overrightarrow{\lambda\, x_{w_{\!j}} \!:\! \widehat{A'_j} \to A .}\, \efftrans {T_2} {A; \overrightarrow{x'_i}; \overrightarrow{x_{w_{\!j}}}; \overrightarrow{V_{\sigalgop}}}\,} {\,\overrightarrow{\Pi x'_i \!:\! \widehat{A_i} .}\, \overrightarrow{\widehat{A'_j} \to A} \to A}
\\[3mm]
(\text{for all } \sigalgop : (x \!:\! I) \longrightarrow O \in \mathcal{S}_{\text{eff}}
\text{ and }
\ljeq {\Gamma \vertbar \Delta} {T_1} {T_2} \in \mathcal{E}_{\text{eff}})
\end{array}
}
\]
and we need to construct a derivation of 
\[
\lj {\Gamma_1,\Gamma_2[W/y]} {\langle A[W/y] , \{V_{\sigalgop}[W/y]\}_{\sigalgop \in \mathcal{S}_{\text{eff}}} \rangle}
\]

To construct this derivation, we first use $(c)$, $(g)$, and $(h)$ on the derivations given by the premises of this rule, in order to get derivations of  
\[
\begin{array}{c}
\lj {\Gamma_1, \Gamma_2[W/y]} A[W/y]
\\[2mm]
\vj {\Gamma_1, \Gamma_2[W/y]} {V_{\sigalgop}[W/y]} {(\Sigma\, x \!:\! I[W/y] . O[W/y] \to A[W/y]) \to A[W/y]}
\\[4mm]
\hspace{-1.6cm} \veq {\Gamma_1, \Gamma_2[W/y]} {\overrightarrow{\lambda\, x'_i \!:\! \widehat{A_i}[W/y] .}\, \overrightarrow{\lambda\, x_{w_{\!j}} \!:\! \widehat{A'_j}[W/y] \to A[W/y] .}\, \efftrans {T_1} {A; \overrightarrow{x'_i}; \overrightarrow{x_{w_{\!j}}}; \overrightarrow{V_{\sigalgop}}}[W/y] \\ \hspace{0.7cm} } {\overrightarrow{\lambda\, x'_i \!:\! \widehat{A_i}[W/y] .}\, \overrightarrow{\lambda\, x_{w_{\!j}} \!:\! \widehat{A'_j}[W/y] \to A[W/y] .}\, \efftrans {T_2} {A; \overrightarrow{x'_i}; \overrightarrow{x_{w_{\!j}}}; \overrightarrow{V_{\sigalgop}}}[W/y]\,} {\\[2mm] \hspace{6.5cm} \overrightarrow{\Pi x'_i \!:\! \widehat{A_i}[W/y] .}\, \overrightarrow{\widehat{A'_j}[W/y] \to A[W/y]} \to A[W/y]}
\end{array}
\]

Next, we use the eMLTT$_{\mathcal{T}_{\text{eff}}}^{\mathcal{H}}$ version of Proposition~\ref{prop:freevariablesofwellformedexpressions} on the given derivations of $\vj {\Gamma_1} W B$, $\lj \diamond I$, $\lj {x \!:\! I} O$, and $\lj \Gamma {A'_j}$, to get the following inclusions:
\[
FVV(W) \subseteq V\!ars(\Gamma_1)
\qquad
FVV(I) = \emptyset
\qquad
FVV(O) \subseteq \{x\}
\qquad
FVV(A'_j) \subseteq V\!ars(\Gamma)
\]

According to our adopted variable conventions, we also know that  
\[
x, x'_1, \ldots, x'_n, x_{w_{1}}, \ldots, x_{w_m} \not\in V\!ars(\Gamma_1, y \!:\! B, \Gamma_2)
\]

As a result, by recalling that the properties of substitution we established for eMLTT in Section~\ref{sect:syntax} extend straightforwardly to  eMLTT$_{\mathcal{T}_{\text{eff}}}^{\mathcal{H}}$, we get the following equations:
\[
I[W/y] = I
\qquad
O[W/y] = O
\qquad
A'_j[W/y] = A'_j
\]

Further, by using the eMLTT$_{\mathcal{T}_{\text{eff}}}^{\mathcal{H}}$ version of Proposition~\ref{prop:effecttermstranslationsubstitution}, we also get that
\[
\begin{array}{c}
\efftrans {T_1} {A; \overrightarrow{x'_i}; \overrightarrow{x_{w_{\!j}}}; \overrightarrow{V_{\sigalgop}}}[W/y] = \efftrans {T_1} {A[W/y]; \overrightarrow{x'_i[W/y]}; \overrightarrow{x_{w_{\!j}}[W/y]}; \overrightarrow{V_{\sigalgop}[W/y]}} = \efftrans {T_1} {A[W/y]; \overrightarrow{x'_i}; \overrightarrow{x_{w_{\!j}}}; \overrightarrow{V_{\sigalgop}[W/y]}}
\\[3mm]
\efftrans {T_2} {A; \overrightarrow{x'_i}; \overrightarrow{x_{w_{\!j}}}; \overrightarrow{V_{\sigalgop}}}[W/y] = \efftrans {T_2} {A[W/y]; \overrightarrow{x'_i[W/y]}; \overrightarrow{x_{w_{\!j}}[W/y]}; \overrightarrow{V_{\sigalgop}[W/y]}} = \efftrans {T_2} {A[W/y]; \overrightarrow{x'_i}; \overrightarrow{x_{w_{\!j}}}; \overrightarrow{V_{\sigalgop}[W/y]}}
\end{array}
\]
where the right-hand equations follow from the properties of substitution.

Finally, by combining all these observations, we get that the derivations we constructed in the beginning of this proof turn out to be in fact derivations of
\[
\hspace{-0.3cm}
\begin{array}{c}
\lj {\Gamma_1, \Gamma_2[W/y]} A[W/y]
\\[2mm]
\vj {\Gamma_1, \Gamma_2[W/y]} {V_{\sigalgop}[W/y]} {(\Sigma\, x \!:\! I . O \to A[W/y]) \to A[W/y]}
\\[4mm]
\hspace{-2.5cm} \veq {\Gamma_1, \Gamma_2[W/y]} {\overrightarrow{\lambda\, x'_i \!:\! \widehat{A_i} .}\, \overrightarrow{\lambda\, x_{w_{\!j}} \!:\! \widehat{A'_j} \to A[W/y] .}\, \efftrans {T_1} {A[W/y]; \overrightarrow{x'_i}; \overrightarrow{x_{w_{\!j}}}; \overrightarrow{V_{\sigalgop}[W/y]}} \\ \hspace{-0.15cm} } {\overrightarrow{\lambda\, x'_i \!:\! \widehat{A_i} .}\, \overrightarrow{\lambda\, x_{w_{\!j}} \!:\! \widehat{A'_j} \to A[W/y] .}\, \efftrans {T_2} {A[W/y]; \overrightarrow{x'_i}; \overrightarrow{x_{w_{\!j}}}; \overrightarrow{V_{\sigalgop}[W/y]}}\,} {\\[2mm] \hspace{8.5cm} \overrightarrow{\Pi x'_i \!:\! \widehat{A_i} .}\, \overrightarrow{\widehat{A'_j} \to A[W/y]} \to A[W/y]}
\end{array}
\]  
Therefore, we can use the formation rule for the user-defined algebra type with these derivations to construct the required derivation of $\lj {\Gamma_1,\Gamma_2[W/y]} {\langle A[W/y] , \{V_{\sigalgop}\}_{\sigalgop \in \mathcal{S}_{\text{eff}}} \rangle}$.

\subsection*{Extending Proposition~\ref{prop:wellformedcomponentsofjudgements} to eMLTT$_{\!\mathcal{T}_{\text{eff}}}^{\mathcal{H}}$}

We begin by recalling that in Proposition~\ref{prop:wellformedcomponentsofjudgements} we showed that the judgements of well-formed expressions and definitional equations only involve well-formed contexts and types, and well-typed terms. For example, given $\ceq \Gamma M N {\ul{C}}$, we showed that
\[
\cj \Gamma M \ul{C}
\qquad
\cj \Gamma N \ul{C}
\]

When extending Proposition~\ref{prop:wellformedcomponentsofjudgements} to eMLTT$_{\mathcal{T}_{\text{eff}}}^{\mathcal{H}}$, we keep the basic proof principle the same: we prove $(a)$--$(j)$ for different kinds of types, terms, and definitional equations simultaneously, by induction on the given derivations, using the eMLTT$_{\mathcal{T}_{\text{eff}}}^{\mathcal{H}}$ versions of the weakening and substitution theorems, where necessary.

The cases for the terms and definitional equations introduced by eMLTT$_{\mathcal{T}_{\text{eff}}}$ are proved as in Section~\ref{sect:fibalgeffectsineMLTT}; this also includes proving Proposition~\ref{prop:welltypednessoftranslatingeffectterms} to show that well-formed effect terms translate into well-typed value terms. Most of the new cases introduced by eMLTT$_{\mathcal{T}_{\text{eff}}}^{\mathcal{H}}$ are proved similarly to other types and terms that involve variable bindings and type annotations. However, in order to be able to account for the congruence rule for the user-defined algebra type, we need to prove the eMLTT$_{\mathcal{T}_{\text{eff}}}^{\mathcal{H}}$ version of Proposition~\ref{prop:wellformedcomponentsofjudgements} simultaneously with Propositions~\ref{prop:repeatedsubstitutioninhandlerssect} and~\ref{prop:replacementinfibeffecttermtranslationindices} below.

\begin{proposition}
\label{prop:repeatedsubstitutioninhandlerssect}
Given a well-typed value term $\vj {\Gamma_1,\Gamma_2,\Gamma_3} V A$ and definitional equations $\veq {\Gamma_1} {V_i} {W_i} {A_i[V_1/x_1, \ldots, V_{i-1}/x_{i-1}]}$, for all $x_i \!:\! A_i$ in $\Gamma_2$, then 
\[
\veq {\Gamma_1,\Gamma_3[\overrightarrow{V_i}/\overrightarrow{x_i}]} {V[\overrightarrow{V_i}/\overrightarrow{x_i}]} {V[\overrightarrow{W_i}/\overrightarrow{x_i}]} {A[\overrightarrow{V_i}/\overrightarrow{x_i}]}
\]
\end{proposition}

\begin{proof}
We prove this proposition by induction on the length of $\Gamma_2$, as discussed below.

\vspace{0.1cm}

\noindent \textit{Base case} (with $\Gamma_2 = \diamond$): In this case, we simply use the reflexivity rule for value terms to prove $\veq {\Gamma_1,\Gamma_3} V V A$.

\vspace{0.1cm}

\noindent \textit{Step case} (with $\Gamma_2 = \Gamma'_2, x_n \!:\! A_n$): In this case, we first use the induction hypothesis on $\vj {\Gamma_1,\Gamma'_2, x_n \!:\! A_n,\Gamma_3} V A$ (with the contexts chosen as $\Gamma_1$ and $\Gamma'_2$ and $x_n \!:\! A_n, \Gamma_3$) to prove
\[
\begin{array}{c}
\hspace{-4cm}
\veq {\Gamma_1, x_n \!:\! A_n[V_1/x_1, \ldots, V_{n-1}/x_{n-1}], \Gamma_3[V_1/x_1, \ldots, V_{n-1}/x_{n-1}]} {\\ \hspace{1cm} V[V_1/x_1, \ldots, V_{n-1}/x_{n-1}]} {V[W_1/x_1, \ldots, W_{n-1}/x_{n-1}]} {A[V_1/x_1, \ldots, V_{n-1}/x_{n-1}]}
\end{array}
\]

Next, by using the simultaneously proved eMLTT$_{\mathcal{T}_{\text{eff}}}^{\mathcal{H}}$ version of Proposition~\ref{prop:wellformedcomponentsofjudgements} on this definitional equation, we get a derivation of
\[
\begin{array}{c}
\hspace{-3.9cm}
\vj {\Gamma_1, x_n \!:\! A_n[V_1/x_1, \ldots, V_{n-1}/x_{n-1}], \Gamma_3[V_1/x_1, \ldots, V_{n-1}/x_{n-1}]} {\\ \hspace{5.6cm} V[W_1/x_1, \ldots, W_{n-1}/x_{n-1}]} {A[V_1/x_1, \ldots, V_{n-1}/x_{n-1}]}
\end{array}
\]

Further, by using the eMLTT$_{\mathcal{T}_{\text{eff}}}^{\mathcal{H}}$ version of the substitution theorem on the definitional equation above, we get a proof of  
\[
\begin{array}{c}
\hspace{-7.6cm}
\veq {\Gamma_1, \Gamma_3[V_1/x_1, \ldots, V_{n-1}/x_{n-1}][V_n/x_n]} {\\ \hspace{-2.55cm} V[V_1/x_1, \ldots, V_{n-1}/x_{n-1}][V_n/x_n]\\ \hspace{2.75cm} } {V[W_1/x_1, \ldots, W_{n-1}/x_{n-1}][V_n/x_n]} {A[V_1/x_1, \ldots, V_{n-1}/x_{n-1}][V_n/x_n]}
\end{array}
\]
and for which we can use the eMLTT$_{\mathcal{T}_{\text{eff}}}^{\mathcal{H}}$ version of Proposition~\ref{prop:simultaneoussubstlemma2} to show that
\[
\begin{array}{c}
\Gamma_1, \Gamma_3[V_1/x_1, \ldots, V_{n-1}/x_{n-1}][V_n/x_n]
=
\Gamma_1, \Gamma_3[V_1/x_1, \ldots, V_{n-1}/x_{n-1}, V_n/x_n]
\\[2mm]
A[V_1/x_1, \ldots, V_{n-1}/x_{n-1}][V_n/x_n]
=
A[V_1/x_1, \ldots, V_{n-1}/x_{n-1}, V_n/x_n]
\end{array}
\]

Finally, we can prove the required equation as follows:
\begin{fleqn}[0.3cm]
\begin{align*}
{\Gamma_1,\Gamma_3[\overrightarrow{V_i}/\overrightarrow{x_i}]} \,\vdash\,\, & 
V[V_1/x_1, \ldots, V_{n-1}/x_{n-1}, V_n/x_n]
\\
=\,\, & V[V_1/x_1, \ldots, V_{n-1}/x_{n-1}][V_n/x_n]
\\
=\,\, & V[W_1/x_1, \ldots, W_{n-1}/x_{n-1}][V_n/x_n]
\\
=\,\, & V[W_1/x_1, \ldots, W_{n-1}/x_{n-1}][W_n/x_n]
\\
=\,\, & V[W_1/x_1, \ldots, W_{n-1}/x_{n-1}, W_n/x_n] : A[\overrightarrow{V_i}/\overrightarrow{x_i}]
\end{align*}
\end{fleqn}
where the first and last equation are proved using the eMLTT$_{\mathcal{T}_{\text{eff}}}^{\mathcal{H}}$\! version of Proposition~\ref{prop:simultaneoussubstlemma2}; the second equation is proved using the definitional 
equation derived above; and the third equation is proved using the replacement rule for value terms.
\end{proof}

\begin{proposition}
\label{prop:replacementinfibeffecttermtranslationindices}
Given a well-formed effect term $\lj {\Gamma \vertbar \Delta} T$ derived from $\mathcal{S}_{\text{eff}}$, value types $A$ and $B$, value terms $V_{i}$ and $W_i$ (for all $x_i \!:\! A_i$ in $\Gamma$), value terms $V'_{j}$ and $W'_j$ (for all $w_j \!:\! A'_j$ in $\Delta$), value terms $V_{\sigalgop}$ and $W_{\sigalgop}$ (for all $\sigalgop : (x \!:\! I) \longrightarrow O$ in $\mathcal{S}_{\text{eff}}$), and a value context $\Gamma'$ such that
\begin{itemize}
\item $\vdash \Gamma'$, 
\item $\ljeq {\Gamma'} A B$, 
\item $\veq {\Gamma'} {V_i} {W_i} {A_i[V_1/x_1, \ldots, V_{i-1}/x_{i-1}]}$, 
\item $\veq {\Gamma'} {V'_j} {W'_j} {A'_j[V_1/x_1, \ldots, V_n/x_n] \to A}$, 
\item $\veq {\Gamma'} {V_{\sigalgop}} {W_{\sigalgop}} {(\Sigma\, x \!:\! I .\, O \to A) \to A}$, 
\end{itemize}
then 
\[
\veq {\Gamma'} {\efftrans T {A; \overrightarrow{V_i}; \overrightarrow{V'_j}; \overrightarrow{V_{\sigalgop}}}} {\efftrans T {B; \overrightarrow{W_i}; \overrightarrow{W'_j}; \overrightarrow{W_{\sigalgop}}}} {A}
\]
\end{proposition}

\begin{proof}
We prove this proposition by induction on the derivation of $\lj {\Gamma \vertbar \Delta} T$.

As a representative example, we consider the case of algebraic operations $\algop_V(y .\, T)$, for which  we need to prove the following definitional equation:
\[
\begin{array}{c}
\hspace{-1cm}
\veq {\Gamma'} {V_{\sigalgop}\,\, \langle V[\overrightarrow{V_i}/\overrightarrow{x_i}] , \lambda\, y \!:\! O[V[\overrightarrow{V_i}/\overrightarrow{x_i}]/x] .\, \efftrans {T} {A; \overrightarrow{V_i},\, y; \overrightarrow{V'_j}; \overrightarrow{V_{\sigalgop}}}\rangle
\\[1mm] \hspace{0.15cm}} { W_{\sigalgop}\,\, \langle V[\overrightarrow{W_i}/\overrightarrow{x_i}] , \lambda\, y \!:\! O[V[\overrightarrow{W_i}/\overrightarrow{x_i}]/x] .\, \efftrans {T} {B; \overrightarrow{W_i},\, y; \overrightarrow{W'_j}; \overrightarrow{W_{\sigalgop}}}\rangle
} {A}
\end{array}
\]

First, we note that we can repeatedly use the replacement rules for value types and value terms with the derivations of $\vj \Gamma V I$ and $\lj {x \!:\! I} O$, in combination with the eMLTT$_{\mathcal{T}_{\text{eff}}}^{\mathcal{H}}$\! version Theorem~\ref{thm:simultaneoussubstitution} (simultaneous value term 
substitution), to prove 
\[
\veq {\Gamma'} {V[\overrightarrow{V_i}/\overrightarrow{x_i}]} {V[\overrightarrow{W_i}/\overrightarrow{x_i}]} {I[\overrightarrow{V_i}/\overrightarrow{x_i}]}
\qquad
\ljeq {\Gamma'} {O[V[\overrightarrow{V_i}/\overrightarrow{x_i}]/x]} {O[V[\overrightarrow{W_i}/\overrightarrow{x_i}]/x]}
\]
However, as $\lj {\diamond} I$, we can use the eMLTT$_{\mathcal{T}_{\text{eff}}}^{\mathcal{H}}$ version of Proposition~\ref{prop:freevariablesofwellformedexpressions} to get that $FVV(I) = \emptyset$, and 
thus we can use the eMLTT$_{\mathcal{T}_{\text{eff}}}$ version of Propopsition~\ref{prop:valuesubstlemma1simultaneous} to get
\[
\veq {\Gamma'} {V[\overrightarrow{V_i}/\overrightarrow{x_i}]} {V[\overrightarrow{W_i}/\overrightarrow{x_i}]} {I}
\]

Next, as $y$ is fresh by our adopted variable conventions, we have a derivation of $\lj {} {\Gamma', y \!:\! O[V[\overrightarrow{V_i}/\overrightarrow{x_i}]/x]}$ 
and thus we can use the induction hypothesis to get
\[
\veq {\Gamma', y \!:\! O[V[\overrightarrow{V_i}/\overrightarrow{x_i}]/x]} {\efftrans T {A; \overrightarrow{V_i} \!, y; \overrightarrow{V'_j}; \overrightarrow{V_{\sigalgop}}}} {\efftrans T {B; \overrightarrow{W_i}, y; \overrightarrow{W'_j}; \overrightarrow{W_{\sigalgop}}}} {A}
\]

Next, by using the congruence rule for lambda abstraction, we get a proof of
\[
\begin{array}{c}
\hspace{-4cm}
\veq {\Gamma'} {\lambda\, y \!:\! O[V[\overrightarrow{V_i}/\overrightarrow{x_i}]/x] .\, \efftrans T {A; \overrightarrow{V_i} \!, y; \overrightarrow{V'_j}; \overrightarrow{V_{\sigalgop}}} \\ \hspace{0.25cm}} { \lambda\, y \!:\! O[V[\overrightarrow{V_i}/\overrightarrow{x_i}]/x].\, \efftrans T {B; \overrightarrow{W_i}, y; \overrightarrow{W'_j}; \overrightarrow{W_{\sigalgop}}}} {O[V[\overrightarrow{V_i}/\overrightarrow{x_i}]/x] \to A}
\end{array}
\]

Finally, using the congruence rules for function application and pairing, in combination with the proofs of definitional equations given above, we can prove 
\[
\begin{array}{c}
\hspace{-1cm}
\veq {\Gamma'} {V_{\sigalgop}\,\, \langle V[\overrightarrow{V_i}/\overrightarrow{x_i}] , \lambda\, y \!:\! O[V[\overrightarrow{V_i}/\overrightarrow{x_i}]/x] .\, \efftrans {T} {A; \overrightarrow{V_i},\, y; \overrightarrow{V'_j}; \overrightarrow{V_{\sigalgop}}}\rangle
\\ \hspace{0.15cm}} { W_{\sigalgop}\,\, \langle V[\overrightarrow{W_i}/\overrightarrow{x_i}] , \lambda\, y \!:\! O[V[\overrightarrow{W_i}/\overrightarrow{x_i}]/x] .\, \efftrans {T} {B; \overrightarrow{W_i},\, y; \overrightarrow{W'_j}; \overrightarrow{W_{\sigalgop}}}\rangle
} {A}
\end{array}
\]
\end{proof}

We now return to the eMLTT$_{\mathcal{T}_{\text{eff}}}^{\mathcal{H}}$ version of Proposition~\ref{prop:wellformedcomponentsofjudgements}. We consider one example case of its proof in detail, so as to  demonstrate how Proposition~\ref{prop:replacementinfibeffecttermtranslationindices} and equational reasoning are used to prove the new cases introduced by eMLTT$_{\mathcal{T}_{\text{eff}}}^{\mathcal{H}}$.

\vspace{0.2cm}

In particular, we consider the case of the congruence rule for the user-defined algebra type. In this case, 
the given derivation ends with
\[
\mkrule
{
\ljeq {\Gamma'} {\langle A , \{V_{\sigalgop}\}_{\sigalgop \in \mathcal{S}_{\text{eff}}} \rangle} {\langle B , \{W_{\sigalgop}\}_{\sigalgop \in \mathcal{S}_{\text{eff}}} \rangle}
}
{
\begin{array}{c}
\ljeq {\Gamma'} A B
\quad
\veq {\Gamma'} {V_{\sigalgop}} {W_{\sigalgop}} {(\Sigma\, x \!:\! I . O \to A) \to A}
\\[2mm]
\hspace{-3.9cm} \veq {\Gamma'} {\overrightarrow{\lambda\, x'_i \!:\! \widehat{A_i} .}\, \overrightarrow{\lambda\, x_{w_{\!j}} \!:\! \widehat{A'_j} \to A .}\, \efftrans {T_1} {A; \overrightarrow{x'_i}; \overrightarrow{x_{w_{\!j}}}; \overrightarrow{V_{\sigalgop}}} \\ \hspace{0.45cm} } {\overrightarrow{\lambda\, x'_i \!:\! \widehat{A_i} .}\, \overrightarrow{\lambda\, x_{w_{\!j}} \!:\! \widehat{A'_j} \to A .}\, \efftrans {T_2} {A; \overrightarrow{x'_i}; \overrightarrow{x_{w_{\!j}}}; \overrightarrow{V_{\sigalgop}}}\,} {\,\overrightarrow{\Pi x'_i \!:\! \widehat{A_i} .}\, \overrightarrow{\widehat{A'_j} \to A} \to A}
\\[3mm]
(\text{for all } \sigalgop : (x \!:\! I) \longrightarrow O \in \mathcal{S}_{\text{eff}}
\text{ and }
\ljeq {\Gamma \vertbar \Delta} {T_1} {T_2} \in \mathcal{E}_{\text{eff}})
\end{array}
}
\]
and we are required to construct derivations of 
\[
\lj {\Gamma'} {\langle A , \{V_{\sigalgop}\}_{\sigalgop \in \mathcal{S}_{\text{eff}}} \rangle}
\qquad
\lj {\Gamma'} {\langle B , \{W_{\sigalgop}\}_{\sigalgop \in \mathcal{S}_{\text{eff}}} \rangle}
\]

We begin by using $(b)$ and $(f)$ on the given derivations of $\ljeq {\Gamma'} A B$ and \linebreak $\veq {\Gamma'} {V_{\sigalgop}} {W_{\sigalgop}} {(\Sigma\, x \!:\! I . O \to A) \to A}$, in order to get derivations of 
\[
\begin{array}{c}
\lj {\Gamma'} A
\qquad
\lj {\Gamma'} B
\\[2mm]
\vj {\Gamma'} {V_{\sigalgop}} {(\Sigma\, x \!:\! I . O \to A) \to A}
\qquad
\vj {\Gamma'} {W_{\sigalgop}} {(\Sigma\, x \!:\! I . O \to A) \to A}
\end{array}
\]
for all $\sigalgop : (x \!:\! I) \longrightarrow O$ in $\mathcal{S}_{\text{eff}}$. 

On the one hand, based on these derivations, we can immediately construct the required derivation of $\lj {\Gamma'} {\langle A , \{V_{\sigalgop}\}_{\sigalgop \in \mathcal{S}_{\text{eff}}} \rangle}$
by using the corresponding formation rule.

On the other hand, more work is needed to construct the required derivation of $\lj {\Gamma'} {\langle B , \{W_{\sigalgop}\}_{\sigalgop \in \mathcal{S}_{\text{eff}}} \rangle}$. To this end, we first use the context and type conversion rule with $\ljeq {} {\Gamma'} {\Gamma'}$ and $\ljeq {\Gamma'} {(\Sigma\, x \!:\! I . O \to A) \to A} {(\Sigma\, x \!:\! I . O \to B) \to B}$ (which we get from $\ljeq {\Gamma'} A B$) on the derivations of $\vj {\Gamma'} {W_{\sigalgop}} {(\Sigma\, x \!:\! I . O \to A) \to A}$ to get derivations of 
\[
\vj {\Gamma'} {W_{\sigalgop}} {(\Sigma\, x \!:\! I . O \to B) \to B}
\]

Next, we prove for all $\ljeq {\Gamma \vertbar \Delta} {T_1} {T_2}$ in $\mathcal{E}_{\text{eff}}$ the following definitional equations: 
\begin{fleqn}[0.3cm]
\begin{align*}
\Gamma' \,\vdash\,\, & \overrightarrow{\lambda\, x'_i \!:\! \widehat{A_i} .}\, \overrightarrow{\lambda\, x_{w_{\!j}} \!:\! \widehat{A'_j} \to B .\,} \efftrans {T_1} {B; \overrightarrow{x'_i}; \overrightarrow{x_{w_{\!j}}}; \overrightarrow{W_{\sigalgop}}}
\\
=\,\, & \overrightarrow{\lambda\, x'_i \!:\! \widehat{A_i} .}\, \overrightarrow{\lambda\, x_{w_{\!j}} \!:\! \widehat{A'_j} \to A .\,} \efftrans {T_1} {A; \overrightarrow{x'_i}; \overrightarrow{x_{w_{\!j}}}; \overrightarrow{V_{\sigalgop}}}
\\
=\,\, & \overrightarrow{\lambda\, x'_i \!:\! \widehat{A_i} .}\, \overrightarrow{\lambda\, x_{w_{\!j}} \!:\! \widehat{A'_j} \to A .\,} \efftrans {T_2} {A; \overrightarrow{x'_i}; \overrightarrow{x_{w_{\!j}}}; \overrightarrow{V_{\sigalgop}}}
\\
=\,\, & \overrightarrow{\lambda\, x'_i \!:\! \widehat{A_i} .}\, \overrightarrow{\lambda\, x_{w_{\!j}} \!:\! \widehat{A'_j} \to B .\,} \efftrans {T_2} {B; \overrightarrow{x'_i}; \overrightarrow{x_{w_{\!j}}}; \overrightarrow{W_{\sigalgop}}} : {\,\overrightarrow{\Pi x'_i \!:\! \widehat{A_i} .}\,\overrightarrow{\widehat{A'_j} \to B} \to B}
\end{align*}
\end{fleqn}
using the assumed definitional equations corresponding to equations $\ljeq {\Gamma \vertbar \Delta} {T_1} {T_2}$ in $\mathcal{E}_{\text{eff}}$ (for the middle equation), in combination with (the repeated use of) the congruence rule for lambda abstraction and Proposition~\ref{prop:replacementinfibeffecttermtranslationindices} (for the first and last equation).

As a result, we can now use the formation rule for the user-defined algebra type to also construct the required derivation of $\lj {\Gamma'} {\langle B , \{W_{\sigalgop}\}_{\sigalgop \in \mathcal{S}_{\text{eff}}} \rangle}$.

\section{Derivable equations}
\label{sect:derivableisomorphismswithhandlers}

In this section we present some useful definitional equations that are derivable in eMLTT$_{\mathcal{T}_{\text{eff}}}^{\mathcal{H}}$. These equations complement those we showed to be derivable in eMLTT and eMLTT$_{\mathcal{T}_{\text{eff}}}$ in Sections~\ref{sect:derivableequations} and~\ref{sect:derivableequationsforeMLTTwithfibalgeffects}, respectively. These derivable equations include unit and associativity equations for the composition operations, and the interaction of the composition operations with other computation and homomorphism terms. 

For better readability, given $\cj {\Gamma, x\!:\! U\ul{C}} M {\ul{D}}$, we abbreviate premises of the form
\[
\begin{array}{c}
\hspace{-0.95cm}
\ceq \Gamma {\lambda\, x \!:\! I .\, \lambda\, x' \!:\! O \to U\ul{C} .\, M[\thunk (\algop^{\ul{C}}_x(y'\!.\, \force {\ul{C}} (x'\, y')))/y] \\[1mm] \hspace{0.25cm}} { \lambda\, x \!:\! I .\, \lambda\, x' \!:\! O \to U\ul{C} .\, \algop^{\ul{D}}_x(y'\!.\, M[x'\, y'/y])} {\Pi\, x \!:\! I .\, (O \to U\ul{C}) \to \ul{D}}
\end{array}
\]
for all $\sigalgop : (x \!:\! I) \longrightarrow O$ in $\mathcal{S}_{\text{eff}}$, 
by writing 
\[
\text{$M$ is a homomorphism in } y 
\]

\begin{proposition}
\label{prop:emlttwithhandlersunitandassoclaws}
The following unit and associativity equations are derivable for the composition operations:
\[
\mkrule
{\ceq \Gamma {\runas M {y \!:\! U\ul{C}} {} {\force {\ul{C}} y}} {M} {\ul{C}}}
{\cj \Gamma M \ul{C}}
\]

\vspace{-0.15cm}

\[
\mkrule
{\heq \Gamma {z \!:\! \ul{C}} {\runas K {y \!:\! U\ul{D}} {} {\force {\ul{D}} y}} {K} {\ul{D}}}
{\hj \Gamma {z \!:\! \ul{C}} K \ul{D}}
\]

\vspace{-0.25cm}

\[
\mkrule
{
\begin{array}{r@{\,\,} l}
\ceq {\Gamma} {& \runas M {y_1 \!:\! U\ul{C}_1} {} {({\runas {N_1} {y_2 \!:\! U\ul{C}_2} {} {N_2}})} \\[-0.5mm]} {& \runas {(\runas {M} {y_1 \!:\! U\ul{C}_1} {} {N_1})} {y_2 \!:\! U\ul{C}_2} {} {N_2}} {\ul{D}}
\end{array}
}
{
\begin{array}{c}
\cj \Gamma M \ul{C}_1 
\\[0.5mm]
\lj \Gamma \ul{C}_2
\quad
\cj {\Gamma, y_1 \!:\! U\ul{C}_1} {N_1} \ul{C}_2
\quad
\text{$N_1$ is a homomorphism in } y_1 
\\[0.5mm]
\lj \Gamma \ul{D}
\quad
\cj {\Gamma, y_2 \!:\! U\ul{C}_2} {N_2} \ul{D}
\quad
\text{$N_2$ is a homomorphism in } y_2
\end{array}
}
\]

\vspace{-0.25cm}

\[
\mkrule
{
\begin{array}{r@{\,\,} l}
\heq \Gamma {z \!:\! \ul{C}} {& \runas K {y_1 \!:\! U\ul{D}_1} {} {({\runas {M} {y_2 \!:\! U\ul{D}_2} {} {N}})} \\[-0.5mm]} { & \runas {(\runas {K} {y_1 \!:\! U\ul{D}_1} {} {M})} {y_2 \!:\! U\ul{D}_2} {} {N}} {\ul{D}_3}
\end{array}
}
{
\begin{array}{c}
\hj \Gamma {z \!:\! \ul{C}} K \ul{D}_1 
\\[0.5mm]
\lj \Gamma \ul{D}_2
\quad
\cj {\Gamma, y_1 \!:\! U\ul{D}_1} {M} \ul{D}_2
\quad
\text{$M$ is a homomorphism in } y_1 
\\[0.5mm]
\lj \Gamma \ul{D}_3
\quad
\cj {\Gamma, y_2 \!:\! U\ul{D}_2} {N} \ul{D}_3
\quad
\text{$N$ is a homomorphism in } y_2
\end{array}
}
\]
\end{proposition}

\begin{proof}
The two unit equations are proved by using the $\eta$-equation for the composition operations, i.e., as
\begin{fleqn}[0.3cm]
\begin{align*}
\Gamma \,\vdash\,\, & \runas M {y \!:\! U\ul{C}} {} {\force {\ul{C}} y}
\\
=\,\, & \runas M {y \!:\! U\ul{C}} {} {z[\force {\ul{C}} y/z]}
\\
=\,\, & z[M/z]
\\
=\,\, & M : \ul{C}
\end{align*}
\end{fleqn}
and similarly for the unit equation for homomorphism terms.

The two associativity equations are proved by using the $\beta$- and $\eta$-equations for the composition operations, i.e., as
\begin{fleqn}[0.1cm]
\begin{align*}
\Gamma \,\vdash\,\, & \runas M {y_1 \!:\! U\ul{C}_1} {} {({\runas {N_1} {y_2 \!:\! U\ul{C}_2} {} {N_2}})}
\\
=\,\, & \runas M {y_1 \!:\! U\ul{C}_1} {} {\big(\runas {\big(\runas {(\force {\ul{C}_1} y_1)} {y'_1 \!:\! U\ul{C}_1} {} {N_1[y_1'/y_1]}\big)} {y_2 \!:\! U\ul{C}_2} {} {N_2}\big)}
\\
=\,\, & \runas M {y_1 \!:\! U\ul{C}_1} {} {\big(\runas {(\runas {z} {y'_1 \!:\! U\ul{C}_1} {} {N_1[y_1'/y_1]})} {y_2 \!:\! U\ul{C}_2} {} {N_2}\big)[\force {\ul{C}_1} y_1/z]}
\\
=\,\, & \runas {(\runas {M} {y'_1 \!:\! U\ul{C}_1} {} {N_1[y'_1/y_1]})} {y_2 \!:\! U\ul{C}_2} {} {N_2}
\\
=\,\, & \runas {(\runas {M} {y_1 \!:\! U\ul{C}_1} {} {N_1})} {y_2 \!:\! U\ul{C}_2} {} {N_2} : \ul{D}
\end{align*}
\end{fleqn}
and similarly for the associativity equation for homomorphism terms.
\end{proof}

\begin{proposition}
Sequential composition commutes with the composition operations from the left:
\[
\mkrule
{
\begin{array}{r@{\,\,} l}
\ceq \Gamma {& \doto {M} {y_1 \!:\! A} {} {(\runas {N_1} {y_2 \!:\! U\ul{C}} {} {N_2})} \\[-0.5mm]} { & \runas {(\doto M {y_1 \!:\! A} {} {N_1})} {y_2 \!:\! U\ul{C}} {} {N_2}} {\ul{D}}
\end{array}
}
{
\begin{array}{c}
\cj \Gamma M FA 
\quad
\lj \Gamma \ul{C}
\quad
\cj {\Gamma, y_1 \!:\! A} {N_1} \ul{C}
\\[0.5mm]
\lj \Gamma \ul{D}
\quad
\cj {\Gamma, y_2 \!:\! U\ul{C}} {N_2} \ul{D}
\quad
\text{$N_2$ is a homomorphism in } y_2 
\end{array}
}
\]

\vspace{-0.15cm}

\[
\mkrule
{
\begin{array}{r@{\,\,} l}
\heq \Gamma {z \!:\! \ul{C}} {& \doto {K} {y_1 \!:\! A} {} {(\runas {M} {y_2 \!:\! U\ul{D}_1} {} {N})} \\[-0.5mm]} { & \runas {(\doto K {y_1 \!:\! A} {} {M})} {y_2 \!:\! U\ul{D}_1} {} {N}} {\ul{D}_2}
\end{array}
}
{
\begin{array}{c@{\qquad} c}
\hj \Gamma {z \!:\! \ul{C}} K FA 
\quad
\lj \Gamma \ul{D}_1
\quad
\cj {\Gamma, y_1 \!:\! A} {M} \ul{D}_1
\\[0.5mm]
\lj \Gamma \ul{D}_2
\quad
\cj {\Gamma, y_2 \!:\! U\ul{D}_1} {N} \ul{D}_2
\quad
\text{$N$ is a homomorphism in } y_2 
\end{array}
}
\]
\end{proposition}

\begin{proof}
Both equations are proved by using the $\beta$- and $\eta$-equations for sequential composition, following the same pattern that we used in Proposition~\ref{prop:seqcompdistributivity} where we showed that sequential composition commutes with other computation term formers from the left.
\end{proof}

\begin{proposition}
Computational pattern-matching commutes with the composition operations from the left:

\[
\mkrule
{
\begin{array}{r@{\,\,} l}
\ceq \Gamma {& \doto {M} {(y_1 \!:\! A, z \!:\! \ul{C}_1)} {} {(\runas {K} {y_2 \!:\! U\ul{C}_2} {} {N})} \\[-0.5mm]} { & \runas {(\doto M {(y_1 \!:\! A, z \!:\! \ul{C}_1)} {} {K})} {y_2 \!:\! U\ul{C}_2} {} {N}} {\ul{D}}
\end{array}
}
{
\begin{array}{c}
\cj \Gamma M \Sigma\, y_1 \!:\! A .\, \ul{C}_1
\quad
\lj \Gamma \ul{C}_2
\quad
\hj {\Gamma, y_1 \!:\! A} {z \!:\! \ul{C}_1} {K} \ul{C}_2
\\[0.5mm]
\lj \Gamma \ul{D}
\quad
\cj {\Gamma, y_2 \!:\! U\ul{C}_2} {N} \ul{D}
\quad
\text{$N$ is a homomorphism in } y_2 
\end{array}
}
\]

\vspace{-0.15cm}

\[
\mkrule
{
\begin{array}{r@{\,\,} l}
\heq \Gamma {z_1 \!:\! \ul{C}} {& \doto {K} {(y_1 \!:\! A, z_2 \!:\! \ul{D}_1)} {} {(\runas {L} {y_2 \!:\! U\ul{D}_2} {} {M})} \\[-0.5mm]} { & \runas {(\doto K {(y_1 \!:\! A, z_2 \!:\! \ul{D}_1)} {} {L})} {y_2 \!:\! U\ul{D}_2} {} {M}} {\ul{D}_3}
\end{array}
}
{
\begin{array}{c}
\hj \Gamma {z_1 \!:\! \ul{C}} K \Sigma\, y_1 \!:\! A .\, \ul{D}_1 
\quad
\lj \Gamma \ul{D}_2
\quad
\hj {\Gamma, y_1 \!:\! A} {z_2 \!:\! \ul{D}_1} {L} \ul{D}_2
\\[0.5mm]
\lj \Gamma \ul{D}_3
\quad
\cj {\Gamma, y_2 \!:\! U\ul{D}_2} {M} \ul{D}_3
\quad
\text{$M$ is a homomorphism in } y_2 
\end{array}
}
\]
\end{proposition}

\begin{proof}
Both equations are proved by using the $\beta$- and $\eta$-equations for computational pattern-matching, following the same common pattern that we used in Proposition~\ref{prop:comppatternmatchingdistributivity} where we showed that computational pattern-matching commutes with other computation term formers from the left.
\end{proof}

\begin{proposition}
The composition operations commute with sequential composition, computational pairing, pattern-matching, lambda abstraction, function application, and homomorphic function application from the left:
\[
\mkrule
{
\begin{array}{r@{\,\,} l}
\ceq \Gamma {& \runas M {y_1 \!:\! U\ul{C}} {} {(\doto {N_1} {y_2 \!:\! A} {} {N_2})} \\[-0.5mm]} { & \doto {(\runas M {y_1 \!:\! U\ul{C}} {} {N_1})} {y_2 \!:\! A} {} {N_2}} {\ul{D}}
\end{array}
}
{
\begin{array}{c}
\cj \Gamma M \ul{C} 
\quad
\cj {\Gamma, y_1 \!:\! U\ul{C}} {N_1} FA
\quad
\text{$N_1$ is a homomorphism in } y_1
\\[0.5mm]
\lj \Gamma \ul{D}
\quad
\cj {\Gamma, y_2 \!:\! A} {N_2} \ul{D}
\end{array}
}
\]

\vspace{0.01cm}

\[
\mkrule
{
\ceq \Gamma {\runas M {y_1 \!:\! U\ul{C}} {} {\langle V , N \rangle}} {\langle V ,  {\runas M {y_1 \!:\! U\ul{C}} {} {N}} \rangle} {\Sigma\, y_2 \!:\! A .\, \ul{D}}
}
{
\begin{array}{c}
\cj \Gamma M \ul{C} 
\quad
\vj \Gamma V A
\\[0.5mm]
\lj {\Gamma, y_2 \!:\! A} \ul{D}
\quad
\cj {\Gamma, y_1 \!:\! U\ul{C}} {N} \ul{D}[V/y_2]
\quad
\text{$N$ is a homomorphism in } y_1 
\end{array}
}
\]

\vspace{0.01cm}

\[
\mkrule
{
\begin{array}{r@{\,\,} l}
\ceq \Gamma {& \runas M {y_1 \!:\! U\ul{C}_1} {} {(\doto {N} {(y_2 \!:\! A, z \!:\! \ul{C}_2)} {} {K})} \\[-0.5mm]} { & \doto {(\runas M {y_1 \!:\! U\ul{C}_1} {} {N})} {(y_2 \!:\! A, z \!:\! \ul{C}_2)} {} {K}} {\ul{D}}
\end{array}
}
{
\begin{array}{c}
\cj \Gamma M \ul{C}_1 
\quad
\cj {\Gamma, y_1 \!:\! U\ul{C}_1} {N} \Sigma\, y_2 \!:\! A .\, \ul{C}_2
\quad
\text{$N$ is a homomorphism in } y_1
\\[0.5mm]
\lj \Gamma \ul{D}
\quad
\hj {\Gamma, y_2 \!:\! A} {z \!:\! \ul{C}_2} {K} \ul{D}
\end{array}
}
\]

\vspace{0.01cm}

\[
\mkrule
{
\begin{array}{r@{\,\,} l}
\ceq \Gamma {& \runas M {y_1 \!:\! U\ul{C}} {} {(\lambda\, y_2 \!:\! A .\, N)} \\[-0.5mm]} { & \lambda\, y_2 \!:\! A .\,  {(\runas M {y_1 \!:\! U\ul{C}} {} {N}})} {\Pi\, y_2 \!:\! A .\, \ul{D}}
\end{array}
}
{
\begin{array}{c@{\qquad} c}
\cj \Gamma M \ul{C} 
\quad
\lj {\Gamma, y_2 \!:\! A} \ul{D}
\\[0.5mm]
\cj {\Gamma, y_1 \!:\! U\ul{C}, y_2 \!:\! A} {N} \ul{D}
\quad
\text{$N$ is a homomorphism in } y_1 
\end{array}
}
\]

\vspace{0.01cm}

\[
\mkrule
{
\ceq \Gamma {\runas M {y_1 \!:\! U\ul{C}} {} {N\, V}} {{(\runas M {y_1 \!:\! U\ul{C}} {} {N})} \, V} {\ul{D}[V/y_2]}
}
{
\begin{array}{c@{\qquad} c}
\cj \Gamma M \ul{C} 
\quad
\vj \Gamma V A
\quad
\lj {\Gamma, y_2 \!:\! A} \ul{D}
\\[0.5mm]
\cj {\Gamma, y_1 \!:\! U\ul{C}} {N} \Pi\, y_2 \!:\! A .\, \ul{D}
\quad
\text{$N$ is a homomorphism in } y_1 
\end{array}
}
\]

\vspace{0.01cm}

\[
\mkrule
{
\ceq \Gamma {\runas M {y_1 \!:\! U\ul{C}} {} {V\, N}} {V\, {(\runas M {y_1 \!:\! U\ul{C}} {} {N})}} {\ul{D}_2}
}
{
\begin{array}{c@{\qquad} c}
\cj \Gamma M \ul{C} 
\quad
\vj \Gamma V {\ul{D}_1 \multimap \ul{D}_2}
\\[0.5mm]
\cj {\Gamma, y_1 \!:\! U\ul{C}} {N} \ul{D}_1
\quad
\text{$N$ is a homomorphism in } y_1 
\end{array}
}
\vspace{0.1cm}
\]
Analogous equations also hold for the composition operation for homomorphism terms. 
\end{proposition}

\begin{proof}
These equations are proved by using the $\beta$- and $\eta$-equations for the composition operations, e.g., the commutativity with computational pairing is proved as follows:
\begin{fleqn}[0.3cm]
\begin{align*}
\Gamma \,\vdash\,\, & \runas M {y_1 \!:\! U\ul{C}} {} {\langle V , N \rangle}
\\
=\,\, & \runas M {y_1 \!:\! U\ul{C}} {} {\langle V , \runas {(\force {\ul{C}} y_1)} {y'_1 \!:\! U\ul{C}} {} {N[y'_1/y_1]} \rangle}
\\
=\,\, & \runas M {y_1 \!:\! U\ul{C}} {} {\langle V , (\runas {z} {y'_1 \!:\! U\ul{C}} {} {N[y'_1/y_1]})\rangle[\force {\ul{C}} y_1/z]}
\\
=\,\, & \langle V ,  {\runas M {y'_1 \!:\! U\ul{C}} {} {N[y'_1/y_1]}} \rangle
\\
=\,\, & \langle V ,  {\runas M {y_1 \!:\! U\ul{C}} {} {N}} \rangle : \Sigma\, y_2 \!:\! A .\, \ul{D}
\end{align*}
\end{fleqn}

Observe that for this proof to be well-formed, we need to lift the homomorphism assumption about $N$ to the pair $\langle V , N \rangle$. We do so by using the congruence rules for computational lambda abstraction and pairing, in combination with the corresponding specialised algebraicity equation we derived in  Proposition~\ref{prop:specialisedalgebraicity}, as shown below.
\begin{fleqn}[0.3cm]
\begin{align*}
\Gamma \,\vdash\,\, & \lambda\, x \!:\! I .\, \lambda\, x' \!:\! O \to U\ul{C} .\, \langle V , N \rangle[\thunk (\algop^{\ul{C}}_x(y'\!.\, \force {\ul{C}} (x'\, y')))/y_1]
\\
=\,\, & \lambda\, x \!:\! I .\, \lambda\, x' \!:\! O \to U\ul{C} .\, \langle V , N[\thunk (\algop^{\ul{C}}_x(y'\!.\, \force {\ul{C}} (x'\, y')))/y_1] \rangle
\\
=\,\, & \lambda\, x \!:\! I .\, \lambda\, x' \!:\! O \to U\ul{C} .\, \langle V , \algop^{\ul{D}[V/y_2]}_x(y'\!.\, N[x'\, y'/y_1]) \rangle
\\
=\,\, & \lambda\, x \!:\! I .\, \lambda\, x' \!:\! O \to U\ul{C} .\, \algop^{\Sigma\, y_2 : A .\, \ul{D}}(y'\!.\, \langle V ,  N[x'\, y'/y_1]\rangle)
\\
=\,\, & \lambda\, x \!:\! I .\, \lambda\, x' \!:\! O \to U\ul{C} .\, \algop^{\Sigma\, y_2 : A .\, \ul{D}}(y'\!.\, \langle V ,  N\rangle[x'\, y'/y_1]) 
\\[-2mm]
& \hspace{7.5cm} : \Pi\, x \!:\! I .\, (O \to U\ul{C}) \to \Sigma\, y_2 \!:\! A .\, \ul{D}
\end{align*}
\vspace{-0.25cm}
\end{fleqn}
\end{proof}

We conclude by noting that analogously to other computation term formers, we can also derive  specialised algebraicity equations for the composition operation.

\index{algebraicity equation!specialised --}
\begin{proposition}
The following specialised algebraicity equation is derivable, for every operation symbol $\sigalgop : (x \!:\! I) \longrightarrow O$ in $\mathcal{S}_{\text{eff}}$:
\[
\mkrule
{\ceq \Gamma {\runas {\sigalgop_V^{\ul{C}}(y.\, M)} {y' \!:\! U\ul{C}} {} {N}} {\sigalgop_V^{\ul{D}}(y.\, \runas M {y' \!:\! U\ul{C}} {} {N})} {\ul{D}}}
{
\begin{array}{c}
\vj \Gamma V I
\quad
\lj \Gamma \ul{C}
\quad
\cj {\Gamma, y \!:\! O[V/x]} {M} {\ul{C}}
\\[0.5mm]
\lj \Gamma \ul{D}
\quad
\cj {\Gamma, y' \!:\! U\ul{C}} {N} {\ul{D}}
\quad
\text{$N$ is a homomorphism in } y' 
\end{array}
}
\]
\end{proposition}

\begin{proof}
This equation is proved by using the general algebraicity equation given in Definition~\ref{def:extensionofeMLTTwithfibalgeffects}, following the same common pattern that we used in Proposition~\ref{prop:specialisedalgebraicity} where we proved specialised algebraicity equations for other computation terms.
\end{proof}

\section{Alternative presentations of eMLTT, eMLTT$_{\!\mathcal{T}_{\text{eff}}}$, and eMLTT$_{\!\mathcal{T}_{\text{eff}}}^{\mathcal{H}}$}
\label{sect:alternativepresentations}

In this section we outline two ways in which we could have defined eMLTT$_{\mathcal{T}_{\text{eff}}}^{\mathcal{H}}$ (and also eMLTT and eMLTT$_{\mathcal{T}_{\text{eff}}}$) differently from the presentation we used in this thesis.

\subsection{Different equational proof obligations}
\label{sect:alternativeauxjudgementaux}

First, we note that instead of using equational proof obligations of the form
\[
\begin{array}{r@{~} c@{~} l}
\Gamma & \vdash & \lambda x \!:\! I . \lambda x' \!:\! O \to U\ul{C} . N[\thunk\! (\algop^{\ul{C}}_x(x''\!. \force {\ul{C}} (x'\, x'')))/y] 
\\[1mm]
& = & \lambda x \!:\! I . \lambda x' \!:\! O \to U\ul{C} . \algop^{\ul{D}}_x(x''\!. N[x'\, x''\!/y]) : {\Pi x \!:\! I . (O \!\to\! U\ul{C}) \!\to\! \ul{D}}
\end{array}
\]
we could have instead used Munch-Maccagnoni's notion of linearity \citep{Munch:Thesis}, i.e.,
\[
\begin{array}{r@{~} c@{~} l}
\Gamma & \vdash & \lambda x \!:\! UFA . \lambda x' \!:\!  A \to U\ul{C} . N[\thunk\! (\doto {(\force {F\!A} x)} {x'' \!:\! A} {\ul{C}} {\force {\ul{C}} {(x'\, x'')}})/y] 
\\[1.5mm]
& = & \lambda x \!:\! UFA . \lambda x' \!:\!  A \to U\ul{C} . \doto {(\force {F\!A} x)} {x'' \!:\! A} {\ul{D}} N[x'\, x''/y] 
\\
&& \hspace{9.1cm} : UFA \to (A \to U\ul{C}) \to \ul{D}
\end{array}
\]

On the one hand, the proof obligations we used in this thesis follow from Munch-Maccagnoni's notion of linearity by straightforward equational reasoning. On the other hand, in a language supporting only algebraic effects (e.g., eMLTT$_{\mathcal{T}_{\text{eff}}}$ and eMLTT$_{\mathcal{T}_{\text{eff}}}^{\mathcal{H}}$), Munch-Maccagnoni's notion of linearity follows from the proof obligations we used in this thesis by appealing to Plotkin and Pretnar's principle of computational induction for algebraic effects, which states that every computation term of type $FA$ is either a returned value or built from computation terms using algebraic operations---see~\cite{Plotkin:Logic}.

While the latter form of proof obligations is also applicable in languages with computational effects other than algebraic (e.g., as used by Levy in \cite{Levy:ContextualIsomorphisms} to characterise general isomorphisms between computation types), we chose the former kind of proof obligations due to their more intuitive reading in the setting of algebraic effects.

\subsection{Omitting homomorphism terms}

Second, we note that we could have omitted computation variables $z$ and homomorphism terms $K$ from eMLTT$_{\mathcal{T}_{\text{eff}}}^{\mathcal{H}}$ (and also eMLTT and eMLTT$_{\mathcal{T}_{\text{eff}}}$) altogether. Instead, we could have used value variables $x$ and an appropriate notion of equational proof obligations to define and type the elimination form for $\Sigma x \!:\! A . \ul{C}$, analogously to the typing rules of composition operations given in Definition~\ref{def:extensionofeMLTTwithhandlers}.
In more detail, this alternative presentation would involve the following elimination form for $\Sigma x \!:\! A . \ul{C}$:
\vspace{0.05cm}
\[
\mkrule
{\cj \Gamma {\doto M {(x \!:\! A, y \!:\! U\ul{C})} {\ul{D}} N} {\ul{D}}}
{
\begin{array}{c}
\cj \Gamma M {\Sigma x \!:\! A . \ul{C}} \quad \lj \Gamma {\ul{D}} 
\quad
\cj {\Gamma, x \!:\! A, y \!:\! U\ul{C}} {N} {\ul{D}}
\\
\text{$N$ is a homomorphism in } y
\end{array}
}
\]
where $M$ would now be eliminated into a pair of values, but with the equational proof obligations (denoted by `$\text{$N$ is a homomorphism in } y$', using the notation of Section~\ref{sect:derivableisomorphismswithhandlers})
ensuring that the computation term $N$ behaves in $y$ as if it was a homomorphism.

While this alternative presentation would have been semantically more precise (because homomorphism terms are 
only an under-approximation of all computation terms that behave as if they were homomorphisms), 
we chose to include both computation and homomorphism terms because the latter enabled us to define  
a structurally cleaner elimination form for the computational $\Sigma$-type in eMLTT, eMLTT$_{\mathcal{T}_{\text{eff}}}$, and eMLTT$_{\mathcal{T}_{\text{eff}}}^{\mathcal{H}}$ .

\section{Interpreting eMLTT$_{\!\mathcal{T}_{\text{eff}}}^{\mathcal{H}}$ in a fibred adjunction model}
\label{sect:interpretingemlttwithhandlers}

In this section we equip eMLTT$_{\mathcal{T}_{\text{eff}}}^{\mathcal{H}}$ with a denotational semantics by showing that it can be soundly interpreted in the fibred adjunction model we used for giving a denotational semantics to eMLTT$_{\mathcal{T}_{\text{eff}}}$ in Section~\ref{sect:fibalgeffectsmodel}. We recall that this fibred adjunction model is built by lifting the adjunction $F_{\!\mathcal{L}_{\mathcal{T}_{\text{eff}}}} \!\dashv\, U_{\!\mathcal{L}_{\mathcal{T}_{\text{eff}}}} : \Mod(\!\mathcal{L}_{\mathcal{T}_{\text{eff}}},\Set) \longrightarrow \Set$ to a split fibred adjunction between $\mathsf{fam}_{\Set}$ and $\mathsf{fam}_{\Mod(\!\mathcal{L}_{\mathcal{T}_{\text{eff}}},\Set)}$, as depicted in the next diagram.
\vspace{-1.6cm}
\[
\xymatrix@C=0.5em@R=5em@M=0.5em{
\ar@{}[dd]^-{\!\!\quad\qquad\qquad\qquad\qquad\perp}
\\
\Fam(\Set) \ar@/_2.5pc/[d]_-{\mathsf{fam}_{\Set}} \ar@{}[d]_-{\dashv\,\,\,\,\,\,\,\,\,\,} \ar@{}[d]^-{\,\,\,\,\,\,\,\,\,\,\dashv} \ar@/^2.5pc/[d]^-{\ia {-}} \ar@/^1.25pc/[rrrrrrrrrr]^-{\widehat{F_{\!\mathcal{L}_{\mathcal{T}_{\text{eff}}}}}} &  &&&&&&&&& \,\,\,\,\,\,\,\,\,\,\,\,\,\,\,\,\,\,\,\, \ar@/^1.25pc/[llllllllll]^-{\widehat{U_{\!\mathcal{L}_{\mathcal{T}_{\text{eff}}}}}}  & \hspace{-1.5cm} \Fam(\Mod(\!\mathcal{L}_{\mathcal{T}_{\text{eff}}},\Set)) \ar@/^2pc/[dlllllllllll]^-{\!\!\!\!\!\!\quad\qquad\mathsf{fam}_{\Mod(\!\mathcal{L}_{\mathcal{T}_{\text{eff}}},\Set)}}
\\
\mathcal{\Set} \ar[u]_-{\!1}
}
\vspace{0.35cm}
\]

We also recall that the countable Lawvere theory $I_{\mathcal{T}_{\text{eff}}} : \aleph^{\text{op}}_{\!1} \longrightarrow \mathcal{L}_{\mathcal{T}_{\text{eff}}}$ is derived from the countable equational theory $\mathbb{T}_{\!\mathcal{T}_{\text{eff}}} = (\mathbb{S}_{\!\mathcal{T}_{\text{eff}}},\mathbb{E}_{\!\mathcal{T}_{\text{eff}}})$, which itself is derived from the given fibred effect theory $\mathcal{T}_{\text{eff}} = (\mathcal{S}_{\text{eff}} , \mathcal{E}_{\text{eff}})$. See Proposition~\ref{prop:lawveretheoryfromequationaltheory} and Definition~\ref{def:countableeqthfromeffth}, respectively, for the detailed definitions of these constructions. 
Analogously to Section~\ref{sect:fibalgeffectsmodel}, we assume that the given fibred effect theory $\mathcal{T}_{\text{eff}}$ is countable.

In order to be able to define the interpretation of the user-defined algebra type and the composition operations in this fibred adjunction model, we first establish that every model of $I_{\mathcal{T}_{\text{eff}}} : \aleph^{\text{op}}_{\!1} \longrightarrow \mathcal{L}_{\mathcal{T}_{\text{eff}}}$ and every morphism between such models are determined by how they behave on operations, i.e., terms of the form $\lj {\overrightarrow{x_o}} \sigalgop_i(x_o)_{1 \,\leq\, o \,\leq\, \vert \sem{x : I; O}_2\,\langle \star , i \rangle \vert}$. 

When proving the above-mentioned property of the models of $I_{\mathcal{T}_{\text{eff}}} : \aleph^{\text{op}}_{\!1} \longrightarrow \mathcal{L}_{\mathcal{T}_{\text{eff}}}$, we use an auxiliary countable Lawvere theory  $I_{\mathcal{T}^d_{\text{eff}}} : \aleph_{\!\!1}^{\text{op}} \longrightarrow \mathcal{L}_{\mathcal{T}^d_{\text{eff}}}$, which we derive from the countable fibred effect theory ${\mathcal{T}^{d}_{\text{eff}}} \defeq (\mathcal{S}_{\text{eff}}, \emptyset)$, as also used in Section~\ref{sect:fibalgeffectsmodel}.

\begin{proposition}
\label{prop:definingmodeloflawthfromops}
\index{ M@$\mathcal{M}_{\langle A , \{f_{\sigalgop_i}\}_{\sigalgop_i \in \mathcal{S}_{\text{eff}}} \rangle}$ (model of a countable Lawvere theory, derived from a set $A$ and functions $f_{\sigalgop_i}$)}
Given a set $A$ and a family of functions 
\[
\begin{array}{c}
f_{\sigalgop_i} : \bigsqcap_{o \in \sem{x : I; O}_2\, \langle \star , i \rangle} A \longrightarrow A
\end{array}
\]
for all $\sigalgop : (x \!:\! I) \longrightarrow O$ in $\mathcal{S}_{\text{eff}}$, then there exists a model 
\[
\mathcal{M}_{\langle A , \{f_{\sigalgop_i}\}_{\sigalgop_i \in \mathcal{S}_{\text{eff}}} \rangle} : \mathcal{L}_{\mathcal{T}^d_{\text{eff}}} \longrightarrow \Set
\]
of the countable Lawvere theory $I_{\mathcal{T}^d_{\text{eff}}} : \aleph_{\!\!1}^{\text{op}} \longrightarrow \mathcal{L}_{\mathcal{T}^d_{\text{eff}}}$.
\end{proposition}

\begin{proof}
We define $\mathcal{M}_{\langle A , \{f_{\sigalgop_i}\}_{\sigalgop_i \in \mathcal{S}_{\text{eff}}} \rangle}$ on objects using countable products, i.e., as\footnote{In the rest of this section, the notation $\bigsqcap_{1 \,\leq\, j \,\leq\, n}\, A$ stands for $A$ when $n = 1$; for $\bigsqcap_{j \in \{1,\ldots,n\}}\, A$ when $n$ is a natural number different from $1$; and for $\bigsqcap_{j \in \mathbb{N}}\, A$ when $n$ is the distinguished symbol $\omega$. In particular, having $\bigsqcap_{1 \,\leq\, j \,\leq\, 1}\, A = A$ allows us to show that the $\beta$-equation for the user-defined algebra type is sound.}
\[
\begin{array}{c}
\mathcal{M}_{\langle A , \{f_{\sigalgop_i}\}_{\sigalgop_i \in \mathcal{S}_{\text{eff}}} \rangle}(n) \defeq \bigsqcap_{1 \,\leq\, j \,\leq\, n}\, A
\end{array}
\]
and on morphisms $(\lj \Delta {t_k})_{1 \,\leq\, k \,\leq\, m} : n \longrightarrow m$ as
\[
\xymatrix@C=15em@R=5em@M=0.5em{
\bigsqcap_{1 \,\leq\, j \,\leq\, n}\, A \ar[r]^-{\langle \mathcal{M}'(\lj {\Delta\,\,} {\,t_k}) \rangle_{1 \,\leq\, k \,\leq\, m}} & \bigsqcap_{1 \,\leq\, k \,\leq\, m}\, A
}
\]
where $\mathcal{M}'(\lj {\Delta} {t_k})$ is defined by recursion on the derivation of $\lj {\Delta} {t_k}$, as follows:
\[
\begin{array}{l c l}
\mathcal{M}'(\lj {\overrightarrow{x_j}} {x_j}) & \defeq & \mathsf{proj}_j
\\[2mm]
\mathcal{M}'(\lj {\Delta} {\sigalgop_i(t_o)_{1 \,\leq\, o \,\leq\, \vert \sem{x : I; O}_2\,\langle \star , i \rangle \vert}}) & \defeq & f_{\sigalgop_i} \comp \langle \mathcal{M}'(\lj {\Delta} {t_o}) \rangle_{1 \,\leq\, o \,\leq\, \vert \sem{x : I; O}_2\,\langle \star , i \rangle \vert}
\end{array}
\]

Next, to show that $\mathcal{M}_{\langle A , \{f_{\sigalgop_i}\}_{\sigalgop_i \in \mathcal{S}_{\text{eff}}} \rangle}$ preserves identity morphisms,  we recall that the identity morphisms are given in $\mathcal{L}_{\mathcal{T}^d_{\text{eff}}}$ by tuples of variables. As a result, we can show
\[
\mathcal{M}_{\langle A , \{f_{\sigalgop_i}\}_{\sigalgop_i \in \mathcal{S}_{\text{eff}}} \rangle}(\id_n) = \langle \mathcal{M}'(\lj {\overrightarrow{x_j}} {x_j}) \rangle_{1 \,\leq\, j \,\leq\, n} = \langle \mathsf{proj}_j \rangle_{1 \,\leq\, j \,\leq\, n} = \id_{\bigsqcap_{1 \,\leq\, j \,\leq\, n}\, A}
\]

Next, we show that given any two morphisms $(\lj \Delta t_k)_{1 \,\leq\, k \,\leq\, n_2} : n_1 \!\longrightarrow\! n_2$ and \linebreak $(\lj {\overrightarrow{x_k}} u_l)_{1 \,\leq\, l \,\leq\, n_3} : n_2 \!\longrightarrow\! n_3$, $\mathcal{M}_{\langle A , \{f_{\sigalgop_i}\}_{\sigalgop_i \in \mathcal{S}_{\text{eff}}} \rangle}$ preserves their composition, i.e.,  
\[
\begin{array}{c}
\mathcal{M}_{\langle A , \{f_{\sigalgop_i}\}_{\sigalgop_i \in \mathcal{S}_{\text{eff}}} \rangle}((\lj \Delta {u_l[\overrightarrow{t_k}/\overrightarrow{x_k}]})_{1 \,\leq\, l \,\leq\, n_3})
\\
=
\\
\mathcal{M}_{\langle A , \{f_{\sigalgop_i}\}_{\sigalgop_i \in \mathcal{S}_{\text{eff}}} \rangle}((\lj {\overrightarrow{x_k}} {u_l})_{1 \,\leq\, l \,\leq\, n_3}) \,\comp\, \mathcal{M}_{\langle A , \{f_{\sigalgop_i}\}_{\sigalgop_i \in \mathcal{S}_{\text{eff}}} \rangle}((\lj \Delta {t_k})_{1 \,\leq\, k \,\leq\, n_2})
\end{array}
\vspace{0.15cm}
\]

This proof proceeds in two steps. First, we show that for all $1 \leq l \leq n_3$, we have 
\[
\begin{array}{c}
\mathcal{M}'(\lj \Delta {u_l[\overrightarrow{t_k}/\overrightarrow{x_k}]})
=
\mathcal{M}'(\lj {\overrightarrow{x_k}} {u_l}) \comp \langle \mathcal{M}'(\lj \Delta {t_k}) \rangle_{1 \,\leq\, k \,\leq\, n_2}
\end{array}
\]
by straightforward induction on the derivation of $\lj {\overrightarrow{x_k}} {u_l}$; we omit the details. Second, the required equation follows by combining these equations with the definition of $\mathcal{M}_{\langle A , \{f_{\sigalgop_i}\}_{\sigalgop_i \in \mathcal{S}_{\text{eff}}} \rangle}$ and the universal property of countable products, i.e., it follows from
\[
\xymatrix@C=8em@R=7em@M=0.5em{
\bigsqcap_{1 \,\leq\, j \,\leq\, n_1}\, A \ar@/^1pc/[rr]^-{\mathcal{M}_{\langle A , \{f_{\sigalgop_i}\}_{\sigalgop_i \in \mathcal{S}_{\text{eff}}} \rangle}((\lj {\Delta\,\,} {\,u_l[\overrightarrow{t_k}/\overrightarrow{x_k}]})_{1 \,\leq\, l \,\leq\, n_3})
}_*+<0.5em>{\dcomment{\text{def.}}} \ar@/_1pc/[rr]_-{\langle \mathcal{M}'(\lj {\Delta\,\,} {\,u_l[\overrightarrow{t_k}/\overrightarrow{x_k}]}) \rangle_{1 \,\leq\, l \,\leq\, n_3}} \ar@/_4.5pc/[rr]_-{\langle \mathcal{M}'(\lj {\overrightarrow{x_k}\,\,} {\,u_l}) \,\,\comp\,\, \langle \mathcal{M}'(\lj {\Delta\,\,} {\,t_k}) \rangle_{1 \,\leq\, k \,\leq\, n_2} \rangle_{1 \,\leq\, l \,\leq\, n_3}}^*+<1em>{\dcomment{\text{equations proved above}}}_*+<3em>{\dcomment{\text{the universal property of countable products}}}
\ar@/_3pc/[dr]_-{\langle \mathcal{M}'(\lj {\Delta\,\,} {\,t_k}) \rangle_{1 \,\leq\, k \,\leq\, n_2}\,\,}
& & 
\bigsqcap_{1 \,\leq\, l \,\leq\, n_3}\, A
\\
& \bigsqcap_{1 \,\leq\, j \,\leq\, n_1}\, A
\ar@/_3pc/[ur]_-{\,\,\,\,\,\,\langle \mathcal{M}'(\lj {\overrightarrow{x_k}\,\,} {\,u_l}) \rangle_{1 \,\leq\, l \,\leq\, n_3}}
}
\]

Finally, we note that $\mathcal{M}_{\langle A , \{f_{\sigalgop_i}\}_{\sigalgop_i \in \mathcal{S}_{\text{eff}}} \rangle}$ preserves countable products because it maps objects of $\mathcal{L}_{\mathcal{T}^d_{\text{eff}}}$ (each of which is itself a countable product) to countable products in $\Set$, variables (i.e., projections in $\mathcal{L}_{\mathcal{T}^d_{\text{eff}}}$) to projections in $\Set$, and tuples of terms (i.e., pairing of morphisms in $\mathcal{L}_{\mathcal{T}^d_{\text{eff}}}$) to pairing in $\Set$. We omit the details of this proof.
\end{proof}

\begin{proposition}
\label{prop:liftingnoeqlawthmodeltoeqlawthmodel}
The model $\mathcal{M}_{\langle A , \{f_{\sigalgop_i}\}_{\sigalgop_i \in \mathcal{S}_{\text{eff}}} \rangle} : \mathcal{L}_{\mathcal{T}^d_{\text{eff}}} \longrightarrow \Set$ of the countable Lawvere theory $I_{\mathcal{T}^d_{\text{eff}}} : \aleph_{\!\!1}^{\text{op}} \longrightarrow \mathcal{L}_{\mathcal{T}^d_{\text{eff}}}$ extends to a model of $I_{\mathcal{T}_{\text{eff}}} : \aleph_{\!\!1}^{\text{op}} \longrightarrow \mathcal{L}_{\mathcal{T}_{\text{eff}}}$ if we have
\[
\mathcal{M}_{\langle A , \{f_{\sigalgop_i}\}_{\sigalgop_i \in \mathcal{S}_{\text{eff}}} \rangle}(\lj {\Delta^\gamma} {T^\gamma_1}) = \mathcal{M}_{\langle A , \{f_{\sigalgop_i}\}_{\sigalgop_i \in \mathcal{S}_{\text{eff}}} \rangle}(\lj {\Delta^\gamma} {T^\gamma_2})
\]
for all $\ljeq {\Gamma \vertbar \Delta} {T_1} {T_2}$ in $\mathcal{E}_{\text{eff}}$ and $\gamma$ in $\sem{\Gamma}$.
\end{proposition}

\begin{proof}
Recalling the definitions of the categories $\mathcal{L}_{\mathcal{T}^d_{\text{eff}}}$ and $\mathcal{L}_{\mathcal{T}_{\text{eff}}}$, the only difference \linebreak between the two is that in the latter the morphisms given by terms $\lj {\Delta^\gamma} {T^\gamma_1}$ and \linebreak $\lj {\Delta^\gamma} {T^\gamma_2}$ are identified, for all $\ljeq {\Gamma \vertbar \Delta} {T_1} {T_2}$ in $\mathcal{E}_{\text{eff}}$ and $\gamma$ in $\sem{\Gamma}$. As a result, for $\mathcal{M}_{\langle A , \{f_{\sigalgop_i}\}_{\sigalgop_i \in \mathcal{S}_{\text{eff}}} \rangle} : \mathcal{L}_{\mathcal{T}^d_{\text{eff}}} \longrightarrow \Set$ to also be a model of the countable Lawvere theory $I_{\mathcal{T}_{\text{eff}}} : \aleph_{\!\!1}^{\text{op}} \longrightarrow \mathcal{L}_{\mathcal{T}_{\text{eff}}}$, it suffices to show that $\mathcal{M}_{\langle A , \{f_{\sigalgop_i}\}_{\sigalgop_i \in \mathcal{S}_{\text{eff}}} \rangle}$ identifies such terms, which follows immediately from the equations we assume in this proposition.
\end{proof}

\begin{proposition}
\label{prop:extendingmorphtohomomorph}
Given models $\mathcal{M}_1 : \mathcal{L}_{\mathcal{T}_{\text{eff}}} \longrightarrow \Set$ and $\mathcal{M}_2 : \mathcal{L}_{\mathcal{T}_{\text{eff}}} \longrightarrow \Set$ of the countable Lawvere theory  $I_{\mathcal{T}_{\text{eff}}} : \aleph_{\!\!1}^{\text{op}} \longrightarrow \mathcal{L}_{\mathcal{T}_{\text{eff}}}$, and a function $f : \mathcal{M}_1(1) \longrightarrow \mathcal{M}_2(1)$ such that
\[
\xymatrix@C=8em@R=3em@M=0.5em{
\bigsqcap_{o \in \vert \sem{x : I; O}_2\, \langle \star , i \rangle \vert} (\mathcal{M}_1(1)) 
\ar[r]^-{\bigsqcap_o (f)}
\ar[d]_-{\cong}
& 
\bigsqcap_{o \in \vert \sem{x : I; O}_2\, \langle \star , i \rangle \vert} (\mathcal{M}_2(1))
\ar[d]^-{\cong}
\\
\mathcal{M}_1(\vert \sem{x \!:\! I; O}_2\, \langle \star , i \rangle \vert)
\ar[d]_-{\mathcal{M}_1(\lj {\overrightarrow{x_o}\,\,} {\,\sigalgop_i(x_o)_{o}})}
&
\mathcal{M}_2(\vert \sem{x \!:\! I; O}_2\, \langle \star , i \rangle \vert)
\ar[d]^-{\mathcal{M}_2(\lj {\overrightarrow{x_o}\,\,} {\,\sigalgop_i(x_o)_{o}})}
\\
\mathcal{M}_1(1)
\ar[r]_-{f}
&
\mathcal{M}_2(1)
}
\]
for all operation symbols $\sigalgop_i : \vert \sem{x \!:\! I; O}_2\, \langle \star , i \rangle \vert$ in $\mathbb{S}_{\!\mathcal{T}_{\text{eff}}}$, then $f$ extends to a morphism 
\[
\mathsf{hom}(f) : \mathcal{M}_1 \longrightarrow \mathcal{M}_2
\]
of models of the countable Lawvere theory $I_{\mathcal{T}_{\text{eff}}} : \aleph_{\!\!1}^{\text{op}} \longrightarrow \mathcal{L}_{\mathcal{T}_{\text{eff}}}$.
\end{proposition}

\begin{proof}
We define the components $\mathsf{hom}(f)_n$ of the natural transformation $\mathsf{hom}(f)$ as
\[
\xymatrix@C=3.5em@R=5em@M=0.5em{
\mathcal{M}_1(n) \ar[r]^-{\cong} & \bigsqcap_{1 \,\leq\, j \,\leq\, n} (\mathcal{M}_1(1)) \ar[r]^-{\bigsqcap_j (f)} & \bigsqcap_{1 \,\leq\, j \,\leq\, n} (\mathcal{M}_2(1)) \ar[r]^-{\cong} & \mathcal{M}_2(n)
}
\]

We prove that $\mathsf{hom}(f)$ is natural in $n$ in two steps. 

First, for any morphism $n \longrightarrow 1$ in $\mathcal{L}_{\mathcal{T}_{\text{eff}}}$, given by a term $\lj \Delta t$, we show that the next diagram commutes, by induction on the given derivation of $\lj \Delta t$.
\[
\xymatrix@C=3.5em@R=5em@M=0.5em{
\mathcal{M}_1(n) \ar[r]^-{\cong} 
\ar[dr]_-{\mathcal{M}_1(\lj {\Delta\,\,} {\,t})}
& 
\bigsqcap_{1 \,\leq\, j \,\leq\, n} (\mathcal{M}_1(1)) \ar[r]^-{\bigsqcap_j (f)} 
& 
\bigsqcap_{1 \,\leq\, j \,\leq\, n} (\mathcal{M}_2(1)) \ar[r]^-{\cong} 
& 
\mathcal{M}_2(n)
\ar[dl]^-{\mathcal{M}_2(\lj {\Delta\,\,} {\,t})}
\\
&
\mathcal{M}_1(1)
\ar[r]_-{f}
&
\mathcal{M}_2(1)
}
\]
We omit the details of the proof and only note that the case for variables follows from the preservation of countable products by $\mathcal{M}_1$ and $\mathcal{M}_2$; and the case for algebraic operations follows from the  commuting squares we assume in the proposition.

Second, using such commuting squares, we show that the naturality square for $\mathsf{hom}(f)$ commutes for any morphism $n \longrightarrow m$ given by a tuple of terms $(\lj \Delta t_k)_{1 \,\leq\, k \,\leq\, m}$:
\[
\xymatrix@C=1.75em@R=7em@M=0.5em{
\mathcal{M}_1(n) \ar[r]^-{\cong} 
\ar@/^1pc/[dr]^>>>>>>>>>>>>>{\langle \mathcal{M}_1(\lj {\Delta\,\,} {\,t_k}) \rangle_k}
\ar[d]_-{\mathcal{M}_1((\lj {\Delta\,\,} {\,t_k})_{k})}^>>>>>{\,\,\,\,\dcomment{\text{pres. of c. prod.}}}^<<<<<<{\,\,\,\qquad\qquad\quad\dcomment{\text{above diagram and the univ. prop. of c. prod.}}}
\ar@/^4pc/[rrr]^-{\mathsf{hom}(f)_n}_<<<<<<<<{\qquad\dcomment{\text{def.}}}
& 
\bigsqcap_{1 \,\leq\, j \,\leq\, n} (\mathcal{M}_1(1)) \ar@/^2pc/[r]^-{\bigsqcap_j (f)} 
& 
\bigsqcap_{1 \,\leq\, j \,\leq\, n} (\mathcal{M}_2(1)) \ar[r]^-{\cong} 
& 
\mathcal{M}_2(n)
\ar@/_1pc/[dl]_>>>>>>>>>>>>>{\langle \mathcal{M}_2(\lj {\Delta\,\,} {\,t_k}) \rangle_k}
\ar[d]^-{\mathcal{M}_2((\lj {\Delta\,\,} {\,t_k})_{k})}_>>>>>{\dcomment{\text{pres. of c. prod.}}\,\,\,\,}
\\
\mathcal{M}_1(m) \ar[r]_-{\cong} 
\ar@/_4pc/[rrr]_-{\mathsf{hom}(f)_m}^>>>>>>>>>{\dcomment{\text{def.}}\qquad}
& 
\bigsqcap_{1 \,\leq\, k \,\leq\, m} (\mathcal{M}_1(1)) \ar@/_2pc/[r]_-{\bigsqcap_k (f)} 
\ar@/^2pc/[r]^-{\langle f \,\comp\, \mathsf{proj}_k \rangle_k}_*+<2.45em>{\!\dcomment{\text{def.}}}
& 
\bigsqcap_{1 \,\leq\, k \,\leq\, m} (\mathcal{M}_2(1)) \ar[r]_-{\cong} 
& 
\mathcal{M}_2(m)
}
\vspace{-0.25cm}
\]
\end{proof}

Using these observations about the models of $I_{\mathcal{T}_{\text{eff}}} : \aleph_{\!\!1}^{\text{op}} \longrightarrow \mathcal{L}_{\mathcal{T}_{\text{eff}}}$ and the morphisms between them, we next show how to interpret eMLTT$_{\mathcal{T}_{\text{eff}}}^{\mathcal{H}}$ in the fibred adjunction model given by the split fibred adjunction $\widehat{F_{\!\mathcal{L}_{\mathcal{T}_{\text{eff}}}}} \dashv \widehat{U_{\!\mathcal{L}_{\mathcal{T}_{\text{eff}}}}}$.

\begin{definition}
\label{def:interpretationofemlttwithhandlers}
\index{interpretation function}
\index{ @$\sem{-}$ (interpretation function)}
We extend the \emph{interpretation} of eMLTT$_{\mathcal{T}_{\text{eff}}}$ in this fibred adjunction model to eMLTT$_{\mathcal{T}_{\text{eff}}}^{\mathcal{H}}$ by defining $\sem{-}$ on the user-defined algebra type as
\vspace{0.1cm}
\[
\mkrule
{
\begin{array}{l c l}
\sem{\Gamma';\langle A , \{V_{\sigalgop}\}_{\sigalgop \in \mathcal{S}_{\text{eff}}} \rangle}_1 & \defeq &\sem{\Gamma'}
\\[0.5mm]
\sem{\Gamma';\langle A , \{V_{\sigalgop}\}_{\sigalgop \in \mathcal{S}_{\text{eff}}} \rangle}_2(\gamma\,') & \defeq & \mathcal{M}_{\langle \sem{\Gamma';A}_2(\gamma\,') , \{f^{\gamma\,'}_{\sigalgop_i}\}_{\sigalgop_i \in \mathcal{S}_{\text{eff}}} \rangle}
\end{array}
}
{
\begin{array}{c}
\sem{\Gamma';A}_1 = \sem{\Gamma'} \in \Set
\quad
\sem{\Gamma';A}_2 : \sem{\Gamma'} \longrightarrow \Set
\quad
\sem{\Gamma';V_{\sigalgop}}_1 = \id_{\sem{\Gamma'}} : \sem{\Gamma'} \longrightarrow \sem{\Gamma'}
\\[2mm]
(\sem{\Gamma';V_{\sigalgop}}_2)_{\gamma\,'} : 1 \longrightarrow \bigsqcap_{\langle i , f \rangle \in \bigsqcup_{i \in \sem{\diamond; I}_2(\star)} \bigsqcap_{o \in \sem{x : I; O}_2\, \langle \star , i \rangle} (\sem{\Gamma';A}_2(\gamma\,'))} (\sem{\Gamma';A}_2(\gamma\,'))
\\[4mm]
\mathcal{M}_{\langle \sem{\Gamma';A}_2(\gamma\,') , \{f^{\gamma\,'}_{\sigalgop_i}\}_{\sigalgop_i \in \mathcal{S}_{\text{eff}}} \rangle}(\lj {\Delta^\gamma} {T^\gamma_1}) = \mathcal{M}_{\langle \sem{\Gamma';A}_2(\gamma\,') , \{f^{\gamma\,'}_{\sigalgop_i}\}_{\sigalgop_i \in \mathcal{S}_{\text{eff}}} \rangle}(\lj {\Delta^\gamma} {T^\gamma_2})
\\[5mm]
\text{for all } \ljeq {\Gamma \vertbar \Delta} {T_1} {T_2} \text{ in } \mathcal{E}_{\text{eff}} \text{ and } \gamma \text{ in } \sem{\Gamma} \text{, where } 
\\[1.5mm]
f^{\gamma\,'}_{\sigalgop_i} \defeq f \mapsto \mathsf{proj}_{\langle i , f \rangle}((\sem{\Gamma';V_{\sigalgop}}_2)_{\gamma\,'}(\star))
\\[0.5mm]
\end{array}
}
\vspace{0.15cm}
\]
on the composition operation for computation terms as
\[
\mkrule
{
\xymatrix@C=5em@R=6em@M=0.5em{
\txt<10pc>{$\sem{\Gamma;\runas M {y \!:\! U\ul{C}} {\ul{D}} N}_1 $\\$ \defeq $\\$ \sem{\Gamma}$}
\ar[dd]_-{\id_{\sem{\Gamma}}}
&
\txt<11pc>{$(\sem{\Gamma;\runas M {y \!:\! U\ul{C}} {\ul{D}} N}_2)_{\gamma} $\\$ \defeq $\\$ 1$}
\ar[d]^-{(\sem{\Gamma;M}_2)_\gamma}
\\
&
U_{\!\mathcal{L}_{\mathcal{T}_{\text{eff}}}}(\sem{\Gamma;\ul{C}}_2(\gamma))
\ar[d]^-{f^\gamma}
\\
\sem{\Gamma}
&
U_{\!\mathcal{L}_{\mathcal{T}_{\text{eff}}}}(\sem{\Gamma;\ul{D}}_2(\gamma))
}
}
{
\begin{array}{c}
\sem{\Gamma;M}_1 = \id_{\sem{\Gamma}} : \sem{\Gamma} \longrightarrow \sem{\Gamma}
\\[2mm]
(\sem{\Gamma;M}_2)_\gamma : 1 \longrightarrow U_{\!\mathcal{L}_{\mathcal{T}_{\text{eff}}}}(\sem{\Gamma;\ul{C}}_2(\gamma))
\\[2mm]
\sem{\Gamma, y \!:\! U\ul{C}; N}_1 = \id_{\bigsqcup_{\gamma \in \sem{\Gamma}} (U_{\!\mathcal{L}_{\mathcal{T}_{\text{eff}}}}(\sem{\Gamma;\ul{C}}_2(\gamma)))} : \sem{\Gamma, y \!:\! U\ul{C}} \longrightarrow \sem{\Gamma, y \!:\! U\ul{C}}
\\[2mm]
(\sem{\Gamma, y \!:\! U\ul{C}; N}_2)_{\langle \gamma , c \rangle} : 1 \longrightarrow U_{\!\mathcal{L}_{\mathcal{T}_{\text{eff}}}}(\sem{\Gamma; \ul{D}}_2(\gamma))
\\[3mm]
\xymatrix@C=10em@R=3em@M=0.5em{
\bigsqcap_o (\mathcal{M}^{\gamma}_1(1)) 
\ar[r]^-{\bigsqcap_{o \in \sem{x : I; O}_2\, \langle \star , i \rangle} (f^\gamma)}
\ar[d]_-{\cong}
& 
\bigsqcap_o (\mathcal{M}^{\gamma}_2(1))
\ar[d]^-{\cong}
\\
\mathcal{M}^{\gamma}_1(\vert \sem{x \!:\! I; O}_2\, \langle \star , i \rangle \vert)
\ar[d]_-{\mathcal{M}^{\gamma}_1(\lj {\overrightarrow{x_o}\,\,} {\,\sigalgop_i(x_o)_{o}})}
&
\mathcal{M}^{\gamma}_2(\vert \sem{x \!:\! I; O}_2\, \langle \star , i \rangle \vert)
\ar[d]^-{\mathcal{M}^{\gamma}_2(\lj {\overrightarrow{x_o}\,\,} {\,\sigalgop_i(x_o)_{o}})}
\\
\mathcal{M}^{\gamma}_1(1)
\ar[r]_-{f^\gamma}
&
\mathcal{M}^{\gamma}_2(1)
}
\\[3mm]
\text{for all } \sigalgop : (x \!:\! I) \longrightarrow O \text{ in } \mathcal{S}_{\text{eff}} \text{, } \gamma \text{ in } \sem{\Gamma} \text{, and } i \text{ in } \sem{\diamond;I}_2 (\star)\text{, where}
\\[1.5mm]
\mathcal{M}^{\gamma}_1 \defeq \sem{\Gamma; \ul{C}}_2(\gamma)
\qquad
\mathcal{M}^{\gamma}_2 \defeq \sem{\Gamma; \ul{D}}_2(\gamma)
\qquad
f^{\gamma} \defeq c \mapsto (\sem{\Gamma, y \!:\! U\ul{C}; N}_2)_{\langle \gamma , c \rangle}(\star)
\\[0.5mm]
\end{array}
}
\]

\pagebreak

\noindent
and on the composition operation for homomorphism terms as
\[
\mkrule
{
\xymatrix@C=3em@R=6em@M=0.5em{
\txt<13pc>{$\sem{\Gamma; z \!:\! \ul{C};\runas K {y \!:\! U\ul{D}_1} {\ul{D}_2} M}_1 $\\$ \defeq $\\$ \sem{\Gamma}$}
\ar[dd]_-{\id_{\sem{\Gamma}}}
&
\txt<14pc>{$(\sem{\Gamma; z \!:\! \ul{C};\runas K {y \!:\! U\ul{D}_1} {\ul{D}_2} M}_2)_{\gamma} $\\$ \defeq $\\$ \sem{\Gamma;\ul{C}}_2(\gamma)$}
\ar[d]^-{(\sem{\Gamma;z : \ul{C};K}_2)_\gamma}
\\
&
\sem{\Gamma;\ul{D}_1}_2(\gamma)
\ar[d]^-{\mathsf{hom}(f^\gamma)}
\\
\sem{\Gamma}
&
\sem{\Gamma;\ul{D}_2}_2(\gamma)
}
}
{
\begin{array}{c}
\sem{\Gamma; z \!:\! \ul{C};K}_1 = \id_{\sem{\Gamma}} : \sem{\Gamma} \longrightarrow \sem{\Gamma}
\\[2mm]
(\sem{\Gamma; z \!:\! \ul{C};K}_2)_\gamma : \sem{\Gamma;\ul{C}}_2(\gamma) \longrightarrow \sem{\Gamma;\ul{D}_1}_2(\gamma)
\\[2mm]
\sem{\Gamma, y \!:\! U\ul{D}_1; M}_1 = \id_{\bigsqcup_{\gamma \in \sem{\Gamma}} (U_{\!\mathcal{L}_{\mathcal{T}_{\text{eff}}}}(\sem{\Gamma;\ul{D}_1}_2(\gamma)))} : \sem{\Gamma, y \!:\! U\ul{D}_1} \longrightarrow \sem{\Gamma, y \!:\! U\ul{D}_1}
\\[2mm]
(\sem{\Gamma, y \!:\! U\ul{D}_1; M}_2)_{\langle \gamma , d \rangle} : 1 \longrightarrow U_{\!\mathcal{L}_{\mathcal{T}_{\text{eff}}}}(\sem{\Gamma; \ul{D}_2}_2(\gamma))
\\[3mm]
\xymatrix@C=10em@R=3em@M=0.5em{
\bigsqcap_o (\mathcal{M}^{\gamma}_1(1)) 
\ar[r]^-{\bigsqcap_{o \in \sem{x : I; O}_2\, \langle \star , i \rangle} (f^\gamma)}
\ar[d]_-{\cong}
& 
\bigsqcap_o (\mathcal{M}^{\gamma}_2(1))
\ar[d]^-{\cong}
\\
\mathcal{M}^{\gamma}_1(\vert \sem{x \!:\! I; O}_2\, \langle \star , i \rangle \vert)
\ar[d]_-{\mathcal{M}^{\gamma}_1(\lj {\overrightarrow{x_o}\,\,} {\,\sigalgop_i(x_o)_{o}})}
&
\mathcal{M}^{\gamma}_2(\vert \sem{x \!:\! I; O}_2\, \langle \star , i \rangle \vert)
\ar[d]^-{\mathcal{M}^{\gamma}_2(\lj {\overrightarrow{x_o}\,\,} {\,\sigalgop_i(x_o)_{o}})}
\\
\mathcal{M}^{\gamma}_1(1)
\ar[r]_-{f^\gamma}
&
\mathcal{M}^{\gamma}_2(1)
}
\\[3mm]
\text{for all } \sigalgop : (x \!:\! I) \longrightarrow O \text{ in } \mathcal{S}_{\text{eff}} \text{, } \gamma \text{ in } \sem{\Gamma} \text{, and } i \text{ in } \sem{\diamond;I}_2 (\star)\text{, where}
\\[1.5mm]
\mathcal{M}^{\gamma}_1 \defeq \sem{\Gamma; \ul{D}_1}_2(\gamma)
\qquad
\mathcal{M}^{\gamma}_2 \defeq \sem{\Gamma; \ul{D}_2}_2(\gamma)
\qquad
f^{\gamma} \defeq d \mapsto (\sem{\Gamma, y \!:\! U\ul{D}_1; M}_2)_{\langle \gamma , d \rangle}(\star)
\\[0.5mm]
\end{array}
}
\]
\end{definition}

\pagebreak

Similarly to eMLTT$_{\mathcal{T}_{\text{eff}}}$, the results from Section~\ref{sect:soundness} that relate  weakening and substitution to reindexing along semantic projection and substitution morphisms extend to eMLTT$_{\mathcal{T}_{\text{eff}}}^{\mathcal{H}}$ straightforwardly. 
However, analogously to eMLTT$_{\mathcal{T}_{\text{eff}}}$, the soundness theorem (Theorem~\ref{thm:soundness}) again needs more attention, as discussed in detail below.

\subsection*{Extending Theorem~\ref{thm:soundness} (Soundness) to eMLTT$_{\!\mathcal{T}_{\text{eff}}}^{\mathcal{H}}$}

\index{soundness theorem}
We begin by recalling that in Theorem~\ref{thm:soundness} we showed that the \emph{a priori} partially defined interpretation function $\sem{-}$  is defined on well-formed contexts and types, and well-typed terms, and that it maps definitionally equal contexts, types, and terms to equal objects and morphisms. For example, given $\ceq \Gamma M N \ul{C}$, we showed that
\[
\sem{\Gamma;M} 
=
\sem{\Gamma;N} 
: 1_{\sem{\Gamma}} \longrightarrow \widehat{U_{\mathcal{L}_{\mathcal{T}_{\text{eff}}}}}(\sem{\Gamma;\ul{C}})
\]

When extending Theorem~\ref{thm:soundness} to eMLTT$_{\mathcal{T}_{\text{eff}}}^{\mathcal{H}}$, we keep the basic proof principle the same: $(a)$--$(l)$ are proved simultaneously, by induction on the given derivations, using the eMLTT$_{\mathcal{T}_{\text{eff}}}^{\mathcal{H}}$ versions of Propositions~\ref{prop:semweakening2},~\ref{prop:semsubstitution2},~\ref{prop:semsubstitution3}, and~\ref{prop:semsubstitution4} to relate syntactic weakening and substitution to their semantic counterparts. We discuss some new cases corresponding to the rules given in Definition~\ref{def:extensionofeMLTTwithhandlers} in detail below.

\vspace{0.2cm}

\noindent
\textbf{Formation rule for the user-defined algebra type:}
In this case, the given derivation ends with 
\[
\mkrule
{
\lj {\Gamma'} {\langle A , \{V_{\sigalgop}\}_{\sigalgop \in \mathcal{S}_{\text{eff}}} \rangle}
}
{
\begin{array}{c}
\lj {\Gamma'} A
\qquad
\vj {\Gamma'} {V_{\sigalgop}} {(\Sigma\, x \!:\! I . O \to A) \to A}
\\[2mm]
\hspace{-3.9cm} \veq {\Gamma'} {\overrightarrow{\lambda\, x'_i \!:\! \widehat{A_i} .}\, \overrightarrow{\lambda\, x_{w_{\!j}} \!:\! \widehat{A'_j} \to A .}\, \efftrans {T_1} {A; \overrightarrow{x'_i}; \overrightarrow{x_{w_{\!j}}}; \overrightarrow{V_{\sigalgop}}} \\ \hspace{0.45cm} } {\overrightarrow{\lambda\, x'_i \!:\! \widehat{A_i} .}\, \overrightarrow{\lambda\, x_{w_{\!j}} \!:\! \widehat{A'_j} \to A .}\, \efftrans {T_2} {A; \overrightarrow{x'_i}; \overrightarrow{x_{w_{\!j}}}; \overrightarrow{V_{\sigalgop}}}\,} {\,\overrightarrow{\Pi x'_i \!:\! \widehat{A_i} .}\, \overrightarrow{\widehat{A'_j} \to A} \to A}
\\[3mm]
(\text{for all } \sigalgop : (x \!:\! I) \longrightarrow O \in \mathcal{S}_{\text{eff}}
\text{ and }
\ljeq {\Gamma \vertbar \Delta} {T_1} {T_2} \in \mathcal{E}_{\text{eff}})
\end{array}
}
\]
and we need to show that
\[
\sem{\Gamma';\langle A , \{V_{\sigalgop}\}_{\sigalgop \in \mathcal{S}_{\text{eff}}} \rangle} \in \Fam_{\sem{\Gamma'}}(\Mod(\!\mathcal{L}_{\mathcal{T}_{\text{eff}}},\Set))
\]
which, for the fibred adjunction model we are working with, is equivalent to showing 
\[
\begin{array}{c}
\sem{\Gamma';\langle A , \{V_{\sigalgop}\}_{\sigalgop \in \mathcal{S}_{\text{eff}}} \rangle}_1 = \sem{\Gamma'} \in \Set
\\[3mm]
\sem{\Gamma';\langle A , \{V_{\sigalgop}\}_{\sigalgop \in \mathcal{S}_{\text{eff}}} \rangle}_2 : \sem{\Gamma'} \longrightarrow \Mod(\!\mathcal{L}_{\mathcal{T}_{\text{eff}}},\Set)
\end{array}
\]

First, we use $(d)$ on the assumed derivations of $\vj {\Gamma'} {V_{\sigalgop}} {(\Sigma\, x \!:\! I . O \to A) \to A}$, in combination with the propositions that relate weakening and substitution to reindexing along semantic projection and substitution morphisms, so as to get
\[
\begin{array}{c}
\sem{\Gamma';V_{\sigalgop}}_1 = \id_{\sem{\Gamma'}} : \sem{\Gamma'} \longrightarrow \sem{\Gamma'}
\\[3mm]
(\sem{\Gamma';V_{\sigalgop}}_2)_{\gamma\,'} : 1 \longrightarrow \bigsqcap_{\langle i , f \rangle \in \bigsqcup_{i \in \sem{\diamond; I}_2(\star)} \bigsqcap_{o \in \sem{x : I; O}_2\, \langle \star , i \rangle} (\sem{\Gamma';A}_2(\gamma\,'))} (\sem{\Gamma';A}_2(\gamma\,'))
\end{array}
\]

Next, we use $(j)$ on the assumed derivations of definitional equations, again in combination with the propositions that relate weakening and substitution to reindexing along semantic projection and substitution morphisms, to get
\[
\begin{array}{c}
\sem{\Gamma'; \overrightarrow{\lambda\, x'_i \!:\! \widehat{A_i} .}\, \overrightarrow{\lambda\, x_{w_{\!j}} \!:\! \widehat{A'_j} \to A .}\, \efftrans {T_1} {A; \overrightarrow{x'_i}; \overrightarrow{x_{w_{\!j}}}; \overrightarrow{V_{\sigalgop}}}}_1 
\\[-1mm]
= 
\\[1mm]
\sem{\Gamma'; \overrightarrow{\lambda\, x'_i \!:\! \widehat{A_i} .}\, \overrightarrow{\lambda\, x_{w_{\!j}} \!:\! \widehat{A'_j} \to A .}\, \efftrans {T_2} {A; \overrightarrow{x'_i}; \overrightarrow{x_{w_{\!j}}}; \overrightarrow{V_{\sigalgop}}}}_1 
\\
= 
\\[-1mm]
\id_{\sem{\Gamma'}}
\end{array}
\]
and
\vspace{3mm}
\[
\begin{array}{c}
(\sem{\Gamma';\overrightarrow{\lambda\, x'_i \!:\! \widehat{A_i} .}\, \overrightarrow{\lambda\, x_{w_{\!j}} \!:\! \widehat{A'_j} \to A .}\, \efftrans {T_1} {A; \overrightarrow{x'_i}; \overrightarrow{x_{w_{\!j}}}; \overrightarrow{V_{\sigalgop}}}}_2)_{\gamma\,'}
\\[-1mm]
=
\\[1mm]
(\sem{\Gamma';\overrightarrow{\lambda\, x'_i \!:\! \widehat{A_i} .}\, \overrightarrow{\lambda\, x_{w_{\!j}} \!:\! \widehat{A'_j} \to A .}\, \efftrans {T_2} {A; \overrightarrow{x'_i}; \overrightarrow{x_{w_{\!j}}}; \overrightarrow{V_{\sigalgop}}}}_2)_{\gamma\,'}
\\[1mm]
\end{array}
\]
as $\sem{\Gamma'} \longrightarrow \sem{\Gamma'}$ and $1 \longrightarrow 
\bigsqcap_{f \in \bigsqcap_{x_{i}} (\sem{\Gamma;A_{i}}_2(\gamma)) }
\bigsqcap_{g \in \bigsqcap_{w_{\!j}} \bigsqcap_{a \in \sem{\Gamma;A'_{\!j}}_2(\gamma)} (\sem{\Gamma';A}_2(\gamma\,')) } (\sem{\Gamma';A}_2(\gamma\,'))$, respectively, 
for all $\ljeq {\Gamma \vertbar \Delta} {T_1} {T_2}$ in $\mathcal{E}_{\text{eff}}$ and $\gamma\,'$ in $\sem{\Gamma'}$. 

Based on the definition of $\sem{-}$ for lambda abstraction, the above equations give us
\[
\begin{array}{c}
(\sem{\Gamma',\overrightarrow{x'_i \!:\! \widehat{A_i}},\overrightarrow{x_{w_{\!j}} \!:\! \widehat{A'_j} \to A};\efftrans {T_1} {A; \overrightarrow{x'_i}; \overrightarrow{x_{w_{\!j}}}; \overrightarrow{V_{\sigalgop}}}}_2)_{\langle \langle \gamma\,', \gamma , \rangle , \overrightarrow{f_{w_{\!j}}} \rangle}
\\
=
\\
(\sem{\Gamma',\overrightarrow{x'_i \!:\! \widehat{A_i}},\overrightarrow{x_{w_{\!j}} \!:\! \widehat{A'_j} \to A};\efftrans {T_2} {A; \overrightarrow{x'_i}; \overrightarrow{x_{w_{\!j}}}; \overrightarrow{V_{\sigalgop}}}}_2)_{\langle \langle \gamma\,', \gamma , \rangle , \overrightarrow{f_{w_{\!j}}} \rangle}
\end{array}
\]
as morphisms $1 \longrightarrow \sem{\Gamma';A}_2(\gamma\,')$, for all $\ljeq {\Gamma \vertbar \Delta} {T_1} {T_2}$ in $\mathcal{E}_{\text{eff}}$, $\gamma$ in $\sem{\Gamma}$, $\gamma\,'$ in $\sem{\Gamma'}$, and $f_{w_{\!j}}$ in $\bigsqcap_{a \in \sem{\Gamma;A'_{\!j}}_2(\gamma)} (\sem{\Gamma';A}_2(\gamma\,'))$, for all $w_{\!j} \!:\! A'_{\!j}$ in $\Delta$.

Next, we show that $\sem{\Gamma';\langle A , \{V_{\sigalgop}\}_{\sigalgop \in \mathcal{S}_{\text{eff}}} \rangle}$ is defined, for which we need to show
\[
\mathcal{M}_{\langle \sem{\Gamma';A}_2(\gamma\,') , \{f^{\gamma\,'}_{\sigalgop_i}\}_{\sigalgop_i \in \mathcal{S}_{\text{eff}}} \rangle}(\lj {\Delta^\gamma} {T^\gamma_1}) = \mathcal{M}_{\langle \sem{\Gamma';A}_2(\gamma\,') , \{f^{\gamma\,'}_{\sigalgop_i}\}_{\sigalgop_i \in \mathcal{S}_{\text{eff}}} \rangle}(\lj {\Delta^\gamma} {T^\gamma_2})
\]
for all $\ljeq {\Gamma \vertbar \Delta} {T_1} {T_2}$ in $\mathcal{E}_{\text{eff}}$ and $\gamma$ in $\sem{\Gamma}$, for which we use Proposition~\ref{prop:relatingsemanticsoffibeffectterms}.

\pagebreak

To be able to use Proposition~\ref{prop:relatingsemanticsoffibeffectterms} to prove the above equations, we first define a family $\mathcal{M}_{\langle \langle \gamma\,' , \gamma \rangle , \overrightarrow{f_{w_{\!j}}} \rangle}$ of models of the countable Lawvere theory $I_{\mathcal{T}^d_{\text{eff}}} : \aleph_{\!\!1}^{\text{op}} \longrightarrow \mathcal{L}_{\mathcal{T}^d_{\text{eff}}}$ by 
\vspace{-0.5cm}
\[
\mathcal{M}_{\langle \langle \gamma\,' , \gamma \rangle , \overrightarrow{f_{w_{\!j}}} \rangle} \defeq \mathcal{M}_{\langle \sem{\Gamma';A}_2(\gamma\,') , \{f^{\gamma\,'}_{\sigalgop_i}\}_{\sigalgop_i \in \mathcal{S}_{\text{eff}}} \rangle}
\vspace{-0.25cm}
\]
where $f_{w_{\!j}}$ is an element of $\bigsqcap_{a \in \sem{\Gamma;A'_{\!j}}_2(\gamma)} (\sem{\Gamma';A}_2(\gamma\,'))$, and $f^{\gamma\,'}_{\sigalgop_i}$ is given as in Definition~\ref{def:interpretationofemlttwithhandlers}.
Observe that by definition we have $\mathcal{M}_{\langle \langle \gamma\,' , \gamma \rangle , \overrightarrow{f_{w_{\!j}}} \rangle}(1) = \sem{\Gamma';A}_2(\gamma\,')$, as discussed in more detail in the footnote in the beginning of this section.

Further, in order to be able to use Proposition~\ref{prop:relatingsemanticsoffibeffectterms}, we also need to show that the next diagram commutes, where we abbreviate $\mathcal{M}_{\langle \langle \gamma\,' , \gamma \rangle , \overrightarrow{f_{w_{\!j}}} \rangle}$ as $\mathcal{M}$. 
\[
\xymatrix@C=5em@R=9em@M=0.5em{
1
\ar[r]^-{\langle \id_1 \rangle_{\langle i , f \rangle}}
\ar@/_10pc/[dddr]_<<<<<<<<<<<<<<<<<<<<<<<<<<<<<<<<<<<<<<<<<<<<<<<<<<{(\sem{\Gamma';V_{\sigalgop}}_2)_{\gamma\,'}}
&
\bigsqcap_{\langle i , f \rangle \in \bigsqcup_{i \in \sem{\diamond; I}_2(\star)} \bigsqcap_{o \in \sem{x : I; O}_2\, \langle \star , i \rangle} (\mathcal{M}(1))} 1
\ar[d]^-{\bigsqcap_{\langle i , f \rangle} (\star \,\mapsto\, f)}_-{\dcomment{\text{the universal property of count. prod.}}\quad\,\,\,}
\\
&
\bigsqcap_{\langle i , f \rangle} \bigsqcap_{o \in \sem{x : I; O}_2\, \langle \star , i \rangle} (\mathcal{M}(1))
\ar@/^7pc/[dd]^-{\bigsqcap_{\langle i , f \rangle} (\sigalgop^{\mathcal{M}}_i)}
\ar[d]_-{\cong}^>>>>>>{\!\!\!\!\quad\dcomment{\text{def. of } \sigalgop^{\mathcal{M}}_i}}
\ar@/_7.5pc/[dd]^<<<<<<<<<<<<<<<<<<<<<<<<<<<<<<<<<<<<<<{\bigsqcap_{\langle i , f \rangle} (f^{\gamma\,'}_{\sigalgop_i})}_-{\dcomment{\text{def.}}\,\,\,\,}
\ar@/_10.5pc/[dd]_<<<<<<{\langle \mathsf{proj}_{\langle i , f \rangle}((\sem{\Gamma';V_{\sigalgop}}_2)_{\gamma\,'}(\star)) \rangle_{\langle i , f \rangle}\,\,\,\,\,\,\,\,\,\,}
\\
&
\bigsqcap_{\langle i , f \rangle} (\mathcal{M}(\vert \sem{x : I; O}_2\, \langle \star , i \rangle \vert))
\ar@/^3.5pc/[d]_-{\bigsqcap_{\langle i , f \rangle} (\mathcal{M}(\lj {\overrightarrow{x_o}\,\,} {\,\sigalgop_i(x_o)_o}))}_<<<<{\dcomment{\text{def. of } \mathcal{M}}\qquad}
\\
&
\bigsqcap_{\langle i , f \rangle} (\mathcal{M}(1))
}
\]

Based on the above, we can now use Proposition~\ref{prop:relatingsemanticsoffibeffectterms} with $\efftrans {T_1} {A; \overrightarrow{x_i}; \overrightarrow{x_{w_{\!j}}}; \overrightarrow{V_{\sigalgop}}}$ and \linebreak $\efftrans {T_2} {A; \overrightarrow{x_i}; \overrightarrow{x_{w_{\!j}}}; \overrightarrow{V_{\sigalgop}}}$ to show that for all $\gamma\,'$, $\gamma$, and $\overrightarrow{f_{w_{\!j}}}$, the following diagram commutes: 

\[
\hspace{-0.45cm}
\xymatrix@C=2em@R=4em@M=0.5em{
1
\ar[d]_>>>>{\langle f_{w_{\!j}} \rangle_{w_{\!j} : A'_{\!j} \in \Delta}}
\ar@/_10.5pc/[dddd]_<<<<<<<<{(\sem{\Gamma',\overrightarrow{x'_i \!:\! \widehat{A_i}},\overrightarrow{x_{w_{\!j}} \!:\! \widehat{A'_j} \to A};\efftrans {T_1} {A; \overrightarrow{x'_i}; \overrightarrow{x_{w_{\!j}}}; \overrightarrow{V_{\sigalgop}}}}_2)_{\langle \langle \gamma\,', \gamma , \rangle , \overrightarrow{f_{w_{\!j}}} \rangle}\qquad\quad\!\!\!\!\!\!\!\!\!\!}
\ar@/^10.5pc/[dddd]^>>>>>>>>>{\!\!\!\!\!\!\!\!\!\!\qquad(\sem{\Gamma',\overrightarrow{x'_i \!:\! \widehat{A_i}},\overrightarrow{x_{w_{\!j}} \!:\! \widehat{A'_j} \to A};\efftrans {T_2} {A; \overrightarrow{x'_i}; \overrightarrow{x_{w_{\!j}}}; \overrightarrow{V_{\sigalgop}}}}_2)_{\langle \langle \gamma\,', \gamma , \rangle , \overrightarrow{f_{w_{\!j}}} \rangle}}
\\
\bigsqcap_{w_{\!j} : A'_{\!j} \in \Delta} \bigsqcap_{a \in {\sem{\Gamma;A'_{\!j}}_2(\gamma)}} (\sem{\Gamma';A}_2(\gamma\,'))
\ar[d]_-{\cong}
\\
\mathcal{M}(\vert \Delta^\gamma \vert)
\ar@/_2pc/[d]_-{\mathcal{M}(\lj {\Delta^\gamma\,\,} {\,T_1^\gamma})}
\ar@/^2pc/[d]^-{\mathcal{M}(\lj {\Delta^\gamma\,\,} {\,T_2^\gamma})}
\\
\mathcal{M}(1)
\ar[d]_-{=}
\\
\sem{\Gamma';A}_2(\gamma\,')
}
\vspace{0.1cm}
\]
where, for better readability, we again write $\mathcal{M}$ for $\mathcal{M}_{\langle \langle \gamma\,' , \gamma \rangle , \overrightarrow{f_{w_{\!j}}} \rangle}$. 

As the above diagram commutes for all $f_{w_{\!j}}$, namely, for all elements of the product $\bigsqcap_{w_{\!j} : A'_{\!j} \in \Delta} \bigsqcap_{a \in {\sem{\Gamma;A'_{\!j}}_2(\gamma)}} (\sem{\Gamma';A}_2(\gamma\,'))$, we can derive the required equations
\vspace{0.1cm}
\[
\mathcal{M}_{\langle \sem{\Gamma';A}_2(\gamma\,') , \{f^{\gamma\,'}_{\sigalgop_i}\}_{\sigalgop_i \in \mathcal{S}_{\text{eff}}} \rangle}(\lj {\Delta^\gamma} {T^\gamma_1}) = \mathcal{M}_{\langle \sem{\Gamma';A}_2(\gamma\,') , \{f^{\gamma\,'}_{\sigalgop_i}\}_{\sigalgop_i \in \mathcal{S}_{\text{eff}}} \rangle}(\lj {\Delta^\gamma} {T^\gamma_2})
\vspace{0.4cm}
\]
from it, for all $\ljeq {\Gamma \vertbar \Delta} {T_1} {T_2}$ in $\mathcal{E}_{\text{eff}}$ and $\gamma$ in $\sem{\Gamma}$, by using the well-pointedness of the category of sets and functions, and the fact that every isomorphism is an epimorphism. 

Finally, as we have shown that $\sem{\Gamma';\langle A , \{V_{\sigalgop}\}_{\sigalgop \in \mathcal{S}_{\text{eff}}} \rangle}$ is defined, and as we know from  Proposition~\ref{prop:liftingnoeqlawthmodeltoeqlawthmodel} that $\mathcal{M}_{\langle \sem{\Gamma';A}_2(\gamma\,') , \{f^{\gamma\,'}_{\sigalgop_i}\}_{\sigalgop_i \in \mathcal{S}_{\text{eff}}} \rangle}$ is a model of $I_{\mathcal{T}_{\text{eff}}} : \aleph_{\!\!1}^{\text{op}} \longrightarrow \mathcal{L}_{\mathcal{T}_{\text{eff}}}$,\vspace{0.1cm}\linebreak for all $\gamma\,'$ in $\sem{\Gamma'}$, then, as required, we have 
\[
\begin{array}{c}
\sem{\Gamma';\langle A , \{V_{\sigalgop}\}_{\sigalgop \in \mathcal{S}_{\text{eff}}} \rangle}_1 = \sem{\Gamma'} \in \Set
\\[3mm]
\sem{\Gamma';\langle A , \{V_{\sigalgop}\}_{\sigalgop \in \mathcal{S}_{\text{eff}}} \rangle}_2 : \sem{\Gamma'} \longrightarrow \Mod(\!\mathcal{L}_{\mathcal{T}_{\text{eff}}},\Set)
\end{array}
\]

\pagebreak

\noindent
\textbf{Typing rule for the composition operation for computation terms:}
 In this case, the given derivation ends with 
\[
\mkrule
{
\cj {\Gamma} {\runas M {y \!:\! U\ul{C}} {\ul{D}} {N}} {\ul{D}}
}
{
\begin{array}{c@{\qquad} c}
\cj \Gamma M \ul{C} 
\quad
\lj \Gamma \ul{D}
\quad
\cj {\Gamma, y \!:\! U\ul{C}} N \ul{D}
\\[2mm]
\hspace{-0.95cm}
\ceq \Gamma {\lambda\, x \!:\! I .\, \lambda\, x' \!:\! O \to U\ul{C} .\, N[\thunk (\algop^{\ul{C}}_x(y'\!.\, \force {\ul{C}} (x'\, y')))/y] \\ \hspace{0.25cm}} { \lambda\, x \!:\! I .\, \lambda\, x' \!:\! O \to U\ul{C} .\, \algop^{\ul{D}}_x(y'\!.\, N[x'\, y'/y])} {\Pi\, x \!:\! I .\, (O \to U\ul{C}) \to \ul{D}}
\\[-1mm]
& \hspace{-3.5cm} (\sigalgop : (x \!:\! I) \longrightarrow O \in \mathcal{S}_{\text{eff}}) 
\end{array}
}
\]
and we need to show that
\[
\sem{\Gamma;{\runas M {y \!:\! U\ul{C}} {\ul{D}} {N}}} : 1_{\sem{\Gamma}} \longrightarrow \widehat{U_{\!\mathcal{L}_{\mathcal{T}_{\text{eff}}}}}(\sem{\Gamma;\ul{D}})
\]
which, for the fibred adjunction model we are working with, is equivalent to showing
\[
\sem{\Gamma;{\runas M {y \!:\! U\ul{C}} {\ul{D}} {N}}}_1 = \id_{\sem{\Gamma}} : \sem{\Gamma} \longrightarrow \sem{\Gamma}
\]
and, for all $\gamma$ in $\sem{\Gamma}$, that
\[
(\sem{\Gamma;{\runas M {y \!:\! U\ul{C}} {\ul{D}} {N}}}_2)_\gamma : 1 \longrightarrow U_{\!\mathcal{L}_{\mathcal{T}_{\text{eff}}}}(\sem{\Gamma;\ul{D}}_2(\gamma))
\]
 
First, we use the induction hypothesis on $\cj \Gamma M \ul{C}$ and $\cj {\Gamma, y \!:\! U\ul{C}} N \ul{D}$
to get
\[
\begin{array}{c}
\sem{\Gamma;M}_1 = \id_{\sem{\Gamma}} : \sem{\Gamma} \longrightarrow \sem{\Gamma}
\qquad
(\sem{\Gamma;M}_2)_\gamma : 1 \longrightarrow U_{\!\mathcal{L}_{\mathcal{T}_{\text{eff}}}}(\sem{\Gamma;\ul{C}}_2(\gamma))
\\[3mm]
\sem{\Gamma, y \!:\! U\ul{C}; N}_1 = \id_{\bigsqcup_{\gamma \in \sem{\Gamma}} (U_{\!\mathcal{L}_{\mathcal{T}_{\text{eff}}}}(\sem{\Gamma; \ul{C}}_2(\gamma)))} : \sem{\Gamma, y \!:\! U\ul{C}} \longrightarrow \sem{\Gamma, y \!:\! U\ul{C}}
\\[3mm]
(\sem{\Gamma, y \!:\! U\ul{C}; N}_2)_{\langle \gamma , c \rangle} : 1 \longrightarrow U_{\!\mathcal{L}_{\mathcal{T}_{\text{eff}}}}(\sem{\Gamma;\ul{D}}_2(\gamma))
 \end{array}
 \]

Next, we use $(k)$ on the assumed derivations of definitional equations
to get
\[
\begin{array}{c}
\sem{\Gamma; {\lambda\, x \!:\! I .\, \lambda\, x' \!:\! O \to U\ul{C} .\, N[\thunk (\algop^{\ul{C}}_x(y'\!.\, \force {\ul{C}} (x'\, y')))/y]}}_1 
\\
=
\\
\sem{\Gamma;{\lambda\, x \!:\! I .\, \lambda\, x' \!:\! O \to U\ul{C} .\, \algop^{\ul{D}}_x(y'\!.\, N[x'\, y'/y])}}_1 
\\
=
\\
\id_{\sem{\Gamma}}
\end{array}
\]
and
\[
\begin{array}{c}
(\sem{\Gamma; {\lambda\, x \!:\! I .\, \lambda\, x' \!:\! O \to U\ul{C} .\, N[\thunk (\algop^{\ul{C}}_x(y'\!.\, \force {\ul{C}} (x'\, y')))/y]}}_2)_\gamma
\\
=
\\
(\sem{\Gamma;{\lambda\, x \!:\! I .\, \lambda\, x' \!:\! O \to U\ul{C} .\, \algop^{\ul{D}}_x(y'\!.\, N[x'\, y'/y])}}_2)_\gamma 
\end{array}
\]
for all $\sigalgop : (x \!:\! I) \longrightarrow O$ in $\mathcal{S}_{\text{eff}}$ and $\gamma$ in $\sem{\Gamma}$. 

Based on the definition of $\sem{-}$ for lambda abstraction, the above equations give us
\vspace{0.1cm}
\[
\begin{array}{c}
(\sem{\Gamma, x \!:\! I, x' \!:\! O \to U\ul{C}; N[\thunk (\algop^{\ul{C}}_x(y'\!.\, \force {\ul{C}} (x'\, y')))/y]}_2)_{\langle \langle \gamma , i \rangle , f \rangle}
\\[-0.5mm]
=
\\
(\sem{\Gamma, x \!:\! I, x' \!:\! O \to U\ul{C}; \algop^{\ul{D}}_x(y'\!.\, N[x'\, y'/y])}_2)_{\langle \langle \gamma , i \rangle , f \rangle}
\end{array}
\vspace{0.1cm}
\]
as morphisms $1 \longrightarrow U_{\!\mathcal{L}_{\mathcal{T}_{\text{eff}}}}(\sem{\Gamma;\ul{D}}_2(\gamma))$, for all $\sigalgop : (x \!:\! I) \longrightarrow O$ in $\mathcal{S}_{\text{eff}}$, $\gamma$ in $\sem{\Gamma}$, $i$ in $\sem{\diamond;I}_2(\star)$, and $f$ in $\bigsqcap_{o \in \sem{x : I; O}_2\, \langle \star , i \rangle}(U_{\!\mathcal{L}_{\mathcal{T}_{\text{eff}}}}(\sem{\Gamma;\ul{C}}_2(\gamma)))$.

\vspace{0.15cm}
Next, we show that $\sem{\Gamma;\runas M {y \!:\! U\ul{C}} {\ul{D}} {N}}$ is defined, by proving that the following diagram commutes:  
\[
\xymatrix@C=12em@R=4em@M=0.5em{
\bigsqcap_{o} (\mathcal{M}^{\gamma}_1(1)) 
\ar@/^3.5pc/[r]^-{\bigsqcap_{o \in \sem{x : I; O}_2\, \langle \star , i \rangle} (f^\gamma)}_*+<0.8em>{\dcomment{\text{def. of } f^\gamma}}
\ar[r]^-{\bigsqcap_o(c \,\mapsto\, (\sem{\Gamma, y : U\ul{C}; N}_2)_{\langle \gamma, c \rangle}(\star))}_*+<0.5em>{\dcomment{\text{the univ. prop. of count. prod.}}}
\ar@/_2.5pc/[r]_-{f \,\mapsto\, \langle (\sem{\Gamma, y : A; N}_2)_{\langle \gamma , \mathsf{proj}_o(f) \rangle}(\star) \rangle_o}
\ar[ddd]_-{\cong}
\ar@{}[dddddd]^>>>>>>>>>>>>>>>>>>>>>>>>>{\!\!\!\!\qquad\dcomment{\text{eMLTT$_{\!\mathcal{T}_{\text{eff}}}^{\mathcal{H}}$ version of Proposition~\ref{prop:semsubstitution2}}}}^>>>>>>>>>>>>>>>>>>>>>>>>>>>>>>>>>>>>>>>>>>{\!\!\!\!\qquad\qquad\qquad\dcomment{\text{def. of } \sem{-}}}
\ar@/^1.15pc/[ddddddr]_>>>>>>>>>{f \,\mapsto\, (\sem{\Gamma, x, x'; N[\thunk (\algop^{\ul{C}}_x(y'\!.\, \force {\ul{C}} (x'\, y')))/y]}_2)_{\langle \langle \gamma , i \rangle , f \rangle}(\star)}
\ar@/^4pc/[ddddddr]^<<<<<<<<<<<<<<<<<<<<<<<<<<<<<<{\!\!\!f \,\mapsto\, (\sem{\Gamma, x, x'; \algop^{\ul{C}}_x(y'\!.\, N[x'\, y'/y])}_2)_{\langle \langle \gamma , i \rangle , f \rangle}(\star)}_-{\dcomment{\text{(k)}}\,\,\,\,}
& 
\bigsqcap_{o} (\mathcal{M}^{\gamma}_2(1))
\ar@/^3pc/[ddd]^-{\cong}_>>>>>>>>>>>>>{\dcomment{\text{def. of } \sem{-}}\qquad\quad}_<<<<<<<<<<<<<<<<{\dcomment{\text{eMLTT$_{\!\mathcal{T}_{\text{eff}}}^{\mathcal{H}}$ version of Proposition~\ref{prop:semsubstitution2}}}\quad}
\\
\\
\\
\mathcal{M}^{\gamma}_1(\vert \sem{x \!:\! I; O}_2\, \langle \star , i \rangle \vert)
\ar@/_2pc/[ddd]_<<<<<<<<<<{\mathcal{M}^{\gamma}_1(\lj {\overrightarrow{x_o}\,\,} {\,\sigalgop_i(x_o)_{o}})}
&
\mathcal{M}^{\gamma}_2(\vert \sem{x \!:\! I; O}_2\, \langle \star , i \rangle \vert)
\ar@/_0.5pc/[ddd]^-{\mathcal{M}^{\gamma}_2(\lj {\overrightarrow{x_o}\,\,} {\,\sigalgop_i(x_o)_{o}})}
\\
\\
\\
\mathcal{M}^{\gamma}_1(1)
\ar@/_3.5pc/[r]_-{f^\gamma}
\ar[r]_-{c \,\mapsto\, (\sem{\Gamma, y : U\ul{C}; N}_2)_{\langle \gamma , c \rangle}(\star)}_*+<2.5em>{\dcomment{\text{def. of } f^\gamma}}
&
\mathcal{M}^{\gamma}_2(1)
}
\]
for all $\sigalgop : (x \!:\! I) \longrightarrow O$ in $\mathcal{S}_{\text{eff}}$, $\gamma$ in $\sem{\Gamma}$, and $i$ in $\sem{\diamond, I}_2(\star)$, and where 
\[
\mathcal{M}^{\gamma}_1 \defeq \sem{\Gamma; \ul{C}}_2(\gamma)
\qquad
\mathcal{M}^{\gamma}_2 = \sem{\Gamma; \ul{D}}_2(\gamma)
\qquad
f^{\gamma} \defeq c \mapsto (\sem{\Gamma, y \!:\! U\ul{C}; N}_2)_{\langle \gamma , c \rangle}(\star)
\]

Finally, as we have shown that $\sem{\Gamma;\runas M {y \!:\! U\ul{C}} {\ul{D}} {N}}$ is defined, then, as required, we have
\[
\sem{\Gamma;\runas M {y \!:\! U\ul{C}} {\ul{D}} {N}}_1 = \id_{\sem{\Gamma}} : \sem{\Gamma} \longrightarrow \sem{\Gamma}
\]
and, for all $\gamma$ in $\sem{\Gamma}$, that
\[
(\sem{\Gamma;\runas M {y \!:\! U\ul{C}} {\ul{D}} {N}}_2)_\gamma : 1 \longrightarrow U_{\!\mathcal{L}_{\mathcal{T}_{\text{eff}}}}(\sem{\Gamma;\ul{D}}_2(\gamma))
\]

\vspace{0.2cm}

\noindent
\textbf{$\beta$-equation for the user-defined algebra type:}
In this case, the given derivation ends with
\[
\mkrule
{\ljeq {\Gamma} {U \langle A , \{V_{\sigalgop}\}_{\sigalgop \in \mathcal{S}_{\text{eff}}} \rangle} {A}}
{
\begin{array}{c}
\lj \Gamma \langle A , \{V_{\sigalgop}\}_{\sigalgop \in \mathcal{S}_{\text{eff}}} \rangle
\end{array}
}
\]
and we need to show 
\[
\sem{\Gamma;U \langle A , \{V_{\sigalgop}\}_{\sigalgop \in \mathcal{S}_{\text{eff}}} \rangle }
=
\sem{\Gamma;A} 
\in \Fam_{\sem{\Gamma}}(\Set)
\]
which, for the fibred adjunction model we are working with, is equivalent to showing
\[
\begin{array}{c}
\sem{\Gamma;U \langle A , \{V_{\sigalgop}\}_{\sigalgop \in \mathcal{S}_{\text{eff}}} \rangle}_1 = \sem{\Gamma;A}_1 = \sem{\Gamma} \in \Set
\\[3mm]
\sem{\Gamma;U \langle A , \{V_{\sigalgop}\}_{\sigalgop \in \mathcal{S}_{\text{eff}}} \rangle}_2 = \sem{\Gamma;A}_2 : \sem{\Gamma} \longrightarrow \Set
\end{array}
\]

First, we use $(c)$ on the derivation of $\lj \Gamma \langle A , \{V_{\sigalgop}\}_{\sigalgop \in \mathcal{S}_{\text{eff}}} \rangle$ to get 
\[
\begin{array}{c}
\sem{\Gamma;\langle A , \{V_{\sigalgop}\}_{\sigalgop \in \mathcal{S}_{\text{eff}}} \rangle}_1 = \sem{\Gamma} \in \Set
\\[3mm]
\sem{\Gamma;\langle A , \{V_{\sigalgop}\}_{\sigalgop \in \mathcal{S}_{\text{eff}}} \rangle}_2 : \sem{\Gamma} \longrightarrow \Mod(\!\mathcal{L}_{\mathcal{T}_{\text{eff}}},\Set)
\end{array}
\]

Next, recalling the definition of $\sem{-}$ for the type of thunked computations, we get
\[
\sem{\Gamma;U \langle A , \{V_{\sigalgop}\}_{\sigalgop \in \mathcal{S}_{\text{eff}}} \rangle } = \widehat{U_{\!\mathcal{L}_{\mathcal{T}_{\text{eff}}}}}(\sem{\Gamma;\langle A , \{V_{\sigalgop}\}_{\sigalgop \in \mathcal{S}_{\text{eff}}} \rangle}) \in \Fam_{\sem{\Gamma}}(\Set)
\]
which, for the fibred adjunction model we are working with, is equivalent to 
\[
\begin{array}{c}
\sem{\Gamma;U \langle A , \{V_{\sigalgop}\}_{\sigalgop \in \mathcal{S}_{\text{eff}}} \rangle }_1 = \sem{\Gamma} \in \Set
\\[3mm]
\sem{\Gamma;U \langle A , \{V_{\sigalgop}\}_{\sigalgop \in \mathcal{S}_{\text{eff}}} \rangle }_2(\gamma) = U_{\!\mathcal{L}_{\mathcal{T}_{\text{eff}}}} \comp \sem{\Gamma; \langle A , \{V_{\sigalgop}\}_{\sigalgop \in \mathcal{S}_{\text{eff}}} \rangle}_2 : \sem{\Gamma} \longrightarrow \Set
\end{array}
\]

Finally, by unfolding the definition of $\sem{-}$ for the user-defined algebra type, and recalling the definitions of $\mathcal{M}_{\langle \sem{\Gamma;A}_2(\gamma) , \{f^{\gamma}_{\sigalgop_i}\}_{\sigalgop_i \in \mathcal{S}_{\text{eff}}} \rangle}$ and $U_{\!\mathcal{L}_{\mathcal{T}_{\text{eff}}}}$, we have
\[
\begin{array}{c}
U_{\!\mathcal{L}_{\mathcal{T}_{\text{eff}}}}(\sem{\Gamma; \langle A , \{V_{\sigalgop}\}_{\sigalgop \in \mathcal{S}_{\text{eff}}} \rangle}_2(\gamma)) 
\\[1mm]
= 
\\[-1mm]
U_{\!\mathcal{L}_{\mathcal{T}_{\text{eff}}}}(\mathcal{M}_{\langle \sem{\Gamma;A}_2(\gamma) , \{f^{\gamma}_{\sigalgop_i}\}_{\sigalgop_i \in \mathcal{S}_{\text{eff}}} \rangle})
\\[1mm]
=
\\[-1mm]
\mathcal{M}_{\langle \sem{\Gamma;A}_2(\gamma) , \{f^{\gamma}_{\sigalgop_i}\}_{\sigalgop_i \in \mathcal{S}_{\text{eff}}} \rangle}(1)
\\[1mm]
=
\\[-1mm]
\sem{\Gamma;A}_2(\gamma)
\end{array}
\]
which means that, as required, we have
\[
\begin{array}{c}
\sem{\Gamma;U \langle A , \{V_{\sigalgop}\}_{\sigalgop \in \mathcal{S}_{\text{eff}}} \rangle}_1 = \sem{\Gamma;A}_1 = \sem{\Gamma} \in \Set
\\[3mm]
\sem{\Gamma;U \langle A , \{V_{\sigalgop}\}_{\sigalgop \in \mathcal{S}_{\text{eff}}} \rangle}_2 = \sem{\Gamma;A}_2 : \sem{\Gamma} \longrightarrow \Set
\end{array}
\]

\vspace{0.2cm}

\noindent
\textbf{$\beta$-equation for the composition operation for computation terms:}
In this case, the given derivation ends with
\[
\mkrule
{
\ceq {\Gamma} {\runas {(\force {\ul{C}} V)} {y \!:\! U\ul{C}} {\ul{D}} {M}} {M[V/y]} {\ul{D}}
}
{
\begin{array}{c@{\qquad} c}
\vj \Gamma V U\ul{C}
\quad
\lj \Gamma \ul{D}
\quad
\cj {\Gamma, y \!:\! U\ul{C}} M \ul{D}
\\[2mm]
\hspace{-0.95cm}
\ceq \Gamma {\lambda\, x \!:\! I .\, \lambda\, x' \!:\! O \to U\ul{C} .\, M[\thunk (\algop^{\ul{C}}_x(y'\!.\, \force {\ul{C}} (x'\, y')))/y] \\ \hspace{0.25cm}} { \lambda\, x \!:\! I .\, \lambda\, x' \!:\! O \to U\ul{C} .\, \algop^{\ul{D}}_x(y'\!.\, M[x'\, y'/y])} {\Pi\, x \!:\! I .\, (O \to U\ul{C}) \to \ul{D}}
\\[-1mm]
& \hspace{-3.5cm} (\sigalgop : (x \!:\! I) \longrightarrow O \in \mathcal{S}_{\text{eff}}) 
\end{array}
}
\]
and we need to show
\[
\sem{\Gamma;{\runas {(\force {\ul{C}} V)} {y \!:\! U\ul{C}} {\ul{D}} {M}}} = \sem{\Gamma;M[V/y]}
: 1_{\sem{\Gamma}} \longrightarrow \widehat{U_{\!\mathcal{L}_{\mathcal{T}_{\text{eff}}}}}(\sem{\Gamma;\ul{D}})
\]
which, for the fibred adjunction model we are working with, is equivalent to showing
\[
\begin{array}{c}
\sem{\Gamma;{\runas {(\force {\ul{C}} V)} {y \!:\! U\ul{C}} {\ul{D}} {M}}}_1 
= 
\sem{\Gamma;M[V/y]}_1 
= 
\id_{\sem{\Gamma}}
: \sem{\Gamma} \longrightarrow \sem{\Gamma}
\end{array}
\]
and, for all $\gamma$ in $\sem{\Gamma}$, that
\[
\begin{array}{c}
(\sem{\Gamma;{\runas {(\force {\ul{C}} V)} {y \!:\! U\ul{C}} {\ul{D}} {M}}}_2)_\gamma
=
(\sem{\Gamma;M[V/y]}_2)_\gamma
: 1 \longrightarrow U_{\!\mathcal{L}_{\mathcal{T}_{\text{eff}}}}(\sem{\Gamma;\ul{D}}_2(\gamma))
\end{array}
\]

First, using the premises of the given $\beta$-equation, we can construct a derivation of
\[
\cj {\Gamma} {\runas {(\force {\ul{C}} V)} {y \!:\! U\ul{C}} {\ul{D}} {M}} {\ul{D}}
\]
which means that $\sem{-}$ is defined on the left-hand side of the required equation, by using $(e)$ on this derivation. By unfolding the definition of $\sem{-}$ for this term, we get
\[
\sem{\Gamma;{\runas {(\force {\ul{C}} V)} {y \!:\! U\ul{C}} {\ul{D}} {M}}}_1
=
\id_{\sem{\Gamma}}
\]
and, for all $\gamma$ in $\sem{\Gamma}$, that
\[
\begin{array}{c}
(\sem{\Gamma;{\runas {(\force {\ul{C}} V)} {y \!:\! U\ul{C}} {\ul{D}} {M}}}_2)_\gamma 
\\
= 
\\[-0.5mm]
U_{\!\mathcal{L}_{\mathcal{T}_{\text{eff}}}}(\mathsf{hom}(f^\gamma)) \comp (\sem{\Gamma;\force {\ul{C}} V}_2)_\gamma
\end{array}
\]
as morphisms $\sem{\Gamma} \longrightarrow \sem{\Gamma}$ and $1 \longrightarrow U_{\!\mathcal{L}_{\mathcal{T}_{\text{eff}}}}(\sem{\Gamma;\ul{D}}_2(\gamma))$, respectively, and where
\[
f^{\gamma} \defeq c \mapsto (\sem{\Gamma, y \!:\! U\ul{C}; M}_2)_{\langle \gamma , c \rangle}(\star)
\]

Next, by unfolding the definition of $\sem{-}$ further (for the forcing of thunked computations) and recalling the definitions of $U_{\!\mathcal{L}_{\mathcal{T}_{\text{eff}}}}$ and $\mathsf{hom}(f^\gamma)$, we get
\[
\begin{array}{c}
U_{\!\mathcal{L}_{\mathcal{T}_{\text{eff}}}}(\mathsf{hom}(f^\gamma)) \comp (\sem{\Gamma;\force {\ul{C}} V}_2)_\gamma 
\\[-1mm]
= 
\\
U_{\!\mathcal{L}_{\mathcal{T}_{\text{eff}}}}(\mathsf{hom}(f^\gamma)) \comp (\sem{\Gamma;V}_2)_\gamma
\\[-1mm]
= 
\\
(\mathsf{hom}(f^\gamma))_1 \comp (\sem{\Gamma;V}_2)_\gamma
\\
=
\\
f^\gamma \comp (\sem{\Gamma;V}_2)_\gamma
\end{array}
\]

Now, by combining these last equations, we get
\[
\begin{array}{c}
(\sem{\Gamma;{\runas {(\force {\ul{C}} V)} {y \!:\! U\ul{C}} {\ul{D}} {M}}}_2)_\gamma (\star)
\\
=
\\
f^\gamma((\sem{\Gamma;V}_2)_\gamma(\star))
\\
=
\\[-1mm]
(\sem{\Gamma, y \!:\! U\ul{C}; M}_2)_{\langle \gamma , (\sem{\Gamma;V}_2)_\gamma(\star) \rangle}(\star)
\end{array}
\]

Next, we use $(d)$ on the derivation of $\vj \Gamma V A$ to get
\[
\begin{array}{c}
\sem{\Gamma;V}_1 = \id_{\sem{\Gamma}} : \sem{\Gamma} \longrightarrow \sem{\Gamma}
\qquad
(\sem{\Gamma;V}_2)_\gamma : 1 \longrightarrow U_{\!\mathcal{L}_{\mathcal{T}_{\text{eff}}}}(\sem{\Gamma;\ul{C}}_2(\gamma))
\end{array}
\]

Next, we can use the eMLTT$_{\!\mathcal{T}_{\text{eff}}}^{\mathcal{H}}$ version of Proposition~\ref{prop:semsubstitution2} to get
\[
\begin{array}{c}
\sem{\Gamma;M[V/y]}_1 = \id_{\sem{\Gamma}} : \sem{\Gamma} \longrightarrow \sem{\Gamma}
\\[3mm]
(\sem{\Gamma;M[V/y]}_2)_\gamma = (\sem{\Gamma, y \!:\! U\ul{C}; M}_2)_{\langle \gamma , (\sem{\Gamma;V}_2)_\gamma (\star) \rangle} : 1 \longrightarrow U_{\!\mathcal{L}_{\mathcal{T}_{\text{eff}}}}(\sem{\Gamma;\ul{D}}_2(\gamma))
\end{array}
\]

Finally, by combining these last two equations with the corresponding two equations we derived by unfolding the definition of $\sem{-}$ earlier, we have, as required, that
\[
\begin{array}{c}
\sem{\Gamma;{\runas {(\force {\ul{C}} V)} {y \!:\! U\ul{C}} {\ul{D}} {M}}}_1 
= 
\sem{\Gamma;M[V/y]}_1 
= 
\id_{\sem{\Gamma}}
: \sem{\Gamma} \longrightarrow \sem{\Gamma}
\end{array}
\]
and, for all $\gamma$ in $\sem{\Gamma}$, that
\[
\begin{array}{c}
(\sem{\Gamma;{\runas {(\force {\ul{C}} V)} {y \!:\! U\ul{C}} {\ul{D}} {M}}}_2)_\gamma
=
(\sem{\Gamma;M[V/y]}_2)_\gamma
: 1 \longrightarrow U_{\!\mathcal{L}_{\mathcal{T}_{\text{eff}}}}(\sem{\Gamma;\ul{D}}_2(\gamma))
\end{array}
\]

\vspace{0.2cm}

\noindent
\textbf{$\eta$-equation for the composition operation for computation terms:}
In this case, the given derivation ends with
\[
\mkrule
{
\ceq {\Gamma} {\runas {M} {y \!:\! U\ul{C}} {\ul{D}} {K[\force {\ul{C}} y/z]}} {K[M/z]} {\ul{D}}
}
{
\cj \Gamma M \ul{C} 
\quad
\hj {\Gamma} {z \!:\! \ul{C}} K \ul{D}
}
\]
and we need to show
\[
\sem{\Gamma;{\runas {M} {y \!:\! U\ul{C}} {\ul{D}} {K[\force {\ul{C}} y/z]}}} = \sem{\Gamma;K[M/z]}
: 1_{\sem{\Gamma}} \longrightarrow \widehat{U_{\!\mathcal{L}_{\mathcal{T}_{\text{eff}}}}}(\sem{\Gamma;\ul{D}})
\]
which, for the fibred adjunction model we are working with, is equivalent to showing
\[
\begin{array}{c}
\sem{\Gamma;{\runas {M} {y \!:\! U\ul{C}} {\ul{D}} {K[\force {\ul{C}} y/z]}}}_1
=
\sem{\Gamma;K[M/z]}_1
=
\id_{\sem{\Gamma}}
: \sem{\Gamma} \longrightarrow \sem{\Gamma}
\end{array}
\]
and, for all $\gamma$ in $\sem{\Gamma}$, that
\[
\hspace{-0.1cm}
\begin{array}{c}
(\sem{\Gamma;{\runas {M} {y \!:\! U\ul{C}} {\ul{D}} {K[\force {\ul{C}} y/z]}}}_2)_\gamma
=
(\sem{\Gamma;K[M/z]}_2)_\gamma
: 1 \longrightarrow U_{\!\mathcal{L}_{\mathcal{T}_{\text{eff}}}}(\sem{\Gamma;\ul{D}}_2(\gamma))
\end{array}
\]

First, using the premises of the given $\eta$-equation, we can construct a derivation of
\[
\cj {\Gamma} {\runas {M} {y \!:\! U\ul{C}} {\ul{D}} {K[\force {\ul{C}} y/z]}} {\ul{D}}
\]
which means that $\sem{-}$ is defined on the left-hand side of the required equation, by using $(e)$ on this derivation. 
By unfolding the definition of $\sem{-}$ for this term, we get
\[
\sem{\Gamma;{\runas {M} {y \!:\! U\ul{C}} {\ul{D}} {K[\force {\ul{C}} y/z]}}}_1 = \id_{\sem{\Gamma}}
\]
and, for all $\gamma$ in $\sem{\Gamma}$, that
\[
\begin{array}{c}
(\sem{\Gamma;{\runas {M} {y \!:\! U\ul{C}} {\ul{D}} {K[\force {\ul{C}} y/z]}}}_2)_\gamma
\\
=
\\
U_{\!\mathcal{L}_{\mathcal{T}_{\text{eff}}}}(\mathsf{hom}(f^\gamma)) \comp (\sem{\Gamma;M}_2)_\gamma
\end{array}
\]
as morphisms $\sem{\Gamma} \longrightarrow \sem{\Gamma}$ and $1 \longrightarrow U_{\!\mathcal{L}_{\mathcal{T}_{\text{eff}}}}(\sem{\Gamma;\ul{C}}_2(\gamma))$, respectively, and where
\[
f^{\gamma} \defeq c \mapsto (\sem{\Gamma, y \!:\! U\ul{C}; K[\force {\ul{C}} y/z]}_2)_{\langle \gamma , c \rangle}(\star)
\]

Next, by recalling the definitions of $U_{\!\mathcal{L}_{\mathcal{T}_{\text{eff}}}}$ and $\mathsf{hom}(f^\gamma)$, we get
\[
\begin{array}{c}
U_{\!\mathcal{L}_{\mathcal{T}_{\text{eff}}}}(\mathsf{hom}(f^\gamma)) \comp (\sem{\Gamma;M}_2)_\gamma
\\
=
\\
(\mathsf{hom}(f^\gamma))_1 \comp (\sem{\Gamma;M}_2)_\gamma
\\
=
\\
f^\gamma \comp (\sem{\Gamma;M}_2)_\gamma
\end{array}
\]

Now, by combining these last equations with the eMLTT$_{\!\mathcal{T}_{\text{eff}}}^{\mathcal{H}}$ version of Propositon~\ref{prop:semweakening2} that relates weakening to reindexing along semantic projection morphisms, and with the eMLTT$_{\!\mathcal{T}_{\text{eff}}}^{\mathcal{H}}$ version of Proposition~\ref{prop:semsubstitution3} that relates substitution of computation terms for computation variables to composition of morphisms, we get
\[
\begin{array}{c}
(\sem{\Gamma;{\runas {M} {y \!:\! U\ul{C}} {\ul{D}} {K[\force {\ul{C}} y/z]}}}_2)_\gamma(\star)
\\
=
\\
f^\gamma((\sem{\Gamma;M}_2)_\gamma(\star))
\\
=
\\[-2mm]
(\sem{\Gamma, y \!:\! U\ul{C}; K[\force {\ul{C}} y/z]}_2)_{\langle \gamma , (\sem{\Gamma;M}_2)_\gamma(\star) \rangle}(\star)
\\[2mm]
=
\\[-1mm]
(U_{\!\mathcal{L}_{\mathcal{T}_{\text{eff}}}}(\sem{\Gamma;z \!:\! \ul{C};K}_2)_{\gamma})((\sem{\Gamma;M}_2)_\gamma(\star))
\end{array}
\]

Next, we use $(e)$ on the assumed derivation of $\cj \Gamma M \ul{C}$ and $(f)$ on the assumed derivation of  $\hj {\Gamma} {z \!:\! \ul{C}} K \ul{D}$ to get
\[
\begin{array}{c}
\sem{\Gamma;M}_1 = \id_{\sem{\Gamma}} : \sem{\Gamma} \longrightarrow \sem{\Gamma}
\qquad
(\sem{\Gamma;M}_2)_\gamma : 1 \longrightarrow U_{\!\mathcal{L}_{\mathcal{T}_{\text{eff}}}}(\sem{\Gamma;\ul{C}}_2(\gamma))
\\[3mm]
\sem{\Gamma;z \!:\! \ul{C};K}_1 = \id_{\sem{\Gamma}} : \sem{\Gamma} \longrightarrow \sem{\Gamma}
\qquad
(\sem{\Gamma;z \!:\! \ul{C};K}_2)_\gamma : \sem{\Gamma;\ul{C}}_2(\gamma) \longrightarrow \sem{\Gamma;\ul{D}}_2(\gamma)
\end{array}
\]
from which we get
\[
\begin{array}{c}
\sem{\Gamma, y \!:\! U\ul{C};z \!:\! \ul{C};K}_1 = \id_{\bigsqcup_{\gamma \in \sem{\Gamma}} (U_{\!\mathcal{L}_{\mathcal{T}_{\text{eff}}}}(\sem{\Gamma;\ul{C}}_2(\gamma)))} : \sem{\Gamma, y \!:\! U\ul{C}} \longrightarrow \sem{\Gamma, y \!:\! U\ul{C}}
\\[3mm]
(\sem{\Gamma, y \!:\! U\ul{C}; z \!:\! \ul{C};K}_2)_{\langle \gamma, c \rangle} : \sem{\Gamma;\ul{C}}_2(\gamma) \longrightarrow \sem{\Gamma;\ul{D}}_2(\gamma)
\end{array}
\]
using the eMLTT$_{\!\mathcal{T}_{\text{eff}}}^{\mathcal{H}}$ version of Propositon~\ref{prop:semweakening2} that relates weakening to reindexing along semantic projection morphisms.

Next, by using the eMLTT$_{\!\mathcal{T}_{\text{eff}}}^{\mathcal{H}}$ version of Proposition~\ref{prop:semsubstitution3}, we get
\[
\begin{array}{c}
\sem{\Gamma;K[M/z]}_1 = \id_{\sem{\Gamma}} : \sem{\Gamma} \longrightarrow \sem{\Gamma}
\\[3mm]
(\sem{\Gamma;K[M/z]}_2)_\gamma = U_{\!\mathcal{L}_{\mathcal{T}_{\text{eff}}}}((\sem{\Gamma;z \!:\! \ul{C};K}_2)_\gamma) \comp (\sem{\Gamma;M}_2)_\gamma
: 1 \longrightarrow U_{\!\mathcal{L}_{\mathcal{T}_{\text{eff}}}}(\sem{\Gamma;\ul{C}}_2(\gamma))
\end{array}
\]

Finally, by combining these last two equations with the corresponding two equations we derived by unfolding the definition of $\sem{-}$ earlier, we have, as required, that
\[
\begin{array}{c}
\sem{\Gamma;{\runas {M} {y \!:\! U\ul{C}} {\ul{D}} {K[\force {\ul{C}} y/z]}}}_1
=
\sem{\Gamma;K[M/z]}_1
=
\id_{\sem{\Gamma}}
: \sem{\Gamma} \longrightarrow \sem{\Gamma}
\end{array}
\]
and, for all $\gamma$ in $\sem{\Gamma}$, that
\[
\hspace{-0.1cm}
\begin{array}{c}
(\sem{\Gamma;{\runas {M} {y \!:\! U\ul{C}} {\ul{D}} {K[\force {\ul{C}} y/z]}}}_2)_\gamma
=
(\sem{\Gamma;K[M/z]}_2)_\gamma
: 1 \longrightarrow U_{\!\mathcal{L}_{\mathcal{T}_{\text{eff}}}}(\sem{\Gamma;\ul{D}}_2(\gamma))
\end{array}
\]

\vspace{0.2cm}

\noindent
\textbf{$\eta$-equation for algebraic operations at the user-defined algebra type:}
In this case, the given derivation ends with
\[
\mkrule
{
\begin{array}{r@{\,\,} l}
\ceq \Gamma {& \algop^{\langle A , \{V_{\sigalgop}\}_{\sigalgop \in \mathcal{S}_{\text{eff}}} \rangle}_V(y.\, M) \\[-0.5mm]} { & \force {\langle A , \{V_{\sigalgop}\}_{\sigalgop \in \mathcal{S}_{\text{eff}}} \rangle} (V_{\sigalgop}\, \langle V , \lambda\, y \!:\! O[V/x] .\, \thunk M \rangle)} {\langle A , \{V_{\sigalgop}\}_{\sigalgop \in \mathcal{S}_{\text{eff}}} \rangle}
\end{array}
}
{
\begin{array}{c}
\vj \Gamma V I 
\quad
\lj \Gamma \langle A , \{V_{\sigalgop}\}_{\sigalgop \in \mathcal{S}_{\text{eff}}} \rangle
\quad
\cj {\Gamma, y \!:\! O[V/x]} M {\langle A , \{V_{\sigalgop}\}_{\sigalgop \in \mathcal{S}_{\text{eff}}} \rangle}
\end{array}
}
\]
and we need to show
\[
\begin{array}{c}
\hspace{-0.5cm}
\sem{\Gamma;\algop^{\langle A , \{V_{\sigalgop}\}_{\sigalgop \in \mathcal{S}_{\text{eff}}} \rangle}_V(y.\, M)}
=
\sem{\Gamma;\force {\langle A , \{V_{\sigalgop}\}_{\sigalgop \in \mathcal{S}_{\text{eff}}} \rangle} (V_{\sigalgop}\, \langle V , \lambda\, y \!:\! O[V/x] .\, \thunk M \rangle)}
\\
\hspace{8.4cm}
: 1 \longrightarrow \widehat{U_{\!\mathcal{L}_{\mathcal{T}_{\text{eff}}}}}(\sem{\Gamma;\langle A , \{V_{\sigalgop}\}_{\sigalgop \in \mathcal{S}_{\text{eff}}} \rangle})
\end{array}
\]
which, for the fibred adjunction model we are working with, is equivalent to showing
\[
\begin{array}{c}
\sem{\Gamma;\algop^{\langle A , \{V_{\sigalgop}\}_{\sigalgop \in \mathcal{S}_{\text{eff}}} \rangle}_V(y.\, M)}_1
\\[-2mm]
=
\\
\sem{\Gamma;\force {\langle A , \{V_{\sigalgop}\}_{\sigalgop \in \mathcal{S}_{\text{eff}}} \rangle} (V_{\sigalgop}\, \langle V , \lambda\, y \!:\! O[V/x] .\, \thunk M \rangle)}_1
\\
=
\\
\id_{\sem{\Gamma}}
\end{array}
\]
and, for all $\gamma$ in $\sem{\Gamma}$, that
\[
\begin{array}{c}
(\sem{\Gamma;\algop^{\langle A , \{V_{\sigalgop}\}_{\sigalgop \in \mathcal{S}_{\text{eff}}} \rangle}_V(y.\, M)}_2)_\gamma
\\[-2mm]
=
\\
(\sem{\Gamma;\force {\langle A , \{V_{\sigalgop}\}_{\sigalgop \in \mathcal{S}_{\text{eff}}} \rangle} (V_{\sigalgop}\, \langle V , \lambda\, y \!:\! O[V/x] .\, \thunk M \rangle)}_2)_\gamma
\end{array}
\]
as morphisms $\sem{\Gamma} \longrightarrow \sem{\Gamma}$ and $1 \longrightarrow U_{\!\mathcal{L}_{\mathcal{T}_{\text{eff}}}}(\sem{\Gamma;\langle A , \{V_{\sigalgop}\}_{\sigalgop \in \mathcal{S}_{\text{eff}}} \rangle}_2(\gamma))$, respectively. 

First, using the eMLTT$_{\!\mathcal{T}_{\text{eff}}}^{\mathcal{H}}$ version of Proposition~\ref{prop:wellformedcomponentsofjudgements}, we get derivations of
\[
\begin{array}{c}
\cj \Gamma {\algop^{\langle A , \{V_{\sigalgop}\}_{\sigalgop \in \mathcal{S}_{\text{eff}}} \rangle}_V(y.\, M)}  {\langle A , \{V_{\sigalgop}\}_{\sigalgop \in \mathcal{S}_{\text{eff}}} \rangle}
\\[2mm]
\cj \Gamma {\force {\langle A , \{V_{\sigalgop}\}_{\sigalgop \in \mathcal{S}_{\text{eff}}} \rangle} (V_{\sigalgop}\, \langle V , \lambda\, y \!:\! O[V/x] .\, \thunk M \rangle)} {\langle A , \{V_{\sigalgop}\}_{\sigalgop \in \mathcal{S}_{\text{eff}}} \rangle}
\end{array}
\]
which means that $\sem{-}$ is defined on both sides of the required equation, by using $(e)$ on these two derivations. 

The equality of the first components of the two sides of the required equation \linebreak (to $\id_{\sem{\Gamma}}$) follows straigthforwardly by unfolding the definition of $\sem{-}$ on both sides.

We prove the equality of the second components of the two sides of the required equation, for all $\gamma$ in $\sem{\Gamma}$, by showing that the next diagram commutes. 

\mbox{}
\[
\hspace{-0.15cm}
\xymatrix@C=1em@R=7em@M=0.5em{
1
\ar@/_1pc/[d]_-{\langle \id_1 \rangle_{o \in \sem{\Gamma;O[V/x]}_2(\gamma)}}^-{\qquad\qquad\dcomment{\text{composition}}}
\ar[r]^-{(\sem{\Gamma;V_{\sigalgop}}_2)_\gamma}
\ar@/_3pc/[ddr]^>>>>>>>>>>>{\!\!\!\!\langle (\sem{\Gamma, y : O[V/x]; M}_2)_{\langle \gamma , o \rangle} \rangle_o}
& 
\bigsqcap_{\langle i , f \rangle} (\sem{\Gamma;A}_2(\gamma))
\ar@/^7pc/[ddddr]^<<<<<<{\!\!\mathsf{proj}_{\langle (\sem{\Gamma;V}_2)_\gamma(\star) , \langle (\sem{\Gamma, y : O[V/x]; \thunk M}_2)_\gamma(\star) \rangle_o \rangle}}_-{\dcomment{\text{def.}}\,\,\,\,\,}
\ar@/^4pc/[ddddr]_<<<<<<<<<<<<<<<<<<<<<<<<<<<{\mathsf{proj}_{\langle (\sem{\Gamma;V}_2)_\gamma(\star) , \langle (\sem{\Gamma, y : O[V/x]; M}_2)_\gamma(\star) \rangle_o \rangle}}
& 
\\
\bigsqcap_o 1
\ar[d]_-{\bigsqcap_o ((\sem{\Gamma, y : O[V/x]; M}_2)_{\langle \gamma , o \rangle})}
\ar[d]^>>>>{\,\,\dcomment{\text{u. prop. of c. pr.}}}
&
&
\\
\bigsqcap_o (U_{\!\mathcal{L}_{\mathcal{T}_{\text{eff}}}}(\mathcal{M}))
\ar[r]^-{=}
\ar@/_5pc/[dddr]_>>>>>>>>>>>>>>>>>>>>>>>>>>>>>>{\sigalgop^{\mathcal{M}}_{(\sem{\Gamma;V}_2)_\gamma(\star)}\!\!\!}
&
\bigsqcap_o(\mathcal{M}(1))
\ar@/_7pc/[d]_-{\cong}
\ar@/^8.5pc/[dd]_<<<<<<<<<<<<<<<<<<<{f \,\mapsto\, \mathsf{proj}_{\langle (\sem{\Gamma;V}_2)_\gamma(\star) , f \rangle}((\sem{\Gamma;V_{\sigalgop}}_2)_\gamma(\star))}
\\
&
\mathcal{M}(\vert \sem{\Gamma; O[V/x]}_2(\gamma) \vert)
\ar[d]_-{\mathcal{M}(\lj {\overrightarrow{x_o}\,\,} {\,\sigalgop_{(\sem{\Gamma;V}_2)_\gamma(\star)}(x_o)_o})}_<<<<<<{\dcomment{\text{def. of } \sigalgop^{\mathcal{M}}_{(\sem{\Gamma;V}_2)_\gamma(\star)}}\quad\,\,\,\,\,\,\,\,}^<<<<<<{\qquad\dcomment{\text{def. of } \mathcal{M}}}
\\
&
\mathcal{M}(1)
\ar[d]_-{=}
&
\sem{\Gamma;A}_2(\gamma)
\ar@/^1.5pc/[dl]^-{=}
\\
&
U_{\!\mathcal{L}_{\mathcal{T}_{\text{eff}}}}\!(\mathcal{M})
&
}
\]
where, for better readability,  we write $\mathcal{M}$ for both $(\sem{\Gamma;\langle A , \{V_{\sigalgop}\}_{\sigalgop \in \mathcal{S}_{\text{eff}}} \rangle}_2(\gamma))$ and $\mathcal{M}_{\langle \sem{\Gamma;A}_2(\gamma) , \{f^{\gamma}_{\sigalgop_i}\}_{\sigalgop_i \in \mathcal{S}_{\text{eff}}} \rangle}$. Recall that these two models of $\mathcal{L}_{\mathcal{T}_{\text{eff}}}$ are equal by definition.

Finally, when we unfold the definition of $\sem{-}$, we see that the two composite 
top-to-bottom morphisms along the outer perimeter of the above diagram are respectively 
equal to 
\vspace{0.15cm}
\[
(\sem{\Gamma;\algop^{\langle A , \{V_{\sigalgop}\}_{\sigalgop \in \mathcal{S}_{\text{eff}}} \rangle}_V(y.\, M)}_2)_\gamma
\]
and
\[
(\sem{\Gamma;\force {\langle A , \{V_{\sigalgop}\}_{\sigalgop \in \mathcal{S}_{\text{eff}}} \rangle} (V_{\sigalgop}\, \langle V , \lambda\, y \!:\! O[V/x] .\, \thunk M \rangle)}_2)_\gamma
\vspace{0.25cm}
\]

As a result, we have, as required, that
\[
\begin{array}{c}
\hspace{-0.5cm}
\sem{\Gamma;\algop^{\langle A , \{V_{\sigalgop}\}_{\sigalgop \in \mathcal{S}_{\text{eff}}} \rangle}_V(y.\, M)}
=
\sem{\Gamma;\force {\langle A , \{V_{\sigalgop}\}_{\sigalgop \in \mathcal{S}_{\text{eff}}} \rangle} (V_{\sigalgop}\, \langle V , \lambda\, y \!:\! O[V/x] .\, \thunk M \rangle)}
\\
\hspace{8.4cm}
: 1 \longrightarrow \widehat{U_{\!\mathcal{L}_{\mathcal{T}_{\text{eff}}}}}(\sem{\Gamma;\langle A , \{V_{\sigalgop}\}_{\sigalgop \in \mathcal{S}_{\text{eff}}} \rangle})
\end{array}
\]


\chapter{Conclusion and future work}
\label{chap:conclusions}

In this thesis we have developed and studied the foundations for combining dependent types and computational effects, two important areas of modern programming language research. In the Introduction, we set out to establish the following claim:
\begin{displayquote}
\vspace{0.15cm}
\textit{Dependent types and computational effects admit a natural combination.}
\end{displayquote}
In retrospect, we can confirm that this is indeed the case. Specifically, we have provided language-based, category-theoretic, and algebraic evidence to support this claim.

\paragraph*{Language-based evidence.}

We have demonstrated that dependent types and computational effects can be naturally combined in a single programming language.
We achieved this by developing a core \emph{effectful dependently typed language}, called eMLTT, that extends intensional MLTT with general computational effects, based on a clear separation between values and computations.
Using eMLTT, we demonstrated that---with minor changes to the typing rules of effectful computations---one can readily use familiar combinators from simply typed languages to program with computational effects in the dependently typed setting, e.g., using sequential composition.
To overcome the limitations caused by these changes to the typing rules of effectful computations,
we introduced eMLTT's distinguishing feature, the \emph{computational $\Sigma$-type}, 
which allows us to uniformly ``close-off" free variables in computation types. 

\paragraph*{Category-theoretic evidence.}

We have also demonstrated that dependent types and computational effects can be naturally combined category-theoretically. To this end, we defined and studied a class of category-theoretic models, called \emph{fibred adjunction models}, suitable for defining a sound and complete interpretation of eMLTT. Specifically, fibred adjunction models naturally combine standard category-theoretic models of dependent types (split closed comprehension categories) and the corresponding generalisation of adjunction-based models of computational effects (split fibred adjunctions).
The naturality of this combination was demonstrated by being able to reuse and generalise various established results about monads and adjunctions, such as the existence of the Eilenberg-Moore resolution, and by showing that the computational $\Sigma$- and $\Pi$-types can be modelled analogously to their value counterparts, namely, as adjoints to weakening functors.
We further presented various examples of fibred adjunction models, ranging from i) those built from models of EEC, to ii) those based on the families of sets fibration, to iii) those built around the fibred Eilenberg-Moore resolutions of split fibred monads, to iv) those based on the fibration of continuous families of $\omega$-complete partial orders. The latter enabled us to extend eMLTT with recursion.

\paragraph*{Algebraic evidence.}

We also investigated the algebraic treatment of computational effects in the presence of dependent types. Specifically, we showed how to extend eMLTT with \emph{fibred algebraic effects} and their \emph{handlers}. To specify such effects, we introduced a dependently typed generalisation of Plotkin and Pretnar's effect theories, whose dependently typed operation symbols enable us to capture precise notions of computation such as state with location-dependent store types and dependently typed update monads. For handlers, we observed that their conventional term-level definition leads to unsound program equivalences becoming derivable in languages that include a notion of homomorphism, such as eMLTT. To solve this problem, we provided a novel type-based treatment of handlers via a new computation type, the \emph{user-defined algebra type}, which pairs a value type (the carrier) with a family of value terms (the operations). This type internalises Plotkin and Pretnar's insight that handlers denote algebras for a given equational theory of computational effects. We demonstrated the generality of this type-based treatment by showing that the conventional presentation of handlers can be routinely derived from it, and that this treatment provides a useful mechanism for reasoning about effectful computations. We also showed that eMLTT with fibred algebraic effects and their handlers can be soundly interpreted in a fibred adjunction model based on the families of sets fibration and models of countable Lawvere theories.

\vspace{0.3cm}

\noindent
In conclusion, the contributions of this thesis can be summed up as follows:
\begin{itemize}
\item one can readily take well-known and established methods for, and results about, programming with computational effects in simply typed languages and successfully adapt them to the dependently typed setting; and 
\item the presence of dependent types, in combination with basing our work on adjunctions rather than monads, provides an opportunity to discover new and interesting language features, and corresponding mathematical structures.
\end{itemize}

\section{Future work directions}
\label{sect:futurework}

There are many directions in which one can take this work forward. 
We discuss some of them in detail, including work on the foundations, improvements to the expressive power of  
eMLTT and its extensions, and developing a (prototype) implementation. 

\subsection{Fibred notions of Lawvere theory}
\label{sect:fiblawveretheories}

In future, we plan to study the denotational semantics of eMLTT$_{\mathcal{T}_{\text{eff}}}$ and eMLTT$_{\mathcal{T}_{\text{eff}}}^{\mathcal{H}}$ at the same level of generality as we did for eMLTT in Chapters~\ref{chap:fibadjmodels} and~\ref{chap:interpretation}. In particular, we plan to extend the denotational semantics of eMLTT$_{\mathcal{T}_{\text{eff}}}$ and eMLTT$_{\mathcal{T}_{\text{eff}}}^{\mathcal{H}}$ from the families of sets fibration to more general fibational models of dependent types. Towards this end, we plan to develop a fibred notion of (countable) Lawvere theory, together with a framework for defining corresponding equational presentations. In particular, we conjecture that our fibred effect theories can be used as a basis for such presentations, by extending them to proper equational theories, i.e., closing the set of equations under reflexivity, symmetry, transitivity, substitution and replacement, and developing the corresponding proof theory. We then plan to study (fibred) local presentability conditions on split closed comprehension categories under which a split fibred free model adjunction exists. This adjunction can then be used as a basis for constructing a fibred adjunction model suitable for defining the interpretations of eMLTT$_{\mathcal{T}_{\text{eff}}}$ and eMLTT$_{\mathcal{T}_{\text{eff}}}^{\mathcal{H}}$.

A related future work direction involves extending eMLTT with local effects, e.g., local names and local state. One possible way forward to account for such computational effects would be to first develop a fibred notion of indexed Lawvere theory~\cite{Power:IndexedLawvereTheories} by working with suitable fibrations of presheaves indexed by names, locations, etc., and then extend eMLTT accordingly. Another way forward could involve developing a fibred version of Staton's parameterised algebraic theories and  their model theory~\cite{Staton:Instances}.

Finally, we also plan to give a general treatment of inequationally presentable computational effects, such as divergence, so as to provide a more general treatment of recursion than our use of the fibration $\mathsf{cfam}_{\CPO} : \CFam(\CPO) \longrightarrow \CPO$ of continuous families of $\omega$-complete partial orders in Section~\ref{sect:continuousfamilies}. Towards this end, we plan to develop a fibred notion of discrete countable enriched Lawvere theory~\cite{Hyland:DiscreteLawTh}, together with a framework for defining inequational presentations corresponding to enrichment in $\omega$-complete partial orders. An important question here involves the exact notion of enrichment one would use for defining such fibred enriched Lawvere theories. As discussed in Section~\ref{sect:shallowenrichment}, there are multiple candidates one could consider using, including those developed in~\cite{Shulman:EnrichedIndexedCategories} and~\cite[Section~8.1]{Vasilakopoulou:Thesis}, and the notion of pre-enrichment we use to model eMLTT's homomorphic function type in Section~\ref{sect:shallowenrichment}.

\subsection{Extending eMLTT with more expressive computation types}
\label{sect:fibredparametrisedeffects}

As it stands, the computation types of eMLTT and its extensions cannot be used to encode detailed specifications about effectful computations except for very basic descriptions of their general shape, e.g., whether a computation is an effectful function.  
To overcome this limitation, we plan to extend eMLTT with dependently typed variants of type-and-effect systems based on, e.g., Katsumata et al.'s graded monads and graded adjunctions~\cite{Katsumata:EffectMonads,Fujii:GradedMonads}, and Atkey's parameterised monads and parameterised adjunctions~\cite{Atkey:ParametrizedNotions,Atkey:Algebras}. 
Specifically, we plan to generalise from grading and parameterising adjunctions by categories to grading and parameterising fibred adjunctions by suitable fibrations, e.g., by a fibred monoidal fibration in the case of graded adjunctions. From a programming language perspective, this means that the gradings and parameters would become first-class citizens, given by value terms of some specified pure value type of ``worlds" of computation, e.g., describing whether a file is open or closed. 

In the case of a type-and-effect system based on parameterised adjunctions, we plan to generalise from working with $\mathcal{W}$-parameterised adjunctions, given by functors
\[
F : \mathcal{W} \times \mathcal{V} \longrightarrow \mathcal{C}
\qquad
U : \mathcal{W}^{\text{op}} \times \mathcal{C} \longrightarrow \mathcal{V}
\]
to working with \emph{split $r$-parameterised fibred adjunctions}, for some given split fibration $r : \mathcal{W} \longrightarrow \mathcal{B}$. We define these to be given by a pair of fibred functors
\[
\xymatrix@C=1.5em@R=2.5em@M=0.5em{
\bigintsss (X \mapsto \mathcal{W}_X \times \mathcal{V}_X) \ar[rrr]^-{F} \ar[dr] &&& \mathcal{C} \ar[dll]^-{q}
\\
& \mathcal{B} &
}
\]
\[
\xymatrix@C=1.5em@R=2.5em@M=0.5em{
\bigintsss (X \mapsto \mathcal{W}^{\text{op}}_X \times \mathcal{C}_X) \ar[rrr]^-{U} \ar[dr] &&& \mathcal{V} \ar[dll]^-{p}
\\
& \mathcal{B} &
}
\vspace{0.2cm}
\]
where $p$ and $q$ are split fibrations used to model value and computation types, respectively, and the domains of these functors are derived from $p$, $q$, and $r$ using the Grothendieck construction. Specifically, based on the discussion in Section~\ref{sect:shallowenrichment}, the domains of these two functors are the product split fibrations $r \times p$ and $r^{\text{op}} \times q$, respectively, thus demonstrating that we indeed have defined a natural fibred generalisation of Atkey's parameterised adjunctions. Of course, we would also generalise the parameterised unit $\eta$ and counit $\varepsilon$ transformations to the fibrational setting.
In addition, there exists an analogous definition of a split $r$-parameterised fibred monad, naturally generalising the notion of a $\mathcal{W}$-parameterised monad $T : \mathcal{W}^{\text{op}} \times \mathcal{W} \times \mathcal{V} \longrightarrow \mathcal{V}$.

Regarding the corresponding extension of eMLTT, such split $r$-parameterised fibred adjoints would give rise to corresponding eMLTT types, namely, $F_W\, A$ and $U_W\, \ul{C}$, where $W$ is a value term of some specified closed pure value type $\mathsf{World}$. For example, to model file-based I/O, $\mathsf{World}$ could be an inductive type with two constructors, called $\mathtt{open}$ and $\mathtt{closed}$. Intuitively, $F_W\,  A$ would be the type of computations that return values of type $A$ and finish evaluating in the world denoted by $W$; and $U_W\, \ul{C}$ would be the type of thunks that can only be forced in a world denoted by $W$. In addition, we plan to develop a fibred generalisation of Atkey's $\mathcal{W}$-parameterised algebraic theories, and investigate the corresponding extensions of eMLTT$_{\mathcal{T}_{\text{eff}}}$ and eMLTT$_{\mathcal{T}_{\text{eff}}}^{\mathcal{H}}$.

As discussed in Section~\ref{sect:relatedwork}, Brady has previously used the corresponding split fibred  parameterised monads $T_{W_1,W_2}\, A$ to extend Idris with computational effects.
As also mentioned in op.~cit., Brady has more recently proposed extending split fibred parameterised monads with additional type-dependency, so as to enable the postcondition world $W_2$ to depend on the return values of the given computation. In more detail, this extension can be illustrated with the following type formation rule:
\vspace{0.25cm}
\[
\mkrule
{\lj \Gamma {T_{W_1,\, x.W_2}\, A}}
{
\vj \Gamma {W_1} {\mathsf{World}}
\quad
\lj \Gamma A
\quad
\vj {\Gamma, x \!:\! A} {W_2} {\mathsf{World}}
}
\]
This additional type-dependency enabled Brady to accommodate generic effects whose postcondition world crucially depends on the outcome of the effect. 
A prototypical example of such an effect is the possibly erroneous file opening operation, typed as
\[
{\cj \Gamma {\mathtt{open\text{-}file}} {T_{\mathtt{closed},x.\,\mathtt{case~} x \mathtt{~of} \mathtt{~} ({\inl {\!} {\!\!(x_1 : 1)} \,\mapsto\, \mathtt{open}}, {\inr {\!} {\!\!(x_2 : 1)} \,\mapsto\, \mathtt{closed}})}(1+1)}}
\]

However, as part of our preliminary work on extending eMLTT with fibred parameterised effects, we have noticed that there does not seem to be a category-theoretically natural notion of adjunction corresponding to $T_{W_1,\, x.W_2}\, A$. In particular, the beautiful symmetries involved in the definition of split $r$-parameterised fibred adjunctions are lost because the functor corresponding to $F_{x.\, W} A$ would be   ``dependently typed", while the functor corresponding to $U_W\, \ul{C}$ remains split fibred as before.
A similar loss of symmetry also affects the unit $\eta$ and the counit $\varepsilon$, where the components of the unit become ``dependently typed" morphisms\footnote{The components of the unit $\eta$ would correspond to terms $\cj {\Gamma,x\!:\!A} {\return x} {T_{W\![x/y],y.\, W}\, A}$, whose type (the codomain of the morphism) crucially depends on the variable $x$ (the domain of the morphism).}.
This has led us to conclude that $T_{W_1,\, x.W_2}\, A$ does not in fact denote some more dependently typed version of a split $r$-parameterised fibred monad. Instead, it can be shown that $T_{W_1,\, x.W_2}\, A$ corresponds to the composition of split $r$-parameterised fibred adjoints (as defined earlier in this section) with the dependent sum functor that models our computational $\Sigma$-type. In particular, in an extension of eMLTT based on a split $r$-parameterised adjunction, we can define  $T_{W_1,\, x.W_2}\, A$ as 
\[
T_{W_1,\, x.W_2}\, A \defeq U_{W_1} (\Sigma\, x \!:\! A .\, (F_{W_2} 1))
\]
and also derive the correspondingly typed combinators for returning values and sequential composition. This is further evidence that the clear distinction between values and computations, together with the computational $\Sigma$-type, have an important and fundamental role to play in combining dependent types and computational effects.

\subsection{Fibrational account of Dijkstra monads}
\label{sect:fibDijkstramonads}

In addition to type-and-effect systems based on graded and parameterised adjunctions, we plan to investigate extending eMLTT's type system with ideas based on how Dijkstra monads are used in F*. To this end, we first need to find an appropriate notion of adjunction corresponding to F*'s Dijkstra monads. As a starting point, we note that in the fibrational setting, Dijkstra monads can be understood as certain relative monads~\cite{Altenkirch:RelMon2}, with respect to the monad of weakest precondition predicate transformers.

Specifically, 
this relative monads based axiomatisation of a Dijkstra monad on a split comprehension category with unit $p : \mathcal{V} \longrightarrow \mathcal{B}$, indexed by a split fibred (weakest preconditions) Kleisli triple $(W\!P, \eta, (-)^\dagger)$ on $p$, involves giving the following data:
\begin{itemize}
\item a functor $T : \mathcal{V} \longrightarrow \mathcal{V}$ such that $T$ strictly preserves Cartesian morphisms and 
\[
\xymatrix@C=2em@R=2em@M=0.5em{
& \mathcal{V} \ar[dl]_-{W\!P} \ar[dr]^-{T}
\\
\mathcal{V} \ar[dr]_-{\ia -} & & \mathcal{V} \ar[dl]^-{p}
\\
& \mathcal{B}
}
\]
\item a unit $\eta^T_A : \ia A \longrightarrow \ia {T(A)}$ in $\mathcal{B}$, for every $A$ in $\mathcal{V}$, such that
\[
\xymatrix@C=7em@R=3em@M=0.5em{
\ia A \ar[d]_-{\id_{\ia A}} \ar[r]^-{\eta^T_A} & \ia {T(A)} \ar[d]^-{\pi_{T(A)}}
\\
\ia A \ar[r]_-{\ia {\eta_A}} & \ia {W\!P(A)}
}
\]
\item and for every commuting square of the form
\[
\xymatrix@C=7em@R=3em@M=0.5em{
\ia A \ar[d]_-{\id_{\ia A}} \ar[r]^-{f} & \ia {T(B)} \ar[d]^-{\pi_{T(B)}}
\\
\ia A \ar[r]_-{\ia {g}} & \ia {W\!P(B)}
}
\]
a Kleisli extension $f_T^\dagger : \ia {T(A)} \longrightarrow \ia {T(B)}$  in $\mathcal{B}$ such that
\[
\xymatrix@C=7em@R=3em@M=0.5em{
\ia {T(A)} \ar[d]_-{\pi_{T(A)}} \ar[r]^-{f_T^\dagger} & \ia {T(B)} \ar[d]^-{\pi_{T(B)}}
\\
\ia {W\!P(A)} \ar[r]_-{\ia {g^\dagger}} & \ia {W\!P(B)}
}
\]
\end{itemize}
such that the natural laws for the interaction of the unit and the Kleisli extension hold.

From a programming language perspective, the functors $W\!P$ and $T$, and their interaction in the  above diagram, can be described using two type formation rules:
\vspace{0.15cm}
\[
\mkrule
{\lj {\Gamma} {W\!P~A}}
{\lj \Gamma A}
\qquad
\mkrule
{\lj {\Gamma, x \!:\! W\!P~A} {T~A}}
{\lj \Gamma A \quad x \not\in V\!ars(\Gamma)}
\]
The unit $\eta^T_A$ and Kleisli extension $f_T^\dagger$ correspond to F*'s typing rules for returning values and sequential composition. For example, the unit $\eta^T_A : \ia {A} \longrightarrow \ia {T(A)}$ in $\mathcal{B}$ can be shown to correspond to a global element $1_{\ia {A}} \longrightarrow \ia {\eta_A}^*(T(A))$ in $\mathcal{V}_{\ia {A}}$, which in turn corresponds to (an idealised version of) F*'s typing rule for returning values:
\vspace{0.15cm}
\[
\mkrule
{\cj \Gamma {\return V} {T~A~[W\!P.\return V/x]}}
{\vj \Gamma V A}
\]

On closer inspection, the above data corresponds exactly to the definition of a relative monad from $\mathcal{V}$ to a certain subcategory of the arrow category $\mathcal{B}^\to$, relative to a functor that maps an object $A$ in $\mathcal{V}$ to the identity morphism $\id_{\ia A} : \ia A \longrightarrow \ia A$.

Based on these observations, we plan to investigate whether the corresponding relative adjunctions can be used to extend eMLTT with weakest precondition based reasoning about computational effects. In addition, we plan to explore an algebraic account of F*'s Dijkstra monads, so as to specify them using operations and equations.

\subsection{Allowing types to depend on effectful computations}
\label{sect:typedependencyonfeffects}

Recall the two key design choices we made when developing eMLTT. These were: \linebreak i) allowing types to depend only on values, and ii) fixing the typing rule for sequential composition by restricting the free variables in the type of the second computation. In future, we plan to extend eMLTT\footnote{In this section, we use eMLTT to jointly refer to eMLTT, eMLTT$_{\mathcal{T}_{\text{eff}}}$, and eMLTT$_{\mathcal{T}_{\text{eff}}}^{\mathcal{H}}$.} so as to lift both these restrictions. In particular, we plan to develop a version of eMLTT in which types could depend on effectful computations directly rather than via thunks. While type-dependency on computations is an intriguing question in itself, these future work plans are also motivated by the problems that arise in the recent work of V{\'{a}}k{\'{a}}r, as discussed in Section~\ref{sect:relatedwork}.

In particular, recall from op.~cit.~that V{\'{a}}k{\'{a}}r investigates a dependently typed version of CBPV built around a dependently typed version of sequential composition:
\[
\mkrule
{\cj {\Gamma_1, \Gamma_2[\thunk M/y]} {\doto {M} {x \!:\! A} {} {N}} {\ul{C}[\thunk M/y]}}
{
\begin{array}{c}
\cj {\Gamma_1} M {FA}
\quad
\lj {\Gamma_1, y \!:\! U\!FA, \Gamma_2} {\ul{C}}
\\[-0.5mm]
\cj {\Gamma_1, x \!:\! A, \Gamma_2[\thunk\! (\return x)/y]} {N} {\ul{C}[\thunk\! (\return x)/y]}
\end{array}
}
\]

While this typing rule solves the problem of simultaneously allowing the type of $N$ to depend on $x$ and restricting it from appearing free in the conclusion, and also enables V{\'{a}}k{\'{a}}r to define call-by-value and call-by-name translations from a dependently typed $\lambda$-calculus into his language, it introduces new problems in the presence of fibred algebraic effects, as discussed in Section~\ref{sect:relatedwork}.
We conjecture that the root cause of these problems is the thunks-based type-dependency on computations.

Consequently, we plan to investigate how to extend eMLTT with computation types that can depend  directly on effectful computations via computation variables. In order to avoid the problems arising from the thunks-based type-dependency in V{\'{a}}k{\'{a}}r's work, we anticipate that the computation variables must be treated in computation types similarly to the way they are currently used in homomorphism terms. Consequently, in addition to computation types $\ul{C}$, $\ul{D}$, $\ldots$ that are dependent on only values, we plan to include \emph{homomorphic} computation types $\ul{\ul{C}}$, $\ul{\ul{D}}$, $\ldots$ that further depend on computation variables, with well-formed such types defined using a judgement $\lj {\Gamma \vertbar z \!:\! \ul{C}} \ul{\ul{D}}$. We can then equip sequential composition with naturally dependent typing rules, given by
\vspace{0.15cm}
\[
\mkrule
{\cj \Gamma {\doto M {x \!:\! A} {} {N}} {\ul{\ul{C}}[M/z]}}
{\cj \Gamma M FA
\quad
\lj {\Gamma \vertbar z \!:\! FA} \ul{\ul{C}}
\quad
\cj {\Gamma, x \!:\! A} N \ul{\ul{C}}[\return x/z]
}
\]
\[
\mkrule
{\hj \Gamma {z_1 \!:\! \ul{C}} {\doto K {x \!:\! A} {} {N}} {\ul{\ul{D}}[K/z_2]}}
{\hj \Gamma {z_1 \!:\! \ul{C}} K FA
\quad
\lj {\Gamma \vertbar z_2 \!:\! FA} \ul{\ul{D}}
\quad
\cj {\Gamma, x \!:\! A} N \ul{\ul{D}}[\return x/z_2]
}
\]
We further speculate that other elimination rules for computation types, such as computational pattern matching, can be given analogous naturally dependent typing rules.

In order that computation variables can be used as they are used in homomorphism terms, we speculate 
that both kinds of computation types will need to include \emph{elimination forms} such as sequential composition, computational pattern matching, and the composition operations. In particular, we anticipate their grammar to be given by
\[
\begin{array}{r c l}
\ul{C} & ::= & \ldots \,\,\,\vertbar\,\,\, \doto M {x \!:\! A} {} {\ul{C}} \,\,\,\vertbar\,\,\, \doto M {(x \!:\! A, z \!:\! \ul{C})} {} \ul{\ul{D}} \,\,\,\vertbar\,\,\, \runas M {x \!:\! U\ul{C}} {} {\ul{D}}
\\[2mm]
\ul{\ul{C}} & ::= & \doto K {x \!:\! A} {} {\ul{C}\,} \,\,\,\vertbar\,\,\, \doto K {(x \!:\! A, z \!:\! \ul{C})} {} \ul{\ul{D}\,} \,\,\,\vertbar\,\,\, \runas K {x \!:\! U\ul{C}} {} {\ul{D}}
\end{array}
\vspace{0.1cm}
\]

While so far it might seem that this extension of eMLTT is going to be straightforward, we expect significant challenges regarding its equational theory and category-theoretic denotational semantics. 
Based on the author's joint work with Plotkin on refinement types for algebraic effects~\cite{Ahman:RefTypes}, we speculate that the natural choice for modelling such computation types is to use families of subsets (more generally, subobjects) of carriers of models of the given fibred effect theory. If we think of these carriers as sets of computation trees, a computation type $\doto {\mathtt{get}^{F\State}(y.\, \return y)} {x \!:\! \State} {} {\ul{C}}$ would denote a family of sets of computation trees each with the following shape:
\vspace{-0.1cm}
\[
\xymatrix@C=1.5em@R=2.5em@M=0.3em{
& & \mathsf{get} \ar@{-}[dll] \ar@{-}[d] \ar@{-}[drr]
\\
c_{s_1} & \ldots & c_{s_i} & \ldots & c_{s_n}
}
\]
where $c_{s_i}$ would be an element of the set of computation trees denoted by $\ul{C}[s_i/x]$. \linebreak
However, it is important to point out that one cannot naively lift all equations from the term level to the type level in this setting. In particular, for general $\lj \diamond \State$ and $\lj \Gamma \ul{C}$, this semantics of computation types would not validate the following definitional equation:
\[
\ljeq \Gamma {\ul{C}} {\doto {\mathtt{get}^{F\State}(y.\, \return y)} {x \!:\! \State} {} {\ul{C}}}
\]
Namely, compared to the corresponding definitional equation between computation terms, a general computation type $\ul{C}$ would denote a non-trivial family of sets of computation trees. As a result, the computation trees $c_{s_i}$ in the above diagram can all be different, meaning that the composite tree might not be in the family denoted by $\ul{C}$.

This is an instance of a general phenomenon that only linear equations can be lifted from a carrier of an algebra to the powerset of the carrier---see the work of Gautam~\cite{gautam:validity} for more details. 
However, it is worth noting that while the semantics in question would not validate the above equation, it would validate the following subtyping inequality:
\[
\Gamma \vdash {\ul{C}} \sqsubseteq {\doto {\mathtt{get}^{F\State}(y.\, \return y)} {x \!:\! \State} {} {\ul{C}}}
\]
This suggests that we probably have to extend eMLTT with a subtyping relation. For refinement types, a general schema for valid such subtyping rules can be found in~\cite{Ahman:RefTypes}.

\subsection{Normalisation and implementation}
\label{sect:normalisationandimplementation}

We also plan to develop a prototype implementation of eMLTT\footnote{In this section, we again use eMLTT to jointly refer to eMLTT, eMLTT$_{\mathcal{T}_{\text{eff}}}$, and eMLTT$_{\mathcal{T}_{\text{eff}}}^{\mathcal{H}}$.} and its extensions. 

As a first step towards implementing a prototype, we plan to develop a normalisation algorithm for the equational theory of eMLTT, using the well-known technique of normalisation-by-evaluation (NBE)~\cite{Dybjer:NBE}. More specifically, we plan to combine the existing work on NBE for dependent types~\cite{Altenkirch:NBEforTT} with the author's previous work on NBE for simply typed languages with algebraic effects~\cite{Ahman:NBE}. Regarding the normalisation algorithm, we could start by normalising the types and terms of eMLTT modulo the given fibred effect theory, and then specialise the normalisation algorithm to specific important computational effects, such as state, analogously to~\cite[Section~5.2]{Ahman:NBE}.

We expect that we would have to weaken the equational theory of eMLTT so as to ensure that its type- and term-equality are decidable. In particular, while part of the equational theory is already set up so as to avoid known sources of undecidability, e.g., we use intensional propositional equality and we omit the $\eta$-equation for primitive recursion\footnote{Already in the simply typed setting, the term-equality in G\"{o}del's System T becomes undecidable when one introduces  the $\eta$-equation (i.e., a uniqueness axiom) for primitive recursion~\cite{Okada:Rewriting}.}, other parts of the equational theory might pose further problems, e.g., the $\eta$-equation for the coproduct type~\cite{Balat:NBE}. Furthermore, the decidability of the equations corresponding to the composition operations is an altogether unknown territory. 

Regarding other sources of undecidability of typechecking, it is worthwhile to recall that checking the correctness of handlers of algebraic effects is an undecidable problem in general~\cite[\S6]{Plotkin:HandlingEffects}. Accordingly, the same would hold for verifying the well-formedness of the user-defined algebra type in eMLTT.
As discussed in Section~\ref{sect:derivingconventionalhandlers}, one way to tackle this problem would be  
to require  
programmers to manually prove equational proof obligations that cannot be established automatically. 
To enable this kind of interaction with programmers, we could replace definitional equations in proof obligations with propositional equalities, and annotate the user-defined algebra type and the composition operations with the corresponding proof terms. 

Note that by changing the proof obligations from definitional equations to propositional equalities, we raise interesting questions regarding the soundness of the denotational semantics of eMLTT. In particular, while definitional and propositional equality conveniently coincide in fibred adjunction models based on the families of sets fibration, the former is usually only included in the latter in more general models of (higher) dependent types, e.g., see~\cite{Bezem:Cubical}. 
Consequently, we have to be careful about which proof terms we allow to witness these proof obligations, so as to ensure that the interpretation of the user-defined algebra type and the composition operations remains sound.

Regarding the implementation, we also need to equip eMLTT 
with a suitable operational semantics. As a starting point, we plan to investigate an operational semantics  based on Lindley and Hillerstr\"{o}m's~\cite{Hillerstrom:Liberating} abstract machine based semantics for handlers of algebraic effects (in the simply typed setting). 
Regarding eMLTT, we expect that the most challenging problem will be accommodating the unfolding of algebraic operations at the user defined algebra type (see the equation in Definition~\ref{def:extensionofeMLTTwithhandlers}).

\appendix

\chapter{Dependently typed parsing example mentioned in Chapter~\ref{chap:introduction}}
\label{chap:appendixC1}

In this appendix we present details of the dependently typed monadic parsing example 
we alluded to in~\cite{Ahman:FibredEffects}, and that we also mentioned in the end of Section~\ref{sect:twoguidingquestions}.
As highlighted in the latter, then similarly to other computationally interesting  examples we are aware of, this example also only requires  
computation types $\Sigma\, x\!:\! A .\, \ul{C}$ where $\ul{C}$ is of the form $FB$, or equivalently, computation types of the form $F(\Sigma\, x \!:\! A .\, B)$.
As a result, as we only need computation types of the form $FB$ for this example, we can present it using a shallow embedding\footnote{The full code is available at 
\url{https://www.github.com/danelahman/Dep-Mon-Parsing/}.} of Moggi's monadic metalanguage in Agda~\cite{Norell:AgdaTutorial}, using the standard parser monad from~\cite{Hutton:MonParsing} (written $\mathtt{P}$ in the code below). By observing that $\mathtt{P}$ is nothing but the tensor product of the global state monad with the lists-based nondeterminism monad (see~\cite{Hyland:CombiningEffects}), we can give the parser combinators we use high-level definitions in terms of the algebraic operations that determine these two monads. 

To keep the example as simple as possible, we consider a very small simply typed language in this appendix, whose terms $t$ are given by the following grammar:
\[
\begin{array}{r c l}
t & ::= & c ~ \vertbar ~ f~t_1~\ldots~t_n
\end{array}
\]
where $c$ and $f$ range over typed constant and function symbols; for simplicity, $n > 0$.

The code for our parser is given below. For better readability, we parameterise it over tokens, types, and constant and function symbols; and conversion functions taking tokens to types, and constant and function symbols. We also assume that the given representation of types has decidable equality ($\mathtt{decTypeEq}$). If it succeeds, the parser will produce as its output a pair of a type $\mathtt{ty:Types}$ and a typed term $\mathtt{tm:Terms~ty}$.

\pagebreak

{\small{$\mathtt{module ~ Parser}$}}

\hspace{0.75cm}
{\small{$\mathtt{(Token : Set) ~ (Types : Set) ~ (ConstSym : Set) ~ (FunSym : Set)}$}}

\hspace{0.75cm}
{\small{$\mathtt{(tokenToType : Token} \to \mathtt{Types} + \mathtt{One)}$}}

\hspace{0.75cm}
{\small{$\mathtt{(decTypeEq : (ty1 ~ ty2 : Types)} \to \mathtt{(Id ~ ty1 ~ ty2)} + \mathtt{(Id ~ ty1 ~ ty2} \to \mathtt{Zero))}$}}

\hspace{0.75cm}
{\small{$\mathtt{(tokenToConstSym : Token} \to \mathtt{ConstSym} + \mathtt{One)}$}}

\hspace{0.75cm}
{\small{$\mathtt{(typeOfConst : ConstSym} \to \mathtt{Types)}$}}

\hspace{0.75cm}
{\small{$\mathtt{(tokenToFunSym : Token} \to \mathtt{FunSym} + \mathtt{One)}$}}

\hspace{0.75cm}
{\small{$\mathtt{(typeOfFun : FunSym} \to \mathtt{(NEList ~ Types)} \times \mathtt{Types) ~where}$}}

\vspace{0.25cm}

\hspace{0.25cm}
{\small{\textcolor{lightgray}{- - $\mathtt{lists ~ of ~ tokens}$}}}

\hspace{0.25cm}
{\small{$\mathtt{Tokens : Set}$}}

\hspace{0.25cm}
{\small{$\mathtt{Tokens = List ~ Token}$}}

\vspace{0.25cm}

\hspace{0.25cm}
{\small{\textcolor{lightgray}{- - $\mathtt{the ~ standard ~(fibred) ~ parser ~ monad ~ with ~ its ~ return ~ and ~ bind}$}}}

\hspace{0.25cm}
{\small{$\mathtt{P : Set} \to \mathtt{Set}$}}

\hspace{0.25cm}
{\small{$\mathtt{P ~ A = Tokens} \to \mathtt{List ~ (Tokens} \times \mathtt{A)}$}}

\vspace{0.25cm}

\hspace{0.25cm}
{\small{$\mathtt{Pf : \{A ~B : Set\}} \to \mathtt{(A} \to \mathtt{B)} \to \mathtt{P ~A} \to \mathtt{P ~B}$}}

\hspace{0.25cm}
{\small{$\mathtt{Pf ~ f ~ p ~ tok = map ~ (\char`\\ \,x \to ((fst ~x) , f~(snd ~x)))~ (p~ tok)}$}}

\vspace{0.25cm}

\hspace{0.25cm}
{\small{$\mathtt{return : \{A : Set\} \to A \to P~ A}$}}

\hspace{0.25cm}
{\small{$\mathtt{return ~a ~tok = listReturn ~(tok , a)}$}}

\vspace{0.25cm}

\hspace{0.25cm}
{\small{$\mathtt{bind : \{A : Set\} ~\{B : Set\} \to P ~A \to (A \to P ~B) \to P ~B}$}}

\hspace{0.25cm}
{\small{$\mathtt{bind ~p ~f ~tok = listBind ~(p ~tok) ~ (\char`\\ \,x \to f ~(snd ~x) ~(fst ~x))}$}}

\vspace{0.25cm}

\hspace{0.25cm}
{\small{\textcolor{lightgray}{- - $\mathtt{generic ~effects ~and ~algebraic ~operations ~for ~the ~parser ~monad}$}}}

\hspace{0.25cm}
{\small{$\mathtt{lkp : P ~Tokens}$}}

\hspace{0.25cm}
{\small{$\mathtt{lkp ~toks = (toks , toks) ~::~ []}$}}

\vspace{0.25cm}

\hspace{0.25cm}
{\small{$\mathtt{put : Tokens \to P ~ One}$}}

\hspace{0.25cm}
{\small{$\mathtt{put ~toks1 ~toks2 = (toks1 , *) ~::~ []}$}}

\vspace{0.25cm}

\hspace{0.25cm}
{\small{$\mathtt{or : \{A : Set\} \to P ~A \to P ~A \to P ~A}$}}

\hspace{0.25cm}
{\small{$\mathtt{or ~p1 ~p2 ~tok = append ~(p1 ~tok) ~(p2 ~tok)}$}}

\vspace{0.25cm}

\hspace{0.25cm}
{\small{$\mathtt{fail : \{A : Set\} \to P ~A}$}}

\hspace{0.25cm}
{\small{$\mathtt{fail ~tok = []}$}}

\newpage

\hspace{0.25cm}
{\small{\textcolor{lightgray}{- - $\mathtt{some ~useful ~combinators ~for ~parsing ~tokens}$}}}

\hspace{0.25cm}
{\small{$\mathtt{parseToken : P ~Token}$}}

\hspace{0.25cm}
{\small{$\mathtt{parseToken = bind ~ lkp ~ (\char`\\ \, \{ [] \to fail;}$}}

\vspace{-0.1cm}

\hspace{4.9cm}
{\small{$\mathtt{(tok ~::~ toks) \to bind ~(put ~toks)}$}}

\vspace{-0.1cm}

\hspace{8.65cm}
{\small{$\mathtt{ (\char`\\ \, \_ \to return ~tok) \})}$}}

\vspace{0.25cm}

\hspace{0.25cm}
{\small{$\mathtt{parseAndConvert : \{X : Set\} \to (Token \to X) \to P ~X}$}}

\hspace{0.25cm}
{\small{$\mathtt{parseAndConvert ~ f = bind ~parseToken ~(\char`\\ \,tok \to return ~(f ~tok))}$}}

\vspace{0.25cm}

\hspace{0.25cm}
{\small{$\mathtt{parseAndTest : \{X : Set\} \to (Token \to X + One) \to P ~X}$}}

\hspace{0.25cm}
{\small{$\mathtt{parseAndTest ~f = bind ~(parseAndConvert ~f)}$}}

\vspace{-0.1cm}

\hspace{4.35cm}
{\small{$\mathtt{(\char`\\ \, b \to +\text{-}elim ~b ~(\char`\\ \, x \to return ~x) ~(\char`\\ \, \_ \to fail))}$}}

\vspace{0.25cm}

\hspace{0.25cm}
{\small{\textcolor{lightgray}{- -$\mathtt{typed ~ASTs~of ~the ~terms ~of ~the ~small ~language}$}}}

\hspace{0.25cm}
{\small{$\mathtt{mutual}$}}

\hspace{0.5cm}
{\small{$\mathtt{data ~Terms : Types \to Set ~where}$}}

\hspace{0.75cm}
{\small{$\mathtt{const : (c : ConstSym) \to Terms ~(typeOfConst ~c)}$}}

\hspace{0.75cm}
{\small{$\mathtt{app : (f : FunSym) \to NEArgumentList ~(fst ~(typeOfFun ~f))}$}}

\vspace{-0.1cm}

\hspace{3.7cm}
{\small{$\mathtt{\to Terms ~(snd ~(typeOfFun ~f))}$}}

\vspace{0.25cm}

\hspace{0.5cm}
{\small{$\mathtt{data ~NEArgumentList : NEList ~Types \to Set ~where}$}}

\hspace{0.75cm}
{\small{$\mathtt{[\_]  : \{ty : Types\} \to Terms ~ty \to NEArgumentList ~[ ~ty~ ]}$}}

\hspace{0.75cm}
{\small{$\mathtt{\_::\!\_ : \{ty : Types\} ~\{tys : NEList ~Types\} \to Terms ~ty}$}}

\vspace{-0.1cm}

\hspace{7.5cm}
{\small{$\mathtt{\to NEArgumentList ~tys}$}}

\vspace{-0.1cm}

\hspace{7.5cm}
{\small{$\mathtt{\to  NEArgumentList ~(ty ~::~ tys)}$}}

\vspace{0.25cm}

\hspace{0.25cm}
{\small{\textcolor{lightgray}{- - $\mathtt{monadic ~parsing ~of ~typed ~ASTs}$}}}

\hspace{0.25cm}
{\small{$\mathtt{mutual}$}}

\vspace{0.25cm}

\hspace{0.5cm}
{\small{$\mathtt{\{\text{-}\# ~TERMINATING~ \#\text{-}\}}$}}

\vspace{0.25cm}

\hspace{0.5cm}
{\small{\textcolor{lightgray}{- - $\mathtt{the ~top\text{-}level ~parser ~for ~the ~language}$}}}

\hspace{0.5cm}
{\small{$\mathtt{parser : P ~(Sigma ~Types ~Terms)}$}}

\hspace{0.5cm}
{\small{$\mathtt{parser = or ~parseConst ~parseFunApp}$}}

\vspace{0.25cm}

\hspace{0.5cm}
{\small{\textcolor{lightgray}{- - $\mathtt{the ~sub\text{-}parser ~for ~constants}$}}}

\hspace{0.5cm}
{\small{$\mathtt{parseConst : P ~(Sigma ~Types ~Terms)}$}}

\hspace{0.5cm}
{\small{$\mathtt{parseConst = bind ~(parseAndTest ~tokenToConstSym)}$}}

\vspace{-0.1cm}

\hspace{3.9cm}
{\small{$\mathtt{(\char`\\ \, c \to return ~(typeOfConst ~c , const ~c))}$}}

\newpage

\hspace{0.5cm}
{\small{\textcolor{lightgray}{- - $\mathtt{the ~sub\text{-}parser ~for ~function ~applications}$}}}

\hspace{0.5cm}
{\small{$\mathtt{parseFunApp : P ~(Sigma ~Types ~Terms)}$}}

\hspace{0.5cm}
{\small{$\mathtt{parseFunApp =}$}}

\vspace{-0.1cm}

\hspace{0.75cm}
{\small{$\mathtt{bind ~(parseAndTest ~tokenToFunSym)}$}}

\vspace{-0.1cm}

\hspace{1.65cm}
{\small{$\mathtt{(\char`\\ \, f \to bind ~(parseNEArgumentList ~(fst ~(typeOfFun ~f)))}$}}

\vspace{-0.1cm}

\hspace{3.7cm}
{\small{$\mathtt{(\char`\\ \, args \to return ~(snd ~(typeOfFun ~f) , app ~f ~args)))}$}}

\vspace{0.25cm}

\hspace{0.5cm}
{\small{\textcolor{lightgray}{- - $\mathtt{parsing ~the ~non\text{-}empty ~lists ~of ~arguments ~in ~function ~applications}$}}}

\hspace{0.5cm}
{\small{$\mathtt{parseNEArgumentList : (tys : NEList ~Types) \to P ~(NEArgumentList ~tys)}$}}

\hspace{0.5cm}
{\small{$\mathtt{parseNEArgumentList ~[ ~ty~ ]      = bind ~(parseTermOfGivenType ~ty)}$}}

\vspace{-0.1cm}

\hspace{6.65cm}
{\small{$\mathtt{(\char`\\ \, tm \to return ~[ ~tm~ ])}$}}

\hspace{0.5cm}
{\small{$\mathtt{parseNEArgumentList ~(ty ~::~ tys) =}$}}

\vspace{-0.1cm}

\hspace{0.75cm}
{\small{$\mathtt{bind ~(parseTermOfGivenType ~ty)}$}}

\vspace{-0.1cm}

\hspace{1.65cm}
{\small{$\mathtt{(\char`\\ \, tm \to bind ~(parseNEArgumentList ~tys)}$}}

\vspace{-0.1cm}

\hspace{3.9cm}
{\small{$\mathtt{(\char`\\ \, tms \to return ~(tm ~::~ tms)))}$}}

\vspace{0.25cm}

\hspace{0.5cm}
{\small{\textcolor{lightgray}{- - $\mathtt{parsing ~a ~term ~of ~given ~type}$}}}

\hspace{0.5cm}
{\small{$\mathtt{parseTermOfGivenType : (ty : Types) \to P ~(Terms ~ty)}$}}

\hspace{0.5cm}
{\small{$\mathtt{parseTermOfGivenType ~ty =}$}}

\vspace{-0.1cm}

\hspace{0.75cm}
{\small{$\mathtt{bind ~parser~ (\char`\\ \, p \to +\text{-}elim ~(decTypeEq ~(fst ~p) ~ty)}$}}

\vspace{-0.1cm}

\hspace{5.45cm}
{\small{$\mathtt{(\char`\\ \, q \to return ~(transport ~q ~(snd ~p)))}$}}

\vspace{-0.1cm}

\hspace{5.45cm}
{\small{$\mathtt{(\char`\\ \, \_ \to fail))}$}}

\vspace{0.5cm}

\noindent
where we highlight that in $\mathtt{parseFunApp}$, the sub-parser for the arguments of a function application ($\mathtt{parseNEArgumentList}$) crucially depends on the type of the particular parsed function. The types and functions we use above are defined as follows:

\vspace{0.5cm}

{\small{\textcolor{lightgray}{- - $\mathtt{propositional ~equality}$}}}

{\small{$\mathtt{data ~Id ~\{A : Set\} ~(a : A) : A \to Set ~where}$}}

\hspace{0.25cm}
{\small{$\mathtt{refl : Id ~a ~a}$}}

\vspace{0.25cm}

{\small{$\mathtt{transport : \{A : Set\} ~\{B : A \to Set\} ~\{a1 ~a2 : A\} \to Id ~a1 ~a2 \to B ~a1 \to B ~a2}$}}

{\small{$\mathtt{transport ~refl ~b = b}$}}

\vspace{0.25cm}

{\small{\textcolor{lightgray}{- - $\mathtt{lists~with~their~monad~structure}$}}}

{\small{$\mathtt{data ~List ~(A : Set) : Set ~where}$}}

\hspace{0.25cm}
{\small{$\mathtt{[] : List ~A}$}}

\hspace{0.25cm}
{\small{$\mathtt{\_::\!\_ : A \to List ~A \to List ~A}$}}

\newpage

{\small{$\mathtt{map : \{X ~Y : Set\} \to (X \to Y) \to List ~X \to List ~Y}$}}

{\small{$\mathtt{map ~f ~[] = []}$}}

{\small{$\mathtt{map ~f ~(x ~::~ xs) = f ~x ~::~ map ~f ~xs}$}}

\vspace{0.25cm}

{\small{$\mathtt{append : \{X : Set\} \to List ~X \to List ~X \to List ~X}$}}

{\small{$\mathtt{append ~[] ~ys = ys}$}}

{\small{$\mathtt{append ~(x ~::~ xs) ~ys = x ~::~ append ~xs ~ys}$}}

\vspace{0.25cm}

{\small{$\mathtt{listReturn : \{X : Set\} \to X \to List ~X}$}}

{\small{$\mathtt{listReturn ~x = x ~::~ []}$}}

\vspace{0.25cm}

{\small{$\mathtt{listBind : \{X ~Y : Set\} \to List ~X \to (X \to List ~Y) \to List ~Y}$}}

{\small{$\mathtt{listBind ~[] ~f = []}$}}

{\small{$\mathtt{listBind ~(x ~::~ xs) ~f = append ~(f ~x) ~(listBind ~xs ~f)}$}}

\vspace{0.25cm}

{\small{\textcolor{lightgray}{- - $\mathtt{non\text{-}empty ~lists}$}}}

{\small{$\mathtt{data ~NEList ~(A : Set) : Set ~where}$}}

\hspace{0.25cm}
{\small{$\mathtt{[\_] : A \to NEList ~A}$}}

\hspace{0.25cm}
{\small{$\mathtt{\_::\!\_ : A \to NEList ~A \to NEList ~A}$}}

\vspace{0.25cm}

{\small{\textcolor{lightgray}{- - $\mathtt{unit ~type}$}}}

{\small{$\mathtt{data ~One : Set ~where}$}}

\hspace{0.25cm}
{\small{$\mathtt{* : One}$}}

\vspace{0.25cm}

{\small{\textcolor{lightgray}{- - $\mathtt{empty ~type}$}}}

{\small{$\mathtt{data ~Zero : Set ~where}$}}

\vspace{0.25cm}

{\small{\textcolor{lightgray}{- - $\mathtt{Sigma\text{-}type}$}}}

{\small{$\mathtt{data ~Sigma ~(A : Set) ~(B : A \to Set) : Set ~where}$}}

\hspace{0.25cm}
{\small{$\mathtt{\_\,,\_ : (a : A) \to (b : B ~a) \to Sigma ~A ~B}$}}

\vspace{0.25cm}

{\small{$\mathtt{fst : \{A : Set\} ~\{B : A \to Set\} \to Sigma ~A ~B \to A}$}}

{\small{$\mathtt{fst ~(a , b) = a}$}}

\vspace{0.25cm}

{\small{$\mathtt{snd : \{A : Set\} \{B : A \to Set\} \to (p : Sigma ~A ~B) \to B ~(fst ~p)}$}}

{\small{$\mathtt{snd ~(a , b) = b}$}}

\vspace{0.25cm}

{\small{\textcolor{lightgray}{- - $\mathtt{product ~type}$}}}

{\small{$\mathtt{\_\times\!\_ : Set \to Set \to Set}$}}

{\small{$\mathtt{A \times B = Sigma ~A ~(\char`\\ \, \_ \to B)}$}}

\newpage

{\small{\textcolor{lightgray}{- - $\mathtt{coproduct ~type}$}}}

{\small{$\mathtt{data ~\_+\!\_~ (A ~B : Set) : Set ~where}$}}

\hspace{0.25cm}
{\small{$\mathtt{inl : A \to A + B}$}}

\hspace{0.25cm}
{\small{$\mathtt{inr : B \to A + B}$}}

\vspace{0.25cm}

{\small{$\mathtt{+\text{-}elim : \{A ~B : Set\} ~\{C : A + B \to Set\} \to (ab : A + B)}$}}

\vspace{-0.1cm}

\hspace{6.2cm}
{\small{$\mathtt{\to ((a : A) \to C ~(inl ~a))}$}}

\vspace{-0.1cm}

\hspace{6.2cm}
{\small{$\mathtt{\to ((b : B) \to C ~(inr ~b))}$}}

\vspace{-0.1cm}

\hspace{6.2cm}
{\small{$\mathtt{\to C ~ab}$}}

{\small{$\mathtt{+\text{-}elim ~(inl ~a) ~f ~g = f ~a}$}}

{\small{$\mathtt{+\text{-}elim ~(inr ~b) ~f ~g = g ~b}$}}

\chapter{Proofs for Chapter~\ref{chap:fibadjmodels}}
\label{chap:appendixC4}

\section{Proof of Proposition~\ref{prop:indexedelimcolimits}}
\label{sect:proofofprop:indexedelimcolimits}

{
\renewcommand{\thetheorem}{\ref{prop:indexedelimcolimits}}
\begin{proposition}
Let us assume a full split comprehension category with unit \linebreak $p : \mathcal{V} \longrightarrow \mathcal{B}$ that has split fibred strong colimits of shape $\mathcal{D}$, a diagram of the form \linebreak $J : \mathcal{D} \longrightarrow \mathcal{V}_X$, and an object $A$ in $\mathcal{V}_{\ia {\mathsf{\ul{colim}}(J)}}$. Then, given a family of vertical morphisms $f_{D} : 1_{\ia {J(D)}} \longrightarrow \ia {\mathsf{\ul{in}}^{J}_D} ^*(A)$, for all objects $D$ in $\mathcal{D}$, such that for all morphisms \linebreak$g : D_i \longrightarrow D_{\!j}$ in $\mathcal{D}$ we have $\ia {J(g)} ^*(f_{D_{\!j}}) = f_{D_i}$, there exists a unique vertical morphism $[f_D]_{D \in \mathcal{D}} : 1_{\ia {\mathsf{\ul{colim}}(J)}} \longrightarrow A$ in $\mathcal{V}_{\ia {\mathsf{\ul{colim}}(J)}}$ satisfying the following ``$\beta$-equations":
\[
\ia {\mathsf{\ul{in}}^{J}_{D_i}}^*([f_D]_{D \in \mathcal{D}}) = f_{D_i}  : 1_{\ia {J(D_i)}} \longrightarrow \ia {\mathsf{\ul{in}}^{J}_{D_i}} ^*(A)
\]
for all objects $D_i$ in $\mathcal{D}$.
\end{proposition}
\addtocounter{theorem}{-1}
}

\begin{proof}
In order to give the definition of $[f_D]_{D \in \mathcal{D}}$, we first define an auxiliary ``pairing \linebreak functor"  $\langle \sigma_f \rangle : ({\underset{\circlearrowleft}{0} \overset{s}{\longrightarrow} \underset{\circlearrowleft}{1}}) \longrightarrow \mathsf{lim}(\widehat{J})$, using the universal property of the limit \linebreak $\mathsf{pr}^{\widehat{J}} : \Delta(\mathsf{lim}(\widehat{J})) \longrightarrow \widehat{J}$ for 
a cone $\sigma_f : \Delta(\underset{\circlearrowleft}{0} \overset{s}{\longrightarrow} \underset{\circlearrowleft}{1}) \longrightarrow \widehat{J}$ that is defined as follows: 
\[
(\sigma_f)_D(0) \defeq 1_{\ia {J(D)}}
\qquad
(\sigma_f)_D(1) \defeq \ia {\mathsf{\ul{in}}^{J}_{D}} ^*(A)
\qquad
(\sigma_f)_D(s) \defeq f_D
\]

In detail, $\langle \sigma_f \rangle$ arises as a unique mediating morphism because for all $g : D_i \longrightarrow D_j$ in $\mathcal{D}$, the outer triangle commutes in the following diagram:
\[
\xymatrix@C=1em@R=2em@M=0.5em{
& & \underset{\circlearrowleft}{0} \overset{s}{\longrightarrow} \underset{\circlearrowleft}{1} \ar@/_2pc/[dddll]_-{(\sigma_f)_{D_j}} \ar@/^2pc/[dddrr]^-{(\sigma_f)_{D_i}} \ar@{-->}[dd]^{\langle \sigma_f \rangle}
\\
\\
& & \mathsf{lim}(\widehat{J}) \ar[dl]_{\mathsf{pr}^{\widehat{J}}_{D_j}} \ar[dr]^{\mathsf{pr}^{\widehat{J}}_{D_i}}
\\
\mathcal{V}_{\ia {J(D_j)}} \ar[r]_-{=} & \widehat{J}(D_j) \ar[rr]_{\ia {J(g)} ^*} & & \widehat{J}(D_i) \ar[r]_-{=} & \mathcal{V}_{\ia {J(D_i)}}
}
\]
because our assumptions about the fibration $p$ and the morphisms $f_D$ give us
\[
\begin{array}{c}
\ia {J(g)} ^*(1_{\ia {J(D_j)}}) = 1_{\ia {J(D_i)}}
\\[2mm]
\ia {J(g)} ^*(\ia {\mathsf{\ul{in}}^{J}_{D_j}} ^*(A)) = \ia {\mathsf{\ul{in}}^{J}_{D_i}} ^*(A)
\\[2mm]
\ia {J(g)} ^*(f_{D_j}) = f_{D_i}
\end{array}
\]

Next, our aim is to define $[f_D]_{D \in \mathcal{D}}$ using the fully-faithfulness of $\langle \ia {\mathsf{\ul{in}}^J_D}^* \rangle_{D \in \mathcal{D}}$ on the morphism $\langle \sigma_f \rangle(s)$. However, before we can do so, we have to show that  
\[
\langle \sigma_f \rangle(0) = \langle \ia {\mathsf{\ul{in}}^J_D}^* \rangle_{D \in \mathcal{D}}(1_{\ia {\mathsf{\ul{colim}}(J)}})
\qquad
\langle \sigma_f \rangle(1) = \langle \ia {\mathsf{\ul{in}}^J_D}^* \rangle_{D \in \mathcal{D}}(A)
\]
in order to ensure that $\langle \sigma_f \rangle(s)$ is in 
\[
\mathsf{lim}(\widehat{J})(\langle \ia {\mathsf{\ul{in}}^J_D}^* \rangle_{D \in \mathcal{D}}(1_{\ia {\mathsf{\ul{colim}}(J)}}) , \langle \ia {\mathsf{\ul{in}}^J_D}^* \rangle_{D \in \mathcal{D}}(A))
\]
To this end, we use the universal property of of the limit $\mathsf{pr}^{\widehat{J}} : \Delta(\mathsf{lim}(\widehat{J})) \longrightarrow \widehat{J}$. 

The left-hand equation follows from observing that for all $g : D_i \longrightarrow D_j$ in $\mathcal{D}$, the following diagram commutes:
\[
\xymatrix@C=1em@R=2em@M=0.5em{
& & \mathbf{1} \ar@/_2pc/[dddll]_-{\star \,\mapsto\, 1_{\ia {J(D_j)}}} \ar@/^2pc/[dddrr]^-{\star \,\mapsto\, 1_{\ia {J(D_i)}}} \ar[dd]
\\
\\
& & \mathsf{lim}(\widehat{J}) \ar[dl]_{\mathsf{pr}^{\widehat{J}}_{D_j}} \ar[dr]^{\mathsf{pr}^{\widehat{J}}_{D_i}}
\\
\mathcal{V}_{\ia {J(D_j)}} \ar[r]_-{=} & \widehat{J}(D_j) \ar[rr]_{\ia {J(g)} ^*} & & \widehat{J}(D_i) \ar[r]_-{=} & \mathcal{V}_{\ia {J(D_i)}}
}
\]
when the unlabelled mediating functor $\mathbf{1} \longrightarrow \mathsf{lim}(\widehat{J})$ is given by either $\star \mapsto \langle \sigma_f \rangle(0)$ or $\star \mapsto \langle \ia {\mathsf{\ul{in}}^J_D}^* \rangle_{D \in \mathcal{D}}(1_{\ia {\mathsf{\ul{colim}}(J)}})$. As a result, these functors must be equal to the unique such functor induced by the universal property of $\mathsf{pr}^{\widehat{J}}$ and, therefore, be equal to each other. 

The right-hand equation is proved analogously. In particular, we observe that for all $g : D_i \longrightarrow D_j$ in $\mathcal{D}$, the following diagram commutes:
\[
\xymatrix@C=1em@R=2em@M=0.5em{
& & \mathbf{1} \ar@/_2pc/[dddll]_-{\star \,\mapsto\, \ia {\mathsf{\ul{in}}^{J}_{D_j}} ^*(A)} \ar@/^2pc/[dddrr]^-{\star \,\mapsto\, \ia {\mathsf{\ul{in}}^{J}_{D_i}} ^*(A)} \ar[dd]
\\
\\
& & \mathsf{lim}(\widehat{J}) \ar[dl]_{\mathsf{pr}^{\widehat{J}}_{D_j}} \ar[dr]^{\mathsf{pr}^{\widehat{J}}_{D_i}}
\\
\mathcal{V}_{\ia {J(D_j)}} \ar[r]_-{=} & \widehat{J}(D_j) \ar[rr]_{\ia {J(g)} ^*} & & \widehat{J}(D_i) \ar[r]_-{=} & \mathcal{V}_{\ia {J(D_i)}}
}
\]
when the unlabelled mediating functor $\mathbf{1} \longrightarrow \mathsf{lim}(\widehat{J})$ is given by either $\star \mapsto \langle \sigma_f \rangle(1)$ or $\star \mapsto \langle \ia {\mathsf{\ul{in}}^J_D}^* \rangle_{D \in \mathcal{D}}(A)$. As a result, these functors must be equal to the unique such functor induced by the universal property of $\mathsf{pr}^{\widehat{J}}$ and, therefore, be equal to each other. 

Now, based on the above observations, we can use the fully-faithfulness of \linebreak $\langle \ia {\mathsf{\ul{in}}^J_D}^* \rangle_{D \in \mathcal{D}}$ and define the required vertical morphism $[f_D]_{D \in \mathcal{D}}$ as
\[
[f_D]_{D \in \mathcal{D}} \defeq \langle \ia {\mathsf{\ul{in}}^J_D}^* \rangle_{D \in \mathcal{D}}^{-1}(\langle \sigma_f \rangle(s)) : 1_{\ia {\mathsf{\ul{colim}}(J)}} \longrightarrow A
\]

Finally, we prove that this morphism $[f_D]_{D \in \mathcal{D}}$ satisfies the required ``$\beta$-equations" $\ia {\mathsf{\ul{in}}^J_{D_i}}^*([f_D]_{D \in \mathcal{D}}) = f_{D_i}$, for all $D_i$ in $\mathcal{D}$, and also that it is a unique such morphism. 

First, the ``$\beta$-equations" are proved as follows:
\begin{fleqn}[0.3cm]
\begin{align*}
& \ia {\mathsf{\ul{in}}^J_{D_i}}^*([f_D]_{D \in \mathcal{D}}) 
\\
=\,\, & \ia {\mathsf{\ul{in}}^J_{D_i}}^*(\langle \ia {\mathsf{\ul{in}}^J_D}^* \rangle_{D \in \mathcal{D}}^{-1}(\langle \sigma_f \rangle(s)))
\\
=\,\, & \mathsf{pr}^{\widehat{J}}_{D_i}(\langle \ia {\mathsf{\ul{in}}^J_{D}}^* \rangle_{D \in \mathcal{D}}(\langle \ia {\mathsf{\ul{in}}^J_D}^* \rangle_{D \in \mathcal{D}}^{-1}(\langle \sigma_f \rangle(s))))
\\
=\,\, & \mathsf{pr}^{\widehat{J}}_{D_i}(\langle \sigma_f \rangle(s))
\\
=\,\, & (\sigma_f)_{D_i}(s)
\\
=\,\, & f_{D_i}
\end{align*}
\end{fleqn}
for all objects $D_i$ in $\mathcal{D}$, using the definitions of $\langle \ia {\mathsf{\ul{in}}^J_D}^* \rangle_{D \in \mathcal{D}}$ and $\langle \sigma_f \rangle$, in combination with the fully-faithfulness of the former. 

In order to show that $[f_D]_{D \in \mathcal{D}}$ is the unique such morphism, we assume that there exists another vertical morphism $h : 1_{\ia {\mathsf{\ul{colim}}(J)}} \longrightarrow A$ in $\mathcal{V}_{\ia {\mathsf{\ul{colim}}(J)}}$, satisfying the ``$\beta$-equations" $\ia {\mathsf{\ul{in}}^{J}_{D_i}}^*(h) = f_{D_i}$ for all objects $D_i$ in $\mathcal{D}$. 

Next, we observe that a functor $\widehat{h} : (\underset{\circlearrowleft}{0} \overset{s}{\longrightarrow} \underset{\circlearrowleft}{1}) \longrightarrow \mathsf{lim}(\widehat(J))$, given by
\[
\widehat{h}(0) \defeq \langle \ia {\mathsf{\ul{in}}^J_D}^* \rangle_{D \in \mathcal{D}}(1_{\ia {\mathsf{\ul{colim}}(J)}})
\quad
\widehat{h}(1) \defeq \langle \ia {\mathsf{\ul{in}}^J_D}^* \rangle_{D \in \mathcal{D}}(A)
\quad
\widehat{h}(s) \defeq \langle \ia {\mathsf{\ul{in}}^J_D}^* \rangle_{D \in \mathcal{D}}(h)
\]
makes the following diagram commute:
\[
\xymatrix@C=1em@R=2em@M=0.5em{
& & \underset{\circlearrowleft}{0} \overset{s}{\longrightarrow} \underset{\circlearrowleft}{1} \ar@/_2pc/[dddll]_-{(\sigma_f)_{D_j}} \ar@/^2pc/[dddrr]^-{(\sigma_f)_{D_i}} \ar[dd]^{\widehat{h}}
\\
\\
& & \mathsf{lim}(\widehat{J}) \ar[dl]_{\mathsf{pr}^{\widehat{J}}_{D_j}} \ar[dr]^{\mathsf{pr}^{\widehat{J}}_{D_i}}
\\
\mathcal{V}_{\ia {J(D_j)}} \ar[r]_-{=} & \widehat{J}(D_j) \ar[rr]_{\ia {J(g)} ^*} & & \widehat{J}(D_i) \ar[r]_-{=} & \mathcal{V}_{\ia {J(D_i)}}
}
\]
for all morphisms $g : D_i \longrightarrow D_j$ in $\mathcal{D}$. 

Therefore, $\widehat{h}$ must be equal to the unique such functor, namely, $\langle \sigma_f \rangle$. 
As a result,
\[
\langle \ia {\mathsf{\ul{in}}^J_D}^* \rangle_{D \in \mathcal{D}}(h) = \langle \sigma_f \rangle(s)
\]
from which it follows that 
\begin{fleqn}[0.3cm]
\begin{align*}
[f_D]_{D \in \mathcal{D}} = \langle \ia {\mathsf{\ul{in}}^J_D}^* \rangle_{D \in \mathcal{D}}^{-1}(\langle \sigma_f \rangle(s)) = \langle \ia {\mathsf{\ul{in}}^J_D}^* \rangle_{D \in \mathcal{D}}^{-1}(\langle \ia {\mathsf{\ul{in}}^J_D}^* \rangle_{D \in \mathcal{D}}(h)) = h
\end{align*}
\end{fleqn}
using the definition of $[f_D]_{D \in \mathcal{D}}$ and the fully-faithfulness of $\langle \ia {\mathsf{\ul{in}}^J_D}^* \rangle_{D \in \mathcal{D}}$. 
\end{proof}

\section{Proof of Proposition~\ref{prop:fibredcolimits}}
\label{sect:proofofprop:fibredcolimits}

{
\renewcommand{\thetheorem}{\ref{prop:fibredcolimits}}
\begin{proposition}
Let us assume a full split comprehension category with unit \linebreak $p : \mathcal{V} \longrightarrow \mathcal{B}$ that has split fibred strong colimits of shape $\mathcal{D}$. Then, given a diagram of the form $J : \mathcal{D} \longrightarrow \mathcal{V}_X$, the cocone $\mathsf{\ul{in}}^{J} : J \longrightarrow \Delta(\mathsf{\ul{colim}}(J))$, induced by the existence of split fibred strong colimits of shape $\mathcal{D}$, is a colimit of $J$ in $\mathcal{V}_X$ in the standard sense, i.e., the cocone $\mathsf{\ul{in}}^{J} : J \longrightarrow \Delta(\mathsf{\ul{colim}}(J))$ is initial amongst the cocones over $J$ in $\mathcal{V}_X$.
\end{proposition}
\addtocounter{theorem}{-1}
}

\begin{proof}
We prove this proposition by appropriately instantiating Proposition~\ref{prop:indexedelimcolimits}. 
Namely, given another cocone $\alpha : J \longrightarrow \Delta(A)$ in $\mathcal{V}_X$, we choose the object in $\mathcal{V}_{\ia {\mathsf{\ul{colim}}(J)}}$ to be $\pi^*_{\mathsf{\ul{colim}}(J)}(A)$ 
and define the morphisms $f_D$ using $\alpha_D$ as the following composites:
\[
\xymatrix@C=3.75em@R=6em@M=0.5em{
1_{\ia{J(D)}} \ar[r]^-{1({\delta_{J(D)}})} & 1_{\ia{\pi^*_{J(D)}(J(D))}} \ar[rr]^-{1(\ia {\pi^*_{J(D)}(\alpha_D)})} && 1_{\ia{\pi^*_{J(D)}(A)}} \ar[d]^-{\varepsilon^{1 \,\dashv\, \ia -}_{\pi^*_{J(D)}(A)}}
\\
\ia{\mathsf{\ul{in}}^J_D}^*(\pi^*_{\mathsf{\ul{colim}}(J)}(A)) &&&  \pi^*_{J(D)}(A) \ar[lll]^-{=}
}
\]
where $\delta_{J(D)}$ is a diagonal morphism, as defined in Definition~\ref{def:diagonalmorphisminbase}; and where the last equality follows from applying $\mathcal{P}$ to $\mathsf{\ul{in}}^J_D$, i.e., from $\pi_{\mathsf{\ul{colim}}(J)} \comp \ia{\mathsf{\ul{in}}^J_D} = \id_X \comp \pi_{J(D)}$.
Furthermore, that $f_D$ is vertical over $\id_{\ia {J(D)}}$ follows from the commutativity of
\[
\xymatrix@C=3.75em@R=6em@M=0.5em{
\ia{J(D)} \ar@/^3pc/[rrr]^-{p(f_D)}_*+<1em>{\dcomment{\text{def. of }f_D}}
\ar[r]_-{\delta_{J(D)}}
\ar@/_1pc/[dr]_-{\id_{\ia{J(D)}}}
& 
\ia {\pi^*_{J(D)}(J(D))}
\ar[r]_-{\ia {\pi^*_{J(D)}(\alpha_D)}} 
\ar[d]_-{\pi_{\pi^*_{J(D)}(J(D))}}^-{\qquad\,\,\,\dcomment{\mathcal{P}({\pi^*_{J(D)}(\alpha_D)})}}_<<<<<{\dcomment{\text{def. of } \delta_{J(D)}}\quad}
& 
\ia {\pi^*_{J(D)}(A)} 
\ar[r]_-{p(\varepsilon^{1 \,\dashv\, \ia -}_{\pi^*_{J(D)}(A)})}
\ar[d]^>>>>>>>>>{\pi_{\pi^*_{J(D)}(A)}}^<<<<<<<<{\!\!\!\!\!\!\!\!\!\quad\dcomment{\text{def. of } \pi_{\pi^*_{J(D)}(A)}}}
& 
\ia {J(D)}
\\
&
\ia{J(D)}
\ar[r]_-{\id_{\ia{J(D)}}}
&
\ia{J(D)}
\ar@/_1pc/[ur]_-{\id_{\ia{J(D)}}}
}
\]

Next, to be able to use Proposition~\ref{prop:indexedelimcolimits}, we also have to check that for all morphisms $g : D_i \longrightarrow D_j$, we have $\ia {J(g)}^*(f_{D_j}) = f_{D_i}$. We do so by recalling that the reindexing $\ia {J(g)}^*(f_{D_j})$ is defined as the unique mediating morphism induced by the Cartesian morphism $\overline{\ia {J(g)}}(\pi^*_{J(D_j)}(A))$. As a consequence, proving that the equation $\ia {J(g)}^*(f_{D_j}) = f_{D_i}$ holds amounts to showing that $f_{D_i}$ satisfies the same universal property as $\ia {J(g)}^*(f_{D_j})$, i.e., we have to show that the following diagram commutes:
\[
\scriptsize
\xymatrix@C=4em@R=4em@M=0.5em{
1_{\ia{J(D_i)}}
\ar[r]_-{1(\delta_{J(D_i)})}
\ar@/^2pc/[rrr]^-{f_{D_i}}_*+<0.75em>{\dscomment{\text{def. of } f_{D_i}}}
\ar@{<-}[d]_-{=}
\ar@/^4pc/[dddd]^-{1(\ia {J(g)})}_-{\dscomment{1 \text{ is s. fib.}}\quad\!\!}
&
1_{\ia {\pi^*_{J(D_i)}(J(D_i))}}
\ar[r]_-{1(\ia {\pi^*_{J(D_i)}(\alpha_{D_i})})}
\ar[d]_-{=}
&
1_{\ia {\pi^*_{J(D_i)}(A)}}
\ar[r]_-{\varepsilon^{1 \,\dashv\, \ia -}_{\pi^*_{J(D_i)}(A)}}
\ar[d]^-{=}_<<<<<{\dscomment{p \text{ is a split fibration}}\qquad\quad}
&
\pi^*_{J(D_i)}(A)
\ar[d]^-{=}
\\
\ia {J(g)}^*(1_{\ia {J(D_j)}})
\ar[ddd]_-{\overline{\ia {J(g)}}(1_{\ia {J(D_j)}})}
&
1_{\ia {\ia {J(g)}^*(\pi^*_{J(D_j)}(J(D_i)))}}
\ar[r]^-{1(\ia {\ia {J(g)}^*(\pi^*_{J(D_j)}(\alpha_{D_i}))})}
\ar[dd]^-{1(\ia {\overline{\ia {J(g)}}(\pi^*_{J(D_j)}(J(D_i)))})}_<<<<<{\dscomment{(a)}\qquad}
&
1_{\ia {\ia {J(g)}^*(\pi^*_{J(D_j)}(A))}}
\ar[ddd]^-{1(\ia {\overline{\ia {J(g)}}(\pi^*_{J(D_j)}(A))})}^<<<<<<{\!\!\!\!\qquad\qquad\dscomment{\text{nat. of } \varepsilon^{1 \,\dashv\, \ia -}}}_<<<<<<{\dscomment{\text{def. of } \ia {J(g)}^*(-)}\qquad\quad}
&
\ia {J(g)}^*(\pi^*_{J(D_j)}(A))
\ar[ddd]^-{\overline{\ia {J(g)}}(\pi^*_{J(D_j)}(A))}
\\
\\
&
1_{\ia {\pi^*_{J(D_j)}(J(D_i))}}
\ar@/^1pc/[dr]^<<<<<<{1(\ia{\pi^*_{J(D_j)}(\alpha_{D_i})})}
\ar[d]_-{1(\ia {\pi^*_{J(D_j)}(J(g))})}^<<<<<{\qquad\dscomment{\text{nat. of } \alpha}}
\\
1_{\ia {J(D_j)}}
\ar[r]^-{1(\delta_{J(D_j)})}
\ar@/_2pc/[rrr]_-{f_{D_j}}^*+<0.75em>{\dscomment{\text{def. of } f_{D_j}}}
&
1_{\ia {\pi^*_{J(D_j)}(J(D_j))}}
\ar[r]^-{1(\ia {\pi^*_{J(D_j)}(\alpha_{D_j})})}
&
1_{\ia {\pi^*_{J(D_j)}(A)}}
\ar[r]^-{\varepsilon^{1 \,\dashv\, \ia -}_{\pi^*_{J(D_j)}(A)}}
&
\pi^*_{J(D_j)}(A)
}
\]
where we prove the commutativity of the subdiagram marked with $(a)$ by showing that 
\[
\hspace{-0.1cm}
\xymatrix@C=3em@R=5em@M=0.5em{
\ia {J(D_i)}
\ar[r]^-{\delta_{J(D_i)}}
&
\ia {\pi^*_{J(D_i)}(J(D_i))}
\ar[r]^-{=}
&
\ia {\ia {J(g)}^(\pi^*_{J(D_j)}(J(D_i)))}
\ar[d]^-{\ia {\overline{\ia {J(g)}}(\pi^*_{J(D_j)}(J(D_i)))}}
\\
\ia {\pi^*_{J(D_j)}(J(D_j))}
&
&
\ia {\pi^*_{J(D_j)}(J(D_i))}
\ar[ll]^-{\ia {\pi^*_{J(D_j)}(J(g))}}
}
\]
and
\[
\xymatrix@C=5em@R=4em@M=0.5em{
\ia {J(D_i)} 
\ar[r]^-{\ia {J(g)}}
& 
\ia {J(D_j)} 
\ar[r]^-{\delta_{J(D_j)}}
& 
\ia {\pi^*_{J(D_j)}(J(D_j))}
}
\]
satisfy the same universal property as the unique 
mediating morphism in the following pullback situation (i.e., they make the two triangles involving $\ia {J(g)}$ commute):
\[
\xymatrix@C=5em@R=5em@M=0.5em{
\ia {J(D_i)} \ar@/_2pc/[dr]_{\ia {J(g)}} \ar@/^5pc/[rr]^{\ia {J(g)}} \ar@{-->}[r] & \ia {\pi^*_{J(D_j)}(J(D_j))} \ar[d]_{\pi_{\pi^*_{J(D_j)}(J(D_j))}}^<{\,\big\lrcorner} \ar[r]^-{\ia {\overline{\pi_{J(D_j)}}(J(D_j))}} & \ia {J(D_j)} \ar[d]^{\pi_{J(D_j)}}_-{\dcomment{\mathcal{P}(\overline{\pi_{J(D_j)}}(J(D_j)))}\quad\,\,\,\,\,\,\,\,\,\,\,\,}
\\
& \ia {J(D_j)} \ar[r]_-{\pi_{J(D_j)}} & X
}
\]
We omit the details of these proofs because they consist of straightforward diagram chasing, based on using the definitions of the diagonal morphisms $\delta_{J(D_i)}$ and $\delta_{J(D_j)}$ as unique mediating morphisms into certain pullback squares (see Definition~\ref{def:diagonalmorphisminbase}).

Now, using Proposition~\ref{prop:indexedelimcolimits}, we get that there exists a unique vertical morphism 
\[
[f_D]_{D \in \mathcal{D}} : 1_{\ia {\mathsf{\ul{colim}}(J)}} \longrightarrow \pi^*_{\mathsf{\ul{colim}}(J)}(A)
\]
such that for all $D_i$ in $\mathcal{D}$ the following ``$\beta$-equation" holds:
\[
\ia {\mathsf{\ul{in}}^J_{D_i}}^*([f_D]_{D \in \mathcal{D}}) = f_{D_i}
\]

Based on this, we define the candidate mediating morphism $[\alpha]$ from $\mathsf{\ul{in}}^J$ to $\alpha$ using the fully-faithfulness of $\mathcal{P}$ on the following morphism in $\mathcal{B}/X$ from $\pi_{\mathsf{\ul{colim}}(J)}$ to $\pi_A$:
\[
\hspace{-0.1cm}
\xymatrix@C=4.5em@R=5em@M=0.5em{
\ia {\mathsf{\ul{colim}}(J)} 
\ar[r]^-{\eta^{1 \,\dashv\, \ia -}_{\ia {\mathsf{\ul{colim}}(J)} }}
\ar@/_2pc/[drr]_-{\id_{\ia {\mathsf{\ul{colim}}(J)}}}
& 
\ia {1_{\ia {\mathsf{\ul{colim}}(J)}}}
\ar[r]^-{\ia {[f_D]_{D \in \mathcal{D}}}}
\ar[dr]_-{\pi_{1_{\ia {\mathsf{\ul{colim}}(J)}}}}_<<<<<{\dcomment{\eta^{1 \,\dashv\, \ia -} \text{ is iso.}}\qquad\quad}
&
\ia {\pi^*_{\mathsf{\ul{colim}}(J)}(A)}
\ar[r]^-{\ia {\overline{\pi_{\mathsf{\ul{colim}}(J)}}(A)}}
\ar[d]^>>>>>>>>{\pi_{\pi^*_{\mathsf{\ul{colim}}(J)}(A)}}_<<<<<{\dcomment{\mathcal{P}([f_D]_{D \in \mathcal{D}})}\quad}
&
\ia A
\ar[d]^-{\pi_A}_<<<<<<{\dcomment{\mathcal{P}(\overline{\pi_{\mathsf{\ul{colim}}(J)}}(A))}\qquad\!\!\!\!}
\\
&
&
\ia {\mathsf{\ul{colim}}(J)} 
\ar[r]_-{\pi_{\mathsf{\ul{colim}}(J)}}
&
X
}
\]
As $p = \mathsf{cod}_{\mathcal{B}} \comp \mathcal{P}$, the fully-faithfulness of $\mathcal{P}$ also gives us that $[\alpha]$ is vertical over $\id_X$.

Next, we check that $[\alpha]$ is indeed a morphism of cocones from $\mathsf{\ul{in}}^J$ to $\alpha$, i.e., we prove that $[\alpha] \comp \mathsf{\ul{in}}^J_D = \alpha_D$ holds, for all $D$ in $\mathcal{D}$. We do so by making use of the fully-faithfulness of $\mathcal{P}$ and instead prove that $\mathcal{P}([\alpha] \comp \mathsf{\ul{in}}^J_D) = \mathcal{P}(\alpha_D)$ holds in $\mathcal{B}/X$, for all $D$ in $\mathcal{D}$, which amounts to showing that the following diagram commutes:
\[
\scriptsize
\xymatrix@C=3em@R=6em@M=0.5em{
\ia {J(D)}
\ar[r]^-{\ia {\mathsf{\ul{in}}^J_D}}
\ar[drr]_-{\mathsf{s}(\ia {\mathsf{\ul{in}}^J_D}^*([f_D]_{D \in \mathcal{D}}))\quad}^>>>>>>{\quad\qquad\dscomment{\text{Proposition~\ref{prop:reindexinginthebasecategory}}}}^-{\quad\dscomment{\mathsf{s}(-) \text{ is iso.}}}_>>>>>>>>>>{\dscomment{\text{Proposition~\ref{prop:indexedelimcolimits}}}\qquad\quad}
\ar@/_3pc/[drr]_-{\mathsf{s}(f_D)}
\ar@/_2pc/[ddr]^>>>>>>>>{\!\!\!\!\!\eta^{1 \,\dashv\, \ia -}_{\ia {J(D)}}}
\ar[ddd]^-{\delta_{J(D)}}
\ar@/_3pc/[dddd]^-{\id_{\ia {J(D)}}}^>>>>>{\qquad\quad\dscomment{\text{def. of } \delta_{J(D)}}}
&
\ia {\mathsf{\ul{colim}}(J)}
\ar[r]^-{\eta^{1 \,\dashv\, \ia -}_{\ia {\mathsf{\ul{colim}}(J)}}}
\ar@/_2pc/[rr]_{\mathsf{s}([f_D]_{D \in \mathcal{D}})}^*+<0.4em>{\dscomment{\text{def. of } \mathsf{s}([f_D]_{D \in \mathcal{D}})}}
&
\ia {1_{\ia {\mathsf{\ul{colim}}(J)}}}
\ar[r]^-{\ia {[f_D]_{D \in \mathcal{D}}}}
&
\ia {\pi^*_{\mathsf{\ul{colim}}(J)}(A)}
\ar[r]^-{\ia {\overline{\pi_{\mathsf{\ul{colim}}(J)}}(A)}}
&
\ia A
\\
&
&
\ia {\ia {\mathsf{\ul{in}}^J_D}^*(\pi^*_{\mathsf{\ul{colim}}(J)}(A))}
\ar[ur]_-{\,\,\,\ia {\overline{\ia {\mathsf{\ul{in}}^J_D}} (\pi^*_{\mathsf{\ul{colim}}(J)}(A))}}
&
\ia {\pi^*_{J(D)}(A)}
\ar[l]_-{=}
\ar[ur]_-{\!\!\!\!\!\!\!\ia {\overline{\pi_{J(D)}}(A)}}^>>>>>>>>>>>>>{\dscomment{p \text{ is a split fib.}}\quad}
\\
&
\ia {1_{\ia {J(D)}}}
\ar[ur]_-{\ia {f_D}}^-{\dscomment{\text{def. of } \mathsf{s}(f_D)}\qquad}_<<<<{\qquad\dscomment{\text{def. of } f_D}}
\ar[d]^-{\!\!\!\ia {1(\delta_{J(D)})}}_-{\dscomment{\text{nat. of } \varepsilon^{1 \,\dashv\, \ia -}}\qquad}
&
\ia {1_{\ia {\pi^*_{J(D)}(A)}}}
\ar[ur]^-{\ia {\varepsilon^{}_{\pi^*_{J(D)}(A)}}\!\!\!\!\!\!\!\!\!}_<<<<<<<<<<{\quad\dscomment{\text{nat. of } \varepsilon^{1 \,\dashv\, \ia -}}}
\\
\ia {\pi^*_{J(D)}(J(D))}
\ar@/_2pc/[rr]_-{\id_{\ia {\pi^*_{J(D)}(J(D))}}}^*+<-0.05em>{\dscomment{1 \,\dashv\, \ia -}}
\ar[r]^-{\eta^{1 \,\dashv\, \ia -}_{\ia {\pi^*_{J(D)}(J(D))}}}
& 
\ia {1_{\ia {\pi^*_{J(D)}(J(D))}}}
\ar[ur]_>>>>>>>{\,\,\,\,\ia {1(\ia {\pi^*_{J(D)}(\alpha_D)})}}
\ar[r]^-{\ia {\varepsilon^{1 \,\dashv\, \ia -}_{\pi^*_{J(D)}(J(D))}}}
&
\ia {\pi^*_{J(D)}(J(D))}
\ar@/_2pc/[uur]^-{\ia {\pi^*_{J(D)}(\alpha_D)}\!\!\!\!\!}_-{\dscomment{\text{def. of } \pi^*_{J(D)}(\alpha_D)}}
\ar@/^1.5pc/[dll]^<<<<<<<<<<{\,\,\,\,\,\,\,\,\ia {\overline{\pi_{J(D)}}(J(D))}}
\\
\ia {J(D)}
\ar@/_10pc/[rrrruuuu]_-{\!\!\!\!\ia {\alpha_D}}
&
&
&
&
}
\]

Finally, we prove that $[\alpha]$ is the unique morphism from $\mathsf{\ul{in}}^J$ to $\alpha$. Namely, assuming another morphism of cocones $h$ from $\mathsf{\ul{in}}^J$ to $\alpha$, we show that $[\alpha] = h$. To this end, we first define a morphism $\widehat{h} : 1_{\ia {\mathsf{\ul{colim}}(J)}} \longrightarrow \pi^*_{\mathsf{\ul{colim}}(J)}(A)$ as the following composite:
\[
\xymatrix@C=3em@R=4em@M=0.5em{
1_{\ia {\mathsf{\ul{colim}}(J)}}
\ar[r]^-{1(\delta_{\mathsf{\ul{colim}}(J)})}
&
1_{\ia {\pi^*_{\mathsf{\ul{colim}}(J)}({\mathsf{\ul{colim}}(J)})}}
\ar[r]^-{1(\ia {\pi^*_{\mathsf{\ul{colim}}(J)}(h)})}
&
1_{\ia {\pi^*_{\mathsf{\ul{colim}}(J)}(A)}}
\ar[r]^-{\varepsilon^{1 \,\dashv\, \ia -}_{\pi^*_{\mathsf{\ul{colim}}(J)}(A)}}
&
\pi^*_{\mathsf{\ul{colim}}(J)}(A)
}
\]
which is vertical over $\id_{\ia {\mathsf{\ul{colim}}(J)}}$ because the following diagram commutes:
\[
\xymatrix@C=5em@R=5em@M=0.5em{
\ia {\mathsf{\ul{colim}}(J)}
\ar[r]^-{\delta_{\mathsf{\ul{colim}}(J)}}
\ar@/_2pc/[dr]_-{\id_{\ia {\mathsf{\ul{colim}}(J)}}}
&
\ia {\pi^*_{\mathsf{\ul{colim}}(J)}(\mathsf{\ul{colim}}(J))}
\ar[r]^-{\ia {\pi^*_{\mathsf{\ul{colim}}(J)}(h)}}
\ar[d]_-{\pi_{\pi^*_{\mathsf{\ul{colim}}(J)}(\mathsf{\ul{colim}}(J))}}_<<<<{\dcomment{\text{def. of } \delta_{\mathsf{\ul{colim}}(J)}}\qquad\,\,\,}
&
\ia {\pi^*_{\mathsf{\ul{colim}}(J)}(A)}
\ar[d]^-{\pi_{\pi^*_{\mathsf{\ul{colim}}(J)}(A)}}_-{\dcomment{\mathcal{P}(\pi^*_{\mathsf{\ul{colim}}(J)}(h))}\qquad\qquad\!\!\!\!}
\\
& 
\ia {\mathsf{\ul{colim}}(J)}
\ar[r]_-{\id_{\ia {\mathsf{\ul{colim}}(J)}}}
&
\ia {\mathsf{\ul{colim}}(J)}
}
\]

Next, as we can show that for all $D$ in $\mathcal{D}$ we have $\ia {\mathsf{\ul{in}}^J_D}^*(\widehat{h}) = f_D$ (we omit the proofs of these equations as they are analogous to the proofs of $\ia {J(g)}^*(f_{D_j}) = f_{D_i}$ given earlier), then the uniqueness of $[f_D]_{D \in \mathcal{D}}$ means that we have $\widehat{h} = [f_D]_{D \in \mathcal{D}}$. Finally, we prove that $[\alpha] = h$ holds by making use of the fully-faithfulness of $\mathcal{P}$ and instead show that $\mathcal{P}([\alpha]) = \mathcal{P}(h)$ holds in $\mathcal{B}/X$, which amounts to proving that the following holds:
\[
\scriptsize
\xymatrix@C=5em@R=5em@M=0.5em{
\ia {\mathsf{\ul{colim}}(J)}
\ar[r]^-{\eta^{1 \,\dashv\, \ia -}_{\ia {\mathsf{\ul{colim}}(J)}}}
\ar[d]^-{\delta_{\mathsf{\ul{colim}}(J)}}^<<<<{\!\!\!\!\qquad\qquad\dscomment{\text{nat. of } \eta^{1 \,\dashv\, \ia -}}}
\ar@/_4pc/[ddd]^-{\id_{\ia {\mathsf{\ul{colim}}(J)}}}
&
\ia {1_{\ia {\mathsf{\ul{colim}}(J)}}}
\ar@/^1pc/[r]^-{\ia {[f_D]_{D \in \mathcal{D}}}}_*+<0.75em>{\dscomment{\text{uniq. of } [f_D]_{D \in \mathcal{D}}}}
\ar@/_1pc/[r]_-{\ia {\widehat{h}}}
\ar[d]_-{\ia {1(\delta_{\mathsf{\ul{colim}}(J)})}}^>>>>>>{\quad\qquad\dscomment{\text{def. of } \widehat{h}}}
&
\ia {\pi^*_{\mathsf{\ul{colim}}(J)}(A)}
\ar[r]^-{\ia {\pi^*_{\mathsf{\ul{colim}}(J)}(A)}}
&
\ia A
\\
\ia {\pi^*_{\mathsf{\ul{colim}}(J)}(\mathsf{\ul{colim}}(J))}
\ar[r]^-{\eta^{1 \,\dashv\, \ia -}_{\ia {\pi^*_{\mathsf{\ul{colim}}(J)}(\mathsf{\ul{colim}}(J))}}}
\ar@/_2pc/[drr]^<<<<<<<<<{\,\,\,\,\id_{\ia {\pi^*_{\mathsf{\ul{colim}}(J)}(\mathsf{\ul{colim}}(J))}}}
\ar@/^1pc/[dd]^-{\ia {\ia {\overline{\pi_{\mathsf{\ul{colim}}(J)}}(\mathsf{\ul{colim}}(J))}}}^>>>>>>>>>>{\qquad\qquad\qquad\qquad\dscomment{\text{id. law}}}_>>>>>>>>>>>>>>>>>>>{\dscomment{\text{def. of } \delta_{\mathsf{\ul{colim}}(J)}}\,\,\,\,}
&
\ia {1_{\ia {\pi^*_{\mathsf{\ul{colim}}(J)}({\mathsf{\ul{colim}}(J)})}}}
\ar[r]_-{\ia {1(\ia {\pi^*_{\mathsf{\ul{colim}}(J)}(h)})}}
\ar[dr]_-{\ia {\varepsilon^{1 \,\dashv\, \ia -}_{\pi^*_{\mathsf{\ul{colim}}(J)}({\mathsf{\ul{colim}}(J)})}}\,\,\,\,}^-{\qquad\qquad\dscomment{\text{nat. of } \varepsilon^{1 \,\dashv\, \ia -}}}_<<<<{\dscomment{1 \,\dashv\, \ia -}\qquad\quad}
&
\ia {1_{\ia {\pi^*_{\mathsf{\ul{colim}}(J)}(A)}}}
\ar[u]^-{\ia {\varepsilon^{1 \,\dashv\, \ia -}_{\pi^*_{\mathsf{\ul{colim}}(J)}(A)}}}
\\
&
&
\ia {\pi^*_{\mathsf{\ul{colim}}(J)}({\mathsf{\ul{colim}}(J)})}
\ar@/_3pc/[uu]_>>>>>>>{\!\!\!\!\ia {\pi^*_{\mathsf{\ul{colim}}(J)}(h)}}
\ar@/^2pc/[dll]_-{\ia {\overline{\pi_{\mathsf{\ul{colim}}(J)}}(\mathsf{\ul{colim}}(J))}\quad}
\\
\ia {\mathsf{\ul{colim}}(J)}
\ar@/_10pc/[uuurrr]_-{\ia {h}}^-{\dscomment{\text{def. of } \pi^*_{\mathsf{\ul{colim}}(J)}(h)}\qquad\qquad\qquad}
}
\]
\end{proof}

\section{Proof of Proposition~\ref{prop:fibredNNO}}
\label{sect:proofofprop:fibredNNO}

{
\renewcommand{\thetheorem}{\ref{prop:fibredNNO}}
\begin{proposition}
Let us assume a full split comprehension category with unit \linebreak $p : \mathcal{V} \longrightarrow \mathcal{B}$ such that $\mathcal{B}$ has a terminal object and $p$ has weak split fibred strong 
natural numbers. Then, each fibre of $p$ has a weak NNO and this structure is preserved on-the-nose by reindexing.
\end{proposition}
\addtocounter{theorem}{-1}
}

\begin{proof}
Given any object $X$ in $\mathcal{B}$, we claim that the diagram
\[
\xymatrix@C=7em@R=6em@M=0.5em{
1_X \ar[r]^-{!_X^*(\mathsf{zero})} & !_X^*(\mathbb{N}) & !_X^*(\mathbb{N}) \ar[l]_-{!_X^*(\mathsf{succ})}
}
\]
defines a weak NNO in $\mathcal{V}_X$. 

To show that this is the case, we assume another diagram in $\mathcal{V}_X$, given by
\[
\xymatrix@C=7em@R=6em@M=0.5em{
1_X \ar[r]^-{f_z} & A & A \ar[l]_-{f_s}
}
\]

\pagebreak

Next, we observe that the morphisms $f_z$ and $f_s$ induce two composite morphisms
\[
\hspace{-0.2cm}
\xymatrix@C=0.7em@R=1em@M=0.5em{
1_{\ia {1_X}} \ar[r]^-{=} & \pi^*_{1_X}(1_X) \ar[rr]^-{\pi^*_{1_X}(f_z)} && \pi^*_{1_X}(A) \ar[r]^-{=} & \ia {!_X^*(\mathsf{zero})}^*(\pi^*_{!_X^*(\mathbb{N})}(A)) \ar[rrrr]^-{\overline{\ia {!_X^*(\mathsf{zero})}}(\pi^*_{!_X^*(\mathbb{N})}(A))} &&&& \pi^*_{!_X^*(\mathbb{N})}(A)
}
\]
\[
\hspace{-0.25cm}
\xymatrix@C=1em@R=1em@M=0.5em{
\pi^*_{!_X^*(\mathbb{N})}(A) \ar[rr]^-{\pi^*_{!_X^*(\mathbb{N})}(f_s)} && \pi^*_{!_X^*(\mathbb{N})}(A) \ar[r]^-{=} & \ia {!_X^*(\mathsf{succ})}^*(\pi^*_{!_X^*(\mathbb{N})}(A)) \ar[rrrr]^-{\overline{\ia {!_X^*(\mathsf{succ})}}(\pi^*_{!_X^*(\mathbb{N})}(A))} &&&& \pi^*_{!_X^*(\mathbb{N})}(A)
}
\]
which we denote in the rest of this proof by $\widehat{f_z}$ and $\widehat{f_s}$, respectively. 

It is easy to see that $\widehat{f_z}$ and $\widehat{f_s}$ are over $\ia {!_X^*(\mathsf{zero})}$ and $\ia {!_X^*(\mathsf{succ})}$, respectively. As a result, we can use the existence of weak split fibred strong natural numbers in $p$ to get a section of $\pi_{\pi^*_{!_X^*(\mathbb{N})}(A)}$, which we denote by $\mathsf{rec}(\widehat{f_z},\widehat{f_s}) : \ia {!_X^*(\mathbb{N})} \longrightarrow \ia {\pi^*_{!_X^*(\mathbb{N})}(A)}$.

Using $\mathsf{rec}(\widehat{f_z},\widehat{f_s})$, we can derive a vertical morphism in $\mathcal{V}_X$, denoted by 
\[
\mathsf{rec}_X(f_z,f_h) :\, !_X^*(\mathbb{N}) \longrightarrow A
\]
and defined using the fully-faithfulness of $\mathcal{P}$ on the next commuting diagram in $\mathcal{B}/X$.
\[
\xymatrix@C=9em@R=4em@M=0.5em{
\ia {!_X^*(\mathbb{N})} \ar[r]^-{\mathsf{rec}(\widehat{f_z},\widehat{f_s})} \ar@/_4pc/[ddr]_-{\pi_{!_X^*(\mathbb{N})}}  \ar@/_2pc/[dr]_-{\id_{\ia {!_X^*(\mathbb{N})}}} & \ia {\pi^*_{!_X^*(\mathbb{N})}(A)} \ar[r]^-{\ia {\overline{\pi_{!_X^*(\mathbb{N})}}(A)}} \ar[d]^-{\pi_{\pi^*_{!_X^*(\mathbb{N})}(A)}}_<<<<<<<{\dcomment{\text{def. of } \mathsf{rec}(\widehat{f_z},\widehat{f_s})}\qquad} \ar@{}[dd]^-{\qquad\qquad\dcomment{\mathcal{P}(\overline{\pi_{!_X^*(\mathbb{N})}}(A))}} & \ia {A} \ar@/^4pc/[ddl]^-{\pi_A}
\\
& \ia {!_X^*(\mathbb{N})} \ar[d]^-{\pi_{!_X^*(\mathbb{N})}}_-{\dcomment{\text{id. law}}\qquad\quad\!\!\!}
\\
& X
}
\]

Next, we proceed by showing that the two standard diagrams describing the interaction of $\mathsf{rec}_X(f_z,f_h)$ with $!_X^*(\mathsf{zero})$ and $!_X^*(\mathsf{succ})$ commute. 

\pagebreak

First, we show the commutativity of
\[
\xymatrix@C=7em@R=5em@M=0.5em{
1_X \ar[r]^-{!_X^*(\mathsf{zero})} \ar[d]_-{\id_{1_X}} & !_X^*(\mathbb{N}) \ar[d]^-{\mathsf{rec}_X(f_z,f_s)}
\\
1_X \ar[r]_-{f_z} & A
}
\]
by using the fully-faithfulness of $\mathcal{P}$ on a diagram in $\mathcal{B}/X$ between $\pi_{1_X} : \ia {1_X} \longrightarrow X$ and $\pi_A : \ia {A} \longrightarrow X$ that is given between the domains of $\pi_{1_X}$ and $\pi_A$ by
\[
\scriptsize
\xymatrix@C=4em@R=3em@M=0.5em{
\ia {1_X} \ar@/_4.5pc/[dddd]_-{\id_{\ia {1_X}}} \ar[rrr]^-{\ia {!_X^*(\mathsf{zero})}} \ar[d]^-{\eta^{1 \,\dashv\, \ia -}_{\ia {1_X}}} &&& \ia {!_X^*(\mathbb{N})} \ar[d]^-{\mathsf{rec}(\widehat{f_z},\widehat{f_s})}_-{\dscomment{\text{def. of } \mathsf{rec}(\widehat{f_z},\widehat{f_s}) }\qquad\qquad\qquad\qquad\qquad\qquad\qquad}
\\
\ia {1_{\ia {1_X}}} \ar[ddr]_-{=}^-{\qquad\qquad\dscomment{\text{def. of } \widehat{f_z} }} \ar[ddd]_-{\pi_{1_{\ia {1_X}}}} \ar@{}[dd]_-{\dscomment{\eta^{1 \,\dashv\, \ia -}_{\ia {1_X}} \text{ is an iso.}}\,\,\,} \ar[rrr]_-{\ia {\widehat{f_z}}} &&& \ia {\pi^*_{!_X^*(\mathbb{N})}(A)} \ar[ddd]^-{\ia {\overline{\pi_{!_X^*(\mathbb{N})}}(A)}}
\\
\ar@{}[dd]^-{\,\,\,\,\,\dscomment{1 \text{ is split fibred}}} && \ia {\ia {!_X^*(\mathsf{zero})}^*(\pi^*_{!_X^*(\mathbb{N})}(A))} \ar[ur]^-{\ia {\overline{\ia {!_X^*(\mathsf{zero})}}(\pi^*_{!_X^*(\mathbb{N})}(A))}} &
\\
& \ia {\pi^*_{1_X}(1_X)} \ar[r]^-{\ia {\pi^*_{1_X}(f_z)}} \ar[dl]^-{\ia {\overline{\pi_{1_X}}(1_X)}}^<<<{\qquad\qquad\qquad\qquad\qquad\dscomment{\text{def. of } \pi^*_{1_X}(f_z)}} & \ia {\pi^*_{1_X}(A)} \ar[u]^-{=}_-{\qquad\quad\dscomment{p \text{ is a split fibration}}} \ar[dr]_-{\ia {\overline{\pi_{1_X}}(A)}} &
\\
\ia {1_X} \ar[rrr]_-{\ia {f_z}} &&& \ia {A}
}
\]

Second, we show the commutativity of 
\[
\xymatrix@C=7em@R=5em@M=0.5em{
!_X^*(\mathbb{N}) \ar[r]^-{!_X^*(\mathsf{succ})} \ar[d]_-{\mathsf{rec}_X(f_z,f_s)} & !_X^*(\mathbb{N}) \ar[d]^-{\mathsf{rec}_X(f_z,f_s)}
\\
A \ar[r]_-{f_s} & A
}
\]
by using the fully-faithfulness of $\mathcal{P}$ on a commuting diagram in $\mathcal{B}/X$ between \linebreak ${\pi_{!_X^*(\mathbb{N})} : \ia {!_X^*(\mathbb{N})} \longrightarrow X}$ and ${\pi_A : \ia {A} \longrightarrow X}$ that is given between the domains of $\pi_{!_X^*(\mathbb{N})}$ and $\pi_A$ by

\pagebreak

\[
\scriptsize
\xymatrix@C=4em@R=3em@M=0.5em{
\ia {!_X^*(\mathbb{N})} \ar[rrr]^-{\ia {!_X^*(\mathsf{succ})}} \ar[d]_-{\mathsf{rec}(\widehat{f_z},\widehat{f_s})}^-{\qquad\qquad\qquad\qquad\qquad\qquad\qquad\dscomment{\text{def. of } \mathsf{rec}(\widehat{f_z},\widehat{f_s})}} &&& \ia {!_X^*(\mathbb{N})} \ar[d]^-{\mathsf{rec}(\widehat{f_z},\widehat{f_s})}
\\
\ia {\pi^*_{!_X^*(\mathbb{N})}(A)} \ar[ddd]_-{\ia {\overline{\pi_{!_X^*(\mathbb{N})}}(A)}} \ar[rrr]_-{\ia {\widehat{f_s}}} \ar[ddr]_-{\id_{\ia {\pi^*_{!_X^*(\mathbb{N})}(A)}}} & \ar@{}[dd]^-{\dscomment{\text{def. of } \widehat{f_s}}} && \ia {\pi^*_{!_X^*(\mathbb{N})}(A)} \ar[ddd]^-{\ia {\overline{\pi_{!_X^*(\mathbb{N})}}(A)}}
\\
\ar@{}[dd]^-{\qquad\dscomment{\text{id. law}}} && \ia {!_X^*(\mathsf{succ})}^*(\pi^*_{!_X^*(\mathbb{N})}(A)) \ar[ur]^-{\ia {\overline{\ia {!_X^*(\mathsf{succ})}}(\pi^*_{!_X^*(\mathbb{N})}(A))}} &
\\
& \ia {\pi^*_{!_X^*(\mathbb{N})}(A)} \ar[r]^-{\ia {\pi^*_{!_X^*(\mathbb{N})}(f_s)}} \ar[dl]^-{\ia {\overline{\pi_{!_X^*(\mathbb{N})}}(A)}}^<<{\qquad\qquad\qquad\qquad\qquad\dscomment{\text{def. of } \pi^*_{!_X^*(\mathbb{N})}(f_s)}} & \ia {\pi^*_{!_X^*(\mathbb{N})}(A)} \ar[u]_-{=}_-{\qquad\quad\dscomment{1 \text{ is a split fibration}}} \ar[dr]_-{\ia {\overline{\pi_{!_X^*(\mathbb{N})}}(A)}} &
\\
\ia {A} \ar[rrr]_-{\ia {f_s}} &&& \ia {A} 
}
\]

Finally, we show that this weak NNO structure is preserved on-the-nose by reindexing. In particular, for all morphisms $f : Y \longrightarrow X$ in $\mathcal{B}$, we have
\[
f^*(!^*_X(\mathbb{N})) = (!_X \comp f)^*(\mathbb{N}) = \,!^*_Y(\mathbb{N})
\]
The proofs of preservation are analogous for ${!_X^*(\mathsf{zero})}$, ${!_X^*(\mathsf{succ})}$, and $\mathsf{rec}_X(f_z,f_h)$.
\end{proof}

\section{Proof of Proposition~\ref{prop:equivalenceofnaturalnumbersinthesisandpaper}}
\label{sect:proofofprop:equivalenceofnaturalnumbersinthesisandpaper}

{
\renewcommand{\thetheorem}{\ref{prop:equivalenceofnaturalnumbersinthesisandpaper}}
\begin{proposition}
Let us assume a full split comprehension category with unit \linebreak $p : \mathcal{V} \longrightarrow \mathcal{B}$ such that $\mathcal{B}$ has a terminal object. Then, $p$ having weak split fibred strong natural numbers is equivalent to $p$ supporting weak natural numbers as in~\cite{Ahman:FibredEffects}, i.e., for every object $X$ in $\mathcal{B}$, every object $A$ in $\mathcal{V}_{\ia {!_X^*(\mathbb{N})}}$, every morphism 
\[
f_z : 1_X \longrightarrow (\funsection(!_X^*(\mathsf{zero})))^*(A)
\]
in $\mathcal{V}_X$, and every morphism 
\[
f_s : 1_{\ia A} \longrightarrow \pi_A^*(\ia {!_X^*(\mathsf{succ})}^* (A))
\]
in $\mathcal{V}_{\ia A}$, there exists a morphism 
\[
\mathsf{i}_A(f_z,f_s) : 1_{\ia {!^*_X(\mathbb{N})}} \longrightarrow A
\]
in $\mathcal{V}_{\ia {!^*_X(\mathbb{N})}}$ such that 
\[
\begin{array}{c}
(\funsection(!_X^*(\mathsf{zero})))^*(\mathsf{i}_A(f_z,f_s)) = f_z
\\[3mm]
\ia {!_X^*(\mathsf{succ})}^* (\mathsf{i}_A(f_z,f_s)) 
=
(\mathsf{s}(\mathsf{i}_A(f_z,f_s)))^*(f_s) 
\end{array}
\vspace{0.2cm}
\]
\end{proposition}
\addtocounter{theorem}{-1}
}

\pagebreak

\begin{proof}
In both directions, we assume an object $X$ in $\mathcal{B}$ and an object $A$ in $\mathcal{V}_{\ia {!_X^*(\mathbb{N})}}$.

\vspace{0.2cm}
\noindent \textbf{Weak natural numbers in~\cite{Ahman:FibredEffects} imply Definition~\ref{def:strongsplitfibredweaknaturals}:}

Given a pair of morphisms
\[
\hspace{-0.5cm}
\xymatrix@C=7em@R=1em@M=0.5em{
1_{\ia {1_X}} \ar[r]^-{f_z} & A & A \ar[l]_-{f_s}
}
\]
in $\mathcal{V}$, with 
\[
p(A) = \ia {!_X^*(\mathbb{N})}
\qquad
p(f_z) = \ia {!_X^*(\mathsf{zero})}
\qquad
p(f_s) = \ia {!_X^*(\mathsf{succ})}
\]
we aim to define a morphism
\[
\mathsf{rec}(f_z,f_s) : \ia {!_X^*(\mathbb{N})} \longrightarrow \ia {A}
\]
that must be a section of $\pi_A : \ia {A} \longrightarrow \ia {!_X^*(\mathbb{N})}$. 

First, we define vertical morphisms
\[
g_z : 1_X \longrightarrow (\mathsf{s}(!^*_X(\mathsf{zero})))^*(A)
\qquad
g_s : 1_{\ia A} \longrightarrow \pi^*_A(\ia {!^*_X(\mathsf{succ})}^*(A))
\]
in $\mathcal{V}_X$ and $\mathcal{V}_{\ia A}$, respectively, by
\[
g_z \defeq (\eta^{1 \,\dashv\, \ia -}_{\ia {1_X}})^*(f_z^\dagger)
\qquad
g_s \defeq \pi^*_A(f_s^\dagger) \comp \pi^*_A(\mathsf{fst}) \comp \eta^{\Sigma_A \,\dashv\, \pi^*_A}_{1_{\ia A}}
\]
where the two vertical morphisms (in $\mathcal{V}_{\ia {1_X}}$ and $\mathcal{V}_{\ia {!_X^*(\mathbb{N})}}$, respectively)
\[
f_z^\dagger : 1_{\ia {1_X}} \longrightarrow \ia {!_X^*(\mathsf{zero})} ^*(A)
\qquad
f_s^\dagger : A \longrightarrow \ia {!_X^*(\mathsf{succ})}^*(A)
\]
arise from using the universal properties of the following two Cartesian morphisms: 
\[
\overline{\ia {!_X^*(\mathsf{zero})}}(A) : \ia {!_X^*(\mathsf{zero})} ^*(A) \longrightarrow A
\qquad
\overline{\ia {!_X^*(\mathsf{succ})}}(A) : \ia {!_X^*(\mathsf{succ})}^*(A) \longrightarrow A
\]
on the given morphisms $f_z$ and $f_s$, respectively, as discussed in Definition~\ref{def:uniquemediatingmorphismforCartesianmorphism}. 

\pagebreak

Next, we recall that the existence of weak natural numbers in 
the sense of~\cite{Ahman:FibredEffects} means that $g_z$ and $g_s$ induce a vertical morphism
\[
\mathsf{i}_A(g_z,g_s) : 1_{\ia {!^*_X(\mathbb{N})}} \longrightarrow A
\]
in $\mathcal{V}_{\ia A}$, which we can in turn use to define $\mathsf{rec}(f_z,f_s)$ by letting
\[
\mathsf{rec}(f_z,f_s) \defeq \mathsf{s}(\mathsf{i}_A(g_z,g_s))
\]

Finally, we prove that $\mathsf{rec}(f_z,f_s)$ makes the next diagram commute in $\mathcal{B}$.
\[
\xymatrix@C=5em@R=5em@M=0.5em{
\ia {1_X} \ar[r]^-{\ia {!_X^*(\mathsf{zero})}} \ar[d]_-{\eta^{1 \,\dashv\, \ia -}_{\ia {1_X}}}^-{\quad\qquad\dcomment{(a)}} & \ia {!_X^*(\mathbb{N})} \ar[d]_-{\mathsf{rec}(f_z,f_s)}^-{\qquad\qquad\dcomment{(b)}} & \ia {!_X^*(\mathbb{N})} \ar[l]_-{\ia {!_X^*(\mathsf{succ})}} \ar[d]^-{\mathsf{rec}(f_z,f_s)}
\\
\ia {1_{\ia {1_X}}} \ar[r]_-{\ia {f_z}} & \ia {A} & \ia {A} \ar[l]^-{\ia {f_s}}
}
\]

In order to show that the square marked with $(a)$ commutes, we recast the equation
\[
(\funsection(!_X^*(\mathsf{zero})))^*(\mathsf{i}_A(g_z,g_s)) = g_z
\]
in $\mathcal{B}$ using Proposition~\ref{prop:reindexinginthebasecategory}. As a result, the outer perimeter of the next diagram (i.e., one of the triangles of the pullback situation in Proposition~\ref{prop:reindexinginthebasecategory}) commutes in $\mathcal{B}$.
\[
\xymatrix@C=4em@R=2em@M=0.5em{
&& \ia {!^*_X(\mathbb{N})} \ar[r]^-{\eta^{1 \,\dashv\, \ia -}_{\ia {!^*_X(\mathbb{N})}}} & \ia {1_{\ia {!^*_X(\mathbb{N})}}} \ar[dd]^-{\ia {\mathsf{i}_A(g_z,g_s)}}
\\
&& \ia {1_X} \ar[d]^-{\ia {g_z}} \ar[u]_-{\,\,\,\ia {!_X^*(\mathsf{zero})}}
\\
X \ar[rr]_-{\mathsf{s}(g_z)} \ar[urr]^-{\eta^{1 \,\dashv\, \ia -}_X\!\!\!\!\!} \ar@/^3pc/[uurr]^-{\funsection(!_X^*(\mathsf{zero}))} \ar@{}[uu]_<<<<{\quad\qquad\qquad\qquad\qquad\dcomment{\text{def. of } \mathsf{s}(g_z)}}_>>>>>>>{\quad\qquad\qquad\qquad\dcomment{\text{def. of } \funsection(!_X^*(\mathsf{zero}))}}_-{\qquad\qquad\qquad\qquad\qquad\qquad\qquad\qquad\qquad\qquad\dcomment{(c)}} && \ia {(\funsection(!_X^*(\mathsf{zero})))^*(A)} \ar[r]_-{\ia {\overline{\funsection(!_X^*(\mathsf{zero}))}(A)}} & \ia A
}
\]

Further, as $\eta^{1 \,\dashv\, \ia -}_X$ is an epimorphism (because it is an isomorphism according to Proposition~\ref{prop:compcatunitiso}), the commutativity of the outer perimeter of this diagram also implies that the square marked with $(c)$ commutes on its own. 

\pagebreak

Based on this last observation, the required commutativity of $(a)$ now follows from 
the commutativity of the following diagram:
\vspace{-1cm}
\[
\scriptsize
\xymatrix@C=6em@R=6em@M=0.5em{
\ar@{}[dddd]^<<<<<<<<<<<<<<{\quad\qquad\qquad\qquad\qquad\qquad\qquad\qquad\qquad\qquad\qquad\dscomment{\text{def. of } \mathsf{rec}(f_z,f_s)}}^<<<<<<<<<<<<<<<<<<<<<<<<<<<<<<{\qquad\qquad\qquad\qquad\qquad\qquad\qquad\dscomment{(c)}}^<<<<<<<<<<<<<<<<<<<<<<<<<<<<<<<<<<<<<<<<<<<<<<<<<<<<{\,\,\,\,\,\,\qquad\qquad\qquad\dscomment{\text{def. of } g_z}}^<<<<<<<<<<<<<<<<<<<<<<<<<<<<<<<<<<<<<<<<<<<<<<<<<{\qquad\qquad\qquad\qquad\qquad\qquad\qquad\qquad\qquad\qquad\dscomment{\text{def. of } \funsection(!_X^*(\mathsf{zero}))}}^<<<<<<<<<<<<<<<<<<<<<<<<<<<<<<<<<<<<<<<<<<<<<<<<<<<<<<<<<<<<<<<<<<<<<<<<<<<{\quad\qquad\qquad\dscomment{\text{def. of } (\eta^{1 \,\dashv\, \ia -}_X)^*(f_z^\dagger)}}^<<<<<<<<<<<<<<<<<<<<<<<<<<<<<<<<<<<<<<<<<<<<<<<<<<<<<<<<<<<<<<<<{\qquad\qquad\qquad\qquad\qquad\qquad\qquad\qquad\qquad\qquad\qquad\qquad\qquad\qquad\dscomment{\text{p is split a fibration}}} &
\\
\ia {!^*_X(\mathbb{N})} \ar[rr]^-{\eta^{1 \,\dashv\, \ia -}_{\ia {!^*_X(\mathbb{N})}}} \ar@/^3pc/[rrr]^-{\mathsf{rec}(f_z,f_s)} && \ia {1_{\ia {!^*_X(\mathbb{N})}}} \ar[r]^-{\ia {\mathsf{i}_A(g_z,g_s)}} & \ia {A}
\\
\ia {1_X} \ar[u]^-{\ia {!_X^*(\mathsf{zero})}} \ar[r]^-{\ia {g_z}} \ar[d]^-{=} & \ia {(\funsection(!_X^*(\mathsf{zero})))^*(A)} \ar[urr]^-{\ia {\overline{\funsection(!_X^*(\mathsf{zero}))}(A)}} \ar[d]_-{=}
\\
\ia {(\eta^{1 \,\dashv\, \ia -}_{X})^*(1_{\ia {1_X}})} \ar[r]^-{\ia {(\eta^{1 \,\dashv\, \ia -}_X)^*(f_z^\dagger)}} \ar[d]_-{\ia {\overline{\eta^{1 \,\dashv\, \ia -}_X}(1_{\ia {1_X}})}} & \ia {(\eta^{1 \,\dashv\, \ia -}_X)^*((\ia {!_X^*(\mathsf{zero})})^*(A))} \ar[d]^-{\ia {\overline{\eta^{1 \,\dashv\, \ia -}_X}((\ia {!_X^*(\mathsf{zero})})^*(A))}} \ar@/_2pc/[uurr]_-{\ia {\overline{\ia {!_X^*(\mathsf{zero})} \,\comp\, \eta^{1 \,\dashv\, \ia -}_X}(A)}}
\\
\ia {1_{\ia {1_X}}} \ar[r]_-{\ia {f_z^\dagger}} & \ia {(\ia {!_X^*(\mathsf{zero})})^*(A)} \ar@/_6pc/[uuurr]_-{\!\!\!\!\!\!\ia {\overline{\ia {!_X^*(\mathsf{zero})}}(A)}}
}
\]
and by observing that we have the following equations:
\[
f_z = \overline{\ia {!_X^*(\mathsf{zero})}}(A) \comp f_z^\dagger
\qquad
\ia {\overline{\eta^{1 \,\dashv\, \ia -}_X}(1_{\ia {1_X}})} = \ia {1(\eta^{1 \,\dashv\, \ia -}_X)} = \eta^{1 \,\dashv\, \ia -}_{\ia {1_X}}
\]
where the two equations on the right follow from $1$ being a split fibred functor, $\eta^{1 \,\dashv\, \ia -}_X$ being an isomorphism, and $\eta^{1 \,\dashv\, \ia -}$ being a natural transformation.

In order to show that the square marked with $(b)$ commutes, we again use Proposition~\ref{prop:reindexinginthebasecategory} to recast the equation
\[
\ia {!_X^*(\mathsf{succ})}^* (\mathsf{i}_A(g_z,g_s)) 
=
(\mathsf{s}(\mathsf{i}_A(f_z,f_s)))^*(g_s) 
\]
in $\mathcal{B}$. As a result, we get that the next diagram, marked with $(d)$, commutes in $\mathcal{B}$.
\[
\xymatrix@C=4em@R=2em@M=0.5em{
&& \ia {!^*_X(\mathbb{N})} \ar[r]^-{\eta^{1 \,\dashv\, \ia -}_{\ia {!^*_X(\mathbb{N})}}} & \ia {1_{\ia {!^*_X(\mathbb{N})}}} \ar[dd]^-{\ia {\mathsf{i}_A(g_z,g_s)}}_-{\dcomment{(d)}\qquad\qquad\qquad\qquad\qquad\,\,\,\,}
\\
&&
\\
\ia {!^*_X(\mathbb{N})} \ar[rr]_-{\mathsf{s}((\mathsf{s}(\mathsf{i}_A(g_z,g_s)))^*(g_s) )} \ar@/^2.5pc/[uurr]^-{\ia {!_X^*(\mathsf{succ})}} && \ia {\ia {!_X^*(\mathsf{succ})}^*(A)} \ar[r]_-{\ia {\overline{\ia {!_X^*(\mathsf{succ})}}(A)}} & \ia A
}
\]
Based on this observation, the commutativity of $(b)$ now follows from that of
\[
\scriptsize
\xymatrix@C=2.75em@R=6em@M=0.5em{
\ia {!^*_X(\mathbb{N})} \ar[rrr]^-{\mathsf{rec}(f_z,f_s)} &&& \ia {A}
\\
\ia {!^*_X(\mathbb{N})} \ar[u]^-{\ia {!_X^*(\mathsf{succ})}}_-{\qquad\qquad\qquad\qquad\qquad\qquad\qquad\qquad\qquad\dscomment{(d)}} \ar[rr]^-{\mathsf{s}((\mathsf{s}(\mathsf{i}_A(g_z,g_s)))^*(g_s) )} \ar[d]^-{\eta^{1 \,\dashv\, \ia -}_{\ia {!^*_X(\mathbb{N})}}}^-{\qquad\qquad\qquad\qquad\qquad\dscomment{\text{def. of } \mathsf{s}((\mathsf{s}(\mathsf{i}_A(g_z,g_s)))^*(g_s) )}} \ar@/_4pc/[ddddddd]^-{\id_{\ia {!^*_X(\mathbb{N})}}}^>>>>>>>>>>>>>>>>>>>>>>>>>>>>>>>>>>>>>>>>>>>>>>>>>>>>>>>>>>>>>>>{\,\,\dscomment{\eta^{1 \,\dashv\, \ia -}_{\ia {!^*_X(\mathbb{N})}} \text{ is iso.}}} &&  \ia {\ia {!_X^*(\mathsf{succ})}^*(A)} \ar[ur]^-{\ia {\overline{\ia {!_X^*(\mathsf{succ})}}(A)}} \ar[d]_-{=} 
\\
\ia {1_{\ia {!^*_X(\mathbb{N})}}} \ar[d]_-{=}^-{\quad\qquad\qquad\dscomment{\text{def. of } (\mathsf{s}(\mathsf{i}_A(g_z,g_s)))^*(g_s)}} \ar[rr]^-{\ia {(\mathsf{s}(\mathsf{i}_A(g_z,g_s)))^*(g_s)}} && \ia {(\mathsf{s}(\mathsf{i}_A(g_z,g_s)))^*(\pi^*_A(\ia {!_X^*(\mathsf{succ})}^*(A)))} \ar[d]_-{\ia {\overline{\mathsf{s}(\mathsf{i}_A(g_z,g_s))}(\pi^*_A(\ia {!_X^*(\mathsf{succ})}^*(A)))}}^<<<<<<<{\quad\dscomment{p \text{ is a s. fib.}}}
\\
\ia {(\mathsf{s}(\mathsf{i}_A(g_z,g_s)))^*(1_{\ia A})} \ar[d]_-{=}^-{\quad\qquad\dscomment{\mathcal{P}(\overline{\mathsf{s}(\mathsf{i}_A(g_z,g_s))}(1_{\ia A}))}} \ar[r]^-{\ia {\overline{\mathsf{s}(\mathsf{i}_A(g_z,g_s))}(1_{\ia A})}} & \ia {1_{\ia A}} \ar[r]^-{\ia {g_s}} \ar[d]_-{=}^-{\qquad\qquad\qquad\dscomment{\text{def. of } g_s}} & \ia {\pi^*_A(\ia {!_X^*(\mathsf{succ})}^*(A))} \ar@/_5.75pc/[uu]_<<<<<<<{\,\,\,\,\ia {\overline{\pi_A}(\ia {!_X^*(\mathsf{succ})}^*(A))}} 
\\
\ia {1_{\ia {!^*_X(\mathbb{N})}}} \ar[dddd]^-{\pi_{1_{\ia {!^*_X(\mathbb{N})}}}} & \ia {\pi^*_A(1_{\ia {!^*_X(\mathbb{N})}})} \ar@/^3pc/[d]^-{\ia {\eta^{\Sigma_A \,\dashv\, \pi^*_A}_{\pi^*_A(1_{\ia {!^*_X(\mathbb{N})}})}}}_>>>>>{\dscomment{\text{def. of } \kappa_{A,\pi^*_A(1_{\ia {!^*_X(\mathbb{N})}})}}\,\,\,\,\,} \ar@/_4pc/[dd]_>>>>>>>{\kappa_{A,\pi^*_A(1_{\ia {!^*_X(\mathbb{N})}})}\!\!\!\!\!} \ar@/_6.5pc/[ddd]_<<<<<<<<<<<<<<<{\id_{\ia {\pi^*_A(1_{\ia {!^*_X(\mathbb{N})}})}}\!\!\!\!\!\!\!\!}
\\
& \ia {\pi^*_A(\Sigma_A(\pi^*_A(1_{\ia {!^*_X(\mathbb{N})}})))} \ar[r]^-{\ia {\pi^*_A(\mathsf{fst})}} \ar[d]^-{\ia {\overline{\pi_A}(\Sigma_A(\pi^*_A(1_{\ia {!^*_X(\mathbb{N})}})))}}^<<<{\qquad\qquad\qquad\dscomment{\text{def. of } \pi^*_A(\mathsf{fst})}} & \ia {\pi^*_A(A)} \ar[uu]^-{\ia {\pi^*_A(f_s^\dagger)}}_-{\qquad\qquad\dscomment{\text{def. of } \pi^*_A(f_s^\dagger)}} \ar[r]^-{\ia {\overline{\pi_A}(A)}} & \ia {A} \ar@/_5.5pc/[uuuul]_>>>>>>>>>>>>>>>>>>>>>>{\ia {f_s^\dagger}} \ar@/_2.75pc/[uuuuu]^>>>>>>>>>>>>>>>>>>>>>>>>>{\ia {f_s}}^>>>>>>>>>>>>>>>>>>>{\dscomment{\text{def. of } f_s^\dagger}\qquad\qquad}
\\
& \ia {\Sigma_A(\pi^*_A(1_{\ia {!^*_X(\mathbb{N})}}))} \ar[urr]^-{\ia {\mathsf{fst}}} \ar@/^1.5pc/[d]^-{\kappa_{A,\pi^*_A(1_{\ia {!^*_X(\mathbb{N})}})}^{-1}}_<<<<{\dscomment{\kappa_{A,\pi^*_A(1_{\ia {!^*_X(\mathbb{N})}})} \text{ is iso.}}\,\,\,}^<<<<{\quad\qquad\dscomment{\text{def. of } \mathsf{fst}}} &
\\
& \ia {\pi^*_A(1_{\ia {!^*_X(\mathbb{N})}})} \ar@/_2pc/[uurr]^-{\pi_{\pi^*_A(1_{\ia {!^*_X(\mathbb{N})}})}}
\\
\ia {!^*_X(\mathbb{N})} \ar@/_4.5pc/[uuurrr]_-{\mathsf{s}(\mathsf{i}_A(g_z,g_s))}
}
\]

\vspace{0.25cm}
\noindent \textbf{Definition~\ref{def:strongsplitfibredweaknaturals} implies weak natural numbers in~\cite{Ahman:FibredEffects}:} 

Given vertical morphisms
\[
f_z : 1_X \longrightarrow (\mathsf{s}(\mathsf{zero}))^*(A)
\qquad
f_s : 1_{\ia A} \longrightarrow \pi^*_A(\ia {!^*_X(\mathsf{succ})}^*(A))
\]
in $\mathcal{V}_X$ and $\mathcal{V}_{\ia A}$, respectively, we aim to define a vertical morphism
\[
\mathsf{i}_A(f_z,f_s) : 1_{\ia {!^*_X(\mathbb{N})}} \longrightarrow A
\]
in $\mathcal{V}_{\ia {!^*_X(\mathbb{N})}}$. 

First, we define morphisms
\[
g_z : 1_{\ia {1_X}} \longrightarrow A
\qquad
g_s : A \longrightarrow A
\]
in $\mathcal{V}$, over $\ia {!^*_X(\mathsf{zero})}$ and $\ia {!^*_X(\mathsf{succ})}$, respectively, by
\[
g_z \defeq \overline{\ia {!^*_X(\mathsf{zero})}}(A) \comp \pi^*_{1_X}(f_z)
\qquad
g_s \defeq \overline{\ia {!^*_X(\mathsf{succ})}}(A) \comp \varepsilon^{\Sigma_A \,\dashv\, \pi^*_A}_{\ia {!^*_X(\mathsf{succ})}^*(A)} \comp \Sigma_A(f_s) \comp \langle \id_A , ! \rangle
\]

Next, according to Definition~\ref{def:strongsplitfibredweaknaturals}, $g_z$ and $g_s$ induce a morphism
\[
\mathsf{rec}(g_z,g_s) : \ia {!^*_X(\mathbb{N})} \longrightarrow \ia {A}
\]
in $\mathcal{B}$, that is the section of $\pi_A$ and makes the following two squares commute:
\[
\xymatrix@C=5em@R=5em@M=0.5em{
\ia {1_X} \ar[r]^-{\ia {!_X^*(\mathsf{zero})}} \ar[d]_-{\eta^{1 \,\dashv\, \ia -}_{\ia {1_X}}} & \ia {!_X^*(\mathbb{N})} \ar[d]_-{\mathsf{rec}(g_z,g_s)} & \ia {!_X^*(\mathbb{N})} \ar[l]_-{\ia {!_X^*(\mathsf{succ})}} \ar[d]^-{\mathsf{rec}(g_z,g_s)}
\\
\ia {1_{\ia {1_X}}} \ar[r]_-{\ia {g_z}} & \ia {A} & \ia {A} \ar[l]^-{\ia {g_s}}
}
\]

As a result, we can now use $\mathsf{rec}(g_z,g_s)$ to define $\mathsf{i}_A(f_z,f_s)$ as
\[
\mathsf{i}_A(f_z,f_s) \defeq \mathsf{s}^{-1}(\mathsf{rec}(g_z,g_s))
\]

Finally, we need to prove that this definition of $\mathsf{i}_A(f_z,f_s)$ satisfies the two equations
\[
(\funsection(!_X^*(\mathsf{zero})))^*(\mathsf{i}_A(f_z,f_s)) = f_z
\qquad
\ia {!_X^*(\mathsf{succ})}^* (\mathsf{i}_A(f_z,f_s)) 
=
(\mathsf{s}(\mathsf{i}_A(f_z,f_s)))^*(f_s) 
\]
in $\mathcal{V}_X$ and $\mathcal{V}_{\ia {!^*_X(\mathbb{N})}}$, respectively, which we use later in Section~\ref{sect:interpretation} to show that the interpretation of eMLTT validates the $\beta$-equations for primitive recursion.

In order to prove that the first of these equations holds in $\mathcal{V}_X$, namely,  
\[
(\funsection(!_X^*(\mathsf{zero})))^*(\mathsf{i}_A(f_z,f_s)) = f_z
\]
we first recast the left-hand side of this equation in $\mathcal{B}$, using Proposition~\ref{prop:reindexinginthebasecategory} and the definition of $\mathsf{i}_A(f_z,f_s)$ from above. In particular, we get that $\mathsf{s}((\funsection(!_X^*(\mathsf{zero})))^*(\mathsf{i}_A(f_z,f_s)))$ is equal to the unique (unnamed) mediating morphism in the next pullback situation.
\[
\xymatrix@C=6em@R=4em@M=0.5em{
& \ia {!_X^*(\mathbb{N})} \ar[r]^-{\eta^{1 \,\dashv\, \ia -}_{\ia {!_X^*(\mathbb{N})}}} & \ia {1_{\ia {!_X^*(\mathbb{N})}}} \ar[d]^{\ia {\mathsf{s}^{-1}(\mathsf{rec}(g_z,g_s))}}_{\dcomment{(e)}\qquad\qquad\qquad\qquad}
\\
X \ar@/_2pc/[dr]_{\id_X} \ar[ur]^{\funsection(!_X^*(\mathsf{zero}))} \ar@{-->}[r] & \ia {(\funsection(!_X^*(\mathsf{zero})))^*(A)} \ar[d]_{\pi_{(\funsection(!_X^*(\mathsf{zero})))^*(A)}}^<{\,\big\lrcorner}_<<<{\dcomment{(f)}\qquad\qquad\qquad} \ar[r]^-{\ia {\overline{\funsection(!_X^*(\mathsf{zero}))}(A)}} & \ia A \ar[d]^{\pi_A}_{\dcomment{\mathcal{P}(\overline{\funsection(!_X^*(\mathsf{zero}))}(A))}\qquad\quad}
\\
& X \ar[r]_-{\funsection(!_X^*(\mathsf{zero}))} & \ia {!_X^*(\mathbb{N})}
}
\]

Now, recalling that $\mathsf{s}$ is an isomorphism, it suffices to show that the equation
\[
\mathsf{s}((\funsection(!_X^*(\mathsf{zero})))^*(\mathsf{i}_A(f_z,f_s))) = \mathsf{s}(f_z)
\]
holds in $\mathcal{B}$, in order to prove that the required equation holds in $\mathcal{V}_X$. We note that this equation  holds because $\mathsf{s}(f_z)$ satisfies the same universal property as the unique unnamed mediating morphism in the above pullback situation, i.e., setting the unnamed morphism to be $\mathsf{s}(f_z)$ makes $(e)$ and $(f)$ commute. 

First, we show the commutativity of $(e)$ by
\[
\scriptsize
\xymatrix@C=5.5em@R=7em@M=0.5em{
\ia {!^*_X(\mathbb{N})} \ar[rrr]^{\eta^{1 \,\dashv\, \ia -}_{\ia {!^*_X(\mathbb{N})}}} \ar@/^8pc/[ddddrrr]^-{\mathsf{rec}(g_z,g_s)} &&& \ia {1_{\ia {!_X^*(\mathbb{N})}}} \ar[dddd]^{\ia {\mathsf{s}^{-1}(\mathsf{rec}(g_z,g_s))}}_<<<<<<<{\dscomment{\text{def. of } \mathsf{s}(\mathsf{s}^{-1}(\mathsf{rec}(g_z,g_s)))}\qquad\qquad}
\\
& & \ia {1_{\ia {1_X}}} \ar[d]^-{=}_>>>{\dscomment{(g)}\qquad} \ar[dddr]^-{\ia {g_z}}_>>>>>>>>>>>>>>>>>>>>>>>>>>>>>>>>>>{\dscomment{\text{def. of } g_z}\qquad}
\\
& & \ia {\pi^*_{1_X}(1_X)} \ar[dl]^<<<<<<<<<<{\ia {\overline{\pi_{1_X}}(1_X)}} \ar[d]^-{\ia {\pi^*_{1_X}(f_z)}}_>>>>{\dscomment{\text{def. of } \pi^*_{1_X}(f_z)}\qquad} &
\\
& \ia {1_X} \ar[dr]_-{\ia {f_z}} \ar[uuul]^-{\ia {!_X^*(\mathsf{zero})}}^<<<<<{\dscomment{\text{def. of } \funsection(!_X^*(\mathsf{zero}))}\quad}_>>>>>>>>>>>>>>>>>>>>>{\qquad\qquad\dscomment{\text{property of } \mathsf{rec}(g_z,g_s)}} \ar[uur]^-{\eta^{1 \,\dashv\, \ia -}_{\ia {1_X}}} & \ia {\pi^*_{1_X}((\funsection(!_X^*(\mathsf{zero})))^*(A))} \ar[d]_<<{\ia {\overline{\pi_{1_X}}((\funsection(!_X^*(\mathsf{zero})))^*(A))}\!\!\!}^-{\!\!\!\!\!\quad\dscomment{p \text{ is a s. fib.}}} \ar[dr]_>>>>>>>>>>{\ia {\overline{\funsection(!_X^*(\mathsf{zero})) \,\comp\, \pi_{1_X} }(A)}\,\,\,\,\,\,\,\,\,}
\\
X \ar[rr]_-{\mathsf{s}(f_z)} \ar[uuuu]^-{\funsection(!_X^*(\mathsf{zero}))} \ar[ur]^-{\eta^{1 \,\dashv\, \ia -}_X\!\!\!\!\!\!}_-{\,\,\,\,\qquad\dscomment{\text{def. of } \mathsf{s}(f_z)}} && \ia {(\funsection(!_X^*(\mathsf{zero})))^*(A)} \ar[r]_-{\ia {\overline{\funsection(!_X^*(\mathsf{zero}))}(A)}} & \ia {A}
}
\]
where we show the commutativity of $(g)$ by
\[
\xymatrix@C=6em@R=6em@M=0.5em{
\ia {1_{\ia {1_X}}} \ar[rr]^-{=} \ar@/^4.5pc/[ddrr]^-{\ia {1(\pi_{1_X})}} &&  \ia {\pi^*_{1_X}(1_X)} \ar[dd]^-{\ia {\overline{\pi_{1_X}}(1_X)}}_<<<<{\dcomment{1 \text{ is split fibred}}\qquad}
\\
& X \ar[dr]^-{\eta^{1 \,\dashv\, \ia -}_X}_-{\dcomment{\eta^{1 \,\dashv\, \ia -}_X \text{ is an iso.}}\qquad} &
\\
\ia {1_X} \ar[uu]^-{\eta^{1 \,\dashv\, \ia -}_{\ia {1_X}}}_>>>>>>>>>>>>>>>>{\qquad\qquad\dcomment{\text{nat. of } \eta^{1 \,\dashv\, \ia -}_{\ia {1_X}}}} \ar[ur]^-{\pi_{1_X}} \ar[rr]_-{\id_{\ia {1_X}}} && \ia {1_X}
}
\]

Second, the commutativity of $(f)$ follows directly from the definition of $\mathsf{s}$. Namely, by definition, the morphism $\mathsf{s}(f_z) : X \longrightarrow \ia {(\funsection(!_X^*(\mathsf{zero})))^*(A)}$ is a section of the projection morphism $\pi_{(\funsection(!_X^*(\mathsf{zero})))^*(A)} : \ia {(\funsection(!_X^*(\mathsf{zero})))^*(A)} \longrightarrow X$, giving us the equation
\[
\pi_{(\funsection(!_X^*(\mathsf{zero})))^*(A)} \comp \mathsf{s}(f_z) =  \id_X
\]

In order to prove that the second required equation holds in $\mathcal{V}_{\ia {!^*_X(\mathbb{N})}}$, namely, 
\[
\ia {!_X^*(\mathsf{succ})}^* (\mathsf{i}_A(f_z,f_s)) 
=
(\mathsf{s}(\mathsf{i}_A(f_z,f_s)))^*(f_s) 
\]
we first recast the left-hand side of this equation in $\mathcal{B}$, using Proposition~\ref{prop:reindexinginthebasecategory} and the definition of $\mathsf{i}_A(f_z,f_s)$ from above. In particular, we get that $\mathsf{s}(\ia {!_X^*(\mathsf{succ})}^* (\mathsf{i}_A(f_z,f_s)))$ is equal to the unique (unnamed) mediating morphism in the next pullback situation.
\[
\xymatrix@C=6em@R=4em@M=0.5em{
& \ia {!_X^*(\mathbb{N})} \ar[r]^-{\eta^{1 \,\dashv\, \ia -}_{\ia {!_X^*(\mathbb{N})}}} & \ia {1_{\ia {!_X^*(\mathbb{N})}}} \ar[d]^{\ia {\mathsf{s}^{-1}(\mathsf{rec}(g_z,g_s))}}_{\dcomment{(h)}\qquad\qquad\qquad\qquad}
\\
\ia {!_X^*(\mathbb{N})} \ar@/_2pc/[dr]_{\id_{\ia {!_X^*(\mathbb{N})}}} \ar[ur]^{\ia {!_X^*(\mathsf{succ})}} \ar@{-->}[r] & \ia {\ia {!_X^*(\mathsf{succ})}^*(A)} \ar[d]_{\pi_{\ia {!_X^*(\mathsf{succ})}^*(A)}}^<{\,\big\lrcorner}_<<<{\dcomment{(i)}\qquad\qquad\qquad} \ar[r]^-{\ia {\overline{\ia {!_X^*(\mathsf{succ})}}(A)}} & \ia A \ar[d]^{\pi_A}_{\dcomment{\mathcal{P}(\overline{\ia {!_X^*(\mathsf{succ})}}(A))}\qquad\quad}
\\
& \ia {!_X^*(\mathbb{N})} \ar[r]_-{\ia {!_X^*(\mathsf{succ})}} & \ia {!_X^*(\mathbb{N})}
}
\]

Next, recalling that $\mathsf{s}$ is an isomorphism and combining this fact with the definition of $\mathsf{i}_A(f_z,f_s)$ from above, it suffices to show that the equation 
\[
\mathsf{s}(\ia {!_X^*(\mathsf{succ})}^* (\mathsf{i}_A(f_z,f_s)))
=
\mathsf{s}((\mathsf{rec}(g_z,g_s))^*(f_s))
\]
holds in $\mathcal{B}$, in order to prove that the required equation holds in $\mathcal{V}_{\ia {!^*_X(\mathbb{N})}}$. We note that this equation holds because $\mathsf{s}((\mathsf{rec}(g_z,g_s))^*(f_s) )$ satisfies the same universal property as the unique unnamed mediating morphism in the above pullback situation, i.e., setting the unnamed morphism to be $\mathsf{s}((\mathsf{rec}(g_z,g_s))^*(f_s))$ makes $(h)$ and $(i)$ commute. 

First, we show the commutativity of  $(h)$ by
\[
\scriptsize
\xymatrix@C=3.5em@R=8em@M=0.5em{
\ia {!^*_X(\mathbb{N})} \ar[rrrr]^-{\eta^{1 \,\dashv\, \ia -}_{\ia {!^*_X(\mathbb{N})}}} \ar@/^8.5pc/[dddddrrrr]^-{\mathsf{rec}(g_z,g_s)} &&&& \ia {1_{\ia {!^*_X(\mathbb{N})}}} \ar[ddddd]_<<<<<<<<<<<<<<<<<<<<<<<{\ia {\mathsf{s}^{-1}(\mathsf{rec}(g_z,g_s))}}_<<<<<<<<<<<{\dscomment{\text{def. of } \mathsf{s}(\mathsf{s}^{-1}(\mathsf{rec}(g_z,g_s)))}\qquad\qquad\qquad\qquad}
\\
& \ia {A} \ar@/^4.5pc/[ddddrrr]^-{\ia {g_s}}
\\
& \ia {1_{\ia {A}}} \ar@/^1pc/[dr]^{\ia {f_s}}
\\
& \ia {(\mathsf{rec}(g_z,g_s))^*(1_{\ia {A}})} \ar[u]_-{\ia {\overline{\mathsf{rec}(g_z,g_s)}(1_{\ia {A}})}} & \ia {\pi^*_A(\ia {!^*_X(\mathsf{succ})} ^*(A))} \ar@/^1pc/[ddrr]_<<<<<<<<<<<<{\ia {\overline{\ia {!^*_X(\mathsf{succ})} \,\comp\, \pi_A}(A)}}
\\
& \ia {1_{\ia {!^*_X(\mathbb{N})}}} \ar[dr]^-{\quad\ia {(\mathsf{rec}(g_z,g_s))^*(f_s)}} \ar[u]^-{=} & \ia {(\mathsf{rec}(g_z,g_s))^*(\pi^*_A(\ia {!^*_X(\mathsf{succ})} ^*(A)))} \ar[u]^-{\ia {\overline{\mathsf{rec}(g_z,g_s)}(\pi^*_A(\ia {!^*_X(\mathsf{succ})} ^*(A)))}}^>>>>>{\dscomment{\text{def. of } (\mathsf{rec}(g_z,g_s))^*(f_s)}\qquad\qquad\,\,} 
\\
\ia {!^*_X(\mathbb{N})} \ar[ur]^-{\!\!\!\!\!\eta^{1 \,\dashv\, \ia -}_{\ia {!^*_X(\mathbb{N})}}}_-{\,\,\,\qquad\quad\dscomment{\text{def. of } \mathsf{s}((\mathsf{rec}(g_z,g_s))^*(f_s))}} \ar[rr]_-{\mathsf{s}((\mathsf{rec}(g_z,g_s))^*(f_s))} \ar[uuuuu]_>>>>>>>>>>>>>>>{\ia {!^*_X(\mathsf{succ})}}_>>>>>>>>>>>>>>>>{\quad\qquad\qquad\dscomment{\text{property of } \mathsf{rec}(g_z,g_s)}} \ar@/^1.8pc/[uuuur]^>>>>>>>>>>>>>>>>>>{\mathsf{rec}(g_z,g_s)}_>>>>>>>>>>>>{\qquad\qquad\dscomment{(j)}} && \ia {\ia {!^*_X(\mathsf{succ})} ^*(A)} \ar[u]_-{=}_-{\qquad\qquad\dscomment{p \text{ is a split fibration}}} \ar[rr]_-{\ia {\overline{\ia {!^*_X(\mathsf{succ})}}(A)}} && \ia {A}
}
\vspace{1cm}
\]
where we show the commutativity of $(j)$ by
\[
\scriptsize
\xymatrix@C=6em@R=4em@M=0.5em{
\ia {!^*_X(\mathbb{N})} \ar[ddd]_-{\eta^{1 \,\dashv\, \ia -}_{\ia {!^*_X(\mathbb{N})}}}^-{\quad\qquad\dscomment{\text{nat. of } \eta^{1 \,\dashv\, \ia -}}} \ar[r]^-{\mathsf{rec}(g_z,g_s)} & \ia {A} \ar[rr]^-{\ia {g_s}} \ar[dddd]_-{\eta^{1 \,\dashv\, \ia -}_{\ia {A}}} \ar[dr]^-{\ia {\langle \id_A , ! \rangle}} && \ia {A}
\\
&& \ia {\Sigma_A(\pi^*_A(1_{\ia {!^*_X(\mathbb{N})}}))} \ar[d]^-{=}^-{\qquad\qquad\qquad\dscomment{\text{def. of } g_s}}_>{\dscomment{(k)}\qquad\qquad\qquad}
\\
&& \ia {\Sigma_A(1_{\ia {A}})} \ar[d]^-{\ia {\Sigma_A(f_s)}}
\\
\ia {1_{\ia {!^*_X(\mathbb{N})}}} \ar[d]_-{=}^>>>>>{\quad\dscomment{1 \text{ is s. fib.}}} \ar[dr]^-{\ia {1(\mathsf{rec}(g_z,g_s))}} && \ia {\Sigma_A(\pi^*_A(\ia {!^*_X(\mathsf{succ})}^*(A)))} \ar[dr]_<<<<<<<<<{\ia {\varepsilon^{\Sigma_A \,\dashv\, \pi^*_A}_{\ia {!^*_X(\mathsf{succ})}^*(A)}}}
\\
\ia {(\mathsf{rec}(g_z,g_s))^*(1_{\ia {A}})} \ar[r]_-{\ia {\overline{\mathsf{rec}(g_z,g_s)}(1_{\ia A})}} & \ia {1_{\ia {A}}} \ar[r]_-{\ia {f_s}} & \ia {\pi^*_A(\ia {!^*_X(\mathsf{succ})} ^*(A))} \ar[r]_-{\ia {\overline{\pi_A}(\ia {!^*_X(\mathsf{succ})} ^*(A))}} & \ia {\ia {!^*_X(\mathsf{succ})} ^*(A)} \ar[uuuu]^-{\overline{\ia {!^*_X(\mathsf{succ})}}(A)}
}
\]
and the commutativity of $(k)$ by
\[
\scriptsize
\xymatrix@C=5.75em@R=6em@M=0.5em{
\ia {A} \ar[rr]^-{\ia {\langle \id_A , ! \rangle}} \ar[dd]_-{\eta^{1 \,\dashv\, \ia -}_{\ia A}}^-{\quad\dscomment{\eta^{1 \,\dashv\, \ia -}_{\ia A} \text{ satisfies same univ. prop. as } h}}^-{\qquad\qquad\qquad\qquad\qquad\qquad\qquad\qquad\qquad\qquad\qquad\qquad\dscomment{\text{def. of } \kappa_{A,\pi^*_A(1_{\ia {!^*_X(\mathbb{N})}})}}} \ar[dr]^-{h} \ar@{}[d]^<<<<<<<<{\!\!\!\!\qquad\qquad\qquad\qquad\qquad\qquad\qquad\qquad\dscomment{\text{def. of } \langle \id_A , ! \rangle}} && \ia {\Sigma_A(\pi^*_A(1_{\ia {!^*_X(\mathbb{N})}}))} \ar[d]_-{\id_{\ia {\Sigma_A(\pi^*_A(1_{\ia {!^*_X(\mathbb{N})}}))}}}
\\
& \ia {\pi^*_A(1_{\ia {!^*_X(\mathbb{N})}})} \ar[ur]^-{\kappa_{A,\pi^*_A(1_{\ia {!^*_X(\mathbb{N})}})}} \ar[dl]^-{=} \ar[d]_>>>>>>>{\ia {\eta^{\Sigma_A \,\dashv\, \pi^*_A}_{\ia {\pi^*_A(1_{\ia {!^*_X(\mathbb{N})}})}}}\!\!\!} \ar@{}[dd]^-{\quad\qquad\qquad\qquad\dscomment{1 \text{ is split fibred}}}_-{\dscomment{1 \text{ is split fibred}}\quad\qquad\qquad\qquad} & \ia {\Sigma_A(\pi^*_A(1_{\ia {!^*_X(\mathbb{N})}}))} \ar[d]^-{=}
\\
\ia {1_{\ia A}} \ar[dd]_-{\ia {f_s}} \ar[dr]_-{\ia {\eta^{\Sigma_A \,\dashv\, \pi^*_A}_{1_{\ia A}}}} & \ia {\pi^*_A(\Sigma_A(\pi^*_A(1_{\ia {!^*_X(\mathbb{N})}})))} \ar[ur]_-{\qquad\ia {\overline{\pi_A}(\Sigma_A(\pi^*_A(1_{\ia {!^*_X(\mathbb{N})}})))}} \ar[d]_-{=} & \ia {\Sigma_A (1_{\ia A})} \ar[ddd]^-{\ia {\Sigma_A (f_s)}}
\\
& \ia {\pi^*_A(\Sigma_A(1_{\ia A}))} \ar[ur]_-{\quad\ia {\overline{\pi_A}(\Sigma_A(1_{\ia A}))}} \ar[d]^-{\ia {\pi^*_A(\Sigma_A(f_s))}}^-{\qquad\qquad\qquad\qquad\dscomment{\text{def. of } \pi^*_A(\Sigma_A(f_s))}}_<<<<<{\dscomment{\text{nat. of } \eta^{\Sigma_A \,\dashv\, \pi^*_A}}\qquad\qquad\qquad\quad} &
\\
\ia {\pi^*_A(\ia {!^*_X(\mathsf{succ})} ^*(A))} \ar[d]_-{\id_{\ia {\pi^*_A(\ia {!^*_X(\mathsf{succ})} ^*(A))}}}^<<<<<{\qquad\quad\dscomment{\Sigma \dashv \pi^*_A}} \ar[r]^-{\ia {\eta^{\Sigma_A \,\dashv\, \pi^*_A}_{\pi^*_A(\ia {!^*_X(\mathsf{succ})} ^*(A))}}} & \ia {\ia {\pi^*_A(\Sigma_A(\pi^*_A(\ia {!^*_X(\mathsf{succ})}^*(A))))}} \ar[dr]_>>>>>>>>>>>>>>>>{\ia {\overline{\pi_A}(\Sigma_A(\pi^*_A(\ia {!^*_X(\mathsf{succ})}^*(A))))}\qquad\quad} \ar[dl]^>>>>>>>>>>>>>>>>>>{\ia {\pi^*_A(\varepsilon^{\Sigma_A \,\dashv\, \pi^*_A}_{\ia {!^*_X(\mathsf{succ})}^*(A)})}}^<<<<<<<{\qquad\qquad\dscomment{\text{def. of } \pi^*_A(\varepsilon^{\Sigma_A \,\dashv\, \pi^*_A}_{\ia {!^*_X(\mathsf{succ})}^*(A)})}} &
\\
\ia {\pi^*_A(\ia {!^*_X(\mathsf{succ})} ^*(A))} \ar[r]_-{\ia {\overline{\pi_A}(\ia {!^*_X(\mathsf{succ})} ^*(A))}} & \ia {\ia {!^*_X(\mathsf{succ})} ^*(A)} & \ia {\Sigma_A(\pi^*_A(\ia {!^*_X(\mathsf{succ})}^*(A)))} \ar[l]^-{\ia {\varepsilon^{\Sigma_A \,\dashv\, \pi^*_A}_{\ia {!^*_X(\mathsf{succ})}^*(A)}}}
}
\]

\pagebreak\noindent
and where $h$ is defined as the unique mediating morphism into the pullback square given by $\mathcal{P}(\overline{\pi_A}(1_{\ia {!^*_X(\mathbb{N})}}))$, for $\ia {!} : \ia {A} \longrightarrow \ia {1_{\ia {!_X^*(\mathbb{N})}}}$ and $\id_{\ia A} : \ia {A} \longrightarrow \ia {A}$. The proof that $\eta^{1 \,\dashv\, \ia -}_{\ia A}$ is equal to $h$ can be found in the proof of Proposition~\ref{prop:semsubstintoweakenedterm}.

Finally, we note that the commutativity of $(i)$ follows directly from the definition of $s$. In particular, we know by definition that the morphism 
\[
\mathsf{s}((\mathsf{rec}(g_z,g_s))^*(f_s)) : \ia {!^*_X(\mathbb{N})} \longrightarrow \ia {\ia {!^*_X(\mathsf{succ})}^*(A)}
\]
is a section of $\pi_{\ia {!^*_X(\mathsf{succ})}^*(A)}$, giving us the required equation
\[
\pi_{\ia {!^*_X(\mathsf{succ})}^*(A)} \comp \mathsf{s}((\mathsf{rec}(g_z,g_s))^*(f_s)) = \id_{\ia {!^*_X(\mathbb{N})}}
\]
\end{proof}

\section{Proof of Proposition~\ref{prop:strengthofsplitfibredmonads}}
\label{sect:proofofprop:strengthofsplitfibredmonads}

{
\renewcommand{\thetheorem}{\ref{prop:strengthofsplitfibredmonads}}
\begin{proposition}
Given a split comprehension category with unit $p : \mathcal{V} \longrightarrow \mathcal{B}$ with strong split dependent sums and a split fibred monad $\mathbf{T} = (T,\eta,\mu)$ on it, then there exists a family of natural transformations 
\[
\sigma_A : \Sigma_A \comp T \longrightarrow T \comp \Sigma_A \qquad\qquad\qquad (A \in \mathcal{V})
\]
collectively called the \emph{dependent strength} of $\mathbf{T}$, satisfying the diagrams $(1)$--$(4)$.
\end{proposition}
\addtocounter{theorem}{-1}
}

\noindent
\textit{Proof.} 
Before proving that the four diagrams $(1)$--$(4)$ given in the proposition commute, we first show that the natural transformation $\alpha_{A,B}$ given in the proposition \linebreak is indeed a natural isomorphism. To this end, we first define its candidate inverse \linebreak 
$\alpha^{-1}_{A,B} : \Sigma_A \comp \Sigma_B \comp \kappa^*_{A,B} \longrightarrow \Sigma_{\Sigma_A(B)}$ as the following composite natural transformation:
\[
\xymatrix@C=13em@R=6em@M=0.5em{
\Sigma_A \comp \Sigma_B \comp \kappa^*_{A,B} \ar[r]^-{\Sigma_A \,\comp\, \Sigma_B \,\comp\, \kappa^*_{A,B} \,\comp\, \eta^{\Sigma_{\Sigma_A(B)} \,\dashv\, \pi^*_{\Sigma_A(B)}}} & \Sigma_A \comp \Sigma_B \comp \kappa^*_{A,B} \comp \pi^*_{\Sigma_A(B)} \comp \Sigma_{\Sigma_A(B)} \ar[d]^-{=}
\\
\Sigma_{\Sigma_A(B)}& \Sigma_A \comp \Sigma_B \comp \pi^*_B \comp \pi^*_A \comp \Sigma_{\Sigma_A(B)} \ar[l]^-{\varepsilon^{\Sigma_A \,\comp \Sigma_B \,\dashv\, \pi^*_B \,\comp\, \pi^*_A} \,\comp\, \Sigma_{\Sigma_A(B)}}
}
\]

\pagebreak

For better readability, we often omit the subscripts from $\kappa^*_{A,B}$ and $(\kappa^{-1}_{A,B})^*$ in the diagrams given below. For the same reason, we also often abbreviate the four functors $\pi^*_B \comp \pi^*_A$, $\Sigma_A \comp \Sigma_B$, $\pi^*_{\Sigma_A(B)}$, and $\Sigma_{\Sigma_A(B)}$ as $\pi^*$, $\Sigma$, $\pi'^*$, and $\Sigma'$, respectively. 

We also note that most of the equality morphisms used in the diagrams given below are induced by reindexing along the following commuting diagram:
\[
\xymatrix@C=6em@R=7em@M=0.5em{
\ia {B} \ar[r]_-{\ia {\eta^{\Sigma_A \,\dashv\, \pi^*_A}_B}} \ar[dr]_-{\pi_B} \ar@/^2.25pc/[rr]^-{\kappa_{A,B}}_{\dcomment{\text{def. of } \kappa_{A,B}}} & \ia {\pi^*_A(\Sigma_A(B))} \ar[r]_-{\ia {\overline{\pi_A}(\Sigma_A(B))}} \ar[d]^-{\pi_{\pi^*_A(\Sigma_A(B))}}_<<<<<<<<<<{\dcomment{\mathcal{P}(\eta^{\Sigma_A \,\dashv\, \pi^*_A}_B)}\,\,\,\,\,\,} & \ia {\Sigma_A(B)} \ar[d]^-{\pi_{\Sigma_A(B)}}_-{\dcomment{\mathcal{P}(\overline{\pi_A}(\Sigma_A(B)))}\quad\,\,\,\,\,} \ar@/_4.8pc/[ll]_-{\kappa^{-1}_{A,B}}^-{\dcomment{\kappa_{A,B} \text{ is an iso.}}}
\\
& \ia A \ar[r]_-{\pi_A} & p(A)
}
\]

We now return to proving that $\alpha_{A,B}$ is a natural isomorphism, by showing that the following two equations hold:
\[
\begin{array}{c}
\alpha^{-1}_{A,B} \comp \alpha_{A,B} = \id_{\Sigma_{\Sigma_A(B)}}
\\[5mm]
\alpha_{A,B} \comp \alpha^{-1}_{A,B} = \id_{\Sigma_A \,\comp\, \Sigma_B \,\comp\, \kappa^*_{A,B}}
\end{array}
\]
We prove these equations by showing that the following two diagrams commute:

\mbox{}

\[
\xymatrix@C=5em@R=6em@M=0.5em{
\Sigma' \ar[r]^-{=} \ar[dddddd]_-{\id_{\Sigma'}}^>>>>>>>>>>>>>>>>>>>>>>>>>>>>>>>>{\qquad\quad\dcomment{\Sigma' \,\dashv\, \pi'^*}} \ar@/_8pc/[dddddr]^<<<<<<<<<<<<<<<<<<<<<<<<{\eta^{\Sigma' \,\dashv\, \pi'^*} \,\comp\, \Sigma'} & \Sigma' \comp (\kappa^{-1})^* \comp \kappa^* \ar[r]^-{\Sigma' \,\comp\, (\kappa^{-1})^* \,\comp\, \eta^{\Sigma \,\dashv\, \pi^*} \,\comp\, \kappa^*} \ar[d]^<<<<<{\Sigma_A \,\comp\, (\kappa^{-1})^* \,\comp\, \kappa^* \comp \eta^{\Sigma' \,\dashv\, \pi'^*}}_-{\dcomment{p \text{ is a split fibration}}\qquad}^-{\qquad\qquad\dcomment{\text{nat. of } \eta^{\Sigma \,\dashv\, \pi^*}}}^>>>>>{\!\!\!\qquad\qquad\Sigma' \,\comp\, (\kappa^{-1})^* \,\comp\, \pi^* \,\comp\, \Sigma \,\comp\, \kappa^* \,\comp\, \eta^{\Sigma' \,\dashv\, \pi'^*}} & \Sigma' \comp (\kappa^{-1})^* \comp \pi^* \comp \Sigma \comp \kappa^* \ar[dd]^-{=} \ar@/^4pc/[ddl]
\\
& \Sigma' \comp (\kappa^{-1})^* \comp \kappa^* \comp \pi'^* \comp \Sigma' \ar@/_0.5pc/[d]^<<<<<<{\Sigma' \,\comp\, (\kappa^{-1})^* \,\comp\, \eta^{\Sigma \,\dashv\, \pi^*} \,\comp\, \kappa^* \,\comp\, \pi'^* \,\comp\, \Sigma'} \ar@/_6pc/[dddd]^-{=} & 
\\
& \txt<7pc>{$\Sigma' \comp (\kappa^{-1})^* \comp \pi^* \comp$\\$\Sigma \comp \kappa^* \comp\pi'^* \comp \Sigma'$}  \ar[d]_{=}_>>>>>>>{\dcomment{p \text{ is a s. fib.}}\quad\!\!\!}^<<<{\qquad\qquad\dcomment{p \text{ is a split fibration}}} & \Sigma' \comp \pi'^* \comp \Sigma \comp \kappa^* \ar[d]^-{\varepsilon^{\Sigma' \,\dashv\, \pi'^*} \,\comp\, \Sigma \,\comp\, \kappa^*}_>>>>>{\dcomment{\text{nat. of } \varepsilon^{\Sigma' \,\dashv\, \pi'^*}}\quad} \ar@/^2pc/[dl]_-{\Sigma' \,\comp\, \pi'^* \,\comp\, \Sigma \,\comp\, \kappa^* \,\comp\, \eta^{\Sigma' \,\dashv\, \pi'^*}}
\\
& \txt<6pc>{$\Sigma' \comp \pi'^* \comp$\\$\Sigma \comp \kappa^* \comp\pi'^* \comp \Sigma'$} \ar@/^1.75pc/[ddr]^<<<<<<<{\,\,\,\,\,\,\,\,\varepsilon^{\Sigma' \,\dashv\, \pi'^*} \,\comp\, \Sigma \,\comp\, \kappa^* \,\comp\, \pi'^* \,\comp\, \Sigma'} \ar[d]_-{=}_<<<<<{\dcomment{\Sigma \,\dashv\, \pi^*}\quad\,\,\,\,\,\,}^>>>{\qquad\dcomment{p \text{ is a s. fib.}}} & \Sigma \comp \kappa^* \ar[dd]^-{\Sigma \,\comp\, \kappa^* \,\comp\, \eta^{\Sigma' \,\dashv\, \pi'^*}}
\\
& \txt<4.5pc>{$\Sigma' \comp \pi'^* \comp$\\$\Sigma \comp \pi^* \comp \Sigma'$} \ar@/^2pc/[ddr]_>>>>>>>>>{\varepsilon^{\Sigma' \,\dashv\, \pi'^*} \,\comp\, \Sigma \,\comp\, \pi^* \,\comp\, \Sigma'} \ar[d]^>>>>>>{\Sigma' \,\comp\, \pi'^* \,\comp\, \varepsilon^{\Sigma \,\dashv\, \pi^*} \,\comp\, \Sigma'} & 
\\
& \Sigma' \comp \pi'^* \comp \Sigma' \ar[dl]^-{\varepsilon^{\Sigma' \,\dashv\, \pi'^*} \,\comp\, \Sigma'}^<<<<<<{\qquad\qquad\quad\dcomment{\text{nat. of } \varepsilon^{\Sigma' \,\dashv\, \pi'^*}}} & \Sigma \comp \kappa^* \comp \pi'^* \comp \Sigma' \ar[d]^-{=}
\\
\Sigma' & & \Sigma \comp \pi^* \comp \Sigma' \ar[ll]^-{\varepsilon^{\Sigma \,\dashv\, \pi^*} \,\comp\, \Sigma'}
}
\]

\pagebreak

\mbox{}

\vspace{-1cm}

\[
\xymatrix@C=5em@R=5em@M=0.5em{
\Sigma \comp \kappa^* \ar[r]^-{\Sigma \,\comp\, \kappa^* \,\comp\, \eta^{\Sigma' \,\dashv\, \pi'^*}} \ar[ddddddd]_-{\id_{\Sigma \,\comp\, \kappa^*}} \ar@/_3pc/[ddddddr]_>>>>>>>>>>>>>>>>>>>>>>>>>>>>>>{\Sigma \,\comp\, \eta^{\Sigma \,\dashv\, \pi^*} \,\comp\, \kappa^*}_>{\dcomment{\Sigma \,\dashv\, \pi^*}\qquad\qquad\quad} & \Sigma \comp \kappa^* \comp \pi'^* \comp \Sigma' \ar[r]^-{=} \ar[d]^-{=}^<<<<{\quad\dcomment{p \text{ is a split fibration}}}_-{\dcomment{(a)}\qquad\qquad\quad\!\!\!\!} & \Sigma \comp \pi^* \comp \Sigma' \ar[d]^{\varepsilon^{\Sigma \,\dashv\, \pi^*} \,\comp\, \Sigma'} \ar[dl]^-{=}
\\
& \Sigma \comp \pi^* \comp \Sigma' \comp (\kappa^{-1})^* \comp \kappa^* \ar@/^2pc/[ddr]_<<<<<<<<<<<<<<<<<<<<<{\varepsilon^{\Sigma \,\dashv\, \pi^*} \,\comp\, \Sigma' \,\comp\, (\kappa^{-1})^* \,\comp\, \kappa^*} \ar[dd]^>>>>>>>>{\Sigma \,\comp\, \pi^* \,\comp\, \Sigma' \,\comp\, (\kappa^{-1})^* \,\comp\, \eta^{\Sigma \,\dashv\, \pi^*} \,\comp\, \kappa^*}^>>>>>>>>>>>>>>>>>{\qquad\quad\dcomment{\text{nat. of } \varepsilon^{\Sigma \,\dashv\, \pi^*}}} & \Sigma' \ar[dd]^{=}_<<<<{\dcomment{p \text{ is a split fibration}}\qquad}
\\
& &
\\
& \txt<7pc>{$\Sigma \comp \pi^* \comp \Sigma' \comp $\\$ (\kappa^{-1})^* \comp \pi^* \comp \Sigma \comp \kappa^*$} \ar@/_2pc/[ddr]^<<<<<<<<<{\varepsilon^{\Sigma \,\dashv\, \pi^*} \,\comp\, \Sigma' \,\comp\, (\kappa^{-1})^* \,\comp\, \pi'^* \,\comp\, \Sigma' \,\comp\, \kappa^*} \ar@/_1pc/[d]^-{=} & \Sigma' \comp (\kappa^{-1})^* \comp \kappa^* \ar[dd]_-{\Sigma' \,\comp\, (\kappa^{-1})^* \,\comp\, \eta^{\Sigma \,\dashv\, \pi^*} \,\comp\, \kappa^*}
\\
& \txt<5pc>{$\Sigma \comp \pi^* \comp \Sigma' \comp $\\$ \pi'^* \comp \Sigma \comp \kappa^*$} \ar@/^1.75pc/[dddr]_>>>>>>>>>>>>{\varepsilon^{\Sigma \,\dashv\, \pi^*} \,\comp\, \Sigma' \,\comp\, \pi'^* \,\comp\, \Sigma \,\comp\, \pi^*} \ar[dd]^>>>>>>>>>>>{\Sigma \,\comp\, \pi^* \,\comp\, \varepsilon^{\Sigma' \,\dashv\, \pi'^*} \,\comp\, \Sigma \,\comp\, \kappa^*} &
\\
& & \Sigma' \comp (\kappa^{-1})^* \comp \pi^* \comp \Sigma \comp \kappa^* \ar[dd]^-{=}_<<<<<{\dcomment{p \text{ is a s. fib.}}\,\,\,\,\,\,\,}
\\
& \Sigma \comp \pi^* \comp \Sigma \comp \kappa^* \ar[dl]^-{\varepsilon^{\Sigma \,\dashv\, \pi^*} \,\comp\, \Sigma \,\comp\, \kappa^*}^<<<<<<<<<<<{\qquad\qquad\quad\dcomment{\text{nat. of } \varepsilon^{\Sigma \,\dashv\, \pi^*}}} & 
\\
\Sigma \comp \kappa^* & & \Sigma' \comp \pi'^* \comp \Sigma \comp \kappa^* \ar[ll]^-{\varepsilon^{\Sigma' \,\dashv\, \pi'^*} \,\comp\, \Sigma \,\comp\, \kappa^*}
}
\vspace{0.5cm}
\]

\noindent
We conclude the proof of these two isomorphism equations by showing that the subdiagram marked with $(a)$ commutes. Its commutativity is proved as follows:

\[
\scriptsize
\xymatrix@C=4em@R=7em@M=0.5em{
\Sigma \comp \kappa^* \comp \pi'^* \comp \Sigma' \ar[rrr]^-{=} \ar[drr]^-{=}^-{\qquad\qquad\qquad\qquad\qquad\qquad\dscomment{p \text{ is a split fibration}}} & & & \txt<4.75pc>{$\Sigma \comp \pi^* \comp \Sigma'  \comp $\\$ (\kappa^{-1})^* \comp \kappa^*$} \ar@/^2pc/[ddddd]_>>>>>>>>>>>>>>>>>>>>>>>>>>>>>>>>>>>>{\Sigma \,\comp\, \pi^* \,\comp\, \Sigma' \,\comp\, (\kappa^{-1})^* \,\comp\, \eta^{\Sigma \,\dashv\, \pi^*} \!\comp\, \kappa^*}
\\
\Sigma \comp \kappa^* \ar[dddddd]_-{\Sigma \,\comp\, \eta^{\Sigma \,\dashv\, \pi^*} \,\comp\, \kappa^*} \ar[u]^-{\Sigma \,\comp\, \kappa^* \,\comp\, \eta^{\Sigma' \,\dashv\, \pi'^*}}_<{\quad\qquad\qquad\qquad\dscomment{p \text{ is a split fibration}}} \ar[dr]_-{=} & & \txt<6pc>{$\Sigma \comp \kappa^* \comp \pi'^* \comp $\\$ \Sigma' \comp (\kappa^{-1})^* \comp \kappa^*$} \ar[ur]^-{=} \ar@/^5pc/[ddd]_<<<<<<<<<<<<<<<<<<<<<<<<<<{\Sigma \,\comp\, \kappa^* \,\comp\, \pi'^* \,\comp\, \Sigma' \,\comp\, (\kappa^{-1})^* \,\comp\, \eta^{\Sigma \,\dashv\, \pi^*} \,\comp\, \kappa^*}^-{\,\,\,\quad\dscomment{p \text{ is a split fibration}}} & 
\\
& \txt<4pc>{$\Sigma \comp \kappa^* \comp $\\$ (\kappa^{-1})^* \comp \kappa^*$} \ar[ur]^>>>>>>>>>>{\Sigma \,\comp\, \kappa^* \,\comp\, \eta^{\Sigma' \,\dashv\, \pi'^*} \,\comp\, (\kappa^{-1})^* \,\comp\, \kappa^*} \ar[d]^>>>>>>>{\Sigma \,\comp\, \kappa^* \,\comp\, (\kappa^{-1})^* \,\comp\, \eta^{\Sigma \,\dashv\, \pi^*} \,\comp\, \kappa^*}_-{\dscomment{p \text{ is a split fibration}}\qquad\,\,\,\,\,\,}^>>>>>>>>>>>>>{\qquad\qquad\qquad\quad\dscomment{\text{nat. of } \eta^{\Sigma' \,\dashv\, \pi'^*}}} & & 
\\
& \txt<4pc>{$\Sigma \comp \kappa^* \comp $\\$ (\kappa^{-1})^* \comp \pi^* \comp \Sigma \comp \kappa^*$} \ar@/_1pc/[dr]^<<<<{\Sigma \,\comp\, \kappa^* \,\comp\, \eta^{\Sigma' \,\dashv\, \pi'^*} \,\comp\, (\kappa^{-1})^*  \,\comp\, \pi^* \,\comp\, \Sigma \,\comp\, \kappa^*} \ar[d]^-{=}^>{\,\,\,\,\,\qquad\dscomment{p \text{ is a s. fib.}}} & &
\\
& \txt<4pc>{$\Sigma \comp \kappa^* \comp $\\$ \pi'^* \comp \Sigma \comp \kappa^*$} \ar@/^1pc/[dr]_>>>>>>{\Sigma \,\comp\, \kappa^* \,\comp\, \eta^{\Sigma' \,\dashv\, \pi'^*} \,\comp\, \pi'^* \,\comp\, \Sigma \,\comp\, \kappa^*\!\!\!\!} \ar@/_4pc/[ddr]_-{\id_{\Sigma \,\comp\, \kappa^* \,\comp\, \pi'^* \,\comp\, \Sigma \,\comp\, \kappa^*}} & \txt<6pc>{$\Sigma \comp \kappa^* \comp \pi'^* \comp $\\$ \Sigma' \comp (\kappa^{-1})^* \comp \pi^* \comp \Sigma \comp \kappa^*$} \ar[dr]^-{=} \ar[d]^-{=} &
\\
& & \txt<5pc>{$\Sigma \comp \kappa^* \comp $\\$ \pi'^* \comp \Sigma' \comp \pi'^* \comp \Sigma \comp \kappa^*$} \ar[d]^-{\Sigma \,\comp\, \kappa^* \,\comp\, \pi'^* \,\comp\, \varepsilon^{\Sigma' \,\dashv\, \pi'^*} \,\comp\, \Sigma \,\comp\, \kappa^*}_-{\dscomment{\Sigma' \,\dashv\, \pi'^*}\qquad} & \txt<6pc>{$\Sigma \comp \pi^* \comp \Sigma' \comp $\\$ (\kappa^{-1})^* \comp \pi^* \comp \Sigma \comp \kappa^*$} \ar[dd]^-{=}
\\
& & \txt<4.5pc>{$\Sigma \comp \kappa^* \comp $\\$ \pi'^* \comp \Sigma \comp \kappa^*$} \ar[dll]^-{=}^-{\qquad\qquad\qquad\qquad\qquad\qquad\qquad\qquad\dscomment{p \text{ is a split fibration}}} & 
\\
\Sigma \comp \pi^* \comp \Sigma \comp \kappa^* & & & \txt<5pc>{$\Sigma \comp \pi^* \comp \Sigma' \comp $\\$ \pi'^* \comp \Sigma \comp \kappa^*$} \ar[lll]^-{\Sigma \,\comp\, \pi^* \,\comp\, \varepsilon^{\Sigma' \,\dashv\, \pi'^*} \,\comp\, \Sigma \,\comp\, \kappa^*}
}
\]

Next, we show that the four diagrams $(1)$--$(4)$ commute. We again omit the subscripts from $\kappa^*_{A,B}$ and $(\kappa^{-1}_{A,B})^*$, and abbreviate the functors $\pi^*_B \comp \pi^*_A$, $\Sigma_A \comp \Sigma_B$, $\pi^*_{\Sigma_A(B)}$, and $\Sigma_{\Sigma_A(B)}$ as $\pi^*$, $\Sigma$, $\pi'^*$, and $\Sigma'$, respectively. In order to further optimise the size of the proof of diagram $(2)$, we instead prove the commutativity of an equivalent diagram, in which we have replaced $\alpha_{A,B,T(C)}$ and $T(\alpha_{A,B,C})$ with their respective inverses.

First, diagram $(1)$ commutes because we have
\[
\scriptsize
\xymatrix@C=8em@R=7em@M=0.5em{
\Sigma_{1_{p(A)}}(\pi^*_{1_{p(A)}}(T(A))) \ar@/_9pc/[ddddrr]_-{\varepsilon^{\Sigma_{1_{p(A)}} \!\!\dashv\, \pi^*_{1_{p(A)}}}_{T(A)}\!\!\!\!\!\!\!\!\!\!} \ar@/_3.5pc/[dddr]^>>>>>>>>>>>>>{\!\!\!\!\!\!\!\!\!\!\!\id_{\Sigma_{1_{p(A)}}(\pi^*_{1_{p(A)}}(T(A)))}} \ar@/_3pc/[ddr]_-{=}_>>>>{\dscomment{T \text{ is s. fib.}}\qquad\quad} \ar[r]^-{=} & \Sigma_{1_{p(A)}}(T(\pi^*_{1_{p(A)}}(A))) \ar[d]^>>>>>>{\Sigma_{1_{p(A)}}(T(\eta^{\Sigma_{1_{p(A)}} \!\!\dashv\, \pi^*_{1_{p(A)}}}_{\pi^*_{1_{p(A)}}(A)}))}^<<<<<{\quad\dscomment{\text{def. of } \sigma_{1_{p(A)},\pi^*_{1_{p(A)}}(A)}}} \ar[r]^-{\sigma_{1_{p(A)},\pi^*_{1_{p(A)}}(A)}} \ar@/_4.5pc/[dd]_<<<<<<<<<<{\id_{\Sigma_{1_{p(A)}}(T(\pi^*_{1_{p(A)}}(A)))}\!\!\!\!}_-{\dscomment{\text{id. law}}\quad}^<<<<<<<<<<<<<<{\,\,\dscomment{\Sigma_{1_{p(A)}} \,\dashv\, \pi^*_{1_{p(A)}}}} & T(\Sigma_{1_{p(A)}}(\pi^*_{1_{p(A)}}(A))) \ar@/^4pc/[dddd]_-{T(\varepsilon^{\Sigma_{1_{p(A)}} \!\!\dashv\, \pi^*_{1_{p(A)}}}_A)}_>>>>>>>>>>>>>>>>>>>>{\dscomment{\text{nat. of } \varepsilon^{\Sigma_{1_{p(A)}} \!\!\dashv\, \pi^*_{1_{p(A)}}}}\qquad\qquad}
\\
& \txt<5pc>{$\Sigma_{1_{p(A)}}(T(\pi^*_{1_{p(A)}}($\\$\Sigma_{1_{p(A)}}(\pi^*_{1_{p(A)}}(A)))))$} \ar[d]^-{\Sigma_{1_{p(A)}}(T(\pi^*_{1_{p(A)}}(\varepsilon^{\Sigma_{1_{p(A)}} \!\!\dashv\, \pi^*_{1_{p(A)}}}_{A})))} \ar[r]_-{=} & \txt<5pc>{$\Sigma_{1_{p(A)}}(\pi^*_{1_{p(A)}}(T($\\$\Sigma_{1_{p(A)}}(\pi^*_{1_{p(A)}}(A)))))$} \ar@/^3pc/[ddl]^>>>>>>>>>>>>>>{\!\!\!\!\Sigma_{1_{p(A)}}(\pi^*_{1_{p(A)}}(T(\varepsilon^{\Sigma_{1_{p(A)}} \!\!\dashv\, \pi^*_{1_{p(A)}}}_{A})))}_>>>>>>>>>>>>>>>>>{\dscomment{T \text{ is split fibred}}\qquad} \ar[u]^>>>>>>{\varepsilon^{\Sigma_{1_{p(A)}} \!\!\dashv\, \pi^*_{1_{p(A)}}}_{T(\Sigma_{1_{p(A)}}(\pi^*_{1_{p(A)}}(A)))}}
\\
& \Sigma_{1_{p(A)}}(T(\pi^*_{1_{p(A)}}(A))) \ar@/^1pc/[d]^-{=} & 
\\
& \Sigma_{1_{p(A)}}(\pi^*_{1_{p(A)}}(T(A))) \ar[dr]^-{\varepsilon^{\Sigma_{1_{p(A)}} \!\!\dashv\, \pi^*_{1_{p(A)}}}_{T(A)}}_<<<<{\dscomment{\text{id. law}}\qquad\quad}
\\
& & T(A)
}
\]

Next, diagram $(3)$, which we prove before diagram $(2)$ for better layout, commutes because we have
\[
\scriptsize
\xymatrix@C=6em@R=7em@M=0.5em{
& \Sigma_A(B) \ar[r]^-{\Sigma_A(\eta_B)} \ar[d]^<<<<<{\Sigma_A(\eta^{\Sigma_A \,\dashv\, \pi^*_A}_B)}^-{\,\,\,\,\,\qquad\qquad\dscomment{\text{nat. of } \eta}} \ar@/_2.5pc/[dddl]_-{\id_{\Sigma_A(B)}} & \Sigma_A(T(B)) \ar@/^2.5pc/[dddr]^-{\sigma_{A,B}} \ar[d]_>>>>>{\Sigma_A(T(\eta^{\Sigma_A \,\dashv\, \pi^*_A}_B))}
\\
& \Sigma_A(\pi^*_A(\Sigma_A(B))) \ar[ddl]^-{\!\!\!\!\varepsilon^{\Sigma_A \,\dashv\, \pi^*_A}_{\Sigma_A(B)}}_<<<<<<<<<<<<<{\dscomment{\Sigma_A \,\dashv\, \pi^*_A}\quad\,\,\,\,} \ar[dr]_-{\Sigma_A(\pi^*_A(\eta_{\Sigma_A(B)}))\,\,\,\,\,} \ar[r]_-{\Sigma_A(\eta_{\pi^*_A(\Sigma_A(B))})} & \Sigma_A(T(\pi^*_A(\Sigma_A(B)))) \ar[d]^-{=}_<<<<<<{\dscomment{\eta \text{ is split fibred}}\quad}^-{\qquad\quad\dscomment{\text{def. of } \sigma_{A,B}}}
\\
& & \Sigma_A(\pi^*_A(T(\Sigma_A(B)))) \ar[dr]_-{\varepsilon^{\Sigma_A \,\dashv\, \pi^*_A}_{T(\Sigma_A(B))}}_<<<<{\dscomment{\text{nat. of } \varepsilon^{\Sigma_A \,\dashv\, \pi^*_A}}\qquad\qquad\qquad\qquad\qquad\qquad} &
\\
\Sigma_A(B) \ar[rrr]_-{\eta_{\Sigma_A(B)}} & & & T(\Sigma_A(B))
}
\]

Next, diagram $(2)$ commutes because we have
\[
\scriptsize
\xymatrix@C=3.9em@R=8.5em@M=0.5em{
\Sigma'(T(C)) \ar[r]^-{\Sigma'(T(\eta^{\Sigma' \,\dashv\, \pi'^*}_C))} & \Sigma'(T(\pi'^*(\Sigma'(C)))) \ar[r]^-{=} & \Sigma'(\pi'^*(T(\Sigma'(C)))) \ar[r]^-{\varepsilon^{\Sigma' \,\dashv\, \pi'^*}_{T(\Sigma'(C))}} &T(\Sigma'(C))
\\
& & & \txt<4pc>{$\Sigma(\pi^*($\\$T(\Sigma'(C))))$} \ar[u]^-{\varepsilon^{\Sigma \,\dashv\, \pi^*}_{T(\Sigma'(C))}\!\!\!}
\\
\txt<3pc>{$\Sigma(\pi^*($\\$\Sigma'(T(C))))$} \ar[r]^-{\Sigma(\pi^*(\Sigma'(T(\eta^{\Sigma' \,\dashv\, \pi'^*}_C))))} \ar@/_1pc/[uu]_-{\varepsilon^{\Sigma \,\dashv\, \pi^*}_{\Sigma'(T(C))}} \ar@{}[uu]^<<<<<<<<<<<{\alpha^{-1}_{A,B,T(C)}} & \txt<4pc>{$\Sigma(\pi^*(\Sigma'(T($\\$\pi'^*(\Sigma'(C))))))$} \ar[uu]_-{\varepsilon^{\Sigma \,\dashv\, \pi^*}_{\Sigma'(T(\pi'^*(\Sigma'(C))))}}^>>>>>>>>>>>>>>{\dscomment{\text{nat. of } \varepsilon^{\Sigma \,\dashv\, \pi^*}}\qquad\quad}_>>>>>>>>>>>>>>{\qquad\qquad\dscomment{T \text{ is split fibred}}} \ar[r]^-{=} & \txt<4pc>{$\Sigma(\pi^*(\Sigma'(\pi'^*($\\$T(\Sigma'(C))))))$} \ar[uu]_-{\varepsilon^{\Sigma \,\dashv\, \pi^*}_{\Sigma'(\pi'^*(T(\Sigma'(C))))}\!\!\!}_>>>>>>>>>>>>>>{\qquad\quad\dscomment{\text{nat. of } \varepsilon^{\Sigma \,\dashv\, \pi^*}}} \ar[ur]^-{\Sigma(\pi^*(\varepsilon^{\Sigma' \,\dashv\, \pi'^*}_{T(\Sigma'(C))}))\!\!\!\!\!\!} & \txt<4pc>{$\Sigma(\kappa^*(\pi'^*($\\$T(\Sigma'(C)))))$} \ar[u]^-{=}^<<<{\dscomment{p \text{ is a split fibration}}\qquad}
\\
\txt<4.5pc>{$\Sigma(\kappa^*(\pi'^*($\\$\Sigma'(T(C)))))$} \ar[r]^>>>>>{\Sigma(\kappa^*(\pi'^*(\Sigma'(T(\eta^{\Sigma' \,\dashv\, \pi'^*}_C)))))} \ar@/_1.75pc/[u]_-{=}^-{\dscomment{\text{def. of } \alpha^{-1}_{A,B,T(C)}}\,} & \txt<5.5pc>{$\Sigma(\kappa^*(\pi'^*($\\$\Sigma'(T(\pi'^*(\Sigma'(C)))))))$} \ar[u]_-{=}^-{\dscomment{p \text{ is a split fibration}}\quad}_-{\qquad\qquad\dscomment{T \text{ is split fibred}}} \ar[r]^-{=} & \txt<4pc>{$\Sigma(\kappa^*(\pi'^*(\Sigma'(\pi'^*($\\$T(\Sigma'(C)))))))$} \ar[u]^-{=} \ar[ur]^>>>>>>>>>>{\Sigma(\kappa^*(\pi'^*(\varepsilon^{\Sigma' \,\dashv\, \pi'^*}_{T(\Sigma'(C))})))\!\!\!\!\!}_<<<<{\qquad\quad\dscomment{\Sigma' \,\dashv\, \pi'^*}}
\\
\Sigma(\kappa^*(T(C))) \ar@{}[d]^-{\qquad\qquad\qquad\qquad\qquad\qquad\qquad\qquad\qquad\dscomment{(b)}} \ar@/_2.15pc/[ddd]^-{=} \ar@/^2.5pc/[uuuu] \ar[r]_-{\Sigma(\kappa^*(T(\eta^{\Sigma' \,\dashv\, \pi'^*}_C)))} \ar[u]_<<<<<<{\Sigma(\kappa^*(\eta^{\Sigma' \,\dashv\, \pi'^*}_{T(C)}))}_-{\qquad\quad\dscomment{\text{nat. of } \eta^{\Sigma' \,\dashv\, \pi'^*}}} & \txt<4pc>{$\Sigma(\kappa^*(T($\\$\pi'^*(\Sigma'(C)))))$} \ar[r]_-{=} \ar[u]^>>>>>>{\Sigma(\kappa^*(\eta^{\Sigma' \,\dashv\, \pi'^*}_{T(\pi'^*(\Sigma'(C)))}))} & \txt<4pc>{$\Sigma(\kappa^*(\pi'^*($\\$T(\Sigma'(C)))))$} \ar[u]^<<<<<<{\Sigma(\kappa^*(\eta^{\Sigma' \,\dashv\, \pi'^*}_{\pi'^*(T(\Sigma'(C)))}))}^-{\dscomment{T \text{ is split fibred}}\qquad\qquad} \ar@/_2pc/[uur]^<<<<<<<<<<<<<<<<<<{\id_{\Sigma(\kappa^*(\pi'^*(T(\Sigma'(C)))))}\!\!\!\!}_-{\,\,\,\,\,\,\,\,\,\,\,\qquad T(\varepsilon^{\Sigma \,\dashv\, \pi^*}_{\Sigma'(C)})} & T(\Sigma(\pi^*( \Sigma'(C)))) \ar@/_2.5pc/[uuuu]
\\
\txt<4pc>{$\Sigma(T(\pi^*_B($\\$\Sigma_B(\kappa^*(C)))))$} \ar[r]^-{=} & \txt<4pc>{$\Sigma(\pi^*_B(T($\\$\Sigma_B(\kappa^*(C)))))$} \ar@/_3pc/[dd]_<<<<<<<{\Sigma_A(\varepsilon^{\Sigma_B \,\dashv\, \pi^*_B}_{T(\Sigma_B(\kappa^*(C)))})} & \Sigma_A(\pi^*_A(T(\Sigma(\kappa^*(C))))) \ar@/_2pc/[ddr]_-{\varepsilon^{\Sigma_A \,\dashv\, \pi^*_A}_{T(\Sigma(\kappa^*(C)))}\!\!\!\!\!} & T(\Sigma(\kappa^*(\pi'^*( \Sigma'(C))))) \ar[u]^-{=}_-{\,\,\,\,\,\,\,\,\,\,\,\,\quad T(\alpha^{-1}_{A,B,C})}_>>>>>{\,\,\,\,\dscomment{\text{def. of } \alpha^{-1}_{A,B,C}}}
\\
& \txt<4pc>{$\Sigma_A(T(\pi^*_A($\\$\Sigma(\kappa^*(C)))))$} \ar[ur]^-{=} &
\\
\Sigma(T(\kappa^*(C))) \ar[uu]_-{\Sigma(T(\eta^{\Sigma_B \,\dashv\, \pi^*_B}_{\kappa^*(C)}))}_<<<<<<<<<{\qquad\dscomment{\text{def. of } \sigma_{B,\kappa^*(C)}}} \ar[r]_-{\Sigma_A(\sigma_{B,\kappa^*(C)})} & \Sigma_A(T(\Sigma_B(\kappa^*(C)))) \ar[u]_-{\Sigma_A(T(\eta^{\Sigma_A \,\dashv\, \pi^*_A}_{\Sigma_B(\kappa^*(C))}))}_-{\qquad\qquad\qquad\qquad\qquad\dscomment{\text{def. of } \sigma_{A,\Sigma_B(\kappa^*(C))}}} \ar[rr]_-{\sigma_{A,\Sigma_B(\kappa^*(C))}} & & T(\Sigma(\kappa^*(C))) \ar[uu]^-{T(\Sigma(\kappa^*(\eta^{\Sigma' \,\dashv\, \pi'^*}_C)))} \ar@/_4.25pc/[uuuuuuu]
}
\]

\pagebreak

Finally, diagram $(4)$ commutes because we have
\[
\scriptsize
\xymatrix@C=5.75em@R=7em@M=0.5em{
T(\Sigma_A(T(B))) \ar@/^1.5pc/[rrr]^-{T(\sigma_{A,B})}_-{\dscomment{\text{def. of } \sigma_{A,B}}} \ar[r]_-{T(\Sigma_A(T(\eta^{\Sigma_A \,\dashv\, \pi^*_A}_B)))} & T(\Sigma_A(T(\pi^*_A(\Sigma_A(B))))) \ar[r]_-{=} & T(\Sigma_A(\pi^*_A(T(\Sigma_A(B))))) \ar[r]_-{T(\varepsilon^{\Sigma_A \,\dashv\, \pi^*_A}_{T(\Sigma_A(B))})} & T(T(\Sigma_A(B))) \ar[dddd]^-{\mu_{\Sigma_A(B)}}
\\
\txt<4pc>{$\Sigma_A(\pi^*_A($\\$T(\Sigma_A(T(B)))))$} \ar[r]^-{\Sigma_A(\pi^*_A(T(\Sigma_A(T(\eta^{\Sigma_A \,\dashv\, \pi^*_A}_B)))))} \ar[u]_-{\varepsilon^{\Sigma_A \,\dashv\, \pi^*_A}_{T(\Sigma_A(T(B)))}}_-{\qquad\qquad\qquad\dscomment{\text{nat. of } \varepsilon^{\Sigma_A \,\dashv\, \pi^*_A}}} & \txt<5pc>{$\Sigma_A(\pi^*_A(T(\Sigma_A($\\$T(\pi^*_A(\Sigma_A(B)))))))$} \ar[r]^-{=} \ar[u]_-{\varepsilon^{\Sigma_A \,\dashv\, \pi^*_A}_{T(\Sigma_A(T(\pi^*_A(\Sigma_A(B)))))}} \ar[d]_-{=}^<<<<<{\quad\dscomment{T \text{ is split fibred}}}_-{\dscomment{T \text{ is split fibred}}\qquad\qquad} & \txt<5pc>{$\Sigma_A(\pi^*_A(T(\Sigma_A($\\$\pi^*_A(T(\Sigma_A(B)))))))$} \ar[u]_-{\varepsilon^{\Sigma_A \,\dashv\, \pi^*_A}_{T(\Sigma_A(\pi^*_A(T(\Sigma_A(B)))))}}^-{\dscomment{T \text{ is split fibred}}\quad\,\,\,\,\,\,} \ar[dd]_-{\Sigma_A(\pi^*_A(T(\varepsilon^{\Sigma_A \,\dashv\, \pi^*_A}_{T(\Sigma_A(B))})))}^<<<<<<<<<<{\qquad\dscomment{\text{nat. of } \varepsilon^{\Sigma_A \,\dashv\, \pi^*_A}}}_>>>>>>>>>>>>>>>{\dscomment{\Sigma_A \,\dashv\, \pi^*_A}\qquad}_>>>>>>>{\dscomment{T \text{ is split fibred}}\quad}
\\
\txt<4pc>{$\Sigma_A(T(\pi^*_A($\\$\Sigma_A(T(B)))))$} \ar[u]_-{=}^-{\dscomment{\text{def.}}\,\,\,\,\,\,\,\,} \ar[r]^-{\Sigma_A(\pi^*_A(T(\Sigma_A(T(\eta^{\Sigma_A \,\dashv\, \pi^*_A}_B)))))} & \txt<5pc>{$\Sigma_A(T(\pi^*_A(\Sigma_A($\\$T(\pi^*_A(\Sigma_A(B)))))))$} \ar[ur]_-{=}
\\
\Sigma_A(T(T(B))) \ar[r]_-{\Sigma_A(T(T(\eta^{\Sigma_A \,\dashv\, \pi^*_A}_B)))} \ar[d]_-{\Sigma_A(\mu_B)} \ar[u]_-{\Sigma_A(T(\eta^{\Sigma_A \,\dashv\, \pi^*_A}_{T(B)}))} \ar@/^2.5pc/[uuu]^<<<<<<<<<<{\sigma_{A,T(B)}\!\!} & \Sigma_A(T(T(\pi^*_A(\Sigma_A(B))))) \ar[u]_-{\Sigma_A(T(\eta^{\Sigma_A \,\dashv\, \pi^*_A}_{T(\pi^*_A(\Sigma_A(B)))}))}^-{\dscomment{\text{nat. of } \eta^{\Sigma_A \,\dashv\, \pi^*_A}}\quad\,\,\,\,\,} \ar[r]_-{=} \ar[d]^-{\Sigma_A(\mu_{\pi^*_A(\Sigma_A(B))})}_-{\dscomment{\text{nat. of } \mu}\qquad\qquad\quad\,\,\,\,\,\,} & \Sigma_A(\pi^*_A(T(T(\Sigma_A(B))))) \ar@/_2pc/[uuur]_<<<<<<<<<<<<<<<<<<<<<<{\!\!\!\!\!\varepsilon^{\Sigma_A \,\dashv\, \pi^*_A}_{T(T(\Sigma_A(B)))}} \ar[d]^-{\Sigma_A(\pi^*_A(\mu_{\Sigma_A(B)}))}^<<<<{\,\,\,\,\qquad\qquad\dscomment{\text{nat. of } \varepsilon^{\Sigma_A \,\dashv\, \pi^*_A}}}_-{\dscomment{\mu \text{ is split fibred}}\qquad\quad}
\\
\Sigma_A(T(B)) \ar[r]^-{\Sigma_A(T(\eta^{\Sigma_A \,\dashv\, \pi^*_A}_B))} \ar@/_1.5pc/[rrr]_-{\sigma_{A,B}}^-{\dscomment{\text{def. of } \sigma_{A,B}}} & \Sigma_A(T(\pi^*_A(\Sigma_A(B)))) \ar[r]^-{=} & \Sigma_A(\pi^*_A(T(\Sigma_A(B)))) \ar[r]^-{\varepsilon^{\Sigma_A \,\dashv\, \pi^*_A}_{T(\Sigma_A(B))}} & T(\Sigma_A(B))
}
\]

We conclude by noting that the subdiagram marked with $(b)$ in the proof of diagram $(2)$ commutes because we have

\pagebreak

\[
\scriptsize
\xymatrix@C=3.5em@R=9em@M=0.5em{
\Sigma(\pi^*(T(\Sigma'(C)))) \ar[rrr]^-{\varepsilon^{\Sigma \,\dashv\, \pi^*}_{T(\Sigma'(C))}} & & & T(\Sigma'(C))
\\
& & & T(\Sigma(\pi^*(\Sigma'(C)))) \ar[u]_-{T(\varepsilon^{\Sigma \,\dashv\, \pi'^*}_{\Sigma'(C)})}^-{\dscomment{\text{nat. of } \varepsilon^{\Sigma \,\dashv\, \pi^*}}\qquad\qquad\qquad\qquad\qquad}
\\
& & \txt<5pc>{$\Sigma(\pi^*(T(\Sigma($\\$\pi^*(\Sigma'(C))))))$} \ar[ur]^-{\varepsilon^{\Sigma \,\dashv\, \pi^*}_{T(\Sigma(\pi^*(\Sigma'(C))))}} \ar[uull]_-{\Sigma(\pi^*(T(\varepsilon^{\Sigma \,\dashv\, \pi^*}_{\Sigma'(C)})))} & T(\Sigma(\kappa^*(\pi'^*(\Sigma'(C))))) \ar[u]_-{=}^<{\dscomment{p \text{ is a split fibration}}\qquad\qquad\quad}
\\
& & \txt<5pc>{$\Sigma(\pi^*(T(\Sigma(\kappa^*($\\$\pi'^*(\Sigma'(C)))))))$} \ar[ur]_-{\varepsilon^{\Sigma \,\dashv\, \pi^*}_{T(\Sigma(\kappa^*(\pi'^*(\Sigma'(C)))))}} \ar[u]^{=}^>{\dscomment{p \text{ is a split fibration}}\qquad\qquad\qquad\qquad}^>>>>>>>>>{\dscomment{T \text{ is split fibred}}\qquad\qquad\qquad\qquad\,\,\,\,\,\,\,}^>>>>>>>>>>>>>>>>>{\dscomment{\Sigma \,\dashv\, \pi^*}\qquad\qquad\qquad\qquad\qquad\!\!} & T(\Sigma(\kappa^*(C))) \ar[u]_-{T(\kappa^*(\eta^{\Sigma' \,\dashv\, \pi'^*}_C))}
\\
& \txt<4pc>{$\Sigma(T(\pi^*(\Sigma(\kappa^*($\\$\pi'^*(\Sigma'(C)))))))$} \ar@/^2pc/[ur]^-{=} & \Sigma(\pi^*(T(\Sigma(\kappa^*(C))))) \ar[r]_-{\Sigma_A(\varepsilon^{\Sigma_B \,\dashv\, \pi^*_B}_{\pi^*_A(T(\Sigma(\kappa^*(C))))})} \ar[u]_-{\Sigma(\pi^*(T(\Sigma(\kappa^*(\eta^{\Sigma' \,\dashv\, \pi'^*}_C)))))}^-{\dscomment{T \text{ is split fibred}}\qquad} & \Sigma_A(\pi^*_A(T(\Sigma_A(\Sigma_B(\kappa^*(C)))))) \ar[u]_-{\varepsilon^{\Sigma_A \,\dashv\, \pi^*_A}_T(\Sigma(\kappa^*(C)))}^>>>>>>>>{\dscomment{\text{def. of } \varepsilon^{\Sigma \,\dashv\, \pi^*}}\qquad\qquad\quad}^<<<<<<{\dscomment{\text{nat. of } \varepsilon^{\Sigma \,\dashv\, \pi^*}}\qquad\qquad\quad}
\\
\Sigma(\kappa^*(\pi'^*(T(\Sigma'(C))))) \ar[uuuuu]^-{=}  & & \txt<3.5pc>{$\Sigma(\pi^*_B(T(\pi^*_A($\\$\Sigma(\kappa^*(C))))))$} \ar[u]_-{=} \ar[r]^-{\Sigma_A(\varepsilon^{\Sigma_B \,\dashv\, \pi^*_B}_{T(\pi^*_A(\Sigma(\kappa^*(C))))})} & \Sigma_A(T(\pi^*_A(\Sigma_A(\Sigma_B(\kappa^*(C)))))) \ar[u]_-{=}^-{\dscomment{T \text{ is split fibred}}\qquad\qquad}
\\
\txt<3.5pc>{$\Sigma(\kappa^*(T($\\$\pi'^*(\Sigma'(C)))))$} \ar[r]_-{=} \ar[u]^-{=}_>>>>>>>>>>>>>>{\qquad\dscomment{T \text{ is split fibred}}}_<<<<<<{\qquad\dscomment{p \text{ is a split fibration}}} & \txt<3.5pc>{$\Sigma(T(\kappa^*($\\$\pi'^*(\Sigma'(C)))))$} \ar@/^5pc/[uuuuuul]^-{=} \ar[uu]_>>>>>>>>>{\Sigma(T(\eta^{\Sigma \,\dashv\, \pi^*}_{\kappa^*(\pi'^*(\Sigma'(C)))}))} & \Sigma(T(\pi^*(\Sigma(\kappa^*(C))))) \ar[u]_-{=} \ar@/^4pc/[uu]^-{=} & \Sigma_A(T(\Sigma_B(\kappa^*(C)))) \ar[u]_-{\Sigma_A(T(\eta^{\Sigma_A \,\dashv\, \pi^*_A}_{\Sigma_B(\kappa^*(C))}))}^-{\dscomment{\text{nat. of } \varepsilon^{\Sigma_B \,\dashv\, \pi^*_B}}\qquad\,\,\,\,\,}
\\
\Sigma(\kappa^*(T(C))) \ar[u]_>>>>>>{\Sigma(\kappa^*(T(\eta^{\Sigma' \,\dashv\, \pi'^*}_C)))}_-{\qquad\quad\dscomment{T \text{ is split fibred}}} \ar[r]_-{=} & \Sigma(T(\kappa^*(C))) \ar[r]_-{\Sigma(T(\eta^{\Sigma_B \,\dashv\, \pi^*_B}_{\kappa^*(C)}))} \ar[u]^<<<<<<{\Sigma(T(\kappa^*(\eta^{\Sigma' \,\dashv\, \pi'^*}_C)))}_>>>>{\,\,\,\,\,\,\,\quad\qquad\dscomment{\text{def. of } \eta^{\Sigma \,\dashv\, \pi^*}}}_<<<<<{\,\,\,\,\,\,\,\qquad\quad\dscomment{\text{nat. of } \eta^{\Sigma \,\dashv\, \pi^*}}} & \Sigma(T(\pi^*_B(\Sigma_B(\kappa^*(C))))) \ar[r]_-{=} \ar[u]^-{\Sigma(T(\pi^*_B(\eta^{\Sigma_A \,\dashv\, \pi^*_A}_{\Sigma_B(\kappa^*(C))})))}_>>>>>>>>>>>{\qquad\dscomment{T \text{ is split fibred}}} & \Sigma(\pi^*_B(T(\Sigma_B(\kappa^*(C))))) \ar[u]_-{\Sigma_A(\varepsilon^{\Sigma_B \,\dashv\, \pi^*_B}_{T(\Sigma_B(\kappa^*(C)))})} \ar[uul]^<<<<<<<<<<<{\Sigma(\pi^*_B(T(\eta^{\Sigma_A \,\dashv\, \pi^*_A}_{\Sigma_B(\kappa^*(C))})))\!\!\!\!}
}
\]

\section{Proof of Theorem~\ref{thm:dependentproductsinEMfibration}}
\label{sect:proofofthm:dependentproductsinEMfibration}

{
\renewcommand{\thetheorem}{\ref{thm:dependentproductsinEMfibration}}
\begin{theorem}
Given a split comprehension category with unit $p : \mathcal{V} \longrightarrow \mathcal{B}$ with split dependent products and a split fibred monad  $\mathbf{T} = (T,\eta,\mu)$ on it, then the corresponding EM-fibration $p^{\mathbf{T}} : \mathcal{V}^{\mathbf{T}} \longrightarrow \mathcal{B}$ has split dependent $p$-products.
\end{theorem}
\addtocounter{theorem}{-1}
}

\begin{proof}
Given an object $A$ in $\mathcal{V}$, the functor $\Pi^{\mathbf{T}}_A : \mathcal{V}^{\mathbf{T}}_{\ia A} \longrightarrow \mathcal{V}^{\mathbf{T}}_{p(A)}$ is given on objects by
\[
\Pi^{\mathbf{T}}_A(B,\beta) \defeq (\Pi_A(B), \beta_{\Pi^{\mathbf{T}}_A})
\]
where the candidate EM-algebra structure map $\beta_{\Pi^{\mathbf{T}}_A} : T(\Pi_A(B)) \longrightarrow \Pi_A(B)$ is defined as the following composite morphism:
\[
\xymatrix@C=5em@R=0.05em@M=0.5em{
T(\Pi_A(B)) \ar[r]^-{\eta^{\pi^*_A \,\dashv\, \Pi_A}_{T(\Pi_A(B))}} & \Pi_A(\pi^*_A(T(\Pi_A(B)))) \ar[dr]^-{=}
\\
& & \Pi_A(T(\pi^*_A(\Pi_A(B)))) \ar[dl]^-{\Pi_A(T(\varepsilon^{\pi^*_A \,\dashv\, \Pi_A}_B))}
\\
\Pi_A(B) & \Pi_A(T(B)) \ar[l]^-{\Pi_A(\beta)}
}
\]
using the split dependent products in $p$, i.e., the adjunction $\pi^*_A \dashv \Pi_A : \mathcal{V}_{\ia A} \longrightarrow \mathcal{V}_{p(A)}$; and making use of Proposition~\ref{prop:verticalEMalgebras} to ensure that $\beta$ is a vertical morphism.

Next, we prove that the morphism $\beta_{\Pi^{\mathbf{T}}_A} : T(\Pi_A(B)) \longrightarrow \Pi_A(B)$ is indeed a structure map of an EM-algebra, by showing that the next two diagrams commute in $\mathcal{V}_{p(A)}$.

\[
\hspace{-0.3cm}
\xymatrix@C=5em@R=4em@M=0.5em{
\Pi_A(B) \ar[rr]^{\eta_{\Pi_A(B)}} \ar[dr]^-{\eta^{\pi^*_A \,\dashv\, \Pi_A}_{\Pi_A(B)}}_>>>>>>>>>>>{\dcomment{\pi^*_A \,\dashv\, \Pi_A}\,\,\,\,\,\,\,\,\,\,\,\,} \ar@/_2pc/[ddr]^-{\!\!\id_{\Pi_A(B)}} \ar@/_7pc/[dddrr]_>>>>>>>>>>>>>>>{\Pi_A(\eta_B)\,\,\,\,}  \ar@/_7pc/[ddddrr]_-{\id_{\Pi_A(B)}\!\!}  & & T(\Pi_A(B)) \ar[d]_-{\eta^{\pi^*_A \,\dashv\, \Pi_A}_{T(\Pi_A(B))}}_-{\dcomment{\text{nat. of } \eta^{\pi^*_A \,\dashv\, \Pi_A}}\qquad\qquad\qquad\quad} \ar@/^6pc/[dddd]^-{\beta_{\Pi^{\mathbf{T}}_A}}
\\
& \Pi_A(\pi^*_A(\Pi_A(B))) \ar[r]^-{\Pi_A(\pi^*_A(\eta_{\Pi_A(B)}))} \ar@/_2pc/[dr]^-{\Pi_A(\eta_{\pi^*_A(\Pi_A(B))})}_>>>{\dcomment{\text{nat. of } \eta}\qquad\qquad} \ar[d]_<<<<<<{\!\!\Pi_A(\varepsilon^{\pi^*_A \,\dashv\, \Pi_A}_B)} & \Pi_A(\pi^*_A(T(\Pi_A(B)))) \ar[d]_-{=}_<<<<{\dcomment{\eta \text{ is split fibred}}\qquad}^-{\,\,\,\,\,\,\dcomment{\text{def. of } \beta_{\Pi^{\mathbf{T}}_A}}}
\\
& \Pi_A(B) \ar@/_2pc/[dr]^<<<<<<<<<{\Pi_A(\eta_B)}_<{\dcomment{\text{id. law}}\,\,\,} & \Pi_A(T(\pi^*_A(\Pi_A(B)))) \ar[d]_-{\Pi_A(T(\varepsilon^{\pi^*_A \,\dashv\, \Pi_A}_B))}
\\
& & \Pi_A(T(B)) \ar[d]_-{\Pi_A(\beta)}_<<<<<{\dcomment{(B,\beta) \text{ is an EM-algebra}}\qquad\,\,\,\,\,\,\,}
\\
& & \Pi_A(B)
}
\]

\pagebreak

\mbox{}

\vspace{0.5cm}

\[
\scriptsize
\xymatrix@C=5em@R=12em@M=0.5em{
T(T(\Pi_A(B))) \ar[rrrr]^-{\mu_{\Pi_A(B)}} \ar[d]^-{T(\eta^{\pi^*_A \,\dashv\, \Pi_A}_{\Pi_A(B)})}^-{\qquad\qquad\qquad\qquad\dscomment{\text{nat. of } \eta^{\pi^*_A \,\dashv\, \Pi_A}}} \ar@/^2.5pc/[drrr]_-{\eta^{\pi^*_A \,\dashv\, \Pi_A}_{T(T(\Pi_A(B)))}} 
& & & & T(\Pi_A(B)) \ar[d]_-{\eta^{\pi^*_A \,\dashv\, \Pi_A}_{T(\Pi_A(B))}}_-{\dscomment{\text{nat. of } \eta^{\pi^*_A \,\dashv\, \Pi_A}}\qquad\qquad\qquad\qquad}
\\
\txt<3.5pc>{
$T(\Pi_A(\pi^*_A($
\\
$T(\Pi_A(B)))))$
}
\ar[rr]^-{\eta^{\pi^*_A \,\dashv\, \Pi_A}_{T(\Pi_A(\pi^*_A(T(\Pi_A(B)))))}}
\ar[d]^-{=}^-{\qquad\qquad\qquad\dscomment{T \text{ is split fibred}}}
 & & 
\txt<3.5pc>{
$\Pi_A(\pi^*_A(T($
\\
$\Pi_A(\pi^*_A(T(\Pi_A(B)))))))$
}
\ar[dl]_-{=}^>>>>>>>>>>>>>{\quad\dscomment{T \text{ is split fibred}}}
\ar[d]_-{=}^<<<<<<<<<<{\!\!\qquad\qquad\dscomment{\pi^*_A \,\dashv\, \Pi_A}}^>>>>>>>>>>{\quad\qquad\dscomment{T \text{ is split fibred}}}
&
\txt<3.5pc>{
$\Pi_A(\pi^*_A(T($
\\
$T(\Pi_A(B)))))$
}
\ar[r]^-{\Pi_A(\pi^*_A(\mu_{\Pi_A(B)}))}
\ar[l]_-{\Pi_A(\pi^*_A(T(\eta^{\pi^*_A \,\dashv\, \Pi_A}_{T(\Pi_A(B))})))}
\ar[d]_-{=}^-{\,\,\,\,\,\,\qquad\dscomment{\mu \text{ is split fibred}}}
& 
\txt<3.5pc>{
$\Pi_A(\pi^*_A($
\\
$T(\Pi_A(B))))$
}
\ar[d]_-{=}
\\
\txt<3pc>{
$T(\Pi_A(T($
\\
$\pi^*_A(\Pi_A(B)))))$
}
\ar[r]^-{\eta^{\pi^*_A \,\dashv\, \Pi_A}_{T(\Pi_A(T(\pi^*_A(\Pi_A(B)))))}}
\ar[d]^>>>>>>>>{T(\Pi_A(T(\varepsilon^{\pi^*_A \,\dashv\, \Pi_A}_B)))}^-{\quad\qquad\dscomment{\text{nat. of } \eta^{\pi^*_A \,\dashv\, \Pi_A}}}
 & 
\txt<4pc>{
$\Pi_A(\pi^*_A(T(\Pi_A($
\\
$T(\pi^*_A(\Pi_A(B)))))))$
}
\ar[r]^-{=}
\ar[d]_<<<<<<<{\Pi_A(\pi^*_A(T(\Pi_A(T(\varepsilon^{\pi^*_A \,\dashv\, \Pi_A}_B)))))}
& 
\txt<4pc>{
$\Pi_A(T(\pi^*_A(\Pi_A($
\\
$T(\pi^*_A(\Pi_A(B)))))))$
}
\ar[r]^-{\Pi_A(T(\varepsilon^{\pi^*_A \,\dashv\, \Pi_A}_{T(\pi^*_A(\Pi_A(B)))}))}
\ar[d]_-{\Pi_A(T(\pi^*_A(\Pi_A(T(\varepsilon^{\pi^*_A \,\dashv\, \Pi_A}_B)))))}_<<<<<<<<<{\dscomment{T \text{ is split fibred}}\qquad\quad}
& 
\txt<3.5pc>{
$\Pi_A(T(T($
\\
$\pi^*_A(\Pi_A(B)))))$
}
\ar[r]^-{\Pi_A(\mu_{\pi^*_A(\Pi_A(B))})}
\ar[d]_-{\Pi_A(T(T(\varepsilon^{\pi^*_A \,\dashv\, \Pi_A}_B)))}_<<<<<<<<<{\dscomment{\text{nat. of } \varepsilon^{\pi^*_A \,\dashv\, \Pi_A}}\qquad\quad}
& 
\txt<3.5pc>{
$\Pi_A(T($
\\
$\pi^*_A(\Pi_A(B))))$
}
\ar[d]_-{\Pi_A(T(\varepsilon^{\pi^*_A \,\dashv\, \Pi_A}_B))}_<<<<<<<<<{\dscomment{\text{nat. of } \mu}\qquad\qquad}
\\
T(\Pi_A(T(B))) \ar[r]_-{\eta^{\pi^*_A \,\dashv\, \Pi_A}_{T(\Pi_A(T(B)))}} \ar[d]^>>>>>>>>>>{T(\Pi_A(\beta))}^-{\qquad\quad\dscomment{\text{nat. of } \eta^{\pi^*_A \,\dashv\, \Pi_A}}}
&
\txt<3.5pc>{
$\Pi_A(\pi^*_A(T($
\\
$\Pi_A(T(B)))))$
}
\ar[r]_-{=}
\ar[d]_<<<<<<<<<{\Pi_A(\pi^*_A(T(\Pi_A(\beta))))}
& 
\txt<3.5pc>{
$\Pi_A(T(\pi^*_A($
\\
$\Pi_A(T(B)))))$
}
\ar[r]_-{\Pi_A(T(\varepsilon^{\pi^*_A \,\dashv\, \Pi_A}_{T(B)}))}
\ar[d]_-{\Pi_A(T(\pi^*_A(\Pi_A(\beta))))}_<<<<<<<<<{\dscomment{T \text{ is split fibred}}\qquad\quad}
&
\Pi_A(T(T(B))) \ar[r]_-{\Pi_A(\mu_B)} \ar[d]_-{\Pi_A(T(\beta))}_<<<<<<<<<<<{\dscomment{\text{nat. of } \varepsilon^{\pi^*_A \,\dashv\, \Pi_A}}\qquad\quad\,\,\,} & \Pi_A(T(B)) \ar[d]_-{\Pi_A(\beta)}_<<<<<<<<<<<{\dscomment{(B,\beta) \text{ is an EM-algebra}}\qquad\!\!\!}
\\
T(\Pi_A(B)) \ar[r]_-{\eta^{\pi^*_A \,\dashv\, \Pi_A}_{T(\Pi_A(B))}} \ar@/_4pc/[rrrr]_-{\beta_{\Pi^{\mathbf{T}}_A}} \ar@{}[d]^<<<<<<<<{\,\,\quad\qquad\qquad\qquad\qquad\qquad\qquad\qquad\qquad\qquad\qquad\dscomment{\text{def. of } \beta_{\Pi^{\mathbf{T}}_A}}} & \txt<3.5pc>{$\Pi_A(\pi^*_A($\\$T(\Pi_A(B))))$} \ar[r]_-{=} & \txt<3.5pc>{$\Pi_A(T($\\$\pi^*_A(\Pi_A(B))))$} \ar[r]_-{\Pi_A(T(\varepsilon^{\pi^*_A \,\dashv\, \Pi_A}_B))} & \Pi_A(T(B)) \ar[r]_-{\Pi_A(\beta)} & \Pi_A(B)
\\
&
}
\]

The functor $\Pi^{\mathbf{T}}_A$ is defined on morphisms $h : (B,\beta) \longrightarrow (B',\beta')$ simply by letting $\Pi^{\mathbf{T}}_A(h) \defeq \Pi_A(h)$. It is easy to see that this gives us an EM-algebra homomorphism:
\[
\xymatrix@C=7em@R=5em@M=0.5em{
T(\Pi_A(B)) \ar[r]^-{T(\Pi_A(h))} \ar[d]^-{\eta^{\pi^*_A \,\dashv\, \Pi_A}_{T(\Pi_A(B))}}^-{\,\,\,\,\quad\qquad\qquad\dcomment{\text{nat. of } \eta^{\pi^*_A \,\dashv\, \Pi_A}}} \ar@/_6pc/[dddd]_{\beta_{\Pi^{\mathbf{T}}_A}} & T(\Pi_A(B')) \ar[d]_-{\eta^{\pi^*_A \,\dashv\, \Pi_A}_{T(\Pi_A(B'))}} \ar@/^6pc/[dddd]^{\beta'_{\Pi^{\mathbf{T}}_A}}
\\
\Pi_A(\pi^*_A(T(\Pi_A(B)))) \ar[r]_-{\Pi_A(\pi^*_A(T(\Pi_A(h))))} \ar[d]^-{=}^-{\,\,\quad\qquad\qquad\dcomment{T \text{ is split fibred}}}_-{\dcomment{\text{def. of } \beta_{\Pi^{\mathbf{T}}_A}}\quad\!\!\!\!} & \Pi_A(\pi^*_A(T(\Pi_A(B')))) \ar[d]_-{=}
\\
\Pi_A(T(\pi^*_A(\Pi_A(B)))) \ar[r]_-{\Pi_A(T(\pi^*_A(\Pi_A(h))))} \ar[d]^>>>>>{\Pi_A(T(\varepsilon^{\pi^*_A \,\dashv\, \Pi_A}_B))}^-{\,\,\,\,\quad\qquad\qquad\dcomment{\text{nat. of } \varepsilon^{\pi^*_A \,\dashv\, \Pi_A}}} & \Pi_A(T(\pi^*_A(\Pi_A(B')))) \ar[d]_<<<<<{\Pi_A(T(\varepsilon^{\pi^*_A \,\dashv\, \Pi_A}_{B'}))}^-{\!\!\!\!\quad\dcomment{\text{def. of } \beta'_{\Pi^{\mathbf{T}}_A}}}
\\
\Pi_A(T(B)) \ar[r]_-{\Pi_A(T(h))} \ar[d]^>>>>>{\Pi_A(\beta)} & \Pi_A(T(B')) \ar[d]_<<<<<{\Pi_A(\beta')}_-{\dcomment{h \text{ is an EM-algebra homomorphism}}\quad\qquad\!\!\!\!\!\!\!\!\!}
\\
\Pi_A(B) \ar[r]_-{\Pi_A(h)} & \Pi_A(B')
}
\]
Further, it is also easy to see that $\Pi^{\mathbf{T}}_A$ preserves identities and composition---these properties follow directly from the functoriality of $\Pi_A$. We therefore omit these proofs.

We proceed by proving that we have an adjunction $\pi^*_A \dashv \Pi^{\mathbf{T}}_A : \mathcal{V}^{\mathbf{T}}_{\ia A} \longrightarrow \mathcal{V}^{\mathbf{T}}_{p(A)}$.

First, we note that the components of the unit and counit natural transformations 
\[
\eta^{\pi^*_A \,\dashv\, \Pi^{\mathbf{T}}_A} : \id_{\mathcal{V}^{\mathbf{T}}_{p(A)}} \longrightarrow \Pi^{\mathbf{T}}_A \comp \pi^*_A
\qquad
\varepsilon^{\pi^*_A \,\dashv\, \Pi^{\mathbf{T}}_A} : \pi^*_A \comp \Pi^{\mathbf{T}}_A \longrightarrow \id_{\mathcal{V}^{\mathbf{T}}_{\ia A}}
\]
are given simply by
\[
\eta^{\pi^*_A \,\dashv\, \Pi^{\mathbf{T}}_A}_{(B,\beta)} \defeq \eta^{\pi^*_A \,\dashv\, \Pi_A}_B
\qquad
\varepsilon^{\pi^*_A \,\dashv\, \Pi^{\mathbf{T}}_A}_{(B,\beta)} \defeq \varepsilon^{\pi^*_A \,\dashv\, \Pi_A}_B
\]

Next, we prove that these components are indeed EM-algebra homomorphisms, by showing that the next two diagrams commute, in $\mathcal{V}^{\mathbf{T}}_{p(A)}$ and $\mathcal{V}^{\mathbf{T}}_{\ia A}$, respectively.
\[
\xymatrix@C=11em@R=5em@M=0.5em{
T(B) \ar[r]^{T(\eta^{\pi^*_A \,\dashv\, \Pi_A}_B)} \ar[ddddd]_-{\beta} \ar@/_5pc/[ddddr]_>>>>>>>>>>>>>>>>>>>>>>>>>>>>>{\eta^{\pi^*_A \,\dashv\, \Pi_A}_{T(B)}\!\!\!} & T(\Pi_A(\pi^*_A(B))) \ar[d]_-{\eta^{\pi^*_A \,\dashv\, \Pi_A}_{T(\Pi_A(\pi^*_A(B)))}}_-{\dcomment{\text{nat. of } \eta^{\pi^*_A \,\dashv\, \Pi_A}}\qquad\qquad\qquad\qquad} \ar@/^9pc/[ddddd]^-{(\pi^*_A(\beta))_{\Pi^{\mathbf{T}}_A}}
\\
& \Pi_A(\pi^*_A(T(\Pi_A(\pi^*_A(B))))) \ar[d]^-{=}^>>>>>{\!\!\quad\dcomment{\text{def. of } (\pi^*_A(\beta))_{\Pi^{\mathbf{T}}_A}}}
\\
& \Pi_A(T(\pi^*_A(\Pi_A(\pi^*_A(B))))) \ar@/^2pc/[d]^-{\Pi_A(T(\varepsilon^{\pi^*_A \,\dashv\, \Pi_A}_{\pi^*_A(B)}))}_-{\dcomment{\pi^*_A \dashv \Pi_A}\,\,}
\\
& \Pi_A(T(\pi^*_A(B))) \ar[d]^-{=} \ar@/^2pc/[u]^-{\Pi_A(T(\pi^*_A(\eta^{\pi^*_A \,\dashv\, \Pi_A}_B)))}^>>>>>{\dcomment{\eta \text{ is split fibred}}\quad}
\\
& \Pi_A(\pi^*_A(T(B))) \ar[d]_-{\Pi_A(\pi^*_A(\beta))}_-{\dcomment{\text{nat. of } \eta^{\pi^*_A \,\dashv\, \Pi_A}}\qquad\qquad\qquad\qquad} \ar@/^9.5pc/[uuu]^>>>>>>>>>>>>{\Pi_A(\pi^*_A(T(\eta^{\pi^*_A \,\dashv\, \Pi_A}_B)))}
\\
B \ar[r]_-{\eta^{\pi^*_A \,\dashv\, \Pi_A}_B} & \Pi_A(\pi^*_A(B))
}
\]

\pagebreak

\mbox{}

\vspace{-1cm}

\[
\xymatrix@C=11em@R=5em@M=0.5em{
T(\pi^*_A(\Pi_A(B))) \ar[r]^-{T(\varepsilon^{\pi^*_A \,\dashv\, \Pi_A}_B)} \ar[d]^-{=} \ar@/_9pc/[ddddd]_-{\pi^*_A(\beta_{\Pi^{\mathbf{T}}_A})}^-{\,\,\,\quad{\dcomment{\text{def. of } \pi^*_A(\beta_{\Pi^{\mathbf{T}}_A})}}} & T(B) \ar[ddddd]^-{\beta}
\\
\pi^*_A(T(\Pi_A(B))) \ar@/_2pc/[d]_-{\pi^*_A(\eta^{\pi^*_A \,\dashv\, \Pi_A}_{T(\Pi_A(B))})}^-{\,\,\,\dcomment{\pi^*_A \dashv \Pi_A}}
\\
\pi^*_A(\Pi_A(\pi^*_A(T(\Pi_A(B))))) \ar[d]_-{=}^<<<<{\,\,\dcomment{T \text{ is split fibred}}} \ar@/_2pc/[u]_-{\varepsilon^{\pi^*_A \,\dashv\, \Pi_A}_{\pi^*_A(T(\Pi_A(B)))}}
\\
\pi^*_A(\Pi_A(T(\pi^*_A(\Pi_A(B))))) \ar[d]_-{\pi^*_A(\Pi_A( T(\varepsilon^{\pi^*_A \,\dashv\, \Pi_A}_B)))} \ar@/_9pc/[uuu]_-{\varepsilon^{\pi^*_A \,\dashv\, \Pi_A}_{T(\pi^*_A(\Pi_A(B)))}}_>>>>>>>>>>>>>{\qquad\qquad\dcomment{\text{nat. of } \varepsilon^{\pi^*_A \,\dashv\, \Pi_A}}}
\\
\pi^*_A(\Pi_A(T(B))) \ar[d]^-{\pi^*_A(\Pi_A(\beta))}^-{\qquad\qquad\qquad\qquad\dcomment{\text{nat. of } \varepsilon^{\pi^*_A \,\dashv\, \Pi_A}}} \ar@/_5pc/[uuuur]_<<<<<<<<<<<<<<<<<<<<<<<<<<<<<{\!\!\!\!\!\!\varepsilon^{\pi^*_A \,\dashv\, \Pi_A}_{T(B)}}
\\
\pi^*_A(\Pi_A(B)) \ar[r]_-{\varepsilon^{\pi^*_A \,\dashv\, \Pi_A}_B} & B
}
\]

The naturality of $\eta^{\pi^*_A \,\dashv\, \Pi^{\mathbf{T}}_A}$ and $\varepsilon^{\pi^*_A \,\dashv\, \Pi^{\mathbf{T}}_A}$, and the two unit-counit laws follow directly from the corresponding properties of the adjunction $\pi^*_A \dashv \Pi_A : \mathcal{V}_{\ia A} \longrightarrow \mathcal{V}_{p(A)}$.

We conclude by noting that the adjunction $\pi^*_A \dashv \Pi^{\mathbf{T}}_A : \mathcal{V}^{\mathbf{T}}_{\ia A} \longrightarrow \mathcal{V}^{\mathbf{T}}_{p(A)}$ also satisfies the split Beck-Chevalley condition from Definition~\ref{def:splitdependentcompproducts}---similarly to the other properties, it also follows directly from the corresponding property of $\pi^*_A \dashv \Pi_A$.
\end{proof}

\newpage

\section{Proof of Theorem~\ref{thm:dependentsumsinEMfibrationwhenmonadpreservesthem}}
\label{sect:proofofthm:dependentsumsinEMfibrationwhenmonadpreservesthem}

{
\renewcommand{\thetheorem}{\ref{thm:dependentsumsinEMfibrationwhenmonadpreservesthem}}
\begin{theorem}
Given a split comprehension category with unit $p : \mathcal{V} \longrightarrow \mathcal{B}$ with strong split dependent sums and a split fibred monad $\mathbf{T} = (T,\eta,\mu)$ on it, then the corresponding EM-fibration $p^{\mathbf{T}} : \mathcal{V}^{\mathbf{T}} \!\longrightarrow\! \mathcal{B}$ has split dependent $p$-sums if the dependent strength of $\mathbf{T}$ is given by a family of natural isomorphisms, i.e., if for every $A$ in $\mathcal{V}$, $\sigma_A : \Sigma_A \comp T \longrightarrow T \comp \Sigma_A$ is a natural isomorphism.
Furthermore, these split dependent $p$-sums are preserved on-the-nose by $U^{\mathbf{T}}$, i.e., we have $U^{\mathbf{T}}(\Sigma^{\mathbf{T}}_A(B,\beta)) = \Sigma_A(U^{\mathbf{T}}(B,\beta))$.
\end{theorem}
\addtocounter{theorem}{-1}
}

\begin{proof}
Given an object $A$ in $\mathcal{V}$, the functor $\Sigma^{\mathbf{T}}_A : \mathcal{V}^{\mathbf{T}}_{\ia A} \longrightarrow \mathcal{V}^{\mathbf{T}}_{p(A)}$ is given on objects by
\[
\Sigma^{\mathbf{T}}_A(B,\beta) \defeq (\Sigma_A(B), \beta_{\Sigma^{\mathbf{T}}_A})
\]
where the candidate EM-algebra structure map $\beta_{\Sigma^{\mathbf{T}}_A} : T(\Sigma_A(B)) \longrightarrow \Sigma_A(B)$ is defined as the following composite morphism:
\[
\xymatrix@C=5em@R=5em@M=0.5em{
T(\Sigma_A(B)) \ar[r]^-{\sigma^{-1}_{A,B}} & \Sigma_A(T(B)) \ar[r]^-{\Sigma_A(\beta)} & \Sigma_A(B)
}
\]
using the split dependent sums in $p$, i.e., the adjunction $\Sigma_A \dashv \pi^*_A : \mathcal{V}_{p(A)} \longrightarrow \mathcal{V}_{\ia A}$; and making use of Proposition~\ref{prop:verticalEMalgebras} to ensure that $\beta$ is a vertical morphism.

Next, we prove that the morphism $\beta_{\Sigma^{\mathbf{T}}_A} : T(\Sigma_A(B)) \longrightarrow \Sigma_A(B)$ is indeed a structure map of an EM-algebra, by showing that the next two diagrams commute in $\mathcal{V}_{p(A)}$.
\[
\xymatrix@C=6em@R=5em@M=0.5em{
\Sigma_A(B) \ar[rrr]^-{\eta_{\Sigma_A(B)}} \ar@/_2pc/[ddrr]^-{\id_{\Sigma_A(B)}} \ar@/_6pc/[dddrrr]_{\id_{\Sigma_A(B)}}^-{\quad\dcomment{\text{id. law}}} &&& T(\Sigma_A(B)) \ar[ddd]^-{\beta_{\Sigma^{\mathbf{T}}_A}} \ar@{}[dd]_-{\dcomment{\text{def. of } \beta_{\Sigma^{\mathbf{T}}_A}}\quad\,\,\,\,\,\,} \ar[dl]_-{\sigma^{-1}_{A,B}}_-{\dcomment{(a)}\qquad\qquad\qquad\qquad\qquad\qquad}
\\
&& \Sigma_A(T(B)) \ar@/^2.25pc/[ddr]^-{\Sigma_A(\beta)} &
\\
&& \Sigma_A(B) \ar[u]^-{\Sigma_A(\eta_B)}_<<<{\!\!\!\!\!\!\!\quad\dcomment{(B,\beta) \text{ is EM-alg.}}} \ar[dr]_-{\id_{\Sigma_A(B)}} &
\\
&&& \Sigma_A(B)
}
\]

\pagebreak

\mbox{}

\vspace{-1cm}

\[
\xymatrix@C=3.5em@R=5em@M=0.5em{
T(T(\Sigma_A(B))) \ar[rrr]^-{T(\beta_{\Sigma^{\mathbf{T}}_A})} \ar[dddd]_-{\mu_{\Sigma_A(B)}}^-{\,\,\,\,\qquad\dcomment{(b)}} \ar[dr]_-{T(\sigma^{-1}_{A,B})} &&& T(\Sigma_A(B)) \ar[dddd]^-{\beta_{\Sigma^{\mathbf{T}}_A}}_-{\dcomment{\text{def. of } \beta_{\Sigma^{\mathbf{T}}_A}}\quad} \ar[ddl]_-{\sigma^{-1}_{A,B}\!\!}
\\
& T(\Sigma_A(T(B))) \ar[d]_-{\sigma^{-1}_{A,{T(B)}}}^-{\qquad\quad\dcomment{\text{nat. of } \sigma^{-1}_A}} \ar[urr]_-{\,\,\,\,\,\,T(\Sigma_A(\beta))}^-{\dcomment{\text{def. of } \beta_{\Sigma^{\mathbf{T}}_A}}\qquad\qquad\qquad\qquad\qquad}
\\
& \Sigma_A(T(T(B))) \ar[d]_-{\Sigma_A(\mu_B)}^-{\quad\dcomment{(B,\beta) \text{ is an EM-algebra}}} \ar[r]^-{\Sigma_A(T(\beta))} & \Sigma_A(T(B)) \ar[ddr]^-{\Sigma_A(\beta)}
\\
& \Sigma_A(T(B)) \ar[drr]_-{\Sigma_A(\beta)}
\\
T(\Sigma_A(B)) \ar[ur]_-{\!\!\!\!\sigma^{-1}_{A,B}}_-{\qquad\qquad\qquad\dcomment{\text{def. of } \beta_{\Sigma^{\mathbf{T}}_A}}} \ar[rrr]_-{\beta_{\Sigma^{\mathbf{T}}_A}} &&& \Sigma_A(B)
}
\]

We conclude the proof that $\beta_{\Sigma^{\mathbf{T}}_A} : T(\Sigma_A(B)) \longrightarrow \Sigma_A(B)$ is an EM-algebra structure map by noting that 
as $\id_{\Sigma_A(B)}$, $\sigma^{-1}_{A,B}$, and $\sigma^{-1}_{A,{T(B)}} \comp T(\sigma^{-1}_{A,B})$ are all isomorphisms, it suffices to show that the subdiagrams marked with $(a)$ and $(b)$ commute when these morphisms are replaced with their inverses. However, after replacing these morphisms with their inverses, we see that the corresponding diagrams are exactly diagrams $(3)$ and $(4)$ from Proposition~\ref{prop:strengthofsplitfibredmonads}, governing the interaction of $\sigma_A$ with $\eta$ and $\mu$.

Next, the functor $\Sigma^{\mathbf{T}}_A$ is defined on morphisms $h : (B,\beta) \longrightarrow (B',\beta')$ simply by letting $\Sigma^{\mathbf{T}}_A(h) \defeq \Sigma_A(h)$. It is easy to see that this gives us an EM-algebra homomorphism:

\[
\xymatrix@C=9em@R=5em@M=0.5em{
T(\Sigma_A(B)) \ar[d]_-{\sigma^{-1}_{A,B}}^-{\,\,\,\,\,\quad\qquad\qquad\dcomment{(c)}} \ar[r]^-{T(\Sigma_A(h))} & T(\Sigma_A(B')) \ar[d]^-{\sigma^{-1}_{A,{B'}}}
\\
\Sigma_A(T(B)) \ar[r]^-{\Sigma_A(T(h))} \ar[d]_-{\Sigma_A(\beta)}^-{\!\!\!\!\!\!\!\!\!\qquad\dcomment{h \text{ is an EM-algebra homomorphism}}} & \Sigma_A(T(B')) \ar[d]^-{\Sigma_A(\beta')}
\\
\Sigma_A(B) \ar[r]_-{\Sigma_A(h)} & \Sigma_A(B')
}
\]

We show that the square marked with $(c)$ commutes by observing that $\sigma^{-1}_{A,B}$ and $\sigma^{-1}_{A,{B'}}$ are both isomorphisms. As a result, it suffices to show that the next diagram commutes, where we have replaced these two morphisms with their respective inverses.
\[
\xymatrix@C=7em@R=5em@M=0.5em{
T(\Sigma_A(B)) \ar[r]^-{T(\Sigma_A(h))} & T(\Sigma_A(B'))
\\
\Sigma_A(\pi^*_A(T(\Sigma_A(B)))) \ar[r]^-{\Sigma_A(\pi^*_A(T(\Sigma_A(h))))} \ar[u]_-{\varepsilon^{\Sigma_A \,\dashv\, \pi^*_A}_{T(\Sigma_A(B))}}_-{\quad\qquad\qquad\dcomment{\text{nat. of } \varepsilon^{\Sigma_A \,\dashv\, \pi^*_A}}} & \Sigma_A(\pi^*_A(T(\Sigma_A(B')))) \ar[u]^-{\varepsilon^{\Sigma_A \,\dashv\, \pi^*_A}_{T(\Sigma_A(B'))}}
\\
\Sigma_A(T(\pi^*_A(\Sigma_A(B)))) \ar[r]^-{\Sigma_A(T(\pi^*_A(\Sigma_A(h))))} \ar[u]_-{=}_-{\!\!\!\!\quad\qquad\qquad\dcomment{T \text{ is split fibred}}}^-{\dcomment{\text{def. of } \sigma_{A,B}}\,\,\,\,\,\,\,\,\,} & \Sigma_A(T(\pi^*_A(\Sigma_A(B')))) \ar[u]^-{=}_-{\,\,\,\,\,\,\,\,\,\dcomment{\text{def. of } \sigma_{A,{B'}}}}
\\
\Sigma_A(T(B)) \ar[r]_-{\Sigma_A(T(h))} \ar@/^6pc/[uuu]^-{\sigma_{A,B}} \ar[u]_>>>>>{\Sigma_A(T(\eta^{\Sigma_A \,\dashv\, \pi^*_A}_B))}_-{\qquad\qquad\quad\dcomment{\text{nat. of } \eta^{\Sigma_A \,\dashv\, \pi^*_A}}} & \Sigma_A(T(B')) \ar@/_6pc/[uuu]_-{\sigma_{A,{B'}}} \ar[u]^<<<<<{\Sigma_A(T(\eta^{\Sigma_A \,\dashv\, \pi^*_A}_B))}
}
\]

The naturality of $\eta^{\Sigma^{\mathbf{T}}_A \,\dashv\, \pi^*_A}$ and $\varepsilon^{\Sigma^{\mathbf{T}}_A \,\dashv\, \pi^*_A}$, and the two unit-counit laws follow directly from the corresponding properties for the adjunction $\Sigma_A \dashv \pi^*_A : \mathcal{V}_{p(A)} \longrightarrow \mathcal{V}_{\ia A}$.

We conclude by noting that the adjunction $\Sigma^{\mathbf{T}}_A \dashv \pi^*_A : \mathcal{V}^{\mathbf{T}}_{p(A)} \longrightarrow \mathcal{V}^{\mathbf{T}}_{\ia A}$ also satisfies the split Beck-Chevalley condition from Definition~\ref{def:splitdependentcompproducts}---similarly to the other properties, it also follows directly from the corresponding property of $\Sigma_A \dashv \pi^*_A$.
\end{proof}

\section{Proof of Theorem~\ref{thm:dependentsumsinEMfibration}}
\label{sect:proofofthm:dependentsumsinEMfibration}

{
\renewcommand{\thetheorem}{\ref{thm:dependentsumsinEMfibration}}
\begin{theorem}
Given a split comprehension category with unit $p : \mathcal{V} \longrightarrow \mathcal{B}$ with strong split dependent sums and a split fibred monad $\mathbf{T} = (T,\eta,\mu)$ on it, then the corresponding EM-fibration $p^{\mathbf{T}} : \mathcal{V}^{\mathbf{T}} \longrightarrow \mathcal{B}$ has split dependent $p$-sums if $p^{\mathbf{T}}$ has split fibred reflexive coequalizers.
\end{theorem}
\addtocounter{theorem}{-1}
}

\begin{proof}
Given an object $A$ in $\mathcal{V}$, the functor $\Sigma^{\mathbf{T}}_A : \mathcal{V}^{\mathbf{T}}_{\ia A} \longrightarrow \mathcal{V}^{\mathbf{T}}_{p(A)}$ is given on an object $(B,\beta)$ of $\mathcal{V}^{\mathbf{T}}_{\ia A}$ as the reflexive coequalizer 
\[
\xymatrix@C=5em@R=5em@M=0.5em{
(T(\Sigma_A(B)), \mu_{\Sigma_A(B)}) \ar[r]^-{e_{A,(B,\beta)}} & \Sigma^{\mathbf{T}}_A(B,\beta)
}
\]
of the following pair of morphisms in $\mathcal{V}^{\mathbf{T}}_{p(A)}$, given by
\[
\xymatrix@C=5em@R=5em@M=0.5em{
(T(\Sigma_A(T(B))), \mu_{\Sigma_A(T(B))}) \ar[r]^-{T(\Sigma_A(\beta))} & (T(\Sigma_A(B)), \mu_{\Sigma_A(B)})
}
\]
and
\[
\xymatrix@C=6em@R=3em@M=0.5em{
(T(\Sigma_A(T(B))), \mu_{\Sigma_A(T(B))}) \ar[r]^-{T(\Sigma_A(T(\eta^{\Sigma_A \,\dashv\, \pi^*_A}_B)))} & (T(\Sigma_A(T(\pi^*_A(\Sigma_A(B))))), \mu_{\Sigma_A(T(\pi^*_A(\Sigma_A(B))))}) \ar[d]^-{=}
\\
& (T(\Sigma_A(\pi^*_A(T(\Sigma_A(B))))), \mu_{\Sigma_A(\pi^*_A(T(\Sigma_A(B))))}) \ar[d]^-{T(\varepsilon^{\Sigma_A \,\dashv\, \pi^*_A}_{T(\Sigma_A(B))})}
\\
(T(\Sigma_A(B)), \mu_{\Sigma_A(B)}) & \ar[l]^-{\mu_{\Sigma_A(B)}} (T(T(\Sigma_A(B))), \mu_{T(\Sigma_A(B))})
}
\]
using the split dependent sums in $p$, i.e., the adjunction $\Sigma_A \dashv \pi^*_A : \mathcal{V}_{p(A)} \longrightarrow \mathcal{V}_{\ia A}$; and making use of Proposition~\ref{prop:verticalEMalgebras} to ensure that $\beta$ is a vertical morphism.
In the rest of this proof, we systematically refer to these two morphisms  as $(1)$ and $(2)$, respectively. Further, as a notational convenience, we often write $(U^{\mathbf{T}}(\Sigma^{\mathbf{T}}_A(B,\beta)), \beta_{\Sigma^{\mathbf{T}}_A})$ for $\Sigma^{\mathbf{T}}_A(B,\beta)$.

It is easy to see that both $(1)$ and $(2)$ are in $\mathcal{V}^{\mathbf{T}}_{p(A)}$. On the one hand, all morphisms used in the definitions of $(1)$ and $(2)$ are vertical over $\id_{p(A)}$. On the other hand, all morphisms used in the definitions of $(1)$ and $(2)$ are EM-algebra homomorphisms because i) we know from the definition of the Eilenberg-Moore resolution that ${F^{\mathbf{T}}(f) = T(f)}$, and ii) it follows from the definition of monads that the components of $\mu$ are EM-algebra homomorphisms, i.e., we have $\mu_{\Sigma_A(B)} \comp \mu_{T(\Sigma_A(B))} = \mu_{\Sigma_A(B)} \comp T(\mu_{\Sigma_A(B)})$. 

The final ingredient we need to construct the reflexive coequalizer $e_{A,(B,\beta)}$ is 
\[
\xymatrix@C=5em@R=5em@M=0.5em{
(T(\Sigma_A(B)), \mu_{\Sigma_A(B)}) \ar[r]^-{T(\Sigma_A(\eta_B))}  & (T(\Sigma_A(T(B))), \mu_{\Sigma_A(T(B))}) 
}
\]
that forms the common section of $(1)$ and $(2)$, as shown by the commutativity of the next diagram. 
In order to minimise the space taken by this diagram, we present it (and other diagrammatic proofs below) in $\mathcal{V}$ rather than in $\mathcal{V}^{\mathbf{T}}$, working directly with the underlying morphisms of the EM-algebra homomorphisms involved. 
\[
\hspace{-0.3cm}
\xymatrix@C=4.5em@R=5em@M=0.5em{
T(\Sigma_A(B)) \ar[rr]^-{T(\Sigma_A(\eta_B))} \ar[ddddd]_-{T(\Sigma_A(\eta_B))} \ar[dr]^-{\,\,\,\,\,T(\Sigma_A(\eta^{\Sigma_A \,\dashv\, \pi^*_A}_B))} \ar@/_8pc/[dddddrr]_>>>>>>>>>>>>>>>>>>>>>>>>>>>>>>>>>>>>>>>>>>>>>>>>>>>{\id_{T(\Sigma_A(B))}\!\!\!} \ar@/_2pc/[dddr]^-{\id_{T(\Sigma_A(B))}} & & T(\Sigma_A(T(B))) \ar[d]^-{T(\Sigma_A(T(\eta^{\Sigma_A \,\dashv\, \pi^*_A}_B)))}_-{\dcomment{\text{nat. of } \eta}\qquad\qquad\qquad}
\\
& T(\Sigma_A(\pi^*_A(\Sigma_A(B)))) \ar[r]^-{T(\Sigma_A(\eta_{\pi^*_A(\Sigma_A(B))}))} \ar@/_2pc/[dr]^-{\,\,\,T(\Sigma_A(\pi^*_A(\eta_{\Sigma_A(B)})))} \ar[dd]^-{T(\varepsilon^{\Sigma_A \,\dashv\, \pi^*_A}_{\Sigma_A(B)})}_<<<<<{\dcomment{\Sigma_A \dashv \pi^*_A}\qquad\,\,\,} & T(\Sigma_A(T(\pi^*_A(\Sigma_A(B))))) \ar[d]^-{=}_<<<<{\dcomment{\eta \text{ is split fibred}}\qquad\quad}
\\
& & T(\Sigma_A(\pi^*_A(T(\Sigma_A(B))))) \ar[d]^-{T(\varepsilon^{\Sigma_A \,\dashv\, \pi^*_A}_{T(\Sigma_A(B))})}_-{\dcomment{\text{nat. of } \eta^{\Sigma_A \,\dashv\, \pi^*_A}}\qquad\qquad}
\\
& T(\Sigma_A(B)) \ar[r]^-{T(\eta_{\Sigma_A(B)})} \ar@/_2pc/[ddr]^-{\!\!\!\!\id_{T(\Sigma_A(B))}}_<<<<<<{\dcomment{\text{id. law}}\quad\!\!\!\!} & T(T(\Sigma_A(B))) \ar[dd]^-{\mu_{\Sigma_A(B)}}_<<<<<<<<<<<{\dcomment{(T,\eta,\mu) \text{ is a monad}}\qquad}
\\
\ar@{}[d]^-{\qquad\quad\dcomment{(B,\beta) \text{ is an EM-algebra}}} &
\\
T(\Sigma_A(T(B))) \ar[rr]_-{T(\Sigma_A(\beta))} & & T(\Sigma_A(B))
}
\]

\pagebreak

Next, we define the action of $\Sigma^{\mathbf{T}}_A$ on morphisms of $\mathcal{V}^{\mathbf{T}}_{\ia A}$ using the universal property of reflexive coequalizers. In detail, given a morphism $h : (B,\beta) \longrightarrow (B',\beta')$, we define the corresponding morphism $\Sigma^{\mathbf{T}}_A(h)$ in $\mathcal{V}^{\mathbf{T}}_{p(A)}$ as the unique mediating morphism in
\[
\xymatrix@C=4.5em@R=4em@M=0.5em{
(T(\Sigma_A(T(B'))),\mu_{\Sigma_A(T(B'))}) \ar@<-0.75ex>[r]_-{(2)'} \ar@<0.75ex>[r]^-{(1)'} & (T(\Sigma_A(B')), \mu_{\Sigma_A(B')}) \ar[r]^-{e_{A,(B',\beta')}} & \Sigma^{\mathbf{T}}_A(B',\beta')
\\
(T(\Sigma_A(T(B))),\mu_{\Sigma_A(T(B))}) \ar[u]^-{T(\Sigma_A(T(h)))} \ar@<-0.75ex>[r]_-{(2)} \ar@<0.75ex>[r]^-{(1)} & (T(\Sigma_A(B)), \mu_{\Sigma_A(B)}) \ar[r]_-{e_{A,(B,\beta)}} \ar[u]_-{T(\Sigma_A(h))} & \Sigma^{\mathbf{T}}_A(B,\beta) \ar@{-->}[u]_-{\Sigma^{\mathbf{T}}_A(h)}
}
\]

In order for $\Sigma^{\mathbf{T}}_A(h)$ to exist and to be the unique such morphism, we need to prove 
\[
e_{A,(B',\beta')} \comp T(\Sigma_A(h)) \comp (1) = e_{A,(B',\beta')} \comp T(\Sigma_A(h)) \comp (2)
\]
We prove this equation by first observing that we have
\[
e_{A,(B',\beta')} \comp (1)' = e_{A,(B',\beta')} \comp (2)'
\]
because $e_{A,(B',\beta')}$ is the reflexive coequalizer of $(1)'$ and $(2)'$.
As a result, it suffices to show that the two left-hand squares given in the above diagram commute.

The left-hand square involving $(1)$ and $(1)'$ commutes because  $h$ is a EM-algebra homo\-morphism---this is best seen when we rotate this square by $90$ degrees:
\[
\xymatrix@C=4.5em@R=3em@M=0.5em{
T(\Sigma_A(T(B))) \ar[r]^-{T(\Sigma_A(T(h)))} \ar[d]_-{T(\Sigma_A(\beta))} & T(\Sigma_A(T(B'))) \ar[d]^-{T(\Sigma_A(\beta'))}
\\
T(\Sigma_A(B)) \ar[r]_-{T(\Sigma_A(h))} & T(\Sigma_A(B'))
}
\]
The other left-hand square, involving $(2)$ and $(2)'$, commutes due to naturality:
\[
\xymatrix@C=7em@R=7.5em@M=0.5em{
T(\Sigma_A(T(B))) \ar[r]^-{T(\Sigma_A(T(h)))} \ar[d]_-{T(\Sigma_A(T(\eta^{\Sigma_A \,\dashv\, \pi^*_A}_B)))} & T(\Sigma_A(T(B'))) \ar[d]^-{T(\Sigma_A(T(\eta^{\Sigma_A \,\dashv\, \pi^*_A}_{B'})))}_-{\dcomment{\text{nat. of } \eta^{\Sigma_A \,\dashv\, \pi^*_A}}\qquad\qquad\qquad\,\,}
\\
T(\Sigma_A(T(\pi^*_A(\Sigma_A(B))))) \ar[r]_-{T(\Sigma_A(T(\pi^*_A(\Sigma_A(h)))))} \ar[d]_-{=} & T(\Sigma_A(T(\pi^*_A(\Sigma_A(B'))))) \ar[d]^-{=}_-{\dcomment{T \text{ is split fibred}}\qquad\qquad\qquad}
\\
T(\Sigma_A(\pi^*_A(T(\Sigma_A(B))))) \ar[r]_-{T(\Sigma_A(\pi^*_A(T(\Sigma_A(h)))))} \ar[d]_-{T(\varepsilon^{\Sigma_A \,\dashv\, \pi^*_A}_{T(\Sigma_A(B))})} & T(\Sigma_A(\pi^*_A(T(\Sigma_A(B'))))) \ar[d]^-{T(\varepsilon^{\Sigma_A \,\dashv\, \pi^*_A}_{T(\Sigma_A(B'))})}_-{\dcomment{\text{nat. of } \varepsilon^{\Sigma_A \,\dashv\, \pi^*_A}}\qquad\qquad\qquad\,\,}
\\
T(T(\Sigma_A(B))) \ar[r]_-{T(T(\Sigma_A(h)))} \ar[d]_-{\mu_{\Sigma_A(B)}} & T(T(\Sigma_A(B'))) \ar[d]^-{\mu_{\Sigma_A(B')}}_-{\dcomment{\text{nat. of } \mu}\qquad\qquad\qquad\quad}
\\
T(\Sigma_A(B)) \ar[r]_-{T(\Sigma_A(h))} & T(\Sigma_A(B'))
}
\]

We omit the proofs showing that $\Sigma^{\mathbf{T}}_A$ preserves identities and composition---both  properties follow directly from using the universal property of reflexive coequalizers.

We proceed by proving that we have an adjunction $ \Sigma^{\mathbf{T}}_A  \dashv \pi^*_A : \mathcal{V}^{\mathbf{T}}_{p(A)} \longrightarrow \mathcal{V}^{\mathbf{T}}_{\ia A}$.

\pagebreak

First, the underlying morphism of a component $\eta^{\Sigma^{\mathbf{T}}_A \,\dashv\, \pi^*_A}_{(B,\beta)}$ of the unit natural transformation
\[
\eta^{\Sigma^{\mathbf{T}}_A \,\dashv\, \pi^*_A} : \id_{\mathcal{V}_{\ia A}} \longrightarrow \pi^*_A \comp \Sigma^{\mathbf{T}}_A
\]
is given by the following composite morphism:
\[
\xymatrix@C=4em@R=4em@M=0.5em{
B \ar[r]^-{\eta^{\Sigma_A \,\dashv\, \pi^*_A}_B} & \pi^*_A(\Sigma_A(B)) \ar[r]^-{\pi^*_A(\eta_{\Sigma_A(B)})} & \pi^*_A(T(\Sigma_A(B))) \ar[r]^-{\pi^*_A(e_{A,(B,\beta)})} & \pi^*_A(U^{\mathbf{T}}(\Sigma^{\mathbf{T}}_A(B,\beta)))
}
\]
which we prove to be a EM-algebra homomorphism from $(B,\beta)$ to  $\pi^*_A(\Sigma^{\mathbf{T}}_A(B,\beta))$ by \linebreak showing that the next diagram commutes in $\mathcal{V}_{\ia A}$.
\[
\scriptsize
\xymatrix@C=5.5em@R=6em@M=0.5em{
T(B) \ar@/_3pc/[dddd]_-{\beta} \ar[dd]^-{\eta^{\Sigma_A \,\dashv\, \pi^*_A}_{T(B)}} \ar[r]^-{T(\eta^{\Sigma_A \,\dashv\, \pi^*_A}_B)} & T(\pi^*_A(\Sigma_A(B))) \ar[r]^-{T(\pi^*_A(\eta_{\Sigma_A(B)}))} \ar[d]_-{=}_-{\dscomment{(b)}\qquad\qquad\quad} & T(\pi^*_A(T(\Sigma_A(B)))) \ar[r]^-{T(\pi^*_A(e_{A,(B,\beta)}))} \ar[d]_-{=}_-{\dscomment{\eta \text{ is split fibred}}\qquad\qquad} & T(\pi^*_A(U^{\mathbf{T}}(\Sigma^{\mathbf{T}}_A(B,\beta)))) \ar[d]^-{=}_-{\dscomment{T \text{ is split fibred}}\qquad\qquad}
\\
 & \pi^*_A(T(\Sigma_A(B))) \ar[r]^-{\pi^*_A(T(\eta_{\Sigma_A(B)}))} \ar@/^2pc/[dddr]_<<<<<<<{\id_{\pi^*_A(T(\Sigma_A(B)))}\!\!\!\!\!\!} & \pi^*_A(T(T(\Sigma_A(B)))) \ar[r]^-{\pi^*_A(T(e_{A,(B,\beta)}))} \ar[ddd]^-{\pi^*_A(\mu_{\Sigma_A(B)})}_<<<<<<{\dscomment{(T,\eta,\mu) \text{ a monad}}\quad} & \pi^*_A(T(U^{\mathbf{T}}(\Sigma^{\mathbf{T}}_A(B,\beta)))) \ar[ddd]^-{\pi^*_A(\beta_{\Sigma^{\mathbf{T}}_A})}_<<<<<<<<<<<<<<<{\dscomment{h \text{ is an EM-algebra homomorphism}}\quad\,\,\,\,\,}
\\
\pi^*_A(\Sigma_A(T(B))) \ar[r]^-{\!\pi^*_A(\eta_{\Sigma_A(T(B))})} \ar@/_2pc/[ddr]^-{\pi^*_A(\Sigma_A(\beta))}_-{\dscomment{\text{nat. of } \eta^{\Sigma_A \,\dashv\, \pi^*_A}}} & \pi^*_A(T(\Sigma_A(T(B)))) \ar@/_2pc/[ddr]_-{\pi^*_A(T(\Sigma_A(\beta)))\!\!\!\!}_<<<<<<{\dscomment{\text{nat. of } \eta}\qquad\qquad} \ar@/^2pc/[ddr]^<<<<<<{\!\!\!\pi^*_A((2))}_-{\dscomment{(a)}\qquad}
\\
&
\\
B \ar[r]_-{\eta^{\Sigma_A \,\dashv\, \pi^*_A}_B} & \pi^*_A(\Sigma_A(B)) \ar[r]_-{\pi^*_A(\eta_{\Sigma_A(B)})} & \pi^*_A(T(\Sigma_A(B))) \ar[r]_-{\pi^*_A(e_{A,(B,\beta)})} & \pi^*_A(U^{\mathbf{T}}(\Sigma^{\mathbf{T}}_A(B,\beta)))
}
\]
Here, the subdiagram marked with $(a)$ commutes because $e_{A,(B,\beta)}$ is the coequalizer of $T(\Sigma_A(\beta))$ and $(2)$; and the subdiagram marked with $(b)$ commutes because we have
\[
\scriptsize
\xymatrix@C=5.5em@R=6em@M=0.5em{
T(B) \ar[rr]^-{T(\eta^{\Sigma_A \,\dashv\, \pi^*_A}_B)} \ar[d]_-{\eta^{\Sigma_A \,\dashv\, \pi^*_A}_{T(B)}}^-{\qquad\qquad\qquad\qquad\qquad\dscomment{\text{nat. of } \eta^{\Sigma_A \,\dashv\, \pi^*_A}}} & & T(\pi^*_A(\Sigma_A(B))) \ar[dd]^-{=} \ar@/_2.5pc/[dddl]_>>>>>>>>>>>>>>>>>>>>>>>>>{\eta_{T(\pi^*_A(\Sigma_A(B)))}\!\!\!}^-{\,\,\,\,\,\,\dscomment{(T,\eta,\mu) \text{ is a monad}}} \ar[dl]_-{\eta^{\Sigma_A \,\dashv\, \pi^*_A}_{T(\pi^*_A(\Sigma_A(B)))}}
\\
\pi^*_A(\Sigma_A(T(B))) \ar[d]_-{\pi^*_A(\eta_{\Sigma_A(T(B))})}^>>>>>{\qquad\quad\dscomment{\text{nat. of } \eta}} \ar[r]^-{\pi^*_A(\Sigma_A(T(\eta^{\Sigma_A \,\dashv\, \pi^*_A}_B)))} & \pi^*_A(\Sigma_A(T(\pi^*_A(\Sigma_A(B))))) \ar[ddl]_<<<<<<<<<<<<{\pi^*_A(\eta_{\Sigma_A(T(\pi^*_A(\Sigma_A(B))))})\!\!\!\!}^>>>>>>>>>>>>>>>>>>>>>>>>>{\qquad\qquad\dscomment{\text{nat. of } \eta}}^>>>>>>>>>>>>>>{\qquad\qquad\dscomment{\eta \text{ is split fibred}}}
\\
\pi^*_A(T(\Sigma_A(T(B)))) \ar[d]_-{\pi^*_A(T(\Sigma_A(T(\eta^{\Sigma_A \,\dashv\, \pi^*_A}_B))))} & & \pi^*_A(T(\Sigma_A(B)))
\\
\pi^*_A(T(\Sigma_A(T(\pi^*_A(\Sigma_A(B)))))) \ar[dd]_-{=}^-{\qquad\quad\dscomment{T \text{ is split fibred}}} & T(T(\pi^*_A(\Sigma_A(B)))) \ar[d]_-{T(\eta^{\Sigma_A \,\dashv\, \pi^*_A}_{T(\pi^*_A(\Sigma_A(B)))})} \ar@/_2.5pc/[uuur]_<<<<<<<<<<<<<<{\!\!\!\mu_{\pi^*_A(\Sigma_A(B))}}
\\
& T(\pi^*_A(\Sigma_A(T(\pi^*_A(\Sigma_A(B)))))) \ar[ul]^-{=} \ar[dl]_-{=}^<<<<<<<{\qquad\qquad\qquad\dscomment{\Sigma_A \dashv \pi^*_A}}^>>>>>>>>>>>>>>>{\!\!\!\qquad\qquad\qquad\qquad\qquad\qquad\qquad\dscomment{T \text{ is split fibred}}}
\\
\pi^*_A(T(\Sigma_A(\pi^*_A(T(\Sigma_A(B)))))) \ar[rr]_-{\pi^*_A(T(\varepsilon^{\Sigma_A \,\dashv\, \pi^*_A}_{T(\Sigma_A(B))}))} & & \pi^*_A(T(T(\Sigma_A(B)))) \ar[uuu]_-{\pi^*_A(\mu_{\Sigma_A(B)})}^>>>>>>>>>>>>>>>>>>>>>>>>{\dscomment{\mu \text{ is split fibred}}\qquad\quad} \ar[uul]_-{=}
}
\]

We omit the proof showing that $\eta^{\Sigma^{\mathbf{T}}_A \,\dashv\, \pi^*_A}$ is natural---it follows directly from the naturality of $\eta$ and $\eta^{\Sigma_A \,\dashv\, \pi^*_A}$, combined with the definition of $\Sigma^{\mathbf{T}}_A$ on morphisms.

Next, the underlying morphism of a component $\varepsilon^{\Sigma^{\mathbf{T}}_A \,\dashv\, \pi^*_A}_{(B,\beta)}$ of the counit natural transformation
\[
\varepsilon^{\Sigma^{\mathbf{T}}_A \,\dashv\, \pi^*_A} : \Sigma^{\mathbf{T}}_A \comp \pi^*_A \longrightarrow \id_{\mathcal{V}_{p(A)}} 
\]
is given by the unique mediating morphism in 

\[
\hspace{-0.05cm}
\xymatrix@C=1.5em@R=3em@M=0.5em{
& & (B,\beta)
\\
(T(\Sigma_A(T(\pi^*_A(B)))),\mu_{\Sigma_A(T(\pi^*_A(B)))}) \ar@<-0.75ex>[r]_-{(2)} \ar@<0.75ex>[r]^-{(1)} & (T(\Sigma_A(\pi^*_A(B))), \mu_{\Sigma_A(\pi^*_A(B))}) \ar[dr]_-{e_{A,\pi^*_A(B,\beta)}\,\,\,\,} \ar[ur]^-{\beta \,\,\comp\,\, T(\varepsilon^{\Sigma_A \,\dashv\, \pi^*_A}_B)\,\,\,\,\,\,\,\,\,\,} & 
\\
& & \Sigma^{\mathbf{T}}_A(\pi^*_A(B,\beta)) \ar@{-->}[uu]_-{\varepsilon^{\Sigma^{\mathbf{T}}_A \,\dashv\, \pi^*_A}_{(B,\beta)}}
}
\]
using the universal property of the reflexive coequalizer $e_{A,(\pi^*_A(B),\pi^*_A(\beta))}$. In order to do so, we first prove that $\beta \comp T(\varepsilon^{\Sigma_A \,\dashv\, \pi^*_A}_B)$ is an EM-algebra homomomorphism, by showing
\[
\xymatrix@C=10em@R=6em@M=0.5em{
T(T(\Sigma_A(\pi^*_A(B)))) \ar[d]_-{\mu_{\Sigma_A(\pi^*_A(B))}} \ar[r]^-{T(T(\varepsilon^{\Sigma_A \,\dashv\, \pi^*_A}_B))} & T(T(B)) \ar[d]^-{\mu_B}_-{\dcomment{\text{nat. of } \mu}\qquad\qquad\qquad\,\,} \ar[r]^-{T(\beta)} & T(B) \ar[d]^-{\beta}_-{\dcomment{(B,\beta) \text{ is an EM-algebra}}\qquad\,\,\,}
\\
T(\Sigma_A(\pi^*_A(B))) \ar[r]_-{T(\varepsilon^{\Sigma_A \,\dashv\, \pi^*_A}_B)} & T(B) \ar[r]_-{\beta} & B
}
\]
Further, for $\varepsilon^{\Sigma^{\mathbf{T}}_A \,\dashv\, \pi^*_A}_{(B,\beta)}$ to exist and be unique such morphism, we also need to show that
\[
\beta \comp T(\varepsilon^{\Sigma_A \,\dashv\, \pi^*_A}_B) \comp (1) 
= 
\beta \comp T(\varepsilon^{\Sigma_A \,\dashv\, \pi^*_A}_B) \comp (2) 
\]
This last equation follows from the commutativity of the next diagram in $\mathcal{V}_{p(A)}$.

\[
\scriptsize
\xymatrix@C=9.5em@R=6em@M=0.5em{
T(\Sigma_A(T(\pi^*_A(B)))) \ar[r]^-{=} \ar[d]_-{T(\Sigma_A(T(\eta^{\Sigma_A \,\dashv\, \pi^*_A}_{\pi^*_A(B)})))}^>>>>>>{\qquad\qquad\dscomment{\Sigma_A \dashv \pi^*_A}} \ar[dr]^-{\id_{T(\Sigma_A(T(\pi^*_A(B))))}}^>>>>>>>>>{\qquad\qquad\quad\dscomment{\text{id. law}}} & T(\Sigma_A(\pi^*_A(T(B)))) \ar[r]^-{T(\Sigma_A(\pi^*_A(\beta)))} \ar@/^4.5pc/[ddd]^-{T(\varepsilon^{\Sigma_A \,\dashv\, \pi^*_A}_{T(B)})} & T(\Sigma_A(\pi^*_A(B))) \ar[ddd]^-{T(\varepsilon^{\Sigma_A \,\dashv\, \pi^*_A}_B)}_<<<<<<<<<<{\dscomment{\text{nat. of } \varepsilon^{\Sigma_A \,\dashv\, \pi^*_A}}\qquad\quad}
\\
T(\Sigma_A(T(\pi^*_A(\Sigma_A(\pi^*_A(B)))))) \ar[d]_-{=} \ar[r]_-{T(\Sigma_A(T(\pi^*_A(\varepsilon^{\Sigma_A \,\dashv\, \pi^*_A}_B))))} & T(\Sigma_A(T(\pi^*_A(B)))) \ar[d]_-{=}_-{\dscomment{T \text{ is split fibred}}\qquad\qquad\qquad\quad\!\!\!\!}
\\
T(\Sigma_A(\pi^*_A(T(\Sigma_A(\pi^*_A(B)))))) \ar[d]_-{T(\varepsilon^{\Sigma_A \,\dashv\, \pi^*_A}_{T(\Sigma_A(\pi^*_A(B)))})} \ar[r]_-{T(\Sigma_A(\pi^*_A(T(\varepsilon^{\Sigma_A \,\dashv\, \pi^*_A}_B))))} & T(\Sigma_A(\pi^*_A(T(B)))) \ar[d]_-{T(\varepsilon^{\Sigma_A \,\dashv\, \pi^*_A}_{T(B)})}_-{\dscomment{\text{nat. of } \varepsilon^{\Sigma_A \,\dashv\, \pi^*_A}}\qquad\qquad\qquad\quad}
\\
T(T(\Sigma_A(\pi^*_A(B)))) \ar[d]_-{\mu_{\Sigma_A(\pi^*_A(B))}} \ar[r]_-{T(T(\varepsilon^{\Sigma_A \,\dashv\, \pi^*_A}_B))} & T(T(B)) \ar[d]_-{\mu_B}_-{\dscomment{\text{nat. of } \mu}\qquad\qquad\qquad\qquad\!\!\!} \ar[r]_-{T(\beta)} & T(B) \ar[d]^-{\beta}_-{\dscomment{(B,\beta) \text{ is an EM-algebra}}\qquad\qquad\,\,\,\,}
\\
T(\Sigma_A(\pi^*_A(B))) \ar[r]_-{T(\varepsilon^{\Sigma_A \,\dashv\, \pi^*_A}_B)} & T(B) \ar[r]_-{\beta} & B
}
\]

\pagebreak

Next, we prove that $\varepsilon^{\Sigma^{\mathbf{T}}_A \,\dashv\, \pi^*_A}$ is a natural transformation: given a morphism \linebreak $h : (B,\beta) \longrightarrow (B',\beta')$ in $\mathcal{V}^{\mathbf{T}}_{p(A)}$, we show that the following diagram commutes in $\mathcal{V}_{p(A)}$:
\[
\xymatrix@C=6em@R=9em@M=0.5em{
U^{\mathbf{T}}(\Sigma^{\mathbf{T}}_A(\pi^*_A(B,\beta))) \ar[rr]^-{\varepsilon^{\Sigma^{\mathbf{T}}_A \,\dashv\, \pi^*_A}_{(B,\beta)}} \ar@/_7pc/[ddd]_-{\Sigma^{\mathbf{T}}_A(\pi^*_A(h))} && B \ar[ddd]^-{h}_<<<<<<<<<<<<<<<<<<<<<<<<<<<<<<<<<<<<<<<<<<<<<<<<<<<<{\dcomment{h \text{ is EM-alg. hom.}}\,\,\,\,\,\,\,\,\,\,\,\,}
\\
T(\Sigma_A(\pi^*_A(B))) \ar[u]_-{e_{A,\pi^*_A(B,\beta)}}_-{\quad\qquad\qquad\qquad\dcomment{\text{def. of } \varepsilon^{\Sigma^{\mathbf{T}}_A \,\dashv\, \pi^*_A}_{(B,\beta)}}} \ar[d]^-{T(\Sigma_A(\pi^*_A(h)))}_-{\dcomment{\text{def. of } \Sigma^{\mathbf{T}}_A(\pi^*_A(h))}\,\,\,} \ar[r]^-{T(\varepsilon^{\Sigma_A \,\dashv\, \pi^*_A}_B)} & T(B) \ar[d]^-{T(h)}_-{\dcomment{\text{nat. of } \varepsilon^{\Sigma_A \,\dashv\, \pi^*_A}}\qquad} \ar[ur]_-{\beta}
\\
T(\Sigma_A(\pi^*_A(B'))) \ar[d]^-{e_{A,\pi^*_A(B',\beta')}}^-{\quad\qquad\qquad\qquad\dcomment{\text{def. of } \varepsilon^{\Sigma^{\mathbf{T}}_A \,\dashv\, \pi^*_A}_{(B',\beta')}}} \ar[r]_-{T(\varepsilon^{\Sigma_A \,\dashv\, \pi^*_A}_{B'})} & T(B') \ar[dr]^-{\beta'}
\\
U^{\mathbf{T}}(\Sigma^{\mathbf{T}}_A(\pi^*_A(B',\beta'))) \ar[rr]_-{\varepsilon^{\Sigma^{\mathbf{T}}_A \,\dashv\, \pi^*_A}_{(B',\beta')}} && B'
}
\]
and then recall an important property of coequalizers---they are epimorphisms. As $e_{A,(\pi^*_A(B),\pi^*_A(\beta))}$ is a coequalizer and thus an epimorphism, the outer square starting \linebreak at $U^{\mathbf{T}}(\Sigma^{\mathbf{T}}_A(\pi^*_A(B),\pi^*_A(\beta)))$ commutes because epimorphisms are right-cancellative.

Next, we prove that the two unit-counit laws hold for $\eta^{\Sigma^{\mathbf{T}}_A \,\dashv\, \pi^*_A}$ and $\varepsilon^{\Sigma^{\mathbf{T}}_A \,\dashv\, \pi^*_A}$, by showing that the next two diagrams commute in $\mathcal{V}_{\ia A}$ and $\mathcal{V}_{p(A)}$, respectively.

\pagebreak

\[
\scriptsize
\xymatrix@C=7em@R=4em@M=0.5em{
\pi^*_A(B) \ar[r]^-{\eta^{\Sigma^{\mathbf{T}}_A \,\dashv\, \pi^*_A}_{\pi^*_A(B,\beta)}} \ar[d]_-{\eta^{\Sigma_A \,\dashv\, \pi^*_A}_{\pi^*_A(B)}}^-{\quad\qquad\dscomment{\text{def. of } \eta^{\Sigma^{\mathbf{T}}_A \,\dashv\, \pi^*_A}}} \ar@/_5pc/[ddd]_-{\id_{\pi^*_A(B)}} & \pi^*_A(U^{\mathbf{T}}(\Sigma^{\mathbf{T}}_A(\pi^*_A(B,\beta)))) \ar@/^5pc/[ddd]^-{\pi^*_A(\varepsilon^{\Sigma^{\mathbf{T}}_A \,\dashv\, \pi^*_A}_{(B,\beta)})}
\\
\pi^*_A(\Sigma_A(\pi^*_A(B))) \ar[r]_-{\pi^*_A(\eta_{\Sigma_A(\pi^*_A(B))})} \ar[dd]_-{\pi^*_A(\varepsilon^{\Sigma_A \,\dashv\, \pi^*_A}_B)} \ar@{}[d]_-{\dscomment{\Sigma_A \dashv \pi^*_A}\quad}^>>>>{\qquad\qquad\dscomment{\text{nat. of } \eta}} & \pi^*_A(T(\Sigma_A(\pi^*_A(B)))) \ar[d]_-{\pi^*_A(T(\varepsilon^{\Sigma_A \,\dashv\, \pi^*_A}_B))}^-{\,\,\,\,\dscomment{\text{def. of } \varepsilon^{\Sigma^{\mathbf{T}}_A \,\dashv\, \pi^*_A}}} \ar[u]^-{\pi^*_A(e_{A,\pi^*_A(B,\beta)})}
\\
& \pi^*_A(T(B)) \ar[d]^-{\pi^*_A(\beta)}_>>>{\dscomment{(B,\beta) \text{ is an EM-algebra}}\qquad\!\!\!\!\!\!\!\!\!\!\!\!\!\!\!\!\!\!\!\!}
\\
\pi^*_A(B) \ar[r]_-{\id_{\pi^*_A(B)}} \ar[ur]^-{\pi^*_A(\eta_B)} & \pi^*_A(B)
}
\]

\[
\scriptsize
\xymatrix@C=3.25em@R=5em@M=0.5em{
U^{\mathbf{T}}(\Sigma^{\mathbf{T}}_A(B,\beta)) \ar[rrr]^-{\Sigma^{\mathbf{T}}_A(\eta^{\Sigma^{\mathbf{T}}_A \,\dashv\, \pi^*_A}_{(B,\beta)})} \ar@/_3.5pc/[dddddd]^<<<<<<<<<<<<{\id_{U^{\mathbf{T}}(\Sigma^{\mathbf{T}}_A(B,\beta))}}^<<<<<<<<<<<<<<<<<<<<<<<<<{\qquad\dscomment{\text{id. law}}} &&& U^{\mathbf{T}}(\Sigma^{\mathbf{T}}_A(\pi^*_A(\Sigma^{\mathbf{T}}_A(B,\beta)))) \ar[dddddd]^-{\varepsilon^{\Sigma^{\mathbf{T}}_A \,\dashv\, \pi^*_A}_{\Sigma^{\mathbf{T}}_A(B,\beta)}}
\\
& T(\Sigma_A(B)) \ar@{}[dd]^-{\quad\qquad\qquad\dscomment{\text{def. of } \eta^{\Sigma^{\mathbf{T}}_A \,\dashv\, \pi^*_A}}} \ar[ul]_-{e_{A,(B,\beta)}}_-{\quad\qquad\qquad\qquad\qquad\qquad\qquad\dscomment{\text{def. of } \Sigma^{\mathbf{T}}_A(\eta^{\Sigma^{\mathbf{T}}_A \,\dashv\, \pi^*_A}_{(B,\beta)})}} \ar[r]^-{T(\Sigma_A(\eta^{\Sigma^{\mathbf{T}}_A \,\dashv\, \pi^*_A}_{(B,\beta)}))} \ar[d]^>>>>>>{T(\Sigma_A(\eta^{\Sigma_A \,\dashv\, \pi^*_A}_B))}_>>>>>{\dscomment{\Sigma_A \dashv \pi^*_A}\,\,\,\,\,\,\,\,\,\,} \ar@/_4pc/[ddddl]_-{\id_{T(\Sigma_A(B))}} & T(\Sigma_A(\pi^*_A(U^{\mathbf{T}}(\Sigma^{\mathbf{T}}_A(B,\beta))))) \ar[ur]^-{e_{A,\pi^*_A(\Sigma^{\mathbf{T}}_A(B,\beta))}\,\,\,\,} \ar[ddd]^-{T(\varepsilon^{\Sigma_A \,\dashv\, \pi^*_A}_{U^{\mathbf{T}}(\Sigma^{\mathbf{T}}_A(B,\beta))})} \ar@{}[dd]^<<<<<<<<<<<<<{\,\,\,\,\,\qquad\qquad\dscomment{\text{def. of } \varepsilon^{\Sigma^{\mathbf{T}}_A \,\dashv\, \pi^*_A}}}
\\
& T(\Sigma_A(\pi^*_A(\Sigma_A(B)))) \ar[d]^-{T(\Sigma_A(\pi^*_A(\eta_{\Sigma_A(B)})))} \ar@/_2pc/[dddl]_<<<<<<<<<<{T(\varepsilon^{\Sigma_A \,\dashv\, \pi^*_A}_{\Sigma_A(B)})\!\!\!\!}
\\
& T(\Sigma_A(\pi^*_A(T(\Sigma_A(B))))) \ar@/_3pc/[uur]^>>>>>{T(\Sigma_A(\pi^*_A(e_{A,(B,\beta)})))} \ar[d]^-{T(\varepsilon^{\Sigma_A \,\dashv\, \pi^*_A}_{T(\Sigma_A(B))})}^-{\qquad\qquad\qquad\dscomment{\text{nat. of } \varepsilon^{\Sigma_A \,\dashv\, \pi^*_A}}}_-{\dscomment{\text{nat. of } \varepsilon^{\Sigma_A \,\dashv\, \pi^*_A}}\qquad} &
\\
& T(T(\Sigma_A(B))) \ar[r]_-{T(e_{A,(B,\beta)})} \ar[dr]_-{\mu_{\Sigma_A(B)}}^>>>>>>>>>{\qquad\qquad\dscomment{e_{A,(B,\beta)} \text{ is an EM-alg. hom.}}} & T(U^{\mathbf{T}}(\Sigma^{\mathbf{T}}_A(B,\beta))) \ar@/^2pc/[ddr]^-{\!\!\!\!\beta_{\Sigma^{\mathbf{T}}_A}}
\\
T(\Sigma_A(B)) \ar[ur]^-{T(\eta_{\Sigma_A(B)})}_-{\qquad\qquad\dscomment{(T,\eta,\mu) \text{ is a monad}}} \ar[rr]_-{\id_{T(\Sigma_A(B))}} && T(\Sigma_A(B)) \ar[dr]_-{e_{A,(B,\beta)}}
\\
U^{\mathbf{T}}(\Sigma^{\mathbf{T}}_A(B,\beta)) \ar[rrr]_-{\id_{U^{\mathbf{T}}(\Sigma^{\mathbf{T}}_A(B,\beta))}} &&& U^{\mathbf{T}}(\Sigma^{\mathbf{T}}_A(B,\beta))
}
\]
Similarly to the naturality proof of $\varepsilon^{\Sigma^{\mathbf{T}}_A \,\dashv\, \pi^*_A}$, the outer square in the second diagram commutes because $e_{A,(B,\beta)}$ is a coequalizer, and thus an epimorphism and right-cancellative.

We conclude our proof of the existence of split dependent $p$-sums by proving that the functors $\Sigma^{\mathbf{T}}_A $ satisfy the split Beck-Chevalley condition. In particular, we show that for any Cartesian morphism $\overline{f}(A) : f^*(A) \longrightarrow A$, the next diagram commutes. 
\[
\scriptsize
\xymatrix@C=5em@R=5.5em@M=0.5em{
U^{\mathbf{T}}(\Sigma^{\mathbf{T}}_{f^*(A)}(\ia {\overline{f}(A)}^*(B,\beta))) \ar[rr]^-{\Sigma^{\mathbf{T}}_{f^*(A)}(\ia {\overline{f}(A)}^*(\eta^{\Sigma^{\mathbf{T}}_A \,\dashv\, \pi^*_A}_{(B,\beta)}))} \ar@/_5pc/[dddddddd]_<<<<<<<{\id_{U^{\mathbf{T}}(\Sigma^{\mathbf{T}}_{f^*(A)}(\ia {\overline{f}(A)}^*(B,\beta)))}} && U^{\mathbf{T}}(\Sigma^{\mathbf{T}}_{f^*(A)}(\ia {\overline{f}(A)}^*(\pi^*_A(\Sigma^{\mathbf{T}}_A(B,\beta))))) \ar@/^4pc/[ddd]^-{=}_-{\dscomment{p \text{ is a s. fib.}}\quad}_-{\dscomment{p \text{ is a split fibration}}\qquad\qquad\qquad\qquad}
\\
\txt<5pc>{$T(\Sigma_{f^*(A)}($\\$\ia {\overline{f}(A)}^*(B)))$}
\ar@/_4pc/[dddd]^-{=}
\ar[d]^<<<<<{T(\Sigma_{f^*(A)}(\ia {\overline{f}(A)}^*(\eta^{\Sigma_A \,\dashv\, \pi^*_A}_B)))}^<<{\qquad\qquad\qquad\qquad\qquad\qquad\dscomment{\text{def. of } \eta^{\Sigma^{\mathbf{T}}_A \,\dashv\, \pi^*_A}}} \ar[u]_-{e_{f^*(A),\ia {\overline{f}(A)}^*(B,\beta)}}_-{\qquad\qquad\qquad\qquad\qquad\dscomment{\text{def. of } \Sigma^{\mathbf{T}}_{f^*(A)}(\ia {\overline{f}(A)}^*(\eta^{\Sigma^{\mathbf{T}}_A \,\dashv\, \pi^*_A}_{(B,\beta)})) }} \ar[rr]^-{T(\Sigma_{f^*(A)}(\ia {\overline{f}(A)}^*(\eta^{\Sigma^{\mathbf{T}}_A \,\dashv\, \pi^*_A}_{(B,\beta)})))} && 
\txt<5pc>{$T(\Sigma_{f^*(A)}(\ia {\overline{f}(A)}^*($\\$\pi^*_A(U^{\mathbf{T}}(\Sigma^{\mathbf{T}}_A(B,\beta))))))$} 
\ar[u]^-{e_{f^*(A),\ia {\overline{f}(A)}^*(\pi^*_A(\Sigma^{\mathbf{T}}_A(B,\beta)))}} \ar[d]_-{=}
\\
\txt<5pc>{$T(\Sigma_{f^*(A)}(\ia {\overline{f}(A)}^*($\\$\pi^*_A(\Sigma_A(B)))))$} \ar[d]^-{=}^-{\qquad\dscomment{p \text{ is a split fibration}}}_-{\dscomment{\text{split BC}}\quad}
\ar@/^1.5pc/[dr]^<<<<<{\,\,\,\,\,\,\,\quad T(\Sigma_{f^*(A)}(\ia {\overline{f}(A)}^*(\pi^*_A(\eta_{\Sigma_A(B)}))))} && 
\txt<5pc>{$T(\Sigma_{f^*(A)}(\pi^*_{f^*(A)}(f^*($\\$U^{\mathbf{T}}(\Sigma^{\mathbf{T}}_A(B,\beta))))))$} \ar@/_3pc/[dd]_-{T(\varepsilon^{\Sigma_{f^*(A)} \,\dashv\, \pi^*_{f^*(A)}}_{f^*(U^{\mathbf{T}}(\Sigma^{\mathbf{T}}_A(B,\beta)))})} 
\ar[d]_>>>>>>{\dshide{e_{A,\pi^*_{f^*(A)}(f^*(\Sigma^{\mathbf{T}}_A(B,\beta)))}}}
\\
\txt<5pc>{$T(\Sigma_{f^*(A)}(\pi^*_{f^*(A)}($\\$f^*(\Sigma_A(B)))))$} \ar@/^1.5pc/[dr]_>>>>>>>{T(\Sigma_{f^*(A)}(\pi^*_{f^*(A)}(f^*(\eta_{\Sigma_A(B)}))))\!\!\!} \ar[dd]^-{T(\varepsilon^{\Sigma_{f^*(A)} \,\dashv\, \pi^*_{f^*(A)}}_{f^*(\Sigma_A(B))})}
& 
\txt<5pc>{$T(\Sigma_{f^*(A)}(\ia {\overline{f}(A)}^*($\\$\pi^*_A(T(\Sigma_A(B))))))$} 
\ar@/^3.5pc/[uur]^-{T(\Sigma_{f^*(A)}(\ia {\overline{f}(A)}^*(\pi^*_A(e_{A,(B,\beta)}))))\!\!\!\!} \ar[d]_-{=} 
\ar@{}[dd]^<<<<<<<<<<<<<<<{\qquad\qquad\qquad\dscomment{\text{nat. of } \varepsilon^{\Sigma_{f^*(A)} \,\dashv\, \pi^*_{f^*(A)}}}}
& 
\txt<5pc>{$U^{\mathbf{T}}(\Sigma^{\mathbf{T}}_{f^*(A)}(\pi^*_{f^*(A)}($\\$f^*(\Sigma^{\mathbf{T}}_A(B,\beta)))))$} \ar@/^3.5pc/[ddddd]^<<<<<<<<<<<<<<<<{\varepsilon^{\Sigma^{\mathbf{T}}_{f^*(A)} \,\dashv\, \pi^*_{f^*(A)}}_{f^*(B,\beta)}}
\\
& 
\txt<5pc>{$T(\Sigma_{f^*(A)}(\pi^*_{f^*(A)}($\\$f^*(T(\Sigma_A(B))))))$} 
\ar@/^1.75pc/[uur]^>>>{T(\Sigma_{f^*(A)}(\pi^*_{f^*(A)}(f^*(e_{A,(B,\beta)}))))\,\,\,\,\,\,} \ar[d]_-{T(\varepsilon^{\Sigma_{f^*(A)} \,\dashv\, \pi^*_{f^*(A)}}_{f^*(T(\Sigma_A(B)))})}_-{\dscomment{\text{nat. of } \varepsilon^{\Sigma_{f^*(A)} \,\dashv\, \pi^*_{f^*(A)}}}\qquad\qquad\qquad\quad} 
\ar@{}[dd]^-{\qquad\qquad\qquad\dscomment{T \text{ is split fibred}}}
&
\txt<5pc>{$T(f^*(U^{\mathbf{T}}($\\$\Sigma^{\mathbf{T}}_{A}(B,\beta))))$} \ar[d]_-{=}
\\
T(f^*(\Sigma_A(B))) \ar[r]_-{T(f^*(\eta_{\Sigma_A(B)}))} \ar[d]_-{=}
& 
T(f^*(T(\Sigma_A(B)))) \ar[d]_-{=}_-{\dscomment{T \text{ is split fibred}}\qquad\qquad\quad} \ar[ur]^-{T(f^*(e_{A,(B,\beta)}))\,\,\,\,}
&
\txt<5pc>{$f^*(T(U^{\mathbf{T}}($\\$\Sigma^{\mathbf{T}}_{A}(B,\beta))))$} \ar@/_2pc/[ddd]_-{f^*(\beta_{\Sigma^{\mathbf{T}}_A})}^-{\dscomment{\text{def. of } \varepsilon^{\Sigma^{\mathbf{T}}_{f^*(A)} \,\dashv\, \pi^*_{f^*(A)}}}}
\\
f^*(T(\Sigma_A(B))) \ar[r]_{f^*(T(\eta_{\Sigma_A(B)}))} \ar[dr]_-{\id_{f^*(T(\Sigma_A(B)))}} 
& 
f^*(T(T(\Sigma_A(B)))) \ar[d]^-{f^*(\mu_{(\Sigma_A(B))})}^<{\!\!\!\!\!\!\!\!\quad\qquad\qquad\dscomment{e_{A,(B,\beta)} \text{ is an EM-alg. hom.}}} \ar[ur]^-{f^*(T(e_{A,(B,\beta)}))\,\,\,\,}
\\
\ar@{}[d]^-{\qquad\quad\dscomment{e_{A,(B,\beta)} \text{ is a split fibred refl. coequalizer}}}
& f^*(T(\Sigma_A(B))) \ar[dr]_-{f^*(e_{A,(B,\beta)})}
\\
U^{\mathbf{T}}(\Sigma^{\mathbf{T}}_{f^*(A)}(\ia {\overline{f}(A)}^*(B,\beta))) \ar[rr]_-{=} && f^*(U^{\mathbf{T}}(\Sigma^{\mathbf{T}}_A(B,\beta)))
}
\]

Analogously to the naturality proof of $\varepsilon^{\Sigma^{\mathbf{T}}_A \,\dashv\, \pi^*_A}$ and the proof of the second unit-counit law, the outer square again commutes because $e_{f^*(A),\ia {\overline{f}(A)}^*(B,\beta)}$ is a coequalizer, and therefore an epimorphism and right-cancellative.
\end{proof}

\renewcommand\thesection{\thechapter.\arabic{section}}

\chapter{Proofs for Chapter~\ref{chap:interpretation}}
\label{chap:appendixC5}

\section{Proof of Proposition~\ref{prop:semweakening2}}
\label{sect:proofofprop:semweakening2}

{
\renewcommand{\thetheorem}{\ref{prop:semweakening2}}
\begin{proposition}[Semantic weakening]
Given value contexts $\Gamma_1$ and $\Gamma_2$, a value type $A$, and a value variable $x$ such that $\sem{\Gamma_1,\Gamma_2} \in \mathcal{B}$ and $\sem{\Gamma_1, x \!:\! A, \Gamma_2} \in \mathcal{B}$, then we have: 
\begin{enumerate}[(a)]
\item Given a value type $B$ such that $\sem{\Gamma_1,\Gamma_2;B} \in \mathcal{V}_{\sem{\Gamma_1,\Gamma_2}}$, then 
\[
\sem{\Gamma_1, x \!:\! A,\Gamma_2;B} = \sproj {\Gamma_1} x A {\Gamma_2}^*(\sem{\Gamma_1,\Gamma_2;B}) \in \mathcal{V}_{\sem{\Gamma_1, x : A,\Gamma_2}}
\]
\item Given a computation type $\ul{C}$ such that $\sem{\Gamma_1,\Gamma_2;\ul{C}} \in \mathcal{C}_{\sem{\Gamma_1,\Gamma_2}}$, then
\[
\sem{\Gamma_1, x \!:\! A,\Gamma_2;\ul{C}} = \sproj {\Gamma_1} x A {\Gamma_2}^*(\sem{\Gamma_1,\Gamma_2;\ul{C}}) \in \mathcal{C}_{\sem{\Gamma_1, x : A,\Gamma_2}}
\]
\item Given a value term $V$ such that $\sem{\Gamma_1,\Gamma_2;V} : 1_{\sem{\Gamma_1,\Gamma_2}} \longrightarrow B$, then
\[
\sem{\Gamma_1, x \!:\! A,\Gamma_2;V} = \sproj {\Gamma_1} x A {\Gamma_2}^*(\sem{\Gamma_1,\Gamma_2;V}) : 1_{\sem{\Gamma_1, x : A,\Gamma_2}} \longrightarrow \sproj {\Gamma_1} x A {\Gamma_2}^*(B)
\]
\item Given a computation term $M$ such that $\sem{\Gamma_1,\Gamma_2;M} : 1_{\sem{\Gamma_1,\Gamma_2}} \longrightarrow U(\ul{C})$, then 
\[
\sem{\Gamma_1, x \!:\! A,\Gamma_2;M} = \sproj {\Gamma_1} x A {\Gamma_2}^*(\sem{\Gamma_1,\Gamma_2;M}) : 1_{\sem{\Gamma_1, x : A,\Gamma_2}} \longrightarrow U(\sproj {\Gamma_1} x A {\Gamma_2}^*(\ul{C}))
\]
\item Given a computation variable $z$, a computation type $\ul{C}$, and a homomorphism term $K$ such that $\sem{\Gamma_1, \Gamma_2; z \!:\! \ul{C}; K} : \sem{\Gamma_1,\Gamma_2;\ul{C}} \longrightarrow \ul{D}$ in $\mathcal{C}_{\sem{\Gamma_1,\Gamma_2}}$, then

\[
\begin{array}{c}
\hspace{-4.5cm}
\sem{\Gamma_1, x \!:\! A, \Gamma_2; z \!:\! \ul{C}; K} = \sproj {\Gamma_1} x A {\Gamma_2}^*(\sem{\Gamma_1, \Gamma_2; z \!:\! \ul{C}; K}) 
\\
\hspace{5.5cm}
: \sproj {\Gamma_1} x A {\Gamma_2}^*(\sem{\Gamma_1, \Gamma_2; \ul{C}}) \longrightarrow \sproj {\Gamma_1} x A {\Gamma_2}^*(\ul{D})
\end{array}
\]
\end{enumerate}
where we use the notation
\[
\sem{\Gamma_1, x \!:\! A,\Gamma_2;B} = \sproj {\Gamma_1} x A {\Gamma_2}^*(\sem{\Gamma_1,\Gamma_2;B}) \in \mathcal{V}_{\sem{\Gamma_1, x : A,\Gamma_2}}
\]
to mean that $\sem{\Gamma_1, x \!:\! A,\Gamma_2;B}$ is defined and that it is equal to $\sproj {\Gamma_1} x A {\Gamma_2}^*(\sem{\Gamma_1,\Gamma_2;B})$ as an object of $\mathcal{V}_{\sem{\Gamma_1, x : A,\Gamma_2}}$. We also use analogous notation for terms and morphisms.
\end{proposition}
\addtocounter{theorem}{-1}
}

\begin{proof}
We prove $(a)$--$(e)$ simultaneously, by induction on the sum of the sizes of the arguments to $\sem{-}$. We omit the cases involving the MLTT fragment of eMLTT because in the setting of contextual categories these proofs can be found in~\cite[Chapter~III]{Streicher:Semantics}.

The proofs of all the cases (both those covered in op.~cit. and the ones discussed below) follow the same general pattern: they rely on the semantic structures we use in the definitions being split, i.e.,  preserved on-the-nose by reindexing functors.

\vspace{0.2cm}
\noindent
\textbf{Type of thunked computations:}
In this case, we assume that 
\[
\sem{\Gamma_1,\Gamma_2;U\ul{C}} \in \mathcal{V}_{\sem{\Gamma_1,\Gamma_2}}
\]
and we need to show that 
\[
\sem{\Gamma_1, x \!:\! A,\Gamma_2;U\ul{C}} = \sproj {\Gamma_1} x A {\Gamma_2}^*(\sem{\Gamma_1,\Gamma_2;U\ul{C}}) \in \mathcal{V}_{\sem{\Gamma_1, x : A,\Gamma_2}}
\] 

First, by inspecting the definition of $\sem{-}$ for $U\ul{C}$, the assumption gives us that 
\[
\sem{\Gamma_1,\Gamma_2;\ul{C}} \in \mathcal{C}_{\sem{\Gamma_1,\Gamma_2}}
\]
on which we can use $(b)$ to get that 
\[
\sem{\Gamma_1, x \!:\! A,\Gamma_2;\ul{C}} = \sproj {\Gamma_1} x A {\Gamma_2}^*(\sem{\Gamma_1,\Gamma_2;\ul{C}}) \in \mathcal{C}_{\sem{\Gamma_1, x : A,\Gamma_2}}
\]

Now, by applying the functor $U$ to both sides of this last equation, we get
\[
U(\sem{\Gamma_1, x \!:\! A,\Gamma_2;\ul{C}}) = U(\sproj {\Gamma_1} x A {\Gamma_2}^*(\sem{\Gamma_1,\Gamma_2;\ul{C}})) \in \mathcal{V}_{\sem{\Gamma_1, x : A,\Gamma_2}}
\]

Next, as $U$ is a split fibred functor, we also have that
\[
U(\sem{\Gamma_1, x \!:\! A,\Gamma_2;\ul{C}}) = \sproj {\Gamma_1} x A {\Gamma_2}^*(U(\sem{\Gamma_1,\Gamma_2;\ul{C}})) \in \mathcal{V}_{\sem{\Gamma_1, x : A,\Gamma_2}}
\]

Finally, by using the definition of $\sem{-}$ for $U\ul{C}$, we get that
\[
\sem{\Gamma_1, x \!:\! A,\Gamma_2;U\ul{C}} = \sproj {\Gamma_1} x A {\Gamma_2}^*(\sem{\Gamma_1,\Gamma_2;U\ul{C}}) \in \mathcal{V}_{\sem{\Gamma_1, x : A,\Gamma_2}}
\]

\vspace{0.2cm}
\noindent
\textbf{Homomorphic function space:}
In this case, we assume that
\[
\sem{\Gamma_1,\Gamma_2;\ul{C} \multimap \ul{D}} \in \mathcal{V}_{\sem{\Gamma_1,\Gamma_2}}
\]
and we need to show that
\[
\sem{\Gamma_1, x \!:\! A, \Gamma_2; \ul{C} \multimap \ul{D}} = \sproj {\Gamma_1} x A {\Gamma_2}^*(\sem{\Gamma_1,\Gamma_2;\ul{C} \multimap \ul{D}}) \in \mathcal{V}_{\sem{\Gamma_1, x : A,\Gamma_2}}
\]

First, by inspecting the definition of $\sem{-}$ for $\ul{C} \multimap \ul{D}$, the assumption gives us that
\[
\sem{\Gamma_1,\Gamma_2;\ul{C}} \in \mathcal{C}_{\sem{\Gamma_1,\Gamma_2}}
\qquad
\sem{\Gamma_1,\Gamma_2;\ul{D}} \in \mathcal{C}_{\sem{\Gamma_1,\Gamma_2}}
\]
on which we can use $(b)$ to get that
\[
\begin{array}{c}
\sem{\Gamma_1, x \!:\! A, \Gamma_2; \ul{C}} = \sproj {\Gamma_1} x A {\Gamma_2}^*(\sem{\Gamma_1,\Gamma_2;\ul{C}}) \in \mathcal{C}_{\sem{\Gamma_1, x : A,\Gamma_2}}
\\[3mm]
\sem{\Gamma_1, x \!:\! A, \Gamma_2; \ul{D}} = \sproj {\Gamma_1} x A {\Gamma_2}^*(\sem{\Gamma_1,\Gamma_2;\ul{D}}) \in \mathcal{C}_{\sem{\Gamma_1, x : A,\Gamma_2}}
\end{array}
\]

Next, by applying the functor $\multimap$ to both sides of these equations, we get 
\[
\begin{array}{c}
\hspace{-6.5cm}
\sem{\Gamma_1, x \!:\! A, \Gamma_2; \ul{C}} \multimap \sem{\Gamma_1, x \!:\! A, \Gamma_2; \ul{D}} = 
\\
\hspace{3cm}
\sproj {\Gamma_1} x A {\Gamma_2}^*(\sem{\Gamma_1,\Gamma_2;\ul{C}}) \multimap \sproj {\Gamma_1} x A {\Gamma_2}^*(\sem{\Gamma_1,\Gamma_2;\ul{D}}) \in \mathcal{V}_{\sem{\Gamma_1, x : A,\Gamma_2}}
\end{array}
\]

Next, as $\multimap$ is a split fibred functor, we also have that
\[
\begin{array}{c}
\hspace{-6.5cm}
\sem{\Gamma_1, x \!:\! A, \Gamma_2; \ul{C}} \multimap \sem{\Gamma_1, x \!:\! A, \Gamma_2; \ul{D}} = 
\\[-1mm]
\hspace{5.25cm}
\sproj {\Gamma_1} x A {\Gamma_2}^*(\sem{\Gamma_1,\Gamma_2;\ul{C}} \multimap \sem{\Gamma_1,\Gamma_2;\ul{D}}) \in \mathcal{V}_{\sem{\Gamma_1, x : A,\Gamma_2}}
\end{array}
\]

Finally, by using the definition of $\sem{-}$ for $\ul{C} \multimap \ul{D}$, we get that
\[
\sem{\Gamma_1, x \!:\! A, \Gamma_2; \ul{C} \multimap \ul{D}} = \sproj {\Gamma_1} x A {\Gamma_2}^*(\sem{\Gamma_1,\Gamma_2;\ul{C} \multimap \ul{D}}) \in \mathcal{V}_{\sem{\Gamma_1, x : A,\Gamma_2}}
\]

\vspace{0.2cm}
\noindent
\textbf{Type of free computations over a value type:}
We omit the proof for this case because it is analogous to the case for the type of thunked computations. 

\vspace{0.2cm}
\noindent
\textbf{Computational $\Sigma$-type:}
In this case, we assume that 
\[
\sem{\Gamma_1,\Gamma_2;\Sigma\, y \!:\! B .\, \ul{C}} \in \mathcal{C}_{\sem{\Gamma_1,\Gamma_2}}
\]
and we need to show that
\[
\sem{\Gamma_1, x \!:\! A,\Gamma_2;\Sigma\, y \!:\! B .\, \ul{C}} = \sproj {\Gamma_1} x A {\Gamma_2}^*(\sem{\Gamma_1,\Gamma_2;\Sigma\, y \!:\! B .\, \ul{C}}) \in \mathcal{C}_{\sem{\Gamma_1, x : A,\Gamma_2}}
\]

First, by inspecting the definition of $\sem{-}$ for $\Sigma\, y \!:\! B .\, \ul{C}$, the assumption gives us that 
\[
\sem{\Gamma_1,\Gamma_2;B} \in \mathcal{V}_{\sem{\Gamma_1,\Gamma_2}} 
\qquad
\sem{\Gamma_1,\Gamma_2, y \!:\! B;\ul{C}} \in \mathcal{C}_{\ia {\sem{\Gamma_1,\Gamma_2;B}}}
\]
on which we can use $(a)$ and the induction hypothesis, respectively, to get that
\[
\begin{array}{c}
\sem{\Gamma_1, x \!:\! A,\Gamma_2;B} = \sproj {\Gamma_1} x A {\Gamma_2}^*(\sem{\Gamma_1,\Gamma_2;B}) \in \mathcal{V}_{\sem{\Gamma_1, x : A,\Gamma_2}}
\\[3mm]
\sem{\Gamma_1, x \!:\! A,\Gamma_2, y \!:\! B;\ul{C}} = \sproj {\Gamma_1} x A {\Gamma_2, y : B}^*(\sem{\Gamma_1,\Gamma_2, y \!:\! B;\ul{C}}) \in \mathcal{C}_{\ia {\sem{\Gamma_1, x : A,\Gamma_2; B}}}
\end{array}
\]

Now, by using the definition of $\sproj {\Gamma_1} x A {\Gamma_2, y : B}$ in the second equation, we get that
\[
\begin{array}{c}
\hspace{-0.25cm}
\sem{\Gamma_1, x \!:\! A,\Gamma_2, y \!:\! B;\ul{C}} = (\ia {\overline{\sproj {\Gamma_1} x A {\Gamma_2}}(\sem{\Gamma_1, \Gamma_2;B})})^*(\sem{\Gamma_1,\Gamma_2, y \!:\! B;\ul{C}}) \in \mathcal{C}_{\ia {\sem{\Gamma_1, x : A,\Gamma_2; B}}}
\end{array}
\]

Next, by applying the functor $\Sigma_{\sem{\Gamma_1, x \!:\! A,\Gamma_2;B}}$ to both sides of this equation, we get that
\[
\begin{array}{c}
\hspace{-8cm}
\Sigma_{\sem{\Gamma_1, x : A,\Gamma_2;B}} (\sem{\Gamma_1, x \!:\! A,\Gamma_2, y \!:\! B;\ul{C}}) = 
\\
\hspace{1.5cm}
\Sigma_{\sem{\Gamma_1, x : A,\Gamma_2;B}} ((\ia {\overline{\sproj {\Gamma_1} x A {\Gamma_2}}(\sem{\Gamma_1, \Gamma_2;B})})^*(\sem{\Gamma_1,\Gamma_2, y \!:\! B;\ul{C}})) \in \mathcal{C}_{\sem{\Gamma_1, x : A,\Gamma_2}}
\end{array}
\]

Next, by using the equation we got from using $(a)$ above, we get that
\[
\begin{array}{c}
\hspace{-8cm}
\Sigma_{\sem{\Gamma_1, x : A,\Gamma_2;B}} (\sem{\Gamma_1, x \!:\! A,\Gamma_2, y \!:\! B;\ul{C}}) = 
\\
\Sigma_{\sproj {\Gamma_1} x A {\Gamma_2}^*(\sem{\Gamma_1,\Gamma_2;B})} ((\ia {\overline{\sproj {\Gamma_1} x A {\Gamma_2}}(\sem{\Gamma_1, \Gamma_2;B})})^*(\sem{\Gamma_1,\Gamma_2, y \!:\! B;\ul{C}})) \in \mathcal{C}_{\sem{\Gamma_1, x : A,\Gamma_2}}
\end{array}
\]

Now, by using the split Beck-Chevalley condition for $\Sigma_{\sem{\Gamma_1, x : A,\Gamma_2;B}}$, we get that
\[
\begin{array}{c}
\hspace{-6.5cm}
\Sigma_{\sem{\Gamma_1, x : A,\Gamma_2;B}} (\sem{\Gamma_1, x \!:\! A,\Gamma_2, y \!:\! B;\ul{C}}) = 
\\
\hspace{4.75cm}
\sproj {\Gamma_1} x A {\Gamma_2}^*(\Sigma_{\sem{\Gamma_1,\Gamma_2;B}} (\sem{\Gamma_1,\Gamma_2, y \!:\! B;\ul{C}})) \in \mathcal{C}_{\sem{\Gamma_1, x : A,\Gamma_2}}
\end{array}
\]

Finally, by using the definition of $\sem{-}$ for $\Sigma\, y \!:\! B .\, \ul{C}$, we get that
\[
\sem{\Gamma_1, x \!:\! A,\Gamma_2;\Sigma\, y \!:\! B .\, \ul{C}} = \sproj {\Gamma_1} x A {\Gamma_2}^*(\sem{\Gamma_1,\Gamma_2;\Sigma\, y \!:\! B .\, \ul{C}}) \in \mathcal{C}_{\sem{\Gamma_1, x : A,\Gamma_2}}
\]

\vspace{0.2cm}
\noindent
\textbf{Computational $\Pi$-type:}
We omit the proof for this case because it is analogous to the case for the computational $\Sigma$-type.

\vspace{0.2cm}
\noindent
\textbf{Thunking a computation:}
In this case, we assume that 
\[
\sem{\Gamma_1,\Gamma_2;\thunk M} : 1_{\sem{\Gamma_1,\Gamma_2}} \longrightarrow U(\ul{C})
\]
and we need to show that 
\[
\begin{array}{c}
\hspace{-3.25cm}
\sem{\Gamma_1, x \!:\! A,\Gamma_2;\thunk M} = \sproj {\Gamma_1} x A {\Gamma_2}^*(\sem{\Gamma_1,\Gamma_2;\thunk M}) 
\\
\hspace{7.5cm}
: 1_{\sem{\Gamma_1, x : A,\Gamma_2}} \longrightarrow \sproj {\Gamma_1} x A {\Gamma_2}^*(U(\ul{C}))
\end{array}
\]

First, by inspecting the definition of $\sem{-}$ for $\thunk M$, the assumption gives us
\[
\sem{\Gamma_1,\Gamma_2;M} : 1_{\sem{\Gamma_1,\Gamma_2}} \longrightarrow \ul{C}
\]

Now, by using $(d)$ on this morphism, we get that
\[
\sem{\Gamma_1, x \!:\! A,\Gamma_2;M} = \sproj {\Gamma_1} x A {\Gamma_2}^*(\sem{\Gamma_1,\Gamma_2;M}) : 1_{\sem{\Gamma_1, x : A,\Gamma_2}} \longrightarrow \sproj {\Gamma_1} x A {\Gamma_2}^*(\ul{C})
\]

Next, by using the definition of $\sem{-}$ for $\thunk M$, we get that
\[
\begin{array}{c}
\hspace{-2.75cm}
\sem{\Gamma_1, x \!:\! A,\Gamma_2;\thunk M} = \sproj {\Gamma_1} x A {\Gamma_2}^*(\sem{\Gamma_1,\Gamma_2;\thunk M}) 
\\
\hspace{7.7cm}
: 1_{\sem{\Gamma_1, x : A,\Gamma_2}} \longrightarrow U(\sproj {\Gamma_1} x A {\Gamma_2}^*(\ul{C}))
\end{array}
\]

Finally, as $U$ is a split fibred functor, we get that
\[
\begin{array}{c}
\hspace{-2.75cm}
\sem{\Gamma_1, x \!:\! A,\Gamma_2;\thunk M} = \sproj {\Gamma_1} x A {\Gamma_2}^*(\sem{\Gamma_1,\Gamma_2;\thunk M}) 
\\
\hspace{7.7cm}
: 1_{\sem{\Gamma_1, x : A,\Gamma_2}} \longrightarrow \sproj {\Gamma_1} x A {\Gamma_2}^*(U(\ul{C}))
\end{array}
\]

\vspace{0.2cm}
\noindent
\textbf{Homomorphic lambda abstraction:}
In this case, we assume that
\[
\sem{\Gamma_1,\Gamma_2;\lambda\, z \!:\! \ul{C} .\, K} : 1_{\sem{\Gamma_1,\Gamma_2}} \longrightarrow \sem{\Gamma_1,\Gamma_2;\ul{C}} \multimap \ul{D}
\]
and we have to show that
\[
\begin{array}{c}
\hspace{-3cm}
\sem{\Gamma_1, x \!:\! A,\Gamma_2;\lambda\, z \!:\! \ul{C} .\, K} = \sproj {\Gamma_1} x A {\Gamma_2}^*(\sem{\Gamma_1,\Gamma_2;\lambda\, z \!:\! \ul{C} .\, K}) 
\\
\hspace{5.5cm}
: 1_{\sem{\Gamma_1, x : A,\Gamma_2}} \longrightarrow \sproj {\Gamma_1} x A {\Gamma_2}^*(\sem{\Gamma_1,\Gamma_2;\ul{C}} \multimap \ul{D})
\end{array}
\]

First, by inspecting the definition of $\sem{-}$ for $\lambda\, z \!:\! \ul{C} .\, K$, the assumption gives us that
\[
\sem{\Gamma_1,\Gamma_2;z \!:\! \ul{C}; K} : \sem{\Gamma_1,\Gamma_2;\ul{C}} \longrightarrow \ul{D}
\]

Now, by using $(e)$ on this morphism, we get that
\[
\begin{array}{c}
\hspace{-3.5cm}
\sem{\Gamma_1, x \!:\! A, \Gamma_2; z \!:\! \ul{C}; K} = \sproj {\Gamma_1} x A {\Gamma_2}^*(\sem{\Gamma_1, \Gamma_2; z \!:\! \ul{C}; K}) 
\\
\hspace{5.5cm}
: \sproj {\Gamma_1} x A {\Gamma_2}^*(\sem{\Gamma_1, \Gamma_2; \ul{C}}) \longrightarrow \sproj {\Gamma_1} x A {\Gamma_2}^*(\ul{D})
\end{array}
\]

Finally, we show that
\[
\begin{array}{c}
\hspace{-3cm}
\sem{\Gamma_1, x \!:\! A,\Gamma_2;\lambda\, z \!:\! \ul{C} .\, K} = \sproj {\Gamma_1} x A {\Gamma_2}^*(\sem{\Gamma_1,\Gamma_2;\lambda\, z \!:\! \ul{C} .\, K}) 
\\
\hspace{5.5cm}
: 1_{\sem{\Gamma_1, x : A,\Gamma_2}} \longrightarrow \sproj {\Gamma_1} x A {\Gamma_2}^*(\sem{\Gamma_1,\Gamma_2;\ul{C}} \multimap \ul{D})
\end{array}
\]
by proving that the next diagram commutes, in which we write $\mathsf{pr}$ for $\sproj {\Gamma_1} x A {\Gamma_2}$. To improve the readability of this diagram, we aggregate some small proof steps.
\[
\vspace{0.3cm}
\xymatrix@C=11.5em@R=6em@M=0.5em{
1_{\sem{\Gamma_1, x : A, \Gamma_2}} \ar[r]^-{\sem{\Gamma_1, x : A,\Gamma_2;\lambda\, z : \ul{C} .\, K}} \ar[dr]_>>>>>>>>>>>>>>>>{\xi^{-1}_{\sem{\Gamma_1, x : A, \Gamma_2},\mathsf{pr}^*(\sem{\Gamma_1, \Gamma_2; \ul{C}}),\mathsf{pr}^*(\ul{D})}(\mathsf{pr}^*(\sem{\Gamma_1, \Gamma_2; z : \ul{C}; K}))\qquad\qquad\qquad} \ar@/_3pc/[dd]_-{=} & \mathsf{pr}^*(\sem{\Gamma_1,\Gamma_2;\ul{C}} \multimap \ul{D})
\\
& \mathsf{pr}^*(\sem{\Gamma_1,\Gamma_2;\ul{C}}) \multimap \mathsf{pr}^*(\ul{D}) \ar[u]_-{=}^>>>>>{\dcomment{\text{def. of } \sem{\Gamma_1, x \!:\! A,\Gamma_2;\lambda\, z \!:\! \ul{C} .\, K}} \quad}^>>>>>>>>>>>>>{\dcomment{(e)}\qquad\quad} \ar@{}[d]_<<<<<<<{\dcomment{\xi^{-1} \text{ is preserved on-the-nose by reindexing}}\qquad} \ar@{}[dd]_>>>>>>>>>>>>>>>>>>>>>>>>>>>>>>{\dcomment{\text{def. of } \sem{\Gamma_1,\Gamma_2;\lambda\, z \!:\! \ul{C} .\, K}}\qquad\qquad\qquad}
\\
\mathsf{pr}^*(1_{\sem{\Gamma_1,\Gamma_2}}) \ar@/^2pc/[r]^-{\mathsf{pr}^*(\xi^{-1}_{\sem{\Gamma_1,\Gamma_2}, \sem{\Gamma_1,\Gamma_2;\ul{C}},\ul{D}}(\sem{\Gamma_1,\Gamma_2; z : \ul{C} ; K}))} \ar@/_2pc/[r]_{\mathsf{pr}^*(\sem{\Gamma_1,\Gamma_2;\lambda\, z : \ul{C} .\, K})} & \mathsf{pr}^*(\sem{\Gamma_1,\Gamma_2;\ul{C}} \multimap \ul{D}) \ar@/_6pc/[uu]_>>>>>>>>>{\!\!\!\!\id_{\mathsf{pr}^*(\sem{\Gamma_1,\Gamma_2;\ul{C}} \multimap \ul{D})}}
\\
&
}
\vspace{-1.5cm}
\]

\vspace{0.1cm}
\noindent
\textbf{Returning a value:}
In this case, we assume that 
\[
\sem{\Gamma_1,\Gamma_2;\return V} : 1_{\sem{\Gamma_1,\Gamma_2}} \longrightarrow U(F(B))
\]
and we need to show that
\[
\begin{array}{c}
\hspace{-3.5cm}
\sem{\Gamma_1, x \!:\! A,\Gamma_2;\return V} = \sproj {\Gamma_1} x A {\Gamma_2}^*(\sem{\Gamma_1,\Gamma_2;\return V}) 
\\
\hspace{7cm}
: 1_{\sem{\Gamma_1, x : A,\Gamma_2}} \longrightarrow U(\sproj {\Gamma_1} x A {\Gamma_2}^*(F(B)))
\end{array}
\]

First, by inspecting the definition of $\sem{-}$ for $\return V$, the assumption gives us
\[
\sem{\Gamma_1,\Gamma_2;V} : 1_{\sem{\Gamma_1,\Gamma_2}} \longrightarrow B
\]

Next, by using $(c)$ on this morphism, we get that
\[
\sem{\Gamma_1, x \!:\! A,\Gamma_2;V} = \sproj {\Gamma_1} x A {\Gamma_2}^*(\sem{\Gamma_1,\Gamma_2;V}) : 1_{\sem{\Gamma_1, x : A,\Gamma_2}} \longrightarrow \sproj {\Gamma_1} x A {\Gamma_2}^*(B)
\]

Finally, we show that
\[
\begin{array}{c}
\hspace{-2.5cm}
\sem{\Gamma_1, x \!:\! A,\Gamma_2;\return V} = \sproj {\Gamma_1} x A {\Gamma_2}^*(\sem{\Gamma_1,\Gamma_2;\return V}) 
\\
\hspace{7cm}
: 1_{\sem{\Gamma_1, x : A,\Gamma_2}} \longrightarrow U(\sproj {\Gamma_1} x A {\Gamma_2}^*(F(B)))
\end{array}
\]
by showing that the next diagram commutes.
\[
\xymatrix@C=11.5em@R=6em@M=0.5em{
1_{\sem{\Gamma_1, x : A,\Gamma_2}} \ar[r]^-{\sem{\Gamma_1, x : A,\Gamma_2;\return V}} \ar[d]_>>>>>>>>>{\sem{\Gamma_1, x : A,\Gamma_2;V}}^-{\quad\qquad\qquad\dcomment{\text{def. of } \sem{\Gamma_1, x \!:\! A,\Gamma_2;\return V}}} \ar@/_7pc/[dd]_-{=} & U(\sproj {\Gamma_1} x A {\Gamma_2}^*(F(B)))
\\
\sproj {\Gamma_1} x A {\Gamma_2}^*(B) \ar[r]^-{\eta^{F \,\dashv\, U}_{\sproj {\Gamma_1} x A {\Gamma_2}^*(B)}} \ar[dr]^>>>>>>>>>>>>>>>>{\quad\sproj {\Gamma_1} x A {\Gamma_2}^*(\eta^{F \,\dashv\, U}_{B})}_>>>>>>>>>>>>>>>>>>>>>>{\dcomment{\text{functoriality of } \sproj {\Gamma_1} x A {\Gamma_2}^*}\qquad\qquad\qquad} & U(F(\sproj {\Gamma_1} x A {\Gamma_2}^*(B))) \ar[u]_-{=}
\\
\sproj {\Gamma_1} x A {\Gamma_2}^*(1_{\sem{\Gamma_1, \Gamma_2}}) \ar[u]_>>>>>>>>>>>>>>{\sproj {\Gamma_1} x A {\Gamma_2}^*(\sem{\Gamma_1,\Gamma_2;V})}^>>>>>>>{\dcomment{(c)}\qquad} \ar[r]_-{\sproj {\Gamma_1} x A {\Gamma_2}^*(\eta^{F \,\dashv\, U} \,\comp\, \sem{\Gamma_1,\Gamma_2;V})} \ar@/_5pc/[r]_-{\sproj {\Gamma_1} x A {\Gamma_2}^*(\sem{\Gamma_1,\Gamma_2;\return V})} & \sproj {\Gamma_1} x A {\Gamma_2}^*(U(F(B))) \ar[u]_-{=}^>>>>>{\dcomment{\text{Proposition~\ref{prop:fibrednaturaltransformationspreserved}}}\qquad\qquad}
\\
& \ar@{}[u]^>>>>>>>{\dcomment{\text{def. of } \sem{\Gamma_1,\Gamma_2;\return V}}\qquad\qquad\qquad\,\,}
}
\]

\noindent
\textbf{Sequential composition for computation terms:}
In this case, we assume that 
\[
\sem{\Gamma_1,\Gamma_2; \doto M {y \!:\! B} {\ul{C}} N} : 1_{\sem{\Gamma_1,\Gamma_2}} \longrightarrow U(\sem{\Gamma_1,\Gamma_2;\ul{C}})
\]
and we need to show that
\[
\begin{array}{c}
\hspace{-1.5cm}
\sem{\Gamma_1, x \!:\! A,\Gamma_2;\doto M {y \!:\! B} {\ul{C}} N} = \sproj {\Gamma_1} x A {\Gamma_2}^*(\sem{\Gamma_1,\Gamma_2;\doto M {y \!:\! B} {\ul{C}} N}) 
\\
\hspace{6cm}
: 1_{\sem{\Gamma_1, x : A,\Gamma_2}} \longrightarrow U(\sproj {\Gamma_1} x A {\Gamma_2}^*(\sem{\Gamma_1,\Gamma_2;\ul{C}}))
\end{array}
\]

First, by inspecting the definition of $\sem{-}$ for $\doto M {y \!:\! B} {\ul{C}} N$, the assumption gives us that
\[
\begin{array}{c}
\sem{\Gamma_1,\Gamma_2; M} : 1_{\sem{\Gamma_1,\Gamma_2}} \longrightarrow U(F(\sem{\Gamma_1,\Gamma_2;B}))
\\[3mm]
\sem{\Gamma_1,\Gamma_2, y \!:\! B; N} : 1_{\sem{\Gamma_1,\Gamma_2, y : B}} \longrightarrow U(\pi^*_{\sem{\Gamma_1,\Gamma_2;B}}(\sem{\Gamma_1,\Gamma_2;\ul{C}}))
\end{array}
\]

Next, by using the induction hypothesis on these morphisms, we get that
\[
\begin{array}{c}
\hspace{-5.4cm}
\sem{\Gamma_1, x \!:\! A,\Gamma_2;M} = \sproj {\Gamma_1} x A {\Gamma_2}^*(\sem{\Gamma_1,\Gamma_2;M}) 
\\
\hspace{5.5cm}
: 1_{\sem{\Gamma_1, x : A,\Gamma_2}} \longrightarrow U(\sproj {\Gamma_1} x A {\Gamma_2}^*(F(\sem{\Gamma_1,\Gamma_2;B})))
\\[3mm]
\hspace{-3.4cm}
\sem{\Gamma_1, x \!:\! A,\Gamma_2, y \!:\! B;N} = \sproj {\Gamma_1} x A {\Gamma_2, y : B}^*(\sem{\Gamma_1,\Gamma_2, y \!:\! B;N}) 
\\
\hspace{3cm}
: 1_{\sem{\Gamma_1, x : A,\Gamma_2, y : B}} \longrightarrow U(\sproj {\Gamma_1} x A {\Gamma_2, y : B}^*(\pi^*_{\sem{\Gamma_1,\Gamma_2;B}}(\sem{\Gamma_1,\Gamma_2;\ul{C}})))
\end{array}
\]

Finally, we show that 
\[
\begin{array}{c}
\hspace{-0.5cm}
\sem{\Gamma_1, x \!:\! A,\Gamma_2;\doto M {y \!:\! B} {\ul{C}} N} = \sproj {\Gamma_1} x A {\Gamma_2}^*(\sem{\Gamma_1,\Gamma_2;\doto M {y \!:\! B} {\ul{C}} N}) 
\\
\hspace{6cm}
: 1_{\sem{\Gamma_1, x : A,\Gamma_2}} \longrightarrow U(\sproj {\Gamma_1} x A {\Gamma_2}^*(\sem{\Gamma_1,\Gamma_2;\ul{C}}))
\end{array}
\]
by proving that the next diagram commutes, in which we write $\mathsf{pr}_{\Gamma_2}$ for $\sproj {\Gamma_1} x A {\Gamma_2}$ and $\mathsf{pr}_{\Gamma_2, y : B}$ for $\sproj {\Gamma_1} x A {\Gamma_2, y : B}$. For better readability, we aggregate small proof steps.

\[
\hspace{-0.3cm}
\scriptsize
\xymatrix@C=2em@R=4.5em@M=0.5em{
1_{\sem{\Gamma_1, x : A,\Gamma_2}} 
\ar[d]_-{\sem{\Gamma_1, x : A, \Gamma_2;M}}^<<<<{\!\qquad\qquad\qquad\dscomment{\text{use of the induction hypothesis on } \sem{\Gamma_1,\Gamma_2;M}}}^>>>>{\qquad\qquad\qquad\qquad\qquad\dscomment{U \text{ is split fibred}}} \ar[r]^-{=} & \mathsf{pr}_{\Gamma_2}^*(1_{\sem{\Gamma_1,\Gamma_2}}) \ar[d]^-{\mathsf{pr}^*_{\Gamma_2}(\sem{\Gamma_1,\Gamma_2;M})}
\\
U(\mathsf{pr}_{\Gamma_2}^*(F(\sem{\Gamma_1,\Gamma_2;B}))) \ar[d]_-{=} \ar[r]_-{=} & \mathsf{pr}^*_{\Gamma_2}(U(F(\sem{\Gamma_1,\Gamma_2;B}))) \ar[dd]^-{\mathsf{pr}^*_{\Gamma_2}(U(F(\langle \id_{\sem{\Gamma_1,\Gamma_2;B}} , ! \rangle)))}_<<<<<<<{\dscomment{F \text{ and } U \text{ are split fibred}} \qquad\qquad\qquad\qquad\quad}_<<<<<<<<<<<<<<<<<<<<<<<<<{\dscomment{\text{the fibred Cartesian products in } p \text{ are split}} \qquad\qquad\qquad\,\,\,\,}
\\
U(F(\mathsf{pr}_{\Gamma_2}^*(\sem{\Gamma_1,\Gamma_2;B}))) \ar[d]_-{U(F(\langle \id_{\mathsf{pr}_{\Gamma_2}^*(\sem{\Gamma_1,\Gamma_2;B})} , ! \rangle))} & 
\\
U(F(\Sigma_{\mathsf{pr}_{\Gamma_2}^*(\sem{\Gamma_1,\Gamma_2;B})}(\pi^*_{\mathsf{pr}_{\Gamma_2}^*(\sem{\Gamma_1,\Gamma_2;B})}(1_{\sem{\Gamma_1, x : A, \Gamma_2}})))) \ar[d]_-{=} \ar[r]_-{=} & \mathsf{pr}^*_{\Gamma_2}(U(F(\Sigma_{\sem{\Gamma_1,\Gamma_2;B}}(\pi^*_{\sem{\Gamma_1,\Gamma_2;B}}(1_{\sem{\Gamma_1,\Gamma_2}}))))) \ar[d]^-{=}_<<<<{\dscomment{\text{split Beck-Chevalley}}\qquad\qquad}_<<<<{\dscomment{\text{def. of } \sproj {\Gamma_1} x A {\Gamma_2, y : B}}\qquad\qquad\qquad\qquad\qquad\qquad\qquad}_<<<<<<<<<<{\dscomment{F \text{ and } U \text{ are split fibred}} \qquad\qquad\qquad\qquad\quad\!\!\!\!}
\\
U(F(\Sigma_{\mathsf{pr}_{\Gamma_2}^*(\sem{\Gamma_1,\Gamma_2;B})}(1_{\sem{\Gamma_1, x : A, \Gamma_2, y : B}}))) \ar[dd]^<<<<<{U(F(\Sigma_{\mathsf{pr}_{\Gamma_2}^*(\sem{\Gamma_1,\Gamma_2;B})}(\sem{\Gamma_1, x : A,\Gamma_2,y : B;N})))} \ar[r]_-{=} & \mathsf{pr}^*_{\Gamma_2}(U(F(\Sigma_{\sem{\Gamma_1,\Gamma_2;B}}(1_{\sem{\Gamma_1,\Gamma_2,y : B}})))) \ar[dd]_>>>>>{\mathsf{pr}^*_{\Gamma_2}(U(F(\Sigma_{\sem{\Gamma_1,\Gamma_2;B}}(\sem{\Gamma_1,\Gamma_2,y:B;N}))))}_>>>>>>>>>>>{\dscomment{\text{split Beck-Chevalley}}\qquad\qquad\qquad\qquad\quad}_>>>>>>>>>>>{\dscomment{\text{def. of } \sproj {\Gamma_1} x A {\Gamma_2, y : B}}\quad\qquad\qquad\qquad\qquad\qquad\qquad\qquad\qquad}_>>>>>>>>>>>{\dscomment{F \text{ and } U \text{ are s. fib.}} \qquad\!\!\!\!}_<<<<<<<<<<<<{\dscomment{\text{use of the induction hypothesis on } \sem{\Gamma_1,\Gamma_2, y \!:\! B;N}}\qquad\qquad\quad}
\\
&
\\
U(F(\Sigma_{\mathsf{pr}_{\Gamma_2}^*(\sem{\Gamma_1,\Gamma_2;B})}(U(\mathsf{pr}^*_{\Gamma_2, y : B}(\pi^*_{\sem{\Gamma_1,\Gamma_2;B}}(\sem{\Gamma_1, \Gamma_2;\ul{C}})))))) \ar[d]_-{=}^-{\qquad\qquad\qquad\qquad\qquad\qquad\qquad\dscomment{\text{split Beck-Chevalley}}}^-{\qquad\qquad\dscomment{\text{def. of } \sproj {\Gamma_1} x A {\Gamma_2, y : B}}} \ar[r]_-{=} & \mathsf{pr}^*_{\Gamma_2}(U(F(\Sigma_{\sem{\Gamma_1,\Gamma_2;B}}(U(\pi^*_{\sem{\Gamma_1,\Gamma_2;B}}(\sem{\Gamma_1,\Gamma_2;\ul{C}})))))) \ar[dd]^-{=}
\\
U(F(\Sigma_{\mathsf{pr}_{\Gamma_2}^*(\sem{\Gamma_1,\Gamma_2;B})}(U(\pi^*_{\mathsf{pr}_{\Gamma_2}^*(\sem{\Gamma_1,\Gamma_2;B})}(\mathsf{pr}^*_{\Gamma_2}(\sem{\Gamma_1,\Gamma_2;\ul{C}})))))) \ar[d]_{=}^-{\qquad\qquad\qquad\qquad\quad\,\,\,\,\dscomment{F \text{ and } U \text{ are split fibred}}}
\\
U(F(\Sigma_{\mathsf{pr}_{\Gamma_2}^*(\sem{\Gamma_1,\Gamma_2;B})}(\pi^*_{\mathsf{pr}_{\Gamma_2}^*(\sem{\Gamma_1,\Gamma_2;B})}(U(\mathsf{pr}^*_{\Gamma_2}(\sem{\Gamma_1,\Gamma_2;\ul{C}})))))) \ar[d]_-{U(F(\varepsilon^{\Sigma_{\mathsf{pr}_{\Gamma_2}^*(\sem{\Gamma_1,\Gamma_2;B})} \,\dashv\, \pi^*_{\mathsf{pr}_{\Gamma_2}^*(\sem{\Gamma_1,\Gamma_2;B})}}_{U(\mathsf{pr}^*_{\Gamma_2}(\sem{\Gamma_1,\Gamma_2;\ul{C}}))}))} \ar[r]_-{=} & \mathsf{pr}^*_{\Gamma_2}(U(F(\Sigma_{\sem{\Gamma_1,\Gamma_2;B}}(\pi^*_{\sem{\Gamma_1,\Gamma_2;B}}(U(\sem{\Gamma_1,\Gamma_2;\ul{C}})))))) \ar[d]^-{\mathsf{pr}^*_{\Gamma_2}(U(F(\varepsilon^{\Sigma_{\sem{\Gamma_1,\Gamma_2;B}} \,\dashv\, \pi^*_{\sem{\Gamma_1,\Gamma_2;B}}}_{U(\sem{\Gamma_1,\Gamma_2;\ul{C}})})))}_>>>>{\dscomment{\text{split Beck-Chevalley}}\qquad\qquad\qquad\qquad\quad}_>>>>{\dscomment{\text{def. of } \sproj {\Gamma_1} x A {\Gamma_2, y : B}}\quad\qquad\qquad\qquad\qquad\qquad\qquad\qquad\qquad}_>>>>{\dscomment{F \text{ and } U \text{ are s. fib.}} \qquad\!\!\!\!}_<<<<{\dscomment{\text{Proposition~\ref{prop:BCfordepsums} for } \varepsilon^{\Sigma_{\mathsf{pr}_{\Gamma_2}^*(\sem{\Gamma_1,\Gamma_2;B})} \,\dashv\, \pi^*_{\mathsf{pr}_{\Gamma_2}^*(\sem{\Gamma_1,\Gamma_2;B})}}}\qquad\qquad\quad}
\\
U(F(U(\mathsf{pr}^*_{\Gamma_2}(\sem{\Gamma_1,\Gamma_2;\ul{C}})))) \ar[d]_-{U(\varepsilon^{F \,\dashv\, U}_{\mathsf{pr}^*_{\Gamma_2}(\sem{\Gamma_1,\Gamma_2;\ul{C}})})} \ar[r]_-{=} & \mathsf{pr}^*_{\Gamma_2}(U(F(U(\sem{\Gamma_1,\Gamma_2;\ul{C}})))) \ar[d]^-{\mathsf{pr}^*_{\Gamma_2}(U(\varepsilon^{F \,\dashv\, U}_{\sem{\Gamma_1,\Gamma_2;\ul{C}}}))}_>>>>{\dscomment{\text{split Beck-Chevalley}}\qquad\qquad\qquad\qquad\quad}_>>>>{\dscomment{\text{def. of } \sproj {\Gamma_1} x A {\Gamma_2, y : B}}\quad\qquad\qquad\qquad\qquad\qquad\qquad\qquad\qquad}_>>>>{\dscomment{F \text{ and } U \text{ are s. fib.}} \qquad\!\!\!\!}_<<<<{\dscomment{\text{Proposition~\ref{prop:fibrednaturaltransformationspreserved} for } \varepsilon^{F \,\dashv\, U}}\qquad\qquad\qquad\qquad\quad}
\\
U(\mathsf{pr}_{\Gamma_2}^*(\sem{\Gamma_1,\Gamma_2;\ul{C}})) \ar[r]_-{=} & \mathsf{pr}^*_{\Gamma_2}(U(\sem{\Gamma_1,\Gamma_2;\ul{C}}))
}
\]

We conclude by observing that the left-hand side top-to-bottom composite morphism is equal to  $\sem{\Gamma_1, x \!:\! A, \Gamma_2;\doto M {y \!:\! B} {\ul{C}} N}$, and that the right-hand side top-to-bottom composite morphism is equal to $\sproj {\Gamma_1} x A {\Gamma_2} ^* (\sem{\Gamma_1,\Gamma_2;\doto M {y \!:\! B} {\ul{C}} N})$. 

\vspace{0.2cm}
\noindent
\textbf{Computational pairing for computation terms:}
In this case, we assume that 
\[
\sem{\Gamma_1,\Gamma_2; \langle V , M \rangle_{(y : B).\, \ul{C}}} : 1_{\sem{\Gamma_1,\Gamma_2}} \longrightarrow U(\Sigma_{\sem{\Gamma_1,\Gamma_2;B}}(\sem{\Gamma_1,\Gamma_2, y \!:\! B ;\ul{C}}))
\]
and we need to show that
\[
\begin{array}{c}
\hspace{-2.5cm}
\sem{\Gamma_1, x \!:\! A,\Gamma_2; \langle V , M \rangle_{(y : B).\, \ul{C}}} = \sproj {\Gamma_1} x A {\Gamma_2}^*(\sem{\Gamma_1,\Gamma_2;\langle V , M \rangle_{(y : B).\, \ul{C}}}) 
\\
\hspace{4cm}
: 1_{\sem{\Gamma_1, x : A,\Gamma_2}} \longrightarrow U(\sproj {\Gamma_1} x A {\Gamma_2}^*(\Sigma_{\sem{\Gamma_1,\Gamma_2}}(\sem{\Gamma_1,\Gamma_2, y \!:\! B ;\ul{C}})))
\end{array}
\]

First, by inspecting the definition of $\sem{-}$ for $\langle V , M \rangle_{(y : B).\, \ul{C}}$, the assumption  gives us
\[
\begin{array}{c}
\sem{\Gamma_1,\Gamma_2;V} : 1_{\sem{\Gamma_1,\Gamma_2}} \longrightarrow \sem{\Gamma_1,\Gamma_2;B}
\\[3mm]
\sem{\Gamma_1,\Gamma_2;M} : 1_{\sem{\Gamma_1,\Gamma_2}} \longrightarrow U((\mathsf{s}(\sem{\Gamma_1,\Gamma_2;V}))^*(\sem{\Gamma_1,\Gamma_2, y \!:\! B ;\ul{C}}))
\end{array}
\]
on which we can use $(c)$ and the induction hypothesis, respectively, to get that
\[
\begin{array}{c}
\sem{\Gamma_1, x \!:\! A,\Gamma_2;V} = \sproj {\Gamma_1} x A {\Gamma_2}^*(\sem{\Gamma_1,\Gamma_2;V}) : 1_{\sem{\Gamma_1, x : A,\Gamma_2}} \longrightarrow \sproj {\Gamma_1} x A {\Gamma_2}^*(B)
\\[3mm]
\hspace{-6cm}
\sem{\Gamma_1, x \!:\! A,\Gamma_2;M} = \sproj {\Gamma_1} x A {\Gamma_2}^*(\sem{\Gamma_1,\Gamma_2;M})
\\
\hspace{2.5cm}
: 1_{\sem{\Gamma_1, x : A, \Gamma_2}} \longrightarrow U(\sproj {\Gamma_1} x A {\Gamma_2}^*((\mathsf{s}(\sem{\Gamma_1,\Gamma_2;V}))^*(\sem{\Gamma_1,\Gamma_2, y \!:\! B ;\ul{C}})))
\end{array}
\]

Finally, we show that 
\[
\begin{array}{c}
\hspace{-1.5cm}
\sem{\Gamma_1, x \!:\! A,\Gamma_2; \langle V , M \rangle_{(y : B).\, \ul{C}}} = \sproj {\Gamma_1} x A {\Gamma_2}^*(\sem{\Gamma_1,\Gamma_2;\langle V , M \rangle_{(y : B).\, \ul{C}}}) 
\\
\hspace{3.5cm}
: 1_{\sem{\Gamma_1, x : A,\Gamma_2}} \longrightarrow U(\sproj {\Gamma_1} x A {\Gamma_2}^*(\Sigma_{\sem{\Gamma_1,\Gamma_2;B}}(\sem{\Gamma_1,\Gamma_2, y \!:\! B ;\ul{C}})))
\end{array}
\]
by proving that the next diagram commutes, in which we write $\mathsf{pr}_{\Gamma_2}$ for $\sproj {\Gamma_1} x A {\Gamma_2}$ and $\mathsf{pr}_{\Gamma_2, y : B}$ for $\sproj {\Gamma_1} x A {\Gamma_2, y : B}$. For better readability, we aggregate small proof steps.
\[
\scriptsize
\xymatrix@C=4em@R=7em@M=0.5em{
1_{\sem{\Gamma_1, x : A,\Gamma_2}} 
\ar[r]^-{=} \ar[d]_-{\sem{\Gamma_1,x:A,\Gamma_2;M}}^<<<<<<{\qquad\qquad\quad\!\!\dscomment{\text{use of the induction hypothesis on } \sem{\Gamma_1,\Gamma_2;M} }}^>>>>>>>{\quad\qquad\qquad\qquad\qquad\dscomment{U \text{ is split fibred}}}
& 
\mathsf{pr}_{\Gamma_2}^*(1_{\sem{\Gamma_1,\Gamma_2}})
\ar[d]^-{\mathsf{pr}_{\Gamma_2}^*(\sem{\Gamma_1,\Gamma_2;M})}
\\
U(\mathsf{pr}_{\Gamma_2}^*((\mathsf{s}(\sem{\Gamma_1,\Gamma_2;V}))^*(\sem{\Gamma_1,\Gamma_2,y \!:\! B;\ul{C}}))) \ar[r]^-{=} \ar[d]_-{=}^{\,\,\,\quad\qquad\qquad\dscomment{\text{Proposition~\ref{prop:BCfordepcompsums} for } \eta^{\Sigma_{\sem{\Gamma_1,\Gamma_2;B}} \,\dashv\, \pi^*_{\sem{\Gamma_1,\Gamma_2;B}}}}} 
&
\mathsf{pr}_{\Gamma_2}^*(U((\mathsf{s}(\sem{\Gamma_1,\Gamma_2;V}))^*(\sem{\Gamma_1,\Gamma_2,y \!:\! B;\ul{C}}))) \ar[ddd]_>>>>>>{\mathsf{pr}_{\Gamma_2}^*(U((\mathsf{s}(\sem{\Gamma_1,\Gamma_2;V}))^*(\eta^{\Sigma_{\sem{\Gamma_1,\Gamma_2;B}} \,\dashv\, \pi^*_{\sem{\Gamma_1,\Gamma_2;B}}}_{\sem{\Gamma_1,\Gamma_2,y \!:\! B;\ul{C}}})))}
\\
U((\mathsf{s}(\mathsf{pr}_{\Gamma_2}^*(\sem{\Gamma_1,\Gamma_2;V})))^*(\mathsf{pr}_{\Gamma_2, y : B}^*(\sem{\Gamma_1,\Gamma_2, y \!:\! B;\ul{C}}))) \ar[dd]^<<<<<{U((\mathsf{s}(\mathsf{pr}_{\Gamma_2}^*(\sem{\Gamma_1,\Gamma_2;V})))^*(\eta^{\Sigma_{\mathsf{pr}_{\Gamma_2}^*(\sem{\Gamma_1,\Gamma_2;B})} \,\dashv\, \pi^*_{\mathsf{pr}_{\Gamma_2}^*(\sem{\Gamma_1,\Gamma_2;B})}}_{\mathsf{pr}_{\Gamma_2, y : B}^*(\sem{\Gamma_1,\Gamma_2, y : B;\ul{C}})}))}^>>>>>>>>>>>>>>>>>>>>>{\qquad\qquad\dscomment{(*)}}^>>>>>>>>>>>>>>>>>>>>>{\quad\qquad\qquad\qquad\dscomment{\text{split Beck-Chevalley}}}^>>>>>>>>>>>>>>>>>>>>>{\quad\qquad\qquad\qquad\qquad\qquad\qquad\qquad\dscomment{U \text{ is split fibred}}}
\\
&
\\
\txt<12pc>{
$U((\mathsf{s}(\mathsf{pr}_{\Gamma_2}^*(\sem{\Gamma_1,\Gamma_2;V})))^*(\pi^*_{\mathsf{pr}_{\Gamma_2}^*(\sem{\Gamma_1,\Gamma_2;B})}($
\\
$\Sigma_{\mathsf{pr}_{\Gamma_2}^*(\sem{\Gamma_1,\Gamma_2;B})}(\mathsf{pr}_{\Gamma_2, y : B}^*(\sem{\Gamma_1,\Gamma_2, y \!:\! B;\ul{C}})))))$
} 
\ar[r]^-{=} \ar[d]_-{=}^<<<<<{\qquad\qquad\dscomment{\mathsf{s}(\mathsf{pr}_{\Gamma_2}^*(\sem{\Gamma_1,\Gamma_2;V})) \text{ is a section of } \pi_{\mathsf{pr}_{\Gamma_2}^*(\sem{\Gamma_1,\Gamma_2;B})}}}^>>>>>>{\,\,\,\,\,\quad\qquad\qquad\dscomment{\mathsf{s}(\sem{\Gamma_1,\Gamma_2;V}) \text{ is a section of } \pi_{\sem{\Gamma_1,\Gamma_2;B}}}}
&
\txt<10.5pc>{
$\mathsf{pr}_{\Gamma_2}^*(U((\mathsf{s}(\sem{\Gamma_1,\Gamma_2;V}))^*(\pi^*_{\sem{\Gamma_1,\Gamma_2;B}}($
\\
$\Sigma_{\Gamma_1,\Gamma_2;B}(\sem{\Gamma_1,\Gamma_2,y \!:\! B;\ul{C}})))))$
}
\ar[dd]^-{=}
\\
U(\Sigma_{\mathsf{pr}_{\Gamma_2}^*(\sem{\Gamma_1,\Gamma_2;B})}(\mathsf{pr}_{\Gamma_2, y : B}^*(\sem{\Gamma_1,\Gamma_2, y \!:\! B;\ul{C}}))) \ar[d]_-{=}^-{\quad\qquad\qquad\dscomment{\text{split Beck-Chevalley}}}^-{\qquad\qquad\qquad\qquad\qquad\qquad\qquad\dscomment{U \text{ is split fibred}}}
\\
U(\mathsf{pr}_{\Gamma_2}^*((\Sigma_{\Gamma_1,\Gamma_2;B}(\sem{\Gamma_1,\Gamma_2,y \!:\! B;\ul{C}}))) \ar[r]_-{=}
&
\mathsf{pr}_{\Gamma_2}^*(U(\Sigma_{\Gamma_1,\Gamma_2;B}(\sem{\Gamma_1,\Gamma_2,y \!:\! B;\ul{C}})))
}
\]

We conclude by observing that the left-hand side top-to-bottom composite morphism is equal to $\sem{\Gamma_1, x \!:\! A,\Gamma_2; \langle V , M \rangle_{(y : B).\, \ul{C}}}$, and that the right-hand side top-to-bottom composite morphism is equal to $\sproj {\Gamma_1} x A {\Gamma_2}^*(\sem{\Gamma_1,\Gamma_2;\langle V , M \rangle_{(y : B).\, \ul{C}}})$. 

In the above diagram, and in other cases of this proof, we use $(*)$ to refer to the following commuting diagram:
\[
\scriptsize
\xymatrix@C=6.5em@R=6em@M=0.5em{
&
\\
\sem{\Gamma_1,\Gamma_2} \ar[r]_-{\eta^{1 \,\dashv\, \ia -}_{\sem{\Gamma_1,\Gamma_2}}} \ar@/^3pc/[rrr]^-{\mathsf{s}(\sem{\Gamma_1,\Gamma_2;V})} \ar@{}[u]_<<<<<<{\!\!\!\qquad\qquad\qquad\qquad\qquad\qquad\qquad\qquad\qquad\dscomment{\text{def. of } \mathsf{s}(\sem{\Gamma_1,\Gamma_2;V})}} & \ia {1_{\sem{\Gamma_1,\Gamma_2}}} \ar[r]^-{\ia {\sem{\Gamma_1,\Gamma_2;V}}} & \ia {\sem{\Gamma_1,\Gamma_2;B}} \ar[r]^-{=} & \sem{\Gamma_1,\Gamma_2,y \!:\! B}
\\
& \ia {\mathsf{pr}_{\Gamma_2}^*(1_{\sem{\Gamma_1,\Gamma_2}})} \ar[u]_<<<<{\ia {\overline{\mathsf{pr}_{\Gamma_2}}(1_{\sem{\Gamma_1,\Gamma_2}})}}^<<<<<<<{\dscomment{1 \text{ is s. fib.}}\,\,\,}_-{\quad\qquad\dscomment{\text{def. of } \mathsf{pr}_{\Gamma_2}^*(\sem{\Gamma_1,\Gamma_2;V})}} \ar[r]_-{\ia {\mathsf{pr}_{\Gamma_2}^*(\sem{\Gamma_1,\Gamma_2;V})}} & \ia {\mathsf{pr}_{\Gamma_2}^*(\sem{\Gamma_1,\Gamma_2;B})} \ar[u]^>>>>{\ia {\overline{\mathsf{pr}_{\Gamma_2}}(\sem{\Gamma_1,\Gamma_2;B})}} &
\\
\sem{\Gamma_1, x \!:\! A, \Gamma_2} \ar[r]^-{\eta^{1 \,\dashv\, \ia -}_{\sem{\Gamma_1, x : A,\Gamma_2}}} \ar[uu]^-{\mathsf{pr}_{\Gamma_2}}_>>>>>>>>>>>{\,\,\,\quad\dscomment{\text{nat. of } \eta^{1 \,\dashv\, \ia -}}} \ar@/_3pc/[rrr]_-{\mathsf{s}(\sem{\Gamma_1, x : A, \Gamma_2;V})} \ar@{}[d]^<<<<<{\quad\qquad\qquad\qquad\qquad\qquad\qquad\qquad\qquad\dscomment{\text{def. of } \mathsf{s}(\sem{\Gamma_1, x \!:\! A,\Gamma_2;V})}} & \ia {1_{\sem{\Gamma_1, x : A, \Gamma_2}}} \ar[r]_-{\ia {\sem{\Gamma_1, x : A,\Gamma_2;V}}} \ar[u]^-{=}_-{\,\,\,\qquad\qquad\qquad\dscomment{(c)}} \ar@/^4pc/[uu]^-{\ia {1(\mathsf{pr}_{\Gamma_2})}} & \ia {\sem{\Gamma_1, x \!:\! A, \Gamma_2;B}} \ar[r]_-{=} \ar[u]_-{=} & \sem{\Gamma_1, x \!:\! A, \Gamma_2, y \!:\! B} \ar[uu]_-{\mathsf{pr}_{\Gamma_2, y : B}}^-{\dscomment{\text{def. of } \mathsf{pr}_{\Gamma_2, y : B}}\qquad\quad}
\\
&
}
\vspace{-0.5cm}
\]

\vspace{0.2cm}
\noindent
\textbf{Computational pattern-matching for computation terms:}
In this case, we assume that
\[
\sem{\Gamma_1,\Gamma_2; \doto M {(y \!:\! B, z \!:\! \ul{C})} {\ul{D}} K} : 1_{\sem{\Gamma_1,\Gamma_2}} \longrightarrow U(\sem{\Gamma_1,\Gamma_2; \ul{D}})
\]
and we need to show that
\[
\begin{array}{c}
\hspace{-6.5cm}
\sem{\Gamma_1, x \!:\! A,\Gamma_2; \doto M {(y \!:\! B, z \!:\! \ul{C})} {\ul{D}} K} = 
\\
\hspace{-3cm}
\sproj {\Gamma_1} x A {\Gamma_2}^*(\sem{\Gamma_1,\Gamma_2; \doto M {(y \!:\! B, z \!:\! \ul{C})} {\ul{D}} K}) 
\\
\hspace{5.5cm}
: 1_{\sem{\Gamma_1, x : A,\Gamma_2}} \longrightarrow U(\sproj {\Gamma_1} x A {\Gamma_2}^*(\sem{\Gamma_1,\Gamma_2; \ul{D}}))
\end{array}
\]

First, by inspecting the definition of $\sem{-}$ for $\doto M {(y \!:\! B, z \!:\! \ul{C})} {\ul{D}} K$, the assumption gives us that
\[
\begin{array}{c}
\sem{\Gamma_1,\Gamma_2; M} : 1_{\sem{\Gamma_1,\Gamma_2}} \longrightarrow U(\Sigma_{\sem{\Gamma_1,\Gamma_2;B}}(\sem{\Gamma_1,\Gamma_2,y \!:\! B;\ul{C}}))
\\[3mm]
\sem{\Gamma_1,\Gamma_2, y \!:\! B; z \!:\! \ul{C}; K} : \sem{\Gamma_1,\Gamma_2,y \!:\! B;\ul{C}} \longrightarrow \pi^*_{\sem{\Gamma_1,\Gamma_2;B}}(\sem{\Gamma_1,\Gamma_2; \ul{D}})
\end{array}
\]
on which we can use the induction hypothesis and $(e)$, respectively, to get that
\[
\begin{array}{c}
\hspace{-6cm}
\sem{\Gamma_1, x \!:\! A,\Gamma_2;M} = \sproj {\Gamma_1} x A {\Gamma_2}^*(\sem{\Gamma_1,\Gamma_2;M}) 
\\
\hspace{3.5cm}
: 1_{\sem{\Gamma_1, x : A,\Gamma_2}} \longrightarrow U(\sproj {\Gamma_1} x A {\Gamma_2}^*(\Sigma_{\sem{\Gamma_1,\Gamma_2;B}}(\sem{\Gamma_1,\Gamma_2,y \!:\! B;\ul{C}})))
\\[4mm]
\hspace{-2.4cm}
\sem{\Gamma_1, x \!:\! A,\Gamma_2, y \!:\! B; z \!:\! \ul{C};K} = \sproj {\Gamma_1} x A {\Gamma_2, y : B}^*(\sem{\Gamma_1,\Gamma_2, y \!:\! B; z \!:\! \ul{C};K})
\\
\hspace{1cm}
: \sproj {\Gamma_1} x A {\Gamma_2, y : B}^*(\sem{\Gamma_1,\Gamma_2,y \!:\! B;\ul{C}}) \longrightarrow \sproj {\Gamma_1} x A {\Gamma_2, y : B}^*(\pi^*_{\sem{\Gamma_1,\Gamma_2;B}}(\sem{\Gamma_1,\Gamma_2; \ul{D}}))
\end{array}
\vspace{0.25cm}
\]

Finally, we show that 
\[
\begin{array}{c}
\hspace{-5.5cm}
\sem{\Gamma_1, x \!:\! A,\Gamma_2; \doto M {(y \!:\! B, z \!:\! \ul{C})} {\ul{D}} K} = 
\\
\hspace{-2cm}
\sproj {\Gamma_1} x A {\Gamma_2}^*(\sem{\Gamma_1,\Gamma_2; \doto M {(y \!:\! B, z \!:\! \ul{C})} {\ul{D}} K}) 
\\
\hspace{5.5cm}
: 1_{\sem{\Gamma_1, x : A,\Gamma_2}} \longrightarrow U(\sproj {\Gamma_1} x A {\Gamma_2}^*(\sem{\Gamma_1,\Gamma_2; \ul{D}}))
\end{array}
\]
by proving that the next diagram commutes, in which we write $\mathsf{pr}_{\Gamma_2}$ for $\sproj {\Gamma_1} x A {\Gamma_2}$ and $\mathsf{pr}_{\Gamma_2, y : B}$ for $\sproj {\Gamma_1} x A {\Gamma_2, y : B}$. For better readability, we aggregate small proof steps.

\vspace{0.3cm}
\[
\scriptsize
\xymatrix@C=2em@R=5em@M=0.5em{
1_{\sem{\Gamma_1, x : A, \Gamma_2}} 
\ar[r]^-{=} \ar[d]_-{\sem{\Gamma_1, x : A, \Gamma_2;M}}^<<<<{\qquad\qquad\dscomment{\text{use of the induction hypothesis on } \sem{\Gamma_1,\Gamma_2;M} }}^>>>>>{\qquad\qquad\qquad\qquad\dscomment{U \text{ is split fibred}}}
& 
\mathsf{pr}_{\Gamma_2}^*(1_{\sem{\Gamma_1,\Gamma_2}})
\ar[d]^-{\mathsf{pr}_{\Gamma_2}^*(\sem{\Gamma_1,\Gamma_2;M})}
\\
U(\mathsf{pr}_{\Gamma_2}^*(\Sigma_{\sem{\Gamma_1,\Gamma_2;B}}(\sem{\Gamma_1,\Gamma_2, y \!:\! B;\ul{C}})))
\ar[r]^-{=} \ar[d]_-{=}^-{\,\,\,\,\,\,\,\qquad\qquad\dscomment{\text{split Beck-Chevalley}}}^-{\,\,\,\,\,\,\,\qquad\qquad\qquad\qquad\qquad\qquad\dscomment{U \text{ is split fibred}}}
&
\mathsf{pr}_{\Gamma_2}^*(U(\Sigma_{\sem{\Gamma_1,\Gamma_2;B}}(\sem{\Gamma_1,\Gamma_2,y \!:\! B;\ul{C}})))
\ar[dd]^-{\mathsf{pr}_{\Gamma_2}^*(U(\Sigma_{\sem{\Gamma_1,\Gamma_2;B}}(\sem{\Gamma_1,\Gamma_2, y : B;z : \ul{C}; K})))}
\\
U(\Sigma_{\mathsf{pr}_{\Gamma_2}^*(\sem{\Gamma_1,\Gamma_2;B})}(\mathsf{pr}^*_{\Gamma_2,y : B}(\sem{\Gamma_1,\Gamma_2,y \!:\! B;\ul{C}})))
\ar[d]_-{U(\Sigma_{\mathsf{pr}_{\Gamma_2}^*(\sem{\Gamma_1,\Gamma_2;B})}(\sem{\Gamma_1, x : A,\Gamma_2, y : B; z : \ul{C}; K}))}^-{\,\,\,\,\,\,\,\qquad\dscomment{\text{use of the induction hypothesis on } \sem{\Gamma_1, x \!:\! A, \Gamma_2; y \!:\! B; K}}}
\\
U(\Sigma_{\mathsf{pr}_{\Gamma_2}^*(\sem{\Gamma_1,\Gamma_2;B})}(\mathsf{pr}^*_{\Gamma_2,y : B}(\pi^*_{\sem{\Gamma_1,\Gamma_2;B}}(\sem{\Gamma_1,\Gamma_2; \ul{D}})))) \ar[r]^-{=} \ar[d]_-{=}^<<<<{\,\,\,\,\,\,\,\qquad\qquad\dscomment{\text{def. of } \mathsf{pr}_{\Gamma_2, y : B}}}^<<<<{\,\,\,\,\,\,\,\qquad\qquad\qquad\qquad\qquad\dscomment{\mathcal{P}(\overline{\mathsf{pr}_{\Gamma_2}}(\sem{\Gamma_1,\Gamma_2;B}))}}^>>>>>{\,\,\,\,\,\,\,\,\,\,\,\quad\qquad\qquad\qquad\dscomment{U \text{ is split fibred}}}
&
\mathsf{pr}_{\Gamma_2}^*(U(\Sigma_{\sem{\Gamma_1,\Gamma_2;B}}(\pi^*_{\sem{\Gamma_1,\Gamma_2;B}}(\sem{\Gamma_1,\Gamma_2; \ul{D}})))) \ar[dd]^-{\mathsf{pr}_{\Gamma_2}^*(U(\varepsilon^{\Sigma_{\sem{\Gamma_1,\Gamma_2;B}} \,\dashv\, \pi^*_{\sem{\Gamma_1,\Gamma_2;B}}}_{\sem{\Gamma_1,\Gamma_2; \ul{D}}}))}
\\
U(\Sigma_{\mathsf{pr}_{\Gamma_2}^*(\sem{\Gamma_1,\Gamma_2;B})}(\pi^*_{\mathsf{pr}_{\Gamma_2}^*(\sem{\Gamma_1,\Gamma_2;B})}(\mathsf{pr}_{\Gamma_2}^*(\sem{\Gamma_1,\Gamma_2; \ul{D}})))) \ar[d]_-{U(\varepsilon^{\Sigma_{\mathsf{pr}_{\Gamma_2}^*(\sem{\Gamma_1,\Gamma_2;B})} \,\dashv\, \pi^*_{\mathsf{pr}_{\Gamma_2}^*(\sem{\Gamma_1,\Gamma_2;B})}}_{\mathsf{pr}^*_{\Gamma_2}(\sem{\Gamma_1,\Gamma_2; \ul{D}})})}^<<<{\,\,\,\,\,\,\,\,\,\,\,\,\,\,\,\qquad\qquad\qquad\dscomment{\text{split Beck-Chevalley}}}^>>>>>{\,\,\,\,\,\,\,\qquad\qquad\dscomment{\text{Proposition~\ref{prop:BCfordepcompsums} for } \varepsilon^{\Sigma_{\sem{\Gamma_1,\Gamma_2;B}} \,\dashv\, \pi^*_{\sem{\Gamma_1,\Gamma_2;B}}}}}
\\
U(\mathsf{pr}_{\Gamma_2}^*(\sem{\Gamma_1,\Gamma_2; \ul{D}}))
\ar[r]_-{=}
&
\mathsf{pr}_{\Gamma_2}^*(U(\sem{\Gamma_1,\Gamma_2; \ul{D}}))
}
\vspace{0.3cm}
\]

To conclude, we observe that the left-hand side top-to-bottom composite morphism is equal to $\sem{\Gamma_1, x \!:\! A,\Gamma_2; \doto M {(y \!:\! B, z \!:\! \ul{C})} {\ul{D}} K}$, and that the right-hand side top-to-bottom composite morphism is equal to $\sproj {\Gamma_1} x A {\Gamma_2}^*(\sem{\Gamma_1,\Gamma_2;\doto M {(y \!:\! B, z \!:\! \ul{C})} {\ul{D}} K})$. 

\vspace{0.2cm}
\noindent
\textbf{Computational lambda abstraction for computation terms:}
In this case, we assume that 
\[
\sem{\Gamma_1,\Gamma_2; \lambda \, y \!:\! B .\, M} : 1_{\sem{\Gamma_1,\Gamma_2}} \longrightarrow U(\Pi_{\sem{\Gamma_1,\Gamma_2;B}}(\ul{C}))
\]

\pagebreak\noindent
and we need to show that
\[
\begin{array}{c}
\hspace{-3.5cm}
\sem{\Gamma_1, x \!:\! A,\Gamma_2; \lambda \, y \!:\! B .\, M} = 
\sproj {\Gamma_1} x A {\Gamma_2}^*(\sem{\Gamma_1,\Gamma_2; \lambda \, y \!:\! B .\, M}) 
\\
\hspace{5.75cm}
: 1_{\sem{\Gamma_1, x : A,\Gamma_2}} \longrightarrow U(\sproj {\Gamma_1} x A {\Gamma_2}^*(\Pi_{\sem{\Gamma_1,\Gamma_2;B}}(\ul{C})))
\end{array}
\]

First, by inspecting the definition of $\sem{-}$ for $\lambda \, y \!:\! B .\, M$, the assumption gives us that
\[
\sem{\Gamma_1,\Gamma_2, y \!:\! B ; M} : 1_{\sem{\Gamma_1,\Gamma_2, y : B}} \longrightarrow U(\ul{C})
\]

Next, by using the induction hypothesis on this morphism, we get that
\[
\begin{array}{c}
\hspace{-3.25cm}
\sem{\Gamma_1, x \!:\! A,\Gamma_2, y \!:\! B ; M} = 
\sproj {\Gamma_1} x A {\Gamma_2, y : B}^*(\sem{\Gamma_1,\Gamma_2, y \!:\! B ; M}) 
\\
\hspace{6.75cm}
: 1_{\sem{\Gamma_1, x : A,\Gamma_2, y : B}} \longrightarrow U(\sproj {\Gamma_1} x A {\Gamma_2, y : B}^*(\ul{C}))
\end{array}
\]

Finally, we show that
\[
\begin{array}{c}
\hspace{-3.15cm}
\sem{\Gamma_1, x \!:\! A,\Gamma_2; \lambda \, y \!:\! B .\, M} = 
\sproj {\Gamma_1} x A {\Gamma_2}^*(\sem{\Gamma_1,\Gamma_2; \lambda \, y \!:\! B .\, M}) 
\\
\hspace{5.75cm}
: 1_{\sem{\Gamma_1, x : A,\Gamma_2}} \longrightarrow U(\sproj {\Gamma_1} x A {\Gamma_2}^*(\Pi_{\sem{\Gamma_1,\Gamma_2;B}}(\ul{C})))
\end{array}
\]
by proving that the next diagram commutes, in which we write $\mathsf{pr}_{\Gamma_2}$ for $\sproj {\Gamma_1} x A {\Gamma_2}$ and $\mathsf{pr}_{\Gamma_2, y : B}$ for $\sproj {\Gamma_1} x A {\Gamma_2, y : B}$. For better readability, we aggregate small proof steps.

\vspace{0.3cm}
\[
\scriptsize
\xymatrix@C=2em@R=7em@M=0.5em{
1_{\sem{\Gamma_1, x : A, \Gamma_2}} 
\ar[r]^{=} \ar[d]_-{\eta^{\pi^*_{\mathsf{pr}_{\Gamma_2}^*(\sem{\Gamma_1,\Gamma_2;B})} \,\dashv\, \Pi_{\mathsf{pr}_{\Gamma_2}^*(\sem{\Gamma_1,\Gamma_2;B})}}_{1_{\sem{\Gamma_1, x : A, \Gamma_2}}}}^-{\!\!\!\!\!\!\!\qquad\qquad\dscomment{\text{Proposition~\ref{prop:BCfordepproducts} for } \eta^{\pi^*_{\sem{\Gamma_1,\Gamma_2;B}} \,\dashv\, \Pi_{\sem{\Gamma_1,\Gamma_2;B}}}}}
& 
\mathsf{pr}_{\Gamma_2}^*(1_{\sem{\Gamma_1,\Gamma_2}})
\ar[d]^-{\mathsf{pr}^*(\eta^{\pi^*_{\sem{\Gamma_1,\Gamma_2;B}} \,\dashv\, \Pi_{\sem{\Gamma_1,\Gamma_2;B}}}_{1_{\sem{\Gamma_1,\Gamma_2}}})}
\\
\Pi_{\mathsf{pr}_{\Gamma_2}^*(\sem{\Gamma_1,\Gamma_2;B})}(\pi^*_{\mathsf{pr}_{\Gamma_2}^*(\sem{\Gamma_1,\Gamma_2;B})}(1_{\sem{\Gamma_1, x : A, \Gamma_2}})) \ar[r]^-{=} \ar[d]_-{=}^-{\quad\qquad\dscomment{\text{split Beck-Chevalley}}}^-{\quad\qquad\qquad\qquad\qquad\qquad\dscomment{1 \text{ is split fibred}}}
&
\mathsf{pr}_{\Gamma_2}^*(\Pi_{\sem{\Gamma_1,\Gamma_2;B}}(\pi^*_{\sem{\Gamma_1,\Gamma_2;B}}(1_{\sem{\Gamma_1,\Gamma_2}}))) \ar[d]^-{=}
\\
\Pi_{\mathsf{pr}_{\Gamma_2}^*(\sem{\Gamma_1,\Gamma_2;B})}(1_{\sem{\Gamma_1, x : A, \Gamma_2, y : B}}) \ar[r]^-{=} \ar[d]_-{\Pi_{\mathsf{pr}_{\Gamma_2}^*(\sem{\Gamma_1,\Gamma_2;B})}(\sem{\Gamma_1, x : A, \Gamma_2, y : B; M})}^<<<<<<{\!\!\!\!\qquad\dscomment{\text{use of the induction hypothesis on } \sem{\Gamma_1, \Gamma_2, y \!:\! B; M}}}^>>>>>>>{\qquad\qquad\qquad\dscomment{\text{split Beck-Chevalley}}}
&
\mathsf{pr}_{\Gamma_2}^*(\Pi_{\sem{\Gamma_1,\Gamma_2;B}}(1_{\sem{\Gamma_1,\Gamma_2,y : B}})) \ar[d]^-{\mathsf{pr}_{\Gamma_2}^*(\Pi_{\sem{\Gamma_1,\Gamma_2;B}}(\sem{\Gamma_1,\Gamma_2, y : B; M}))}
\\
\Pi_{\mathsf{pr}_{\Gamma_2}^*(\sem{\Gamma_1,\Gamma_2;B})}(U(\mathsf{pr}_{\Gamma_2, y : B}^*(\ul{C}))) \ar[r]^-{=} \ar[d]_-{(\zeta^{-1}_{\Pi,\mathsf{pr}^*_{\Gamma_2}(\sem{\Gamma_1,\Gamma_2;B})})_{\mathsf{pr}_{\Gamma_2, y : B}^*(\ul{C})}}^>>>>>>>{\quad\qquad\dscomment{\text{split Beck-Chevalley}}}^>>>>>>>{\quad\qquad\qquad\qquad\qquad\qquad\dscomment{U \text{ is split fibred}}}^<<<<<<{\quad\dscomment{\text{expanding the defs. of } \zeta^{-1}_{\Pi,\mathsf{pr}^*_{\Gamma_2}(\sem{\Gamma_1,\Gamma_2;B})} \text{ and } \zeta^{-1}_{\Pi,\sem{\Gamma_1,\Gamma_2;B}}}}
&
\mathsf{pr}_{\Gamma_2}^*(\Pi_{\sem{\Gamma_1,\Gamma_2;B}}(U(\ul{C}))) \ar[d]^-{\mathsf{pr}_{\Gamma_2}^*((\zeta^{-1}_{\Pi,\sem{\Gamma_1,\Gamma_2;B}})_{\ul{C}})}
\\
U(\Pi_{\mathsf{pr}_{\Gamma_2}^*(\sem{\Gamma_1,\Gamma_2;B})}(\mathsf{pr}_{\Gamma_2, y : B}^*(\ul{C}))) \ar[r]_-{=}
&
\mathsf{pr}_{\Gamma_2}^*(U(\Pi_{\sem{\Gamma_1,\Gamma_2;B}}(\ul{C})))
}
\vspace{0.3cm}
\]

\pagebreak
We conclude by observing that the left-hand side top-to-bottom composite morphism is equal to $\sem{\Gamma_1, x \!:\! A,\Gamma_2; \lambda \, y \!:\! B .\, M}$, and that the right-hand side top-to-bottom composite morphism
is equal to $\sproj {\Gamma_1} x A {\Gamma_2}^*(\sem{\Gamma_1,\Gamma_2;\lambda \, y \!:\! B .\, M})$. 

\vspace{0.2cm}
\noindent
\textbf{Computational function application for computation terms:}
In this case, we assume that 
\[
\sem{\Gamma_1,\Gamma_2; M(V)_{(y : B).\,\ul{C}}} : 1_{\sem{\Gamma_1,\Gamma_2}} \longrightarrow U((\mathsf{s}(\sem{\Gamma_1,\Gamma_2;V}))^*(\sem{\Gamma_1,\Gamma_2,y \!:\! B;\ul{C}}))
\]
and we need to show that
\[
\begin{array}{c}
\hspace{-3.5cm}
\sem{\Gamma_1, x \!:\! A,\Gamma_2; M(V)_{(y : B).\,\ul{C}}} = 
\sproj {\Gamma_1} x A {\Gamma_2}^*(\sem{\Gamma_1,\Gamma_2; M(V)_{(y : B).\,\ul{C}}}) 
\\
\hspace{2.5cm}
: 1_{\sem{\Gamma_1, x : A,\Gamma_2}} \longrightarrow U(\sproj {\Gamma_1} x A {\Gamma_2}^*((\mathsf{s}(\sem{\Gamma_1,\Gamma_2;V}))^*(\sem{\Gamma_1,\Gamma_2,y \!:\! B;\ul{C}})))
\end{array}
\]

First, by inspecting the definition of $\sem{-}$ for $M(V)_{(y : B).\,\ul{C}}$, the assumption gives us that
\[
\begin{array}{c}
\sem{\Gamma_1,\Gamma_2; V} : 1_{\sem{\Gamma_1,\Gamma_2}} \longrightarrow \sem{\Gamma_1,\Gamma_2;B}
\\[3mm]
\sem{\Gamma_1,\Gamma_2; M} : 1_{\sem{\Gamma_1,\Gamma_2}} \longrightarrow U(\Pi_{\sem{\Gamma_1,\Gamma_2;B}}(\sem{\Gamma_1,\Gamma_2,y \!:\! B;\ul{C}}))
\end{array}
\]
on which we can use $(c)$ and the induction hypothesis, respectively, to get that
\[
\begin{array}{c}
\sem{\Gamma_1, x \!:\! A,\Gamma_2; V} = 
\sproj {\Gamma_1} x A {\Gamma_2}^*(\sem{\Gamma_1,\Gamma_2; V}) 
: 1_{\sem{\Gamma_1, x : A,\Gamma_2}} \longrightarrow \sproj {\Gamma_1} x A {\Gamma_2}^*(\sem{\Gamma_1,\Gamma_2;B})
\\[3mm]
\hspace{-6.8cm}
\sem{\Gamma_1, x \!:\! A,\Gamma_2; M} = 
\sproj {\Gamma_1} x A {\Gamma_2}^*(\sem{\Gamma_1,\Gamma_2; M}) 
\\
\hspace{3.8cm}
: 1_{\sem{\Gamma_1, x : A,\Gamma_2}} \longrightarrow U(\sproj {\Gamma_1} x A {\Gamma_2}^*(\Pi_{\sem{\Gamma_1,\Gamma_2;B}}(\sem{\Gamma_1,\Gamma_2,y \!:\! B;\ul{C}})))
\end{array}
\]

Finally, we show that 
\[
\begin{array}{c}
\hspace{-2.5cm}
\sem{\Gamma_1, x \!:\! A,\Gamma_2; M(V)_{(y : B).\,\ul{C}}} = 
\sproj {\Gamma_1} x A {\Gamma_2}^*(\sem{\Gamma_1,\Gamma_2; M(V)_{(y : B).\,\ul{C}}}) 
\\
\hspace{2.5cm}
: 1_{\sem{\Gamma_1, x : A,\Gamma_2}} \longrightarrow U(\sproj {\Gamma_1} x A {\Gamma_2}^*((\mathsf{s}(\sem{\Gamma_1,\Gamma_2;V}))^*(\sem{\Gamma_1,\Gamma_2,y \!:\! B;\ul{C}})))
\end{array}
\]
by proving that the next diagram commutes, in which we write $\mathsf{pr}_{\Gamma_2}$ for $\sproj {\Gamma_1} x A {\Gamma_2}$ and $\mathsf{pr}_{\Gamma_2, y : B}$ for $\sproj {\Gamma_1} x A {\Gamma_2, y : B}$. For better readability, we aggregate small proof steps.
\[
\scriptsize
\xymatrix@C=5em@R=6.5em@M=0.5em{
1_{\sem{\Gamma_1,x:A,\Gamma_2}} 
\ar[r]^-{=} \ar[d]_-{=}^<<<<<<{\!\!\!\!\!\!\!\!\qquad\qquad\qquad\dscomment{\mathsf{s}(\sem{\Gamma_1, x \!:\! A, \Gamma_2;V}) \text{ is a section of } \pi_{\mathsf{pr}_{\Gamma_2}^*(\sem{\Gamma_1,\Gamma_2;B})}}}^>>>>>>>{\!\!\!\!\quad\qquad\qquad\qquad\dscomment{\mathsf{s}(\sem{\Gamma_1, \Gamma_2;V}) \text{ is a section of } \pi_{\sem{\Gamma_1,\Gamma_2;B}}}}
& 
\mathsf{pr}_{\Gamma_2}^*(1_{\sem{\Gamma_1,\Gamma_2}})
\ar[d]^-{=}
\\
(\mathsf{s}(\sem{\Gamma_1, x \!:\! A, \Gamma_2;V}))^*(\pi^*_{\mathsf{pr}_{\Gamma_2}^*(\sem{\Gamma_1,\Gamma_2;B})}(1_{\sem{\Gamma_1,x:A,\Gamma_2}})) 
\ar[r]^-{=} \ar[dd]^<<<<<<{(\mathsf{s}(\sem{\Gamma_1, x : A, \Gamma_2;V}))^*(\pi^*_{\mathsf{pr}_{\Gamma_2}^*(\sem{\Gamma_1,\Gamma_2;B})}(\sem{\Gamma_1,x : A, \Gamma_2;M}))}^<<<<<<<<<<<<<{\!\!\!\!\!\!\!\!\qquad\qquad\qquad\dscomment{\text{use of the induction hypothesis on } \sem{\Gamma_1,\Gamma_2;M}}}^<<<<<<<<<<<<<<<<<<<{\,\,\,\,\,\,\,\qquad\dscomment{(*) \text{ from above}}}^<<<<<<<<<<<<<<<<<<<{\,\,\,\,\,\,\,\,\,\,\,\,\,\qquad\qquad\qquad\qquad\dscomment{\text{def. of } \mathsf{pr}_{\Gamma_2, y : B}}}^<<<<<<<<<<<<<<<<<<<{\,\,\,\,\,\,\,\,\,\,\,\,\,\,\,\qquad\qquad\qquad\qquad\qquad\qquad\qquad\dscomment{\mathcal{P}(\overline{\mathsf{pr}_{\Gamma_2}}(\sem{\Gamma_1,\Gamma_2;B}))}}^<<<<<<<<<<<<<<<<<<<<<<<<<{\,\,\,\,\,\,\,\,\,\,\,\qquad\qquad\qquad\qquad\dscomment{U \text{ is split fibred}}}
&
\mathsf{pr}_{\Gamma_2}^*((\mathsf{s}(\sem{\Gamma_1, \Gamma_2;V}))^*(\pi^*_{\sem{\Gamma_1,\Gamma_2;B}}(1_{\sem{\Gamma_1,\Gamma_2}})))
\ar[dd]_>>>>>>{\mathsf{pr}_{\Gamma_2}^*((\mathsf{s}(\sem{\Gamma_1, \Gamma_2;V}))^*(\pi^*_{\sem{\Gamma_1,\Gamma_2;B}}(\sem{\Gamma_1,\Gamma_2,y : B;M})))}
\\
&
\\
\txt<12pc>{
$(\mathsf{s}(\sem{\Gamma_1, x \!:\! A, \Gamma_2;V}))^*(\pi^*_{\mathsf{pr}_{\Gamma_2}^*(\sem{\Gamma_1,\Gamma_2;B})}($
\\
$U(\mathsf{pr}_{\Gamma_2}^*(\Pi_{\sem{\Gamma_1,\Gamma_2;B}}(\sem{\Gamma_1,\Gamma_2,y \!:\! B;\ul{C}})))))$
}
\ar[r]^-{=} \ar[d]_-{=}^<<<<<<{\,\,\,\,\,\,\,\qquad\dscomment{(*) \text{ from above}}}^<<<<<<{\,\,\,\,\,\,\,\,\,\,\,\,\,\qquad\qquad\qquad\qquad\dscomment{\text{def. of } \mathsf{pr}_{\Gamma_2, y : B}}}^<<<<<<{\,\,\,\,\,\,\,\,\,\,\,\,\,\,\,\qquad\qquad\qquad\qquad\qquad\qquad\qquad\dscomment{\mathcal{P}(\overline{\mathsf{pr}_{\Gamma_2}}(\sem{\Gamma_1,\Gamma_2;B}))}}^>>>>>>>{\,\,\,\,\,\,\,\,\,\,\,\qquad\qquad\qquad\qquad\dscomment{U \text{ is split fibred}}}
&
\txt<12pc>{
$\mathsf{pr}_{\Gamma_2}^*((\mathsf{s}(\sem{\Gamma_1, \Gamma_2;V}))^*(\pi^*_{\sem{\Gamma_1,\Gamma_2;B}}($
\\
$U(\Pi_{\sem{\Gamma_1,\Gamma_2;B}}(\sem{\Gamma_1,\Gamma_2,y \!:\! B;\ul{C}})))))$
}
\ar[d]^-{=}
\\
\txt<12pc>{
$U((\mathsf{s}(\sem{\Gamma_1, x \!:\! A, \Gamma_2;V}))^*(\pi^*_{\mathsf{pr}_{\Gamma_2}^*(\sem{\Gamma_1,\Gamma_2;B})}($
\\
$\mathsf{pr}_{\Gamma_2}^*(\Pi_{\sem{\Gamma_1,\Gamma_2;B}}(\sem{\Gamma_1,\Gamma_2,y \!:\! B;\ul{C}})))))$
}
\ar[r]^-{=} \ar[d]_-{=}^<<<<<<{\,\,\,\,\,\,\,\qquad\dscomment{(*) \text{ from above}}}^<<<<<<{\,\,\,\,\,\,\,\,\,\,\,\,\,\qquad\qquad\qquad\qquad\dscomment{\text{def. of } \mathsf{pr}_{\Gamma_2, y : B}}}^<<<<<<{\,\,\,\,\,\,\,\,\,\,\,\,\,\,\,\qquad\qquad\qquad\qquad\qquad\qquad\qquad\dscomment{\mathcal{P}(\overline{\mathsf{pr}_{\Gamma_2}}(\sem{\Gamma_1,\Gamma_2;B}))}}^>>>>>>>{\,\,\,\,\,\,\,\,\,\,\,\qquad\qquad\qquad\qquad\dscomment{U \text{ is split fibred}}}
&
\txt<12pc>{
$\mathsf{pr}_{\Gamma_2}^*(U((\mathsf{s}(\sem{\Gamma_1, \Gamma_2;V}))^*(\pi^*_{\sem{\Gamma_1,\Gamma_2;B}}($
\\
$\Pi_{\sem{\Gamma_1,\Gamma_2;B}}(\sem{\Gamma_1,\Gamma_2,y \!:\! B;\ul{C}})))))$
}
\ar[ddd]_>>>>>>>{\mathsf{pr}_{\Gamma_2}^*(U((\mathsf{s}(\sem{\Gamma_1, \Gamma_2;V}))^*(\varepsilon^{\pi^*_{\sem{\Gamma_1,\Gamma_2;B}} \,\dashv\, \Pi_{\sem{\Gamma_1,\Gamma_2;B}}}_{\sem{\Gamma_1,\Gamma_2,y \!:\! B;\ul{C}}})))}
\\
\txt<12pc>{
$U((\mathsf{s}(\sem{\Gamma_1, x \!:\! A, \Gamma_2;V}))^*(\mathsf{pr}_{\Gamma_2, y : B}^*($
\\
$\pi^*_{\sem{\Gamma_1,\Gamma_2;B}}(\Pi_{\sem{\Gamma_1,\Gamma_2;B}}(\sem{\Gamma_1,\Gamma_2,y \!:\! B;\ul{C}})))))$
}
\ar[dd]^<<<<<{U((\mathsf{s}(\sem{\Gamma_1, x \!:\! A, \Gamma_2;V}))^*(\mathsf{pr}_{\Gamma_2, y : B}^*(\varepsilon^{\pi^*_{\sem{\Gamma_1,\Gamma_2;B}} \,\dashv\, \Pi_{\sem{\Gamma_1,\Gamma_2;B}}}_{\sem{\Gamma_1,\Gamma_2,y \!:\! B;\ul{C}}})))}^-{\qquad\qquad\qquad\dscomment{\text{Proposition~\ref{prop:BCfordepproducts} for } \varepsilon^{\pi^*_{\sem{\Gamma_1,\Gamma_2;B}} \,\dashv\, \Pi_{\sem{\Gamma_1,\Gamma_2;B}}}}}
\\
&
\\
U((\mathsf{s}(\sem{\Gamma_1, x \!:\! A, \Gamma_2;V}))^*(\mathsf{pr}_{\Gamma_2, y : B}^*(\sem{\Gamma_1,\Gamma_2,y \!:\! B;\ul{C}})))
\ar[r]_-{=}
&
\mathsf{pr}_{\Gamma_2}^*(U((\mathsf{s}(\sem{\Gamma_1, \Gamma_2;V}))^*(\sem{\Gamma_1,\Gamma_2,y \!:\! B;\ul{C}})))
}
\]

We conclude by observing that the left-hand side top-to-bottom composite morphism is equal to $\sem{\Gamma_1, x \!:\! A,\Gamma_2; M(V)_{(y : B).\,\ul{C}}}$, and that the right-hand side top-to-bottom\linebreak

\pagebreak\noindent 
composite morphism is equal to $\sproj {\Gamma_1} x A {\Gamma_2}^*(\sem{\Gamma_1,\Gamma_2;M(V)_{(y : B).\,\ul{C}}})$. 

\vspace{0.2cm}
\noindent
\textbf{Forcing a thunked computation:}
We omit the proof for this case because it is analogous to the case for thunking a computation.

\vspace{0.2cm}
\noindent
\textbf{Homomorphic function application for computation terms:}
In this case, we assume that
\[
\sem{\Gamma_1,\Gamma_2; V(M)_{\ul{C}, \ul{D}}} : 1_{\sem{\Gamma_1,\Gamma_2}} \longrightarrow U(\sem{\Gamma_1,\Gamma_2;\ul{D}})
\]
and we need to show that
\[
\begin{array}{c}
\hspace{-4cm}
\sem{\Gamma_1, x \!:\! A,\Gamma_2; V(M)_{\ul{C}, \ul{D}}} = \sproj {\Gamma_1} x A {\Gamma_2}^*(\sem{\Gamma_1,\Gamma_2; V(M)_{\ul{C}, \ul{D}}}) 
\\
\hspace{6.5cm}
: 1_{\sem{\Gamma_1, x : A,\Gamma_2}} \longrightarrow U(\sproj {\Gamma_1} x A {\Gamma_2}^*(\sem{\Gamma_1,\Gamma_2;\ul{D}}))
\end{array}
\]

First, by inspecting the definition of $\sem{-}$ for $V(M)_{\ul{C}, \ul{D}}$, the assumption gives us
\[
\begin{array}{c}
\sem{\Gamma_1,\Gamma_2; V} : 1_{\sem{\Gamma_1,\Gamma_2}} \longrightarrow \sem{\Gamma_1,\Gamma_2;\ul{C}} \multimap \sem{\Gamma_1,\Gamma_2;\ul{D}}
\\[3mm]
\sem{\Gamma_1,\Gamma_2; M} : 1_{\sem{\Gamma_1,\Gamma_2}} \longrightarrow U(\sem{\Gamma_1,\Gamma_2;\ul{C}})
\end{array}
\]
on which we can use $(c)$ and the induction hypothesis, respectively, to get that
\[
\begin{array}{c}
\sem{\Gamma_1, x : A,\Gamma_2; V} : 1_{\sem{\Gamma_1, x : A,\Gamma_2}} \longrightarrow \sproj {\Gamma_1} x A {\Gamma_2}^*(\sem{\Gamma_1,\Gamma_2;\ul{C}} \multimap \sem{\Gamma_1,\Gamma_2;\ul{D}})
\\[3mm]
\sem{\Gamma_1, x : A,\Gamma_2; M} : 1_{\sem{\Gamma_1, x : A,\Gamma_2}} \longrightarrow U(\sproj {\Gamma_1} x A {\Gamma_2}^*(\sem{\Gamma_1,\Gamma_2;\ul{C}}))
\end{array}
\]

Finally, we show that 
\[
\begin{array}{c}
\hspace{-3cm}
\sem{\Gamma_1, x \!:\! A,\Gamma_2; V(M)_{\ul{C}, \ul{D}}} = \sproj {\Gamma_1} x A {\Gamma_2}^*(\sem{\Gamma_1,\Gamma_2; V(M)_{\ul{C}, \ul{D}}}) 
\\
\hspace{6.5cm}
: 1_{\sem{\Gamma_1, x : A,\Gamma_2}} \longrightarrow U(\sproj {\Gamma_1} x A {\Gamma_2}^*(\sem{\Gamma_1,\Gamma_2;\ul{D}}))
\end{array}
\]

\pagebreak\noindent
by proving that the next diagram commutes, in which we write $\mathsf{pr}$ for $\sproj {\Gamma_1} x A {\Gamma_2}$. To improve the readability of this diagram, we aggregate small proof steps.

\[
\hspace{0.5cm}
\xymatrix@C=11.5em@R=8em@M=0.5em{
1_{\sem{\Gamma_1, x : A, \Gamma_2}} \ar[r]^-{\sem{\Gamma_1, x : A,\Gamma_2; V(M)_{\ul{C}, \ul{D}}}} \ar[d]^-{\sem{\Gamma_1,x : A, \Gamma_2;M}}^<<<{\,\,\,\quad\qquad\dcomment{\text{def. of } \sem{\Gamma_1, x \!:\! A,\Gamma_2; V(M)_{\ul{C}, \ul{D}}}}}^<<<<<<<<<<<<{\qquad\qquad\qquad\dcomment{(c)}} \ar@/_6pc/[ddd]^-{=} & U(\mathsf{pr}^*(\sem{\Gamma_1,\Gamma_2;\ul{D}}))
\\
U(\mathsf{pr}^*(\sem{\Gamma_1,\Gamma_2;\ul{C}})) \ar[ur]_<<<<<<<<<<<<<<<<{\quad\qquad\qquad\qquad U(\xi_{\sem{\Gamma_1,\Gamma_2},\mathsf{pr}^*(\sem{\Gamma_1,\Gamma_2;\ul{C}}),\mathsf{pr}^*(\sem{\Gamma_1,\Gamma_2;\ul{D}})}(\mathsf{pr}^*(\sem{\Gamma_1,\Gamma_2;V})))} \ar[d]_-{=}_-{\dcomment{\text{use of i.h.}}\quad}^-{\qquad\dcomment{U \text{ is split fibred}}}^-{\qquad\qquad\qquad\qquad\qquad\dcomment{\xi \text{ is preserved on-the-nose by reindexing}}}
\\
\mathsf{pr}^*(U(\sem{\Gamma_1,\Gamma_2;\ul{C}})) \ar[dr]^<<<<<<<<<<<<<<<<{\qquad\qquad\qquad\mathsf{pr}^*(U(\xi_{\sem{\Gamma_1,\Gamma_2}, \sem{\Gamma_1,\Gamma_2;\ul{C}}, \sem{\Gamma_1,\Gamma_2;\ul{D}}}(\sem{\Gamma_1,\Gamma_2;V})))}
\\
\mathsf{pr}^*(1_{\sem{\Gamma_1,\Gamma_2}}) \ar[r]_-{\mathsf{pr}^*(\sem{\Gamma_1,\Gamma_2; V(M)_{\ul{C}, \ul{D}}})} \ar[u]_-{\mathsf{pr}^*(\sem{\Gamma_1,\Gamma_2;M})}_<<<<<<{\quad\qquad\dcomment{\text{def. of } \sem{\Gamma_1,\Gamma_2; V(M)_{\ul{C}, \ul{D}}}}} & \mathsf{pr}^*(U(\sem{\Gamma_1,\Gamma_2;\ul{D}})) \ar@/_6pc/[uuu]^-{=}
}
\]

\vspace{0.2cm}
\noindent
\textbf{Computation variables:}
In this case, we assume that
\[
\sem{\Gamma_1,\Gamma_2;z \!:\! \ul{C};z} : \sem{\Gamma_1,\Gamma_2;\ul{C}} \longrightarrow \sem{\Gamma_1,\Gamma_2;\ul{C}}
\]
and we need to show that 
\[
\begin{array}{c}
\hspace{-4.5cm}
\sem{\Gamma_1, x \!:\! A,\Gamma_2;z \!:\! \ul{C};z} = \sproj {\Gamma_1} x A {\Gamma_2}^*(\sem{\Gamma_1,\Gamma_2;z \!:\! \ul{C};z}) 
\\
\hspace{4.5cm}
: \sproj {\Gamma_1} x A {\Gamma_2}^*(\sem{\Gamma_1,\Gamma_2;\ul{C}}) \longrightarrow \sproj {\Gamma_1} x A {\Gamma_2}^*(\sem{\Gamma_1,\Gamma_2;\ul{C}})
\end{array}
\]

\pagebreak 

First, by inspecting the definition of $\sem{-}$ for $z$, the assumption also gives us that 
\[
\sem{\Gamma_1,\Gamma_2;\ul{C}} \in \mathcal{C}_{\sem{\Gamma_1,\Gamma_2}}
\]

Next, by using $(b)$ on this object, we get that
\[
\sem{\Gamma_1, x \!:\! A, \Gamma_2; \ul{C}} = \sproj {\Gamma_1} x A {\Gamma_2}^*(\sem{\Gamma_1,\Gamma_2;\ul{C}}) \in \mathcal{C}_{\sem{\Gamma_1, x : A,\Gamma_2}}
\]

Next, by using the functoriality of $\sproj {\Gamma_1} x A {\Gamma_2}^*$, we get that
\[
\begin{array}{c}
\hspace{-4.5cm}
\id_{\sem{\Gamma_1, x : A, \Gamma_2; \ul{C}}} 
=
\sproj {\Gamma_1} x A {\Gamma_2}^*(\id_{\sem{\Gamma_1,\Gamma_2;\ul{C}}})
\\
\hspace{4cm}
: \sproj {\Gamma_1} x A {\Gamma_2}^*(\sem{\Gamma_1,\Gamma_2;\ul{C}}) \longrightarrow \sproj {\Gamma_1} x A {\Gamma_2}^*(\sem{\Gamma_1,\Gamma_2;\ul{C}})
\end{array}
\]

Finally, by using the definition of $\sem{-}$ for $z$, we get that
\[
\begin{array}{c}
\hspace{-4cm}
\sem{\Gamma_1, x \!:\! A,\Gamma_2;z \!:\! \ul{C};z} = \sproj {\Gamma_1} x A {\Gamma_2}^*(\sem{\Gamma_1,\Gamma_2;z \!:\! \ul{C};z}) 
\\
\hspace{4.5cm}
: \sproj {\Gamma_1} x A {\Gamma_2}^*(\sem{\Gamma_1,\Gamma_2;\ul{C}}) \longrightarrow \sproj {\Gamma_1} x A {\Gamma_2}^*(\sem{\Gamma_1,\Gamma_2;\ul{C}})
\end{array}
\]

\vspace{0.1cm}
\noindent
\textbf{Other cases for homomorphism terms:}
We omit the cases for sequential composition, computational pairing, computational pattern-matching, computational lambda abstraction, and homomorphic function application for homomorphism terms because they are analogous to the corresponding cases for computation terms discussed above. 
\end{proof}

\newpage

\section{Proof of Proposition~\ref{prop:semweakeningandsubstitutioncommuting}}
\label{sect:proofofprop:semweakeningandsubstitutioncommuting}

{
\renewcommand{\thetheorem}{\ref{prop:semweakeningandsubstitutioncommuting}}
\begin{proposition}
Given value contexts $\Gamma_1$ and $\Gamma_2$, value variables $x$ and $y$, value types $A$ and $B$, and a value term $V$ such that $\sem{\Gamma_1,\Gamma_2[V/y]} \in \mathcal{B}$, $\sem{\Gamma_1, y \!:\! B, \Gamma_2} \in \mathcal{B}$, $\sem{\Gamma_1, x \!:\! A,\Gamma_2[V/y]} \in \mathcal{B}$, $\sem{\Gamma_1, x \!:\! A, y \!:\! B,\Gamma_2} \in \mathcal{B}$, and $\sem{\Gamma_1; V} : 1_{\sem{\Gamma_1}} \longrightarrow \sem{\Gamma_1;B}$, then 
\[
\ssubst {\Gamma_1} {y} {B} {\Gamma_2} {V} \comp \sproj {\Gamma_1} {x} {A} {\Gamma_1[V/y]}
=
\sproj {\Gamma_1} {x} {A} {y : B, \Gamma_2} \comp \ssubst {\Gamma_1, x : A} y B {\Gamma_2} V 
\]
\end{proposition}
\addtocounter{theorem}{-1}
}

\begin{proof}
We prove this equation by induction on the length of $\Gamma_2$. Both the base case and the step case of induction are proved similarly, by straightforward diagram chasing.

\noindent
\textit{Base case (with $\Gamma_2 = \diamond$):}
\[
\scriptsize
\xymatrix@C=5em@R=6em@M=0.5em{
\sem{\Gamma_1, x \!:\! A} 
\ar[r]^-{\sproj {\Gamma_1} {x} {A} {\diamond}}
\ar@/^0.5pc/[dr]^<<<<<{=}
\ar@/_1.5pc/[ddr]_>>>>>>>>>>>>>>{\eta^{1 \,\dashv\, \ia -}_{\sem{\Gamma_1, x \!:\! A}}\!\!\!\!\!}^-{\,\,\,\,\,\,\dscomment{\text{iso.}}}
\ar@/_4.75pc/[dddd]_<<<<<<<{\ssubst {\Gamma_1, x : A} {y} {B} {\diamond} {V}}
\ar@/^1pc/[ddd]_-{\mathsf{s}(\sem{\Gamma_1, x : A; V})}_>>>>>>>>>>>>>{\dscomment{\text{def. of } \ssubst {\Gamma_1, x : A} {y} {B} {\diamond} {V}}}^>>>>>>>>>>>>{\!\!\!\!\quad\dscomment{\text{def. of } \mathsf{s}(\sem{\Gamma_1, x : A; V})}}
& 
\sem{\Gamma_1} 
\ar[rr]^-{\ssubst {\Gamma_1} {y} {B} {\diamond} {V}}
\ar@/^1pc/[dr]^>>>>>>>{\!\!\!\!\!\!\!\eta^{1 \,\dashv\, \ia -}_{\sem{\Gamma_1}}}_-{\dscomment{\text{iso.}}\,\,\,\,\,\,\,}^>{\dscomment{\text{def. of } \mathsf{s}(\sem{\Gamma_1;V})}}
\ar@/^1pc/[drr]^>>>>>>>>>>>{\,\,\,\,\,\,\mathsf{s}(\sem{\Gamma_1;V})}
&&
\sem{\Gamma_1, y \!:\! B}
\ar[d]^-{=}_<<<{\dscomment{\text{def. of } \ssubst {\Gamma_1} {y} {B} {\diamond} {V}}\qquad}
\\
&
\ia {\sem{\Gamma_1; A}}
\ar[u]^-{\pi_{\sem{\Gamma_1;A}}\!\!\!}^>>>{\dscomment{\text{def. of } {\sproj {\Gamma_1} {x} {A} {\diamond}}}\,\,\,\,\,}
&
\ia {1_{\sem{\Gamma_1}}}
\ar[r]_-{\ia {\sem{\Gamma_1; V}}}
\ar@/^1pc/[ul]^-{\pi_{1_{\sem{\Gamma_1}}}\!\!\!\!}
&
\ia {\sem{\Gamma_1; B}}
\ar@/^0.5pc/[dddd]_-{\id_{\ia {\sem{\Gamma_1; B}}}}
\\
&
\ia {1_{\sem{\Gamma_1, x : A}}}
\ar@/_1pc/[uul]_<<<<<<<<<<<<{\!\!\!\pi_{1_{\sem{\Gamma_1, x : A}}}}
\ar[dl]^-{\,\,\,\,\ia {\sem{\Gamma_1, x : A; V}}}^>>>>>{\qquad\qquad\qquad\dscomment{\text{Proposition~\ref{prop:semweakening2}}}}^>>>>>{\qquad\qquad\qquad\qquad\qquad\qquad\qquad\qquad\dscomment{\text{def. of } \sproj {\Gamma_1} {x} {A} {\diamond}}}
\ar[r]_-{=}
&
\ia {\pi^*_{\sem{\Gamma_1;A}}(1_{\sem{\Gamma_1}})}
\ar[d]_-{\ia {\pi^*_{\sem{\Gamma_1;A}}(\sem{\Gamma_1;V})}}
\ar[u]^-{\ia {\overline{\pi_{\sem{\Gamma_1;A}}}(1_{\sem{\Gamma_1}})}}^>{\dscomment{\mathcal{P}(\overline{\pi_{\sem{\Gamma_1;A}}}(1_{\sem{\Gamma_1}}))}\qquad\quad}_>>>>>>>>>{\!\!\!\!\quad\dscomment{\text{def. of } \pi^*_{\sem{\Gamma_1;A}}(\sem{\Gamma_1;V})}}
\\
\ia {\sem{\Gamma_1, x \!:\! A; B}}
\ar[dr]_-{=}
&
&
\ia {\pi^*_{\sem{\Gamma_1;A}}(\sem{\Gamma_1;B})}
\ar[ddr]_-{\ia {\overline{\pi_{\sem{\Gamma_1;A}}}(\sem{\Gamma_1;B})}}^<<<<<<<<<<{\quad\qquad\dscomment{\text{id. law}}}
\ar@/_1.5pc/[uur]_<<<<<<<<<<{\!\!\!\!\!\ia {\overline{\pi_{\sem{\Gamma_1;A}}}(\sem{\Gamma_1;B})}}
\\
\sem{\Gamma_1, x \!:\! A, y \!:\! B}
\ar[u]^-{=}
\ar[d]_-{\sproj {\Gamma_1} {x} {A} {y : B}}^<<<<{\,\,\,\,\dscomment{\text{def. of } \sproj {\Gamma_1} {x} {A} {y : B}}}
&
\ia {\sproj {\Gamma_1} {x} {A} {\diamond}^*(\sem{\Gamma_1; B})}
\ar[dl]^-{\,\,\,\,\,\,\,\,\,\,\,\ia {\overline{\sproj {\Gamma_1} {x} {A} {\diamond}}(\sem{\Gamma_1;B})}}^<<<{\qquad\qquad\qquad\qquad\qquad\dscomment{\text{def. of } \sproj {\Gamma_1} {x} {A} {\diamond}}}
\ar[ur]_-{=}
\\
\sem{\Gamma_1, y \!:\! B}
\ar[rrr]_-{=}
&
&
&
\ia {\sem{\Gamma_1;B}}
}
\vspace{0.25cm}
\]

\pagebreak

\noindent
\textit{Step case (with $\Gamma_2 = \Gamma'_2, x_n \!:\! A_n$):}
\[
\hspace{0.65cm}
\scriptsize
\xymatrix@C=2em@R=6em@M=0.5em{
\sem{\Gamma_1, x \!:\! A, \Gamma'_2[V/y], x_n \!:\! A_n[V/y]}
\ar@/_5.5pc/[dddd]_-{=}
\ar@/_8.5pc/[ddddddd]^>>>>>>>>>>>>>>>>>>>>>>>>>>>>>>>>>>>>>>>>>>>>>>>>>{\ssubst {\Gamma_1, x : A} {y} {B} {\Gamma'_2, x_n : A_n} {V}}
\ar[rr]^-{\sproj {\Gamma_1} {x} {A} {\Gamma'_2, x_n : A_n}}
\ar[d]^-{=}^>>>>>{\quad\qquad\qquad\qquad\qquad\dscomment{\text{def. of } \sproj {\Gamma_1} {x} {A} {\Gamma'_2, x_n : A_n}}}^<<<<<{\!\!\quad\quad\qquad\qquad\qquad\qquad\dscomment{\text{Proposition~\ref{prop:semweakening2}}}}
&
&
\sem{\Gamma_1, \Gamma'_2[V/y], x_n \!:\! A_n[V/y]]}
\ar@/^4.5pc/[ddd]_-{=}
\ar@/^9pc/[ddddddd]_-{\ssubst {\Gamma_1} {y} {B} {\Gamma'_2, x_n : A_n} {V}}
\\
\txt<7pc>{
$\{\sproj {\Gamma_1} {x} {A} {\Gamma'_2[V/y]}^*($\\$\sem{\Gamma_1, \Gamma'_2[V/y]; A_n[V/y]})\}$
}
\ar[d]^-{=}^-{\qquad\qquad\qquad\qquad\qquad\qquad\dscomment{\text{Proposition~\ref{prop:semsubstitution2}}}}_-{\dscomment{p \text{ is a s. fib.}}\quad}
\ar[rr]_-{\ia {\overline{\sproj {\Gamma_1} {x} {A} {\Gamma'_2[V/y]}}(\sem{\Gamma_1, \Gamma'_2[V/y]; A_n[V/y]})}}
&
&
\ia {\sem{\Gamma_1, \Gamma'_2[V/y]; A_n[V/y]}}
\ar[u]_-{=}
\\
\txt<7pc>{
$\{\sproj {\Gamma_1} {x} {A} {\Gamma'_2[V/y]}^*($\\$\ssubst {\Gamma_1} {y} {B} {\Gamma_2} {V}^*($\\$\sem{\Gamma_1, \Gamma'_2[V/y]; A_n[V/y]}))\}$
}
\ar[d]^-{=}_-{\dscomment{\text{i.h.}}\qquad\,\,\,}
\ar[drr]^<<<<<<<<<<<<<{\quad\qquad\qquad\qquad\ia {\overline{\sproj {\Gamma_1} {x} {A} {\Gamma'_2[V/y]}}(\ssubst {\Gamma_1} {y} {B} {\Gamma_2} {V}^*(\sem{\Gamma_1, \Gamma'_2[V/y]; A_n[V/y]}))}}
\\
\txt<7pc>{
$\{\ssubst {\Gamma_1, x : A} {y} {B} {\Gamma'_2} {V}^*($\\$\sproj {\Gamma_1} {x} {A} {y : B, \Gamma'_2}^*($\\$\sem{\Gamma_1, y \!:\! B, \Gamma'_2;A_n}))\}$
}
\ar[d]^-{=}^-{\qquad\qquad\dscomment{p \text{ is a split fibration}}}^-{\qquad\qquad\qquad\qquad\qquad\qquad\qquad\dscomment{\text{induction hypothesis}}}
&
&
\txt<7pc>{
$\{\ssubst {\Gamma_1} {y} {B} {\Gamma'_2} {V}^*($\\$\sem{\Gamma_1, y \!:\! B, \Gamma'_2; A_n})\}$
}
\ar[dd]_-{\ia {\overline{\ssubst {\Gamma_1} {y} {B} {\Gamma'_2} {V}}(\sem{\Gamma_1, y : B, \Gamma'_2; A_n})}}^-{\qquad\dscomment{\text{def. of } \ssubst {\Gamma_1} {y} {B} {\Gamma'_2, x_n : A_n} {V}}}
\\
\txt<7pc>{
$
\{\ssubst {\Gamma_1, x : A} {y} {B} {\Gamma'_2} {V}^*($\\$\sem{\Gamma_1, x \!:\! A, y \!:\! B, \Gamma_2;A_n})\}
$}
\ar[dd]^<<<<<<<<<<<{\ia {\overline{\ssubst {\Gamma_1, x : A} {y} {B} {\Gamma'_2} {V}}(\sem{\Gamma_1, x : A, \Gamma'_2; A_n})}}_<<<<<<{\dscomment{\text{def. of } \ssubst {\Gamma_1, x : A} {y} {B} {\Gamma'_2, x_n : A_n} {V}}\,\,\,\,\,}
\\
&&
\ia {\sem{\Gamma_1, y \!:\! B, \Gamma'_2; A_n}}
\ar[dd]^-{=}
\\
\ia {\sem{\Gamma_1, x \!:\! A, y \!:\! B, \Gamma'_2; A_n}}
\ar[d]^-{=}^>>>>>{\qquad\qquad\qquad\qquad\dscomment{\text{def. of } \sproj {\Gamma_1} {x} {A} {y : B, \Gamma'_2, x_n : A_n}}}^<<<<<<<<{\,\,\,\,\quad\qquad\qquad\qquad\qquad\dscomment{\text{Proposition~\ref{prop:semweakening2}}}}
\ar[r]_-{=}
&
\txt<5.5pc>{
$\{\sproj {\Gamma_1} {x} {A} {y : B, \Gamma'_2}^*($\\$\sem{\Gamma_1, y \!:\! B, \Gamma'_2;A_n})\}$
}
\ar[ur]^-{\ia {\overline{\sproj {\Gamma_1} {x} {A} {y : B, \Gamma'_2}}(\sem{\Gamma_1, y : B, \Gamma'_2; A_n})}\,\,\,\,\,\,\,\,\,}
&
\\
\sem{\Gamma_1, x \!:\! A, y \!:\! B, \Gamma'_2, x_n \!:\! A_n}
\ar[rr]_-{\sproj {\Gamma_1} {x} {A} {y : B, \Gamma'_2, x_n : A_n}}
&
&
\sem{\Gamma_1, y \!:\! B, \Gamma'_2; x_n : A_n}
}
\]
\end{proof}

\newpage 

\section{Proof of Proposition~\ref{prop:reindexingalongkappaandpairing}}
\label{sect:proofofprop:reindexingalongkappaandpairing}

{
\renewcommand{\thetheorem}{\ref{prop:reindexingalongkappaandpairing}}
\begin{proposition}
Given a value context $\Gamma$, value variables $x_1$, $x_2$, and $y$, and \linebreak value types $A_1$, $A_2$, and $B$ such that $x_2 \not\in V\!ars(\Gamma) \cup \{y\}$, $\sem{\Gamma} \in \mathcal{B}$, $\sem{\Gamma;A} \in \mathcal{V}_{\sem{\Gamma}}$, \linebreak $\sem{\Gamma, x_1 \!:\! A_1;A_2} \in \mathcal{V}_{\sem{\Gamma, x_1 \!:\! A_1}}$, and $\sem{\Gamma, y \!:\! (\Sigma\, x_1 \!:\! A_1 .\, A_2) ; B} \in \mathcal{V}_{\sem{\Gamma, y : (\Sigma\, x_1 : A_1 .\, A_2)}}$, then we have
\[
\sem{\Gamma, x_1 \!:\! A_1, x_2 \!:\! A_2, B[\langle x_1, x_2 \rangle/y]} = \kappa_{\sem{\Gamma; A_1},\sem{\Gamma, x_1 : A_1; A_2}}^*(\sem{\Gamma, y \!:\! (\Sigma\, x_1 \!:\! A_1 .\, A_2); B}) 
\]
\end{proposition}
\addtocounter{theorem}{-1}
}

\begin{proof}
We begin by noting that both sides of this equation can be rewritten as follows.

On the one hand, the left-hand side of this equation can be rewritten as
\[
\begin{array}{c}
\hspace{-8cm}
(\mathsf{s}(\sem{\Gamma, x_1 \!:\! A_1, x_2 \!:\! A_2; \langle x_1 , x_2 \rangle}))^*(
\\
\hspace{-4cm}
\ia{\overline{\pi_{\sem{\Gamma, x_1 : A_1 ; A_2}}}(\sem{\Gamma, x_1 \!:\! A_1; \Sigma\, x_1 \!:\! A_1 .\, A_2})}^*(
\\
\hspace{4cm}
\ia{\overline{\pi_{\sem{\Gamma;A_1}}}(\sem{\Gamma; \Sigma\, x_1 \!:\! A_1 .\, A_2})}^*(\sem{\Gamma, y \!:\! (\Sigma\, x_1 \!:\! A_1 .\, A_2); B})))
\end{array}
\]
based on Propositions~\ref{prop:semweakening2} and~\ref{prop:semsubstitution2}, and the definition of morphisms $\sproj {\Gamma_1} {x} {A} {\Gamma_2}$.

On the other hand, the right-hand side of this equation can be rewritten as
\[
\ia {\eta^{\Sigma_{\sem{\Gamma; A_1}} \,\dashv\, \pi^*_{\sem{\Gamma; A_1}}}_{\sem{\Gamma, x_1 : A_1; A_2}}}^*(\ia {\overline{\pi_{\sem{\Gamma;A_1}}}(\sem{\Gamma; \Sigma\, x_1 \!:\! A_1 .\, A_2})}^*(\sem{\Gamma, y \!:\! (\Sigma\, x_1 \!:\! A_1 .\, A_2); B}))
\]
based on the definitions of $\kappa_{\sem{\Gamma; A_1},\sem{\Gamma, x_1 : A_1; A_2}}$ and $\sem{\Gamma;\Sigma\, x_1 \!:\! A_1 .\, A_2}$.

Now, as a result of $p : \mathcal{V} \longrightarrow \mathcal{B}$ being a split fibration, it suffices to show 
\[
\begin{array}{c}
\ia{\overline{\pi_{\sem{\Gamma, x_1 : A_1 ; A_2}}}(\sem{\Gamma, x_1 \!:\! A_1; \Sigma\, x_1 \!:\! A_1 .\, A_2})} 
\comp \mathsf{s}(\sem{\Gamma, x_1 \!:\! A_1, x_2 \!:\! A_2; \langle x_1 , x_2 \rangle})
\\
=
\\
\ia {\eta^{\Sigma_{\sem{\Gamma; A_1}} \,\dashv\, \pi^*_{\sem{\Gamma; A_1}}}_{\sem{\Gamma, x_1 : A_1; A_2}}}
\end{array}
\]
for the required equation to be true, which follows from the commutativity of the following diagram:
\[
\scriptsize
\xymatrix@C=5em@R=5em@M=0.5em{
\sem{\Gamma, x_1, x_2}
\ar[d]^-{\eta^{1 \,\dashv\, \ia -}_{\sem{\Gamma, x_1, x_2}}}^>>>>{\quad\qquad\qquad\qquad\qquad\dscomment{(*)}}
\ar@/_8pc/[dddddd]_<<<<<<<<<<<<{\mathsf{s}(\sem{\Gamma, x_1, x_2; \langle x_1, x_2 \rangle})}
\ar[r]^-{=}
&
\ia {\sem{\Gamma, x_1; A_2}}
\ar[ddddddd]^-{\ia {\eta^{\Sigma_{\sem{\Gamma;A_1}} \,\dashv\, \pi^*_{\sem{\Gamma; A_1}}}_{\sem{\Gamma, x_1 ; A_2}}}}
\\
\ia {1_{\sem{\Gamma, x_1, x_2}}}
\ar[d]^-{\ia {\eta^{\Sigma_{\sem{\Gamma, x_1;A_2}} \,\dashv\, \pi^*_{\sem{\Gamma, x_1; A_2}}}_{1_{\sem{\Gamma, x_1, x_2}}}}}
\\
\ia {\pi^*_{\sem{\Gamma, x_1;A_2}}(\Sigma_{\sem{\Gamma, x_1; A_2}}(1_{\sem{\Gamma, x_1, x_2}}))}
\ar[d]_-{\ia {\pi^*_{\sem{\Gamma, x_1;A_2}}(\mathsf{fst})}}
\\
\ia {\pi^*_{\sem{\Gamma, x_1;A_2}}(\sem{\Gamma, x_1;A_2})}
\ar[d]_-{=}_-{\dscomment{\text{def. of } \mathsf{s}(\sem{\Gamma, x_1, x_2; \langle x_1, x_2 \rangle})}\quad\,\,\,\,}
\ar@/_2pc/[uuur]_<<<<<<<<<<<<<{\,\,\,\,\,\ia {\overline{\pi_{\sem{\Gamma, x_1;A_2}}}(\sem{\Gamma, x_1;A_2})}}
\\
\txt<10pc>{
$\{(\mathsf{s}(\sem{\Gamma, x_1, x_2; x_1}))^*( $\\$ \sem{\Gamma, x_1, x_2, x'_1; A_2[x'_1/x_1]})\}$
}
\ar[d]_-{\dshide{\ia {(\mathsf{s}(\sem{\Gamma, x_1, x_2; x_1}))^*(\eta^{\Sigma_{\sem{\Gamma, x_1, x_2;A_1}} \,\dashv\, \pi^*_{\sem{\Gamma, x_1, x_2;A_1}}}_{\sem{\Gamma, x_1, x_2, x'_1; A_2[x'_1/x_1]}})}}}^-{\qquad\qquad\qquad\qquad\qquad\dscomment{(**)}}
\\
\txt<10pc>{
$\{(\mathsf{s}(\sem{\Gamma, x_1, x_2; x_1}))^*($\\$ \pi^*_{\sem{\Gamma, x_1, x_2;A_1}}( $\\$ \Sigma_{\Gamma, x_1, x_2; A_1}( $\\$ \hspace{0.5cm}\sem{\Gamma, x_1, x_2, x'_1; A_2[x'_1/x_1]})))\}$
}
\ar[d]_-{=}
\\
\ia {\Sigma_{\sem{\Gamma, x_1, x_2; A_1}}(\sem{\Gamma, x_1, x_2, x'_1; A_2[x'_1/x_1]})}
\ar[d]_-{=}
\\
\txt<10pc>{
$\{\pi^*_{\sem{\Gamma, x_1; A_2}}(\Sigma_{\sem{\Gamma, x_1; A_1}}( $\\$ \sem{\Gamma, x_1, x'_1; A_2[x'_1/x_1]}))\}$
}
\ar[d]_-{\ia {\overline{\pi_{\sem{\Gamma, x_1; A_2}}}(\Sigma_{\sem{\Gamma, x_1; A_1}}(\sem{\Gamma, x_1, x'_1; A_2[x'_1/x_1]}))}}
& 
\ia {\pi^*_{\sem{\Gamma;A_1}}(\Sigma_{\sem{\Gamma;A_1}}(\sem{\Gamma, x_1 ; A_2}))}
\\
\ia {\Sigma_{\sem{\Gamma, x_1; A_1}}(\sem{\Gamma, x_1, x'_1; A_2[x'_1/x_1]})}
\ar[r]_-{=}
&
\ia {\pi^*_{\sem{\Gamma;A_1}}(\Sigma_{\sem{\Gamma;A_1}}(\sem{\Gamma, x'_1 ; A_2[x'_1/x_1]}))}
\ar[u]_-{=}
}
\]

\pagebreak

\noindent 
where the subdiagram marked with $(*)$ commutes because we have
\[
\scriptsize
\xymatrix@C=7em@R=3.5em@M=0.5em{
\sem{\Gamma, x_1, x_2}
\ar[d]_-{\eta^{1 \,\dashv\, \ia -}_{\sem{\Gamma, x_1, x_2}}}^-{\qquad\qquad\qquad\dscomment{\eta^{1 \,\dashv\, \ia -}_{\sem{\Gamma, x_1, x_2}} \text{ and } \pi_{1_{\sem{\Gamma, x_1, x_2}}} \text{ form an isomorphism (Proposition~\ref{prop:compcatunitiso})}}}
\ar[rr]^-{=}
&&
\ia {\sem{\Gamma, x_1; A_2}}
\\
\ia {1_{\sem{\Gamma, x_1, x_2}}}
\ar@/^2pc/[ddr]^-{\!\!\!\!\!\!\!\!\id_{\ia {1_{\sem{\Gamma, x_1, x_2}}}}}
\ar[dd]_-{\ia {\eta^{\Sigma_{\sem{\Gamma, x_1;A_2}} \,\dashv\, \pi^*_{\sem{\Gamma, x_1; A_2}}}_{1_{\sem{\Gamma, x_1, x_2}}}}}^>>>>>>>{\qquad\dscomment{\text{def. of } \kappa}}
\ar[dddr]^-{\!\!\kappa_{\sem{\Gamma; A_1},\sem{\Gamma, x_1; A_2}}}^<<<<<<<<<<<<<<<<<<<<{\,\,\,\,\,\,\,\,\,\,\,\dscomment{\kappa \text{ is an iso.}}}
\\
\\
\ia {\pi^*_{\sem{\Gamma, x_1;A_2}}(\Sigma_{\sem{\Gamma, x_1; A_2}}(1_{\sem{\Gamma, x_1, x_2}}))}
\ar[dd]_-{\ia {\pi^*_{\sem{\Gamma, x_1;A_2}}(\mathsf{fst})}}
\ar[dr]_>>>>>>>>>>>>>>{\ia{\overline{\pi_{\sem{\Gamma, x_1;A_2}}}(\Sigma_{\sem{\Gamma, x_1; A_2}}(1_{\sem{\Gamma, x_1, x_2}}))}\qquad\qquad\quad}
&
\ia {1_{\sem{\Gamma, x_1,x_2}}}
\ar[uuur]_-{\!\!\!\!\pi_{1_{\sem{\Gamma, x_1, x_2}}}}_<<<<<{\quad\qquad\qquad\dscomment{\text{def. of } \mathsf{fst}}}
\\
&
\ia {\Sigma_{\sem{\Gamma, x_1; A_2}}(1_{\sem{\Gamma, x_1, x_2}})}
\ar[dr]_-{\ia {\mathsf{fst}}}_<<<<<<{\dscomment{\text{def. of } \pi^*_{\sem{\Gamma, x_1;A_2}}(\mathsf{fst})}\qquad\qquad\qquad\qquad\qquad\qquad\qquad\qquad}
\ar[u]_-{\kappa^{-1}_{\sem{\Gamma; A_1},\sem{\Gamma, x_1; A_2}}}
\\
\ia {\pi^*_{\sem{\Gamma, x_1;A_2}}(\sem{\Gamma, x_1;A_2})}
\ar[rr]_-{\ia {\overline{\pi_{\sem{\Gamma, x_1;A_2}}}(\sem{\Gamma, x_1;A_2})}}
&&
\ia {\sem{\Gamma, x_1;A_2}}
\ar[uuuuu]_-{\id_{\ia {\sem{\Gamma, x_1;A_2}}}}
}
\]
and the subdiagram marked with $(**)$ commutes because we have
\[
\scriptsize
\xymatrix@C=1em@R=4em@M=0.5em{
\ia {\pi^*_{\sem{\Gamma, x_1;A_2}}(\sem{\Gamma, x_1;A_2})}
\ar[d]_-{=}
\ar[rr]^-{\ia {\overline{\pi_{\sem{\Gamma, x_1;A_2}}}(\sem{\Gamma, x_1;A_2})}}
\ar@/^10pc/[dddddr]^<<<<<<<<<<<<<<<<<<<<<<<<<<<<<<<{\!\!\!\!\ia {\pi^*_{\sem{\Gamma, x_1; A_2}}(\eta^{\Sigma_{\sem{\Gamma;A_1}} \,\dashv\, \pi^*_{\sem{\Gamma; A_1}}}_{\sem{\Gamma, x_1 ; A_2}})}}^<<<<<<<<<<<<<<<<<{\qquad\qquad\qquad\dscomment{\text{def. of } \pi^*_{\sem{\Gamma, x_1; A_2}}(\eta^{\Sigma_{\sem{\Gamma;A_1}} \,\dashv\, \pi^*_{\sem{\Gamma; A_1}}}_{\sem{\Gamma, x_1 ; A_2}})}}
&
&
\ia {\sem{\Gamma, x_1; A_2}}
\ar[dddd]^-{\ia {\eta^{\Sigma_{\sem{\Gamma;A_1}} \,\dashv\, \pi^*_{\sem{\Gamma; A_1}}}_{\sem{\Gamma, x_1 ; A_2}}}}
\\
\txt<10pc>{
$\{(\mathsf{s}(\sem{\Gamma, x_1, x_2; x_1}))^*( $\\$ \sem{\Gamma, x_1, x_2, x'_1; A_2[x'_1/x_1]})\}$
}
\ar[d]^-{{\ia {(\mathsf{s}(\sem{\Gamma, x_1, x_2; x_1}))^*(\eta^{\Sigma_{\sem{\Gamma, x_1, x_2;A_1}} \,\dashv\, \pi^*_{\sem{\Gamma, x_1, x_2;A_1}}}_{\sem{\Gamma, x_1, x_2, x'_1; A_2[x'_1/x_1]}})}}}
\\
\txt<10pc>{
$\{(\mathsf{s}(\sem{\Gamma, x_1, x_2; x_1}))^*($\\$ \pi^*_{\sem{\Gamma, x_1, x_2;A_1}}(\Sigma_{\Gamma, x_1, x_2; A_1}( $\\$ \sem{\Gamma, x_1, x_2, x'_1; A_2[x'_1/x_1]})))\}$
}
\ar[d]_-{=}^-{\qquad\qquad\dscomment{\text{Proposition~\ref{prop:semweakening2}}}}^-{\qquad\qquad\qquad\qquad\qquad\qquad\dscomment{\text{Proposition~\ref{prop:semsubstitution2}}}}
\\
\ia {\Sigma_{\sem{\Gamma, x_1, x_2; A_1}}(\sem{\Gamma, x_1, x_2, x'_1; A_2[x'_1/x_1]})}
\ar[d]_-{=}^-{\quad\dscomment{\text{split Beck-Chevalley}}}^{\quad\qquad\qquad\qquad\qquad\dscomment{\text{defs. of } \sproj {\Gamma_1} {x} {A} {\Gamma_2} \text{ and } \ssubst {\Gamma_1} {x} {A} {\Gamma_2} V}}
\\
\txt<10pc>{
$\{\pi^*_{\sem{\Gamma, x_1; A_2}}(\Sigma_{\sem{\Gamma, x_1; A_1}}( $\\$ \sem{\Gamma, x_1, x'_1; A_2[x'_1/x_1]}))\}$
}
\ar[dd]^>>>>>{\ia {\overline{\pi_{\sem{\Gamma, x_1; A_2}}}(\Sigma_{\sem{\Gamma, x_1; A_1}}(\sem{\Gamma, x_1, x'_1; A_2[x'_1/x_1]}))}}^>>>>>>>>>>>{\qquad\qquad\dscomment{\text{Proposition~\ref{prop:semweakening2}}}}^>>>>>>>>>>>{\quad\qquad\qquad\qquad\qquad\qquad\qquad\qquad\dscomment{\text{split Beck-Chevalley}}}^>>>>>>>>>>>{\qquad\qquad\qquad\qquad\qquad\qquad\qquad\qquad\qquad\qquad\qquad\qquad\qquad\dscomment{\text{def. of } \sproj {\Gamma_1} {x} {A} {\Gamma_2}}}
\ar[dr]_-{=}
&
& 
\ia {\pi^*_{\sem{\Gamma;A_1}}(\Sigma_{\sem{\Gamma;A_1}}(\sem{\Gamma, x_1 ; A_2}))}
\\
& \txt<10pc>{
$\{\pi^*_{\sem{\Gamma, x_1; A_2}}(\pi^*_{\sem{\Gamma;A_1}}($\\$\Sigma_{\sem{\Gamma;A_1}}(\sem{\Gamma, x_1 ; A_2})))\}$
}
\ar[ur]_<<<<<<<<<<<{\!\!\!\!\qquad\qquad\ia {\overline{\pi_{\sem{\Gamma, x_1; A_2}}}(\pi^*_{\sem{\Gamma;A_1}}(\Sigma_{\sem{\Gamma;A_1}}(\sem{\Gamma, x_1 ; A_2})))}}
\\
\ia {\Sigma_{\sem{\Gamma, x_1; A_1}}(\sem{\Gamma, x_1, x'_1; A_2[x'_1/x_1]})}
\ar[rr]_-{=}
&&
\txt<5pc>{
$\{\pi^*_{\sem{\Gamma;A_1}}(\Sigma_{\sem{\Gamma;A_1}}($\\$\sem{\Gamma, x'_1 ; A_2[x'_1/x_1]}))\}$
}
\ar@/_3.75pc/[uu]_-{=}
}
\]

To improve the readability of the above proofs, we have omitted the types in value contexts, writing ${\Gamma, x_1, x_2}$ for ${\Gamma, x_1 \!:\! A_1, x_2 \!:\! A_2}$. We have also omitted details of equalities that follow from the use of the split Beck-Chevalley conditions, Propositions~\ref{prop:semweakening2} and~\ref{prop:semsubstitution2}, and the definitions of semantic projection and substitution morphisms, e.g., 
\[
\Sigma_{\sem{\Gamma, x_1,x_2;A_1}}(\sem{\Gamma, x_1, x_2, x'_1; B[x'_1/x_1]}) = \pi^*_{\sem{\Gamma, x_1; B}}(\Sigma_{\sem{\Gamma, x_1;A_1}}(\sem{\Gamma, x_1, x'_1; B[x'_1/x_1]}))
\]
\end{proof}

\renewcommand\thesection{\thechapter.\arabic{section}}

\chapter{Proofs for Chapter~\ref{chap:fibalgeffects}}
\label{chap:appendixC6}

\section{Proof of Proposition~\ref{prop:relatingsemanticsoffibeffectterms}}
\label{sect:proofofprop:relatingsemanticsoffibeffectterms}

{
\renewcommand{\thetheorem}{\ref{prop:relatingsemanticsoffibeffectterms}}
\begin{proposition}
Given a well-formed effect term $\lj {\Gamma \vertbar \Delta} T$ derived from $\mathcal{S}_{\text{eff}}$, a computation type $\ul{C}$, value terms $V_{i}$ (for all $x_i \!:\! A_i$ in $\Gamma$), value terms $V'_{\!j}$ (for all $w_{\!j} \!:\! A'_{\!j}$ in $\Delta$), value terms $W_{\sigalgop}$ (for all $\sigalgop : (x \!:\! I) \longrightarrow O$ in $\mathcal{S}_{\text{eff}}$), and a value context $\Gamma'$ such that
\begin{itemize}
\item $\sem{\Gamma'} \in \Set$, 
\item $\sem{\Gamma';V_i}_1 = \id_{\sem{\Gamma'}}  : \sem{\Gamma'} \longrightarrow \sem{\Gamma'}$, and 
\item $(\sem{\Gamma';V_i}_2)_{\gamma\,'} : 1 \longrightarrow $
\\[-7.5mm]

\hspace{1cm} $\sem{x_1 \!:\! A_1, \ldots, x_{i-1} \!:\! A_{i - 1};A_i}_2\, \langle \langle \langle \star , (\sem{\Gamma'; V_1}_2)_{\gamma\,'}(\star) \rangle , \ldots \rangle , (\sem{\Gamma'; V_{i - 1}}_2)_{\gamma\,'}(\star) \rangle$, 
\end{itemize}
together with a value type $A$ and a family of models $\mathcal{M}_{\gamma\,'} : \mathcal{L}_{\mathcal{T}^{d}_{\text{eff}}} \longrightarrow \Set$ (for all $\gamma\,'$ in $\sem{\Gamma'}$) of the Lawvere theory $I_{\mathcal{T}^{d}_{\text{eff}}} : \aleph_{\!\!1}^{\text{op}} \longrightarrow \mathcal{L}_{\mathcal{T}^{d}_{\text{eff}}}$ (where ${\mathcal{T}^{d}_{\text{eff}}} \defeq (\mathcal{S}_{\text{eff}}, \emptyset)$) such that
\index{ T@${\mathcal{T}^{d}_{\text{eff}}}$ (fibred effect theory $(\mathcal{S}_{\text{eff}}, \emptyset)$)}
\begin{itemize}
\item $\sem{\Gamma';A}_1 = \sem{\Gamma'}$, 
\item $\sem{\Gamma';A}_2(\gamma\,') = \mathcal{M}_{\gamma\,'}(1)$,  
\item $\sem{\Gamma';V'_{\!j}}_1 = \id_{\sem{\Gamma'}}  : \sem{\Gamma'} \longrightarrow \sem{\Gamma'}$, 
\item $(\sem{\Gamma';V'_{\!j}}_2)_{\gamma\,'} : 1 \longrightarrow \bigsqcap_{a \in \sem{\Gamma;A'_{\!j}}_2(\gamma)}(\sem{\Gamma';A}_2(\gamma\,'))$, 
\item $\sem{\Gamma';W_{\sigalgop}}_1 = \id_{\sem{\Gamma'}} : \sem{\Gamma'} \longrightarrow \sem{\Gamma'}$, and
\item $(\sem{\Gamma';W_{\sigalgop}}_2)_{\gamma\,'} = \bigsqcap_{\langle i , f \rangle} \sigalgop^{\mathcal{M}_{\gamma\,'}}_{i} \,\comp\,\, \bigsqcap_{\langle i , f \rangle} (\star \mapsto f) \,\comp\,\, \langle \id_1 \rangle_{\langle i , f \rangle}$
\\[-7.5mm]

\hspace{3.5cm} $: 1 \longrightarrow \bigsqcap_{\langle i , f \rangle \in \bigsqcup_{i \in \sem{\diamond;I}_2(\star)} \bigsqcap_{o \in {\sem{x : I ; O}_2\, \langle \star , i \rangle}} (\sem{\Gamma';A}_2(\gamma\,'))} (\sem{\Gamma';A}_2(\gamma\,'))$, 
\end{itemize}
then 
\[
\sem{\Gamma'; \efftrans T {\!\!A; \overrightarrow{V_i}; \overrightarrow{V'_{\!j}}; \overrightarrow{W_{\sigalgop}}}}_1 = \id_{\sem{\Gamma'}} : \sem{\Gamma'} \longrightarrow \sem{\Gamma'}
\]
and, for all $\gamma\,'$ in $\sem{\Gamma'}$, the function 
\[
(\sem{\Gamma'; \efftrans T {\!\!A; \overrightarrow{V_i}; \overrightarrow{V'_{\!j}}; \overrightarrow{W_{\sigalgop}}}}_2)_{\gamma\,'} : 1 \longrightarrow \sem{\Gamma';A}_2(\gamma\,')
\]
is defined and equal to the following composite function:
\[
\xymatrix@C=4em@R=5em@M=0.5em{
1 
\ar[rr]^-{\langle (\sem{\Gamma';V'_{\!j}}_2)_{\gamma\,'} \rangle_{w_{\!j} : A'_{\!j} \in \Delta}}
&&
\bigsqcap_{w_{\!j} : A'_{\!j} \in \Delta} \bigsqcap_{a \in {\sem{\Gamma;A'_{\!j}}_2(\gamma)}} (\sem{\Gamma';A}_2(\gamma\,'))
\ar[d]^-{\cong}
\\
\sem{\Gamma';A}_2(\gamma\,')
&
\mathcal{M}_{\gamma\,'}(1)
\ar[l]^-{=}
&
\mathcal{M}_{\gamma\,'}(\vert \Delta^\gamma \vert)
\ar[l]^-{\mathcal{M}_{\gamma\,'}(\lj {\Delta^\gamma\,\,} {\,T^\gamma})}
}
\]
where $\vert \Delta^\gamma \vert$ denotes the length of the context $\Delta^\gamma$; and where we use the abbreviation
\[
\gamma \,\defeq \langle \langle \langle \star , (\sem{\Gamma'; V_1}_2)_{\gamma\,'}(\star) \rangle , \ldots \rangle , (\sem{\Gamma'; V_n}_2)_{\gamma\,'}(\star) \rangle
\]
\end{proposition}
\addtocounter{theorem}{-1}
}

\begin{proof}
We prove this proposition by induction on the given derivation of $\lj {\Gamma \vertbar \Delta} T$, using the eMLTT$_{\mathcal{T}_{\text{eff}}}$ versions of Propositions~\ref{prop:semweakening2},~\ref{prop:semsubstitution2},~\ref{prop:semweakening5}, and~\ref{prop:semsubstitution5} to relate syntactic weakening and substitution to reindexing along semantic projection and substitution morphisms. 

We discuss all cases of this proof in detail below. In each of the cases, the proof of
\[
\sem{\Gamma'; \efftrans T {\!\!A; \overrightarrow{V_i}; \overrightarrow{V'_{\!j}}; \overrightarrow{W_{\sigalgop}}}}_1 = \id_{\sem{\Gamma'}} : \sem{\Gamma'} \longrightarrow \sem{\Gamma'}
\]
is straightforward, by combining our assumptions with the definitions of $\sem{-}$ and $\efftrans {-} {}$.

\vspace{0.2cm}

\noindent \textbf{Effect variables:}
In this case, the given derivation ends with 
\[
\mkrule
{\lj {\Gamma \vertbar \Delta_1, w_{\!j} \!:\! A'_{\!j}, \Delta_2} {w_{\!j}\,(V)}}
{
\lj \Gamma {\Delta_1, w_{\!j} \!:\! A'_{\!j}, \Delta_2}
\quad
\vj \Gamma V A'_{\!j}
}
\]

\pagebreak

First, according to the definition of $\efftrans {-} {}$ for effect variables, we have that
\[
(\sem{\Gamma'; \efftrans {w_{\!j}\,(V)} {\!\!A; \overrightarrow{V_i}; \overrightarrow{V'_{\!j}}; \overrightarrow{W_{\sigalgop}}}}_2)_{\gamma\,'}
\]
is equal to
\[
(\sem{\Gamma'; V'_{\!j}\, (V[\overrightarrow{V_i}/\overrightarrow{x_i}])}_2)_{\gamma\,'}
\]
which, by unfolding the definition of $\sem -$ for function application, is equal to
\[
\xymatrix@C=5em@R=4em@M=0.5em{
1
\ar[r]^-{(\sem{\Gamma'; V'_{\!j}}_2)_{\gamma\,'}}
&
\bigsqcap_{a \in {\sem{\Gamma;A'_{\!j}}_2(\gamma)}} (\sem{\Gamma';A}_2(\gamma\,'))
\ar[r]^-{\mathsf{proj}_{(\sem{\Gamma;V}_2)_\gamma(\star)}}
&
\sem{\Gamma';A}_2(\gamma\,')
}
\]
where we implicitly make use of the fact that $\sem{\Gamma;A'_{\!j}}_2(\gamma) = \sem{\Gamma';A'_{\!j}[\overrightarrow{V_i}/\overrightarrow{x_i}]}_2(\gamma\,')$ \linebreak and $(\sem{\Gamma;V}_2)_\gamma = (\sem{\Gamma';V[\overrightarrow{V_i}/\overrightarrow{x_i}]}_2)_{\gamma\,'}$, both of which follow from the definition of $\gamma$ and the eMLTT$_{\mathcal{T}_{\text{eff}}}$ versions of the results that relate syntactic substitution to its semantic counterpart (see Propositions~\ref{prop:semsubstitution2} and~\ref{prop:semsubstitution5}).

The required equation then follows from the commutativity of the following diagram:
\[
\scriptsize
\xymatrix@C=12em@R=5em@M=0.5em{
1
\ar[r]^-{\langle (\sem{\Gamma';V'_{\!j}}_2)_{\gamma\,'} \rangle_{w_{\!j} : A'_{\!j} \in \Delta}}
\ar@/_2pc/[ddr]^<<<<<<<<<<<<<<<<<<{\!\!\!(\sem{\Gamma';V'_{\!j}}_2)_{\gamma\,'}}
\ar@/_3.5pc/[dddr]^-{\!\!\!(\sem{\Gamma';V'_{\!j}}_2)_{\gamma\,'}}
\ar@/_4.5pc/[ddddr]_>>>>>>>>>>>>>{(\sem{\Gamma'; V'_{\!j}\, (V[\overrightarrow{V_i}/\overrightarrow{x_i}])}_2)_{\gamma\,'}}
&
\bigsqcap_{w_{\!j} : A'_{\!j} \in \Delta} \bigsqcap_{a \in {\sem{\Gamma;A'_{\!j}}_2(\gamma)}} (\sem{\Gamma';A}_2(\gamma\,'))
\ar[r]^-{\cong}
\ar[d]_-{=}_<<<<<<{\dscomment{w_{\!j}\text{'th projection}}\qquad\qquad\quad}^<<<<<{\quad\dscomment{\text{preservation of count. prod.}}}
&
\mathcal{M}_{\gamma\,'}(\vert \Delta^\gamma \vert)
\ar@/^2pc/[ddd]_-{\mathcal{M}_{\gamma\,'}(\lj {\Delta^\gamma\,\,} {\,(w_{\!j}\, (V))^\gamma})}_>>>>>>>>>>>>>>>>>>>>>>>>>>>>>{\dscomment{\text{preservation of count. prod.}}\qquad\qquad\qquad\qquad\qquad\qquad}
\ar@/^1pc/[dl]^-{\cong}
\ar@/_3.5pc/[ddd]_<<<<<<<<<<<<<<<<<<<<<<<<<<<<<<<<<<{\mathcal{M}_{\gamma\,'}(\lj {\Delta^\gamma\,\,} {\,x^{(\sem{\Gamma;V}_2)_\gamma(\star)}_{w_{\!j}}})}^>>>>>>>>>>>>>>>>{\,\,\,\,\,\,\qquad\dscomment{\text{def.}}}
\\
&
\bigsqcap_{w_{\!j} : A'_{\!j} \in \Delta} \bigsqcap_{a \in {\sem{\Gamma;A'_{\!j}}_2(\gamma)}} (\mathcal{M}_{\gamma\,'}(1))
\ar[d]_-{{\mathsf{proj}_{w_{\!j}}}}
&
\\
&
\bigsqcap_{a \in {\sem{\Gamma;A'_{\!j}}_2(\gamma)}} (\mathcal{M}_{\gamma\,'}(1))
\ar[d]^-{=}_<<<<<{\dscomment{\text{assumption about } \mathcal{M}_{\gamma\,'}}\quad\!\!\!\!\!}
\ar[dr]_-{\mathsf{proj}_{(\sem{\Gamma;V}_2)_\gamma(\star)}}
&
\\
&
\bigsqcap_{a \in {\sem{\Gamma;A'_{\!j}}_2(\gamma)}} (\sem{\Gamma';A}_2(\gamma\,'))
\ar[d]^>>>>>{\mathsf{proj}_{(\sem{\Gamma;V}_2)_\gamma(\star)}}^<<<<{\qquad\dscomment{\text{assumption about } \sem{\Gamma';A}_2(\gamma\,')}}_{\dscomment{\text{def.}}\quad\,\,\,\,}
&
\mathcal{M}_{\gamma\,'}(1)
\ar@/^2pc/[dl]^-{=}
\\
&
\sem{\Gamma';A}_2(\gamma\,')
}
\]

\vspace{0.75cm}

\pagebreak

\noindent \textbf{Algebraic operations:}
In this case, the given derivation ends with
\vspace{0.2cm}
\[
\mkrulelabel
{\lj {\Gamma \vertbar \Delta} {\sigalgop_V(y.\, T)}}
{
\lj \Gamma \Delta
\quad
\vj \Gamma V I
\quad
\lj {\Gamma, y \!:\! O[V/x] \vertbar \Delta} {T}
}
{(\sigalgop : (x \!:\! I) \longrightarrow O \in \mathcal{S}_{\text{eff}})}
\vspace{0.05cm}
\]

First, according to the definition of $\efftrans - {}$ for algebraic operations, we have that
\[
(\sem{\Gamma'; \efftrans {\sigalgop_V(y.\, T)} {\!\!A; \overrightarrow{V_i}; \overrightarrow{V'_{\!j}}; \overrightarrow{W_{\sigalgop}}}}_2)_{\gamma\,'} 
\]
is equal to
\[
(\sem{\Gamma'; {W_{\sigalgop}\,\, \langle V[\overrightarrow{V_i}/\overrightarrow{x_i}] , \lambda\, y \!:\! O[V[\overrightarrow{V_i}/\overrightarrow{x_i}]/x] .\, \efftrans {T} {\!\!A; \overrightarrow{V_i}, y; \overrightarrow{V'_{\!j}}; \overrightarrow{W_{\sigalgop}}}}}_2)_{\gamma\,'} 
\]
which, by unfolding the definition of $\sem -$ for function application, pairing, and lambda abstraction, is equal to the following composite function:
\[
\xymatrix@C=6em@R=6em@M=0.5em{
1
\ar[d]_-{(\sem{\Gamma'; W_{\sigalgop}}_2)_{\gamma\,'}} 
\\
\bigsqcap_{\langle i , f \rangle \in \bigsqcup_{i \in \sem{\diamond;I}_2(\star)} \bigsqcap_{o \in {\sem{x : I ; O}_2\, \langle \star , i \rangle}} (\sem{\Gamma';A}_2(\gamma\,'))} (\sem{\Gamma';A}_2(\gamma\,'))
\ar[d]_-{\mathsf{proj}_{\langle (\sem{\Gamma;V}_2)_\gamma(\star) , \langle (\sem{\Gamma'\!, y : O[V[\overrightarrow{V_i}/\overrightarrow{x_i}]/x] ; \efftrans T {}}_2)_{\langle \gamma\,' \!, o \rangle} \rangle_{o \in \sem{x : I ; O}_2\,\langle \star , (\sem{\Gamma;V}_2)_\gamma(\star) \rangle} \rangle}}
\\
\sem{\Gamma';A}_2(\gamma\,')
}
\]
where, analogously to the case of effect variables, we again implicitly make use of the fact that $\sem{\Gamma;A'_{\!j}}_2(\gamma) = \sem{\Gamma';A'_{\!j}[\overrightarrow{V_i}/\overrightarrow{x_i}]}_2(\gamma\,')$ and $(\sem{\Gamma;V}_2)_\gamma = (\sem{\Gamma';V[\overrightarrow{V_i}/\overrightarrow{x_i}]}_2)_{\gamma\,'}$.

The required equation then follows from the commutativity of the following diagram: 
\[
\scriptsize
\xymatrix@C=6em@R=5em@M=0.5em{
1
\ar[dd]^>>>>>>>>>>>{\langle \id_1 \rangle_{\langle i , f \rangle}}
\ar@/_2pc/[ddddddddd]^-{(\sem{\Gamma'; W_{\sigalgop}}_2)_{\gamma\,'}}^>>>>>>>>>>>>>>>>>>>>>>>>>>>>>>>>>>>>>>>>>>>>>>>>>>>>>>>>>>>>>>>>>{\quad\qquad\dscomment{\text{assumption about } W_{\sigalgop}}}
\ar[dr]_>>>>>>>>>>>>>>>>>>{\langle \id_1 \rangle_{o \in \sem{x : I ; O}_2\,\langle \star , (\sem{\Gamma;V}_2)_\gamma(\star) \rangle}\qquad}
\ar[r]^-{\langle (\sem{\Gamma'; V'_{\!j}}_2)_{\gamma\,'} \rangle_{w_{\!j} : A'_{\!j} \in \Delta}}
&
\bigsqcap_{w_{\!j}} \bigsqcap_{a \in \sem{\Gamma;A'_{\!j}}_2(\gamma)}
(\sem{\Gamma';A}_2(\gamma\,'))x
\ar[r]^-{\cong}
&
\mathcal{M}_{\gamma\,'}(\vert \Delta^\gamma \vert)
\ar@/^1pc/[ddddl]_<<<<<<<<{\mathcal{M}_{\gamma\,'}((\lj {\Delta^{\gamma}\,\,} {\,T^{\langle \gamma , o \rangle}})_{1 \,\leq\, o \,\leq\, \vert \sem{x : I ; O}_2\,\langle \star , (\sem{\Gamma;V}_2)_\gamma(\star) \rangle \vert})}
\ar[ddddd]_>>>>>>>>>>>>>>>>>>>>>>>>>>>>>{\mathcal{M}_{\gamma\,'}(\lj {\Delta^{\gamma}\,\,} {\,\sigalgop_{(\sem{\Gamma;V}_2)_\gamma(\star)}(T^{\langle \gamma , o \rangle})_o})}_>>>>>>>>>>>>>>>>>>>>>>>>>>>>>>>>>>>>>{\dscomment{\text{functoriality of } \mathcal{M}_{\gamma\,'}}\quad\,\,\,\,\,\,\,}
\\
&
\bigsqcap_{o \in \sem{x : I ; O}_2\,\langle \star , (\sem{\Gamma;V}_2)_\gamma(\star) \rangle} 1
\ar[d]_-{\bigsqcap_o ((\sem{\Gamma'\!, y; \efftrans T {}}_2)_{\langle \gamma\,'\!, o \rangle})}^-{\quad\qquad\qquad\dscomment{(*)}}
&
\\
\bigsqcap_{\langle i , f \rangle} 1
\ar@/_2.5pc/[dddr]^<<<<<<<<<<<<<<<<<<<<<<<<<<{\bigsqcap_{\langle i , f \rangle} (\star \,\mapsto\, f)}
&
\bigsqcap_{o} (\sem{\Gamma';A}_2(\gamma\,'))
\ar[d]_-{=}
&
\\
&
\bigsqcap_{o} (\mathcal{M}_{\gamma\,'}(1))
\ar[d]_-{\cong}
&
\\
&
\mathcal{M}_{\gamma\,'}(\vert \sem{x \!:\! I ; O}_2\,\langle \star , (\sem{\Gamma;V}_2)_\gamma(\star) \rangle \vert)
\ar@/^2pc/[dr]_-{\mathcal{M}_{\gamma\,'}(\lj {\overrightarrow{x_o}\,\,} {\,\sigalgop_{(\sem{\Gamma;V}_2)_\gamma(\star)}(x_o)_o})\quad}
&
\\
&
\bigsqcap_{\langle i , f \rangle} \bigsqcap_{o \in \sem{x : I; O}_2\, \langle \star , i \rangle} (\sem{\Gamma';A}_2(\gamma\,'))
\ar[d]^-{=}
\ar@/_3.5pc/[ddddl]_-{\bigsqcap_{\langle i , f \rangle} \sigalgop^{\mathcal{M}_{\gamma\,'}}_i\!\!\!\!\!}
&
\mathcal{M}_{\gamma\,'}(1)
\ar[dddd]^-{=}
\\
&
\bigsqcap_{\langle i , f \rangle} \bigsqcap_{o} (\mathcal{M}_{\gamma\,'}(1))
\ar[d]^-{\cong}
&
\\
&
\bigsqcap_{\langle i , f \rangle} (\mathcal{M}_{\gamma\,'}(\vert \sem{x \!:\! I ; O}_2\,\langle \star , i \rangle \vert))
\ar[d]^-{\bigsqcap_{\langle i , f \rangle} (\mathcal{M}_{\gamma\,'}(\lj {\overrightarrow{x_o}\,\,} {\,\sigalgop_i(x_o)_o}))}_-{\dscomment{\text{def. of } \sigalgop^{\mathcal{M}_{\gamma\,'}}_i}\qquad\qquad\quad}
&
\\
&
\bigsqcap_{\langle i , f \rangle} (\mathcal{M}_{\gamma\,'}(1))
\ar[dl]_-{=}^<<<<<<<<<<{\quad\qquad\qquad\qquad\qquad\qquad\qquad\qquad\qquad\dscomment{\text{projections from countable products}}}
&
\\
\bigsqcap_{\langle i , f \rangle} (\sem{\Gamma';A}_2(\gamma\,'))
\ar[rr]_-{\mathsf{proj}_{\langle (\sem{\Gamma;V}_2)_\gamma(\star) , \langle (\sem{\Gamma'\!, y : O[V[\overrightarrow{V_i}/\overrightarrow{x_i}]/x] ; \efftrans T {}}_2)_{\langle \gamma\,' \!, o \rangle} \rangle_{o \in \sem{x : I ; O}_2\,\langle \star , (\sem{\Gamma;V}_2)_\gamma(\star) \rangle} \rangle}}
&
&
\sem{\Gamma';A}_2(\gamma\,')
}
\]
where, for better readability, we abbreviate some of the indices of countable products. For the same reason, we also write the value context $\Gamma'\!, y \!:\! O[V[\overrightarrow{V_i}/\overrightarrow{x_i}]/x]$ as $\Gamma'\!, y$.

In the previous diagram, $(*)$ refers to the following commuting diagram:
\[
\scriptsize
\xymatrix@C=3.5em@R=6em@M=0.5em{
1
\ar[d]^-{\langle \id_1 \rangle_{o \in \sem{x : I ; O}_2\,\langle \star , (\sem{\Gamma;V}_2)_\gamma(\star) \rangle}\qquad}^<<<<<<<<{\quad\qquad\qquad\qquad\qquad\qquad\qquad\qquad\dscomment{\text{pairing for countable products}}}
\ar[rr]^-{\langle (\sem{\Gamma'; V'_{\!j}}_2)_{\gamma\,'} \rangle_{w_{\!j} : A'_{\!j} \in \Delta}}
&
&
\bigsqcap_{w_{\!j}} \bigsqcap_{a \in \sem{\Gamma;A'_{\!j}}_2(\gamma)}
(\sem{\Gamma';A}_2(\gamma\,'))
\ar[ddd]^-{\cong}_-{\dscomment{\text{preservation of countable products}}\quad}
\ar[ddl]_-{\langle \id_{\bigsqcap_{w_{\!j}} \bigsqcap_{a \in \sem{\Gamma;A'_{\!j}}_2(\gamma)}
(\sem{\Gamma';A}_2(\gamma\,'))} \rangle_{o}\!\!\!\!}
\\
\bigsqcap_{o \in \sem{x : I ; O}_2\,\langle \star , (\sem{\Gamma;V}_2)_\gamma(\star) \rangle} 1
\ar@/_1.5pc/[ddr]_>>>>>>>>>>>>>>>{\bigsqcap_{o} (\langle (\sem{\Gamma'\!, y ; V'_{\!j}}_2)_{\langle \gamma\,'\!, o \rangle} \rangle_{w_{\!j}})}
\ar@/_1pc/[dr]^-{\,\,\,\,\quad\bigsqcap_{o} (\langle (\sem{\Gamma'; V'_{\!j}}_2)_{\gamma\,'} \rangle_{w_{\!j}})}
\ar@/_2pc/[ddd]^>>>>>>>>>>>{\bigsqcap_o ((\sem{\Gamma'\!, y; \efftrans T {}}_2)_{\langle \gamma\,'\!, o \rangle})}
&
\\
&
\bigsqcap_{o} \bigsqcap_{w_{\!j}} \bigsqcap_{a} (\sem{\Gamma';A}_2(\gamma\,'))
\ar@/^5.5pc/[dd]^-{\cong}
&
\\
&
\bigsqcap_{o} \bigsqcap_{w_{\!j}} \bigsqcap_{a} (\sem{\Gamma'\!, y;A}_2\, \langle \gamma\,' , o \rangle))
\ar@/_4pc/[dd]_-{\cong}_>>>>>>>>>{\dscomment{\text{induction hypothesis}}\qquad\qquad\qquad\qquad\qquad\qquad}^<<<<<{\qquad\dscomment{\text{weakening}}}
\ar[u]_-{=}^>>>>>{\dscomment{\text{weakening}}\quad}
&
\mathcal{M}_{\gamma\,'}(\vert \Delta^{\gamma} \vert)
\ar@/^1pc/[dl]^>>>>>>>>{\quad\langle \id_{\mathcal{M}_{\gamma\,'}(\vert \Delta^{\gamma} \vert)} \rangle_{o}}
\ar[ddddddd]_>>>>>>>>>>>>>>>>>>>>>>>>>>>>>>>>>>>>>>>>>>>>>{\mathcal{M}_{\gamma\,'}((\lj {\Delta^{\gamma}\,\,} {\,T^{\langle \gamma , o \rangle}})_o)}_<<<<<<<<<<<<<<<<<<<<<<<<<<<<{\dscomment{\text{preservation of countable products}}\quad}
\\
\bigsqcap_{o} (\sem{\Gamma';A}_2(\gamma\,'))
\ar@/_1pc/[ddd]_-{=} \ar@{}[ddd]^>>>>>>>>{\quad\dscomment{\text{weakening}}}
&
\bigsqcap_{o} (\mathcal{M}_{\gamma\,'}(\vert \Delta^{\gamma} \vert))
\ar@/^8pc/[dddd]_-{\bigsqcap_o (\mathcal{M}_{\gamma\,'}(\lj {\Delta^{\gamma}\,\,} {\,T^{\langle \gamma , o \rangle}}))}
&
\\
&
\bigsqcap_{o} (\mathcal{M}_{\langle \gamma\,' , o \rangle}(\vert \Delta^{\langle \gamma , o \rangle} \vert))
\ar[u]_-{=}
\ar[dd]_<<<<<<<<<<<<{\bigsqcap_o (\mathcal{M}_{\langle \gamma\,' , o \rangle}(\lj {\Delta^{\langle \gamma , o \rangle}\,\,} {\,T^{\langle \gamma , o \rangle}}))}^<<<<<<<<{\qquad\dscomment{\text{weakening}}}_<{\dscomment{\text{weakening}}\qquad\qquad\qquad\quad}
\\
&
&
\\
\bigsqcap_o (\mathcal{M}_{\gamma\,'}(1))
\ar@/_7.5pc/[dddrr]_-{\cong}^<<<<<<<<{\quad\qquad\dscomment{\text{weakening}}}
&
\bigsqcap_{o} (\mathcal{M}_{\langle \gamma\,' , o \rangle}(1))
\ar@/^3.5pc/[uuul]^-{=}
\ar[l]_-{=}
\ar[d]_-{=}
\ar@/_4pc/[dd]_-{\cong}^>>>>>{\qquad\dscomment{\text{weakening}}}
\\
&
\bigsqcap_o (\mathcal{M}_{\gamma\,'}(1))
\ar@/^1pc/[ddr]_-{\cong}
\\
&
\mathcal{M}_{\langle \gamma\,' , o \rangle}(\vert \sem{x : I ; O}_2\,\langle \star , (\sem{\Gamma;V}_2)_\gamma(\star) \rangle \vert)
\ar[dr]_-{=}
&
\\
&
&
\mathcal{M}_{\gamma\,'}(\vert \sem{x : I ; O}_2\,\langle \star , (\sem{\Gamma;V}_2)_\gamma(\star) \rangle \vert)
}
\]

Observe that throughout the last diagram, we have extensively used the eMLTT$_{\mathcal{T}_{\text{eff}}}$ version of Propositions~\ref{prop:semweakening2} and~\ref{prop:semweakening5} to relate syntactic weakening to reindexing along semantic projection morphisms (marked with ``weakening"), e.g., to show that
\[
\sem{\Gamma', y \!:\! O[V[\overrightarrow{V_i}/\overrightarrow{x_i}]/x]; A}_2\, \langle \gamma\,'\!, o \rangle = \sem{\Gamma'; A}_2(\gamma\,')
\]

Further, as we know from our assumptions that $\lj \Gamma \Delta$, analogous equations for all $w_{\!j} \!:\! A'_{\!j}$ in $\Delta$ also enable us extend weakening to derived contexts $\Delta^{\langle \gamma , o \rangle}$, as follows:
\[
\vert \Delta^{\langle \gamma , o \rangle} \vert = \vert \Delta^\gamma \vert
\]

\vspace{0.2cm}

\noindent \textbf{Pattern-matching:} In this case, the given derivation ends with 
\[
\mkrule
{\lj {\Gamma \vertbar \Delta} {\pmatchsf V {(y_1 \!:\! B_1, y_2 \!:\! B_2)} {} T}}
{
\lj \Gamma \Delta 
\quad
\vj \Gamma V \Sigma\, y_1 \!:\! B_1 .\, B_2
\quad
\lj {\Gamma, y_1 \!:\! B_1, y_2 \!:\! B_2 \vertbar \Delta} T
}
\]

First, according to the definition of $\efftrans - {}$ for pattern-matching, we have that
\[
(\sem{\Gamma'; \efftrans {\pmatchsf V {(y_1 \!:\! B_1, y_2 \!:\! B_2)} {} T} {\!\!A; \overrightarrow{V_i}; \overrightarrow{V'_{\!j}}; \overrightarrow{W_{\sigalgop}}}}_2)_{\gamma\,'}
\]
is equal to
\[
(\sem{\Gamma'; \pmatch {V[\overrightarrow{V_i}/\overrightarrow{x_i}]} {(y_1 \!:\! B_1[\overrightarrow{V_i}/\overrightarrow{x_i}], y_2 \!:\! B_2[\overrightarrow{V_i}/\overrightarrow{x_i}])} {} {\efftrans T {\!\!A; \overrightarrow{V_i}, y_1, y_2; \overrightarrow{V'_{\!j}}; \overrightarrow{W_{\sigalgop}}}}}_2)_{\gamma\,'}
\]
which, by unfolding the definition of $\sem -$ for pattern-matching and assuming that
\[
(\sem{\Gamma';V[\overrightarrow{V_i}/\overrightarrow{x_i}]}_2)_{\gamma\,'}(\star) = (\sem{\Gamma;V})_\gamma (\star) = \langle b_1 , b_2 \rangle
\]
is equal to
\[
(\sem{\Gamma'\!, y_1 \!:\! B_1[\overrightarrow{V_i}/\overrightarrow{x_i}], y_2 \!:\! B_2[\overrightarrow{V_i}/\overrightarrow{x_i}]; \efftrans T {\!\!A; \overrightarrow{V_i}, y_1, y_2; \overrightarrow{V'_{\!j}}; \overrightarrow{W_{\sigalgop}}}}_2)_{\langle \langle \gamma\,' \! ,\, b_1 \rangle ,\, b_2 \rangle}
\]

The required equation then follows from the commutativity of the following diagram:
\[
\scriptsize
\xymatrix@C=3em@R=4.5em@M=0.5em{
1
\ar[rr]^-{\langle (\sem{\Gamma'; V'_{\!j}}_2)_{\gamma\,'} \rangle_{w_{\!j} : A'_{\!j} \in \Delta}}
\ar[dddddd]^<<<<<<<<<<<<<<<<<<<<<<<<<<<<<<<<<<<<<<<<<<<<<{(\sem{\Gamma'\!, y_1 , y_2 ; \efftrans T {\!\!A; \overrightarrow{V_i}, y_1, y_2; \overrightarrow{V'_{\!j}}; \overrightarrow{W_{\sigalgop}}}}_2)_{\langle \langle \gamma\,' \! ,\, b_1 \rangle ,\, b_2 \rangle}}
\ar@/^2pc/[ddr]^<<<<<<<<<<<<<<<<<<<<<<{\!\!\!\!\langle (\sem{\Gamma'\!, y_1, y_2; V'_{\!j}}_2)_{\langle \langle \gamma\,' \! ,\, b_1 \rangle ,\, b_2\rangle} \rangle_{w_{\!j} : A'_{\!j} \in \Delta}}^<<<<<<<<<<<<<<{\quad\qquad\qquad\qquad\qquad\qquad\dscomment{\text{weakening}}}
&
&
\bigsqcap_{w_{\!j}} \bigsqcap_{a \in \sem{\Gamma;A'_{\!j}}_2(\gamma)}
(\sem{\Gamma';A}_2(\gamma\,'))
\ar[d]^-{\cong}
\\
&
&
\mathcal{M}_{\gamma\,'}(\vert \Delta^\gamma \vert)
\ar@/^1pc/[dddd]_>>>>>>>{\mathcal{M}_{\gamma\,'}(\lj {\Delta^\gamma\,\,} {\,T^{\langle \langle \gamma ,\, b_1 \rangle ,\, b_2 \rangle}})}
\\
&
\bigsqcap_{w_{\!j}} \bigsqcap_{a \in \sem{\Gamma;A'_{\!j}}_2(\gamma)}
(\sem{\Gamma'\!, y_1, y_2;A}_2\, \langle \langle \gamma\,' \!, b_1 \rangle , b_2\rangle)
\ar@/^2pc/[uur]_-{=}_<<<<<<<<<<<<{\qquad\qquad\dscomment{\text{weakening}}}
\ar[d]^-{\cong}
&
\\
&
\mathcal{M}_{\langle \langle \gamma\,' \!, b_1 \rangle , b_2 \rangle}(\vert \Delta^{\langle \langle \gamma ,\, b_1 \rangle ,\, b_2 \rangle} \vert)
\ar@/_2.5pc/[uur]_-{=}
\ar[dd]^-{\mathcal{M}_{\langle \langle \gamma\,' \!, b_1 \rangle , b_2 \rangle}(\lj {\Delta^{\langle \langle \gamma,\, b_1 \rangle ,\, b_2 \rangle}\,\,} {\,T^{\langle \langle \gamma,\, b_1 \rangle ,\, b_2 \rangle}})}_-{\dscomment{\text{induction hypothesis}}\qquad\qquad\qquad}^<<<<{\quad\qquad\qquad\qquad\qquad\dscomment{\text{weakening}}}
\\
&
&
\\
&
\mathcal{M}_{\langle \langle \gamma\,' , b_1 \rangle , b_2 \rangle}(1)
\ar[r]_-{=}
\ar[dl]_-{=}
&
\mathcal{M}_{\gamma\,'}(1)
\ar[d]^-{=}_-{\dscomment{\text{weakening}}\qquad\qquad\qquad\qquad\qquad}
\\
\sem{\Gamma'\!, y_1, y_2;A}_2\, \langle \langle \gamma\,' \! , b_1 \rangle , b_2 \rangle
\ar[rr]_-{=}
&&
\sem{\Gamma';A}_2(\gamma\,')
}
\]

Similarly to the earlier case for algebraic operations, we have again abbreviated some of the value contexts and indices of countable products for better readability.

\vspace{0.2cm}

\noindent \textbf{Case analysis:} In this case, the given derivation ends with 
\[
\mkrule
{\lj {\Gamma \vertbar \Delta} {\mathsf{case~} V \mathsf{~of}_{} \mathsf{~} ({\seminl {\!} {\!\!(y_1 \!:\! B_1)} \mapsto T_1}, {\seminr {\!} {\!\!(y_2 \!:\! B_2)} \mapsto T_2})}}
{
\lj \Gamma \Delta 
\quad
\vj \Gamma V B_1 + B_2
\quad
\lj {\Gamma, y_1 \!:\! B_1 \vertbar \Delta} T_1
\quad
\lj {\Gamma, y_2 \!:\! B_2 \vertbar \Delta} T_2
}
\]

First, according to the definition of $\efftrans - {}$ for case analysis, we have that
\[
(\sem{\Gamma'; \efftrans {\mathsf{case~} V \mathsf{~of}_{} \mathsf{~} ({\seminl {\!} {\!\!(y_1 \!:\! B_1)} \mapsto T_1}, {\seminr {\!} {\!\!(y_2 \!:\! B_2)} \mapsto T_2})} {\!\!A; \overrightarrow{V_i}; \overrightarrow{V'_{\!j}}; \overrightarrow{W_{\sigalgop}}}}_2)_{\gamma\,'}
\]
is equal to 
\[
(\sem{\Gamma'; \mathtt{case~} V[\overrightarrow{V_i}/\overrightarrow{x_i}] \mathtt{~of}_{} \mathtt{~} ({\inl {\!} {\!\!(y_1 \!:\! B_1[\overrightarrow{V_i}/\overrightarrow{x_i}])} \mapsto \efftrans {T_1} {}}, {\inr {\!} {\!\!(y_2 \!:\! B_2[\overrightarrow{V_i}/\overrightarrow{x_i}])} \mapsto \efftrans {T_2} {}})}_2)_{\gamma\,'}
\]

\pagebreak
\noindent
which, by unfolding the definition of $\sem -$ for case analysis, is either equal to 
\[
(\sem{\Gamma'\!, y_1 \!:\! B_1[\overrightarrow{V_i}/\overrightarrow{x_i}]; \efftrans {T_1} {\!\!A; \overrightarrow{V_i}, y_1; \overrightarrow{V'_{\!j}}; \overrightarrow{W_{\sigalgop}}}}_2)_{\langle \gamma\,' \! ,\, b_1 \rangle}
\]
or 
\[
(\sem{\Gamma'\!, y_2 \!:\! B_2[\overrightarrow{V_i}/\overrightarrow{x_i}]; \efftrans {T_2} {\!\!A; \overrightarrow{V_i}, y_2; \overrightarrow{V'_{\!j}}; \overrightarrow{W_{\sigalgop}}}}_2)_{\langle \gamma\,' \! ,\, b_2 \rangle}
\]
depending on whether 
\[
(\sem{\Gamma';V[\overrightarrow{V_i}/\overrightarrow{x_i}]}_2)_{\gamma\,'}(\star) = (\sem{\Gamma;V})_\gamma (\star) = \seminl {} {b_1}
\]
or
\[
(\sem{\Gamma';V[\overrightarrow{V_i}/\overrightarrow{x_i}]}_2)_{\gamma\,'}(\star) = (\sem{\Gamma;V})_\gamma (\star) = \seminr {} {b_2}
\vspace{0.1cm}
\]

The required equation for the $\seminl {} {b_1}$ case then follows from the commutativity of the next diagram, where, similarly to the case for algebraic operations, we again abbreviate some of the value contexts and indices of countable products for better readability.
\[
\scriptsize
\xymatrix@C=4em@R=4.5em@M=0.5em{
1
\ar[rr]^-{\langle (\sem{\Gamma'; V'_{\!j}}_2)_{\gamma\,'} \rangle_{w_{\!j} : A'_{\!j} \in \Delta}}
\ar[dddddd]^<<<<<<<<<<<<<<<<<<<<<<<<<<<<<<<<<<<<<<<<<<<<<<{(\sem{\Gamma'\!, y_1; \efftrans T {\!\!A; \overrightarrow{V_i}, y_1; \overrightarrow{V'_{\!j}}; \overrightarrow{W_{\sigalgop}}}}_2)_{\langle \gamma\,' \! ,\, b_1 \rangle}}
\ar@/^2pc/[ddr]^<<<<<<<<<<<<<<<<<<<<<<{\!\!\!\!\!\!\!\!\!\!\!\langle (\sem{\Gamma'\!, y_1; V'_{\!j}}_2)_{\langle \gamma\,' \! ,\, b_1 \rangle} \rangle_{w_{\!j} : A'_{\!j} \in \Delta}}^<<<<<<<<<<<<<{\qquad\qquad\qquad\qquad\dscomment{\text{weakening}}}
&
&
\bigsqcap_{w_{\!j}} \bigsqcap_{a \in \sem{\Gamma;A'_{\!j}}_2(\gamma)}
(\sem{\Gamma';A}_2(\gamma\,'))
\ar[d]^-{\cong}
\\
&
&
\mathcal{M}_{\gamma\,'}(\vert \Delta^\gamma \vert)
\ar@/^1pc/[dddd]_>>>>>>>>{\mathcal{M}_{\gamma\,'}(\lj {\Delta^\gamma\,\,} {\,T^{\langle  \gamma ,\, b_1 \rangle}})}
\\
&
\bigsqcap_{w_{\!j}} \bigsqcap_{a \in \sem{\Gamma;A'_{\!j}}_2(\gamma)}
(\sem{\Gamma'\!, y_1;A}_2\, \langle \gamma\,' \!, b_1 \rangle)
\ar@/^2pc/[uur]_-{=}_<<<<<<<<{\qquad\qquad\dscomment{\text{weakening}}}
\ar[d]^-{\cong}
&
\\
&
\mathcal{M}_{\langle \gamma\,' \!, b_1 \rangle}(\vert \Delta^{\langle \gamma ,\, b_1 \rangle} \vert)
\ar@/_2.5pc/[uur]_-{=}
\ar[dd]^-{\mathcal{M}_{\langle \gamma\,' \!, b_1 \rangle}(\lj {\Delta^{\langle  \gamma,\, b_1 \rangle}\,\,} {\,T^{\langle \gamma,\, b_1 \rangle}})}_-{\dscomment{\text{induction hypothesis}}\qquad\qquad}^<<<<{\qquad\qquad\qquad\qquad\dscomment{\text{weakening}}}
\\
&
&
\\
&
\mathcal{M}_{\langle \gamma\,' , b_1 \rangle}(1)
\ar[r]_-{=}
\ar[dl]_-{=}
&
\mathcal{M}_{\gamma\,'}(1)
\ar[d]^-{=}_-{\dscomment{\text{weakening}}\qquad\qquad\qquad\qquad\qquad}
\\
\sem{\Gamma'\!, y_1;A}_2\, \langle \gamma\,' \! , b_1 \rangle
\ar[rr]_-{=}
&&
\sem{\Gamma';A}_2(\gamma\,')
}
\vspace{0.25cm}
\]

We omit the proof of the other case (for $\seminr {} {b_2}$) because it is proved analogously.
\end{proof}

\renewcommand\thesection{\thechapter.\arabic{section}}

\bibliographystyle{abbrv}


\bibliography{references}

\begin{thebibliography}{100}

\bibitem{CompCert}
{The CompCert project. Website:} \url{http://compcert.inria.fr/}.

\bibitem{Everest:Project}
{The Everest project. Website:} \url{https://project-everest.github.io/}.

\bibitem{MulticoreOCaml}
{The multicore OCaml project. Website:}
  \url{https://github.com/ocamllabs/ocaml-multicore/}.

\bibitem{Abbott:Containers}
M.~G. Abbott, T.~Altenkirch, and N.~Ghani.
\newblock Categories of containers.
\newblock In A.~D. Gordon, editor, {\em Proc. of 6th Int. Conf. on Foundations
  of Software Science and Computational Structures, FOSSACS 2003}, volume 2620
  of {\em LNCS}, pages 23--38. Springer, 2003.

\bibitem{Saleh:OpSemantics}
F.~Abou{-}Saleh and D.~Pattinson.
\newblock Comodels and effects in mathematical operational semantics.
\newblock In F.~Pfenning, editor, {\em Proc. of 16th Int. Conf. on Foundations
  of Software Science and Computational Structures, FoSSaCS 2013}, volume 7794
  of {\em LNCS}, pages 129--144. Springer, 2013.

\bibitem{Abramsky:DomainTheory}
S.~Abramsky and A.~Jung.
\newblock Domain theory.
\newblock In S.~Abramsky, D.~M. Gabbay, and T.~S.~E. Maibaum, editors, {\em
  Handbook of Logic in Computer Science}, volume~3, pages 1--168. Clarendon
  Press, 1994.

\bibitem{Adamek:LocallyPresentableCats}
J.~Adamek and J.~Rosicky.
\newblock {\em Locally Presentable and Accessible Categories}.
\newblock Number 189 in London Mathematical Society Lecture Note Series.
  Cambridge Univ. Press, 1994.

\bibitem{Ahman:Dcontainers}
D.~Ahman, J.~Chapman, and T.~Uustalu.
\newblock When is a container a comonad?
\newblock {\em Logical Methods in Computer Science}, 10(3), 2014.

\bibitem{Ahman:FibredEffects}
D.~Ahman, N.~Ghani, and G.~D. Plotkin.
\newblock Dependent types and fibred computational effects.
\newblock In B.~Jacobs and C.~L\"{o}ding, editors, {\em Proc. of 19th Int.
  Conf. on Foundations of Software Science and Computation Structures, FoSSaCS
  2016}, volume 9634 of {\em LNCS}, pages 1--19. Springer, 2016.

\bibitem{Ahman:DM4Free}
D.~Ahman, C.~Hri{\c t}cu, K.~Maillard, G.~Mart\'inez, G.~Plotkin, J.~Protzenko,
  A.~Rastogi, and N.~Swamy.
\newblock Dijkstra monads for free.
\newblock In A.~D. Gordon, editor, {\em Proc. of 44th {ACM} {SIGPLAN} Symp. on
  Principles of Programming Languages, POPL 2017}, pages 515--529. ACM, 2017.

\bibitem{Ahman:RefTypes}
D.~Ahman and G.~D. Plotkin.
\newblock Refinement types for algebraic effects (extended abstract).
\newblock In T.~Uustalu, editor, {\em Book of abstracts of the 21th Meeting
  "Types for Proofs and Programs", TYPES 2015}, pages 10--11. Inst. of
  Cybernetics at TUT, 2015.

\bibitem{Ahman:NBE}
D.~Ahman and S.~Staton.
\newblock Normalization by evaluation and algebraic effects.
\newblock In D.~Kozen and M.~Mislove, editors, {\em Proc. of 29th Conf. on the
  Mathematical Foundations of Programming Semantics, MFPS XXIX}, volume 298 of
  {\em ENTCS}, pages 51--69. Elsevier, 2013.

\bibitem{Ahman:UpdateMonads}
D.~Ahman and T.~Uustalu.
\newblock Update monads: Cointerpreting directed containers.
\newblock In R.~Matthes and A.~Schubert, editors, {\em Post-proc. of the 19th
  Meeting ``Types for Proofs and Programs", TYPES 2013}, volume~26 of {\em
  LIPIcs}, pages 1--23. Schloss Dagstuhl -- Leibniz-Zentrum f{\"u}r Informatik,
  Dagstuhl Publishing, 2014.

\bibitem{Altenkirch:RelMon2}
T.~Altenkirch, J.~Chapman, and T.~Uustalu.
\newblock Monads need not be endofunctors.
\newblock {\em Logical Methods in Computer Science}, 11(1), 2015.

\bibitem{Altenkirch:NBEforTT}
T.~Altenkirch and A.~Kaposi.
\newblock Normalisation by evaluation for dependent types.
\newblock In D.~Kesner and B.~Pientka, editors, {\em Proc. of 1st Int. Conf. on
  Formal Structures for Computation and Deduction, FSCD 2016}, volume~52 of
  {\em LIPIcs}, pages 6:1--6:16. Schloss Dagstuhl - Leibniz-Zentrum fuer
  Informatik, 2016.

\bibitem{Atkey:Algebras}
R.~Atkey.
\newblock Algebras for parameterised monads.
\newblock In A.~Kurz, M.~Lenisa, and A.~Tarlecki, editors, {\em Proc. of 3rd
  Int. Conf. on Algebra and Coalgebra in Computer Science, CALCO 2009}, volume
  5728 of {\em LNCS}, pages 3--17. Springer, 2009.

\bibitem{Atkey:ParametrizedNotions}
R.~Atkey.
\newblock Parameterised notions of computation.
\newblock {\em J. Funct. Program.}, 19(3 \& 4):335--376, 2009.

\bibitem{Atkey:DepTypes}
R.~Atkey, N.~Ghani, and P.~Johann.
\newblock A relationally parametric model of dependent type theory.
\newblock In S.~Jagannathan and P.~Sewell, editors, {\em Proc. of 41st Ann.
  {ACM} {SIGPLAN-SIGACT} Symp. on Principles of Programming Languages, {POPL}
  2014}, pages 503--516. ACM, 2014.

\bibitem{Balat:NBE}
V.~Balat, R.~D. Cosmo, and M.~P. Fiore.
\newblock Extensional normalisation and type-directed partial evaluation for
  typed lambda calculus with sums.
\newblock In N.~D. Jones and X.~Leroy, editors, {\em Proc. of 31st {ACM}
  {SIGPLAN-SIGACT} Symp. on Principles of Programming Languages, POPL 2004},
  pages 64--76. ACM, 2004.

\bibitem{Barr:Toposes}
M.~Barr and C.~Wells.
\newblock {\em Toposes, Triples and Theories}.
\newblock Number 278 in Grundlehren der mathematischen Wissenschaften.
  Springer-Verlag, 1985.

\bibitem{Barr:CatThForCS}
M.~Barr and C.~Wells.
\newblock {\em Category theory for computing science}.
\newblock Prentice Hall International Series in Computer Science. Prentice
  Hall, 1990.

\bibitem{Bauer:EffectSystem}
A.~Bauer and M.~Pretnar.
\newblock An effect system for algebraic effects and handlers.
\newblock In R.~Heckel and S.~Milius, editors, {\em Proc. of 5th Int. Conf. on
  Algebra and Coalgebra in Computer Science, CALCO 2013}, volume 8089 of {\em
  LNCS}, pages 1--16. Springer, 2013.

\bibitem{Bauer:AlgebraicEffects}
A.~Bauer and M.~Pretnar.
\newblock Programming with algebraic effects and handlers.
\newblock {\em J. Log. Algebr. Meth. Program.}, 84(1):108--123, 2015.

\bibitem{Bezem:Cubical}
M.~Bezem, T.~Coquand, and S.~Huber.
\newblock A model of type theory in cubical sets.
\newblock In R.~Matthes and A.~Schubert, editors, {\em Post-proc. of the 19th
  Meeting ``Types for Proofs and Programs", TYPES 2013}, volume~26 of {\em
  LIPIcs}, pages 107--128. Schloss Dagstuhl - Leibniz-Zentrum fuer Informatik,
  2013.

\bibitem{Borceux:HandbookVol2}
F.~Borceux.
\newblock {\em Handbook of Categorical Algebra, vol. 2: Categories and
  Structures}.
\newblock Cambridge Univ. Press, 1994.

\bibitem{Brady:Idris}
E.~Brady.
\newblock Idris, a general-purpose dependently typed programming language:
  Design and implementation.
\newblock {\em J. Funct. Program.}, 23(5):552--593, 2013.

\bibitem{Brady:Effects}
E.~Brady.
\newblock Programming and reasoning with algebraic effects and dependent types.
\newblock In G.~Morrisett and T.~Uustalu, editors, {\em Proc. of 18th {ACM}
  {SIGPLAN} Int. Conf. on Functional Programming, ICFP 2013}, pages 133--144.
  ACM, 2013.

\bibitem{Brady:ResourceDependentEffects}
E.~Brady.
\newblock Resource-dependent algebraic effects.
\newblock In J.~Hage, editor, {\em Proc. of 15th Symp. on Trends in Functional
  Programming, TFP 2014}, volume 8843 of {\em LNCS}, pages 18--33. Springer,
  2015.

\bibitem{Casinghino:CombiningProofs}
C.~Casinghino, V.~Sj\"{o}bergberg, and S.~Weirich.
\newblock Combining proofs and programs in a dependently typed language.
\newblock In S.~Jagannathan and P.~Sewell, editors, {\em Proc. of 41st Ann.
  {ACM} {SIGPLAN-SIGACT} Symp. on Principles of Programming Languages, POPL
  2014}, pages 33--45. ACM, 2014.

\bibitem{Castellan:Report}
S.~Castellan.
\newblock Dependent type theory as the initial category with families.
\newblock Technical report, Chalmers University of Technology, 2014.

\bibitem{Cenciarelli:Modularity}
P.~Cenciarelli and E.~Moggi.
\newblock A syntactic approach to modularity in denotational semantics.
\newblock In {\em Proc. of 5th. Biennial Meeting on Category Theory and
  Computer Science, CTCS 1993}. CWI Technical report, 1993.

\bibitem{Cheney:DepNomTypeTheory}
J.~Cheney.
\newblock A dependent nominal type theory.
\newblock {\em Logical Methods in Computer Science}, 8(1), 2012.

\bibitem{Dijkstra:GCommands}
E.~W. Dijkstra.
\newblock Guarded commands, nondeterminacy and formal derivation of programs.
\newblock {\em Commun. {ACM}}, 18(8):453--457, 1975.

\bibitem{Dybjer:NBE}
P.~Dybjer and A.~Filinski.
\newblock Normalization and partial evaluation.
\newblock In G.~Barthe, P.~Dybjer, L.~Pinto, and J.~Saraiva, editors, {\em
  Proc. of Int. Summer School on Applied Semantics, APPSEM 2000}, volume 2395
  of {\em LNCS}, pages 137--192. Springer, 2002.

\bibitem{Egger:EnrichedEffectCalculus}
J.~Egger, R.~E. M{\o}gelberg, and A.~Simpson.
\newblock The enriched effect calculus: syntax and semantics.
\newblock {\em J. Log. Comput.}, 24(3):615--654, 2014.

\bibitem{Fujii:GradedMonads}
S.~Fujii, S.~Katsumata, and P.~Melli{\`{e}}s.
\newblock Towards a formal theory of graded monads.
\newblock In B.~Jacobs and C.~L\"{o}ding, editors, {\em Proc. of 19th Int.
  Conf. on Foundations of Software Science and Computation Structures, FoSSaCS
  2016}, volume 9634 of {\em LNCS}, pages 513--530. Springer, 2016.

\bibitem{gautam:validity}
N.~D. Gautam.
\newblock The validity of equations of complex algebras.
\newblock {\em Arch. Math. Logic}, 3(3-4), 1957.

\bibitem{Ghani:FibredInduction}
N.~Ghani, P.~Johann, and C.~Fumex.
\newblock Indexed induction and coinduction, fibrationally.
\newblock {\em Logical Methods in Computer Science}, 9(3), 2013.

\bibitem{Gierz:ContinuousLattices}
G.~Gierz, K.~H. Hofmann, K.~Keimel, J.~D. Lawson, M.~Mislove, and D.~S. Scott.
\newblock {\em Continuous Lattices and Domains}.
\newblock Number~93 in Encyclopedia of Mathematics and its Applications.
  Cambridge Univ. Press, 2003.

\bibitem{Gratzer:UniversalAlgebra}
G.~A. Gr{\"a}tzer.
\newblock {\em Universal Algebra}.
\newblock Springer, 2nd edition, 1979.

\bibitem{Hancock:InteractivePrograms}
P.~Hancock and A.~Setzer.
\newblock Interactive programs in dependent type theory.
\newblock In P.~Clote and H.~Schwichtenberg, editors, {\em Proc. of 14th Ann.
  Conf. of the EACSL on Computer Science Logic, CSL 2000}, volume 1862 of {\em
  LNCS}, pages 317--331. Springer, 2000.

\bibitem{Hermida:fibinduction}
C.~Hermida and B.~Jacobs.
\newblock Structural induction and coinduction in a fibrational setting.
\newblock {\em Inf. Comput.}, 145(2):107 -- 152, 1998.

\bibitem{Hillerstrom:Liberating}
D.~Hillerstr\"{o}m and S.~Lindley.
\newblock Liberating effects with rows and handlers.
\newblock In J.~Chapman and W.~Swierstra, editors, {\em Proc. of 1st Wksh. on
  Type-Driven Development, TyDe 2016}, pages 15--27. ACM, 2016.

\bibitem{Hofmann:Thesis}
M.~Hofmann.
\newblock {\em Extensional concepts in intensional type theory}.
\newblock PhD thesis, Laboratory for Foundations in Computer Science,
  University of Edinburgh, 1995.

\bibitem{Hofmann:SyntaxAndSemantics}
M.~Hofmann.
\newblock Syntax and semantics of dependent types.
\newblock In A.~M. Pitts and P.~Dybjer, editors, {\em Semantics and Logics of
  Computation}, pages 79--130. Cambridge Univ. Press, 1997.

\bibitem{Honda:LangPrimitives}
K.~Honda, V.~T. Vasconcelos, and M.~Kubo.
\newblock Language primitives and type discipline for structured
  communication-based programming.
\newblock In C.~Hankin, editor, {\em Proc. of 7th European Symp. on
  Programming, ESOP 1998}, volume 1381 of {\em LNCS}, pages 122--138. Springer,
  1998.

\bibitem{Hutton:MonParsing}
G.~Hutton and E.~Meijer.
\newblock Monadic parsing in {Haskell}.
\newblock {\em J. Funct. Program.}, 8(4):437--444, 1998.

\bibitem{Hyland:Continuations}
M.~Hyland, P.~B. Levy, G.~Plotkin, and J.~Power.
\newblock Combining algebraic effects with continuations.
\newblock {\em Theor. Comput. Sci.}, 375(1-3):20--40, 2007.

\bibitem{Hyland:CombiningEffects}
M.~Hyland, G.~Plotkin, and J.~Power.
\newblock Combining effects: Sum and tensor.
\newblock {\em Theor. Comput. Sci.}, 357(1--3):70--99, 2006.

\bibitem{Hyland:DiscreteLawTh}
M.~Hyland and J.~Power.
\newblock Discrete {Lawvere} theories and computational effects.
\newblock {\em Theor. Comput. Sci.}, 366(1-2):144--162, 2006.

\bibitem{Jacobs:Book}
B.~Jacobs.
\newblock {\em Categorical Logic and Type Theory}.
\newblock Number 141 in Studies in Logic and the Foundations of Mathematics.
  North Holland, 1999.

\bibitem{Kammar:Thesis}
O.~Kammar.
\newblock {\em An Algebraic Theory of Type-and-Effect Systems}.
\newblock PhD thesis, School of Informatics, University of Edinburgh, 2014.

\bibitem{Kammar:Handlers}
O.~Kammar, S.~Lindley, and N.~Oury.
\newblock Handlers in action.
\newblock In G.~Morrisett and T.~Uustalu, editors, {\em Proc. of 18th {ACM}
  {SIGPLAN} Int. Conf. on Functional Programming, ICFP 2013}, pages 145--158.
  ACM, 2013.

\bibitem{Kammar:AlgebraicFoundations}
O.~Kammar and G.~D. Plotkin.
\newblock Algebraic foundations for effect-dependent optimisations.
\newblock In M.~Hicks, editor, {\em Proc. of 39th {ACM} {SIGPLAN-SIGACT} Symp.
  on Principles of Programming Languages, POPL 2012}, pages 349--360. ACM,
  2012.

\bibitem{Katsumata:EffectMonads}
S.~Katsumata.
\newblock Parametric effect monads and semantics of effect systems.
\newblock In S.~Jagannathan and P.~Sewell, editors, {\em Proc. of 41st Ann.
  {ACM} {SIGPLAN-SIGACT} Symp. on Principles of Programming Languages, {POPL}
  2014}, pages 633--646. ACM, 2014.

\bibitem{Kelly:EnrichedCats}
G.~Kelly.
\newblock {\em Basic Concepts of Enriched Category Theory}.
\newblock Number~64 in Lecture Notes in Mathematics. Cambridge Univ. Press,
  1982.

\bibitem{Kernighan:CLanguage}
B.~W. Kernighan and D.~Ritchie.
\newblock {\em The C Programming Language}.
\newblock Prentice Hall Software Series, 2nd edition, 1988.

\bibitem{Kock:StrongMonads}
A.~Kock.
\newblock Strong functors and monoidal monads.
\newblock {\em Arch. Math.}, 23:113--120, 1972.

\bibitem{Krishnaswami:LinearDependentTypes}
N.~R. Krishnaswami, P.~Pradic, and N.~Benton.
\newblock Integrating linear and dependent types.
\newblock In D.~Walker, editor, {\em Proc. of 42nd Ann. {ACM} {SIGPLAN-SIGACT}
  Symp. on Principles of Programming Languages, POPL 2015}, pages 17--30. ACM,
  2015.

\bibitem{Leijen:Handlers}
D.~Leijen.
\newblock Type directed compilation of row-typed algebraic effects.
\newblock In A.~D. Gordon, editor, {\em Proc. of 44th {ACM} {SIGPLAN} Symp. on
  Principles of Programming Languages, POPL 2017}, pages 486--499. ACM, 2017.

\bibitem{Levy:CBPV}
P.~B. Levy.
\newblock {\em Call-By-Push-Value: A Functional/Imperative Synthesis}, volume~2
  of {\em Semantics Structures in Computation}.
\newblock Springer, 2004.

\bibitem{Levy:MonadsForExceptions}
P.~B. Levy.
\newblock Monads and adjunctions for global exceptions.
\newblock In S.~Brookes and M.~Mislove, editors, {\em Proc. of 22nd Conf. on
  Mathematical Foundations of Programming Semantics, MFPS XXII}, volume 158 of
  {\em ENTCS}, pages 261--287. Elsevier, 2006.

\bibitem{Levy:ContextualIsomorphisms}
P.~B. Levy.
\newblock Contextual isomorphisms.
\newblock In A.~D. Gordon, editor, {\em Proc. of 44th {ACM} {SIGPLAN} Symp. on
  Principles of Programming Languages, POPL 2017}, pages 400--414. ACM, 2017.

\bibitem{Lindley:DoBeDoBeDo}
S.~Lindley, C.~McBride, and C.~McLaughlin.
\newblock Do be do be do.
\newblock In A.~D. Gordon, editor, {\em Proc. of 44th {ACM} {SIGPLAN} Symp. on
  Principles of Programming Languages, POPL 2017}, pages 500--514. ACM, 2017.

\bibitem{Linton:Coequalizers}
F.~Linton.
\newblock Coequalizers in categories of algebras.
\newblock In B.~Eckmann, editor, {\em Proc. of Seminar on Triples and
  Categorical Homology Theory}, volume~80 of {\em LNCS}, pages 75--90.
  Springer, 1969.

\bibitem{MacLane:CatWM}
S.~Mac~Lane.
\newblock {\em Categories for the Working Mathematician}.
\newblock Number~5 in Graduate Texts in Mathematics. Springer-Verlag, 1971.

\bibitem{Manes:AlgTheories}
E.~G. Manes.
\newblock {\em Algebraic Theories}, volume~26 of {\em Graduate Texts in
  Mathematics}.
\newblock Springer-Verlag, 1976.

\bibitem{Marlow:HaskellReport}
S.~Marlow.
\newblock {Haskell 2010 Language Report}.
\newblock April 2010.

\bibitem{MartinLof:IntuitionisticTT}
P.~Martin-L{\"o}f.
\newblock An intuitionisitc theory of types, predicative part.
\newblock In E.~Rose and S.~J.C., editors, {\em Proc. of Logic Colloquium
  1973}, pages 73--118. North-Holland, 1975.

\bibitem{MartinLof:Bibliopolis}
P.~Martin-L\"{o}f.
\newblock {\em Intuitionistic Type Theory}.
\newblock Bibliopolis, 1984.

\bibitem{Coq:Manual}
\mbox{The {Coq} Development Team}.
\newblock {\em The {Coq} Proof Assistant Reference Manual}.
\newblock $\pi r^2$ Project, Version 8.5pl3, October 26, 2016.

\bibitem{McBride:Kleisli}
C.~McBride.
\newblock Functional pearl: Kleisli arrows of outrageous fortune.
\newblock {\em J. Funct. Program.}
\newblock (To appear).

\bibitem{ML:Standard}
R.~Milner, M.~Tofte, R.~Harper, and D.~MacQueen.
\newblock {\em The Definition of Standard ML (Revised)}.
\newblock The MIT Press, 1997.

\bibitem{Moggi:ComputationalLambdaCalculus}
E.~Moggi.
\newblock Computational lambda-calculus and monads.
\newblock In R.~Parikh, editor, {\em Proc. of 4th Ann. Symp. on Logic in
  Computer Science, LICS 1989}, pages 14--23. IEEE, 1989.

\bibitem{Moggi:NotionsofComputationandMonads}
E.~Moggi.
\newblock Notions of computation and monads.
\newblock {\em Inf. Comput.}, 93(1):55--92, 1991.

\bibitem{Munch:Thesis}
G.~Munch-Maccagnoni.
\newblock {\em {S}yntax and {M}odels of a non-{A}ssociative {C}omposition of
  {P}rograms and {P}roofs}.
\newblock PhD thesis, Univ. Paris Diderot, 2013.

\bibitem{Nanevski:HTT}
A.~Nanevski, G.~Morrisett, and L.~Birkedal.
\newblock {Hoare Type Theory}, polymorphism and separation.
\newblock {\em J. Funct. Program.}, 18(5-6):865--911, 2008.

\bibitem{Norell:Thesis}
U.~Norell.
\newblock {\em Towards a practical programming language based on dependent type
  theory}.
\newblock PhD thesis, Department of Computer Science and Engineering, Chalmers
  University of Technology, 2007.

\bibitem{Norell:AgdaTutorial}
U.~Norell and J.~Chapman.
\newblock Dependently typed programming in agda.
\newblock Tutorial.

\bibitem{Okada:Rewriting}
M.~Okada and P.~J. Scott.
\newblock A note on rewriting theory for uniqueness of iteration.
\newblock {\em Theory Appl. Categ.}, 6(4):47--64, 1999.

\bibitem{Palmgren:Universes}
E.~Palmgren.
\newblock On universes in type theory.
\newblock In G.~Sambin and J.~M. Smith, editors, {\em Twenty-five Years of
  Constructive Type Theory: Proc. of Congress Held in Venice, October 1995},
  volume~36 of {\em Oxford Logic Guides}, pages 191--204. Clarendon Press,
  Oxford, 1998.

\bibitem{Palmgren:DomainInterpretation}
E.~Palmgren and V.~Stoltenberg-Hansen.
\newblock Domain interpretations of {Martin-L\"{o}f's} partial type theory.
\newblock {\em Ann. Pure Appl. Logic}, 48(2):135 -- 196, 1990.

\bibitem{Petersen:HTT}
R.~L. Petersen, L.~Birkedal, A.~Nanevski, and G.~Morrisett.
\newblock A realizability model for impredicative {Hoare Type Theory}.
\newblock In S.~Drossopoulou, editor, {\em Proc. of 17th European Symp. on
  Programming, ESOP 2008}, pages 337--352. Springer-Verlag, 2008.

\bibitem{PittsAM:evall}
A.~M. Pitts.
\newblock Evaluation logic.
\newblock In G.~Birtwistle, editor, {\em IVth Higher Order Workshop, {Banff}
  1990}, Workshops in Computing, pages 162--189. Springer-Verlag, Berlin, 1991.

\bibitem{Pitts:CategoricalLogic}
A.~M. Pitts.
\newblock {Categorical Logic}.
\newblock In S.~Abramsky, D.~M. Gabbay, and T.~S.~E. Maibaum, editors, {\em
  Handbook of Logic in Computer Science, Volume 5. Algebraic and Logical
  Structures}, pages 39--128. Oxford University Press, 2000.

\bibitem{Pitts:NominalMLTT}
A.~M. Pitts, J.~Matthiesen, and J.~Derikx.
\newblock A dependent type theory with abstractable names.
\newblock In I.~Mackie and M.~Ayala-Rincon, editors, {\em Proc. of 9th Wksh. on
  Logical and Semantic Frameworks, with Applications, LSFA 2014}, volume 312 of
  {\em ENTCS}, pages 19--50. Elsevier, 2015.

\bibitem{Plotkin:PisaNotes}
G.~Plotkin.
\newblock Pisa notes (on domain theory).
\newblock Available at:
  \url{http://homepages.inf.ed.ac.uk/gdp/publications/Domains_a4.ps}, 1983.

\bibitem{Plotkin:SemanticsForAlgOperations}
G.~Plotkin and J.~Power.
\newblock Semantics for algebraic operations.
\newblock In S.~Brookes and M.~Mislove, editors, {\em Proc. of 17th Conf. on
  the Mathematical Foundations of Programming Semantics, MFPS XVII}, volume~45
  of {\em ENTCS}, pages 332--345. Elsevier, 2001.

\bibitem{Plotkin:AlgOperations}
G.~Plotkin and J.~Power.
\newblock Algebraic operations and generic effects.
\newblock {\em Appl. Categor. Struct.}, 11(1):69--94, 2003.

\bibitem{Plotkin:TensorsOfModels}
G.~Plotkin and J.~Power.
\newblock Tensors of comodels and models for operational semantics.
\newblock In A.~Bauer and M.~Mislove, editors, {\em Proc. of 24th Conf. on
  Mathematical Foundations of Programming Semantics, MFPS XXIV}, volume 218 of
  {\em ENTCS}, pages 295--311. Elsevier, 2008.

\bibitem{Plotkin:BinaryHandlers}
G.~D. Plotkin.
\newblock Concurrency and the algebraic theory of effects.
\newblock Invited talk at CONCUR 2012.

\bibitem{Plotkin:NotionsOfComputation}
G.~D. Plotkin and J.~Power.
\newblock Notions of computation determine monads.
\newblock In M.~Nielsen, editor, {\em Proc. of 5th Int. Conf. on Foundations of
  Software Science and Computation Structures, FOSSACS 2002}, volume 2303 of
  {\em LNCS}, pages 342--356. Springer, 2002.

\bibitem{Plotkin:Logic}
G.~D. Plotkin and M.~Pretnar.
\newblock A logic for algebraic effects.
\newblock In F.~Pfenning, editor, {\em Proc. of 23th Ann. {IEEE} Symp. on Logic
  in Computer Science, LICS 2008}, pages 118--129. IEEE, 2008.

\bibitem{Plotkin:Handlers}
G.~D. Plotkin and M.~Pretnar.
\newblock Handlers of algebraic effects.
\newblock In G.~Castagna, editor, {\em Proc. of 18th European Symp. on
  Programming, {ESOP} 2009}, volume 5502 of {\em LNCS}, pages 80--94. Springer,
  2009.

\bibitem{Plotkin:HandlingEffects}
G.~D. Plotkin and M.~Pretnar.
\newblock Handling algebraic effects.
\newblock {\em Logical Methods in Computer Science}, 9(4:23), 2013.

\bibitem{Power:EnrichedLawvereTheories}
J.~Power.
\newblock Enriched {Lawvere} theories.
\newblock {\em Theory Appl. Categ}, 6(7):83--93, 1999.

\bibitem{Power:CountableTheories}
J.~Power.
\newblock Countable {Lawvere} theories and computational effects.
\newblock In A.~K. Seda, T.~Hurley, M.~Schellekens, M.~M. an~Airchinnigh, and
  G.~Strong, editors, {\em Proc. of 3rd Irish Conf. on the Mathematical
  Foundations of Computer Science and Information Technology, MFCSIT 2004},
  volume 161 of {\em ENTCS}, pages 59--71. Elsevier, 2006.

\bibitem{Power:IndexedLawvereTheories}
J.~Power.
\newblock Indexed {Lawvere} theories for local state.
\newblock In B.~Hart, T.~G. Kucera, A.~Pillay, P.~J. Scott, and R.~A.~G. Seely,
  editors, {\em Models, Logics and Higher-Dimensional Categories: A Tribute to
  the Work of Mih{\'a}ly Makkai}, volume~53 of {\em CRM Proceedings \& Lecture
  Notes}, pages 213--229. Amer. Math. Soc., 2011.

\bibitem{Pretnar:Thesis}
M.~Pretnar.
\newblock {\em The Logic and Handling of Algebraic Effects}.
\newblock PhD thesis, School of Informatics, University of Edinburgh, 2010.

\bibitem{Pretnar:Tutorial}
M.~Pretnar.
\newblock An introduction to algebraic effects and handlers. {Invited tutorial
  paper}.
\newblock In D.~Ghica, editor, {\em Proc. of 31st Conf. on the Mathematical
  Foundations of Programming Semantics, MFPS XXXI}, volume 319 of {\em ENTCS},
  pages 19--35. Elsevier, 2015.

\bibitem{Pym:BunchedImplications}
D.~Pym.
\newblock {\em The Semantics and Proof Theory of the Logic of Bunched
  Implications}, volume~26 of {\em Applied Logic Series}.
\newblock Springer Netherlands, 2002.

\bibitem{Schopp:Thesis}
U.~Sch\"{o}pp.
\newblock {\em Names and Binding in Type Theory}.
\newblock PhD thesis, School of Informatics, University of Edinburgh, 2006.

\bibitem{Schopp:DTTWithNames}
U.~Sch\"{o}pp and I.~Stark.
\newblock A dependent type theory with names and binding.
\newblock In J.~Marcinkowski and A.~Tarlecki, editors, {\em Proc. of 18th Ann.
  Conf. of the EACSL on Computer Science Logic, CSL 2004}, volume 3210 of {\em
  LNCS}. Springer, 2004.

\bibitem{Shulman:EnrichedIndexedCategories}
M.~Shulman.
\newblock Enriched indexed categories.
\newblock {\em Theory Appl. Categ}, 28:616--695, 2013.

\bibitem{Smyth:RecDomEqs}
M.~B. Smyth and G.~D. Plotkin.
\newblock The category-theoretic solution of recursive domain equations.
\newblock {\em {SIAM} J. Comput.}, 11(4):761--783, 1982.

\bibitem{Staton:Instances}
S.~Staton.
\newblock Instances of computational effects: An algebraic perspective.
\newblock In O.~Kupferman, editor, {\em Proc. of 28th Ann. {ACM/IEEE} Symp. on
  Logic in Computer Science, LICS 2013}, pages 519--519. IEEE, 2013.

\bibitem{Streicher:Semantics}
T.~Streicher.
\newblock {\em Semantics of Type Theory. Correctness, Completeness and
  Independence Results}.
\newblock Progress in Theoretical Computer Science. Birkh\"{a}user, 1991.

\bibitem{Swamy:FStar}
N.~Swamy, C.~Hri{\c t}cu, C.~Keller, A.~Rastogi, A.~Delignat-Lavaud, S.~Forest,
  K.~Bhargavan, C.~Fournet, P.-Y. Strub, M.~Kohlweiss, J.-K. Zinzindohoue, and
  S.~Zanella-B\'eguelin.
\newblock Dependent types and multi-monadic effects in {F*}.
\newblock In R.~Majumdar, editor, {\em Proc. of 43rd Ann. {ACM}
  {SIGPLAN-SIGACT} Symp. on Principles of Programming Languages, POPL 2016},
  pages 256--270. ACM, 2016.

\bibitem{Tarski:ConceptOfTruth}
A.~Tarski.
\newblock The concept of truth in formalized languages.
\newblock In J.~Corcoran, editor, {\em Logic, Semantics, Metamathematics},
  pages 152--278. Hackett Publishing, 1983.
\newblock (Translation of the original 1933 paper published in Polish.
  Translated by J.~H.~Woodger based on a 1936 translation to German.).

\bibitem{Vakar:LinearLF}
M.~V{\'a}k{\'a}r.
\newblock A categorical semantics for linear logical frameworks.
\newblock In A.~Pitts, editor, {\em Proc. of 18th Int. Conf. on Foundations of
  Software Science and Computation Structures, FoSSaCS 2015}, volume 9034 of
  {\em LNCS}, pages 102--116. Springer, 2015.

\bibitem{Vakar:FrameworkForDepEffs}
M.~V{\'{a}}k{\'{a}}r.
\newblock A framework for dependent types and effects.
\newblock {\em CoRR}, abs/1512.08009, 2015.

\bibitem{Vakar:EffectfulDepTypes}
M.~V{\'{a}}k{\'{a}}r.
\newblock An effectful treatment of dependent types.
\newblock {\em CoRR}, abs/1603.04298, 2016.

\bibitem{Vasilakopoulou:Thesis}
C.~Vasilakopoulou.
\newblock {\em Generalization of Algebraic Operations via Enrichment}.
\newblock PhD thesis, Department of Pure Mathematics and Mathematical
  Statistics, University of Cambridge, 2014.

\bibitem{Voevodsky:CSystems}
V.~{Voevodsky}.
\newblock {Subsystems and regular quotients of C-systems}.
\newblock {\em CoRR}, abs/1406.7413, June 2014.

\bibitem{Wadler:Monads}
P.~Wadler.
\newblock The essence of functional programming.
\newblock In R.~Sethi, editor, {\em Proc. of 19th Ann. {ACM} {SIGPLAN-SIGACT}
  Symp. on Principles of Programming Languages, POPL 1992}, pages 1--14. ACM,
  1992.

\end{thebibliography}



\renewcommand\indexname{Notation and Subject Index}

\cleardoublepage
\phantomsection
\addcontentsline{toc}{chapter}{Notation and Subject Index}
\printindex

\end{document}